\documentclass[11pt,a4paper,openany]{book}

\usepackage[english]{babel}       
\usepackage[utf8]{inputenc}       
\usepackage[T1]{fontenc}          
\usepackage{lmodern}              

\usepackage[left=1.3cm, right=1.3cm, bottom=2cm, top=2cm]{geometry} 
\usepackage{setspace}             
\usepackage{titlesec}             
\usepackage{microtype}  
\usepackage[bottom]{footmisc} 
\usepackage[Glenn]{fncychap}      
\usepackage{fancyhdr}             
\usepackage{graphicx}             
\usepackage{subcaption}          
\usepackage{tikz}                
\usepackage{eso-pic}
\usepackage{placeins}            
\usepackage{float}               
\usepackage{longtable}           

\usepackage{amsfonts}             
\usepackage{amsmath}              
\allowdisplaybreaks[4]               
\usepackage{amssymb}              
\usepackage{amsthm}               
\usepackage{bm}                   
\usepackage{mathrsfs}             
\usepackage{dsfont}               

\theoremstyle{plain}
\newtheorem{theorem}{Theorem}
\newtheorem{lemma}[theorem]{Lemma}

\theoremstyle{definition}
\newtheorem{definition}{Definition}
\newtheorem{example}{Example}

\theoremstyle{remark}
\newtheorem{remark}{Remark}
\newtheorem*{solution*}{Solution}

\newcommand{\angstrom}{\textup{\AA}} 
\DeclareMathOperator{\sgn}{sgn}
\DeclareMathOperator{\perm}{perm}
\DeclareMathOperator{\grad}{grad}
\DeclareMathOperator{\dive}{div}

\DeclareMathOperator{\Tr}{Tr}
\DeclareMathOperator{\Span}{Span}
\DeclareMathOperator{\re}{Re}
\DeclareMathOperator{\im}{Im}
\DeclareMathOperator{\pv}{P.V.}
\DeclareMathOperator{\const}{const.}

\makeatletter
\renewcommand{\thepart}{\Roman{part}}
\renewcommand\part{%
  \if@openright
    \cleardoublepage
  \else
    \clearpage
  \fi
  \thispagestyle{plain}%
  \null\vfil
  \secdef\@part\@spart}

\def\@part[#1]#2{%
  \ifnum \c@secnumdepth > -2\relax
    \refstepcounter{part}%
    \addcontentsline{toc}{part}{Part \thepart: #1}%
  \else
    \addcontentsline{toc}{part}{#1}%
  \fi
  {\centering
   \interlinepenalty \@M
   \normalfont\Huge\bfseries
   Part \thepart: #2\par}%
  \vfil\null
  \clearpage}

\def\@spart#1{%
  {\centering
   \interlinepenalty \@M
   \normalfont\Huge\bfseries
   #1\par}%
  \vfil\null
  \clearpage}
\makeatother

\usepackage[
    colorlinks=false,         
    linkbordercolor=red,      
    pdfborder={0 0 1}         
]{hyperref}

\usepackage{pdfpages}
\usepackage{fancyhdr}
\usepackage{ifthen}

\begin{document}
\frontmatter
\let\cleardoublepage\clearpage
\fancypagestyle{tocstyle}{
  \fancyhf{}
  \fancyhead[OL,ER]{\contentsname}
  \fancyfoot[C]{\thepage}
}

\renewcommand{\thefootnote}{\fnsymbol{footnote}}

\begin{center}
  \vspace*{4cm}
  
  {\LARGE\bfseries Lecture Notes on Quantum Many-Body Theory:\\
  A Pedagogical Introduction}\\[0.7cm]
  
  {\large
    Fabrizio Tafuri\textsuperscript{1},\quad
    Carmine Antonio Perroni\textsuperscript{1,2,3,}\footnotemark[1],\quad
    Giulio De Filippis\textsuperscript{1,2,3,}\footnotemark[2]
  }\\[0.5cm]
  
  \begin{minipage}{0.9\textwidth}
    \footnotesize
    \textsuperscript{1} Università degli Studi di Napoli Federico II, Dipartimento di Fisica “E. Pancini”, Complesso Universitario di Monte Sant’Angelo, Via Cinthia 21, I‑80126 Napoli, Italia\\[0.1cm]
    \textsuperscript{2} Istituto Nazionale di Fisica Nucleare (INFN), Sezione di Napoli, Complesso Universitario di Monte Sant'Angelo, Via Cinthia 21, I‑80126 Napoli, Italia\\[0.1cm]
    \textsuperscript{3} CNR‑SPIN – Istituto Superconduttori, Materiali Innovativi e Dispositivi, Università degli Studi di Napoli Federico II, Dipartimento di Fisica “E. Pancini”, Via Cinthia 21, I‑80126 Napoli, Italia
  \end{minipage}\\[2cm]
  
 \begin{minipage}{0.9\textwidth}
  {\large\bfseries Abstract:} {\large
  In these notes, we present a rigorous and self-contained introduction to the fundamental concepts and methods of quantum many-body theory. The text is designed to provide a solid theoretical foundation for the study of interacting quantum systems, combining clarity with mathematical precision. Core topics are developed systematically, with detailed derivations and comprehensive proofs that aim to make the material accessible to graduate students and beginning PhD students. Special attention is given to formal consistency and pedagogical structure, so as to guide the reader through both the conceptual and technical aspects of the subject. This work is intended as a reliable starting point for further exploration and research in modern quantum many-body physics.}
\end{minipage}

\end{center}

  \vfill

  \footnotetext[1]{\texttt{carmine.perroni@unina.it}}
  \footnotetext[2]{\texttt{giulio.defilippis@unina.it}}
  
  \thispagestyle{empty}

\cleardoublepage
\thispagestyle{tocstyle}
\tableofcontents

\cleardoublepage
\pagestyle{plain}

\mainmatter

\chapter*{Introduction}
\addcontentsline{toc}{chapter}{Introduction} 
\markboth{Introduction}{Introduction}       
This book is a compilation of the lecture notes of the Many-Body Course held at the Department of Physics of the University of Naples “Federico II”. They are written with MSc Physics students in mind. Clearly, they are not verbatim: they have been edited by Fabrizio Tafuri, who was a master's student at the time and attended the course during the 2018-2019 academic year. The course itself has been offered regularly since 2009 by Professors Carmine
Antonio Perroni and Giulio De Filippis. The aim is to introduce the students to the basic approaches used at finite temperature in the calculations of physical features of systems containing many interacting non-relativistic particles. \newline
After a brief review of the principles of quantum mechanics, the second quantization formalism is introduced in Part I. It enormously simplifies the treatment of many interacting particles: it allows one to avoid treating directly many-particle wavefunctions. Any system turns out to be described in terms of the field operators that replace the single-particle wavefunction in the Schrödinger representation. Field operators incorporate the Bose-Einstein and Fermi-Dirac statistics and allow to prove that any bosonic or fermionic system of identical particles can be described in terms of quanta. \newline
The next step in Part II is the description of a solid by using the second quantization formalism. The student is gradually introduced to the full Hamiltonian: 1) the starting point is the band theory, based on the assumption that nuclei/ions are fixed at the lattice positions; 2) later, the lattice vibrations around the equilibrium positions are taken into account, leading to the concept of normal modes in classical mechanics: they are quantized, demonstrating that oscillations in the classical theory are described in terms of non-interacting bosons in quantum mechanics, i.e., phonons, shedding further light on the wave-particle duality; 3) finally, the electron-phonon interaction is included in the Hamiltonian: it plays a crucial role in describing the onset of superconductivity, in metallic compounds with low critical temperature, and transport properties of metals. \newline
The next important steps in Part III are the introduction of the retarded Green's functions and the Feynman Dyson perturbation theory based on the famous Feynman diagrams. The linear response theory allows us to highlight the relevance of the Green's functions: the response of a system to a weak external perturbation can be deduced by investigating the fluctuations at thermal equilibrium in the absence of the perturbation. By merging linear response theory and retarded Green's functions, one can obtain the dielectric function, the magnetic susceptibility and the optical absorption, i.e., all quantities experimentally accessible. \newline
In Part IV, the lecture notes contain: 1) a simple description of the Bose-Einstein condensation both in the absence and in the presence of interactions, by using canonical transformations; 2) the quantization of the electromagnetic field, that leads to photons. The lecture notes end with Part V providing a toolkit of group theory and symmetries in physics and Part VI dedicated to appendices. \newline
The book also contains a simple model, describing an impurity interacting with phonons, exactly solved through unitary transformations in Chapter 6. The aim is to introduce the student to the most used approaches in the many-body theory, emphasizing that, for describing a system of interacting particles, one can choose among many and different techniques, each of them shedding light on specific features of the physical system under investigation. \newline
Finally, a particular mention deserves the concept of quasiparticle, a brilliant idea proposed by Landau in 1941. Quasiparticles represent low energy long lived excitations that behave as weakly interacting particles even if the system contains many strongly interacting bodies. Examples are phonons and plasmons in metals, Cooper pairs in superconductors, excitons in semiconductors, polarons in systems with strong electron-phonon interaction. In some cases, they are collective excitations of the original interacting particles, in other cases they coincide with the original particles dressed by the interaction either with the other particles or other fields. In the book there are a lot of applications of this crucial physical concept. \newline
We hope that the book will prompt the students to appreciate many-body physics, helping them grasp advanced approaches typically used in the treatment of systems with strong interactions.
\chapter*{General references}
\addcontentsline{toc}{chapter}{General references}
\markboth{General references}{General references}
The development of these lecture notes has been informed by a selection of foundational texts in quantum many-body theory, which have shaped both the structure and the perspective adopted here. While the exposition is self-contained, the interested reader will find in the following references a broader context, rigorous derivations, and further insights that go beyond the scope of these pages.
\begin{itemize}
  \item[\textbf{[1]}] A. A. Abrikosov, L. P. Gorkov, and I. E. Dzyaloshinski, \emph{Methods of Quantum Field Theory in Statistical Physics}, 1st Dover edition, Dover Publications, New York, 1975. ISBN: 978-0486632285.

  \item[\textbf{[2]}] A. L. Fetter and J. D. Walecka, \emph{Quantum Theory of Many-Particle Systems}, McGraw-Hill, New York, 1971.

  \item[\textbf{[3]}] G. D. Mahan, \emph{Many-Particle Physics}, 3rd ed., Kluwer Academic/Plenum Publishers, New York, 2000. ISBN: 0-306-45423-7.

  \item[\textbf{[4]}] W. Nolting, \emph{Theoretical Physics 9 – Fundamentals of Many-Body Physics}, 2nd ed., Springer, Cham, 2018. ISBN: 978-3-319-75794-6.

  \item[\textbf{[5]}] H. Bruus and K. Flensberg, \emph{Many-Body Quantum Theory in Condensed Matter Physics: An Introduction}, Oxford Graduate Texts, Oxford University Press, Oxford, 2004. ISBN: 978-0-19-856633-5.

  \item[\textbf{[6]}] A. Zagoskin, \emph{Quantum Theory of Many-Body Systems: Techniques and Applications}, Springer, New York, 1998. ISBN: 978-0-387-98275-5.

  \item[\textbf{[7]}] M. Fabrizio, \emph{A Course in Quantum Many-Body Theory: From Conventional Fermi Liquids to Strongly Correlated Systems}, Springer, Cham, 2023. ISBN: 978-3-031-11496-2.

  \item[\textbf{[8]}] D. J. Thouless, \emph{The Quantum Mechanics of Many-Body Systems}, 2nd ed., Dover Publications, Mineola, 2014 (1st ed. 1961). ISBN: 978-0-486-49344-8.

  \item[\textbf{[9]}] W. H. Dickhoff and D. Van Neck, \emph{Many-Body Theory Exposed! Propagator Description of Quantum Mechanics in Many-Body Systems}, 2nd ed., World Scientific, Singapore, 2008. ISBN: 978-981-281-380-4.

  \item[\textbf{[10]}] N. H. March, W. H. Young, and S. Sampanthar, \emph{The Many-Body Problem in Quantum Mechanics}, Dover Publications, New York, 1995 (originally published 1967). ISBN: 978-0-486-68812-7.

  \item[\textbf{[11]}] T. D. Schultz, \emph{Quantum Field Theory and the Many-Body Problem}, Academic Press, New York, 1964. ISBN: 978-0126326509.

  \item[\textbf{[12]}] R. D. Mattuck, \emph{A Guide to Feynman Diagrams in the Many-Body Problem}, 2nd ed., Dover Publications, New York, 1992 (1st ed. 1967). ISBN: 978-0-486-41938-7.

  \item[\textbf{[13]}] L. Peliti, \emph{Statistical Mechanics in a Nutshell}, Princeton University Press, Princeton, 2011. ISBN: 978-0-691-14519-8.

  \item[\textbf{[14]}] O. Madelung, \emph{Introduction to Solid State Theory}, Springer Series in Solid-State Sciences, vol. 2, Springer, Berlin, 1996 (1st ed. 1978). ISBN: 978-3-540-60443-3.

  \item[\textbf{[15]}] C. Cohen-Tannoudji, J. Dupont-Roc, and G. Grynberg, \emph{Photons and Atoms: Introduction to Quantum Electrodynamics}, Wiley-Interscience, New York, 1989. ISBN: 0-471-62181-0.

  \item[\textbf{[16]}] F. H. L. Essler, H. Frahm, F. Göhmann, A. Klümper, and V. E. Korepin, \emph{The One-Dimensional Hubbard Model}, Cambridge University Press, Cambridge, 2005. ISBN: 978-0-521-86018-8.

  \item[\textbf{[17]}] M. Born and V. A. Fock, “Beweis des Adiabatensatzes”, \emph{Zeitschrift für Physik A}, vol. 51, no. 3–4, pp. 165–180, 1928.

  \item[\textbf{[18]}] T. Kato, “On the Adiabatic Theorem of Quantum Mechanics”, \emph{Journal of the Physical Society of Japan}, vol. 5, no. 6, pp. 435–439, 1950.

  \item[\textbf{[19]}] A. Altland and B. D. Simons, \emph{Condensed Matter Field Theory}, 2nd ed., Cambridge University Press, Cambridge, 2010. ISBN-13: 978-0521769754; ISBN-10: 0521769752. 

  \item[\textbf{[20]}] J. W. Negele and H. Orland, \emph{Quantum Many-Particle Systems}, Addison-Wesley, Reading, MA, 1988. ISBN-13: 978-0201125931; ISBN-10: 0201125935. 

  \item[\textbf{[21]}] P. Coleman, \emph{Introduction to Many-Body Physics}, Cambridge University Press, Cambridge, 2015. ISBN-13: 978-0521864886; ISBN-10: 0521864887.

  \item[\textbf{[22]}] N. Nagaosa, \emph{Quantum Field Theory in Condensed Matter Physics}, Springer, Berlin, 2010. ISBN-13: 978-3642084850; ISBN-10: 3642084850. 

  \item[\textbf{[23]}] A. M. Tsvelik, \emph{Quantum Field Theory in Condensed Matter Physics}, 2nd ed., Cambridge University Press, Cambridge, 2007. ISBN-13: 978-0521529808; ISBN-10: 0521529808.

   \item[\textbf{[24]}] Dupuis, N.: \textit{Field Theory of Condensed Matter and Ultracold Gases — Volume 1}. World Scientific, 2023. ISBN 978-1-80061-390-4.

  \item[\textbf{[25]}] J. F. Cornwell, \emph{Group Theory in Physics: An Introduction}, Vol. 1, Academic Press, London, 1997. ISBN: 978-0121898001.
\end{itemize}
\chapter*{Notation and symbols}
\addcontentsline{toc}{chapter}{Notation and symbols}
\markboth{Notation and symbols}{Notation and symbols}
This text covers topics ranging from classical mechanics to quantum mechanics in second quantization. As a consequence, many mathematical symbols can take on different meanings depending on the specific context in which they are used. \\
To reduce notational complexity and improve readability, we adopt the convention that only fundamental operators (such as position and momentum operators, as well as unitary or antiunitary symmetry operators, e.g. Wigner operators) are explicitly marked with a hat symbol "\(\hat{\ }\)". Generic interaction operators, including the Hamiltonian in first quantization, are instead left without a hat, while their corresponding second-quantized forms are consistently denoted with one. Less standard operators or scalar quantities are also left without a hat, unless otherwise specified. In addition, for the sake of notational economy, creation and annihilation operators, due to their pervasive use throughout the text, are likewise written without a hat. \\
Moreover, many symbols used throughout the text may carry multiple meanings; the Notation and Symbols table specifies their primary interpretation, while specific cases are clarified within the main text. This approach allows us to maintain a compact notation without compromising the formal precision required for the theoretical development.
\begin{longtable}{p{4cm} p{13.5cm}}
\caption{Notation and symbols} \\
\hline
\textbf{Symbol} & \textbf{Meaning / Description} \\
\hline
\endfirsthead
\hline
\textbf{Symbol} & \textbf{Meaning / Description} \\
\hline
\endhead
\hline
\endfoot

$H$ & Hilbert space. \\

$\mathcal{H}$ & First-quantized Hamiltonian. \\

$\hat{\mathcal{H}}$ & Second-quantized Hamiltonian. Operator on Fock space. \\

$U$, $V$, $\ldots$ & Semiclassical interaction functions or one-/two-body interaction operators in first quantization. \\

$\hat{U}$, $\hat{V}$, $\ldots$ & One-/two-body interaction operators in second quantization. \\

$a$, $A$, $b$, $B$, $\ldots$ & Annihilation operators. \\
& \small{(Uppercase/lowercase letters are chosen for clarity and do not carry physical meaning.)} \\

$a^\dagger$, $A^\dagger$, $b^\dagger$, $B^\dagger$, $\ldots$ & Creation operators. \\
& \small{(Uppercase/lowercase letters are chosen for clarity and do not carry physical meaning.)} \\

$\psi$, $\varphi$, $\ldots$ & Wave vectors. \\

$\hat{\psi}$ & Field operator. \\

$\hat{S}_i$ ($i = 1,2,3$ or $x,y,z$) & Components of the spin operator in first quantization. \\

$s_x$, $s_y$, $s_z$ & Spin densities in first quantization. \\

$\sigma_x$, $\sigma_y$, $\sigma_z$, $\bm{\sigma}$ & Pauli matrices. \\

$\sigma$, $\sigma'$, $\sigma_1$, $\sigma_2$, $\ldots$ & These symbols may refer to different quantities (e.g. Pauli matrices, spin components, permutation operators, etc.); its precise meaning will be specified case by case. \\

$\chi_{\sigma}(s)$ & Component of the eigenfunction of the spin operator \(\hat{S}_z\) with eigenvalue \(\sigma\) in the spin basis labeled by \(s\). \\

$x$ & This symbols may refer to different quantities (compact notation $x=(\mathbf{r},s)$, combining spatial position $\mathbf{r}$ and spin $s$ for a single particle, etc.); its precise meaning will be specified case by case. \\

$\textbf{r}$ & Spatial position. \\

$\hat{\textbf{r}}$ & Position operator. \\

$\textbf{p}$ & Momentum. \\

$\hat{\textbf{p}}$ & Momentum operator. \\

$\mathbf{k}$ & This symbol may refer to different quantities (e.g., wave vector, crystal momentum, etc.); its precise meaning will be specified case by case. In the following, we will generally refer to this symbol as the momentum, as it is related to the physical momentum $\mathbf{p}$ by a factor of $\hbar$. \\

$\textbf{k}_F$ & Fermi momentum. \\

$\bar{a}$ & Generic $n$-dimensional vector. This notation is used in the Part Toolkit of Group Theory and Symmetries in Physics. \\

$\textbf{a}$, $\textbf{A}$, $\textbf{b}$, $\textbf{B}$, $\ldots$ & Generic three-dimensional vectors; their precise meaning will be specified case by case. \\

$|\textbf{a}|$, $a$, $|\textbf{b}|$, $b$, $\ldots$ & Modulus of a three-dimensional vector. Depending on the context, the modulus of a vector (e.g., $\textbf{a}$) may be denoted either by absolute value symbols ($|\textbf{a}|$) or by the corresponding non-bold letter ($a$), to simplify notation. This dual notation also reflects the fact that in some cases it is important to emphasize the vectorial nature of the quantity, such as in the case of eigenvalues of the kinetic energy operator, which depend on the modulus squared of a three-dimensional wavevector, e.g., $\mathcal{E}_{\textbf{k}} = \frac{\hbar^2 \textbf{k}^2}{2m}$. \newline
Additionally, it should be noted that a non-bold symbol (e.g., $a$) may sometimes indicate a scalar quantity that is not necessarily positive (e.g., a component of a vector), rather than the strictly positive modulus of a vector. In any case, the intended meaning is clear from the context. \\

$\alpha$ & This symbol may refer to different quantities (e.g. generical quantum state label, the quantum state label ($\textbf{k},\sigma$), etc.); its precise meaning will be specified case by case. \\

$\rho$ & Symbol that may denote different quantities (density matrix, reduced density matrix, particle density (fluctuation) operator in first quantization, density states, etc.); its precise interpretation will be specified case by case. \\

$\hat{\rho}(\textbf{r})$, $\hat{\rho}_{\textbf{q}}$ & Particle-density operator (and its Fourier components) in second quantization. \\

$\mathcal{E}$, $\mathcal{E}_{\alpha}$, $\mathcal{E}_{n,\mathbf{k},\sigma}$, $\ldots$ & Energies. \\

$\mathcal{E}_F$ & Fermi energy. \\

$t$ & Real time. \\

$\tau$ & This symbols may refer to different quantities (real time in the imaginary-time formalism, number of molecules in a solid, etc.); its precise meaning will be specified case by case. \\

$T$ & Absolute temperature (in Kelvin). \\

$k_B$ & Boltzmann constant, $k_B = 1.380649 \cdot 10^{-23} \,\frac{J}{K}$. \\

$\beta$ & Beta parameter, defined as \(\beta = \frac{1}{k_B T}\). \\

$\mu$ & This symbols may refer to different quantities (chemical potential, atomic species index, etc.); its precise meaning will be specified case by case. \\

$\mu_B$ & Bohr magneton, defined as $\mu_B = \frac{\hslash |q_e|}{2 m_e c}$. \\

$q_e$ & Electronic charge, $q_e = - 1.602 \cdot 10^{-19} \text{C}$. \\

$m$ & This symbol may refer to different quantities (e.g., particle mass, electron mass, electronic band index, etc.); its precise meaning will be specified case by case. \\

$m_e$ & Electron mass, \( m_e = 9.109 \cdot 10^{-31} \,\mathrm{kg} \). We will explicitly use the symbol \( m_e \) (rather than \( m \)) to avoid confusion with other uses of \( m \). \\

$\varepsilon$ & This symbol may refer to different quantities (e.g. statistical index: $\varepsilon = +1$ for bosons, $\varepsilon = -1$ for fermions, and it is used to express quantities that depend on particle statistics in a unified formalism; arbitrarily small constant, etc.); its precise meaning will be specified case by case. \\

$\varepsilon_{\textbf{q}}(\omega)$ & Dielectric function. \\

$Z$ & Partition function (canonical or grand-canonical depending on context); its precise meaning will be specified case by case. \\

$N$ & This symbols may refer to different quantities (number of particles, number of unit cells in a three-dimensional crystal, etc.); its precise meaning will be specified case by case. \\

$\hat{N}$, $\hat{N}_\alpha$, $\ldots$ & Number operator. \\

$n_{-1}(\mathcal{E})$, $n_{-1}(\mathcal{E}_{\alpha})$, $\ldots$ & Fermi-Dirac statistics (or distribution). \\

$n_{1}(\mathcal{E})$, $n_{1}(\mathcal{E}_{\alpha})$, $\ldots$ & Bose-Einstein statistics (or distribution). \\

$n_{\varepsilon}(\mathcal{E})$, $n_{\varepsilon}(\mathcal{E}_{\alpha})$, $\ldots$ & Unified Bose-Einstein or Fermi-Dirac statistics (or distribution). \\

$G^{(r)}_{AB}(t,t')$, $G^{(m)}_{AB}(\tau',\tau'')$, $G^{(r)}$, $G^{(m)}$, $G$, $\ldots$ & Green's functions in real or imaginary time. \\

$\nabla$ & Nabla operator used to denote gradient, divergence, curl, and
Laplacian. \\

$P_{\sigma}$ & Operator corresponding to the permutation $\sigma$. \\

$U(t)$, $U(\tau)$ & Time evolution operator in real and imaginary time formalism. \\

$\hat{R}$, $\hat{R}_{\textbf{u}}$, $\ldots$ & Rotation operator. \\

$\hat{\Omega}$, $\hat{\Omega}(R)$, $\hat{\Omega}(R,\mathbf{a})$, $\ldots$ & Representation operators. \\

$\hat{\Theta}$ & Time-reversal operator. \\

$\hat{T}(\mathbf{a})$ & Spatial translation operator by vector $\mathbf{a}$. \\

$\hat{T}_n$ & Lattice translation operator by vector $\mathbf{R}_n$. \\

$\hat{T}_{\tau}$ & Wick time-ordering operator. \\

$\hat{T}_D$ & Dyson time-ordering operator. \\

$\Theta(\ldots)$ & Heaviside theta function. \\

$\mathds{1}$ & Identity operator or matrix; its precise meaning will be clear from the context. \\

$\langle \ldots \rangle$ & This symbol may refer to different quantities (thermal average, expectation value in a quantum state, statistical average, etc.); its precise meaning will be specified case by case. \\

$0$ & Zero scalar or null operator; its precise meaning will be clear from the context. \\

$\delta_{ij}$ & Kronecker delta, equal to 1 if $i = j$, and 0 otherwise. \\

$\delta(x - x')$ & Generic Dirac delta distribution; its precise dimensionality  (e.g., 1D, 3D) will be clear from the context. \\
\end{longtable}
\pagestyle{fancy}
\fancyhf{}
\fancyhead[EL,OR]{\thepage}
\fancyhead[OL,ER]{%
  \ifthenelse{\value{chapter}>0}{\chaptername\ \thechapter}{}%
}
\part{Foundations of Quantum Many-Body Theory}
\chapter{Quantum mechanics review}
This chapter provides a concise overview of the fundamental properties of quantum mechanics necessary for a rigorous understanding of the topics developed later. We begin with the formalism of a single particle, including the description of spin as an intrinsic degree of freedom, essential for characterizing quantum systems with internal angular momentum. The chapter continues with a discussion of plane waves as examples of momentum eigenstates, which serve as essential tools in momentum-space representations and in the analysis of free particle dynamics. This is followed by the development of a general abstract framework based on creation and annihilation operators, within which coherent states are introduced and analyzed. These states play a fundamental role in various physical contexts, such as the quantum harmonic oscillator and quantum field theory, due to their close connection with classical behavior and their minimal quantum uncertainty. The chapter concludes with a treatment of the quantum adiabatic theorem, which offers key insights into the slow evolution of quantum systems under gradually changing external parameters. \newline
Then, the different pictures of time evolution will be introduced, in particular the Schrödinger picture, the Heisenberg picture and the interaction picture, which represent alternative but equivalent ways to treat time and the evolution operator in a quantum system. After that, we introduce the one-particle Green's function, a fundamental concept that encapsulates the evolution of a quantum state over time. The propagator provides a complete description of how the wavefunction changes from an initial to a final state, incorporating all possible quantum paths. Its properties, such as causality and unitarity, ensure the consistency of the quantum dynamics and allow for a direct connection between the formalism and physical observables. \newline
Finally, the concluding part of the chapter focuses on essential tools from statistical mechanics, beginning with the formalism of the density matrix, which generalizes the description of quantum states to include mixed states. This framework enables a comprehensive treatment of phenomena such as decoherence, partial measurements, and open quantum systems. In addition, foundational concepts like the Maxwell-Boltzmann distribution are introduced to connect quantum mechanics with thermodynamic behavior. \newline
This formal recall lays a solid foundation for the systematic use of quantum mechanics tools throughout the subsequent sections, serving as a compact reference rather than an exhaustive treatment.
\section{One particle formalism}
The physical states of a single quantum particle with spin, confined in a volume $V$, are represented by elements of a Hilbert space $H_1$ defined as the tensor product as follows
\begin{align}
H_{1} &= H_{orbital} \otimes H_{spin} \equiv \notag \\
&\equiv H_o \otimes H_s,
\end{align}
i.e., it is the tensor product of the Hilbert space of the orbital degrees of freedom $H_o$ for the finite Hilbert space $H_s$ of the spin degree of freedom. Given $\Psi, \Phi \in \mathcal{L}^2\left(\Omega\right)$, we define the scalar product in $H_o$ as
\begin{equation}
\left( \Psi,\Phi \right)_{H_o} = \int_{V} d^{3} \textbf{r} \Psi^{*}(\textbf{r}) \Phi(\textbf{r}).
\end{equation} 
By assumption, $H_o$ is separable, that is, there exists a basis, i.e., a numerable, orthonormal and complete set of vectors. For example, a possible orthonormal basis \(\{ \phi_n(\textbf{r}) \}\) in \(H_o\) can be chosen as the eigenfunctions of the Hamiltonian. Such eigenfunctions satisfy the orthonormality property
\begin{equation}
\left( \phi_{n},\phi_{n'} \right)_{H_o} = \delta_{n,n'},
\end{equation}
and the completeness property
\begin{equation}
\sum_{n} \phi^{*}_{n}(\textbf{r}) \phi_{n}(\textbf{r}') = \delta \left( \textbf{r}-\textbf{r}' \right).
\end{equation}
In many contexts, it is convenient to adopt the so-called Dirac notation, in which state vectors are denoted by symbols such as \(|\psi\rangle\), while their duals (linear functionals) are represented by \(\langle\phi|\). The inner product between two states is written as $\langle \Psi | \Phi \rangle$, which, in the position representation, corresponds to
\begin{equation}
\langle \Psi | \Phi \rangle = \int_V d^3\mathbf{r}\; \Psi^*(\mathbf{r}) \Phi(\mathbf{r}).
\end{equation}
This notation is purely formal and does not imply a specific choice of representation: states cannot always be described as functions, but the notation is particularly useful for expressing general relations in a compact form, regardless of the basis used. For example, in the position representation, a state \(|\psi\rangle\) is described by the wavefunction \(\psi(\mathbf{r})\), defined as
\begin{align}
\psi(\textbf{r}) &= \langle \mathbf{r} | \psi \rangle = \notag \\
&= \sum_n c_n \phi_n(\textbf{r}),
\end{align}
where \(|\mathbf{r}\rangle\) is a generalized eigenstate of the position operator \(\hat{\mathbf{r}}\). According to the probabilistic interpretation of quantum mechanics, $\left| c_n \right|^2$ is the probability that, immediately after a position measurement on the state \(|\psi\rangle\), the system collapses into the eigenstate \(\phi_n(\mathbf{r})\). \newline
Regarding the spin degrees of freedom, the spin Hilbert space \(H_s\) is a complex vector space generated by the \((2S+1)\) eigenfunctions of the spin operator \(\hat{S}_z\), where \(S\) is a fixed quantum number, integer or half-integer depending on the nature of the particle. Let \(\{\chi_{\sigma}\}\) denote the eigenfunctions of \(\hat{S}_z\), corresponding to eigenvalues $\sigma$, with
\begin{equation}
\sigma \in \{-S, -S+1, \ldots, S-1, S\}.
\end{equation}
The index \(\sigma\) labels the basis vectors in \(H_s\), and the values \(s\) represent the possible measurement outcomes of the operator \(\hat{S}_z\).  
Given \(\chi, \chi' \in H_s\), we define the scalar product in \(H_s\) as
\begin{equation}
\left( \chi , \chi' \right)_{H_s} = \sum_{s=-S}^{S} \chi^*(s) \chi'(s).
\end{equation}
Similarly to the position representation, in the spin representation any state \(\chi\) can be expanded as
\begin{align}
\chi(s) &= \langle s | \chi \rangle = \notag \\
&= \sum_{\sigma=-S}^{S} c_{\sigma} \chi_{\sigma}(s),
\end{align}
where \(|s\rangle\) is an eigenstate of \(\hat{S}_z\) and \(\{\chi_\sigma\}\) is a basis. The eigenstates \(\chi_{\sigma}(s)\) form an orthonormal basis of \(H_s\), satisfying the orthonormality relation
\begin{align}
\left( \chi_{\sigma}, \chi_{\sigma'} \right)_{H_s} &= \sum_{s=-S}^S \chi^*_{\sigma}(s) \chi_{\sigma'}(s) = \notag \\
&= \delta_{\sigma,\sigma'},
\end{align}
and the completeness relation
\begin{equation}
\sum_{\sigma=-S}^{S} \chi_{\sigma}^*(s) \chi_{\sigma}(s') = \delta_{s,s'}.
\end{equation}
According to the probabilistic interpretation of quantum mechanics, \(|c_{\sigma}|^2\) is the probability that immediately after a measurement of \(\hat{S}_z\) on the spin state \(|\chi\rangle\), the state collapses into the eigenstate \(\chi_\sigma\). \newline
For the canonical basis vectors $\chi_\sigma$, the components are given by
\begin{align}
\chi_{\sigma}(s) &= \langle s | \sigma \rangle = \notag \\
&= \delta_{s,\sigma}.
\end{align}
\begin{example}[Spin-$\frac{1}{2}$ particles] 
The Pauli matrices form a set of three Hermitian, traceless matrices that play a central role in the description of intrinsic angular momentum for spin-$\frac{1}{2}$ particles, as well as in the representation of the Lie algebra (see Section \ref{Lie algebra}). They are defined as
\begin{align}
\sigma_x &\equiv \sigma_1 = 
\begin{pmatrix}
0 & 1 \\
1 & 0
\end{pmatrix}
,
\end{align}
\begin{align}
\sigma_y &\equiv \sigma_2 = 
\begin{pmatrix}
0 & -i \\
i & 0
\end{pmatrix}
,
\end{align}
\begin{align}
\sigma_z &\equiv \sigma_3 = 
\begin{pmatrix}
1 & 0 \\
0 & -1
\end{pmatrix}
.
\end{align}
Their fundamental properties are:
\begin{itemize}
\item Hermitian: \( \sigma_i^\dagger = \sigma_i \);
\item Traceless: \( \operatorname{Tr}(\sigma_i) = 0 \);
\item Determinant: \( \det(\sigma_i) = -1 \);
\item Squared: \( \sigma_i^2 = \mathds{1} \);
\item Commutators:
\begin{equation}
\left[ \sigma_i, \sigma_j \right] = 2i \, \varepsilon_{ijk} \sigma_k;
\end{equation}
\item Anticommutators:
\begin{equation}
\left\lbrace \sigma_i, \sigma_j \right\rbrace = 2\delta_{ij};
\end{equation}
\item Pauli identity:
\begin{equation}
(\bm{\sigma} \cdot \textbf{a} ) ( \bm{\sigma} \cdot \textbf{b} ) = (\textbf{a} \cdot \textbf{b}) \mathds{1} + i \bm{\sigma} \cdot (\textbf{a} \times \textbf{b}).
\label{eq: Pauliidentity}
\end{equation}
\end{itemize}
The Pauli matrices are used to represent spin-$\frac{1}{2}$ operators $\hat{S}_i$ as follows
\begin{equation}
\hat{S}_i = \frac{\hslash}{2} \sigma_i.
\end{equation}
As a particular case, we now consider the eigenvalue equations for the spin eigenfunctions of a spin-$\frac{1}{2}$ particle, such as the electron. Let \(\chi_{\sigma}(s)\) denote the components of the eigenvector \(\chi_{\sigma}\) of the operator \(\hat{S}_z\) in the spin basis, that is,
\begin{equation}
\chi_{-\frac{1}{2}}(s) =
\begin{pmatrix}
0 \\
1
\end{pmatrix},
\end{equation}
\begin{equation}
\chi_{+\frac{1}{2}}(s) =
\begin{pmatrix}
1 \\
0
\end{pmatrix},
\end{equation}
these results can be compactly expressed as
\begin{equation}
\hat{S}_z \chi_{\sigma}(s) = \hslash \sigma \chi_{\sigma}(s), \ \sigma = \pm \dfrac{1}{2},
\label{eq: equazioneautovaloriSzconbasechi}
\end{equation}
As for the action of the operator $\hat{S}_x$, we have
\begin{align}
\hat{S}_x \begin{pmatrix}
1 \\
0
\end{pmatrix} &=
\dfrac{\hslash}{2}
\begin{pmatrix}
0 & 1 \\
1 & 0
\end{pmatrix}
\begin{pmatrix}
1 \\
0
\end{pmatrix}
= \notag \\
&= \dfrac{\hslash}{2}
\begin{pmatrix}
0 \\
1
\end{pmatrix},
\end{align}
\begin{align}
\hat{S}_x \begin{pmatrix}
0 \\
1
\end{pmatrix} &=
\dfrac{\hslash}{2}
\begin{pmatrix}
0 & 1 \\
1 & 0
\end{pmatrix}
\begin{pmatrix}
0 \\
1
\end{pmatrix}
= \notag \\
&= \dfrac{\hslash}{2}
\begin{pmatrix}
1 \\
0
\end{pmatrix},
\end{align}
these results can be compactly expressed as
\begin{equation}
\hat{S}_x \chi_{\sigma}(s) = \hslash \left| \sigma \right| \chi_{- \sigma}(s), \ \sigma = \pm \dfrac{1}{2},
\label{eq: equazioneautovaloriSxconbasechi}
\end{equation}
and similarly
\begin{align}
\hat{S}_y \begin{pmatrix}
1 \\
0
\end{pmatrix} &=
\dfrac{\hslash}{2}
\begin{pmatrix}
0 & -i \\
i & 0
\end{pmatrix}
\begin{pmatrix}
1 \\
0
\end{pmatrix}
= \notag \\
&= \dfrac{\hslash}{2}
\begin{pmatrix}
0 \\
i
\end{pmatrix} = \notag \\
&=
i \dfrac{\hslash}{2}
\begin{pmatrix}
0 \\
1
\end{pmatrix},
\end{align}
\begin{align}
\hat{S}_y \begin{pmatrix}
0 \\
1
\end{pmatrix} &=
\dfrac{\hslash}{2}
\begin{pmatrix}
0 & -i \\
i & 0
\end{pmatrix}
\begin{pmatrix}
0 \\
1
\end{pmatrix}
= \notag \\
&= \dfrac{\hslash}{2}
\begin{pmatrix}
-i \\
0
\end{pmatrix} = \notag \\
&= - i \dfrac{\hslash}{2}
\begin{pmatrix}
1 \\
0
\end{pmatrix},
\end{align}
these results can be compactly expressed as
\begin{equation}
\hat{S}_y \chi_{\sigma}(s) = i \hslash \sigma \chi_{- \sigma}(s), \ \sigma = \pm \dfrac{1}{2}.
\label{eq: equazioneautovaloriSyconbasechi}
\end{equation}
\end{example}
Since $H_1 = H_o \otimes H_s$, we show that a basis for a particle with spin is given by the tensor product of a basis of $H_o$ by a basis of $H_s$, i.e.,
\begin{align}
\left\lbrace \varphi_{n,\sigma}(\textbf{r},s) \right\rbrace_{n,\sigma} &\equiv \left\lbrace f_n(\textbf{r}) \otimes \chi_\sigma(s) \right\rbrace_{n,\sigma} \equiv \notag \\
&\equiv \left\lbrace f_n(\textbf{r}) \chi_\sigma(s) \right\rbrace_{n,\sigma}.
\end{align}
Indeed, the scalar product in $H_{1}$ is defined as
\begin{equation}
\left( \phi(\textbf{r},s),\psi(\textbf{r},s) \right)_{H_1} = \sum_{s=-S}^{S} \int_{V} d^3\textbf{r} \phi^\dag(\textbf{r},s) \psi(\textbf{r},s) ,
\end{equation}
whose form is derived from the scalar products defined in the spaces $H_o$ and $H_s$. Such a scalar product satisfies the orthonormality property
\begin{align}
(\varphi_{n,\sigma},\varphi_{n',\sigma'})_{H_1} &= \sum_{s=-S}^{S} \int_V d^3\textbf{r} f_n^*(\textbf{r}) f_{n'}(\textbf{r}) \chi^*_\sigma(s) \chi_{\sigma'}(s) = \notag \\
&= (f_{n},f_{n'})_{H_o} \ (\chi_{\sigma},\chi_{\sigma'})_{H_s} = \notag \\
&= \delta_{n,n'} \delta_{\sigma,\sigma'},
\end{align}
and completeness property
\begin{align}
\sum_n \sum_\sigma \psi^\dag(\textbf{r},s) \psi(\textbf{r}',s') &= \sum_{n} f^{*}_{n}(\textbf{r}) f_{n}(\textbf{r}') \sum_\sigma \chi^*_{\sigma}(s) \chi_{\sigma}(s') = \notag \\
&= \delta \left( \textbf{r}-\textbf{r}' \right) \delta_{s,s'},
\end{align}
that is, $\left\lbrace \varphi_{n,\sigma}(\textbf{r},s) \right\rbrace_{n,\sigma}$ is a basis. The square modulus of a state of a particle with spin, say $\psi$, is
\begin{equation}
||\psi||^2 = \sum_{s=-S}^{S} \int_V d^3\textbf{r} \psi^\dag(\textbf{r},s) \psi(\textbf{r},s).
\end{equation} 
Throughout this text, unless otherwise specified, we shall use the following compact notation:
\begin{equation}
\left( \textbf{r},s \right) \equiv x,
\end{equation}
\begin{equation}
(n,\sigma) \equiv \alpha,
\end{equation}
\begin{equation}
\sum_{s=-S}^{S} \int_V d^3\textbf{r} \ \equiv \int dx,
\end{equation}
\begin{equation}
\delta(x-x') = \delta\left(\textbf{r}-\textbf{r}'\right) \delta_{s,s'},
\end{equation}
which, in turn, implies
\begin{equation}
\varphi(x) = \sum_{\alpha} c_{\alpha} \varphi_{\alpha}(x),
\end{equation}
\begin{equation}
\varphi_{\alpha}(x) = f_{n}(\textbf{r}) \chi_{\sigma}(s),
\end{equation}
\begin{equation}
\left( \varphi_{\alpha}, \varphi_{\alpha'} \right) = \delta_{\alpha,\alpha'},
\end{equation}
\begin{equation}
\int dx \Psi^\dag(x) \Phi(x) \equiv \sum_{s=-S}^{S} \int_V d^{3}\textbf{r} \Psi^\dag(\textbf{r},s) \Phi(\textbf{r},s),
\end{equation}
for each state vector in the space $H_{1}$.
\subsection{Eigenfunctions of a free particle with spin in a finite volume}
Consider a free particle with spin \(S\) confined in a rectangular box of volume \(V = L_x L_y L_z\), subject to periodic boundary conditions. The single-particle Hamiltonian for a spinless free particle is
\begin{equation}
\mathcal{H} = \frac{\hat{\mathbf{p}}^2}{2m},
\end{equation}
where \(\hat{\mathbf{p}} = -i\hbar \nabla\) is the momentum operator and \(m\) the particle mass. The eigenvalue problem
\begin{equation}
\mathcal{H} \phi(\mathbf{r}) = \mathcal{E} \phi(\mathbf{r}),
\end{equation}
with the periodic boundary conditions $\phi(\mathbf{r} + L_i \mathbf{e}_i) = \phi(\mathbf{r}), \quad i = x,y,z$, selects the allowed spatial wavefunctions. The solutions to this problem are the so-called (spatial) plane waves
\begin{equation}
\phi_{\mathbf{k}}(\mathbf{r}) = \frac{1}{\sqrt{V}} e^{i \mathbf{k} \cdot \mathbf{r}},
\label{eq: ondepiane}
\end{equation}
where the quantized wavevectors are given by
\begin{equation}
k_i = \frac{2\pi}{L_i} n_i, \quad n_i \in \mathbb{Z}, \quad i = x,y,z,
\end{equation}
ensuring periodicity (see Chapter \ref{The thermodynamic limit}). The corresponding eigenenergies are given by
\begin{equation}
\mathcal{E}_{\mathbf{k}} = \frac{\hbar^2 \mathbf{k}^2}{2m}.
\label{eq: autovaloriondepiane}
\end{equation}
These functions form a complete orthonormal basis of the spatial Hilbert space $H=\mathcal{L}^2(V)$, satisfying
\begin{equation}
\int_V d^3 \textbf{r}\, \phi_{\mathbf{k}}^*(\mathbf{r}) \phi_{\mathbf{k}'}(\mathbf{r}) = \delta_{\mathbf{k}, \mathbf{k}'}.
\end{equation}
Now, introducing the spin degrees of freedom, since the Hamiltonian $\mathcal{H}$ acts trivially on spin, the eigenfunctions of the full Hamiltonian are tensor products
\begin{align}
\psi_{\mathbf{k}, \sigma}(\mathbf{r}, s) &= \phi_{\mathbf{k}}(\mathbf{r}) \chi_{\sigma}(s) = \notag \\
&= \frac{1}{\sqrt{V}} e^{i \mathbf{k} \cdot \mathbf{r}} \chi_{\sigma}(s),
\label{eq: ondepianeperautofunzionispin}
\end{align}
where $\chi_{\sigma}(s)$ denote the components of the spin eigenfunctions in the $\hat{S}_z$ basis. These eigenfunctions satisfy the orthonormality relation
\begin{align}
\langle \psi_{\mathbf{k}, \sigma} | \psi_{\mathbf{k}', \sigma'} \rangle 
&= \sum_{s=-S}^S \int_V d^3 \textbf{r} \psi_{\mathbf{k}, \sigma}^*(\mathbf{r}, s) \psi_{\mathbf{k}', \sigma'}(\mathbf{r}, s) 
= \notag \\
&= \delta_{\mathbf{k}, \mathbf{k}'} \delta_{\sigma, \sigma'}.
\end{align}
The emergence of plane waves as spatial eigenfunctions follows naturally from solving the free-particle Schrödinger equation with periodic boundary conditions, which enforce the discretization of momentum eigenvalues. The spin degrees of freedom, being unaffected by the kinetic energy operator, simply multiply these spatial solutions and add a finite-dimensional internal space structure to the state vector. Thus, the full state of a free particle with spin in a volume \(V\) is described by a discrete set of plane wave momenta combined with spin eigenstates, forming a complete orthonormal basis of the Hilbert space $H_1$. \newline
This basis of momentum-spin eigenfunctions is particularly useful in quantum many-body theory, where it forms the starting point for both numerical methods (such as exact diagonalization or Monte Carlo techniques) and analytical approaches based on perturbation theory. Throughout this text, we will make extensive use of this basis to formulate and analyze interacting systems in second quantization.
\section{The creation and annihilation operator algebra}
In the previous section, we described quantum states using the first-quantization formalism, working within the position and spin representations and employing Dirac notation to characterize single-particle states. In this section, we shift to a more abstract and algebraic point of view, where the focus is placed directly on the operators that create and annihilate particles in given quantum states. To set the stage, we will first discuss coherent states, which naturally give rise to the creation and annihilation operators. This will allow us to build the algebra of these operators in a way that is independent of the specific statistical nature of the particles involved. After establishing this general framework, we will specialize the treatment to the bosonic and fermionic cases, where the algebraic properties of creation and annihilation operators take distinct forms. By introducing a unified compact notation, we will derive a set of commutation relations that generalize to both bosons and fermions, providing a robust foundation for the many-body formalism developed in the following chapters.
\subsection{Coherent states}
The quantum harmonic oscillator plays a central role in quantum mechanics due to its solvability and its appearance as an approximation near equilibrium in many systems. In this section, we rigorously define and study the so-called coherent states, which are special quantum states of the harmonic oscillator with classical-like behavior. \newline
Consider a one-dimensional quantum harmonic oscillator with Hamiltonian
\begin{equation}
\mathcal{H} = \hbar \omega \left( a^\dagger a + \frac{1}{2} \right),
\end{equation}
where \( a \) and \( a^\dagger \) are the annihilation and creation operators, respectively, defined by
\begin{equation}
a = \sqrt{\frac{m\omega}{2\hbar}} \left( \hat{x} + \frac{i}{m\omega} \hat{p} \right), 
\end{equation}
\begin{equation}
a^\dagger = \sqrt{\frac{m\omega}{2\hbar}} \left( \hat{x} - \frac{i}{m\omega} \hat{p} \right),
\end{equation}
with $\hat{x}$ and $\hat{p}$ being the position and momentum operators satisfying \( [\hat{x},\hat{p}] = i\hbar \). The operators \( a \), \( a^\dagger \) satisfy the canonical commutation relation
\begin{equation}
\left[ a, a^\dagger \right] = \mathds{1}.
\end{equation}
The states \( |n\rangle \), known as the number states, are the eigenstates of the number operator 
\begin{equation}
\hat{N} = a^\dagger a, 
\end{equation}
which satisfies
\begin{equation}
\hat{N} |n\rangle = n |n\rangle, \quad n = 0, 1, 2, \ldots
\end{equation}
with
\begin{equation}
a |n\rangle = \sqrt{n} |n-1\rangle, \qquad \text{for } n \geq 1,
\label{eq: azioneoperatoredistruzioneastratto}
\end{equation}
\begin{equation}
a |0\rangle = 0,
\label{eq: azioneoperatoredistruzioneastrattosuvuoto}
\end{equation}
\begin{equation}
a^\dagger |n\rangle = \sqrt{n+1} |n+1\rangle, \qquad \text{for } n \geq 0.
\label{eq: azioneoperatorecreazioneastratto}
\end{equation}
A coherent state \( |\alpha\rangle \) is defined as an eigenstate of the annihilation operator, that is,
\begin{equation}
a |\alpha\rangle = \alpha |\alpha\rangle, \quad \alpha \in \mathbb{C}.
\label{eq: statocoerente}
\end{equation}
These states are not eigenstates of the Hamiltonian (unless \( \alpha = 0 \)), but they have important semiclassical properties. We have
\begin{theorem}[Construction of coherent states]
A coherent state \( |\alpha\rangle \) can be written in the following form
\begin{equation}
|\alpha\rangle = e^{-\frac{|\alpha|^2}{2}} \sum_{n=0}^\infty \frac{\alpha^n}{\sqrt{n!}} |n\rangle.
\label{eq: espansionestatocoerente}
\end{equation}
\end{theorem}
\begin{proof}
We will now derive this result by explicitly constructing the coherent state from the properties of the annihilation operator. To construct the coherent state, we begin by assuming that \( |\alpha\rangle \) can be expressed as a superposition of number states \( |n\rangle \), i.e.,
\begin{equation}
|\alpha\rangle = \sum_{n=0}^\infty c_n |n\rangle.
\end{equation}
We apply the annihilation operator \( a \) to this expression, using the $\eqref{eq: azioneoperatoredistruzioneastratto}$, then
\begin{equation}
a |\alpha\rangle = \sum_{n=1}^\infty c_n \sqrt{n} |n-1\rangle.
\end{equation}
Next, we relabel the sum by setting \( m = n - 1 \), so we get
\begin{equation}
a |\alpha\rangle = \sum_{m=0}^\infty c_{m+1} \sqrt{m+1} |m\rangle.
\end{equation}
From the eigenvalue equation $\eqref{eq: statocoerente}$, we also have
\begin{equation}
a |\alpha\rangle = \alpha \sum_{m=0}^\infty c_m |m\rangle.
\end{equation}
By equating the coefficients of \( |m\rangle \) on both sides, we derive the recurrence relation
\begin{equation}
c_{m+1} \sqrt{m+1} = \alpha c_m.
\end{equation}
This simplifies to
\begin{equation}
c_{m+1} = \frac{\alpha}{\sqrt{m+1}} c_m.
\end{equation}
By iterating this relation starting from \( c_0 \), we obtain
\begin{equation}
c_n = \frac{\alpha^n}{\sqrt{n!}} c_0.
\end{equation}
Finally, we normalize \( |\alpha\rangle \) by requiring that \( \langle \alpha | \alpha \rangle = 1 \), which gives
\begin{align}
1 &= |c_0|^2 \sum_{n=0}^\infty \frac{|\alpha|^{2n}}{n!} = \notag \\
&= |c_0|^2 e^{|\alpha|^2},
\end{align}
\begin{equation}
c_0 = e^{-\frac{|\alpha|^2}{2}},
\end{equation}
which leads to the thesis.
\end{proof}
Ultimately, the expectation values of position and momentum in a coherent state are
\begin{align}
\langle \hat{x} \rangle &= \sqrt{\frac{2\hbar}{m\omega}} \, \mathrm{Re}(\alpha), \\
\langle \hat{p} \rangle &= \sqrt{2\hbar m\omega} \, \mathrm{Im}(\alpha),
\end{align}
and their uncertainties satisfy
\begin{equation}
\Delta x \Delta p = \frac{\hbar}{2},
\end{equation}
showing that coherent states are minimum uncertainty states.
\subsection{The creation, annihilation and number operators}
A fundamental way to distinguish between bosons and fermions is by examining the algebra satisfied by their respective annihilation and creation operators. These operators encapsulate the quantum statistical behavior of the particles. In the bosonic case, the annihilation and creation operators \( a_\alpha \) and \( a^\dagger_\beta \) obey the commutation relation, i.e.,
\begin{equation}
\left[ a_\alpha , a^\dagger_\beta \right] = \delta_{\alpha,\beta} \mathds{1},
\label{eq: algebrabosonica}
\end{equation}
while the fermionic annihilation and creation operators \( a_\alpha \) and \( a^\dagger_\beta \) satisfy the anticommutation relation, i.e.,
\begin{equation}
\left\lbrace a_\alpha , a^\dagger_\beta \right\rbrace = \delta_{\alpha,\beta} \mathds{1}.
\label{eq: algebrafermionica}
\end{equation}
In general, however, we will deal with operators whose statistical nature may be arbitrary. For this purpose, let us introduce an operator that accounts for this possibility: given two generic operators $A$ and $B$, both bosonic or both fermionic, we define the parentheses
\begin{equation}
\left[\cdot,\cdot\right]^{(\varepsilon)} \ : \ \left[ A,B \right]^{(\varepsilon)} \equiv A B - \varepsilon B A,
\label{eq: parentesicommutatoreanticommutatore}
\end{equation}
which takes the form of a commutator in the case of bosons and an anticommutator in the case of fermions, reflecting their underlying quantum statistics. This distinction is governed by the value of the statistical index, defined as
\begin{equation}
\varepsilon = 
\begin{cases}
+1, & \text{bosons} \\ 
-1, & \text{fermions}
\end{cases} ,
\label{eq: epsilonfermionibosoni}
\end{equation}
which takes into account the bosonic or fermionic nature of the operators. Consequently, thanks to the parentheses $\eqref{eq: parentesicommutatoreanticommutatore}$, we can write the bosonic $\eqref{eq: algebrabosonica}$ and fermionic $\eqref{eq: algebrafermionica}$ algebras in a compact form as follows
\begin{equation}
a_\alpha a^\dagger_\beta - \varepsilon a^\dagger_\beta a_\alpha = \delta_{\alpha,\beta} \mathds{1},
\label{eq: algebracommutatoreanticommutatorecreazionedistruzione1}
\end{equation}
which will often be useful in the form
\begin{equation}
a_\alpha a^\dagger_\beta = \delta_{\alpha,\beta} \mathds{1} + \varepsilon a^\dagger_\beta a_\alpha.
\label{eq: algebracommutatoreanticommutatorecreazionedistruzione2}
\end{equation}
In particular, for a creation and annihilation operators, bosonic or fermionic, referring to different labels, the commutation or anticommutation rules are both described in compact form by
\begin{equation}
a_\alpha a^\dagger_\beta = \varepsilon a^\dagger_\beta a_\alpha, \ \alpha \neq \beta.
\label{eq: algebracommutatoreanticommutatorecreazionedistruzione3}
\end{equation}
\begin{remark}[Notation]
In many applications, the identity operator \( \mathds{1} \) appearing in commutation or anticommutation relations is understood and thus often omitted for brevity. This simplification is particularly common when the algebraic structure is the main focus, and the presence of the identity operator does not affect the physical interpretation. Whenever omitted, its presence should be considered implicit.
\end{remark}
Now, given a one-particle state labeled by $\alpha$, the number operator, associated with this state, denoted by $\hat{N}_\alpha$, is defined as
\begin{equation}
\hat{N}_\alpha = a^\dag_\alpha a_\alpha.
\end{equation}
We have
\begin{theorem}
Regardless of the statistical nature of creation and annihilation operators, the number operator satisfies the following commutators
\begin{equation}
\left[ a_\alpha , \hat{N}_\beta \right] = \delta_{\alpha,\beta} a_{\beta},
\label{eq: commutatoreoperatorenumeroconoperatoredistruzione}
\end{equation}
\begin{equation}
\left[ a^\dag_\alpha , \hat{N}_\beta \right] = - \delta_{\alpha,\beta} a^\dag_{\beta}.
\label{eq: commutatoreoperatorenumeroconoperatorecreazione}
\end{equation}
\begin{proof}
For both computations, we will make use of the unified algebraic relation $\eqref{eq: algebracommutatoreanticommutatorecreazionedistruzione2}$. We get
\begin{align}
\left[a_\alpha,\hat{N}_\beta\right] &= a_\alpha \hat{N}_\beta - \hat{N}_\beta a_\alpha = \notag \\
&=a_\alpha a^\dag_\beta a_\beta - a^\dag_\beta a_\beta a_\alpha = \notag \\
&= \left( \delta_{\alpha,\beta} a_\beta + \varepsilon a_\beta^\dag a_\alpha \right) a_\beta - a^\dag_\beta a_\beta a_\alpha = \notag \\
&= \delta_{\alpha,\beta} a_{\beta} + \varepsilon a_\beta^\dag a_\alpha a_\beta - a^\dag_\beta a_\beta a_\alpha = \notag \\
&= \delta_{\alpha,\beta} a_{\beta} + \varepsilon^2 a_\beta^\dag a_\beta a_\alpha - a^\dag_\beta a_\beta a_\alpha = \notag \\
&= \delta_{\alpha,\beta} a_{\beta}.
\end{align}
Similarly, for the second commutator, we have
\begin{align}
\left[ a^\dag_\alpha , \hat{N}_\beta \right] &= a^\dag_\alpha \hat{N}_\beta - \hat{N}_\beta a^\dag_\alpha = \notag \\
&= a^\dag_\alpha a^\dag_\beta a_\beta - a^\dag_\beta a_\beta a^\dag_\alpha = \notag \\
&= a^\dag_\alpha a^\dag_\beta a_\beta - a^\dag_\beta \left( \delta_{\beta,\alpha} + \varepsilon a^\dag_\alpha a_\beta \right) = \notag \\
&= a^\dag_\alpha a^\dag_\beta a_\beta - \delta_{\beta,\alpha} a^\dag_\beta - \varepsilon a^\dag_\beta a^\dag_\alpha a_\beta = \notag \\
&= a^\dag_\alpha a^\dag_\beta a_\beta - \delta_{\alpha,\beta} a^\dag_\beta - \varepsilon^2 a^\dag_\alpha a^\dag_\beta a_\beta = \notag \\
&= - \delta_{\alpha,\beta} a^\dag_{\beta}.
\end{align}
\end{proof}
\end{theorem}
These relations will be particularly useful when dealing with Hamiltonians written in second quantization, such as those involving interaction terms or mean-field approximations. In addition, we have
\begin{theorem}
Regardless of the statistical nature of creation and annihilation operators, they satisfy the following commutators
\begin{equation}
\left[ a_\alpha , a^\dagger_\beta a_\gamma \right] = \delta_{\alpha,\beta} a_{\gamma},
\label{eq: commutatoreaalphaadaggerbetaagamma}
\end{equation}
\begin{equation}
\left[ a^\dagger_\alpha , a^\dagger_\beta a_\gamma \right] = - \delta_{\alpha,\gamma} a^\dag_{\beta}.
\label{eq: commutatoreadaggeralphaadaggerbetaagamma}
\end{equation}
\begin{proof}
Also in this case, for both computations, we will make use of the unified algebraic relation $\eqref{eq: algebracommutatoreanticommutatorecreazionedistruzione2}$. We get
\begin{align}
\left[ a_\alpha , a^\dagger_\beta a_\gamma \right] &= a_\alpha a^\dagger_\beta a_\gamma - a^\dagger_\beta a_\gamma a_\alpha = \notag \\
&= \left( \delta_{\alpha,\beta} + \varepsilon a^\dagger_\beta a_\alpha \right) a_\gamma - a^\dagger_\beta a_\gamma a_\alpha = \notag \\
&= \delta_{\alpha,\beta} a_\gamma + \varepsilon a^\dagger_\beta a_\alpha a_\gamma - a^\dagger_\beta a_\gamma a_\alpha = \notag \\
&= \delta_{\alpha,\beta} a_\gamma + \varepsilon^2 a^\dagger_\beta a_\gamma a_\alpha - a^\dagger_\beta a_\gamma a_\alpha = \notag \\
&= \delta_{\alpha,\beta} a_\gamma.
\end{align}
Similarly, for the second commutator, we have
\begin{align}
\left[ a^\dagger_\alpha , a^\dagger_\beta a_\gamma \right] &= a^\dagger_\alpha a^\dagger_\beta a_\gamma - a^\dagger_\beta a_\gamma a^\dagger_\alpha = \notag \\
&= a^\dagger_\alpha a^\dagger_\beta a_\gamma - a^\dagger_\beta \left( \delta_{\gamma,\alpha} + \varepsilon a^\dagger_\alpha a_\gamma \right) = \notag \\
&= a^\dagger_\alpha a^\dagger_\beta a_\gamma - \delta_{\alpha,\gamma} a^\dagger_\beta - \varepsilon a^\dagger_\beta a^\dagger_\alpha a_\gamma = \notag \\
&= a^\dagger_\alpha a^\dagger_\beta a_\gamma - \delta_{\alpha,\gamma} a^\dagger_\beta - \varepsilon^2 a^\dagger_\alpha a^\dagger_\beta a_\gamma = \notag \\
&= - \delta_{\alpha,\gamma} a^\dagger_\beta.
\end{align}
\end{proof}
\end{theorem}
\section{Quantum mechanics pictures}
In quantum mechanics, dynamical pictures (or pictures or representations) are the multiple equivalent ways to mathematically formulate the dynamics of a quantum system. The most famous are the Schrödinger and Heisenberg pictures, which are referred to a generical Hamiltonian. On the other hand, the interaction and thermal interaction pictures are particularly useful when the Hamiltonian \( \mathcal{H} \) decomposes into the sum of a diagonalizable term \( \mathcal{H}_0 \) and a perturbative term \( \mathcal{H}_I \) (e.g., a Hamiltonian including interactions between particles), i.e.,
\begin{equation}
\mathcal{H} = \mathcal{H}_0 + \mathcal{H}_I.
\label{eq: HamiltonianascompostainH_0eH_I}
\end{equation}
These representations allow us to treat the effect of interactions as perturbations evolving over a solvable background. This separation is especially advantageous in quantum field theory and many-body physics, where exact solutions are rarely available, and perturbative techniques are essential for practical computations.
\subsection{Schrödinger picture}
In the Schrödinger picture the state vectors of the system depend on time while the operators do not, and the time evolution of the system is described by the equation
\begin{equation}
i \hslash \partial_t \psi(t) = \mathcal{H} \psi(t),
\end{equation}
which is called Schrödinger's equation. For systems in which energy is conserved, the Hamiltonian operator $\mathcal{H}$ does not depend explicitly on time and the solution of Schrödinger's equation is
\begin{align}
\psi(t) &= U(t) \psi(0) = \notag \\
&= e^{- i \frac{\mathcal{H}}{\hslash} t} \psi(0),
\end{align}
and since $\mathcal{H}$ is a self-adjoint operator, $e^{-i \frac{\mathcal{H}}{\hslash} t}$ is a unitary operator.
\subsection{Heisenberg picture}
In the Heisenberg picture, the time dependence of a quantum system is transferred from the state vectors to the operators. That is, the state vectors remain time-independent, while the operators evolve in time according to the Heisenberg equation of motion. The time-evolved operator \( A(t) \) is given by
\begin{equation}
A(t) = e^{i \frac{\mathcal{H}}{\hslash} t} A e^{- i \frac{\mathcal{H}}{\hslash} t},
\label{eq: Heisenbergpicture}
\end{equation}
where \( \mathcal{H} \) is the Hamiltonian of the system. This transformation ensures that the expectation values of observables remain the same as in the Schrödinger picture.
\subsection{Interaction picture}\label{Interaction picture}
In the interaction (or Dirac's) picture both operators and states depend explicitly on time and the dynamical evolution of the operator $A$ involve a unitary transformation related to $\mathcal{H}_0$ as follows
\begin{equation}
A^{(0)}(t) = e^{i \frac{\mathcal{H}_0}{\hslash} t} A e^{- i \frac{\mathcal{H}_0}{\hslash} t} ,
\label{eq: interactionpicture}
\end{equation}
where the superscript $0$ emphasizes that the evolution of the operators in the interaction representation is with respect to the Hamiltonian $\mathcal{H}_0$; on the other hand, the time evolution of the states is given by
\begin{align}
\psi(t) &= e^{i \frac{\mathcal{H}_0}{\hslash} t} e^{- i \frac{\mathcal{H}}{\hslash} t} \psi(0) \equiv \notag \\
&\equiv U(t) \psi(0) ,
\end{align}
where we have defined the unitary transformation
\begin{equation}
U(t) = e^{i \frac{\mathcal{H}_0}{\hslash} t} e^{- i \frac{\mathcal{H}}{\hslash} t}.
\end{equation}
Note that
\begin{align}
\langle A(t) \rangle 
&= \langle \psi(t) | A | \psi(t) \rangle = \notag \\
&= \left\langle \psi(0) \left| e^{i \frac{\mathcal{H}}{\hbar} t} e^{- i \frac{\mathcal{H}_0}{\hbar} t} e^{i \frac{\mathcal{H}_0}{\hbar} t} A e^{- i \frac{\mathcal{H}_0}{\hbar} t} e^{i \frac{\mathcal{H}_0}{\hbar} t} e^{- i \frac{\mathcal{H}}{\hbar} t} \right| \psi(0) \right\rangle = \notag \\
&= \left\langle \psi(0) \left| e^{i \frac{\mathcal{H}}{\hbar} t} A e^{- i \frac{\mathcal{H}}{\hbar} t} \right| \psi(0) \right\rangle,
\end{align}
as it must be. 
\begin{theorem}
The unitary transformation \( U(t) \) associated with the picture representation $\eqref{eq: interactionpicture}$ satisfies the relation
\begin{equation}
\dfrac{\partial U(t)}{\partial t} = - \dfrac{i}{\hslash} \mathcal{H}_I^{(0)}(t) U(t),
\end{equation}
with
\begin{equation}
U(0) = \mathds{1}.
\end{equation}
\begin{proof}
We have
\begin{align}
\dfrac{\partial U}{\partial t} &= \dfrac{i}{\hslash} \mathcal{H}_0 e^{i \frac{\mathcal{H}_0}{\hslash} t} e^{- i \frac{\mathcal{H}}{\hslash} t} - \dfrac{i}{\hslash} e^{i \frac{\mathcal{H}_0}{\hslash} t} \mathcal{H} e^{- i \frac{\mathcal{H}}{\hslash} t} = \notag \\
&= \dfrac{i}{\hslash} e^{i \frac{\mathcal{H}_0}{\hslash} t} \mathcal{H}_0 e^{- i \frac{\mathcal{H}}{\hslash} t} - \dfrac{i}{\hslash} e^{i \frac{\mathcal{H}_0}{\hslash} t} \mathcal{H} e^{- i \frac{\mathcal{H}}{\hslash} t} = \notag \\
&= \dfrac{i}{\hslash} e^{i \frac{\mathcal{H}_0}{\hslash} t} \left( \mathcal{H}_0 - \mathcal{H} \right) e^{- i \frac{\mathcal{H}}{\hslash} t} = \notag \\
&= - \dfrac{i}{\hslash} e^{i \frac{\mathcal{H}_0}{\hslash} t} \mathcal{H}_I e^{- i \frac{\mathcal{H}}{\hslash} t} = \notag \\
&= - \dfrac{i}{\hslash} e^{i \frac{\mathcal{H}_0}{\hslash} t} \mathcal{H}_I e^{- i \frac{\mathcal{H}_0}{\hslash} t} e^{i \frac{\mathcal{H}_0}{\hslash} t} e^{- i \frac{\mathcal{H}}{\hslash} t} = \notag \\
&= - \dfrac{i}{\hslash} \mathcal{H}_I^{(0)}(t) U(t),
\end{align}
where we have used the commutator $\left[ \mathcal{H}_0 , e^{i \frac{\mathcal{H}_0}{\hslash} t} \right]=0$, after which we have used the identity $e^{- i \frac{\mathcal{H}_0}{\hslash} t} e^{i \frac{\mathcal{H}_0}{\hslash} t}$ to bring out the interaction representation of $\mathcal{H}_I$ with respect to $\mathcal{H}_0$.
\end{proof}
\end{theorem}
Together with the transformation $U(t)$, we define the matrix $S(t,t')$ such that
\begin{equation}
\psi(t) = S(t,t') \psi(t').
\end{equation}
From
\begin{equation}
\psi(t) = U(t) \psi(0),
\end{equation}
\begin{align}
\psi(t) &= S(t,t') \psi(t') = \notag \\
&= S(t,t') U(t') \psi(0) ,
\end{align}
we get
\begin{equation}
S(t,t') U(t') = U(t), \ \forall \ \psi(0),
\end{equation}
which implies
\begin{align}
S(t,t') &= U(t) U^{-1}(t') \equiv \notag \\
&\equiv U(t) U^{\dag}(t').
\end{align}
By definition it follows
\begin{equation}
S(t,0) = U(t),
\end{equation}
\begin{align}
S(0,t) &= U^{-1}(t) = \notag \\
&= U^{\dag}(t).
\end{align}
\begin{theorem}
The matrix $S(t,t')$ of the picture representation satisfies:
\begin{itemize}
\item [i)]
\begin{equation}
S(t,t) = \mathds{1},
\end{equation}
which is the initial condition for the matrix $S(t,t')$;
\item [ii)]
\begin{equation}
S(t_1,t_2) S(t_2,t_3) = S(t_1,t_3);
\end{equation}
\item [iii)]
\begin{equation}
S^{\dag}(t,t') = U(t') U^{\dag}(t).
\end{equation}
\end{itemize}
\begin{proof}
\begin{itemize}
From the definition we have
\item [i)]
\begin{align}
S(t,t) &=U(t) U^{\dag}(t) = \notag \\
&= \mathds{1};
\end{align}
\item [ii)]
\begin{align}
S(t_1,t_2) S(t_2,t_3) &= U(t_1) U^{\dag}(t_2) U(t_2) U^{\dag}(t_3) = \notag \\
&= U(t_1) U^{\dag}(t_3) \equiv \notag \\
&\equiv S(t_1,t_3);
\end{align}
\item [iii)]
\begin{align}
S^{\dag}(t,t') &= \left( U(t) U^{\dag}(t') \right)^{\dag} = \notag \\
&= U^{\dag}{}^{\dag}(t') U^{\dag}(t) = \notag \\
&= U(t') U^{\dag}(t).
\end{align}
\end{itemize}
\end{proof}
\end{theorem}
Finally, note that matrix $S(t,t')$ can be written as a function of the Hamiltonian $\mathcal{H}$ as
\begin{align}
S(t,t') &= U(t) U^{\dag}(t') = \notag \\
&= e^{i \frac{\mathcal{H}_0}{\hslash} t} e^{- i \frac{\mathcal{H}}{\hslash} t} \left( e^{i \frac{\mathcal{H}_0}{\hslash} t'} e^{- i \frac{\mathcal{H}}{\hslash} t'} \right)^{\dag} = \notag \\
&= e^{i \frac{\mathcal{H}_0}{\hslash} t} e^{- i \frac{\mathcal{H}}{\hslash} t} e^{i \frac{\mathcal{H}}{\hslash} t'} e^{- i \frac{\mathcal{H}_0}{\hslash} t'} = \notag \\
&= e^{i \frac{\mathcal{H}_0}{\hslash} t} e^{- i \frac{\mathcal{H}}{\hslash} (t-t')} e^{- i \frac{\mathcal{H}_0}{\hslash} t'}.
\end{align}
\section{The quantum adiabatic theorem}\label{The quantum adiabatic theorem}
The quantum adiabatic theorem is a fundamental result that describes the behavior of quantum many-particle systems undergoing slow (adiabatic) evolution, i.e., when the Hamiltonian varies smoothly and sufficiently slowly in time. Suppose a quantum system is initially prepared in a given state, and then the system Hamiltonian is varied slowly in time. It is known that the system’s state evolves according to the Schrödinger equation with a time-dependent Hamiltonian. Remarkably, if the evolution is adiabatic and the spectrum of the Hamiltonian is discrete and nondegenerate, the system remains in the instantaneous eigenstate corresponding to its initial eigenstate throughout the evolution. Here we present a simple quantitative characterization of this principle.
\begin{theorem}[The quantum adiabatic theorem]
Let $\mathcal{H}(t)$ be a Hamiltonian that varies very slowly with time, and assume its spectrum is discrete and nondegenerate. Then, given two times $t_1$ and $t_2$ with $t_2 > t_1$, if at time $t_1$ the system is in the $n$-th eigenstate of $\mathcal{H}(t_1)$, it will be in the $n$-th eigenstate of $\mathcal{H}(t_2)$ at time $t_2$.
\begin{proof}
At each instant $t$, the eigenfunctions and eigenvalues satisfy the eigenvalue problem
\begin{equation}
\mathcal{H}(t) \varphi_m(t) = \mathcal{E}_m(t) \varphi_m(t),
\end{equation}
with orthonormality condition
\begin{equation}
\langle \varphi_n(t) | \varphi_m(t) \rangle = \delta_{n,m}.
\end{equation}
The wave function can be expanded in the instantaneous eigenbasis as
\begin{equation}
\Psi(t) = \sum_m c_m(t) \varphi_m(t) e^{-\frac{i}{\hbar} \theta_m(t)},
\end{equation}
where each phase is defined by
\begin{equation}
\theta_m(t) = \int_0^t \mathcal{E}_m(t') \, \mathrm{d}t'.
\end{equation}
Inserting this expansion into the time-dependent Schrödinger equation, we obtain
\begin{equation}
i \hbar \sum_m \left[ \partial_t c_m(t) \varphi_m(t) + c_m(t) \partial_t \varphi_m(t) - \frac{i}{\hbar} c_m(t) \varphi_m(t) \partial_t \theta_m(t) \right] e^{-\frac{i}{\hbar} \theta_m(t)} = \sum_m c_m(t) \mathcal{H}(t) \varphi_m(t) e^{-\frac{i}{\hbar} \theta_m(t)}.
\end{equation}
Since $\partial_t \theta_m(t) = \mathcal{E}_m(t)$, the terms containing $\mathcal{E}_m(t)$ cancel out on both sides, leaving
\begin{equation}
\sum_m \left[ \partial_t c_m(t) \varphi_m(t) + c_m(t) \partial_t \varphi_m(t) \right] e^{-\frac{i}{\hbar} \theta_m(t)} = 0.
\end{equation}
Multiplying from the left by $\langle \varphi_n(t) |$ and using orthonormality, we find
\begin{equation}
\partial_t c_n(t) = - \sum_m c_m(t) \langle \varphi_n(t) | \partial_t \varphi_m(t) \rangle e^{-\frac{i}{\hbar} \left( \theta_m(t) - \theta_n(t) \right)}.
\end{equation}
Separating the term with $m = n$,
\begin{equation}
\partial_t c_n(t) = - c_n(t) \langle \varphi_n(t) | \partial_t \varphi_n(t) \rangle - \sum_{m \neq n} c_m(t) \langle \varphi_n(t) | \partial_t \varphi_m(t) \rangle e^{-\frac{i}{\hbar} \left( \theta_m(t) - \theta_n(t) \right)}.
\end{equation}
Because the spectrum is nondegenerate, the phase factors $e^{-\frac{i}{\hbar}(\theta_m - \theta_n)}$ oscillate rapidly and average out, making the object
\begin{equation}
\sum_{m \neq n} c_m(t) \langle \varphi_n(t) | \partial_t \varphi_m(t) \rangle e^{-\frac{i}{\hbar} \left( \theta_m(t) - \theta_n(t) \right)}
\label{eq: terminediapprossimazioneadiabatica}
\end{equation} 
negligible. Moreover, since $|c_n(t)|^2 \leq 1$, it remains to evaluate the terms
\begin{equation}
\langle \varphi_n(t) | \partial_t \varphi_m(t) \rangle.
\end{equation}
Differentiating the eigenvalue equation with respect to time,
\begin{equation}
\partial_t \mathcal{H}(t) \varphi_m(t) + \mathcal{H}(t) \partial_t \varphi_m(t) = \partial_t \mathcal{E}_m(t) \varphi_m(t) + \mathcal{E}_m(t) \partial_t \varphi_m(t),
\end{equation}
and projecting on $\varphi_n(t)$ with $n \neq m$, we get
\begin{equation}
\langle \varphi_n(t) | \partial_t \mathcal{H}(t) | \varphi_m(t) \rangle + \mathcal{E}_n(t) \langle \varphi_n(t) | \partial_t \varphi_m(t) \rangle = \mathcal{E}_m(t) \langle \varphi_n(t) | \partial_t \varphi_m(t) \rangle,
\end{equation}
which yields
\begin{equation}
\left( \mathcal{E}_n(t) - \mathcal{E}_m(t) \right) \langle \varphi_n(t) | \partial_t \varphi_m(t) \rangle = - \langle \varphi_n(t) | \partial_t \mathcal{H}(t) | \varphi_m(t) \rangle.
\label{eq: terminenonadiabaticosviluppocn}
\end{equation}
Since
\begin{equation}
0 < \left| \mathcal{E}_n(t) - \mathcal{E}_m(t) \right| < +\infty, \quad \forall \ n \neq m,
\end{equation}
the quantity $\langle \varphi_n(t) | \partial_t \varphi_m(t) \rangle$ is negligible if the matrix elements of $\partial_t \mathcal{H}(t)$ are sufficiently small. Under this condition, it is legitimate to neglect the non-adiabatic term in the evolution of $c_n(t)$; this is known as the adiabatic approximation. The expression $\eqref{eq: terminenonadiabaticosviluppocn}$ is called the non-adiabatic term. In the adiabatic approximation, the time evolution of the coefficient $c_n$ reduces to
\begin{equation}
\partial_t c_n(t) = - c_n(t) \langle \varphi_n(t) | \partial_t \varphi_n(t) \rangle.
\end{equation}
From the orthonormality condition, it follows that
\begin{equation}
\langle \varphi_n(t) | \partial_t \varphi_n(t) \rangle + \langle \partial_t \varphi_n(t) | \varphi_n(t) \rangle = 0,
\end{equation}
which implies
\begin{equation}
2 \, \mathrm{Re} \left\lbrace \langle \varphi_n(t) | \partial_t \varphi_n(t) \rangle \right\rbrace = 0.
\end{equation}
Therefore, $\langle \varphi_n(t) | \partial_t \varphi_n(t) \rangle$ is purely imaginary. Solving the first-order differential equation with separable variables, we have
\begin{equation}
\int \frac{d c_n(t)}{c_n(t)} = - \int \langle \varphi_n(t) | \partial_t \varphi_n(t) \rangle \, \mathrm{d}t,
\end{equation}
which gives
\begin{equation}
\ln c_n(t) = - \int_0^t \langle \varphi_n(t') | \partial_{t'} \varphi_n(t') \rangle \, \mathrm{d}t' + \const
\end{equation}
Exponentiating and fixing the integration constant by initial conditions yields
\begin{equation}
c_n(t) = c_n(0) \, e^{ - \int_0^t \langle \varphi_n(t') | \partial_{t'} \varphi_n(t') \rangle \, \mathrm{d}t' }.
\end{equation}
The adiabatic phase, given by the exponent, is purely imaginary and is called the Berry phase acquired by the eigenstate during the adiabatic evolution. This completes the proof that, under adiabatic evolution, the system remains in its instantaneous eigenstate up to a phase factor.
\end{proof}
\end{theorem}
\section{One particle Green's function}
Let \( U(t, t_1) \) be the unitary time-evolution operator defined by
\begin{equation}
|\psi(t)\rangle = U(t, t_1) |\psi(t_1)\rangle,
\end{equation}
where the state \( |\psi(t)\rangle \) satisfies the Schrödinger equation. Then, the time-evolution operator \( U(t, t_1) \) itself obeys the corresponding Schrödinger equation,
\begin{equation}
i \hbar \frac{\partial}{\partial t} U(t, t_1) = \mathcal{H} \, U(t, t_1).
\end{equation}
Consequently, the operator \( U(t, t_1) \) must satisfy
\begin{equation}
\lim_{t_{2} \rightarrow t_{1}} U(t_{2},t_{1}) = \mathds{1},
\end{equation}
\begin{equation}
U^{\dagger}(t_{2},t_{1}) = U^{-1}(t_{2},t_{1}),
\end{equation}
\begin{equation}
U(t_{3},t_{1}) = U(t_{3},t_{2}) U(t_{2},t_{1}), \ t_3>t_2>t_1.
\end{equation}
The last property shows that the time evolution operator must satisfy a semigroup relation. The Schrödinger representation for the temporal evolution operator is
\begin{equation}
\langle \mathbf{r}_2 | \psi(t_2) \rangle = \langle \mathbf{r}_2 | U(t_2, t_1) | \psi(t_1) \rangle,
\end{equation}
then we insert $\mathds{1} = \int d\mathbf{r}_1 \, |\mathbf{r}_1 \rangle \langle \mathbf{r}_1 |$ as follows
\begin{equation}
\langle \mathbf{r}_2 | \psi(t_2) \rangle = \int d\mathbf{r}_1 \, \langle \mathbf{r}_2 | U(t_2, t_1) | \mathbf{r}_1 \rangle \, \langle \mathbf{r}_1 | \psi(t_1) \rangle,
\end{equation}
and we have
\begin{equation}
\psi(\textbf{r}_{2},t_{2}) = \int d\textbf{r}_{1} \ G(\textbf{r}_{1},t_{1},\textbf{r}_{2},t_{2}) \ \psi(\textbf{r}_{1},t_{1}),
\end{equation}
where
\begin{equation}
G(\mathbf{r}_1, t_1, \mathbf{r}_2, t_2) = \langle \mathbf{r}_2 | U(t_2, t_1) | \mathbf{r}_1 \rangle
\end{equation}
is called the Green's function. If $\mathcal{H}$ does not depend on time,
\begin{equation}
U(t_{2},t_{1}) = e^{- i \frac{\mathcal{H}}{\hslash} (t_{2} - t_{1})}
\end{equation}
then the Green's function is
\begin{equation}
G(\mathbf{r}_1, t_1, \mathbf{r}_2, t_2) = \langle \mathbf{r}_2 | e^{-i \frac{\mathcal{H}}{\hbar} (t_2 - t_1)} | \mathbf{r}_1 \rangle.
\end{equation}
Assume that the Hamiltonian has a discrete, non-degenerate spectrum, i.e.,
 $\mathcal{H} |\psi_n\rangle = \mathcal{E}_n |\psi_n\rangle$, then
 \begin{align}
G(\mathbf{r}_1, t_1, \mathbf{r}_2, t_2) &= \sum_{n,m} \langle \mathbf{r}_2 | \psi_n \rangle \langle \psi_n | e^{-i \frac{\mathcal{H}}{\hbar} (t_2 - t_1)} | \psi_m \rangle \langle \psi_m | \mathbf{r}_1 \rangle = \notag \\
&= \sum_{n,m} \langle \mathbf{r}_2 | \psi_n \rangle \langle \psi_n | e^{-i \frac{\mathcal{E}_n}{\hbar} (t_2 - t_1)} | \psi_m \rangle \langle \psi_m | \mathbf{r}_1 \rangle = \notag \\
&= \sum_{n,m} \langle \mathbf{r}_2 | \psi_n \rangle e^{-i \frac{\mathcal{E}_n}{\hbar} (t_2 - t_1)} \langle \psi_n | \psi_m \rangle \langle \psi_m | \mathbf{r}_1 \rangle = \notag \\
&= \sum_{n,m} \langle \mathbf{r}_2 | \psi_n \rangle e^{-i \frac{\mathcal{E}_n}{\hbar} (t_2 - t_1)} \delta_{n m} \langle \psi_m | \mathbf{r}_1 \rangle = \notag \\
&= \sum_{n} \psi_n(\mathbf{r}_2) e^{-i \frac{\mathcal{E}_n}{\hbar} (t_2 - t_1)} \psi_n^*(\mathbf{r}_1) = \notag \\
&= \sum_{n} \psi_n^*(\mathbf{r}_1) \psi_n(\mathbf{r}_2) e^{-i \frac{\mathcal{E}_n}{\hbar} (t_2 - t_1)}.
\end{align}
\subsection{Meaning of one-particle Green's function}
Consider the one-particle Schrödinger equation with initial condition $\psi(t')$. If the Hamiltonian does not depend on time, at time $t$, $t>t'$, the wave function is given by
\begin{align}
|\psi(t)\rangle = e^{-i \frac{\mathcal{H}}{\hbar} (t - t')} |\psi(t')\rangle, \quad \forall \ t > t'.
\end{align}
Then, in the representation of configuration, we project on $|\mathbf{r}\rangle$ both members, use in the second member the completeness of $|\mathbf{r}'\rangle$ (i.e., we insert the decomposition of the identity $\mathds{1} = \int d\mathbf{r}' \, |\mathbf{r}'\rangle \langle \mathbf{r}'|$) and multiply both members by Heaviside's $\Theta(t-t')$ (to comply with the principle of temporal causality)
\begin{equation}
\langle \mathbf{r} | \psi(t) \rangle = \langle \mathbf{r} | e^{-i \frac{\mathcal{H}}{\hbar} (t - t')} | \psi(t') \rangle,
\end{equation}
\begin{equation}
\psi(\mathbf{r}, t) = \int d\mathbf{r}' \, \langle \mathbf{r} | e^{-i \frac{\mathcal{H}}{\hbar} (t - t')} | \mathbf{r}' \rangle \langle \mathbf{r}' | \psi(t') \rangle,
\end{equation}
\begin{equation}
\psi(\mathbf{r}, t) = \int d\mathbf{r}' \, \langle \mathbf{r} | e^{-i \frac{\mathcal{H}}{\hbar} (t - t')} | \mathbf{r}' \rangle \, \psi(\mathbf{r}', t'),
\end{equation}
\begin{equation}
\Theta(t - t') \, \psi(\mathbf{r}, t) = i \hbar \int \frac{d\mathbf{r}'}{i \hbar} \, \Theta(t - t') \, \langle \mathbf{r} | e^{-i \frac{\mathcal{H}}{\hbar} (t - t')} | \mathbf{r}' \rangle \, \psi(\mathbf{r}', t'),
\end{equation}
\begin{equation}
\Theta(t - t') \, \psi(\mathbf{r}, t) = i \hbar \int d\mathbf{r}' \, G^{(r)}(\mathbf{r},t,\mathbf{r}',t') \, \psi(\mathbf{r}', t').
\label{eq: costruzionepropagatorefunzioned'ondasingolaparticella}
\end{equation}
We define retarded Green's function as
\begin{equation}
G^{(r)}(\mathbf{r}, t, \mathbf{r}', t') = - \frac{i}{\hbar} \, \Theta(t - t') \, \langle \mathbf{r} | e^{-i \frac{\mathcal{H}}{\hbar} (t - t')} | \mathbf{r}' \rangle,
\end{equation}
which satisfies several properties. First, it does not depend on the initial conditions, since it does not contain \(\psi(t')\). Given the initial condition and the retarded Green's function, we can calculate the solution of the Schrödinger equation at each instant \(t\) with \(t > t'\). Moreover, it is itself a solution of the Schrödinger equation with the particular initial condition $\psi(\mathbf{r}', t') = \delta(\mathbf{r}' - \mathbf{r}_0), \ \forall \ t > t', \ \forall \ \mathbf{r}_0$. Indeed, from $\eqref{eq: costruzionepropagatorefunzioned'ondasingolaparticella}$ it follows
\begin{equation}
\Theta(t - t') \psi(\mathbf{r}, t) = i \hbar \int d\mathbf{r}' \, G^{(r)}(\mathbf{r},t,\mathbf{r}',t') \, \delta(\mathbf{r}' - \mathbf{r}_0),
\end{equation}
which reduces to
\begin{equation}
\Theta(t - t') \psi(\mathbf{r}, t) = i \hbar \, G^{(r)}(\mathbf{r},t,\mathbf{r}_0, t'),
\end{equation}
and the Green's function satisfies the equation
\begin{equation}
i \hbar \, \partial_t G^{(r)}(\mathbf{r},t,\mathbf{r}_0,t') = \mathcal{H} \, G^{(r)}(\mathbf{r},t,\mathbf{r}_0,t').
\end{equation}
Equivalently,
\begin{equation}
i \hbar \, \partial_t \left( \frac{1}{i \hbar} \Theta(t - t') \psi(\mathbf{r}, t) \right) = \frac{1}{i \hbar} \mathcal{H} \, \Theta(t - t') \psi(\mathbf{r}, t),
\end{equation}
which implies
\begin{equation}
i \hbar \, \partial_t \left( \Theta(t - t') \psi(\mathbf{r}, t) \right) = \mathcal{H} \, \Theta(t - t') \psi(\mathbf{r}, t).
\end{equation}
Since \( t > t' \), the statement follows. The object $G^{(r)}(\textbf{r},t,\textbf{r}',t')$ is also called the one-particle, or single-particle, propagator, for a reason we will see shortly. We compute $G^{(r)}(\textbf{r},t,\textbf{r}',t')$ as a function of the eigenstates $\lbrace \varphi_\alpha \rbrace$ as follows
\begin{align}
G^{(r)}(\mathbf{r}, t, \mathbf{r}', t') &= - \frac{i}{\hbar} \Theta(t - t') \sum_{\alpha, \beta} \langle \mathbf{r} | \varphi_{\alpha} \rangle \langle \varphi_{\alpha} | e^{-i \frac{\mathcal{H}}{\hbar} (t - t')} | \varphi_{\beta} \rangle \langle \varphi_{\beta} | \mathbf{r}' \rangle = \notag \\
&= - \frac{i}{\hbar} \Theta(t - t') \sum_{\alpha, \beta} \varphi_\alpha(\mathbf{r}) \langle \varphi_{\alpha} | e^{-i \frac{\mathcal{E}_{\beta}}{\hbar} (t - t')} | \varphi_{\beta} \rangle \varphi_\beta^*(\mathbf{r}') = \notag \\
&= - \frac{i}{\hbar} \Theta(t - t') \sum_{\alpha, \beta} \varphi_\alpha(\mathbf{r}) e^{-i \frac{\mathcal{E}_\beta}{\hbar} (t - t')} \langle \varphi_{\alpha} | \varphi_{\beta} \rangle \varphi_\beta^*(\mathbf{r}') = \notag \\
&= - \frac{i}{\hbar} \Theta(t - t') \sum_{\alpha, \beta} \varphi_\alpha(\mathbf{r}) e^{-i \frac{\mathcal{E}_{\beta}}{\hbar} (t - t')} \delta_{\alpha, \beta} \varphi_\beta^*(\mathbf{r}') = \notag \\
&= - \frac{i}{\hbar} \Theta(t - t') \sum_\alpha \varphi_\alpha(\mathbf{r}) e^{-i \frac{\mathcal{E}_\alpha}{\hbar} (t - t')} \varphi_\alpha^*(\mathbf{r}').
\end{align}
Note that
\begin{align}
\lim_{t \rightarrow t'^{ +}} G^{(r)}(\textbf{r},t,\textbf{r}',t') = - \dfrac{i}{\hslash}\delta(\textbf{r}-\textbf{r}') ,
\end{align}
whereas for $t \rightarrow t'^{ -}$ there is no dynamics and the wave function is obviously the initial one. \newline
Now we illustrate the meaning of a single-particle propagator. Consider a particle and suppose that in a time interval $t-t'$ it transits from point $\textbf{r}'$ to point $\textbf{r}$, then $\langle \textbf{r} | e^{- i \frac{\mathcal{H}}{\hbar} (t - t')} | \textbf{r}' \rangle$ represents the probability amplitude for the particle to be found at position $\mathbf{r}$ at time $t$, given that it was at position $\mathbf{r}'$ at time $t'$.
\subsection{One-dimensional free particle Green's function}
Let us compute the Green's function $G_{0}(x',t',x,t)$ associated with the Hamiltonian of a free particle moving along a line, that is,
\begin{equation}
G_{0}(x', t', x, t) = \langle x' | e^{- \frac{i}{\hbar} \frac{\hat{p}^2}{2m} (t' - t)} | x \rangle,
\end{equation}
where $x$ denotes the position of the particle along the one-dimensional space. Since $\mathcal{H}_{0}$ does depend solely on the momentum, we insert the identity $\mathds{1} = \int dp \ |p\rangle \langle p|$ in $G_{0}$ as follows
\begin{equation}
G_{0}(x', t', x, t) = \int dp \int dp' \langle x' | p' \rangle \langle p' | e^{- \frac{i}{\hbar} \frac{\hat{p}^2}{2m} (t' - t)} | p \rangle \langle p | x \rangle.
\end{equation}
From the identity
\begin{equation}
\langle x | p \rangle = \frac{1}{\sqrt{2 \pi \hbar}} e^{\frac{i}{\hbar} p x},
\end{equation}
we have
\begin{align}
G_{0}(x',t',x,t) &= \int \dfrac{dp}{\sqrt{2 \pi \hbar}} \int \dfrac{dp'}{\sqrt{2 \pi \hbar}} \ e^{\frac{i}{\hbar} (p'x'-px)} \langle p'|e^{- \frac{i}{\hbar} \frac{\hat{p}^2}{2m} (t'-t)}|p \rangle = \notag \\ 
&= \int \dfrac{dp}{\sqrt{2 \pi \hbar}} \int \dfrac{dp'}{\sqrt{2 \pi \hbar}} \ e^{\frac{i}{\hbar} (p'x'-px)} e^{- \frac{i}{\hbar} \frac{p^2}{2m} (t'-t)} \delta(p'-p) = \notag \\ 
&= \int \dfrac{dp}{\sqrt{2 \pi \hbar}} \int \dfrac{dp'}{\sqrt{2 \pi \hbar}} \ e^{\frac{i}{\hbar} (p'x'-px) - \frac{i}{\hbar} \frac{p^2}{2m} (t'-t)} \delta(p'-p) = \notag \\ 
&= \int \dfrac{dp}{2 \pi \hbar} \ e^{\frac{i}{\hbar} p (x'-x) - \frac{i}{\hbar} \frac{p^2}{2m} (t'-t)}.
\end{align}
We introduce an arbitrary phase factor $\delta$ to ensure the convergence of the series, i.e.,
\begin{equation}
G_{0}(x',t',x,t) = \int \dfrac{dp}{2 \pi \hslash} \ e^{\frac{i}{\hslash} p (x'-x) - \frac{i}{\hslash} \frac{p^2}{2m} (t'-t-i \delta)} 
\end{equation}
and note that in this form, if we set
\begin{equation}
A = i \ \dfrac{t' - t - i \delta}{m \hslash},
\end{equation}
\begin{equation}
B = \dfrac{i}{\hslash} (x'-x),
\end{equation}
$G_{0}$ is a gaussian integral, i.e.,
\begin{equation}
\int_{-\infty}^{+\infty} dy \ e^{- \frac{A}{2} y^{2} + B y} = \sqrt{\dfrac{2 \pi}{A}} \ e^{- \frac{B^{2}}{2} A}, \ A,B \in \mathbb{C}, \ \re(A)>0
\end{equation}
then
\begin{align}
G_{0}(x',t',x,t) &= \dfrac{1}{2 \pi \hslash} \sqrt{\dfrac{2 \pi m \hslash}{i (t'-t-i \delta)}} \ e^{\frac{i m (x'-x)^{2}}{2 \hslash (t'-t-i \delta)}} = \notag \\
&= \sqrt{\dfrac{m}{2 \pi i \hslash (t'-t-i \delta)}} \ e^{\frac{i m (x'-x)^{2}}{2 \hslash (t'-t-i \delta)}} 
\end{align}
is a pseudo-Gaussian, because of the imaginary unit.
\subsubsection{Analytical properties}
\begin{itemize}
\item
We show that in the limit for $t'\rightarrow t^{+}$, $G_{0}$ is a Dirac delta with respect to positions.
\begin{proof}
Being interested in the limit of a difference of times, we set $i (t'-t)=- i \delta$ in $G_{0}$ and write
\begin{align}
G_{0} &= \sqrt{\dfrac{m}{2 \pi i \hslash (-i \delta)}} \ e^{\frac{i m (x'-x)^{2}}{2 \hslash (-i \delta)}} = \notag \\
&= \sqrt{\dfrac{m}{2 \pi \hslash \delta}} \ e^{- \frac{m (x'-x)^{2}}{2 \hslash \delta}} 
\end{align}
and we study the limit for $\delta \rightarrow 0^{+}$ of the integral on the real axis of $G_{0}$ multiplied a rapidly decreasing function $f(x)$, i.e.,
\begin{equation}
\lim_{\delta \rightarrow 0^{+}} \int_{-\infty}^{+\infty} dx \ G_{0} f(x) = \lim_{\delta \rightarrow 0^{+}} \int_{-\infty}^{+\infty} dx \sqrt{\dfrac{m}{2 \pi \hslash \delta}} \ e^{- \frac{m (x'-x)^{2}}{2 \hslash \delta}} f(x).
\end{equation}
We set
\begin{equation}
R=\frac{m}{\hslash}, 
\end{equation}
then
\begin{equation}
\lim_{\delta \rightarrow 0^{+}} \int_{-\infty}^{+\infty} dx G_{0} f(x) = \lim_{\delta \rightarrow 0^{+}} \int_{-\infty}^{+\infty} dx \sqrt{\dfrac{R}{2 \pi \delta}} \ e^{- \frac{R (x'-x)^{2}}{2 \delta}} f(x) ,
\end{equation}
and we set
\begin{equation}
z=\sqrt{\dfrac{R}{\delta}} \left(x'-x\right) ,
\end{equation}
\begin{equation}
x = x' - \sqrt{\dfrac{\delta}{R}} z,
\end{equation}
then
\begin{equation}
\lim_{\delta \rightarrow 0^{+}} \int_{-\infty}^{+\infty} dx \ G_{0} f(x) = \lim_{\delta \rightarrow 0^{+}} \int_{-\infty}^{+\infty} \left( - \dfrac{dz}{\sqrt{2 \pi}} \right) e^{- \frac{z^2}{2}} f\left(x' - \sqrt{\dfrac{\delta}{R}} z\right) ,
\end{equation}
\begin{equation}
\lim_{\delta \rightarrow 0^{+}} \int_{-\infty}^{+\infty} dx \ G_{0} f(x) = \lim_{\delta \rightarrow 0^{+}} \int_{+\infty}^{-\infty} \dfrac{dz}{\sqrt{2 \pi}} e^{- \frac{z^2}{2}} f\left(x' - \sqrt{\dfrac{\delta}{R}} z\right) ,
\end{equation}
\begin{align}
\lim_{\delta \rightarrow 0^{+}} \int_{-\infty}^{+\infty} dx \ G_{0} f(x) &= f(x') \int_{+\infty}^{-\infty} \dfrac{dz}{\sqrt{2 \pi}} \ e^{- \frac{z^2}{2}} = \notag \\
&= f(x') \int_{-\infty}^{+\infty} \dfrac{dz}{\sqrt{2 \pi}} \ e^{- \frac{z^2}{2}} = \notag \\
&= f(x') ,
\end{align}
and finally we get
\begin{equation}
\lim_{t' \rightarrow t^{+}} G_{0}(x',t',x,t) = \delta(x'-x) ,
\end{equation}
where the limit is understood in the sense of distributions.
\end{proof}
\item
Under the assumption $\delta < |t'-t| \ll 1$, we show that the support of $G_0$, i.e., the spacetime region where $G_0$ is significantly nonzero, is given by
\begin{align}
|x'-x| &\simeq \sqrt{\dfrac{\hslash |t'-t|}{m}} \simeq \notag \\
&\simeq (\sqrt{|t'-t|}).
\end{align}
\begin{proof}
Without loss of generality, we assume $t'>t$. Consider
\begin{equation}
\int_{-\infty}^{+\infty} dx \ G_{0} f(x),
\end{equation}
where $f(x)$ is a rapidly decreasing function. Neglecting $\delta$, which by assumption is several orders of magnitude less than $t'-t$, we have
\begin{equation}
\int_{-\infty}^{+\infty} dx \sqrt{\dfrac{m}{2 \pi i \hslash (t'-t)}} \ e^{\frac{i m (x'-x)^{2}}{2 \hslash (t'-t)}} f(x),
\end{equation}
we set
\begin{equation}
z^{2} = \dfrac{m (x'-x)^{2}}{\hslash (t'-t)} ,
\end{equation}
then
\begin{equation}
|x' - x| = \sqrt{\dfrac{\hslash (t'-t)}{m}} z,
\end{equation}
and for $x > x'$, with
\begin{equation}
x = x' + \sqrt{\dfrac{\hslash (t'-t)}{m}} z,
\end{equation}
we get
\begin{equation}
\dfrac{1}{2 \pi i} \int_{-\infty}^{+\infty} dz \ e^{- \frac{i z^{2}}{2}} f\left( x' + \sqrt{\dfrac{\hslash (t'-t)}{m}} z \right).
\end{equation}
Note that for $|z| \gg 1$, the positive and negative contributions of the integral tend to elide due to the oscillating factor $e^{- \frac{i z^{2}}{2}}$, so the terms significantly different from zero belong to the neighborhood of $|z|=1$, i.e.,
\begin{equation}
x \simeq x' \pm \sqrt{\dfrac{\hslash (t'-t)}{m}}
\end{equation}
\begin{equation}
|x-x'| \simeq \sqrt{\dfrac{\hslash (t'-t)}{m}}.
\end{equation}
\end{proof}
\end{itemize}
\section{Density matrix formalism}
A density matrix (or density operator) is a matrix that describes an ensemble of physical systems as quantum states, even if the ensemble contains only one system. The development of such a formalism is due to Landau, and its applications involve quantum mechanics to one particle as well as to many bodies. Density matrices are thus crucial tools in areas of quantum mechanics such as quantum statistical mechanics and open quantum systems.
\subsection{Density matrix}
Recall that the probability that an observable $G$ takes the value $\alpha_r$ at time $t$ in the state $\psi(t)$, and the expectation value $\langle G \rangle$ of the observable, are given respectively by
\begin{equation}
P(G = \alpha_{r} | t) = \langle \psi(t) | P_{r} | \psi(t) \rangle,
\end{equation}
\begin{equation}
\langle G \rangle (t) = \langle \psi(t) | G | \psi(t) \rangle,
\end{equation}
where $P_r$ is the projection operator onto the eigenspace corresponding to the eigenvalue $\alpha_r$. We now wish to rewrite the equations above using the trace operator. To this end, we define the density operator as
\begin{equation}
\rho(t) = |\psi(t)\rangle\langle\psi(t)|,
\label{eq: statopuro}
\end{equation}
which projects arbitrary states onto the direction of the state vector $\psi(t)$. Using the definition $\eqref{eq: statopuro}$, we have
\begin{theorem}[Probability and expectation value in terms of the trace]
Let $G$ be an observable with spectral decomposition $G = \sum_r \alpha_r P_r$, where $P_r$ is the projection operator onto the eigenspace associated with eigenvalue $\alpha_r$. Then, the probability of obtaining the outcome $\alpha_r$ at time $t$, and the expectation value of $G$ at time $t$, are given by
\begin{equation}
P(G=\alpha_{r}|t) = \Tr \left( \rho(t) P_{r} \right) ,
\label{eq: probabilitybymeansofthetrace}
\end{equation}
\begin{equation}
\langle G \rangle (t) = \text{Tr} \left( \rho(t) G \right).
\label{eq: expectationvaluebymeansofthetrace}
\end{equation}
\begin{proof}
Let $\{ |\varphi_i\rangle \}_i$ be an orthonormal basis of the Hilbert space. Then:
\begin{align}
\operatorname{Tr} \left[ \rho(t) P_r \right] &= \sum_i \langle \varphi_i | \rho(t) P_r | \varphi_i \rangle = \notag \\
&= \sum_i \langle \varphi_i | \psi(t) \rangle \langle \psi(t) | P_r | \varphi_i \rangle = \notag \\
&= \sum_i \langle \psi(t) | P_r | \varphi_i \rangle \langle \varphi_i | \psi(t) \rangle = \notag \\
&= \langle \psi(t) | P_r | \psi(t) \rangle,
\end{align}
which proves equation $\eqref{eq: probabilitybymeansofthetrace}$. Similarly,
\begin{align}
\operatorname{Tr} \left[ \rho(t) G \right] &= \sum_i \langle \varphi_i | \rho(t) G | \varphi_i \rangle = \notag \\
&= \sum_i \langle \varphi_i | \psi(t) \rangle \langle \psi(t) | G | \varphi_i \rangle = \notag \\
&= \sum_i \langle \psi(t) | G | \varphi_i \rangle \langle \varphi_i | \psi(t) \rangle = \notag \\
&= \langle \psi(t) | G | \psi(t) \rangle,
\end{align}
which proves equation $\eqref{eq: expectationvaluebymeansofthetrace}$.
\end{proof}
\end{theorem}
Now suppose that the initial state of the system is not known precisely, but is known to be one of the states $|\psi_1(0)\rangle, \, |\psi_2(0)\rangle, \, \ldots, \, |\psi_n(0)\rangle$, where each state is normalized, but the set is not necessarily orthonormal. Associated with these states is a set of probabilities $\{p_1, \ldots, p_n\}$, satisfying $\sum_{i=1}^{n} p_i = 1$. Assume that a measurement of the observable $G$ is performed on a system prepared in one of these states. Then, at time $t$, the corresponding evolved states are $|\psi_1(t)\rangle, \, |\psi_2(t)\rangle, \, \ldots, \, |\psi_n(t)\rangle$. The probability of obtaining a given eigenvalue $\alpha_r$ of $G$, averaged over the ensemble of possible initial states, is given by
\begin{equation}
P_{i}(G=\alpha_{r}|t) = \langle \psi_{i}(t) | P^{r}_{i} | \psi(t) \rangle.
\end{equation}
However, if the state of the system is not precisely known, then the probability of observing a given eigenvalue of the observable $\hat{G}$ must be computed as a weighted average over all possible states. That is,
\begin{equation}
P(G = \alpha_{r} | t) = \sum_{i=1}^{n} p_{i} \langle \psi_{i}(t) | P^{r}_{i} | \psi(t) \rangle,
\end{equation}
and similarly, the expectation value will be provided by
\begin{equation}
\langle G \rangle (t) = \sum_{i=1}^{n} p_{i} \langle \psi_{i}(t) | G | \psi(t) \rangle.
\end{equation}
From the above considerations, we define the density matrix for mixed states, i.e., the general density matrix, as
\begin{equation}
\rho(t) = \sum_{i} p_{i} |\psi(t)\rangle \langle \psi(t)|.
\label{eq: miscelastatistica}
\end{equation}
This expression reduces to the projector defined in equation $\eqref{eq: statopuro}$ in the special case where the system is in a well-defined quantum state. In this case, the density operator describes a pure state, which encodes the maximal possible information about the system. On the other hand, when the density operator takes the form $\eqref{eq: miscelastatistica}$, with more than one nonzero term in the sum, the state is referred to as a mixed state. Here, $\rho(t)$ represents a statistical mixture of the states $\{|\psi_i(t)\rangle\}$, which are not necessarily orthonormal. In this case, the information about the system is incomplete, reflecting a lack of precise knowledge about its actual state.
\begin{theorem}
From the linearity of the trace and density matrix definition, $\eqref{eq: probabilitybymeansofthetrace}$ and $\eqref{eq: expectationvaluebymeansofthetrace}$ can also be applied to mixed states.
\end{theorem}
\begin{theorem} The operator $\eqref{eq: miscelastatistica}$ satisfies:
\begin{itemize}
\item[1)] it is hermitian;
\item[2)] $\Tr(\rho(t))=1, \ \forall \ t$;
\item[3)] it is positive definite;
\item[4)] a convex sum of density matrices is a density matrix. 
\end{itemize}
\begin{proof}
\begin{itemize}
\item[1)] The coefficients $p_{i}$ are real and the matrix $|\psi_{i}(t)\rangle \langle \psi_{i}(t)|$ is hermitian, then the density matrix is hermitian.
\item[2)] We have
\begin{align}
\Tr(\rho(t)) &= \sum_{j=1}^{+\infty} \sum_{i=1}^{n} p_{i} \langle \varphi_{j}|\psi_{i}(t)\rangle \langle \psi_{i}(t)|\varphi_{j} \rangle = \notag \\
&= \sum_{j=1}^{+\infty} \sum_{i=1}^{n} p_{i} \langle \psi_{i}(t)|\varphi_{j} \rangle \langle \varphi_{j}|\psi_{i}(t) \rangle = \notag \\
&= \sum_{i=1}^{n} p_{i} \langle \psi_{i}(t)|\psi_{i}(t)\rangle = \notag \\
&= \sum_{i=1}^{n} p_{i}.
\end{align}
\item[3)] We have
\begin{align}
\langle \varphi|\rho(t)|\varphi \rangle &= \left\langle \varphi \left| \sum_{i=1}^{n} p_{i} |\psi_{i}(t)\rangle \langle \psi_{i}(t)| \right| \varphi \right\rangle = \notag \\
&= \sum_{i=1}^{n} p_{i} |\langle \varphi | \psi_{i}(t) \rangle|^{2} \geq 0, \quad \forall \ \varphi \in \mathcal{H}.
\end{align}
\item[4)] For the sake of simplicity, let us consider case of two density matrices, $\rho_{1}$ and $\rho_{2}$, and let be
\begin{equation}
\rho = \lambda \rho_{1} + (1-\lambda) \rho_{2}, \ \lambda \in [0,1]
\end{equation}
the convex sum. Since $\lambda$ is real and $\rho_{1}$, $\rho_{2}$ are hermitians, then $\rho$ is hermitian. Since $\lambda \in [0,1]$ and $\rho_{1}$, $\rho_{2}$ are positive definite, then $\rho$ is positive definite. Finally
\begin{align}
\Tr(\rho) &= \lambda \Tr(\rho_{1}) + (1-\lambda) \Tr(\rho_{2}) = \notag \\
&= \lambda + 1 - \lambda = \notag \\
&= 1.
\end{align}
\end{itemize}
\end{proof}
\end{theorem}
\subsection{Entropy}
A quantum system described by a mixed state exhibits greater uncertainty, or disorder, compared to one in a pure state. To quantify this degree of disorder, we define the quantity
\begin{equation}
S = - k_{B} \Tr \left\lbrace \rho \ln{\rho} \right\rbrace,
\label{eq: entropy}
\end{equation}
which is called the entropy.  
\begin{theorem}
The entropy $\eqref{eq: entropy}$ satisfies:
\begin{itemize}
\item[1)] it is positive definite;
\item[2)] it is minimum for a pure state and maximum for a mixed state with equal coefficients. 
\end{itemize}
\begin{proof}
\begin{itemize}
\item[1)] Given a basis of eigenvectors of $\rho(t)$, then
\begin{align}
S &= - k_{B} \Tr \left\lbrace \rho \ln{\rho} \right\rbrace = \notag \\
&= - k_{B} \sum_{i=1}^{N} p_{i} \ln{p_{i}} \geq 0.
\end{align}
\item[2)] For a pure state we have
\begin{align}
S &= - k_{B} \Tr \left\lbrace \rho \ln{\rho} \right\rbrace = \notag \\
&= - k_{B} \ln{1} = \notag \\
&= 0.
\end{align}
For mixed states, since entropy depends on probabilities $p_{i}$, maximizing $S$ implies solving constrained optimization problem, i.e.,
\begin{equation}
\max \left\lbrace - k_{B} \sum_{i=1}^{n} p_{i} \ln{p_{i}} \right\rbrace,
\end{equation}
\begin{equation}
\sum_{i=1}^{n} p_{i} = 1.
\end{equation}
We apply the method of Lagrange multipliers and consider the function
\begin{align}
f(p_{i}) &= S - \lambda \sum_{i} p_{i} = \notag \\
&= - \sum_{i} \left( k_{B} p_{i} \ln{p_{i}} + \lambda p_{i} \right) ,
\end{align}
where $\lambda$ is a Lagrange multiplier. It must be
\begin{align}
\dfrac{\partial f}{\partial p_{j}} &= - \left[ k_{B} \left( \ln{p_{j}} + 1 \right) + \lambda \right] = \notag \\
&= 0 ,
\end{align}
which implies
\begin{align}
p_{j} &= \const = \notag \\
&= \dfrac{1}{N}, \ \forall \ j ,
\end{align}
that is,
\begin{equation}
\rho = \dfrac{1}{N} \mathds{1},
\end{equation}
and finally it has shown
\begin{align}
S &= - k_{B} \Tr \left\lbrace \rho \ln{\rho} \right\rbrace = \notag \\
&= - k_{B} \sum_{i=1}^{N} \dfrac{1}{N} \ln{\left(\dfrac{1}{N}\right)} = \notag \\
&= k_{B} \sum_{i=1}^{N} \dfrac{1}{N} \ln{N} = \notag \\
&= k_{B} \ln{N}.
\end{align}
\end{itemize}
\end{proof}
\end{theorem}
\subsection{Maxwell-Boltzmann distribution}
Consider a quantum system described within the framework of the canonical ensemble, i.e., in thermodynamic equilibrium at temperature $T$ and characterized by an average energy $\langle \mathcal{E} \rangle$, then
\begin{theorem}
Maxwell-Boltzmann distribution
\begin{equation}
\rho = \dfrac{e^{-\beta \mathcal{H}}}{Z}
\end{equation}
makes entropy maximum and therefore from the postulate of maximum entropy it is configured as the density matrix of the system in a canonical ensemble.
\begin{proof}
Since the system is at thermodynamic equilibrium, we look for a time-independent density matrix operator, i.e.,
\begin{equation}
\dfrac{\partial \rho}{\partial t} = 0,
\end{equation}
from Liouville's theorem in quantum mechanics it then follows
\begin{equation}
\left[ \mathcal{H} , \rho \right] = 0,
\end{equation}
and thus Hamiltonian and density matrix share a common eigenbasis. Given
\begin{equation}
\mathcal{H} = \sum_{ru} \mathcal{E}_{r} |\varphi_{ru}\rangle \langle \varphi_{ru}|,
\end{equation}
where $u=1,...g_r$ labels the degeneracy of the eigenvalue $\mathcal{E}_r$, and we set $g_{r} = \dim(H_{r})$. Since the density matrix commutes with the Hamiltonian, then
\begin{equation}
\rho = \sum_{ru} \ p_{r} |\varphi_{ru}\rangle \langle \varphi_{ru}|.
\end{equation}
We are interested in determining the coefficients $p_{r}$ that make the entropy maximum: this is a constrained optimization problem with constraints
\begin{align}
\sum_{ru} p_{r} &= \sum_{r} g_{r} p_{r} = \notag \\
&= 1,
\end{align}
\begin{align}
\sum_{ru} p_{r} \mathcal{E}_{r} &= \sum_{r} g_{r} p_{r} \mathcal{E}_{r} = \notag \\
&= \langle \mathcal{E} \rangle.
\end{align}
We use the method of Lagrange multipliers and we define
\begin{equation}
h = S - \lambda \sum_{r} g_{r} p_{r} - \mu \sum_{r} g_{r} p_{r} \mathcal{E}_{r} ,
\end{equation}
then
\begin{align}
\dfrac{\partial h}{\partial p_{j}} &= - \left[ k_{B} \left( \ln{p_{j}} + 1 \right) + \lambda + \mu \mathcal{E}_{j} \right] g_{j} = \notag \\
&= 0,
\end{align}
\begin{equation}
\ln{p_{j}} = - k_{1} \mathcal{E}_{j} + k_{2},
\end{equation}
where $k_{1}$ and $k_{2}$ are constant. Finally we get
\begin{equation}
p_{j} = k_3 e^{- \beta \mathcal{E}_{j}} ,
\end{equation}
which, after normalization, is equal to the Maxwell-Boltzmann distribution, with $\beta = \frac{1}{k_{B} T}$.
\end{proof}
\end{theorem}
\subsection{Thermal average}
In many-body physics, systems whose interactions cannot be fully incorporated into a Hamiltonian or whose equations of motion do not admit exact solutions are said to be described by incomplete information. In such cases, the system is no longer characterized by a single pure state $|\psi(t)\rangle$, but rather by a statistical mixture of normalized states $\left\{ |\psi_i\rangle \right\}_{i=1,\ldots,N}$, each occurring with probability $p_i$. The corresponding density operator (or density matrix) for a pure state and for a statistical mixture are given in equations $\eqref{eq: statopuro}$ and $\eqref{eq: miscelastatistica}$, respectively. According to statistical mechanics, the density operator $\rho$ describing a quantum system in thermal equilibrium at temperature $T$ with a heat bath is defined as
\begin{equation}
\rho = \frac{e^{-\beta \mathcal{H}}}{Z},
\end{equation}
where the partition function $Z$ is given by
\begin{equation}
Z = \Tr \left(e^{-\beta \mathcal{H}} \right).
\end{equation}
After diagonalizing the Hamiltonian, the eigenvalues $\{ \mathcal{E}_r \}$ define the energy spectrum of the system, and the probability of occupying the state with energy $\mathcal{E}_r$ is
\begin{equation}
p_r = \frac{e^{-\beta \mathcal{E}_r}}{Z}.
\end{equation}
The expectation value of a generic observable $G$ at time $t$ is given by equation $\eqref{eq: expectationvaluebymeansofthetrace}$. At thermal equilibrium, this expression takes the form
\begin{equation}
\langle G \rangle = \Tr \left(\rho G \right), \quad \rho = \frac{e^{-\beta \mathcal{H}}}{Z},
\label{eq: mediatermica}
\end{equation}
where $\rho$ is the density operator for the many-body system. Equation $\eqref{eq: mediatermica}$ defines the thermal average, which holds for systems at thermodynamic equilibrium.
\chapter{Second quantization for non-relativistic identical particles}
In this chapter, we introduce the formalism of second quantization, a fundamental tool for the clear and efficient description of quantum many-body systems. Unlike the traditional quantum mechanical approach, which treats particles as individual entities, second quantization adopts a perspective centered on occupation states: one works in Fock space, where the particle number is not fixed, and quantum states are constructed using creation and annihilation operators. \newline
By distinguishing between fermions and bosons, according to the statistics they obey, we will describe the structure of Fock space for each case, highlighting the fundamental algebraic properties of the relevant operators. In particular, we will examine the commutation and anticommutation relations that characterize the behavior of bosons and fermions, respectively. This formalism is especially well-suited for dealing with systems containing a very large number of particles, as is often the case in condensed matter problems or field theories. \newline
A crucial aspect will be the systematic rewriting of observable operators in the language of second quantization. Specifically, we will show how one-body operators (such as kinetic energy or external potentials), two-body operators (such as particle interactions), and more generally N-body operators, can be expressed in terms of creation and annihilation operators. This compact and powerful formulation allows for a unified treatment of both free and interacting terms, and it forms the basis for developing perturbative theories and advanced numerical approaches. \newline
A central role will be played by the particle density operator, both in real space and in momentum space. This object provides a direct connection to physically observable quantities, such as density fluctuations and spatial correlations. Importantly, the particle density operator can be expressed elegantly in terms of the field operators, which create and annihilate particles at specific points in space. This formalism not only simplifies the mathematical description but also provides a natural language to incorporate interactions and to analyze quantum many-body effects systematically. In this framework, interactions often appear as density-dependent terms, and the study of density correlations becomes a powerful method to uncover the collective properties of the system. \newline
Then, we will discuss in detail the jellium model, which will serve throughout this book as an ideal testing ground for applying many-body techniques within the second quantization framework. The jellium model describes a system of interacting electrons embedded in a uniform, positively charged background that ensures overall charge neutrality. Despite its apparent simplicity, this model captures the essential physics of electron-electron interactions, making it a powerful tool for exploring the microscopic origin of collective behavior in many-body systems, providing a clean and conceptually transparent setting in which to test approximations and analytical techniques, including perturbation theory, Hartree-Fock method, diagrammatic methods and linear response theory. Within the second quantization formalism, the jellium model reveals its full utility. Its translation-invariant structure allows for a natural formulation in momentum space, where creation and annihilation operators facilitate an elegant and compact representation of the kinetic and interaction terms. Moreover, the model lends itself well to the study of fundamental quantities such as the density operator and response functions, all of which can be expressed and analyzed efficiently using second quantized operators. For these reasons, the jellium model appears repeatedly throughout this book, not only as a pedagogical example but also as a practical tool for developing intuition and verifying theoretical predictions. 
\section{Bose-Einstein and Fermi-Dirac statistics}
In quantum statistical mechanics, particles are classified into two fundamental categories based on their statistical behavior: fermions and bosons. Fermions obey the Pauli exclusion principle, which implies that each quantum state can be occupied by at most one particle. In contrast, bosons do not follow this principle, and multiple particles can occupy the same quantum state simultaneously. These intrinsic properties are reflected in the different statistical distributions that govern the average occupation number of a quantum state with energy $\mathcal{E}$, at temperature $T$, and chemical potential $\mu$. For fermions (e.g., electrons, protons, neutrons), the average occupation number is given by
\begin{equation}
n_{-1}(\mathcal{E}) = \frac{1}{e^{\beta(\mathcal{E} - \mu)} + 1},
\end{equation}
where $\beta = \frac{1}{k_B T}$. The $+$ sign in the denominator reflects the fermionic nature and the Pauli exclusion principle. For bosons (e.g., photons, phonons, bosonic atoms, ecc.), the average occupation number becomes
\begin{equation}
n_{+1}(\mathcal{E}) = \frac{1}{e^{\beta(\mathcal{E} - \mu)} - 1}.
\end{equation}
The $-$ sign indicates bosonic statistics, allowing multiple particles to occupy the same state, leading to collective phenomena such as Bose-Einstein condensation. Both distributions can be compactly written as
\begin{equation}
n_{\varepsilon}(\mathcal{E}) = \frac{1}{e^{\beta(\mathcal{E} - \mu)} - \varepsilon},
\label{eq: distribuzioneBoseFermiunificata}
\end{equation}
where $\varepsilon$ is given by $\eqref{eq: epsilonfermionibosoni}$. This unified form is particularly useful in field-theoretic approaches, such as in the Matsubara formalism and in many-body perturbation theory, where it appears naturally in contour integrations and thermal averages. 
\section{System of non-relativistic identical particles of fixed number}
Given a system of $N$ non-relativistic identical particles, the related Hilbert space $H_{N}$ is given by the tensor product $N-1$ times of $H_{1}$ with itself, i.e.,
\begin{equation}
H_{N} = H_{1} \otimes H_{1} \otimes \ldots \otimes \, H_{1}.
\end{equation}
Fixed $s_1,\ldots,s_N$ for the spins, the functions of $H_{N}$ are elements of $\mathcal{L}^2\left(V^{N}\right)$, so the scalar product for many particles is automatically generalized as
\begin{align}
\left( \psi,\phi \right)_{H_{N}} &= \int dx_{1} \int dx_{2} \ldots \int dx_{N} \psi^{*}(x_{1},\ldots,x_{N}) \phi(x_{1},\ldots,x_{N}) \equiv \notag \\
&\equiv \sum_{s_1=-S}^{S} \ldots \sum_{s_N=-S}^{S} \int_{V} d^{3}\textbf{r}_{1} \ldots \int_{V} d^{3}\textbf{r}_{N} \psi^{*}(\textbf{r}_{1},s_1,\ldots,\textbf{r}_{N},s_N) \phi(\textbf{r}_{1},s_1,\ldots,\textbf{r}_{N},s_N).
\end{align}
A basis in $H_{N}$ is given by the tensor product of the single particle bases, i.e.,
\begin{equation}
\lbrace \varphi_{\alpha_{1}}(x_{1}) \varphi_{\alpha_{2}}(x_{2}) \ldots \varphi_{\alpha_{N}}(x_{N}) \rbrace_{\alpha_{1},\ldots,\alpha_{N}} ,
\end{equation}
which consists of all possible permutations of the set of indices $\lbrace \alpha_{1},\ldots,\alpha_{N} \rbrace$ for fixed $x_1,\ldots,x_N$. Because of the principle of indistinguishability of identical particles, physical states whose wave functions differ by a permutation of the variables $x_{1},\ldots,x_{N}$, that is, by a different assignment of the same set of individual degrees of freedom to the $N$ particles, coincide. However, not all functions in $H_N$ can describe systems of identical particles: one must select those that have a definite characterization under the action of permutations. A permutation $\sigma$ is an application between a finite subset of the natural numbers in itself such that $\sigma$ takes the elements of the set $\left\lbrace 1,\ldots,N \right\rbrace$ and maps them into itself with an ordering $\left\lbrace \sigma(1),\ldots,\sigma(N) \right\rbrace$. The action of a permutation is generally made explicit by a matrix notation as follows
\begin{equation}
\sigma(1,\ldots,N) =
\begin{pmatrix}
1  & \ldots & N \\ 
\sigma(1) & \ldots & \sigma(N)
\end{pmatrix}.
\end{equation}
If the permutation involves only two elements, say of positions $i$ and $j$, it is called transposition and is denoted by
\begin{equation}
\sigma_{ij} =
\begin{pmatrix}
1 & \ldots & i & \ldots & j & \ldots & N \\ 
1 & \ldots & j & \ldots & i & \ldots & N
\end{pmatrix}.
\end{equation}
The set of permutations, together with the composition operation "$\cdot$'' defined by $\sigma \cdot \tau \equiv \sigma \tau$, forms a group (see Chapter \ref{Foundations of group theory}). In this group:
\begin{itemize}
\item the inverse element of a permutation $\sigma$, denoted by $\sigma^{-1}$, is the permutation obtained by applying the transpositions that compose $\sigma$ in reverse order;
\item the identity element is the permutation that leaves all elements unchanged, and is given by
\begin{equation}
\sigma_{\text{Id}} =
\begin{pmatrix}
1 & 2 & \ldots & i & \ldots &j& \ldots & N-1 & N \\ 
1 & 2 & \ldots & i & \ldots &j& \ldots & N-1 & N
\end{pmatrix}.
\end{equation}
\end{itemize}
By definition, every permutation $\sigma$ is a composition of transpositions: we distinguish between odd and even permutations, realized by an odd and even number of permutations, respectively, so we define 
\begin{equation}
\sgn(\sigma) = 
\begin{cases}
+1, \ \sigma \ \text{even} \\ 
-1, \ \sigma \ \text{odd}
\end{cases}.
\end{equation}
If $\sigma$ is even (odd) then $\sigma^{-1}$ is also even (odd), since the numbers of transpositions composing them are equal. Each permutation $\sigma$ can be related to an operator $P_{\sigma}$ in the space $H_N$ and whose action is defined as
\begin{equation}
P_{\sigma} \Psi(x_{1}, \ldots, x_{N}) = \Psi(x_{\sigma(1)}, \ldots, x_{\sigma(N)}).
\end{equation}
If the wave function does not describe a system of identical particles, in general the new wave function does not have a definite relation with respect to the old wave function. The permutation operator forms a group, that is, the permutation operator group, with the inverse element 
\begin{equation}
P^{-1}_{\sigma} = P_{\sigma^{-1}} ,
\end{equation}
and with the identity element
\begin{equation}
P_{\mathrm{Id}} = \mathds{1}.
\end{equation}
Consequently, the map
\begin{equation}
\sigma \longrightarrow P_{\sigma}
\end{equation}
induces homomorphism between groups, i.e.,
\begin{align}
P_{\sigma_{1} \sigma_{2}} \Psi(x_{1}, \ldots, x_{N}) &= \Psi(x_{(\sigma_{1} \sigma_{2}) (1)}, \ldots, x_{(\sigma_{1} \sigma_{2})(N)}) = \notag \\
&= P_{\sigma_{1}} \Psi(x_{\sigma_{2}(1)}, \ldots, x_{\sigma_{2}(N)}) = \notag \\
&= P_{\sigma_{1}} P_{\sigma_{2}} \Psi(x_{1}, \ldots, x_{N}) ,
\end{align}
that is,
\begin{equation}
P_{\sigma_{1} \sigma_{2}} = P_{\sigma_{1}} P_{\sigma_{2}}.
\end{equation}
The operator of a transposition of the individual degrees of freedom $x_{i}$, $x_{j}$ will be denoted by $P_{ij}$. Permutation operators satisfy the properties:
\begin{itemize}
\item[1.] $P_\sigma^\dag = P^{-1}_{\sigma}$: $P_{\sigma}$ is unitary.
\item[2.] Transposition operators are self-adjoint and have eigenvalues $+1$ and $-1$.
\end{itemize}
\begin{proof}
\begin{itemize}
\item[1.] Given
\begin{equation}
\left( \psi , P_{\sigma} \phi \right) = \int dx_{1} \ldots \int dx_{N} \psi^{*}(x_{1}, \ldots, x_{N}) \phi(x_{\sigma(1)},\ldots,x_{\sigma(N)}) ,
\end{equation}
the map between two sequences induced by a permutation $\sigma$, i.e.,
\begin{equation}
y_i = x_{\sigma(i)} \quad \Leftrightarrow \quad x_i = y_{\sigma^{-1}(i)}
\end{equation}
leaves the measurement element unchanged, that is,
\begin{equation}
dy_{1} dy_{2} \ldots dy_{N} = dx_{\sigma(1)} dx_{\sigma(2)} \ldots dx_{\sigma(N)}, \ \forall \ \sigma ,
\end{equation}
then
\begin{align}
\left( \psi , P_{\sigma} \phi \right) &= \int dy_{1} \ldots \int dy_{N} \psi^{*}(y_{\sigma^{-1}(1)}, \ldots, y_{\sigma^{-1}(N)}) \phi(y_{1},\ldots,y_{N}) = \notag \\
&= \int dy_{1} \ldots \int dy_{N} P_{\sigma^{-1}} \psi^{*}(y_{1}, \ldots, y_{N}) \phi(y_{1},\ldots,y_{N}) = \notag \\
&= \left( P_{\sigma^{-1}} \psi , \phi \right).
\end{align}
\item[2.] The transposition operators satisfy
\begin{align}
P_{ij}^{\dagger} &= P_{ji} = \notag \\
&= P_{ij},
\end{align}
where in the first equality we applied the definition of adjoint, while in the second we used the property that a transposition is symmetric with respect to degrees of freedom, that is, the exchange of $x_{i}$ with $x_{j}$ coincides with the exchange of $x_{j}$ with $x_{i}$. Then, 
\begin{align}
P_{ij}^{\dagger} P_{ij} &= P_{ij} P_{ij}^{\dagger} = \notag \\
&= \mathds{1} .
\end{align}
From $P_{ij}^{\dagger}=P_{ij}$ and $P_{ij}^{\dagger}P_{ij}=\mathds{1}$ it follows that the transposition operator is self-adjoint and unitary, and therefore diagonalizable with real eigenvalues. Moreover, since
\begin{equation}
P_{ij}^{2}=\mathds{1},
\end{equation}
any eigenvalue $\lambda$ of $P_{ij}$ must satisfy $\lambda^{2}=1$. Hence, the only possible eigenvalues of the transposition operator are
\begin{equation}
\lambda = \pm 1.
\end{equation}
\end{itemize}
\end{proof}
According to the principle of indistinguishability, an operator can describe an observable of a system of identical particles only if it is invariant under the action of particle exchange, that is, such that
\begin{equation}
\left [ A ,P_{\sigma} \right] = 0, \ \forall \ \sigma.
\end{equation}
In addition, the states $\psi(x_1,\ldots,x_N)$ describing the system of identical particles must have square moduli invariant under permutation 
\begin{equation}
|\psi|^2 = \left|P_{\sigma} \psi \right|^2.
\end{equation}
In particular, under the action of a transposition $P_{ij}$, there may be at most a difference in a phase factor, i.e.,
\begin{equation}
\psi(x_1,\ldots,x_j,\ldots,x_i,\ldots,x_N) = e^{i\theta} \psi(x_1,\ldots,x_i,\ldots,x_j,\ldots,x_N) .
\end{equation}
Let us apply $P_{ij}$ to both sides, so that
\begin{equation}
P_{ij} \psi(x_1,\ldots,x_j,\ldots,x_i,\ldots,x_N) = P_{ij} e^{i\theta} \psi(x_1,\ldots,x_i,\ldots,x_j,\ldots,x_N),
\end{equation}
\begin{equation}
P_{ij} \psi(x_1,\ldots,x_j,\ldots,x_i,\ldots,x_N) = e^{i\theta} P_{ij} \psi(x_1,\ldots,x_i,\ldots,x_j,\ldots,x_N),
\end{equation}
\begin{equation}
\psi(x_1,\ldots,x_i,\ldots,x_j,\ldots,x_N) = e^{i\theta} \psi(x_1,\ldots,x_j,\ldots,x_i,\ldots,x_N),
\end{equation}
\begin{equation}
\psi(x_1,\ldots,x_{i},\ldots,x_{j},\ldots,x_{N}) = e^{2i\theta} \psi(x_1,\ldots,x_{i},\ldots,x_{j},\ldots,x_{N}) ,
\end{equation}
\begin{equation}
\theta = 0, 
\end{equation}
\begin{equation}
\theta = \pi.
\end{equation}
Only the wave functions in the Hilbert space $H_N$ describing systems of identical particles exhibit a definite behavior under the action of a transposition: they are mapped either to themselves or to their negative. Accordingly, we distinguish
\begin{itemize}
\item totally symmetric functions are such that $\theta = 0$ for any $sigma$, they are therefore invariant for any permutation, i.e.,
\begin{equation}
P_\sigma \psi(x_1,\ldots,x_N) = \psi(x_1,\ldots,x_N), \ \forall \ \sigma;  
\end{equation}
\item totally antisymmetric functions are such that $\theta = 0$ with even $\sigma$ and $\theta = \pi$ with odd $\sigma$, i.e.,
\begin{equation}
P_{\sigma} \psi(x_{1},\ldots,x_{N}) = 
\begin{cases}
+ \psi(x_{1},\ldots,x_{N}), \ \sigma \ \text{even} \\
- \psi(x_{1},\ldots,x_{N}), \ \sigma \ \text{odd}
\end{cases}.
\end{equation}
\end{itemize}
Let us define two subspaces of $H_{N}$: the space $H_{N}^{A}$ of antisymmetric functions and the space $H_{N}^{S}$ of symmetric functions. We then define the antisymmetrization and symmetrization operators as follows
\begin{equation}
P^{(A,N)} = \dfrac{1}{N!} \sum_\sigma \sgn(\sigma) P_{\sigma},
\label{eq: operatoreantisimmetrizzazione}
\end{equation}
\begin{equation}
P^{(S,N)} = \dfrac{1}{N!} \sum_\sigma P_{\sigma},
\label{eq: operatoresimmetrizzazione}
\end{equation}
respectively. The factor $N!$ is used to normalize the action of the operators, being the number of total permutations. We have
\begin{theorem}
The antisymmetrization and symmetrization operators, i.e., $\eqref{eq: operatoreantisimmetrizzazione}$ and $\eqref{eq: operatoresimmetrizzazione}$ respectively, are projectors and in particular they project generic functions of $H_{N}$ into two orthogonal eigenspaces, i.e., $H_{N}^{A}$ and $H_{N}^{S}$, with eigenvalues $-1$ and $+1$, respectively.
\begin{proof}
First, we must verify that they are self-adjoint and idempotent. Consider the antisymmetrization operator and apply to it the adjoint operation, i.e.,
\begin{align}
\left[ P^{(A,N)}\right]^{\dagger} &= \dfrac{1}{N!} \sum_\sigma \sgn(\sigma)  P_{\sigma}^{\dagger} = \notag \\
&= \dfrac{1}{N!} \sum_\sigma \sgn(\sigma) P_{\sigma^{-1}} = \notag \\
&= \dfrac{1}{N!} \sum_{\sigma} \sgn(\sigma^{-1}) P_{\sigma^{-1}} = \notag \\
&= \dfrac{1}{N!} \sum_{\sigma^{-1}} \sgn(\sigma^{-1}) P_{\sigma^{-1}} = \notag \\
&= P^{(A,N)},
\end{align}
where we made use of $\sgn(\sigma^{-1})= \sgn(\sigma)$ and the fact that we are summing over all possible permutations, so the summation index is a dummy index. It remains to verify the idempotency of $P^{(A,N)}$: let us apply to it a permutation operator 
\begin{align}
P_{\sigma'} P^{(A,N)} &= \dfrac{1}{N!} \sum_{\sigma} \sgn(\sigma) P_{\sigma'} P_{\sigma} = \notag \\
&= \dfrac{1}{N!} \sum_{\sigma} \sgn^2(\sigma') \sgn(\sigma) P_{\sigma'} P_{\sigma} = \notag \\
&= \dfrac{1}{N!} \sgn(\sigma') \sum_{\sigma} \sgn(\sigma') \sgn(\sigma) P_{\sigma'} P_{\sigma} = \notag \\
&= \dfrac{1}{N!} \sgn(\sigma') \sum_{\sigma} \sgn(\sigma' \sigma) P_{\sigma' \sigma} = \notag \\ 
&= \sgn(\sigma') P^{(A,N)},
\end{align} 
where we used the identity \(1 = \sgn^{2}(\sigma')\) and exploited the dummy summation index, with \(\sigma'\) fixed and \(\sigma\) running over all permutations. It follows that
\begin{align}
P^{(A,N)} P^{(A,N)} &= \dfrac{1}{N!} \sum_{\sigma} \sgn(\sigma) P_{\sigma} P^{(A,N)} = \notag \\
&= \dfrac{1}{N!} \sum_{\sigma} \sgn(\sigma) \sgn(\sigma) P^{(A,N)} = \notag \\
&= \dfrac{1}{N!} \sum_{\sigma} \sgn^{2}(\sigma) P^{(A,N)} = \notag \\
&= \dfrac{1}{N!} \left( \sum_{\sigma} 1 \right) P^{(A,N)} = \notag \\
&= \dfrac{1}{N!} N! P^{(A,N)} = \notag \\
&= P^{(A,N)} ,
\end{align}
so it is proved that $P^{(A,N)}$ is a projection operator. \newline
It is similarly shown that $P^{(S,N)}$ is a projection operator, with the difference that it satisfies the property
\begin{equation}
P_{\sigma'} P^{(S,N)} = P^{(S,N)}, \ \forall \ \sigma'.
\end{equation}
Let us verify that $P^{(S,N)}$ and $P^{(A,N)}$ project elements of $H_N$ into symmetric and antisymmetric spaces, respectively. From 
\begin{equation}
P^{(A,N)} \psi = \phi, \ \psi \in H_N,
\end{equation}
it follows
\begin{equation}
P_{\sigma} \phi = \sgn(\sigma) \phi = 
\begin{cases}
\ \ \phi \ , \ \sigma \ \text{even} \\
- \phi \ , \ \sigma \ \text{odd}
\end{cases} ,
\end{equation} 
\begin{equation}
P^{(A,N)} \psi \in H^{A}_{N}, \ \forall \ \psi \in H_{N} ,
\end{equation}
and from
\begin{equation}
P^{(S,N)} \psi' = \phi', \ \psi' \in H_N,
\end{equation}
it follows
\begin{equation}
P_{\sigma} \phi' = \phi', \ \forall \ \sigma ,
\end{equation}
\begin{equation}
P^{(S,N)} \psi' \in H^{S}_{N}, \ \forall \ \psi' \in H_{N}.
\end{equation}
The operators \(P^{(S,N)}\) and \(P^{(A,N)}\) project functions from \(H_N\) onto the physically relevant subspaces. We now show that the two subspaces \(H_{N}^{A}\) and \(H_{N}^{S}\) have an empty intersection. By applying a transposition operator, an odd permutation, we observe that
\begin{equation}
P_{ij} P^{(A,N)} \psi = - P^{(A,N)} \psi,
\end{equation}
\begin{equation}
P_{ij} P^{(S,N)} \psi' = P^{(S,N)} \psi',
\end{equation}
that is, the antisymmetric and symmetric wave functions are eigenvectors of the transposition operator, which is self-adjoint: the eigenvalues $-1$ and $+1$ correspond to orthogonal eigenspaces.
\end{proof}
\end{theorem}
From phenomenological evidence, the following classification is well established:
\begin{itemize}
\item particles whose quantum states lie in the antisymmetric subspace \(H_{N}^{A}\) exhibit half-integer spin and are identified as fermions;
\item particles whose quantum states lie in the symmetric subspace \(H_{N}^{S}\) exhibit integer spin and are identified as bosons.
\end{itemize}
A special case of a system of identical particles is that of two particles, in which case a state is written as the sum of a completely symmetrical and a completely antisymmetrical state, i.e.,
\begin{equation}
\psi(x_1,x_2) \equiv \dfrac{1}{2} \left[ \psi^S(x_1,x_2) + \psi^S(x_2,x_1) \right] + \dfrac{1}{2} \left[ \psi^A(x_1,x_2) - \psi^A(x_2,x_1) \right],
\end{equation}
then the space $H_2$ is equal to the soace given by the direct sum of the symmetric and antisymmetric spaces, i.e.,
\begin{equation}
H_2 = H^S_2 \oplus H^A_2.
\end{equation}
For $N>2$ the situation becomes more complicated: we have the direct sum of three spaces, symmetrical, antisymmetrical and an additional space, which includes the functions that do not have definite symmetry with respect to the group of permutations, but which has no physical relevance.
\section{Fermionic basis and bosonic basis}
Given a $N \times N$ matrix $A$, the determinant is defined by 
\begin{equation}
\det A = \sum_\sigma \sgn(\sigma) a_{1,\sigma(1)} \ldots a_{N,\sigma(N)},
\label{eq: determperrighe}
\end{equation}
that is, for each permutation $\sigma$ we have to take $N$ elements from the matrix $A$, one for each row, from different columns and multiply them, then add up the products obtained from all possible permutations. We observe that the $\eqref{eq: determperrighe}$ is written by rows, obviously, the same determinant is obtained through a definition that is a function of the columns
\begin{equation}
\det A = \sum_\tau \sgn(\tau) a_{\tau(1),1} \ldots a_{\tau(N),N},
\label{eq: determpercol}
\end{equation}
and as can be seen from the $\eqref{eq: determperrighe}$, or equivalently from the $\eqref{eq: determpercol}$, there is an analogy with the antisymmetrization operator. \newline 
The definition of permanent follows that of determinant, except that the signs of permutations are missing, i.e.,
\begin{equation}
\perm A = \sum_\sigma a_{1,\sigma(1)} \ldots a_{N,\sigma(N)}.
\label{eq: permanente}
\end{equation}
Starting from $H_{N} = H_{1} \otimes H_{2} \otimes \ldots \otimes H_{1}$ with basis $\lbrace \varphi_{\alpha_{1}}(x_{1}) \ldots \varphi_{\alpha_{N}}(x_{N}) \rbrace_{\alpha_{1},\ldots, \alpha_{N}}$, we determine a basis for the antisymmetric subspace $H^A_N$ and a basis for the symmetric subspace $H^S_N$.
\subsection{Fermionic basis}
Let us consider the action of the antisymmetrization operator to a basis of $H_{N}$, i.e.,
\begin{align}
\varphi^{(A,N)}_{\alpha_1,\ldots,\alpha_N}(x_1,\ldots,x_N) &\equiv P^{(A,N)} \left\lbrace \varphi_{\alpha_1}(x_1) \ldots \varphi_{\alpha_N}(x_N) \right\rbrace = \notag \\
&= \dfrac{1}{N!} \sum_{\sigma} \sgn(\sigma) P_{\sigma} \left\lbrace \varphi_{\alpha_1}(x_{1}) \ldots \varphi_{\alpha_N}(x_{N}) \right\rbrace = \notag \\
&= \dfrac{1}{N!} \sum_{\sigma} \sgn(\sigma) \varphi_{\alpha_1}(x_{\sigma(1)}) \ldots \varphi_{\alpha_N}(x_{\sigma(N)}) ,
\end{align}
where the permutation acts on the index of the degree of freedom. If we consider the states $\lbrace \alpha_i \rbrace$ as row indices and the degrees of freedom $\lbrace x_j \rbrace$ as column indices, the function is obtained by multiplying $(N!)^{-1}$ by the determinant (see $\eqref{eq: determperrighe}$) of the matrix whose element is given by $\lbrace \varphi_{\alpha_{i}}(x_{j}) \rbrace$. However, if we swap two indices in the determinant formula, e.g. give $\alpha_1$ to the second particle and $\alpha_2$ to the first particle, the determinant changes at most by one sign, but it is the same state, whereas the basis must consist of states orthogonal to each other. Accordingly, given $\varphi^{(A,N)}_{\alpha_1,\ldots,\alpha_N}(x_1,\ldots,x_N)$, one must arrange the states without repetition and respecting the experimental evidence that the lowest-energy states are occupied first. One possible way is to arrange them in ascending order
\begin{equation}
\alpha_1 < \alpha_2 < \ldots < \alpha_N,
\end{equation}
where the chain of inequalities is to be interpreted as follows: the states are arranged in ascending order according to the quantum number \(n\); whenever two particles possess the same \(n\), they necessarily differ in their spin quantum number \(\sigma\). This reconstruction captures the essence of the fermionic phenomenology embodied in the Pauli exclusion principle. The object
\begin{equation}
\varphi^{(A,N)}_{\alpha_1,\ldots,\alpha_N}(x_1,\ldots,x_N) = 
\dfrac{1}{N!} \det
\begin{pmatrix}
\varphi_{\alpha_1}(x_{1}) & \ldots & \varphi_{\alpha_1}(x_{N}) \\
\vdots & \ddots & \vdots \\
\varphi_{\alpha_N}(x_{1}) & \ldots & \varphi_{\alpha_N}(x_{N}) \\
\end{pmatrix} 
\label{eq: basefermionicainiz}
\end{equation}
turns out to be an element of $H^A_N$. Let us check whether $\varphi^{(A,N)}_{\alpha_1,\ldots,\alpha_N}$ can be a basis in $H_N^A$ or whether we need to change its definition. Its completeness derives from the completeness of $\varphi_{\alpha_i}(x_1)$ in $H_1$: it is complete in $H_N$ the product of bases and the projection of the product in $H_N^A$ is also complete. Regarding orthonormality, we have
\begin{align}
(\varphi^{(A,N)}_{(\alpha_1',\ldots,\alpha_N')} , \varphi^{(A,N)}_{(\alpha_1,\ldots,\alpha_N)}) &= \left( P^{(A,N)}\varphi_{\alpha_1'}(x_1) \ldots \varphi_{\alpha_N'}(x_N),P^{(A,N)}\varphi_{\alpha_1}(x_1) \ldots \varphi_{\alpha_N}(x_N) \right) = \notag \\
&= \left(\varphi_{\alpha_1'}(x_1) \ldots \varphi_{\alpha_N'}(x_N),\left[\hat{P}^{(A,N)}\right]^{\dagger} P^{(A,N)}\varphi_{\alpha_1}(x_1) \ldots \varphi_{\alpha_N}(x_N)\right) = \notag \\
&= \left(\varphi_{\alpha_1'}(x_1) \ldots \varphi_{\alpha_N'}(x_N),P^{(A,N)} P^{(A,N)}\varphi_{\alpha_1}(x_1) \ldots \varphi_{\alpha_N}(x_N)\right) = \notag \\
&= \left(\varphi_{\alpha_1'}(x_1) \ldots \varphi_{\alpha_N'}(x_N), P^{(A,N)}\varphi_{\alpha_1}(x_1) \ldots \varphi_{\alpha_N}(x_N)\right)  ,
\end{align}
where we have used the properties of the projection operator $\left[P^{(A,N)}\right]^\dag = P^{(A,N)}$ and $\left[P^{(A,N)}\right]^2 = P^{(A,N)}$. From
\begin{equation}
\dfrac{1}{N!} \sum_\sigma \sgn(\sigma) \int dx_1 \ldots \int dx_N \varphi^*_{\alpha'_1}(x_1) \ldots \varphi^*_{\alpha'_{N}}(x_N) \varphi_{\alpha_{\sigma(1)}}(x_1) \ldots \varphi_{\alpha_{\sigma(N)}}(x_N) ,
\end{equation}
where the determinant is written explicitly so that the permutation acts on the row indices (see $\eqref{eq: determpercol}$), with the row and column indices identified with the subscripts of $\varphi$ and $x$, respectively. By setting $\alpha_{\sigma(i)}=\alpha_i$, we observe that the integral is the product of $N$ integrals of the form $\int dx \varphi^*_{\alpha'_i}(x) \varphi_{\alpha_j}(x) \equiv (\alpha'_i,\alpha_j)$, then the scalar product of the $\varphi^{(A,N)}_{(\alpha_1,\ldots,\alpha_N)}$ has been linked to 
\begin{equation}
\left( \varphi^{(A,N)}_{(\alpha_1',\ldots,\alpha_N')} , \varphi^{(A,N)}_{(\alpha_1,\ldots,\alpha_N)} \right) = \dfrac{1}{N!} \det
\begin{pmatrix}
&(\alpha_1',\alpha_1) & \ldots & (\alpha'_1,\alpha_N) & \\
& \vdots & \ddots & \vdots \\
&(\alpha_N',\alpha_1)& \ldots & (\alpha'_N,\alpha_N) &
\end{pmatrix}.
\label{eq: prodottoscalarebasefermionicasbagliata}
\end{equation}
We now show that if the two sets of indices differ by at least one particle state, the determinant vanishes. Suppose, for instance, that only $\alpha_1 \neq \alpha'_1$. In this case, the first row and the first column of the matrix $\eqref{eq: prodottoscalarebasefermionicasbagliata}$ are entirely zero, which makes the determinant identically zero. Similarly, whenever any other pair of indices does not coincide, the corresponding rows and columns are both zero, so that the determinant vanishes. It follows that a nonzero determinant arises only when all one-particle states coincide pairwise. Consequently, the Slater determinants constructed from distinct sets of single-particle states are mutually orthogonal. \newline
Let us now examine the normalization and reconsider the matrix $\eqref{eq: prodottoscalarebasefermionicasbagliata}$, whose generic elements $(\alpha'_i,\alpha_j)$ represent scalar products in the one-particle Hilbert space $\mathcal{H}_1$. When the sets of indices coincide pairwise, the matrix reduces to the identity, its determinant equals $1$, and the corresponding scalar product evaluates to $(N!)^{-1}$. In order for the vector to be normalized, the definition of $\varphi^{(A,N)}_{\alpha_1,\ldots,\alpha_N}(x_1,\ldots,x_N)$ in equation $\eqref{eq: basefermionicainiz}$ must include a factor of $\sqrt{N!}$, i.e.,
\begin{equation}
\varphi^{(A,N)}_{\alpha_1,\ldots,\alpha_N}(x_1,\ldots,x_N) = \sqrt{N!} P^{(A,N)} \left\lbrace \varphi_{\alpha_1}(x_1) \ldots \varphi_{\alpha_N}(x_N) \right\rbrace.
\end{equation}
From
\begin{align}
\varphi^{(A,N)}_{\alpha_1,\ldots,\alpha_N}(x_1,\ldots,x_N) &= \sqrt{N!} \dfrac{1}{N!} \sum_{\sigma} \sgn(\sigma) P_{\sigma} \left\lbrace \varphi_{\alpha_1}(x_{1}) \ldots \varphi_{\alpha_N}(x_{N}) \right\rbrace = \notag \\
&= \dfrac{1}{\sqrt{N!}} \sum_{\sigma} \sgn(\sigma) \varphi_{\alpha_1}(x_{\sigma(1)}) \ldots \varphi_{\alpha_N}(x_{\sigma(N)}),
\end{align}
the scalar product becomes
\begin{align}
\left( \varphi^{(A,N)}_{(\alpha_1',\ldots,\alpha_N')} , \varphi^{(A,N)}_{(\alpha_1,\ldots,\alpha_N)} \right) &= \left( \sqrt{N!} P^{(A,N)}\varphi_{\alpha_1'}(x_1) \ldots \varphi_{\alpha_N'}(x_N), \sqrt{N!} P^{(A,N)}\varphi_{\alpha_1}(x_1) \ldots \varphi_{\alpha_N}(x_N) \right) = \notag \\
&= N! \left(\varphi_{\alpha_1'}(x_1) \ldots \varphi_{\alpha_N'}(x_N), P^{(A,N)}\varphi_{\alpha_1}(x_1) \ldots \varphi_{\alpha_N}(x_N)\right) = \notag \\
&= \det
\begin{pmatrix}
&(\alpha_1',\alpha_1) & \ldots & (\alpha'_1,\alpha_N) & \\
& \vdots & \ddots & \vdots \\
&(\alpha_N',\alpha_1) & \ldots & (\alpha'_N,\alpha_N) &
\end{pmatrix}.
\end{align}
Finally, the fermionic basis is given by
\begin{equation}
\varphi^{(A,N)}_{\alpha_1,\ldots,\alpha_N}(x_1,\ldots,x_N) = \dfrac{1}{\sqrt{N!}} \det
\begin{pmatrix}
\varphi_{\alpha_1}(x_{1}) & \ldots & \varphi_{\alpha_1}(x_{N}) \\
\vdots & \ddots & \vdots \\
\varphi_{\alpha_N}(x_{1}) & \ldots & \varphi_{\alpha_N}(x_{N}) \\
\end{pmatrix} ,
\label{eq: basefermionica}
\end{equation}
with $\alpha_1 < \alpha_2 < \ldots < \alpha_N$. 
\subsection{Bosonic basis}
Let us consider the action of the symmetrization operator on a basis of $H_{N}$, i.e.,
\begin{align}
\varphi^{(S,N)}_{\alpha_1,\ldots,\alpha_N}(x_1,\ldots,x_N) &\equiv P^{(S,N)} \left\lbrace \varphi_{\alpha_1}(x_1) \ldots \varphi_{\alpha_N}(x_N) \right\rbrace = \notag \\
&= \dfrac{1}{N!} \sum_{\sigma} P_{\sigma} \left\lbrace \varphi_{\alpha_1}(x_{1}) \ldots \varphi_{\alpha_N}(x_{N}) \right\rbrace = \notag \\
&= \dfrac{1}{N!} \sum_{\sigma} \varphi_{\alpha_1}(x_{\sigma(1)}) \ldots \varphi_{\alpha_N}(x_{\sigma(N)}) .
\end{align}
We note that, in the case of bosons, the symmetrized wavefunction remains non-zero even when two or more indices are equal, i.e., the permanent does not vanish. The indices are arranged according to the convention 
\begin{equation}
\alpha_1 \leq \alpha_2 \leq \ldots \leq \alpha_N ,
\end{equation}
which reproduces the experimental evidence that lower-energy states are occupied first. According to this convention, we define
\begin{equation}
\varphi^{(S,N)}_{\alpha_1,\ldots,\alpha_N}(x_1,\ldots,x_N) = 
\dfrac{1}{N!} \perm
\begin{pmatrix}
\varphi_{\alpha_1}(x_{1}) & \ldots & \varphi_{\alpha_1}(x_{N}) \\
\vdots & \ddots & \vdots \\
\varphi_{\alpha_N}(x_{1}) & \ldots & \varphi_{\alpha_N}(x_{N}) \\
\end{pmatrix} ,
\label{eq: basebosonicainiz}
\end{equation}
which implies that the function lies in the symmetric subspace \(H_N^S\). Let us now verify whether equation $\eqref{eq: basebosonicainiz}$ can serve as a basis for \(H_N^S\), or if its definition requires modification. Since the indices \(\alpha_i\) can be repeated, consider the case where a particular index \(\alpha_k\) appears \(N_{\alpha_k}\) times, with \(\sum_k N_{\alpha_k} = N\). In this case, instead of the matrix in equation $\eqref{eq: prodottoscalarebasefermionicasbagliata}$, we obtain a block-diagonal matrix composed of blocks of size \(N_{\alpha_k} \times N_{\alpha_k}\), each filled entirely with ones. Because the permanent of a block-diagonal matrix factorizes as the product of the permanents of its blocks, and since the permanent of each such block equals \(N_{\alpha_k}!\), the total permanent is given by $N_{\alpha_1}! \ldots N_{\alpha_M}!$, where $M$ denotes the number of distinct one-particle states. Therefore, each function $\varphi^{(S,N)}_{\alpha_1,\ldots,\alpha_N}(x_1,\ldots,x_N)$ must be multiplied by
\begin{equation}
\dfrac{1}{\sqrt{N_{\alpha_1}!} \cdots \sqrt{N_{\alpha_M}!}}
\end{equation}
to compensate for the overcounting due to repeated indices in the symmetrized sum (permanent). In addition, a factor of \(\sqrt{N!}\) is needed to cancel the prefactor \((N!)^{-1}\) appearing in equation $\eqref{eq: basebosonicainiz}$. Putting everything together, the properly normalized bosonic basis functions are given by
\begin{equation}
\varphi^{(S,N)}_{\alpha_1,\ldots,\alpha_N}(x_1,\ldots,x_N) = \dfrac{1}{\sqrt{N!} \sqrt{N_{\alpha_1}!} \ldots \sqrt{N_{\alpha_M}!}} \perm
\begin{pmatrix}
\varphi_{\alpha_1}(x_{1}) & \ldots & \varphi_{\alpha_1}(x_{N}) \\
\vdots & \ddots & \vdots \\
\varphi_{\alpha_N}(x_{1}) & \ldots & \varphi_{\alpha_N}(x_{N}) \\
\end{pmatrix},
\label{eq: basebosonica}
\end{equation}
with $\alpha_1 \leq \alpha_2 \leq \ldots \leq \alpha_M$, where $M$ denotes the number of distinct one-particle states.
\section{Fock space}
We remove the constraint that the number of identical particles $N$ be fixed. A Hilbert space with a variable number of particles is called Fock space, we denote it by $\mathcal{F}$ and it is formed by a direct sum of Hilbert spaces associated with particles in a fixed number, i.e.,
\begin{align}
\mathcal{F} &= H_0 \bigoplus H_1 \bigoplus H_2 \bigoplus \ldots = \notag \\
&= \bigoplus_{N=0}^{\infty} \ H_N ,
\end{align}
where the space $H_0$ has dimension $1$ and is associated with the particle vacuum. The vacuum vector $|0\rangle$ is given by
\begin{equation}
|0\rangle \equiv \psi^{(0)} = 
\begin{pmatrix}
1 \\
0 \\
0 \\
\vdots \\
\end{pmatrix} ,
\end{equation}
and every Hilbert space $H_N$ with $N \geq 1$ is $\infty$-dimensional. What does a vector $|\psi>$ look like in Fock space? It has projections in all Hilbert subspaces with a fixed number of particles, i.e.,
\begin{equation}
|\psi \rangle = 
\begin{pmatrix}
&\psi^{(0)} \\
&\psi^{(1)}(x_1) \\
&\psi^{(2)}(x_1,x_2) \\
& \vdots \\
\end{pmatrix} 
=
\begin{pmatrix}
&\psi^{(0)}\\
&0\\
&0\\
&\vdots
\end{pmatrix}
+
\begin{pmatrix}
&0\\
&\psi^{(1)}(x_1)\\
&0\\
&\vdots
\end{pmatrix}
+
\begin{pmatrix}
&0\\
&0\\
&\psi^{(2)}(x_1,x_2)\\
&\vdots
\end{pmatrix}
+
\ldots
\ .
\end{equation}
Fock space is a Hilbert space if provided with the scalar product 
\begin{equation}
\left(\psi,\phi\right)_{\mathcal{F}} = \sum_{N=0}^{\infty} \left(\psi^{(N)},\phi^{(N)}\right)_{H_{N}} ,
\end{equation}
where the general term of the sequence is the scalar product of the $N$-th components of the vector. Obviously, the vector $\psi^{(N)}$ describing the state of a system of $N$ particles reduces in Fock space to the vector with elements all null except the $N$-th element
\begin{equation}
\psi^{(N)} = 
\begin{pmatrix}
& 0 \\ 
& \vdots \\ 
& 0 \\
& \psi^{(N)}(x_1,\ldots,x_N) \\ 
& 0 \\
& \vdots \\
& \vdots \\
\end{pmatrix}.
\end{equation}
As in the case of systems with a fixed particle number, in Fock space we are likewise interested in the following relevant subspaces
\begin{equation}
F^A = H_0 \bigoplus H_1 \bigoplus_{N=2}^\infty H_N^A,
\end{equation}
\begin{equation}
F^S = H_0 \bigoplus H_1 \bigoplus_{N=2}^\infty H_N^S,
\end{equation}
since the vectors in such subspaces are the only ones describing systems of identical particles. The product of the square modulus of the $N$-th component by the $N$-dimensional measuring element
\begin{equation}
|\psi^{(N)}(x_1,\ldots,x_N)|^2 dx_1 \ldots dx_N
\end{equation}
denotes the probability of finding the system with $N$ particles of spin $s_1,\ldots,s_N$ in the volume elements $d^3\textbf{r}_1,\ldots,d^3\textbf{r}_N$. If we integrate the square modulus of the state vector with respect to spin and positions, we have the probability that the system is of $N$ particles
\begin{equation}
\int dx_1 \ldots \int dx_N |\psi^{(N)}(x_1,\ldots,x_N)|^2 = P_N,
\end{equation}
being $\sum_{N=1}^\infty P_N = 1$.
\section{Creation and annihilation operators}
Having introduced the Hilbert space suitable for the description of systems with a fluctuating number of particles, namely the Fock space $\mathcal{F}$, we now require operators that connect sectors with different particle numbers. Specifically, we need operators that map from $H_N$ to $H_{N \pm 1}$. Within the framework of quantum many-body theory, we define the annihilation operator associated with a single-particle state $\varphi \in H_1$ as the operator that, when acting on a many-body state $\psi \in \mathcal{F}$, removes one particle in the state $\varphi$, that it, it acts as
\begin{equation}
a_\varphi : \ \left[a_\varphi |\psi\rangle \right]^{(N)}(x_1,\ldots,x_N) \equiv \sqrt{N+1} \int dx \varphi^*(x) \psi^{(N+1)}(x,x_1,\ldots,x_N),
\label{eq: operatoredistruzionemanybody}
\end{equation}
where $\left[a_\varphi |\psi\rangle \right]^N(x_1,\ldots,x_N)$ is the $N$-th component of the vector $\psi \in \mathcal{F}$: integration with respect to $dx$ eliminates the dependence on any one among the $(N+1)$ variables of $\psi^{(N+1)}$. \newline
For each $\varphi \in H_1$ and $\psi \in \mathcal{F}$ we define the creation operator $a^\dag_\varphi$ as the operator that acts as
\begin{equation}
a^\dag_\varphi: \left[a^\dag_\varphi |\psi\rangle \right]^{(N)}(x_1,\ldots,x_N) = \frac{1}{\sqrt{N}}\sum_{j=1}^{N} \varepsilon^{j+1} \varphi(x_j) \psi^{(N-1)}(x_1,\ldots,x_{j-1},x_{j+1},\ldots,x_N), \ N \geq 1,
\label{eq: operatorecreazionemanybody}
\end{equation}
\begin{equation}
\left[a_\varphi^\dag |\psi\rangle \right]^{(0)} = 0,
\end{equation}
indeed, $a_{\varphi}^\dag$ has no projection on $H_0$, since it must create particles.
\begin{remark}[Notation]
Given a basis $\{\varphi_\alpha\}$ of $H_1$, we denote by $a_\alpha \equiv a_{\varphi_\alpha}$ the annihilation operator associated with the single-particle state $\varphi_\alpha$, and similarly by $a^\dagger_\alpha \equiv a^\dagger_{\varphi_\alpha}$ the corresponding creation operator.
\end{remark}
An annihilation operator $a_{\varphi}$ can be written in terms of $\lbrace a_{\varphi_\alpha} \rbrace$: indeed, since a generic function $\varphi \in H_1$ is written as a linear combination of the basis, $\varphi = \sum_\alpha c_\alpha \varphi_\alpha$, follows
\begin{align}
\left[ a_\varphi |\psi\rangle \right]^{(N)}(x_1,\ldots,x_N) &= \sqrt{N+1} \int dx \varphi^*(x) \psi^{(N+1)}(x,x_1,\ldots,x_N) = \notag \\
&= \sqrt{N+1} \int dx \sum_\alpha c_\alpha^* \varphi^*_\alpha(x) \psi^{(N+1)}(x,x_1,\ldots,x_N) = \notag \\
&= \sum_\alpha c_\alpha^* \sqrt{N+1} \int dx \varphi^*_\alpha(x) \psi^{(N+1)}(x,x_1,\ldots,x_N) = \notag \\
&= \sum_\alpha c_\alpha^* \left[ a_{\varphi_{\alpha}} |\psi\rangle \right]^{(N)}(x_1,\ldots,x_N) ,
\end{align}
that is,
\begin{equation}
a_\varphi = \sum_\alpha c^*_\alpha a_{\varphi_\alpha}.
\end{equation}
We have
\begin{theorem}
The operator $\eqref{eq: operatorecreazionemanybody}$ is the adjoint of the operator $\eqref{eq: operatoredistruzionemanybody}$, and vice versa. That is, for all \( \phi, \psi \in \mathcal{F} \) and \( \varphi \in H_1 \), the following identity holds
\begin{equation}
\langle \phi | a_\varphi | \psi \rangle = \langle \psi | a_\varphi^\dagger | \phi \rangle^*.
\end{equation}
\begin{proof}
Let us focus on the second term and compute the scalar product between the two wave functions in Fock space, namely
\begin{equation}
\langle \psi | a_\varphi^\dagger | \phi \rangle^* = \sum_{N=1}^\infty \left( \left[ a_\varphi^\dagger | \phi \rangle \right]^{(N)}, \psi^{(N)} \right)_{H_N},
\end{equation}
with the series starting at $1$ instead of $0$, since in the vacuum (particle-free) vector the creation operator has no components. Now, consider the generic term in the $N$-particle Hilbert space $H_N$, i.e.,
\begin{align}
& \left(\left[a^\dag_\varphi | \phi \rangle \right]^{(N)},\psi^{(N)}\right)_{H_{N}} = \notag \\
&= \dfrac{1}{\sqrt{N}} \sum_{j=1}^N \varepsilon^{j+1} \int dx_1 \ldots \int dx_j \ldots \int dx_N \varphi^*(x_j) \left( \phi^{(N-1)}(x_1,\ldots,x_{j-1},x_{j+1},\ldots,x_N) \right)^* \psi^{(N)}(x_1,\ldots,x_N) = \notag \\
&= \dfrac{1}{\sqrt{N}} \sum_{j=1}^N \varepsilon^{j+1} \int dx_1 \ldots \int dx_{j-1} \int dx_{j+1} \ldots \int dx_N \notag \\
& \ \ \ \ \ \ \ \ \ \ \ \ \ \ \ \ \left( \phi^{(N-1)}(x_1,\ldots,x_{j-1},x_{j+1},\ldots,x_N) \right)^* \int dx_j \varphi^*(x_j) \psi^{(N)}(x_1,\ldots,x_N) ,
\end{align}
\begin{equation}
I_j = \int dx_j \varphi^*(x_j) \psi^{(N)}(x_1,\ldots,x_N).
\end{equation}
Now we compute the integral $I_j$ in the cases $j=1$, $j=2$ and $j=3$ by rearranging the degrees of freedom of the function $\psi^{(N)}$. We want to bring out the action of the annihilation operator on the state vector. For $j=1$
\begin{equation}
I_1 = \int dx_1 \varphi^*(x_1) \psi^{(N)}(x_1,\ldots,x_N),
\end{equation}
for $j=2$ we have
\begin{align}
I_2 &= \int dx_2 \varphi^*(x_2)\psi^{(N)}(x_1,x_2,\ldots,x_N) = \notag \\
&= \varepsilon \int dx_2 \varphi^*(x_2) \psi^{(N)}(x_2,x_1,\ldots,x_N),
\end{align}
for $j=3$ we have
\begin{align}
I_3 &= \int dx_3 \varphi^*(x_3)\psi^{(N)}(x_1,x_2,x_3,\ldots,x_N) = \notag \\
&= \varepsilon^2 \int dx_3 \varphi^*(x_3) \psi^{(N)}(x_3,x_1,x_2,\ldots,x_N),
\end{align}
we want to derive a general pattern from it. Given the set $\left\lbrace x_1,\ldots,x_N \right\rbrace$, we take the j-th term, $x_j$, replace it with $x_1$, so the transposition produces a factor $\varepsilon$. Now, $x_1$ is in position $j$, but we want to take $x_1$ to position $2$, since we want to get the sequence $x_j,x_1,\ldots,x_{j-1},x_{j+1},\ldots,x_N$. $j-2$ exchanges are needed to order the sub-sequence $x_2,\ldots,x_{j-1},x_{j+1},\ldots,x_N$. Together with the initial permutation, we have $j-1$ exchanges, which originate the factor $\varepsilon^{j-1}$. The integral corresponding to a generic \(j\)-th term is therefore
\begin{equation}
I_j = \varepsilon^{j-1} \int dx_j \varphi^*(x_j) \psi^{(N)}(x_j,x_1,\ldots,x_{j-1},x_{j+1},\ldots,x_N).
\end{equation}
Recall that in the definition of the creation operator there is a factor \(\varepsilon^{j+1}\). Multiplying this by the factor \(\varepsilon^{j-1}\) arising from the permutation, we obtain $\varepsilon^{j+1}\varepsilon^{j-1} = \varepsilon^{2j}= 1$, which shows that the dependence on \(\varepsilon\) cancels out in the scalar product, yielding
\begin{align}
\dfrac{1}{\sqrt{N}} & \sum_{j=1}^N \int dx_j \varphi^*(x_j) \int dx_1 \ldots \int dx_{j-1} \int dx_{j+1} \ldots \int dx_N \notag \\
& \left( \phi^{(N-1)}(x_1,\ldots,x_{j-1},x_{j+1},\ldots,x_N) \right)^* \psi^{(N)}(x_{j},x_1,\ldots,x_{j-1},x_{j+1},\ldots,x_N). 
\end{align}
Now the choice of the degree of freedom over which to integrate does not imply any sign changes in the scalar product, since for every permutation in $\psi^{(N)}$ there will be the same one in the product $\varphi^* \phi^{(N-1)}$, so the integral with respect to $x_j$ is on a dummy index. $x_j$ is renamed and replaced with $x$, the particle index is lost, and the remaining variables are renamed as $x_1,\ldots,x_{N-1}$
\begin{align}
& \dfrac{1}{\sqrt{N}} \sum_{j=1}^N \int dx \varphi^*(x) \int dx_1 \ldots \int dx_{j-1} \int dx_{j+1} \ldots \int dx_N \notag \\
& \ \ \ \ \ \left( \phi^{(N-1)}(x_1,\ldots,x_{j-1},x_{j+1},\ldots,x_N) \right)^* \psi^{(N)}(x,x_1,\ldots,x_{j-1},x_{j+1},\ldots,x_N) = \notag \\
&= \dfrac{1}{\sqrt{N}} \left( \sum_{j=1}^N 1 \right) \int dx \varphi^*(x) \int dx_1 \ldots \int dx_{N-1} \left( \phi^{(N-1)}(x_1,\ldots,x_{N-1}) \right)^* \psi^{(N)}(x,x_1,\ldots,x_{N-1)} = \notag \\
&= \sqrt{N} \int dx \varphi^*(x) \int dx_1 \ldots \int dx_{N-1} \left( \phi^{(N-1)}(x_1,\ldots,x_{N-1}) \right)^* \psi^{(N)}(x,x_1,\ldots,x_{N-1}).
\end{align}
Finally, rearranging the terms and first performing the integral on $x$, it follows
\begin{align}
\left(\left[a^\dag_\varphi |\phi\rangle \right]^{(N)},\psi^{(N)}\right)_{H_{N}} &= \int dx_1 \ldots \int dx_{N-1} \left( \phi^{(N-1)}(x_1,\ldots,x_{N-1}) \right)^* \sqrt{N} \int dx \varphi^{*}(x) \psi^{(N)}(x,x_1,\ldots,x_{N-1}) = \notag \\
&= \int dx_1 \ldots \int dx_{N-1} \left( \phi^{(N-1)}(x_1,\ldots,x_{N-1}) \right)^* \left[a_{\varphi} |\phi\rangle \right]^{(N-1)} = \notag \\
&= \left( \phi^{(N-1)} , \left[a_{\varphi} |\phi\rangle \right]^{(N-1)} \right)_{H_{N-1}}, 
\end{align}
then
\begin{align}
\langle \psi | a_\varphi^\dagger | \phi \rangle^* &= \sum_{N=1}^{\infty} \left( \phi^{(N-1)} , \left[a_{\varphi} |\phi\rangle \right]^{(N-1)} \right)_{H_{N-1}} = \notag \\
&= \sum_{N=0}^{\infty} \left( \phi^{(N)} , \left[a_{\varphi} |\phi\rangle \right]^{(N)} \right)_{H_{N}},
\end{align}
which is the thesis.
\end{proof}
\end{theorem}
\section{Algebra of many-body creation and annihilation operators}
\begin{theorem}
The creation and annihilation operators satisfy the following algebras
\begin{equation}
\left[a_{\varphi},a_{\varphi'}\right]^{(\varepsilon)} = 0 = \left[ a^\dag_{\varphi} , a^\dag_{\varphi'} \right]^{(\varepsilon)}, \ \forall \ \varphi, \varphi' \in \mathcal{F},
\end{equation}
\begin{equation}
\left[a_{\varphi},a^\dag_{\varphi'}\right]^{(\varepsilon)} = (\varphi,\varphi')_{H_1}, \ \forall \ \varphi, \varphi' \in \mathcal{F},
\label{eq: algebraoperatoricreazdistruz}
\end{equation}
from which it follows
\begin{equation}
\left[a_\alpha,a^\dag_{\alpha'}\right]^{(\varepsilon)} = \delta_{\alpha,\alpha'}.
\end{equation}
Let us prove $\eqref{eq: algebraoperatoricreazdistruz}$. 
\begin{proof}
Given $\left[a_\varphi, a^\dag_{\varphi'}\right]^{(\varepsilon)} |\psi^{(N)}\rangle = a_\varphi a^\dag_{\varphi'} |\psi^{(N)}\rangle - \varepsilon\, a^\dag_{\varphi'} a_\varphi |\psi^{(N)}\rangle$, we compute separately the action of each operator product on the \(N\)-th component of the wave function in Fock space. We have
\begin{align}
\varepsilon \left[ a^\dag_{\varphi'} a_\varphi |\psi\rangle \right]^{(N)}(x_1,\ldots,x_N) &= \dfrac{\varepsilon}{\sqrt{N}} \sum_{j=1}^N \varepsilon^{j+1} \varphi'(x_j) \left[a_\varphi |\psi\rangle \right]^{(N-1)}(x_1,\ldots,x_{j-1},x_{j+1},\ldots,x_N) = \notag \\ 
&= \dfrac{1}{\sqrt{N}} \sum_{j=1}^N \varepsilon^{j+2} \varphi'(x_j) \left[a_\varphi |\psi\rangle \right]^{(N-1)}(x_1,\ldots,x_{j-1},x_{j+1},\ldots,x_N) = \notag \\ 
&= \dfrac{1}{\sqrt{N}} \sum_{j=1}^N \varepsilon^{j+2} \varphi'(x_j) \sqrt{N} \int dx \varphi^*(x) \psi^{(N)}(x,x_1,\ldots,x_{j-1},x_{j+1},\ldots,x_N) \equiv \notag \\
&\equiv \sum_{j=1}^N \varepsilon^{j} \varphi'(x_j) \int dx \varphi^*(x) \psi^{(N)}(x,x_1,\ldots,x_{j-1},x_{j+1},\ldots,x_N) ,
\end{align}
and given that $\varphi'(x_j)$ does not depend on the integration variable, we have
\begin{equation}
\sum_{j=1}^N \int dx \varepsilon^{j} \varphi'(x_j) \varphi^*(x) \psi^{(N)}(x,x_1,\ldots,x_{j-1},x_{j+1},\ldots,x_N).
\end{equation}
Regarding the second term, we have
\begin{align}
& \left[a_\varphi a^\dag_{\varphi'} |\psi\rangle \right]^{(N)}(x_1,\ldots,x_N) = \sqrt{N+1}\int dx \varphi^*(x) \left[a^\dag_{\varphi'} |\psi\rangle \right]^{(N+1)}(x,x_1,\ldots,x_N) ,
\end{align}
where the order of the operators is now reversed: a particle is created before one is annihilated. We proceed as follows
\begin{align}
\left[a_\varphi a^\dag_{\varphi'} |\psi\rangle \right]^{(N)}(x_1,\ldots,x_N) &= \sqrt{N+1} \int dx \varphi^*(x) \left[a^\dag_{\varphi'} |\psi\rangle \right]^{(N+1)}(x,x_1,\ldots,x_N) = \notag \\
&= \int dx \varphi^*(x) \bigg\{ \varepsilon^2 \varphi'(x) \psi^{(N)}(x_1,\ldots,x_N) + \notag \\
&+\varepsilon^3 \varphi'(x_1)\psi^{(N)}(x,x_2,\ldots,x_N) \ + \varepsilon^4 \varphi'(x_2)\psi^{(N)}(x,x_1,x_3,\ldots,x_N) + \ldots\bigg\}.
\end{align}
By summing the two terms, we obtain
\begin{align}
\left[ \left( a_\varphi a^\dag_{\varphi'} - \varepsilon a^\dag_{\varphi'} a_\varphi \right) |\psi\rangle \right]^{(N)} &= \left( \int dx \varphi^*(x) \varphi'(x) \right) \psi^{(N)}(x_1,\ldots,x_N) = \notag \\
&= (\varphi,\varphi')_{H_1} |\psi^{(N)}\rangle, \ \forall \ N,
\end{align}
which is the thesis.
\end{proof}
\end{theorem}
Let us consider $\varphi_\alpha \in H_1$, $a_{\varphi_\alpha}$, $a^\dag_{\varphi_\alpha}$: from the orthonormality we get
\begin{equation}
\left[a_{\varphi_{\alpha}},a^\dag_{\varphi_\alpha'}\right]^{(\varepsilon)} = \delta_{\alpha,\alpha'},
\end{equation}
which defines the algebra of creation and annihilation operators and from which the fundamental properties of fermions and bosons are derived, as we shall see next. \newline
To conclude this section, we present a theorem that enables the creation and annihilation operators to be expressed explicitly in terms of the transformation relating the two bases.
\begin{theorem}
Given two one-particle bases \( \left\lbrace \varphi_{\alpha}(x) \right\rbrace \) and \( \left\lbrace \omega_{\beta}(x) \right\rbrace \), related by a linear transformation
\begin{equation}
\omega_{\beta}(x) = \sum_{\alpha} m_{\alpha,\beta} \varphi_{\alpha}(x),
\end{equation}
the annihilation $c_{\beta}$ and creation $c_\beta^\dagger$ operators associated with $\omega_{\beta}(x)$ are written as
\begin{equation}
c_{\beta} = \sum_{\alpha} m^{*}_{\alpha,\beta} b_{\alpha},
\label{eq: operatoredistruzionenellabasedialtrooperatoredistruzione}
\end{equation}
\begin{equation}
c^\dagger_{\beta} = \sum_{\alpha} m_{\alpha,\beta} b^\dagger_{\alpha},
\label{eq: operatorecreazionenellabasedialtrooperatorecreazione}
\end{equation}
being $b_\alpha$ and $b^\dagger_\alpha$ the annihilation and creation operators, respectively, associated with the basis $\varphi_{\alpha}(x)$. 
\begin{proof}
To prove the statement, we consider the action of the annihilation operator \( c_{\omega_\beta} \) on an arbitrary state \( |\psi\rangle \in H^{(N+1)} \), that is,
\begin{align}
\left[c_{{\omega}_{\beta}} |\psi\rangle \right]^{(N)}(x_1,\ldots,x_N) &= \sqrt{N+1} \int dx \omega_\beta^*(x) \psi^{(N+1)}(x,x_1,\ldots,x_N) = \notag \\
&= \sqrt{N+1} \int dx \left( \sum_{\alpha} m_{\alpha,\beta} \varphi_{\alpha}(x) \right)^* \psi^{(N+1)}(x,x_1,\ldots,x_N) = \notag \\
&= \sqrt{N+1} \int dx \left( \sum_{\alpha} m^*_{\alpha,\beta} \varphi^*_{\alpha}(x) \right) \psi^{(N+1)}(x,x_1,\ldots,x_N) = \notag \\
&= \sum_\alpha m^*_{\alpha,\beta} \sqrt{N+1} \int dx \varphi^*_{\alpha}(x) \psi^{(N+1)}(x,x_1,\ldots,x_N) = \notag \\
&= \sum_\alpha m^*_{\alpha,\beta} \left[b_{{\varphi}_{\alpha}} |\psi\rangle \right]^{(N)}(x_1,\ldots,x_N), \ \forall \ |\psi\rangle,
\end{align}
and the expression for the creation operator follows analogously.
\end{proof}
\end{theorem}
\section{Fermionic basis and bosonic basis in Fock space}
Starting from a one-particle basis, we aim to construct a basis for the Fock space. The creation and annihilation operators serve as fundamental tools for generating fermionic and bosonic bases within this space. According to the definition of the creation operator \(a^\dagger_\alpha\), its action on the vacuum state yields
\begin{equation}
\left[a^\dagger_\alpha |0\rangle \right]^{(1)}(x) = \varphi_\alpha(x),
\end{equation}
meaning that the creation operator, when applied to the vacuum, creates a single particle in the state \(\varphi_\alpha(x) \in H_1\).
\subsection{Fermionic basis in Fock space}
We apply to the vacuum state the fermionic creation operators associated with the subscripts $\alpha_1$, $\alpha_2$ , with $\alpha_1 \neq \alpha_2$, i.e.,
\begin{align}
\left[a_{\alpha_1}^\dag a^\dag_{\alpha_2} |0\rangle \right]^{(2)}(x_1,x_2) &= \frac{1}{\sqrt{2}} \left(\varphi_{\alpha_1}(x_1) \varphi_{\alpha_2}(x_2) - \varphi_{\alpha_1}(x_2) \varphi_{\alpha_2}(x_1)\right) = \notag \\
&= \frac{1}{\sqrt{2}} \det 
\begin{pmatrix}
\varphi_{\alpha_1}(x_1) & \varphi_{\alpha_1}(x_2)\\
\varphi_{\alpha_2}(x_1) & \varphi_{\alpha_2}(x_2) 
\end{pmatrix},
\end{align}
which coincides with the basis of the antisymmetric space \(H_2^A\), provided that \(\alpha_1 < \alpha_2\). Conversely, if we compute $\left[a^\dagger_{\alpha_2} a^\dagger_{\alpha_1} |0\rangle \right]^{(2)}$, the rows of the corresponding matrix are reversed. Indeed, due to the fermionic anticommutation relations, $a^\dagger_{\alpha_1} a^\dagger_{\alpha_2} = -\, a^\dagger_{\alpha_2} a^\dagger_{\alpha_1}$, with $\alpha_1 \neq \alpha_2$, the determinant changes sign, introducing only a global phase factor between the two states, which does not affect the physical state. Next, we consider the case of \(N\) fermions and construct a basis of the antisymmetric subspace \(H^A_{N+1}\). This is done by applying the creation operator \(a^\dagger_\alpha\) to the basis elements of \(H^A_N\), i.e.,
\begin{equation}
\left[ a^\dag_\alpha \varphi^{(A,N)}_{{\alpha_1, \ldots,\alpha_N}} \right]^{(N+1)}(x_1,\ldots,x_{N+1}) = \dfrac{1}{\sqrt{N+1}}\sum_{j=1}^{N+1} \varepsilon^{j+1}\varphi_\alpha(x_j) \varphi^{(A,N)}_{\alpha_1,\ldots,\alpha_N}(x_1,\ldots,x_{j-1},x_{j+1},\ldots,x_{N+1}) ,
\end{equation}
\begin{align}
\varphi^{(A,N)}_{\alpha_1,\ldots,\alpha_N}(&x_1,\ldots,x_{j-1},x_{j+1},\ldots,x_{N+1}) = \notag \\
&= \frac{1}{\sqrt{N!}} \det
\begin{pmatrix}
\varphi_{\alpha_1}(x_{1}) & \ldots & \varphi_{\alpha_1}(x_{j-1}) & \varphi_{\alpha_1}(x_{j+1}) & \ldots & \varphi_{\alpha_1}(x_{N+1}) \\
\vdots & \ddots & \vdots & \vdots & \ddots & \vdots \\
\varphi_{\alpha_N}(x_{1}) & \ldots & \varphi_{\alpha_N}(x_{j-1}) & \varphi_{\alpha_N}(x_{j+1}) & \ldots & \varphi_{\alpha_N}(x_{N+1}) \\
\end{pmatrix} ,
\end{align}
being $\varepsilon=-1$. So, it follows
\begin{align}
\left[ a^\dag_\alpha \varphi^{(A,N)}_{{\alpha_1,\ldots,\alpha_N}} \right]^{(N+1)}(x_1,\ldots,x_{N+1}) &= \dfrac{1}{\sqrt{(N+1)!}} \sum_{j=1}^{N+1} (-1)^{j+1} \varphi_\alpha(x_j) \notag \\
& \ \ \ \ \ \det 
\begin{pmatrix}
\varphi_{\alpha_1}(x_{1}) & \ldots & \varphi_{\alpha_1}(x_{j-1}) & \varphi_{\alpha_1}(x_{j+1}) & \ldots & \varphi_{\alpha_1}(x_{N+1}) \\
\vdots & \ddots & \vdots & \vdots & \ddots & \vdots \\
\varphi_{\alpha_N}(x_{1}) & \ldots & \varphi_{\alpha_N}(x_{j-1}) & \varphi_{\alpha_N}(x_{j+1}) & \ldots & \varphi_{\alpha_N}(x_{N+1}) \\
\end{pmatrix} ,
\end{align}
which coincides with the Laplace expansion of the determinant. Recall that given a $(N+1)\times(N+1)$ matrix $A$, Laplace's rule provides a method of calculating the determinant given by
\begin{equation}
\det A = \sum_{j=1}^{N+1} (-1)^{i+j} A_{ij} \det M_{ij},
\end{equation}
where $M_{ij}$ is the matrix obtained by deleting the $i$-th row and the $j$-th column of $A$ and is called the minor of row $i$ and column $j$. $\left[ a^\dag_\alpha \varphi^{(A,N)}_{{\alpha_1, \ldots,\alpha_N}} \right]^{(N+1)}(x_1,\ldots,x_{N+1})$ has the structure of a Laplace expansion, where the identifications are assumed to be that the index of $\alpha_i$ is the row index, the index of the degree of freedom is a column index, and the expansion is performed on the first row, at subscript $\alpha$. It follows
\begin{equation}
\left[a^\dag_\alpha \varphi^{(A,N)}_{{\alpha_1,\ldots,\alpha_N}}\right]^{(N+1)}(x_1,\ldots,x_{N+1}) = \dfrac{1}{\sqrt{(N+1)!}} \sum_{j=1}^{N+1} (-1)^{1+j} A_{1j} \det \ M_{1j} ,
\end{equation}
\begin{equation}
A = 
\begin{pmatrix}
\varphi_{\alpha}(x_{1}) & \ldots & \varphi_{\alpha}(x_{j-1}) & \varphi_{\alpha}(x_j) & \varphi_{\alpha}(x_{j+1}) & \ldots & \varphi_{\alpha}(x_{N+1}) \\
\varphi_{\alpha_1}(x_{1}) & \ldots & \varphi_{\alpha_1}(x_{j-1}) & \varphi_{\alpha_1}(x_{j}) & \varphi_{\alpha_1}(x_{j+1}) & \ldots & \varphi_{\alpha_1}(x_{N+1}) \\
\vdots & \ddots & \vdots & \vdots & \vdots & \ddots & \vdots \\
\varphi_{\alpha_N}(x_{1}) & \ldots & \varphi_{\alpha_N}(x_{j-1}) & \varphi_{\alpha_N}(x_{j}) & \varphi_{\alpha_N}(x_{j+1}) & \ldots & \varphi_{\alpha_N}(x_{N+1}) \\
\end{pmatrix} ,
\end{equation}
and the minor with respect to row $1$ and column $j$ is the matrix 
\begin{equation}
M_{1j} = 
\begin{pmatrix}
\varphi_{\alpha_1}(x_{1}) & \ldots & \varphi_{\alpha_1}(x_{j-1}) & \varphi_{\alpha_1}(x_{j+1}) & \ldots & \varphi_{\alpha_1}(x_{N+1}) \\
\vdots & \ddots & \vdots & \vdots & \ddots & \vdots \\
\varphi_{\alpha_N}(x_{1}) & \ldots & \varphi_{\alpha_N}(x_{j-1}) & \varphi_{\alpha_N}(x_{j+1}) & \ldots & \varphi_{\alpha_N}(x_{N+1}) \\
\end{pmatrix}.
\end{equation}
Recall that it must be $\alpha_1 < \alpha_2 < \ldots < \alpha_N$, therefore, in order for the element $\left[a^\dag_\alpha \varphi^{(A,N)}_{\alpha_1,\ldots,\alpha_N}\right]^{(N+1)}(x_1,\ldots,x_{N+1})$ to be a basis in $H^A_{N+1}$, we need to arrange the row of subscript $\alpha$ according to the above convention, assuming of course that it does not coincide with any of the subscripts already present, otherwise Pauli's principle for fermions would imply a null determinant. If, on the other hand, $\alpha \neq \alpha_i$ $\forall i$, let $S_\alpha$ be the number $S_\alpha$ of exchanges needed to move the row $\alpha$ to the right position. Depending on whether $S_\alpha$ is even or odd, we will have a sign change or not. Since the sign of the permutation can be written as $(-1)^{S_{\alpha}}$, we have
\begin{equation}
\left[a^\dag_\alpha \varphi^{(A,N)}_{{\alpha_1,\ldots,\alpha_N}}\right]^{(N+1)}(x_1,\ldots,x_{N+1}) = (-1)^{S_{\alpha}} \varphi^{(A,N+1)}_{\alpha_1,\ldots,\alpha_{i-1},\alpha,\alpha_{i},\ldots,\alpha_N}(x_1,\ldots,x_{N+1}).
\end{equation}
In the occupation-number basis the action of the creation operator can be expressed as
\begin{equation}
a^\dagger_\alpha |N_1, \ldots, N_\alpha, \ldots, N_\infty; N \rangle = (1 - N_\alpha) (-1)^{S_\alpha} |N_1, \ldots, 1 + N_\alpha, \ldots, N_\infty; N + 1 \rangle,
\end{equation}
where $(1-N_\alpha)$ is justified as follows: we are adding the row associated with $\lbrace \varphi_{\alpha}(x_j) \rbrace$, if it was already present, $N_{\alpha}=1$ and the resulting vector coincides with the null vector, conversely if $\alpha$ is not present, then $N_\alpha = 0$ and $a^\dag_\alpha$ adds a fermion in a new state. On the other hand, the phase factor $(-1)^{S_\alpha}$ is due to the fact that, a priori, $\alpha$ is not added respecting the ordering convention. Generalizing, a basis in the space $H_N^A$ is obtainable by means of
\begin{equation}
a^\dagger_{\alpha_1} \ldots a^\dagger_{\alpha_N} |0\rangle.
\end{equation} 
Let us now analyze how to go from \(H^A_N\) to \(H^A_{N-1}\), studying the action of the annihilation operator \(a_{\alpha}\) on a string of \(N\) creation operators, i.e., $a_{\alpha} a^\dagger_{\alpha_1} \cdots a^\dagger_{\alpha_N} |0\rangle$. Suppose that particles occupy the states \(\alpha_1, \ldots, \alpha_N\) but not the state \(\alpha\). From the fermionic algebra, we have the anticommutation relation $a_{\alpha} a^\dagger_{\alpha_i} = - a^\dagger_{\alpha_i} a_{\alpha}$, which means that \(a_{\alpha}\) anticommutes with all \(a^\dagger_{\alpha_i}\) and therefore acts directly on the vacuum state \(|0\rangle\), that is
\begin{align}
a_{\alpha} a^\dagger_{\alpha_1} \ldots a^\dagger_{\alpha_N} |0\rangle \ 
&= (-1)^{S_{\alpha}} a^\dagger_{\alpha_1} \ldots a^\dagger_{\alpha_N} a_{\alpha} |0\rangle \ = \notag \\
&= 0.
\end{align}
On the other hand, if $\alpha$ is present in the string, $a_{\alpha}$ anticommutes with all $a^\dag_{\alpha_i}$ with $\alpha \neq \alpha_i$, up to $a_\alpha^\dag$. We use $a_\alpha a^\dag_\alpha = 1-a^\dag_\alpha a_\alpha$ as follows
\begin{align}
a_{\alpha} a^\dagger_{\alpha_1} \ldots a^\dagger_{\alpha_N} |0\rangle &= (-1)^{S_{\alpha}} a_{\alpha_1}^\dagger \ldots a_\alpha a^\dagger_\alpha \ldots a_{\alpha_N}^\dagger |0\rangle \ = \notag \\
&= (-1)^{S_{\alpha}} a_{\alpha_1}^\dagger \ldots \left( 1 - a^\dagger_\alpha a_\alpha \right) \ldots a_{\alpha_N}^\dagger |0\rangle \ = \notag \\
&= (-1)^{S_{\alpha}} a_{\alpha_1}^\dagger \ldots 1 \ldots a_{\alpha_N}^\dagger |0\rangle - (-1)^{S_{\alpha}} (-1)^{S'_{\alpha}} a_{\alpha_1}^\dagger \ldots a^\dagger_\alpha \ldots a_{\alpha_N}^\dagger a_\alpha |0\rangle \ = \notag \\
&= (-1)^{S_{\alpha}} a_{\alpha_1}^\dagger \ldots 1 \ldots a_{\alpha_N}^\dagger |0\rangle \ = \notag \\
&= (-1)^{S_{\alpha}} |N_1, \ldots, N_\alpha = 0, \ldots, N_N, 0, 0, \ldots, N_\infty; N \rangle,
\end{align}
where $S'_{\alpha}$ denotes the number of exchanges required to move $a_\alpha$ past $a_\alpha^\dagger$ until it acts on the vacuum $|0\rangle$, resulting in the null vector. This process leaves us with the vector $(-1)^{S_{\alpha}} a_{\alpha_1}^\dagger \ldots 1 \ldots a_{\alpha_N}^\dagger |0\rangle$. Note that the operator $a_\alpha^\dagger$ is replaced by the identity operator $1$, which represents the absence of a particle in the state $\alpha$. \newline
In the occupation-number basis, this can be succinctly expressed as
\begin{equation}
a_\alpha |N_1, \ldots, N_\infty; N\rangle = (-1)^{S_\alpha} N_\alpha \, |N_1, \ldots, N_\alpha - 1, \ldots, N_\infty; N - 1\rangle.
\end{equation}
In the occupation-number basis, for fermions we have obtained
\begin{equation}
\begin{cases}
a^\dagger_{\alpha} |N_1,\ldots,N_\alpha,\ldots,N_\infty;N\rangle = (-1)^{S_\alpha} (1 - N_\alpha) |N_1,\ldots,1 + N_\alpha,\ldots;N+1\rangle \\
a_{\alpha} |N_1,\ldots,N_\alpha,\ldots,N_\infty;N\rangle = (-1)^{S_\alpha} N_\alpha |N_1,\ldots,1 - N_\alpha,\ldots;N-1\rangle.
\end{cases}
\end{equation}
\subsection{Bosonic basis in Fock space}
Consider the action of $a_\alpha^\dag$ on a basis of $H^S_N$ (recall that $M$ denotes the number of distinct one-particle states), i.e.,
\begin{align}
& \left[a_\alpha^\dag \varphi^{(S,N)}_{\alpha_1,\ldots,\alpha_N}\right]^{(N+1)}(x_1,\ldots,x_{N+1}) = \notag \\
&= \dfrac{1}{\sqrt{N+1}}\sum_{j=1}^{N+1} \varphi_\alpha(x_j) \dfrac{1}{\sqrt{N! N_{\alpha_1}! \ldots N_{\alpha_M}!}} \perm
\begin{pmatrix}
& \varphi_{\alpha_1}(x_{1}) & \ldots & \varphi_{\alpha_1}(x_{j-1}) & \varphi_{\alpha_1}(x_{j+1}) & \ldots & \varphi_{\alpha_1}(x_{N+1}) \\
& \vdots & \ddots & \vdots & \vdots & \ddots & \vdots \\
& \varphi_{\alpha_N}(x_{1}) & \ldots & \varphi_{\alpha_N}(x_{j-1}) & \varphi_{\alpha_N}(x_{j+1}) & \ldots & \varphi_{\alpha_N}(x_{N+1}) \\
\end{pmatrix},
\end{align}
in order for it to be a basis, the indices of $\left[a_\alpha^\dag \varphi^{(S,N)}_{\alpha_1,\ldots,\alpha_N}\right]^{(N+1)}(x_1,\ldots,x_{N+1})$ must be ordered in an ascending direction, and since $\alpha$ has been added, its row must be shifted until an ascending order of the indices is obtained again. Unlike the fermionic case, however, the inversion of the rows in the permanent does not alter the sign, and thus any phase factors between the matrices are not present. As for normalization, suppose that the index $\alpha$ is present $N_\alpha$ times, now it will be present $N_\alpha + 1$ times, so the normalization constant of the basis of $H_{N+1}^S$ is $\left( \sqrt{(N_\alpha + 1)!} \right)^{-\frac{1}{2}}$, since $N_\alpha !$ is in the denominator, we multiply and divide by $\sqrt{N_\alpha + 1}$, i.e.,
\begin{align}
\left[a_\alpha^\dag \varphi^{(S,N)}_{\alpha_1,\ldots,\alpha_N}\right]^{(N+1)} (x_1,\ldots,x_{N+1}) &= \dfrac{1}{\sqrt{(N+1)!}} \frac{1}{\sqrt{N_{\alpha_1}!} \ldots \sqrt{N_{\alpha_M}!}} \frac{\sqrt{N_\alpha + 1}}{\sqrt{N_\alpha + 1}} \sum_{j=1}^{N+1}A_{1j} \ \perm M_{1j} = \notag \\
&= \sqrt{N_\alpha + 1} \varphi^{(S,N+1)}_{\{\alpha_1,\ldots,\alpha_N\}}(x_1,\ldots,x_{N+1}),
\end{align}
\begin{equation}
A =
\begin{pmatrix}
&\varphi_{\alpha}(x_{1}) & \ldots & \varphi_{\alpha}(x_{j-1}) & \varphi_{\alpha}(x_{j}) & \varphi_{\alpha}(x_{j+1}) & \ldots & \varphi_{\alpha}(x_{N+1}) \\
&\varphi_{\alpha_1}(x_{1}) & \ldots & \varphi_{\alpha_1}(x_{j-1}) & \varphi_{\alpha_1}(x_{j}) & \varphi_{\alpha_1}(x_{j+1}) & \ldots & \varphi_{\alpha_1}(x_{N+1}) \\
&\vdots & \ddots & \vdots & \vdots & \vdots & \ddots & \vdots \\
&\varphi_{\alpha_N}(x_{1}) & \ldots & \varphi_{\alpha_N}(x_{j-1}) & \varphi_{\alpha_N}(x_{j}) & \varphi_{\alpha_N}(x_{j+1}) & \ldots & \varphi_{\alpha_N}(x_{N+1}) \\
\end{pmatrix} ,
\end{equation}
and the minor with respect to row $1$ and column $j$ is given by the matrix 
\begin{equation}
M_{1j} =
\begin{pmatrix}
&\varphi_{\alpha_1}(x_{1}) & \ldots & \varphi_{\alpha_1}(x_{j-1}) & \varphi_{\alpha_1}(x_{j+1}) & \ldots & \varphi_{\alpha_1}(x_{N+1}) \\
&\vdots & \ddots & \vdots & \vdots & \ddots & \vdots \\
&\varphi_{\alpha_N}(x_{1}) & \ldots & \varphi_{\alpha_N}(x_{j-1}) & \varphi_{\alpha_N}(x_{j+1}) & \ldots & \varphi_{\alpha_N}(x_{N+1}) \\
\end{pmatrix}.
\end{equation}
Let us make explicit the action of the creation operator on the occupation-number states
\begin{equation}
a_\alpha^\dagger |N_1, \ldots, N_\alpha, \ldots, N_\infty; N \rangle = \sqrt{N_\alpha + 1} \, |N_1, \ldots, N_\alpha + 1, \ldots, N_\infty; N + 1 \rangle,
\end{equation}
note that the multiplicative factor is the same as that arising from the action of the creation operator on the eigenstates of the harmonic oscillator. To construct the bosonic basis, we start from the vacuum state and generate particles through successive applications of creation operators. Unlike the fermionic case, here it is possible to have terms of the form $\left( a_{\alpha_k}^\dag \right)^{N_k}$, representing $N_k$ particles occupying the state $\alpha_k$. In general the strings of operators have the form
\begin{equation}
\left(a^\dagger_{\alpha_1}\right)^{N_1} \ldots \left(a^\dagger_{\alpha_M}\right)^{N_M} |0\rangle,
\end{equation}
with $\sum N_i = N$, and a basis in the space $H_N^S$ is obtained as
\begin{equation}
\dfrac{(a^\dag_{\alpha_1})^{N_1}}{\sqrt{N_{\alpha_1}!}} \ldots \dfrac{(a^\dag_{\alpha_M})^{N_{\alpha_M}}}{\sqrt{N_{\alpha_M}!}} |0\rangle.
\end{equation}
To move from $H_N^S$ to $H_{N-1}^S$, we need to study the action of the annihilation operator on the basis. Let us consider the two possible cases: the state $\alpha$ is not in the string or the state is present and appears $N_\alpha \geq 1$ times. In the first case, $a_\alpha$ commutes with all other $a^\dag_{\alpha_i}$, up to the state $|0 \rangle$ and gives the empty vector; in the second case, the annihilation operator commutes with all creation operators with different indices up to the $N_\alpha$ creation operators $a^\dag_{\alpha}$. Ultimately, we need to calculate the commutator $\left[ a, \left( a^\dag \right)^N \right]$. We have
\begin{theorem}
Let $a$ and $a^\dagger$ be bosonic annihilation and creation operators, then
\begin{equation}
\left[a, \left( a^\dag \right) ^N \right] = N \left( a^\dag \right)^{N-1}, \ \forall \ N.
\label{eq: commutatoreoperatoredistruzioneecreazionenbosonici}
\end{equation}
\begin{proof}
We proceed by induction. For $N=1$, the left-hand side coincides with the bosonic commutator between the annihilation and creation operators, so the statement holds. Assume it holds for $N-1$, i.e.,
\begin{equation}
\left[a,\left( a^\dag \right)^{N-1}\right] = \left( N-1 \right) \left( a^\dag \right)^{N-2} ,
\end{equation}
and we check the validity for $N$. Since the operators are bosonic, we can use $\eqref{eq: Jacobiidentity}$ as follows
\begin{align}
\left[a,\left(a^\dag\right)^N\right] &= \left[a,a^\dag \left(a^\dag\right)^{N-1}\right] = \notag \\
&= \left[ a,a^\dagger \right] \left(a^\dag\right)^{N-1} +  a^\dag\left[a,\left(a^\dag\right)^{N-1}\right] = \notag \\
&= \left( a^\dag\right)^{n-1} + a^\dag\left[a,\left(a^\dag\right)^{N-1}\right] = \notag \\
&= \left( a^\dag\right)^{N-1} + a^\dag\left(N-1\right)\left(a^\dag\right)^{N-2} = \notag \\
&= N \left(a^\dag\right)^{N-1} ,
\end{align}
so, $\eqref{eq: commutatoreoperatoredistruzioneecreazionenbosonici}$ is thus proven.
\end{proof}
\end{theorem}
From 
\begin{equation}
a_\alpha \left(a^\dag_\alpha\right)^{N_\alpha} = N_\alpha \left(a^\dag_\alpha\right)^{N_\alpha - 1} + a^\dag_\alpha a_\alpha, 
\end{equation}
the term $a^\dag_\alpha a_\alpha$ causes the annihilation operator to act on $|0\rangle$, yielding the vacuum vector, whereas the first term yields
\begin{align}
a_\alpha \dfrac{(a^\dag_{\alpha_1})^{N_1}}{\sqrt{N_{\alpha_1}!}} \ldots \dfrac{(a^\dag_{\alpha})^{N_\alpha}}{\sqrt{N_{\alpha}!}} \ldots \dfrac{(a^\dag_{\alpha_M})^{N_M}}{\sqrt{N_{\alpha_M}!}} |0\rangle \ &= \dfrac{(a^\dag_{\alpha_1})^{N_1}}{\sqrt{N_{\alpha_1}!}} \ldots N_\alpha \dfrac{(a^\dag_{\alpha})^{N_\alpha-1}}{\sqrt{N_{\alpha}!}} \ldots \dfrac{(a^\dag_{\alpha_M})^{N_M}}{\sqrt{N_{\alpha_M}!}} |0\rangle = \notag \\
&= \dfrac{(a^\dag_{\alpha_1})^{N_1}}{\sqrt{N_{\alpha_1}!}} \ldots N_\alpha \dfrac{(a^\dag_{\alpha})^{N_\alpha-1}}{\sqrt{N_{\alpha}} \sqrt{(N_{\alpha}-1)!}} \ldots \dfrac{(a^\dag_{\alpha_M})^{N_M}}{\sqrt{N_{\alpha_M}!}} |0\rangle = \notag \\
&= \dfrac{(a^\dag_{\alpha_1})^{N_1}}{\sqrt{N_{\alpha_1}!}} \ldots \sqrt{N_\alpha} \dfrac{(a^\dag_{\alpha})^{N_\alpha-1}}{\sqrt{(N_{\alpha}-1)!}} \ldots \dfrac{(a^\dag_{\alpha_M})^{N_M}}{\sqrt{N_{\alpha_M}!}} |0\rangle
\end{align}
which in the occupation-number basis is written as
\begin{equation}
a_\alpha |N_1,\ldots,N_\infty;N \rangle = \sqrt{N_\alpha}|N_1,\ldots,N_\alpha -1,\ldots,N_\infty;N-1 \rangle.
\end{equation}
$a_\alpha$ destroys a boson and has the same structure as the annihilation operator for the harmonic oscillator. \newline
In the occupation-number basis, for bosons we have obtained
\begin{equation}
\begin{cases}
a^\dag_\alpha |N_1,\ldots,N_\alpha,\ldots,N_\infty;N\rangle = \sqrt{N_\alpha + 1}\, |N_1,\ldots,N_\alpha + 1,\ldots,N_\infty;N+1\rangle \\
a_\alpha |N_1,\ldots,N_\alpha,\ldots,N_\infty;N\rangle = \sqrt{N_\alpha}\, |N_1,\ldots,N_\alpha - 1,\ldots,N_\infty;N-1\rangle
\end{cases}.
\end{equation}
\subsection{The number operator in many-body systems}
Let us analyze the action of the operator \(a^\dag_\alpha a_\alpha\) on a system of fermions and bosons. Intuitively, since this operator first annihilates a particle in the state \(\alpha\) and then creates one back in the same state, the initial and final Hilbert spaces coincide. However, an important question arises: how does the state change under this operation? This operator is fundamental, as it counts the number of particles occupying the state \(\alpha\), it is precisely the number operator for the mode \(\alpha\). Understanding its action is key to describing the occupation properties of quantum many-body systems. In the fermionic case, its action reads
\begin{align}
a_\alpha^\dag a_\alpha |N_1, \ldots, N_\alpha, \ldots, N_\infty; N \rangle 
\ &= (-1)^{S_\alpha} N_\alpha a^\dag_\alpha |N_1, \ldots, 1 - N_\alpha, \ldots, N_\infty; N - 1 \rangle \ = \notag \\
&= (-1)^{S_\alpha} (-1)^{S_\alpha} N_\alpha \left[ 1 - \left( 1 - N_\alpha \right) \right] |N_1, \ldots, 1 + \left( 1 - N_\alpha \right), \ldots, N_\infty; N \rangle \ = \notag \\
&= N_\alpha^2 |N_1, \ldots, 2 - N_\alpha, \ldots, N_\infty; N \rangle \ = \notag \\
&= N_\alpha |N_1, \ldots, 2 - N_\alpha, \ldots, N_\infty; N \rangle \ = \notag \\
&= N_\alpha |N_1, \ldots, N_\alpha, \ldots, N_\infty; N \rangle.
\end{align}
The substitution \(2 - N_\alpha \rightarrow N_\alpha\) in the last step is justified by considering the possible values of \(N_\alpha\). If the state \(\alpha\) was initially unoccupied (\(N_\alpha = 0\)), the multiplicative factor \(N_\alpha = 0\) annihilates the vector, yielding the null vector. In this case, whether we write \(2 - N_\alpha\) or \(0\) at the \(\alpha\)-th position makes no difference. Conversely, if the state was occupied (\(N_\alpha = 1\)), then \(2 - N_\alpha = 1\) matches the occupation number correctly. Therefore, the final result is either the original state (if the particle was present) or the zero vector (if it was not). For the bosonic case, we have instead
\begin{align}
a_\alpha^\dag a_\alpha |N_1, \ldots, N_\alpha, \ldots, N_\infty; N \rangle \ &= \sqrt{N_\alpha} \, a^\dag_\alpha |N_1, \ldots, N_\alpha - 1, \ldots, N_\infty; N - 1 \rangle \ = \notag \\
&= \sqrt{N_\alpha} \sqrt{N_\alpha - 1 + 1} \, |N_1, \ldots, N_\alpha - 1 + 1, \ldots, N_\infty; N \rangle \ = \notag \\
&= N_\alpha |N_1, \ldots, N_\alpha, \ldots, N_\infty; N \rangle.
\end{align}
In both cases, the operator $\hat{N}_\alpha = a_\alpha^\dag a_\alpha$ acts as a number operator, counting the number of particles in the state $\varphi_\alpha$.
\section{Field operators}
Given a basis $\varphi_\alpha \in H_1$, we define the annihilation and creation operators field as
\begin{equation}
\hat{\psi}(x) \equiv \sum_\alpha \varphi_\alpha(x) a_\alpha,
\label{eq: operatorecampodistruzione}
\end{equation}
\begin{equation}
\hat{\psi}^{\dag}(x) \equiv \sum_\alpha \varphi^{*}_\alpha(x) a^{\dag}_\alpha,
\label{eq: operatorecampocreazione}
\end{equation}
respectively, which satisfies the algebras
\begin{equation}
\left[\hat{\psi}(x),\hat{\psi}(x')\right]^{(\varepsilon)} = 0, 
\end{equation}
\begin{equation}
\left[\hat{\psi}^\dag(x),\hat{\psi}^\dag(x')\right]^{(\varepsilon)} = 0,
\end{equation}
\begin{align}
\left[\hat{\psi}(x),\hat{\psi}^\dag(x')\right]^{(\varepsilon)} &= \sum_\alpha \sum_{\alpha'} \varphi_\alpha(x) \varphi^*_{\alpha'}(x')\left[a_\alpha,a^\dag_{\alpha'}\right]^{(\varepsilon)} = \notag \\
&= \sum_\alpha \varphi_\alpha(x) \varphi_\alpha^*(x') = \notag \\
&= \delta(x-x').
\end{align}
Due to the orthonormality of $\left\lbrace \varphi_\alpha(x) \right\rbrace$, we can write $a_{\alpha'}$ in terms of $\hat{\psi}(x)$ 
\begin{align}
\int dx \varphi_{\alpha'}^*(x) \hat{\psi}(x) &= \int dx \sum_\alpha \varphi_{\alpha'}^*(x) \varphi_\alpha(x) a_\alpha = \notag \\
&= \sum_{\alpha} \left( \int dx \varphi_{\alpha'}^*(x) \varphi_\alpha(x) \right) a_{\alpha} = \notag \\
&= \sum_{\alpha} \delta_{\alpha,\alpha'} a_\alpha = \notag \\
&= a_{\alpha'}.
\end{align}
We let the annihilation field operator $\hat{\psi}(x)$ act on a vector $|\phi\rangle \in \mathcal{F}$. Let us consider its $N$-th component
\begin{align}
\left[ \hat{\psi}(x) |\phi \rangle \right]^{(N)} (x_1, \ldots, x_N) &= \sum_\alpha \varphi_\alpha(x) \left[ a_\alpha |\phi \rangle \right]^{(N)} (x_1, \ldots, x_N) = \notag \\
&= \sum_\alpha \varphi_\alpha(x) \sqrt{N+1} \int dx' \, \varphi_\alpha^*(x') \, \phi^{(N+1)} (x', x_1, \ldots, x_N) = \notag \\
&= \sqrt{N+1} \int dx' \left( \sum_\alpha \varphi_\alpha(x) \varphi_\alpha^*(x') \right) \phi^{(N+1)} (x', x_1, \ldots, x_N) = \notag \\
&= \sqrt{N+1} \int dx' \, \delta(x - x') \, \phi^{(N+1)} (x', x_1, \ldots, x_N) = \notag \\
&= \sqrt{N+1} \, \phi^{(N+1)} (x, x_1, \ldots, x_N).
\end{align}
Note that while the annihilation operator \(a_\varphi\) is defined with respect to the wavefunction \(\varphi(x')\), the annihilation field operator \(\hat{\psi}(x)\) is defined using the Dirac delta function \(\delta(x - x')\). In this sense, \(\hat{\psi}(x)\) can be viewed as an annihilation operator associated with the generalized function \(\delta(x - x')\). Just as \(a_\varphi\) annihilates a boson or fermion in the state described by the wavefunction \(\varphi\), the field operator \(\hat{\psi}(x)\) annihilates a particle localized at the point \(\mathbf{r}\) with spin \(s\). In other words, \(\hat{\psi}(x)\) annihilates bosons or fermions at a specific position in space and with a well-defined spin. \newline
Let the creation field operator \(\hat{\psi}^\dagger(x)\) act on a vector \(|\phi \rangle \in \mathcal{F}\). Consider its \(N\)-particle component, that is
\begin{align}
\left[\hat{\psi}^\dagger(x) |\phi \rangle \right]^{(N)} (x_1,\ldots,x_N) &= \sum_\alpha \varphi_\alpha^*(x) \left[a_\alpha^\dagger |\phi \rangle \right]^{(N)}(x_1,\ldots,x_N) = \notag \\
&= \sum_{\alpha} \varphi_{\alpha}^{*}(x) \frac{1}{\sqrt{N}} \sum_{j=1}^N \varepsilon^{j+1} \varphi_{\alpha}(x_j) \phi^{(N-1)}(x_1,\ldots,x_{j-1},x_{j+1},\ldots,x_N) = \notag \\
&= \frac{1}{\sqrt{N}} \sum_{j=1}^N \left( \sum_{\alpha} \varphi_{\alpha}^{*}(x) \varphi_{\alpha}(x_j) \right) \varepsilon^{j+1} \phi^{(N-1)}(x_1,\ldots,x_{j-1},x_{j+1},\ldots,x_N) = \notag \\
&= \frac{1}{\sqrt{N}} \sum_{j=1}^N \delta(x - x_j) \varepsilon^{j+1} \phi^{(N-1)}(x_1,\ldots,x_{j-1},x_{j+1},\ldots,x_N).
\end{align}
Note that, whereas the creation operator $a^{\dag}_\varphi$ involves the function $\varphi(x_j)$ in its definition, the creation field operator $\hat{\psi}^\dag(x)$ involves the Dirac delta function $\delta(x - x_j)$: in this sense, $\hat{\psi}^\dag(x)$ acts like a creation operator associated with the generalized function $\delta(x - x_j)$. Just as $a^{\dag}_\varphi$ creates a boson or fermion in the quantum state described by the wavefunction $\varphi$, the field operator $\hat{\psi}^\dag(x)$ creates a particle at the spatial point $\textbf{r}$ with spin $s$, that is, it creates bosons or fermions at a specific position in space and with a definite spin. \newline
Ultimately, note that, in the construction of $\left[\hat{\psi}(x)|\phi\rangle\right]^{(N)}$ and $\left[\hat{\psi}^{\dag}(x)|\phi\rangle\right]^{(N)}$, completeness has been used, a property shared by all bases: consequently, the field operators are defined independently of the choice of basis in the one-particle Hilbert space. In this text, however, field operators will often be expanded in terms of plane waves multiplied by spin eigenfunctions, namely according to equation $\eqref{eq: ondepianeperautofunzionispin}$, and we have
\begin{equation}
\hat{\psi}(\textbf{r},s) = \sum_{\textbf{k},\sigma} \dfrac{e^{i \textbf{k} \cdot \textbf{r}}}{\sqrt{V}} \chi_{\sigma}(s) C_{\textbf{k},\sigma},
\label{eq: operatorecampodistruzionebaseondepianeperspin}
\end{equation}
\begin{equation}
\hat{\psi}^\dagger(\textbf{r},s) = \sum_{\textbf{k},\sigma} \dfrac{e^{-i \textbf{k} \cdot \textbf{r}}}{\sqrt{V}} \chi^*_{\sigma}(s) C^\dagger_{\textbf{k},\sigma}.
\label{eq: operatorecampocreazionebaseondepianeperspin}
\end{equation}
\section{Representation in Fock space of one-body operators}
Let us consider an operator \(F_N\) defined on the \(N\)-particle Hilbert space \(H_N\), which in first quantization is expressed as the sum of single-particle operators, i.e.,
\begin{equation}
F_N \equiv \sum_{i=1}^N f(x_i),
\end{equation}
i.e., a one-body operator. The one-body operator $F_N$ naturally induces an operator $\hat F$ on Fock space, which is defined by its action on each $N$-particle space $H_N$, that is
\begin{equation}
\langle \phi | \hat{F} | \phi' \rangle_{\mathcal{F}} \ \equiv \sum_{N=1}^\infty \left(\phi^{(N)}, F_N \phi'^{(N)}\right)_{H_N},
\end{equation}
let us analyze the scalar product in the \(N\)-particle space \(H_N\):
\begin{align}
I &\equiv \left(\phi^{(N)}, F_N \phi'^{(N)}\right)_{H_N} = \notag \\
&= \int dx_1 \ldots \int dx_N \left( \phi^{(N)}(x_1, \ldots, x_N) \right)^* \left( \sum_{i=1}^N f(x_i) \right) \phi'^{(N)}(x_1, \ldots, x_N) = \notag \\
&= \sum_{i=1}^N \int dx_1 \ldots \int dx_N \left( \phi^{(N)}(x_1, \ldots, x_N) \right)^* f(x_i) \phi'^{(N)}(x_1, \ldots, x_N).
\end{align}
Let us calculate the integral with respect to $dx_i$, i.e.,
\begin{equation}
I = \sum_{i=1}^N \int dx_1 \ldots \int dx_{i-1} \int dx_{i+1} \ldots \int dx_N \int dx_i \left( \phi^{(N)}(x_1,\ldots,x_i,\ldots,x_N) \right)^* f(x_i) \phi'^{(N)}(x_1,\ldots,x_i,\ldots,x_N) ,
\end{equation}
we move $x_i$ to the first position in both $\phi$ and $\phi'$, the total number of permutations will be even, so the calculations to follow do not depend on the statistical nature of the particles. We set $x_i = x$ and rename the variables on which we do not integrate , i.e.,
\begin{align}
& \sum_{i=1}^N \int dx_1 \ldots \int dx_{N-1}\int dx \left( \phi^{(N)}(x,x_1,\ldots,x_{N-1}) \right)^* f(x) \phi'^{(N)}(x,x_1,\ldots,x_{N-1}) = \notag \\
& = \int dx_1 \ldots \int dx_{N-1} \int dx \left\lbrace \sqrt{N} \left( \phi^{(N)}(x,x_1,\ldots,x_{N-1}) \right)^* \right\rbrace f(x) \left\lbrace \sqrt{N}\phi'^{(N)}(x,x_1,\ldots,x_{N-1}) \right\rbrace ,
\end{align}
where we used the fact that the summand does not depend on the index \(i\) and the identity
\begin{align}
\sum_{i=1}^N 1 &= N = \notag \\
&= \sqrt{N} \sqrt{N} .
\end{align}
From $\eqref{eq: operatorecampodistruzione}$, we write
\begin{align}
I &= \int dx_1 \ldots \int dx_{N-1} \int dx \left(\left[\hat{\psi}(x)|\phi\rangle \right]^{(N-1)}(x_1,\ldots,x_{N-1})\right)^* f(x) \left(\left[\hat{\psi}(x)|\phi'\rangle \right]^{(N-1)}(x_1,\ldots,x_{N-1})\right) = \notag \\
&= \int dx \int dx_1 \ldots \int dx_{N-1} \left(\left[\hat{\psi}(x) |\phi\rangle \right]^{(N-1)}(x_1,\ldots,x_{N-1})\right)^* f(x) \left(\left[\hat{\psi}(x) |\phi'\rangle \right]^{(N-1)}(x_1,\ldots,x_{N-1})\right) = \notag \\
&= \int dx \left( \left[ \hat{\psi}(x) |\phi\rangle \right]^{(N-1)} , f(x) \left[ \hat{\psi}(x) |\phi'\rangle \right]^{(N-1)} \right)_{H_{N-1}},
\end{align}
and the series becomes
\begin{align}
\sum_{N=1}^\infty \left(\phi^{(N)}, F_N \phi'^{(N)}\right)_{H_N} &= \sum_{N=1}^{\infty} \int dx \left( \left[ \hat{\psi}(x) |\phi\rangle \right]^{(N-1)} , f(x) \left[ \hat{\psi}(x) |\phi'\rangle \right]^{(N-1)} \right)_{H_{N-1}} = \notag \\
&= \sum_{N=0}^{\infty} \int dx \left( \left[ \hat{\psi}(x) |\phi\rangle \right]^{(N)} , f(x) \left[ \hat{\psi}(x) |\phi'\rangle \right]^{(N)} \right)_{H_{N}}.
\end{align}
Assume the order of summation and integration can be interchanged, that is,
\begin{align}
\langle \phi | \hat{F} | \phi' \rangle_{\mathcal{F}} &= \sum_{N=1}^\infty \left(\phi^{(N)}, F_N \phi'^{(N)}\right)_{H_N} = \notag \\
&= \int dx \sum_{N=0}^{\infty} \left( \left[ \hat{\psi}(x) |\phi\rangle \right]^{(N)} , f(x) \left[ \hat{\psi}(x) |\phi'\rangle \right]^{(N)} \right)_{H_{N}} = \notag \\
&= \int dx \left( \hat{\psi}(x)|\phi\rangle , f(x) \hat{\psi}(x) |\phi'\rangle \right)_{\mathcal{F}} = \notag \\
&= \int dx \langle\phi| \hat{\psi}^{\dag}(x) f(x) \hat{\psi}(x) |\phi'\rangle_{\mathcal{F}} \ , \ \forall \ \phi, \phi' \in \mathcal{F}^{A} \ \left( \forall \ \phi, \phi' \in \mathcal{F}^{S} \right).
\end{align}
In second quantization the one-body operator $F_N$ is replaced in Fock space by
\begin{align}
\hat{F} = \int dx \hat{\psi}^\dag(x) f(x) \hat{\psi}(x).
\end{align}
The particle index is lost: two field operators create and annihilate particles at a given point in space with a specific spin component. The statistical nature of the particles is encoded in the creation and annihilation operators. Moreover, the operator $\hat{F}$ plays a role analogous to an average, where the field operators take the place of the wave function and its complex conjugate in defining the expectation value of an observable. In first quantization, the wave function of a single particle is a vector in a Hilbert space; in second quantization, it becomes an operator.
\subsection{Kinetic term and one-body potential}
Let us consider a system of $N$ non-interacting particles (fermions or bosons), possibly subject to an external one-body potential $V(x)$. The Hamiltonian in first quantization is
\begin{align}
\mathcal{H} = \sum_i^N \left( \dfrac{\hat{\textbf{p}}_i^2}{2m} + V(x_i) \right).
\end{align}
Given the restriction of $\mathcal{H}$ to each subspace $H_N$ of Fock space $\mathcal{F}$, our goal is to express $\mathcal{H}$ in second quantization. The second-quantized Hamiltonian is
\begin{align}
\hat{\mathcal{H}} = \int dx \, \hat{\psi}^\dag(x) \left( - \dfrac{\hslash^2 \nabla^2}{2m} + V(x) \right) \hat{\psi}(x).
\end{align}
Expanding the field operators in a single-particle basis $\lbrace \varphi_\alpha(x) \rbrace$
\begin{equation}
\hat{\mathcal{H}} = \sum_{\alpha,\alpha'} \mathcal{H}_{\alpha , \alpha'} a_{\alpha'}^\dag a_\alpha, 
\end{equation}
\begin{equation}
\mathcal{H}_{\alpha, \alpha'} = \int dx \, \varphi^*_{\alpha'}(x) \left( -\dfrac{\hslash^2 \nabla^2}{2m} + V(x) \right) \varphi_\alpha(x).
\end{equation}
By choosing the basis of eigenfunctions of the one-body Hamiltonian
\begin{equation}
\left( - \dfrac{\hslash^2 \nabla^2}{2m} + V(x) \right) \varphi_\alpha(x) = \mathcal{E}_\alpha \varphi_\alpha(x),
\end{equation}
the Hamiltonian becomes diagonal, i.e.,
\begin{equation}
\hat{\mathcal{H}} = \sum_\alpha \mathcal{E}_\alpha a_\alpha^\dag a_\alpha.
\label{eq: Hamiltonianasecondaquantizzazioneparticellelibere}
\end{equation}
If the potential $V(x)$ is zero, we recover the free-particle case described by the plane waves in $\eqref{eq: ondepiane}$, with eigenvalues given by $\eqref{eq: autovaloriondepiane}$, with $\mathcal{E}_\alpha = \mathcal{E}_{\textbf{k}}$. The eigenstates in the $N$-particle sector $H_N$ are (anti)symmetrized products of the single-particle eigenstates, i.e., Slater determinants or permanents. The problem remains exactly solvable due to the absence of interactions among the particles. Since $\hat{N}_{\alpha} = a_\alpha^\dag a_\alpha$ counts the number of particles, the Hamiltonian $\hat{\mathcal{H}}$ is written in terms of energy quanta. We have quantized the system; it exchanges energy only for multiples of $\mathcal{E}_\alpha$. 
\section{Average occupation number for non-interacting particles}
We consider a system of non-interacting particles described by the Hamiltonian 
$\hat{\mathcal{H}} = \hat{\mathcal{H}}_0 - \mu \hat{N}$, where $\hat{\mathcal{H}_0}$ is given by $\eqref{eq: Hamiltonianasecondaquantizzazioneparticellelibere}$, \(\mu\) is the chemical potential, and \(\hat{N} = \sum_\alpha C^{\dagger}_\alpha C_\alpha\) is the total number operator. This corresponds to working within the grand canonical ensemble, where the total number of particles is allowed to fluctuate. The thermal average occupation number of a single-particle state \(\alpha\) is defined as
\begin{align}
\left\langle \hat{N}_\alpha \right\rangle &= \langle C^{\dagger}_\alpha C_\alpha \rangle = \notag \\ 
&= \dfrac{1}{Z} \, \mathrm{Tr} \left[ e^{-\beta(\hat{\mathcal{H}}_0 - \mu \hat{N})} C^{\dagger}_\alpha C_\alpha \right],
\end{align}
where \( Z = \mathrm{Tr}[e^{-\beta(\hat{\mathcal{H}} - \mu \hat{N})}] \) is the grand canonical partition function. Since the system is non-interacting, the Hamiltonian can be written as a sum of independent terms, i.e.,
\begin{equation}
\hat{\mathcal{H}}_0 - \mu \hat{N} = \sum_{\alpha} (\mathcal{E}_\alpha - \mu) \, C^{\dagger}_\alpha C_\alpha,
\end{equation}
As a consequence, the trace factorizes into a product over all single-particle states, and we can restrict ourselves to considering a single state \( \alpha \). We obtain
\begin{align}
\langle C^{\dagger}_\alpha C_\alpha \rangle &= \frac{1}{Z_\alpha} \sum_{N_\alpha} N_\alpha \, e^{-\beta(\mathcal{E}_\alpha - \mu) N_\alpha}, \\
Z_\alpha &= \sum_{N_\alpha} e^{-\beta(\mathcal{E}_\alpha - \mu) N_\alpha},
\end{align}
where the sum over the values of \( N_\alpha \) depends on the statistics of the particles.
\begin{itemize}
\item Fermions. The possible occupation numbers are \( N_\alpha = 0, 1 \), therefore
\begin{equation}
Z_\alpha^{(f)} = 1 + e^{-\beta(\mathcal{E}_\alpha - \mu)}, 
\end{equation}
\begin{align}
\langle C^{\dagger}_\alpha C_\alpha \rangle &= \frac{e^{-\beta(\mathcal{E}_\alpha - \mu)}}{1 + e^{-\beta(\mathcal{E}_\alpha - \mu)}} = \notag \\
&= \frac{1}{e^{\beta(\mathcal{E}_\alpha - \mu)} + 1}.
\end{align}
\item Bosons. The possible values of \( N_\alpha \) are \( 0, 1, 2, \dots \), therefore
\begin{align}
Z_\alpha^{(b)} &= \sum_{N_\alpha=0}^{\infty} e^{-\beta(\mathcal{E}_\alpha - \mu) N_\alpha} = \notag \\
&= \frac{1}{1 - e^{-\beta(\mathcal{E}_\alpha - \mu)}} \quad (\mu < \mathcal{E}_\alpha), 
\end{align}
\begin{equation}
\sum_{N_\alpha=0}^{\infty} N_\alpha \, e^{-\beta(\mathcal{E}_\alpha - \mu) N_\alpha} = \frac{e^{-\beta(\mathcal{E}_\alpha - \mu)}}{\left(1 - e^{-\beta(\mathcal{E}_\alpha - \mu)}\right)^2}, 
\end{equation}
\begin{align}
\langle C^{\dagger}_\alpha C_\alpha \rangle &= \frac{e^{-\beta(\mathcal{E}_\alpha - \mu)}}{(1 - e^{-\beta(\mathcal{E}_\alpha - \mu)})^2} (1 - e^{-\beta(\mathcal{E}_\alpha - \mu)}) = \notag \\
&= \frac{1}{e^{\beta(\mathcal{E}_\alpha - \mu)} - 1}.
\end{align}
\end{itemize}
Finally, we can write the result in a unified form as
\begin{align}
\left\langle \hat{N}_\alpha \right\rangle &= \dfrac{1}{e^{\beta(\mathcal{E}_\alpha - \mu)} \pm 1} = \notag \\ 
&= n_{\varepsilon}(\mathcal{E}_\alpha),
\label{eq: mediatermicanumerodioccupazioneedistribuzioneFermiBose}
\end{align}
where $n_{\varepsilon}(\mathcal{E}_\alpha)$ is given by $\eqref{eq: distribuzioneBoseFermiunificata}$, that is, it is equal to the Fermi-Dirac or Bose-Einstein statistics, depending on the nature of the particles. Finally, note that from $\eqref{eq: mediatermicanumerodioccupazioneedistribuzioneFermiBose}$ we can write
\begin{equation}
\left\langle a^\dagger_{\alpha'} a_{\alpha} \right\rangle = \delta_{\alpha,\alpha'} n_{\varepsilon}(\mathcal{E}_\alpha),
\label{eq: mediatermicaadaggeraconalphadiversi}
\end{equation}
which expresses that the thermal average is diagonal in the one-particle basis.
\section{Time evolution of creation and annihilation operators in the interaction picture}
Here we compute the interaction picture evolution of the annihilation and creation operators with respect to the Hamiltonian $\eqref{eq: Hamiltonianasecondaquantizzazioneparticellelibere}$, as defined in $\eqref{eq: interactionpicture}$, within the framework of second quantization. Let us begin with the annihilation operator \( a_\alpha \). Its time evolution in the interaction picture is given by
\begin{equation}
a^{(0)}_\alpha(t) = e^{i \frac{\hat{\mathcal{H}}_0}{\hslash} t} a_\alpha e^{- i \frac{\hat{\mathcal{H}}_0}{\hslash} t},
\end{equation}
and let us compute its time first derivative, i.e.,
\begin{align}
\dfrac{da^{(0)}_{\alpha}(t)}{dt} &= \dfrac{d\left( e^{i \frac{\hat{\mathcal{H}}_0}{\hslash} t} a_\alpha e^{- i \frac{\hat{\mathcal{H}}_0}{\hslash} t} \right)}{dt} = \notag \\
&= \dfrac{de^{i \frac{\hat{\mathcal{H}}_0}{\hslash} t}}{dt} a_\alpha e^{- i \frac{\hat{\mathcal{H}}_0}{\hslash} t} + e^{i \frac{\hat{\mathcal{H}}_0}{\hslash} t} \dfrac{d a_\alpha}{dt} e^{- i \frac{\hat{\mathcal{H}}_0}{\hslash} t} + e^{i \frac{\hat{\mathcal{H}}_0}{\hslash} t} a_\alpha \dfrac{d e^{- i \frac{\hat{\mathcal{H}}_0}{\hslash} t}}{dt} = \notag \\
&= \dfrac{de^{i \frac{\hat{\mathcal{H}}_0}{\hslash} t}}{dt} a_\alpha e^{- i \frac{\hat{\mathcal{H}}_0}{\hslash} t} + e^{i \frac{\hat{\mathcal{H}}_0}{\hslash} t} a_\alpha \dfrac{d e^{- i \frac{\hat{\mathcal{H}}_0}{\hslash} t}}{dt} = \notag \\
&= \dfrac{i}{\hslash} \hat{\mathcal{H}}_0 e^{i \frac{\hat{\mathcal{H}}_0}{\hslash} t} a_\alpha e^{- i \frac{\hat{\mathcal{H}}_0}{\hslash} t} - \dfrac{i}{\hslash} e^{i \frac{\hat{\mathcal{H}}_0}{\hslash} t} a_\alpha \hat{\mathcal{H}}_0 e^{-i \frac{\hat{\mathcal{H}}_0}{\hslash} t} = \notag \\
&= \dfrac{i}{\hslash} \hat{\mathcal{H}}_0 e^{i \frac{\hat{\mathcal{H}}_0}{\hslash} t} a_\alpha e^{- i \frac{\hat{\mathcal{H}}_0}{\hslash} t} - \dfrac{i}{\hslash} e^{i \frac{\hat{\mathcal{H}}_0}{\hslash} t} a_\alpha e^{-i \frac{\hat{\mathcal{H}}_0}{\hslash} t} \hat{\mathcal{H}}_0 = \notag \\
&= \dfrac{i}{\hslash} \hat{\mathcal{H}}_0 a^{(0)}_{\alpha}(t) - \dfrac{i}{\hslash} a^{(0)}_{\alpha}(t) \hat{\mathcal{H}}_0,
\end{align}
therefore, we find the Heisenberg equation of motion for the annihilation operator
\begin{equation}
\dfrac{da^{(0)}_{\alpha}(t)}{dt} = - \dfrac{i}{\hslash} \left[a_\alpha(t),\hat{\mathcal{H}}_0 \right],
\label{eq: evoluzionetemporaleoperatoredistruzione}
\end{equation}
that is, the evolution time of the annihilation operator is a commutator regardless of the statistical nature of the particle. Now, the time evolution equation of the annihilation operator $\eqref{eq: evoluzionetemporaleoperatoredistruzione}$ has an exact solution, indeed, from the equation $\eqref{eq: commutatoreoperatorenumeroconoperatoredistruzione}$ it follows that the time evolution of the annihilation operator of non-interacting particles is given by
\begin{equation}
\dfrac{d {a}^{(0)}_\alpha(t)}{dt} = - \dfrac{i}{\hslash} \mathcal{E}_\alpha a_\alpha,
\end{equation}
\begin{equation}
a^{(0)}_\alpha(t) = e^{-i \frac{\mathcal{E}_\alpha}{\hslash} t} a_\alpha.
\label{eq: evoluzionetemporaleoperatoredistruzione1}
\end{equation}
Now, consider the interaction picture of the creation operator $a_\alpha$, i.e.,
\begin{equation}
a^{\dagger (0)}_\alpha(t) = e^{i \frac{\hat{\mathcal{H}}_0}{\hslash} t} a^\dag_\alpha e^{- i \frac{\hat{\mathcal{H}}_0}{\hslash} t},
\end{equation}
then, similarly to the annihilation operator, its time evolution is described by
\begin{equation}
\dfrac{da^{\dagger (0)}_{\alpha}(t)}{dt} = - \dfrac{i}{\hslash} \left[ a^\dag_\alpha(t),\hat{\mathcal{H}}_0 \right],
\label{eq: evoluzionetemporaleoperatorecreazione}
\end{equation}
and again, from the equation $\eqref{eq: commutatoreoperatorenumeroconoperatorecreazione}$ it follows that the time evolution of the creation operator of non-interacting particles is given by
\begin{equation}
\dfrac{d {a}^{\dagger (0)}_\alpha(t)}{dt} = \dfrac{i}{\hslash} \mathcal{E}_\alpha a^\dag_\alpha,
\end{equation}
\begin{equation}
a^{\dagger (0)}_\alpha(t) = e^{i \frac{\mathcal{E}_\alpha}{\hslash} t} a^\dag_\alpha.
\label{eq: evoluzionetemporaleoperatorecreazione1}
\end{equation}
\section{Particle density (fluctuation) operator}
The particle density fluctuation operator is defined as follows
\begin{equation}
\rho(\textbf{r}) = \sum_{i=1}^N \delta \left( \textbf{r} - \textbf{r}_{i} \right),
\label{eq: operatorefluttuazionedensitàparticelle1quantizzazione}
\end{equation}
and we are now interested in expressing it in the formalism of second quantization. We get
\begin{equation}
\hat{\rho}(\textbf{r}) = \sum_{s'=-S}^{S} \int d\textbf{r}' \hat{\psi}^\dag(\textbf{r}',s') \delta \left( \textbf{r}' - \textbf{r} \right) \hat{\psi}(\textbf{r}',s'),
\end{equation}
that is,
\begin{equation}
\hat{\rho}(\textbf{r}) = \sum_{s=-S}^{S} \hat{\psi}^\dag(\textbf{r},s) \hat{\psi}(\textbf{r},s).
\label{eq: operatorefluttuazionedensitàparticelle2quantizzazione}
\end{equation}
The particle density fluctuation operator counts how many particles are in $\textbf{r}$, regardless of the spin. Now, we calculate the Fourier transform of the particle density operator, that is
\begin{equation}
\hat{\rho}_{\textbf{q}} = \int d^3 \textbf{r} e^{i \textbf{q} \cdot \textbf{r}} \hat{\rho}(\textbf{r}) ,
\label{eq: definizionegeneraletrasformataFourieroperatorefluttuazionedensitaparticelle}
\end{equation}
which thanks to $\eqref{eq: operatorefluttuazionedensitàparticelle2quantizzazione}$ becomes
\begin{equation}
\hat{\rho}_{\textbf{q}} = \sum_{s=-S}^{S} \int d^3 \textbf{r} e^{i \textbf{q} \cdot \textbf{r}} \hat{\psi}^\dag(\textbf{r},s) \hat{\psi}(\textbf{r},s).
\end{equation}
By means of the expansion $\eqref{eq: operatorecampodistruzionebaseondepianeperspin}$, $\eqref{eq: operatorecampocreazionebaseondepianeperspin}$, we have
\begin{align}
\hat{\rho}_{\textbf{q}} &= \dfrac{1}{V} \sum_{s=-S}^{S} \int d^3 \textbf{r} e^{i \textbf{q} \cdot \textbf{r}} \sum_{\textbf{k}_1,\sigma_1} \sum_{\textbf{k}_2,\sigma_2} e^{- i \textbf{k}_1 \cdot \textbf{r}} e^{i \textbf{k}_2 \cdot \textbf{r}} \chi^*_{\sigma_1}(s) \chi_{\sigma_2}(s) C^\dagger_{\textbf{k}_1,\sigma_1} C_{\textbf{k}_2,\sigma_2} = \notag \\
&= \dfrac{1}{V} \sum_{\textbf{k}_1,\sigma_1} \sum_{\textbf{k}_2,\sigma_2} \int d^3 \textbf{r} e^{- i (\textbf{k}_1 - \textbf{k}_2 - \textbf{q}) \cdot \textbf{r}} \sum_{s=-S}^{S} \chi^*_{\sigma_1}(s) \chi_{\sigma_2}(s) C^\dagger_{\textbf{k}_1,\sigma_1} C_{\textbf{k}_2,\sigma_2} = \notag \\
&= \dfrac{1}{V} \sum_{\textbf{k}_1,\sigma_1} \sum_{\textbf{k}_2,\sigma_2} V \delta_{\textbf{k}_1 , \textbf{k}_2 + \textbf{q}} \delta_{\sigma_1,\sigma_2} C^\dagger_{\textbf{k}_1,\sigma_1} C_{\textbf{k}_2,\sigma_2} = \notag \\
&= \sum_{\textbf{k},\sigma} C^\dagger_{\textbf{k} + \textbf{q},\sigma} C_{\textbf{k},\sigma},
\label{eq: trasformatadiFourieroperatoredensitadifluttuazioneparticelle}
\end{align}
where in the last steps, by means of the sums with respect to $\textbf{k}_1$, $\sigma_1$, the deltas imply $\sigma_1=\sigma_2$ and $\textbf{k}_1=\textbf{k}_2=\textbf{q}$, and we renamed the dummy indexes $\textbf{k}_2 \rightarrow \textbf{k}$, $\sigma_2 \rightarrow \sigma$. Finally, note that
\begin{equation}
\hat{\rho}^\dag_{\textbf{q}} = \sum_{\textbf{k},\sigma} C^\dag_{\textbf{k},\sigma} C_{\textbf{k}+\textbf{q},\sigma} ,
\end{equation}
and if we set $\textbf{k}'=\textbf{k}+\textbf{q}$, we finally have
\begin{equation}
\hat{\rho}^\dag_{\textbf{q}} = \sum_{\textbf{k}',\sigma} C^\dagger_{\textbf{k}' - \textbf{q},\sigma} C_{\textbf{k}',\sigma},
\end{equation}
which implies
\begin{equation}
\hat{\rho}^\dag_{\textbf{q}} = \hat{\rho}_{-\textbf{q}}.
\label{eq: aggiuntotrasformatadiFourieroperatoredensitadifluttuazioneparticelle}
\end{equation}
\begin{remark}[Why particle density "fluctuation" operator?] Here, we explain the motivation behind the naming of the particle density fluctuation operator $\eqref{eq: operatorefluttuazionedensitàparticelle2quantizzazione}$. In particular, given the equation $\eqref{eq: definizionegeneraletrasformataFourieroperatorefluttuazionedensitaparticelle}$, we define
\begin{equation}
\Delta \hat{\rho}(\mathbf{r}) = \hat{\rho}(\mathbf{r}) - \langle \hat{\rho}(\mathbf{r}) \rangle.
\end{equation}
For a homogeneous medium, the expectation value is independent of the position \(\mathbf{r}\), so that
\begin{equation}
\langle \hat{\rho}(\mathbf{r}) \rangle = \langle \hat{\rho} \rangle,
\end{equation}
and consequently, we have
\begin{align}
\Delta \hat{\rho}_{\textbf{q}} &= \int d^3 \textbf{r} \Delta \hat{\rho}(\textbf{r}) e^{i \textbf{q} \cdot \textbf{r}} = \notag \\
&= \hat{\rho}_{\textbf{q}} - \langle \hat{\rho} \rangle \int d^3 \textbf{r} e^{i \textbf{q} \cdot \textbf{r}} ,
\end{align}
but the integral $\int d^3 \textbf{r} e^{i \textbf{q} \cdot \textbf{r}}$ is nonzero only for $\textbf{q}=\textbf{0}$, then
\begin{equation}
\Delta \hat{\rho}_{\textbf{q}} = \hat{\rho}_{\textbf{q}}, \ \textbf{q} \neq \textbf{0}.
\end{equation}
This is the reason why the Fourier component \(\hat{\rho}_{\mathbf{q}}\) (for \(\mathbf{q} \neq 0\)) is also referred to as the particle density fluctuation operator.
\end{remark}
Regardless of the fermionic or bosonic nature of the creation and annihilation operators in $\eqref{eq: trasformatadiFourieroperatoredensitadifluttuazioneparticelle}$, $\eqref{eq: aggiuntotrasformatadiFourieroperatoredensitadifluttuazioneparticelle}$, $\hat{\rho}_{\textbf{q}}$ and $\hat{\rho}_{\textbf{q}}^\dag$ are bosonic operators, since they are given by a product of an even number of operators, then they do not change sign under the action of a permutation. \newline
We now present some useful relations involving the fluctuation operator \(\hat{\rho}_{\mathbf{q}}\), which will be instrumental in various applications.
\begin{theorem}[Calculation of some commutators involving the Fourier transform of the particle density operator] Let $\mathcal{E}_{\mathbf{k}}$ be an eigenvalue of the non-interacting many-body Hamiltonian, for instance $\mathcal{E}_{\mathbf{k}} = \frac{\hbar^2 \textbf{k}^2}{2m} - \mu$, then, the Fourier transform of particle density, the operator $\eqref{eq: trasformatadiFourieroperatoredensitadifluttuazioneparticelle}$ satisfies
\begin{itemize}
\item[1)]
\begin{equation}
\left[ \hat{\rho}_{\textbf{q}} , \hat{\rho}_{\textbf{q}}^\dag \right] = 0 ;
\label{eq: trasformatadiFourieroperatoredensitadifluttuazioneparticellecommutatore1}
\end{equation}
\item[2)]
\begin{equation}
\left[ \hat{\rho}_{\textbf{q}} , \sum_{\textbf{k},\sigma} \mathcal{E}_{\textbf{k}} C^\dag_{\textbf{k},\sigma} C_{\textbf{k},\sigma} \right] = \sum_{\textbf{k},\sigma} \left( \mathcal{E}_{\textbf{k}} - \mathcal{E}_{\textbf{k}+\textbf{q}} \right) C^\dag_{\textbf{k}+\textbf{q},\sigma} C_{\textbf{k},\sigma} ;
\label{eq: trasformatadiFourieroperatoredensitadifluttuazioneparticellecommutatore2}
\end{equation}
\item[3)]
\begin{equation}
\left[ \left[ \hat{\rho}_{\textbf{q}} , \sum_{\textbf{k},\sigma} \mathcal{E}_{\textbf{k}} C^\dag_{\textbf{k},\sigma} C_{\textbf{k},\sigma} \right], \hat{\rho}_{\textbf{q}}^\dag \right] = \sum_{\textbf{k},\sigma} \left( \mathcal{E}_{\textbf{k}-\textbf{q}} - 2 \mathcal{E}_{\textbf{k}} + \mathcal{E}_{\textbf{k}+\textbf{q}} \right) C^\dag_{\textbf{k},\sigma} C_{\textbf{k},\sigma} ;
\label{eq: trasformatadiFourieroperatoredensitadifluttuazioneparticellecommutatore3}
\end{equation}
\item[4)]
\begin{equation}
\left[ \hat{\rho}_{\textbf{q}} , \sum_{\textbf{k},\sigma} C^\dag_{\textbf{k},\sigma} C_{\textbf{k},\sigma} \right] = 0 ,
\label{eq: trasformatadiFourieroperatoredensitadifluttuazioneparticellecommutatore4}
\end{equation}
which has a clear physical interpretation: the density operator $\hat{\rho}_{\textbf{q}}$ conserves the total number of particles, it represents a fluctuation in the spatial distribution of the particle number, rather than a change in the total particle count;
\item[5)]
\begin{equation}
\left[ \hat{\rho}_{\textbf{q}} , \sum_{\textbf{q}'} \hat{\rho}_{\textbf{q}'} \hat{\rho}^{\dagger}_{\textbf{q}'} \right] = 0.
\label{eq: trasformatadiFourieroperatoredensitadifluttuazioneparticellecommutatore5}
\end{equation}
\end{itemize}
\begin{proof}
Each proof makes use of the Jacobi identity $\eqref{eq: Jacobiidentity}$, and/or one of the commutation relations $\eqref{eq: commutatoreaalphaadaggerbetaagamma}$, $\eqref{eq: commutatoreadaggeralphaadaggerbetaagamma}$.
\begin{itemize}
\item[1)]
\begin{align}
\left[ \hat{\rho}_{\textbf{q}} , \hat{\rho}_{\textbf{q}}^\dag \right] &= \sum_{\substack{\textbf{k}',\textbf{k}'' \\ \sigma',\sigma''}} \left[ C^\dag_{\textbf{k}'+\textbf{q},\sigma'} C_{\textbf{k}',\sigma'} , C^\dag_{\textbf{k}''-\textbf{q},\sigma''} C_{\textbf{k}'',\sigma''} \right] = \notag \\
&= \sum_{\substack{\textbf{k}',\textbf{k}'' \\ \sigma',\sigma''}} \left( C^\dag_{\textbf{k}'+\textbf{q},\sigma'} \left[ C_{\textbf{k}',\sigma'} , C^\dag_{\textbf{k}''-\textbf{q},\sigma''} C_{\textbf{k}'',\sigma''} \right] + \left[ C^\dag_{\textbf{k}'+\textbf{q},\sigma'} , C^\dag_{\textbf{k}''-\textbf{q},\sigma''} C_{\textbf{k}'',\sigma''} \right] C_{\textbf{k}',\sigma'} \right) = \notag \\
&= \sum_{\substack{\textbf{k}',\textbf{k}'' \\ \sigma',\sigma''}} \left( C^\dag_{\textbf{k}'+\textbf{q},\sigma'} \delta_{\textbf{k}',\textbf{k}''-\textbf{q}} \delta_{\sigma',\sigma''} C_{\textbf{k}'',\sigma''} - \delta_{\textbf{k}'+\textbf{q},\textbf{k}''} \delta_{\sigma',\sigma''} C^\dag_{\textbf{k}'' - \textbf{q},\sigma''} C_{\textbf{k}',\sigma'} \right),
\end{align}
we sum with respect to $\textbf{k}'$, $\sigma'$ in the first term so that deltas imply $\textbf{k}'=\textbf{k}'' - \textbf{q}$, $\sigma'=\sigma''$, and we sum with respect to $\textbf{k}''$, $\sigma''$ in the second term, so that deltas imply $\textbf{k}''=\textbf{k}' + \textbf{q}$, $\sigma'' = \sigma'$, then
\begin{equation}
\left[ \hat{\rho}_{\textbf{q}} , \hat{\rho}_{\textbf{q}}^\dag \right] =  \sum_{\textbf{k}'',\sigma''} C^\dag_{\textbf{k}'',\sigma''} C_{\textbf{k}'',\sigma''} - \sum_{\textbf{k}',\sigma'} C^\dag_{\textbf{k}',\sigma'} C_{\textbf{k}',\sigma'} ,
\end{equation}
which is the thesis.
\item[2)]
\begin{align}
\left[ \hat{\rho}_{\textbf{q}} , \sum_{\textbf{k},\sigma} \mathcal{E}_{\textbf{k}} C^\dag_{\textbf{k},\sigma} C_{\textbf{k},\sigma} \right] &= \sum_{\textbf{k},\sigma} \mathcal{E}_{\textbf{k}} \left[ \hat{\rho}_{\textbf{q}} , C^\dag_{\textbf{k},\sigma} C_{\textbf{k},\sigma} \right] = \notag \\
&= \sum_{\textbf{k},\sigma} \mathcal{E}_{\textbf{k}} \left[ \sum_{\textbf{k}',\sigma'} C^\dag_{\textbf{k}'+\textbf{q},\sigma'} C_{\textbf{k}',\sigma'},C^\dag_{\textbf{k},\sigma} C_{\textbf{k},\sigma} \right] = \notag \\
&= \sum_{\textbf{k},\sigma} \mathcal{E}_{\textbf{k}} \sum_{\textbf{k}',\sigma'} \left( C^\dag_{\textbf{k}'+\textbf{q},\sigma'} \left[ C_{\textbf{k}',\sigma'},C^\dag_{\textbf{k},\sigma} C_{\textbf{k},\sigma} \right] + \left[ C^\dag_{\textbf{k}'+\textbf{q},\sigma'},C^\dag_{\textbf{k},\sigma}C_{\textbf{k},\sigma} \right] C_{\textbf{k}',\sigma'} \right) = \notag \\
&= \sum_{\textbf{k},\sigma} \mathcal{E}_{\textbf{k}} \sum_{\textbf{k}',\sigma'} \left( C^\dag_{\textbf{k}'+\textbf{q},\sigma'} \delta_{\textbf{k}',\textbf{k}} \delta_{\sigma',\sigma} C_{\textbf{k},\sigma} - \delta_{\textbf{k}'+\textbf{q},\textbf{k}} \delta_{\sigma',\sigma} C^\dag_{\textbf{k},\sigma} C_{\textbf{k}',\sigma'} \right),
\end{align}
we sum with respect to $\textbf{k}'$, $\sigma'$ in the first term so that deltas imply $\textbf{k}'=\textbf{k}$, $\sigma'=\sigma$, and we sum with respect to $\textbf{k}$, $\sigma$ in the second term, so that deltas imply $\textbf{k}=\textbf{k}' + \textbf{q}$, $\sigma = \sigma'$, then
\begin{equation}
\left[ \hat{\rho}_{\textbf{q}} , \sum_{\textbf{k},\sigma} \mathcal{E}_{\textbf{k}} C^\dag_{\textbf{k},\sigma} C_{\textbf{k},\sigma} \right] = \sum_{\textbf{k},\sigma} \mathcal{E}_{\textbf{k}} C^\dag_{\textbf{k}+\textbf{q},\sigma} C_{\textbf{k},\sigma} - \sum_{\textbf{k}',\sigma'} \mathcal{E}_{\textbf{k}'+\textbf{q}} C^\dag_{\textbf{k}'+\textbf{q},\sigma'} C_{\textbf{k}',\sigma'},
\end{equation}
and finally renaming the dummy indexes, we get the thesis.
\item[3)]
From $\eqref{eq: trasformatadiFourieroperatoredensitadifluttuazioneparticellecommutatore2}$ we have
\begin{align}
& \left[ \left[ \hat{\rho}_{\textbf{q}} , \sum_{\textbf{k},\sigma} \mathcal{E}_{\textbf{k}} C^\dag_{\textbf{k},\sigma} C_{\textbf{k},\sigma} \right], \hat{\rho}_{\textbf{q}}^\dag \right] = \notag \\
&= \left[ \sum_{\textbf{k}',\sigma'} \left( \mathcal{E}_{\textbf{k}'} - \mathcal{E}_{\textbf{k}'+\textbf{q}} \right) C^\dag_{\textbf{k}'+\textbf{q},\sigma'} C_{\textbf{k}',\sigma'} , \sum_{\textbf{k}'',\sigma''} C^\dag_{\textbf{k}''-\textbf{q},\sigma''} C_{\textbf{k}'',\sigma''} \right] = \notag \\
&= \sum_{\substack{\textbf{k}',\textbf{k}'' \\ \sigma',\sigma''}} \left( \mathcal{E}_{\textbf{k}'} - \mathcal{E}_{\textbf{k}'+\textbf{q}} \right) \left[ C^\dag_{\textbf{k}'+\textbf{q},\sigma'} C_{\textbf{k}',\sigma'} , C^\dag_{\textbf{k}''-\textbf{q},\sigma''} C_{\textbf{k}'',\sigma''} \right] = \notag \\
&= \sum_{\substack{\textbf{k}',\textbf{k}'' \\ \sigma',\sigma''}} \left( \mathcal{E}_{\textbf{k}'} - \mathcal{E}_{\textbf{k}'+\textbf{q}} \right) \left( C^\dag_{\textbf{k}'+\textbf{q},\sigma'} \left[ C_{\textbf{k}',\sigma'} , C^\dag_{\textbf{k}''-\textbf{q},\sigma''} C_{\textbf{k}'',\sigma''} \right] + \left[ C^\dag_{\textbf{k}'+\textbf{q},\sigma'} , C^\dag_{\textbf{k}''-\textbf{q},\sigma''} C_{\textbf{k}'',\sigma''} \right] C_{\textbf{k}',\sigma'} \right) = \notag \\
&= \sum_{\substack{\textbf{k}',\textbf{k}'' \\ \sigma',\sigma''}} \left( \mathcal{E}_{\textbf{k}'} - \mathcal{E}_{\textbf{k}'+\textbf{q}} \right) \left( C^\dag_{\textbf{k}'+\textbf{q},\sigma'} \delta_{\textbf{k}',\textbf{k}''-\textbf{q}} \delta_{\sigma',\sigma''} C_{\textbf{k}'',\sigma''} - \delta_{\textbf{k}'+\textbf{q},\textbf{k}''} \delta_{\sigma',\sigma''} C^\dag_{\textbf{k}''-\textbf{q},\sigma''} C_{\textbf{k}',\sigma'} \right).
\end{align}
we sum with respect to $\textbf{k}'$, $\sigma'$ in the first term so that deltas imply $\textbf{k}'=\textbf{k}'' - \textbf{q}$, $\sigma'=\sigma''$, and we sum with respect to $\textbf{k}''$, $\sigma''$ in the second term, so that deltas imply $\textbf{k}''=\textbf{k}' + \textbf{q}$, $\sigma'' = \sigma'$, then
\begin{align}
\left[ \left[ \hat{\rho}_{\textbf{q}} , \sum_{\textbf{k},\sigma} \mathcal{E}_{\textbf{k}} C^\dag_{\textbf{k},\sigma} C_{\textbf{k},\sigma} \right], \hat{\rho}_{\textbf{q}}^\dag \right] &= \sum_{\textbf{k}'',\sigma''} \left( \mathcal{E}_{\textbf{k}''-\textbf{q}} - \mathcal{E}_{\textbf{k}''} \right) C^\dag_{\textbf{k}'',\sigma''} C_{\textbf{k}'',\sigma''} - \sum_{\textbf{k}',\sigma'} \left( \mathcal{E}_{\textbf{k}'} - \mathcal{E}_{\textbf{k}'+\textbf{q}} \right) C^\dag_{\textbf{k}',\sigma'} C_{\textbf{k}',\sigma'},
\end{align}
and finally renaming the dummy indexes, we get the thesis.
\item[4)] Following steps analogous to those in proof of $\eqref{eq: trasformatadiFourieroperatoredensitadifluttuazioneparticellecommutatore2}$, except that energies are absent in this case, we find
\begin{align}
\left[ \hat{\rho}_{\textbf{q}} , \sum_{\textbf{k},\sigma} \mathcal{E}_{\textbf{k}} C^\dag_{\textbf{k},\sigma} C_{\textbf{k},\sigma} \right] &= \sum_{\substack{\textbf{k}',\textbf{k}'' \\ \sigma',\sigma''}} \left( C^\dag_{\textbf{k}'+\textbf{q},\sigma'} \delta_{\textbf{k}',\textbf{k}''-\textbf{q}} \delta_{\sigma',\sigma''} C_{\textbf{k}'',\sigma''} - \delta_{\textbf{k}'+\textbf{q},\textbf{k}''} \delta_{\sigma',\sigma''} C^\dag_{\textbf{k}''-\textbf{q},\sigma''} C_{\textbf{k}',\sigma'} \right) = \notag \\
&= \sum_{\textbf{k}'',\sigma''} C^\dag_{\textbf{k}'',\sigma''} C_{\textbf{k}'',\sigma''} - \sum_{\textbf{k}',\sigma'} C^\dag_{\textbf{k}',\sigma'} C_{\textbf{k}',\sigma'} ,
\end{align}
and finally renaming the dummy indexes, we get the thesis.
\item[5)]
\begin{align}
\left[ \hat{\rho}_{\textbf{q}} , \sum_{\textbf{q}'} \hat{\rho}_{\textbf{q}'} \hat{\rho}^{\dagger}_{\textbf{q}'} \right] &= \sum_{\textbf{q}'} \left[ \hat{\rho}_{\textbf{q}} , \hat{\rho}_{\textbf{q}'} \hat{\rho}^{\dagger}_{\textbf{q}'} \right] = \notag \\
&= \sum_{\textbf{q}'} \hat{\rho}_{\textbf{q}'} \left[ \hat{\rho}_{\textbf{q}} , \hat{\rho}^{\dagger}_{\textbf{q}'} \right] + \sum_{\textbf{q}'} \left[ \hat{\rho}_{\textbf{q}} , \hat{\rho}_{\textbf{q}'} \right] \hat{\rho}^{\dagger}_{\textbf{q}'},
\end{align}
from equation $\eqref{eq: trasformatadiFourieroperatoredensitadifluttuazioneparticellecommutatore1}$ and from $\left[ \hat{\rho}_{\textbf{q}} , \hat{\rho}_{\textbf{q}'} \right] = 0$, the thesis follows.
\end{itemize}
\end{proof}
\end{theorem}
\section{Representation in Fock space of two-body operators}
Consider the two-body operator
\begin{equation}
U_N = \frac{1}{2} \sum_{i \neq j}^N U(x_i, x_j),
\end{equation}
where the interaction potential satisfies the symmetry condition
\begin{equation}
U(x_i, x_j) = U(x_j, x_i).
\end{equation}
In complete analogy with the case of one-body operators, the operator \(U_N\) naturally induces an operator \(\hat U\) on Fock space \(\mathcal F\), defined through its action on each \(N\)-particle sector. More precisely, for two states \(\phi,\phi' \in \mathcal F\) we set
\begin{align}
\langle \phi | \hat{U} | \phi' \rangle_{\mathcal{F}} \equiv \sum_{N=2}^\infty \left(\phi^{(N)},U_N \phi'^{(N)}\right)_{H_{N}}.
\end{align}
Let us study the scalar product in the space $H_N$, that is,
\begin{equation}
\left(\phi^{(N)},U_N \phi'^{(N)}\right)_{H_N} = \dfrac{1}{2} \sum_{i \neq j}^{N} \int dx_1 \ldots \int dx_N \left( \phi^{(N)}(x_1,\ldots,x_N) \right)^{*} U(x_i,x_j) \phi'^{(N)}(x_1,\ldots,x_N),
\end{equation}
we fix an index $i$ and divide the sum in the following way $\dfrac{1}{2} \sum_{i=1}^N \left\lbrace \sum_{j<i} \ldots + \sum_{j>i} \ldots \right\rbrace$. For example, for $j<i$
\begin{align}
& \int dx_1 \ldots \int dx_{j-1} \int dx_{j+1} \ldots \int dx_{i-1} \int dx_{i+1} \ldots \int dx_N \notag \\
& \ \ \ \ \ \ \ \int dx_j \int dx_i \left( \phi^{(N)}(x_1,\ldots,x_{j},\ldots,x_i,\ldots,x_N) \right)^{*} U(x_i,x_j) \left( \phi'^{(N)}(x_1,\ldots,x_j,\ldots,x_i,\ldots,x_N) \right).
\end{align}
We place $x_j \equiv x$, $x_i \equiv x'$, rename the other $N-2$ variables as the sequence $x_1,\ldots,x_{N-2}$ and finally reorder the total sequence as $x,x',x_1,\ldots,x_{N-2}$. We get
\begin{equation}
\int dx_1 \ldots \int dx_{N-2} \int dx \int dx' \left( \phi^{(N)}(x,x',x_1,\ldots,x_{N-2}) \right)^{*} U(x',x) \phi'^{(N)}(x,x',x_1,\ldots,x_{N-2}).
\end{equation}
We proceed in the same way for $j>i$, since the variables are dummy and $U(x,x') = U(x',x)$. The two sums do not depend on the indices and return
\begin{equation}
\sum_{j < i} 1 = \frac{N(N-1)}{2},
\label{eq: sommaiminorejdi1ugualeN_N_1_diviso_2}
\end{equation}
then
\begin{align}
\left(\phi^{(N)},U_N\phi'^{(N)}\right)_{H_N} &= \dfrac{N(N-1)}{2} \int dx_1 \ldots \int dx_{N-2} \int dx \int dx' \notag \\
&\left( \phi^{(N)}(x,x',x_1,\ldots,x_{N-2}) \right)^{*} U(x,x')\phi'^{(N)}(x,x',x_1,\ldots,x_{N-2}) = \notag \\
&= \dfrac{1}{2} \int dx_1 \ldots \int dx_{N-2} \int dx \int dx' \sqrt{N}\sqrt{N-1} \left( \phi^{(N)}(x,x',x_1,\ldots,x_{N-2}) \right)^{*} \notag \\
&U(x,x') \sqrt{N} \sqrt{N-1} \phi'^{(N)}(x,x',x_1,\ldots,x_{N-2}) ,
\end{align}
where we have used
\begin{equation}
\dfrac{N(N-1)}{2} = \dfrac{\sqrt{N-1}\sqrt{N}\sqrt{N-1}\sqrt{N}}{2}.
\end{equation}
We insert
\begin{equation}
\left[ \hat{\psi}(x) \hat{\psi}(x') \phi \right]^{(N-2)}(x_1,\ldots,x_{N-2}) = \sqrt{N}\sqrt{N-1}\phi^{(N)}(x,x',x_1,\ldots,x_{N-2}),
\end{equation}
into the scalar product in \( H_N \), exchanging the order of integration between \(\int dx_1 \ldots dx_{N-2}\) and \(\int dx \int dx'\), thereby obtaining a scalar product in \( H_{N-2} \), that is,
\begin{align}
\langle \phi | \hat{U} | \phi' \rangle_{\mathcal{F}} &= \sum_{N=2}^\infty (\phi^{(N)},U_N \phi'^{(N)})_N = \notag \\
&= \dfrac{1}{2} \sum_{N=2}^\infty \int dx \int dx' \left( \left[\hat{\psi}(x)\hat{\psi}(x') |\phi\rangle \right]^{(N-2)} , U(x,x') \left[\hat{\psi}(x)\hat{\psi}(x') |\phi'\rangle \right]^{(N-2)}\right)_{H_{N-2}} = \notag \\
&= \sum_{N=0}^\infty \dfrac{1}{2} \int dx \int dx' \left( \left[\hat{\psi}(x)\hat{\psi}(x') |\phi\rangle \right]^{(N)} , U(x,x')\left[\hat{\psi}(x)\hat{\psi}(x')||\phi'\rangle\right]^{(N)}\right)_{H_{N}}.
\end{align}
Assuming that it is possible to reverse the order between $\sum_{N=0}^{\infty}$ and $\int dx \int dx'$, we have
\begin{align}
\langle \phi | \hat{U} | \phi' \rangle_{\mathcal{F}} &= \dfrac{1}{2} \int dx \int dx' \sum_{N=0}^\infty \left( \left[\hat{\psi}(x)\hat{\psi}(x') |\phi\rangle \right]^{(N)} , U(x,x')\left[\hat{\psi}(x)\hat{\psi}(x') |\phi'\rangle \right]^{(N)}\right)_{H_{N}} = \notag \\
&= \dfrac{1}{2} \int dx \int dx' \left( \hat{\psi}(x)\hat{\psi}(x') |\phi\rangle , U(x,x')\hat{\psi}(x)\hat{\psi}(x') |\phi'\rangle \right)_{\mathcal{F}} = \notag \\
&= \dfrac{1}{2} \int dx \int dx' \langle\phi| \hat{\psi}^{\dag}(x') \hat{\psi}^{\dag}(x) U(x,x') \hat{\psi}(x) \hat{\psi}(x') |\phi'\rangle_{\mathcal{F}},
\end{align}
we formed a scalar product in $\mathcal{F}$ and the obtained relation holds $\forall \phi, \phi' \in \mathcal{F}$. The two-body operator $U_N = \frac{1}{2} \sum_{i \neq j}^N U(x_i,x_j)$ with $N \geq 2$ and $U(x_i,x_j) = U(x_j,x_i)$ in second quantization in Fock space becomes
\begin{align}
\hat{U} = \dfrac{1}{2} \int dx \int dx' \ \hat{\psi}^\dag(x')\hat{\psi}^\dag(x) U(x,x')\hat{\psi}(x)\hat{\psi}(x').
\label{eq: operatoreaduecorpisecondaquantizzazione}
\end{align}
Note that the order of action of the creation field operators is mirrored by that of the annihilation field operators, as shown by $\eqref{eq: operatoreaduecorpisecondaquantizzazione}$. Given a basis $\lbrace \varphi_\alpha \rbrace \in H_1$, according to equation $\eqref{eq: operatoreaduecorpisecondaquantizzazione}$, we have
\begin{align}
\hat{U} &= \dfrac{1}{2} \sum_{\substack{\alpha_1,\alpha_2, \\ \alpha_3,\alpha_4}} \left[\int dx \int dx' \varphi_{\alpha_1}^*(x)\varphi_{\alpha_2}^*(x')U(x,x')\varphi_{\alpha_3}(x')\varphi_{\alpha_4}(x)\right]a_{\alpha_1}^\dag a_{\alpha_2}^\dag a_{\alpha_3}a_{\alpha_4} = \notag \\
&= \dfrac{1}{2} \sum_{\substack{\alpha_1,\alpha_2, \\ \alpha_3,\alpha_4}} \langle \alpha_1\alpha_2|U|\alpha_3\alpha_4 \rangle a_{\alpha_1}^\dag a_{\alpha_2}^\dag a_{\alpha_3}a_{\alpha_4},
\end{align}
which is independent of both the basis and the statistical nature of the particles.
\subsection{Two-body Hamiltonian}
In first quantization, a typical Hamiltonian of matter physics is given by
\begin{align}
\mathcal{H} &= \sum_{i=1}^N \left[ \dfrac{\hat{\textbf{p}}_i^2}{2m} + V(x_i) \right] + \dfrac{1}{2} \sum_{i \neq j} U(x_i,x_j) \equiv \notag \\
&\equiv \mathcal{H}_0 + \dfrac{1}{2}\sum_{i \neq j} U(x_i,x_j),
\end{align}
where $\sum_{i=1}^N V(x_i)$ is an external field and $\frac{1}{2} \sum_{i \neq j}^N U(x_i,x_j)$ is a two-body operator. In $H_1$ we use the basis of the eigenstates of $\mathcal{H}_0$, from which
\begin{equation}
\hat{\mathcal{H}} = \int dx \hat{\psi}^\dag(x) \mathcal{H}_0(x) \hat{\psi}(x) + \dfrac{1}{2} \int dx \int dx' \ \hat{\psi}^\dag(x)\hat{\psi}^\dag(x') U(x,x') \hat{\psi}(x')\hat{\psi}(x),
\label{eq: hamiltonianamateriainterazioneaduecorpi}
\end{equation}
\begin{align}
\hat{\mathcal{H}} = \sum_{\alpha_1,\alpha_2} \langle \alpha_1|\mathcal{H}_0|\alpha_2 \rangle a^\dag_{\alpha_1}a_{\alpha_2} + \dfrac{1}{2} \sum_{\substack{\alpha_1,\alpha_2, \\ \alpha_3,\alpha_4}} \langle \alpha_1 \alpha_2|U|\alpha_3 \alpha_4 \rangle a_{\alpha_1}^\dag a_{\alpha_2}^\dag a_{\alpha_3} a_{\alpha_4}.
\end{align}
The first term is diagonalizable; however, the eigenstates of $\hat{\mathcal{H}}_0$ are not eigenstates of the two-body operator. One therefore works in the basis of the eigenstates of $\hat{\mathcal{H}}_0$ and treats the two-body term as a perturbation. When considering interactions between particles, the eigenstates of the full Hamiltonian $\hat{\mathcal{H}}$ are expected to become linear combinations of Slater determinants or Slater permanents. The coefficients of these linear combinations are not known exactly, since the eigenvalue problem cannot be solved exactly, but they are the subject of study in Feynman's perturbation theory.
\section{Reduced density matrices}
Let us consider a system of $N$ particles and let $\eqref{eq: statopuro}$ be the pure state describing this system. The generic matrix element is given by
\begin{equation}
\rho(x_1,\ldots,x_N;x'_1,\ldots,x'_N) = \psi(x_1,\ldots,x_N) \psi^*(x'_1,\ldots,x'_N),
\end{equation}
whose diagonal elements are given by
\begin{equation}
\rho(x_1,\ldots,x_N) = \left| \psi(x_1,\ldots,x_N) \right|^2 .
\end{equation}
We define the first-order reduced density matrix, or single-particle reduced density matrix, as the object
\begin{equation}
\rho_1(x) = N \int dx_2 \int dx_3 \ldots \int d x_N \left| \psi(x,x_2,\ldots,x_N) \right|^2,
\label{eq: matricedensitaridottasingolaparticellaprimaquantizzazione}
\end{equation}
which represents the probability of finding one of the $N$ particles at $x$, that is, at position $\textbf{r}$ with spin component $s$. We have
\begin{theorem}[Representation of the single-particle reduced density matrix]
Let $\psi(x_1, \ldots, x_N)$ be a normalized wavefunction of a system of $N$ indistinguishable particles, then
\begin{equation}
\rho_1(x) = \dfrac{1}{N} \left\langle \sum_{i=1}^N \delta(x - x_i) \right\rangle,
\label{eq: rappresentazionedensitaridottasingolaparticella}
\end{equation}
where the expectation value $\langle \ldots \rangle$ is taken with respect to the state $\psi$.
\begin{proof}
Given the quantity
\begin{equation}
\left\langle \sum_{i=1}^N \delta(x - x_i) \right\rangle = \int dx_1 \cdots dx_N \left(\sum_{i=1}^N \delta(x - x_i)\right) |\psi(x_1, \ldots, x_N)|^2,
\end{equation}
since the particles are indistinguishable, the wavefunction is symmetric (for bosons) or antisymmetric (for fermions), and in both cases the squared absolute value is symmetric with respect to any permutation of the indices, consequently, the object
\begin{equation}
\int dx_1 \cdots dx_N \delta(x - x_i) |\psi(x_1, \ldots, x_N)|^2
\end{equation}
does not depend on $i$, so we can fix $i=1$ and write
\begin{align}
\left\langle \sum_{i=1}^N \delta(x - x_i) \right\rangle &= \sum_{i=1}^N \int dx_1 \cdots dx_N \delta(x - x_i) |\psi(x_1, \ldots, x_N)|^2 = \notag \\
&= N \int dx_1 \cdots dx_N \delta(x - x_1) |\psi(x_1, \ldots, x_N)|^2.
\end{align}
Integrating over the variable $x_1$, we obtain
\begin{equation}
\left\langle \sum_{i=1}^N \delta(x - x_i) \right\rangle = N \int dx_2 \cdots dx_N |\psi(x, x_2, \ldots, x_N)|^2,
\end{equation}
then
\begin{align}
\rho_1(x) &= N \int dx_2 \cdots dx_N |\psi(x, x_2, \ldots, x_N)|^2 = \notag \\
&= \left\langle \sum_{i=1}^N \delta(x - x_i) \right\rangle,
\end{align}
from which the statement follows.
\end{proof}
\end{theorem}
Note that the particle density operator in first quantization, i.e., equation $\eqref{eq: operatorefluttuazionedensitàparticelle1quantizzazione}$, does not appear in the theorem, since it is defined differently, that is, it is independent of spin. Thanks to $\eqref{eq: rappresentazionedensitaridottasingolaparticella}$, we define the projection of the single-particle reduced density matrix onto the $N$-particle Fock subspace as
\begin{align}
\tilde{\rho}_1(x) &= N \rho_1(x) = \notag \\
&= \left\langle \sum_{i=1}^N \delta(x - x_i) \right\rangle.
\end{align}
Similarly to the matrix $\eqref{eq: matricedensitaridottasingolaparticellaprimaquantizzazione}$, we define the two-body reduced density matrix, or pair reduced density matrix, as the object
\begin{equation}
\rho_2(x,x') = N (N-1) \int dx_3 \int dx_4 \ldots \int d x_N \left| \psi(x,x',x_3,\ldots,x_N) \right|^2,
\label{eq: matricedensitaridottacoppiaprimaquantizzazione}
\end{equation}
which is the probability of finding a particle at $x$, that is, at position $\textbf{r}$ with spin component $s$, and a particle at $x'$, that is, at position $\textbf{r}'$ with spin component $s'$. We have
\begin{theorem}[Representation of the two-body reduced density matrix]
Let $\psi(x_1, \ldots, x_N)$ be a normalized wavefunction of a system of $N$ indistinguishable particles, then
\begin{equation}
\rho_2(x,x') = \dfrac{1}{N(N-1)} \left\langle \sum_{i \ne j} \delta(x - x_i)\, \delta(x' - x_j) \right\rangle,
\label{eq: rappresentazionedensitaridottacoppia}
\end{equation}
where the expectation value $\langle \ldots \rangle$ is taken with respect to the state $\psi$.
\begin{proof}
Given the quantity
\begin{equation}
\left\langle \sum_{i \ne j} \delta(x - x_i)\, \delta(y - x_j) \right\rangle = \int dx_1 \cdots dx_N \left( \sum_{i \ne j} \delta(x - x_i)\, \delta(x' - x_j) \right) |\psi(x_1, \ldots, x_N)|^2,
\end{equation}
Since the particles are indistinguishable, the wavefunction is symmetric (for bosons) or antisymmetric (for fermions), and in both cases the squared absolute value is symmetric with respect to any permutation of the variables, it follows that the object
\begin{equation}
\int dx_1 \cdots dx_N\, \delta(x - x_i)\, \delta(x' - x_j)\, |\psi(x_1, \ldots, x_N)|^2
\end{equation}
does not depend on the pair of indices $i \neq j$, and given that the total number of ordered pairs $(i, j)$ with $i \ne j$ is $N(N - 1)$, we can choose an arbitrary pair, for example $i = 1$, $j = 2$, and we write
\begin{align}
\left\langle \sum_{i \ne j} \delta(x - x_i)\, \delta(x' - x_j) \right\rangle &= \sum_{i \ne j} \int dx_1 \cdots dx_N \delta(x - x_i)\, \delta(x' - x_j) |\psi(x_1, \ldots, x_N)|^2 = \notag \\
&= N(N - 1) \int dx_1 \cdots dx_N\, \delta(x - x_1)\, \delta(x' - x_2)\, |\psi(x_1, \ldots, x_N)|^2.
\end{align}
Integrating over the variables $x_1$, $x_2$, we obtain
\begin{equation}
\left\langle \sum_{i \ne j} \delta(x - x_i)\, \delta(x' - x_j) \right\rangle = N (N-1) \int dx_3 \cdots dx_N |\psi(x,x', x_3, \ldots, x_N)|^2,
\end{equation}
then
\begin{align}
\rho_2(x,x') &= N (N-1) \int dx_3 \cdots dx_N |\psi(x,x', x_3, \ldots, x_N)|^2 = \notag \\
&= \left\langle \sum_{i \ne j} \delta(x - x_i)\, \delta(x' - x_j) \right\rangle,
\end{align}
from which the statement follows.
\end{proof}
\end{theorem}
Thanks to $\eqref{eq: rappresentazionedensitaridottacoppia}$, we define the projection of the two-body reduced density matrix onto the $N$-particle Fock subspace as
\begin{align}
\tilde{\rho}_2(x,x') &= N (N-1) \rho_2(x,x') = \notag \\
&= \left\langle \sum_{i \ne j} \delta(x - x_i)\, \delta(x' - x_j) \right\rangle.
\end{align}
We now want to express the reduced matrices $\eqref{eq: matricedensitaridottasingolaparticellaprimaquantizzazione}$, $\eqref{eq: matricedensitaridottacoppiaprimaquantizzazione}$ in second quantization. From $\eqref{eq: rappresentazionedensitaridottasingolaparticella}$ and $\eqref{eq: rappresentazionedensitaridottacoppia}$, we have
\begin{align}
\tilde{\rho}_1(x) &= \int dx_1\, \hat{\psi}^\dagger(x_1) \left\langle \delta(x - x_1) \right\rangle \hat{\psi}(x_1) = \notag \\
&= \left\langle \int dx_1\, \hat{\psi}^\dagger(x_1) \delta(x - x_1) \hat{\psi}(x_1) \right\rangle = \notag \\
&= \left\langle \hat{\psi}^\dagger(x)\, \hat{\psi}(x) \right\rangle,
\label{eq : matricedensitaridottasingolaparticellasecondaquantizzazione}
\end{align}
\begin{align}
\tilde{\rho}_2(x,x') &= \dfrac{1}{2} \int dx_1 \int dx_2\, \hat{\psi}^\dagger(x_1)\, \hat{\psi}^\dagger(x_2)\, \left\langle \delta(x - x_1)\, \delta(x' - x_2) \right\rangle\, \hat{\psi}(x_2)\, \hat{\psi}(x_1) = \notag \\
&= \dfrac{1}{2} \left\langle \int dx_1 \int dx_2\, \hat{\psi}^\dagger(x_1)\, \hat{\psi}^\dagger(x_2)\, \delta(x - x_1)\, \delta(x' - x_2)\, \hat{\psi}(x_2)\, \hat{\psi}(x_1) \right\rangle = \notag \\
&= \dfrac{1}{2} \left\langle \hat{\psi}^\dagger(x)\, \hat{\psi}^\dagger(x')\, \hat{\psi}(x')\, \hat{\psi}(x) \right\rangle.
\label{eq: matricedensitaridottacoppiasecondaquantizzazione}
\end{align}
The quantities introduced above allow us to define the pair correlation function
\begin{equation}
g(x,x') = \dfrac{\left\langle \psi^\dagger(x) \psi^\dagger(x') \psi(x') \psi(x) \right\rangle}{\left\langle \psi^\dagger(x) \psi(x) \right\rangle \left\langle \psi^\dagger(x') \psi(x') \right\rangle},
\label{eq: funzionecorrelazionedicoppia}
\end{equation}
which is directly proportional to the joint probability of finding one particle at position \( \textbf{r} \) and another at \( \textbf{r}' \). In particular, when the two particles are sufficiently far apart, the numerator factorizes as
\begin{equation}
\left\langle \hat{\psi}^\dagger(\textbf{r}, s)\, \hat{\psi}^\dagger(\textbf{r}', s')\, \hat{\psi}(\textbf{r}', s')\, \hat{\psi}(\textbf{r}, s) \right\rangle \xrightarrow[\left| \textbf{r} - \textbf{r}' \right| \rightarrow +\infty]{} \left\langle \hat{\psi}^\dagger(\textbf{r}, s)\, \hat{\psi}(\textbf{r}, s) \right\rangle \left\langle \hat{\psi}^\dagger(\textbf{r}', s')\, \hat{\psi}(\textbf{r}', s') \right\rangle,
\end{equation}
then
\begin{equation}
\lim_{\left| \textbf{r}-\textbf{r}' \right| \rightarrow + \infty} g(\textbf{r},s,\textbf{r}',s') = 1.
\end{equation}
\section{Representation in Fock space of $n$-body operators}
An $n$-body operator in first quantization is expressed as
\begin{equation}
M_N(n) = \dfrac{1}{n!} \sum_{i_1 \neq i_2 \neq \ldots \neq i_n}^N W \left( x_{i_1}, \ldots , x_{i_n} \right),
\end{equation}
where the factor \( \frac{1}{n!} \) accounts for the \( n! \) identical terms
\begin{equation}
W \left( x_1, x_2, \ldots , x_n \right) = W \left( x_2, x_1, \ldots , x_n \right) = \ldots = W \left( x_n, \ldots , x_1 \right),
\end{equation}
which arise from the symmetry under permutations of the degrees of freedom. By generalizing the result obtained for the two-body case, an $n$-body operator with \( n \geq 2 \) can be written in second quantization in Fock space as
\begin{align}
\hat{M} = \dfrac{1}{n!} \int dx_1 \ldots \int dx_n \ \hat{\psi}^\dag(x_1) \ldots \hat{\psi}^\dag(x_n) W(x_1,\ldots,x_n)\hat{\psi}(x_n) \ldots \hat{\psi}(x_1).
\label{eq: operatoreancorpisecondaquantizzazione}
\end{align}
\section{Jellium model}
\subsection{The model in the first quantization}
Throughout this text, we will make extensive use of the jellium model, one of the simplest and most useful idealizations for describing electronic systems in solids. In this model, the ions are replaced by a uniformly distributed positive background that neutralizes the charge of the valence electrons. The absence of an explicit crystal structure allows for a clearer analysis of electron–electron interactions. Since the ionic mass is much greater than the electronic one, the positive background is assumed to be static. The jellium model thus provides a minimal yet effective theoretical framework, which will be used frequently in the analyses that follow. In first quantization, for a system of $N$ electrons in a uniform background, the Hamiltonian takes the form
\begin{align}
\mathcal{H} &= \sum_{i=1}^N \dfrac{\hat{\textbf{p}}_i^2}{2m_e} + \dfrac{1}{2}\sum_{i \neq j}^N \dfrac{q_e^2}{|\textbf{r}_i - \textbf{r}_j|} e^{-k_s|\textbf{r}_i - \textbf{r}_j|} + \dfrac{1}{2}\int d^3\textbf{R} \int d^3\textbf{R}' \dfrac{q_e^2}{|\textbf{R} - \textbf{R}'|}\rho(\textbf{R})\rho(\textbf{R}') e^{-k_s|\textbf{R}-\textbf{R}'|} + \notag \\
&- q_e^2 \sum_{i=1}^N \int d^3 \textbf{R} \dfrac{\rho(\textbf{R})}{|\textbf{R}-\textbf{r}_i|} e^{-k_s|\textbf{R}-\textbf{r}_i|} .
\end{align}
In this form, the Hamiltonian includes a screened Yukawa-type interaction instead of the bare Coulomb potential. The screening is controlled by the parameter $k_s >0$, which represents the inverse screening length. Larger values of \( k_s \) correspond to stronger screening (i.e., shorter-range interactions), while in the limit \( k_s \to 0 \), one recovers the standard unscreened Coulomb interaction, i.e.,
\begin{equation}
q_e^2 \frac{e^{-k_s|\textbf{r} - \textbf{r}'|}}{|\textbf{r} - \textbf{r}'|} \xrightarrow[k_s \to 0]{} \dfrac{q_e^2}{|\textbf{r} - \textbf{r}'|}.
\end{equation}
The function \( \rho(\mathbf{R}) \) denotes the positive background charge density, the so-called jellium background, that neutralizes the electron charge. Assuming a uniform distribution, the background density is taken to be constant, that is
\begin{align}
\rho(\textbf{R}) &= \rho = \notag \\
&= \dfrac{N}{V},
\end{align}
where \( V \) is the volume of the system. This condition ensures overall charge neutrality, which is essential to avoid infrared divergences and to guarantee that the total energy per unit volume remains finite. We consider the thermodynamic limit, where both the number \( N \) of electrons and the volume \( V \) tend to infinity, while keeping the electronic density constant. Finally, we take the limit \( k_s \rightarrow 0 \) to recover the long-range Coulomb interaction. The order of these limits is crucial and cannot be reversed. Due to periodic boundary conditions, the system is invariant under spatial translations. Consequently, the integrand in
\begin{equation}
\dfrac{1}{2} \int d^3\textbf{R} \int d^3\textbf{R}' \dfrac{q_e^2}{|\textbf{R} - \textbf{R}'|} \rho(\textbf{R})\rho(\textbf{R}') e^{-k_s|\textbf{R}-\textbf{R}'|} 
\end{equation}
depends only on the difference \( \textbf{R} - \textbf{R}' \), and not on the individual positions. Using \( \int d^3\textbf{R}' = V \), we find
\begin{equation}
\dfrac{1}{2} \int d^3\textbf{R} \int d^3\textbf{R}' \dfrac{q_e^2}{|\textbf{R} - \textbf{R}'|}\rho(\textbf{R})\rho(\textbf{R}')e^{-k_s|\textbf{R}-\textbf{R}'|} = \dfrac{1}{2} q_e^2 \left( \dfrac{N}{V} \right)^2 V \int d^3\textbf{R} \dfrac{e^{-k_s |\textbf{R}|}}{|\textbf{R}|}.
\end{equation}
Switching to spherical coordinates and using the identity
\begin{equation}
\int_0^{+\infty} x e^{-a x} \, dx = \dfrac{1}{a^2}, \quad a > 0,
\end{equation}
we evaluate the integral as follows
\begin{align}
\dfrac{1}{2} q_e^2 \left( \dfrac{N}{V} \right)^2 V \int d^3\textbf{R} \dfrac{e^{-k_s |\textbf{R}|}}{|\textbf{R}|} &= \dfrac{1}{2} q_e^2 \left( \dfrac{N}{V} \right)^2 V \int_0^\infty 4 \pi R^2 \dfrac{e^{-k_s R}}{R} \, dR = \notag \\
&= \dfrac{q_e^2}{2} \left( \dfrac{N}{V} \right)^2 V 4\pi \dfrac{1}{k_s^2} = \notag \\
&= \dfrac{q_e^2}{2} \dfrac{N^2}{V} \dfrac{4\pi}{k_s^2}.
\end{align}
Now consider the interaction between electrons and the background
\begin{equation}
- q_e^2 \sum_{i=1}^N \int d^3\textbf{R} \dfrac{\rho(\textbf{R})}{|\textbf{R}-\textbf{r}_i|} e^{-k_s|\textbf{R}-\textbf{r}_i|} = - q_e^2 \left( \dfrac{N}{V} \right) \sum_{i=1}^N \int d^3\textbf{R} \dfrac{e^{-k_s|\textbf{R}-\textbf{r}_i|}}{|\textbf{R}-\textbf{r}_i|}.
\end{equation}
The integral does not explicitly depend on the electronic position \( \textbf{r}_i \), due to translational invariance. Therefore, we may shift coordinates and write
\begin{align}
- q_e^2 \sum_{i=1}^N \int d^3\textbf{R} \dfrac{\rho(\textbf{R})}{|\textbf{R}-\textbf{r}_i|} e^{-k_s|\textbf{R}-\textbf{r}_i|} &= - q_e^2 \left( \dfrac{N}{V} \right) \sum_{i=1}^N \int d^3\textbf{R} \dfrac{e^{-k_s |\textbf{R}|}}{|\textbf{R}|} = \notag \\
&= - q_e^2 \dfrac{N^2}{V} \dfrac{4\pi}{k_s^2}.
\end{align}
Combining the two background-related terms, we obtain
\begin{equation}
\dfrac{1}{2}\int d^3\textbf{R} \int d^3\textbf{R}' \dfrac{q_e^2}{|\textbf{R} - \textbf{R}'|} \rho(\textbf{R}) \rho(\textbf{R}') e^{-k_s|\textbf{r}-\textbf{r}'|} - q_e^2 \sum_{i=1}^N \int d^3\textbf{R} \frac{\rho(\textbf{R})}{|\textbf{R}-\textbf{r}_i|}e^{-k_s|\textbf{R}-\textbf{r}_i|} = -\dfrac{q_e^2}{2} \dfrac{N^2}{V} \dfrac{4\pi}{k_s^2}.
\end{equation}
Therefore, the jellium Hamiltonian in first quantization is give by
\begin{equation}
\mathcal{H} = \sum_{i=1}^N \dfrac{\hat{\textbf{p}}_i^2}{2m_e} + \dfrac{1}{2} \sum_{i \neq j}^N \dfrac{q_e^2}{|\textbf{r}_i - \textbf{r}_j|} e^{-k_s|\textbf{r}_i - \textbf{r}_j|} - \dfrac{q_e^2}{2} \dfrac{N^2}{V} \dfrac{4\pi}{k_s^2}.
\end{equation}
\subsection{The model in the second quantization}
After expressing the Hamiltonian in first quantization, we aim to rewrite it in second quantization, and given that it is a two-body operator, we need to compute
\begin{equation}
\hat{\mathcal{H}} = \int dx \hat{\psi}^\dag(x) \dfrac{\hat{\textbf{p}}^2}{2m_e} \hat{\psi}(x) + \dfrac{1}{2} q_e^2 \int dx \int dx' \hat{\psi}^\dag(x) \hat{\psi}^\dag(x') \dfrac{e^{-k_s|\textbf{r}-\textbf{r}'|}}{|\textbf{r} - \textbf{r}'|} \hat{\psi}(x') \hat{\psi}(x) - \dfrac{q_e^2}{2} \dfrac{N^2}{V} \dfrac{4\pi}{k_s^2}.
\end{equation}
By means of the expansion $\eqref{eq: operatorecampodistruzionebaseondepianeperspin}$ $\eqref{eq: operatorecampocreazionebaseondepianeperspin}$, the kinetic operator is diagonal, then
\begin{align}
\hat{\mathcal{H}} &= \sum_{\textbf{k},\sigma} \dfrac{\hslash^2 \textbf{k}^2}{2m_e} C^\dag_{\textbf{k},\sigma} C_{\textbf{k},\sigma} + \dfrac{1}{2} \dfrac{q_e^2}{V^2} \sum_{\substack{\textbf{k}_1,\textbf{k}_2, \\ \textbf{k}_3,\textbf{k}_4}}\sum_{\substack{\sigma_1,\sigma_2, \\ \sigma_3,\sigma_4}} \sum_{s} \sum_{s'} \int d^3\textbf{r} \int d^3\textbf{r}' e^{-i \textbf{k}_1 \cdot \textbf{r}} e^{-i \textbf{k}_2 \cdot \textbf{r}'} e^{i \textbf{k}_3 \cdot \textbf{r}'} e^{i \textbf{k}_4 \cdot \textbf{r}} \notag \\
&\chi_{\sigma_1}^*(s) \chi_{\sigma_2}^*(s') \chi_{\sigma_3}(s') \chi_{\sigma_4}(s) \dfrac{e^{-k_s|\textbf{r}-\textbf{r}'|}}{|\textbf{r}-\textbf{r}'|} C_{\textbf{k}_1,\sigma_1}^\dag C_{\textbf{k}_2,\sigma_2}^\dag C_{\textbf{k}_3,\sigma_3} C_{\textbf{k}_4,\sigma_4} - \dfrac{q_e^2}{2} \dfrac{N^2}{V} \dfrac{4\pi}{k_s^2},
\end{align}
and given that the electron-electron interaction does not depend on spin variables, we can use the following orthonormality relations
\begin{equation}
\sum_{s} \chi_{\sigma_1}^*(s) \chi_{\sigma_4}(s) = \delta_{\sigma_1,\sigma_4},
\end{equation}
\begin{equation}
\sum_{s'} \chi^*_{\sigma_2}(s') \chi_{\sigma_3}(s') = \delta_{\sigma_2,\sigma_3},
\end{equation}
and we sum over $\sigma_3$, $\sigma_4$, from which it follows that $\sigma_1 = \sigma_4$ e $\sigma_2=\sigma_3$, that is,
\begin{align}
\hat{\mathcal{H}} &= \sum_{\textbf{k},\sigma} \dfrac{\hslash^2 \textbf{k}^2}{2m_e} C^\dag_{\textbf{k},\sigma} C_{\textbf{k},\sigma} + \dfrac{1}{2} \dfrac{q_e^2}{V^2} \sum_{\substack{\textbf{k}_1,\textbf{k}_2, \\ \textbf{k}_3,\textbf{k}_4}}\sum_{\sigma_1,\sigma_2} \int d^3\textbf{r} \int d^3\textbf{r}' e^{-i \textbf{k}_1 \cdot \textbf{r}} e^{-i \textbf{k}_2 \cdot \textbf{r}'} e^{i \textbf{k}_3 \cdot \textbf{r}'} e^{i \textbf{k}_4 \cdot \textbf{r}} \notag \\
& \dfrac{e^{-k_s|\textbf{r}-\textbf{r}'|}}{|\textbf{r}-\textbf{r}'|} C_{\textbf{k}_1,\sigma_1}^\dag C_{\textbf{k}_2,\sigma_2}^\dag C_{\textbf{k}_3,\sigma_2} C_{\textbf{k}_4,\sigma_1} - \dfrac{q_e^2}{2} \dfrac{N^2}{V} \dfrac{4\pi}{k_s^2} .
\end{align}
Thanks to translational invariance, we set $\textbf{r} = \textbf{r}' + \textbf{r}''$, which implies $d\textbf{r}=d\textbf{r}''$, and we integrate with respect to $\textbf{r}''$. Every function depending on $\textbf{r}$ becomes a function of $\textbf{r}' + \textbf{r}''$, thus
\begin{equation}
\dfrac{1}{2} \dfrac{q_e^2}{V^2} \sum_{\substack{\textbf{k}_1, \textbf{k}_2 \\ \textbf{k}_3, \textbf{k}_4}} \sum_{\sigma_1, \sigma_2} \int d^3\textbf{r}' \int d^3\textbf{r}'' e^{-i(\textbf{k}_1 + \textbf{k}_2 - \textbf{k}_3 - \textbf{k}_4) \cdot \textbf{r}'}  e^{-i(\textbf{k}_1 - \textbf{k}_4) \cdot \textbf{r}''} \frac{e^{-k_s |\textbf{r}''|}}{|\textbf{r}''|} C^\dagger_{\textbf{k}_1, \sigma_1} C^\dagger_{\textbf{k}_2, \sigma_2} C_{\textbf{k}_3, \sigma_2} C_{\textbf{k}_4, \sigma_1}.
\end{equation}
The integral over $\textbf{r}'$ yields a Kronecker delta showing the conservation of total momentum
\begin{equation}
\int d^3\textbf{r}' \; e^{-i(\textbf{k}_1 + \textbf{k}_2 - \textbf{k}_3 - \textbf{k}_4) \cdot \textbf{r}'} = V \delta_{\textbf{k}_1 + \textbf{k}_2,\; \textbf{k}_3 + \textbf{k}_4}.
\end{equation}
Thus, the electron-electron interaction can be rewritten as
\begin{equation}
\dfrac{1}{2} \dfrac{q_e^2}{V} \sum_{\substack{\textbf{k}_1, \textbf{k}_2 \\ \textbf{k}_3, \textbf{k}_4}} \sum_{\sigma_1, \sigma_2} \delta_{\textbf{k}_1 + \textbf{k}_2,\; \textbf{k}_3 + \textbf{k}_4} \left[ \int d^3\textbf{r}'' \frac{e^{-k_s |\textbf{r}''|}}{|\textbf{r}''|} e^{-i(\textbf{k}_1 - \textbf{k}_4) \cdot \textbf{r}''} \right] C^\dagger_{\textbf{k}_1, \sigma_1} C^\dagger_{\textbf{k}_2, \sigma_2} C_{\textbf{k}_3, \sigma_2} C_{\textbf{k}_4, \sigma_1}.
\end{equation}
Now we use the conservation of total momentum, that is, we set $\textbf{k}_1 = \textbf{k}_4 + \textbf{q}$, $\textbf{k}_2 = \textbf{k}_3 - \textbf{q}$, so that $\textbf{k}_3$, $\textbf{k}_4$ and $\textbf{q}$ are independent, while $\textbf{k}_1$ and $\textbf{k}_2$ are functions of $\textbf{k}_3$, $\textbf{k}_4$, $\textbf{q}$. We sum with respect to $\textbf{k}_1$ and $\textbf{k}_2$, and use $\delta_{\textbf{k}_1+\textbf{k}_2,\textbf{k}_3+\textbf{k}_4}$,
\begin{equation}
\dfrac{1}{2} \dfrac{q_e^2}{V} \sum_{\sigma_1 \sigma_2} \sum_{\substack{\textbf{k}_3,\textbf{k}_4, \\\textbf{q}}} C^\dag_{\textbf{k}_4 + \textbf{q},\sigma_1} C^\dag_{\textbf{k}_3 - \textbf{q},\sigma_2} C_{\textbf{k}_3,\sigma_2} C_{\textbf{k}_4,\sigma_1} \int d^3 \textbf{r} \frac{e^{-k_s|\textbf{r}|}}{|\textbf{r}|}e^{-i\textbf{q}\cdot\textbf{r}},
\end{equation}
and we use the Fourier transform of the Yukawa potential, i.e.,
\begin{equation}
\int d^3 \textbf{r} e^{-i\textbf{q}\cdot\textbf{r}} \dfrac{e^{-k_s|\textbf{r}|}}{|\textbf{r}|} = \dfrac{4\pi}{q^2+k_s^2},
\label{eq: trasformataFourierpotenzialeYukawa}
\end{equation}
and we have
\begin{equation}
\dfrac{1}{2} \dfrac{q_e^2}{V} \sum_{\sigma_1 \sigma_2} \sum_{\substack{\textbf{k}_3,\textbf{k}_4, \\\textbf{q}}} \dfrac{4\pi}{q^2+k_s^2} C^\dag_{\textbf{k}_4 + \textbf{q},\sigma_1} C^\dag_{\textbf{k}_3 - \textbf{q},\sigma_2} C_{\textbf{k}_3,\sigma_2} C_{\textbf{k}_4,\sigma_1}.
\end{equation}
Now, we decompose $\sum_{\textbf{q}}$ into a term with $\textbf{q}=\textbf{0}$ and one with $\textbf{q} \neq \textbf{0}$, we rename $\textbf{k}_3$ as $\textbf{k}_1$ and $\textbf{k}_4$ as $\textbf{k}_2$, 
\begin{align}
\hat{\mathcal{H}} &= \sum_{\textbf{k},\sigma} \frac{\hslash^2 \textbf{k}^2}{2m_e}C^\dag_{\textbf{k},\sigma}C_{\textbf{k},\sigma} + \dfrac{q_e^2}{2V}\sum_{\substack{\textbf{k}_1,\textbf{k}_2, \\ \textbf{q}\neq \textbf{0}}} \sum_{\sigma_1 \sigma_2} \dfrac{4\pi}{q^2 + k_s^2}C^\dag_{\textbf{k}_1+\textbf{q},\sigma_1}C^\dag_{\textbf{k}_2-\textbf{q},\sigma_2}C_{\textbf{k}_2,\sigma_2}C_{\textbf{k}_1,\sigma_1} + \notag \\
&+ \dfrac{q_e^2}{2V} \sum_{\textbf{k}_1,\textbf{k}_2} \sum_{\sigma_1,\sigma_2}\frac{4\pi}{k^2_s} C^\dag_{\textbf{k}_1,\sigma_1}C^\dag_{\textbf{k}_2,\sigma_2}C_{\textbf{k}_2,\sigma_2}C_{\textbf{k}_1,\sigma_1} - \dfrac{q_e^2}{2} \dfrac{N^2}{V} \frac{4\pi}{k_s^2}.
\end{align}
From
\begin{align}
C^\dag_{\textbf{k}_1,\sigma_1}C^\dag_{\textbf{k}_2,\sigma_2}C_{\textbf{k}_2,\sigma_2}C_{\textbf{k}_1,\sigma_1} &= -C_{\textbf{k}_2,\sigma_2}^\dag C_{\textbf{k}_1,\sigma_1}^\dag C_{\textbf{k}_2,\sigma_2}C_{\textbf{k}_1,\sigma_1} = \notag \\
&= - C^\dag_{\textbf{k}_2,\sigma_2} \left[ \delta_{\textbf{k}_1,\textbf{k}_2}\delta_{\sigma_1,\sigma_2} - C_{\textbf{k}_2,\sigma_2}C_{\textbf{k}_1,\sigma_1}^\dag \right] C_{\textbf{k}_1,\sigma_1},
\end{align}
we have
\begin{equation}
\dfrac{q_e^2}{2V}\sum_{\textbf{k}_1,\textbf{k}_2} \sum_{\sigma_1,\sigma_2} \dfrac{4\pi}{k^2_s} C^\dag_{\textbf{k}_1,\sigma_1} C^\dag_{\textbf{k}_2,\sigma_2} C_{\textbf{k}_2,\sigma_2} C_{\textbf{k}_1,\sigma_1} = - \dfrac{q_e^2}{2V}\frac{4\pi}{k_s^2} \left[ \sum_{\textbf{k},\sigma}C_{\textbf{k},\sigma}^\dag C_{\textbf{k},\sigma} - \left(\sum_{\textbf{k},\sigma}C^\dag_{\textbf{k},\sigma}C_{\textbf{k},\sigma}\right)^2 \right].
\end{equation}
We work within the Fock space, however, since the particle number is fixed, the description shifts from the grand canonical ensemble to the canonical ensemble. To ensure conservation of the average electron number, when operating in a fixed-particle-number subspace, we can replace the operator by its eigenvalue, that is, $\sum_{\textbf{k},\sigma}C_{\textbf{k},\sigma}^\dag C_{\textbf{k},\sigma} \rightarrow N$. As a result, the corresponding term in the Hamiltonian simplifies to
\begin{equation}
- \dfrac{q_e^2}{2} \dfrac{N^2}{V} \dfrac{4\pi}{k_s^2},
\end{equation}
then
\begin{align}
-\dfrac{q_e^2}{2V}\frac{4\pi}{k^2_s}\sum_{\textbf{k},\sigma}C_{\textbf{k},\sigma}^\dag C_{\textbf{k},\sigma} &= - \dfrac{q_e^2}{2}\dfrac{4\pi}{k^2_s} \sum_{\textbf{k},\sigma} \dfrac{C_{\textbf{k},\sigma}^\dag C_{\textbf{k},\sigma}}{V} = \notag \\
&= - \dfrac{q_e^2}{2}\dfrac{4\pi}{k^2_s} \dfrac{N}{V} = \notag \\
&= - \dfrac{q_e^2}{2}\dfrac{4\pi}{k^2_s} \rho,
\end{align}
which is called constant background term and is typically neglected when studying dynamic properties of a system, as we show. Indeed, taking the limits for $N \rightarrow \infty$ and $V \rightarrow \infty$, so that $\rho$ remains constant, we expect the energy to grow with the number $N$, but we are interested in the energy per particle, then the object
\begin{equation}
- \dfrac{\dfrac{q_e^2}{2} \dfrac{4\pi}{k^2_s} \rho}{N}
\end{equation}
vanishes for $N \rightarrow \infty$. The remaining terms make a contribution proportional to the number of particles, and so if these terms are divided by $N$, we have a finite asymptotic trend. After the limits $N, V \rightarrow \infty$, we can make $k_s \rightarrow 0$ and obtain the hamiltonian of the jellium model in second quantization, i.e.,
\begin{equation}
\hat{\mathcal{H}} = \sum_{\textbf{k},\sigma} \dfrac{\hslash^2 \textbf{k}^2}{2m_e} C^\dag_{\textbf{k},\sigma}C_{\textbf{k},\sigma} + \dfrac{1}{2V}\sum_{\substack{\textbf{k}_1,\textbf{k}_2, \\ \textbf{q} \neq \textbf{0}}} \sum_{\sigma_1,\sigma_2} \dfrac{4 \pi q_e^2}{q^2} C^\dag_{\textbf{k}_1 + \textbf{q},\sigma_1} C^\dag_{\textbf{k}_2 - \textbf{q},\sigma_2} C_{\textbf{k}_2,\sigma_2} C_{\textbf{k}_1,\sigma_1}.
\label{eq: hamiltonianamodellojellium1}
\end{equation}
In the limit $k_s \rightarrow 0$, the Fourier transform of the Yukawa potential, multiplied by $q_e^2$ reduces to $\frac{4 \pi q_e^2}{q^2}$, which is precisely the Fourier transform of the Coulomb potential. This shows that when the screening parameter $k_s$, vanishes, the Yukawa potential naturally recovers the long-range Coulomb interaction. \newline
Note that the jellium Hamiltonian $\eqref{eq: hamiltonianamodellojellium1}$ can be rewritten in a more insightful form, namely in terms of the Fourier-transformed particle density operator, equation $\eqref{eq: trasformatadiFourieroperatoredensitadifluttuazioneparticelle}$, its adjoint, and the number operator. Indeed, writing $C^{\dagger}_{\textbf{k}_2 - \textbf{q},\sigma_2} C_{\textbf{k}_1,\sigma_1} = \delta_{\textbf{k}_1,\textbf{k}_2 - \textbf{q}} \delta_{\sigma_1,\sigma_2} - C_{\textbf{k}_1,\sigma_1} C^{\dagger}_{\textbf{k}_2 - \textbf{q},\sigma_2}$, we invert the order of the annihilation operators in the jellium Hamiltonian and obtain
\begin{align}
\hat{\mathcal{H}} &= \sum_{\textbf{k},\sigma} \dfrac{\hslash^2 \textbf{k}^2}{2m_e} C^\dag_{\textbf{k},\sigma}C_{\textbf{k},\sigma} + \dfrac{1}{2V}\sum_{\substack{\textbf{k}_1,\textbf{k}_2, \\ \textbf{q} \neq \textbf{0}}} \sum_{\sigma_1,\sigma_2} \dfrac{4 \pi q_e^2}{q^2}C^\dag_{\textbf{k}_1 + \textbf{q},\sigma_1} C^\dag_{\textbf{k}_2 - \textbf{q},\sigma_2} C_{\textbf{k}_1,\sigma_1} C_{\textbf{k}_2,\sigma_2} = \notag \\
&= \sum_{\textbf{k},\sigma} \dfrac{\hslash^2 \textbf{k}^2}{2m_e} C^\dag_{\textbf{k},\sigma}C_{\textbf{k},\sigma} + \dfrac{1}{2V}\sum_{\substack{\textbf{k}_1,\textbf{k}_2, \\ \textbf{q} \neq \textbf{0}}} \sum_{\sigma_1,\sigma_2} \dfrac{4 \pi q_e^2}{q^2} C^\dag_{\textbf{k}_1 + \textbf{q},\sigma_1} \left( \delta_{\textbf{k}_1,\textbf{k}_2 - \textbf{q}} \delta_{\sigma_1,\sigma_2} - C_{\textbf{k}_1,\sigma_1} C^{\dagger}_{\textbf{k}_2 - \textbf{q},\sigma_2} \right) C_{\textbf{k}_2,\sigma_2}.
\end{align}
We manipulate as follows
\begin{align}
& \sum_{\textbf{k}_1,\textbf{k}_2} \sum_{\sigma_1,\sigma_2} C^\dag_{\textbf{k}_1 + \textbf{q},\sigma_1} \left( \delta_{\textbf{k}_1,\textbf{k}_2 - \textbf{q}} \delta_{\sigma_1,\sigma_2} - C_{\textbf{k}_1,\sigma_1} C^{\dagger}_{\textbf{k}_2 - \textbf{q},\sigma_2} \right) C_{\textbf{k}_2,\sigma_2} = \notag \\
&= \sum_{\textbf{k}_1,\textbf{k}_2} \sum_{\sigma_1,\sigma_2} C^\dag_{\textbf{k}_1 + \textbf{q},\sigma_1} \delta_{\textbf{k}_1,\textbf{k}_2 - \textbf{q}} \delta_{\sigma_1,\sigma_2} C_{\textbf{k}_2,\sigma_2} - \sum_{\textbf{k}_1,\textbf{k}_2} \sum_{\sigma_1,\sigma_2} C^\dag_{\textbf{k}_1 + \textbf{q},\sigma_1} C_{\textbf{k}_1,\sigma_1} C^{\dagger}_{\textbf{k}_2 - \textbf{q},\sigma_2} C_{\textbf{k}_2,\sigma_2} = \notag \\
&= \sum_{\textbf{k}_1,\textbf{k}_2} \sum_{\sigma_1,\sigma_2} C^\dag_{\textbf{k}_1 + \textbf{q},\sigma_1} \delta_{\textbf{k}_1,\textbf{k}_2 - \textbf{q}} \delta_{\sigma_1,\sigma_2} C_{\textbf{k}_2,\sigma_2} - \hat{\rho}_{\textbf{q}} \hat{\rho}^{\dagger}_{\textbf{q}},
\end{align}
and summing over $\textbf{k}_1$, $\sigma_1$, the deltas imply $\textbf{k}_1 = \textbf{k}_2 - \textbf{q}$, $\sigma_1 = \sigma_2$, that is, the first term becomes
\begin{align}
\sum_{\textbf{k}_2,\sigma_2} C^\dag_{\textbf{k}_2,\sigma_2} C_{\textbf{k}_2,\sigma_2} &= \sum_{\textbf{k}_2,\sigma_2} \hat{N}_{\textbf{k}_2,\sigma_2} = \notag \\
&= \hat{N},
\end{align}
and the jellium Hamiltonian $\eqref{eq: hamiltonianamodellojellium1}$ can be equivalently written as
\begin{equation}
\hat{\mathcal{H}} = \sum_{\textbf{k},\sigma} \dfrac{\hslash^2 \textbf{k}^2}{2m_e} C^\dag_{\textbf{k},\sigma}C_{\textbf{k},\sigma} + \dfrac{1}{2V}\sum_{\textbf{q} \neq \textbf{0}} \dfrac{4 \pi q_e^2}{q^2} \left( \hat{N} - \hat{\rho}_{\textbf{q}} \hat{\rho}^{\dagger}_{\textbf{q}} \right).
\label{eq: hamiltonianamodellojellium2}
\end{equation}
\subsubsection{Extension of the model to the grand canonical ensemble}
In the jellium model, the ensemble of electrons immersed in a neutralizing background is traditionally described by a Hamiltonian that conserves the total particle number. However, to treat systems in which the number of electrons can fluctuate, as often occurs in many experimental and theoretical situations, it is convenient to use the grand canonical formalism. One then introduces the chemical potential \(\mu\) and considers the grand Hamiltonian operator
\begin{equation}
\hat{\mathcal{H}} - \mu \hat{N},
\end{equation}
where \(\hat{\mathcal{H}}\) is the system Hamiltonian, $\mu$ is the chemical potential and \(\hat{N}\) is the total particle number operator. The introduction of the chemical potential in the grand canonical jellium model allows for a more realistic and flexible description of various physical phenomena, including:
\begin{itemize}
\item {Green's functions in the grand canonical formalism.} \newline
In the grand canonical context, the finite-temperature Green's function is computed by taking the statistical average with chemical potential \(\mu\). This implies that the energy of the electronic states is shifted by a term \(-\mu\), thus modifying the spectral structure of the operators and influencing the system’s response properties. This approach is fundamental to correctly describe the electronic dynamics in systems with a variable particle number.
\item {Hartree-Fock theory and Fermi energy calculation.} \newline  
Within the grand canonical formalism, the chemical potential determines the Fermi level (see Chapter \ref{The Fermi momentum}), and consequently the average number of particles in the system. The self-consistent solution of the Hartree-Fock equation is performed by imposing the value of \(\mu\) so that the resulting electron density matches the desired value. The Fermi energy \(\mathcal{E}_F = \mu\) is thus the parameter governing the electronic properties of the jellium.
\item {Calculation of partition functions and thermodynamic properties.} \newline
The grand canonical partition function
\begin{equation}
Z = \mathrm{Tr}\left[e^{-\beta \left( \hat{\mathcal{H}} - \mu \hat{N} \right)}\right]
\end{equation}
is fundamental for calculating thermodynamic quantities such as free energy, particle density, and susceptibilities. In jellium, this formalism allows the study of particle number fluctuations and phase transitions induced by variations of \(\mu\).
\item {Description of open systems and particle exchange.} \newline  
The jellium model with chemical potential is particularly useful for describing electronic systems in contact with a particle reservoir, such as metal junctions, double-layer capacitors, or low-dimensional materials where the electron density can vary in response to an external bias.
\end{itemize}
In the grand canonical ensemble, the second-quantized Hamiltonian of the jellium model is straightforwardly adapted as follows
\begin{equation}
\hat{\mathcal{H}} = \sum_{\textbf{k},\sigma} \left( \dfrac{\hslash^2 \textbf{k}^2}{2m_e} - \mu \right) C^\dag_{\textbf{k},\sigma}C_{\textbf{k},\sigma} + \dfrac{1}{2V}\sum_{\substack{\textbf{k}_1,\textbf{k}_2, \\ \textbf{q} \neq \textbf{0}}} \sum_{\sigma_1,\sigma_2} \dfrac{4 \pi q_e^2}{q^2} C^\dag_{\textbf{k}_1 + \textbf{q},\sigma_1} C^\dag_{\textbf{k}_2 - \textbf{q},\sigma_2} C_{\textbf{k}_2,\sigma_2} C_{\textbf{k}_1,\sigma_1}.
\label{eq: hamiltonianamodellojellium3}
\end{equation}

\part{A Quantum Many-Body Model of a Solid}

\fancypagestyle{nochapterhead}{
  \fancyhf{}
  \fancyhead[EL,OR]{\thepage}  
}

\cleardoublepage
\phantomsection 
\addcontentsline{toc}{chapter}{Three steps to model a solid – Preliminary discussion}
\markboth{Three steps to model a solid – Preliminary discussion}{}
\pagestyle{nochapterhead}
\section*{Three steps to model a solid - Preliminary discussion}
A typical example in quantum many-body theory is the description of the dynamics of a solid. The simplest model assumes atoms arranged with a specific spatial periodicity. Each atom consists of a nucleus (ion) and valence electrons, treated collectively as point-like entities. Each atom contains \( Z_V \) valence electrons, which are responsible for forming chemical bonds between atoms, in addition to electrons in closed inner shells that remain tightly bound to the nucleus. When atoms are brought close together to form a crystal lattice, the orbitals associated with the inner closed shells remain largely unaltered, whereas the outermost orbitals are strongly affected by the presence of neighboring atoms. In other words, the atomic wave functions overlap. For each positive ion corresponding to the valence electrons, there are \( Z_V \) valence electrons present. The Hamiltonian describing such a solid within the formalism of first quantization is highly complex, that is,
\begin{equation}
\mathcal{H} = T^{(e)} + T^{(i)} + V^{(e-e)} + V^{(i-i)} + V^{(e-i)},
\end{equation}
where
\begin{equation}
T^{(e)} = \sum_{j=1}^{N_e} \frac{\hat{\mathbf{p}}_j^{2}}{2 m_e},
\end{equation}
is the kinetic energy of the valence electrons, where \(N_e\) is the total number of valence electrons in the crystal,
\begin{equation}
T^{(i)} = \sum_{j=1}^{N_i} \frac{\hat{\mathbf{P}}_j^{2}}{2 M_j},
\end{equation}
is the kinetic energy of the ions, where \(N_i\) is the total number of ions,
\begin{equation}
V^{(e-e)} = \frac{1}{2} \sum_{\substack{i,j=1 \\ i \neq j}}^{N_e} \frac{q_e^{2}}{|\mathbf{r}_i - \mathbf{r}_j|},
\end{equation}
is the Coulomb repulsion energy between valence electrons, \(\mathbf{r}_i\) being the electron position,
\begin{equation}
V^{(i-i)} = \frac{1}{2} \sum_{\substack{i,j=1 \\ i \neq j}}^{N_i} \frac{Z_{V,i} Z_{V,j} q_e^{2}}{|\mathbf{R}_i - \mathbf{R}_j|},
\end{equation}
is the Coulomb repulsion energy between ions, where \(\mathbf{R}_i\) are the ion positions, and \(Z_{V,i}\), \(Z_{V,j}\) are the valence electron numbers of atoms \(i\) and \(j\), respectively,
\begin{equation}
V^{(e-i)} = - \sum_{i=1}^{N_e} \sum_{j=1}^{N_i} \frac{Z_{V,j} q_e^{2}}{|\mathbf{r}_i - \mathbf{R}_j|},
\end{equation}
is the Coulomb attraction between ions and valence electrons. In general, ion and electron populations are strongly coupled at densities of order \(10^{23} \text{cm}^{-3}\), then the eigenvalue problem is impossible to solve exactly, and approximations are necessary. First, the mass of ions is several orders of magnitude larger than that of electrons, so the dominant dynamics is governed by the electrons. Typical electron response times are on the order of femtoseconds (\(\sim \text{fs}\)), whereas ion response times are much slower, on the order of picoseconds or longer (\(\gtrsim \text{ps}\)). In the limit \( M_j \to \infty \), ions can be treated as fixed classical points, and the problem reduces to describing approximately \(10^{23}\) electrons interacting among themselves and with fixed nuclei according to the Hamiltonian
\begin{equation}
\mathcal{H}^{(e)} = T^{(e)} + V^{(e-e)} + V^{(e-i)},
\end{equation}
which represents the electronic energy in the fixed nuclei (Born-Oppenheimer) approximation. Despite this simplification, the problem remains intractable for exact solutions. Suppose we have computes an eigenfunction of this fixed nuclei Hamiltonian, denoted by \(\varphi_n(\{ \mathbf{r} \}; \{ \mathbf{R} \})\), i.e.,
\begin{equation}
\mathcal{H}^{(e)} \varphi_{n}\left( \lbrace \textbf{r} \rbrace; \lbrace \textbf{R} \rbrace \right) = \mathcal{E}_{n}\left( \lbrace \textbf{R} \rbrace \right)) \varphi_{n}\left( \lbrace \textbf{r} \rbrace;\lbrace \textbf{R} \rbrace \right),
\end{equation}
we then pose the question whether this electronic solution can be related to an eigenstate of the full Hamiltonian
\begin{equation}
\mathcal{H} = \mathcal{H}^{(e)} + V^{(i-i)} + T^{(i)}.
\end{equation}
Considering the total wavefunction expanded in the eigenbasis of $\mathcal{H}$, we investigate the possibility of expressing it as a superposition of the ionic eigenstates coupled with the electronic eigenfunctions as follows
\begin{equation}
\psi \left( \lbrace \textbf{r} \rbrace,\lbrace \textbf{R} \rbrace \right) = \sum_{n} c_n \left( \lbrace \textbf{R} \rbrace \right) \varphi_n \left( \lbrace \textbf{r} \rbrace;\lbrace \textbf{R} \rbrace \right).
\label{eq: vettorestatosolidopassaggio1e2}
\end{equation} 
The goal is to determine the coefficient functions \( c_n(\{ \mathbf{R} \}) \) for each ionic configuration \(\{ \mathbf{R} \}\). We impose that \(|\psi\rangle\) is an eigenstate of the full Hamiltonian \(\mathcal{H}\). The electronic Hamiltonian \(\mathcal{H}^{(e)}\) acts on \(\varphi_n(\{ \mathbf{r} \}; \{ \mathbf{R} \})\) yielding the eigenvalues \(\mathcal{E}_n(\{ \mathbf{R} \})\). The ion-ion interaction term \(V^{(i-i)}\) acts multiplicatively with respect to the ionic coordinates, whereas the ionic kinetic energy operator \(T^{(i)}\), being a Laplacian in \(\{ \mathbf{R} \}\), acts on both the coefficient functions \(c_n(\{ \mathbf{R} \})\) and on the parametric dependence of \(\varphi_n(\{ \mathbf{r} \}; \{ \mathbf{R} \})\) as follows
\begin{equation}
\nabla^2_{\textbf{R}_j} \left( c_n \varphi_n \right) = \left( \nabla^2_{\textbf{R}_j} c_n \right) \varphi_n + c_n \left( \nabla^2_{\textbf{R}_j} \varphi_n \right) + 2 \nabla_{\textbf{R}_j} c_n \cdot \nabla_{\textbf{R}_j} \varphi_n.
\end{equation}
The eigenvalue equation then becomes
\begin{equation}
A + \sum_n \left\lbrace \left[ \mathcal{E}_n(\lbrace \textbf{R} \rbrace) + V^{(i-i)} \left( \lbrace \textbf{R} \rbrace \right) \right] c_n \varphi_n - \sum_j \dfrac{\hbar^{2}}{2 M_j} (\nabla^2_{\textbf{R}_j} c_n) \varphi_n \right\rbrace = \mathcal{E} \sum_n c_n \varphi_n,
\end{equation} 
\begin{equation}
A = - \sum_n \sum_j \dfrac{\hbar^{2}}{2 M_j} \left[ c_n \left( \nabla^2_{\textbf{R}_j} \varphi_n \right) + 2 \nabla_{\textbf{R}_j} c_n \cdot \nabla_{\textbf{R}_j} \varphi_n \right] ,
\end{equation}
where we have omitted the explicit dependence of the eigenfunctions on both the electronic and ionic position variables to simplify the notation. We now multiply both sides from the left by \( \varphi^{*}_m \) and integrate over the electronic coordinates. Using the orthonormality of the set \( \lbrace \varphi_n \rbrace \), we obtain
\begin{equation}
D + \left[ \mathcal{E}_m \left( \lbrace \textbf{R} \rbrace \right) + V^{(i-i)} \left( \lbrace \textbf{R} \rbrace \right) - \sum_j \dfrac{\hbar^{2}}{2 M_j} \nabla^2_{\textbf{R}_j} \right] c_m = \mathcal{E} c_m,
\end{equation}
where
\begin{equation}
D = - \sum_j \sum_n \dfrac{\hbar^{2}}{2 M_j} \left\lbrace \langle \varphi_m | \nabla^2_{\textbf{R}_j} | \varphi_n \rangle c_n + \langle \varphi_m | \nabla_{\textbf{R}_j} | \varphi_n \rangle \cdot \nabla_{\textbf{R}_j} c_n \right\rbrace
\end{equation}
is called the non-adiabatic term. Since the ions move, the electrons do not remain confined in the same electronic state \( \varphi_m \), but undergo transitions from state \( m \) to other states \( n \). Electrons and ions exchange energy and influence one another. Then, the term \( D \) represents an electron-ion interaction contribution, involving both first and second derivatives of the electronic wavefunctions \( \lbrace \varphi_n \rbrace \) with respect to the ionic coordinates. In order to compute such derivatives, one must account for the variations of the electronic wavefunctions with respect to infinitesimal displacements of \( \textbf{R}_j \). However, since electrons are much faster than ions, it is generally expected that these variations are negligible and can be safely omitted. \newline
To describe the dynamics of a solid, we proceed through three steps. 
\begin{itemize}
\item[1.] The fixed-ion electronic problem
\begin{equation}
\mathcal{H}^{(e)} \varphi_n = \mathcal{E}_n \varphi_n 
\end{equation}
should be solved for each $\lbrace \textbf{R} \rbrace$ configuration of the nuclei. Such a problem is not solvable, so we fix the ions at the vertices of a lattice. The main difficulty now lies in treating approximately $10^{23} \text{cm}^{-3}$ electrons interacting both with one another and with fixed ions. We apply to $\mathcal{H}^{(e)}$ the Hartree-Fock method and ultimately have to study the problem of a particle in an effective potential $V(\textbf{r})$, which is periodic on the crystal with respect to each of its elementary cells. After applying the Hartree-Fock method, the Hamiltonian 
\begin{equation}
\mathcal{H}^{(e)} \xrightarrow[\text{Hartree Fock}]{} \mathcal{H}^{(e)}_{H.F.} = \dfrac{\hat{\textbf{p}}^2}{2 m} + V(\textbf{r})
\label{eq: HamiltonianaperiodicacampomedioHartreeFock}
\end{equation}
is periodic. Ultimately, we must study the problem of a particle subjected to a periodic potential, the so-called Bloch's problem, i.e.,
\begin{equation}
\mathcal{H}^{(e)}_{H.F.} \varphi_n = \mathcal{E}_n \varphi_n,
\label{eq: problemadiBloch}
\end{equation}
and we will compute the so-called Bloch eigenfunctions.
\begin{remark}
Although the Bloch problem employs a one-particle Hartree-Fock potential, residual electronic interactions remain and can be significant. For example, in models of superconductivity, an additional local interaction term, such as the Hubbard Hamiltonian, is often incorporated into the full solid-state Hamiltonian.
\end{remark}
\item[2.] Given the solutions of the fixed-ion electronic problem, we need to solve the ion problem. We expand the effective potential $V^{(i-i)} (\lbrace \textbf{R} \rbrace) + \mathcal{E}_n(\lbrace \textbf{R} \rbrace)$ to second order with respect to its minimum, i.e., we consider its harmonic expansion. Now, there are on the order of $10^{23} \text{cm}^{-3}$ ionic variables coupled together, but because of the harmonic expansion, such coupling is simple to deal with. Finally, we consider the classical normal modes, find the normal coordinates and quantize them: in this way, we obtain the energy quantities of the lattice vibration field, which are called phonons.
\item[3.] Our focus is on electron-phonon coupling. 
\end{itemize}
\begin{remark}[Born-Oppenheimer approximation]
Step 1 and 2 give the Born-Oppenheimer approximation, that is, $D=0$ and the state vector is given by $\eqref{eq: vettorestatosolidopassaggio1e2}$. The set \( \lbrace \varphi_n \rbrace \) consists of eigenfunctions of the electronic problem with fixed ionic positions, whereas the set \( \lbrace c_n \rbrace \) comprises eigenfunctions of the ionic problem, in which the ions possess their own dynamics and interact through an effective potential
\begin{equation}
V_{eff}^{(i-i)} = V^{(i-i)} \left( \lbrace \textbf{R} \rbrace \right) + \mathcal{E}_{n} \left( \lbrace \textbf{R} \rbrace \right) ,
\end{equation}
which includes not only the repulsive coulombic potential $V^{(i-i)} (\lbrace \textbf{R} \rbrace)$ but also a potential $\mathcal{E}_{n} (\lbrace \textbf{R} \rbrace)$ that originates from electron dynamics: electrons are shielding ions. Neglecting the non-adiabatic term makes it impossible to describe some phenomena such as superconductivity and transport in metals.
\end{remark}
At the end of the three steps, the Hamiltonian of a solid in second quantization has the form
\begin{equation}
\hat{\mathcal{H}} = \sum_{n,\textbf{k},\sigma} \mathcal{E}_{n,\textbf{k}} C^{\dag}_{n,\textbf{k},\sigma} C_{n,\textbf{k},\sigma} + \sum_{\textbf{q},s} \hslash \omega_{\textbf{q},s} \left( a^{\dag}_{\textbf{q},s} a_{\textbf{q},s} + \dfrac{1}{2} \right) \ + \sum_{\substack{m,m',\sigma,s, \\ \textbf{k},\textbf{G},\textbf{q}}} T_{\textbf{k},\textbf{q},\textbf{G}}^s \ C^{\dag}_{m',\textbf{k}+\textbf{q}+\textbf{G},\sigma} C_{m,\textbf{k},\sigma} \left( a_{\textbf{q},s} + a^{\dag}_{-\textbf{q},s} \right),
\label{eq: Hamiltonianadiunsolido}
\end{equation}
where $C$, $C^\dagger$ are fermionic operators and $a$, $a^\dagger$ are bosonic operators. \newline
Ultimately, in place of the Hamiltonian $\eqref{eq: Hamiltonianadiunsolido}$, a simplified form of the solid Hamiltonian is given by
\begin{equation}
\hat{\mathcal{H}} = \sum_{\textbf{k},\sigma} \mathcal{E}_{\textbf{k}} C^{\dag}_{\textbf{k},\sigma} C_{\textbf{k},\sigma} + \sum_{\textbf{q}} \hslash \omega_{\textbf{q}} a^{\dag}_{\textbf{q}} a_{\textbf{q}} + \sum_{\textbf{k},\textbf{q},\sigma} M_{\textbf{q}} C^{\dag}_{\textbf{k}+\textbf{q},\sigma} C_{\textbf{k},\sigma} \left( a_{\textbf{q}} + a^{\dag}_{-\textbf{q}} \right).
\label{eq: Hamiltonianasemplificatadiunsolido}
\end{equation}
This preliminary discussion provides a conceptual foundation that will guide the detailed development of each step in the following three chapters. In this way, the reader gains a comprehensive overview before delving into each individual phase.
\clearpage
\pagestyle{fancy}
\fancyhf{}
\fancyhead[EL,OR]{\thepage}
\fancyhead[OL,ER]{%
  \ifthenelse{\value{chapter}>0}{\chaptername\ \thechapter}{}%
}

\chapter{Step I of the solid model: Bloch electrons}
Here we show the fundamental concept of the crystal lattice, which forms the geometric basis for describing solids. Lattices represent the periodic arrangement of atoms in space, and their symmetry is crucial for understanding the electronic and structural properties of materials. The intrinsic periodicity of the lattice allows the use of specific mathematical tools, such as Bloch's theory, which provides a general solution to the quantum problem of an electron in a periodic potential. \newline
Bloch's theorem states that the electron’s eigenfunctions in a crystal can be expressed as products of periodic functions and plane waves, giving rise to the so-called Bloch functions, which are essential for studying the energy band structure. These functions reflect the crystal's periodicity and form the foundation for understanding electronic behavior in solids. However, for certain applications, it is more convenient to use another basis of eigenfunctions called Wannier functions, which offer a complementary viewpoint to the Bloch functions. Wannier functions are particularly useful for studying local interactions and for numerical methods in material simulations. \newline
This chapter aims to provide a clear and rigorous overview of these concepts, starting from the geometric definition of lattices, moving through the formulation of Bloch's problem, and culminating in the construction and properties of Bloch and Wannier functions. Understanding these tools is essential for effectively addressing more complex problems in solid-state physics. Finally, based on the ideas developed in this chapter, we will derive the expression
\begin{equation}
\sum_{n,\textbf{k},\sigma} \mathcal{E}_{n,\textbf{k}} C^{\dag}_{n,\textbf{k},\sigma} C_{n,\textbf{k},\sigma} 
\end{equation}
in the Hamiltonian $\eqref{eq: Hamiltonianadiunsolido}$.
\section{Bravais lattices and their properties}
In this section, we introduce geometric concepts from statistical physics that are essential for proving Bloch's theorem.
\subsection{Bravais lattices}
In geometry and crystallography, a Bravais lattice, named after Auguste Bravais, is an infinite array of discrete points generated by a set of discrete translation operations. The Bravais lattice concept is used to formally define a crystalline arrangement and its frontiers: a crystal is made up of one or more atoms, called the basis or motif, at each lattice point. The basis may consist of atoms, molecules, or polymer strings of solid matter, and the lattice provides the locations of the basis. Placed the origin of the Cartesian axes on any point of the lattice, each point is identified by a vector. A Bravais lattice is generated by translation operations in the space of a set of vectors, called primitive vectors. The primitive vectors are linearly independent and their choice is not unique: the general definition of a Bravais lattice vector $\textbf{R}$ in $d$ dimension is
\begin{equation}
\textbf{R} = \sum_{i=1}^d n_i \textbf{a}_i,
\end{equation}
where $n_1$ are integers and $\textbf{a}_i$ are the primitive vectors of the lattice. Group theory allows defining the number of possible Bravais lattices for each dimension of space. Two Bravais lattices are often considered equivalent if they have isomorphic symmetry groups. In this sense, there are 5 possible Bravais lattices in 2-dimensional space and 14 possible Bravais lattices in 3-dimensional space. In particular, the lattice vectors in three dimensions have form
\begin{equation}
\textbf{R} = n_1 \textbf{a}_1 + n_2 \textbf{a}_2 + n_3 \textbf{a}_3,
\end{equation}
where the primitive vectors $\textbf{a}_i$ are not complanar. \newline
Primitive unit cells are defined as unit cells with the smallest volume for a given lattice. There can be more than one way to choose a primitive cell for a given crystal and each choice will have a different primitive cell shape, but the primitive cell volume is the same for every choice and each choice will have the property that a one-to-one correspondence can be established between primitive unit cells and discrete lattice points over the associated lattice. Among all possible primitive cells for a given crystal, an obvious primitive cell may be the parallelepiped formed by a chosen set of primitive translation vectors, that is, the set of all points $x_1 \textbf{a}_1 + x_2 \textbf{a}_2 + x_3 \textbf{a}_3$ where $0 \leq x_i < 1$ and $\textbf{a}_i$ are the chosen primitive vector. This primitive cell does not always show the clear symmetry of a given crystal. The unique property of a crystal is that its atoms are arranged in a regular three-dimensional array called a lattice. All the properties attributed to crystalline materials stem from this highly ordered structure. Such a structure exhibits discrete translational symmetry. In order to model and study such a periodic system, one needs a mathematical "handle" to describe the symmetry and hence draw conclusions about the material properties consequent to this symmetry. The Wigner–Seitz cell is a means to achieve this. To construct a Wigner–Seitz cell, let us start by first picking a lattice point. After a point is chosen, lines are drawn to all nearby lattice points. At the midpoint of each line, another line is drawn normal to each of the first set of lines. The smallest area enclosed in this way is called the Wigner–Seitz primitive cell. A Wigner–Seitz cell is an example of a primitive cell, which is a unit cell containing exactly one lattice point. For any given lattice, there is only one Wigner–Seitz cell for any given lattice. It is the locus of points in space that are closer to that lattice point than to any of the other lattice points.
\subsection{Direct and reciprocal Bravais lattices}
Recall that the vectors \( \mathbf{R} \in \mathbb{R}^3 \) characterizing the translational symmetry of the lattice are of the form
\begin{align}
\mathbf{R} &\equiv \mathbf{R}_n = \notag \\
&= n_1 \mathbf{a}_1 + n_2 \mathbf{a}_2 + n_3 \mathbf{a}_3, \quad n_i \in \mathbb{Z},
\end{align}
where \( \mathbf{a}_1, \mathbf{a}_2, \mathbf{a}_3 \) are the primitive lattice vectors of the direct lattice. Each triplet of integers \( n = (n_1, n_2, n_3) \in \mathbb{Z}^3 \) uniquely and bijectively corresponds to a lattice vector \( \mathbf{R}_n \), which identifies a point in the lattice. In what follows, we will use the notations \( \mathbf{R} \) and \( \mathbf{R}_n \) interchangeably, depending on whether or not it is necessary to make the dependence on \( n \) explicit. The translational invariance of lattice-defined quantities is naturally expressed in terms of these vectors. Now, let $f(\textbf{r})$ be a periodic function on a Bravais lattice, i.e., such that
\begin{equation}
f\left(\textbf{r}+\textbf{R}\right) = f(\textbf{r}),
\end{equation}
where $\textbf{r}$ is defined in the unit cell. To write a Fourier expansion of such a function, we must determine plane waves that satisfy
\begin{equation}
e^{i \textbf{G} \cdot \textbf{r}} = e^{i \textbf{G} \cdot \left(\textbf{r}+\textbf{R}\right)} ,
\end{equation}
i.e., we need a vector $\textbf{G}$ such that
\begin{equation}
e^{i \textbf{G} \cdot \textbf{R}} = 1, \ \forall \ \textbf{R} ,
\end{equation} 
\begin{equation}
\textbf{G} \cdot \textbf{R} = 2 \pi l, \ l \in \mathbb{Z}.
\label{eq: condizionevettoreGreciproco}
\end{equation}
It can be shown that the $\eqref{eq: condizionevettoreGreciproco}$ is satisfied if and only if we consider $3$ vectors $\textbf{b}_1$, $\textbf{b}_2$, $\textbf{b}_3$ such that
\begin{equation}
\textbf{a}_i \cdot \textbf{b}_j = 2 \pi \delta_{ij}.
\end{equation}
One solution is represented by
\begin{equation}
\textbf{b}_1 = 2 \pi \dfrac{\textbf{a}_2 \times \textbf{a}_3}{\textbf{a}_1 \cdot \left[ \textbf{a}_2 \times \textbf{a}_3 \right]},
\end{equation}
\begin{equation}
\textbf{b}_2 = 2 \pi \dfrac{\textbf{a}_3 \times \textbf{a}_1}{\textbf{a}_1 \cdot \left[ \textbf{a}_2 \times \textbf{a}_3 \right]},
\end{equation}
\begin{equation}
\textbf{b}_3 = 2 \pi \dfrac{\textbf{a}_1 \times \textbf{a}_2}{\textbf{a}_1 \cdot \left[ \textbf{a}_2 \times \textbf{a}_3 \right]},
\end{equation}
whose dimensions are given by
\begin{equation}
[b_i] = [\text{length}]^{-1}.
\end{equation}
The vectors $\textbf{G}$ which satisfy the $\eqref{eq: condizionevettoreGreciproco}$ are of the form
\begin{equation}
\textbf{G} = m_1 \textbf{b}_1 + m_2 \textbf{b}_2 + m_3 \textbf{b}_3, \ m_i \in \mathbb{Z} ,
\end{equation}
and given that the vectors $\lbrace \textbf{b}_i \rbrace$ form a linearly independent set, they constitute a basis. That is, they span a vector space and belong to it. Note that both because of the presence of the numerical factor $2 \pi$ in the product $\textbf{a}_i \cdot \textbf{b}_j$ and because of the physical dimensions of the $\lbrace \textbf{b}_i \rbrace$, we are naturally induced to interpret these vectors as wave vectors. The three $\textbf{b}_i$ vectors form a new lattice, called the reciprocal lattice. In addition, the quantity
\begin{equation}
V_d = \textbf{a}_1 \cdot \left[ \textbf{a}_2 \times \textbf{a}_3 \right]
\end{equation}
has a physical interpretation, i.e., it is the volume of the elementary cell of the direct lattice, i.e., $V_{\text{direct}} \equiv V_d$. The cell of the primitive vectors $\textbf{a}_1$, $\textbf{a}_2$, $\textbf{a}_3$ on the Bravais lattice locates a reciprocal cell of vectors $\textbf{b}_1$, $\textbf{b}_2$, $\textbf{b}_3$ that belong to a lattice that has the dimensions of the reciprocal of a length and is then called a reciprocal lattice. The primitive cell of the reciprocal lattice in momentum space is called the first Brillouin zone (1$^\text{st}$ BZ). The volume of the elementary cell of the reciprocal lattice, i.e., $V_{\text{reciprocal}} \equiv V_r$, which dimensionally is the reciprocal of a volume ($[V_r] = [L]^{-3}$), is given by
\begin{equation}
V_r = \textbf{b}_1 \cdot \left[ \textbf{b}_2 \times \textbf{b}_3 \right].
\end{equation}
It satisfies
\begin{equation}
V_r = \dfrac{(2 \pi)^{3}}{V_d} ,
\end{equation}
and is equal to the volume of the first Brillouin zone. A periodic function on Bravais lattice, say $f\left(\textbf{r}\right)$, can be expanded by means of the reciprocal lattice vectors as follows
\begin{equation}
f\left(\textbf{r}\right) = \sum_{\textbf{G}} c_{\textbf{G}} \ \dfrac{e^{i \textbf{G} \cdot \textbf{r}}}{\sqrt{V_d}},
\end{equation}
\begin{equation}
c_{\textbf{G}} = \dfrac{1}{\sqrt{V_d}} \int_{V_d} d^3 \textbf{r} f(\textbf{r}) e^{- i \textbf{G} \cdot \textbf{r}},
\end{equation}
where $\left\lbrace \frac{e^{i \mathbf{G} \cdot \mathbf{r}}}{\sqrt{V_d}} \right\rbrace$ denotes a set of spatial plane waves that form an orthonormal and complete basis on the domain $V_d$, i.e., within each elementary cell. Let $f(\textbf{k})$ be defined on the reciprocal lattice and such that
\begin{equation}
f\left(\textbf{k}\right) = f\left(\textbf{k}+\textbf{G}\right), \ \forall \ \textbf{G}, 
\end{equation}
where $\textbf{k}$ takes values in the first Brillouin zone. Such a function can be expanded in Fourier series on the vectors of the direct lattice, i.e.,
\begin{equation}
f(\textbf{k}) = \sum_{\textbf{R}} c_{\textbf{R}} \ \dfrac{e^{i \textbf{k} \cdot \textbf{R}}}{\sqrt{V_r}},
\end{equation}
with
\begin{equation}
c_{\textbf{R}} = \dfrac{1}{\sqrt{V_r}} \int_{V_r} d^3 \textbf{k} f\left(\textbf{k}\right) e^{- i \textbf{k} \cdot \textbf{R}}.
\end{equation}
\subsection{Simple cubic lattice}
The simplest Bravais lattice is the simple cubic lattice. Given a site, all nearest neighbors are at distance $a$ from each other. A basis in the direct simple cubic lattice is given by the vectors
\begin{equation}
\textbf{a}_1 = a (1,0,0),
\end{equation}
\begin{equation}
\textbf{a}_2 = a (0,1,0),
\end{equation}
\begin{equation}
\textbf{a}_3 = a (0,0,1),
\end{equation}
with volume $V_d = a^3$. The reciprocal lattice of a simple cubic lattice is a simple cubic lattice, with primitive vectors
\begin{equation}
\textbf{b}_1 = \dfrac{2 \pi}{a} (1,0,0),
\end{equation}
\begin{equation}
\textbf{b}_2 = \dfrac{2 \pi}{a} (0,1,0),
\end{equation}
\begin{equation}
\textbf{b}_3 = \dfrac{2 \pi}{a} (0,0,1) ,
\end{equation}
with volume $V_r = \frac{8 \pi^3}{a^3}$. Finally, a convenient basis for the first Brillouin zone of the simple cubic lattice consists of the wave vectors $\textbf{k}=(k_z,k_y,k_z)$, whose components satisfy
\begin{equation}
k_i \in \left[ - \dfrac{\pi}{a} , \dfrac{\pi}{a} \right], \ i=x,y,z.
\label{eq: componentibasefirstBrillouinreticolocubico}
\end{equation}
This defines a cubic region in reciprocal space centered at the origin, which constitutes the first Brillouin zone for the simple cubic lattice.
\section{The quasiparticle concept in quantum many-body theory}
In the context of quantum many-body theory, the concept of a quasiparticle represents one of the most powerful theoretical tools for describing complex many-body systems. A quasiparticle is not an elementary particle in the strict sense, but rather a collective excitation of the system that behaves as if it were an independent particle, with well-defined mass, charge, and energy. When many particles interact, their individual behavior becomes intractable; however, it is observed that the system responds effectively as if composed of "renormalized" entities: the quasiparticles. A classic example is that of electrons in a solid, where interactions with the lattice and with other electrons modify the dynamical properties of a free electron, giving rise to an effective electron with a different mass. Other examples include phonons, which represent the collective vibrations of the crystal lattice, plasmons, associated with collective oscillations of the electronic density. \newline
The electrons in solids and phonons are described in detail in the chapters on a quantum many-body model of a solid, namely, in this chapter and the next, respectively, while other quasiparticles, such as plasmons, are discussed in the subsequent chapters, with thorough explanations of their physical significance and role in many-body systems.
\section{Bloch's theorem}
Assume we have applied the mean-field Hartree-Fock method to the electronic Hamiltonian within the fixed nuclei approximation, then the resulting eigenvalue equation takes the form given in $\eqref{eq: problemadiBloch}$. In this framework, each electron no longer interacts directly with the others, but rather experiences an average effective potential that approximately accounts for their presence. In particular, $\mathcal{H}^{(e)}_{H.F.}$ in $\eqref{eq: HamiltonianaperiodicacampomedioHartreeFock}$ represents the Hamiltonian of a single electron subject to a periodic effective potential \(V(\mathbf{r})\), which reflects the translational symmetry of the crystal lattice, namely
\begin{equation}
V\left(\textbf{r}+\textbf{R}\right) = V(\textbf{r}),
\label{eq: potenzialeperiodicoreticolo}
\end{equation}
or equivalently
\begin{equation}
V\left(\textbf{r}+\textbf{R}_{\textbf{n}}\right) = V(\textbf{r}).
\end{equation}
In the context of a lattice system, we adapt the notion of spatial translation to reflect the discrete periodicity of the structure. Accordingly, we adapt the action of the translation operator on the lattice, as given in equation $\eqref{eq: operatoretraslazionespaziale}$, as follows
\begin{equation}
\hat{T}_{\textbf{n}} : \ \hat{T}_{\textbf{n}} \psi(\textbf{r}) = \psi\left(\textbf{r}+\textbf{R}_{\textbf{n}}\right) .
\end{equation}
It is observed that
\begin{equation}
\left[ \hat{T}_{\textbf{n}} , \hat{T}_{\textbf{n}'} \right] = 0 
\end{equation}
since the order of the translations does not matter. In addition, the equations
\begin{equation}
\left[ \hat{T}_{\textbf{n}} , \dfrac{\hat{\textbf{p}}^2}{2m} \right] = 0 ,
\end{equation}
\begin{equation}
\left[ \hat{T}_{\textbf{n}} , V(\textbf{r}) \right] = 0
\end{equation} 
imply
\begin{equation}
\left[ \hat{T}_{\textbf{n}} , \mathcal{H}^{(e)}_{H.F.} \right] = 0.
\end{equation}
So, the Hamiltonian $\mathcal{H}^{(e)}_{H.F.}$ and the translation operator $\hat{T}_{\textbf{n}}$ have a set of eigenfunctions and the goal of this section is to compute such eigenfunctions, the Bloch eigenfunctions or Bloch waves.
\begin{theorem}[Bloch's theorem] Let $V(\mathbf{r})$ be a potential that is periodic with respect to a crystal lattice, i.e., of the form $\eqref{eq: potenzialeperiodicoreticolo}$, then the solutions to the one-particle Schrödinger equation with respect to the Hamiltonian $\eqref{eq: HamiltonianaperiodicacampomedioHartreeFock}$ can be chosen in the form
\begin{equation}
\varphi_{n,\textbf{k}}(\textbf{r}) = e^{i \textbf{k} \cdot \textbf{r}} \ U_{n,\textbf{k}}(\textbf{r}),
\label{eq: teoremadiBloch1}
\end{equation}
where
\begin{itemize}
\item $\mathbf{k}$ is a vector in the first Brillouin zone;
\item $n \in \mathbb{N}$ is the band index;
\item $U_{n,\textbf{k}}(\textbf{r})$ is a function that is periodic on the lattice, i.e.,
\begin{equation}
U_{n,\textbf{k}}\left(\textbf{r}+\textbf{R}\right) = U_{n,\textbf{k}}(\textbf{r}).
\label{eq: teoremadiBloch2}
\end{equation}
\end{itemize}
\begin{proof}
Since $V(\textbf{r})$ is periodic on the direct lattice with respect to the unit cell, it can be expanded as Fourier series, i.e.,
\begin{equation}
V(\textbf{r}) = \sum_{\textbf{G}} V_{\textbf{G}} \ \dfrac{e^{i \textbf{G} \cdot \textbf{r}}}{\sqrt{V_d}},
\end{equation}
\begin{equation}
V_{\textbf{G}} = \dfrac{1}{\sqrt{V_d}} \int_{V_d} d^3 \textbf{r} \ V(\textbf{r}) \ e^{- i \textbf{G} \cdot \textbf{r}},
\end{equation}
where
\begin{equation}
\textbf{G} = m_1 \textbf{b}_1 + m_2 \textbf{b}_2 + m_3 \textbf{b}_3, \ m_i \in \mathbb{Z},
\end{equation}
is a vector of the reciprocal lattice of the crystal lattice. Moreover, $V(\textbf{r})$ is a real function, then
\begin{align}
V^*(\textbf{r}) &= \sum_{\textbf{G}} V^*_{\textbf{G}} \ \dfrac{e^{- i \textbf{G} \cdot \textbf{r}}}{\sqrt{V_d}} = \notag \\
&= \sum_{- \textbf{G}} V^*_{- \textbf{G}} \ \dfrac{e^{i \textbf{G} \cdot \textbf{r}}}{\sqrt{V_d}} ,
\end{align}
\begin{equation}
V_{\textbf{G}}^{*} = V_{-\textbf{G}}.
\end{equation}
If $\textbf{G}=\textbf{0}$, that is, the wave vector becomes a point on the reciprocal lattice, then $V_{\textbf{G}=\textbf{0}}$ becomes the mean value of $V(\textbf{r})$ with respect to an elementary cell of the direct lattice, i.e., it is a constant. The object $\mathcal{H}^{(e)}_{H.F.}$, which is an energy, is defined up to an arbitrary constant, so can set  
\begin{equation}
V_{\textbf{G}=\textbf{0}} = 0.
\end{equation}
The lattice model is based on the assumption that, in each spatial direction, the system size tends to infinity, even though real crystals have finite dimensions. Near the surface, lattice properties are no longer meaningful. However, if the volume of the crystal is sufficiently large and the number of atoms at the surface is negligible compared to those in the bulk, the crystal can be approximated by an infinite lattice. Real boundary conditions are thus replaced by analytically more convenient ones. Periodic Born-von Karman boundary conditions are assumed. Consider a crystal with $N_1$ cells along the $\mathbf{a}_1$ direction, $N_2$ cells along the $\mathbf{a}_2$ direction, and $N_3$ cells along the $\mathbf{a}_3$ direction, with $N_i \sim 10^{23}$ (recall that the typical atomic density is $\sim 10^{23}\,\mathrm{cm}^{-3}$), for $i=1,2,3$. Now, let us form a larger crystal, a supercrystal, as follows. For each direction, the sequence $N_i \textbf{a}_i$ is repeated nfinitely many times. The elementary cell of the supercrystal coincides with the entire crystal. Since the elementary cell of the crystal is defined by the lattice vectors $\textbf{a}_1$, $\textbf{a}_2$, and $\textbf{a}_3$, the elementary cell of the supercrystal is characterized by the vectors
\begin{equation}
\begin{cases}
\tilde{\textbf{a}}_1 = N_1 \textbf{a}_1 \\
\tilde{\textbf{a}}_2 = N_2 \textbf{a}_2 \\
\tilde{\textbf{a}}_3 = N_3 \textbf{a}_3 \\
\end{cases},
\end{equation}
where $\tilde{\textbf{a}}_i$ are the primitive vectors of the supercrystal, and $N_1$, $N_2$, and $N_3$ are the number of unit cells along each lattice direction. The structure of the reciprocal lattice corresponding to this supercrystal can be determined, and it can be shown that the primitive vectors of the reciprocal lattice are given by
\begin{equation}
\tilde{\textbf{b}}_i = \dfrac{\textbf{b}_i}{N_i}, \ i = 1,2,3.
\end{equation}
If $\textbf{q}$ is a vector of the reciprocal lattice of the supercrystal, it is a linear combination of the primitive vectors with rational coefficients, i.e.,
\begin{equation}
\textbf{q} = \dfrac{m_1}{N_1} \textbf{b}_1 + \dfrac{m_2}{N_2} \textbf{b}_2 + \dfrac{m_3}{N_3} \textbf{b}_3, \ m_i \in \mathbb{Z}, \ N_i \in \mathbb{N}.
\label{eq: vettorereticoloreciprocosupercristallo}
\end{equation}
Given the Bloch eigenvalue problem, the wave function is periodic on the supercrystal, i.e.,
\begin{equation}
\psi \left( \textbf{r} + \tilde{\textbf{R}} \right) = \psi(\textbf{r}),
\end{equation}
i.e., the wave function is replicated in each crystal translated by a vector of the supercrystal lattice of the form
\begin{equation}
\tilde{\textbf{R}} = c_1 \tilde{\textbf{a}}_1 + c_2 \tilde{\textbf{a}}_2 + c_3 \tilde{\textbf{a}}_3, \ c_i \in \mathbb{Z} .
\end{equation}
Consequently, $\psi(\textbf{r})$ can be expanded as Fourier series by means of the vector $\textbf{q}$ of the reciprocal lattice of the supercrystal, i.e.,
\begin{equation}
\psi(\textbf{r}) = \sum_{\textbf{q}} \varphi_{\textbf{q}} \ \dfrac{e^{i \textbf{q} \cdot \textbf{r}}}{\sqrt{V}},
\label{eq: formainizialeautofunzioniBloch1}
\end{equation}
with
\begin{equation}
\varphi_{\textbf{q}} = \dfrac{1}{\sqrt{V}} \int_{V} d^3 \textbf{r} \ \psi(\textbf{r}) \ e^{- i \textbf{q} \cdot \textbf{r}}.
\label{eq: formainizialeautofunzioniBloch2}
\end{equation}
Consequently, the computation of $\varphi_{\textbf{q}}$ is required, as it leads to the solution of the Bloch eigenvalue problem and ultimately to the proof of Bloch's theorem. We require that $\psi(\textbf{r})$ be a solution of the Bloch eigenvalue problem, that is,
\begin{equation}
\left[ \dfrac{\hat{\textbf{p}}^2}{2 m_e} + \sum_{\textbf{G}} V_{\textbf{G}} \ \dfrac{e^{i \textbf{G} \cdot \textbf{r}}}{\sqrt{V_d}} \right] \sum_{\textbf{q}} \varphi_{\textbf{q}} \ \dfrac{e^{i \textbf{q} \cdot \textbf{r}}}{\sqrt{V}} = \mathcal{E} \sum_{\textbf{q}} \varphi_{\textbf{q}} \ \dfrac{e^{i \textbf{q} \cdot \textbf{r}}}{\sqrt{V}},
\end{equation}
\begin{equation}
\dfrac{\hat{\textbf{p}}^2}{2m_e} \sum_{\textbf{q}} \varphi_{\textbf{q}} \ \dfrac{e^{i \textbf{q} \cdot \textbf{r}}}{\sqrt{V}} = \sum_{\textbf{q}} \varphi_{\textbf{q}} \dfrac{\hslash^2 \textbf{q}^2}{2m_e} \dfrac{e^{i \textbf{q} \cdot \textbf{r}}}{\sqrt{V}},
\end{equation}
\begin{equation}
\sum_{\textbf{G}} V_{\textbf{G}} \dfrac{e^{i \textbf{G} \cdot \textbf{r}}}{\sqrt{V_d}} \sum_{\textbf{q}} \varphi_{\textbf{q}} \dfrac{e^{i \textbf{q} \cdot \textbf{r}}}{\sqrt{V}} = \sum_{\textbf{G}} \sum_{\textbf{q}} \dfrac{V_{\textbf{G}}}{\sqrt{V_d}} \varphi_{\textbf{q}} \dfrac{e^{i (\textbf{q}+\textbf{G}) \cdot \textbf{r}}}{\sqrt{V}}.
\end{equation}
By definition, the components of the vectors $\textbf{q}$ are rational, while those of the vectors $\textbf{G}$ are integers. Therefore, the components of $\textbf{q} + \textbf{G}$ are also rational. Since the sum is taken over the $\textbf{q}$ vectors, we perform the change of variables $\textbf{q} \rightarrow \textbf{q} - \textbf{G}$ in $\varphi_{\textbf{q}}$, and accordingly set $\textbf{q} + \textbf{G} \rightarrow \textbf{q}$ in the plane-wave element $\frac{e^{i (\textbf{q}+\textbf{G}) \cdot \textbf{r}}}{\sqrt{V}}$. Given
\begin{equation}
\sum_{\textbf{G}} \sum_{\textbf{q}} \dfrac{V_{\textbf{G}}}{\sqrt{V_d}} \varphi_{\textbf{q}} \dfrac{e^{i (\textbf{q}+\textbf{G}) \cdot \textbf{r}}}{\sqrt{V}} = \sum_{\textbf{G}} \sum_{\textbf{q}} \dfrac{V_{\textbf{G}}}{\sqrt{V_d}} \varphi_{\textbf{q}-\textbf{G}} \dfrac{e^{i \textbf{q} \cdot \textbf{r}}}{\sqrt{V}},
\end{equation}
we have
\begin{equation}
\sum_{\textbf{q}} \left[ \dfrac{\hslash^{2} \textbf{q}^{2}}{2 m_e} \varphi_{\textbf{q}} + \sum_{\textbf{G}} \dfrac{V_{\textbf{G}}}{\sqrt{V_d}} \varphi_{\textbf{q}-\textbf{G}} - \mathcal{E} \varphi_{\textbf{q}} \right] \dfrac{e^{i \textbf{q} \cdot \textbf{r}}}{\sqrt{V}} = 0,
\end{equation}
which, since $\left\lbrace \frac{e^{i \textbf{q} \cdot \textbf{r}}}{\sqrt{V}} \right\rbrace$ forms a basis for functions defined on the crystal, implies
\begin{equation}
\left( \dfrac{\hslash^{2} \textbf{q}^{2}}{2 m_e} - \mathcal{E} \right) \varphi_{\textbf{q}} + \sum_{\textbf{G}} \dfrac{V_{\textbf{G}}}{\sqrt{V_d}} \varphi_{\textbf{q}-\textbf{G}} = 0, \ \forall \ \textbf{q}.
\end{equation}
Now, we write the division between an integer $m_1$ and a natural $N_1$ as the sum of an integer $d_1$ and the remainder natural $S_1$ divided by $N_1$, i.e.,
\begin{equation}
\dfrac{m_1}{N_1} = d_1 + \dfrac{S_1}{N_1}, \ m_1 \in \mathbb{Z}, \ N_1 \in \mathbb{N}, \ d_1 \in \mathbb{Z}, \ S_1 \in \mathbb{N}, \ 0 \leq S_1 \leq N_1 - 1.
\end{equation}
From $\eqref{eq: vettorereticoloreciprocosupercristallo}$ we have
\begin{equation}
\textbf{q} = \left( d_1 + \dfrac{S_1}{N_1} \right) \textbf{b}_1 + \left( d_2 + \dfrac{S_2}{N_2} \right) \textbf{b}_2 + \left( d_3 + \dfrac{S_3}{N_3} \right) \textbf{b}_3, \ d_i \in \mathbb{Z}, \ S_i \in \mathbb{N}, \ \dfrac{S_i}{N_i} < 1.
\label{eq: vettorereticoloreciprocosupercristallo2}
\end{equation}
A vector $\textbf{q}$ of the reciprocal lattice of the supercrystal can be written as
\begin{equation}
\textbf{q} = \textbf{G}' + \textbf{k},
\end{equation}
that is, it is the sum of a vector $\textbf{G}'$ of the original reciprocal lattice and a vector $\textbf{k}$ of the elementary cell of the original reciprocal lattice. The definitions of the vectors $\textbf{k}$ depend on the rests $\frac{S_i}{N_i}$, with $0 \leq S_i \leq N_i - 1$, so there exist $N_i$ such vectors. We write the vector $\textbf{q}$ taking into account the previous considerations, i.e.,
\begin{equation}
\left[ \dfrac{\hslash^{2}}{2 m_e} \left( \textbf{k}+\textbf{G}' \right)^{2} - \mathcal{E} \right] \varphi_{\textbf{k}+\textbf{G}'} + \sum_{\textbf{G}} \dfrac{V_{\textbf{G}}}{\sqrt{V_d}} \varphi_{\textbf{k}+\textbf{G}'-\textbf{G}} = 0.
\end{equation}
Now, all the wave vectors of the Bloch eigenvalue problem lie either in the reciprocal lattice of the crystal or in the first Brillouin zone. We set $\textbf{G}'-\textbf{G}=\textbf{G}''$ in the sum with respect to $\textbf{G}$, which becomes a sum over $\textbf{G}''$, i.e.,
\begin{equation}
\left[ \dfrac{\hslash^{2}}{2 m_e} \left( \textbf{k}+\textbf{G}' \right)^{2} - \mathcal{E} \right] \varphi_{\textbf{k}+\textbf{G}'} + \sum_{\textbf{G}''} \dfrac{V_{\textbf{G}'-\textbf{G}''}}{\sqrt{V_d}} \varphi_{\textbf{k}+\textbf{G}''} = 0.
\end{equation}
We rename the dummy index $\textbf{G}''=\textbf{G}$ in the second term as follows
\begin{equation}
\left[ \dfrac{\hslash^{2}}{2 m_e} \left( \textbf{k}+\textbf{G}' \right)^{2} - \mathcal{E} \right] \varphi_{\textbf{k}+\textbf{G}'} + \sum_{\textbf{G}} \dfrac{V_{\textbf{G}'-\textbf{G}}}{\sqrt{V_d}} \varphi_{\textbf{k}+\textbf{G}} = 0 ,
\end{equation}
and we obtain the Bloch eigenvalue problem in explicit form
\begin{equation}
\sum_{\textbf{G}} \left\lbrace \left[ \dfrac{\hslash^{2}}{2 m_e} \left( \textbf{k}+\textbf{G} \right)^{2} - \mathcal{E} \right] \delta_{\textbf{G},\textbf{G}'} + \dfrac{V_{\textbf{G}'-\textbf{G}}}{\sqrt{V_d}} \right\rbrace \varphi_{\textbf{k}+\textbf{G}} = 0.
\label{eq: equazioneautovaloriproblemaBloch}
\end{equation}
The diagonal terms are kinetic, i.e., $\frac{\hslash^{2}(\textbf{k}+\textbf{G})^{2}}{2 m_e}$, while the off-diagonal terms are $\frac{V_{\textbf{G}'-\textbf{G}}}{\sqrt{V_d}}$, being $V_{\textbf{0}} = 0$. The matrix $\eqref{eq: equazioneautovaloriproblemaBloch}$ is block diagonal: for each $\textbf{k}$ there is a block. Each block has the same size and the non-vanishing matrix elements depend on how many $\textbf{G}$ are involved in the expansion of the periodic effective potential $V(\textbf{r})$. There exist as many blocks as there are $\textbf{k}$ vectors, i.e., there are $N= N_1 N_2 N_3$ blocks. By fixing a particular \(\textbf{k}\), we diagonalize its corresponding block to obtain the eigenvectors $\varphi_{n,\textbf{k}+\textbf{G}}$. Here, \(\textbf{k}\) labels the block (i.e., the crystal momentum), while $n$ indexes the eigenstates within the block, naturally associated with the energy band index for each \(\textbf{k}\). Thus, \(\textbf{k}\) and \(n\) are known as Bloch indices. It follows that the corresponding eigenvalues of the Hamiltonian can be denoted by \(\mathcal{E}_{n,\textbf{k}}\), representing the energy of the \(n\)-th band at crystal momentum \(\textbf{k}\). Now, we have an explicit form of $\psi(\textbf{r})$ of the $\eqref{eq: formainizialeautofunzioniBloch1}$, in particular it is a superposition of Bloch waves, eigenfunctions of a block matrix, i.e.,
\begin{equation}
\varphi_{n,\textbf{k}}(\textbf{r}) = \sum_{\textbf{G}} \dfrac{e^{i \left(\textbf{k}+\textbf{G}\right) \cdot \textbf{r}}}{\sqrt{V}} \varphi_{n,\textbf{k}+\textbf{G}}.
\label{eq: autofunzionidiBloch}
\end{equation}
Now, if we rewrite equation $\eqref{eq: autofunzionidiBloch}$ in the form of $\eqref{eq: teoremadiBloch1}$, that is, if we define
\begin{equation}
U_{n,\textbf{k}}(\textbf{r}) = \sum_{\textbf{G}} \dfrac{e^{i \textbf{G} \cdot \textbf{r}}}{\sqrt{V}} \psi_{n,\textbf{k}+\textbf{G}},
\label{eq: formaesplicitafunzioniperiodicheteoremaBloch}
\end{equation}
then
\begin{align}
U_{n,\textbf{k}}\left(\textbf{r}+\textbf{R}\right) &= \sum_{\textbf{G}} \dfrac{e^{i \textbf{G} \cdot \left( \textbf{r} + \textbf{R} \right)}}{\sqrt{V}} \varphi_{n,\textbf{k}+\textbf{G}} = \notag \\
&= \sum_{\textbf{G}} \dfrac{e^{i \textbf{G} \cdot \textbf{r}}}{\sqrt{V}} \varphi_{n,\textbf{k}+\textbf{G}} = \notag \\
&= U_{n,\textbf{k}}(\textbf{r}),
\end{align}
that is, the function $U_{n,\textbf{k}}(\textbf{r})$ is periodic with the periodicity of the crystal lattice (i.e., with respect to unit cells), and the theorem is thus proved.
\end{proof}
\end{theorem}
Given the action of the momentum operator $\hat{\textbf{p}}=-i \hslash \nabla$ on a Bloch wave, one finds 
\begin{equation}
\hat{\textbf{p}} \varphi_{n,\textbf{k}}(\textbf{r}) = \hslash \textbf{k} \varphi_{n,\textbf{k}}(\textbf{r}) - i \hbar e^{i \textbf{k} \cdot \textbf{r}} \nabla U_{n,\textbf{k}}.
\end{equation}
In the case of many free particles, the Bloch eigenfunctions reduce to plane waves $\eqref{eq: ondepiane}$, which are eigenfunctions of the momentum operator. In a crystal lattice, however, the vector $\textbf{k}$ is no longer the physical momentum, but rather the crystal quasi-momentum, and Bloch waves are not eigenfunctions of the momentum operator with eigenvalue $\hslash \textbf{k}$. \newline
Electrons in the lattice become quasiparticles whose properties are modified by the periodic potential. In particular, its inertial response is described by an effective mass, different from the bare electron mass. To describe these quasiparticles within Bloch's formalism, we refer to them as Bloch quasi-electrons. The effective mass of a Bloch quasi-electron depends on the wave vector $\textbf{k}$ and is generally anisotropic. It is therefore characterized by a mass tensor defined as
\begin{equation}
\dfrac{1}{m_{ij}\left(n,\textbf{k}\right)} = \dfrac{1}{\hslash^2} \dfrac{\partial^{2} \mathcal{E}_{n,\textbf{k}}}{\partial k_j \partial k_i}.
\end{equation}
In the case of many free particles, $\mathcal{E}_{n,\textbf{k}} = \mathcal{E}_{\textbf{k}} = \frac{\hslash^2 \textbf{k}^2}{2m_e}$, $\forall \ n$, then the quantity $m_{ij}^{-1}\left(n,\textbf{k}\right)$ is a diagonal matrix and the diagonal terms are equal to $m^{-1}_e$. De Broglie's relation $|\textbf{p}|=\hslash |\textbf{k}|$ is no longer valid; it must be generalized, just as the definition of group velocity must be generalized: the group velocity of a Bloch wave is given by
\begin{equation}
v_{n,\textbf{k}} = \dfrac{1}{\hslash} \nabla_{\textbf{k}} \mathcal{E}_{n,\textbf{k}}.
\end{equation}
To conclude the section, we now prove that the Bloch eigenstates are also eigenfunctions of the spatial translation operator. Indeed,
\begin{align}
\hat{T}_{\textbf{m}} \varphi_{n,\textbf{k}}(\textbf{r}) &= \varphi_{n,\textbf{k}}\left(\textbf{r}+\textbf{R}_{\textbf{m}} \right) = \notag \\
&= e^{i \textbf{k} \cdot \left( \textbf{r} + \textbf{R}_{\textbf{m}} \right)} U_{n,\textbf{k}}\left(\textbf{r} + \textbf{R}_{\textbf{m}} \right) = \notag \\
&= e^{i \textbf{k} \cdot \left( \textbf{r} + \textbf{R}_{\textbf{m}} \right)} U_{n,\textbf{k}}(\textbf{r}) = \notag \\
&= e^{i \textbf{k} \cdot \textbf{r}} e^{i \textbf{k} \cdot \textbf{R}_{\textbf{m}}} U_{n,\textbf{k}}(\textbf{r}) = \notag \\
&= e^{i \textbf{k} \cdot \textbf{R}_{\textbf{m}}} \varphi_{n,\textbf{k}}(\textbf{r}),
\label{eq: azionedellatraslazionesulleautofunzionidiBloch}
\end{align}
that is, $\varphi_{n,\textbf{k}}(\textbf{r})$ are eigenfunctions of the translation operator $\hat{T}_{\textbf{m}}$. It is worth noting that the action of the translation operator on Bloch functions, given in equation $\eqref{eq: azionedellatraslazionesulleautofunzionidiBloch}$, is perfectly consistent with the action of the spatial translation operator defined in equation $\eqref{eq: azioneoperatoretraslazionespazialesufunzionedonda}$, as the exponential terms differ only by a sign.
\begin{remark}
In what follows, we will adopt the convention of writing the wave vectors in the first Brillouin zone of the crystal as
\begin{equation}
\dfrac{S_1}{N_1} \textbf{b}_1 + \dfrac{S_2}{N_2} \textbf{b}_2 + \dfrac{S_3}{N_3} \textbf{b}_3, \ S_i \in \mathbb{N}, \ 0 \leq S_i \leq N_i - 1.
\end{equation}
\label{eq: vettoridondadellaprimaBZdelcristallo}
\end{remark}
\section{Bloch waves}
Valence electrons are responsible for the formation of chemical bonds in a crystal, so we are particularly interested in determining their corresponding eigenfunctions, denoted by $\psi_{n,\textbf{k}+\textbf{G}}$. To this end, we begin by solving the many-electron problem in the absence of an external potential, treating the ionic potential as a perturbation. This approach is known as the Bloch model or the nearly-free electron model. We first solve the eigenvalue problem for the case $V(\textbf{r})=0$. From equation $\eqref{eq: equazioneautovaloriproblemaBloch}$, note that the eigenvalues correspond to the kinetic energies, since in this limit the Hamiltonian matrix becomes diagonal, that is
\begin{align}
\mathcal{E}_{n,\textbf{k}} &= \dfrac{\hslash^{2}}{2 m_e} \left[ \textbf{k} + \textbf{G} \right]^{2} = \notag \\
&= \dfrac{\hslash^{2}}{2 m_e} \left[ \textbf{k} + m_1 \textbf{b}_1 + m_2 \textbf{b}_2 + m_3 \textbf{b}_3 \right]^{2}.
\end{align}
Let us consider a simple cubic lattice, $\textbf{k}$ is defined in the first Brillouin zone, which is itself a simple cubic lattice of linear size $\frac{2 \pi}{a}$ and components givem by $\eqref{eq: componentibasefirstBrillouinreticolocubico}$. Bloch's theorem returns the following eigenfunctions
\begin{equation}
\varphi^{(0)}_{n,\textbf{k}}(\textbf{r}) = \dfrac{e^{i (\textbf{k}+\textbf{G}) \cdot \textbf{r}}}{\sqrt{V}} ,
\end{equation}
where the superscript $(0)$ denotes the unperturbed (free-electron) eigenfunction, i.e., for $V(\mathbf{r}) = 0$. The functions $\varphi^{(0)}_{n,\textbf{k}+\textbf{G}}$ are eigenvectors of a diagonal Hamiltonian; they are proportional to the $N$ unit vectors of the standard basis. In other words, in the absence of an ionic potential, the Bloch eigenfunctions reduce to plane waves. It is important to note that the wavevectors within the first Brillouin zone are sufficient to fully describe the dynamics of many non-interacting free particles. For example, in one dimension, on a simple cubic lattice, \(\textbf{k}\) is replaced by the scalar \(k\) and $\textbf{r}$ is replaced by $x$, and we have
\begin{equation}
\mathcal{E}_{0,k} = \dfrac{\hslash^{2} k^2}{2 m_e},
\end{equation}
\begin{equation}
\mathcal{E}_{-1,k} = \dfrac{\hslash^{2}}{2 m_e} \left( k - \dfrac{2 \pi}{a} \right)^{2},
\end{equation}
\begin{equation}
\mathcal{E}_{+1,k} = \dfrac{\hslash^{2}}{2 m_e} \left( k + \dfrac{2 \pi}{a} \right)^{2},
\end{equation}
which corresponds to three parabolic dispersion curves centered at \(0\), \(-\frac{2\pi}{a}\), and \(\frac{2\pi}{a}\), respectively. The free-electron parabola is originally defined over the entire wavevector domain. However, when \(k\) is restricted to lie within the first Brillouin zone, the dispersion relation becomes confined to that zone. The remaining parts of the free-particle spectrum are recovered by translating the central parabola by reciprocal lattice vectors. Therefore, it is sufficient to study the energy bands within the first Brillouin zone. What is the effect of the periodic potential? In general, a perturbing potential shifts energy levels and may lift degeneracies. In this case, the parabolas centered at \(-\frac{2\pi}{a}\) and \(\frac{2\pi}{a}\) intersect the central one (centered at \(0\)), resulting in two doubly degenerate energy levels. For example, at \(k = \frac{\pi}{a}\), the degenerate eigenfunctions correspond to \(n = 0\) and \(n = -1\), i.e., $\frac{e^{i \frac{\pi}{a} x}}{\sqrt{L}}$, $\frac{e^{- i \frac{\pi}{a} x}}{\sqrt{L}}$. However, the presence of the periodic potential lifts this degeneracy, as we shall now see. According to perturbation theory, we must diagonalize a \(2 \times 2\) matrix at \(k = \frac{\pi}{a}\). For a system of free particles, the eigenvalues correspond to the diagonal terms. The off-diagonal elements, instead, are determined by the matrix elements of the potential between the two degenerate unperturbed states, i.e.,
\begin{equation}
\langle n = 0 | V(x) | n = -1 \rangle, \ k = \frac{\pi}{a}.
\end{equation}
The resulting matrix takes the form
\begin{equation}
\begin{pmatrix}
\dfrac{\hslash^{2}}{2m_e} \left( \dfrac{\pi}{a} \right)^{2} & \tilde{V} \\
\tilde{V}^{*} & \dfrac{\hslash^{2}}{2m_e} \left( - \dfrac{\pi}{a} \right)^{2} 
\end{pmatrix},
\end{equation}
where
\begin{equation}
\tilde{V} = \dfrac{1}{L} \int dx e^{- i \frac{\pi}{a} x} V(x) e^{i \frac{\pi}{a} x}.
\end{equation}
From the eigenvalue equation
\begin{equation}
\left[ \dfrac{\hslash^{2}}{2m_e} \left( \dfrac{\pi}{a} \right)^{2} - \mathcal{E} \right]^{2} - \left|\tilde{V}\right|^{2} = 0,
\end{equation}
we get 
\begin{equation}
\mathcal{E}_{\pm} = \dfrac{\hslash^{2}}{2m_e} \left( \dfrac{\pi}{a} \right)^{2} \pm \left|\tilde{V}\right|^{2} .
\end{equation}
In other words, the presence of the ionic potential is responsible for the opening of energy gaps in a solid. To summarize: the free-electron spectrum, originally continuous, is folded into the first Brillouin zone due to the periodicity of the crystal lattice. This folding creates points of degeneracy where different energy branches intersect. The periodic potential lifts these degeneracies, resulting in the formation of energy gaps. Near the degenerate points, the energy bands split and slightly deviate from the free-electron parabolas, creating a gap. This behavior repeats itself at higher energies, giving rise to the characteristic band structure of solids. Bloch’s theorem provides the fundamental explanation for the existence and structure of these energy bands.
\section{Wannier functions}
The tight-binding model is suited for describing valence electrons that are strongly localized around their respective nuclei, as is the case, for example, with the valence electrons of transition metals in the periodic table. The model begins with isolated atoms. Suppose we have \(N\) such atoms initially placed at infinite separation. These atoms are then gradually brought together to form a crystal, reducing the inter-atomic distance until it equals the lattice constant \(a\). At infinite separation, each atomic energy level is \(N\)-fold degenerate. As atoms approach one another, electrons can tunnel to neighboring nuclei, a process known as hybridization, which lifts the degeneracy and causes the energy levels to split. This splitting gives rise to energy bands. Hence, in the tight-binding picture, the electronic band structure of the crystal emerges from the quantum‑mechanical tunneling of electrons between the periodic potential wells created by the atoms. This phenomenon is called electron hopping. From $\eqref{eq: azionedellatraslazionesulleautofunzionidiBloch}$ it follows
\begin{equation}
\left|\hat{T}_m \varphi_{n,\textbf{k}}\left(\textbf{r}\right)\right|^{2} = \left|\varphi_{n,\textbf{k}}(\textbf{r})\right|^{2}.
\end{equation}
Bloch eigenfunctions are extended waves: the probability of finding the electron at a point within an elementary cell is independent of the cell. As discussed earlier, any point in the reciprocal lattice space of a crystal can be obtained by a translation from the elementary cell, then
\begin{equation}
\varphi_{n,\textbf{k}+\textbf{G}}(\textbf{r}) = \varphi_{n,\textbf{k}}(\textbf{r}),
\end{equation}
in other words, the elementary cell where we compute $\varphi_{n,\textbf{k}}(\textbf{r})$ is arbitrary. Then $\varphi_{n,\textbf{k}}(\textbf{r})$ can be expanded as Fourier series on the direct lattice as
\begin{equation}
\varphi_{n,\textbf{k}}(\textbf{r}) = \sum_{\textbf{R}} \dfrac{e^{i \textbf{k} \cdot \textbf{R}}}{\sqrt{N}} \omega_{n}\left(\textbf{r},\textbf{R}\right) ,
\end{equation}
where $\omega_{n}\left(\textbf{r},\textbf{R}\right)$ are the Wannier functions. 
\begin{theorem}
Let \(\{ \mathbf{R} \}\) be a finite set of vectors of the direct lattice, and let \(\{ \mathbf{k} \}\) be the corresponding discrete set of wavevectors in the first Brillouin zone. Then, the matrix
\begin{equation}
\dfrac{e^{i \textbf{k} \cdot \textbf{R}}}{\sqrt{N}}
\label{eq: esponenzialediunatrasformatadiscretadiFourier}
\end{equation}
is a unitary transformation.
\begin{proof}
From equation $\eqref{eq: vettoridondadellaprimaBZdelcristallo}$, the components of $\mathbf{k}$ are given by $\frac{S_i}{N_i}$, with integers satisfying $0 \leq S_i \leq N_i - 1$. Therefore, there are $N = N_1 N_2 N_3$ such vectors. On the other hand, the vectors $\mathbf{R}$ belong to the direct lattice, where the correspondence between lattice points and Wigner-Seitz cells is one-to-one, yielding the same number $N = N_1 N_2 N_3$ of \(\mathbf{R}\) vectors. Hence, a one-to-one mapping exists between the sets \(\{\mathbf{k}\}\) and \(\{\mathbf{R}\}\), ensuring that the scalar product \(\mathbf{k} \cdot \mathbf{R}\) is well defined. Consequently,
\begin{align}
\textbf{R} \cdot \textbf{k} &= \left( n_1 \textbf{a}_1 + n_2 \textbf{a}_2 + n_3 \textbf{a}_3 \right) \cdot \left( \dfrac{S_1}{N_1} \textbf{b}_1 + \dfrac{S_2}{N_2} \textbf{b}_2 + \dfrac{S_3}{N_3} \textbf{b}_3 \right) = \notag \\
&= 2 \pi \left[ \dfrac{n_1 S_1}{N_1} + \dfrac{n_2 S_2}{N_2} + \dfrac{n_3 S_3}{N_3} \right], \ n_i, S_i \in \mathbb{N}, \ 0 \leq n_i \leq N_i - 1, \  0 \leq S_i \leq N_i - 1,
\end{align}
where we used $\textbf{a}_i \cdot \textbf{b}_j = 2 \pi \delta_{ij}$. Consider
\begin{align}
\sum_{\textbf{R}} \dfrac{e^{i \textbf{k} \cdot \textbf{R}}}{\sqrt{N}} &= \sum_{n_1 , n_2 , n_3} \dfrac{e^{i 2 \pi n_1 \frac{S_1}{N_1}}}{\sqrt{N_1}} \dfrac{e^{i 2 \pi n_2 \frac{S_2}{N_2}}}{\sqrt{N_2}} \dfrac{e^{i 2 \pi n_3 \frac{S_3}{N_3}}}{\sqrt{N_3}} = \notag \\
&= \left[ \sum_{n_1 = 0}^{N_1 - 1} \dfrac{e^{i 2 \pi n_1 \frac{S_1}{N_1}}}{\sqrt{N_1}} \right] \left[ \sum_{n_2 = 0}^{N_2 - 1} \dfrac{e^{i 2 \pi n_2 \frac{S_2}{N_2}}}{\sqrt{N_2}} \right] \left[ \sum_{n_3 = 0}^{N_3 - 1} \dfrac{e^{i 2 \pi n_3 \frac{S_3}{N_3}}}{\sqrt{N_3}} \right].
\end{align}
Each of these three terms is a function of $\textbf{k}$, that is, of $S_i$, $i=1,2,3$. Consider the first component, if $S_1=0$ (or equivalently if $k_1 = 0$), we have
\begin{align}
\sum_{n_1 = 0}^{N_1 - 1} \left. \dfrac{e^{i 2 \pi n_1 \frac{S_1}{N_1}}}{\sqrt{N_1}} \right|_{S_1 = 0} &= \dfrac{1}{\sqrt{N_1}} \sum_{n_1 = 0}^{N_1 - 1} 1 = \notag \\
&= \sqrt{N_1},
\end{align}
and we want to prove that it is zero for $S_1 \neq 0$. If we assume that $S_1 \neq 0$ and that $e^{i 2 \pi n_1 \frac{S_1}{N_1}} \neq 1$, we have
\begin{align}
\sum_{n_1 = 0}^{N_1 - 1} \dfrac{e^{i 2 \pi n_1 \frac{S_1}{N_1}}}{\sqrt{N_1}} &= \sum_{n_1 = 0}^{N_1 - 1} \dfrac{e^{\left( i 2 \pi \frac{S_1}{N_1} \right) n_1}}{\sqrt{N_1}} = \notag \\
&= \dfrac{1}{\sqrt{N_1}} \dfrac{1-\left( e^{i 2 \pi \frac{S_1}{N_1}} \right)^{N_1}}{1-e^{i 2 \pi \frac{S_1}{N_1}}} = \notag \\
&= \dfrac{1}{\sqrt{N_1}} \dfrac{1-e^{i 2 \pi S_1}}{1-e^{i 2 \pi \frac{S_1}{N_1}}}.
\end{align}
Now, the quantity $S_1$ is a natural, so the above ratio is zero except for $S_1 \neq 0$, which we ruled out a priori. From
\begin{equation}
\sum_{\textbf{R}} \dfrac{e^{i \left( \textbf{k}_i - \textbf{k}_j \right) \cdot \textbf{R}}}{N} = \delta_{\textbf{k}_i,\textbf{k}_j},
\end{equation} 
the matrix is unitary.
\end{proof}
\end{theorem}
\begin{remark}[Discrete Fourier transform]
Given a set of $N$ complex numbers $\left\lbrace c_n \right\rbrace$, with \(n = 0, \ldots, N-1\), its discrete Fourier transform (DFT) is defined as the set
\begin{equation}
\tilde{c}_{k} = \sum_{n=0}^{N-1} \dfrac{e^{i \frac{2 \pi k n}{N}}}{\sqrt{N}} c_n, \ k = 0,\ldots,N-1.
\end{equation}
\label{oss: trasformatadiscretadiFourier}
\end{remark}
According to Remark $\eqref{oss: trasformatadiscretadiFourier}$, Bloch eigenfunctions are the discrete Fourier transforms of Wannier functions. Since $\eqref{eq: esponenzialediunatrasformatadiscretadiFourier}$ is a unitary transformation, is invertible and the inverse Fourier transform is indexed on the $\textbf{k}$ vectors of the first Brillouin zone. Then, Wannier functions are defined by a discrete inverse Fourier transform of Bloch eigenfunctions, i.e.,
\begin{equation}
\omega_{n}\left(\textbf{r},\textbf{R}\right) = \sum_{\textbf{k} \in \text{1}^{\text{st}} \, \text{BZ}} \dfrac{e^{- i \textbf{k} \cdot \textbf{R}}}{\sqrt{N}} \psi_{n,\textbf{k}}(\textbf{r}).
\end{equation}
For each site $\textbf{R}$ there exists a Wannier function, unlike de-localized and extended Bloch functions. The unitary transformation allows one to go from the extended basis of Bloch functions to the localized basis of Wannier functions. A very useful property is the following 
\begin{equation}
\omega_{n}\left(\textbf{r}+\textbf{R}',\textbf{R}+\textbf{R}'\right) = \omega_n\left(\textbf{r},\textbf{R}\right),
\end{equation}
where $\textbf{R}'$ is any vector of the direct lattice. For $\textbf{R}'=-\textbf{R}$,
\begin{align}
\omega_{n}\left(\textbf{r}+\textbf{R}',\textbf{R}+\textbf{R}'\right) &= \omega_n\left(\textbf{r}-\textbf{R},\textbf{0}\right) \equiv \notag \\
&\equiv \omega_{n}\left(\textbf{r}-\textbf{R}\right) ,
\end{align}
that is, it does not depend on the two indices separately but only on their difference vector. Wannier functions depend on \(n\) and \(\mathbf{r} - \mathbf{R}\), hence the relation between the two functions can be written as
\begin{equation}
\omega_{n}\left(\textbf{r}-\textbf{R}\right) = \sum_{\textbf{k} \in \text{1}^{\text{st}} \, \text{BZ}} \dfrac{e^{- i \textbf{k} \cdot \textbf{R}}}{\sqrt{N}} \varphi_{n,\textbf{k}}(\textbf{r}),
\end{equation}
\begin{equation}
\varphi_{n,\textbf{k}}(\textbf{r}) = \sum_{\textbf{R}} \dfrac{e^{i \textbf{k} \cdot \textbf{R}}}{\sqrt{N}} \omega_{n}\left(\textbf{r}-\textbf{R}\right).
\end{equation}
Ultimately, it can be shown that the Wannier functions form a complete orthonormal set with respect to both the band index and the lattice site.
\section{Tight-binding model}
Given an atom, an electron of one of the outermost orbitals, i.e., a valence electron, is immersed in an atomic potential $V^{at.}(\textbf{r})$, and thus satisfies the eigenvalue equation
\begin{equation}
\left[ \dfrac{\hat{\textbf{p}}^{2}}{2 m_e} + V^{at.}(\textbf{r}) \right] \varphi_n^{at.}(\textbf{r}) = \mathcal{E}_n^{at.} \varphi_n^{at.}(\textbf{r}) ,
\end{equation}
where $\varphi_n^{at.}(\textbf{r})$ are the eigenfunctions of the atomic orbitals and $n$ is referred to the orbital. To form the crystal, many atoms must be arranged in the lattice, if the atoms are $N$, the energy $\mathcal{E}_n^{at.}$ is $N$ times degenerate. When the atoms are in the crystal, the degeneracy is removed because of hybridization and a band is created. Now, we replace the Wannier functions by localized atomic orbitals. The eigenfunctions of $\mathcal{H}^{(e)}_{H.F.}$, the Bloch functions $\varphi_{n,\textbf{k}}(\textbf{r})$, can be written in terms of Wannier functions, so if we rewrite
\begin{equation}
\varphi_{n,\textbf{k}}(\textbf{r}) = \sum_{\textbf{R}} \dfrac{e^{i \textbf{k} \cdot \textbf{R}}}{\sqrt{N}} \varphi_n^{at.}\left(\textbf{r}-\textbf{R}\right),
\end{equation}
the eigenvalue equation for $\mathcal{H}^{(e)}_{H.F.}$ can be written as
\begin{equation}
\left[ \dfrac{\hat{\textbf{p}}^{2}}{2 m_e} + V(\textbf{r}) \right] \left[ \sum_{\textbf{R}} \dfrac{e^{i \textbf{k} \cdot \textbf{R}}}{\sqrt{N}} \varphi_n^{at.}\left(\textbf{r}-\textbf{R}\right) \right] = \mathcal{E} \sum_{\textbf{R}} \dfrac{e^{i \textbf{k} \cdot \textbf{R}}}{\sqrt{N}} \varphi_n^{at.}\left(\textbf{r}-\textbf{R}\right) ,
\end{equation}
\begin{equation}
\sum_{\textbf{R}} e^{i \textbf{k} \cdot \textbf{R}} \left[ \dfrac{\hat{\textbf{p}}^{2}}{2 m_e} + V(\textbf{r}) \right] \varphi_n^{at.}\left(\textbf{r}-\textbf{R}\right) = \mathcal{E} \sum_{\textbf{R}} e^{i \textbf{k} \cdot \textbf{R}} \varphi_n^{at.} \left(\textbf{r}-\textbf{R}\right).
\end{equation}
We multiply both members to the left by $\left( \varphi_n^{at.}(\textbf{r}) \right)^*$, integrate both members with respect to the measure $d^3 \textbf{r}$, i.e.,
\begin{equation}
\sum_{\textbf{R}} e^{i \textbf{k} \cdot \textbf{R}} \int d^3 \textbf{r} \left( \varphi_n^{at.}(\textbf{r}) \right)^* \left[ \dfrac{\hat{\textbf{p}}^{2}}{2 m_e} + V(\textbf{r}) \right] \varphi_n^{at.}\left(\textbf{r}-\textbf{R}\right) = \mathcal{E} \sum_{\textbf{R}} e^{i \textbf{k} \cdot \textbf{R}} \int d^3 \textbf{r} \ \left( \varphi_n^{at.}(\textbf{r}) \right)^* \varphi_n^{at.}\left(\textbf{r}-\textbf{R}\right),
\end{equation}
\begin{equation}
\sum_{\textbf{R}} e^{i \textbf{k} \cdot \textbf{R}} \int d^3 \textbf{r} \left( \varphi_n^{at.}(\textbf{r}) \right)^* \left[ \dfrac{\hat{\textbf{p}}^{2}}{2 m_e} - \mathcal{E} \right] \varphi_n^{at.}\left(\textbf{r}-\textbf{R}\right) = \sum_{\textbf{R}} e^{i \textbf{k} \cdot \textbf{R}} \int d^3 \textbf{r} \ \left( \varphi_n^{at.}(\textbf{r}) \right)^* \left[ - V(\textbf{r}) \right] \varphi_n^{at.}\left(\textbf{r}-\textbf{R}\right).
\end{equation}
We add $V^{at.}\left(\textbf{r}-\textbf{R}\right)$ to both members in the integrand function and recall the eigenvalue equation for the atomic potential, that is
\begin{equation}
\left[ \dfrac{\hat{\textbf{p}}^{2}}{2 m_e} + V^{at.}\left(\textbf{r}-\textbf{R}\right) \right] \varphi_n^{at.}\left(\textbf{r}-\textbf{R}\right) = \mathcal{E}_n^{at.} \varphi_n^{at.}\left(\textbf{r}-\textbf{R}\right),
\end{equation}
then
\begin{equation}
\sum_{\textbf{R}} e^{i \textbf{k} \cdot \textbf{R}} \int d^3 \textbf{r} \left( \varphi_n^{at.}(\textbf{r}) \right)^* \left[ \mathcal{E}_n^{at.} - \mathcal{E} \right] \varphi_n^{at.}\left(\textbf{r}-\textbf{R}\right) = \sum_{\textbf{R}} e^{i \textbf{k} \cdot \textbf{R}} \int d^3 \textbf{r} \ \left( \varphi_n^{at.}(\textbf{r}) \right)^* \left[ V^{at.}\left(\textbf{r}-\textbf{R}\right) - V(\textbf{r}) \right] \varphi_n^{at.}\left(\textbf{r}-\textbf{R}\right).
\end{equation}
On the left-hand side, since the orbital functions form a complete orthonormal set, the scalar products between the atomic orbital centered at \(\mathbf{0}\) and those centered at \(\mathbf{r} - \mathbf{R}\) vanish for all \(\mathbf{R} \neq \mathbf{0}\), while for \(\mathbf{R} = \mathbf{0}\) the scalar product equals 1, yielding the term \(\mathcal{E}_n^{at.} - \mathcal{E}\). On the right-hand side, when \(\mathbf{R} = \mathbf{0}\), the integral reduces to a constant independent of \(\mathbf{k}\), but dependent on the orbital \(n\); let us denote this constant by \(-\tilde{c}_n\). Conversely, for \(\mathbf{R} \neq \mathbf{0}\), the quantity \(\left[ V^{at.}(\mathbf{r} - \mathbf{R}) - V(\mathbf{r}) \right]\) enables the interaction and overlap between \(\left( \varphi^{at.}_n(\mathbf{r}) \right)^*\) and \(\varphi^{at.}_n(\mathbf{r} - \mathbf{R})\), thereby permitting electron transfer between an atom and its nearest neighbors. We then write
\begin{equation}
\mathcal{E}_n^{at.} - \mathcal{E} = - \tilde{c}_n + \sum_{\bm{\delta}} e^{i \textbf{k} \cdot \bm{\delta}} \ t_n,
\end{equation}
\begin{equation}
\mathcal{E} = \mathcal{E}_n^{at.} + \tilde{c}_n - \sum_{\bm{\delta}} e^{i \textbf{k} \cdot \bm{\delta}} \ t_n,
\end{equation}
where
\begin{equation}
t_n = \int d^3 \textbf{r} \ \left( \varphi_n^{at.}(\textbf{r}) \right)^* \left[ V^{at.}\left(\textbf{r}-\bm{\delta}\right) - V(\textbf{r}) \right] \varphi_n^{at.}\left(\textbf{r}-\bm{\delta}\right)
\label{eq: terminedihopping}
\end{equation}
is the integral hopping and $\bm{\delta}$ is the vector connecting the atom in $\textbf{r}$ with its nearest neighbors. From $\eqref{eq: terminedihopping}$, if we add and subtract the kinetic term as follows
\begin{equation}
t_n = \int d^3 \textbf{r} \ \left( \varphi_n^{at.}(\textbf{r}) \right)^* \left[ V^{at.}\left(\textbf{r}-\bm{\delta}\right) + \dfrac{\hat{\textbf{p}}^2}{2m_e} - \dfrac{\hat{\textbf{p}}^2}{2m_e} - V(\textbf{r}) \right] \varphi_n^{at.}\left(\textbf{r}-\bm{\delta}\right),
\end{equation}
and we make use of the eigenvalue equation for the atomic orbital, then
\begin{align}
t_n &= \int d^3 \textbf{r} \ \left( \varphi_n^{at.}(\textbf{r}) \right)^* \mathcal{E}_n^{at.} \varphi_n^{at.}\left(\textbf{r}-\bm{\delta}\right) - \int d^3 \textbf{r} \ \left( \varphi_n^{at.}(\textbf{r}) \right)^* \left[ \dfrac{\hat{\textbf{p}}^2}{2m_e} + V(\textbf{r}) \right] \varphi_n^{at.}\left(\textbf{r}-\bm{\delta}\right) = \notag \\
&= \mathcal{E}_n^{at.} \int d^3 \textbf{r} \ \left( \varphi_n^{at.}(\textbf{r}) \right)^* \varphi_n^{at.}\left(\textbf{r}-\bm{\delta}\right) - \int d^3 \textbf{r} \ \left( \varphi_n^{at.}(\textbf{r}) \right)^* \left[ \dfrac{\hat{\textbf{p}}^2}{2m_e} + V(\textbf{r}) \right] \varphi_n^{at.}\left(\textbf{r}-\bm{\delta}\right).
\end{align}
The first integral is null, due to the completeness of the eigenfunctions $\varphi_n^{at.}(\textbf{r})$, while the second contains the Hamiltonian $\mathcal{H}^{(e)}_{H.F.}$, so
\begin{equation}
- t_n = \int d^3 \textbf{r} \ \left( \varphi_n^{at.}(\textbf{r}) \right)^* \mathcal{H}^{(e)}_{H.F.} \varphi_n^{at.}\left(\textbf{r}-\bm{\delta}\right).
\end{equation}
The hopping term is responsible for the motion of an electron from one site to its nearest neighbors and depends on the $n$ orbital. Of course, to describe the actual interactions, the contributions of the interactions of one atom with all atoms must be added. However, if the electrons are localized, such terms are negligible. Here, the energy depends on the quantum number $n$ and the crystalline momentum $\textbf{k}$. We write
\begin{equation}
\mathcal{E}_{n,\textbf{k}} = \mathcal{E}_n^{at.} + \tilde{c}_n - \sum_{\bm{\delta}} e^{i \textbf{k} \cdot \bm{\delta}} \ t_n.
\end{equation}
Now, if the sum $\sum_{\bm{\delta}} e^{i \textbf{k} \cdot \bm{\delta}} \ t_n$ vanishes, the energy reduces to the atomic energy plus a constant $\tilde{c}_n$, which arises from the presence of all atoms in the lattice and represents the crystal field effect, the potential exerted on the electron by the surrounding atoms. Conversely, if the hopping integral is nonzero, the energy becomes $\textbf{k}$-dependent, resulting in the formation of an energy band.
\subsection{Application on a simple cubic lattice}
In the simple cubic lattice model, atoms are positioned at the vertices of a cube with side length \(a\), where each site has six nearest neighbors, and the displacement vector \(\bm{\delta}\) can take one of the following values: \((\pm a, 0, 0)\), \((0, \pm a, 0)\), or \((0, 0, \pm a)\). Our goal is to compute the energy band \(\mathcal{E}_{n,\mathbf{k}}\) and to analyze its behavior in the limiting cases. In particular, we are interested in its expansion near the center of the Brillouin zone (\(\mathbf{k} \rightarrow \mathbf{0}\)). We have
\begin{align}
\mathcal{E}_{n,\textbf{k}} &= \mathcal{E}_n^{at.} + \tilde{c}_n - \sum_{\bm{\delta}} e^{i \textbf{k} \cdot \bm{\delta}} \ (+t_n) = \notag \\
&= \mathcal{E}_n^{at.} + \tilde{c}_n - t_n \left[ e^{i k_x a} + e^{- i k_x a} + e^{i k_y a} + e^{- i k_y a} + e^{i k_z a} + e^{- i k_z a} \right] = \notag \\
&= \mathcal{E}_n^{at.} + \tilde{c}_n - 2 t_n \left[ \cos(k_x a) + \cos(k_y a) + \cos(k_z a) \right], \ k_x, k_y, k_z \in \left[- \dfrac{\pi}{a} , \dfrac{\pi}{a}  \right].
\end{align}
In the one-dimensional case, that is, for a linear lattice chain, the vector \(\mathbf{k}\) is replaced by the scalar \(k\), and the energy band becomes
\begin{equation}
\mathcal{E}_{n,k} = \mathcal{E}_n^{at.} + \tilde{c}_n - 2 t_n \cos(k a).
\end{equation}
Since $k = \frac{2 \pi}{\lambda}$, in the limit $k \rightarrow 0$, the wavelength $\lambda$ of the Bloch function is strictly greater than the lattice step, that is
\begin{equation}
\cos(k a) \sim 1 - \dfrac{(k a)^2}{2},
\end{equation}
\begin{equation}
\mathcal{E}_n(k) = t_n (k a)^2 + \const,
\end{equation}
in other words for small wave vectors, the energy band is approximated by a parabola. The previous equation can be rewritten as
\begin{equation}
\mathcal{E}_{n,k} \sim \dfrac{\hslash^2 k^2}{2 m^*},
\end{equation}
up to a constant, where 
\begin{equation}
m^{*}(k = 0) = \dfrac{\hslash^2}{2 t_n a^2}
\end{equation}
is the effective mass for $k =0$. If $k \rightarrow 0$, then
\begin{equation}
\mathcal{E}_{n,k} \underset{k \rightarrow 0}{\sim} \dfrac{\hslash^2 k^2}{2 m^*},
\end{equation}
that is, the energy spectrum approaches that of a free particle, with the electron mass effectively replaced by a renormalized effective mass. When the hopping amplitude vanishes, the energy band becomes completely flat, the effective mass diverges, and the electron is strictly localized at an atomic site. In contrast, increasing the hopping amplitude reduces the effective mass, allowing the electron to delocalize and propagate more easily through the lattice. The parameter \(t_n\) thus governs both the width of the energy band and the strength of the inter-site electron hopping. In three dimensions, the group velocity is given by
\begin{equation}
\textbf{v}_{n,\textbf{k}} = \dfrac{1}{\hslash} 2 t_n a \left( \sin(k_x a), \sin(k_y a), \sin(k_z a) \right),
\end{equation}
which satisfies
\begin{equation}
\textbf{v}_{n,\textbf{k}} \underset{\textbf{k} \rightarrow \textbf{0}}{\sim} \dfrac{\hslash \textbf{k}}{m^*},
\end{equation}
i.e., the system behaves as a free particle system, where the electron mass is replaced by an effective mass that accounts for the effects of the periodic potential.
\section{Bloch Hamiltonian in second quantization}
The Hamiltonian $\mathcal{H}^{(e)}_{H.F.}$ is a one-body operator; in second quantization, the Bloch Hamiltonian must be computed as follows
\begin{equation}
\hat{\mathcal{H}}^{(e)}_{H.F.} = \int dx \hat{\psi}^{\dag}(x) \mathcal{H}^{(e)}_{H.F.}(x) \hat{\psi}(x),
\end{equation}
and we examine the equivalent representations in the Bloch and Wannier bases.
\subsection{Bloch's description}
We use Bloch's eigenfunctions as the basis, so the annihilation and creation field operators are written as
\begin{equation}
\hat{\psi}(x) = \sum_{n,\textbf{k},\sigma} \varphi_{n,\textbf{k}}(\textbf{r}) \chi_{\sigma}(s) C_{n,\textbf{k},\sigma},
\label{eq: operatorecampodistruzioneinfunzioniBloch}
\end{equation}
\begin{equation}
\hat{\psi}^\dagger(x) = \sum_{n,\textbf{k},\sigma} \varphi^*_{n,\textbf{k}}(\textbf{r}) \chi^*_{\sigma}(s) C^\dagger_{n,\textbf{k},\sigma},
\label{eq: operatorecampocreazioneinfunzioniBloch}
\end{equation}
where $\textbf{k}$ and $n$ are the Bloch indices, so Bloch's description in second quantization is given by
\begin{align}
\hat{\mathcal{H}}^{(e)}_{H.F.} &= \int dx \left( \sum_{n,\mathbf{k},\sigma} \varphi^*_{n,\mathbf{k}}(\mathbf{r})\, \chi^*_{\sigma}(s)\, C^\dagger_{n,\mathbf{k},\sigma} \right) \mathcal{H}^{(e)}_{H.F.} \left( \sum_{n',\mathbf{k}',\sigma'} \varphi_{n',\mathbf{k}'}(\mathbf{r})\, \chi_{\sigma'}(s)\, C_{n',\mathbf{k}',\sigma'} \right) \notag \\
&= \int dx \left( \sum_{n,\mathbf{k},\sigma} \varphi^*_{n,\mathbf{k}}(\mathbf{r})\, \chi^*_{\sigma}(s)\, C^\dagger_{n,\mathbf{k},\sigma} \right) 
\mathcal{E}_{n',\mathbf{k}'} \left( \sum_{n',\mathbf{k}',\sigma'} \varphi_{n',\mathbf{k}'}(\mathbf{r})\, \chi_{\sigma'}(s)\, C_{n',\mathbf{k}',\sigma'} \right) \notag \\
&= \sum_{n,\mathbf{k},\sigma} \mathcal{E}_{n,\mathbf{k}}\, C^\dagger_{n,\mathbf{k},\sigma} C_{n,\mathbf{k},\sigma},
\end{align}
which represents the first term of the Hamiltonian $\eqref{eq: Hamiltonianadiunsolido}$. Each eigenvector is a Slater determinant of the Bloch wave functions with eigenvalues the occupation numbers $0$ and $1$. 
\subsection{Wannier's description}
Let us expand the field operators in the basis of Wannier functions. The indices \(\{n, \mathbf{R}, \sigma\}\) refer respectively to: \(n\), the band or orbital index; \(\mathbf{R}\), the position of the lattice site (i.e., the center of the corresponding unit cell); and \(\sigma\), the spin projection. Under the approximation that atomic orbitals are replaced by Wannier functions, the field operators take the form
\begin{equation}
\hat{\psi}(x) = \sum_{n,\textbf{R},\sigma} \varphi^{at.}_n\left(\textbf{r}-\textbf{R}\right) \chi_{\sigma}(s) C_{n,\textbf{R},\sigma},
\end{equation}
\begin{equation}
\hat{\psi}^\dagger(x) = \sum_{n,\textbf{R},\sigma} \left( \varphi^{at.}_n\left(\textbf{r}-\textbf{R}\right) \right)^* \chi^*_{\sigma}(s) C^\dagger_{n,\textbf{R},\sigma},
\end{equation}
where $C_{n,\textbf{R},\sigma}$ ($C^\dagger_{n,\textbf{R},\sigma}$) destroys (creates) a fermion on the atomic orbital $n$ centered on the $\textbf{R}$ site with spin component $\sigma$. Wannier operators do not act on delocalized waves, like Bloch's, but on a specific lattice atom. We consider only contributions from the same site as $\textbf{R}$ or its nearest neighbors, so, substituting the Hamiltonian operator, we have
\begin{equation}
\hat{\mathcal{H}}^{(e)}_{H.F.} = \sum_{n,\textbf{R},\sigma} \left(\mathcal{E}_n^{at.} + \tilde{c}_n\right) C^{\dag}_{n,\textbf{R},\sigma} C_{n,\textbf{R},\sigma} - \sum_{\substack{n,\textbf{R}, \\ \bm{\delta},\sigma}} t_n C^{\dag}_{n,\textbf{R} - \bm{\delta},\sigma} C_{n,\textbf{R},\sigma}.
\end{equation}
The first term is diagonal if $\textbf{R}= \textbf{R}'$, while the second term is off-diagonal, for $\textbf{R}' = \textbf{R} - \bm{\delta}$: the Hamiltonian operator is not diagonal in the Wannier basis, consistent with the fact that it is the Bloch functions that are eigenfunctions of the Bloch Hamiltonian and not the Wannier functions. From equations $\eqref{eq: operatoredistruzionenellabasedialtrooperatoredistruzione}$, $\eqref{eq: operatorecreazionenellabasedialtrooperatorecreazione}$, with the following identifications
\begin{equation}
\left\lbrace
\begin{array}{ll}
b_{n,\mathbf{k},\sigma}, \quad b^\dagger_{n,\mathbf{k},\sigma} & \text{Bloch operators} \\[6pt]
c_{n,\mathbf{R},\sigma}, \quad c^\dagger_{n,\mathbf{R},\sigma} & \text{Wannier operators} \\[6pt]
m_{\mathbf{k}} = \dfrac{e^{-i \mathbf{k} \cdot \mathbf{R}}}{\sqrt{N}} & \text{Coefficients of the unitary transformation}
\end{array}
\right.
,
\end{equation}
the annihilation operator in Wannier's basis is
\begin{equation}
c_{n,\textbf{R},\sigma} = \sum_{\textbf{k} \in \text{1}^{\text{st}} \, \text{BZ}} \dfrac{e^{i \textbf{k} \cdot \textbf{R}}}{\sqrt{N}} b_{n,\textbf{k},\sigma} ,
\end{equation}
and the creation operator in Wannier's basis is
\begin{equation}
c^{\dagger}_{n,\textbf{R},\sigma} = \sum_{\textbf{k} \in \text{1}^{\text{st}} \, \text{BZ}} \dfrac{e^{- i \textbf{k} \cdot \textbf{R}}}{\sqrt{N}} b^{\dagger}_{n,\textbf{k},\sigma}.
\label{eq: operatoricreazionediWannierinfunzionedeglioperatoricreazioneBloch}
\end{equation}
The hamiltonian becomes
\begin{equation}
\hat{\mathcal{H}}^{(e)}_{H.F.} = \sum_{\substack{n,\textbf{k}, \\ \textbf{R},\sigma}} \left(\mathcal{E}_n^{at.} + \tilde{c}_n\right) b^{\dag}_{n,\textbf{k},\sigma} b_{n,\textbf{k},\sigma} - \sum_{\substack{n,\textbf{k}, \\ \bm{\delta},\sigma}} t_n e^{i \textbf{k} \cdot \bm{\delta}} b^{\dag}_{n,\textbf{k},\sigma} b_{n,\textbf{k},\sigma}.
\end{equation}
The first and second terms involve the following sums
\begin{equation}
\dfrac{1}{N} \sum_{\textbf{R}} e^{i \textbf{k} \cdot \textbf{R}} e^{-i \textbf{k}' \cdot \textbf{R}} = \delta_{\textbf{k},\textbf{k}'},
\end{equation}
\begin{equation}
\dfrac{1}{N} \sum_{\textbf{R}} e^{i \textbf{k} \cdot (\textbf{R}-\bm{\delta})} e^{-i \textbf{k}' \cdot \textbf{R}} = \delta_{\textbf{k},\textbf{k}'} e^{- i \textbf{k} \cdot \bm{\delta}},
\end{equation}
respectively. In particular, the exponential $e^{- i \textbf{k} \cdot \bm{\delta}}$ cancels out the corresponding positive exponential term present in the Hamiltonian operator, effectively removing the summation over $\bm{\delta}$. As a result, the Hamiltonian $\hat{\mathcal{H}}^{(e)}_{H.F.}$ becomes diagonal, and the total contribution is given by
\begin{equation}
\hat{\mathcal{H}}^{(e)}_{H.F.} = \sum_{n,\textbf{k},\sigma} \mathcal{E}_{n,\textbf{k}} b^{\dag}_{n,\textbf{k},\sigma} b_{n,\textbf{k},\sigma} ,
\end{equation}
that is, the Wannier and Bloch descriptions are equivalent within the framework of second quantization.
\section{One-dimensional electronic crystal structure}
Given a one-dimensional chain of lattice step $a$, at the center of each pair of sites is placed an atom of a species different from the first: if we add a repeated basis of atoms to each lattice point, we obtain what is called crystal structure. Assume that each atom has only one valence electron, we refer to the outermost orbital for each atom, say $\mathcal{E}_1$ for the first species and $\mathcal{E}_2$ for the second, and construct a model for the Hamiltonian. We assume that $\mathcal{E}_2>\mathcal{E}_1$. From the two atomic levels of the valence electrons of the two types of isolated atoms, the two atoms will be coupled by a jump term. In second quantization,
\begin{equation}
\hat{\mathcal{H}} = \sum_{\tau,\sigma} \mathcal{E}_1 c^{\dag}_{\tau,\sigma} c_{\tau,\sigma} + \sum_{\tau,\sigma} \mathcal{E}_2 d^{\dag}_{\tau + \frac{1}{2},\sigma} d_{\tau + \frac{1}{2},\sigma} - t \sum_{\tau,\sigma} \left[ c^{\dag}_{\tau,\sigma} d_{\tau + \frac{1}{2},\sigma} + c^{\dag}_{\tau,\sigma} d_{\tau - \frac{1}{2},\sigma} + d^{\dag}_{\tau + \frac{1}{2},\sigma} c_{\tau,\sigma} + d^{\dag}_{\tau - \frac{1}{2},\sigma} c_{\tau,\sigma} \right],
\end{equation}
which we write briefly as
\begin{equation}
\hat{\mathcal{H}} = \sum_{\tau,\sigma} \mathcal{E}_1 c^{\dag}_{\tau,\sigma} c_{\tau,\sigma} + \sum_{\tau,\sigma} \mathcal{E}_2 d^{\dag}_{\tau + \frac{1}{2},\sigma} d_{\tau + \frac{1}{2},\sigma} - t \sum_{\tau,\sigma} \left[ c^{\dag}_{\tau,\sigma} d_{\tau + \frac{1}{2},\sigma} + c^{\dag}_{\tau,\sigma} d_{\tau - \frac{1}{2},\sigma} + \text{h.c.} \right],
\end{equation}
where the abbreviation $\text{h.c.}$ stands for the Hermitian conjugate. The operators $c^{\dag}_{\tau,\sigma}$ and $c_{\tau,\sigma}$ correspond to the species $\mathcal{E}1$, whereas $d^{\dag}_{\tau + \frac{1}{2} \sigma}$ and $d_{\tau + \frac{1}{2},\sigma}$ correspond to the species $\mathcal{E}_2$. Here, $\tau$ denotes the site index, and $\sigma$ labels the spin. The Hamiltonian $\hat{\mathcal{H}}$ is composed of three terms: the first represents the contribution of the first species, the second corresponds to the second species, and the third describes the hopping between the two species. Our objective is to diagonalize $\hat{\mathcal{H}}$ in order to find its eigenvalues and eigenvectors. We now turn to Bloch’s formalism. Recall that the Bloch and Wannier annihilation operators are related by
\begin{equation}
c_{\tau,\sigma} = \sum_{k \in \text{1}^{\text{st}} \, \text{BZ}} \dfrac{e^{i k \tau a}}{\sqrt{N}} c_{k,\sigma},
\end{equation}
\begin{equation}
d_{\tau + \frac{1}{2},\sigma} = \sum_{k \in \text{1}^{\text{st}} \, \text{BZ}} \dfrac{e^{i k \left( \tau + \frac{1}{2} \right) a}}{\sqrt{N}} d_{k,\sigma} ,
\end{equation}
where, in the one-dimensional case, the vector \(\mathbf{k}\) is replaced by the scalar \(k\). We know that the operators $c^{\dag}_{\tau,\sigma} c_{\tau,\sigma}$ and $d^{\dag}_{\tau + \frac{1}{2},\sigma} d_{\tau + \frac{1}{2},\sigma}$ are diagonal in both bases. Then,
\begin{equation}
\sum_{\tau,\sigma} c^{\dag}_{\tau,\sigma} d_{\tau + \frac{1}{2},\sigma} = \dfrac{1}{N} \sum_{\tau,\sigma} \sum_k \sum_{k'} e^{- i k \tau a} e^{i k' \left( \tau + \frac{1}{2} \right) a} c^{\dag}_{k,\sigma} d_{k',\sigma}.
\end{equation}
From the completeness 
\begin{equation}
\dfrac{1}{N} \sum_{\tau} e^{- i k \tau a} e^{i k' \left( \tau + \frac{1}{2} \right) a} = e^{i k' \frac{a}{2}} \dfrac{1}{N} \sum_{\tau} e^{- i k \tau a} e^{i k' \tau a} = e^{i k' \frac{a}{2}} \delta_{k',k},
\end{equation}
we have
\begin{equation}
\sum_{\tau,\sigma} c^{\dag}_{\tau,\sigma} d_{\tau + \frac{1}{2},\sigma} = \sum_{\sigma} \sum_k e^{i k \frac{a}{2}} c^{\dag}_{k,\sigma} d_{k,\sigma} ,
\end{equation}
and similarly
\begin{equation}
\sum_{\tau,\sigma} c^{\dag}_{\tau,\sigma} d_{\tau - \frac{1}{2},\sigma} = \sum_{\sigma} \sum_k e^{- i k \tau \frac{a}{2}} c^{\dag}_{k,\sigma} d_{k,\sigma}.
\end{equation}
The Hamiltonian can be written as
\begin{align}
\hat{\mathcal{H}} &= \sum_{k,\sigma} \mathcal{E}_1 c^{\dag}_{k,\sigma} c_{k,\sigma} + \sum_{k,\sigma} \mathcal{E}_2 d^{\dag}_{k,\sigma} d_{k,\sigma} + \sum_{k,\sigma} \alpha_k \left[ c^{\dag}_{k,\sigma} d_{k,\sigma} + \text{h.c.} \right] ,
\end{align}
with
\begin{equation}
\alpha_k = - 2 t \cos \left( \dfrac{ka}{2} \right).
\end{equation}
The hopping term between the two atomic species makes the matrix non-diagonal. We write the Hamiltonian in matrix form
\begin{equation}
\hat{\mathcal{H}} = \sum_{k,\sigma} 
\begin{pmatrix}
c^{\dag}_{k,\sigma} & d^{\dag}_{k,\sigma}
\end{pmatrix}
\begin{pmatrix}
\mathcal{E}_1 & \alpha_k \\
\alpha_k & \mathcal{E}_2 
\end{pmatrix}
\begin{pmatrix}
c_{k,\sigma} \\
d_{k,\sigma}
\end{pmatrix},
\end{equation}
we use the identity $P P^{-1}$ as follows
\begin{equation}
\mathcal{H} = \sum_{k,\sigma} 
\begin{pmatrix}
c^{\dag}_{k,\sigma} & d^{\dag}_{k,\sigma}
\end{pmatrix}
P P^{-1}
\begin{pmatrix}
\mathcal{E}_1 & \alpha_k \\
\alpha_k & \mathcal{E}_2 
\end{pmatrix}
P P^{-1}
\begin{pmatrix}
c_{k,\sigma} \\
d_{k,\sigma}
\end{pmatrix},
\end{equation}
where $P$ is a $2 \times 2$ matrix whose columns are the eigenvectors of the matrix 
\begin{equation}
M(k) =
\begin{pmatrix}
\mathcal{E}_1 & \alpha_k \\
\alpha_k & \mathcal{E}_2 
\end{pmatrix},
\end{equation}
which, being symmetrical with respect to the real numbers, i.e., it is a self-adjoint matrix, can be diagonalized. The columns of $P$ are real and orthogonal to each other, i.e., the object
\begin{equation}
P^{-1} M P
\end{equation}
is diagonal and the diagonal elements are the eigenvalues, say $\mathcal{E}^{(+)}(k)$ and $\mathcal{E}^{(-)}(k)$. 
\begin{equation}
P^{-1} M P =
\begin{pmatrix}
\mathcal{E}^{(+)}(k) & 0 \\
0 & \mathcal{E}^{(-)}(k)
\end{pmatrix},
\end{equation}
\begin{equation}
\hat{\mathcal{H}} = \sum_{k,\sigma} 
\begin{pmatrix}
c^{\dag}_{k,\sigma} & d^{\dag}_{k,\sigma}
\end{pmatrix}
P 
\begin{pmatrix}
\mathcal{E}^{(+)}(k) & 0 \\
0 & \mathcal{E}^{(-)}(k)
\end{pmatrix}
P^{-1}
\begin{pmatrix}
c_{k,\sigma} \\
d_{k,\sigma}
\end{pmatrix},
\end{equation}
\begin{equation}
\mathcal{E}^{(\pm)}(k) = \dfrac{\mathcal{E}_1 + \mathcal{E}_2}{2} \pm \dfrac{1}{2} \sqrt{(\mathcal{E}_1 - \mathcal{E}_2)^{2} + 4 \alpha_k^2}.
\end{equation}
We define two operators as
\begin{equation}
\begin{pmatrix}
\tilde{c}_{k,\sigma} \\
\tilde{d}_{k,\sigma}
\end{pmatrix}
=
P^{-1} 
\begin{pmatrix}
c_{k,\sigma} \\
d_{k,\sigma}
\end{pmatrix},
\end{equation}
$\tilde{c}_{k,\sigma}$ and $\tilde{d}_{k,\sigma}$ are linear combination of the annihilation operators of both atomic species, so $\tilde{c}_{k,\sigma}$ destroys the quasi-momentum in the Bloch wave of both the first orbital and the second orbital. We apply the added operator and have
\begin{equation}
\begin{pmatrix}
\tilde{c}^{\dag}_{k,\sigma} & \tilde{d}^{\dag}_{k,\sigma}
\end{pmatrix}
= 
\begin{pmatrix}
c^{\dag}_{k,\sigma} & d^{\dag}_{k,\sigma}
\end{pmatrix}
\left( P^{-1} \right)^{*}_{t},
\end{equation}
given that $P$ is an orthogonal matrix, the transpose of the inverse of $P$ is equal to $P$ and the complex conjugate operation leaves the matrix unchanged, i.e.,
\begin{align}
\left( P^{-1} \right)^{*}_{t} &= P^{*} = \notag \\
&= P,
\end{align}
then
\begin{equation}
\begin{pmatrix}
\tilde{c}^{\dag}_{k,\sigma} & \tilde{d}^{\dag}_{k,\sigma}
\end{pmatrix}
= 
\begin{pmatrix}
c^{\dag}_{k,\sigma} & d^{\dag}_{k,\sigma}
\end{pmatrix}
P,
\end{equation}
which is just the term to the left of the matrix $P^{-1} M P$. Substituting in $\hat{\mathcal{H}}$ as follows
\begin{align}
\hat{\mathcal{H}} &= \sum_{k,\sigma} 
\begin{pmatrix}
\tilde{c}^{\dag}_{k,\sigma} & \tilde{d}^{\dag}_{k,\sigma}
\end{pmatrix}
\begin{pmatrix}
\mathcal{E}^{(+)}(k) & 0 \\
0 & \mathcal{E}^{(-)}(k)
\end{pmatrix}
\begin{pmatrix}
\tilde{c}_{k,\sigma} \\
\tilde{d}_{k,\sigma}
\end{pmatrix}
= \notag \\
&= \sum_{k,\sigma} \left\lbrace \mathcal{E}^{(+)}(k) \tilde{c}^{\dag}_{k,\sigma} \tilde{c}_{k,\sigma} + \mathcal{E}^{(-)}(k) \tilde{d}^{\dag}_{k,\sigma} \tilde{d}_{k,\sigma} \right\rbrace,
\end{align}
i.e., the Hamiltonian $\hat{\mathcal{H}}$ has been diagonalized. A hybrid level has formed between the two individual energy levels: the electrons of the two atoms transfer to the nearest neighbors and the $\mathcal{E}^{(\pm)}(k)$ bands originate. We note that the function $\alpha_k$ has a minimum at $k=0$ and two maxima at $k=\pm \frac{\pi}{a}$, so from  
\begin{equation}
\mathcal{E}^{(\pm)}(k) = \dfrac{\mathcal{E}_1 + \mathcal{E}_2}{2} \pm \dfrac{1}{2} \sqrt{(\mathcal{E}_1 - \mathcal{E}_2)^{2} + 4 \alpha_k^2} ,
\end{equation}
we infer that a gap of energy $\mathcal{E}_2 - \mathcal{E}_1 > 0$ is formed. Suppose that each of the atoms has one valence electron. All electrons will occupy the lower band while the upper band will remain empty.
\section{Hubbard's model}
In this section, we introduce a widely used model in the study of strongly correlated electron systems: the Hubbard model. This model captures essential features of electron interactions in solids and serves as a fundamental framework in condensed matter physics. In Hubbard's solid model, ions are attached to the vertices of a lattice, and electrons interact with each other and with ions through the Coulomb force. Under these assumptions, electron dynamics is described in first quantization by the Hamiltonian
\begin{equation}
\mathcal{H} = \sum_{i=1}^N h_1\left(\hat{\textbf{r}}_i,\hat{\textbf{p}}_i\right) + \sum_{1 \leq i < j \leq N} h_2(\hat{\textbf{r}}_i,\hat{\textbf{r}}_j).
\label{eq: Hubbardprimaquantizzazione}
\end{equation}
The one-body Hamiltonian 
\begin{align}
\mathcal{H}_1 &= \sum_{i=1}^N h_1\left(\hat{\textbf{r}}_i,\hat{\textbf{p}}_i\right) = \notag \\
&= \sum_{i=1}^N \dfrac{\hat{\textbf{p}}_i^2}{2m} + V_I(\textbf{r}_i) 
\end{align}
is given by the kinetic term and the ion-induced potential $V_I(\mathbf{r}_i)$, which is periodic with respect to the elementary cell of the crystal. The two-body Hamiltonian 
\begin{align}
\mathcal{H}_2 &= \sum_{1 \leq i < j \leq N} h_2(\hat{\textbf{r}}_i,\hat{\textbf{r}}_j) = \notag \\
&= \sum_{1 \leq i < j \leq N} h_2(\left|\textbf{r}_i-\textbf{r}_j\right|)
\end{align}
describes the Coulomb interaction. 
\subsection{One-band Hubbard's model}
We use the basis of the eigenstates of the one-body Hamiltonian $\tilde{h}_1$ to write $\eqref{eq: Hubbardprimaquantizzazione}$ in the second quantization. Recall that the interaction potential with ions, $V_I$, is periodic, consequently the eigenvalue equation $\tilde{h}_1 \varphi_{n,\textbf{k}}(\textbf{r}) = \mathcal{E}_{n,\textbf{k}} \varphi_{n,\textbf{k}}(\textbf{r})$ is a Bloch eigenvalue problem. Since the Wannier functions are localized, the field operators are given by
\begin{equation}
\hat{\psi}(x) = \sum_{n,i,\sigma} \omega_n(\textbf{r}-\textbf{R}_i) \chi_\sigma(s) C_{n,i,\sigma},
\label{eq: operatorecampodistruzionefunzioniWannier}
\end{equation}
\begin{equation}
\hat{\psi}^\dagger(x) = \sum_{n,i,\sigma} \omega^*_n(\textbf{r}-\textbf{R}_i) \chi^*_\sigma(s) C^\dagger_{n,i,\sigma},
\label{eq: operatorecampocreazionefunzioniWannier}
\end{equation}
and we have
\begin{align}
\hat{\mathcal{H}} &= \int dx \ \hat{\psi}^\dagger(x) h_1 \hat{\psi}(x) + \dfrac{1}{2} \int dx \int dx' \ \hat{\psi}^\dagger(x) \hat{\psi}^\dagger(x') h_2(x,x') \hat{\psi}(x') \hat{\psi}(x) = \notag \\
&= \sum_s \sum_{\substack{n,i,\sigma \\ n',j,\sigma'}} \int d^3\textbf{r} \ \omega^*_n(\textbf{r}-\textbf{R}_i) \chi^*_\sigma(s) h_1(\textbf{r}) \omega_{n'}(\textbf{r}-\textbf{R}_j) \chi_{\sigma'}(s) C^\dagger_{n,i,\sigma} C_{n',j,\sigma'} + \notag \\
&+ \dfrac{1}{2} \sum_{s,s'} \sum_{\substack{n_1,i,\sigma_1 \\ n_2,j,\sigma_2}} \sum_{\substack{n_3,k,\sigma_3 \\ n_4,l,\sigma_4}} \int d^3\textbf{r} \int d^3\textbf{r}' \ \omega^*_{n_1}(\textbf{r}-\textbf{R}_i) \chi^*_{\sigma_1}(s) \omega^*_{n_2}(\textbf{r}'-\textbf{R}_j) \chi^*_{\sigma_2}(s') h_2(\textbf{r},\textbf{r}') \notag \\
& \ \ \ \ \omega_{n_3}(\textbf{r}'-\textbf{R}_l) \chi_{\sigma_3}(s') \omega_{n_4}(\textbf{r}-\textbf{R}_l) \chi_{\sigma_4}(s) C^\dagger_{n_1,i,\sigma_1} C^\dagger_{n_2,j,\sigma_2} C_{n_3,k,\sigma_3} C_{n_4,l,\sigma_4}.
\end{align}
Regarding the one-body operator, we use $\sum_s \chi^*_\sigma(s) \chi_{\sigma'}(s) = \delta_{\sigma,\sigma'}$ and we sum with respect to $\sigma'$ 
\begin{equation}
\hat{\mathcal{H}}_1 = \sum_{\substack{n,i,\sigma \\ n',j}} \int d^3\textbf{r} \ \omega^*_n(\textbf{r}-\textbf{R}_i) h_1(\textbf{r}) \omega_{n'}(\textbf{r}-\textbf{R}_j) C^\dagger_{n,i,\sigma} C_{n',j,\sigma}.
\end{equation}
Since the one-body operator is diagonal in the basis of Bloch eigenfunctions rather than in the Wannier basis, we express the Wannier functions in terms of the Bloch functions and we replace the operator with its eigenvalue, as obtained from the solution of the corresponding eigenvalue problem as follows
\begin{align}
\hat{\mathcal{H}}_1 &= \sum_{\substack{n,i,\sigma \\ n',j}} \sum_{\textbf{k},\textbf{k'}} \int d^3\textbf{r} \ \varphi^*_{n,\textbf{k}}(\textbf{r}) h_1(\textbf{r}) \varphi_{n',\textbf{k}'}(\textbf{r}) \dfrac{e^{i \textbf{k} \cdot \textbf{R}_i}}{\sqrt{N}} \dfrac{e^{- i \textbf{k}' \cdot \textbf{R}_j}}{\sqrt{N}} C^\dagger_{n,i,\sigma} C_{n',j,\sigma} = \notag \\
&= \sum_{\substack{n,i,\sigma \\ n',j}} \sum_{\textbf{k},\textbf{k'}} \int d^3\textbf{r} \ \mathcal{E}_{n',\textbf{k}'} \varphi^*_{n,\textbf{k}}(\textbf{r}) \varphi_{n',\textbf{k}'}(\textbf{r}) \dfrac{e^{i \textbf{k} \cdot (\textbf{R}_i-\textbf{R}_j)}}{N} C^\dagger_{n,i,\sigma} C_{n',j,\sigma}.
\end{align}
The Bloch eigenfunctions constitute a complete orthonormal basis; therefore, upon integration over the spatial variable $\textbf{r}$, all cross terms with $\textbf{k} \neq \textbf{k}'$ or $n \neq n'$ vanish, and only the diagonal contributions remain, that is
\begin{align}
\hat{\mathcal{H}}_1 &= \sum_{n,\textbf{k},i,j,\sigma} \mathcal{E}_{n,\textbf{k}} \dfrac{e^{i \textbf{k} \cdot (\textbf{R}_i-\textbf{R}_j)}}{N} C^\dagger_{n,i,\sigma} C_{n,j,\sigma} \equiv \notag \\
&\equiv - \sum_{n,i,j,\sigma} t_{ij}^n C^\dagger_{n,i,\sigma} C_{n,j,\sigma},
\end{align}
with
\begin{align}
t_{ij}^n &= - \int d^3\textbf{r} \ \omega^*_n(\textbf{r}-\textbf{R}_i) h_1(\textbf{r}) \omega_{n}(\textbf{r}-\textbf{R}_j) = \notag \\
&= - \sum_{\textbf{k}} \mathcal{E}_{n,\textbf{k}} \dfrac{e^{i \textbf{k} \cdot (\textbf{R}_i-\textbf{R}_j)}}{N}.
\end{align}
Regarding the two-body Hamiltonian, we use
\begin{equation}
\sum_{s} \chi^*_{\sigma_1}(s) \chi_{\sigma_4}(s) = \delta_{\sigma_1,\sigma_4},
\end{equation}
\begin{equation}
\sum_{s'} \chi^*_{\sigma_2}(s') \chi_{\sigma_3}(s') = \delta_{\sigma_2,\sigma_3},
\end{equation}
then we sum over $\sigma_3$ and $\sigma_4$, and define $\sigma_1 \equiv \sigma$, $\sigma_2 \equiv \sigma'$,
\begin{align}
\hat{\mathcal{H}}_2 &= \dfrac{1}{2} \sum_{\substack{n_1,n_2 \\ n_3,n_4}} \sum_{i,j,k,l} \sum_{\sigma,\sigma'} \int d^3\textbf{r} \int d^3\textbf{r}' \notag \\
& \ \ \ \ \ \omega^*_{n_1}(\textbf{r}-\textbf{R}_i) \omega^*_{n_2}(\textbf{r}'-\textbf{R}_j) U(\textbf{r},\textbf{r}') \omega_{n_3}(\textbf{r}'-\textbf{R}_l) \omega_{n_4}(\textbf{r}-\textbf{R}_l) C^\dagger_{n_1,i,\sigma} C^\dagger_{n_2,j,\sigma'} C_{n_3,k,\sigma'} C_{n_4,l,\sigma} ,
\end{align}
and we set
\begin{equation}
h_{ijkl}^{n_1 n_2 n_3 n_4} = \int d^3\textbf{r} \int d^3\textbf{r}' \ \omega^*_{n_1}(\textbf{r}-\textbf{R}_i) \omega^*_{n_2}(\textbf{r}'-\textbf{R}_j) h_2(\textbf{r},\textbf{r}') \omega_{n_3}(\textbf{r}'-\textbf{R}_l) \omega_{n_4}(\textbf{r}-\textbf{R}_l).
\end{equation}
The Hamiltonian $\eqref{eq: Hubbardprimaquantizzazione}$ becomes in second quantization
\begin{equation}
\hat{\mathcal{H}} = - \sum_{n,i,j,\sigma} t_{ij}^n C^\dagger_{n,i,\sigma} C_{n,j,\sigma} + \dfrac{1}{2} \sum_{\substack{n_1,n_2 \\ n_3,n_4}} \sum_{i,j,k,l} \sum_{\sigma,\sigma'} h_{ijkl}^{n_1 n_2 n_3 n_4} C^\dagger_{n_1,i,\sigma} C^\dagger_{n_2,j,\sigma'} C_{n_3,k,\sigma'} C_{n_4,l,\sigma}.
\label{eq: Hamiltonianasolidopotenzialeausiliariosecondaquantizzazione}
\end{equation}
When the Fermi level (see Chapter \ref{The Fermi momentum}) lies within a single conduction band, coupling to energetically distant bands can be neglected. One therefore adopts the single-band approximation, assuming that Wannier states belonging to different bands are orthonormal. Within this framework, electron motion is governed mainly by hopping processes between nearby lattice sites, since the hopping amplitude decreases rapidly with distance. As for two-body interactions, only the dominant contribution is retained, corresponding to the interaction between two electrons occupying the same lattice site. This leads to the definition of the on-site Hubbard interaction, while all other longer-range interaction terms are neglected. The resulting effective description is a single-band model with short-range hopping and local interactions. From
\begin{equation}
t_{ij}^{nn'} = t_{ij} \delta_{n,n'},
\end{equation}
\begin{equation}
h_{ijkl}^{n_1 n_2 n_3 n_4} = h_{ijkl} \delta_{n_1,n} \delta_{n_2,n} \delta_{n_3,n} \delta_{n_4,n} ,
\end{equation}
\begin{equation}
h_{ijkl}^{nnnn} = U \delta_{i,j} \delta_{j,k} \delta_{k,l},
\end{equation}
we obtain the one-band Hubbard's model
\begin{equation}
\hat{\mathcal{H}} = - \sum_{\langle i,j \rangle,\sigma} t_{ij} C^\dagger_{i,\sigma} C_{j,\sigma} + \dfrac{U}{2} \sum_{i,\sigma \neq \sigma'} C^\dagger_{i,\sigma} C^\dagger_{i,\sigma'} C_{i,\sigma'} C_{i,\sigma}.
\label{eq: HamiltonianaHubbardunabandaforma1}
\end{equation}
The interaction term includes a sum over spin indices with the constraint $\sigma \neq \sigma'$, since the Pauli exclusion principle forbids two electrons with identical spin from occupying the same site $i$. The factor $\frac{1}{2}$ accounts for the fact that the sum includes both spin configurations (e.g., $\uparrow\downarrow$ and $\downarrow\uparrow$), thereby avoiding double counting of the interaction energy. Finally, we use fermionic algebra with $\sigma \neq \sigma'$ to write the two-body operator in a more physically suggestive form, indeed, from
\begin{align}
C^\dagger_{i,\sigma} C^\dagger_{i,\sigma'} C_{i,\sigma'} C_{i,\sigma} &= - C^\dagger_{i,\sigma} C^\dagger_{i,\sigma'} C_{i,\sigma} C_{i,\sigma'} = \notag \\
&= C^\dagger_{i,\sigma} C_{i,\sigma} C^\dagger_{i,\sigma'} C_{i,\sigma'} = \notag \\
&= \hat{N}_{i,\sigma} \hat{N}_{i,\sigma'},
\end{align}
and finally
\begin{equation}
\hat{\mathcal{H}} = - \sum_{\langle i,j \rangle,\sigma} t_{ij} C^\dagger_{i,\sigma} C_{j,\sigma} + \dfrac{U}{2} \sum_{i,\sigma \neq \sigma'} \hat{N}_{i,\sigma} \hat{N}_{i,\sigma'},
\label{eq: HamiltonianaHubbardunabandaforma2}
\end{equation}
which is the one-band Hubbard Hamiltonian in second quantization. The first term describes the hopping of electrons between neighboring sites $i$ and $j$ with amplitude $t_{ij}$, conserving spin $\sigma$. The second term accounts for the on-site Coulomb repulsion, where $U$ is the energy cost for having two electrons with opposite spins on the same site. The factor of $\frac{1}{2}$ ensures that each pairwise interaction is counted only once, thereby avoiding double-counting of the interaction energy.
\newpage
\section{Figures}
\begin{figure}[H]
\centering
\renewcommand{\arraystretch}{1.5}
\setlength{\tabcolsep}{10pt}

\newlength{\imagewidth}
\setlength{\imagewidth}{2.53cm}

\begin{tabular}{|c|c|c|c|c|}
\hline
\textbf{Cell} & \textbf{Simple} & \textbf{Base-centered} & \textbf{Body-centered} & \textbf{Face-centered} \\
\hline
\textbf{Cubic} 
  & \includegraphics[width=\imagewidth]{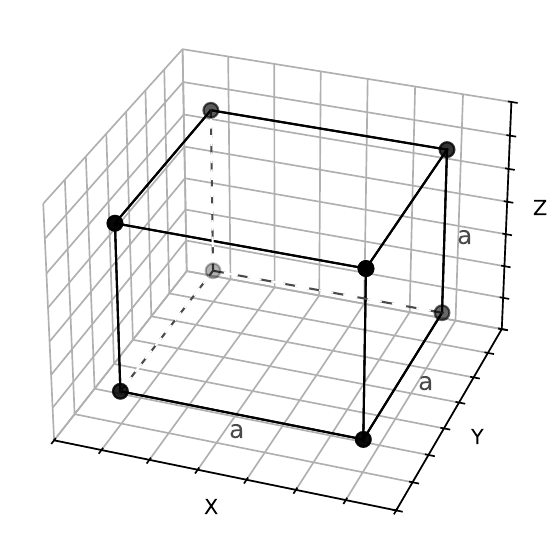} 
  & & \includegraphics[width=\imagewidth]{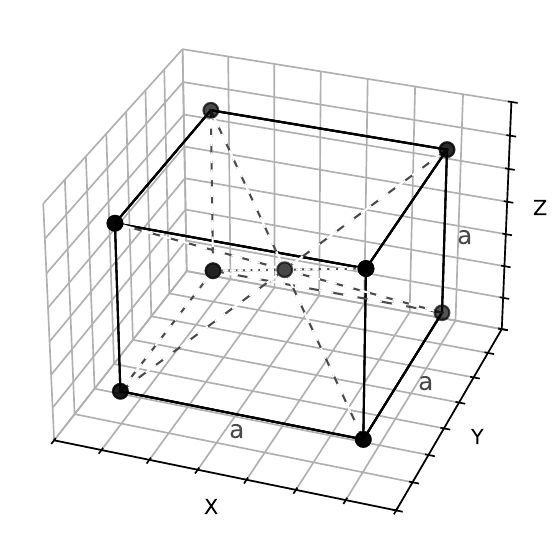} 
  & \includegraphics[width=\imagewidth]{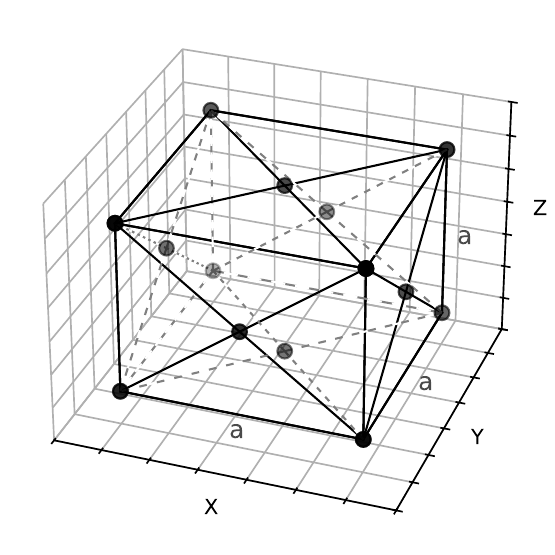} \\
\hline
\textbf{Tetragonal} 
  & \includegraphics[width=\imagewidth]{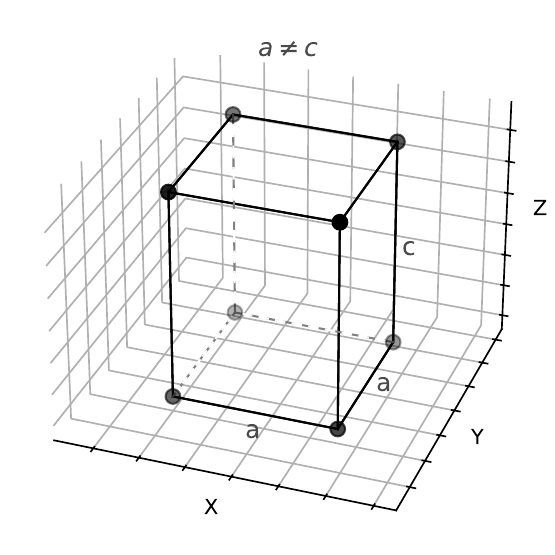} 
  & & \includegraphics[width=\imagewidth]{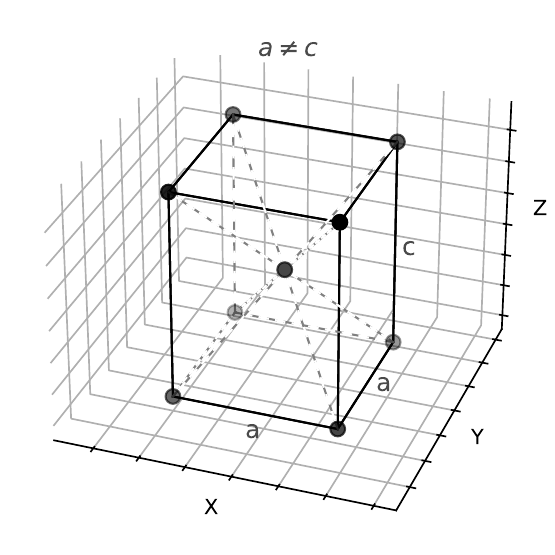} 
  & \\
\hline
\textbf{Orthorhombic} 
  & \includegraphics[width=\imagewidth]{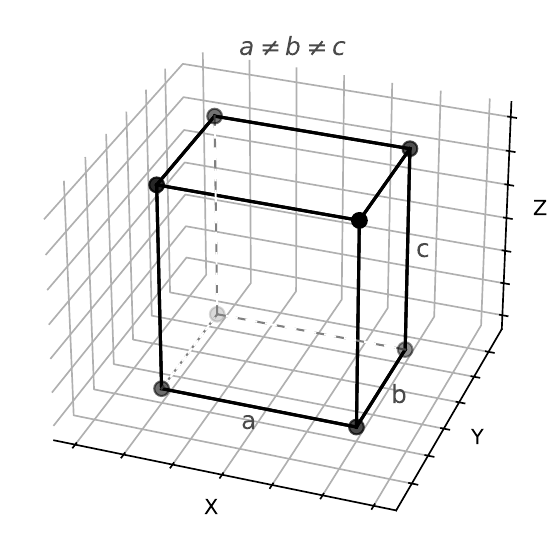} 
  & \includegraphics[width=\imagewidth]{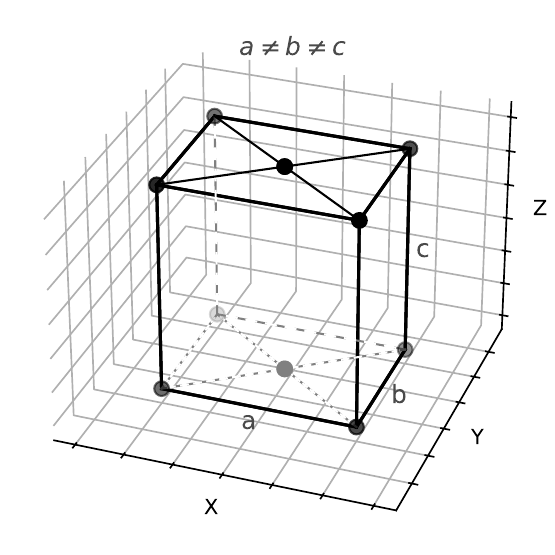} 
  & \includegraphics[width=\imagewidth]{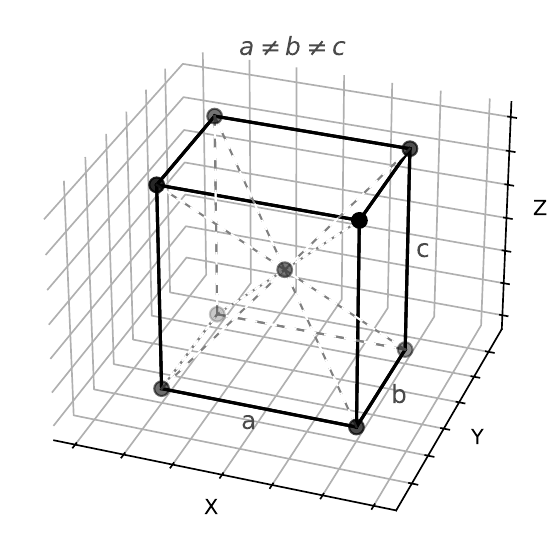} 
  & \includegraphics[width=\imagewidth]{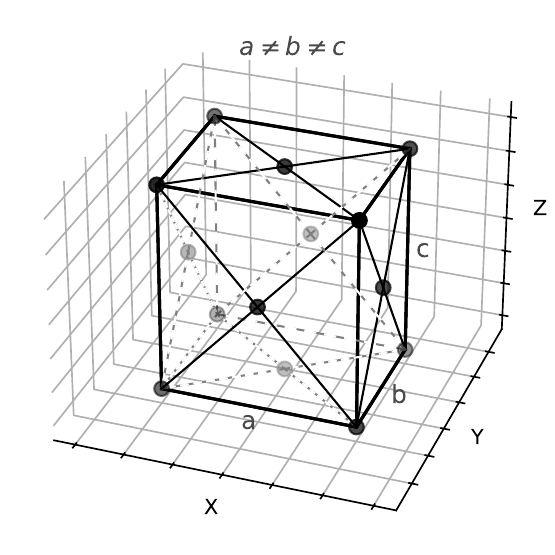} \\
\hline
\textbf{Monoclinic} 
  & \includegraphics[width=\imagewidth]{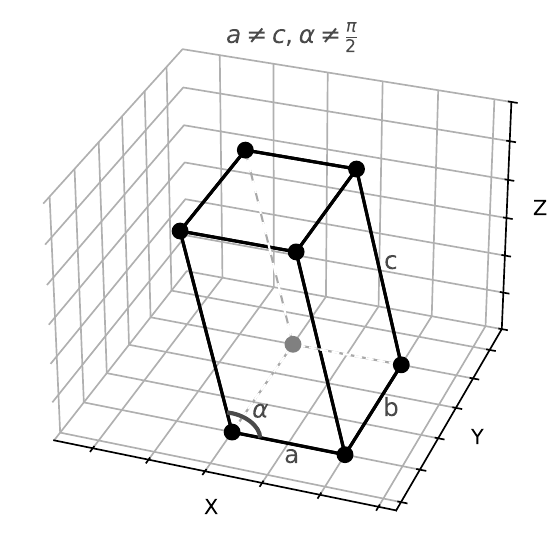} 
  & \includegraphics[width=\imagewidth]{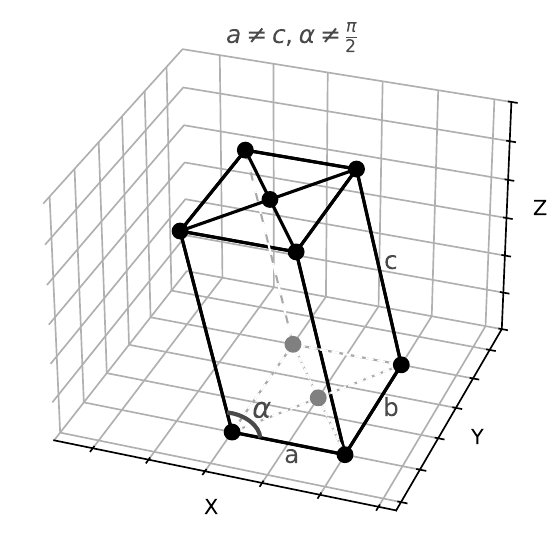} 
  & & \\
\hline
\textbf{Triclinic} 
  & \includegraphics[width=\imagewidth]{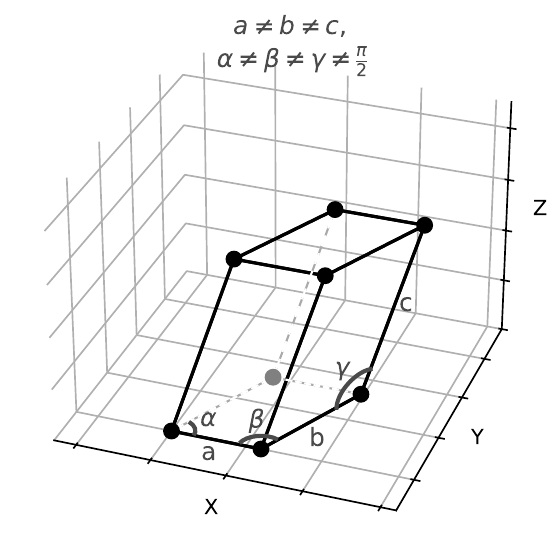} 
  & & & \\
\hline
\textbf{Hexagonal} 
  & \includegraphics[width=\imagewidth]{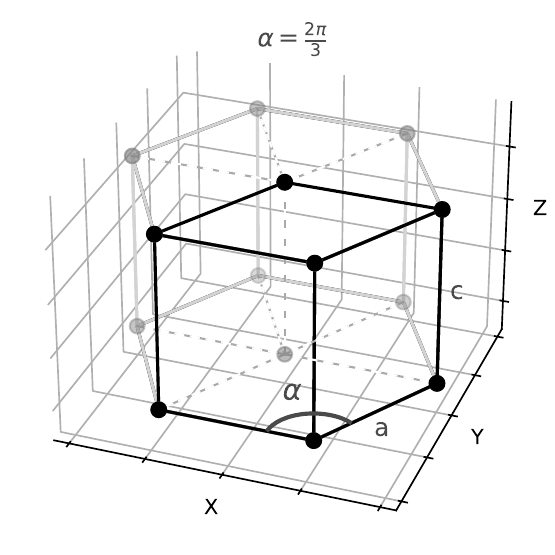} 
  & & & \\
\hline
\textbf{Rhombohedral} 
  & \includegraphics[width=\imagewidth]{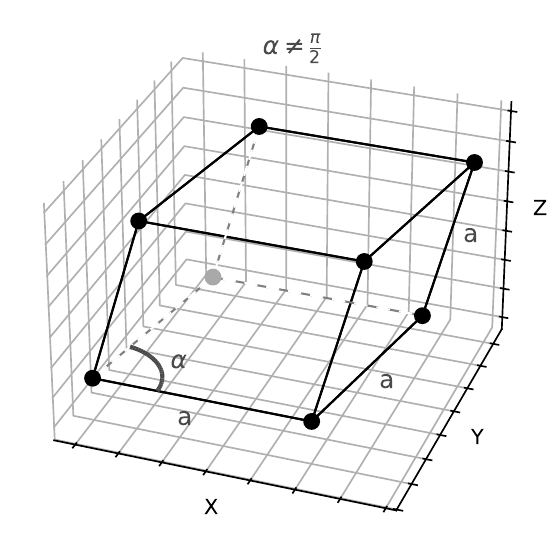} 
  & & & \\
\hline
\end{tabular}
\caption{The 14 3-dimensional Bravais lattices, grouped into seven crystal systems: triclinic, monoclinic, orthorhombic, tetragonal, trigonal, hexagonal, and cubic. Each lattice is defined by its symmetry, the lengths and angles of its primitive vectors, and can occur in primitive (P), body-centered (I), face-centered (F), or base-centered (C) forms. This classification encompasses all possible periodic arrangements of points in three-dimensional space consistent with crystallographic symmetry.}
\label{fig: Unit_cells_of_14_Bravais_lattices}
\end{figure}
\chapter{Step II of the solid model: phonons}
In a crystalline solid, atoms are not fixed at their equilibrium positions but oscillate around them due to quantum and thermal fluctuations. These collective vibrations can be described, as a first approximation, as harmonic oscillations of nuclei connected by elastic forces. From a classical point of view, this naturally leads to the analysis of a system of coupled equations of motion, which describe the elastic waves propagating through the lattice. The solutions to these equations are normal modes, representing the collective vibrational patterns of the crystal. However, to fully understand the physical implications of these vibrations, it is necessary to introduce the quantization of motion. In analogy with the quantum harmonic oscillator, each normal mode is associated with a quantum of energy: the phonon. Phonons thus represent the elementary quantum excitations of lattice vibrations, analogous to photons for the electromagnetic field. \newline
This transition from classical to quantum formalism not only clarifies the particle-like nature of vibrations, but also provides a powerful and elegant language for describing phenomena such as the heat capacity of solids and the electron-phonon interaction. In particular, we will present the Einstein and Debye models for calculating the lattice specific heat, which are especially valuable from a pedagogical perspective. The electron–phonon interaction, on the other hand, will be addressed in a later chapter. Moreover, the quantum formalism allows us to introduce phonon creation and annihilation operators, which greatly simplify the analysis of scattering processes and enable a compact description of the system's total vibrational energy. \newline
In this chapter, we will explore this transition in detail, starting from the classical description of lattice vibrations and proceeding to their quantization, with the goal of showing how the concept of the phonon emerges naturally from the theoretical treatment of the crystal lattice. In particular, we will derive the expression
\begin{equation}
\sum_{\textbf{q},s} \hslash \omega_{\textbf{q},s} \left( a^{\dag}_{\textbf{q},s} a_{\textbf{q},s} + \dfrac{1}{2} \right)
\end{equation}
in the Hamiltonian $\eqref{eq: Hamiltonianadiunsolido}$.
\section{Quantization of the lattice vibrations}
\subsection{The harmonic approximation}
Ionic dynamics is described by an effective ion-ion potential of the form
\begin{equation}
V^{eff.}_{i-i} = V_{i-i} + \mathcal{E}_{n} \left( \lbrace \textbf{R} \rbrace \right) ,
\end{equation}
where $V_{i-i}$ is the Coulomb repulsion and $\mathcal{E}_{n} \left( \lbrace \textbf{R} \right)$ is the screening energy due to the electrons and resulting from solving the fixed-ion electron problem. We are in the low-energy physics scenario, so we replace $\mathcal{E}_{n}\left( \lbrace \textbf{R} \rbrace \right)$ with the minimum electronic eigenvalue $\mathcal{E}_{0} \left( \lbrace \textbf{R} \rbrace \right)$
\begin{equation}
V^{eff.}_{i-i} = V_{i-i} + \mathcal{E}_{0} \left( \lbrace \textbf{R} \rbrace \right).
\label{eq: potenzialeionicoeffettivo}
\end{equation}
We aim to minimize the effective interionic potential $V^{eff.}_{i-i}$. However, this is a highly nontrivial problem, as many different crystal configurations correspond to energy values that are very close to one another. To make progress, we consider the second-order Taylor expansion of \( V^{eff.}_{i-i} \), as given in equation $\eqref{eq: potenzialeionicoeffettivo}$, around the stable equilibrium positions of the ions. This leads to the harmonic approximation, in which $V^{eff.}_{i-i}$ is replaced by a harmonic potential. Such an approximation is well justified by experimental evidence indicating that ions typically undergo only small displacements from their equilibrium positions at the lattice sites. We now consider a crystal lattice defined by the primitive vectors \( \textbf{a}_1 \), \( \textbf{a}_2 \), and \( \textbf{a}_3 \), with ions positioned at the lattice vertices and \( \tau \) distinct atomic species per unit cell. The entire lattice comprises \( N = N_1 N_2 N_3 \) cells, each labelled by a triplet of integers \( \textbf{n} = (n_1, n_2, n_3) \). Let \( \textbf{R}_{\textbf{n},\mu}(t) \) denote the position at time \( t \) of an ion of species \( \mu \) located in cell \( \textbf{n} \). This position can be expressed as
\begin{equation}
\textbf{R}_{\textbf{n},\mu}(t) = \textbf{R}_{\textbf{n}} + \textbf{d}_{\mu} + \textbf{S}_{\textbf{n},\mu}(t), \left|\textbf{S}_{\textbf{n},\mu}(t)\right| \ll a_i, \ i=1,2,3 ,
\end{equation} 
where \(\textbf{R}_{\textbf{n}}\) denotes the position of the elementary cell labeled by \(\textbf{n}\); \(\textbf{d}_{\mu}\) is the position vector of the atom of species \(\mu\) (\(\mu = 1, \ldots, \tau\)) within the unit cell indexed by \(\textbf{0}\); and \(\textbf{S}_{\textbf{n},\mu}(t)\) represents the displacement from the equilibrium position \(\textbf{R}_{\textbf{n}} + \textbf{d}_{\mu}\) at time \(t\). Accordingly, the second-order Taylor expansion of $V^{eff.}_{i-i}$ is performed with respect to the variables \(\{\textbf{R}_{\textbf{n},\mu}\}\), around the equilibrium positions \(\textbf{R}_{\textbf{n}} + \textbf{d}_{\mu}\). In particular, one needs to compute: the Hessian matrix, the matrix of first derivatives evaluated at the minimum point, which is identically zero, and a constant corresponding to the value of the potential at the ions equilibrium positions. Since the potential is defined up to an additive constant, we assume, without loss of generality, that this constant is zero. This expansion of the potential is therefore expressed as
\begin{equation}
V^{eff.}_{i-i}\left(\lbrace \textbf{R} \rbrace\right) \ = \ \dfrac{1}{2} \sum_{\substack{\textbf{n} \\ \mu,i}} \sum_{\substack{\textbf{m} \\ \nu,k}} W^{\textbf{n},\mu,i}_{\textbf{m},\nu,k} S_{\textbf{n},\mu,i} S_{\textbf{m},\nu,k} + o \left( |S|^{3} \right),
\end{equation}
where  
\begin{equation}
W^{\textbf{n},\mu,i}_{\textbf{m},\nu,k} = \dfrac{\partial^2 V^{eff.}_{i-i}}{\partial R_{\textbf{m},\nu,k} \partial R_{\textbf{n},\mu,i}}
\end{equation}
is the multi-index object of the second derivatives of the potential with respect to the positions $R_{\textbf{n},\mu,i}$, $R_{\textbf{m},\nu,k}$ and is calculated for $\textbf{S}=\textbf{0}$. Since $V^{eff.}_{i-i}$ has a minimum at $\textbf{S}=\textbf{0}$, it is a positive definite form. Let us now study the classical Lagrangian for ion dynamics. The time-dependent quantity in $\textbf{R}_{\textbf{n},\mu}$ is $\textbf{S}_{\textbf{n},\mu}$, then the kinetic term is
\begin{align}
T &= \dfrac{1}{2} \sum_{\substack{\textbf{n} \\ \mu,i}} M_{\mu} \dot{R}^{2}_{\textbf{n},\mu,i} \equiv \notag \\
&\equiv \dfrac{1}{2} \sum_{\substack{\textbf{n} \\ \mu,i}} M_{\mu} \dot{S}^{2}_{\textbf{n},\mu,i},
\end{align}
and the Lagrangian function is
\begin{align}
\mathcal{L} &= T - V = \notag \\
&= \dfrac{1}{2} \sum_{\substack{\textbf{n} \\ \mu,i}} M_{\mu} \dot{S}^{2}_{\textbf{n},\mu,i} - \dfrac{1}{2} \sum_{\substack{\textbf{n} \\ \mu,i}} \sum_{\substack{\textbf{m} \\ \nu,k}} W^{\textbf{n},\mu,i}_{\textbf{m},\nu,k} S_{\textbf{n},\mu,i} S_{\textbf{m},\nu,k}.
\end{align}
The equations of motion follow from the Euler-Lagrange equations
\begin{equation}
\dfrac{d}{dt} \dfrac{\partial \mathcal{L}}{\partial \dot{q}^{h}} - \dfrac{\partial \mathcal{L}}{\partial q^{h}} = 0,
\end{equation}
with identifications $h \rightarrow (\textbf{n},\mu,i)$, $q \rightarrow S$, $q^{h} \rightarrow S_{\textbf{n},\mu,i}$, the Euler-Lagrange equations for the ionic dynamics become
\begin{equation}
M_{\mu} \ddot{S}_{\textbf{n},\mu,i} = - \sum_{\substack{\textbf{m} \\ \nu,k}} W^{\textbf{n},\mu,i}_{\textbf{m},\nu,k} S_{\textbf{m},\nu,k} .
\end{equation}
The equation describes a system of many coupled harmonic oscillators. Specifically, the total number of vectors $\mathbf{n}$ corresponds to the number of elementary cells in the crystal lattice, namely $N_1 N_2 N_3$, where $N_i$ is the number of cells along the $i$-th crystallographic direction (with $i = 1, 2, 3$). Since there are three spatial directions and $\tau$ distinct atomic species, the total number of coupled oscillators is
\begin{equation}
3 \tau N_1 N_2 N_3.
\end{equation}
Our goal is to diagonalize the Lagrangian. Because the Lagrangian is a quadratic form, this diagonalization can be carried out exactly. Through this process, we will determine the normal modes of the system. Let us first consider the quantities $W^{\textbf{n},\mu,i}_{\textbf{m},\nu,k}$. If, instead of a system of coupled harmonic oscillators, we were studying a single harmonic oscillator, the quantity \( W \) would simply correspond to the elastic constant of the oscillator. In the general case, these quantities are referred to as elastic material constants or force constants. The force constants \( W^{\mathbf{n}, \mu, i}_{\mathbf{m}, \nu, k} \) satisfy the following properties.
\begin{itemize}
\item [1.] Lattice translation invariance, i.e.,
\begin{equation}
W^{\textbf{n}+\textbf{h},\mu,i}_{\textbf{m}+\textbf{h},\nu,k} = W^{\textbf{n},\mu,i}_{\textbf{m},\nu,k}, \ \forall \ \textbf{R}_{\textbf{h}} \in \text{direct lattice},
\end{equation}
indeed, translating the cells labeled by $\textbf{n}$ and $\textbf{m}$ by a lattice translation vector $\textbf{R}_{\textbf{h}}$ does not affect the derivatives, as these are evaluated with respect to the equilibrium positions within the unit cell. In particular, when $\textbf{R}_{\textbf{h}} = - \textbf{R}_{\textbf{m}}$, then
\begin{equation}
W^{\textbf{n},\mu,i}_{\textbf{m},\nu,k} = W^{\textbf{n}-\textbf{m},\mu,i}_{\textbf{0},\nu,k},
\label{eq: proprietàcoefficientiWdifferenzavettori}
\end{equation} 
therefore the objects $W^{\textbf{n},\mu,i}_{\textbf{m},\nu,k}$ depend only on the difference $\textbf{R}_{\textbf{n}} - \textbf{R}_{\textbf{m}}$ and not individually on the vectors $\textbf{R}_{\textbf{n}}$ and $\textbf{R}_{\textbf{m}}$.
\item [2.] Index symmetry. From the analyticity of the potential, the order of the derivatives can be exchanged, leading to the symmetry
\begin{equation}
W^{\textbf{n},\mu,i}_{\textbf{m},\nu,k} = W^{\textbf{m},\nu,k}_{\textbf{n},\mu,i}.
\label{eq: proprietàcoefficientiWinversioneindici}
\end{equation}
\item [3.] Reality. The coefficients \( W^{\mathbf{n}, \mu, i}_{\mathbf{m}, \nu, k} \) are real, since they arise as second derivatives of an effective potential, which is a real function of real variables.
\item [4.] Force balance under rigid translation. The quantity
\begin{equation}
M_{\mu} \ddot{S}_{\textbf{n},\mu,i}
\end{equation}
represents the total force acting on the atom of species \( \mu \) in cell \( \mathbf{n} \), along direction \( i \), due to the interactions with all other ions in the lattice. If we displace an atom \( (\mathbf{m}, \nu) \) by one lattice constant in direction \( k \), i.e., \( S_{\mathbf{m}, \nu, k} = 1 \), then the term \( W^{\mathbf{n}, \mu, i}_{\mathbf{m}, \nu, k} S_{\mathbf{n}, \mu, i} \) represents (minus) the force that this displacement exerts on the atom \( (\mathbf{n}, \mu) \) along direction \( i \). According to this interpretation,
\begin{equation}
\sum_{\textbf{m},\nu} \left. W^{\textbf{n},\mu,i}_{\textbf{m},\nu,k} \right|_{k \ \text{fixed}} \ = 0,
\label{eq: proprietàcoefficientiWsommasucellaesuspecieatomicadirezionefissata}
\end{equation}
that is, a rigid translation of all atoms by one lattice constant in direction \( k \) leaves all relative positions unchanged, and therefore all elastic forces cancel.
\end{itemize}
\subsection{Classical formalism}
From this point onward, in order to solve the Euler-Lagrange equations, we apply a series of transformations aimed at diagonalizing the Lagrangian and ultimately obtaining an equation that governs atomic oscillations. As a first step, we perform a change of variables from the original coordinates \( S \) to a new set of coordinates \( U \), chosen to simplify the treatment of the atomic masses \( M_{\mu} \). Let \( M = \sum_{\mu} M_{\mu} \) denote the total mass of the system, then we define
\begin{equation}
U_{\textbf{n},\mu,i} = \sqrt{\dfrac{M_{\mu}}{M}} S_{\textbf{n},\mu,i},
\end{equation}
then
\begin{equation}
\mathcal{L} = \dfrac{1}{2} M \sum_{\substack{\textbf{n} \\ \mu,i}} \dot{U}^{2}_{\textbf{n},\mu,i} - \dfrac{1}{2} \sum_{\substack{\textbf{n} \\ \mu,i}} \sum_{\substack{\textbf{m} \\ \nu,k}} D^{\textbf{n},\mu,i}_{\textbf{m},\nu,k} U_{\textbf{n},\mu,i} U_{\textbf{m},\nu,k},
\end{equation}
where
\begin{equation}
D^{\textbf{n},\mu,i}_{\textbf{m},\nu,k} = \dfrac{M}{\sqrt{M_{\mu} M_{\nu}}} W^{\textbf{n},\mu,i}_{\textbf{m},\nu,k}
\end{equation}
are the rescaled force constants. The next goal is to decouple the system of coupled harmonic oscillators. To achieve this, we first aim to eliminate the double summation over the lattice indices \( \mathbf{n} \) and \( \mathbf{m} \). This is conveniently done by transforming to reciprocal space, introducing wave vectors \( \mathbf{q} \) defined on the first Brillouin zone of the reciprocal lattice. Since the displacements \( U_{\mathbf{n}, \mu, i} \) are defined on the \( N \) sites of the direct lattice, we express \( U_{\mathbf{n}, \mu, i} \) as the action of a unitary transformation $\eqref{eq: esponenzialediunatrasformatadiscretadiFourier}$, with matrix elements $\frac{e^{i \textbf{q} \cdot \textbf{R}_{\textbf{n}}}}{\sqrt{N}}$, on the variables \( C_{\mathbf{q}, \mu, i} \), specifically
\begin{equation}
U_{\textbf{n},\mu,i} = \sum_{\textbf{q}} \dfrac{e^{i \textbf{q} \cdot \textbf{R}_{\textbf{n}}}}{\sqrt{N}} C_{\textbf{q},\mu,i} ,
\end{equation}
where the wave vectors $\textbf{q}$ are defined in equation $\eqref{eq: vettoridondadellaprimaBZdelcristallo}$. It is worth noting that the transformation between the real-space displacements \( U_{\mathbf{n}, \mu, i} \) and the reciprocal-space variables \( C_{\mathbf{q}, \mu, i} \) is formally analogous to the relation between Wannier functions and Bloch waves. The Lagrangian is modified as
\begin{equation}
\mathcal{L} = \dfrac{1}{2} M \sum_{\textbf{n},\mu,i} \sum_{\textbf{q}} \sum_{\textbf{q}'} \dfrac{1}{N} e^{i \textbf{q} \cdot \textbf{R}_{\textbf{n}}} e^{i \textbf{q}' \cdot \textbf{R}_{\textbf{n}}} \dot{C}_{\textbf{q},\mu,i} \dot{C}_{\textbf{q}',\mu,i} - \dfrac{1}{2} \sum_{\substack{\textbf{n} \\ \mu,i}} \sum_{\substack{\textbf{m} \\ \nu,k}} D^{\textbf{n},\mu,i}_{\textbf{m},\nu,k} \sum_{\textbf{q}} \sum_{\textbf{q}'} \dfrac{1}{N} e^{i \textbf{q} \cdot \textbf{R}_{\textbf{n}}} e^{i \textbf{q}' \cdot \textbf{R}_{\textbf{m}}} C_{\textbf{q},\mu,i} C_{\textbf{q}',\nu,k}.
\end{equation}
In the first term, the quantities \(\dot{C}_{\mathbf{q}, \mu, i} \dot{C}_{\mathbf{q}, \mu, i}\) do not depend on the lattice index \(\mathbf{n}\), therefore
\begin{equation}
\dfrac{1}{N} \sum_{\textbf{n}} e^{i \textbf{q} \cdot \textbf{R}_{\textbf{n}}} e^{i \textbf{q}' \cdot \textbf{R}_{\textbf{n}}} = \delta_{\textbf{q},-\textbf{q}'} ,
\end{equation}
\begin{align}
\mathcal{L} &= \dfrac{1}{2} M \sum_{\mu,i} \sum_{\textbf{q}} \sum_{\textbf{q}'} \delta_{\textbf{q},-\textbf{q}'} \dot{C}_{\textbf{q},\mu,i} \dot{C}_{\textbf{q}',\mu,i} - \dfrac{1}{2} \sum_{\substack{\textbf{n} \\ \mu,i}} \sum_{\substack{\textbf{m} \\ \nu,k}} D^{\textbf{n},\mu,i}_{\textbf{m},\nu,k} \sum_{\textbf{q}} \sum_{\textbf{q}'} \dfrac{1}{N} e^{i \textbf{q} \cdot \textbf{R}_{\textbf{n}}} e^{i \textbf{q}' \cdot \textbf{R}_{\textbf{m}}} C_{\textbf{q},\mu,i} C_{\textbf{q}',\nu,k} = \notag \\
&= \dfrac{1}{2} M \sum_{\substack{\textbf{q} \\ \mu,i}} \dot{C}_{\textbf{q},\mu,i} \dot{C}_{-\textbf{q},\mu,i} - \dfrac{1}{2} \sum_{\substack{\textbf{n} \\ \mu,i}} \sum_{\substack{\textbf{m} \\ \nu,k}} D^{\textbf{n},\mu,i}_{\textbf{m},\nu,k} \sum_{\textbf{q}} \sum_{\textbf{q}'} \dfrac{1}{N} e^{i \textbf{q} \cdot \textbf{R}_{\textbf{n}}} e^{i \textbf{q}' \cdot \textbf{R}_{\textbf{m}}} C_{\textbf{q},\mu,i} C_{\textbf{q}',\nu,k}.
\end{align}
We cannot apply the relation \(\delta_{\mathbf{q}, -\mathbf{q}'}\) to the second term because the force constants depend explicitly on the lattice index \(\mathbf{n}\). Instead, we utilize the symmetry properties of \(D\) (and equivalently \(W\)) and insert the identity operator in the form $\mathds{1}=e^{- i \textbf{q} \cdot \textbf{R}_{\textbf{m}}} e^{i \textbf{q} \cdot \textbf{R}_{\textbf{m}}}$ to rewrite the Lagrangian as follows
\begin{align}
\mathcal{L} &= \dfrac{1}{2} M \sum_{\substack{\textbf{q} \\ \mu,i}} \dot{C}_{\textbf{q},\mu,i} \dot{C}_{-\textbf{q},\mu,i} - \dfrac{1}{2} \sum_{\substack{\textbf{n} \\ \mu,i}} \sum_{\substack{\textbf{m} \\ \nu,k}} D^{\textbf{n},\mu,i}_{\textbf{m},\nu,k} \sum_{\textbf{q}} \sum_{\textbf{q}'} \dfrac{1}{N} e^{i \textbf{q} \cdot \textbf{R}_{\textbf{n}}} e^{- i \textbf{q} \cdot \textbf{R}_{\textbf{m}}} e^{i \textbf{q} \cdot \textbf{R}_{\textbf{m}}} e^{i \textbf{q}' \cdot \textbf{R}_{\textbf{m}}} C_{\textbf{q},\mu,i} C_{\textbf{q}',\nu,k} = \notag \\
&= \dfrac{1}{2} M \sum_{\substack{\textbf{q} \\ \mu,i}} \dot{C}_{\textbf{q},\mu,i} \dot{C}_{-\textbf{q},\mu,i} - \dfrac{1}{2} \sum_{\substack{\textbf{n} \\ \mu,i}} \sum_{\substack{\textbf{m} \\ \nu,k}} D^{\textbf{n}-\textbf{m},\mu,i}_{\textbf{0},\nu,k} \sum_{\textbf{q}} \sum_{\textbf{q}'} \dfrac{1}{N} e^{i \textbf{q} \cdot \left(\textbf{R}_{\textbf{n}} - \textbf{R}_{\textbf{m}}\right)} e^{i \left(\textbf{q}+\textbf{q}'\right) \cdot \textbf{R}_{\textbf{m}}} C_{\textbf{q},\mu,i} C_{\textbf{q}',\nu,k} ,
\end{align}
where, in the last step, we have made explicit that the force constants depend only on the difference between the lattice vectors. We set $\textbf{R}_{\textbf{n}} \rightarrow \textbf{R}_{\textbf{m}} + \textbf{R}_{\textbf{h}}, \ \forall \ \textbf{R}_{\textbf{m}}$, then
\begin{equation}
\mathcal{L} = \dfrac{1}{2} M \sum_{\substack{\textbf{q} \\ \mu,i}} \dot{C}_{\textbf{q},\mu,i} \dot{C}_{-\textbf{q},\mu,i} - \dfrac{1}{2} \sum_{\substack{\textbf{h} \\ \mu,i}} \sum_{\nu,k} \sum_{\textbf{q}} D^{\textbf{h},\mu,i}_{\textbf{0},\nu,k} e^{i \textbf{q} \cdot \textbf{R}_{\textbf{h}}} \sum_{\textbf{m}} \sum_{\textbf{q}'} \dfrac{1}{N} e^{i \left(\textbf{q}+\textbf{q}'\right) \cdot \textbf{R}_{\textbf{m}}} C_{\textbf{q},\mu,i} C_{\textbf{q}',\nu,k} ,
\end{equation}
and we define
\begin{equation}
\left[ \tilde{D}(\textbf{q}) \right]_{\substack{\mu,\nu \\ i,k}} \ = \ \sum_{\textbf{h}} D^{\textbf{h},\mu,i}_{\textbf{0},\nu,k} \ e^{i \textbf{q} \cdot \textbf{R}_{\textbf{h}}} ,
\label{eq: costantidiforzaDtilde}
\end{equation}
so that the dependence of the force constants on the positional index of the cell disappears, allowing us to write
\begin{equation}
\dfrac{1}{N} \sum_{\textbf{m}} e^{i \left(\textbf{q}+\textbf{q}'\right) \cdot \textbf{R}_{\textbf{m}}} = \delta_{\textbf{q},-\textbf{q}'} ,
\end{equation}
\begin{align}
\mathcal{L} &= \dfrac{1}{2} M \sum_{\substack{\textbf{q} \\ \mu,i}} \dot{C}_{\textbf{q},\mu,i} \dot{C}_{-\textbf{q},\mu,i} - \dfrac{1}{2} \sum_{\textbf{q}} \sum_{\substack{\mu,i \\ \nu,k}} \left[ \tilde{D}(\textbf{q}) \right]_{\substack{\mu,\nu \\ i,k}} C_{\textbf{q},\mu,i} C_{-\textbf{q},\nu,k}.
\end{align}
Now, each term indexed by \((\mathbf{q}, \mu, i)\) couples only with terms indexed by \(\{(-\mathbf{q}, \mu, i)\}_{\mu,i}\), so the Lagrangian becomes block diagonal. For each wave vector \(\mathbf{q}\), there is a block of dimension \(3 \tau \times 3 \tau\), and the total number of such blocks equals the number of \(\mathbf{q}\)-vectors, i.e., \(N\) blocks. Thus, the Lagrangian decomposes into \(N\) independent blocks. The matrices defined in $\eqref{eq: costantidiforzaDtilde}$ are Hermitian; indeed,
\begin{align}
\left[ \left[ \tilde{D}(\textbf{q}) \right]_{\substack{\mu,\nu \\ i,k}} \right]^{\dag} &= \sum_{\textbf{h}} D^{\textbf{h},\nu,k}_{\textbf{0},\mu,i} \ e^{- i \textbf{q} \cdot \textbf{R}_{\textbf{h}}} = \notag \\
&= \sum_{\textbf{h}} D^{-\textbf{h},\nu,k}_{\textbf{0},\mu,i} \ e^{i \textbf{q} \cdot \textbf{R}_{\textbf{h}}} = \notag \\
&= \sum_{\textbf{h}} D^{\textbf{0},\nu,k}_{\textbf{h},\mu,i} \ e^{i \textbf{q} \cdot \textbf{R}_{\textbf{h}}} = \notag \\
&= \sum_{\textbf{h}} D^{\textbf{h},\mu,i}_{\textbf{0},\nu,k} \ e^{i \textbf{q} \cdot \textbf{R}_{\textbf{h}}} \equiv \notag \\
&\equiv \left[ \tilde{D}(\textbf{q}) \right]_{\substack{\mu,\nu \\ i,k}}.
\end{align}
The matrices \(D\) are real, so the adjoint operation coincides with the transpose, which in our notation corresponds to exchanging the order of the index pairs \((\mu, \nu)\) and \((i, k)\) in the definition of \(\tilde{D}(\mathbf{q})\). We substituted \(\mathbf{R}_{\mathbf{h}} \to -\mathbf{R}_{\mathbf{h}}\) since \(\mathbf{h}\) is a dummy summation index, then applied property \eqref{eq: proprietàcoefficientiWdifferenzavettori}, and finally reordered the indices using property $\eqref{eq: proprietàcoefficientiWdifferenzavettori}$. Therefore, for each \(\mathbf{q}\), the matrix \(\tilde{D}(\mathbf{q})\) has \(3 \tau\) real eigenvalues and \(3 \tau\) corresponding eigenvectors. Let us now define \(\mathbf{C}_{\mathbf{q}}\) as the column vector composed of the \(3 \tau\) components \(C_{\mathbf{q}, \mu, i}\) for each fixed \(\mathbf{q}\). Under this notation, the Lagrangian becomes
\begin{align}
\mathcal{L} &= \dfrac{1}{2} M \sum_{\textbf{q}} \left[ \dot{\textbf{C}}_{- \textbf{q}} \right]_t \cdot \ \dot{\textbf{C}}_{\textbf{q}} \ - \ \dfrac{1}{2} \sum_{\textbf{q}} \left[ \textbf{C}_{- \textbf{q}} \right]_t \tilde{D}(\textbf{q}) \ \textbf{C}_{\textbf{q}},
\end{align}
where we have rearranged the order of \(\dot{\mathbf{C}}_{\mathbf{q}}\), \(\dot{\mathbf{C}}_{-\mathbf{q}}\), \(\mathbf{C}_{-\mathbf{q}}\), and \(\mathbf{C}_{\mathbf{q}}\) to write the first term as a row-by-column scalar product and the second term as a matrix product. The matrix \(\tilde{D}(\mathbf{q})\) is Hermitian by construction. Fixing \(\mathbf{q}\), we denote by \(P(\mathbf{q})\) the \(3 \tau \times 3 \tau\) matrix whose columns are the eigenvectors of \(\tilde{D}(\mathbf{q})\). The object
\begin{equation}
P^{-1}(\textbf{q}) \tilde{D}(\textbf{q}) P(\textbf{q}) = K(\textbf{q})
\end{equation}
is a diagonal matrix whose diagonal elements are the \(3 \tau\) real eigenvalues of \(\tilde{D}(\mathbf{q})\), and we have
\begin{align}
\mathcal{L} &= \dfrac{1}{2} M \sum_{\textbf{q}} \left[ \dot{\textbf{C}}_{- \textbf{q}} \right]_t \cdot \ \dot{\textbf{C}}_{\textbf{q}} \ - \ \dfrac{1}{2} \sum_{\textbf{q}} \left[ \textbf{C}_{- \textbf{q}} \right]_t P(\textbf{q}) P^{-1}(\textbf{q}) \tilde{D}(\textbf{q}) P(\textbf{q}) P^{-1}(\textbf{q}) \textbf{C}_{\textbf{q}} = \notag \\
&= \dfrac{1}{2} M \sum_{\textbf{q}} \left[ \dot{\textbf{C}}_{- \textbf{q}} \right]_t \cdot \ \dot{\textbf{C}}_{\textbf{q}} \ - \ \dfrac{1}{2} \sum_{\textbf{q}} \left[ \textbf{C}_{- \textbf{q}} \right]_t P(\textbf{q}) K(\textbf{q}) P^{-1}(\textbf{q}) \textbf{C}_{\textbf{q}}.
\end{align}
Due to the similarity transformation $P^{-1}(\textbf{q}) \tilde{D}(\textbf{q}) P(\textbf{q})$, the matrix \(\tilde{D}(\mathbf{q})\) becomes diagonal in the basis formed by the vectors
\begin{equation}
\left[ \textbf{C}_{- \textbf{q}} \right]_t P(\textbf{q}), 
\end{equation}
\begin{equation}
P^{-1}(\textbf{q}) \textbf{C}_{\textbf{q}},
\end{equation}
which are respectively row and column vectors. We then define the column vector
\begin{equation}
\textbf{Q}_{\textbf{q}} = P^{-1}(\textbf{q}) \textbf{C}_{\textbf{q}} ,
\end{equation}
and we aim to compute
\begin{align}
\left[ \textbf{Q}_{-\textbf{q}} \right]_t &= \left[ P^{-1}(-\textbf{q}) \textbf{C}_{-\textbf{q}} \right]_t = \notag \\
&= \left[ \textbf{C}_{-\textbf{q}} \right]_t \left[ P^{-1}(-\textbf{q}) \right]_t .
\end{align}
Since \(P(\mathbf{q})\) is a change-of-basis matrix, i.e., a unitary transformation, we have
\begin{equation}
P^\dag(-\textbf{q}) = P^{-1}(-\textbf{q}) ,
\end{equation}
\begin{equation}
\left[ P^{*}(-\textbf{q}) \right]_{t} = P^{-1}(-\textbf{q}),
\end{equation}
\begin{equation}
P^*(-\textbf{q}) = \left[ P^{-1}(-\textbf{q}) \right]_t,
\end{equation}
\begin{equation}
\left[ \textbf{Q}_{-\textbf{q}} \right]_t = \left[ \textbf{C}_{-\textbf{q}} \right]_t P^*(-\textbf{q}).
\end{equation}
The quantity \(\left[ \mathbf{Q}_{-\mathbf{q}} \right]^T\) can be manipulated by exploiting the properties of \(\tilde{D}(\mathbf{q})\). From equation $\eqref{eq: costantidiforzaDtilde}$, the matrices \(\tilde{D}(\mathbf{q})\) satisfy
\begin{equation}
\left[ \tilde{D}(\textbf{q}) \right]^{*} = \tilde{D}(-\textbf{q}).
\end{equation}
Then, we manipulate the eigenvalue equation as follows
\begin{equation}
\tilde{D}(\textbf{q}) V = \lambda V ,
\end{equation}
\begin{equation}
\left[ \tilde{D}(\textbf{q}) \right]^{*} V^{*} = \lambda V^{*},
\end{equation}
\begin{equation}
\tilde{D}(-\textbf{q}) V^{*} = \lambda V^{*},
\end{equation}
\begin{equation}
\left[ \tilde{D}(-\textbf{q})\right]^{*} V = \lambda V ,
\end{equation}
therefore, \(\tilde{D}(\mathbf{q})\) and \(\left[ \tilde{D}(-\mathbf{q}) \right]^*\) share the same eigenvalues, which are real, as well as the eigenvectors, consequently we have
\begin{equation}
P^*(-\textbf{q}) = P(\textbf{q}).
\end{equation}
The object
\begin{equation}
\left[ \textbf{Q}_{-\textbf{q}} \right]_t = \left[ \textbf{C}_{-\textbf{q}} \right]_t P(\textbf{q}) 
\end{equation}
is the row vector appearing in the Lagrangian
\begin{align}
\mathcal{L} &= \dfrac{1}{2} M \sum_{\textbf{q}} \left[ \dot{\textbf{C}}_{- \textbf{q}} \right]_t P(\textbf{q}) P^{-1}(\textbf{q}) \dot{\bar{C}}_{\textbf{q}} \ - \ \dfrac{1}{2} \sum_{\textbf{q}} \left[ \textbf{C}_{- \textbf{q}} \right]_t P(\textbf{q}) K(\textbf{q}) P^{-1}(\textbf{q}) \textbf{C}_{\bar{q}} = \notag \\
&= \dfrac{1}{2} M \sum_{\textbf{q}} \left[ \dot{\textbf{Q}}_{-\textbf{q}} \right]_t \cdot \ \dot{\textbf{Q}}_{\textbf{q}} \ - \ \dfrac{1}{2} \sum_{\textbf{q}} \left[ \textbf{Q}_{-\textbf{q}} \right]_t K(\textbf{q}) \ \textbf{Q}_{\textbf{q}}.
\end{align}
Since $K(\mathbf{q})$ is diagonal, we can express the scalar products in the Lagrangian in terms of the components of $\left[ \dot{\textbf{Q}}_{-\textbf{q}} \right]_t$, $\left[ \dot{\textbf{Q}}_{-\textbf{q}} \right]_t$, $\left[ \textbf{Q}_{-\textbf{q}} \right]_t$ and $\left[ \textbf{Q}_{-\textbf{q}} \right]_t$, i.e.,
\begin{equation}
\mathcal{L} = \dfrac{1}{2} M \sum_{\textbf{q}} \sum_{s=1}^{3 \tau} \dot{Q}_{-\textbf{q},s} \dot{Q}_{\textbf{q},s} \ - \ \dfrac{1}{2} \sum_{\textbf{q}} \sum_{s=1}^{3 \tau} Q_{-\textbf{q},s} \left[ K(\textbf{q}) \right]_{s,s} Q_{\textbf{q},s} .
\end{equation}
The vectors $\mathbf{Q}_{\mathbf{q}}$ and $\mathbf{Q}_{-\mathbf{q}}$ are indexed by a single index $s$ in both the kinetic and potential terms. \newline
The displacements $S_{\mathbf{n},\mu,i}$ can now be rewritten in terms of the variables $Q_{\mathbf{q},s}$. For convenience, we group the indices $(\mu,i)$ into a single index $l$. The vector $\mathbf{C}_{\mathbf{q}}$ can then be written as a function of $\mathbf{Q}_{\mathbf{q}}$ as
\begin{equation}
\textbf{C}_{\textbf{q}} = P(\textbf{q}) \textbf{Q}_{\textbf{q}}.
\end{equation}
Now, recalling the transformations between $S_{\textbf{n},l}$ and $U_{\textbf{n},l}$, and between $U_{\textbf{n},l}$ and $C_{\textbf{q},l}$, we have 
\begin{equation}
S_{\textbf{n},l} = \sqrt{\dfrac{M}{M_{\mu}}} \sum_{\textbf{q}} \dfrac{e^{i \textbf{q} \cdot \textbf{R}_{\textbf{n}}}}{\sqrt{N}} \sum_{s=1}^{3 \tau} \left[ P(\textbf{q}) \right]_{l,s} Q_{\textbf{q},s}.
\end{equation}
In other words, the displacement in the cell \(\mathbf{n}\) is a linear combination of the variables \(Q_{\mathbf{q},s}\). The displacements \(S_{\mathbf{n},\mu,i}\) are real quantities, while the \(Q_{\mathbf{q},s}\) are generally complex; thus, the number of \(Q_{\mathbf{q},s}\) variables is twice that of the \(S_{\mathbf{n},\mu,i}\). Consequently, there must be a constraint on the \(Q_{\mathbf{q},s}\) to ensure they form a linearly independent system. Since \(S_{\mathbf{n},l}\) is real, it follows that
\begin{equation}
S^{*}_{\textbf{n},l} = \sqrt{\dfrac{M}{M_{\mu}}} \sum_{\textbf{q}} \dfrac{e^{- i \textbf{q} \cdot \textbf{R}_{\textbf{n}}}}{\sqrt{N}} \sum_{s=1}^{3 \tau} \left[ P^{*}(\textbf{q}) \right]_{l,s} Q^{*}_{\textbf{q},s} ,
\end{equation}
we rename the dummy index $\textbf{q} \rightarrow - \textbf{q}$ as follows
\begin{equation}
S^{*}_{\textbf{n},l} = \sqrt{\dfrac{M}{M_{\mu}}} \sum_{\textbf{q}} \dfrac{e^{i \textbf{q} \cdot \textbf{R}_{\textbf{n}}}}{\sqrt{N}} \sum_{s=1}^{3 \tau} \left[ P^{*}(-\textbf{q}) \right]_{l,s} Q^{*}_{-\textbf{q},s} ,
\end{equation}
and this must be equal to \( S_{\mathbf{n},l} \). Recall that
\begin{equation}
\left[ P^{*}(-\textbf{q}) \right]_{l,s} = \left[ P(\textbf{q}) \right]_{l,s},
\end{equation}
then
\begin{equation}
Q^{*}_{-\textbf{q},s} = Q_{\textbf{q},s}.
\label{eq: vincolomodinormalifononi}
\end{equation}
Among the \(6 N \tau\) variables \(Q_{\mathbf{q},s}\), only \(3 N \tau\) are linearly independent. The Lagrangian becomes
\begin{equation}
\mathcal{L} = \dfrac{1}{2} M \sum_{\textbf{q}} \sum_{s=1}^{3 \tau} |\dot{Q}_{\textbf{q},s}|^{2} \ - \ \dfrac{1}{2} \sum_{\textbf{q}} \sum_{s=1}^{3 \tau} \left[ K(\textbf{q}) \right]_{s,s} |Q_{\textbf{q},s}|^{2} ,
\end{equation}
where \(Q_{\mathbf{q},s}\) denote the classical normal modes. The Euler-Lagrange equations corresponding to \(\mathcal{L}\) are
\begin{equation}
\dfrac{d}{dt} \dfrac{\partial \mathcal{L}}{\partial \dot{Q}^{*}_{\textbf{q},s}} - \dfrac{\partial \mathcal{L}}{\partial Q^{*}_{\textbf{q},s}} = 0, \ \forall \ \textbf{q}, \ \forall s ,
\end{equation}
where we used the complex conjugates of \( Q_{\mathbf{q},s} \) and \( \dot{Q}_{\mathbf{q},s} \). To properly handle terms like \( Q_{\mathbf{q},s} Q^*_{\mathbf{q},s} \) and \( \dot{Q}_{\mathbf{q},s} \dot{Q}^*_{\mathbf{q},s} \), we must consider both the components at \(\mathbf{q}\) and \(-\mathbf{q}\). These components are related and can be expressed in terms of each other, which removes the factor of \(\frac{1}{2}\). Consequently, the Euler-Lagrange equations take the form
\begin{equation}
M \ddot{Q}_{\textbf{q},s} = - \left[ K(\textbf{q}) \right]_{s,s} Q_{\textbf{q},s}.
\end{equation}
The Lagrangian we have derived is positive definite; hence, the eigenvalues of \(\tilde{D}(\mathbf{q})\) and \(K(\mathbf{q})\) are all positive. We define
\begin{equation}
\left[ K(\textbf{q}) \right]_{s,s} = M \omega^{2}_{\textbf{q},s},
\end{equation}
so that the equations become
\begin{equation}
\ddot{Q}_{\textbf{q},s} = - \omega^{2}_{\textbf{q},s} Q_{\textbf{q},s},
\end{equation}
which corresponds to a system of \(3 \tau N\) independent second-order differential equations, each describing a simple harmonic motion with solutions given by
\begin{equation}
Q_{\textbf{q},s}(t) = A_{\textbf{q},s} e^{i \omega_{\textbf{q},s} t} + B_{\textbf{q},s} e^{- i \omega_{\textbf{q},s} t}.
\end{equation}
The displacement field 
\begin{equation}
S_{\textbf{n},l} = \sqrt{\dfrac{M}{M_{\mu}}} \sum_{\textbf{q}} \dfrac{e^{i \textbf{q} \cdot \textbf{R}_{\textbf{n}}}}{\sqrt{N}} \sum_{s=1}^{3 \tau} \left[ P(\textbf{q}) \right]_{l,s} Q_{\textbf{q},s}
\end{equation}
is thus a superposition of \(3 \tau N\) normal modes, each of which satisfies the equation of simple harmonic motion. Recall that the index \(l\) groups the indices \((\mu, i)\), and therefore takes on \(3 \tau\) values. For each atomic species \(\mu\), we now collect the three components \(S_{\textbf{n}, \mu, i}\), corresponding to the Cartesian directions \(i = 1, 2, 3\), into a vector \(\textbf{S}_{\textbf{n}, \mu}\). This vector represents the displacement of the atom of species \(\mu\) in the cell labeled by \(\textbf{n}\). Thus, on the left-hand side we have the substitution \(S_{\textbf{n}, \mu, i} \rightarrow \textbf{S}_{\textbf{n}, \mu}\). On the right-hand side, only the quantity \(\left[ P(\textbf{q}) \right]_{l, s}\) depends on the index \(l\). Here, \(\left[ P(\textbf{q}) \right]_{l, s}\) denotes the entry in the \(l\)-th row and \(s\)-th column of the matrix \(P(\textbf{q})\), which is a \(3 \tau \times 3 \tau\) unitary matrix whose columns are the eigenvectors of the dynamical matrix \(\tilde{D}(\textbf{q})\), i.e.,
\begin{equation}
P(\textbf{q})
=
\begin{pmatrix}
\varepsilon_{1,1}^{(1)}(\textbf{q}) & \varepsilon_{1,1}^{(2)}(\textbf{q}) & \ldots & \varepsilon_{1,1}^{(3 \tau)}(\textbf{q}) \\
\varepsilon_{1,2}^{(1)}(\textbf{q}) & \varepsilon_{1,2}^{(2)}(\textbf{q}) & \ldots & \varepsilon_{1,2}^{(3 \tau)}(\textbf{q}) \\
\varepsilon_{1,3}^{(1)}(\textbf{q}) & \varepsilon_{1,3}^{(2)}(\textbf{q}) & \ldots & \varepsilon_{1,3}^{(3 \tau)}(\textbf{q}) \\
\vdots & \vdots & \ddots & \vdots \\
\varepsilon_{\tau,1}^{(1)}(\textbf{q}) & \varepsilon_{\tau,1}^{(2)}(\textbf{q}) & \ldots & \varepsilon_{\tau,1}^{(3\tau)}(\textbf{q}) \\
\varepsilon_{\tau,2}^{(1)}(\textbf{q}) & \varepsilon_{\tau,2}^{(2)}(\textbf{q}) & \ldots & \varepsilon_{\tau,2}^{(3\tau)}(\textbf{q}) \\
\varepsilon_{\tau,3}^{(1)}(\textbf{q}) & \varepsilon_{\tau,3}^{(2)}(\textbf{q}) & \ldots & \varepsilon_{\tau,3}^{(3\tau)}(\textbf{q}) 
\end{pmatrix}.
\end{equation}
Each column of \(P(\textbf{q})\) consists of \(3 \tau\) components, where the index \(s\) spans both the atomic species \(\mu\) and the Cartesian direction \(i\). We group the three scalar components \(\varepsilon_{\mu,1}^{(s)}(\textbf{q})\), \(\varepsilon_{\mu,2}^{(s)}(\textbf{q})\), and \(\varepsilon_{\mu,3}^{(s)}(\textbf{q})\), which correspond to the displacements of atom \(\mu\) along the three directions, into a single vector \(\bm{\varepsilon}_{\mu}^{(s)}(\textbf{q})\). Then, the displacement of the atom of species \( \mu \) in the unit cell located at \( \textbf{n} \) is given by
\begin{align}
\textbf{S}_{\textbf{n},\mu} &= \sqrt{\dfrac{M}{M_{\mu}}} \sum_{\textbf{q}} \sum_{s=1}^{3 \tau} \dfrac{e^{i \textbf{q} \cdot \textbf{R}_{\textbf{n}}}}{\sqrt{N}} \ \bm{\varepsilon}_{\mu}^{(s)}(\textbf{q}) \ Q_{\textbf{q},s} = \notag \\
&= \sqrt{\dfrac{M}{M_{\mu}}} \sum_{\textbf{q}} \sum_{s=1}^{3 \tau} \dfrac{e^{i \textbf{q} \cdot \textbf{R}_{\textbf{n}}}}{\sqrt{N}} \ \bm{\varepsilon}_{\mu}^{(s)}(\textbf{q}) \left( A_{\textbf{q},s} e^{i \omega_{\textbf{q},s} t} + B_{\textbf{q},s} e^{- i \omega_{\textbf{q},s} t} \right).
\end{align}
Due to this structure, where the factor \( e^{i \textbf{q} \cdot \textbf{R}_{\textbf{n}}} \) encodes spatial oscillations and \( Q_{\textbf{q},s} \) describes temporal oscillations, the displacement \( \textbf{S}_{\textbf{n},\mu}(t) \) results as a linear combination of forward and backward propagating plane waves. Each term in the sum corresponds to a mode characterized by a wavevector \( \textbf{q} \), a polarization vector \( \bm{\varepsilon}_{\mu}^{(s)}(\textbf{q}) \), and a time-dependent amplitude \( Q_{\textbf{q},s} \). The polarization vector \( \bm{\varepsilon}_{\mu}^{(s)}(\textbf{q}) \) specifies the direction and relative amplitude of the displacement of atom \( \mu \) in the unit cell for the \( s \)-th normal mode with wavevector \( \textbf{q} \). The atomic displacements are composed of superpositions of normal modes, each with a well-defined wavevector and polarization, forming a complete basis for the vibrational motion of the lattice. From the expression
\begin{align}
\mathcal{L} &\equiv \mathcal{L}\left(Q_{\textbf{q},s},\dot{Q}_{\textbf{q},s}\right) = \notag \\
&= \dfrac{1}{2} M \sum_{\textbf{q}} \sum_{s=1}^{3 \tau} |\dot{Q}_{\textbf{q},s}|^{2} \ - \ \dfrac{1}{2} \sum_{\textbf{q}} \sum_{s=1}^{3 \tau} \left[ K(\textbf{q}) \right]_{s,s} |Q_{\textbf{q},s}|^{2},
\label{eq: Lagrangianaclassicavibrazionireticolari}
\end{align}
we compute the related Hamiltonian, that is,
\begin{equation}
\mathcal{H} = \sum_h P_h \dot{Q}_h - \mathcal{L} \left( Q_h,\dot{Q}_h(P_h) \right) ,
\end{equation}
where $h \equiv (\textbf{q},s)$ and $P_h$ is the momentum and is defined as
\begin{equation}
P_h = \dfrac{\partial \mathcal{L}}{\partial \dot{Q}^*_h}.
\end{equation} 
We have
\begin{align}
P_{\textbf{q},s} &= \dfrac{\partial \mathcal{L}}{\partial \dot{Q}^*_{\textbf{q},s}} = \notag \\
&= M \dot{Q}_{\textbf{q},s},
\end{align}
then
\begin{equation}
\dot{Q}_{\textbf{q},s} = \dfrac{P_{\textbf{q},s}}{M},
\end{equation}
consequently the \( P_{\textbf{q},s} \) fulfill $\eqref{eq: vincolomodinormalifononi}$, in the same way as the \( \dot{Q}_{\textbf{q},s} \). The variables \( P_{\textbf{q},s} \) form a set of \( 3 \tau N\) independent canonical momenta, and the object
\begin{equation}
\mathcal{H} = \sum_{\textbf{q}} \sum_{s=1}^{3\tau} \dfrac{|P_{\textbf{q},s}|^2}{2M} + \sum_{\textbf{q}} \sum_{s=1}^{3\tau} \dfrac{1}{2} \left[ K(\textbf{q}) \right]_{s,s} |Q_{\textbf{q},s}|^2
\label{eq: hamiltonianavibrazionireticolariclassica}
\end{equation}
is the Hamiltonian of $3 \tau N$ independent simple harmonic motions. 
\subsection{Quantum formalism}
In quantum mechanics, the objects \( Q_{\textbf{q},s} \) and \( P_{\textbf{q},s} \) are promoted to operators, so the constraints $\eqref{eq: vincolomodinormalifononi}$ are naturally replaced by the canonical commutation relations
\begin{equation}
\hat{Q}^\dag_{\mathbf{q},s} = \hat{Q}_{-\mathbf{q},s},
\end{equation}
\begin{equation}
\hat{P}^\dag_{\mathbf{q},s} = \hat{P}_{-\mathbf{q},s},
\end{equation}
and they must satisfy the algebras of the position and momentum operators, i.e.,
\begin{equation}
\left[ \hat{Q}_{\mathbf{q},s} , \hat{Q}_{\mathbf{q}',s'} \right] = \left[ \hat{P}_{\mathbf{q},s} , \hat{P}_{\mathbf{q}',s'} \right] = 0,
\end{equation}
\begin{equation}
\left[ \hat{Q}^\dagger_{\mathbf{q},s} , \hat{P}_{\mathbf{q}',s'} \right] = i \hbar \delta_{\mathbf{q},\mathbf{q}'} \delta_{s,s'}.
\end{equation}
or equivalently
\begin{equation}
\left[ \hat{Q}_{\mathbf{q},s} , \hat{P}_{\mathbf{q}',s'} \right] = i \hbar \delta_{\mathbf{q},-\mathbf{q}'} \delta_{s,s'}.
\end{equation}
We define 
\begin{equation}
\hat{Q}_{\textbf{q},s} = \sqrt{\dfrac{\hslash}{2 M \omega_{\textbf{q},s}}} \left( a_{\textbf{q},s} + a_{-\textbf{q},s}^{\dag} \right),
\end{equation}
\begin{equation}
\hat{P}_{\textbf{q},s} = i \sqrt{\dfrac{M \hslash \omega_{\textbf{q},s}}{2}} \left( a^{\dag}_{\textbf{q},s} - a_{-\textbf{q},s} \right) ,
\end{equation}
whose adjoint are given by
\begin{equation}
\hat{Q}_{-\textbf{q},s} = \sqrt{\dfrac{\hslash}{2 M \omega_{\textbf{q},s}}} \left( a^{\dag}_{\textbf{q},s} + a_{-\textbf{q},s} \right),
\end{equation}
\begin{equation}
\hat{P}_{-\textbf{q},s} = - i \sqrt{\dfrac{M \hslash \omega_{\textbf{q},s}}{2}} \left( a_{\textbf{q},s} - a^{\dag}_{-\textbf{q},s} \right).
\end{equation}
We write
\begin{equation}
a_{\textbf{q},s} + a_{-\textbf{q},s}^{\dag} = \sqrt{\dfrac{2 M \omega_{\textbf{q},s}}{\hslash}} \hat{Q}_{\textbf{q},s} ,
\label{eq: aacrocefononi1}
\end{equation}
\begin{equation}
a^{\dag}_{\textbf{q},s} - a_{-\textbf{q},s} = - i \sqrt{\dfrac{2}{M \hslash \omega_{\textbf{q},s}}} \hat{P}_{\textbf{q},s} ,
\label{eq: aacrocefononi2}
\end{equation}
\begin{equation}
a^{\dag}_{\textbf{q},s} + a_{-\textbf{q},s} = \sqrt{\dfrac{2 M \omega_{\textbf{q},s}}{\hslash}} \hat{Q}_{-\textbf{q},s} ,
\label{eq: aacrocefononi3}
\end{equation}
\begin{equation}
a_{\textbf{q},s} - a^{\dag}_{-\textbf{q},s} = i \sqrt{\dfrac{2}{M \hslash \omega_{\textbf{q},s}}} \hat{P}_{-\textbf{q},s} ,
\label{eq: aacrocefononi4}
\end{equation}
we sum equations $\eqref{eq: aacrocefononi1}$ and $\eqref{eq: aacrocefononi4}$ 
\begin{align}
a_{\textbf{q},s} &= \dfrac{1}{2} \sqrt{\dfrac{2 M \omega_{\textbf{q},s}}{\hslash}} \hat{Q}_{\textbf{q},s} + \dfrac{i}{2} \sqrt{\dfrac{2}{M \hslash \omega_{\textbf{q},s}}} \hat{P}_{-\textbf{q},s} = \notag \\
&= \sqrt{\dfrac{M \omega_{\textbf{q},s}}{2 \hslash}} \hat{Q}_{\textbf{q},s} + i \sqrt{\dfrac{1}{2 M \hslash \omega_{\textbf{q},s}}} \hat{P}_{-\textbf{q},s},
\end{align}
and we sum equations $\eqref{eq: aacrocefononi3}$ and $\eqref{eq: aacrocefononi2}$,
\begin{align}
a^{\dag}_{\textbf{q},s} &= \dfrac{1}{2} \sqrt{\dfrac{2 M \omega_{\textbf{q},s}}{\hslash}} \hat{Q}_{-\textbf{q},s} - \dfrac{i}{2} \sqrt{\dfrac{2}{M \hslash \omega_{\textbf{q},s}}} \hat{P}_{\textbf{q},s} = \notag \\
&= \sqrt{\dfrac{M \omega_{\textbf{q},s}}{2 \hslash}} \hat{Q}_{-\textbf{q},s} - i \sqrt{\dfrac{1}{2 M \hslash \omega_{\textbf{q},s}}} \hat{P}_{\textbf{q},s},
\end{align}
that is, the operators $a_{\textbf{q},s}$ and $a^{\dag}_{\textbf{q},s}$ are not adjoint operators of each other. Nevertheless, from the commutation relations of the position and momentum operators, it immediately follows that $a_{\textbf{q},s}$ and $a^{\dag}_{\textbf{q},s}$ close a bosonic algebra: $a_{\textbf{q},s}$ and $a^{\dag}_{\textbf{q},s}$ are annihilation and creation operators, respectively. The Hamiltonian $\eqref{eq: hamiltonianavibrazionireticolariclassica}$ is written as a function of the operators $a_{\textbf{q},s}$ and $a^{\dag}_{\textbf{q},s}$ as
\begin{equation}
\hat{\mathcal{H}} = \sum_{\textbf{q}} \sum_{s=1}^{3\tau} \hslash \omega_{\textbf{q},s} \left( a^{\dag}_{\textbf{q},s} a_{\textbf{q},s} + \dfrac{1}{2} \right)
\label{eq: hamiltonianavibrazionireticolariquantizzate}
\end{equation}
which is the Hamiltonian in second quantization of $3 \tau N$ independent simple harmonic oscillators and describes the oscillations of ions in the electronic effective potential, and it represents the second term of the Hamiltonian $\eqref{eq: Hamiltonianadiunsolido}$. The field \( \mathbf{S}_{\mathbf{n},\mu} \) of lattice vibrations has been quantized, and lattice energy exchanges occur for integer multiples of \( \hslash \omega_{\mathbf{q},s} \). The field energy quanta of lattice vibrations are named phonons, and the atomic displacement field is given by the operator
\begin{align}
\textbf{S}_{\textbf{n},\mu} &= \sqrt{\dfrac{M}{M_{\mu}}} \sum_{\textbf{q}} \sum_{s=1}^{3 \tau} \dfrac{e^{i \textbf{q} \cdot \textbf{R}_{\textbf{n}}}}{\sqrt{N}} \ \bm{\varepsilon}_{\mu}^{(s)}(\textbf{q}) \hat{Q}_{\textbf{q},s} = \notag \\
&= \sqrt{\dfrac{M}{M_{\mu}}} \sum_{\textbf{q}} \sum_{s=1}^{3 \tau} \dfrac{e^{i \textbf{q} \cdot \textbf{R}_{\textbf{n}}}}{\sqrt{N}} \ \bm{\varepsilon}_{\mu}^{(s)}(\textbf{q}) \sqrt{\dfrac{\hslash}{2 M \omega_{\textbf{q},s}}} \left( a_{\textbf{q},s} e^{- i \omega_{\textbf{q},s} t} + a_{-\textbf{q},s}^{\dag} e^{i \omega_{\textbf{q},s} t} \right) = \notag \\
&= \sum_{\textbf{q}} \sum_{s=1}^{3 \tau} \dfrac{e^{i \textbf{q} \cdot \textbf{R}_{\textbf{n}}}}{\sqrt{N}} \ \bm{\varepsilon}_{\mu}^{(s)}(\textbf{q}) \sqrt{\dfrac{\hslash}{2 M_{\mu} \omega_{\textbf{q},s}}} \left( a_{\textbf{q},s} e^{- i \omega_{\textbf{q},s} t} + a_{-\textbf{q},s}^{\dag} e^{i \omega_{\textbf{q},s} t} \right).
\label{eq: spostamentoquantizzatofononi}
\end{align}
For simplicity, we will use the same symbol to denote both the classical field and the corresponding operator. \newline
The amplitudes of the classical oscillations, $A_{\textbf{q},s}$ and $B_{\textbf{q},s}$, have become operators. If we fix $s$, that is, fix the atomic species and direction, we obtain a so-called branch. For a given branch, varying the wavevector $\textbf{q}$ produces the ionic oscillation bands of the solid. Just as the quantum number $n$ indexes an electronic Bloch band, the index $s$ labels the phonon bands. These bands are classified as acoustic or optical depending on whether the phonon frequency $\omega_{\textbf{q},s}$ tends to zero or remains finite, respectively, as $\textbf{q} \rightarrow \textbf{0}$. If the polarization vector of the phonon wave, that is $\bm{\varepsilon}_{\mu}(\textbf{q})$, is parallel to $\textbf{q}$, the phonon mode is longitudinal; if perpendicular, it is transverse. Unlike photons, which are exclusively transverse electromagnetic waves, phonons can be either longitudinal or transverse elastic waves.
\section{1D lattice with one atom per cell: acoustic branch}
Consider a linear chain with lattice constant \(a\) and a single atom per unit cell, i.e., \(\tau = 1\). We impose Born-von Karman boundary conditions, meaning \(S(N) = S(0)\), so that the ends of the chain are joined, and the \(N\) atoms are arranged along a circumference. The eigenvalues \(M \omega_{\textbf{q},s} \equiv M \omega_\textbf{q}\) of the matrix \(K(\textbf{q})\), which diagonalizes \(\tilde{D}(\textbf{q})\), are determined by calculating the force constants \(W^{\mathbf{n},\mu,i}_{\mathbf{m},\nu,k}\). Since there is only one atom per cell and the lattice is one-dimensional, the indices simplify: the lattice vectors \(\mathbf{n}\), \(\mathbf{m}\), and wavevector \(\mathbf{q}\) become scalars \(n\), \(m\), and \(q\), and the atomic and directional indices \((\mu,i)\) and \((\nu,k)\) disappear. The force constants reduce to the form \(W^n_m\). In the simplest model, each atom interacts only with its nearest neighbors via an elastic potential with elastic constant $k_{el}$, i.e.,
\begin{equation}
V = \dfrac{k_{el}}{2} \sum_{n=0}^{N-1} \left( S_{n} - S_{n+1} \right)^2.
\end{equation}
\begin{remark}[Elastic potential in a linear chain.] 
The potential $V$ can be written as
\begin{equation}
V = \dfrac{k_{el}}{2} \sum_{n=0}^{N-1} \left( 2 S_n^2 - S_n S_{n+1} - S_n S_{n-1} \right).
\label{eq: potenzialeelasticoinunacatenalineare}
\end{equation}
Indeed, from 
\begin{align}
\left( S_{n} - S_{n+1} \right)^2 &= \left( S_{n} - S_{n+1} \right) \left( S_{n} - S_{n+1} \right) = \notag \\
&= S^2_{n} - S_{n} S_{n+1} - S_{n+1} S_{n} + S^2_{n+1},
\end{align}
we now consider the last term. Since the atoms in the chain are identical simple harmonic oscillators and periodic boundary conditions are assumed, the system is translationally invariant. As a consequence, quantities associated with a given lattice site do not depend on the specific choice of the site index. In particular, nearest-neighbour products are invariant under a relabelling of the lattice sites. By performing a shift of the summation index (made valid by the periodic boundary conditions), one obtains
\begin{equation}
\sum_{n=0}^{N-1} S_{n+1}S_{n} = \sum_{n=0}^{N-1} S_{n}S_{n-1}.
\end{equation}
\end{remark}
From $\eqref{eq: potenzialeelasticoinunacatenalineare}$, we write
\begin{align}
V &= \dfrac{k_{el}}{2} \sum_{n=0}^{N-1} \left( 2 S_n^2 - S_n S_{n+1} - S_n S_{n-1} \right) = \notag \\
&= \dfrac{k_{el}}{2} \sum_{n=0}^{N-1} \sum_{m=0}^{N-1} \left( 2 S_n S_m \delta_{n,m} - S_n S_m \delta_{m,n+1} - S_n S_m \delta_{m,n-1} \right).
\end{align}
Indeed,
\begin{equation}
V = \dfrac{1}{2} W^n_m S_n S_m,
\end{equation}
with
\begin{equation}
W^n_m = 2 k_{el} \delta_{n,m} - k_{el} \delta_{n,m+1} - k_{el} \delta_{n,n-1}.
\end{equation}
Let us calculate $\tilde{D}(q)$, we have
\begin{align}
\tilde{D}(q) &= \sum_h W^h_0 e^{i q h a} = \notag \\
&= 2k_{el} - k_{el} \left(e^{iqa} + e^{-iqa}\right) = \notag \\
&= 2k_{el} - 2k_{el} \cos(qa) = \notag \\
&= 2k_{el} \left(1-\cos(qa)\right) \equiv \notag \\
&\equiv k_q ,
\end{align}
where $k_q$ is the eigenvalue of $K(q)$, since $\tilde{D}(q)$ is a $1 \times 1$ matrix. Note that $k_q \geq 0$, as it must be. From $k_q = M \omega_q^2$, we get
\begin{align}
\omega_q^2 &= \dfrac{2k_{el}}{M} \left( 1-\cos(qa) \right) = \notag \\
&= \dfrac{4k_{el}}{M} \sin^2\left( \dfrac{qa}{2} \right),
\end{align}
\begin{equation}
\omega_q = \sqrt{\dfrac{4k_{el}}{M}} \left|\sin\left( \frac{qa}{2} \right)\right|, \ - \dfrac{\pi}{a} \leq q \leq \dfrac{\pi}{a},
\end{equation}
where $q$ is defined in the first Brillouin zone. In the limit $qa \rightarrow 0$, we get
\begin{equation}
\omega_q \sim \sqrt{\dfrac{2k_{el}}{M}} |q| \dfrac{a}{2}.
\end{equation}
Since \( |q| = \frac{2 \pi}{\lambda} \), in the limit \( q a \rightarrow 0 \) we have \( \lambda \gg a \): the lattice is replaced by a continuous structure and \( \omega_q \) becomes linear with respect to \( |q| \). We write
\begin{align}
\omega_q &\sim \sqrt{\dfrac{2k_{el}}{M}} |q| \dfrac{a}{2} = \notag \\
&= V_s |q|,
\end{align}
where 
\begin{equation}
V_s = \sqrt{\dfrac{2k_{el}}{M}} \dfrac{a}{2}
\end{equation}
 is a velocity, indicating the existence of a dispersion relation. \newline
Let us consider the displacement vector \( \mathbf{S}_{n,\mu} \). The vector \( \bm{\varepsilon}_\mu(q) \) characterizes the relative displacement of an atom with respect to its nearest neighbors, representing longitudinal compression and rarefaction waves. This interpretation is consistent with modeling the solid as a one-dimensional ring. Sound waves propagate through the lattice with velocity \( V_s \), corresponding to the acoustic modes. When \( |q| = \frac{\pi}{a} \), the associated wavelength \( \lambda = \frac{2\pi}{|q|} = 2a \) becomes comparable to the lattice constant \( a \). In this regime, the continuum approximation fails, and the discrete nature of the lattice must be taken into account. Here, atoms oscillate out of phase with their nearest neighbors, giving rise to standing wave patterns. Conversely, at \( q = 0 \), all atoms oscillate in phase, corresponding to a uniform translation of the lattice. In one dimension, the spatial modulation of the displacement is captured by the phase factor \( e^{i q n a} \), which satisfies periodic boundary conditions and reflects the translational symmetry of the lattice, indeed it satisfies
\begin{equation}
\lim_{q \rightarrow 0} e^{i q n a} = 1 ,
\end{equation}
corresponding to in-phase waves, and
\begin{equation}
\lim_{|q| \rightarrow \frac{\pi}{a}} e^{i q n a} = e^{i n \pi},
\end{equation}
which implies nearest neighbors oscillate in opposite phase.
\section{1D lattice with two atoms per cell: acoustic branches and optical branches}
Given a linear chain of lattice step $a$ with two atomic species per cell, of mass $M_1$ and $M_2$, we use a model analogous to the previous section. The potential is
\begin{equation}
V = \dfrac{k_{el}}{2} \sum_{n=0}^{N-1} \left( S_n^{(1)} - S_n^{(2)} \right)^2 + \dfrac{k_{el}}{2} \sum_{n=0}^{N-1} \left( S_n^{(2)} - S_{n+1}^{(1)} \right)^2 ,
\end{equation}
where the first and second terms refer to pairs of nearest-neighbor atoms in the same cell (intra-cell) or in different cells (inter-cell), respectively. From
\begin{align}
V &= \dfrac{k_{el}}{2} \sum_{n=0}^{N-1} \left[ \left( S_n^{(1)} \right)^2 + \left( S_n^{(2)} \right)^2 - 2 S_n^{(1)} S_n^{(2)} + \left( S_n^{(2)} \right)^2 + \left( S_{n+1}^{(1)} \right)^2 - 2 S_n^{(2)} S_{n+1}^{(1)} \right] = \notag \\
&= \dfrac{k_{el}}{2} \sum_{n=0}^{N-1} \left[ \left( S_n^{(1)} \right)^2 + 2 \left( S_n^{(2)} \right)^2 - 2 S_n^{(1)} S_n^{(2)} + \left( S_{n+1}^{(1)} \right)^2 - 2 S_n^{(2)} S_{n+1}^{(1)} \right] ,
\end{align}
we use
\begin{equation}
\sum_{n=0}^{N-1} \left( S_{n+1}^{(1)} \right)^2 = \sum_{n=0}^{N-1} \left( S_n^{(1)} \right)^2,
\end{equation}
\begin{equation}
\sum_{n=0}^{N-1} 2 S_n^{(2)} S_{n+1}^{(1)} = \sum_{n=0}^{N-1} \left( S_n^{(2)} S_{n+1}^{(1)} + S_n^{(1)} S_{n-1}^{(2)} \right) ,
\end{equation}
then
\begin{align}
V &= \dfrac{k_{el}}{2} \sum_{n=0}^{N-1} \left[ \left( S_n^{(1)} \right)^2 + 2 \left( S_n^{(2)} \right)^2 - 2 S_n^{(1)} S_n^{(2)} + \left( S_{n}^{(1)} \right)^2 - S_n^{(2)} S_{n+1}^{(1)} - S_n^{(1)} S_{n-1}^{(2)} \right] = \notag \\
&= \dfrac{k_{el}}{2} \sum_{n=0}^{N-1} \left[ 2 \left( S_n^{(1)} \right)^2 + 2 \left( S_n^{(2)} \right)^2 - 2 S_n^{(1)} S_n^{(2)} - S_n^{(2)} S_{n+1}^{(1)} - S_n^{(1)} S_{n-1}^{(2)} \right].
\end{align}
Here, $\textbf{n}$ and $\textbf{m}$ are replaced by scalars, $\tau=2$, $i=k=1$ and $\mu, \nu = 1,2$, then the force constants are
\begin{align}
W^{n-m,\mu}_{0,\nu} &\equiv W^{h,\mu}_{0,\nu} = \notag \\
&= 2 k_{el} \delta_{h,0} \delta_{\mu,\nu} - k_{el} \delta_{h,0} \delta_{\mu,1} \delta_{\nu,2} - k_{el} \delta_{h,0} \delta_{\mu,2} \delta_{\nu,1} - k_{el} \delta_{h,-a} \delta_{\mu,2} \delta_{\nu,1} - k_{el} \delta_{h,a} \delta_{\mu,1} \delta_{\nu,2}.
\end{align}
We have
\begin{equation}
D^{n,\mu}_{m,\nu} = \sqrt{\dfrac{M}{M_{\mu} M_{\nu}}} W^{n,\mu}_{m,\nu} ,
\end{equation}
\begin{equation}
\left[ \tilde{D}(q) \right]^{\mu}_{\nu} = \sum_h D^{h,\mu}_{0,\nu} e^{i q R_{h}} ,
\end{equation}
and $\mu, \ \nu=1,2$, then $\tilde{D}(q)$ is a $2 \times 2$ matrix given by
\begin{equation}
\tilde{D}(q) 
=
\begin{pmatrix}
k_{el} \dfrac{2M}{M_1} & - k_{el} \dfrac{M}{\sqrt{M_1 M_2}} (1+e^{iqa}) \\ \\
- k_{el} \dfrac{M}{\sqrt{M_1 M_2}} (1+e^{-iqa}) & k_{el} \dfrac{2M}{M_2}
\end{pmatrix} ,
\end{equation}
and it is hermitian. Let us compute the eigenvalues $M \omega_q^2$ of $\tilde{D}(q)$. We divide the matrix $\tilde{D}(q)$ by $M$ and have
\begin{equation}
\det 
\begin{pmatrix}
k_{el} \dfrac{2}{M_1} - \omega^2 & - k_{el} \dfrac{1}{\sqrt{M_1 M_2}} (1+e^{iqa}) \\ \\
- k_{el} \dfrac{1}{\sqrt{M_1 M_2}} (1+e^{-iqa}) & k_{el} \dfrac{2}{M_2} - \omega^2
\end{pmatrix}
= 0,
\end{equation}
\begin{equation}
\left( \dfrac{2k_{el}}{M_1} - \omega^2 \right) \left( \dfrac{2k_{el}}{M_2} - \omega^2 \right) - \dfrac{k_{el}^2}{M_1 M_2} (1+e^{iqa}) (1+e^{-iqa}) = 0 ,
\end{equation}
\begin{equation}
\omega^4 - \left( \dfrac{2k_{el}}{M_1} + \dfrac{2k_{el}}{M_2} \right) \omega^2 + \dfrac{k_{el}^2}{M_1 M_2} - \dfrac{k_{el}^2}{M_1 M_2} (2 + 2 \cos(qa)) = 0,
\end{equation}
\begin{equation}
\omega^4 - 2 k_{el} \left( \dfrac{1}{M_1} + \dfrac{1}{M_2} \right) \omega^2 + \dfrac{2k_{el}^2}{M_1 M_2} (1 - \cos(qa)) = 0,
\end{equation}
so
\begin{equation}
\omega^2_{\pm} = k_{el} \left( \dfrac{1}{M_1} + \dfrac{1}{M_2} \right) \pm k_{el} \sqrt{\left( \dfrac{1}{M_1} + \dfrac{1}{M_2} \right)^2 - \dfrac{2}{M_1 M_2} (1-\cos(qa))} .
\label{eq: omega+-reticolo1Dduespecieatomicheforma1}
\end{equation}
From $\eqref{eq: omega+-reticolo1Dduespecieatomicheforma1}$ it follows
\begin{equation}
\omega^2_{-}(q=0) = 0,
\end{equation}
\begin{equation}
\omega^2_{+}(q=0) = 2 k_{el} \dfrac{(M_1+M_2)}{M_1M_2},
\end{equation}
\begin{align}
\omega^2_{\pm} \left( q = \dfrac{\pi}{a} \right) &= k_{el} \left( \dfrac{1}{M_1} + \dfrac{1}{M_2} \right) \pm k_{el} \sqrt{\left( \dfrac{1}{M_1} + \dfrac{1}{M_2} \right)^2 - \dfrac{2}{M_1 M_2} (1-\cos(\pi))} = \notag \\
&= k_{el} \left( \dfrac{1}{M_1} + \dfrac{1}{M_2} \right) \pm \dfrac{k_{el}}{M_1M_2} \left|M_2 - M_1 \right| .
\end{align}
For instance, assuming \( M_2 > M_1 \), then $\omega^2_{+} = \frac{2k_{el}}{M_1}$, $\omega^2_{-} = \frac{2k_{el}}{M_2}$.
\section{3D lattice with many atoms per cell}
Here, we study the general case of $\tau \geq 2$ atomic species per cell in three dimensions, specifically in a simple cubic lattice, and show that three acoustic frequencies exist. From
\begin{align}
\left[ \tilde{D}(\textbf{q}) \right]_{\substack{\mu,\nu \\ i,k}} \ &= \ \sum_{\textbf{h}} D^{\textbf{h},\mu,i}_{\textbf{0},\nu,k} \ e^{i \textbf{q} \cdot \textbf{R}_{\textbf{h}}} \ = \notag \\
&= \sum_{\textbf{h}} \sqrt{\dfrac{M^2}{M_{\mu}M_{\nu}}} W^{\textbf{h},\mu,i}_{\textbf{0},\nu,k} \ e^{i \textbf{q} \cdot \textbf{R}_{\textbf{h}}},
\end{align}
we have
\begin{align}
\left[ \tilde{D}(\textbf{0}) \right]_{\substack{\mu,\nu \\ i,k}} \ &= \ \sum_{\textbf{h}} \sqrt{\dfrac{M^2}{M_{\mu}M_{\nu}}} W^{\textbf{h},\mu,i}_{\textbf{0},\nu,k} \ = \notag \\
&= \sum_{\textbf{h}} \sqrt{\dfrac{M^2}{M_{\mu}M_{\nu}}} W^{\textbf{0},\mu,i}_{-\textbf{h},\nu,k} \ = \notag \\
&= \sum_{\textbf{h}} \sqrt{\dfrac{M^2}{M_{\mu}M_{\nu}}} W^{\textbf{0},\mu,i}_{\textbf{h},\nu,k} .
\label{eq: matriceD3Dlatticewithmanyatoms}
\end{align}
In the last step we used that $\mathbf{h}$ is a dummy index. 
\begin{theorem}
The matrix $\eqref{eq: matriceD3Dlatticewithmanyatoms}$ has three eigenvalues equal to zero.
\begin{proof}
We seek vectors $\mathbf{V}$ that satisfy the eigenvalue equation, i.e.,
\begin{equation}
\sum_{\nu,k} \left[ \tilde{D}(\textbf{0}) \right]_{\substack{\mu,\nu \\ i,k}} V_{\nu,k} = 0,
\end{equation}
\begin{equation}
\sum_{\textbf{h},\nu,k} \sqrt{\dfrac{M^2}{M_{\mu}M_{\nu}}} W^{\textbf{0},\mu,i}_{\textbf{h},\nu,k} V_{\nu,k} = 0,
\end{equation}
\begin{equation}
\sum_{\textbf{h},\nu,k} \sqrt{\dfrac{1}{M_{\nu}}} W^{\textbf{0},\mu,i}_{\textbf{h},\nu,k} V_{\nu,k} = 0.
\end{equation}
We now insert the following three vectors
\begin{equation}
V^{(1)}_{\nu,k}= \sqrt{M_{\nu}} \delta_{k,1},
\end{equation}
\begin{equation}
V^{(2)}_{\nu,k}= \sqrt{M_{\nu}} \delta_{k,2} ,
\end{equation}
\begin{equation}
V^{(3)}_{\nu,k}= \sqrt{M_{\nu}} \delta_{k,3},
\end{equation}
into the eigenvalue equation
\begin{equation}
\sum_{\textbf{h},\nu,k} \sqrt{\dfrac{1}{M_{\nu}}} W^{\textbf{0},\mu,i}_{\textbf{h},\nu,k} V^{(j)}_{\nu,k} = 0, \ \forall \ j = 1,2,3.
\end{equation}
Indeed, from $\eqref{eq: proprietàcoefficientiWsommasucellaesuspecieatomicadirezionefissata}$, it follows that
\begin{align}
\sum_{\textbf{h},\nu,k} \sqrt{\dfrac{1}{M_{\nu}}} W^{\textbf{0},\mu,i}_{\textbf{h},\nu,k} \sqrt{M_{\nu}} \delta_{k,j} &= \sum_{\textbf{h},\nu,k} W^{\textbf{0},\mu,i}_{\textbf{h},\nu,k} \delta_{k,j} = \notag \\
&= \sum_{\textbf{h},\nu} \left. W^{\textbf{0},\mu,i}_{\textbf{h},\nu,j} \right|_{j \ fixed} \ = 0, \ \forall \ j = 1,2,3.
\end{align}
We have shown that among the $3 \tau$ branches, at least three correspond to acoustic modes.
\end{proof}
\end{theorem}
It can be shown that only three branches are acoustic, one corresponding to longitudinal motion and the other two to transverse motion. Let us calculate the displacements \( S^{(\mathrm{ac.})}_{\mathbf{n}, \mu, i} \) of these acoustic branches. By definition, the eigenvectors \( V^{(j)}_{\nu,k} \) are equal to the displacements \( U_{\mathbf{n}, \mu, i} \), hence
\begin{align}
S^{(ac.)}_{\textbf{n},\mu,i} &= \sqrt{\dfrac{M}{M_{\mu}}} U_{\textbf{n},\mu,i} = \notag \\
&= \sqrt{\dfrac{M}{M_{\mu}}} V^{(i)}_{\nu,k} = \notag \\
&= \sqrt{\dfrac{M}{M_{\mu}}} \sqrt{M_{\mu}} \delta_{k,i} = \notag \\
&= \sqrt{M} \delta_{k,i}, \ i = 1,2,3.
\end{align}
In other words, the displacements of the acoustic branches depend neither on the atom nor on the cell: all atoms move in phase, and these longitudinal motions are responsible for the propagation of sound waves in the solid. Let us turn to the optical branches. The matrix \(\tilde{D}(\mathbf{q})\) is Hermitian, so the eigenvectors \( V_{\nu,k}^{(s)} \) corresponding to different eigenvalues are orthogonal to each other and therefore satisfy
\begin{equation}
\sum_{\nu,k} V_{\nu,k}^{(s)*}(\textbf{q}) V_{\nu,k}^{(s')}(\textbf{q}) = \delta_{s,s'}.
\end{equation}
Consider an optical branch and any of the three acoustic branches, i.e.,
\begin{equation}
\sum_{\nu,k} V_{\nu,k}^{(o)*}(\textbf{q}) V_{\nu,k}^{(a)}(\textbf{q}) = 0.
\end{equation}
If \( \mathbf{q} = \mathbf{0} \), the eigenvectors of the acoustic branches are known. Moreover, the matrix \( \tilde{D}(\mathbf{0}) \) is real and symmetric, so its eigenvectors can be chosen to be real.
\begin{align}
0 &= \sum_{\nu,k} V_{\nu,k}^{(o)}(\textbf{0}) V_{\nu,k}^{(a)}(\textbf{0}) = \notag \\
&= \sum_{\nu,k} V_{\nu,k}^{(o)}(\textbf{0}) \sqrt{M_{\mu}} \delta_{k,j} ,
\end{align}
\begin{equation}
\sum_{\nu} V_{\nu,j}^{(o)}(\textbf{0}) \sqrt{M_{\mu}} = 0,
\end{equation}
\begin{equation}
\sum_{\nu} V_{\nu,j}^{(o)}(\textbf{0}) \sqrt{\dfrac{M}{M_{\mu}}} \sqrt{M_{\mu}} \sqrt{\dfrac{M_{\mu}}{M}} = 0 ,
\end{equation}
\begin{equation}
\sum_{\nu} S_{\textbf{n}=\textbf{0},\mu,j}(\textbf{q}=\textbf{0}) \sqrt{M_{\mu}} \sqrt{\dfrac{M_{\mu}}{M}} = 0 ,
\end{equation}
\begin{equation}
\dfrac{1}{\sqrt{M}} \sum_{\nu} S_{\textbf{n}=\textbf{0},\mu,j}(\textbf{q}=\textbf{0}) M_{\nu} = 0 ,
\end{equation}
\begin{equation}
\dfrac{1}{M} \sum_{\nu} S_{\textbf{n}=\textbf{0},\mu,j}(\textbf{q}=\textbf{0}) M_{\nu} = 0.
\label{eq: proprietàcentrodimassadeglispostamentidellebrancheottiche}
\end{equation}
The equation $\eqref{eq: proprietàcentrodimassadeglispostamentidellebrancheottiche}$ states that the center of mass of the optical branches remains stationary. This implies that atoms in the optical modes move out of phase with respect to their neighbors, effectively forming oscillating electric dipoles within the unit cell. The optical branch is so named because the frequency it attains at \( \mathbf{q} = \mathbf{0} \) typically lies in the infrared region, and can interact with light. When electromagnetic radiation in the visible or infrared spectrum impinges on the crystal, it can couple to these vibrational modes. The lattice then becomes excited and vibrates collectively; each dipole oscillates and interacts with the radiation. This coupling leads to the formation of hybrid light-matter quasiparticles known as polaritons. These quasiparticles exhibit mixed properties and play an important role in the optical response of polar crystals, a topic that we will explore in detail in one of the following chapters.
\section{Lattice specific heat}
The phonon Hamiltonian $\eqref{eq: hamiltonianavibrazionireticolariquantizzate}$ provides an adequate description of the lattice contribution to the specific heat in metals. Within the Born-Oppenheimer approximation, the total specific heat arises from both electronic and ionic degrees of freedom; the lattice specific heat specifically accounts for the ionic part. We have
\begin{equation}
\langle \hat{\mathcal{H}}_{ret.} \rangle = \sum_{s} \int D_{s}(\omega) \dfrac{\hslash \omega}{e^{\beta \hslash \omega_{\textbf{q},s}} - 1} d\omega,
\end{equation}
where 
\begin{equation}
D_{s}(\omega) = \sum_{\textbf{q}} \delta\left( \omega - \omega_{\textbf{q},s} \right)
\end{equation}
is the density of modes or density of states of the $s$ branch and it takes account for the number of vibrations for each energy. 
\subsection{Einstein's approximation of the density of modes}
Einstein's approximation of the density of modes consists of setting 
\begin{equation}
\omega_{\textbf{q},s} = \omega_0 > 0, \ \forall \ \textbf{q}, \ s
\end{equation}
that is, all branches become optical. We have 
\begin{equation}
D_{s}(\omega) = N \delta(\omega - \omega_0),
\end{equation}
\begin{align}
\langle \hat{\mathcal{H}}_{ret.} \rangle &= \sum_{s} \int D_{s}(\omega) \dfrac{\hslash \omega}{e^{\beta \hslash \omega_{\textbf{q},s}} - 1} d\omega = \notag \\
&= \sum_s N \dfrac{\hslash \omega_0}{e^{\beta \hslash \omega_0} - 1} = \notag \\
&= 3 \tau N \dfrac{\hslash \omega_0}{e^{\frac{\hslash \omega_0}{k_B T}} - 1}.
\end{align}
Let us compute the energy of a mole: for each species we consider an Avogadro number of atoms (recall $N=N_A$, $N_A k_B = R$), then
\begin{align}
\mathcal{E}_{mol} &= 3 \tau N_A \dfrac{\hslash \omega_0}{e^{\frac{\hslash \omega_0}{k_B T}} - 1} \equiv \notag \\
&\equiv 3 \tau N_A \dfrac{\hslash \omega_0}{e^{\frac{T_E}{T}} - 1} ,
\end{align}
and the molar specific heat at constant volume is
\begin{align}
C_V &= \dfrac{\partial \mathcal{E}_{mol}}{\partial T} = \notag \\
&= \dfrac{3 \tau N_A \hslash \omega_0}{\left( e^{\frac{T_E}{T}} -1 \right)^2} e^{\frac{T_E}{T}} \dfrac{T_E}{T} = \notag \\ 
&= 3 \tau N_A \hslash \omega_0 \dfrac{1}{\left( e^{\frac{T_E}{T}} -1 \right)^2} e^{\frac{T_E}{T}} \dfrac{T_E}{T} \dfrac{k_B}{k_B} = \notag \\
&= \dfrac{3 \tau R}{\left( e^{\frac{T_E}{T}} -1 \right)^2} \left( \dfrac{T_E}{T} \right)^2 e^{\frac{T_E}{T}}.
\end{align}
Let us analyze the behavior of the specific heat in the limits of high and low temperature. At high temperature, i.e., \( T \gg T_E \), the specific heat approaches $C_V \to 3 \tau R$, which is consistent with the equipartition theorem: classically, each quadratic degree of freedom contributes an average energy of \(\frac{k_B T}{2}\). Therefore, the total classical energy is given by $3 \tau \frac{k_B T}{2} 2 N$, where the factor $2$ accounts for the two degrees of freedom of the harmonic oscillator. Differentiating this energy with respect to temperature yields $C_V = 3 \tau N k_B$, matching the Dulong-Petit law. On the other hand, Einstein's approximation fails to correctly describe the low-temperature behavior. For \( T \ll T_E \), it predicts an exponential decay of the specific heat, $C_V \sim e^{-\frac{T_E}{T}}$, which contradicts experimental observations. This discrepancy arises because Einstein's model neglects the low-energy excitations associated with the acoustic branches near \(\mathbf{q} \to \mathbf{0}\). Only by including these acoustic modes can the isochoric specific heat be accurately reproduced. Indeed, as temperature decreases below a certain threshold, the excitation of these modes becomes energetically inaccessible, causing them to cease contributing to the specific heat. Thus, to correctly capture the low-temperature behavior, we must go beyond Einstein's approximation and explicitly consider the contributions from the acoustic branches.
\subsection{Debye's approximation of the density of modes}
Debye assumes that the frequencies are linear with respect to $\textbf{q}$. Since in the limit $qa \rightarrow 0$ we have $\lambda \gg a$, Debye's approximation is equivalent to replacing the crystal lattice with a continuous elastic medium and the entire spectrum becomes continuous. The dispersion relation becomes linear
\begin{equation}
\omega_{\textbf{q},s} = \left|\textbf{q}\right| V_s,
\end{equation}
where the velocity $V_s$ can have only two values: $V_t$ for the two transverse modes and $V_l$ for the longitudinal mode. From spatial periodicity
\begin{align}
q_x (x+L_x) &= q_x x + q_x L_x = \notag \\
&= qx + q_x n_x \lambda = \notag \\
&= q_x x + \dfrac{2\pi}{\lambda} n_x \lambda = \notag \\
&= q_x x + 2 \pi n_x,
\end{align}
where $n_x$ is the number of waves with wave vector $\textbf{q}$ and in the direction $x$ of the lattice, of length $L_x$. Given the wave vector $(q_x,q_y,q_z)$, the density of the number of waves with wave vector between $\textbf{q}$ and $\textbf{q}+d\textbf{q}$ is
\begin{equation}
dn_x dn_y dn_z = \dfrac{L_x L_y L_z}{(2 \pi)^3} dq_x dq_y dq_z.
\end{equation}
Given $D_{s}(\omega)$, in the thermodynamic limit approximation $\eqref{eq: illimitetermodinamico}$, we have
\begin{align}
D_{s}(\omega) &= \dfrac{L_x L_y L_z}{(2 \pi)^3} \int d^3\textbf{q} \ \delta \left( \omega - |\textbf{q}|V_s \right) \equiv \notag \\
&\equiv \dfrac{V}{8 \pi^3} \int d^3\textbf{q} \ \delta \left( \omega - |\textbf{q}|V_s \right),
\end{align}
where $V$ is the volume of the crystal. Using polar coordinates, 
\begin{equation}
D_{s}(\omega) = 4 \pi \dfrac{V}{8 \pi^3} \int \textbf{q}^2 dq \ \delta \left( \omega - |\textbf{q}|V_s \right) ,
\end{equation}
\begin{equation}
D_{s}(\omega) = 4 \pi \dfrac{V}{8 \pi^3} \left( \dfrac{\omega_{\textbf{q},s}}{V_s} \right)^2 \int dq \ \delta \left( \omega - |\textbf{q}|V_s \right).
\end{equation}
Now, we use the following property of Dirac's delta
\begin{equation}
\int \delta\left[ f(x) \right] dx = \int \sum_i \dfrac{\delta(x-x_i)}{\left|f'(x_i)\right|} dx, \ f(x_i) \neq 0, \ \forall \ i ,
\end{equation}
$x_i$ being the zeros of $f$. In our case, the function is $\omega(\textbf{q})$ and the first derivative is equal to $V_s$. The density of modes is written as
\begin{equation}
D_s(\omega) = 4 \pi \dfrac{V}{8 \pi^3} \left( \dfrac{\omega}{V_s} \right)^2 \dfrac{1}{V_s} ,
\end{equation}
then the energy becomes 
\begin{align}
\mathcal{E} &= \sum_s \int D_{s}(\omega) \dfrac{\hslash \omega}{e^{\frac{\hslash \omega}{k_B T}} - 1} d\omega = \notag \\
&= \sum_s \int 4 \pi \dfrac{V}{8 \pi^3} \left( \dfrac{\omega}{V_s} \right)^2 \dfrac{1}{V_s}  \dfrac{\hslash \omega}{e^{\frac{\hslash \omega}{k_B T}} - 1} d\omega = \notag \\
&= 4 \pi \dfrac{V}{8 \pi^3} \sum_s \left( \dfrac{1}{V_s} \right)^3 \int \dfrac{\hslash \omega^3}{e^{\frac{\hslash \omega}{k_B T}} - 1} d\omega = \notag \\
&= 4 \pi \dfrac{V}{8 \pi^3} \sum_s \left( \dfrac{1}{V_s} \right)^3 \int \dfrac{\hslash \omega^3}{e^{\frac{\hslash \omega}{k_B T}} - 1} d\omega \notag \\
&= 4 \pi \dfrac{V}{8 \pi^3} \left[ \dfrac{2}{V^3_t} + \dfrac{1}{V^3_l} \right] \int \dfrac{\hslash \omega^3}{e^{\frac{\hslash \omega}{k_B T}} - 1} d\omega.
\end{align}
The definition of $D_s(\omega)$ implies a constraint, indeed
\begin{align}
\sum_s \int d\omega D_{s}(\omega) &= \sum_{\textbf{q},s} \int d\omega \delta\left( \omega - \omega_{\textbf{q},s} \right) = \notag \\
&= \sum_{\textbf{q},s} 1 = \notag \\
&= 3 N ,
\end{align}
i.e., the total number of acoustic phonons must be equal to the number of elementary cells in the lattice, so
\begin{align}
\sum_s \int d\omega D_{s}(\omega) &= \int_0^{\omega_D} d\omega \ 4 \pi \dfrac{V}{8 \pi^3} \left[ \dfrac{2}{V^3_t} + \dfrac{1}{V^3_l} \right] \omega^2 = \notag \\
&= 3 N.
\end{align}
Debye's approximation implies that a maximum frequency $\omega_D$, the Debye frequency, exists. The integral with respect to the measure $d\omega$ gives
\begin{equation}
4 \pi \dfrac{V}{8 \pi^3} \left[ \dfrac{2}{V^3_t} + \dfrac{1}{V^3_l} \right] \dfrac{\omega_D^3}{3} = 3 N ,
\end{equation}
from which we compute $\omega_D$. The Debye frequency is related to a temperature 
\begin{equation}
T_D = \dfrac{\hslash \omega_D}{k_B},
\end{equation}
which is the Debye temperature. We insert into the energy $\mathcal{E}$ the quantity
\begin{equation}
4 \pi \dfrac{V}{8 \pi^3} \left[ \dfrac{2}{V^3_t} + \dfrac{1}{V^3_l} \right] = \dfrac{9N}{\omega_D^3} ,
\end{equation}
then 
\begin{equation}
\mathcal{E} = \dfrac{9N\hslash}{\omega_D^3} \int_0^{\omega_D} d\omega \dfrac{\omega^3}{e^{\frac{\hslash \omega}{k_B T}} - 1}.
\end{equation}
We multiply and divide by $\hslash^3$
\begin{equation}
\mathcal{E} = \dfrac{9N}{(\hslash \omega_D)^3} \hslash \int_0^{\omega_D} d\omega \dfrac{(\hslash \omega)^3}{e^{\frac{\hslash \omega}{k_B T}} - 1} ,
\end{equation}
and we set $x = \dfrac{\hslash \omega}{k_B T}$, then
\begin{align}
\mathcal{E} &= \dfrac{9N}{(\hslash \omega_D)^3} \hslash \int_0^{\frac{\hslash \omega_D}{k_B T}} dx \dfrac{k_B T}{\hslash} (k_B T)^3 \dfrac{x^3}{e^x - 1} = \notag \\
&= \dfrac{9N(k_B T)^4}{(\hslash \omega_D)^3} \int_0^{\frac{\hslash \omega_D}{k_B T}} dx \dfrac{x^3}{e^x - 1} = \notag \\
&= \dfrac{9N(k_B T)^4}{(\hslash \omega_D)^3} \int_0^{\frac{T_D}{T}} dx \dfrac{x^3}{e^x - 1} = \notag \\
&= 9Nk_B T \left( \dfrac{T}{T_D} \right)^3 \int_0^{\frac{T_D}{T}} dx \dfrac{x^3}{e^x - 1}.
\end{align}
For a mole, we have
\begin{equation}
\mathcal{E}_{mol} = 9 R T \left( \dfrac{T}{T_D} \right)^3 \int_0^{\frac{T_D}{T}} dx \dfrac{x^3}{e^x - 1},
\end{equation}
and the molar specific heat is
\begin{equation}
C_{V,mol} = \dfrac{\partial \mathcal{E}_{mol}}{\partial T}.
\end{equation}
We now examine the behavior of the molar internal energy and specific heat in the high- and low-temperature limits, relative to the Debye temperature \( T_D \).
In the high-temperature regime, where \( T \gg T_D \), we have
\begin{equation}
\dfrac{T}{T_D} \gg 1.
\end{equation}
In this limit, the upper limit of integration in the Debye model becomes small, so the integrand is significant only for \( x \ll 1 \), and we can use the approximation \( e^x \approx 1 + x \). The molar internal energy then becomes
\begin{align}
\mathcal{E}_{mol}(T \gg T_D) &\sim 9 R T \left( \dfrac{T}{T_D} \right)^3 \int_0^{\frac{T_D}{T}} dx \ x^2 = \notag \\
&= 9 R T \left( \dfrac{T}{T_D} \right)^3 \left( \dfrac{T_D}{T} \right)^3 \dfrac{1}{3} = \notag \\
&= 3 R T.
\end{align}
As a result, the molar specific heat at constant volume is
\begin{equation}
C_{V,mol} = 3 R,
\end{equation}
which coincides with the classical Dulong-Petit law. In the low-temperature regime, where \( T \ll T_D \), the upper limit of integration becomes large, and the full form of the Bose-Einstein distribution must be retained. In this case, the molar internal energy behaves as $\mathcal{E}_{mole}(T \ll T_D) \propto T^4$, so the molar specific heat scales as
\begin{equation}
C_{V,mol} \propto T^{3}.
\end{equation}
This \(T^3\) dependence agrees with experimental observations and confirms the success of Debye's model at low temperatures.
\chapter{Step III of the solid model: electron-phonon interaction}
In this chapter, we derive the electron-phonon interaction Hamiltonian, a key element in the study of the electronic properties of solids. The interaction between electrons and phonons lies at the heart of many physical phenomena, including electrical resistance in metals, the formation of Cooper pairs in superconductors, and the renormalization of the electronic mass in complex materials. Our goal is to construct an effective model that describes how electrons couple to phonons in a crystal, starting from microscopic considerations and arriving at a compact expression suitable for use within the framework of second quantization. In particular, we will derive the expression
\begin{equation}
\sum_{\substack{m,m',\sigma,s, \\ \textbf{k},\textbf{G},\textbf{q}}} T_{\textbf{k},\textbf{q},\textbf{G}}^s \ C^{\dag}_{m',\textbf{k}+\textbf{q}+\textbf{G},\sigma} C_{m,\textbf{k},\sigma} \left( a_{\textbf{q},s} + a^{\dag}_{-\textbf{q},s} \right)
\end{equation}
in the Hamiltonian $\eqref{eq: Hamiltonianadiunsolido}$.
\section{Electron-phonon interaction}
The electron-ion interaction is given by
\begin{equation}
V^{(e-i)} = \sum_j \sum_{\textbf{n},\mu} V_{\mu} \left[ \textbf{r}_j - \textbf{R}_{\textbf{n}} - \textbf{d}_{\mu} - \textbf{S}_{\textbf{n},\mu} \right] ,
\end{equation}
where the sums are indexed by the number of electrons \( j \), the number of nuclei \( \mathbf{n} \), and the number of atomic species \( \mu \). We add and subtract the potential energy evaluated at \(\mathbf{S}_{\mathbf{n},\mu} = \mathbf{0}\), i.e., at the ionic equilibrium positions, that is
\begin{equation}
V^{(e-i)} = \sum_j \sum_{\textbf{n},\mu} V_{\mu} \left[ \textbf{r}_j - \textbf{R}_{\textbf{n}} - \textbf{d}_{\mu} \right] + \sum_j \sum_{\textbf{n},\mu} \left\lbrace V_{\mu} \left[ \textbf{r}_j - \textbf{R}_{\textbf{n}} - \textbf{d}_{\mu} - \textbf{S}_{\textbf{n},\mu} \right] - V_{\mu} \left[ \textbf{r}_j - \textbf{R}_{\textbf{n}} - \textbf{d}_{\mu} \right] \right\rbrace.
\end{equation}
The first term originates from the fixed ions at the equilibrium positions, and thus it is included in the electronic problem studied in the previous chapter. The second term corresponds to the interaction between electrons and ions when the latter are displaced from their equilibrium positions. Therefore, the quantity that characterizes the electron-photon interaction, which we will now analyze, is
\begin{equation}
\sum_j \sum_{\textbf{n},\mu} \left\lbrace V_{\mu} \left[ \textbf{r}_j - \textbf{R}_{\textbf{n}} - \textbf{d}_{\mu} - \textbf{S}_{\textbf{n},\mu} \right] - V_{\mu} \left[ \textbf{r}_j - \textbf{R}_{\textbf{n}} - \textbf{d}_{\mu} \right] \right\rbrace.
\end{equation}
Since the ions oscillate with small amplitude around their equilibrium positions, we expand the potential \( V_{\mu} \left( \mathbf{r}_j - \mathbf{R}_{\mathbf{n}} - \mathbf{d}_{\mu} - \mathbf{S}_{\mathbf{n},\mu} \right) \) as a Taylor series with respect to \(\mathbf{S}_{\mathbf{n},\mu}\). We have
\begin{equation}
V_{\mu} \left[ \textbf{r}_j - \textbf{R}_{\textbf{n}} - \textbf{d}_{\mu} - \textbf{S}_{\textbf{n},\mu} \right] - V_{\mu} \left[ \textbf{r}_j - \textbf{R}_{\textbf{n}} - \textbf{d}_{\mu} \right] \simeq - \sum_j \sum_{\textbf{n},\mu} \textbf{S}_{\textbf{n},\mu} \cdot \left. \nabla_{\textbf{S}_{\textbf{n},\mu}} V_{\mu} \left[ \textbf{r}_j - \textbf{R}_{\textbf{n}} - \textbf{d}_{\mu} - \textbf{S}_{\textbf{n},\mu} \right] \right|_{\textbf{S}_{\textbf{n},\mu}=\textbf{0}} ,
\end{equation}
where the minus sign in the second term arises from the functional dependence of \( V^{(e-i)} \) on \(-\mathbf{S}_{\mathbf{n},\mu} \). We aim to express this quantity in second quantization. Since \( V_{\mu} \) (and consequently \( \nabla_{\mathbf{S}_{\mathbf{n},\mu}} V_{\mu} \)) is a one-body potential, we employ the basis of Bloch waves multiplied by the standard spin eigenfunctions, that is $\eqref{eq: operatorecampodistruzioneinfunzioniBloch}$, $\eqref{eq: operatorecampocreazioneinfunzioniBloch}$. Then
\begin{align}
\hat{\mathcal{H}}_{e-ph} &= - \sum_{\textbf{n},\mu} \textbf{S}_{\textbf{n},\mu} \cdot \int dx \ \hat{\psi}^\dag(x) \left. \nabla_{\textbf{S}_{\textbf{n},\mu}} V_{\mu} \left[ \textbf{r} - \textbf{R}_{\textbf{n}} - \textbf{d}_{\mu} - \textbf{S}_{\textbf{n},\mu} \right] \right|_{\textbf{S}_{\textbf{n},\mu}=\textbf{0}} \hat{\psi}(x) = \notag \\
&= - \sum_{\textbf{n},\mu} \textbf{S}_{\textbf{n},\mu} \cdot \sum_{\sigma,\sigma'} \sum_{n,n'} \sum_{\textbf{k},\textbf{k}'} C^{\dag}_{n',\textbf{k}',\sigma'} C_{n,\textbf{k},\sigma} \notag \\
& \ \ \ \ \ \ \ \ \ \ \sum_s \int d^3\textbf{r} \ \varphi^*_{n',\textbf{k}'}(\textbf{r}) \chi^*_{\sigma'}(s) \left. \nabla_{\textbf{S}_{\textbf{n},\mu}} V_{\mu} \left[ \textbf{r} - \textbf{R}_{\textbf{n}} - \textbf{d}_{\mu} - \textbf{S}_{\textbf{n},\mu} \right] \right|_{\textbf{S}_{\textbf{n},\mu}=\textbf{0}} \varphi_{n,\textbf{k}}(\textbf{r}) \chi_{\sigma}(s) = \notag \\
&= - \sum_{\textbf{n},\mu} \textbf{S}_{\textbf{n},\mu} \cdot \sum_{\sigma,\sigma'} \sum_{n,n'} \sum_{\textbf{k},\textbf{k}'} C^{\dag}_{n',\textbf{k}',\sigma} C_{n,\textbf{k},\sigma} \delta_{\sigma,\sigma'} \notag \\
& \ \ \ \ \ \ \ \ \ \ \int d^3\textbf{r} \ \varphi^*_{n',\textbf{k}'}(\textbf{r}) \left. \nabla_{\textbf{S}_{\textbf{n},\mu}} V_{\mu} \left[ \textbf{r} - \textbf{R}_{\textbf{n}} - \textbf{d}_{\mu} - \textbf{S}_{\textbf{n},\mu} \right] \right|_{\textbf{S}_{\textbf{n},\mu}=\textbf{0}} \varphi_{n,\textbf{k}}(\textbf{r}) = \notag \\
&= - \sum_{\textbf{n},\mu} \textbf{S}_{\textbf{n},\mu} \cdot \sum_{\sigma} \sum_{n,n'} \sum_{\textbf{k},\textbf{k}'} C^{\dag}_{n',\textbf{k}',\sigma} C_{n,\textbf{k},\sigma} \int d^3\textbf{r} \ \varphi^*_{n',\textbf{k}'}(\textbf{r}) \left. \nabla_{\textbf{S}_{\textbf{n},\mu}} V_{\mu} \left[ \textbf{r} - \textbf{R}_{\textbf{n}} - \textbf{d}_{\mu} - \textbf{S}_{\textbf{n},\mu} \right] \right|_{\textbf{S}_{\textbf{n},\mu}=\textbf{0}} \varphi_{n,\textbf{k}}(\textbf{r}).
\end{align}
From the structure of \(\hat{\mathcal{H}}_{e-f}\), it follows that an electron is annihilated in band \(n\) with quasi-momentum \(\mathbf{k}\) and spin component \(\sigma\), and simultaneously created in band \(n'\) with quasi-momentum \(\mathbf{k}'\) and the same spin component \(\sigma\). The band index \(n'\) may coincide with \(n\) or differ from it: thus, the electron-phonon interaction can induce both intra-band and inter-band transitions. The Hamiltonian \(\hat{\mathcal{H}}_{e-ph}\) involves the quantized displacement \(\mathbf{S}_{\mathbf{n},\mu}\) defined in equation $\eqref{eq: spostamentoquantizzatofononi}$, which is expressed in terms of the phonon annihilation and creation operators \(a_{\mathbf{q},s}\) and \(a^{\dagger}_{-\mathbf{q},s}\). This corresponds respectively to the annihilation of a phonon with momentum \(\mathbf{q}\) or the creation of one with momentum \(-\mathbf{q}\). Through interaction with a phonon of momentum \(\pm \mathbf{q}\), the electron undergoes a transition from state \(\mathbf{k}\) to \(\mathbf{k}'\), with conservation of total momentum requiring that
\begin{equation}
\textbf{k}' = \textbf{k} + \textbf{q}.
\end{equation}
Given the expression for $\textbf{S}_{\textbf{n},\mu}$, the task is to compute the matrix element of the electron-phonon interaction Hamiltonian. Recall that we have imposed periodic boundary conditions: the crystal, with volume $V$, is repeated infinitely many times along each crystallographic direction $i$. As a result, the potential $V_{\mu}$ is periodic over the entire supercrystal, that is,
\begin{equation}
V_{\mu}\left[ \textbf{r} - \textbf{R}_{\textbf{n}} - \textbf{d}_{\mu} - \textbf{S}_{\textbf{n},\mu} \right] = \sum_{\textbf{q}_1} \dfrac{e^{i \textbf{q}_1 \cdot \left( \textbf{r} - \textbf{R}_{\textbf{n}} - \textbf{d}_{\mu} - \textbf{S}_{\textbf{n},\mu} \right)}}{\sqrt{N}} V_{\mu}(\textbf{q}_1),
\end{equation}
where  
\begin{equation}
\textbf{q}_1 = \dfrac{\tilde{n}_1}{N_1} \textbf{b}_1 + \dfrac{\tilde{n}_2}{N_2} \textbf{b}_2 + \dfrac{\tilde{n}_3}{N_3} \textbf{b}_3, \ \tilde{n}_i \in \mathbb{Z}
\end{equation}
denote the wave vectors for each cell in the reciprocal lattice of the supercrystal. We define
\begin{equation}
\tilde{V}_{\mu}(\textbf{q}_1) = V_{\mu}(\textbf{q}_1) e^{- i \textbf{q}_1 \cdot \textbf{d}_{\mu}} ,
\end{equation}
then
\begin{equation}
V_{\mu}\left[ \textbf{r} - \textbf{R}_{\textbf{n}} - \textbf{d}_{\mu} - \textbf{S}_{\textbf{n},\mu} \right] = \sum_{\textbf{q}_1} \dfrac{e^{i \textbf{q}_1 \cdot \left( \textbf{r} - \textbf{R}_{\textbf{n}} - \textbf{S}_{\textbf{n},\mu} \right)}}{\sqrt{N}} \tilde{V}_{\mu}(\textbf{q}_1),
\end{equation}
\begin{equation}
\left. \nabla_{\textbf{S}_{\textbf{n},\mu}} V_{\mu} \left[ \textbf{r} - \textbf{R}_{\textbf{n}} - \textbf{d}_{\mu} - \textbf{S}_{\textbf{n},\mu} \right] \right|_{\textbf{S}_{\textbf{n},\mu}=\textbf{0}} \ = \sum_{\textbf{q}_1} (- i \textbf{q}_1) \dfrac{e^{i \textbf{q}_1 \cdot \left( \textbf{r} - \textbf{R}_{\textbf{n}} \right)}}{\sqrt{N}} \tilde{V}_{\mu}(\textbf{q}_1) ,
\end{equation}
where $(-i \textbf{q}_1)$ in the last equation is derived from the gradient operation. From the previous results, we have
\begin{align}
\hat{\mathcal{H}}_{e-ph} &= - \sum_{\textbf{n},\mu} \textbf{S}_{\textbf{n},\mu} \cdot \sum_{\sigma} \sum_{n,n'} \sum_{\textbf{k},\textbf{k}'} C^{\dag}_{n',\textbf{k}',\sigma} C_{n,\textbf{k},\sigma} \int d^3\textbf{r} \ \varphi^*_{n',\textbf{k}'}(\textbf{r}) \sum_{\textbf{q}_1} (- i \textbf{q}_1) \dfrac{e^{i \textbf{q}_1 \cdot \left( \textbf{r} - \textbf{R}_{\textbf{n}} \right)}}{\sqrt{N}} \tilde{V}_{\mu}(\textbf{q}_1) \varphi_{n,\textbf{k}}(\textbf{r}) = \notag \\
&= \sum_{\textbf{n},\mu} \textbf{S}_{\textbf{n},\mu} \cdot \sum_{\sigma} \sum_{n,n'} \sum_{\textbf{k},\textbf{k}'} C^{\dag}_{n',\textbf{k}',\sigma} C_{n,\textbf{k},\sigma} \int_V d^3\textbf{r} \ \varphi^*_{n',\textbf{k}'}(\textbf{r}) \sum_{\textbf{q}_1} i \textbf{q}_1 \dfrac{e^{i \textbf{q}_1 \cdot \left( \textbf{r} - \textbf{R}_{\textbf{n}} \right)}}{\sqrt{N}} \tilde{V}_{\mu}(\textbf{q}_1) \varphi_{n,\textbf{k}}(\textbf{r}).
\end{align}
Now, $\frac{e^{i \textbf{q} \cdot \textbf{R}_{\textbf{n}}}}{\sqrt{N}}$ (appearing in $\textbf{S}_{\textbf{n},\mu}$) and $\frac{e^{i \textbf{q}_1 \cdot \left( \textbf{r} - \textbf{R}_{\textbf{n}} \right)}}{\sqrt{N}}$ (appearing in $\nabla_{\textbf{S}_{\textbf{n},\mu}} V_{\mu} \mid_{\textbf{S}_{\textbf{n},\mu}=\textbf{0}}$) both depend on $\mathbf{n}$. Since the wave vectors \(\mathbf{q}\) and \(\mathbf{q}_1\) belong to different domains, namely, \(\mathbf{q}\) varies within the first Brillouin zone, whereas \(\mathbf{q}_1\) takes values in the reciprocal lattice vectors of the supercrystal, the sum over \(\mathbf{n}\) does not yield a Kronecker delta. Therefore, no immediate momentum selection occurs, and more elaborate algebraic manipulations are required to evaluate the interaction terms.
\begin{equation}
\sum_{\textbf{n}} \dfrac{e^{i \textbf{q} \cdot \textbf{R}_{\textbf{n}}}}{\sqrt{N}} \dfrac{e^{- i \textbf{q}_1 \cdot \textbf{R}_{\textbf{n}}}}{\sqrt{N}} = \sum_{n_1,n_2,n_3} \dfrac{e^{i (\textbf{q} - \textbf{q}_1) \cdot \textbf{R}_{\textbf{n}}}}{N},
\end{equation} 
we have
\begin{align}
\dfrac{1}{N} \sum_{n_1,n_2,n_3} e^{i (\textbf{q}-\textbf{q}_1) \cdot \textbf{R}_{\textbf{n}}} &= \dfrac{1}{N} \left( \sum_{n_1=0}^{N_1 - 1} e^{i \frac{n_1}{N_1} (\tilde{\tilde{n}}_1-\tilde{n}_1) \textbf{a}_1 \cdot \textbf{b}_1} \right) \left( \sum_{n_2=0}^{N_2 - 1} e^{i \frac{n_2}{N_2} (\tilde{\tilde{n}}_2-\tilde{n}_2) \textbf{a}_2 \cdot \textbf{b}_2} \right) \left( \sum_{n_3=0}^{N_3 - 1} e^{i \frac{n_3}{N_3} (\tilde{\tilde{n}}_3-\tilde{n}_3) \textbf{a}_3 \cdot \textbf{b}_3} \right) = \notag \\
&= \dfrac{1}{N_1} \left( \sum_{n_1=0}^{N_1 - 1} e^{i 2 \pi \frac{n_1}{N_1} (\tilde{\tilde{n}}_1-\tilde{n}_1)} \right) \dfrac{1}{N_2} \left( \sum_{n_2=0}^{N_2 - 1} e^{i 2 \pi \frac{n_2}{N_2} (\tilde{\tilde{n}}_2-\tilde{n}_2)} \right) \dfrac{1}{N_3} \left( \sum_{n_3=0}^{N_3 - 1} e^{i 2 \pi \frac{n_3}{N_3} (\tilde{\tilde{n}}_3-\tilde{n}_3)} \right) ,
\end{align}
which is the product of three truncated geometric progressions of complex common ratio of unit modulus. Let us consider the first one
\begin{equation}
\sum_{n_1=0}^{N_1 - 1} \left( e^{i \frac{2 \pi}{N_1} (\tilde{\tilde{n}}_1-\tilde{n}_1)} \right)^{n_1} = 
\begin{cases}
N_1, \ e^{i \frac{2 \pi}{N_1} (\tilde{\tilde{n}}_1-\tilde{n}_1)} = 1 \\
\dfrac{1- \left(e^{i \frac{2 \pi}{N_1} (\tilde{\tilde{n}}_1-\tilde{n}_1)} \right)^{N_1}}{1 - e^{i \frac{2 \pi}{N_1} (\tilde{\tilde{n}}_1-\tilde{n}_1)}}, \ e^{i \frac{2 \pi}{N_1} (\tilde{\tilde{n}}_1-\tilde{n}_1)} \neq 1 
\end{cases}.
\end{equation}
From
\begin{align}
\left(e^{i \frac{2 \pi}{N_1} (\tilde{\tilde{n}}_1-\tilde{n}_1)} \right)^{N_1} &= e^{i 2 \pi (\tilde{\tilde{n}}_1-\tilde{n}_1)} = \notag \\
&= e^{i 2 \pi m} = \notag \\
&= 1, \ m \in \mathbb{Z} ,
\end{align}
the quantity above vanishes and the truncated geometric progression is nonnull only if
\begin{equation}
e^{i \frac{2 \pi}{N_1} (\tilde{\tilde{n}}_1-\tilde{n}_1)} = 1 ,
\end{equation}
\begin{equation}
\dfrac{\tilde{\tilde{n}}_1-\tilde{n}_1}{N_1} \in \mathbb{Z},
\end{equation}
but $\tilde{\tilde{n}}_1 \in \left[ 0,N_1 -1 \right]$, while $\tilde{n}_1 \in \mathbb{Z}$, so 
\begin{equation}
\tilde{n}_1 = \tilde{\tilde{n}}_1 + m_1 N_1, \ m_1 \in \mathbb{Z}.
\end{equation}
We insert the previous results into
\begin{align}
\textbf{q}_1 &= \left( \dfrac{\tilde{\tilde{n}}_1}{N_1} + m_1 \right) \textbf{b}_1 + \left( \dfrac{\tilde{\tilde{n}}_2}{N_2} +m_2 \right) \textbf{b}_2 + \left( \dfrac{\tilde{\tilde{n}}_3}{N_3} + m_3 \right) \textbf{b}_3 \ \equiv \notag \\
&\equiv \textbf{q} + \textbf{G},
\end{align}
where
\begin{equation}
\textbf{q} = \dfrac{\tilde{\tilde{n}}_1}{N_1} \textbf{b}_1 + \dfrac{\tilde{\tilde{n}}_2}{N_2} \textbf{b}_2 + \dfrac{\tilde{\tilde{n}}_3}{N_3} \textbf{b}_3, \ \tilde{\tilde{n}}_i \in \left[ 0,N_i - 1 \right]
\end{equation}
belongs to the first Brillouin zone of the reciprocal lattice of the crystal, and $\textbf{G}$ is an arbitrary vector of the reciprocal lattice. Therefore, combining the results of the three truncated geometric series, the sum over $\textbf{n}$ can be expressed as
\begin{equation}
\sum_{\textbf{n}} \dfrac{e^{i (\textbf{q}-\textbf{q}_1) \cdot \textbf{R}_{\textbf{n}}}}{N} = \delta_{\textbf{q}_1,\textbf{q}+\textbf{G}}.
\end{equation}
We get
\begin{align}
\hat{\mathcal{H}}_{e-ph} &= \sum_{\textbf{n},\mu} \sum_{\textbf{q}} \sum_{s=1}^{3 \tau} \dfrac{e^{i \textbf{q} \cdot \textbf{R}_{\textbf{n}}}}{\sqrt{N}} \ \bm{\varepsilon}_{\mu}^{(s)}(\textbf{q}) \sqrt{\dfrac{\hslash}{2 M_{\mu} \omega_{\textbf{q},s}}} \left( a_{\textbf{q},s} + a_{-\textbf{q},s}^{\dag} \right) \sum_{\sigma} \sum_{n,n'} \sum_{\textbf{k},\textbf{k}'} C^{\dag}_{n',\textbf{k}',\sigma} C_{n,\textbf{k},\sigma} \notag \\
& \ \ \ \ \ \ \ \int d^3\textbf{r} \ \varphi^*_{n',\textbf{k}'}(\textbf{r}) \sum_{\textbf{q}_1} i \textbf{q}_1 \dfrac{e^{i \textbf{q}_1 \cdot \left( \textbf{r} - \textbf{R}_{\textbf{n}} \right)}}{\sqrt{N}} \tilde{V}_{\mu}(\textbf{q}_1) \varphi_{n,\textbf{k}}(\textbf{r}) = \notag \\
&= \sum_{\mu} \sum_{\textbf{q}} \sum_{s=1}^{3 \tau} \ \bm{\varepsilon}_{\mu}^{(s)}(\textbf{q}) \sqrt{\dfrac{\hslash}{2 M_{\mu} \omega_{\textbf{q},s}}} \left( a_{\textbf{q},s} + a_{-\textbf{q},s}^{\dag} \right) \sum_{\sigma} \sum_{n,n'} \sum_{\textbf{k},\textbf{k}'} C^{\dag}_{n',\textbf{k}',\sigma} C_{n,\textbf{k},\sigma} \notag \\
& \ \ \ \ \ \ \ \int d^3\textbf{r} \ \varphi^*_{n',\textbf{k}'}(\textbf{r}) \sum_{\textbf{q}_1} i \textbf{q}_1 \delta_{\textbf{q}_1,\textbf{q}+\textbf{G}} e^{i \textbf{q}_1 \cdot \textbf{r}} \tilde{V}_{\mu}(\textbf{q}_1) \varphi_{n,\textbf{k}}(\textbf{r}) = \notag \\
&= i \sum_{\mu} \sum_{\textbf{q}} \sum_{s=1}^{3 \tau} \ \bm{\varepsilon}_{\mu}^{(s)}(\textbf{q}) \sqrt{\dfrac{\hslash}{2 M_{\mu} \omega_{\textbf{q},s}}} \left( a_{\textbf{q},s} + a_{-\textbf{q},s}^{\dag} \right) \sum_{\sigma} \sum_{n,n'} \sum_{\textbf{k},\textbf{k}'} C^{\dag}_{n',\textbf{k}',\sigma} C_{n,\textbf{k},\sigma} \notag \\
& \ \ \ \ \ \ \ \int d^3\textbf{r} \ \varphi^*_{n',\textbf{k}'}(\textbf{r}) \sum_{\textbf{q}_1} \textbf{q}_1 \delta_{\textbf{q}_1,\textbf{q}+\textbf{G}} e^{i \textbf{q}_1 \cdot \textbf{r}} \tilde{V}_{\mu}(\textbf{q}_1) \varphi_{n,\textbf{k}}(\textbf{r}),
\end{align}
and we sum over $\textbf{q}_1$, then
\begin{align}
\hat{\mathcal{H}}_{e-ph} &= i \sum_{\mu} \sum_{\textbf{q}} \sum_{s=1}^{3 \tau} \ \bm{\varepsilon}_{\mu}^{(s)}(\textbf{q}) \sqrt{\dfrac{\hslash}{2 M_{\mu} \omega_{\textbf{q},s}}} \left( a_{\textbf{q},s} + a_{-\textbf{q},s}^{\dag} \right) \sum_{\sigma} \sum_{n,n'} \sum_{\textbf{k},\bar{k}'} C^{\dag}_{n',\textbf{k}',\sigma} C_{n,\textbf{k},\sigma} \notag \\
& \ \ \ \ \ \ \ \int d^3\textbf{r} \ \varphi^*_{n',\textbf{k}'}(\textbf{r}) \left( \textbf{q}+\textbf{G} \right) e^{i (\textbf{q}+\textbf{G}) \cdot \textbf{r}} \tilde{V}_{\mu}\left(\textbf{q}+\textbf{G}\right) \varphi_{n,\textbf{k}}(\textbf{r}).
\end{align}
Given
\begin{align}
\int d^3\textbf{r} & \ \varphi^*_{n',\textbf{k}'}(\textbf{r}) e^{i (\textbf{q}+\textbf{G}) \cdot \textbf{r}} \varphi_{n,\textbf{k}}(\textbf{r}) = \int_V d^3\textbf{r} \ U^*_{n',\textbf{k}'}(\textbf{r}) e^{- i \textbf{k}' \cdot \textbf{r}} e^{i (\textbf{q}+\textbf{G}) \cdot \textbf{r}} U_{n,\textbf{k}}(\textbf{r}) e^{i \textbf{k} \cdot \textbf{r}},
\end{align}
we decompose the integral over the total volume $V$ as a sum of integrals over the volumes $V_{\textbf{n}}$ of the individual unit cells labeled by $\textbf{n}$. To do so, we express the position vector $\textbf{r}$ as the sum of a position $\textbf{r}_{\textbf{0}}$ within the reference cell $\textbf{0}$, denoted by $V_d$, and a lattice translation vector $\textbf{R}_{\textbf{n}}$, i.e., we set $\textbf{r} \rightarrow \textbf{r}_{\textbf{0}} + \textbf{R}_{\textbf{n}}$. The functions $U_{n',\textbf{k}'}\left(\textbf{r}_{\textbf{0}}+\textbf{R}_{\textbf{n}}\right)$ are periodic with respect to lattice translations, that is, they satisfy $U_{n',\textbf{k}'}\left(\textbf{r}_{\textbf{0}}+\textbf{R}_{\textbf{n}}\right) = U_{n',\textbf{k}'}\left(\textbf{r}_{\textbf{0}}\right)$. Therefore, this change of variables does not alter the form of the periodic functions, but it introduces a phase factor due to the Bloch wave nature of the full electron wavefunctions, then
\begin{align}
\int d^3\textbf{r} \ \varphi^*_{n',\textbf{k}'}(\textbf{r}) & e^{i (\textbf{q}+\textbf{G}) \cdot \textbf{r}} \varphi_{n,\textbf{k}}(\textbf{r}) = \notag \\
&= \sum_{\textbf{n}} \int_{V_d} d^3\textbf{r}_{\textbf{0}} \ U^*_{n',\textbf{k}'}(\textbf{r}_{\textbf{0}}+\textbf{R}_{\textbf{n}}) e^{- i \textbf{k}' \cdot (\textbf{r}_{\textbf{0}}+\textbf{R}_{\textbf{n}})} e^{i (\textbf{q}+\textbf{G}) \cdot (\textbf{r}_{\textbf{0}}+\textbf{R}_{\textbf{n}})} U_{n,\textbf{k}}(\textbf{r}_{\textbf{0}}+\textbf{R}_{\textbf{n}}) e^{i \textbf{k} \cdot (\textbf{r}_{\textbf{0}}+\textbf{R}_{\textbf{n}})} = \notag \\
&= \sum_{\textbf{n}} \int_{V_d} d^3\textbf{r}_{\textbf{0}} \ U^*_{n',\textbf{k}'}(\textbf{r}_{\textbf{0}}) e^{- i \textbf{k}' \cdot (\textbf{r}_{\textbf{0}}+\textbf{R}_{\textbf{n}})} e^{i (\textbf{q}+\textbf{G}) \cdot (\textbf{r}_{\textbf{0}}+\textbf{R}_{\textbf{n}})} U_{n,\textbf{k}}(\textbf{r}_{\textbf{0}}) e^{i \textbf{k} \cdot (\textbf{r}_{\textbf{0}}+\textbf{R}_{\textbf{n}})} =  \notag \\
&= \sum_{\textbf{n}} \int_{V_d} d^3\textbf{r}_{\textbf{0}} \ U^*_{n',\textbf{k}'}(\textbf{r}_{\textbf{0}}) e^{- i \left( \textbf{k}' - \textbf{k} - \textbf{q} - \textbf{G} \right) \cdot \textbf{r}_{\textbf{0}}} e^{- i \left( \textbf{k}' - \textbf{k} - \textbf{q} - \textbf{G} \right) \cdot \textbf{R}_{\textbf{n}}} U_{n,\textbf{k}}(\textbf{r}_{\textbf{0}}) \equiv \notag \\
&\equiv N \sum_{\textbf{n}} \int_{V_d} d^3\textbf{r}_{\textbf{0}} \ U^*_{n',\textbf{k}'}(\textbf{r}_{\textbf{0}}) e^{- i \left( \textbf{k}' - \textbf{k} - \textbf{q} - \textbf{G} \right) \cdot \textbf{r}_{\textbf{0}}} \frac{e^{- i \left( \textbf{k}' - \textbf{k} - \textbf{q} - \textbf{G} \right) \cdot \textbf{R}_{\textbf{n}}}}{N} U_{n,\textbf{k}}(\textbf{r}_{\textbf{0}}) ,
\end{align}
where at the last step we multiplied and divided by $N$ to get 
\begin{equation}
\sum_{\textbf{n}} \dfrac{e^{- i \left( \textbf{k}' - \textbf{k} - \textbf{q} - \textbf{G} \right) \cdot \textbf{R}_{\textbf{n}}}}{N} = \delta_{\textbf{k}',\textbf{k}+\textbf{q}+\textbf{G}},
\end{equation}
and recall that $\textbf{k}'$, $\textbf{k}$ and $\textbf{q}$ are wave vectors of the first Brillouin zone, while $\textbf{G}$ is any vector of the reciprocal crystal lattice. We have
\begin{equation}
\int_V d^3\textbf{r} \varphi^*_{n',\textbf{k}'}(\textbf{r}) e^{i (\textbf{q}+\textbf{G}) \cdot \textbf{r}} \varphi_{n,\textbf{k}}(\textbf{r}) = N \int_{V_d} d^3\textbf{r}_{\textbf{0}} \ U^*_{n',\textbf{k}'}(\textbf{r}_{\textbf{0}}) e^{- i \left( \textbf{k}' - \textbf{k} - \textbf{q} - \textbf{G} \right) \cdot \textbf{r}_{\textbf{0}}} \delta_{\textbf{k}',\textbf{k}+\textbf{q}+\textbf{G}} U_{n,\textbf{k}}(\textbf{r}_{\textbf{0}}),
\end{equation}
\begin{align}
\hat{\mathcal{H}}_{e-ph} &= i \sum_{\mu} \sum_{\textbf{q}} \sum_{s=1}^{3 \tau} \ \bm{\varepsilon}_{\mu}^{(s)}(\textbf{q}) \sqrt{\dfrac{\hslash}{2 M_{\mu} \omega_{\textbf{q},s}}} \left( a_{\textbf{q},s} + a_{-\textbf{q},s}^{\dag} \right) \cdot \sum_{\sigma} \sum_{n,n'} \sum_{\textbf{k},\textbf{k}'} C^{\dag}_{n',\textbf{k}',\sigma} C_{n,\textbf{k},\sigma} \notag \\
& \ \ \ \ \ \ \ \left( \textbf{q}+\textbf{G} \right) \left[ N \int_{V_d} d^3\textbf{r}_{\textbf{0}} \ U^*_{\textbf{k}',n'}(\textbf{r}_{\textbf{0}}) e^{- i \left( \textbf{k}' - \textbf{k} - \textbf{q} - \textbf{G} \right) \cdot \textbf{r}_{\textbf{0}}} \delta_{\textbf{k}',\textbf{k}+\textbf{q}+\textbf{G}} U_{\textbf{k},n}(\textbf{r}_{\textbf{0}}) \right] .
\end{align}
We perform the sum over $\textbf{k}'$, which activates the delta function and enforces momentum conservation. As a result, the electron-phonon interaction Hamiltonian takes the form
\begin{align}
\hat{\mathcal{H}}_{e-ph} &= i \sum_{\substack{n,n',\sigma, \\ s,\mu, \\ \textbf{k},\textbf{G},\textbf{q}}} C^{\dag}_{n',\textbf{k}+\textbf{q}+\textbf{G},\sigma} C_{n,\textbf{k},\sigma} \left( a_{\textbf{q},s} + a^{\dag}_{-\textbf{q},s} \right) \left( \sqrt{\dfrac{\hslash}{2 M_{\mu} \omega_{\textbf{q},s}}} \right) \notag \\
& \ \ \ \ \ \ \ \ \ \left\lbrace \bm{\varepsilon}_{\mu}^{(s)}(\textbf{q}) \cdot \left( \textbf{q}+\textbf{G} \right) \right\rbrace \tilde{V}_{\mu}\left(\textbf{q}+\textbf{G}\right) \left[ N \int_{V_d} d^3\textbf{r} \ U^*_{n',\textbf{k}+\textbf{q}+\textbf{G}}(\textbf{r}) U_{n,\textbf{k}}(\textbf{r}) \right] \equiv \notag \\
&\equiv \sum_{\substack{n,n',\sigma,s, \\ \textbf{k},\textbf{G},\textbf{q}}} T_{\textbf{k},\textbf{q},\textbf{G}}^s \ C^{\dag}_{n',\textbf{k}+\textbf{q}+\textbf{G},\sigma} C_{n,\textbf{k},\sigma} \left( a_{\textbf{q},s} + a^{\dag}_{-\textbf{q},s} \right) ,
\end{align}
which represents the third term of the Hamiltonian $\eqref{eq: Hamiltonianadiunsolido}$, and we have defined
\begin{equation}
T_{\textbf{k},\textbf{q},\textbf{G}}^s = i \sum_\mu \left( \sqrt{\dfrac{\hslash}{2 M_{\mu} \omega_{\textbf{q},s}}} \right) \left\lbrace \bm{\varepsilon}_{\mu}^{(s)}(\textbf{q}) \cdot \left( \textbf{q}+\textbf{G}) \right) \right\rbrace \tilde{V}_{\mu}\left(\textbf{q}+\textbf{G}\right) \left[ N \int_{V_d} d^3\textbf{r} \ U^*_{n',\textbf{k}+\textbf{q}+\textbf{G}}(\textbf{r}) U_{n,\textbf{k}}(\textbf{r}) \right].
\end{equation}
An electron is annihilated in the state with momentum $\textbf{k}$ and created in the state with momentum $\textbf{k} + \textbf{q} + \textbf{G}$, where $\textbf{q}$ is the phonon momentum. Let us now consider the constraint $\textbf{k}' = \textbf{k} + \textbf{q} + \textbf{G}$. The resulting vector $\textbf{k}'$ must lie within the first Brillouin zone, since $\textbf{k}$ and $\textbf{q}$ are both defined within the first Brillouin zone of the reciprocal lattice. However, the sum $\textbf{k} + \textbf{q}$ may fall outside the first Brillouin zone. In such cases, a reciprocal lattice vector $\textbf{G}$ is introduced to map the momentum back into the first Brillouin zone. This gives rise to what are known as Umklapp processes. If instead $\textbf{k} + \textbf{q}$ already lies within the first Brillouin zone, then $\textbf{G} = \textbf{0}$, and we refer to these as normal processes. In the case of normal processes, the vibrational modes contributing to the electron-phonon interaction and satisfying
\begin{equation}
\bm{\varepsilon}_{\mu}(\textbf{q}) \cdot \textbf{q} \neq 0
\end{equation}
are longitudinal. Transverse phonons contribute to the dynamics only for Umklapp processes, with $\textbf{G} \neq \textbf{0}$. \newline
The possible processes are two. First one is given by $C^{\dag}_{n',\textbf{k}',\sigma} C_{n,\textbf{k},\sigma} a_{\textbf{q},s}$, that is, a phonon with momentum $\textbf{q}$ and an electron with momentum $\textbf{k}$ are destroyed in the $n$ band; an electron is created with momentum $\textbf{k}'$, in the $n'$ band, with the unchanged spin component. Second process is given by $C^{\dag}_{n',\textbf{k}',\sigma} C_{n,\textbf{k},\sigma} a^{\dag}_{-\textbf{q},s}$, that is, an electron with momentum $\textbf{k}$ in the $n$ band is destroyed; a phonon with momentum $-\textbf{q}$ and an electron with momentum $\textbf{k}$ in the $n'$ band with the unchanged spin component are created.
\section{Simplification of the Hamiltonian of electron-phonon interaction}
At low energies, only a single conduction band and the most relevant phonon branch need to be considered. Neglecting interband, Umklapp, and momentum-dependent electronic effects further simplifies the description. In this regime, the electron–phonon coupling depends only on the phonon momentum, leading to a compact interaction where electrons scatter within the same band by emitting or absorbing phonons. The simplified electron-phonon Hamiltonian is thus given by
\begin{equation}
\hat{\mathcal{H}}_{e-ph} = \sum_{\textbf{q},\textbf{k},\sigma} M_{\textbf{q}} C^{\dag}_{\textbf{k}+\textbf{q},\sigma} C_{\textbf{k},\sigma} \left( a_{\textbf{q}} + a^{\dag}_{-\textbf{q}} \right) 
\label{eq: interazioneparticellebagnotermico}
\end{equation}
with a single electronic band ($n=n'=1$) and a single phonon branch ($s=1$).

\part{Developments of Quantum Many-Body Theory}

\chapter{Many-body Green's functions}
Many-body Green's functions are among the most fundamental tools for studying interacting quantum systems. In this chapter, we will formally introduce Green's functions for both fermions and bosons, highlighting how they allow us to describe the time evolution, correlations, and response of a system in the presence of interactions. Green's functions incorporate interaction effects in a systematic way, providing access to physically relevant observables through their analytical structure. We will then illustrate how these functions can be used to extract fundamental thermodynamic and state quantities, such as the average particle number in the presence of interactions. This connection makes Green’s functions not just formal tools, but a direct bridge between quantum field theory and experimental observation. \newline
In particular, we will study spectral functions, which encode the system's essential energetic information and determine its excitation properties. These quantities are intimately connected to real-time Green's functions, which are necessary for capturing dynamical behavior and accessing physically measurable spectra. Spectral functions are also directly related to the density of states, and they allow one to extract observable quantities via the so-called sum rules. These rules act as constraints on the information contained in Green's and spectral functions, ensuring that the formalism is consistent with fundamental conservation laws. We will show how these rules are useful both for interpreting theoretical results and for testing the validity of numerical or analytical approximations. A paradigmatic example illustrating the use of Green’s functions in the context of interactions is the Independent Boson Model. This model describes a quantum particle coupled to a bath of independent bosonic modes, providing an exactly solvable case that highlights the effect of coupling to a bosonic environment on the particle’s dynamics and excitation properties. \newline
Moreover, many of thermal averages and perturbative expansions are most naturally expressed in terms of imaginary-time Green's functions, which are fundamental for describing equilibrium properties at finite temperature. The imaginary-time formalism simplifies the treatment of thermal states by mapping the quantum statistical mechanics problem onto a field theory in imaginary time, enabling powerful computational techniques such as Matsubara frequency summations. On the other hand, to access real-time dynamics, spectral properties, and response functions directly comparable to experiments, one must work with real-time Green's functions, often obtained through analytic continuation from the imaginary-time domain. 
\section{The retarded one particle Green's function in quantum many-body theory}
Here we want to generalize the definition of retarded Green's function to a quantum many-body system. Let \( \mathcal{H} \) be a one-body Hamiltonian with eigenstates \( \{ \varphi_\alpha(x) \} \) and corresponding eigenvalues \( \{ \mathcal{E}_\alpha \} \), satisfying the time-independent Schrödinger equation. The general solution of the time-dependent Schrödinger equation for a single particle can be written as
\begin{equation}
\psi(x,t) = \sum_\alpha c_\alpha\, e^{-i \frac{\mathcal{E}_\alpha}{\hbar} t} \varphi_\alpha(x),
\label{eq: operatorecampodistruzioneevoluzionetemporale}
\end{equation}
where \( c_\alpha \) are complex coefficients determined by the initial condition. In the second quantization formalism, the state of the system is described in Fock space, while observables and time evolution are represented by operators. The field annihilation operator \( \hat{\psi}(x,t) \), which removes a particle at position \( \mathbf{r} \) and time \( t \) (recall that \( x = (\mathbf{r}, s) \)), is defined in $\eqref{eq: operatorecampodistruzione}$. In the non-interacting case, the time evolution of the annihilation operators \( a_\alpha \) is given by $\eqref{eq: evoluzionetemporaleoperatoredistruzione1}$, so that the field operator becomes
\begin{equation}
\hat{\psi}(x,t) = \sum_\alpha e^{- i \frac{\mathcal{E}_\alpha}{\hslash} t} \varphi_\alpha (x) a_\alpha.
\end{equation}
Similarly, from $\eqref{eq: operatorecampocreazione}$, $\eqref{eq: evoluzionetemporaleoperatorecreazione1}$, the creation field operator evolves as
\begin{equation}
\hat{\psi}^\dag(x,t) = \sum_\alpha e^{i \frac{\mathcal{E}_\alpha}{\hslash} t} \varphi^*_\alpha (x) a^\dag_\alpha.
\label{eq: operatorecampocreazioneevoluzionetemporale}
\end{equation}
Similarly, the field operators create a particle in $x'$ at time $t'$ and then destroy it in $x$ at time $t$, so we might expect the many-body Green's function to be
\begin{equation}
- \frac{i}{\hslash} \Theta(t-t') \langle \hat{\psi}(x,t) \hat{\psi}^\dagger(x',t') \rangle,
\label{eq: propagatoremanybodieserrato}
\end{equation}
where $\langle \ldots \rangle$ denotes the thermal average operation. The $\eqref{eq: propagatoremanybodieserrato}$ is an incorrect equation, indeed 
\begin{equation}
\lim_{t \rightarrow t'^{+}} - \frac{i}{\hslash} \Theta(t-t') \left\langle \hat{\psi}(x,t) \hat{\psi}^\dagger(x',t') \right\rangle \neq - \dfrac{i}{\hslash} \delta(x-x').
\end{equation}
Since $\left[\hat{\psi}(x),\hat{\psi}^\dag(x')\right]^{(\varepsilon)} = \delta(x-x')$, the retarded many-body Green's function associated with the field operators (see Eqq. \ref{eq: operatorecampodistruzione}, \ref{eq: operatorecampocreazione}), expressed in the Heisenberg picture (see Eq. \ref{eq: Heisenbergpicture}), is defined as
\begin{equation}
G^{(r)}(x,t;x',t') = - \frac{i}{\hslash} \Theta(t-t') \left\langle [\hat{\psi}(x,t), \hat{\psi}^\dagger(x',t')]^{(\varepsilon)} \right\rangle. 
\label{eq: propagatoremanybodies}
\end{equation}
The object in the thermal average in $\eqref{eq: propagatoremanybodies}$ is explicitly
\begin{equation}
\left[\hat{\psi}(x,t),\hat{\psi}^\dag(x',t')\right]^{(\varepsilon)} = \hat{\psi}(x,t) \hat{\psi}^{\dag}(x',t') - \varepsilon \hat{\psi}^{\dag}(x',t') \hat{\psi}(x,t), \ t > t'
\end{equation}
which are respectively called the particle term (first create and then destroy) and the hole term (first destroy and then create): we have both particle and hole propagations, and this is conceptually the correct generalization of the propagator to a many-body system. We prove that indeed the $\eqref{eq: propagatoremanybodies}$ satisfies the properties of a propagator. First, we verify that for non-interacting particles the many-body propagator is proportional to the single-particle propagator: given the Hamiltonian of free non-interacting particles, i.e., $\eqref{eq: Hamiltonianasecondaquantizzazioneparticellelibere}$, we have
\begin{align}
G^{(r)}(x,t,x',t') &= - \dfrac{i}{\hslash} \Theta(t-t') \sum_{\alpha,\alpha'} \varphi_\alpha(x) e^{- i \frac{\mathcal{E}_\alpha}{\hslash} (t-t')} \varphi^*_{\alpha'}(x') \left\langle \left[ a_\alpha, a^\dagger_{\alpha'} \right]^{(\varepsilon)} \right\rangle = \notag \\
&= - \dfrac{i}{\hslash} \Theta(t-t') \sum_{\alpha,\alpha'} \varphi_\alpha(x) e^{- i \frac{\mathcal{E}_\alpha}{\hslash} (t-t')} \varphi^*_{\alpha'}(x') \delta_{\alpha,\alpha'} = \notag \\
&= - \dfrac{i}{\hslash} \Theta(t-t') \sum_{\alpha} \varphi_\alpha(x) e^{- i \frac{\mathcal{E}_\alpha}{\hslash} (t-t')} \varphi^*_{\alpha}(x'),
\label{eq: propagatoreperparticellenoninteragenti}
\end{align}
which is the thesis. Then, for both non-interacting and interacting particles the $\eqref{eq: propagatoremanybodies}$ satisfies
\begin{equation}
\lim_{t \rightarrow t'^+} - \dfrac{i}{\hslash} \Theta(t-t') \left\langle \left[\psi(x,t),\psi^\dag(x',t')\right]^{(\varepsilon)} \right\rangle = - \dfrac{i}{\hslash} \delta(x-x').
\end{equation}
Finally, given two operators $A$ and $B$, we define Green's function of operators $A$ and $B$ at times $t$ and $t'$ as
\begin{align}
G^{(r)}_{AB}(t,t') &= - \dfrac{i}{\hslash} \Theta(t-t') \left\langle \left[ A(t), B(t') \right]^{(\varepsilon)} \right\rangle \equiv \notag \\
&\equiv - \dfrac{i}{\hslash} \Theta(t-t') \left[  \langle A(t)B(t') \rangle - \varepsilon  \langle B(t')A(t) \rangle \right],
\label{eq: definizionefunzionediGreenritardata}
\end{align} 
which quantifies the correlation between operators $A$ and $B$ at times $t$ and $t'$. Many functions of physical interest are written in the form $G^{(r)}_{AB}(t,t')$.

At the conclusion of this section, we present an important property of Green's functions that will be extensively used in the following. In many-body physics and quantum statistical mechanics, the concept of thermal equilibrium plays a fundamental role in determining the properties of correlation functions. When a system is coupled to a thermal reservoir at temperature $T$, it reaches a stationary state characterized by invariance under time translations. This time-translation invariance implies that dynamical quantities such as the retarded Green's function depend only on the elapsed time between events, rather than on the absolute times themselves. The following theorem formalizes this property.
\begin{theorem}[Time-translation invariance of Green's functions in thermal equilibrium]\label{Time-translation invariance of Green's functions in thermal equilibrium} At thermal equilibrium, the retarded many-body Green's function $\eqref{eq: definizionefunzionediGreenritardata}$ exhibits time-translation invariance, i.e.,
\begin{equation}
G^{(r)}_{AB}(t,t') = G^{(r)}_{AB}(t-t').
\end{equation}
\begin{proof}
Given the retarded Green's function $\eqref{eq: definizionefunzionediGreenritardata}$, regarding the first thermal average, we have
\begin{align}
\langle A(t)B(t') \rangle &= \dfrac{1}{Z} \Tr \left[ e^{-\beta \hat{\mathcal{H}}} e^{i \frac{\hat{\mathcal{H}}}{\hslash} t} A e^{- i \frac{\hat{\mathcal{H}}}{\hslash} t} e^{i \frac{\hat{\mathcal{H}}}{\hslash} t'} B e^{- i \frac{\hat{\mathcal{H}}}{\hslash} t'} \right] = \notag \\
&= \dfrac{1}{Z} \Tr \left[ e^{-\beta \hat{\mathcal{H}}} e^{i \frac{\hat{\mathcal{H}}}{\hslash} t} A e^{- i \frac{\hat{\mathcal{H}}}{\hslash} (t-t')} B e^{- i \frac{\hat{\mathcal{H}}}{\hslash} t'} \right] = \notag \\
&= \dfrac{1}{Z} \Tr \left[ e^{- i \frac{\hat{\mathcal{H}}}{\hslash} t'} e^{-\beta \hat{\mathcal{H}}} e^{i \frac{\hat{\mathcal{H}}}{\hslash} t} A e^{- i \frac{\hat{\mathcal{H}}}{\hslash} (t-t')} B \right] = \notag \\
&= \dfrac{1}{Z} \Tr \left[ e^{-\beta \hat{\mathcal{H}}} e^{- i \frac{\hat{\mathcal{H}}}{\hslash} t'} e^{i \frac{\hat{\mathcal{H}}}{\hslash} t} A e^{- i \frac{\hat{\mathcal{H}}}{\hslash} (t-t')} B \right] = \notag \\
&= \dfrac{1}{Z} \Tr \left[ e^{-\beta \hat{\mathcal{H}}} e^{i \frac{\hat{\mathcal{H}}}{\hslash} (t-t')} A e^{- i \frac{\hat{\mathcal{H}}}{\hslash} (t-t')} B \right] = \notag \\
&= \langle A(t-t')B(0) \rangle,
\end{align}
where we have used the cyclic property $\eqref{eq: proprietàciclicatraccia}$ of the trace, then we inverted the product $e^{- i \frac{\hat{\mathcal{H}}}{\hslash} t'} e^{-\beta \hat{\mathcal{H}}}$, which is admissible, since the exponent operators commute. Similarly, we have
\begin{align}
\langle B(t')A(t) \rangle &= \dfrac{1}{Z} \Tr \left[ e^{-\beta \hat{\mathcal{H}}} e^{i \frac{\hat{\mathcal{H}}}{\hslash} t'} B e^{- i \frac{\hat{\mathcal{H}}}{\hslash} t'} e^{i \frac{\hat{\mathcal{H}}}{\hslash} t} A e^{- i \frac{\hat{\mathcal{H}}}{\hslash} t} \right] = \notag \\
&= \dfrac{1}{Z} \Tr \left[ e^{-\beta \hat{\mathcal{H}}} e^{i \frac{\hat{\mathcal{H}}}{\hslash} t'} B e^{i \frac{\hat{\mathcal{H}}}{\hslash} (t-t')} A e^{- i \frac{\hat{\mathcal{H}}}{\hslash} t} \right] = \notag \\
&= \dfrac{1}{Z} \Tr \left[ e^{i \frac{\hat{\mathcal{H}}}{\hslash} t'} e^{-\beta \hat{\mathcal{H}}} B e^{i \frac{\hat{\mathcal{H}}}{\hslash} (t-t')} A e^{- i \frac{\hat{\mathcal{H}}}{\hslash} t} \right] = \notag \\
&= \dfrac{1}{Z} \Tr \left[ e^{-\beta \hat{\mathcal{H}}} B e^{i \frac{\hat{\mathcal{H}}}{\hslash} (t-t')} A e^{- i \frac{\hat{\mathcal{H}}}{\hslash} t} e^{i \frac{\hat{\mathcal{H}}}{\hslash} t'} \right] = \notag \\
&= \dfrac{1}{Z} \Tr \left[ e^{-\beta \hat{\mathcal{H}}} B e^{i \frac{\hat{\mathcal{H}}}{\hslash} (t-t')} A e^{- i \frac{\hat{\mathcal{H}}}{\hslash} (t-t')} \right] = \notag \\
&= \langle B(0)A(t-t') \rangle.
\end{align}
By combining both results, we observe that the thermal averages depend only on the time difference \( t - t' \), confirming time-translation invariance of the thermal expectation values in the Heisenberg picture with a time-independent Hamiltonian, that is, establishing the thesis.
\end{proof}
\end{theorem}
\section{Thermal interaction picture}\label{Thermal interaction picture}
\begin{figure}[H]
\centering
\includegraphics[scale=0.4]{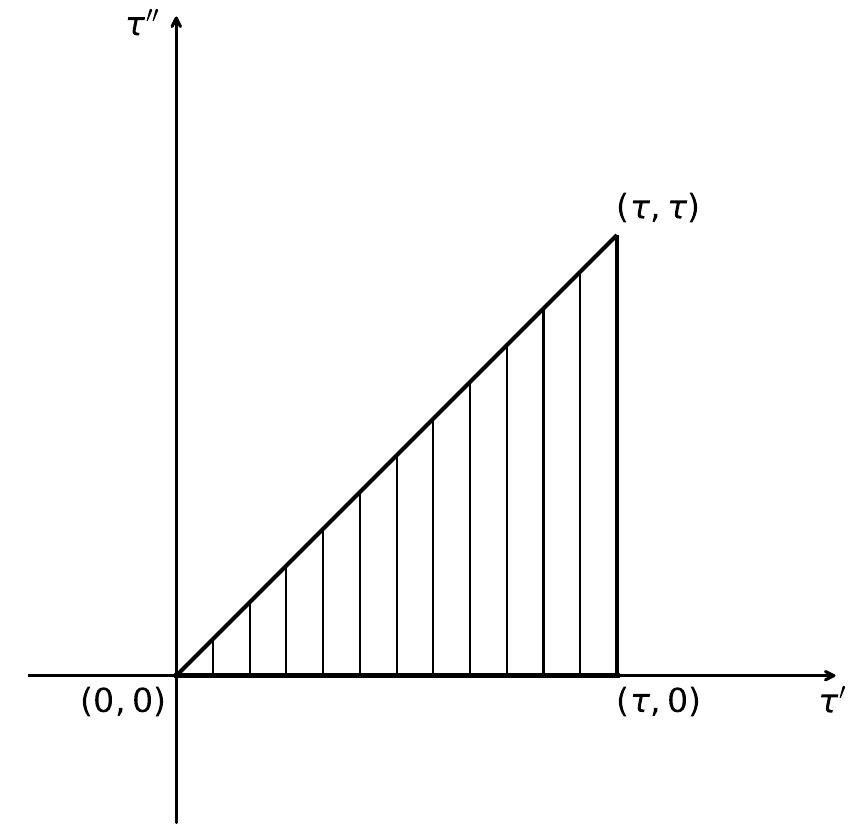}
\caption{Integration region corresponding to $\eqref{eq: primaregioneintegrazioneordinesecondointerazioneserieDyson}$, where $0 \leq \tau'' \leq \tau' \leq \tau$. This domain reflects the time ordering $\tau'' \leq \tau'$ required by the Dyson series at second order. The integration is performed first over $\tau''$ (from $0$ to $\tau'$), then over $\tau'$ (from $0$ to $\tau$), and matches the natural ordering of operators in $\hat{T}_D \lbrace \hat{\mathcal{H}}_I^{(0)}(\tau') \hat{\mathcal{H}}_I^{(0)}(\tau'') \rbrace$.}
\label{fig: primaregioneintegrazioneHtauprimoHtausecondo}
\end{figure}

\begin{figure}[H]
\centering
\includegraphics[scale=0.4]{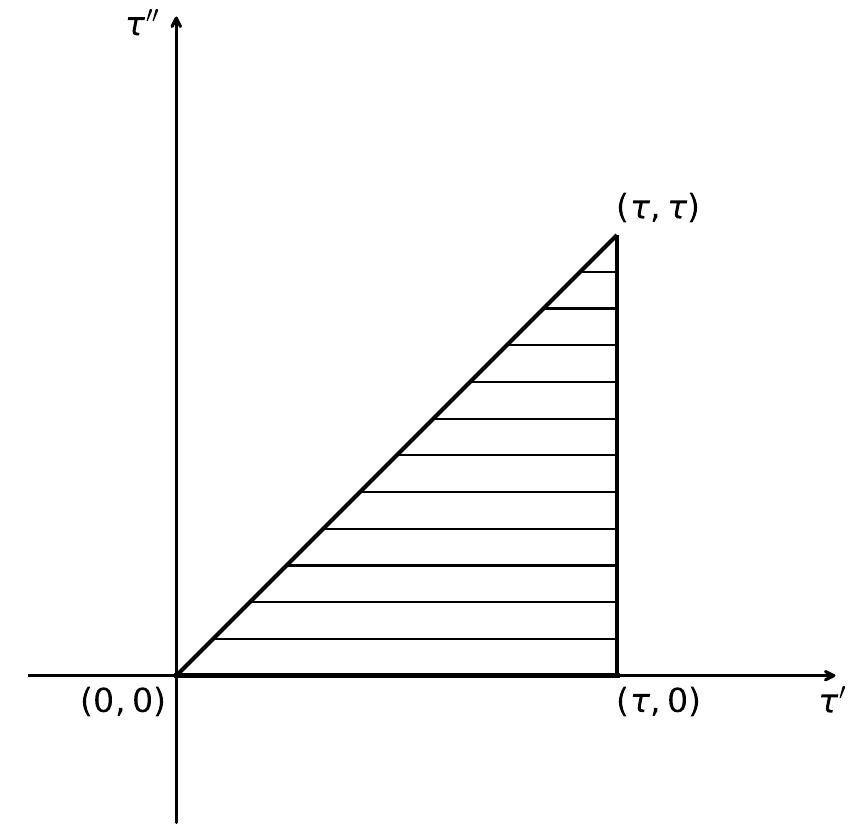}
\caption{Integration region corresponding to $\eqref{eq: secondaregioneintegrazioneordinesecondointerazioneserieDyson}$, where $0 \leq \tau'' \leq \tau$ and $\tau'' \leq \tau' \leq \tau$. This is an equivalent rewriting of the domain in Figure~\ref{fig: primaregioneintegrazioneHtauprimoHtausecondo}, preserving the time ordering $\tau'' \leq \tau'$. The integration is performed first over $\tau''$, and for each fixed $\tau''$, over $\tau'$ from $\tau''$ to $\tau$.}
\label{fig: secondaregioneintegrazioneHtauprimoHtausecondo}
\end{figure}

\begin{figure}[H]
\centering
\includegraphics[scale=0.4]{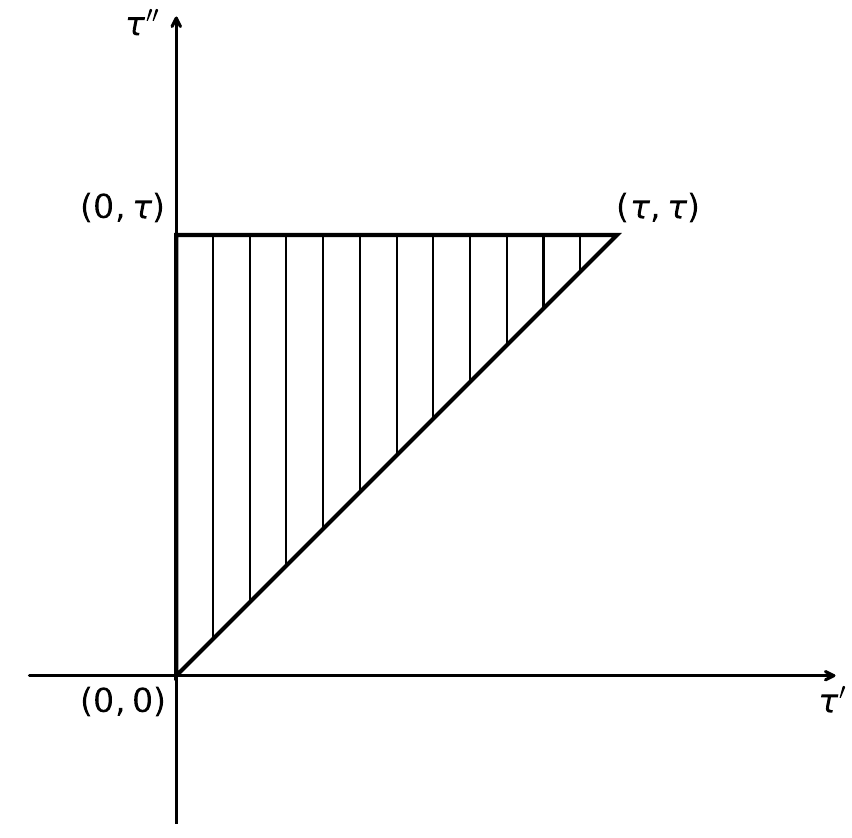}
\caption{Integration region corresponding to $\eqref{eq: terzaregioneintegrazioneordinesecondointerazioneserieDyson}$, where $0 \leq \tau' \leq \tau'' \leq \tau$. This region corresponds to the opposite time ordering, $\tau' \leq \tau''$, and appears in the second-order Dyson term when operators are explicitly reordered by $\hat{T}_D$. The integration is carried out first over $\tau'$, and then over $\tau''$ from $\tau'$ to $\tau$.}
\label{fig: terzaregioneintegrazioneHtauprimoHtausecondo}
\end{figure}

Here, we adapt the formalism of the interaction picture for imaginary times. Let the system be in thermal equilibrium at temperature $T$, we define the thermal interaction picture of an operator $A$ with respect to the Hamiltonian $\hat{\mathcal{H}}_0$ from $\eqref{eq: HamiltonianascompostainH_0eH_I}$ within the framework of second quantization, as
\begin{equation}
A^{(0)}(\tau) = e^{\frac{\hat{\mathcal{H}}_0}{\hslash} \tau} A e^{- \frac{\hat{\mathcal{H}}_0}{\hslash} \tau}, \ \tau \in \left[- \beta \hslash , \beta \hslash \right], \ \tau = i t \in \mathbb{R}, \ t \in \im \mathbb{C}.
\label{eq: thermalinteractionpicture}
\end{equation}
Such a picture is called thermal since the temperature is finite and nonzero. On the other hand, the states are transformed by means of a time evolution operator
\begin{equation}
U(\tau) = e^{\frac{\hat{\mathcal{H}}_0}{\hslash} \tau} e^{- \frac{\hat{\mathcal{H}}}{\hslash} \tau}.
\end{equation}
From
\begin{equation}
U^\dag(\tau) = e^{- \frac{\hat{\mathcal{H}}}{\hslash} \tau} e^{\frac{\hat{\mathcal{H}}_0}{\hslash} \tau},
\end{equation}
and
\begin{equation}
U^{-1}(\tau) = e^{\frac{\hat{\mathcal{H}}}{\hslash} \tau} e^{- \frac{\hat{\mathcal{H}}_0}{\hslash} \tau},
\end{equation}
the operator $U(\tau)$ is not a unitary transformation, indeed $U^\dag(\tau) \neq U^{-1}(\tau)$. \newline
We now introduce two time-ordering operators that will be used in the subsequent discussion. First, the Wick time-ordering operator, denoted by $\hat{T}_\tau$, is defined follows
\begin{equation}
\hat{T}_\tau A(\tau') B(\tau'') =
\begin{cases}
A(\tau') B(\tau''), & \tau' > \tau'', \\
\varepsilon \, B(\tau'') A(\tau'), & \tau'' > \tau',
\end{cases}
\end{equation}
that is, it arranges products of operators from left to right in decreasing order with respect to the time variable $\tau$, taking into account the statistical nature of the operators. Second, the Dyson time-ordering operator, denoted by \(\hat{T}_D\), is defined as
\begin{equation}
\hat{T}_D A(\tau') B(\tau'') =
\begin{cases}
A(\tau') B(\tau''), & \tau' > \tau'' \\[6pt]
B(\tau'') A(\tau'), & \tau'' > \tau'
\end{cases},
\end{equation}
where \(\hat{T}_D\) orders operators from left to right in decreasing order of the time variable \(t\), regardless of their statistics. We are now able to prove
\begin{theorem}[The Dyson's series]
Given an Hamiltonian of the form $\eqref{eq: HamiltonianascompostainH_0eH_I}$, within the framework of the second quantizazion, the time evolution of the operator $U(\tau)$ involves the thermal interaction representations of $\hat{\mathcal{H}_I}$, i.e.,
\begin{equation}
U(\tau) = \hat{T}_\tau \exp \left\lbrace - \dfrac{1}{\hslash} \int_0^\tau d\tau_1 \hat{\mathcal{H}}_I^{(0)}(\tau_1) \right\rbrace ,
\label{eq: seriediDysonconesponenziale}
\end{equation}
which is known as Dyson's series.
\begin{proof}
Let us compute the first time derivative of the operator $U(\tau)$, i.e.,
\begin{equation}
\dfrac{\partial U}{\partial \tau} = - \dfrac{1}{\hslash} \hat{\mathcal{H}}_I^{(0)}(\tau) U(\tau),
\label{eq: equazioneoperatoreevoluzionetemporaleinterazionetermica}
\end{equation}
with the initial condition $U(0) = \mathds{1}$, indeed
\begin{align}
\dfrac{\partial U}{\partial \tau} &= \dfrac{1}{\hslash} \hat{\mathcal{H}}_0 e^{\frac{\hat{\mathcal{H}}_0}{\hslash} \tau} e^{- \frac{\hat{\mathcal{H}}}{\hslash} \tau} - \dfrac{1}{\hslash} e^{\frac{\hat{\mathcal{H}}_0}{\hslash} \tau} \hat{\mathcal{H}} e^{- \frac{\hat{\mathcal{H}}}{\hslash} \tau} = \notag \\
&= \dfrac{1}{\hslash} e^{\frac{\hat{\mathcal{H}}_0}{\hslash} \tau} \hat{\mathcal{H}}_0 e^{- \frac{\hat{\mathcal{H}}}{\hslash} \tau} - \dfrac{1}{\hslash} e^{\frac{\hat{\mathcal{H}}_0}{\hslash} \tau} \hat{\mathcal{H}} e^{- \frac{\hat{\mathcal{H}}}{\hslash} \tau} = \notag \\
&= - \dfrac{1}{\hslash} e^{\frac{\hat{\mathcal{H}}_0}{\hslash} \tau} \hat{\mathcal{H}}_I e^{- \frac{\hat{\mathcal{H}}}{\hslash} \tau} = \notag \\
&= - \dfrac{1}{\hslash} e^{\frac{\hat{\mathcal{H}}_0}{\hslash} \tau} \hat{\mathcal{H}}_I e^{- \frac{\hat{\mathcal{H}}_0}{\hslash} \tau} e^{\frac{\hat{\mathcal{H}}_0}{\hslash} \tau} e^{- \frac{\hat{\mathcal{H}}}{\hslash} \tau} = \notag \\
&= - \dfrac{1}{\hslash} \hat{\mathcal{H}}_I^{(0)}(\tau) U(\tau) ,
\end{align}
where we used the commutator $\left[ \hat{\mathcal{H}}_0 , e^{\frac{\hat{\mathcal{H}}_0}{\hslash} \tau} \right]=0$, after which we used the identity $ \mathds{1} = e^{- \frac{\hat{\mathcal{H}}_0}{\hslash} \tau} e^{\frac{\hat{\mathcal{H}}_0}{\hslash} \tau}$ to bring out the thermal interaction representation of $\hat{\mathcal{H}}_I$ with respect to $\hat{\mathcal{H}}_0$. Let us integrate both members of $\eqref{eq: equazioneoperatoreevoluzionetemporaleinterazionetermica}$ between $0$ and $\tau$, i.e.,
\begin{equation}
\int_0^{\tau} \dfrac{\partial U}{\partial \tau'} d\tau' = - \dfrac{1}{\hslash} \int_0^{\tau} \hat{\mathcal{H}}_I^{(0)}(\tau') U(\tau') d\tau',
\end{equation}
\begin{equation}
U(\tau) - U(0) = - \dfrac{1}{\hslash} \int_0^{\tau} \hat{\mathcal{H}}_I^{(0)}(\tau') U(\tau') d\tau',
\end{equation}
\begin{equation}
U(\tau) = U(0) - \dfrac{1}{\hslash} \int_0^{\tau} \hat{\mathcal{H}}_I^{(0)}(\tau') U(\tau') d\tau',
\end{equation}
\begin{equation}
U(\tau) = \mathds{1} - \dfrac{1}{\hslash} \int_0^{\tau} \hat{\mathcal{H}}_I^{(0)}(\tau') U(\tau') d\tau',
\end{equation}
and we insert the above integral relation of $U(\tau)$ into $U(\tau')$ as follows
\begin{align}
U(\tau) &= \mathds{1} - \dfrac{1}{\hslash} \int_0^{\tau} \hat{\mathcal{H}}_I^{(0)}(\tau') \left[ \mathds{1} - \dfrac{1}{\hslash} \int_0^{\tau'} \hat{\mathcal{H}}_I^{(0)}(\tau'') U(\tau'') d\tau'' \right] d\tau' = \notag \\
&=  \mathds{1} - \dfrac{1}{\hslash} \int_0^{\tau} \hat{\mathcal{H}}_I^{(0)}(\tau') d\tau' + \left( - \dfrac{1}{\hslash} \right)^2 \int_0^{\tau} d\tau' \hat{\mathcal{H}}_I^{(0)}(\tau') \int_0^{\tau'} d\tau'' \hat{\mathcal{H}}_I^{(0)}(\tau'') U(\tau'') ,
\end{align}
and iterating, we construct a series. We now want to write the integral
\begin{equation}
J_2 = \int_0^{\tau} d\tau' \int_0^{\tau'} d\tau'' \hat{\mathcal{H}}_I^{(0)}(\tau') \hat{\mathcal{H}}_I^{(0)}(\tau'') 
\end{equation}
in a physically more expressive form in which the integration limits are decoupled. By means of the Dyson time-ordering operator, we write
\begin{equation}
J_2 = \int_0^{\tau} d\tau' \int_0^{\tau'} d\tau'' \hat{\mathcal{H}}_I^{(0)}(\tau') \hat{\mathcal{H}}_I^{(0)}(\tau'') = \int_0^{\tau} d\tau' \int_0^{\tau'} d\tau'' \hat{T}_D \left\lbrace \hat{\mathcal{H}}_I^{(0)}(\tau') \hat{\mathcal{H}}_I^{(0)}(\tau'') \right\rbrace.
\label{eq: primaregioneintegrazioneordinesecondointerazioneserieDyson}
\end{equation}
In Figure $\eqref{fig: primaregioneintegrazioneHtauprimoHtausecondo}$ the region and direction of the above integration are shown. This integral can also be calculated by reversing the order of integration, that is, according to the Figure $\eqref{fig: secondaregioneintegrazioneHtauprimoHtausecondo}$, i.e.,
\begin{equation}
J_2 = \int_0^{\tau} d\tau' \int_0^{\tau'} d\tau'' \hat{\mathcal{H}}_I^{(0)}(\tau') \hat{\mathcal{H}}_I^{(0)}(\tau'') = \int_0^{\tau} d\tau'' \int_{\tau''}^{\tau} d\tau' \hat{\mathcal{H}}_I^{(0)}(\tau') \hat{\mathcal{H}}_I^{(0)}(\tau''),
\label{eq: secondaregioneintegrazioneordinesecondointerazioneserieDyson}
\end{equation}
and given that $\tau'$ and $\tau''$ are dummy indices, they can be inverted in the last member of the $\eqref{eq: secondaregioneintegrazioneordinesecondointerazioneserieDyson}$ as follows
\begin{equation}
J_2 = \int_0^{\tau} d\tau' \int_{\tau'}^{\tau} d\tau'' \hat{\mathcal{H}}_I^{(0)}(\tau'') \hat{\mathcal{H}}_I^{(0)}(\tau') = \int_0^{\tau} d\tau' \int_{\tau'}^{\tau} d\tau'' \hat{T}_D \left\lbrace \hat{\mathcal{H}}_I^{(0)}(\tau') \hat{\mathcal{H}}_I^{(0)}(\tau'') \right\rbrace ,
\label{eq: terzaregioneintegrazioneordinesecondointerazioneserieDyson}
\end{equation}
where we have inserted the operator $\hat{T}_D$ since $\tau'' \geq \tau'$. Now the integral is computed in the triangle above the diagonal $\tau'=\tau''$, as in the Figure $\eqref{fig: terzaregioneintegrazioneHtauprimoHtausecondo}$. Consequently, the integral $J_2$ can be calculated in the following two equivalent ways
\begin{equation}
J_2 = \int_0^{\tau} d\tau' \int_0^{\tau'} d\tau'' \hat{T}_D \left\lbrace \hat{\mathcal{H}}_I^{(0)}(\tau') \hat{\mathcal{H}}_I^{(0)}(\tau'') \right\rbrace,
\label{eq: primointegraleJ2Dyson}
\end{equation}
\begin{equation}
J_2 = \int_0^{\tau} d\tau' \int_{\tau'}^{\tau} d\tau'' \hat{T}_D \left\lbrace \hat{\mathcal{H}}_I^{(0)}(\tau') \hat{\mathcal{H}}_I^{(0)}(\tau'') \right\rbrace,
\label{eq: secondointegraleJ2Dyson}
\end{equation}
and summing $\eqref{eq: primointegraleJ2Dyson}$ and $\eqref{eq: secondointegraleJ2Dyson}$, it follows
\begin{equation}
J_2 = \dfrac{1}{2} \int_0^{\tau} d\tau' \int_0^{\tau} d\tau'' \hat{T}_D \left\lbrace \hat{\mathcal{H}}_I^{(0)}(\tau') \hat{\mathcal{H}}_I^{(0)}(\tau'') \right\rbrace.
\end{equation}
Regarding $U(\tau)$, we perform iteration for $n$ integrals and send $n$ to infinity to compose the series
\begin{equation}
U(\tau) = \mathds{1} + \sum_{n=1}^{\infty} \left( - \dfrac{1}{\hslash} \right)^n \dfrac{1}{n!} \int_0^\tau d\tau_1 \ldots \int_0^\tau d\tau_n \ \hat{T}_\tau \left\lbrace \hat{\mathcal{H}}_I^{(0)}(\tau_1) \ldots \hat{\mathcal{H}}_I^{(0)}(\tau_n) \right\rbrace,
\label{eq: seriediDysonconsommatoria}
\end{equation}
which is equivalent to equation $\eqref{eq: seriediDysonconesponenziale}$. Note that the factor \(\frac{1}{2}\) generalizes to \(\frac{1}{n!}\) because the set of operators \(\left\lbrace \hat{\mathcal{H}}_I^{(0)}(\tau_1), \ldots, \hat{\mathcal{H}}_I^{(0)}(\tau_n) \right\rbrace\) can be reordered in \(n!\) different ways. Ultimately, note the Dyson time-ordering operator \(\hat{T}_D\) has been replaced by the Wick time-ordering operator \(\hat{T}_\tau\). 
\end{proof}
\end{theorem}
\begin{remark}[Wick and Dyson time ordering operators in interacting quantum many-body theory Hamiltonians]
For many-body Hamiltonians with two-body interactions, the time-ordering operators \(\hat{T}_D\) and \(\hat{T}_\tau\) coincide. This is because, in second quantization, each interaction term \(\hat{\mathcal{H}}_I\) includes an even number of creation and annihilation operators. When exchanging two interaction terms evaluated at different times, the total number of transpositions between operators is even. As a result, any signs arising from particle statistics cancel out. Therefore, the time-ordering procedure is independent of whether the particles are bosons or fermions, and we have \(\hat{T}_D \equiv \hat{T}_\tau\).
\end{remark}
Together with the transformation $U(\tau)$, we define the matrix $S(\tau,\tau')$ such that
\begin{equation}
\psi(\tau) = S(\tau,\tau') \psi(\tau').
\end{equation}
From $\psi(\tau) = U(\tau) \psi(0)$ and from
\begin{align}
\psi(\tau) &= S(\tau,\tau') \psi(\tau') = \notag \\
&= S(\tau,\tau') U(\tau') \psi(0) ,
\end{align}
we get
\begin{equation}
S(\tau,\tau') U(\tau') = U(\tau), \ \forall \ \psi(0) ,
\end{equation}
which implies
\begin{equation}
S(\tau,\tau') = U(\tau) U^{-1}(\tau').
\end{equation}
By definition it follows
\begin{equation}
S(\tau,0) = U(\tau),
\end{equation}
\begin{equation}
S(0,\tau) = U^{-1}(\tau).
\end{equation}
\begin{theorem}
The $S(\tau,\tau')$ matrix satisfies:
\begin{itemize}
\item [i)]
\begin{equation}
S(\tau,\tau) = \mathds{1},
\end{equation}
which is the initial condition for the matrix $S(\tau,\tau')$;
\item [ii)]
\begin{equation}
S(\tau_1,\tau_2) S(\tau_2,\tau_3) = S(\tau_1,\tau_3).
\end{equation}
\end{itemize}
\begin{proof}
Both statements follow directly from the definition of \( S(\tau,\tau') \). \newline
For $i)$, we have $S(\tau,\tau) = U(\tau) U^{-1}(\tau) = \mathds{1}$. \newline
For $ii)$, we compute
\begin{align}
S(\tau_1,\tau_2) S(\tau_2,\tau_3) &= U(\tau_1) U^{-1}(\tau_2) U(\tau_2) U^{-1}(\tau_3) = \notag \\
&= U(\tau_1) U^{-1}(\tau_3) \equiv \notag \\
&\equiv S(\tau_1,\tau_3).
\end{align}
\end{proof}
\end{theorem}
As a function of the Hamiltonian, the matrix $S(\tau,\tau')$ is written as
\begin{align}
S(\tau,\tau') &= U(\tau) U^{-1}(\tau') = \notag \\
&= e^{\frac{\hat{\mathcal{H}}_0}{\hslash} \tau} e^{- \frac{\hat{\mathcal{H}}}{\hslash} \tau} e^{\frac{\hat{\mathcal{H}}}{\hslash} \tau'} e^{- \frac{\hat{\mathcal{H}}_0}{\hslash} \tau'} = \notag \\
&= e^{\frac{\hat{\mathcal{H}}_0}{\hslash} \tau} e^{- \frac{\hat{\mathcal{H}}}{\hslash} (\tau-\tau')} e^{- \frac{\hat{\mathcal{H}}_0}{\hslash} \tau'}.
\end{align}
Analogously to the evolution operator \( U(\tau) \), the matrix \( S(\tau,\tau') \) satisfies the differential equation
\begin{equation}
\dfrac{\partial S(\tau,\tau')}{\partial \tau} = - \dfrac{1}{\hslash} \hat{\mathcal{H}}_I^{(0)}(\tau) S(\tau,\tau'),
\end{equation}
and, upon integration between \( \tau' \) and \( \tau \), it admits the formal series expansion
\begin{equation}
S(\tau,\tau') = \mathds{1} + \sum_{n=1}^{\infty} \left( - \dfrac{1}{\hslash} \right)^n \dfrac{1}{n!} \int_{\tau'}^{\tau} d\tau_1 \ldots \int_{\tau'}^{\tau} d\tau_n \ \hat{T}_\tau \left\lbrace \hat{\mathcal{H}}_I^{(0)}(\tau_1) \ldots \hat{\mathcal{H}}_I^{(0)}(\tau_n) \right\rbrace.
\end{equation}
In conclusion, we manipulate the thermal Heisenberg picture with respect to a generic Hamiltonian $\eqref{eq: HamiltonianascompostainH_0eH_I}$ by inserting the identity $\mathds{1} = e^{- \frac{\hat{\mathcal{H}}_0}{\hslash} \tau} e^{\frac{\hat{\mathcal{H}}_0}{\hslash} \tau}$ as follows
\begin{align}
A(\tau) &= e^{\frac{\hat{\mathcal{H}}}{\hslash} \tau} A e^{- \frac{\hat{\mathcal{H}}}{\hslash} \tau} = \notag \\[6pt]
&= e^{\frac{\hat{\mathcal{H}}}{\hslash} \tau} e^{- \frac{\hat{\mathcal{H}}_0}{\hslash} \tau} e^{\frac{\hat{\mathcal{H}}_0}{\hslash} \tau} A e^{-\frac{\hat{\mathcal{H}}_0}{\hslash} \tau} e^{\frac{\hat{\mathcal{H}}_0}{\hslash} \tau} e^{- \frac{\hat{\mathcal{H}}}{\hslash} \tau} = \notag \\[6pt]
&= U^{-1}(\tau) A^{(0)}(\tau) U(\tau) \equiv \notag \\[6pt]
&\equiv S(0,\tau) A^{(0)}(\tau) S(\tau,0).
\end{align}
The thermal Heisenberg representation $A(\tau)$ with respect to a generic Hamiltonian that includes interactions is written as a function of its thermal interaction representation $A^{(0)}(\tau)$ and matrix $S$.
\section{Time evolution of creation and annihilation operators in the thermal interaction picture}
Here we compute the thermal interaction picture of the annihilation and creation operators with respect to the Hamiltonian $\eqref{eq: HamiltonianascompostainH_0eH_I}$, as defined by $\eqref{eq: thermalinteractionpicture}$, within the framework of second quantization, i.e.,
\begin{equation}
a_\alpha^{(0)}(\tau) = e^{\frac{\hat{\mathcal{H}}_0}{\hslash} \tau} a_\alpha e^{- \frac{\hat{\mathcal{H}}_0}{\hslash} \tau} ,
\end{equation}
and let us compute its time first derivative, i.e.,
\begin{align}
\dfrac{d a_\alpha^{(0)}(\tau)}{dt} &= \dfrac{d \left( e^{\frac{\hat{\mathcal{H}}_0}{\hslash} \tau} a_\alpha e^{- \frac{\hat{\mathcal{H}}_0}{\hslash} \tau} \right)}{dt} = \notag \\
&= \dfrac{d e^{\frac{\hat{\mathcal{H}}_0}{\hslash} \tau}}{dt} a_\alpha e^{- \frac{\hat{\mathcal{H}}_0}{\hslash} \tau} + e^{\frac{\hat{\mathcal{H}}_0}{\hslash} \tau} \dfrac{d a_\alpha}{dt} e^{- \frac{\hat{\mathcal{H}}_0}{\hslash} \tau} + e^{\frac{\hat{\mathcal{H}}_0}{\hslash} \tau} a_\alpha \dfrac{d e^{- \frac{\hat{\mathcal{H}}_0}{\hslash} \tau}}{dt} = \notag \\
&= \dfrac{d e^{\frac{\hat{\mathcal{H}}_0}{\hslash} \tau}}{dt} a_\alpha e^{- \frac{\hat{\mathcal{H}}_0}{\hslash} \tau} + e^{\frac{\hat{\mathcal{H}}_0}{\hslash} \tau} a_\alpha \dfrac{d e^{- \frac{\hat{\mathcal{H}}_0}{\hslash} \tau}}{dt} = \notag \\
&= \dfrac{\hat{\mathcal{H}}_0}{\hslash} e^{\frac{\hat{\mathcal{H}}_0}{\hslash} \tau} a_\alpha e^{- \frac{\hat{\mathcal{H}}_0}{\hslash} \tau} - e^{\frac{\hat{\mathcal{H}}_0}{\hslash} \tau} a_\alpha \dfrac{\hat{\mathcal{H}}_0}{\hslash} e^{- \frac{\hat{\mathcal{H}}_0}{\hslash} \tau},
\end{align}
then
\begin{equation}
\dfrac{d a_\alpha^{(0)}(\tau)}{d\tau} = - \dfrac{1}{\hslash} \left[ a_\alpha,\hat{\mathcal{H}}_0 \right] ,
\end{equation}
and from the equation $\eqref{eq: commutatoreoperatorenumeroconoperatoredistruzione}$ it follows that the time evolution of the annihilation operator for imaginary times is given by
\begin{equation}
a_\alpha^{(0)}(\tau) = e^{- \frac{\mathcal{E}_\alpha}{\hslash} \tau} a_\alpha.
\label{eq: evoluzionetemporaleraprresentazioneHeisenbergtermicaoperatoredistruzione}
\end{equation}
Similarly, given the termal interaction picture of the creation operator $a^\dag_\alpha$ with respect to an Hamiltonian $\hat{\mathcal{H}}$ of the form $\eqref{eq: HamiltonianascompostainH_0eH_I}$ within the framework of the second quantization, i.e.,
\begin{equation}
a^{\dagger (0)}_\alpha(\tau) = e^{\frac{\hat{\mathcal{H}}_0}{\hslash} \tau} a^\dag_\alpha e^{- \frac{\hat{\mathcal{H}}_0}{\hslash} \tau},
\end{equation}
then, similarly to the annihilation operator, its time evolution for imaginary times is described by
\begin{equation}
\dfrac{d a^{\dagger (0)}_\alpha(\tau)}{d\tau} = - \dfrac{1}{\hslash} \left[ a^\dag_\alpha,\hat{\mathcal{H}}_0 \right],
\end{equation}
and again, from the equation $\eqref{eq: commutatoreoperatorenumeroconoperatorecreazione}$ it follows that the time evolution of the creation operator for imaginary times is given by
\begin{equation}
a_\alpha^{\dagger (0)}(\tau) = e^{\frac{\mathcal{E}_\alpha}{\hslash} \tau} a^\dag_\alpha.
\label{eq: evoluzionetemporaleraprresentazioneHeisenbergtermicaoperatorecreazione}
\end{equation}
Consequently, in the thermal interaction representation, because of the real exponentials, the annihilation and creation operators are no longer self-adjoint to each other.
\begin{remark}[Finite average life times]
For interacting particles, the time evolution of the annihilation and creation operators \(a_\alpha\) and \(a_\alpha^\dagger\) in the thermal interaction picture involves exponential factors containing generally complex energies of the form \(\mathcal{E}_\alpha + i \delta\). The real part \(\mathcal{E}_\alpha\) corresponds to the excitation energy, while the imaginary part \(i \delta\) introduces a damping factor \(e^{-\delta t}\). Physically, this damping represents the finite lifetime of quasiparticle excitations, arising from interaction-induced decay processes and scattering events within the system. Such a finite lifetime implies that excitations are not stable but decay over time, leading to a broadening of energy levels. Quantitatively, the decay rate \(\delta\) and related lifetime can be computed systematically by means of Feynman’s perturbation theory, which accounts for the interactions through self-energy corrections in the Green's functions framework.
\end{remark}
\section{Matsubara Green's function}
The thermal Heisenberg representation of $A$ at time $\tau$, that is $A(\tau)$, is defined as
\begin{equation}
A(\tau) = e^{\frac{\hat{\mathcal{H}}}{\hslash} \tau} A e^{- \frac{\hat{\mathcal{H}}}{\hslash} \tau}, \ \tau \in \left[ - \beta \hslash , \beta \hslash \right], \ \tau = i t \in \mathbb{R}, \ t \in \im \ \mathbb{C} ,
\end{equation}
which is a Heisenberg representation for imaginary times $t$. For reasons that will become clear shortly, we define the Matsubara Green's function (or thermal Green's function) of the operators $A$ and $B$ at times $\tau'$ and $\tau'$, i.e.,
\begin{equation}
G^{(m)}_{AB}(\tau',\tau'') = - \dfrac{1}{\hslash} \left\langle \hat{T}_\tau A(\tau')B(\tau'') \right\rangle .
\label{eq: funzioneGreenMastubara1}
\end{equation}
Theorem \ref{Time-translation invariance of Green's functions in thermal equilibrium} also applies to the Matsubara Green's function. Consequently, from now on, whenever the system is in thermodynamic equilibrium, we will write it concisely as
\begin{align}
G^{(m)}_{AB}(\tau,0) &\equiv G^{(m)}_{AB}(\tau - 0) \equiv \notag \\
&\equiv G^{(m)}_{AB}(\tau) = \notag \\
&= - \frac{1}{\hslash} \left\langle \hat{T}_\tau A(\tau) B(0) \right\rangle = \notag \\
&= - \frac{1}{\hslash} \left\langle \Theta(\tau) A(\tau) B(0) + \varepsilon \, \Theta(-\tau) B(0) A(\tau) \right\rangle = \notag \\
&= - \frac{1}{\hslash} \left[ \Theta(\tau) \left\langle A(\tau) B(0) \right\rangle + \varepsilon \, \Theta(-\tau) \left\langle B(0) A(\tau) \right\rangle \right].
\label{eq: funzioneGreenMastubara2}
\end{align}
where the action of \(\hat{T}_{\tau}\) has been made explicit.
\subsection{Properties of bosonic and fermionic Matsubara Green's functions}
Consider the Matsubara Green's function for generic operators \(A\) and \(B\). Given the instant \(\tau - \beta \hslash\), with \(\tau \in (0, \beta \hslash]\), we evaluate the thermal average of the second term in equation $\eqref{eq: funzioneGreenMastubara2}$, that is,
\begin{align}
G_{AB}^{(m)}\left(\tau - \beta \hslash\right) &= -\dfrac{\varepsilon}{\hslash} \Tr \left[\dfrac{e^{-\beta \hat{\mathcal{H}}}}{Z} B \, e^{\frac{\hat{\mathcal{H}}}{\hslash}(\tau - \beta \hslash)} A \, e^{-\frac{\hat{\mathcal{H}}}{\hslash}(\tau - \beta \hslash)}\right] = \notag \\
&= -\dfrac{\varepsilon}{\hslash} \Tr \left[\dfrac{e^{-\beta \hat{\mathcal{H}}}}{Z} \, e^{\frac{\hat{\mathcal{H}}}{\hslash} \tau} A \, e^{-\frac{\hat{\mathcal{H}}}{\hslash} \tau} \, e^{\beta \hat{\mathcal{H}}} \, e^{- \beta \hat{\mathcal{H}}} B \right] = \notag \\
&= -\dfrac{\varepsilon}{\hslash} \Tr \left[\dfrac{e^{-\beta \hat{\mathcal{H}}}}{Z} \, e^{\frac{\hat{\mathcal{H}}}{\hslash} \tau} A \, e^{-\frac{\hat{\mathcal{H}}}{\hslash} \tau} B \right] = \notag \\
&= \varepsilon G_{AB}^{(m)}(\tau).
\label{eq: traslazionetaudiGMatsubara}
\end{align}
Then, if $\tau - \beta \hslash \in (-\beta \hslash,0]$, i.e., if $\tau \in (0,\beta \hslash]$, the thermal Green's function computed at time $\tau-\beta\hslash$ is equal to the thermal Green's function evaluated at $\tau \in (0,\beta \hslash]$, multiplied by $\varepsilon$. It follows
\begin{equation}
G_{AB}^{(m)}\left(\tau - \beta \hslash\right) = G_{AB}^{(m)}(\tau), \ \tau \in (0,\beta \hslash], \ \ \ \text{(bosons)}
\end{equation}
that is, the bosonic thermal Green's function is a periodic function of period $\beta \hslash$, and
\begin{equation}
G_{AB}^{(m)}(\tau - \beta \hslash) = - G_{AB}^{(m)}(\tau), \ \tau \in (0,\beta \hslash], \ \ \ \text{(fermions)}
\end{equation}
that is, the fermionic thermal Green's function is an antisymmetric, periodic function of period $\beta \hslash$. The calculations involving Matsubara Green's functions can be consistently carried out by restricting the variable $\tau$ to the interval \((0, \beta \hslash]\), where the formalism is well-defined and periodic (or antiperiodic) boundary conditions apply. \newline
Starting from the Fourier series expansion formula $\eqref{eq: svilupposerieFourier}$, adapted to the interval $[-L, L]$ instead of $[-\frac{L}{2}, \frac{L}{2}]$, and taking into account that the Matsubara Green's function $G_{AB}^{(m)}(\tau)$ is periodic on the interval $\left[-\beta \hslash, \beta \hslash\right]$, we can expand it in a Fourier series over any interval of width $2 \beta \hslash$ as
\begin{equation}
G_{AB}^{(m)}(\tau) = \dfrac{1}{\beta \hslash} \sum_{n \in \mathbb{Z}} G_{AB}^{(m)}(i\omega_n) e^{-i \omega_n \tau},
\end{equation}
\begin{equation}
G_{AB}^{(m)}(i\omega_n) = \dfrac{1}{2} \int_{-\beta \hslash}^{\beta \hslash} d\tau G_{AB}^{(m)}(\tau) e^{i \omega_n \tau},
\end{equation}
\begin{equation}
\omega_n = \dfrac{n \pi}{\beta \hslash}, \ n \in \mathbb{Z}.
\end{equation}
From
\begin{equation}
G_{AB}(i\omega_n) = \dfrac{1}{2} \left[\int_{-\beta \hslash}^0 d\tau G_{AB}^{(m)}(\tau) e^{i \omega_n \tau} + \int_0^{\beta \hslash} d\tau G_{AB}^{(m)}(\tau) e^{i \omega_n \tau} \right] ,
\end{equation}
we compute the first integral. We set $\tau' = \tau + \beta \hslash$ ($\tau' \in ]0,\beta \hslash[$) and use equation $\eqref{eq: traslazionetaudiGMatsubara}$, then
\begin{align}
G_{AB}(i\omega_n) &= \dfrac{1}{2} \left[\int_0^{\beta \hslash} d\tau' G_{AB}^{(m)}(\tau'-\beta \hslash) e^{i \omega_n (\tau' - \beta \hslash)} + \int_0^{\beta \hslash} d\tau G_{AB}^{(m)}(\tau) e^{i \omega_n \tau} \right] = \notag \\
&= \dfrac{1}{2} \left[\int_0^{\beta \hslash} d\tau' \varepsilon G_{AB}^{(m)}(\tau') e^{i \omega_n (\tau' - \beta \hslash)} + \int_0^{\beta \hslash} d\tau G_{AB}^{(m)}(\tau) e^{i \omega_n \tau} \right] ,
\end{align}
and we rename the dummy index $\tau'$, i.e.,
\begin{align}
G_{AB}^{(m)}(i\omega_n) &= \dfrac{1}{2} \int_0^{\beta \hslash} d \tau G_{AB}^{(m)}(\tau) e^{i \omega_n \tau}\left[1 + \varepsilon e^{-i\omega_n \beta \hslash} \right] \equiv \notag \\
& \equiv \dfrac{1}{2} \int_0^{\beta \hslash} d \tau G_{AB}^{(m)}(\tau) e^{i \omega_n \tau}\left[1 + \varepsilon e^{-i n \pi} \right].
\end{align}
There are two cases, depending on the statistical nature of the particles. For bosons, if $n$ is even, the factor $\frac{1}{2}$ is simplified by $2$, if $n$ is odd the Green's function $G_{AB}^{(m)}(i\omega_n)$ is null: only the even frequencies are non-null and the factor $\frac{1}{2}$ disappears. For fermions, if $n$ is odd, the factor $\frac{1}{2}$ simplifies to $2$, if $n$ is even, the Green's function $G_{AB}^{(m)}(i\omega_n)$ is null: only the odd frequencies are non-null and the factor $\frac{1}{2}$ disappears. Any Matsubara Green's function can be written as
\begin{equation}
G_{AB}^{(m)}(\tau) = \dfrac{1}{\beta \hslash} \sum_{n \in \mathbb{Z}} G_{AB}^{(m)}(i \omega_n) e^{- i \omega_n \tau},
\label{eq: funzioneGreenMatsubaraserieFourier1}
\end{equation}
with
\begin{equation}
G_{AB}^{(m)}(i\omega_n) = \int_0^{\beta \hslash}d \tau G_{AB}^{(m)}(\tau) e^{i \omega_n \tau},
\label{eq: funzioneGreenMatsubaraserieFourier2}
\end{equation}
where the frequencies
\begin{equation}
i \omega_n = 
\begin{cases}
\displaystyle \frac{i 2n\pi}{\beta \hslash}, & \text{bosons} \\
\displaystyle \frac{i(2n+1)\pi}{\beta \hslash}, & \text{fermions}
\end{cases}
\end{equation}
belong to the imaginary axis. Finally, note that the statistical index $\eqref{eq: epsilonfermionibosoni}$ satisfies
\begin{equation}
e^{i \omega_n \beta \hslash} \equiv \varepsilon .
\label{eq: varepsilonefrequenzebosonfermion}
\end{equation}
\section{Spectral function and analytical extension of Green's function}
The thermal Green function $G_{AB}^{(m)}(i\omega_n)$ is defined on the imaginary frequency axis, but our interest lies in its behavior on the real frequency axis. Let $\hat{\mathcal{H}}$ be a Hamiltonian that includes interactions among particles, and suppose that $\hat{\mathcal{H}}$ has been diagonalized, i.e., we assume its eigenvalues $\lbrace \mathcal{E}_n \rbrace$ and corresponding eigenstates $\lbrace |\psi_n\rangle \rbrace \equiv \lbrace |n\rangle \rbrace$ are known. We express the retarded Green's function for two generic operators $A$ and $B$ in terms of this eigenbasis as follows
\begin{align}
G_{AB}^{(r)}(t) &= - \dfrac{i}{\hslash} \, \Theta(t) \, \dfrac{1}{Z} \, \Tr \left\lbrace e^{-\beta \hat{\mathcal{H}}} \left[ e^{i \frac{\hat{\mathcal{H}}}{\hslash} t} A \, e^{-i \frac{\hat{\mathcal{H}}}{\hslash} t} \, B - \varepsilon \, B \, e^{i \frac{\hat{\mathcal{H}}}{\hslash} t} A \, e^{-i \frac{\hat{\mathcal{H}}}{\hslash} t} \right] \right\rbrace = \notag \\
&= - \dfrac{i}{\hslash} \, \Theta(t) \, \dfrac{1}{Z} \sum_{n,m} \Big[ \langle n | e^{-\beta \hat{\mathcal{H}}} e^{i \frac{\hat{\mathcal{H}}}{\hslash} t} A e^{-i \frac{\hat{\mathcal{H}}}{\hslash} t} | m \rangle \langle m | B | n \rangle - \varepsilon \, \langle n | B | m \rangle \langle m | e^{-\beta \hat{\mathcal{H}}} e^{i \frac{\hat{\mathcal{H}}}{\hslash} t} A e^{-i \frac{\hat{\mathcal{H}}}{\hslash} t} | n \rangle \Big] = \notag \\
&= - \dfrac{i}{\hslash} \, \Theta(t) \, \dfrac{1}{Z} \sum_{n,m} \left[ \langle n | A | m \rangle \langle m | B | n \rangle \, e^{-\beta \mathcal{E}_n} \, e^{i \frac{\mathcal{E}_n - \mathcal{E}_m}{\hslash} t} \right. \left. - \, \varepsilon \, \langle n | B | m \rangle \langle m | A | n \rangle \, e^{-\beta \mathcal{E}_m} \, e^{i \frac{\mathcal{E}_m - \mathcal{E}_n}{\hslash} t} \right].
\end{align}
we exchange the dummy indices $n$, $m$ in the second term and we get
\begin{align}
G_{AB}^{(r)}(t) &= -\dfrac{i}{\hslash} \, \Theta(t) \, \dfrac{1}{Z} \sum_{n,m} \langle n|A|m\rangle \langle m|B|n\rangle \, e^{i \frac{\mathcal{E}_n - \mathcal{E}_m}{\hslash} t} \left[ e^{-\beta \mathcal{E}_n} - \varepsilon \, e^{-\beta \mathcal{E}_m} \right] \equiv \notag \\
&\equiv -\dfrac{i}{\hslash} \, \Theta(t) \, \dfrac{1}{Z} \sum_{n,m} A_{nm} B_{mn} \, e^{i \frac{\mathcal{E}_n - \mathcal{E}_m}{\hslash} t} \left[ e^{-\beta \mathcal{E}_n} - \varepsilon \, e^{-\beta \mathcal{E}_m} \right],
\end{align}
The retarded Green's function at the thermodynamic equilibrium of $A$ and $B$ can be written in the form 
\begin{equation}
G_{AB}^{(r)}(t) = \Theta(t) C(t),
\label{eq: funzionedigreencomeprodottothetaHeavisideefunzionecorrelazione}
\end{equation}
where
\begin{equation}
C(t) = -\dfrac{i}{\hslash} \left\langle [A(t),B(0)]^{(\varepsilon)} \right\rangle
\label{eq: funzionedicorrelazioneinpropagatoremanybodies}
\end{equation}
is the correlation function of $A$ and $B$ at thermodynamic equilibrium. For $t \rightarrow \infty$, $C(t)$ is zero or constant. Let us pose $z = \omega + i \delta$, $\omega, \delta \in \mathbb{R}$ and apply a Fourier transform, i.e.,
\begin{align}
G_{AB}^{(r)}(z) &= \int_{-\infty}^{+\infty} dt \ G_{AB}^{(r)}(t) e^{i z t} \ \equiv \notag \\
&\equiv \int_{0}^{+\infty} dt \ G_{AB}^{(r)}(t) e^{i\omega t}e^{- \delta t},
\end{align}
where at the last step we used Heaviside's function \( \Theta(t) \). Conversely, if we know \( G_{AB}^{(\mathrm{ret})}(z) \), to calculate \( G_{AB}^{(\mathrm{ret})}(t) \) it is sufficient to apply the Fourier antitransform, i.e.,
\begin{equation}
G_{AB}^{(r)}(t) = \int_{-\infty}^{+\infty} \dfrac{d\omega}{2\pi} G_{AB}^{(r)}(\omega + i \delta) e^{-i(\omega + i \delta)t}, \ \forall \ \delta > 0 ,
\end{equation}
where the prescription \( \omega + i \delta \) ensures the correct analytic structure in the upper half of the complex frequency plane. If the eigenstates and eigenvalues of \( \hat{\mathcal{H}} \) were known, the Green's function could be explicitly expressed in the energy eigenbasis as
\begin{align}
G_{AB}^{(r)}(z) &= \int_{0}^{\infty} dt \ G_{AB}^{(r)}(t) e^{i z t} = \notag \\
&= - \dfrac{i}{\hslash} \dfrac{1}{Z} \sum_{n,m} A_{nm}B_{mn} \left( e^{-\beta \mathcal{E}_n}-\varepsilon e^{-\beta \mathcal{E}_m} \right) \int_0^\infty dt \ e^{i z t} e^{i \frac{\mathcal{E}_n-\mathcal{E}_m}{\hslash} t}.
\end{align}
For every $\delta > 0$, $e^{- \delta t}$ guarantees that $G_{AB}^{(r)}(z)$ is analytic in the upper complex half-plane, indeed
\begin{align}
G_{AB}^{(r)}(z) &= - \dfrac{i}{\hslash Z} \sum_{n,m} A_{nm}B_{mn} \left( e^{-\beta \mathcal{E}_n}-\varepsilon e^{-\beta \mathcal{E}_m} \right) \int_0^\infty dt \ e^{i \left( z + \frac{\mathcal{E}_n-\mathcal{E}_m}{\hslash} \right) t} = \notag \\
&= - \dfrac{i}{\hslash Z} \sum_{n,m} A_{nm}B_{mn} \left( e^{-\beta \mathcal{E}_n}-\varepsilon e^{-\beta \mathcal{E}_m} \right) \int_0^\infty dt \ \dfrac{i \left( z + \frac{\mathcal{E}_n-\mathcal{E}_m}{\hslash} \right)}{i \left( z + \frac{\mathcal{E}_n-\mathcal{E}_m}{\hslash} \right)} e^{i \left( z + \frac{\mathcal{E}_n-\mathcal{E}_m}{\hslash} \right) t} = \notag \\
&= - \dfrac{i}{\hslash Z} \sum_{n,m} A_{nm}B_{mn} \left( e^{-\beta \mathcal{E}_n}-\varepsilon e^{-\beta \mathcal{E}_m} \right) \dfrac{1}{i \left( z + \frac{\mathcal{E}_n-\mathcal{E}_m}{\hslash} \right)} \left[ e^{i \left( z + \frac{\mathcal{E}_n-\mathcal{E}_m}{\hslash} \right) t} \right]_0^{\infty} = \notag \\
&= - \dfrac{i}{\hslash Z} \sum_{n,m} A_{nm}B_{mn} \left( e^{-\beta \mathcal{E}_n}-\varepsilon e^{-\beta \mathcal{E}_m} \right) \dfrac{1}{i \left( z + \frac{\mathcal{E}_n-\mathcal{E}_m}{\hslash} \right)} \left[ e^{i \left( \omega + \frac{\mathcal{E}_n-\mathcal{E}_m}{\hslash} \right) t} e^{- \delta t} \right]_0^{\infty} = \notag \\
&= - \dfrac{1}{\hslash Z} \sum_{n,m} A_{nm}B_{mn} \left( e^{-\beta \mathcal{E}_n}-\varepsilon e^{-\beta \mathcal{E}_m} \right) \dfrac{1}{\left( z + \frac{\mathcal{E}_n-\mathcal{E}_m}{\hslash} \right)} \left[ e^{i \left( \omega + \frac{\mathcal{E}_n-\mathcal{E}_m}{\hslash} \right) t} e^{- \delta t} \right]_0^{\infty}.
\end{align}
The last term is the product of an oscillating factor for a decreasing monotone function, in the limit for $t \rightarrow \infty$ tends to zero, so
\begin{align}
G_{AB}^{(r)}(z) &= - \dfrac{1}{\hslash Z} \sum_{n,m} A_{nm}B_{mn} \left( e^{-\beta \mathcal{E}_n}-\varepsilon e^{-\beta \mathcal{E}_m} \right) \dfrac{1}{\left( z + \frac{\mathcal{E}_n-\mathcal{E}_m}{\hslash} \right)} (0-1) = \notag \\
&= \dfrac{1}{\hslash Z} \sum_{n,m} A_{nm}B_{mn} \left( e^{-\beta \mathcal{E}_n}-\varepsilon e^{-\beta \mathcal{E}_m} \right) \dfrac{1}{\left( z + \frac{\mathcal{E}_n-\mathcal{E}_m}{\hslash} \right)} = \notag \\
&= \dfrac{1}{Z} \sum_{n,m} A_{nm}B_{mn} \dfrac{e^{-\beta \mathcal{E}_n} - \varepsilon e^{-\beta \mathcal{E}_m}}{\hslash \left( z + \frac{\mathcal{E}_n - \mathcal{E}_m}{\hslash}\right)} ,
\label{eq: trasformatadiFourierdellafunzionediGreenritardatanellabasedegliautostati}
\end{align}
and the Green's function is complex or not depending on the denominator $\left( z + \frac{\mathcal{E}_n - \mathcal{E}_m}{\hslash} \right)$. We write $G_{AB}^{(r)}(z)$ in integral form
\begin{align}
G_{AB}^{(r)}(z) = \int_{-\infty}^\infty \dfrac{d\omega'}{2 \pi}\frac{A_{AB}(\omega ')}{\hslash(z - \omega')},
\end{align}
and by comparison we get
\begin{equation}
A_{AB}(\omega') = \dfrac{2\pi}{Z} \sum_{n,m} A_{nm}B_{mn}(e^{-\beta \mathcal{E}_n } - \varepsilon e^{-\beta \mathcal{E}_m}) \delta \left[ \omega' - \frac{\mathcal{E}_m - \mathcal{E}_n}{\hslash} \right],
\label{eq: funzionespettrale}
\end{equation}
which is called the spectral function of the operators $A$ and $B$. In the most physically relevant case, one considers \( B = A^\dagger \), yielding
\begin{equation}
A_{AA^\dag}(\omega') = \dfrac{2\pi}{Z} \sum_{n,m} |A_{nm}|^2 (e^{-\beta \mathcal{E}_n } - \varepsilon e^{-\beta \mathcal{E}_m}) \delta \left[ \omega' - \frac{\mathcal{E}_m - \mathcal{E}_n}{\hslash} \right].
\end{equation}
If \( A \) is taken to be the annihilation operator \( C_{\mathbf{k}} \) and \( B = C_{\mathbf{k}}^\dagger \), the resulting function is the one-particle spectral function \( A(\mathbf{k}, \omega) \), which describes the probability of removing (for \( \omega < \mu \)) or adding (for \( \omega > \mu \)) a particle with momentum \( \mathbf{k} \) and energy \( \omega \). The spectral function \( A(\mathbf{k}, \omega) \) plays a central role in many-body theory, as it encodes the probability of adding or removing a particle with momentum \( \mathbf{k} \) and energy \( \omega \). It reflects the intrinsic electronic structure of the system, including renormalizations due to interactions and the finite lifetime of quasiparticles. Importantly, the spectral function is directly connected to experimental observables. In particular, it is accessible via Angle-Resolved Photoemission Spectroscopy (ARPES), a powerful technique for probing the electronic structure of materials. ARPES involves illuminating a material with ultraviolet or X-ray photons to eject electrons from its surface. By measuring the kinetic energy and emission angle of the photoemitted electrons, one reconstructs their in-plane momentum and binding energy. The resulting photoemission intensity is, under suitable approximations, proportional to the product \( A(\mathbf{k}, \omega) n_{-1}(\hbar \omega) \), where \( n_{-1}(\hbar \omega) \) is the Fermi-Dirac distribution. This enables direct access to the occupied part of the spectral function, making ARPES an invaluable tool for studying dispersions, energy gaps, correlation effects, and emergent many-body phenomena in quantum materials. \newline
As for the thermal Green's function, suppose we know eigenvalues and eigenfunctions of the Hamiltonian with interactions and consider times $\tau>0$: the time ordering operator can be omitted and we get
\begin{align}
G_{AB}^{(m)}(\tau) &= - \dfrac{1}{\hslash} \left\langle A(\tau)B(0) \right\rangle = \notag \\
&= - \dfrac{1}{\hslash} \text{Tr} \left[ \dfrac{e^{-\beta \mathcal{H}}e^{\frac{\mathcal{H}}{\hslash} \tau} A e^{- \frac{\mathcal{H}}{\hslash} \tau} B}{Z}\right] = \notag \\
&= - \dfrac{1}{\hslash Z} \sum_{n,m} A_{nm}B_{mn} e^{-\beta \mathcal{E}_n}e^{\frac{\mathcal{E}_n-\mathcal{E}_m}{\hslash} \tau}.
\end{align} 
From \( G_{AB}^{(r)}(t) \), we obtained \( G_{AB}^{(r)}(z) \) in the upper half of the complex plane. We now apply the Fourier transform to the Matsubara Green's function \( G_{AB}^{(m)}(\tau) \) as follows
\begin{align}
G_{AB}^{(m)}(i\omega_n) &= \int_0^{\beta \hslash} G_{AB}^{(m)}(\tau) e^{i \omega_n \tau}d\tau = \notag \\
&= - \dfrac{1}{\hslash Z} \sum_{n,m} A_{nm}B_{mn} e^{- \beta \mathcal{E}_n}\int_0^{\beta \hslash} d\tau \ e^{i\omega_n \tau}e^{\frac{\mathcal{E}_n-\mathcal{E}_m}{\hslash} \tau} = \notag \\
&= - \dfrac{1}{\hslash Z} \sum_{n,m} A_{nm}B_{mn} e^{-\beta \mathcal{E}_n} \dfrac{e^{i \omega_n \beta \hslash}e^{\beta(\mathcal{E}_n-\mathcal{E}_m)}-1}{i\omega_n + \frac{\mathcal{E}_n - \mathcal{E}_m}{\hslash}} = \notag \\
&= - \dfrac{1}{\hslash Z} \sum_{n,m} A_{nm}B_{mn} e^{-\beta \mathcal{E}_n} \dfrac{\varepsilon e^{\beta(\mathcal{E}_n-\mathcal{E}_m)}-1}{i\omega_n + \frac{\mathcal{E}_n - \mathcal{E}_m}{\hslash}} = \notag \\
&= - \dfrac{1}{\hslash Z} \sum_{n,m} A_{nm}B_{mn} \dfrac{\varepsilon e^{-\beta \mathcal{E}_n}e^{\beta(\mathcal{E}_n-\mathcal{E}_m)} - e^{-\beta \mathcal{E}_n}}{i\omega_n + \frac{\mathcal{E}_n - \mathcal{E}_m}{\hslash}} = \notag \\
&= \dfrac{1}{\hslash Z} \sum_{n,m} A_{nm}B_{mn} \dfrac{e^{-\beta \mathcal{E}_n} - \varepsilon e^{-\beta \mathcal{E}_n}e^{\beta(\mathcal{E}_n-\mathcal{E}_m)}}{i\omega_n + \frac{\mathcal{E}_n - \mathcal{E}_m}{\hslash}} \equiv \notag \\
&\equiv \dfrac{1}{Z} \sum_{n,m} A_{nm}B_{mn} \dfrac{e^{-\beta \mathcal{E}_n} - \varepsilon e^{-\beta \mathcal{E}_m}}{\hslash \left( i\omega_n + \frac{\mathcal{E}_n - \mathcal{E}_m}{\hslash} \right)} ,
\end{align}
where we used equation $\eqref{eq: varepsilonefrequenzebosonfermion}$. We now express \( G^{(m)}_{AB}(i\omega_n) \) in integral form:
\begin{align}
G_{AB}^{(m)}(i \omega_n) = \int_{-\infty}^\infty \frac{d\omega'}{2\pi} \dfrac{\tilde{A}_{AB}(\omega')}{\hslash( i\omega_n - \omega')},
\end{align}
from which by comparison, $\tilde{A}_{AB}(\omega')$ is the spectral function found above, then
\begin{equation}
G_{AB}^{(m)}(i\omega_n) = \int_{-\infty}^\infty \frac{d\omega'}{2\pi}\dfrac{A_{AB}(\omega')}{\hslash( i\omega_n - \omega')}.
\end{equation}
On the other hand, the Fourier transform of the retarded many-body Green's function is given by
\begin{equation}
G_{AB}^{(r)}(z) =  \int_{-\infty}^{+\infty} \dfrac{d\omega'}{2\pi} \dfrac{A_{AB}(\omega')}{\hslash (z - \omega')}, \ z = \omega + i \delta, \ \delta>0 ,
\label{eq: integraledellafunzionediGreeninterminidellafunzionespettrale}
\end{equation}
that is, the spectral function of the Fourier transform of the retarded Green's function \( G_{AB}^{(r)}(z) \) coincides with that of the Fourier transform of the Matsubara Green's function \( G_{AB}^{(m)}(i\omega_n) \), as given by equation $\eqref{eq: funzionespettrale}$. According to the Abrikosov-Gor'kov theorem, if \( G_{AB}^{(m)}(i\omega_n) \) is known on the imaginary frequency axis \( i\omega_n \), it can be analytically continued to the upper half of the complex plane. We do not provide a proof of this theorem here. Our focus is on the Fourier transform of the Matsubara Green's function evaluated along the real frequency axis. To connect the Matsubara Green's function defined on the imaginary axis to values on the real axis, we consider complex frequencies of the form $z = \omega + i \delta$, $\omega, \delta \in \mathbb{R}$, $\delta > 0$ and then take the limit
\begin{equation}
G_{AB}^{(r)}(\omega) = \lim_{\delta \rightarrow 0^+} G^{(m)}_{AB}(\omega + i \delta).
\end{equation}
\begin{theorem}[Exact relation for the spectral function] Let $A$ and $B$ be two generic operators. Then the spectral function \( A_{AB}(\omega) \) satisfies the following exact identity
\begin{equation}
\int_{-\infty}^{+\infty} \dfrac{d\omega}{2\pi} \dfrac{A_{AB}(\omega)}{1-\varepsilon e^{-\beta \hslash \omega}} = \ \langle AB \rangle,
\label{eq: mediaprodottooperatoricalcolataconfunzionespettrale}
\end{equation}
where \(\langle \ldots \rangle\) denotes the thermal average.
\begin{proof}
We compute the integral on the left-hand side of equation $\eqref{eq: mediaprodottooperatoricalcolataconfunzionespettrale}$. Using the explicit expression of the spectral function, that is, equation $\eqref{eq: funzionespettrale}$, we obtain,
\begin{align}
\int_{-\infty}^{+\infty} \dfrac{d\omega}{2\pi} \dfrac{A_{AB}(\omega)}{1-\varepsilon e^{-\beta \hslash \omega}} &= \dfrac{2\pi}{Z} \sum_{n,m} A_{nm} B_{mn}\left(e^{-\beta \mathcal{E}_n} - \varepsilon e^{-\beta \mathcal{E}_m} \right) \int_{-\infty}^{+\infty} \dfrac{d\omega}{2\pi} \dfrac{1}{1-\varepsilon e^{-\beta \hslash \omega}} \delta \left[ \omega - \dfrac{\mathcal{E}_m - \mathcal{E}_n}{\hslash} \right] = \notag \\
&= \dfrac{2\pi}{Z} \sum_{n,m} A_{nm} B_{mn} \left( e^{-\beta \mathcal{E}_n} - \varepsilon e^{-\beta \mathcal{E}_m}\right) \dfrac{1}{2\pi} \dfrac{1}{1-\varepsilon e^{-\beta \hslash \left(\frac{\mathcal{E}_m - \mathcal{E}_n}{\hslash}\right)}} = \notag \\
&= \dfrac{1}{Z} \sum_{n,m} A_{nm} B_{mn} e^{-\beta \mathcal{E}_n} \left(1 - \varepsilon e^{-\beta(\mathcal{E}_m - \mathcal{E}_n)} \right) \dfrac{1}{1-\varepsilon e^{-\beta \left(\mathcal{E}_m - \mathcal{E}_n\right)}} \equiv \notag \\
&\equiv \sum_{n,m} \dfrac{e^{-\beta \mathcal{E}_n}}{Z} \langle n|A|m \rangle \langle m|B|n \rangle \ = \notag \\
&= \sum_{n} \dfrac{e^{-\beta \mathcal{E}_n}}{Z} \langle n|AB|n \rangle,
\end{align}
which is the thesis.
\end{proof}
\end{theorem}
\section{Many-body Green's functions in real and frequency space for translationally invariant systems}
Given a system of free particles described by the Hamiltonian \(\hat{\mathcal{H}}_0\) at thermodynamic equilibrium, coupled to an external bath represented by the interaction Hamiltonian \(\hat{\mathcal{H}}_I\), we compute the free Green's function for the annihilation and creation field operators, namely
\begin{equation}
G^{(m)(0)}_{x,x'}(\tau) = - \dfrac{1}{\hslash} \left\langle \hat{T}_\tau \hat{\psi}^{(0)}(x,\tau) \hat{\psi}^\dagger(x',0) \right\rangle_0.
\end{equation}
We expand the field operators in the basis of the eigenfunctions of $\mathcal{H}_0$ and assume for simplicity $\tau>0$, then
\begin{align}
G^{(m)(0)}_{x,x'}(\tau) &= - \dfrac{1}{\hslash} \sum_{\alpha,\alpha'} \varphi_\alpha(x) \varphi^*_{\alpha'}(x') \left\langle \hat{T}_\tau a^{(0)}_\alpha(\tau) a^\dagger_{\alpha'} \right\rangle_0 \ = \notag \\
&= - \dfrac{1}{\hslash} \sum_{\alpha,\alpha'} \varphi_\alpha(x) \varphi^*_{\alpha'}(x') e^{- \frac{\mathcal{E}_\alpha \tau}{\hslash}} \langle a_\alpha a^\dagger_{\alpha'} \rangle_0.
\end{align}
The terms obtained by summing over $\alpha$ with the condition $\alpha' \neq \alpha$ are given by
\begin{equation}
\sum_{\substack{\alpha' \\ \alpha' \neq \alpha}} \varphi_\alpha(x) \varphi^*_{\alpha'}(x') e^{- \frac{\mathcal{E}_\alpha \tau}{\hslash}} \langle a_\alpha a^\dagger_{\alpha'} \rangle_0 = \sum_{\substack{\alpha' \\ \alpha' \neq \alpha}} \varphi_\alpha(x) \varphi^*_{\alpha'}(x') e^{- \frac{\mathcal{E}_\alpha \tau}{\hslash}} \langle \varepsilon a^\dagger_{\alpha'} a_\alpha \rangle_0,
\end{equation}
where we used $\eqref{eq: algebracommutatoreanticommutatorecreazionedistruzione3}$, given that $\alpha \neq \alpha'$. From $\eqref{eq: mediatermicaadaggeraconalphadiversi}$, we have
\begin{align}
\sum_{\substack{\alpha' \\ \alpha' \neq \alpha}} \varphi_\alpha(x) \varphi^*_{\alpha'}(x') e^{- \frac{\mathcal{E}_\alpha \tau}{\hslash}} \langle \varepsilon a^\dagger_{\alpha'} a_\alpha \rangle_0 &= \sum_{\substack{\alpha' \\ \alpha' \neq \alpha}} \varphi_\alpha(x) \varphi^*_{\alpha'}(x') e^{- \frac{\mathcal{E}_\alpha \tau}{\hslash}} \varepsilon \delta_{\alpha,\alpha'} n_{\varepsilon}(\mathcal{E}_\alpha) = \notag \\
&= 0.
\end{align}
Consequently, the free many-body Green's function at thermodynamic equilibrium becomes
\begin{equation}
G^{(m)(0)}_{x,x'}(\tau) = - \dfrac{1}{\hslash} \sum_{\alpha} \varphi_\alpha(x) \varphi^*_\alpha(x') e^{- \frac{\mathcal{E}_\alpha \tau}{\hslash}} \left[ 1 + \dfrac{\varepsilon}{e^{\beta \mathcal{E}_{\alpha}} - \varepsilon} \right].
\end{equation}
We apply a time Fourier transform, that is
\begin{equation}
G^{(m)(0)}_{x,x'}(i \omega_n) = \int_0^{\beta \hslash} d\tau e^{i \omega_n \tau} G^{(m)(0)}_{x,x'}(\tau) ,
\end{equation}
and since only $e^{- \frac{\mathcal{E}_\alpha \tau}{\hslash}}$ depends explicitly on $\tau$, we have
\begin{align}
G^{(m)(0)}_{x,x'}(i \omega_n) &= \left( - \dfrac{1}{\hslash} \sum_{\alpha} \varphi_\alpha(x) \varphi^*_\alpha(x') \left[ 1 + \dfrac{\varepsilon}{e^{\beta \mathcal{E}_{\alpha}} - \varepsilon} \right] \right) \int_0^{\beta \hslash} d\tau e^{i \omega_n \tau} e^{- \frac{\mathcal{E}_{\alpha}}{\hslash} \tau} = \notag \\
&= \left( - \dfrac{1}{\hslash} \sum_{\alpha} \varphi_\alpha(x) \varphi^*_\alpha(x') \left[ \dfrac{e^{\beta \mathcal{E}_{\alpha}} - \varepsilon + \varepsilon}{e^{\beta \mathcal{E}_{\alpha}} - \varepsilon} \right] \right) \left( \dfrac{1}{i \omega_n - \frac{\mathcal{E}_{\alpha}}{\hslash}} \left[ e^{i \omega_n - \frac{\mathcal{E}_{\alpha}}{\hslash}} \right]_0^{\beta \hslash} \right) = \notag \\
&= - \dfrac{1}{\hslash} \sum_{\alpha} \varphi_\alpha(x) \varphi^*_\alpha(x') \dfrac{e^{\beta \mathcal{E}_{\alpha}}}{e^{\beta \mathcal{E}_{\alpha}} - \varepsilon} \dfrac{1}{i \omega_n - \frac{\mathcal{E}_{\alpha}}{\hslash}} \left[ e^{i \omega_n \beta \hslash} e^{- \beta \hslash} - 1 \right] = \notag \\
&= - \dfrac{1}{\hslash} \sum_{\alpha} \varphi_\alpha(x) \varphi^*_\alpha(x') \dfrac{e^{\beta \mathcal{E}_{\alpha}}}{e^{\beta \mathcal{E}_{\alpha}} - \varepsilon} \dfrac{1}{i \omega_n - \frac{\mathcal{E}_{\alpha}}{\hslash}} \left( \varepsilon e^{- \beta \mathcal{E}_\alpha} - 1 \right) = \notag \\
&= - \dfrac{1}{\hslash} \sum_{\alpha} \varphi_\alpha(x) \varphi^*_\alpha(x') \dfrac{1}{i \omega_n - \frac{\mathcal{E}_{\alpha}}{\hslash}} \dfrac{\varepsilon - e^{\beta \mathcal{E}_{\alpha}}}{e^{\beta \mathcal{E}_{\alpha}} - \varepsilon} = \notag \\
&= \sum_{\alpha} \dfrac{\varphi_\alpha(x) \varphi^*_\alpha(x')}{\hslash \left[ i \omega_n - \frac{\mathcal{E}_{\alpha}}{\hslash} \right]}.
\label{eq: propagatoreparticelleliberescrittocomeprodottiautostatidivisoenergie}
\end{align}
This final expression highlights that the Matsubara Green's function in frequency space depends solely on the spectrum \( \mathcal{E}_\alpha \) and the eigenfunctions \( \varphi_\alpha(x) \), and it is valid for both bosons and fermions through the appropriate choice of the Matsubara frequencies \( \omega_n \).
\section{Equations-of-motion method}
Given the retarded Green's function $G_{AB}^{(r)}(t)$ at the thermodynamic equilibrium of two generic operators $A$ and $B$, i.e., the equations $\eqref{eq: funzionedigreencomeprodottothetaHeavisideefunzionecorrelazione}$, $\eqref{eq: funzionedicorrelazioneinpropagatoremanybodies}$, we define $z=\omega + i\delta$, $\omega, \delta \in \mathbb{R}$ e $\delta > 0$ and integrate by parts the quantity
\begin{equation}
\int_{0}^\infty dt \ \dot{C}(t) e^{i z t} = \left. C(t)e^{izt} \right|_0^{\infty} - iz \int_{0}^{\infty} dt \ C(t) e^{izt} ,
\end{equation}
with
\begin{equation}
\left. C(t) e^{izt} \right|_0^{\infty} \ = -C(0).
\end{equation}
We have
\begin{equation}
\int_{0}^\infty dt \ \dot{C}(t) e^{i z t} = - C(0) - iz\int_{0}^{\infty} dt \ C(t) e^{izt},
\end{equation}
\begin{equation}
\int_{0}^\infty dt \ \dot{C}(t) e^{i z t} = - C(0) - iz\int_{0}^{\infty} dt \ \Theta(t) C(t) e^{izt},
\end{equation}
\begin{equation}
\int_{0}^\infty dt \ \dot{C}(t) e^{i z t} = - C(0) - iz\int_{0}^{\infty} dt \ G^{(r)}_{AB}(t) e^{izt},
\end{equation}
\begin{equation}
\int_{0}^\infty dt \ \dot{C}(t) e^{i z t} = - C(0) - iz G^{(r)}_{AB}(z),
\end{equation}
\begin{equation}
G^{(r)}_{AB}(z) = - \dfrac{1}{iz} C(0) - \dfrac{1}{iz} \int_{0}^\infty dt \ \dot{C}(t) e^{i z t},
\end{equation}
\begin{equation}
G^{(r)}_{AB}(z) = \left( - \dfrac{1}{iz} \right) \left( - \dfrac{i}{\hslash} \left\langle \left[ A,B\right]^{(\varepsilon)} \right\rangle \right) - \dfrac{1}{iz} \int_{0}^\infty dt \ \dot{C}(t) e^{i z t},
\end{equation}
\begin{equation}
G^{(r)}_{AB}(z) = \dfrac{1}{\hslash z} \left\langle \left[ A,B \right]^{(\varepsilon)} \right\rangle - \dfrac{1}{iz} \int_{0}^\infty dt \ \dot{C}(t) e^{i z t},
\end{equation}
\begin{equation}
G^{(r)}_{AB}(z) = \dfrac{1}{\hslash z} \left\langle \left[ A,B \right]^{(\varepsilon)} \right\rangle - \dfrac{1}{iz} \int_{0}^\infty dt \left( \left( - \dfrac{i}{\hslash} \right) \dfrac{1}{i\hslash} \left\langle \left[\left[A(t),\hat{\mathcal{H}}\right],B\right]^{(\varepsilon)} \right\rangle \right) e^{i z t},
\end{equation}
\begin{equation}
G^{(r)}_{AB}(z) = \dfrac{1}{\hslash z} \left\langle \left[ A,B \right]^{(\varepsilon)} \right\rangle + \dfrac{1}{iz} \int_{0}^\infty dt \dfrac{1}{\hslash^2} \left\langle \left[\left[A(t),\hat{\mathcal{H}}\right],B\right]^{(\varepsilon)} \right\rangle e^{i z t},
\end{equation}
\begin{equation}
G^{(r)}_{AB}(z) = \dfrac{1}{\hslash z} \left\langle \left[ A,B \right]^{(\varepsilon)} \right\rangle + \dfrac{1}{i \hslash^2 z} \int_{0}^\infty dt \left\langle \left[\left[A(t),\hat{\mathcal{H}}\right],B\right]^{(\varepsilon)} \right\rangle e^{i z t},
\end{equation}
\begin{equation}
G^{(r)}_{AB}(z) = \dfrac{1}{\hslash z} \left\langle \left[ A,B \right]^{(\varepsilon)} \right\rangle - \dfrac{i}{\hslash^2 z} \int_{0}^\infty dt \left\langle \left[\left[A(t),\hat{\mathcal{H}}\right],B\right]^{(\varepsilon)} \right\rangle e^{i z t},
\end{equation}
\begin{equation}
G^{(r)}_{AB}(z) = \dfrac{1}{\hslash z} \left\langle \left[ A,B \right]^{(\varepsilon)} \right\rangle + \dfrac{1}{\hslash z} \int_{0}^\infty dt \left( - \dfrac{i}{\hslash} \left\langle \left[\left[A(t),\hat{\mathcal{H}}\right],B\right]^{(\varepsilon)} \right\rangle \right) e^{i z t},
\end{equation}
\begin{equation}
G_{AB}^{(r)}(z) = \dfrac{1}{\hslash z} \left\langle \left[A,B\right]^{(\varepsilon)} \right\rangle + \dfrac{1}{\hslash z} G_{\left[A,\hat{\mathcal{H}} \right],B}^{(r)}(z) ,
\label{eq: tecnicaequazionidelmoto}
\end{equation}
i.e., we obtain a new retarded Green's function: apart from the term $\frac{1}{\hslash z} \left\langle \left[A,B\right]^{(\varepsilon)} \right\rangle$, the $G_{AB}^{(r)}(z)$ is related to a new Green's function, $G_{\left[A,\mathcal{H}\right],B}^{(r)}(z)$. From the equation connecting $G_{AB}^{(r)}(z)$ and $G_{\left[A,\hat{\mathcal{H}} \right],B}^{(r)}(z)$, integrating by parts, $G_{AB}^{(r)}(z)$ is related not only to $G_{\left[A,\hat{\mathcal{H}}\right],B}^{(r)}(z)$, but also to another Green's function, which is the $G_{\left[\left[A,\hat{\mathcal{H}}\right],\hat{\mathcal{H}}\right],B}^{(r)}(z)$ and so on for each integration. Consider a system of free particles, described by $\eqref{eq: Hamiltonianasecondaquantizzazioneparticellelibere}$, then the retarded Green's function at thermodynamic equilibrium with respect to the operators $A = a_\alpha$, $B = a_\alpha^\dag$ is
\begin{align}
G_{a_{\alpha} a_{\alpha}^{\dag}}^{(r)(0)}(t) &\equiv G_{\alpha}^{(r)(0)}(t) = \notag \\
&= - \dfrac{i}{\hslash}\Theta(t) \left\langle \left[ a^{(0)}_\alpha(t),a_\alpha^\dag(0) \right]^{(\varepsilon)} \right\rangle.
\end{align}
The many-body Green's function $G_{\alpha}^{(r)(0)}(t)$ describes the time evolution of the $\alpha$ state. Starting from the generalized commutator $\left[a^{(0)}_\alpha(t), a^\dag_\alpha(0)\right]^{(\varepsilon)} = a^{(0)}_\alpha(t)a^\dag_\alpha(0) - \varepsilon a^\dag_\alpha(0)a^{(0)}_\alpha(t)$, we observe that the first term, $a^{(0)}_\alpha(t)a^\dag_\alpha(0)$, represents a process in which a particle is created in state $\alpha$ at time $0$ and annihilated at time $t$: this is a particle term, and it propagates forward in time. The second term, $-\varepsilon a^\dag_\alpha(0)a^{(0)}_\alpha(t)$, represents a process in which a particle in state $\alpha$ is annihilated at time $t$ and then created at time $0$: this is a hole term, and it propagates backward in time. Recalling that $\left\langle \left[a_\alpha, a^\dag_\alpha\right]^{(\varepsilon)} \right\rangle = \left\langle \delta_{\alpha,\alpha} \, \mathds{1} \right\rangle = 1$, and that the equation of motion reads $[a_\alpha, \hat{\mathcal{H}}_0] = \mathcal{E}_\alpha a_\alpha$, we proceed to determine the propagator using the equations-of-motion method, as follows
\begin{align}
G_\alpha^{(r)(0)}(z) &= \dfrac{1}{\hslash z} \left\langle \left[a_\alpha,a_\alpha^\dag\right]^{(\varepsilon)} \right\rangle + \dfrac{1}{\hslash z} G_{\left[a_\alpha,\hat{\mathcal{H}}_0 \right],\alpha^\dag}^{(r)}(z) = \notag \\
&= \dfrac{1}{\hslash z} + \dfrac{1}{\hslash z} G_{\mathcal{E}_\alpha a_\alpha,a_\alpha^\dag}^{(r)}(z) = \notag \\
&= \dfrac{1}{\hslash z} + \dfrac{1}{\hslash z}\mathcal{E}_\alpha G_\alpha^{(r)(0)}(z),
\end{align}
which can be rearranged as an algebraic equation in \( G_\alpha^{(r)(0)}(z) \), that is,
\begin{align}
G_\alpha^{(r)(0)}(z) &= \dfrac{1}{\hslash z} \left[ \dfrac{1}{1 - \frac{\mathcal{E}_\alpha}{\hslash z}} \right] = \notag \\
&= \frac{1}{\hslash} \left[ \dfrac{1}{z - \frac{\mathcal{E}_\alpha}{\hslash}} \right] ,
\end{align}
and we directly derive the rational expression for the Green's function, i.e.,
\begin{align}
G_\alpha^{(r)(0)}(\omega + i\delta) &= \dfrac{1}{\hslash}\left[\frac{1}{\omega + i\delta - \frac{\mathcal{E}_\alpha}{\hslash}}\right] \equiv \notag \\
&\equiv \dfrac{1}{\hslash}\left[\frac{1}{\left( \omega - \frac{\mathcal{E}_\alpha}{\hslash} \right) + i\delta}\right],
\end{align}
and multiply and divide by the complex conjugate of the denominator, i.e.,
\begin{equation}
G_\alpha^{(r)}(\omega + i\delta) = \dfrac{1}{\hslash} \left[\frac{\omega - \frac{\mathcal{E}_\alpha}{\hslash} - i\delta}{\left(\omega - \frac{\mathcal{E}_\alpha}{\hslash}\right)^2 + \delta^2}\right].
\end{equation}
The imaginary part of the Fourier transform of the retarded Green's function
\begin{align}
\im G_\alpha^{(r)(0)}(\omega + i \delta) &= \dfrac{1}{\hslash} \im \left[\frac{\omega - \frac{\mathcal{E}_\alpha}{\hslash} - i\delta}{\left(\omega - \frac{\mathcal{E}_\alpha}{\hslash}\right)^2 + \delta^2}\right] = \notag \\
&= - \dfrac{1}{\hslash} \dfrac{\delta}{\left(\omega - \frac{\mathcal{E}_\alpha}{\hslash} \right)^2 + \delta^2}
\end{align}
is a measurable quantity. The imaginary part of the retarded Green's function for free particles is a Lorentzian function and in the limit $\delta \rightarrow 0^+$ tends to a Dirac's delta. Indeed, from $\eqref{eq: BreitWignerformula}$ we have
\begin{equation}
\lim_{\delta \rightarrow 0^+} \im G^{(r)(0)}_{\alpha}(\omega + i \delta) = - \lim_{\delta \rightarrow 0^+} \dfrac{1}{\hslash} \dfrac{\pi}{\pi} \dfrac{\delta}{\left(\omega - \frac{\mathcal{E}_\alpha}{\hslash} \right)^2 + \delta^2},
\end{equation}
\begin{equation}
\im G^{(r)(0)}_{\alpha}(\omega) = - \dfrac{\pi}{\hslash} \delta\left[\omega - \frac{\mathcal{E}_\alpha}{\hslash}\right],
\end{equation}
\begin{equation}
-2 \hslash \im G_\alpha^{(r)(0)}(\omega) = 2 \pi \delta\left[\omega - \frac{\mathcal{E}_\alpha}{\hslash}\right],
\end{equation}
where in the last step we multiplied both members by $-2$. Let us now determine the spectral function of free particles. Consider first the general case, from $\eqref{eq: trasformatadiFourierdellafunzionediGreenritardatanellabasedegliautostati}$ with $B=A^{\dag}$ and $z=\omega+i \delta$, we have
\begin{align}
G_{AA^\dag}^{(r)}(\omega + i \delta) &= \dfrac{1}{Z} \sum_{n,m} |A_{nm}|^2 \dfrac{e^{-\beta \mathcal{E}_n} - \varepsilon e^{-\beta \mathcal{E}_m}}{\hslash \left( \omega + i \delta + \frac{\mathcal{E}_n - \mathcal{E}_m}{\hslash} \right)} = \notag \\
&= \dfrac{1}{Z} \sum_{n,m} |A_{nm}|^2 \frac{e^{-\beta \mathcal{E}_n} - \varepsilon e^{-\beta \mathcal{E}_m}}{\hslash \left( \omega + i \delta + \frac{\mathcal{E}_n - \mathcal{E}_m}{\hslash} \right) \left( \omega + \frac{\mathcal{E}_n - \mathcal{E}_m}{\hslash} - i \delta \right)} \left( \omega + \frac{\mathcal{E}_n - \mathcal{E}_m}{\hslash} - i \delta \right) = \notag \\
&= \dfrac{1}{Z} \sum_{n,m} |A_{nm}|^2 \frac{e^{-\beta \mathcal{E}_n} - \varepsilon e^{-\beta \mathcal{E}_m}}{\hslash \left( \left( \omega + \frac{\mathcal{E}_n - \mathcal{E}_m}{\hslash} \right)^2 + \delta^2 \right)} \left( \omega + \frac{\mathcal{E}_n - \mathcal{E}_m}{\hslash} - i \delta \right) .
\end{align}
From this expression, we extract the imaginary part as follows
\begin{align}
- 2 \hslash \im G_{AA^\dag}^{(r)}(\omega) &= \lim_{\delta \rightarrow 0^+} \left( -2 \hslash \im G_{AA^\dag}^{(r)}(\omega + i \delta) \right) = \notag \\
&= \lim_{\delta \rightarrow 0^+} \left( -2 \hslash \dfrac{1}{Z} \sum_{n,m} |A_{nm}|^2 \dfrac{\left( - \delta \right)}{\hslash \left( \left( \omega + \frac{\mathcal{E}_n - \mathcal{E}_m}{\hslash} \right)^2 + \delta^2 \right)} \left( e^{-\beta \mathcal{E}_n} - \varepsilon e^{-\beta \mathcal{E}_m} \right) \right) = \notag \\
&= \lim_{\delta \rightarrow 0^+} \left( 2 \dfrac{1}{Z} \sum_{n,m} |A_{nm}|^2 \dfrac{\delta}{\left( \left( \omega + \frac{\mathcal{E}_n - \mathcal{E}_m}{\hslash} \right)^2 + \delta^2 \right)} \left( e^{-\beta \mathcal{E}_n} - \varepsilon e^{-\beta \mathcal{E}_m} \right) \right) = \notag \\
&= \dfrac{1}{Z} \sum_{n,m} |A_{nm}|^2 2 \left(e^{-\beta \mathcal{E}_n} - \varepsilon e^{-\beta \mathcal{E}_m} \right) \pi \delta\left[\omega + \frac{\mathcal{E}_n - \mathcal{E}_m}{\hslash}\right] \equiv \notag \\
&\equiv \dfrac{2 \pi}{Z} \sum_{n,m} |A_{nm}|^2 \left(e^{-\beta \mathcal{E}_n} - \varepsilon e^{-\beta \mathcal{E}_m} \right) \delta\left[\omega + \frac{\mathcal{E}_n - \mathcal{E}_m}{\hslash} \right] ,
\end{align}
that is equal to the spectral function $\eqref{eq: funzionespettrale}$ computed for $B=A^\dag$. Finally, we get
\begin{align}
A_{AA^\dag}(\omega) = - 2 \hslash \im G_{AA^\dag}^{(r)}(\omega) ,
\label{eq: funzionespettraleparteimmaginariapropagatore}
\end{align}
that is, the spectral function of the operators \( A \) and \( A^\dagger \), calculated as a function of \( \omega \), is related to the imaginary part of the retarded Green's function of the same operators evaluated at \( \omega \). In the case of free particles, the spectral function is a piecewise-defined function concentrated around the excitation energies of the system. Indeed, by comparison with the imaginary part of the retarded Green's function derived above, we find
\begin{equation}
A_{a a^\dag}(\omega) = 2 \pi \delta \left[\omega - \frac{\mathcal{E}_\alpha}{\hslash}\right]. 
\end{equation}
\subsection{Application of the equations-of-motion method to Anderson's single impurity model}
Anderson's single impurity model describes fermions interacting with an impurity, such as the case of electrons in a metal interacting with a single discrete state of fixed energy. Anderson's model neglects the spin of the particles. We will apply equations-of-motion method, showing that for this model they close exactly to the second order. The Hamiltonian of Anderson's model is given by
\begin{align}
\hat{\mathcal{H}} = \sum_{\textbf{k}} \mathcal{E}_{\textbf{k}} C^\dag_{\textbf{k}} C_{\textbf{k}} + \mathcal{E} b^\dag b + \sum_{\textbf{k}} A_{\textbf{k}} \left( C_{\textbf{k}}^\dag b + b^\dag C_{\textbf{k}} \right) ,
\end{align}
where the first term is the continuous level of free electrons; $\mathcal{E} b^\dag b$ is the impurity state; $A_{\textbf{k}}$ quantifies the interaction between electrons and impurity for each momentum $\textbf{k}$. The operators in $\hat{\mathcal{H}}$ satisfy $\left\lbrace C_{\textbf{k}},b \right\rbrace = 0$, $\left\lbrace C_{\textbf{k}},b^\dag \right\rbrace = 0$, $\left\lbrace C_{\textbf{k}}^\dag,b \right\rbrace = 0$, $\left\lbrace C_{\textbf{k}}^\dag,b^\dag \right\rbrace = 0$. Anderson's model approximation describes a solid: $\mathcal{E}_{\textbf{k}}$ is the highest energy band, with no interactions between particles. Impurity, e.g., an atom that is not aligned with others, is embedded in the continuous energy level: a flat, dispersion-free energy level arises, which is the energy level of impurity, $\mathcal{E} b^\dag b$. In terms of interaction: $b^\dag C_{\textbf{k}}$ is an electron that goes from the continuum to the impurity, while $C_{\textbf{k}}^\dag b$ is an electron that goes from the impurity to the continuum. Given the Green's function of the impurity, that is,
\begin{align}
G_{bb^{\dag}}^{(r)}(t) &= - \dfrac{i}{\hslash} \Theta(t) \left\langle \left[ b(t),b^\dag(0)\right]^{(\varepsilon)} \right\rangle \ = \notag \\
&= - \dfrac{i}{\hslash} \Theta(t) \left\langle \left\lbrace b(t),b^\dag(0)\right\rbrace \right\rangle,
\end{align}
we apply the equations-of-motion method and we have 
\begin{align}
z G_{bb^\dag}^{(r)}(z) &= \dfrac{1}{\hslash} \left\langle \left\lbrace b, b^\dag \right\rbrace \right\rangle + \dfrac{1}{\hslash} G^{(r)}_{\left[b,\hat{\mathcal{H}}\right], b^\dag}(z) \notag \\
&= \dfrac{1}{\hslash} + \dfrac{1}{\hslash} G^{(r)}_{\left[b,\hat{\mathcal{H}}\right], b^\dag}(z),
\end{align}
from fermionic algebra, it follows
\begin{align}
\left[ b,\hat{\mathcal{H}} \right] &= \left[ b, \sum_{\textbf{k}} \mathcal{E}_{\textbf{k}} C^\dag_{\textbf{k}} C_{\textbf{k}} + \mathcal{E} b^\dag b + \sum_{\textbf{k}} A_{\textbf{k}} \left( C_{\textbf{k}}^\dag b + b^\dag C_{\textbf{k}} \right) \right] = \notag \\
&= \left[ b, \mathcal{E} b^\dag b + \sum_{\textbf{k}} A_{\textbf{k}} \left( C_{\textbf{k}}^\dag b + b^\dag C_{\textbf{k}} \right) \right] = \notag \\
&= \left[ b, \mathcal{E} b^\dag b \right] + \left[ b, \sum_{\textbf{k}} A_{\textbf{k}} \left( C_{\textbf{k}}^\dag b + b^\dag C_{\textbf{k}} \right) \right] = \notag \\
&= \mathcal{E} b + \sum_{\textbf{k}} A_{\textbf{k}} C_{\textbf{k}} ,
\end{align}
then
\begin{equation}
zG_{bb^{\dag}}^{(r)}(z) = \dfrac{1}{\hslash} + \dfrac{\mathcal{E}}{\hslash}G^{(r)}_{bb^{\dag}}(z) + \dfrac{1}{\hslash} \sum_{\textbf{k}} A_{\textbf{k}} G^{(r)}_{C_{\textbf{k}},b^\dag}(z).
\end{equation}
Regarding the second order of the technique of equations-of-motion method for $G^{(r)}_{C_{\textbf{k}},b^\dag}(z)$, i.e.,
\begin{align}
zG^{(r)}_{C_{\textbf{k}},b^\dag}(z) &= \dfrac{1}{\hslash} \left\langle \left\lbrace C_{\textbf{k}},b^\dag\right\rbrace \right\rangle + \dfrac{1}{\hslash} G^{(r)}_{[C_{\textbf{k}},\hat{\mathcal{H}}],b^\dag}(z) = \notag \\
&= \dfrac{1}{\hslash} G^{(r)}_{[C_{\textbf{k}},\hat{\mathcal{H}}],b^\dag}(z),
\end{align}
we have to compute $\left[C_{\textbf{k}},\hat{\mathcal{H}}\right]$, then
\begin{align}
\left[ C_{\textbf{k}} , \hat{\mathcal{H}} \right] &= \left[ C_{\textbf{k}} , \sum_{\textbf{k}'} \mathcal{E}_{\textbf{k}'} C^\dag_{\textbf{k}'} C_{\textbf{k}'} + \mathcal{E} b^\dag b + \sum_{\textbf{k}'} A_{\textbf{k}'} \left(C_{\textbf{k}'}^\dag b + b^\dag C_{\textbf{k}'}\right) \right] = \notag \\
&= \left[ C_{\textbf{k}} , \sum_{\textbf{k}'} \mathcal{E}_{\textbf{k}'} C^\dag_{\textbf{k}'} C_{\textbf{k}'} + \sum_{\textbf{k}'} A_{\textbf{k}'} \left(C_{\textbf{k}'}^\dag b + b^\dag C_{\textbf{k}'}\right) \right] = \notag \\
&= \sum_{\textbf{k}'} \mathcal{E}_{\textbf{k}'} \left[ C_{\textbf{k}} , C^\dag_{\textbf{k}'} C_{\textbf{k}'} \right] + \sum_{\textbf{k}'} A_{\textbf{k}'} \left[ C_{\textbf{k}} , \left(C_{\textbf{k}'}^\dag b + b^\dag C_{\textbf{k}'} \right) \right] = \notag \\
&= \sum_{\textbf{k}'} \mathcal{E}_{\textbf{k}'} \delta_{\textbf{k}',\textbf{k}} C_{\textbf{k}'} + \sum_{\textbf{k}'} A_{\textbf{k}'} \left( C_{\textbf{k}} C_{\textbf{k}'}^\dag b - C_{\bar{k}'}^\dag b C_{\textbf{k}} + C_{\bar{k}} b^\dag C_{\textbf{k}'} - b^\dag C_{\textbf{k}'} C_{\textbf{k}} \right) = \notag \\
&= \mathcal{E}_{\textbf{k}} C_{\textbf{k}} + \sum_{\textbf{k}'} A_{\textbf{k}'} \left( C_{\textbf{k}} C_{\textbf{k}'}^\dag b - C_{\textbf{k}'}^\dag C_{\textbf{k}} b + C_{\textbf{k}} C_{\textbf{k}'} b^\dag - C_{\textbf{k}'} C_{\textbf{k}} b^\dag \right) = \notag \\
&= \mathcal{E}_{\textbf{k}} C_{\textbf{k}} + A_{\textbf{k}} b ,
\end{align}
and we have
\begin{align}
zG_{C_{\textbf{k}},b^\dag}^{(r)}(z) &= \dfrac{1}{\hslash} G^{(r)}_{\left[C_{\textbf{k}},\hat{\mathcal{H}}\right],b^\dag}(z) = \notag \\
&= \dfrac{1}{\hslash} \mathcal{E}_{\textbf{k}} G^{(r)}_{C_{\textbf{k}},b^\dag}(z) + \dfrac{A_{\textbf{k}}}{\hslash}G^{(r)}_{bb^\dag}(z),
\end{align}
\begin{equation}
G_{C_{\textbf{k}},b^\dag}^{(r)}(z) = \dfrac{A_{\textbf{k}}}{\hslash}G^{(r)}_{bb^\dag}(z) \dfrac{1}{z - \frac{\mathcal{E}_\alpha}{\hslash}}.
\end{equation}
We insert $G_{C_{\textbf{k}},b^\dag}^{(r)}(z)$ in the expression for $G_{C_{\textbf{k}},b^\dag}^{(r)}(z)$ as follows
\begin{align}
zG_{bb^{\dag}}^{(r)}(z) &= \frac{1}{\hslash} + \dfrac{\mathcal{E}}{\hslash}G^{(r)}_{bb^{\dag}}(z) + \frac{1}{\hslash} \sum_{\textbf{k}} A_{\textbf{k}} G^{(r)}_{C_{\textbf{k}},b^\dag}(z) = \notag \\
&= \dfrac{1}{\hslash} + \dfrac{\mathcal{E}}{\hslash} G^{(r)}_{bb^{\dag}}(z) + \sum_{\textbf{k}} \dfrac{A_{\textbf{k}}^2}{\hslash^2} G^{(r)}_{bb^\dag}(z)\dfrac{1}{z - \frac{\mathcal{E}_{\bar{k}}}{\hslash}},
\end{align}
and we get
\begin{equation}
G_{bb^{\dag}}^{(r)}(z) = \dfrac{1}{\hslash}\dfrac{1}{z - \frac{\mathcal{E}}{\hslash} - \sum_{\textbf{k}} \frac{A_{\textbf{k}}^2}{\hslash^2} \dfrac{1}{z - \frac{\mathcal{E}_{\textbf{k}}}{\hslash}}}.
\end{equation}
In the non-interacting limit, i.e., when \( A_{\textbf{k}} = 0 \), the Green's function reduces to its free form, characterized by a simple pole on the real axis. The corresponding spectral function is then given by $2 \pi \delta\left(\omega - \frac{\mathcal{E}}{\hslash}\right)$. We define the improper self-energy as the complex function
\begin{equation}
\Sigma^*(z) \equiv \sum_{\textbf{k}} \frac{A_{\textbf{k}}^2}{\hslash} \dfrac{1}{z - \frac{\mathcal{E}_{\textbf{k}}}{\hslash}}.
\end{equation}
This quantity will be further analyzed in the chapter dedicated to Feynman's perturbative expansion. The Green's function of an impurity in a metal is
\begin{equation}
G_{bb^{\dag}}^{(r)}(z)  = \dfrac{1}{\hslash} \dfrac{1}{z - \frac{\mathcal{E}}{\hslash} - \frac{\Sigma^*(z)}{\hslash}}.
\label{eq: funzionediGreensistemainvariantetraslspazialeeallequilibriotermod}
\end{equation}
By means of Feynman's perturbative theory (see Chapter $\ref{Feynman's perturbative theory of the thermal Green's function}$, equation $\eqref{eq: funzionediGreensistemainvariantetraslspazialeeallequilibriotermod2}$) we will demonstrate that, for any system in thermodynamic equilibrium, the corresponding many-body Green's function necessarily takes the form given in equation $\eqref{eq: funzionediGreensistemainvariantetraslspazialeeallequilibriotermod}$. In this expression, the energy \(\mathcal{E}\) of the unperturbed level explicitly appears, while the function \(\Sigma^*(z)\) contains the infinite sum of all Feynman diagrams representing interaction effects. How does \(\Sigma^*(z)\) modify the free Green's function? The Green's function of the impurity, \(G_{bb^{\dag}}^{(r)}(z)\), may exhibit a pole \(z_0\) either on the real axis or in the complex lower half-plane. Suppose that \(z_0\) lies in the complex lower half-plane: if \(G_{bb^{\dag}}^{(r)}(z)\) has a pole at \(z_0\), then necessarily \(\Sigma^*(z_0) \neq 0\). Indeed, if \(\Sigma^*(z_0) = 0\), the pole of the Green's function would be located at \(z = \frac{\mathcal{E}}{\hslash}\), as in the non-interacting case, which contradicts the assumption that the pole lies in the complex plane. Then
\begin{align}
G_{bb^{\dag}}^{(r)}(z) &= \frac{1}{\hslash} \dfrac{1}{z - \frac{\mathcal{E}}{\hslash} - \frac{\Sigma^*(z)}{\hslash}} \approx \notag \\
&\approx \dfrac{1}{\hslash} \dfrac{1}{z - \frac{\mathcal{E}}{\hslash} - \frac{\Sigma^*(z_0)}{\hslash}},
\end{align}
where $\Sigma^*(z_0)$ has real and imaginary parts. We set $\Sigma^*(z_0) = \widetilde{\mathcal{E}} + i \gamma$. If it were worth $\gamma>0$
\begin{equation}
G_{b b^\dagger}^{(r)}(z) \approx \dfrac{1}{\hslash} \dfrac{1}{z - \frac{\mathcal{E}}{\hslash} - \frac{\widetilde{\mathcal{E}}}{\hslash} - i \frac{\gamma}{\hslash}} ,
\end{equation}
that is, the Green's function would have the pole $z_0=\frac{\mathcal{E}}{\hslash} + \frac{\widetilde{\mathcal{E}}}{\hslash} + i \frac{\gamma}{\hslash}$ in the complex upper half-plane, but this is not possible, so it must be $\gamma < 0$. We set $\Sigma^*(z_0) = \widetilde{\mathcal{E}} - i\gamma$, $\gamma > 0$
\begin{align}
G_{bb^{\dag}}^{(r)}(z) &\approx \dfrac{1}{\hslash} \dfrac{1}{z - \frac{\mathcal{E}}{\hslash} - \frac{\widetilde{\mathcal{E}}}{\hslash} + i \frac{\gamma}{\hslash}} = \notag \\
&= \dfrac{1}{\hslash} \dfrac{z - \frac{\mathcal{E}}{\hslash} - \frac{\widetilde{\mathcal{E}}}{\hslash} - i \frac{\gamma}{\hslash}}{\left( z - \frac{\mathcal{E}}{\hslash} - \frac{\widetilde{\mathcal{E}}}{\hslash} \right)^2 + \frac{\gamma^2}{\hslash^2}}.
\end{align}
Thus, the role of the improper self-energy becomes clear: in the case of free particles, the Green's function exhibits a pole on the real axis, while in the presence of interactions, this pole is shifted into the complex lower half-plane. This behavior is observed in the frequency domain. Naturally, one may wonder how this modification manifests itself in the time domain. We will employ the residue theorem (see Theorem $\eqref{thm: teoremaresidui}$) to evaluate the inverse Fourier transform of the Green’s function and thereby elucidate its temporal behavior. We have
\begin{align}
G_{bb^{\dag}}^{(r)}(t) &= \int_{\Gamma} \frac{dz}{2\pi}e^{-izt} G^{(r)}_{b b^\dagger}(z) = \notag \\
&= \frac{1}{\hslash} \int_{\Gamma} \frac{dz}{2\pi} \frac{e^{-izt}}{z - \frac{\mathcal{E}}{\hslash} - \frac{\widetilde{\mathcal{E}}}{\hslash} + i \frac{\gamma}{\hslash}} ,
\end{align}
where $\Gamma$ is a line parallel to the real axis and belongs to the upper half-plane. Assume $t > 0$. The pole is $z_0=\frac{\mathcal{E}}{\hslash} + \frac{\widetilde{\mathcal{E}}}{\hslash} - i \frac{\gamma}{\hslash}$, then to obtain a circuit containing $z_0$, we consider the lower half-plane. The line integral of the function around the circuit is equal to $2 \pi i$ times the sum of residues of the function at the points. Since for $R \rightarrow +\infty$, we apply Jordan's lemma, it remains to calculate the integral along the line. For the simple pole of the first order we have
\begin{align}
G_{bb^{\dag}}^{(r)}(t) &= \dfrac{1}{\hslash} \dfrac{1}{2\pi} \left( - 2 \pi i \right) \Theta(t) e^{-i \left[ \frac{\mathcal{E} + \widetilde{\mathcal{E}}}{\hslash} - i \frac{\gamma}{\hslash} \right] t} = \notag \\
&= - \dfrac{i}{\hslash} \Theta(t) e^{-i \left[ \frac{\mathcal{E} + \widetilde{\mathcal{E}}}{\hslash} - i \frac{\gamma}{\hslash} \right] t} ,
\end{align}
where we used Heaviside's $\Theta(t)$ so that it is $t>0$. If $t<0$, the circuit must belong to the upper half-plane, where there are no poles, and the integral is null, consequently the propagator is also null. The improper self-energy has two effects on the Green's function: the real part produces renormalization of energy levels and the imaginary part generates decays in time. We are interested in $z = \omega + i \delta$ in the limit $\delta \rightarrow 0^+$. How is the spectral function modified?
\begin{align}
A_{bb^\dagger}(\omega) &= \lim_{\delta \rightarrow 0^+} \left( -2 \hslash \im G^{(r)}_{b b^\dagger}(\omega + i \delta) \right) = \notag \\
&= - 2 \left[\dfrac{\frac{\gamma}{\hslash}}{\left(\omega - \frac{\mathcal{E} + \widetilde{\mathcal{E}}}{\hslash} \right)^2 + \frac{\gamma^2}{\hslash^2}}\right],
\end{align}
which is a Lorentzian function. In conclusion, if the approximation \( \Sigma^*(z) \simeq \Sigma^*(z_0) \) holds, the improper self-energy leads to several important physical consequences. First, it induces a renormalization of the energy levels, shifting them from their unperturbed positions on the real axis. Secondly, it introduces a finite decay rate \( \gamma \), associated with the imaginary part of the self-energy. As a result, the spectral function acquires a Lorentzian profile with a width given by \( \frac{\gamma}{\hslash} = \frac{1}{\tau} \), where \( \tau \) is the lifetime of the excitation. In the limit \( \gamma \rightarrow 0^+ \), the Lorentzian function tends toward a Dirac delta. However, this delta is no longer centered at the unperturbed energy \( \mathcal{E} \), but rather at \( \mathcal{E} + \widetilde{\mathcal{E}} \), due to the real part of the self-energy remaining finite. The smaller the value of \( \gamma \), the larger the corresponding lifetime \( \tau \): as \( \gamma \rightarrow 0 \), we have \( \tau \rightarrow \infty \), indicating that the excitation becomes stable, with an infinite lifetime. Conversely, for finite \( \gamma \), the lifetime \( \tau \) is also finite, and the excitation with energy \( \mathcal{E} + \widetilde{\mathcal{E}} \) decays into other states after a finite time.
\section{Green's function of free particles}
In the case of free particles, we have seen by the equations-of-motion method that
\begin{align}
G_\alpha^{(r)(0)}(z) = \dfrac{1}{\hslash}\dfrac{1}{z - \frac{\mathcal{E}_\alpha}{\hslash}},
\end{align}
where the pole $z = \frac{\mathcal{E}_\alpha}{\hslash}$ is on the real axis and the spectral function is given by
\begin{align}
-2 \hslash \im G_\alpha^{(r)(0)}(\omega) = 2\pi \delta \left[\omega - \dfrac{\mathcal{E}_\alpha}{\hslash}\right].
\end{align}
In general, $A_{AB}$ is sum of several Dirac deltas, but in the case of free particles it includes only one delta, centered in $\frac{\mathcal{E}_\alpha}{\hslash}$, with respect to which we are calculating the Green's function. To compute $G_{\alpha}^{(r)}(t)$, one would have to perform the Fourier antitransform. For free particles, the time evolution of the $a_\alpha$ operator is given in \eqref{eq: evoluzionetemporaleoperatoredistruzione1}, then
\begin{align}
G_\alpha^{(r)(0)}(t) &= -\dfrac{1}{\hslash}\Theta(t) \left\langle \left[a^{(0)}_\alpha(t),a^\dag_\alpha(0)\right]^{(\varepsilon)} \right\rangle= \notag \\
&= -\dfrac{1}{\hslash}\Theta(t) e^{- i \frac{\mathcal{E}_\alpha}{\hslash} t}.
\end{align}
In conclusion, the free Green's function evolves in time by a phase factor $e^{- i \frac{\mathcal{E}_\alpha}{\hslash} t}$, in Fourier transform it has a pole on the real axis at the point $z = \frac{\mathcal{E}_\alpha}{\hslash}$ and its spectral function is $\delta\left[\omega - \frac{\mathcal{E}_\alpha}{\hslash}\right]$.
\section{Independent boson model}
Given the Hamiltonian
\begin{equation}
\hat{\mathcal{H}} = \mathcal{E} b^\dagger b + \sum_{\mathbf{q}} \hbar \omega_{\mathbf{q}}\, a^\dagger_{\mathbf{q}} a_{\mathbf{q}} + \sum_{\mathbf{q}} M_{\mathbf{q}} \left( a_{\mathbf{q}} + a^\dagger_{\mathbf{q}} \right) b^\dagger b,
\label{eq: singleimpuritymodel}
\end{equation}
the three terms describe, respectively: the energy of a fermionic impurity, the phonon bath, and the interaction between the impurity and the phonon modes. The operators satisfy the following relations
\begin{equation}
\left[ b, a_{\mathbf{q}} \right] = 0,
\end{equation}
\begin{equation}
\left[ b^\dagger, a_{\mathbf{q}} \right] = 0,
\end{equation}
\begin{equation}
\left[ b, a^\dagger_{\mathbf{q}} \right] = 0,
\end{equation}
\begin{equation}
\left[ b^\dagger, a^\dagger_{\mathbf{q}} \right] = 0.
\end{equation}
Since \( \hat{\mathcal{{H}}}^\dagger = \hat{\mathcal{H}} \), it must be \( M_{\mathbf{q}} \in \mathbb{R} \). The term \( \mathcal{E}\, b^\dagger b \) represents a single nondispersive energy level, e.g., an impurity in a metal, and it interacts with a set of independent harmonic oscillators with energy \( \hbar \omega_{\mathbf{q}} \). The impurity, through the interaction term, can either create or destroy a phonon. Equation $\eqref{eq: singleimpuritymodel}$ is the simplest Hamiltonian model containing an interaction term, since it is linear and exactly diagonalizable, as we shall now prove. To diagonalize the Hamiltonian, we employ a suitable unitary operator \( U \). Given a generic operator \( S \), the condition for \( U = e^{-S} \) to be unitary is that \( S^\dagger = -S \), i.e., \( S \) must be anti-Hermitian. We diagonalize the initial Hamiltonian according to the transformation
\begin{equation}
e^S \hat{\mathcal{H}} e^{-S} = \hat{\tilde{\mathcal{H}}},
\label{eq: diagonalization_unitary}
\end{equation}
where, for the Hamiltonian $\eqref{eq: singleimpuritymodel}$, the generator \( S \) is given by
\begin{equation}
S = \sum_{\mathbf{q}'} r_{\mathbf{q}'}\, b^\dagger b \left( a^\dagger_{\mathbf{q}'} - a_{\mathbf{q}'} \right),
\label{eq: generator_S}
\end{equation}
with
\begin{equation}
r_{\mathbf{q}'} = \frac{M_{\mathbf{q}'}}{\hbar \omega_{\mathbf{q}'}}
\label{eq: rq_coefficient}
\end{equation}
a dimensionless constant. We observe that \( S \) involves the ratio between the interaction energy and the oscillator energy; the modulus of the operator grows with the interaction strength. To apply the exponential transformation of \( S \), we make use of the BCH formula $\eqref{eq: BCHformula}$. Note that the phononic term transforms as
\begin{equation}
e^S b^\dagger b\, e^{-S} = \left( e^S b^\dagger e^{-S} \right) \left( e^S b\, e^{-S} \right),
\label{eq: transformed_bdaggerb}
\end{equation}
that is, the operator \( b^\dagger b \) is mapped into the product of the individually transformed operators. Now consider the transformation of the bosonic annihilation operator \( a_{\mathbf{q}} \). We first compute the commutator
\begin{align}
[S, a_{\mathbf{q}}] &= \left[ \sum_{\mathbf{q}'} r_{\mathbf{q}'} b^\dagger b \left( a^\dagger_{\mathbf{q}'} - a_{\mathbf{q}'} \right),\, a_{\mathbf{q}}
\right] = \notag \\
&= b^\dagger b \left( \sum_{\mathbf{q}'} r_{\mathbf{q}'} \left[a^\dagger_{\mathbf{q}'}, a_{\mathbf{q}} \right] - \sum_{\mathbf{q}'} r_{\mathbf{q}'} \left[a_{\mathbf{q}'}, a_{\mathbf{q}}\right] \right) =\notag \\
&= b^\dagger b \left( \sum_{\mathbf{q}'} r_{\mathbf{q}'} \left[ a^\dagger_{\mathbf{q}'}, a_{\mathbf{q}} \right] - \sum_{\mathbf{q}'} r_{\mathbf{q}'} \left[ a_{\mathbf{q}'}, a_{\mathbf{q}} \right] \right) = \notag \\
&= b^\dagger b \sum_{\mathbf{q}'} r_{\mathbf{q}'} \left[ a^\dagger_{\mathbf{q}'}, a_{\mathbf{q}} \right] = \notag \\
&= b^\dagger b \sum_{\mathbf{q}'} r_{\mathbf{q}'} (-\delta_{\mathbf{q}', \mathbf{q}}) = \notag \\
&= -r_{\mathbf{q}}\, b^\dagger b,
\label{eq: commutator_S_aq}
\end{align}
\begin{align}
\left[ S, a^\dagger_{\mathbf{q}} \right] &= \left[ \sum_{\mathbf{q}'} r_{\mathbf{q}'} b^\dagger b \left( a^\dagger_{\mathbf{q}'} - a_{\mathbf{q}'} \right),\, a^\dagger_{\mathbf{q}} \right] = \notag \\
&= b^\dagger b \left( \sum_{\mathbf{q}'} r_{\mathbf{q}'} \left[ a^\dagger_{\mathbf{q}'}, a^\dagger_{\mathbf{q}}\right] - \sum_{\mathbf{q}'} r_{\mathbf{q}'} \left[a_{\mathbf{q}'}, a^\dagger_{\mathbf{q}}\right] \right) = \notag \\
&= b^\dagger b \left( \sum_{\mathbf{q}'} r_{\mathbf{q}'} \left[ a^\dagger_{\mathbf{q}'}, a^\dagger_{\mathbf{q}}\right] - \sum_{\mathbf{q}'} r_{\mathbf{q}'} \left[a_{\mathbf{q}'}, a^\dagger_{\mathbf{q}}\right] \right) = \notag \\
&= b^\dagger b \left( 0 - \sum_{\mathbf{q}'} r_{\mathbf{q}'} \delta_{\mathbf{q}', \mathbf{q}} \right) = \notag \\
&= -r_{\mathbf{q}}\, b^\dagger b,
\label{eq: commutator_S_adaggerq}
\end{align}
then we find that the bosonic operators transform as
\begin{align}
e^{S} a_{\mathbf{q}} e^{-S} &= a_{\mathbf{q}} - r_{\mathbf{q}} b^\dagger b, \\
e^{S} a^\dagger_{\mathbf{q}} e^{-S} &= a^\dagger_{\mathbf{q}} - r_{\mathbf{q}} b^\dagger b, 
\end{align}
since all higher-order commutators vanish identically. For the impurity operators, we compute the commutators
\begin{align}
\left[ S, b \right] 
&= \sum_{\mathbf{q}'} r_{\mathbf{q}'} \left( a^\dagger_{\mathbf{q}'} - a_{\mathbf{q}'} \right) \left[ b^\dagger b, b \right] = \notag \\
&= -b \sum_{\mathbf{q}'} r_{\mathbf{q}'} \left( a^\dagger_{\mathbf{q}'} - a_{\mathbf{q}'} \right),
\end{align}
\begin{align}
\left[ S, b^\dagger \right] 
&= \sum_{\mathbf{q}'} r_{\mathbf{q}'} \left( a^\dagger_{\mathbf{q}'} - a_{\mathbf{q}'} \right) \left[ b^\dagger b, b^\dagger \right] = \notag \\
&= b^\dagger \sum_{\mathbf{q}'} r_{\mathbf{q}'} \left( a^\dagger_{\mathbf{q}'} - a_{\mathbf{q}'} \right).
\end{align}
As a result, the impurity operators transform as
\begin{equation}
e^{S} b\, e^{-S} = b \left( \mathds{1} - S' + \frac{1}{2!} S'^2 - \cdots \right),
\end{equation}
\begin{equation}
e^{S} b^\dagger\, e^{-S} = b^\dagger \left( \mathds{1} + S' + \frac{1}{2!} S'^2 + \cdots \right), 
\end{equation}
where the exponential series follows from the BCH expansion $\eqref{eq: BCHformula}$, and the operator $S'$ is defined as
\begin{equation}
S' = \sum_{\mathbf{q}'} r_{\mathbf{q}'}\, \left( a^\dagger_{\mathbf{q}'} - a_{\mathbf{q}'} \right).
\label{eq: generator_S'}
\end{equation}
The operators \( b \) and \( b^\dagger \) satisfy
\begin{equation}
e^{S} b\, e^{-S} = b\, e^{-\sum_{\mathbf{q}} r_{\mathbf{q}} \left( a^\dagger_{\mathbf{q}} - a_{\mathbf{q}} \right)},
\label{eq: transformed_b_exponential}
\end{equation}
\begin{equation}
e^{S} b^\dagger\, e^{-S} = b^\dagger\, e^{\sum_{\mathbf{q}} r_{\mathbf{q}} \left( a^\dagger_{\mathbf{q}} - a_{\mathbf{q}} \right)}.
\label{eq: transformed_bdagger_exponential}
\end{equation}
\begin{align}
e^{S} b^\dagger b\, e^{-S} &= \left( e^{S} b^\dagger e^{-S} \right) \left( e^{S} b\, e^{-S} \right) \nonumber \\
&= b^\dagger b.
\label{eq:transformed_bdaggerb}
\end{align}
Consequently the Hamiltonian $\eqref{eq: singleimpuritymodel}$ transforms as
\begin{align}
\hat{\tilde{\mathcal{H}}} &= \mathcal{E} e^{S} b^\dagger b\, e^{-S} + \sum_{\mathbf{q}} \hbar \omega_{\mathbf{q}}\, e^{S} a^\dagger_{\mathbf{q}} a_{\mathbf{q}}\, e^{-S} + \sum_{\mathbf{q}} M_{\mathbf{q}}\, e^{S} (a_{\mathbf{q}} + a^\dagger_{\mathbf{q}})\, b^\dagger b\, e^{-S} = \notag \\
&= \mathcal{E} b^\dagger b + \sum_{\mathbf{q}} \hbar \omega_{\mathbf{q}}\, e^{S} a^\dagger_{\mathbf{q}}\, e^{-S} e^{S} a_{\mathbf{q}}\, e^{-S} + \sum_{\mathbf{q}} M_{\mathbf{q}}\, e^{S} (a_{\mathbf{q}} + a^\dagger_{\mathbf{q}})\, e^{-S} \cdot e^{S} b^\dagger b\, e^{-S} = \notag \\
&= \mathcal{E} b^\dagger b
+ \sum_{\mathbf{q}} \hbar \omega_{\mathbf{q}}\, \left( a^\dagger_{\mathbf{q}} - \frac{M_{\mathbf{q}}}{\hbar \omega_{\mathbf{q}}} b^\dagger b \right) \left( a_{\mathbf{q}} - \frac{M_{\mathbf{q}}}{\hbar \omega_{\mathbf{q}}} b^\dagger b \right) + \sum_{\mathbf{q}} M_{\mathbf{q}}\, e^{S} (a_{\mathbf{q}} + a^\dagger_{\mathbf{q}})\, e^{-S} b^\dagger b = \notag \\
&= \mathcal{E} b^\dagger b + \sum_{\mathbf{q}} \hbar \omega_{\mathbf{q}}\, \left( a^\dagger_{\mathbf{q}} - \frac{M_{\mathbf{q}}}{\hbar \omega_{\mathbf{q}}} b^\dagger b \right) \left( a_{\mathbf{q}} - \frac{M_{\mathbf{q}}}{\hbar \omega_{\mathbf{q}}} b^\dagger b \right) + b^\dagger b \sum_{\mathbf{q}} M_{\mathbf{q}}\, e^{S} (a_{\mathbf{q}} + a^\dagger_{\mathbf{q}})\, e^{-S} = \notag \\
&= \mathcal{E} b^\dagger b + \sum_{\mathbf{q}} \hbar \omega_{\mathbf{q}} \left[ a^\dagger_{\mathbf{q}} a_{\mathbf{q}} + \frac{M_{\mathbf{q}}^2}{(\hbar \omega_{\mathbf{q}})^2} b^\dagger b - \frac{M_{\mathbf{q}}}{\hbar \omega_{\mathbf{q}}} b^\dagger b\, a^\dagger_{\mathbf{q}} - \frac{M_{\mathbf{q}}}{\hbar \omega_{\mathbf{q}}} b^\dagger b\, a_{\mathbf{q}} \right] + b^\dagger b \sum_{\mathbf{q}} M_{\mathbf{q}} \left( a^\dagger_{\mathbf{q}} + a_{\mathbf{q}} - \frac{2 M_{\mathbf{q}}}{\hbar \omega_{\mathbf{q}}} b^\dagger b \right) = \notag \\
&= \mathcal{E} b^\dagger b + \sum_{\mathbf{q}} \left[ \hbar \omega_{\mathbf{q}}\, a^\dagger_{\mathbf{q}} a_{\mathbf{q}} + \frac{M_{\mathbf{q}}^2}{\hbar \omega_{\mathbf{q}}} b^\dagger b - M_{\mathbf{q}} b^\dagger b\, a^\dagger_{\mathbf{q}} - M_{\mathbf{q}} b^\dagger b\, a_{\mathbf{q}} \right] + \sum_{\mathbf{q}} \left[ M_{\mathbf{q}} b^\dagger b\, a^\dagger_{\mathbf{q}} + M_{\mathbf{q}} b^\dagger b\, a_{\mathbf{q}} - \frac{2 M_{\mathbf{q}}^2}{\hbar \omega_{\mathbf{q}}} (b^\dagger b)^2 \right] = \notag \\
&= \mathcal{E} b^\dagger b + \sum_{\mathbf{q}} \left[ \hbar \omega_{\mathbf{q}}\, a^\dagger_{\mathbf{q}} a_{\mathbf{q}} + \frac{M_{\mathbf{q}}^2}{\hbar \omega_{\mathbf{q}}} b^\dagger b - \frac{2 M_{\mathbf{q}}^2}{\hbar \omega_{\mathbf{q}}} (b^\dagger b)^2 \right]. 
\label{eq: singleimpuritymodel2}
\end{align}
Since the operator \( b^\dagger b \) is fermionic, we have
\begin{equation}
(b^\dagger b)^2 = b^\dagger b,
\end{equation}
then
\begin{align}
\hat{\tilde{\mathcal{H}}} &= \mathcal{E}\, b^\dagger b + \sum_{\mathbf{q}} \left( 
\hbar \omega_{\mathbf{q}}\, a^\dagger_{\mathbf{q}} a_{\mathbf{q}} +
\frac{M_{\mathbf{q}}^2}{\hbar \omega_{\mathbf{q}}} b^\dagger b -
\frac{2 M_{\mathbf{q}}^2}{\hbar \omega_{\mathbf{q}}} b^\dagger b 
\right) = \notag \\
&= \mathcal{E}\, b^\dagger b + \sum_{\mathbf{q}} \left( 
\hbar \omega_{\mathbf{q}}\, a^\dagger_{\mathbf{q}} a_{\mathbf{q}} -
\frac{M_{\mathbf{q}}^2}{\hbar \omega_{\mathbf{q}}} b^\dagger b 
\right) .
\label{eq: singleimpuritymodel3}
\end{align}
We can thus finally express the Hamiltonian $\eqref{eq: singleimpuritymodel}$ in diagonal form as follows
\begin{align}
\hat{\tilde{\mathcal{H}}} &= \sum_{\mathbf{q}} \hbar \omega_{\mathbf{q}}\, a^\dagger_{\mathbf{q}} a_{\mathbf{q}} + \left( \mathcal{E} - \sum_{\mathbf{q}} \frac{M_{\mathbf{q}}^2}{\hbar \omega_{\mathbf{q}}} \right) b^\dagger b \equiv \notag \\
&\equiv \sum_{\mathbf{q}} \hbar \omega_{\mathbf{q}}\, a^\dagger_{\mathbf{q}} a_{\mathbf{q}} + \tilde{\mathcal{E}}\, b^\dagger b,
\label{eq: singleimpuritymodel4}
\end{align}
with
\begin{equation}
\tilde{\mathcal{E}} = \mathcal{E} - \sum_{\mathbf{q}} \frac{M_{\mathbf{q}}^2}{\hbar \omega_{\mathbf{q}}} .
\label{eq: autovalorifermionicisingleimpuritymodeldiagonalizzata}
\end{equation}
By means of the new operators, the Hamiltonian is given by the sum of a bosonic and a fermionic term, so the eigenstates are given by the tensor product of the bosonic and fermionic eigenstates. In the occupation-number basis, the bosonic
basis is
\begin{equation}
|N_{\textbf{q}_1}, \ldots, N_{\textbf{q}_N} \rangle = \frac{(a^\dagger_{\textbf{q}_1})^{N_{\textbf{q}_1}}}{\sqrt{N_{\textbf{q}_1}!}} \cdots \frac{(a^\dagger_{\textbf{q}_N})^{N_{\textbf{q}_N}}}{\sqrt{N_{\textbf{q}_N}!}} |0 \rangle_b,
\end{equation}
with eigenvalues
\begin{equation}
N_{\mathbf{q}_1} \hbar \omega_{\mathbf{q}_1} + \cdots + N_{\mathbf{q}_N} \hbar \omega_{\mathbf{q}_N}.
\end{equation}
Due to the impurity, the fermionic state in the occupation-number basis is
\begin{equation}
|N_b \rangle = b^\dagger |0 \rangle_f,
\end{equation}
and the corresponding eigenvalue is given by $\eqref{eq: autovalorifermionicisingleimpuritymodeldiagonalizzata}$. In the diagonalized Hamiltonian $\hat{\tilde{\mathcal{H}}}$, the impurity becomes dressed by its interaction with the phonon field and acquires an energy $\tilde{\mathcal{E}}$. This dressed impurity, a fermionic quasiparticle resulting from the coupling to the phonons, is known as a single-level polaron. This model, describing a single impurity linearly coupled to a bosonic environment, is commonly referred to as the independent boson model (IBM) in the literature. The Green's function of the single-level polaron corresponds to a delta function, reflecting the fact that the system is exactly solvable and the single-level polaron is a well-defined quasiparticle with no decay. However, in experiments, one does not directly probe the Green's function of the dressed impurity. Instead, measurements access the Green's function of the bare impurity, i.e., the original fermionic operator before the unitary transformation. The central objective, therefore, is to determine the form of the Green's function associated with the original impurity operator the one that is physically relevant and experimentally observable. To summarize, the eigenstates and eigenvalues of $\hat{\tilde{\mathcal{H}}}$ are given by
\begin{equation}
|N_{\textbf{q}_1}, \ldots, N_{\textbf{q}_N} \rangle b^\dagger |0 \rangle_f,
\label{eq: autostatisingleimpuritymodeldiagonalizzata}
\end{equation}
\begin{equation}
N_{\mathbf{q}_1} \hbar \omega_{\mathbf{q}_1} + \cdots + N_{\mathbf{q}_N} \hbar \omega_{\mathbf{q}_N} + \mathcal{E} - \sum_{\mathbf{q}} \frac{M_{\mathbf{q}}^2}{\hbar \omega_{\mathbf{q}}}.
\label{eq: autovalorisingleimpuritymodeldiagonalizzata}
\end{equation}
respectively. Now, it is natural to ask what the eigenstates and eigenvalues of the Hamiltonian \( \hat{\mathcal{H}} \) in the original basis. We multiply to the left by \( e^{-S} \) the eigenvalue equation for the diagonalized Hamiltonian as follows
\begin{equation}
e^{-S} \hat{\tilde{\mathcal{H}}} |\psi_n \rangle = \mathcal{E}_n e^{-S} |\psi_n \rangle, 
\end{equation}
and from
\begin{align}
e^{-S} \hat{\tilde{\mathcal{H}}} &= e^{-S} e^{S} \hat{\mathcal{H}} e^{-S} = \notag \\
&= \hat{\mathcal{H}} e^{-S},
\end{align}
we find 
\begin{equation}
\hat{\mathcal{H}} e^{-S} |\psi_n \rangle = \mathcal{E}_n e^{-S} |\psi_n \rangle ,
\end{equation}
that is, \( e^{-S} |\psi_n \rangle \) is an eigenstate of the original Hamiltonian \( \hat{\mathcal{H}} \) with the same eigenvalue \( \mathcal{E}_n \) as \( |\psi_n \rangle \). This is expected, since \( e^{-S} \) represents a unitary transformation and thus preserves the spectrum of the Hamiltonian. The eigenstates of \( \hat{\mathcal{H}} \) are given by  
\begin{equation}
e^{-S} |N_{\mathbf{q}_1}, \ldots, N_{\mathbf{q}_N} \rangle b^\dagger | 0 \rangle_f,
\end{equation}
which, in the original basis, are written as
\begin{equation}
e^{-S} |N_{\mathbf{q}_1}, \ldots, N_{\mathbf{q}_N} \rangle b^\dagger |0 \rangle_f = e^{- \sum_{\mathbf{q}} r_{\mathbf{q}} (a^\dagger_{\mathbf{q}} - a_{\mathbf{q}}) b^\dagger b} |N_{\mathbf{q}_1}, \ldots, N_{\mathbf{q}_N} \rangle b^\dagger |0 \rangle_f .
\end{equation}
Since \( b^\dagger b \) acts as a projector onto the occupied impurity state, on the subspace spanned by \( b^\dagger |0\rangle_f \) it behaves like the identity operator. Therefore, when acting on this state, we can replace \( b^\dagger b \) by the identity operator \( \mathds{1} \), and we write
\begin{equation}
e^{-S} |N_{\mathbf{q}_1}, \ldots, N_{\mathbf{q}_N} \rangle b^\dagger |0\rangle_f = e^{- \sum_{\mathbf{q}} r_{\mathbf{q}} (a^\dagger_{\mathbf{q}} - a_{\mathbf{q}})} |N_{\mathbf{q}_1}, \ldots, N_{\mathbf{q}_N} \rangle b^\dagger |0 \rangle_f .
\end{equation}
The exponential operator on the right-hand side is of the form \( e^{A + B} \), so we can use Glauber's formula $\eqref{eq: Glauberformula}$. We set
\begin{equation}
A = -\sum_{\mathbf{q}} r_{\mathbf{q}}\, a^\dagger_{\mathbf{q}}, 
\end{equation}
\begin{equation}
B = \sum_{\mathbf{q}} r_{\mathbf{q}}\, a_{\mathbf{q}},
\end{equation}
and we find that the eigenfunctions in the original basis are  
\begin{equation}
e^{-S} |N_{\mathbf{q}_1}, \ldots, N_{\mathbf{q}_N} \rangle b^\dagger |0 \rangle_f = e^{-\frac{1}{2} \sum_{\mathbf{q}} r_{\mathbf{q}}^2} e^{-\sum_{\mathbf{q}} r_{\mathbf{q}} a^\dagger_{\mathbf{q}}}
\, e^{\sum_{\mathbf{q}} r_{\mathbf{q}} a_{\mathbf{q}}} \, |N_{\mathbf{q}_1}, \ldots, N_{\mathbf{q}_N} \rangle b^\dagger |0 \rangle_f .
\end{equation}
In the transformed basis, all harmonic oscillators occupy their ground state, so the dressed ground state reads
\begin{equation}
| \tilde{G}_s \rangle = | 0, 0, 0, \ldots, 0 \rangle\, b^\dagger | 0 \rangle_f,
\end{equation}
whereas in the original basis, this state is expressed as
\begin{equation}
| G_s \rangle = e^{- \frac{1}{2} \sum_{\mathbf{q}} r_{\mathbf{q}}^2} e^{- \sum_{\mathbf{q}} r_{\mathbf{q}} a^\dagger_{\mathbf{q}}} e^{\sum_{\mathbf{q}} r_{\mathbf{q}} a_{\mathbf{q}}} | 0, \ldots, 0 \rangle\, b^\dagger | 0 \rangle.
\end{equation}
Since
\begin{equation}
e^{\sum_{\mathbf{q}} r_{\mathbf{q}} a_{\mathbf{q}}} | 0, \ldots, 0 \rangle = | 0, \ldots, 0 \rangle,
\end{equation}
the exponential acts as the identity on the vacuum. Thus, the ground state becomes
\begin{align}
| G_s \rangle &= e^{- \frac{1}{2} \sum_{\mathbf{q}} r_{\mathbf{q}}^2}
\, e^{- \sum_{\mathbf{q}} r_{\mathbf{q}} a^\dagger_{\mathbf{q}}}
| 0, \ldots, 0 \rangle\, b^\dagger | 0 \rangle_f = \notag \\
&= e^{- \frac{1}{2} \sum_{\mathbf{q}} r_{\mathbf{q}}^2}
\left( \prod_{\mathbf{q}} e^{- r_{\mathbf{q}} a^\dagger_{\mathbf{q}}} | 0 \rangle_{b,\mathbf{q}} \right) b^\dagger | 0 \rangle_f.
\end{align}
Note that the fundamental state of $\eqref{eq: singleimpuritymodel}$ is a coherent state, being the product of coherent states. Now, referring to the old basis, we calculate the number of phonons with momentum $\mathbf{q}$. If the interaction is null, the average number of phonons $\left\langle a^\dagger_{\mathbf{q}} a_{\mathbf{q}} \right\rangle$ in the ground state vanishes. Otherwise, the average number of phonons can be calculated as
\begin{align}
\sum_{\mathbf{q}} \left\langle G_s \left| a^\dagger_{\mathbf{q}} a_{\mathbf{q}} \right| G_s \right\rangle &= \sum_{\mathbf{q}} \left\langle \tilde{G}_s \left| e^S a^\dagger_{\mathbf{q}} e^{-S} \, e^S a_{\mathbf{q}} e^{-S} \right| \tilde{G}_s \right\rangle = \notag \\
&= \sum_{\mathbf{q}} \frac{M_{\mathbf{q}}^2}{(\hbar \omega_{\mathbf{q}})^2} .
\end{align}
The greater the interaction strength $M_{\mathbf{q}}$, the greater the number of phonons excited.
\subsection{Green's function in the independent boson model}
Here, we aim to compute the retarded Green's function of the single-level polaron, that is
\begin{align}
G^{(r)}_{bb^\dagger}(t) &= -\frac{i}{\hbar} \Theta(t) \left\langle \left\lbrace b(t), b^\dagger(0) \right\rbrace \right\rangle = \notag \\
&= -\frac{i}{\hbar} \Theta(t) \left\langle b(t) b^\dagger(0) + b^\dagger(0) b(t) \right\rangle = \notag \\
&= -\frac{i}{\hbar} \Theta(t) \frac{1}{Z} \mathrm{Tr} \left[ e^{-\beta \hat{\mathcal{H}}} \left( b(t) b^\dagger(0) + b^\dagger(0) b(t) \right) \right] = \notag \\
&= -\frac{i}{\hbar} \Theta(t) \frac{1}{Z} \mathrm{Tr} \left[ e^{-\beta \hat{\mathcal{H}}} \left( e^{\frac{i}{\hbar} \hat{\mathcal{H}} t} b\, e^{-\frac{i}{\hbar} \hat{\mathcal{H}} t} b^\dagger + b^\dagger\, e^{\frac{i}{\hbar} \hat{\mathcal{H}} t} b\, e^{-\frac{i}{\hbar} \hat{\mathcal{H}} t} \right) \right].
\end{align}
Since we can diagonalize \( \hat{\tilde{\mathcal{H}}} = e^{S} \hat{\mathcal{H}} e^{-S} \), but not \( \hat{\mathcal{H}} \) directly, we apply the similarity transformation to each operator: we multiply to the left by \( e^{S} \), and to the right by \( e^{-S} \), and use the cyclic property of the trace to move the final \( e^{S} \) to the front
\begin{align}
G^{(r)}_{bb^\dagger}(t) &= -\frac{i}{\hbar} \Theta(t) \frac{1}{Z} \mathrm{Tr} \Big[ e^{S} e^{-\beta \hat{\mathcal{H}}} e^{-S} \Big( e^{S} e^{\frac{i}{\hbar} \hat{\mathcal{H}} t} e^{-S} e^{S} b\, e^{-S} e^{S} e^{-\frac{i}{\hbar} \hat{\mathcal{H}} t} e^{-S} e^{S} b^\dagger e^{-S} + \notag \\
&\quad + e^{S} b^\dagger e^{-S} e^{S} e^{\frac{i}{\hbar} \hat{\mathcal{H}} t} e^{-S} e^{S} b\, e^{-S} e^{S} e^{-\frac{i}{\hbar} \hat{\mathcal{H}} t} e^{-S} \Big) \Big].
\end{align}
From the relations
\begin{equation}
e^{S} e^{-\beta \hat{\mathcal{H}}} e^{-S} = e^{-\beta \hat{\tilde{\mathcal{H}}}},
\end{equation}
\begin{equation}
e^{S} e^{i \frac{\hat{\mathcal{H}}}{\hbar} t} e^{-S} = e^{i \frac{\hat{\tilde{\mathcal{H}}}}{\hbar} t},
\end{equation}
\begin{equation}
e^{S} b e^{-S} = b \, e^{-\sum_{\mathbf{q}} r_{\mathbf{q}} \left(a^\dagger_{\mathbf{q}} - a_{\mathbf{q}}\right)},
\end{equation}
\begin{equation}
e^{S} b^\dagger e^{-S} = b^\dagger \, e^{\sum_{\mathbf{q}} r_{\mathbf{q}} \left(a^\dagger_{\mathbf{q}} - a_{\mathbf{q}}\right)},
\end{equation}
we obtain the expression
\begin{align}
G^{(r)}_{bb^\dagger}(t) &= -\frac{i}{\hbar} \, \Theta(t) \, \frac{1}{Z} \, \text{Tr} \left[ e^{-\beta \hat{\tilde{\mathcal{H}}}} \left( e^{i \frac{\hat{\tilde{\mathcal{H}}}}{\hbar} t} \, b \, e^{-\sum_{\mathbf{q}} r_{\mathbf{q}} (a^\dagger_{\mathbf{q}} - a_{\mathbf{q}})} \, e^{-i \frac{\hat{\tilde{\mathcal{H}}}}{\hbar} t} \, b^\dagger \, e^{\sum_{\mathbf{q}} r_{\mathbf{q}} (a^\dagger_{\mathbf{q}} - a_{\mathbf{q}})} \right. \right. \notag \\
&\left. \left. + \, b^\dagger \, e^{\sum_{\mathbf{q}} r_{\mathbf{q}} (a^\dagger_{\mathbf{q}} - a_{\mathbf{q}})} \, e^{i \frac{\hat{\tilde{\mathcal{H}}}}{\hbar} t} \, b \, e^{-\sum_{\mathbf{q}} r_{\mathbf{q}} (a^\dagger_{\mathbf{q}} - a_{\mathbf{q}})} \, e^{-i \frac{\hat{\tilde{\mathcal{H}}}}{\hbar} t} \right) \right] .
\end{align}
Since the Hamiltonian $\hat{\tilde{\mathcal{H}}}$ is diagonal, Under Heisenberg time evolution, the operators $a_{\textbf{q}}$, $a^\dagger_{\textbf{q}}$, $b$, $b^\dagger$ evolve multiplicatively by a phase factor determined by their respective eigenfrequencies. Therefore, the Green's function becomes
\begin{align}
G^{(r)}_{bb^\dagger}(t) &= -\frac{i}{\hbar} \, \Theta(t) \, \frac{1}{Z} \, \text{Tr} \left[ e^{-\beta \hat{\tilde{\mathcal{H}}}} \left( e^{- i \frac{\tilde{\mathcal{E}}}{\hbar} t} \, b \, e^{-\sum_{\mathbf{q}} r_{\mathbf{q}} (a^\dagger_{\mathbf{q}} e^{i \omega_{\mathbf{q}} t} - a_{\mathbf{q}} e^{-i \omega_{\mathbf{q}} t})} b^\dagger \, e^{\sum_{\mathbf{q}} r_{\mathbf{q}} (a^\dagger_{\mathbf{q}} - a_{\mathbf{q}})} \right.\right. = \notag \\
& \left.\left. + \, b^\dagger \, e^{\sum_{\mathbf{q}} r_{\mathbf{q}} (a^\dagger_{\mathbf{q}} - a_{\mathbf{q}})} e^{- i \frac{\tilde{\mathcal{E}}}{\hbar} t} \, b \, e^{-\sum_{\mathbf{q}} r_{\mathbf{q}} (a^\dagger_{\mathbf{q}} e^{i \omega_{\mathbf{q}} t} - a_{\mathbf{q}} e^{-i \omega_{\mathbf{q}} t})}
\right) \right].
\end{align}
If we define the operator
\begin{equation}
d(t) \equiv e^{\sum_{\mathbf{q}} r_{\mathbf{q}} \left( a^\dagger_{\mathbf{q}} e^{i \omega_{\mathbf{q}} t} - a_{\mathbf{q}} e^{-i \omega_{\mathbf{q}} t} \right)},
\end{equation}
we can express the time-evolved dressing operator in a compact form as follows
\begin{equation}
G^{(r)}_{bb^\dagger}(t) = -\frac{i}{\hbar} \, \Theta(t) \, \frac{1}{Z} \, \mathrm{Tr} \left\lbrace e^{- i \frac{\tilde{\mathcal{E}}}{\hbar} t} e^{-\beta \hat{\tilde{\mathcal{H}}}} \left[ (b d)(t) \, \left( b^\dagger d^\dagger \right)(0) + \left( b^\dagger d^\dagger\right)(0) \, (b d)(t) \right] \right\rbrace.
\end{equation}
The second term vanishes: indeed, since the operator combination \( (b^\dagger d^\dagger)(0) (b d)(t) \) acting on the impurity vacuum state yields the null state, it can be identified as the null operator over the entire interval \([0, t]\). Consequently, we have
\begin{equation}
G^{(r)}_{bb^\dagger}(t) = -\frac{i}{\hbar} \, \Theta(t) \, \frac{1}{Z} \, \mathrm{Tr} \left[ e^{- i \frac{\tilde{\mathcal{E}}}{\hbar} t} e^{-\beta \hat{\tilde{\mathcal{H}}}} (b d)(t) \, \left( b^\dagger d^\dagger \right)(0) \right] .
\end{equation}
We now explicitly insert the time-evolved operators and write
\begin{equation}
G^{(r)}_{bb^\dagger}(t) = -\frac{i}{\hbar} \, \Theta(t) \, \frac{1}{Z} \,
\mathrm{Tr} \left[ e^{- i \frac{\tilde{\mathcal{E}}}{\hbar} t} e^{-\beta \hat{\tilde{\mathcal{H}}}} b \, e^{-\sum_{\mathbf{q}} r_{\mathbf{q}} (a^\dagger_{\mathbf{q}} e^{i \omega_{\mathbf{q}} t} - a_{\mathbf{q}} e^{-i \omega_{\mathbf{q}} t})} b^\dagger \, e^{\sum_{\mathbf{q}} r_{\mathbf{q}} (a^\dagger_{\mathbf{q}} - a_{\mathbf{q}})} \right] .
\end{equation}
We evaluate the above trace with respect to the basis $\eqref{eq: autostatisingleimpuritymodeldiagonalizzata}$, in which the Hamiltonian $\eqref{eq: singleimpuritymodel4}$ is diagonal: the bosonic and fermionic operators act independently on the eigenstates of $\hat{\tilde{\mathcal{H}}}$, that is,
\begin{align}
G^{(r)}_{bb^\dagger}(t) &= - \frac{i}{\hbar} \, \Theta(t) \, \frac{1}{Z_f} \, \mathrm{Tr} \left[ e^{- i \frac{\tilde{\mathcal{E}}}{\hbar} t} e^{-\beta \tilde{\mathcal{E}} b^\dagger b} b b^\dagger \right] \frac{1}{Z_b} \, \mathrm{Tr} \left[ e^{-\beta \sum_{\mathbf{q}} \hbar \omega_{\mathbf{q}} a^\dagger_{\mathbf{q}} a_{\mathbf{q}}} e^{-\sum_{\mathbf{q}} r_{\mathbf{q}} (a^\dagger_{\mathbf{q}} e^{i \omega_{\mathbf{q}} t} - a_{\mathbf{q}} e^{-i \omega_{\mathbf{q}} t})} e^{\sum_{\mathbf{q}} r_{\mathbf{q}} (a^\dagger_{\mathbf{q}} - a_{\mathbf{q}})} \right] = \notag \\
&= - \frac{i}{\hbar} \, \Theta(t) \, e^{- i \frac{\tilde{\mathcal{E}}}{\hbar} t} \frac{1}{Z_f} \, \mathrm{Tr} \left[ e^{-\beta \tilde{\mathcal{E}} b^\dagger b} b b^\dagger \right] \frac{1}{Z_b} \, \mathrm{Tr} \left[ e^{-\beta \sum_{\mathbf{q}} \hbar \omega_{\mathbf{q}} a^\dagger_{\mathbf{q}} a_{\mathbf{q}}} e^{-\sum_{\mathbf{q}} r_{\mathbf{q}} (a^\dagger_{\mathbf{q}} e^{i \omega_{\mathbf{q}} t} - a_{\mathbf{q}} e^{-i \omega_{\mathbf{q}} t})} e^{\sum_{\mathbf{q}} r_{\mathbf{q}} (a^\dagger_{\mathbf{q}} - a_{\mathbf{q}})} \right],
\end{align}
where the partition function factorizes as a product of the bosonic and fermionic partition functions, i.e.,
\begin{equation}
Z = Z_f Z_b,
\end{equation}
with
\begin{equation}
Z_f = \mathrm{Tr} \left[ e^{-\beta \tilde{\mathcal{E}} b^\dagger b} \right],
\end{equation}
\begin{equation}
Z_b = \mathrm{Tr} \left[ e^{-\beta \sum_{\mathbf{q}} \hbar \omega_{\mathbf{q}} a^\dagger_{\mathbf{q}} a_{\mathbf{q}}} \right].
\end{equation}
Working in the canonical ensemble with fermionic occupation restricted to 0 and 1, the fermionic trace can be evaluated explicitly. Assuming that the impurity level is initially unoccupied, the fermionic contribution reduces to unity. As a result, only the bosonic contribution remains, that is
\begin{align}
& \frac{1}{Z_b} \, \mathrm{Tr} \Bigg[ e^{-\beta \sum_{\mathbf{q}} \hbar \omega_{\mathbf{q}} a^\dagger_{\mathbf{q}} a_{\mathbf{q}}} \, e^{-\sum_{\mathbf{q}} r_{\mathbf{q}} (a^\dagger_{\mathbf{q}} e^{i \omega_{\mathbf{q}} t} - a_{\mathbf{q}} e^{-i \omega_{\mathbf{q}} t})} e^{\sum_{\mathbf{q}} r_{\mathbf{q}} (a^\dagger_{\mathbf{q}} - a_{\mathbf{q}})} \Bigg] = \frac{1}{Z_b} \sum_{N_{\mathbf{q}_1}, \dots, N_{\mathbf{q}_N}, \dots, N_{\infty}} \notag \\
& \Big\langle N_{\mathbf{q}_1}, \dots, N_{\mathbf{q}_N}, \dots, N_{\infty} \Big| e^{-\beta \sum_{\mathbf{q}} \hbar \omega_{\mathbf{q}} a^\dagger_{\mathbf{q}} a_{\mathbf{q}}} e^{-\sum_{\mathbf{q}} r_{\mathbf{q}} (a^\dagger_{\mathbf{q}} e^{i \omega_{\mathbf{q}} t} - a_{\mathbf{q}} e^{-i \omega_{\mathbf{q}} t})}
\, e^{\sum_{\mathbf{q}} r_{\mathbf{q}} (a^\dagger_{\mathbf{q}} - a_{\mathbf{q}})}
\Big| N_{\mathbf{q}_1}, \dots, N_{\mathbf{q}_N}, \dots N_{\infty} \Big\rangle.
\end{align}
Now, each exponential function is a sum of operators that individually act on the state labeled by $N_{\mathbf{q}}$, so the trace over many indices can be replaced by a product over $\mathbf{q}$ of series indexed by the infinite admissible values of each $N_{\mathbf{q}}$. That is,
\begin{align}
&\frac{1}{Z_b} \, \mathrm{Tr} \left[ e^{-\beta \sum_{\mathbf{q}} \hbar \omega_{\mathbf{q}} a^\dagger_{\mathbf{q}} a_{\mathbf{q}}} \, e^{-\sum_{\mathbf{q}} r_{\mathbf{q}} (a^\dagger_{\mathbf{q}} e^{i \omega_{\mathbf{q}} t} - a_{\mathbf{q}} e^{-i \omega_{\mathbf{q}} t})} \, e^{\sum_{\mathbf{q}} r_{\mathbf{q}} (a^\dagger_{\mathbf{q}} - a_{\mathbf{q}})} \right] = \notag \\
&= \prod_{\mathbf{q}} \frac{1}{Z_{b,\mathbf{q}}} \sum_{N_{\mathbf{q}} = 0}^\infty \left\langle N_{\mathbf{q}} \left| e^{-\beta \hbar \omega_{\mathbf{q}} a^\dagger_{\mathbf{q}} a_{\mathbf{q}}} \, e^{-r_{\mathbf{q}} (a^\dagger_{\mathbf{q}} e^{i \omega_{\mathbf{q}} t} - a_{\mathbf{q}} e^{-i \omega_{\mathbf{q}} t})} \, e^{r_{\mathbf{q}} (a^\dagger_{\mathbf{q}} - a_{\mathbf{q}})} \right| N_{\mathbf{q}} \right\rangle = \notag \\
&= \prod_{\mathbf{q}} \frac{1}{Z_{b,\mathbf{q}}} \sum_{N_{\mathbf{q}} = 0}^\infty e^{-\beta \hbar \omega_{\mathbf{q}} N_{\mathbf{q}}} \left\langle N_{\mathbf{q}} \left| e^{-r_{\mathbf{q}} (a^\dagger_{\mathbf{q}} e^{i \omega_{\mathbf{q}} t} - a_{\mathbf{q}} e^{-i \omega_{\mathbf{q}} t})} \, e^{r_{\mathbf{q}} (a^\dagger_{\mathbf{q}} - a_{\mathbf{q}})} \right| N_{\mathbf{q}} \right\rangle.
\end{align}
We made the density operator exponent explicit using the number operator $a^\dagger_{\mathbf{q}} a_{\mathbf{q}}$, and replaced it by its eigenvalue $n_{\mathbf{q}}$. The bosonic partition function for each mode $\mathbf{q}$ factorizes as follows
\begin{align}
Z_{b, \mathbf{q}} &= \mathrm{Tr} \left[ e^{-\beta \hbar \omega_{\mathbf{q}} a^\dagger_{\mathbf{q}} a_{\mathbf{q}}} \right] = \notag \\
&= \sum_{N_{\mathbf{q}} = 0}^{\infty} e^{-\beta \hbar \omega_{\mathbf{q}} N_{\mathbf{q}}} = \notag \\
&= \frac{1}{1 - e^{-\beta \hbar \omega_{\mathbf{q}}}}.
\end{align}
Therefore, the bosonic trace becomes
\begin{equation}
\prod_{\mathbf{q}} \left(1 - e^{-\beta \hbar \omega_{\mathbf{q}}} \right) 
\sum_{N_{\mathbf{q}} = 0}^{\infty} e^{-\beta \hbar \omega_{\mathbf{q}} N_{\mathbf{q}}} 
\left\langle N_{\mathbf{q}} \left| 
e^{-r_{\mathbf{q}} (a^\dagger_{\mathbf{q}} e^{i \omega_{\mathbf{q}} t} - a_{\mathbf{q}} e^{-i \omega_{\mathbf{q}} t})} 
\, e^{r_{\mathbf{q}} (a^\dagger_{\mathbf{q}} - a_{\mathbf{q}})} 
\right| N_{\mathbf{q}} \right\rangle.
\end{equation}
Now, from the Glauber formula $\eqref{eq: Glauberformula}$ in the form 
\begin{align}
e^{-A + B} &= e^{-A} e^{B} e^{\frac{1}{2} [A, B]} ,
\end{align}
we set 
\begin{equation}
A = r_{\mathbf{q}} a^\dagger_{\mathbf{q}} e^{i \omega_{\mathbf{q}} t}, 
\end{equation}
\begin{equation}
B = r_{\mathbf{q}} a_{\mathbf{q}} e^{-i \omega_{\mathbf{q}} t}, 
\end{equation}
and we get
\begin{align}
e^{-r_{\mathbf{q}} \left( a^\dagger_{\mathbf{q}} e^{i \omega_{\mathbf{q}} t} - a_{\mathbf{q}} e^{-i \omega_{\mathbf{q}} t} \right)}
&= e^{-r_{\mathbf{q}} a^\dagger_{\mathbf{q}} e^{i \omega_{\mathbf{q}} t}} \, e^{r_{\mathbf{q}} a_{\mathbf{q}} e^{-i \omega_{\mathbf{q}} t}} \, e^{- \frac{1}{2} r_{\mathbf{q}}^2} = \notag \\
&= e^{- \frac{1}{2} r_{\mathbf{q}}^2} \, e^{-r_{\mathbf{q}} a^\dagger_{\mathbf{q}} e^{i \omega_{\mathbf{q}} t}} \, e^{r_{\mathbf{q}} a_{\mathbf{q}} e^{-i \omega_{\mathbf{q}} t}} .
\end{align}
Similarly, from the Glauber formula $\eqref{eq: Glauberformula}$ in the form
\begin{align}
e^{A - B} &= e^{A} e^{-B} e^{\frac{1}{2} [A, B]} ,
\end{align}
we set
\begin{equation}
A = r_{\mathbf{q}} a^\dagger_{\mathbf{q}} , 
\end{equation}
\begin{equation}
B = r_{\mathbf{q}} a_{\mathbf{q}} ,
\end{equation}
and we get
\begin{align}
e^{r_{\mathbf{q}} (a^\dagger_{\mathbf{q}} - a_{\mathbf{q}})} 
&= e^{r_{\mathbf{q}} a^\dagger_{\mathbf{q}}} \, e^{-r_{\mathbf{q}} a_{\mathbf{q}}} \, e^{- \frac{1}{2} r_{\mathbf{q}}^2} = \notag \\
&= e^{- \frac{1}{2} r_{\mathbf{q}}^2} \, e^{r_{\mathbf{q}} a^\dagger_{\mathbf{q}}} \, e^{-r_{\mathbf{q}} a_{\mathbf{q}}},
\end{align}
consequently the bosonic trace becomes
\begin{equation}
\prod_{\mathbf{q}} \left( 1 - e^{-\beta \hbar \omega_{\mathbf{q}}} \right)
\sum_{N_{\mathbf{q}} = 0}^{\infty}
e^{-\beta \hbar \omega_{\mathbf{q}} N_{\mathbf{q}}} \,
\left\langle N_{\mathbf{q}} \left| \,
e^{-r_{\mathbf{q}}^2} \,
e^{-r_{\mathbf{q}} a^\dagger_{\mathbf{q}} e^{i \omega_{\mathbf{q}} t}} \,
e^{r_{\mathbf{q}} a_{\mathbf{q}} e^{-i \omega_{\mathbf{q}} t}} \,
e^{r_{\mathbf{q}} a^\dagger_{\mathbf{q}}} \,
e^{-r_{\mathbf{q}} a_{\mathbf{q}}} \,
\right| N_{\mathbf{q}} \right\rangle .
\end{equation}
Now, regarding the product \( e^{r_{\mathbf{q}} a_{\mathbf{q}} e^{-i \omega_{\mathbf{q}} t}} \, e^{r_{\mathbf{q}} a^\dagger_{\mathbf{q}}} \), we want to rewrite it so that the product \( e^{r_{\mathbf{q}} a_{\mathbf{q}} e^{-i \omega_{\mathbf{q}} t}} \) appears to the right of the operator \( e^{r_{\mathbf{q}} a^\dagger_{\mathbf{q}}} \). We write
\begin{equation}
e^{r_{\mathbf{q}} a_{\mathbf{q}} e^{-i \omega_{\mathbf{q}} t}} \, e^{r_{\mathbf{q}} a^\dagger_{\mathbf{q}}}
\equiv
e^{r_{\mathbf{q}} a^\dagger_{\mathbf{q}}}
\left(
e^{-r_{\mathbf{q}} a^\dagger_{\mathbf{q}}}
\, e^{r_{\mathbf{q}} a_{\mathbf{q}} e^{-i \omega_{\mathbf{q}} t}}
\, e^{r_{\mathbf{q}} a^\dagger_{\mathbf{q}}}
\right) ,
\end{equation}
then consider the product \( e^{r_{\mathbf{q}} a_{\mathbf{q}} e^{-i \omega_{\mathbf{q}} t}} \, e^{r_{\mathbf{q}} a^\dagger_{\mathbf{q}}} \). Using the Glauber formula $\eqref{eq: Glauberformula}$, we set
\begin{equation}
A = r_{\mathbf{q}} a_{\mathbf{q}} e^{-i \omega_{\mathbf{q}} t}, 
\end{equation}
\begin{equation}
B = r_{\mathbf{q}} a^\dagger_{\mathbf{q}}, 
\end{equation}
and we get
\begin{align}
e^{r_{\mathbf{q}} a_{\mathbf{q}} e^{-i \omega_{\mathbf{q}} t}} \, e^{r_{\mathbf{q}} a^\dagger_{\mathbf{q}}} &= e^{\frac{1}{2} r_{\mathbf{q}}^2 \left[ a_{\mathbf{q}} e^{-i \omega_{\mathbf{q}} t}, a^\dagger_{\mathbf{q}} \right]} \, e^{r_{\mathbf{q}} a_{\mathbf{q}} e^{-i \omega_{\mathbf{q}} t} + r_{\mathbf{q}} a^\dagger_{\mathbf{q}}} = \notag \\
&= e^{\frac{1}{2} r_{\mathbf{q}}^2 e^{-i \omega_{\mathbf{q}} t} \left[ a_{\mathbf{q}}, a^\dagger_{\mathbf{q}} \right]}
\, e^{r_{\mathbf{q}} a_{\mathbf{q}} e^{-i \omega_{\mathbf{q}} t} + r_{\mathbf{q}} a^\dagger_{\mathbf{q}}} = \notag \\
&= e^{\frac{1}{2} r_{\mathbf{q}}^2 e^{-i \omega_{\mathbf{q}} t}} \,
e^{r_{\mathbf{q}} a_{\mathbf{q}} e^{-i \omega_{\mathbf{q}} t} + r_{\mathbf{q}} a^\dagger_{\mathbf{q}}} ,
\end{align}
then
\begin{equation}
e^{-r_{\mathbf{q}} a^\dagger_{\mathbf{q}}} e^{r_{\mathbf{q}} a_{\mathbf{q}} e^{-i \omega_{\mathbf{q}} t}} e^{r_{\mathbf{q}} a^\dagger_{\mathbf{q}}} = e^{\frac{1}{2} r_{\mathbf{q}}^2 e^{-i \omega_{\mathbf{q}} t}} e^{-r_{\mathbf{q}} a^\dagger_{\mathbf{q}}}  e^{r_{\mathbf{q}} a_{\mathbf{q}} e^{-i \omega_{\mathbf{q}} t} + r_{\mathbf{q}} a^\dagger_{\mathbf{q}}} .
\end{equation}
Now we repeat the same procedure for the product $e^{-r_{\mathbf{q}} a^\dagger_{\mathbf{q}}} e^{r_{\mathbf{q}} a_{\mathbf{q}} e^{-i \omega_{\mathbf{q}} t} + r_{\mathbf{q}} a^\dagger_{\mathbf{q}}}$. Applying the Glauber formula $\eqref{eq: Glauberformula}$, we set
\begin{equation}
A = -r_{\mathbf{q}} a^\dagger_{\mathbf{q}}, 
\end{equation}
\begin{equation}
B = r_{\mathbf{q}} a_{\mathbf{q}} e^{-i \omega_{\mathbf{q}} t} + r_{\mathbf{q}} a^\dagger_{\mathbf{q}},
\end{equation}
and we get
\begin{align}
e^{-r_{\mathbf{q}} a^\dagger_{\mathbf{q}}} \, e^{r_{\mathbf{q}} a_{\mathbf{q}} e^{-i \omega_{\mathbf{q}} t} + r_{\mathbf{q}} a^\dagger_{\mathbf{q}}} &= e^{\frac{1}{2} r_{\mathbf{q}}^2 \left[ -a^\dagger_{\mathbf{q}}, a_{\mathbf{q}} e^{-i \omega_{\mathbf{q}} t} + a^\dagger_{\mathbf{q}} \right]} \, e^{-r_{\mathbf{q}} a^\dagger_{\mathbf{q}} + r_{\mathbf{q}} a_{\mathbf{q}} e^{-i \omega_{\mathbf{q}} t} + r_{\mathbf{q}} a^\dagger_{\mathbf{q}}} = \notag \\
&= e^{\frac{1}{2} r_{\mathbf{q}}^2 \left[ -a^\dagger_{\mathbf{q}}, a_{\mathbf{q}} e^{-i \omega_{\mathbf{q}} t} \right]}
\, e^{r_{\mathbf{q}} a_{\mathbf{q}} e^{-i \omega_{\mathbf{q}} t}} = \notag \\
&= e^{\frac{1}{2} r_{\mathbf{q}}^2 e^{-i \omega_{\mathbf{q}} t} \left[ -a^\dagger_{\mathbf{q}}, a_{\mathbf{q}} \right]}
\, e^{r_{\mathbf{q}} a_{\mathbf{q}} e^{-i \omega_{\mathbf{q}} t}} = \notag \\
&= e^{\frac{1}{2} r_{\mathbf{q}}^2 e^{-i \omega_{\mathbf{q}} t}} \,
e^{r_{\mathbf{q}} a_{\mathbf{q}} e^{-i \omega_{\mathbf{q}} t}} .
\end{align}
We group the results as follows
\begin{align}
e^{r_{\mathbf{q}} a_{\mathbf{q}} e^{-i \omega_{\mathbf{q}} t}} e^{r_{\mathbf{q}} a_{\mathbf{q}}^\dagger} &\equiv e^{r_{\mathbf{q}} a_{\mathbf{q}}^\dagger} \left( e^{-r_{\mathbf{q}} a_{\mathbf{q}}^\dagger} \, e^{r_{\mathbf{q}} a_{\mathbf{q}} e^{-i \omega_{\mathbf{q}} t}} \, e^{r_{\mathbf{q}} a_{\mathbf{q}}^\dagger} \right) = \notag \\
&= e^{r_{\mathbf{q}} a_{\mathbf{q}}^\dagger} \left( e^{\frac{1}{2} r_{\mathbf{q}}^2 e^{-i \omega_{\mathbf{q}} t}} \, e^{-r_{\mathbf{q}} a_{\mathbf{q}}^\dagger} \, e^{r_{\mathbf{q}} a_{\mathbf{q}} e^{-i \omega_{\mathbf{q}} t} + r_{\mathbf{q}} a_{\mathbf{q}}^\dagger}
\right) = \notag \\
&= e^{r_{\mathbf{q}} a_{\mathbf{q}}^\dagger} \left( e^{\frac{1}{2} r_{\mathbf{q}}^2 e^{-i \omega_{\mathbf{q}} t}} \, e^{\frac{1}{2} r_{\mathbf{q}}^2 e^{-i \omega_{\mathbf{q}} t}} \, e^{r_{\mathbf{q}} a_{\mathbf{q}} e^{-i \omega_{\mathbf{q}} t}} \right) = \notag \\
&= e^{r_{\mathbf{q}}^2 e^{-i \omega_{\mathbf{q}} t}} \,
e^{r_{\mathbf{q}} a_{\mathbf{q}}^\dagger} \,
e^{r_{\mathbf{q}} a_{\mathbf{q}} e^{-i \omega_{\mathbf{q}} t}},
\end{align}
that is, we have thus obtained the desired ordering, and we can reorder the operators in the bosonic trace as follows
\begin{align}
& \prod_{\mathbf{q}} \left(1 - e^{-\beta \hbar \omega_{\mathbf{q}}} \right) \sum_{N_{\mathbf{q}} = 0}^{\infty} e^{-\beta \hbar \omega_{\mathbf{q}} N_{\mathbf{q}}} \left\langle N_{\mathbf{q}} \left| e^{-r_{\mathbf{q}}^2} \, e^{-r_{\mathbf{q}} a^\dagger_{\mathbf{q}} e^{i \omega_{\mathbf{q}} t}} \, e^{r_{\mathbf{q}} a_{\mathbf{q}} e^{-i \omega_{\mathbf{q}} t}} \, e^{r_{\mathbf{q}} a^\dagger_{\mathbf{q}}} \, e^{-r_{\mathbf{q}} a_{\mathbf{q}}} \right| N_{\mathbf{q}} \right\rangle = \notag \\
&= \prod_{\mathbf{q}} \left(1 - e^{-\beta \hbar \omega_{\mathbf{q}}} \right) \sum_{N_{\mathbf{q}} = 0}^{\infty} e^{-\beta \hbar \omega_{\mathbf{q}} N_{\mathbf{q}}} \left\langle N_{\mathbf{q}} \left| e^{-r_{\mathbf{q}}^2} \, e^{-r_{\mathbf{q}} a^\dagger_{\mathbf{q}} e^{i \omega_{\mathbf{q}} t}} \, e^{r_{\mathbf{q}}^2 e^{-i \omega_{\mathbf{q}} t}} \, e^{r_{\mathbf{q}} a^\dagger_{\mathbf{q}} e^{i \omega_{\mathbf{q}} t}} \, e^{r_{\mathbf{q}} a_{\mathbf{q}} e^{-i \omega_{\mathbf{q}} t}} \, e^{-r_{\mathbf{q}} a_{\mathbf{q}}} \right| N_{\mathbf{q}} \right\rangle = \notag \\
&= \prod_{\mathbf{q}} \left(1 - e^{-\beta \hbar \omega_{\mathbf{q}}} \right) \sum_{N_{\mathbf{q}} = 0}^{\infty} e^{-\beta \hbar \omega_{\mathbf{q}} N_{\mathbf{q}}} \left\langle N_{\mathbf{q}} \left| e^{-r_{\mathbf{q}}^2} \, e^{r_{\mathbf{q}}^2 e^{-i \omega_{\mathbf{q}} t}} \, e^{-r_{\mathbf{q}} a^\dagger_{\mathbf{q}} e^{i \omega_{\mathbf{q}} t}} \, e^{r_{\mathbf{q}} a^\dagger_{\mathbf{q}} e^{i \omega_{\mathbf{q}} t}} \, e^{r_{\mathbf{q}} a_{\mathbf{q}} e^{-i \omega_{\mathbf{q}} t}} \, e^{-r_{\mathbf{q}} a_{\mathbf{q}}} \right| N_{\mathbf{q}} \right\rangle = \notag \\ 
&= \prod_{\mathbf{q}}
\left(1 - e^{-\beta \hbar \omega_{\mathbf{q}}} \right)
\sum_{N_{\mathbf{q}} = 0}^{\infty}
e^{-\beta \hbar \omega_{\mathbf{q}} N_{\mathbf{q}}}
\left\langle N_{\mathbf{q}} \left| 
e^{-r_{\mathbf{q}}^2 \left(1 - e^{-i \omega_{\mathbf{q}} t}\right)} \,
e^{-r_{\mathbf{q}} a^\dagger_{\mathbf{q}} e^{i \omega_{\mathbf{q}} t}} \,
e^{r_{\mathbf{q}} a^\dagger_{\mathbf{q}} e^{i \omega_{\mathbf{q}} t}} \,
e^{r_{\mathbf{q}} a_{\mathbf{q}} e^{-i \omega_{\mathbf{q}} t}} \,
e^{-r_{\mathbf{q}} a_{\mathbf{q}}} \right| N_{\mathbf{q}} \right\rangle = \notag \\
&= \prod_{\mathbf{q}} e^{-r_{\mathbf{q}}^2 \left(1 - e^{-i \omega_{\mathbf{q}} t} \right)} \left(1 - e^{-\beta \hbar \omega_{\mathbf{q}}} \right) \sum_{N_{\mathbf{q}} = 0}^{\infty} e^{-\beta \hbar \omega_{\mathbf{q}} N_{\mathbf{q}}} \left\langle N_{\mathbf{q}} \left|  e^{r_{\mathbf{q}} \left(1 - e^{i \omega_{\mathbf{q}} t} \right) a^\dagger_{\mathbf{q}}} \, e^{-r_{\mathbf{q}} \left(1 - e^{-i \omega_{\mathbf{q}} t} \right) a_{\mathbf{q}}} \right| N_{\mathbf{q}} \right\rangle .
\end{align}
By temporarily ignoring the dependence on \( \textbf{q} \), consider the expression
\begin{equation}
e^{-r \left (1 - e^{-i \omega t} \right) a} |N\rangle,
\end{equation}
We now expand the exponential as a power series
\begin{equation}
e^{-r \left(1 - e^{-i \omega t}\right) a} |N\rangle = \sum_{m=0}^{N} \frac{(-1)^m}{m!} \left[ r \left(1 - e^{-i \omega t} \right) \right]^m a^m |N\rangle .
\end{equation}
Here the sum terminates at \( m = N \), since \( a^m |N\rangle = 0 \) for \( m > N \), because the annihilation operator cannot remove more quanta than present in the state. The action of \( a^m \) on the number state \( |N\rangle \) gives
\begin{equation}
a^m |N\rangle = \sqrt{\frac{N!}{(N - m)!}}\, |N - m\rangle ,
\end{equation}
thus, we obtain
\begin{equation}
e^{-r a \left(1 - e^{-i \omega t} \right)} |N \rangle = \sum_{m=0}^{N} \frac{(-1)^m}{m!} \left[ r \left( 1 - e^{-i \omega t} \right) \right]^m \sqrt{\frac{N!}{(N - m)!}}\, |N - m\rangle .
\end{equation}
Similarly, consider
\begin{align}
\langle N | e^{r a^\dagger (1 - e^{i \omega t})} &= \sum_{m'=0}^{N} \frac{1}{m'!} \left[ r \left(1 - e^{i \omega t} \right) \right]^{m'} \langle N | (a^\dagger)^{m'} = \notag \\
&= \sum_{m'=0}^{N} \frac{1}{m'!} \left[ r \left(1 - e^{i \omega t} \right) \right]^{m'} \sqrt{\frac{N!}{(N - m')!}}\, \langle N - m' | .
\end{align}
The scalar products between terms of the two series are proportional to 
\(\langle N - m' | N - m \rangle = \delta_{m, m'}\). Therefore, the full matrix element becomes
\begin{align}
\left\langle N \left| e^{r \left(1 - e^{i \omega t} \right) a^\dagger } \, e^{-r \left(1 - e^{-i \omega t} \right) a} \right| N \right\rangle &= \sum_{m=0}^{N}  \frac{(-1)^m}{(m!)^2} \left[ r^2 (1 - e^{i \omega t})(1 - e^{-i \omega t}) \right]^m \frac{N!}{(N - m)!} = \notag \\
&= \sum_{m=0}^{N} \frac{(-1)^m}{(m!)^2} \left[ r^2 |1 - e^{-i \omega t}|^2 \right]^m \frac{N!}{(N - m)!} = \notag \\
&= \sum_{m=0}^{N} \frac{(-1)^m}{m!} \left( r^2 |1 - e^{-i \omega t}|^2 \right)^m \frac{N!}{m!(N - m)!} = \notag \\
&= L_N \left( | \lambda |^2 \right) ,
\end{align}
where \( L_N \) is the Laguerre polynomial of order \( N \), and we define the complex variable
\begin{equation}
\lambda \equiv r (1 - e^{-i \omega t}) ,
\end{equation}
so that \( |\lambda|^2 = r^2 |1 - e^{-i \omega t}|^2 \). We set
\begin{equation}
z = e^{-\beta \hbar \omega} ,
\end{equation}
then, from the generating function for the Laguerre polynomials
\begin{equation}
\sum_{N=0}^{\infty} z^N L_N \left( |\lambda|^2 \right) = \frac{1}{1 - z} \exp\left( -\frac{z}{1 - z} |\lambda|^2 \right),
\end{equation}
we can compute the weighted sum
\begin{align}
\left(1 - e^{-\beta \hbar \omega} \right) \sum_{N=0}^{\infty} e^{-\beta \hbar \omega N} L_N \left( |\lambda|^2 \right) &= (1 - z) \sum_{N=0}^{\infty} z^N L_N \left( |\lambda|^2 \right) = \notag \\
&= (1 - z) \frac{1}{1 - z} \exp\left( -\frac{z}{1 - z} |\lambda|^2 \right) = \notag \\
&= \exp\left( -\frac{z}{1 - z} |\lambda|^2 \right) .
\end{align}
We also note the identity
\begin{align}
\frac{z}{1 - z} &= \frac{e^{-\beta \hbar \omega}}{1 - e^{-\beta \hbar \omega}} = \notag \\
&= \frac{1}{e^{\beta \hbar \omega} - 1} \equiv \notag \\
&\equiv n_{1}(\hslash \omega),
\end{align}
where \( n_{1}(\hslash \omega) \) is the Bose-Einstein statistics, then we get
\begin{equation}
\left(1 - e^{-\beta \hbar \omega} \right) \sum_{N=0}^{\infty} e^{-\beta \hbar \omega N} L_N \left( |\lambda|^2 \right) = \exp \left( - N_{1}(\hslash \omega) |\lambda|^2 \right).
\end{equation}
Restoring the $\mathbf{q}$ subscript and including the multiplicative factor $e^{-r_{\mathbf{q}}^2 \left(1 - e^{-i \omega_{\mathbf{q}} t} \right)}$, the bosonic trace finally becomes
\begin{equation}
\prod_{\mathbf{q}} \exp\left[ - r_{\mathbf{q}}^2 \left( \left(1 - e^{-i \omega_{\mathbf{q}} t} \right) + \left|1 - e^{-i \omega_{\mathbf{q}} t}\right|^2 n_1(\hslash \omega_{\mathbf{q}}) \right) \right].
\end{equation}
Note that from
\begin{align}
\left|1 - e^{-i \omega_{\mathbf{q}} t}\right|^2 &= \left( 1 - e^{-i \omega_{\mathbf{q}} t} \right) \left( 1 - e^{i \omega_{\mathbf{q}} t} \right) = \notag \\
&= 2 - e^{i \omega_{\mathbf{q}} t} - e^{-i \omega_{\mathbf{q}} t} ,
\end{align}
\begin{align}
\left(1 - e^{-i \omega_{\mathbf{q}} t}\right) 
+ \left|1 - e^{-i \omega_{\mathbf{q}} t}\right|^2 n_1\left(\hslash \omega_{\mathbf{q}}\right) &= \left(1 - e^{-i \omega_{\mathbf{q}} t}\right) 
+ \left(2 - e^{i \omega_{\mathbf{q}} t} - e^{-i \omega_{\mathbf{q}} t}\right) n_1\left(\hslash \omega_{\mathbf{q}}\right) = \notag \\
&= 1 - e^{-i \omega_{\mathbf{q}} t} + 2 n_1\left(\hslash \omega_{\mathbf{q}}\right) - n_1\left(\hslash \omega_{\mathbf{q}}\right) e^{i \omega_{\mathbf{q}} t} - n_1\left(\hslash \omega_{\mathbf{q}}\right) e^{-i \omega_{\mathbf{q}} t} = \notag \\
&= 1 + 2 n_1\left(\hslash \omega_{\mathbf{q}}\right) - \left(1 + n_1\left(\hslash \omega_{\mathbf{q}}\right)\right) e^{-i \omega_{\mathbf{q}} t} - n_1\left(\hslash \omega_{\mathbf{q}}\right) e^{i \omega_{\mathbf{q}} t} = \notag \\
&= \left(n_1\left(\hslash \omega_{\mathbf{q}}\right) + 1\right)\left(1 - e^{-i \omega_{\mathbf{q}} t}\right) + n_1\left(\hslash \omega_{\mathbf{q}}\right)\left(1 - e^{i \omega_{\mathbf{q}} t}\right) ,
\end{align}
the bosonic trace finally becomes
\begin{equation}
\prod_{\mathbf{q}} \exp\left[ - r_{\mathbf{q}}^2 \left( (n_1(\hslash \omega_{\mathbf{q}}) + 1)(1 - e^{-i \omega_{\mathbf{q}} t}) + n_1(\hslash \omega_{\mathbf{q}})(1 - e^{i \omega_{\mathbf{q}} t}) \right) \right].
\end{equation}
By grouping the bosonic and fermionic traces, the retarded Green's function in the independent boson model at thermodynamic equilibrium is given by
\begin{equation}
G^{(r)}_{bb^\dagger}(t) = -\frac{i}{\hbar} \, \Theta(t) \, e^{- i \frac{\tilde{\mathcal{E}}}{\hbar} t} \prod_q \exp\left[ - r_q^2 \left(n_1\left(\hslash \omega_{\mathbf{q}}\right) + 1\right)\left(1 - e^{-i \omega_{\mathbf{q}} t}\right) + n_1\left(\hslash \omega_{\mathbf{q}}\right)\left(1 - e^{i \omega_{\mathbf{q}} t}\right) \right] .
\end{equation}
\section{Sum rules for the moments of the spectral function}
We now present a set of exact sum rules for the spectral function, which follow from fundamental properties of the Green's function.
\begin{theorem}[Zeroth moment sum rule of the spectral function]
The zeroth-order moment of the spectral function $\eqref{eq: funzionespettrale}$ satisfy
\begin{equation}
\int_{-\infty}^{+\infty}\frac{d\omega}{2\pi}A_{AB}(\omega) = \  \left\langle [A,B]^{(\varepsilon)} \right\rangle.
\label{eq: momentonullodifunzionespettrale}
\end{equation}
\begin{proof}
\begin{align}
\int_{-\infty}^{+\infty} \dfrac{d\omega}{2\pi} A_{AB}(\omega) &= \int_{-\infty}^{+\infty} \dfrac{d\omega}{2\pi} \frac{2\pi}{Z} \sum_{n,m}A_{nm}B_{mn}\left(e^{-\beta \mathcal{E}_n} - \varepsilon e^{-\beta \mathcal{E}_m}\right) \delta \left[ \omega - \dfrac{\mathcal{E}_m - \mathcal{E}_n}{\hslash} \right] = \notag \\
&= \dfrac{2\pi}{Z} \sum_{n,m} A_{nm}B_{mn}\left(e^{-\beta \mathcal{E}_n} - \varepsilon e^{-\beta \mathcal{E}_m}\right) \int_{-\infty}^{+\infty} \dfrac{d\omega}{2\pi} \delta \left[ \omega - \dfrac{\mathcal{E}_m - \mathcal{E}_n}{\hslash} \right] = \notag \\
&= \dfrac{2\pi}{Z} \sum_{n,m} A_{nm}B_{mn}\left(e^{-\beta \mathcal{E}_n} - \varepsilon e^{-\beta \mathcal{E}_m}\right) \dfrac{1}{2\pi} = \notag \\
&= \dfrac{1}{Z} \sum_{n,m} A_{nm}B_{mn}\left(e^{-\beta \mathcal{E}_n} - \varepsilon e^{-\beta \mathcal{E}_m}\right). 
\end{align}
We compute the product
\begin{equation}
\dfrac{1}{Z} \sum_{n,m} A_{nm}B_{mn} e^{-\beta \mathcal{E}_n} - \dfrac{1}{Z} \sum_{n,m} A_{nm}B_{mn} \varepsilon e^{-\beta \mathcal{E}_m},
\end{equation}
we write the two traces as 
\begin{equation}
\dfrac{1}{Z} \sum_{n,m} \langle n | A | m \rangle \langle m | B | n \rangle \, e^{- \beta \mathcal{E}_n} = \dfrac{1}{Z} \sum_n \langle n | AB | n \rangle \, e^{- \beta \mathcal{E}_n} ,
\end{equation}
\begin{align}
- \dfrac{1}{Z} \sum_{n,m} \langle n | A | m \rangle \langle m | B | n \rangle \, \varepsilon \, e^{- \beta \mathcal{E}_m} &= - \dfrac{1}{Z} \sum_{n,m} \langle m | B | n \rangle \langle n | A | m \rangle \, \varepsilon \, e^{- \beta \mathcal{E}_m} = \notag \\
&= - \dfrac{1}{Z} \sum_m \langle m | B A | m \rangle \, \varepsilon \, e^{- \beta \mathcal{E}_m}.
\end{align}
We sum with respect to $n$ and $m$, respectively, and we obtain the thermal average of the operator $\left[ A,B \right]^{(\varepsilon)}$, that is, the thesis.
\end{proof}
\end{theorem}
\begin{theorem}[First moment sum rule of the spectral function]
The first-order moment of the spectral function $\eqref{eq: funzionespettrale}$ satisfy
\begin{equation}
\int_{-\infty}^{+\infty} \dfrac{d\omega}{2\pi} \omega A_{AB}(\omega) = \dfrac{1}{\hslash} \left\langle [[A,\hat{\mathcal{H}}],B]^{(\varepsilon)}\right\rangle,
\label{eq: momentoprimodifunzionespettrale}
\end{equation}
\begin{proof}
\begin{align}
\int_{-\infty}^{+\infty} \dfrac{d\omega}{2\pi} \omega A_{AB}(\omega) &= \int_{-\infty}^{+\infty} \dfrac{d\omega}{2\pi} \dfrac{2\pi}{Z} \omega \sum_{n,m}A_{nm}B_{mn}\left(e^{-\beta \mathcal{E}_n} - \varepsilon e^{-\beta \mathcal{E}_m}\right) \delta \left[ \omega - \dfrac{\mathcal{E}_m - \mathcal{E}_n}{\hslash} \right] = \notag \\
&= \dfrac{2\pi}{Z} \sum_{n,m} A_{nm}B_{mn}\left(e^{-\beta \mathcal{E}_n} - \varepsilon e^{-\beta \mathcal{E}_m}\right) \int_{-\infty}^{+\infty} \dfrac{d\omega}{2\pi} \omega \delta \left[ \omega - \dfrac{\mathcal{E}_m - \mathcal{E}_n}{\hslash} \right] = \notag \\
&= \frac{2 \pi}{Z}\sum_{n,m} A_{nm}B_{mn}\left(e^{-\beta \mathcal{E}_n} - \varepsilon e^{-\beta \mathcal{E}_m}\right) \dfrac{1}{2 \pi} \left( \dfrac{\mathcal{E}_m - \mathcal{E}_n}{\hslash} \right) = \notag \\
&= \frac{1}{Z}\sum_{n,m} A_{nm}B_{mn}\left(e^{-\beta \mathcal{E}_n} - \varepsilon e^{-\beta \mathcal{E}_m}\right)\left( \dfrac{\mathcal{E}_m - \mathcal{E}_n}{\hslash} \right),
\end{align}
note that this time we cannot directly use the trace property used earlier because of the product for $(\mathcal{E}_m - \mathcal{E}_n)$. We write
\begin{equation}
\dfrac{1}{Z} \sum_{n,m} A_{nm}B_{mn} e^{-\beta \mathcal{E}_n} \dfrac{\mathcal{E}_m - \mathcal{E}_n}{\hslash} - \dfrac{\varepsilon}{Z} \sum_{n,m} A_{nm}B_{mn} e^{-\beta \mathcal{E}_m} \dfrac{\mathcal{E}_m - \mathcal{E}_n}{\hslash} ,
\end{equation}
and we consider the first term, and we get
\begin{align}
\dfrac{1}{Z} \sum_{n,m} A_{nm} B_{mn} e^{-\beta \mathcal{E}_n} \dfrac{\mathcal{E}_m - \mathcal{E}_n}{\hslash} &= \dfrac{1}{\hslash Z} \sum_{n,m} A_{nm}B_{mn} e^{-\beta \mathcal{E}_n} \mathcal{E}_m - \dfrac{1}{\hslash Z} \sum_{n,m} A_{nm}B_{mn} e^{-\beta \mathcal{E}_n} \mathcal{E}_n \equiv \notag \\
& \equiv \dfrac{1}{\hslash Z} \sum_{n,m} \langle n | A | m \rangle \langle m | B | n \rangle \, e^{-\beta \mathcal{E}_n} \, \mathcal{E}_m 
- \dfrac{1}{\hslash Z} \sum_{n,m} \langle n | A | m \rangle \langle m | B | n \rangle \, e^{-\beta \mathcal{E}_n} \, \mathcal{E}_n = \notag \\
&= \dfrac{1}{\hslash Z} \sum_{n,m} \langle n | A \mathcal{E}_m | m \rangle \langle m | B | n \rangle \, e^{-\beta \mathcal{E}_n}
- \dfrac{1}{\hslash Z} \sum_{n,m} \langle n | \mathcal{E}_n A | m \rangle \langle m | B | n \rangle \, e^{-\beta \mathcal{E}_n} = \notag \\
&= \dfrac{1}{\hslash Z} \sum_{n,m} \langle n | A \hat{\mathcal{H}} | m \rangle \langle m | B | n \rangle \, e^{-\beta \mathcal{E}_n}
- \dfrac{1}{\hslash Z} \sum_{n,m} \langle n | \hat{\mathcal{H}} A | m \rangle \langle m | B | n \rangle \, e^{-\beta \mathcal{E}_n} = \notag \\
&= \dfrac{1}{\hslash Z} \sum_{n,m} \langle n | [A, \hat{\mathcal{H}}] | m \rangle \langle m | B | n \rangle \, e^{-\beta \mathcal{E}_n}.
\end{align}
Similarly for the second term, we have
\begin{equation}
- \dfrac{\varepsilon}{\hslash Z} \sum_{n,m} \langle n | [A, \hat{\mathcal{H}}] | m \rangle \langle m | B | n \rangle \, e^{-\beta \mathcal{E}_m} = - \dfrac{\varepsilon}{\hslash Z} \sum_{n,m} \langle m | B | n \rangle \langle n | [A, \hat{\mathcal{H}}] | m \rangle \, e^{-\beta \mathcal{E}_m}.
\end{equation}
We sum with respect to $m$ in the first term and with respect to $n$ in the second term, combine the sums under one index, and obtain the thesis.
\end{proof}
\end{theorem}
In principle, it is possible to compute moments of all orders: each increase in the power of \( \omega \) corresponds to an additional commutator with the Hamiltonian \( \hat{\mathcal{H}} \). This approach is known as the moments method.
\section{Application of Kramers-Krönig relations to retarded Green's function}
The first Kramers-Krönig relation (see Chapter \ref{Kramers-Krönig relations}) can be used to calculate the real part of Green's function, after the imaginary part has been measured. From $\eqref{eq: integraledellafunzionediGreeninterminidellafunzionespettrale}$ 
\begin{align}
G^{(r)}_{AB}(z) &= \int_{-\infty}^{+\infty} \dfrac{d\omega'}{2\pi} \dfrac{A_{AB}(\omega')}{\hslash (z - \omega')} = \notag \\
&= \int_{-\infty}^{+\infty} \dfrac{d\omega'}{2\pi} \dfrac{A_{AB}(\omega')}{\hslash (\omega - \omega' + i \delta)} = \notag \\
&= \dfrac{1}{\hslash} \int_{-\infty}^{+\infty} \frac{d\omega'}{2 \pi}\frac{A_{AB}(\omega')}{(\omega - \omega')^2 + \delta^2}(\omega - \omega' - i \delta) ,
\end{align}
we apply the limit $\delta \rightarrow 0^+$. In the case of physical interest $B=A^\dag$ the spectral function is a real function, then the contributions to the integral from the real and imaginary parts are
\begin{align}
G_{AB}^{(r)}(\omega) &= - \dfrac{i}{\hslash} \lim_{\delta \rightarrow 0^+} \int_{-\infty}^{+\infty} \dfrac{d\omega'}{2 \pi} \dfrac{\delta}{(\omega - \omega')^2 + \delta^2} A_{AB}(\omega') + \dfrac{1}{\hslash}\lim_{\delta \rightarrow 0^+}\int_{-\infty}^{+\infty} \frac{d\omega'}{2\pi}A_{AB}(\omega') \dfrac{\omega - \omega'}{(\omega - \omega')^2 + \delta^2} = \notag \\
&= - \dfrac{i}{\hslash} \int_{-\infty}^{+\infty} \dfrac{d\omega'}{2}  \left( \lim_{\delta \rightarrow 0^+} \dfrac{1}{\pi} \dfrac{\delta }{(\omega - \omega')^2 + \delta^2} \right) A_{AB}(\omega') + \dfrac{1}{\hslash} \int_{-\infty}^{+\infty} \dfrac{d\omega'}{2\pi} \dfrac{A_{AB}(\omega')}{\omega - \omega'} \left( \lim_{\delta \rightarrow 0^+} \dfrac{(\omega - \omega')^2}{(\omega - \omega')^2 + \delta^2} \right) = \notag \\ 
&= - \dfrac{i}{\hslash} \int_{-\infty}^{+\infty} \dfrac{d\omega'}{2} \delta(\omega-\omega') A_{AB}(\omega') + \dfrac{1}{\hslash} \int_{-\infty}^{+\infty} \frac{d\omega'}{2\pi} \dfrac{A_{AB}(\omega')}{\omega - \omega'} \left( \lim_{\delta \rightarrow 0^+} \dfrac{(\omega - \omega')^2}{(\omega - \omega')^2 + \delta^2} \right) = \notag \\
&= - \dfrac{i}{2 \hslash} A_{AB}(\omega) + \dfrac{1}{\hslash} \int_{-\infty}^{+\infty} \frac{d\omega'}{2\pi} \dfrac{A_{AB}(\omega')}{\omega - \omega'} \left( \lim_{\delta \rightarrow 0^+} \dfrac{(\omega - \omega')^2}{(\omega - \omega')^2 + \delta^2} \right),
\end{align}
where in the first term we triggered Dirac's delta and in the second term we multiplied and divided by $\omega - \omega'$, finally in both we have reversed the limit and integral operations. The quantity
\begin{equation}
\lim_{\delta \rightarrow 0^+} \dfrac{(\omega - \omega')^2}{(\omega - \omega')^2 + \delta^2}
\end{equation}
approaches $1$ for $|\omega - \omega'| \gg \delta$, and tends to $0$ for $|\omega-\omega'| \ll \delta$. Then, in the limit $\delta \rightarrow 0^+$ the singularity is for $\omega = \omega'$, in which case the function
\begin{equation}
\dfrac{A_{AB}(\omega')}{\omega - \omega'} 
\end{equation}
has a first-order pole. Accordingly, we can rewrite the second term in \( G^{(r)}_{AB}(\omega) \) as a principal value integral over frequency. This yields the representation:
\begin{align}
G_{AB}^{(r)}(\omega) = - \dfrac{i}{2 \hslash} A_{AB}(\omega) + \dfrac{1}{\hslash} \, \pv \int_{-\infty}^{+\infty} \frac{d\omega'}{2\pi} \, \frac{A_{AB}(\omega')}{\omega - \omega'}.
\end{align}
In the physically relevant case where \( B = A^\dag \), the spectral function \( A_{AA^\dag}(\omega) \) is real-valued for real arguments. Consequently, we have
\begin{align}
G_{AA^\dag}^{(r)}(\omega) = - \dfrac{i}{2 \hslash} A_{AA^\dag}(\omega) + \dfrac{1}{\hslash} \, \pv \int_{-\infty}^{+\infty} \frac{d\omega'}{2\pi} \, \frac{A_{AA^\dag}(\omega')}{\omega - \omega'}.
\end{align}
Taking the imaginary part of both sides gives
\begin{align}
\im G_{AA^\dag}^{(r)}(\omega) &= \im \left( - \dfrac{i}{2 \hslash} A_{AA^\dag}(\omega) \right) + \im \left( \dfrac{1}{\hslash} \, \pv \int_{-\infty}^{+\infty} \frac{d\omega'}{2\pi} \, \frac{A_{AA^\dag}(\omega')}{\omega - \omega'} \right) = \notag \\
&= \dfrac{1}{2\hslash} A_{AA^\dag}(\omega),
\end{align}
which can be inverted to express the spectral function in terms of the imaginary part of the Green's function:
\begin{equation}
A_{AA^\dag}(\omega) = -2 \hslash \im G_{AA^\dag}^{(r)}(\omega).
\end{equation}
By substituting this result into the principal value integral, we explicitly recover the first Kramers-Krönig relation. This fundamental relation expresses the real part of the retarded Green's function as a transform of its imaginary part. This identity reflects the analyticity of \( G^{(r)}(\omega) \) in the upper half of the complex plane and ultimately encodes the causality of the physical system. The relation shows that the dispersive (real) part of the response is completely determined by its absorptive (imaginary) part, in accordance with the general principles of linear response theory. 
\section{Average occupation number for interacting particles}
Here we want to show an important application of many-body Green's function.
In quantum many-body theory, understanding how particles occupy available quantum states is fundamental. The number operator plays a crucial role in this context, as it counts the occupation of a given many particles state. The thermal average occupation number of fermions or bosons in the state $(\textbf{k},\sigma)$ with respect to a Hamiltonian of free particles, i.e., $\eqref{eq: Hamiltonianasecondaquantizzazioneparticellelibere}$, is given by $\eqref{eq: mediatermicanumerodioccupazioneedistribuzioneFermiBose}$ in absence of interaction. Let us now consider a Hamiltonian of the form $\eqref{eq: HamiltonianascompostainH_0eH_I}$ within the framework of the second quantization, and let us compute how the average occupancy number changes in the state $(\textbf{k},\sigma)$ in an interacting system. Let us use the property $\eqref{eq: mediaprodottooperatoricalcolataconfunzionespettrale}$ of the spectral function. We set $A=C_{\textbf{k},\sigma}$, $B=C^{\dagger}_{\textbf{k},\sigma}$, and we use $C_{\textbf{k},\sigma} C^{\dagger}_{\textbf{k},\sigma} - \varepsilon C^{\dagger}_{\textbf{k},\sigma} C_{\textbf{k},\sigma} = \mathds{1}$, then
\begin{equation}
\left\langle C^{\dagger}_{\textbf{k},\sigma} C_{\textbf{k},\sigma} \right\rangle = \int_{-\infty}^{+\infty} \dfrac{d\omega}{2\pi} \dfrac{A_{\textbf{k},\sigma}(\omega)}{1-\varepsilon e^{-\beta \hslash \omega}} ,
\end{equation}
\begin{equation}
\varepsilon \left\langle C^{\dagger}_{\textbf{k},\sigma} C_{\textbf{k},\sigma} \right\rangle = \int_{-\infty}^{+\infty} \dfrac{d\omega}{2\pi} \dfrac{A_{\textbf{k},\sigma}(\omega)}{1-\varepsilon e^{-\beta \hslash \omega}} - 1.
\end{equation}
From $\eqref{eq: momentonullodifunzionespettrale}$, we write
\begin{align}
1 &= \left\langle \left[C_{\textbf{k},\sigma},C{^\dagger}_{\textbf{k},\sigma}\right]^{(\varepsilon)} \right\rangle = \notag \\
&= \int_{-\infty}^{+\infty}\frac{d\omega}{2\pi}A_{\textbf{k},\sigma}(\omega) ,
\end{align}
and we have
\begin{align}
\varepsilon \left\langle C^{\dagger}_{\textbf{k},\sigma} C_{\textbf{k},\sigma} \right\rangle &= \int_{-\infty}^{+\infty} \dfrac{d\omega}{2\pi} \dfrac{A_{\textbf{k},\sigma}(\omega)}{1-\varepsilon e^{-\beta \hslash \omega}} - \int_{-\infty}^{+\infty}\frac{d\omega}{2\pi}A_{\textbf{k},\sigma}(\omega) = \notag \\
&= \int_{-\infty}^{+\infty} \dfrac{d\omega}{2\pi} A_{\textbf{k},\sigma}(\omega) \left[ \dfrac{\varepsilon e^{- \beta \hslash \omega}}{1 - \varepsilon e^{- \beta \hslash \omega}} \right],
\end{align}
that is,
\begin{align}
\left\langle \hat{N}_{\mathbf{k},\sigma} \right\rangle &= \int_{-\infty}^{+\infty} \frac{d\omega}{2\pi} \frac{A_{\mathbf{k},\sigma}(\omega)}{e^{\beta \omega} - \varepsilon} = \notag \\
&= \int_{-\infty}^{+\infty} \frac{d\omega}{2\pi} A_{\mathbf{k},\sigma}(\omega) \, n_{\varepsilon}(\omega).
\label{eq: mediatermicanumerodioccupazionecomeintegralefunzionespettrale}
\end{align}
Consequently, the many-body propagator, by means of the spectral function, allows calculation of the average occupation number and such object, due to interaction, is written as an integral of the Fermi-Dirac or Bose-Einstein statistics.
\section{Pole close to the real axis: quasiparticle picture}
As previously discussed, in Chapter $\ref{Feynman's perturbative theory of the thermal Green's function}$ we will demonstrate that any retarded Green's function can be expressed in the form of equation $\eqref{eq: funzionediGreensistemainvariantetraslspazialeeallequilibriotermod}$, where $\Sigma^*(z)$ is a complex function, known as the improper self-energy, that shifts the poles of the free Green's function from the real axis (in the non-interacting case) into the lower half of the complex plane (in the presence of interactions). The imaginary part of the self-energy determines the damping factor $\gamma$: the larger the shift of the pole into the complex plane, the stronger the damping. Let us denote by $z_0$ a pole of the retarded Green's function $G^{(r)}(z)$ lying in the lower half-plane. We define $z_0 \equiv \Omega - i \gamma$, $\gamma > 0$, and impose the condition for $z_0$ to be a pole, that is,
\begin{align}
\Omega - i\gamma - \dfrac{\mathcal{E}}{\hslash} - \dfrac{1}{\hslash} \Sigma^*(\Omega - i \gamma) = 0.
\end{align}
To analyze the behavior of the Green's function near the pole, we assume that the pole is close to the real axis, i.e., $\gamma \ll |\Omega|$. Under this assumption, we perform a first-order Taylor expansion of $\Sigma^*(z)$ around the real point $\Omega$, i.e.,
\begin{align}
\Sigma^*(\Omega-i \gamma) & \simeq \Sigma^*(\Omega) - i \gamma \left. \dfrac{\partial \Sigma^*}{\partial z}\right|_{z=\Omega} = \notag \\
&= \re \Sigma^*(\Omega) + i \im \Sigma^*(\Omega) -i\gamma \left[ \left. \dfrac{\partial \re \Sigma^*(\omega)}{\partial \omega}\right|_{\omega = \Omega} + i \left. \dfrac{\partial \im \Sigma^*(\omega)}{\partial \omega}\right|_{\omega=\Omega}\right],
\end{align}
where we used that $\Sigma^*(z)$ is analytic and so its derivative does not depend on direction and we can calculate it on the real axis. 
\begin{equation}
\Omega - i\gamma - \dfrac{\mathcal{E}}{\hslash} - \dfrac{1}{\hslash} \left( \re \Sigma^*(\Omega) + i \im \Sigma^*(\Omega) - i\gamma \left. \frac{\partial \re \Sigma^*(\omega)}{\partial \omega}\right|_{\omega = \Omega} + \gamma \left. \frac{\partial \im \Sigma^*(\omega)}{\partial \omega}\right|_{\omega = \Omega} \right) = 0,
\end{equation}
\begin{equation}
\Omega - i\gamma - \dfrac{\mathcal{E}}{\hslash} - \dfrac{1}{\hslash} \left( \re \Sigma^*(\Omega) +i \im \Sigma^*(\Omega) - i\gamma \left. \frac{\partial \re \Sigma^*(\omega)}{\partial \omega}\right|_{\omega = \Omega} \right) = 0,
\end{equation}
we have neglected $\gamma \left. \frac{\partial \im \Sigma^*(\omega)}{\partial \omega}\right|_{\omega = \Omega}$ because the partial derivative of the imaginary part is already proportional to $\gamma$, so the product for $\gamma$ is a second-order term and can be neglected. By setting equal $0$ imaginary part and real part, we get respectively
\begin{equation}
\gamma = - \dfrac{\frac{1}{\hslash} \im \Sigma^*(\Omega)}{1 - \frac{1}{\hslash} \left. \frac{\partial \re \Sigma^*(\omega)}{\partial \omega}\right|_{\omega = \Omega}} ,
\label{eq: parteimmaginariapoloprossimoall'assereale}
\end{equation}
\begin{equation}
\Omega = \dfrac{\mathcal{E}}{\hslash} + \dfrac{\re \Sigma^*(\Omega)}{\hslash}.
\label{eq: parterealepoloprossimoall'assereale}
\end{equation}
From the first order expansion in $\gamma$ it is deduced that the imaginary part of the improper self-energy determines the decay times, while the real part implies the shift of energy levels, which is to be determined self-consistently as zero of the $\eqref{eq: parterealepoloprossimoall'assereale}$. The energy is no longer $\frac{\mathcal{E}}{\hslash}$. Given $G^{(r)}(\omega)$, we take the limit for $\delta \rightarrow 0^+$ and obtain $G^{(r)}(\omega)$. Here, the goal is to study Green's function $G^{(r)}(\omega)$ for $\omega \approx \Omega$ and to do this, $\Sigma^*(z)$ must be expanded respect to $\Omega$. We get
\begin{align}
G^{(r)}(\omega) &= \dfrac{1}{\hslash} \dfrac{1}{\omega - \frac{\mathcal{E}}{\hslash} - \frac{\re \Sigma^*(\Omega)}{\hslash} - i \frac{\im  \Sigma^*(\Omega)}{\hslash} - \frac{1}{\hslash} \left. \frac{\partial \re  \Sigma^*(\omega)}{\partial \omega}\right|_{\omega = \Omega}(\omega - \Omega)} = \notag \\
&= \dfrac{1}{\hslash} \dfrac{1}{\omega - \Omega - i \frac{\im \Sigma^*(\Omega)}{\hslash} - \frac{1}{\hslash} \left. \frac{\partial \re \Sigma^*(\omega)}{\partial \omega}\right|_{\omega = \Omega}(\omega - \Omega)},
\label{eq: Greenfrequenzarealiinterazionideboli}
\end{align}
where in the denominator we used $z = \omega + i \delta$ and omitted the second-order term containing the partial derivative of the imaginary part. In the limit $\delta \rightarrow 0^+$, $i \delta$ does not imply singularity since $\Sigma^*(z)$ includes a finite imaginary part. The form obtained for $G^{(r)}(\omega)$ is general, it holds for all Green's functions as long as the imaginary part of $\Sigma^*(\Omega)$ is small, but $\Sigma^*(\Omega)$ is related to the intensity of interactions: this expansion applies only to weak interactions. In other words, we are developing the propagator with weak interactions around the propagator of the noninteracting system. It follows
\begin{equation}
G^{(r)}(\omega) = \dfrac{1}{\hslash} \dfrac{1}{(\omega - \Omega)\left(1 - \frac{1}{\hslash} \left. \frac{\partial \re \Sigma^*(\omega)}{\partial \omega}\right|_{\omega = \Omega}\right) - i \frac{\im \Sigma^*(\Omega)}{\hslash}},
\end{equation}
we use $\eqref{eq: parteimmaginariapoloprossimoall'assereale}$ as follows
\begin{equation}
\dfrac{1}{\hslash} \im \Sigma^*(\Omega) = - \gamma \left( 1 - \frac{1}{\hslash} \left. \frac{\partial \re \Sigma^*(\omega)}{\partial \omega}\right|_{\omega = \Omega} \right),
\end{equation}
and we get
\begin{align}
\im G^{(r)}(\omega) \approx \frac{1}{\hslash}\left[\dfrac{1}{1 - \frac{1}{\hslash} \left. \frac{\partial \re \Sigma^*(\omega)}{\partial \omega}\right|_{\omega = \Omega}} \right] \left[-\dfrac{\gamma}{(\omega - \Omega)^2 + \gamma^2}\right].
\end{align}
Multiplying by $- 2 \hslash \pi$, we have
\begin{equation}
A(\omega) \simeq 2 \pi \left( \dfrac{1}{1 - \frac{1}{\hslash} \left. \frac{\partial \re \Sigma^*(\omega)}{\partial \omega}\right|_{\omega = \Omega}} \right) \left[-\dfrac{\gamma}{(\omega - \Omega)^2 + \gamma^2}\right] ,
\label{eq: funzionespettraleGreendeboliinterazioni}
\end{equation}
\begin{equation}
A(\omega) \simeq 2 \pi r \left[-\dfrac{\gamma}{(\omega - \Omega)^2 + \gamma^2}\right],
\end{equation}
with
\begin{equation}
r = \left( \frac{1}{1 - \frac{1}{\hslash} \left. \frac{\partial \re \Sigma^*(\omega)}{\partial \omega}\right|_{\omega = \Omega}} \right).
\end{equation}
The equation $\eqref{eq: funzionespettraleGreendeboliinterazioni}$ holds for $\gamma$ small and it is a Lorentzian function normalized to $1$, with a peak around $\Omega$ and a width of $\gamma$. In a neighborhood of $\Omega$, the spectral function becomes a Lorentzian that has a peak at the real part $\Omega$, with width bound to the imaginary part $\gamma$. The factor $r$ is definitely less than $1$. Indeed, consider an Hamiltonian of the form $\eqref{eq: HamiltonianascompostainH_0eH_I}$ within the second quantization formalism, with $\hat{\mathcal{H}}_0$ given by $\eqref{eq: Hamiltonianasecondaquantizzazioneparticellelibere}$. If the intensity of the interaction is weak, the Green's function at real frequencies at the lowest order has the form given by $\eqref{eq: Greenfrequenzarealiinterazionideboli}$. From the equation $\eqref{eq: momentonullodifunzionespettrale}$ adapted for a spectral function of noninteracting fermions, we have
\begin{equation}
\int_{-\infty}^{+\infty}d\omega \frac{A_\alpha(\omega)}{2\pi} = 1.
\end{equation}
For fermions, $A_\alpha(\omega)$ is a real and positive-defined function. Its integral with respect to the measure $\omega$ should equal 1, and around $\Omega$, the function $A_\alpha(\omega)$ is a Lorentzian normalized to 1. Substituting equation $\eqref{eq: funzionespettraleGreendeboliinterazioni}$ into the integral, the result is $r$. Then, if the interactions are weak, we have
\begin{equation}
A(\omega) = 2 \pi r \left[-\dfrac{\gamma}{(\omega - \Omega)^2 + \gamma^2}\right] + f(\omega) ,
\end{equation}
where $f(\omega)$ is a function whose integral must give $1-r$. We expect the parameter $r$ to be close to $1$, since in the free (non-interacting) case we have exactly $r = 1$, and we are currently considering a perturbative regime. In summary, when the interactions are weak, the Dirac delta peak characterizing the spectral function of a free particle is broadened into a Lorentzian profile, and the residue $r$ of the Green's function at the pole becomes strictly less than $1$.
\newpage
\section{Figures}
\FloatBarrier
\chapter{Feynman's perturbative theory of the thermal Green's function}\label{Feynman's perturbative theory of the thermal Green's function}
Here, we begin with the Dyson equation for Green's functions of field operators, which serves as a cornerstone for the study of interacting quantum systems. The Dyson equation expresses the full Green's function in terms of the non-interacting one and a self-energy term that encapsulates all interaction effects, providing a compact and systematic framework for perturbation theory. \newline
As part of this discussion, we will also cover the thermal Wick theorem, an essential result for constructing the perturbative expansion around the thermal averages of the non-interacting system. This expansion is expressed in terms of Feynman diagrams, which graphically represent the various interaction contributions. The theorem provides the formal basis for expressing such thermal averages of operator products as combinations of contractions, enabling a systematic treatment of interaction effects. \newline
Building on Feynman formalism, we introduce Feynman diagrammatics, a powerful graphical and intuitive tool that simplifies the representation and calculation of perturbative contributions. Feynman diagrams visually encode the complex interactions between particles and fields, making the analysis of quantum processes and the physical interpretation of computed quantities more accessible. \newline
Moreover, we will consider the summations over Matsubara frequencies, a crucial step in finite-temperature field theory that allows the transition from time-dependent operators to frequency space and facilitates the evaluation of thermal Green’s functions. \newline
Finally, we discuss how, through Green’s functions and the associated diagrammatics, it is possible to compute the expectation value of the particle density operator in interacting systems. This connection bridges abstract theoretical frameworks with physically measurable quantities, making the chapter both theoretically significant and practically relevant for condensed matter physics, quantum field theory, and beyond.
\section{Dyson expansion of the thermal Green's function}
In this section, we follow Feynman's approach to the study of Green's functions at finite temperature. The starting point is the evaluation of the partition function within the framework of second quantization, which allows us to reformulate thermodynamic and dynamical quantities in terms of the interaction picture. This leads to an expression of the Matsubara Green's function as a perturbative series in powers of the interaction. \newline
A fundamental step in the computation of the thermal Green's function is the evaluation of the partition function \( Z \), associated with a generic Hamiltonian of the form given in $\eqref{eq: HamiltonianascompostainH_0eH_I}$, within the framework of second quantization. Specifically, we proceed with the following sequence of manipulations
\begin{align}
Z &= \Tr e^{- \beta \hat{\mathcal{H}}} = \notag \\
&= \Tr e^{- \beta \hat{\mathcal{H}}_0} e^{\beta \hat{\mathcal{H}}_0} e^{- \beta \hat{\mathcal{H}}} = \notag \\
&= \Tr e^{- \beta \hat{\mathcal{H}}_0} \ U(\beta \hslash) = \notag \\
&= \dfrac{\Tr e^{- \beta \hat{\mathcal{H}}_0}}{\Tr e^{- \beta \hat{\mathcal{H}}_0}} \Tr e^{- \beta \hat{\mathcal{H}}_0} \ U(\beta \hslash) = \notag \\
&= \Tr e^{- \beta \hat{\mathcal{H}}_0} \ \dfrac{\Tr e^{- \beta \hat{\mathcal{H}}_0} \ U(\beta \hslash)}{\Tr e^{- \beta \hat{\mathcal{H}}_0}} = \notag \\
&= \Tr e^{- \beta \hat{\mathcal{H}}_0} \langle U(\beta \hslash) \rangle_0 \ \equiv \notag \\
&\equiv Z_0 \ \langle U(\beta \hslash) \rangle_0,
\end{align}
where $Z_0$ and $\langle \ldots \rangle_0$ denote the partition function and the thermal average with respect to the Hamiltonian $\hat{\mathcal{H}}_0$, respectively, and $U(\beta \hslash)$ is evolution operator in the thermal interaction picture. Given two general operators $A$ and $B$, the Matsubara Green's function at positive times becomes
\begin{align}
G^{(m)}_{AB}(\tau) &= - \dfrac{1}{\hslash} \left\langle \hat{T}_\tau A(\tau) B(0) \right\rangle = \notag \\
&= - \dfrac{1}{\hslash} \langle A(\tau) B(0) \rangle \ = \notag \\
&= - \dfrac{1}{\hslash} \dfrac{\Tr e^{- \beta \hat{\mathcal{H}}} A(\tau) B(0)}{\Tr e^{- \beta \hat{\mathcal{H}}}} = \notag \\
&= - \dfrac{1}{\hslash} \dfrac{\Tr e^{- \beta \hat{\mathcal{H}}} A(\tau) B(0)}{\Tr e^{- \beta \hat{\mathcal{H}}_0} \langle U(\beta \hslash) \rangle_0} = \notag \\
&= - \dfrac{1}{\hslash} \dfrac{1}{\langle U(\beta \hslash) \rangle_0 } \dfrac{\Tr e^{- \beta \hat{\mathcal{H}}} A(\tau) B(0)}{\Tr e^{- \beta \hat{\mathcal{H}}_0}} = \notag \\
&= - \dfrac{1}{\hslash} \dfrac{1}{\langle U(\beta \hslash) \rangle_0 } \dfrac{\Tr e^{- \beta \hat{\mathcal{H}}_0} e^{\beta \hat{\mathcal{H}}_0} e^{- \beta \hat{\mathcal{H}}} A(\tau) B(0)}{\Tr e^{- \beta \hat{\mathcal{H}}_0}} = \notag \\
&= - \dfrac{1}{\hslash} \dfrac{1}{\langle U(\beta \hslash) \rangle_0 } \dfrac{\Tr e^{- \beta \hat{\mathcal{H}}_0} U(\beta \hslash) A(\tau) B(0)}{\Tr e^{- \beta \hat{\mathcal{H}}_0}}, \quad \tau > 0.
\end{align}
Using $U(\beta \hslash) = S(\beta \hslash,0)$ e $A(\tau)=S(0,\tau) A^{(0)}(\tau) S(\tau,0)$, we write
\begin{align}
G^{(m)}_{AB}(\tau) &= - \dfrac{1}{\hslash} \dfrac{1}{\langle U(\beta \hslash) \rangle_0} \dfrac{\Tr \left( e^{- \beta \hat{\mathcal{H}}_0} U(\beta \hslash) A(\tau) B(0) \right)}{\Tr \left( e^{- \beta \hat{\mathcal{H}}_0} \right)} = \notag \\
&= - \dfrac{1}{\hslash} \dfrac{1}{\langle U(\beta \hslash) \rangle_0} \dfrac{\Tr \left( e^{- \beta \hat{\mathcal{H}}_0} S(\beta \hslash,0) S(0,\tau) A^{(0)}(\tau) S(\tau,0) B(0) \right)}{\Tr \left( e^{- \beta \hat{\mathcal{H}}_0} \right)} = \notag \\
&= - \dfrac{1}{\hslash} \dfrac{1}{\langle U(\beta \hslash) \rangle_0} \dfrac{\Tr \left( e^{- \beta \hat{\mathcal{H}}_0} S(\beta \hslash,\tau) A^{(0)}(\tau) S(\tau,0) B(0) \right)}{\Tr \left( e^{- \beta \hat{\mathcal{H}}_0} \right)} = \notag \\
&\equiv - \dfrac{1}{\hslash} \dfrac{\langle S(\beta \hslash,\tau) A^{(0)}(\tau) S(\tau,0) B(0) \rangle_0}{\langle U(\beta \hslash) \rangle_0}, \quad \tau > 0.
\label{eq: GreentermicainfunzionediU(betahslash)1}
\end{align}
If the order of $A^{(0)}(\tau)$ and $S(\tau,0)$ can be reversed, the thermal Green's function is
\begin{align}
G^{(m)}_{AB}(\tau) &= - \dfrac{1}{\hslash} \dfrac{\langle S(\beta \hslash,\tau) S(\tau,0) A^{(0)}(\tau) B(0) \rangle_0}{\langle U(\beta \hslash) \rangle_0} = \notag \\
&= - \dfrac{1}{\hslash} \dfrac{\langle S(\beta \hslash,0) A^{(0)}(\tau) B(0) \rangle_0}{\langle U(\beta \hslash) \rangle_0} = \notag \\
&= - \dfrac{1}{\hslash} \dfrac{\langle U(\beta \hslash) A^{(0)}(\tau) B(0) \rangle_0}{\langle U(\beta \hslash) \rangle_0}, \quad \tau > 0,
\end{align}
which, upon expanding \( U(\beta \hslash) \) in a Dyson series, becomes
\begin{align}
G^{(m)}_{AB}(\tau) &= \dfrac{
    -\dfrac{1}{\hslash} \langle A^{(0)}(\tau) B(0) \rangle_0
}{
    1 + \sum_{n=1}^{\infty} \left( -\dfrac{1}{\hslash} \right)^n \dfrac{1}{n!}
    \int_0^{\beta \hslash} d\tau_1 \cdots \int_0^{\beta \hslash} d\tau_n\,
    \left\langle \hat{T}_\tau \left\lbrace \hat{\mathcal{H}}_I^{(0)}(\tau_1) \cdots \hat{\mathcal{H}}_I^{(0)}(\tau_n) \right\rbrace \right\rangle_0 
} + \notag \\
&\quad + \dfrac{
    -\dfrac{1}{\hslash} \sum_{n=1}^{\infty} \left( -\dfrac{1}{\hslash} \right)^n \dfrac{1}{n!}
    \int_0^{\beta \hslash} d\tau_1 \cdots \int_0^{\beta \hslash} d\tau_n\,
    \left\langle \hat{T}_\tau \left\lbrace \hat{\mathcal{H}}_I^{(0)}(\tau_1) \cdots \hat{\mathcal{H}}_I^{(0)}(\tau_n) A^{(0)}(\tau) B(0) \right\rbrace \right\rangle_0
}{
    1 + \sum_{n=1}^{\infty} \left( -\dfrac{1}{\hslash} \right)^n \dfrac{1}{n!}
    \int_0^{\beta \hslash} d\tau_1 \cdots \int_0^{\beta \hslash} d\tau_n\,
    \left\langle \hat{T}_\tau \left\lbrace \hat{\mathcal{H}}_I^{(0)}(\tau_1) \cdots \hat{\mathcal{H}}_I^{(0)}(\tau_n) \right\rbrace \right\rangle_0
} ,
\label{eq: GreentermicainfunzionediU(betahslash)2}
\end{align}
and recall $\tau>0$. We have
\begin{theorem}
Equations $\eqref{eq: GreentermicainfunzionediU(betahslash)1}$ and $\eqref{eq: GreentermicainfunzionediU(betahslash)2}$ are equivalent.
\begin{proof}
Consider the $n$-th term of the series in the numerator of equation $\eqref{eq: GreentermicainfunzionediU(betahslash)2}$, i.e.,
\begin{equation}
\int_0^{\beta \hslash} d\tau_1 \ldots \int_0^{\beta \hslash} d\tau_n \left\langle \hat{T}_\tau \left\lbrace \hat{\mathcal{H}}_I^{(0)}(\tau_1) \ldots \hat{\mathcal{H}}_I^{(0)}(\tau_n) \right\rbrace A^{(0)}(\tau) B(0) \right\rangle_0.
\end{equation}
Since $\tau>0$, we write
\begin{equation}
\left( \int_0^\tau d\tau_1 + \int_\tau^{\beta \hslash} d\tau_1 \right) \left( \int_0^\tau d\tau_2 + \int_\tau^{\beta \hslash} d\tau_2 \right) \ldots \left( \int_0^\tau d\tau_n + \int_\tau^{\beta \hslash} d\tau_n \right),
\label{eq: ennesimointegraleserieDysonGreentermicainfunzionediU(betahslash)2}
\end{equation}
then onsider $n_1$ integrals of the form $\int_\tau^{\beta \hslash}$ and $m_1 = n - n_1$ integrals of the form $\int_0^\tau$. How can we explicitly order such a product? The correct time ordering, decreasing from left to right, is
\begin{align}
& \int_\tau^{\beta \hslash} d\tau_1 \ldots \int_\tau^{\beta \hslash} d\tau_{n_1} \notag \\
& \left\langle 
\hat{T}_\tau \left\lbrace \hat{\mathcal{H}}_I^{(0)}(\tau_1) \ldots \hat{\mathcal{H}}_I^{(0)}(\tau_{n_1}) \right\rbrace 
A^{(0)}(\tau) 
\int_0^\tau d\tau_{m_1+1} \ldots \int_0^{\tau} d\tau_n \,
\hat{T}_\tau \left\lbrace \hat{\mathcal{H}}_I^{(0)}(\tau_{m_1+1}) \ldots \hat{\mathcal{H}}_I^{(0)}(\tau_n) \right\rbrace 
B(0)
\right\rangle_0,
\end{align}
which corresponds to one of the $\frac{n!}{n_1! m_1!}$ possible combinations obtained by selecting \(n_1\) integrals out of a total of \(n\). However, all these products are equivalent since the operator \(\hat{T}_\tau\) reorders them identically within each integral. Therefore, we can represent the entire sum by the above expression multiplied by the combinatorial factor $\frac{n!}{n_1! m_1!}$, i.e.,
\begin{align}
& \dfrac{n!}{n_1! m_1!} \int_\tau^{\beta \hslash} d\tau_1 \ldots \int_\tau^{\beta \hslash} d\tau_{n_1} \notag \\
& \left\langle 
\hat{T}_\tau \left\lbrace \hat{\mathcal{H}}_I^{(0)}(\tau_1) \ldots \hat{\mathcal{H}}_I^{(0)}(\tau_{n_1}) \right\rbrace 
A^{(0)}(\tau) 
\int_0^\tau d\tau_{m_1+1} \ldots \int_0^{\tau} d\tau_n \,
\hat{T}_\tau \left\lbrace \hat{\mathcal{H}}_I^{(0)}(\tau_{m_1+1}) \ldots \hat{\mathcal{H}}_I^{(0)}(\tau_n) \right\rbrace 
B(0)
\right\rangle_0.
\end{align}
To generate all terms of the integral in $\eqref{eq: ennesimointegraleserieDysonGreentermicainfunzionediU(betahslash)2}$, both \(n_1\) and \(m_1\) must vary from \(0\) to \(n\), subject to the constraint \(n_1 + m_1 = n\). Hence, we have
\begin{align}
& \sum_{n_1=0}^{\infty} \sum_{m_1=0}^{\infty} \delta_{n_1+m_1,n} \dfrac{n!}{n_1! m_1!} \int_\tau^{\beta \hslash} d\tau_1 \ldots \int_\tau^{\beta \hslash} d\tau_{n_1} \notag \\
& \left\langle 
\hat{T}_\tau \left\lbrace \hat{\mathcal{H}}_I^{(0)}(\tau_1) \ldots \hat{\mathcal{H}}_I^{(0)}(\tau_{n_1}) \right\rbrace 
A^{(0)}(\tau) 
\int_0^\tau d\tau_{m_1+1} \ldots \int_0^{\tau} d\tau_n \,
\hat{T}_\tau \left\lbrace \hat{\mathcal{H}}_I^{(0)}(\tau_{m_1+1}) \ldots \hat{\mathcal{H}}_I^{(0)}(\tau_n) \right\rbrace 
B(0)
\right\rangle_0,
\end{align}
where \(\delta_{n_1 + m_1, n}\) enforces the constraint on \(n\) and allows the summations to be formally extended to infinity. Consequently, the series in the numerator of $\eqref{eq: GreentermicainfunzionediU(betahslash)2}$ becomes
\begin{align}
& \sum_{n=1}^{\infty} \sum_{n_1=0}^{\infty} \sum_{m_1=0}^{\infty} \left( - \dfrac{1}{\hslash} \right)^n \dfrac{1}{n!} \delta_{n_1+m_1,n} \dfrac{n!}{n_1! m_1!} \int_\tau^{\beta \hslash} d\tau_1 \ldots \int_\tau^{\beta \hslash} d\tau_{n_1} \notag \\
& \left\langle \hat{T}_\tau \left\lbrace \hat{\mathcal{H}}_I^{(0)}(\tau_1) \ldots \hat{\mathcal{H}}_I^{(0)}(\tau_{n_1}) \right\rbrace A^{(0)}(\tau) \int_0^\tau d\tau_{m_1+1} \ldots \int_0^{\tau} d\tau_n \,\hat{T}_\tau \left\lbrace \hat{\mathcal{H}}_I^{(0)}(\tau_{m_1+1}) \ldots \hat{\mathcal{H}}_I^{(0)}(\tau_n) \right\rbrace B(0) \right\rangle_0 = \notag \\
&= \sum_{n=0}^{\infty} \sum_{n_1=0}^{\infty} \sum_{m_1=0}^{\infty} \left( - \dfrac{1}{\hslash} \right)^{n_1} \left( - \dfrac{1}{\hslash} \right)^{m_1} \delta_{n_1+m_1,n} \dfrac{1}{n_1! m_1!} \int_\tau^{\beta \hslash} d\tau_1 \ldots \int_\tau^{\beta \hslash} d\tau_{n_1} \notag \\
& \left\langle \hat{T}_\tau \left\lbrace \hat{\mathcal{H}}_I^{(0)}(\tau_1) \ldots \hat{\mathcal{H}}_I^{(0)}(\tau_{n_1}) \right\rbrace A^{(0)}(\tau) \int_0^\tau d\tau_{m_1+1} \ldots \int_0^{\tau} d\tau_n \,\hat{T}_\tau \left\lbrace \hat{\mathcal{H}}_I^{(0)}(\tau_{m_1+1}) \ldots \hat{\mathcal{H}}_I^{(0)}(\tau_n) \right\rbrace B(0) \right\rangle_0,
\end{align}
where $n!$ has been simplified, and the power $\left( - \frac{1}{\hslash} \right)^n$ is expressed as a function of $n_1$ and $m_1$. Summing over $n$, the $\delta_{n_1+m_1,n}$ is triggered and removed, and the two sums factorize as
\begin{equation}
\sum_{n_1=0}^{\infty} \left( - \dfrac{1}{\hslash} \right)^{n_1} \dfrac{1}{n_1!} \int_\tau^{\beta \hslash} d\tau_1 \ldots \int_\tau^{\beta \hslash} d\tau_{n_1} \hat{T}_\tau \left\lbrace \hat{\mathcal{H}}_I^{(0)}(\tau_1) \ldots \hat{\mathcal{H}}_I^{(0)}(\tau_{n_1}) \right\rbrace = S(\beta \hslash,\tau),
\end{equation}
\begin{equation}
\sum_{m_1=0}^{\infty} \left( - \dfrac{1}{\hslash} \right)^{m_1} \dfrac{1}{m_1!} \int_0^\tau d\tau_{m_1+1} \ldots \int_0^\tau d\tau_n \hat{T}_\tau \left\lbrace \hat{\mathcal{H}}_I^{(0)}(\tau_{m_1+1}) \ldots \hat{\mathcal{H}}_I^{(0)}(\tau_n) \right\rbrace = S(\tau,0),
\end{equation}
that is, the operator
\begin{equation}
S(\beta \hslash,\tau) A^{(0)}(\tau) S(\tau,0) B(0)
\end{equation}
has been reconstructed. Therefore, $\eqref{eq: GreentermicainfunzionediU(betahslash)1}$ and $\eqref{eq: GreentermicainfunzionediU(betahslash)2}$ are equal.
\end{proof}
\end{theorem}
Any many-body Green's function can be written as a function of thermal averages, with respect to non-interacting $\hat{\mathcal{H}}_0$ Hamiltonians, of operators that are expressible as powers of the interaction Hamiltonian $\hat{\mathcal{H}}_I$, whose time evolution depends only on $\hat{\mathcal{H}}_0$. Ultimately, any Green's function is an expansion of non-interacting Green's functions. Then, $\eqref{eq: GreentermicainfunzionediU(betahslash)2}$ is equivalent to the Green's function $\eqref{eq: funzioneGreenMastubara2}$ adapted for $\tau >0$, and computed with respect to a Hamiltonian $\eqref{eq: HamiltonianascompostainH_0eH_I}$. The $\eqref{eq: GreentermicainfunzionediU(betahslash)2}$ includes Hamiltonians that evolve according to $\hat{\mathcal{H}}_0$.
\section{Wick's thermal theorem}
Consider thermal averages of the form $\langle A^{(1)}_{\alpha_1} \ldots A^{(2N)}_{\alpha_{2N}} \rangle_0$, where: each $A$ is proportional to a creation or annihilation operator and does not depend explicitly on time; the subscript $0$ emphasizes that the averages must be evaluated on the non-interacting Hamiltonian $\hat{\mathcal{H}_0}$ of the form $\eqref{eq: Hamiltonianasecondaquantizzazioneparticellelibere}$; the superscript $(i)$ on $A^{(i)}$ indicates the operator’s position in the ordered product, while the subscript $\alpha_i$ labels the one-particle quantum state. The calculation of these thermal averages leads to Wick's thermal theorem, whose proof we present in this section. Previously, we introduce a definition preparatory to Wick's theorem: given two operators $A^{(i)}_{\alpha_i}$ and $A^{(j)}_{\alpha_j}$, we define their contraction operator as
\begin{equation}
C^{ij} \equiv A^{(i)}_{\alpha_i} A^{(j)}_{\alpha_j} - \varepsilon A^{(j)}_{\alpha_j} A^{(i)}_{\alpha_i},
\end{equation}
and we simply refer to the thermal average of this operator as the contraction:
\begin{equation}
\langle C^{ij} \rangle_0 = \langle A^{(i)}_{\alpha_i} A^{(j)}_{\alpha_j} - \varepsilon A^{(j)}_{\alpha_j} A^{(i)}_{\alpha_i} \rangle_0.
\end{equation}
\begin{theorem}[Wick's thermal theorem]\label{Wick's thermal theorem}
Let $A^{(1)}_{\alpha_1} \ldots A^{(2N)}_{\alpha_{2N}}$ be operators, each of which is either a creation or an annihilation operator, and assume they are time-independent. Then, the thermal average of their product, taken with respect to the non-interacting Hamiltonian \( \hat{\mathcal{H}}_0 \), can be expressed entirely in terms of two-point thermal averages (contractions) as follows
\begin{equation}
\langle A^{(1)}_{\alpha_1} \ldots A^{(2N)}_{\alpha_{2N}} \rangle_0 = \sum_\sigma \varepsilon^{\mathrm{sign}(\sigma)} \langle A^{(\sigma(1))}_{\alpha_{\sigma(1)}} A^{(\sigma(2))}_{\alpha_{\sigma(2)}} \rangle_0 \, \langle A^{(\sigma(3))}_{\alpha_{\sigma(3)}} A^{(\sigma(4))}_{\alpha_{\sigma(4)}} \rangle_0 \, \ldots \, \langle A^{(\sigma(2N-1))}_{\alpha_{\sigma(2N-1)}} A^{(\sigma(2N))}_{\alpha_{\sigma(2N)}} \rangle_0,
\label{eq: teoremaWickgeneralizzato1}
\end{equation}
where the sum is taken over all possible complete pairings \( \sigma \) of the indices \( 1, \ldots, 2N \), such that
\begin{equation}
\sigma : \ \sigma(1) < \sigma(2), \ \sigma(3) < \sigma(4), \ \ldots, \ \sigma(2N-1) < \sigma(2N), \quad \text{and} \quad \sigma(1) < \sigma(3) < \ldots < \sigma(2N-1).
\label{eq: teoremaWickgeneralizzato2}
\end{equation}
Each thermal contraction \( \langle A^{(i)}_{\alpha_i} A^{(j)}_{\alpha_j} \rangle_0 \) is non-zero only when \( A^{(i)}_{\alpha_i} \) and \( A^{(j)}_{\alpha_j} \) are a creation and an annihilation operator referring to the same quantum state.
\begin{proof}
For sake of simplicity, here we consider the case with $4$ operators, i.e., $\langle A^{(1)}_{\alpha_1} A^{(2)}_{\alpha_2} A^{(3)}_{\alpha_3} A^{(4)}_{\alpha_4} \rangle_0$. Regarding the contraction $C^{ij}$, we consider the following cases:

\begin{itemize}
\item {Case 1:} $\alpha_i = \alpha_j \equiv \alpha$, $A^{(i)}_{\alpha} = a_{\alpha}$, $A^{(j)}_{\alpha} = a^{\dagger}_{\alpha}$. Then
\begin{align}
C^{ij} &= a_\alpha a_\alpha^\dag - \varepsilon a_\alpha^\dag a_\alpha = \notag \\
&= \left[ a_\alpha , a_\alpha^\dagger \right]^{(\varepsilon)} = \notag \\
&= \mathds{1}.
\end{align}
\item {Case 2:} $\alpha_i = \alpha_j \equiv \alpha$, $A^{(i)}_{\alpha} = a^{\dagger}_{\alpha}$, $A^{(j)}_{\alpha} = a_{\alpha}$. Then
\begin{align}
C^{ij} &= a_\alpha^\dag a_\alpha - \varepsilon\, a_\alpha a_\alpha^\dag = \notag \\
&= \varepsilon \left( a_\alpha a_\alpha^\dag - \mathds{1} \right) - \varepsilon\, a_i a_i^\dag = \notag \\
&= \varepsilon a_\alpha a_\alpha^\dag - \varepsilon \mathds{1} - \varepsilon a_\alpha a_\alpha^\dag = \notag \\
&= -\varepsilon \mathds{1}.
\end{align}
\item Case 3: $\alpha_i \neq \alpha_j$. In this case, the creation and annihilation operators associated with different quantum states commute (bosons) or anticommute (fermions), then
\begin{equation}
C^{ij} = 0.
\end{equation}
\end{itemize}
In summary, the contraction \( \langle C^{ij} \rangle_0 \) takes the form
\begin{equation}
\left\langle C^{ij} \right\rangle_0 = 
\begin{cases}
+1 \ \ \ , & \alpha_i = \alpha_j \equiv \alpha, \ \ A^{(i)}_{\alpha} = a_\alpha, \ \ A^{(j)}_{\alpha} = a_\alpha^\dag \\
-\varepsilon \ \ \ , & \alpha_i = \alpha_j \equiv \alpha, \ \ A^{(i)}_{\alpha} = a_\alpha^\dag, \ \ A^{(j)}_{\alpha} = a_\alpha \\
0 \ \ \ , & \text{otherwise}
\end{cases} .
\end{equation}
Now, we use several times $A^{(i)}_{\alpha_i} A^{(j)}_{\alpha_j} = C^{ij} + \varepsilon A^{(j)}_{\alpha_j} A^{(i)}_{\alpha_i}$ in the thermal average as follows
\begin{align}
& \left\langle A^{(1)}_{\alpha_1} A^{(2)}_{\alpha_2} A^{(3)}_{\alpha_3} A^{(4)}_{\alpha_4} \right\rangle_0 = \notag \\
&= \left\langle \left( C^{12} + \varepsilon A^{(2)}_{\alpha_2} A^{(1)}_{\alpha_1} \right) A^{(3)}_{\alpha_3} A^{(4)}_{\alpha_4} \right\rangle_0 = \notag \\
&= \left\langle C^{12} \right\rangle_0 \left\langle A^{(3)}_{\alpha_3} A^{(4)}_{\alpha_4} \right\rangle_0 + \varepsilon \left\langle A^{(2)}_{\alpha_2} A^{(1)}_{\alpha_1} A^{(3)}_{\alpha_3} A^{(4)}_{\alpha_4} \right\rangle_0 = \notag \\
&= \left\langle C^{12} \right\rangle_0 \left\langle A^{(3)}_{\alpha_3} A^{(4)}_{\alpha_4} \right\rangle_0 + \varepsilon \left\langle A^{(2)}_{\alpha_2} \left( C^{13} + \varepsilon A^{(3)}_{\alpha_3} A^{(1)}_{\alpha_1} \right) A^{(4)}_{\alpha_4} \right\rangle_0 = \notag \\
&= \left\langle C^{12} \right\rangle_0 \left\langle A^{(3)}_{\alpha_3} A^{(4)}_{\alpha_4} \right\rangle_0 + \varepsilon \left\langle C^{13} \right\rangle_0 \left\langle A^{(2)}_{\alpha_2} A^{(4)}_{\alpha_4} \right\rangle_0 + \varepsilon^2 \left\langle A^{(2)}_{\alpha_2} A^{(3)}_{\alpha_3} A^{(1)}_{\alpha_1} A^{(4)}_{\alpha_4} \right\rangle_0 = \notag \\
&\equiv \left\langle C^{12} \right\rangle_0 \left\langle A^{(3)}_{\alpha_3} A^{(4)}_{\alpha_4} \right\rangle_0 + \varepsilon \left\langle C^{13} \right\rangle_0 \left\langle A^{(2)}_{\alpha_2} A^{(4)}_{\alpha_4} \right\rangle_0 + \left\langle A^{(2)}_{\alpha_2} A^{(3)}_{\alpha_3} A^{(1)}_{\alpha_1} A^{(4)}_{\alpha_4} \right\rangle_0 = \notag \\
&= \left\langle C^{12} \right\rangle_0 \left\langle A^{(3)}_{\alpha_3} A^{(4)}_{\alpha_4} \right\rangle_0 + \varepsilon \left\langle C^{13} \right\rangle_0 \left\langle A^{(2)}_{\alpha_2} A^{(4)}_{\alpha_4} \right\rangle_0 + \left\langle A^{(2)}_{\alpha_2} A^{(3)}_{\alpha_3} \left( C^{14} + \varepsilon A^{(4)}_{\alpha_4} A^{(1)}_{\alpha_1} \right) \right\rangle_0 = \notag \\
&= \left\langle C^{12} \right\rangle_0 \left\langle A^{(3)}_{\alpha_3} A^{(4)}_{\alpha_4} \right\rangle_0 + \varepsilon \left\langle C^{13} \right\rangle_0 \left\langle A^{(2)}_{\alpha_2} A^{(4)}_{\alpha_4} \right\rangle_0 + \left\langle C^{14} \right\rangle_0 \left\langle A^{(2)}_{\alpha_2} A^{(3)}_{\alpha_3} \right\rangle_0 + \varepsilon \left\langle A^{(2)}_{\alpha_2} A^{(3)}_{\alpha_3} A^{(4)}_{\alpha_4} A^{(1)}_{\alpha_1} \right\rangle_0 .
\end{align}
Now we want to check whether it is possible to manipulate the term $\varepsilon \left\langle A^{(2)}_{\alpha_2} A^{(3)}_{\alpha_3} A^{(4)}_{\alpha_4} A^{(1)}_{\alpha_1} \right\rangle_0$ so that it becomes proportional to the first member; we write
\begin{align}
\left\langle A^{(2)}_{\alpha_2} A^{(3)}_{\alpha_3} A^{(4)}_{\alpha_4} A^{(1)}_{\alpha_1} \right\rangle_0 &\equiv \dfrac{1}{Z_0} \operatorname{Tr} \left( e^{-\beta \hat{\mathcal{H}}_0} A_2 A_3 A_4 A_1 \right) = \notag \\
&= \dfrac{1}{Z_0} \operatorname{Tr} \left( A^{(2)}_{\alpha_2} A^{(3)}_{\alpha_3} A^{(4)}_{\alpha_4} A^{(1)}_{\alpha_1} e^{-\beta \hat{\mathcal{H}}_0} \right) = \notag \\
&= \dfrac{1}{Z_0} \operatorname{Tr} \left( A^{(2)}_{\alpha_2} A^{(3)}_{\alpha_3} A^{(4)}_{\alpha_4} e^{-\beta \hat{\mathcal{H}}_0} e^{\beta \hat{\mathcal{H}}_0} A^{(1)}_{\alpha_1} e^{-\beta \hat{\mathcal{H}}_0} \right) = \notag \\
&= \dfrac{1}{Z_0} \operatorname{Tr} \left( A^{(2)}_{\alpha_2} A^{(3)}_{\alpha_3} A^{(4)}_{\alpha_4} e^{-\beta \hat{\mathcal{H}}_0} e^{\delta_1 \beta \mathcal{E}_{\alpha_1}} A^{(1)}_{\alpha_1} \right) \equiv \notag \\
&\equiv \dfrac{e^{\delta_1 \beta \mathcal{E}_{\alpha_1}}}{Z_0} \operatorname{Tr} \left( A^{(2)}_{\alpha_2} A^{(3)}_{\alpha_3} A^{(4)}_{\alpha_4} e^{-\beta \hat{\mathcal{H}}_0} A^{(1)}_{\alpha_1} \right) = \notag \\
&= \dfrac{e^{\delta_1 \beta \mathcal{E}_{\alpha_1}}}{Z_0} \operatorname{Tr} \left( e^{-\beta \hat{\mathcal{H}}_0} A^{(1)}_{\alpha_1} A^{(2)}_{\alpha_2} A^{(3)}_{\alpha_3} A^{(4)}_{\alpha_4} \right) \equiv \notag \\
&\equiv e^{\delta_1 \beta \mathcal{E}_{\alpha_1}} \left\langle A^{(1)}_{\alpha_1} A^{(2)}_{\alpha_2} A^{(3)}_{\alpha_3} A^{(4)}_{\alpha_4} \right\rangle_0 .
\end{align}
We used the property $\eqref{eq: proprietàciclicatraccia}$ of the trace by shifting $e^{- \beta \hat{\mathcal{H}}_0}$ to the right; we inserted the identity $\mathds{1}= e^{-\beta \hat{\mathcal{H}}_0} e^{\beta \hat{\mathcal{H}}_0}$ between $A^{(4)}_{\alpha_4}$ and $A^{(1)}_{\alpha_1}$; by means of the thermal interaction picture, we wrote the evolute of $A^{(1)}_{\alpha_1}$ at time $\beta \hslash$ in a form that includes $\eqref{eq: evoluzionetemporaleraprresentazioneHeisenbergtermicaoperatorecreazione}$, $\eqref{eq: evoluzionetemporaleraprresentazioneHeisenbergtermicaoperatoredistruzione}$, that is,
\begin{align}
e^{\beta \hat{\mathcal{H}}_0} A^{(1)}_{\alpha_1} e^{-\beta \hat{\mathcal{H}}_0} &= A^{(1)(0)}_{\alpha_1}(\beta \hbar) = \notag \\
&= e^{\delta_1 \beta \mathcal{E}_{\alpha_1}} A^{(1)(0)}_{\alpha_1}(0) \equiv \notag \\
&\equiv e^{\delta_1 \beta \mathcal{E}_{\alpha_1}} A^{(1)}_{\alpha_1} ,
\end{align}
where $\mathcal{E}_{\alpha_1}$ is the one-particle energy of the quantum state associated with $A^{(1)}_{\alpha_1}$, and $\delta_1$ is $+1$ or $-1$ depending on whether the $A^{(1)}_{\alpha_1}$ operator is a creation or annihilation operator, respectively, i.e.,
\begin{equation}
\delta_1 =
\begin{cases}
+1, \ A^{(1)}_{\alpha_1} = a_{\alpha_1}^\dag \\
-1, \ A^{(1)}_{\alpha_1} = a_{\alpha_1}
\end{cases} ,
\end{equation}
Finally, we have used the cyclic property of the trace, see $\eqref{eq: proprietàciclicatraccia}$, first to bring \(A^{(1)}_{\alpha_1}\) to the left and then to move \(e^{-\beta \hat{\mathcal{H}}_0}\) back to the right. As a result, we obtain
\begin{align}
& \left\langle A^{(1)}_{\alpha_1} A^{(2)}_{\alpha_2} A^{(3)}_{\alpha_3} A^{(4)}_{\alpha_4} \right\rangle_0 = \notag \\
&= \left\langle C^{12} \right\rangle_0 \left\langle A^{(3)}_{\alpha_3} A^{(4)}_{\alpha_4} \right\rangle_0 + \varepsilon \left\langle C^{13} \right\rangle_0 \left\langle A^{(2)}_{\alpha_2} A^{(4)}_{\alpha_4} \right\rangle_0 + \left\langle C^{14} \right\rangle_0 \left\langle A^{(2)}_{\alpha_2} A^{(3)}_{\alpha_3} \right\rangle_0 + \varepsilon e^{\delta_1 \beta \mathcal{E}_{\alpha_1}} \left\langle A^{(1)}_{\alpha_1} A^{(2)}_{\alpha_2} A^{(3)}_{\alpha_3} A^{(4)}_{\alpha_4} \right\rangle_0 ,
\end{align}
\begin{align}
& \left(1 - \varepsilon e^{\delta_1 \beta \mathcal{E}_{\alpha_1}} \right) \left\langle A^{(1)}_{\alpha_1} A^{(2)}_{\alpha_2} A^{(3)}_{\alpha_3} A^{(4)}_{\alpha_4} \right\rangle_0 = \left\langle C^{12} \right\rangle_0 \left\langle A^{(3)}_{\alpha_3} A^{(4)}_{\alpha_4} \right\rangle_0 + \varepsilon \left\langle C^{13} \right\rangle_0 \left\langle A^{(2)}_{\alpha_2} A^{(4)}_{\alpha_4} \right\rangle_0 + \left\langle C^{14} \right\rangle_0 \left\langle A^{(2)}_{\alpha_2} A^{(3)}_{\alpha_3} \right\rangle_0 ,
\end{align}
and finally
\begin{equation}
\left\langle A^{(1)}_{\alpha_1} A^{(2)}_{\alpha_2} A^{(3)}_{\alpha_3} A^{(4)}_{\alpha_4} \right\rangle_0 = \dfrac{\left\langle C^{12} \right\rangle_0 \left\langle A^{(3)}_{\alpha_3} A^{(4)}_{\alpha_4} \right\rangle_0 + \varepsilon \left\langle C^{13} \right\rangle_0 \left\langle A^{(2)}_{\alpha_2} A^{(4)}_{\alpha_4} \right\rangle_0 + \left\langle C^{14} \right\rangle_0 \left\langle A^{(2)}_{\alpha_2} A^{(3)}_{\alpha_3} \right\rangle_0}{1 - \varepsilon e^{\delta_1 \beta \mathcal{E}_{\alpha_1}}} .
\label{eq: teoremaWick1}
\end{equation}
This expression explicitly shows how the thermal average of a four-operator product reduces to a combination of two-point thermal contractions, in accordance with the generalized Wick’s theorem. Equation $\eqref{eq: teoremaWick1}$ represents the thermal Wick's theorem specialized to the case of four operators.
\end{proof}
\end{theorem}
Note that Wick's theorem is referred to the thermal average of an even number of operators, since such an object is different from $0$ only when the number of operators is half creation and half annihilation. Now we show an equivalent form of Wick's thermal theorem. By definition, we have $\left\langle A^{(1)}_{\alpha_1} A^{(2)}_{\alpha_2} \right\rangle_0 = \left\langle C^{12} \right\rangle_0 + \varepsilon \left\langle A^{(2)}_{\alpha_2} A^{(1)}_{\alpha_1} \right\rangle_0$, and
\begin{align}
\langle A^{(2)}_{\alpha_2} A^{(1)}_{\alpha_1} \rangle_0 &= \Tr \left( \dfrac{e^{-\beta \hat{\mathcal{H}}_0} A^{(2)}_{\alpha_2} A^{(1)}_{\alpha_1}}{Z_0} \right) = \notag \\
&= \Tr \left( \dfrac{A^{(2)}_{\alpha_2} A^{(1)}_{\alpha_1} e^{-\beta \hat{\mathcal{H}}_0}}{Z_0} \right) = \notag \\
&= \Tr \left( \dfrac{A^{(2)}_{\alpha_2} e^{-\beta \hat{\mathcal{H}}_0} e^{\beta \hat{\mathcal{H}}_0} A^{(1)}_{\alpha_1} e^{-\beta \hat{\mathcal{H}}_0}}{Z_0} \right) = \notag \\
&= \Tr \left( \dfrac{A^{(2)}_{\alpha_2} e^{-\beta \hat{\mathcal{H}}_0} e^{\delta_1 \beta \mathcal{E}_{\alpha_1}} A^{(1)}_{\alpha_1}}{Z_0} \right) = \notag \\
&= e^{\delta_1 \beta \mathcal{E}_{\alpha_1}} \langle A^{(1)}_{\alpha_1} A^{(2)}_{\alpha_2} \rangle_0,
\end{align}
that is
\begin{equation}
\left\langle A^{(1)}_{\alpha_1} A^{(2)}_{\alpha_2} \right\rangle_0 = \dfrac{\left\langle C^{12} \right\rangle_0}{1 - \varepsilon e^{\delta_1 \beta \mathcal{E}_{\alpha_1}}} .
\end{equation}
Now, given the thermal average of the product of the 4 operators, we use
\begin{equation}
\left\langle C^{12} \right\rangle_0 = \left( 1 - \varepsilon e^{\delta_1 \beta \mathcal{E}_{\alpha_1}} \right) \left\langle A^{(1)}_{\alpha_1} A^{(2)}_{\alpha_2} \right\rangle_0
\end{equation}
for each $\langle C^{1j} \rangle_0$, then we finally arrive at the alternative form of Wick's theorem, namely
\begin{align}
\left\langle A^{(1)}_{\alpha_1} A^{(2)}_{\alpha_2} A^{(3)}_{\alpha_3} A^{(4)}_{\alpha_4} \right\rangle_0 &= \left\langle A^{(1)}_{\alpha_1} A^{(2)}_{\alpha_2} \right\rangle_0 \left\langle A^{(3)}_{\alpha_3} A^{(4)}_{\alpha_4} \right\rangle_0 + \varepsilon \left\langle A^{(1)}_{\alpha_1} A^{(3)}_{\alpha_3} \right\rangle_0 \left\langle A^{(2)}_{\alpha_2} A^{(4)}_{\alpha_4} \right\rangle_0 + \notag \\
&+ \left\langle A^{(1)}_{\alpha_1} A^{(4)}_{\alpha_4} \right\rangle_0 \left\langle A^{(2)}_{\alpha_2} A^{(3)}_{\alpha_3} \right\rangle_0 .
\label{eq: teoremaWick2}
\end{align}
Equation $\eqref{eq: teoremaWick2}$ is the form commonly used in practice, where each term of the type $\langle A_i A_j \rangle_0$ is often referred to as a contraction. In this formulation, the contraction $\langle C_{ij} \rangle_0$ is effectively replaced by the corresponding thermal expectation value, and the sign factors (including $\varepsilon = \pm 1$ for bosons or fermions) are incorporated into the properly permuted terms. \newline
Consider the contractions of the form $\langle A^{(i)}_{\alpha_i} A^{(j)}_{\alpha_j} \rangle_0$: these give non-zero contributions onlyonly when
the two operators refer to the same quantum state and one of them is a creation
operator while the other is an annihilation operator (see equation $\eqref{eq: mediatermicaadaggeraconalphadiversi}$). In particular, for the number operator, we recover equation $\eqref{eq: mediatermicanumerodioccupazioneedistribuzioneFermiBose}$, which expresses the average occupation number in terms of the Fermi-Dirac or Bose-Einstein distribution. \newline
Wick's thermal theorem admits a straightforward extension to time-dependent operators. Although we do not derive it here, the proof follows the same lines as in the static case, with contractions replaced by Matsubara Green's functions, apart from physical constants.
\section{Dyson equation for thermal Green's function of field operators in diagrammatic and integral form}
\begin{figure}[H]
\centering
\includegraphics[scale=0.7]{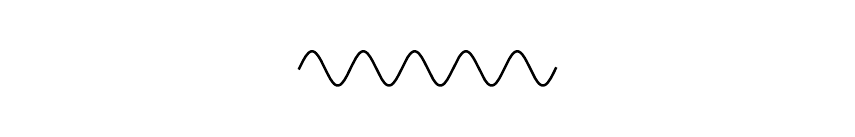}
\caption{Schematic representation of a basic interaction in Feynman diagrammatics. In particular, bosonic lines are represented as wavy curves.}
\label{fig: interazione}
\end{figure}

\begin{figure}[H]
\centering
\includegraphics[scale=0.7]{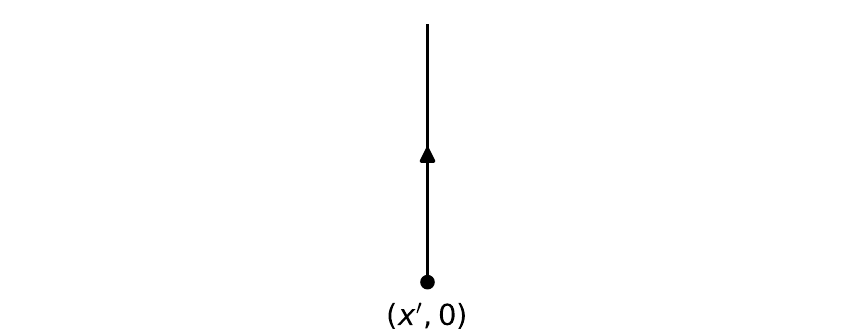}
\caption{Graphical representation of an annihilation operator in Feynman diagrammatics. The operator is depicted as the starting point of a fermionic line, corresponding to the removal of a particle from the quantum state.}
\label{fig: operatoredistruzionediagrammatica}
\end{figure}

\begin{figure}[H]
\centering
\includegraphics[scale=0.7]{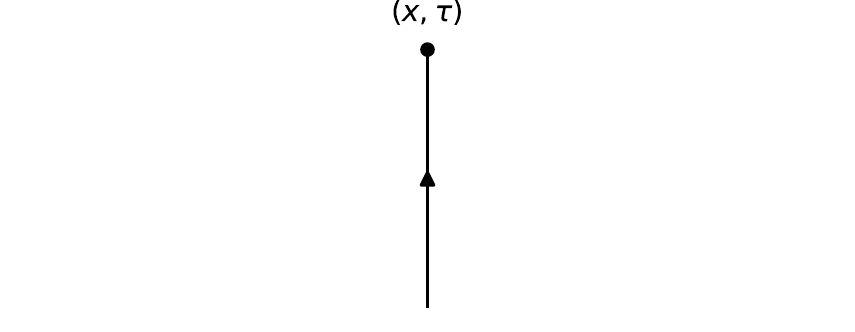}
\caption{Graphical representation of a creation operator in Feynman diagrammatics. The operator is depicted as the endpoint of a fermionic line, corresponding to the addition of a particle to the quantum state.}
\label{fig: operatorecreazionediagrammatica}
\end{figure}

\begin{figure}[H]
\centering
\includegraphics[scale=0.8]{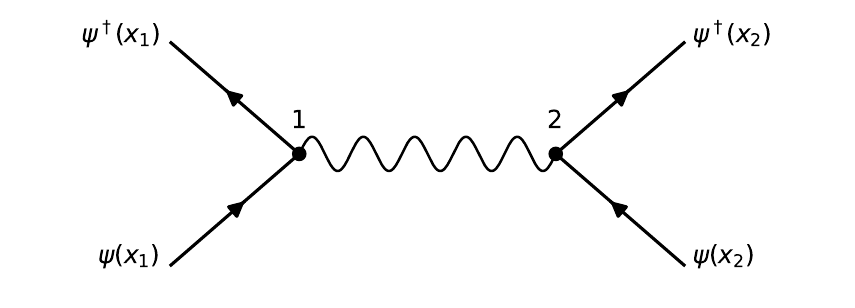}
\caption{Graphical representation of a two-body interaction in Feynman diagrammatics. Two fermionic lines exchange a bosonic mediator, illustrating a typical interaction between two particles.}
\label{fig: interazioneaduecorpi}
\end{figure}

\begin{figure}[H]
\centering
\includegraphics[scale=0.8]{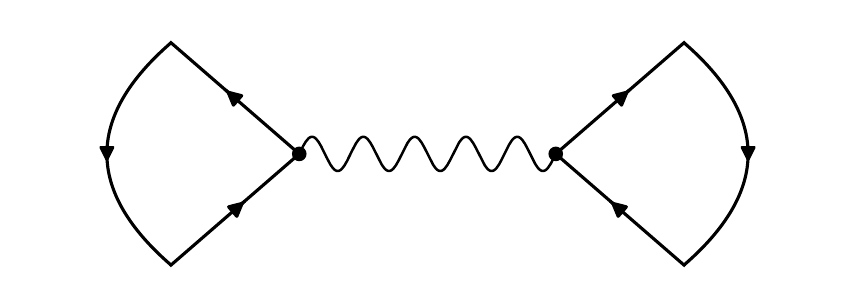}
\caption{Schematic representation of the first possible contraction contributing to the first-order denominator of the many-body Green's function of field operators, expanded in the Dyson series; see equation $\eqref{eq: seriediDysonperfunzionediGreentermicadeglioperatoricampo}$.}
\label{fig: primapossibilitàcontrazioneinterazioneprimoordine1}
\end{figure}

\begin{figure}[H]
\centering
\includegraphics[scale=0.8]{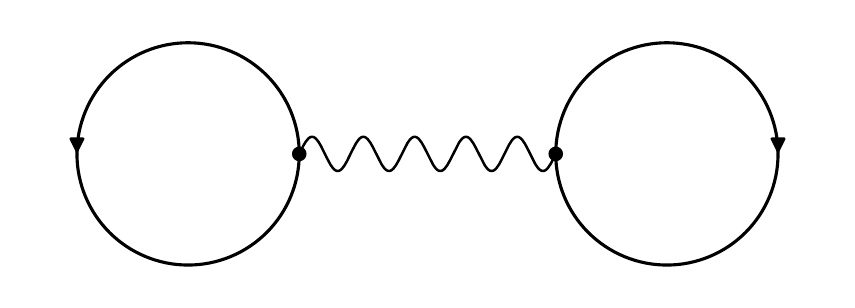}
\caption{Feynman diagrammatic representation of the Figure $\ref{fig: primapossibilitàcontrazioneinterazioneprimoordine1}$.}
\label{fig: primapossibilitàcontrazioneinterazioneprimoordine2}
\end{figure}

\begin{figure}[H]
\centering
\includegraphics[scale=0.8]{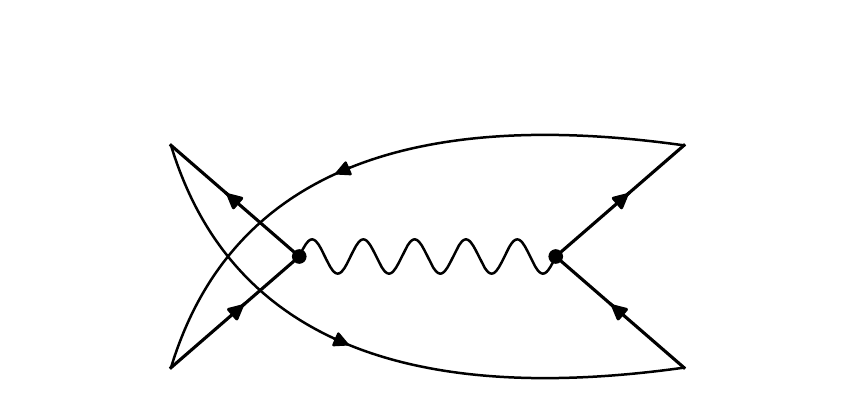}
\caption{Schematic representation of the second possible contraction contributing to the first-order denominator of the many-body Green's function of field operators, expanded in the Dyson series; see equation $\eqref{eq: seriediDysonperfunzionediGreentermicadeglioperatoricampo}$.}
\label{fig: secondapossibilitàcontrazioneinterazioneprimoordine1}
\end{figure}

\begin{figure}[H]
\centering
\includegraphics[scale=0.8]{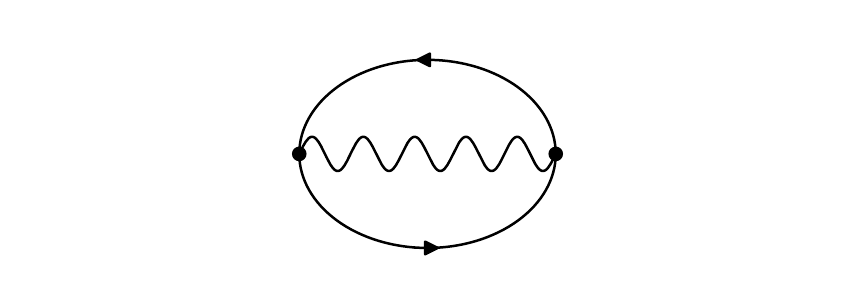}
\caption{Feynman diagrammatic representation of the Figure $\ref{fig: secondapossibilitàcontrazioneinterazioneprimoordine1}$.}
\label{fig: secondapossibilitàcontrazioneinterazioneprimoordine2}
\end{figure}

\begin{figure}[H]
\centering
\includegraphics[scale=0.8]{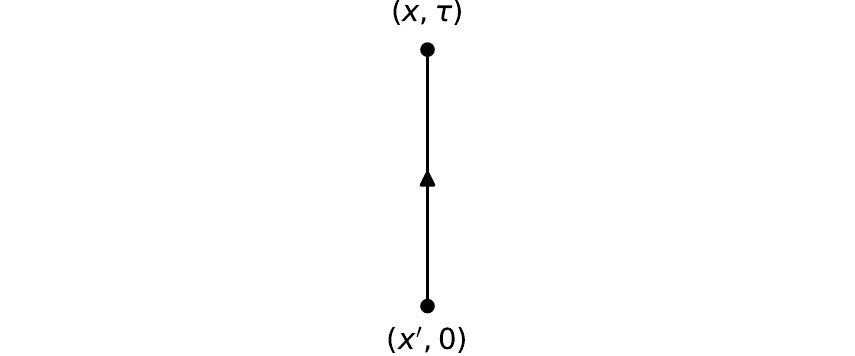}
\caption{Feynman diagrammatic representation of the free, or bare, many-body Green's function of field operators, illustrating the propagation of a non-interacting particle between two spacetime points.}
\label{fig: propagatorenudo}
\end{figure}

\begin{figure}[H]
\centering
\includegraphics[scale=1]{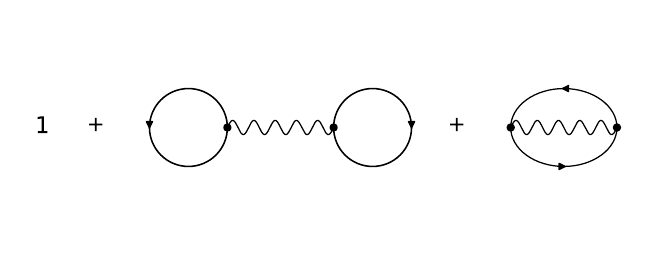}
\caption{Feynman diagrammatic representation of the first-order denominator of the many-body Green's function of field operators, expanded in the Dyson series; see equation $\eqref{eq: seriediDysonperfunzionediGreentermicadeglioperatoricampo}$.}
\label{fig: primoordinedenominatore}
\end{figure}

\begin{figure}[H]
\centering
\includegraphics[scale=0.6]{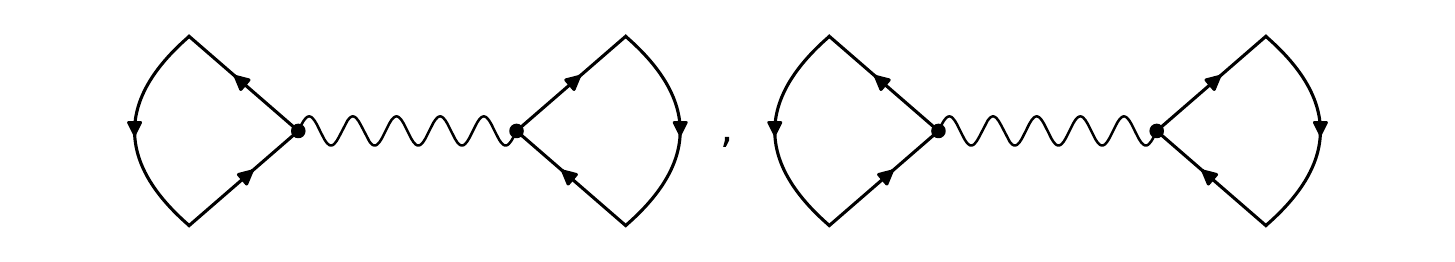}
\caption{Schematic representation of the first possible contraction contributing to the second-order denominator of the many-body Green's function of field operators, expanded in the Dyson series; see equation $\eqref{eq: seriediDysonperfunzionediGreentermicadeglioperatoricampo}$.}
\label{fig: primapossibilitàcontrazionedenominatoresecondoordine}
\end{figure}

\begin{figure}[H]
\centering
\includegraphics[scale=0.8]{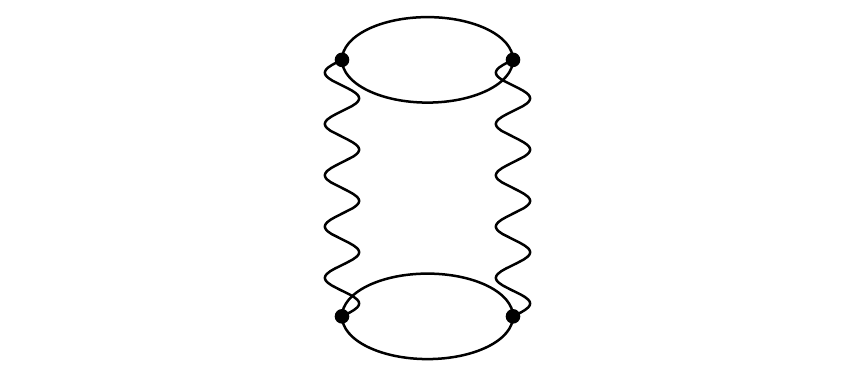}
\caption{Feynman diagrammatic representation of the second possible contraction contributing to the second-order denominator of the many-body Green's function of field operators, expanded in the Dyson series; see equation $\eqref{eq: seriediDysonperfunzionediGreentermicadeglioperatoricampo}$.}
\label{fig: secondapossibilitàcontrazionedenominatoresecondoordine}
\end{figure}

\begin{figure}[H]
\centering
\includegraphics[scale=0.8]{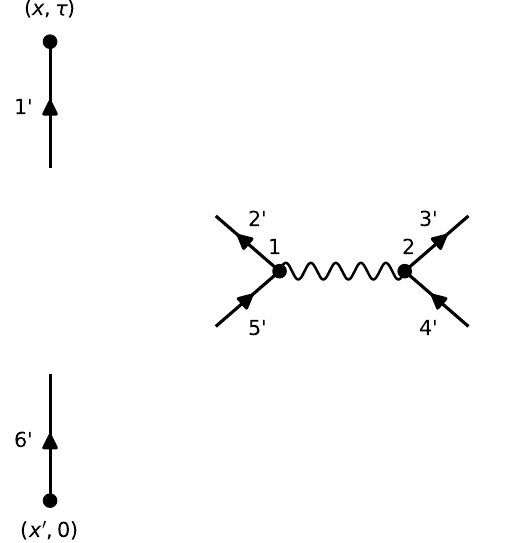}
\caption{Feynman diagrammatic representation of the first-order numerator contribution in the expansion of the many-body Green's function of field operators, expanded in the Dyson series; see equation $\eqref{eq: seriediDysonperfunzionediGreentermicadeglioperatoricampo}$.}
\label{fig: numeratorealprimoordine}
\end{figure}

\begin{figure}[H]
\centering
\includegraphics[scale=0.8]{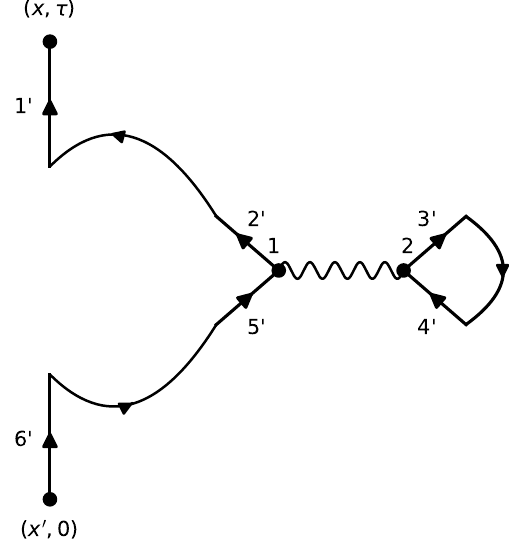}
\caption{Schematic representation of the first possible contraction contributing to the first-order numerator of the many-body Green's function of field operators, expanded in the Dyson series; see equation $\eqref{eq: seriediDysonperfunzionediGreentermicadeglioperatoricampo}$.}
\label{fig: primapossibilitàcontrazionenumeratoreprimoordine1}
\end{figure}

\begin{figure}[H]
\centering
\includegraphics[scale=0.63]{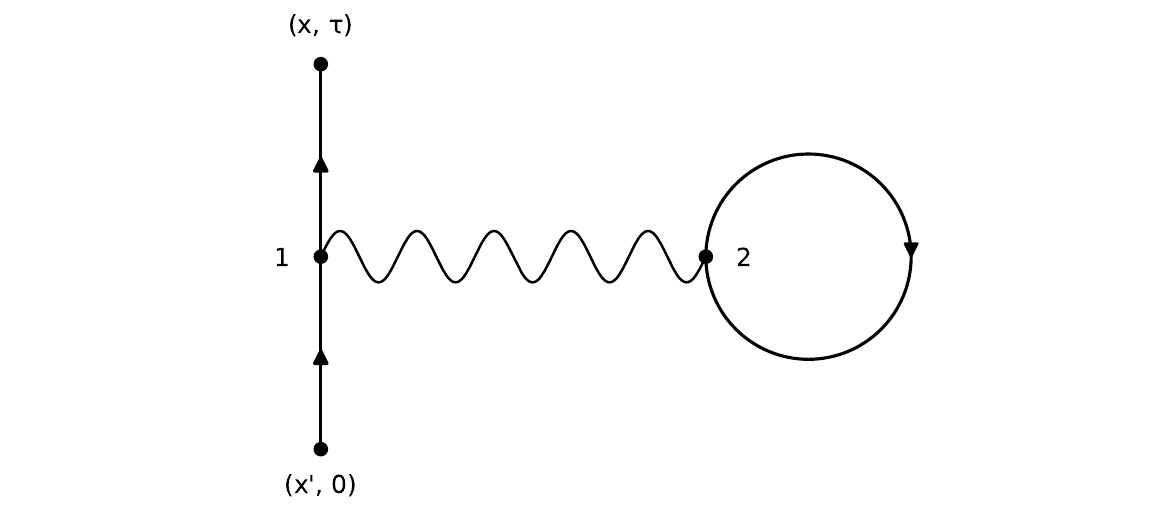}
\caption{Feynman diagrammatic representation of the Figure $\ref{fig: primapossibilitàcontrazionenumeratoreprimoordine1}$.}
\label{fig: primapossibilitàcontrazionenumeratoreprimoordine2}
\end{figure}

\begin{figure}[H]
\centering
\includegraphics[scale=0.7]{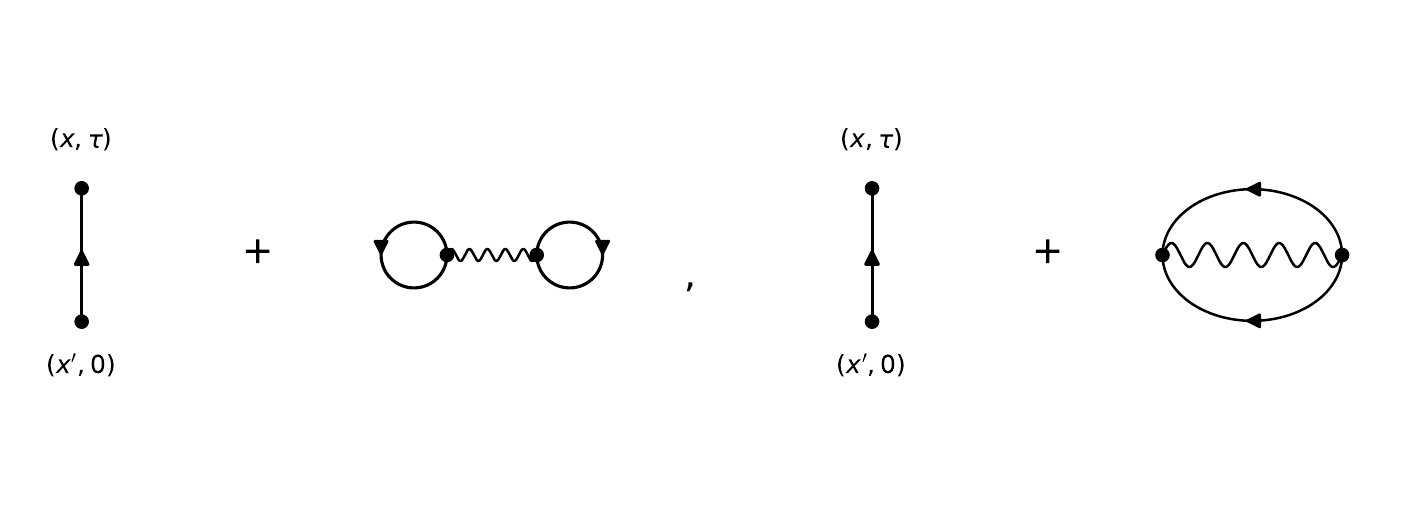}
\caption{Feynman diagrammatic representation of the third possible contraction contributing to first-order numerator contribution in the expansion of the many-body Green's function of field operators, expanded in the Dyson series; see equation $\eqref{eq: seriediDysonperfunzionediGreentermicadeglioperatoricampo}$.}
\label{fig: terzapossibilitàcontrazionenumeratoreprimoordine}
\end{figure}

\begin{figure}[H]
\centering
\includegraphics[scale=0.43]{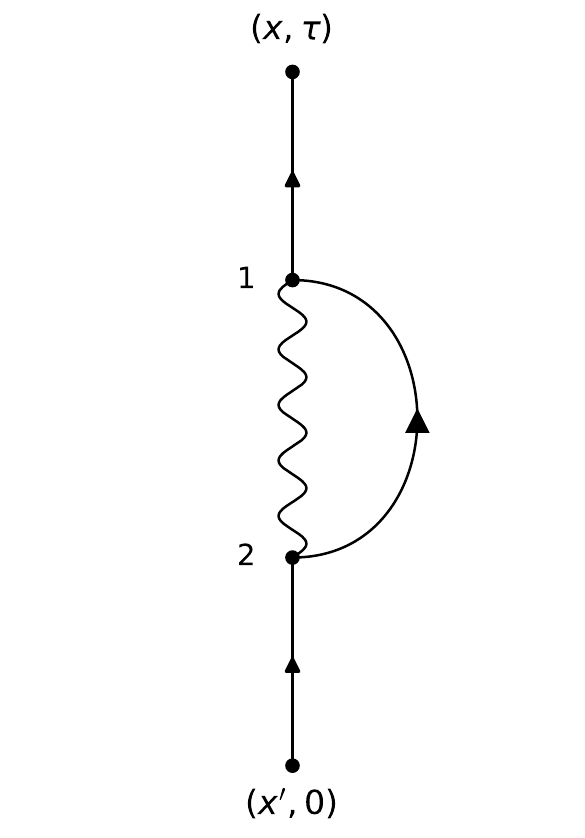}
\caption{Feynman diagrammatic representation of the fourth possible contraction contributing to first-order numerator contribution in the expansion of the many-body Green's function of field operators, expanded in the Dyson series; see equation $\eqref{eq: seriediDysonperfunzionediGreentermicadeglioperatoricampo}$.}
\label{fig: quartapossibilitàcontrazionenumeratoreprimoordine}
\end{figure}

\begin{figure}[H]
\centering
\includegraphics[scale=0.65]{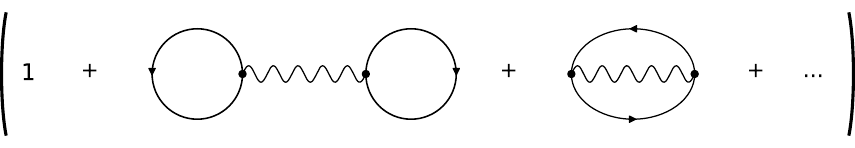}
\caption{Feynman diagrammatic representation of the denominator in the Dyson expansion of the many-body Green's function of field operators, expanded in the Dyson series; see equation $\eqref{eq: seriediDysonperfunzionediGreentermicadeglioperatoricampo}$.}
\label{fig: denominatorefunzionediGreennelladiagrammatica}
\end{figure}

\begin{figure}[H]
\centering
\makebox[\textwidth][c]{%
  \includegraphics[scale=0.7]{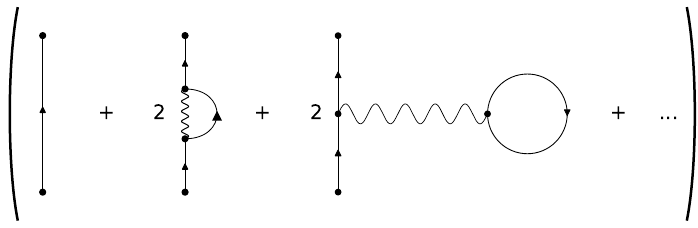}%
  \hspace{1mm}%
  \raisebox{0.5cm}{\includegraphics[scale=0.65]{Denominator_of_Dyson_expansion_of_Green_function.pdf}}%
}
\caption{Feynman diagrammatic representation of the numerator in the Dyson expansion of the many-body Green's function of field operators, expanded in the Dyson series; see equation $\eqref{eq: seriediDysonperfunzionediGreentermicadeglioperatoricampo}$.}
\label{fig: numeratorefunzionediGreennelladiagrammatica}
\end{figure}

Here, our goal is to specialize the Dyson equation to the case in which the system is governed by a Hamiltonian consisting of a free part, given by equation $\eqref{eq: Hamiltonianasecondaquantizzazioneparticellelibere}$, and a two-body interaction term described by equation $\eqref{eq: operatoreaduecorpisecondaquantizzazione}$. That is, we consider the full Hamiltonian in the form of equation $\eqref{eq: hamiltonianamateriainterazioneaduecorpi}$. Consider the Matsubara Green's function of the field operators, i.e.,
\begin{equation}
G^{(m)}_{x,x'}(\tau,0) = - \dfrac{1}{\hslash} \left\langle \hat{T}_\tau \, \hat{\psi}(x,\tau) \, \hat{\psi}^\dagger(x',0) \right\rangle,
\label{eq: funzioneGreenMastubaraoperatoricampo}
\end{equation}
where the subscript $(x,x')$ is a shortened form of $\left( \hat{\psi}(x),\hat{\psi}^\dag(x')\right)$. We have previously shown that any Green's function of Matsubara can be written in the form $\eqref{eq: GreentermicainfunzionediU(betahslash)2}$, which we use for the field operators, $A=\hat{\psi}(x,\tau)$ and $B=\hat{\psi}^\dag(x',0)$. We aim to calculate the object
\begin{align}
G^{(m)}_{x,x'}(\tau,0) &= 
\frac{
    -\frac{1}{\hslash} \left\langle \hat{T}_\tau \hat{\psi}^{(0)}(x,\tau) \hat{\psi}^\dag(x',0) \right\rangle_0
}{
    1 + \sum_{n=1}^{\infty} \left( -\frac{1}{\hslash} \right)^n \frac{1}{n!} 
    \int_0^{\beta \hslash} d\tau_1 \ldots \int_0^{\beta \hslash} d\tau_n 
    \left\langle \hat{T}_\tau \left\{ \hat{\mathcal{H}}_I^{(0)}(\tau_1) \ldots \hat{\mathcal{H}}_I^{(0)}(\tau_n) \right\} \right\rangle_0
}
\notag \\
&\quad + 
\frac{
    -\frac{1}{\hslash} \sum_{n=1}^{\infty} \left( -\frac{1}{\hslash} \right)^n \frac{1}{n!} 
    \int_0^{\beta \hslash} d\tau_1 \ldots \int_0^{\beta \hslash} d\tau_n 
    \left\langle \hat{T}_\tau \left\{ \hat{\mathcal{H}}_I^{(0)}(\tau_1) \ldots \hat{\mathcal{H}}_I^{(0)}(\tau_n) \right\} 
    \hat{\psi}^{(0)}(x,\tau) \hat{\psi}^\dag(x',0) \right\rangle_0
}{
    1 + \sum_{n=1}^{\infty} \left( -\frac{1}{\hslash} \right)^n \frac{1}{n!} 
    \int_0^{\beta \hslash} d\tau_1 \ldots \int_0^{\beta \hslash} d\tau_n 
    \left\langle \hat{T}_\tau \left\{ \hat{\mathcal{H}}_I^{(0)}(\tau_1) \ldots \hat{\mathcal{H}}_I^{(0)}(\tau_n) \right\} \right\rangle_0
}.
\label{eq: seriediDysonperfunzionediGreentermicadeglioperatoricampo}
\end{align}
Note that the general form of an interaction Hamiltonian in second quantization always involves field operators, which are expressed as combinations of creation and annihilation operators. Therefore, since every interaction Hamiltonian is written in the thermal representation, from Wick's thermal theorem we are ultimately required to compute quantities of the form
\begin{equation}
\left\langle \hat{T}_\tau \, a^{(0)}_{\alpha_1}(\tau_1) \, a^{\dagger(0)}_{\alpha_2}(\tau_2) \right\rangle_0 
= - \hslash \, G^{(m)(0)}_{\alpha_1,\alpha_2}(\tau_1,\tau_2),
\end{equation}
which can be rewritten as
\begin{equation}
\left\langle \hat{T}_\tau \, a^{\dagger(0)}_{\alpha_2}(\tau_2) \, a^{(0)}_{\alpha_1}(\tau_1) \right\rangle_0 
= - \hslash \, \varepsilon \, G^{(m)(0)}_{\alpha_1,\alpha_2}(\tau_1,\tau_2).
\end{equation}
Consider the denominator of $\eqref{eq: seriediDysonperfunzionediGreentermicadeglioperatoricampo}$. For $n=0$, it gives unity. For $n=1$, the term of the series is
\begin{equation}
1 - \dfrac{1}{\hslash} \int_0^{\beta \hslash} d\tau_1 \, \left\langle \hat{T}_\tau \, \hat{\mathcal{H}}_I^{(0)}(\tau_1) \right\rangle_0,
\end{equation}
where for an interaction Hamiltonian given by a two-body operator, that is, equation $\eqref{eq: operatoreaduecorpisecondaquantizzazione}$, we have
\begin{equation}
1 - \dfrac{1}{\hslash} \int_0^{\beta \hslash} d\tau_1 \, \dfrac{1}{2} \int dx_1 \int dx_2 \, U(x_1, x_2) \,
\left\langle \hat{T}_\tau \, 
\hat{\psi}^{\dagger (0)}(x_1, \tau_1) \, 
\hat{\psi}^{\dagger (0)}(x_2, \tau_1) \, 
\hat{\psi}^{(0)}(x_2, \tau_1) \, 
\hat{\psi}^{(0)}(x_1, \tau_1) 
\right\rangle_0.
\end{equation}
We apply Wick's thermal theorem by noting that $\hat{\psi}^{\dag (0)}(x_1,\tau_1)$ can only contract with $\hat{\psi}^{(0)}(x_2,\tau_1)$ (there is one exchange and a $\varepsilon$ appears) and with $\hat{\psi}(x_1,\tau_1)$ (there are two exchanges), therefore it follows
\begin{align}
& 1 - \dfrac{1}{\hslash} \int_0^{\beta \hslash} d\tau_1 \, \dfrac{1}{2} \int dx_1 \int dx_2 \, U(x_1, x_2) \Big( 
\left\langle \hat{T}_\tau \, \hat{\psi}^{\dagger (0)}(x_1, \tau_1) \, \hat{\psi}^{(0)}(x_1, \tau_1) \right\rangle_0 \,
\left\langle \hat{T}_\tau \, \hat{\psi}^{\dagger (0)}(x_2, \tau_1) \, \hat{\psi}^{(0)}(x_2, \tau_1) \right\rangle_0 \notag \\
& \quad + \, \varepsilon \,
\left\langle \hat{T}_\tau \, \hat{\psi}^{\dagger (0)}(x_1, \tau_1) \, \hat{\psi}^{(0)}(x_2, \tau_1) \right\rangle_0 \,
\left\langle \hat{T}_\tau \, \hat{\psi}^{\dagger (0)}(x_2, \tau_1) \, \hat{\psi}^{(0)}(x_1, \tau_1) \right\rangle_0 \Big) .
\end{align}
Feynman invented diagrams and rules to visualize a term in the series, such as the one above. A wavy line, see Figure $\eqref{fig: interazione}$, represents an interaction between particles. The extremes of a wavy line are called vertices: each interaction line carries with it $2$ vertices, which we denote by $1$ and $2$, or $x_1$ and $x_2$, respectively. For a Hamiltonian that includes two-body interaction, two annihilation field operators and two creation field operators are associated with the interaction line: an annihilation field operator is represented by a line entering a vertex (see Figure $\eqref{fig: operatoredistruzionediagrammatica}$), while a creation field operator is represented by a line leaving a vertex (see Figure $\eqref{fig: operatorecreazionediagrammatica}$). At first order, the denominator of $\eqref{eq: seriediDysonperfunzionediGreentermicadeglioperatoricampo}$ includes the interaction only once, so a graph of the type in Figure $\eqref{fig: interazioneaduecorpi}$ appears only once. Recall that in Wick's thermal theorem, nonzero contractions are those of the form $\langle a a^\dag \rangle_0$ and $\langle a^\dag a \rangle_0$, so the graph can only be completed in $2$ ways.
\begin{itemize}
\item
$I$: Figure $\eqref{fig: primapossibilitàcontrazioneinterazioneprimoordine1}$ or equivalently $\eqref{fig: primapossibilitàcontrazioneinterazioneprimoordine2}$. $\hat{\psi}^{\dag (0)}(x_1)$ is contracted with $\hat{\psi}^{(0)}(x_1)$ and $\hat{\psi}^{\dag (0)}(x_2)$ is contracted with $\hat{\psi}^{(0)}(x_2)$ or briefly $1$ with $1$ and $2$ with $2$ and two loops are formed.
\item
$II$: Figure $\eqref{fig: secondapossibilitàcontrazioneinterazioneprimoordine1}$ or equivalently $\eqref{fig: secondapossibilitàcontrazioneinterazioneprimoordine2}$. $\hat{\psi}^{\dag (0)}(x_1)$ contracted with $\hat{\psi}^{(0)}(x_2)$ and $\hat{\psi}^{\dag (0)}(x_2)$ contracted with $\hat{\psi}^{(0)}(x_1)$ or briefly $1$ with $2$ and $2$ with $1$. 
\end{itemize}
By Wick's thermal theorem, when an operator $\hat{\psi}^{\dag (0)}(x,\tau)$ is contracted with an operator $\hat{\psi}^{(0)}(x',0)$, we must calculate the average of that product with respect to a Hamiltonian $\hat{\mathcal{H}}_0$ without interactions, that is, we must calculate the free propagator
\begin{equation}
G^{(m)(0)}_{x,x'}(\tau,0) = - \dfrac{1}{\hslash} \left\langle \hat{T}_\tau \, \hat{\psi}^{(0)}(x,\tau) \, \hat{\psi}^{\dagger (0)}(x',0) \right\rangle_0.
\end{equation}
Which is a the representation of a propagator in Feynman diagrammatics? Consider two points, the point $(x',0)$ where the particle is created and the point $(x,\tau)$ where the particle is destroyed, representing as in Figure $\eqref{fig: propagatorenudo}$. Such line is half creation and half annihilation and it represents the propagation from $(x',0)$ to $(x,\tau)$. A solid line represents a free propagator, that is, a particle whose dynamics is described by a Hamiltonian $\mathcal{H}_0$. The solid line is always read from top to bottom. Regarding the first-order denominator of the $\eqref{eq: seriediDysonperfunzionediGreentermicadeglioperatoricampo}$, there are two possibilities to close the graph. Two solid lines are closed on themselves and originate two loops, that is, two $G^{(m)(0)}$, in which case there is no particle exchange; or the particles are exchanged. The first order of the denominator is shown in Figure $\eqref{fig: primoordinedenominatore}$. Now, consider the second order of the denominator of the $\eqref{eq: seriediDysonperfunzionediGreentermicadeglioperatoricampo}$. Again there are two possibilities for closing the graph.
\begin{itemize}
\item $I$: Figure $\eqref{fig: primapossibilitàcontrazionedenominatoresecondoordine}$. We obtain disconnected diagrams, meaning that the two interactions are not connected to each other and remain independent.
\item $II$: Figure $\eqref{fig: secondapossibilitàcontrazionedenominatoresecondoordine}$. We get a connected diagram: in this case the diagrams can cross each other and each vertex of one interaction is also connected to the other interaction.
\end{itemize}
The first order of the $\eqref{eq: seriediDysonperfunzionediGreentermicadeglioperatoricampo}$ includes only connected diagrams, while the second order originates both connected and disconnected diagrams. \newline
Now, consider the numerator of the $\eqref{eq: seriediDysonperfunzionediGreentermicadeglioperatoricampo}$. For $n=0$ there are no integrals and we have
\begin{equation}
- \dfrac{1}{\hslash} \left\langle \hat{T}_\tau \, \hat{\psi}^{(0)}(x, 0) \, \hat{\psi}^{\dagger (0)}(x', \tau) \right\rangle_0 
= G^{(m)(0)}_{x,x'}(\tau),
\end{equation} 
which is the free thermal Green's function. On the other hand, for $n=0$ the denominator is worth $1$. Consequently, Feynman's approach is a perturbative expansion. We observe that the integrand function of the $n$-th term of the Dyson series of Green's function for two generic operators $A$ and $B$, i.e., the object
\begin{equation}
\left\langle \hat{T}_\tau \left\lbrace \hat{\mathcal{H}}_I^{(0)}(\tau_1) \ldots \hat{\mathcal{H}}_I^{(0)}(\tau_n) \right\rbrace A^{(0)}(\tau) \, B^{(0)}(0) \right\rangle_0
\end{equation}
can be written as
\begin{equation}
\left\langle \hat{T}_\tau \left\lbrace A^{(0)}(\tau) \, \hat{\mathcal{H}}_I^{(0)}(\tau_1) \ldots \hat{\mathcal{H}}_I^{(0)}(\tau_n) \, B(0) \right\rbrace \right\rangle_0.
\end{equation}
In fact, if we insert $A^{(0)}(\tau)$ and $B(0)$ at the beginning and end of the product, respectively, the action of $\hat{T}_\tau$ is unchanged. Since $B$ is evaluated at time $0$, it remains to the right of all other operators. Regarding $A$, for each place reversal with $\hat{\mathcal{H}}_I$ an even number of exchanges are required, so the order in which it figures does not matter. Then, for $A=\hat{\psi}^{(0)}(x,\tau)$ and $B=\hat{\psi}^{\dag (0)}(x',0)$, the first-order development of the numerator is given by
\begin{align}
& G^{(m)(0)}_{x,x'}(\tau,0) 
- \frac{1}{\hslash} \left( \frac{1}{\hslash} \right) \int_0^{\beta \hslash} d\tau_1 \, \frac{1}{2} \int dx_1 \int dx_2 \, U(x_1,x_2) \notag \\
& \quad  
\left\langle \hat{T}_\tau \,
\hat{\psi}^{(0)}(x,\tau) \,
\hat{\psi}^{\dagger (0)}(x_1,\tau_1) \,
\hat{\psi}^{\dagger (0)}(x_2,\tau_1) \,
\hat{\psi}^{(0)}(x_2,\tau_1) \,
\hat{\psi}^{(0)}(x_1,\tau_1) \,
\hat{\psi}^{\dagger (0)}(x',0) 
\right\rangle_0
\end{align}
to which we want to apply the rules of diagrammatics. We use the identifications $\hat{\psi}^{(0)}(x,\tau)=1'$, $\hat{\psi}^{\dag (0)}(x_1,\tau_1)=2'$, $\hat{\psi}^{\dag (0)}(x_2,\tau_1)=3'$, $\hat{\psi}^{(0)}(x_2,\tau_1)=4'$, $\hat{\psi}^{(0)}(x_1,\tau_1)=5'$, $\hat{\psi}^{\dag (0)}(x',0)=6'$, where operators $2'$, $3'$, $4'$ and $5'$ are originated by the interaction, while $1'$ is a half-line that is annihilated in $(x,\tau)$ and $6'$ is a half-line that is created in $(x',0)$ (see Figure $\eqref{fig: numeratorealprimoordine}$). We study the possible cases. 
\begin{itemize}
\item $I$: Figure $\eqref{fig: primapossibilitàcontrazionenumeratoreprimoordine2}$. $1'$ is closed with $2'$, $3'$ with $4'$, $5'$ with $6'$ as in Figure $\eqref{fig: primapossibilitàcontrazionenumeratoreprimoordine1}$, and we get the $\eqref{fig: primapossibilitàcontrazionenumeratoreprimoordine2}$, which represents a direct term, since there are no particle exchanges.
\item $II$. $1'$ is closed with $3'$, $4'$ with $6'$, $2'$ with $5'$. The diagram we obtain is equivalent to that in Case I, two vertices of the interaction being equivalent to each other: since by the principle of indistinguishability $V(x_1,x_2)=V(x_2,x_1)$ holds, the direct term in Figure $\eqref{fig: primapossibilitàcontrazionenumeratoreprimoordine2}$ appears twice.
\item $III$. $1'$ is closed with $6'$: we obtain a free thermal Green's function $G^{(m) (0)}_{x,x'}(\tau,0)$. For the interaction diagram, the two cases examined for the denominator at first order apply. We then obtain the two disconnected diagrams in Figure $\eqref{fig: terzapossibilitàcontrazionenumeratoreprimoordine}$.
\item $IV$: $1'$ is closed with $2'$, $6'$ with $4'$ and $3'$ with $5'$. The $\eqref{fig: quartapossibilitàcontrazionenumeratoreprimoordine}$ represents an exchange term, as operators $2$ and $1$ have been exchanged with each other. The particle is created in $(x',0)$, propagates to $2$, interacts, then propagates from $1$ to $(x,\tau)$. See Figure $\eqref{fig: quartapossibilitàcontrazionenumeratoreprimoordine}$.
\item $V$. $1'$ is closed with $3'$, $6'$ with $5'$ and $2'$ with $4'$: $1$ exchanges with $2$ and by the principle of indistinguishability we get the previous diagram.
\end{itemize}
Ultimately, given the thermal Green function of the field operators in the $\eqref{eq: seriediDysonperfunzionediGreentermicadeglioperatoricampo}$, the denominator and numerator are shown in Figure $\eqref{fig: denominatorefunzionediGreennelladiagrammatica}$ and Figure $\eqref{fig: numeratorefunzionediGreennelladiagrammatica}$, respectively. It can be seen that some numerator diagrams simplify denominator diagrams, at the numerator there are only connected diagrams: it can be shown that it is possible to eliminate the denominator as long as only connected diagrams are considered at the numerator. \newline
Finally, the Feynman perturbative expansion of a thermal Green's function is given by
\begin{equation}
G^{(m)}_{AB}(\tau,0) = G^{(m)(0)}_{AB}(\tau,0) - \frac{1}{\hslash} \sum_{n=1}^\infty \left( - \frac{1}{\hslash} \right)^n \frac{1}{n!} \int_0^{\beta \hslash} d\tau_1 \cdots \int_0^{\beta \hslash} d\tau_n \, \left\langle \hat{T}_\tau \, A^{(0)}(\tau) \, \hat{\mathcal{H}}_I^{(0)}(\tau_1) \cdots \hat{\mathcal{H}}_I^{(0)}(\tau_n) \, B(0) \right\rangle_{0,conn.}
\label{eq: sviluppoperturbativodifeynmandiunafunzionediGreentermica}
\end{equation}
We have expressed \( G^{(m)}_{AB}(\tau, 0) \) as a power series in \( \hat{\mathcal{H}}^{(0)}_I \). Since the power series appears only in the numerator, \( G^{(m)}_{AB}(\tau, 0) \) is itself a genuine power series, rather than a ratio of two series.
\subsection{First order analytic in Feynman's perturbative theory}
\begin{figure}[H]
\centering
\includegraphics[scale=0.63]{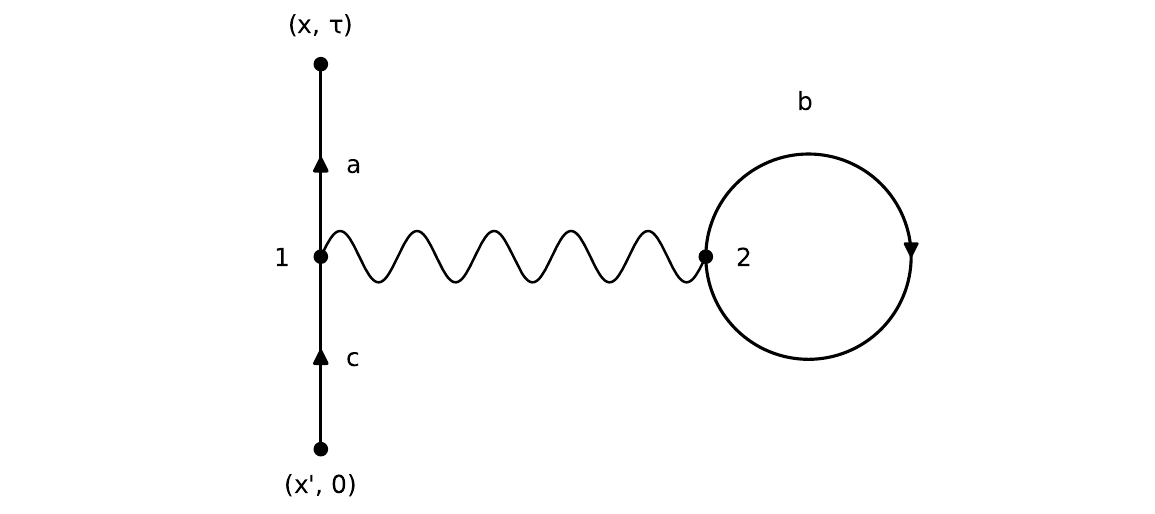}
\caption{Feynman diagrammatic representation of the first possible contraction contributing to the first-order term in the perturbative expansion of the many-body Green's function of field operators; see equation $\eqref{eq: sviluppoperturbativodifeynmandiunafunzionediGreentermica}$.}
\label{fig: primapossibilitàcontrazionesviluppoFeynmanprimoordine}
\end{figure}

\begin{figure}[H]
\centering
\includegraphics[scale=0.53]{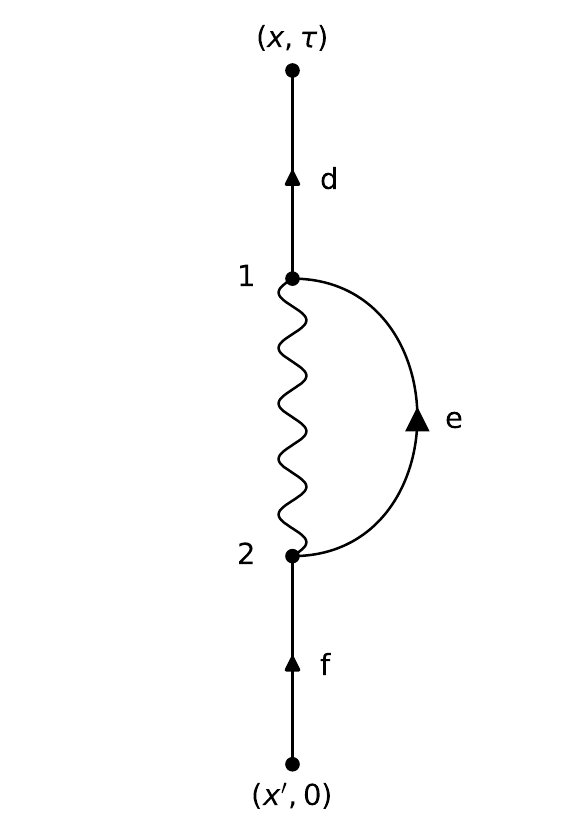}
\caption{Feynman diagrammatic representation of the second possible contraction contributing to the first-order term in the perturbative expansion of the many-body Green's function of field operators; see equation $\eqref{eq: sviluppoperturbativodifeynmandiunafunzionediGreentermica}$.}
\label{fig: secondapossibilitàcontrazionesviluppoFeynmanprimoordine}
\end{figure}

We study the first-order of $\eqref{eq: sviluppoperturbativodifeynmandiunafunzionediGreentermica}$.
\begin{itemize}
\item $I$ diagram, see Figure $\eqref{fig: primapossibilitàcontrazionesviluppoFeynmanprimoordine}$. Such a diagram arises from contractions $a \equiv 1'-2'$, $b \equiv 3'-4'$, and $c \equiv 5'-6'$. $a$ corresponds to $-\hslash G^{(m)(0)}{x,x_1}(\tau,\tau_1)$ because, reading from top to bottom, we first have $\hat{\psi}^{(0)}(x,\tau)$ and then $\hat{\psi}^{\dagger (0)}(x_1,\tau_1)$. $b$ corresponds to $-\hslash \varepsilon G^{(m)(0)}{x_2,x_2}(\tau_1,\tau_1 + \delta)$, as $\delta \rightarrow 0^+$. The Green's function is calculated at negative times because the second time instant follows the first. Since the operators are evaluated at the same time, we take $\delta \rightarrow 0^+$. Finally, $c$ corresponds to $-\hslash G^{(m)(0)}_{x_1,x'}(\tau_1,0)$.
\item $II$ diagram, see Figure $\eqref{fig: secondapossibilitàcontrazionesviluppoFeynmanprimoordine}$. Such a diagram arises from contractions $d \equiv 1'-2'$, $e \equiv 3'-5'$, $f \equiv 4'-6'$. $d$ corresponds to $- \hslash G^{(m)(0)}_{x,x_1}(\tau,\tau_1)$, $e$ corresponds to $- \hslash \varepsilon G^{(m)(0)}_{x_1,x_2}(\tau_1,\tau_1 + \delta)$ and $f$ corresponds to $- \hslash  G^{(m)(0)}_{x_2,x'}(\tau_1,0)$.
\end{itemize} 
Each of these diagrams appears twice, because of the principle of indistinguishability. The analytic first order in $\eqref{eq: sviluppoperturbativodifeynmandiunafunzionediGreentermica}$ evaluated at $\tau>0$ is
\begin{align}
G^{(m)(1)}_{x,x'}(\tau,0) &= G^{(m)(0)}_{x,x'}(\tau,0) 
+ \left(- \frac{1}{\hslash} \right) \left(- \frac{1}{\hslash} \right) 
\int_0^{\beta \hslash} d\tau_1 \, \frac{1}{2} \int dx_1 \int dx_2 \, U(x_1,x_2) \notag \\
& \quad \left\langle 
\hat{T}_\tau \hat{\psi}^{(0)}(x,\tau) \hat{\psi}^{\dag (0)}(x_1,\tau_1) \hat{\psi}^{\dag (0)}(x_2,\tau_1) 
\hat{\psi}^{(0)}(x_2,\tau_1) \hat{\psi}^{(0)}(x_1,\tau_1) \hat{\psi}^{\dag (0)}(x',0) 
\right\rangle_{0,conn.} = \notag \\
&= G^{(m)(0)}_{x,x'}(\tau,0) + \left(- \frac{1}{\hslash} \right)^2 \int_0^{\beta \hslash} d\tau_1 \, \frac{1}{2} \int dx_1 \int dx_2 \, U(x_1,x_2) \notag \\
& \quad \Big\lbrace 
2 (-\hslash) G^{(m)(0)}_{x,x_1}(\tau,\tau_1) (-\hslash) \varepsilon G^{(m)(0)}_{x_2,x_2}(\tau_1,\tau_1 + \delta) (-\hslash) G^{(m)(0)}_{x_1,x'}(\tau_1,0) \notag \\
& \quad \quad + 2 \varepsilon (-\hslash) G^{(m)(0)}_{x,x_1}(\tau,\tau_1) (-\hslash) \varepsilon G^{(m)(0)}_{x_1,x_2}(\tau_1,\tau_1+\delta) \cdot (-\hslash) G^{(m)(0)}_{x_2,x'}(\tau_1,0) 
\Big\rbrace = \notag \\
&= G^{(m)(0)}_{x,x'}(\tau,0) + \left(- \frac{1}{\hslash} \right)^2 \int_0^{\beta \hslash} d\tau_1 \int dx_1 \int dx_2 \, U(x_1,x_2) \notag \\
& \quad \Big\lbrace 
(-\hslash) G^{(m)(0)}_{x,x_1}(\tau,\tau_1) \cdot (-\hslash) \varepsilon G^{(m)(0)}_{x_2,x_2}(\tau_1,\tau_1 + \delta) (-\hslash) G^{(m)(0)}_{x_1,x'}(\tau_1,0) \notag \\
& \quad \quad + \varepsilon (-\hslash) G^{(m)(0)}_{x,x_1}(\tau,\tau_1) \cdot (-\hslash) \varepsilon G^{(m)(0)}_{x_1,x_2}(\tau_1,\tau_1+\delta) \cdot (-\hslash) G^{(m)(0)}_{x_2,x'}(\tau_1,0) 
\Big\rbrace.
\end{align}
Note that the spatial integrations are with respect to the interaction vertices. We associate a spacetime variable with each vertex over which we integrate. We will refer to the vertices as $(x_1, \tau_1)$ and $(x_2, \tau_2)$ instead of $1$ and $2$, respectively. We have two spatial integrals and one temporal integral. To obtain two temporal integrals, we multiply by $\int_0^{\beta \hslash} d\tau_2 \delta (\tau_1-\tau_2)$ and the terms $b$, $e$, and $f$ should be rewritten as
\begin{equation}
b \longrightarrow - \hslash \varepsilon G^{(m)(0)}_{x_2,x_2}(\tau_1,\tau_1 + \delta) \rightarrow - \hslash \varepsilon G^{(m)(0)}_{x_2,x_2}(\tau_2,\tau_2 + \delta),
\end{equation}
\begin{equation}
e \longrightarrow - \hslash \varepsilon G^{(m)(0)}_{x_1,x_2}(\tau_1,\tau_1 + \delta) \rightarrow - \hslash \varepsilon G^{(m)(0)}_{x_1,x_2}(\tau_1,\tau_2 + \delta),
\end{equation}
\begin{equation}
f \longrightarrow - \hslash G^{(m)(0)}_{x_2,x'}(\tau_1,0) \rightarrow - \hslash G^{(m)(0)}_{x_2,x'}(\tau_2,0),
\end{equation}
and finally
\begin{align}
G^{(m)(1)}_{x,x'}(\tau,0) &= G^{(m)(0)}_{x,x'}(\tau,0) + \left(- \frac{1}{\hslash} \right)^2 
\int_0^{\beta \hslash} d\tau_1 \int_0^{\beta \hslash} d\tau_2 \int dx_1 \int dx_2 \, U(x_1,x_2) \, \delta(\tau_1 - \tau_2) \notag \\
& \quad \biggl\{ (-\hslash) G^{(m)(0)}_{x,x_1}(\tau, \tau_1) (-\hslash) \varepsilon G^{(m)(0)}_{x_2,x_2}(\tau_2, \tau_2 + \delta) (-\hslash) G^{(m)(0)}_{x_1,x'}(\tau_1,0) \notag \\
& \quad + \varepsilon (-\hslash) G^{(m)(0)}_{x,x_1}(\tau, \tau_1) (-\hslash) \varepsilon G^{(m)(0)}_{x_1,x_2}(\tau_1, \tau_2 + \delta) (-\hslash) G^{(m)(0)}_{x_2,x'}(\tau_2, 0) \biggr\} \notag \\
&= G^{(m)(0)}_{x,x'}(\tau,0) - \hslash \int_0^{\beta \hslash} d\tau_1 \int_0^{\beta \hslash} d\tau_2 \int dx_1 \int dx_2 \, U(x_1,x_2) \, \delta(\tau_1 - \tau_2) \notag \\
& \quad \biggl\{ \varepsilon \, G^{(m)(0)}_{x,x_1}(\tau, \tau_1) \, G^{(m)(0)}_{x_2,x_2}(\tau_2, \tau_2 + \delta) \, G^{(m)(0)}_{x_1,x'}(\tau_1, 0) \notag \\
& \quad + G^{(m)(0)}_{x,x_1}(\tau, \tau_1) \, G^{(m)(0)}_{x_1,x_2}(\tau_1, \tau_2 + \delta) \, G^{(m)(0)}_{x_2,x'}(\tau_2, 0) \biggr\}.
\label{eq: sviluppoperturbativoalprimoordinedellaGreentermica}
\end{align}
Let us examine what $\eqref{eq: sviluppoperturbativoalprimoordinedellaGreentermica}$ tells us about the structure of the \(n\)-th order term. At order \(n\), the diagram contains \(2n+1\) free propagators \(G^{(m)(0)}\). This follows from the fact that each interaction line contributes four half-particle lines, which pair up to form two free propagators. The overall numerical prefactor can be identified as follows:
\begin{itemize}
\item The factor $-\frac{1}{\hslash}$ in front of the entire series is the conventional normalization.
\item Each interaction vertex contributes a factor $-\frac{1}{\hslash}$, so at order \(n\) we get $\left(-\frac{1}{\hslash}\right)^n$.
\item Every time a free propagator \(G^{(m)(0)}\) appears, it contributes a factor \(-\hslash\). Since there are \(2n+1\) propagators, this gives \((- \hslash)^{2n+1}\).
\item Finally, we include a factor \(\varepsilon\) for each closed loop, where \(\varepsilon = -1\) for fermions and \(\varepsilon = +1\) for bosons.
\end{itemize}
Putting everything together, the total numerical factor at order \(n\) becomes
\begin{equation}
\left( - \dfrac{1}{\hslash} \right) \left( - \dfrac{1}{\hslash} \right)^n (-\hslash)^{2n+1} \varepsilon^{\text{number of closed loops}} = \left( - \hslash \right)^n \varepsilon^{\text{number of closed loops}}.
\end{equation}
\section{Self-energy}
\begin{figure}[H]
\centering
\includegraphics[scale=0.8]{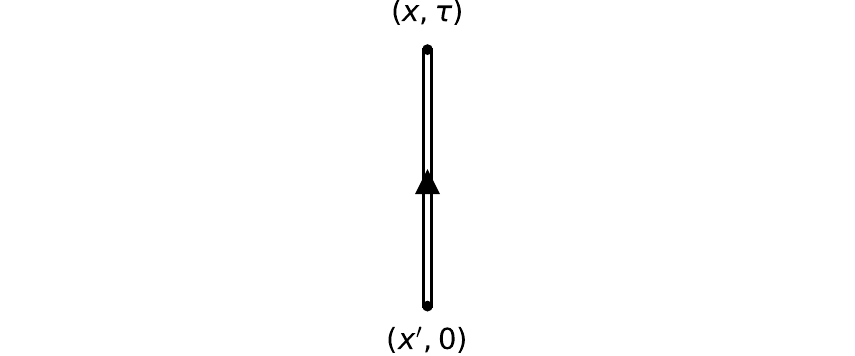}
\caption{Feynman diagrammatic representation of the full (interacting) many-body Green's function of field operators, illustrating the propagation of a non-interacting particle between two spacetime points.}
\label{fig: propagatore}
\end{figure}

\begin{figure}[H]
\centering
\includegraphics[scale=1]{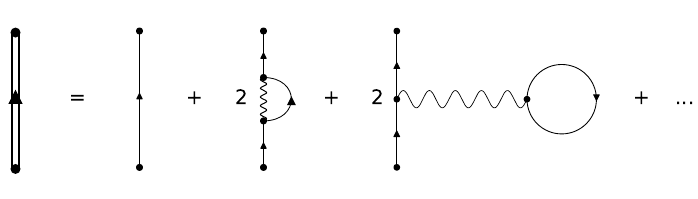}
\caption{Feynman diagrammatic representation of the perturbative expansion of the many-body Green's function of field operators; see equation $\eqref{eq: sviluppoperturbativodifeynmandiunafunzionediGreentermica}$.}
\label{fig: sviluppoperturbativoFeynmanindiagrammatica}
\end{figure}

\begin{figure}[H]
\centering
\includegraphics[scale=0.47]{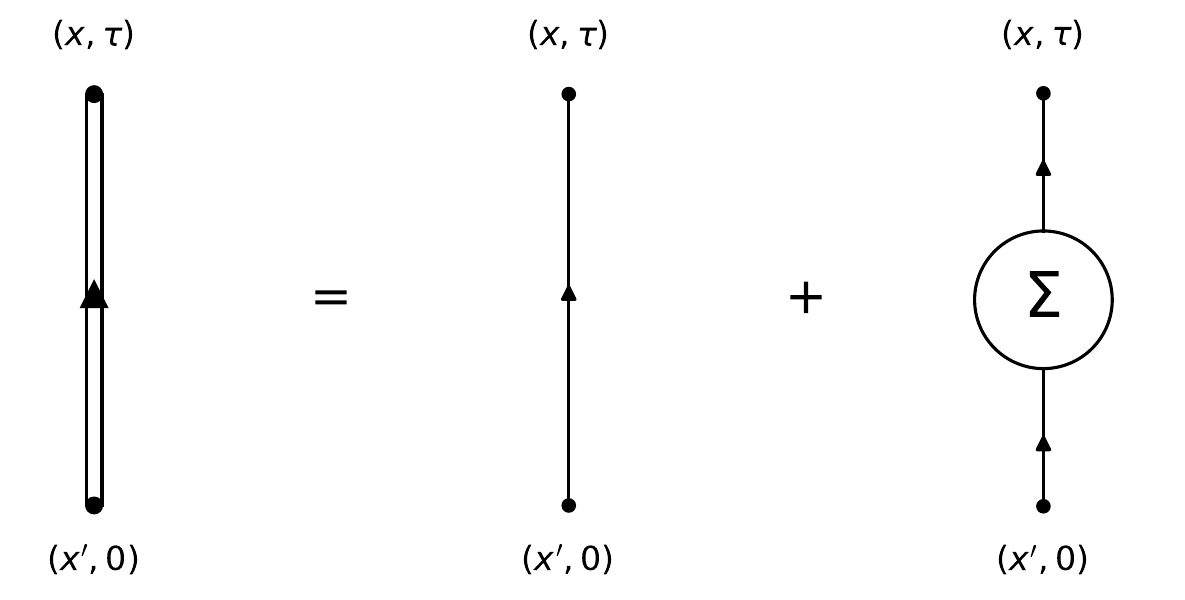}
\caption{Feynman diagrammatic representation of the many-body Green's function in terms of the self-energy.}
\label{fig: sviluppoperturbativoFeynmanindiagrammaticaconself-energy}
\end{figure}

\begin{figure}[H]
\centering
\includegraphics[scale=0.47]{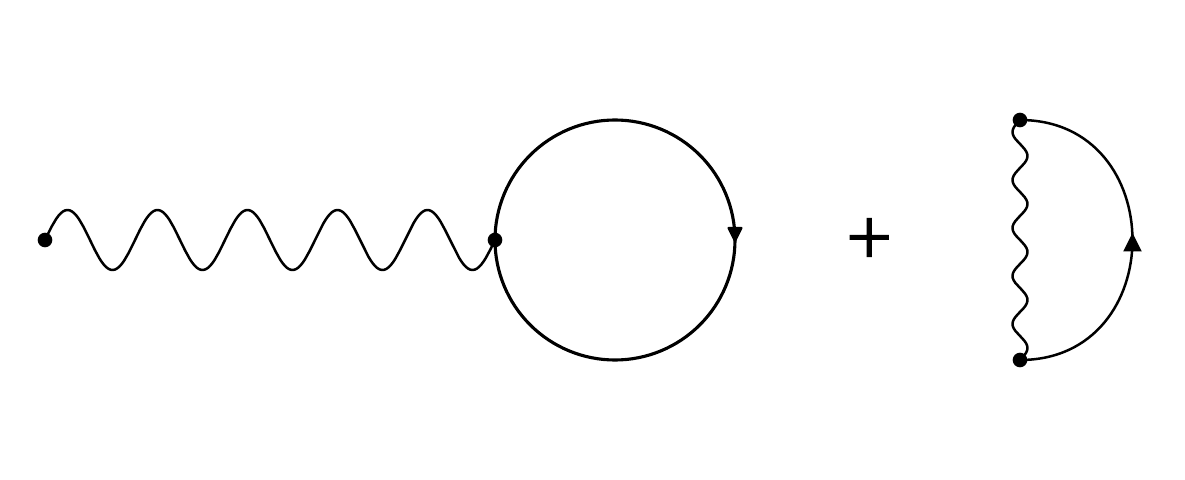}
\caption{Feynman diagrammatic representation of the first-order self-energy; see equation $\eqref{eq: self-energyprimoordineanalitico}$.}
\label{fig: self-energyprimoordinediagrammatica}
\end{figure}

\begin{figure}[H]
\centering
\includegraphics[scale=0.3]{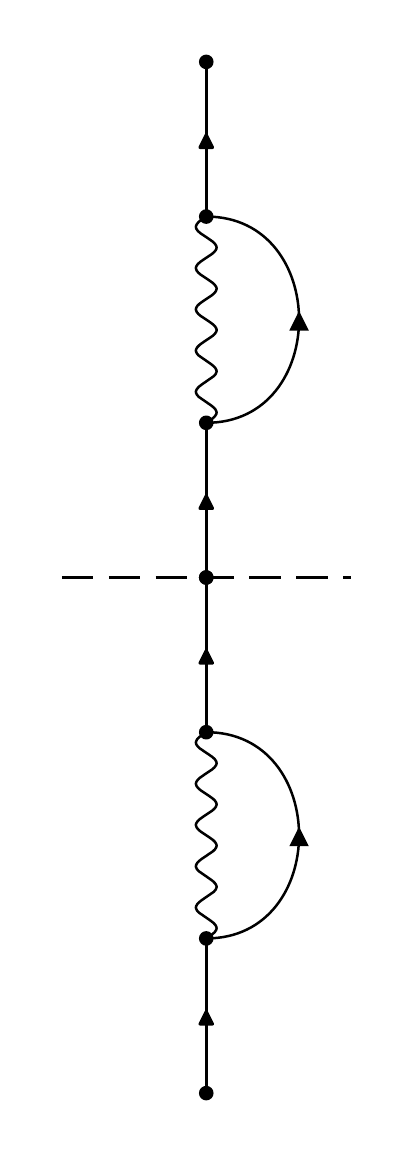}
\caption{Feynman diagrammatic representation of a second-order reducible connected diagram.}
\label{fig: diagrammaconnessosecondoordineriducibile}
\end{figure}

\begin{figure}[H]
\centering
\includegraphics[scale=0.5]{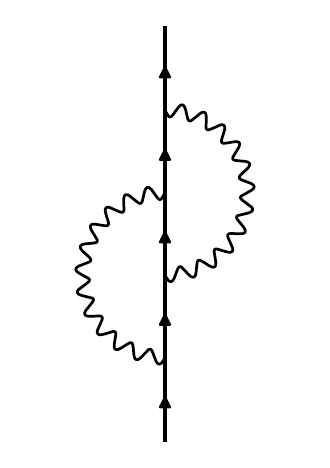}
\caption{Feynman diagrammatic representation of a second-order irreducible connected diagram.}
\label{fig: diagrammaconnessosecondoordineirriducibile}
\end{figure}

\begin{figure}[H]
\centering
\includegraphics[scale=0.52]{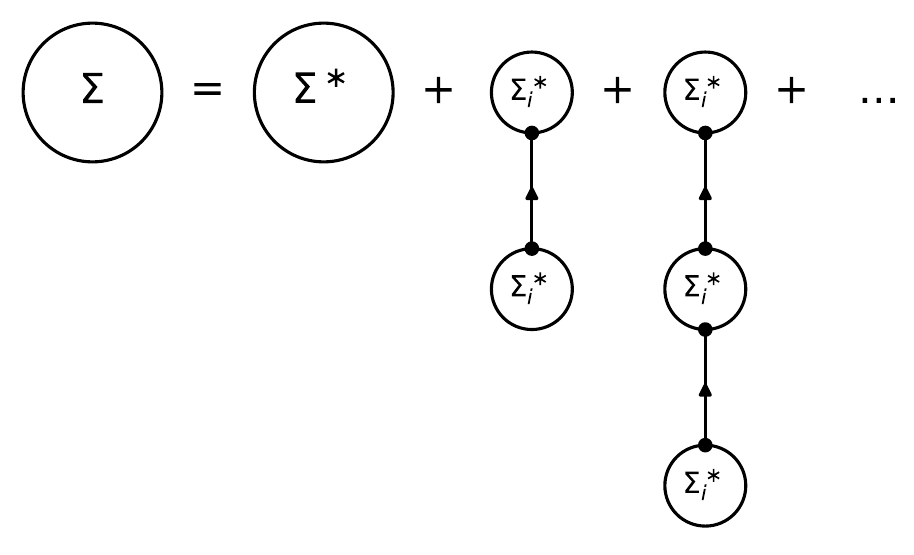}
\caption{Feynman diagrammatic representation of the decomposition of the self-energy $\Sigma$ into the improper self-energy $\Sigma^*$ and the reducible contributions built from improper self-energy insertions $\Sigma_i^*$.}
\label{fig: scomposizioneself-energy}
\end{figure}

\begin{figure}[H]
\centering
\includegraphics[scale=0.37]{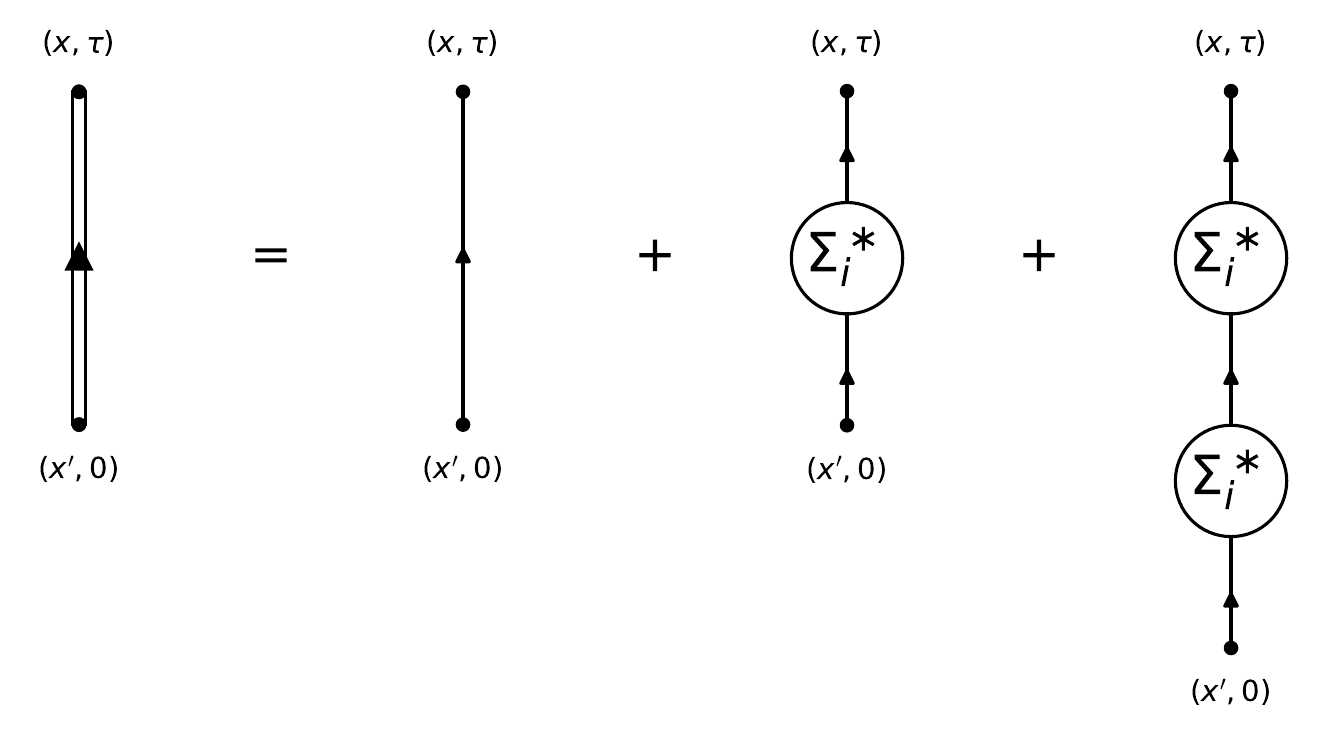}
\caption{Feynman diagrammatic representation of the decomposition of the full many-body Green's function into a sum of reducible self-energy contributions.}
\label{fig: propagatoreinfunzionedellaself-energypropria}
\end{figure}

\begin{figure}[H]
\centering
\includegraphics[scale=0.47]{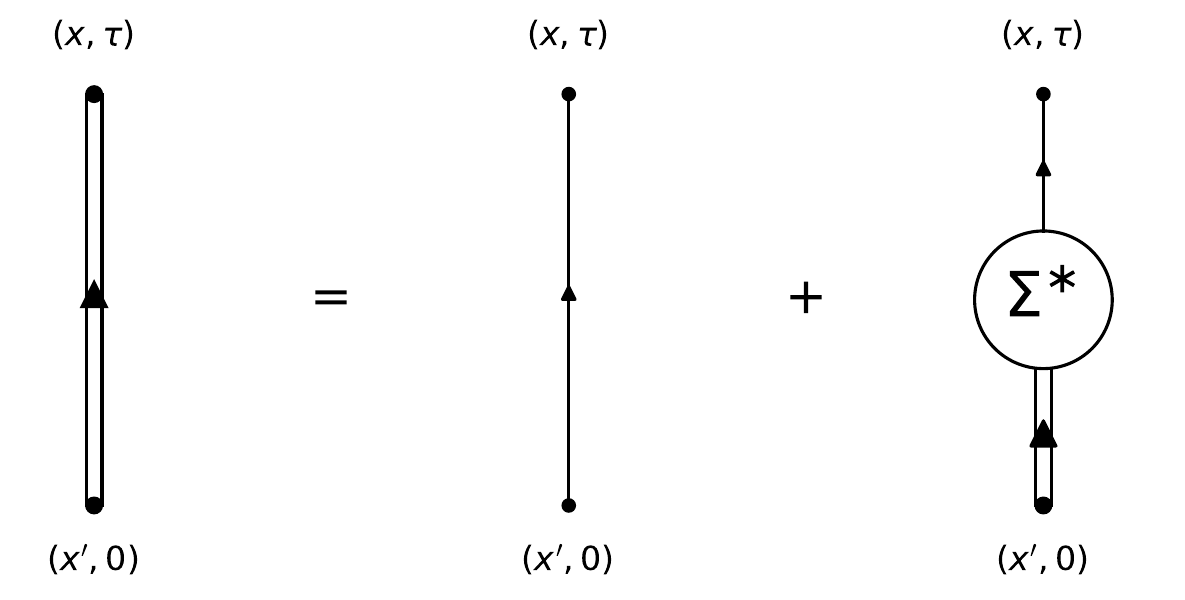}
\caption{Feynman diagrammatic representation of Dyson equation of many-body Green's function; see equation $\eqref{eq: equazioneDysonformaintegrale}$.}
\label{fig: equazioneDysonformadiagrammatica}
\end{figure}

The Green's function $G^{(m)}_{x,x'}(\tau,0)$ is depicted in Figure $\eqref{fig: propagatore}$, while its perturbative expansion is shown in Figure $\eqref{fig: sviluppoperturbativoFeynmanindiagrammatica}$. As we include higher-order terms, the resulting diagrams exhibit a common structure: an "outgoing free propagator", some "intermediate interaction structure", followed by an "incoming free propagator". This organization allows Feynman's perturbative expansion to be recast in the form shown in Figure $\eqref{fig: sviluppoperturbativoFeynmanindiagrammaticaconself-energy}$, where $\Sigma$ denotes the self-energy, a function that depends on the interaction vertices $x_1$, $x_2$ and the times $\tau_1$, $\tau_2$. At first order, the self-energy can be obtained by cutting the incoming and outgoing propagator lines of a diagram; its diagrammatic representation is given in Figure $\eqref{fig: self-energyprimoordinediagrammatica}$. Figure $\eqref{fig: sviluppoperturbativoFeynmanindiagrammaticaconself-energy}$ corresponds to an equation that can be recast in integral form, namely
\begin{equation}
G^{(m)}_{x,x'}(\tau,0) = G^{(m)(0)}_{x,x'}(\tau,0) + \int_0^{\beta \hslash} d\tau_1 \int_0^{\beta \hslash} d\tau_2 \int dx_1 \int dx_2 G^{(m)(0)}_{x,x_1}(\tau,\tau_1) \Sigma_{x_1,x_2}(\tau_1,\tau_2) G^{(m)(0)}_{x_2,x'}(\tau_2,0),
\end{equation}
that is, we obtained via an integral equation the exact expression of $G^{(m)}_{x,x'}(\tau,0)$ in terms of the free Green's functions $G^{(m)(0)}$. In practice, integrals are difficult to compute and one typically proceeds numerically or by making a careful choice of higher order diagrams to include. Comparing the equation above with the first-order perturbative expansion $\eqref{eq: sviluppoperturbativoalprimoordinedellaGreentermica}$, we derive the analytical expression of the first-order self-energy in Figure $\eqref{fig: self-energyprimoordinediagrammatica}$, that is
\begin{align}
\Sigma^{(1)}_{x_1,x_2}(\tau_1,\tau_2) &= - \varepsilon \hslash \delta(x_1-x_2) \delta(\tau_1 - \tau_2) \int dx'_2 \int_0^{\beta \hslash} d\tau'_2 U(x_1,x'_2) \delta(\tau_1-\tau'_2) G^{(m)(0)}_{x'_2,x'_2}(\tau'_2,\tau'_2+\delta) \ + \notag \\
&- \hslash U(x_1,x_2) \delta(\tau_1-\tau_2) G^{(m)(0)}_{x_1,x_2}(\tau_1,\tau_2+\delta).
\label{eq: self-energyprimoordineanalitico}
\end{align}
We define a self-energy insertion (or self-energy contribution) as the part of a Feynman diagram that results from removing one incoming and one outgoing propagator line. The self-energy $\Sigma$ is then defined as the sum over all such self-energy insertions. A connected diagram is said to be reducible if its associated self-energy insertion is reducible, i.e., if it can be decomposed into diagrams of lower order by cutting any internal free propagator line $G^{(m)(0)}$. Otherwise, the diagram is called irreducible. Note that all first-order diagrams are irreducible. As illustrative second-order examples, in Figure $\eqref{fig: diagrammaconnessosecondoordineriducibile}$ we show a reducible connected diagram, while in Figure $\eqref{fig: diagrammaconnessosecondoordineirriducibile}$ we present an irreducible connected diagram. In general, for all orders higher than one, both reducible and irreducible connected diagrams appear. Irreducible self-energy insertions are also referred to as proper, while reducible ones are called improper. Let $\Sigma^*$ denote the sum of all improper self-energy insertions, as previously introduced in the context of the Anderson model. The full self-energy $\Sigma$ can be expressed as the sum of $\Sigma^*$ and all diagrams that contribute to improper (reducible) self-energy terms; see Figure $\eqref{fig: scomposizioneself-energy}$. We can now express the diagram in Figure $\eqref{fig: sviluppoperturbativoFeynmanindiagrammaticaconself-energy}$ as a function of the proper self-energy insertions. In Figure $\eqref{fig: propagatoreinfunzionedellaself-energypropria}$, we isolate a single outgoing free propagator $G^{(m)(0)}$ and one proper self-energy insertion at each order, and sum over the remaining (excluded) diagrams. The resulting expression corresponds to the full propagator $G^{(m)}_{x_2,x'}(\tau_2,0)$. What we obtain is the Dyson equation in diagrammatic form, shown in Figure $\eqref{fig: equazioneDysonformadiagrammatica}$. In summary, the many-body Green function can be written either in terms of the self-energy $\Sigma$ and two free propagators $G^{(m)(0)}$, or alternatively in terms of the self-energy $\Sigma^*$ and the full (interacting) Green function $G^{(m)}$. \newline
Since exact evaluations are not feasible, one typically resorts to perturbation theory. It is important to note that these two expressions lead to different perturbative expansions. In particular, the Dyson equation involving $\Sigma^*$ is generally preferred: indeed, if we compute $\Sigma$ to first order, as in Figure $\eqref{fig: sviluppoperturbativoFeynmanindiagrammaticaconself-energy}$, the Green's function is expanded as $G^{(m)(0)}$ plus first-order diagrams. Conversely, if we compute $\Sigma^*$ to first order in the Dyson equation of Figure $\eqref{fig: equazioneDysonformadiagrammatica}$, the Green's function corresponds to a resummation that already includes infinitely many reducible diagrams. We finally rewrite the Dyson equation in its integral form, i.e.,
\begin{equation}
G^{(m)}_{x,x'}(\tau,0) = G^{(m)(0)}_{x,x'}(\tau,0) + \int_0^{\beta \hslash} d\tau_1 \int_0^{\beta \hslash} d\tau_2 \int dx_1 \int dx_2 G^{(m)(0)}_{x,x_1}(\tau,\tau_1) \Sigma_{x_1,x_2}^*(\tau_1,\tau_2)G^{(m)}_{x_2,x'}(\tau_2,0).
\label{eq: equazioneDysonformaintegrale}
\end{equation}
\section{Dyson algebraic equation}
The Dyson equation of field operators in integral form, i.e., equation $\eqref{eq: equazioneDysonformaintegrale}$, is greatly simplified for systems at thermodynamic equilibrium, i.e., invariant by time translation, indeed
\begin{equation}
G^{(m)}_{x,x'}(\tau) = G^{(m)(0)}_{x,x'}(\tau) + \int_0^{\beta \hslash} d\tau_1 \int_0^{\beta \hslash} d\tau_2 \int dx_1 \int dx_2 G^{(m)(0)}_{x,x_1}(\tau-\tau_1) \Sigma_{x_1,x_2}^*(\tau_1-\tau_2)G^{(m)}_{x_2,x'}(\tau_2). 
\label{eq: equazioneDysonformaintegraleequilibriotermodinamico}
\end{equation}
Recalling $\eqref{eq: funzioneGreenMatsubaraserieFourier1}$, $\eqref{eq: funzioneGreenMatsubaraserieFourier2}$, we multiply both members of the $\eqref{eq: equazioneDysonformaintegraleequilibriotermodinamico}$ by $e^{i \omega_n \tau}$ and integrate with respect to time in the range $[ 0 , \beta \hslash [$, i.e.,
\begin{align}
G^{(m)}_{x,x'}(i \omega_n) &= G^{(m)(0)}_{x,x'}(i \omega_n) + \int_0^{\beta \hslash} d\tau \int_0^{\beta \hslash} d\tau_1 \int_0^{\beta \hslash} d\tau_2 \int dx_1 \int dx_2 \ e^{i \omega_n \tau} \left( \dfrac{1}{\beta \hslash} \right)^3 \notag \\
& \ \ \ \sum_{n_1} \sum_{n_2} \sum_{n_3} G^{(m)(0)}_{x,x_1}(i \omega_{n_1}) \Sigma_{x_1,x_2}^*(i \omega_{n_2})G^{(m)}_{x_2,x'}(i \omega_{n_3}) e^{- i \omega_{n_1} (\tau-\tau_1)} e^{- i \omega_{n_2} (\tau_1 - \tau_2)} e^{- i \omega_{n_3} \tau_2}.
\end{align}
With respect to the measure $d\tau$, the only nonzero contribution arises from the term $e^{(i\omega_n - i\omega_{n_1})\tau}$. Let us analyze the corresponding integral. Depending on whether the propagators in $G$ are fermionic or bosonic, the Matsubara frequencies $i\omega_{n_1}$, $i\omega_{n_2}$, and $i\omega_{n_3}$ are odd or even, respectively. As for the self-energy, in the case $n = 1$, it consists of two diagrams, each given by the product of an interaction line (bosonic) and a particle propagator (either bosonic or fermionic). Therefore, the frequency structure of the self-energy ultimately depends on the nature of the internal propagator. As a consequence, the difference $\omega_n - \omega_{n_1} = \omega_m$ is always a bosonic frequency, i.e., it is always even. Hence,
\begin{align}
\int_0^{\beta \hslash} e^{(i \omega_n - i \omega_{n_1}) \tau} \, d\tau &= \int_0^{\beta \hslash} e^{i \omega_m \tau} \, d\tau = \notag \\
&= \int_0^{\beta \hslash} e^{\frac{2 \pi i m}{\beta \hslash} \tau} \, d\tau = \notag \\
&= 
\begin{cases}
\beta \hslash, & m = 0 \\[8pt]
\dfrac{e^{i \omega_m \beta \hslash} - 1}{i \omega_m}, & m \neq 0
\end{cases} .
\end{align}
Since $i\omega_m$ is a bosonic frequency, it is an even multiple of $\frac{\pi}{\beta\hbar}$, then, $e^{i \omega_ \beta \hslash} = 1$. As a result, the integral over $\tau$ yields a nonzero contribution only when $m = 0$. Ultimately, we obtain
\begin{equation}
\int_0^{\beta \hslash} e^{(i \omega_n - i \omega_{n_1}) \tau} = \delta_{n,n_1} \beta \hslash,
\end{equation}
and similarly
\begin{equation}
\int_0^{\beta \hslash} e^{(i \omega_{n_1} - i \omega_{n_2}) \tau} = \delta_{n_1,n_2} \beta \hslash,
\end{equation}
\begin{equation}
\int_0^{\beta \hslash} e^{(i \omega_{n_2} - i \omega_{n_3}) \tau} = \delta_{n_2,n_3} \beta \hslash.
\end{equation}
We simplify the prefactor $(\beta \hslash)^{-3}$ in the Dyson equation, and by summing over the indices $n_1$, $n_2$, and $n_3$, the Kronecker deltas enforce the conditions $n = n_1 = n_2 = n_3$. Consequently, the frequencies also coincide: $\omega_n = \omega_{n_1} = \omega_{n_2} = \omega_{n_3}$. This reflects the conservation of energy at each vertex of the corresponding Feynman diagram. The Dyson equation at thermodynamic equilibrium in frequency space can thus be rewritten as
\begin{equation}
G^{(m)}_{x,x'}(i \omega_n) = G^{(m)(0)}_{x,x'}(i \omega_n) + \int dx_1 \int dx_2  G^{(m)(0)}_{x,x_1}(i \omega_n) \Sigma_{x_1,x_2}^*(i \omega_n) G^{(m)}_{x_2,x'}(i \omega_n).
\label{eq: equazioneDysonformaintegraleequilibriotermodinamicospaziodellefrequenze}
\end{equation}
We now consider whether the Dyson integral equation can be further simplified in real space. In the thermodynamic limit, boundary conditions become irrelevant, and we adopt periodic boundary conditions to ensure invariance under spatial translations. Due to this translational invariance, the Green's functions and the self-energy depend only on the spatial difference $\mathbf{r}_i - \mathbf{r}_j$. We therefore decompose the index $x$ into its spatial part $\mathbf{r}$ and its spin component $s$, and apply a spatial Fourier transform to equation $\eqref{eq: equazioneDysonformaintegraleequilibriotermodinamicospaziodellefrequenze}$. Given the standard forms of the Fourier transform in real and frequency space, we obtain:
\begin{equation}
G_{\textbf{r}-\textbf{r}',s,s'}(i \omega_n) = \int \dfrac{d^3\textbf{k}}{(2 \pi)^3} e^{- i \textbf{k} \cdot (\textbf{r}-\textbf{r}')} G_{\textbf{k},s,s'}(i \omega_n),
\end{equation}
\begin{equation}
G_{\textbf{k},s,s'}(i \omega_n) = \int d^3(\textbf{r}-\textbf{r}') e^{i \textbf{k} \cdot (\textbf{r}-\textbf{r}')} G_{\textbf{r}-\textbf{r}',s,s'}(i \omega_n),
\end{equation}
we multiply by $e^{i \textbf{k} \cdot (\textbf{r}-\textbf{r}')}$ both members of $\eqref{eq: equazioneDysonformaintegraleequilibriotermodinamicospaziodellefrequenze}$ and integrate with respect to the measure $d^3(\textbf{r}-\textbf{r}')$ as follows
\begin{align}
& G^{(m)}_{\textbf{k},s,s'}(i \omega_n) = G^{(m)(0)}_{\textbf{k},s,s'}(i \omega_n) + \sum_{s_1,s_2} \int d^3(\textbf{r}-\textbf{r}') e^{i \textbf{k} \cdot (\textbf{r}-\textbf{r}')} \int d^3\textbf{r}_1 \int d^3\textbf{r}_2 \notag \\
& \left( \int \dfrac{d^3\textbf{k}_1}{(2 \pi)^3} e^{- \textbf{k}_1 \cdot (\textbf{r}-\textbf{r}_1)} G^{(m)(0)}_{\textbf{k}_1,s,s_1} \right) \left( \int \dfrac{d^3\textbf{k}_2}{(2 \pi)^3} e^{- \textbf{k}_2 \cdot (\textbf{r}_1-\textbf{r}_2)} \Sigma^*_{\mathbf{k},s_1,s_2}(i\omega_n) \right) \left( \int \dfrac{d^3\textbf{k}_3}{(2 \pi)^3} e^{- \textbf{k}_3 \cdot (\textbf{r}_2-\textbf{r}')} G^{(m)}_{\textbf{k}_3,s_2,s'} \right) = \notag \\
&= G^{(m)(0)}_{\textbf{k},s,s'}(i \omega_n) + \sum_{s_1,s_2} \int d^3(\textbf{r}-\textbf{r}') \int d^3\textbf{r}_1 \int d^3\textbf{r}_2 \int \dfrac{d^3\textbf{k}_1}{(2 \pi)^3} \int \dfrac{d^3\textbf{k}_2}{(2 \pi)^3} \int \dfrac{d^3\textbf{k}_3}{(2 \pi)^3} \notag \\
& G^{(m)(0)}_{\textbf{k}_1,s,s_1} \Sigma^*_{\mathbf{k},s_1,s_2}(i\omega_n) G^{(m)}_{\textbf{k}_3,s_2,s'} e^{i \textbf{k} \cdot \textbf{r}} e^{- i \textbf{k} \cdot \textbf{r}'} e^{- i \textbf{k}_1 \cdot \textbf{r}} e^{i \textbf{k}_1 \cdot \textbf{r}_1} e^{- i \textbf{k}_2 \cdot \textbf{r}_1} e^{i \textbf{k}_2 \cdot \textbf{r}_2} e^{- i \textbf{k}_3 \cdot \textbf{r}_2} e^{i \textbf{k}_3 \cdot \textbf{r}'}.
\end{align}
If we integrate with respect to the measure $d\mathbf{r}_2$, we observe that
\begin{equation}
\int d^3\textbf{r}_2 e^{i (\textbf{k}_2 - \textbf{k}_3) \cdot \textbf{r}_2} = (2 \pi)^3 \delta(\textbf{k}_2-\textbf{k}_3),
\end{equation}
which allows us to perform
\begin{equation}
\int \dfrac{d^3\textbf{k}_3}{(2 \pi)^3} (2 \pi)^3 \delta(\textbf{k}_2-\textbf{k}_3) G^{(m)}_{\textbf{k}_3,s_2,s'} e^{i \textbf{k}_3 \cdot \textbf{r}'} = G^{(m)}_{\textbf{k}_2,s_2,s'} e^{i \textbf{k}_2 \cdot \textbf{r}'},
\end{equation}
and the Dyson equation is written as
\begin{align}
G^{(m)}_{\textbf{k},s-s'}(i \omega_n) &= G^{(m)(0)}_{\textbf{k},s,s'}(i \omega_n) + \sum_{s_1,s_2} \int d^3(\textbf{r}-\textbf{r}') \int d^3\textbf{r}_1 \int \dfrac{d^3\textbf{k}_1}{(2 \pi)^3} \int \dfrac{d^3\textbf{k}_2}{(2 \pi)^3} \notag \\
& \ \ \ \ \ \ G^{(m)(0)}_{\textbf{k}_1,s,s_1} \Sigma^*_{\mathbf{k},s_1,s_2}(i\omega_n) G^{(m)}_{\textbf{k}_2,s_2,s'} e^{i \textbf{k} \cdot \textbf{r}} e^{- i \textbf{k} \cdot \textbf{r}'} e^{- i \textbf{k}_1 \cdot \textbf{r}} e^{i \textbf{k}_1 \cdot \textbf{r}_1} e^{- i \textbf{k}_2 \cdot \textbf{r}_1} e^{i \textbf{k}_2 \cdot \textbf{r}'}.
\end{align}
Similarly
\begin{equation}
\int d^3\textbf{r}_1 e^{i (\textbf{k}_1 - \textbf{k}_2) \cdot \textbf{r}_1} = (2 \pi)^3 \delta(\textbf{k}_1-\textbf{k}_2),
\end{equation}
\begin{equation}
\int \dfrac{d^3\textbf{k}_2}{(2 \pi)^3} (2 \pi)^3 \delta(\textbf{k}_1-\textbf{k}_2) e^{i \textbf{k}_2 \cdot \textbf{r}'} \Sigma^*_{\textbf{k}_2,s_1,s_2} G^{(m)}_{\textbf{k}_2,s_2,s'} = e^{i \textbf{k}_1 \cdot \textbf{r}'} \Sigma^*_{\mathbf{k},s_1,s_2}(i\omega_n) G^{(m)}_{\textbf{k}_1,s_2,s'},
\end{equation}
\begin{equation}
G^{(m)}_{\textbf{k},s,s'}(i \omega_n) = G^{(m)(0)}_{\textbf{k},s,s'}(i \omega_n) + \sum_{s_1,s_2} \int d^3(\textbf{r}-\textbf{r}') \int \dfrac{d^3\textbf{k}_1}{(2 \pi)^3} G^{(m)(0)}_{\textbf{k}_1,s-s_1} \Sigma^*_{\mathbf{k},s_1,s_2}(i\omega_n) G^{(m)}_{\textbf{k}_1,s_2,s'} e^{i \textbf{k} \cdot \textbf{r}} e^{- i \textbf{k} \cdot \textbf{r}'} e^{- i \textbf{k}_1 \cdot \textbf{r}} e^{i \textbf{k}_1 \cdot \textbf{r}'}.
\end{equation}
Finally, we use
\begin{equation}
\int d^3(\textbf{r}-\textbf{r}') e^{i \textbf{k} \cdot (\textbf{r}-\textbf{r}')} e^{- i \textbf{k}_1 \cdot (\textbf{r}-\textbf{r}')} = (2 \pi)^3 \delta(\textbf{k}-\textbf{k}_1),
\end{equation}
\begin{equation}
\int \dfrac{d^3\textbf{k}_1}{(2 \pi)^3} (2 \pi)^3 \delta(\textbf{k}-\textbf{k}_1) G^{(m)(0)}_{\textbf{k}_1,s,s_1} \Sigma^*_{\mathbf{k},s_1,s_2}(i\omega_n) G^{(m)}_{\textbf{k}_1,s_2,s'} = G^{(m)(0)}_{\textbf{k},s,s_1} \Sigma^*_{\textbf{k},s_1,s_2} G^{(m)}_{\textbf{k},s_2,s'},
\end{equation}
and we write the Dyson equation as
\begin{equation}
G^{(m)}_{\mathbf{k},s,s'}(i\omega_n) = G^{(m)(0)}_{\mathbf{k},s,s'}(i\omega_n) + \sum_{s_1, s_2} G^{(m)(0)}_{\mathbf{k},s,s_1}(i\omega_n) \Sigma^*_{\mathbf{k},s_1,s_2}(i\omega_n) G^{(m)}_{\mathbf{k},s_2,s'}(i\omega_n),
\end{equation}
which expresses the Dyson equation in frequency and momentum space, with full spin structure. This form reflects the conservation of momentum at each vertex of the diagram, as all propagators and self-energies share the same momentum label $\mathbf{k}$. \newline
In many physical systems, in particular, those that are spin-rotation invariant or in the absence of spin-dependent interactions (such as spin-orbit coupling or external magnetic fields), the Green's function and self-energy become diagonal in the spin basis. That is,
\begin{equation}
G^{(m)(0)}_{\mathbf{k},s,s_1}(i\omega_n) = \delta_{s,s_1} G^{(m)(0)}_{\mathbf{k},\sigma}(i\omega_n),
\end{equation}
\begin{equation}
\Sigma^*_{\mathbf{k},s_1,s_2}(i\omega_n) = \delta_{s_1,s_2} \Sigma^*_{\mathbf{k},\sigma}(i\omega_n),
\end{equation}
\begin{equation}
G^{(m)}_{\mathbf{k},s_2,s'}(i\omega_n) = \delta_{s_2,s'} G^{(m)}_{\mathbf{k},\sigma}(i\omega_n).
\end{equation}
Consequently, under this further assumption, the Dyson equation simplifies to a scalar form
\begin{equation}
G^{(m)}_{\mathbf{k},\sigma}(i\omega_n) = G^{(m)(0)}_{\mathbf{k},\sigma}(i\omega_n) + G^{(m)(0)}_{\mathbf{k},\sigma}(i\omega_n) \, \Sigma^*_{\mathbf{k},\sigma}(i\omega_n) \, G^{(m)}_{\mathbf{k},\sigma}(i\omega_n),
\label{eq: equazioneDysonalgebrica}
\end{equation}
which is an algebraic equation for the Green's function of given spin and momentum. Consider a system of non-interacting particles and let us compute $G^{(m)(0)}_{\textbf{k},\sigma}(i \omega_n)$ starting from
\begin{equation}
G^{(m)}_{\textbf{k},\sigma}(i \omega_n) = \dfrac{G^{(m)(0)}_{\textbf{k},\sigma}(i \omega_n)}{1 - G^{(m)(0)}_{\textbf{k},\sigma}(i \omega_n)\Sigma_{\textbf{k},\sigma}^*(i \omega_n)}.
\end{equation}
Given
\begin{align}
G^{(m)(0)}_{\textbf{k},\sigma}(\tau) &= - \dfrac{1}{\hslash} \left\langle C^{(0)}_{\textbf{k},\sigma}(\tau)C^\dagger_{\textbf{k},\sigma} \right\rangle_0 = \notag \\
&= - \dfrac{1}{\hslash} \left[ e^{- \frac{\mathcal{E}_{\textbf{k},\sigma} \tau}{\hslash}} \left( 1 + \varepsilon n_{\varepsilon}(\mathcal{E}_{\textbf{k},\sigma}) \right) \Theta(\tau) + \varepsilon e^{- \frac{\mathcal{E}_{\textbf{k},\sigma} \tau}{\hslash}} n_{\varepsilon}(\mathcal{E}_{\textbf{k},\sigma}) \Theta(- \tau) \right] ,
\label{eq: propagatorenudoMatsubaraspaziomomentietempiimmaginaricomemedia}
\end{align}
where $n_{\varepsilon}(\mathcal{E}_{\textbf{k},\sigma})$ is given by $\eqref{eq: distribuzioneBoseFermiunificata}$, then
\begin{align}
G^{(m)(0)}_{\textbf{k},\sigma}(i \omega_n) &= \int_0^{\beta \hslash} d\tau e^{i \omega_n \tau} G^{(m)(0)}_{\textbf{k},\sigma}(\tau) = \notag \\
&= \dfrac{1}{\hslash \left[ i \omega_n - \frac{\mathcal{E}_{\textbf{k},\sigma}}{\hslash} \right]},
\end{align}
and the solution of the Dyson equation can then be written as follows
\begin{align}
G^{(m)}_{\textbf{k},\sigma}(i \omega_n) &= \dfrac{G^{(m)(0)}_{\textbf{k},\sigma}(i \omega_n)}{1 - G^{(m)(0)}_{\textbf{k},\sigma}(i \omega_n)\Sigma_{\textbf{k},\sigma}^*(i \omega_n)} = \notag \\
&= \dfrac{1}{\left[ G^{(m)(0)}_{\textbf{k},\sigma}(i \omega_n) \right]^{-1} - \Sigma_{\textbf{k},\sigma}^*(i \omega_n)} = \notag \\
&= \dfrac{1}{\hslash \left[ i \omega_n - \frac{\mathcal{E}_{\textbf{k},\sigma}}{\hslash} \right]- \Sigma_{\textbf{k},\sigma}^*(i \omega_n)} \equiv \notag \\
&\equiv \dfrac{1}{\hslash \left[ i \omega_n - \frac{\mathcal{E}_{\textbf{k},\sigma}}{\hslash} - \frac{\Sigma_{\textbf{k},\sigma}^*(i \omega_n)}{\hslash} \right]}.
\label{eq: funzionediGreensistemainvariantetraslspazialeeallequilibriotermod2}
\end{align}
As anticipated, we have recovered an equation of the form $\eqref{eq: funzionediGreensistemainvariantetraslspazialeeallequilibriotermod}$: for systems exhibiting invariance under both spatial and temporal translations, the many-body Green's function reduces to a particularly simple functional form: it depends only on the Matsubara frequency \( i\omega_n \), the crystal momentum \( \textbf{k} \), reflecting the underlying translational symmetries, and the improper self-energy $\Sigma^*$. \newline
In the absence of interactions, that is, when the self-energy vanishes identically, i.e., \( \Sigma_{\textbf{k},\sigma}^*(i \omega_n) = 0 \), the full (interacting) Green's function \( G^{(m)}_{\textbf{k},\sigma}(i \omega_n) \) exactly reduces to the non-interacting one \( G^{(m)(0)}_{\textbf{k},\sigma}(i \omega_n) \), as expected. This confirms the formal consistency of the perturbative framework: interactions modify the system's dynamics exclusively through the self-energy, which effectively introduces both an energy shift and a possible damping of the spectral features.
\section{A Matsubara sum}
\begin{figure}[H]
\centering
\includegraphics[scale=0.7]{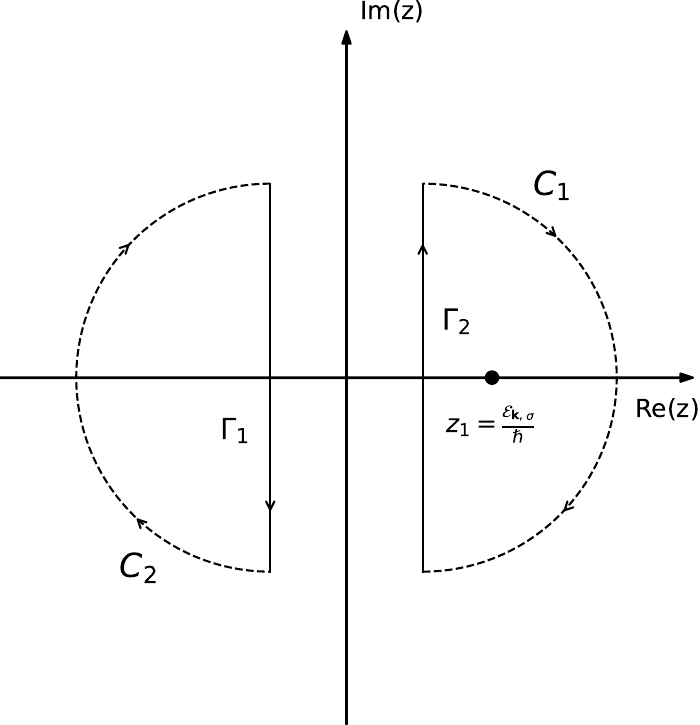}
\caption{Integration contour used for the evaluation of the Matsubara sum in equation $\eqref{eq: primarelazioneesattafrequenzeMatsubara2}$ via the residue theorem.}
\label{fig: graficointegraleresiduiprimasommamatsubara}
\end{figure}

From the definition of time Fourier transform we can derive an equation for Matsubara frequencies. Indeed, the propagator $\eqref{eq: propagatorenudoMatsubaraspaziomomentietempiimmaginaricomemedia}$ can be written in Fourier time transform as
\begin{align}
G^{(m)(0)}_{\textbf{k},\sigma}(\tau) &= \dfrac{1}{\beta \hslash} \sum_n e^{- i \omega_n} G^{(m)(0)}_{\textbf{k},\sigma}(i \omega_n) = \notag \\
&= \dfrac{1}{\beta \hslash} \sum_n \dfrac{e^{- i \omega_n}}{\hslash \left[ i \omega_n - \frac{\mathcal{E}_{\textbf{k},\sigma}}{\hslash} \right]} ,
\label{eq: propagatorenudoMatsubaraspaziomomentietempiimmaginaricometrasformata}
\end{align}
and from comparison between $\eqref{eq: propagatorenudoMatsubaraspaziomomentietempiimmaginaricomemedia}$ and $\eqref{eq: propagatorenudoMatsubaraspaziomomentietempiimmaginaricometrasformata}$ we have
\begin{equation}
\dfrac{1}{\beta} \sum_n \dfrac{e^{- i \omega_n}}{\hslash \left[ i \omega_n - \frac{\mathcal{E}_{\textbf{k},\sigma}}{\hslash} \right]} = 
\begin{cases}
- e^{- \frac{\mathcal{E}_{\textbf{k},\sigma} \tau}{\hslash}} \left( 1 + \varepsilon n_{\varepsilon}(\mathcal{E}_{\textbf{k},\sigma}) \right), \ \tau>0 \\
- \varepsilon e^{- \frac{\mathcal{E}_{\textbf{k}} \tau}{\hslash}} n_{\varepsilon}(\mathcal{E}_{\textbf{k},\sigma}), \ \ \ \ \ \ \ \ \ \ \ \ \tau < 0
\end{cases}.
\label{eq: primarelazioneesattafrequenzeMatsubara1}
\end{equation}
Due to the discontinuity of the Heaviside theta function, we need to examine the limit as $\tau \rightarrow 0^+$ and $\tau \rightarrow 0^-$, we would like to interchange the sum and the limit operators. However, this is not allowed. In fact, we know that the bosonic and fermionic frequencies satisfy $\omega_n \sim n$, so for infinitesimal $\tau$ we have
\begin{equation}
\dfrac{1}{\beta} \sum_n \dfrac{e^{- i \omega_n}}{\hslash \left[ i \omega_n - \frac{\mathcal{E}_{\textbf{k},\sigma}}{\hslash} \right]} \sim \sum_n \dfrac{1}{n},
\end{equation}
which is divergent. We must therefore introduce a factor $e^{- i \omega_n \delta}$, with $\delta \in (0, \beta \hslash)$ arbitrarily small, in order to ensure the convergence of the series. Consequently, the series is replaced by
\begin{equation}
\dfrac{1}{\beta} \sum_n \dfrac{e^{- i \omega_n \delta}}{\hslash \left[ i \omega_n - \frac{\mathcal{E}_{\textbf{k},\sigma}}{\hslash} \right]},
\label{eq: seriemodificatapropagatoreMatsubara}
\end{equation}
and from equation $\eqref{eq: primarelazioneesattafrequenzeMatsubara1}$ it follows
\begin{theorem}\label{thm: primasommaMatsubara}
The Matsubara frequencies satisfy
\begin{equation}
\dfrac{1}{\beta} \sum_n \dfrac{e^{- i \omega_n \delta}}{\hslash \left[ i \omega_n - \frac{\mathcal{E}_{\textbf{k},\sigma}}{\hslash} \right]} = 
\begin{cases}
- e^{- \frac{\mathcal{E}_{\textbf{k},\sigma} \delta}{\hslash}} \left( 1 + \varepsilon n_{\varepsilon}(\mathcal{E}_{\textbf{k},\sigma}) \right), \ \tau>0 \\
- \varepsilon e^{- \frac{\mathcal{E}_{\textbf{k},\sigma} \delta}{\hslash}} n_{\varepsilon}(\mathcal{E}_{\textbf{k},\sigma}), \ \ \ \ \ \ \ \ \ \tau < 0
\end{cases}.
\label{eq: primarelazioneesattafrequenzeMatsubara2}
\end{equation}
where $n_{\varepsilon}(\mathcal{E}_{\textbf{k},\sigma})$ is given by $\eqref{eq: distribuzioneBoseFermiunificata}$.
\begin{proof}
Here, we aim to compute the sum of the series $\eqref{eq: seriemodificatapropagatoreMatsubara}$ independently of the propagators and we use the residue theorem $\eqref{thm: teoremaresidui}$ (see Figure $\eqref{fig: graficointegraleresiduiprimasommamatsubara}$). We will consider the case of fermions, then $\omega_n = \frac{(2n+1) \pi}{\beta \hslash}$ and the Matsubara frequencies $\omega_n$ are the poles of the Fermi-Dirac function
\begin{equation}
n_{-1}(z) = \dfrac{1}{e^{\beta z} + 1}.
\end{equation}
Such poles $z = i \omega_n$ lie on the imaginary axis and they are simple poles, indeed,
\begin{align}
\operatorname{Res}_{i \omega_n}(n_{-1}) &= \lim_{z \rightarrow i \omega_n} (z - i \omega_n) n_{-1}(z) = \notag \\
&= \lim_{z \rightarrow i \omega_n} \dfrac{z - i \omega_n}{e^{\beta z} + 1} = \notag \\
&= \lim_{z \rightarrow i \omega_n} \dfrac{0 + 1 (z - i \omega_n)}{0 + \beta e^{\beta z} (z - i \omega_n)} = \notag \\
&= \lim_{z \rightarrow i \omega_n} \dfrac{1}{\beta e^{\beta z}} = \notag \\
&= - \dfrac{1}{\beta}.
\end{align}
In the complex plane of frequencies, we draw small squares around each pole, oriented counterclockwise. Since the horizontal sides are traversed in both directions, their contributions cancel out when summing over the squares. Moreover, since the sum runs over the set $\mathbb{Z}$, the resulting contour $\Gamma$ consists of two lines parallel to the imaginary axis and closes at infinity. We now show that the Matsubara sum can be computed by means of the integral
\begin{equation}
I = \int_{\Gamma} dz \left( - \dfrac{\beta}{2 \pi i} \right) n_{-1}(z) g(z),
\end{equation}
with
\begin{equation}
g(z) = \dfrac{e^{z \delta}}{z - \frac{\mathcal{E}_{\textbf{k},\sigma}}{\hslash}},
\end{equation}
so that
\begin{equation}
\dfrac{1}{\beta} \sum_n \dfrac{e^{- i \omega_n \delta}}{\hslash \left[ i \omega_n - \frac{\mathcal{E}_{\textbf{k},\sigma}}{\hslash} \right]} = \int_{\Gamma} dz \left( - \dfrac{\beta}{2 \pi i} \right) n_{-1}(z) \dfrac{e^{z \delta}}{z - \frac{\mathcal{E}_{\textbf{k},\sigma}}{\hslash}}.
\end{equation}
We fix the pole $z_1 = \frac{\mathcal{E}_{\textbf{k},\sigma}}{\hslash}$ of the function $g(z)$ on the positive real semi-axis, assuming it to be positive. We construct a contour consisting of two semicircles, $C_1$ and $C_2$, for each line. If we show that the contribution of the integrand along the arcs vanishes in the limit as the radius tends to infinity, we obtain the integral along the contour $\Gamma$. Let us assume $\re z > 0$ and $z \rightarrow \infty$, i.e., we consider the curve $C_1$: in order for Jordan's lemma $\eqref{lmm: lemmaJordan}$ to apply, the integrand must vanish faster than $z^{-1}$. For $z \rightarrow \infty$, we have $e^{z \delta} \sim e^{\re z \ \delta}$, and the Fermi-Dirac statistics $n_{-1}(z) \sim e^{- \beta \re z}$ ensures convergence, since the exponential $e^{\re z \delta}$ is well-defined for $\delta \in (0, \beta \hslash)$, with the endpoints excluded. Jordan's lemma $\eqref{lmm: lemmaJordan}$ can be applicated and the integral along $C_1$ vanishes. Now suppose $\re z < 0$ and $z \rightarrow \infty$, i.e., we consider the curve $C_2$. In this case, we have $n_{-1}(z) \sim 1$. However, the exponential $e^{\re z \ \delta}$ prevents the divergence of the integral, since we must perform the integral first and then take the limit $\delta \rightarrow 0$, exactly as for the series. We have shown that the integrals along the semicircles vanish, so the integrals over the two contours $\Gamma_1 + C_2$ and $\Gamma_2 + C_1$ reduce to the residue term only, and the only pole lies in the upper half-plane at $z_1 = \frac{\mathcal{E}_{\textbf{k},\sigma}}{\hslash}$, and it lies outside the contour $\Gamma$. By the residue theorem $\eqref{thm: teoremaresidui}$, we have
\begin{equation}
\dfrac{1}{\beta} \sum_n \dfrac{e^{- i \omega_n \delta}}{\hslash \left[ i \omega_n - \frac{\mathcal{E}_{\textbf{k},\sigma}}{\hslash} \right]} = e^{\frac{\mathcal{E}_\textbf{k} \delta}{\hslash}} n_{-1} ( \mathcal{E}_{\textbf{k}} ) ,
\end{equation}
which is equal to $\eqref{eq: primarelazioneesattafrequenzeMatsubara2}$ adapted for fermions ($\varepsilon = -1$) and for $\tau < 0$.
\end{proof}
\end{theorem}
\section{Thermal average of the particle density operator}
We show here another useful application of many-body propagator. Given a non-interacting Hamiltonian $\hat{\mathcal{H}}_0$, the thermal average of the particle density operator
\begin{equation}
\langle \hat{\rho}(\textbf{r}) \rangle_0 = \sum_{s=-S}^{S} \left\langle \hat{\psi}^\dag(\textbf{r},s) \hat{\psi}(\textbf{r},s) \right\rangle_0
\end{equation}
can be manipulated as follows
\begin{align}
\langle \hat{\rho}(\textbf{r}) \rangle_0 &= \varepsilon \sum_{s=-S}^S \left\langle \hat{T}_\tau \, \hat{\psi}(\textbf{r},s,\tau) \, \hat{\psi}^\dag(\textbf{r},s,\tau+\delta) \right\rangle_0 = \notag \\
&= - \hslash \varepsilon \sum_{s=-S}^S G^{(m)(0)}(\textbf{r},s,-\delta),
\end{align}
where $G^{(m)(0)}$ denotes the non-interacting many-body Green's function. Note that the value $\delta$ is due to the fact that the operators are arranged in ascending order with respect to time, since $\delta>0$. Given a free propagator of the form $\eqref{eq: propagatoreparticelleliberescrittocomeprodottiautostatidivisoenergie}$, we apply a Fourier transform to the density operator with respect to time
\begin{equation}
\langle \hat{\rho}(\textbf{r}) \rangle_0 = - \hslash \varepsilon \sum_{s=-S}^S \dfrac{1}{\beta \hslash} \sum_n G^{(m)(0)}(\textbf{r},s,i \omega_n) e^{i \omega_n \delta} ,
\end{equation}
and we insert the explicit form of the free many-body Green's function, i.e.,
\begin{equation}
\langle \hat{\rho}(\textbf{r}) \rangle_0 = - \hslash \varepsilon \sum_{s=-S}^S \dfrac{1}{\beta \hslash} \sum_n \sum_{j} \dfrac{\left| \varphi_{j}(\textbf{r},s) \right|^2}{\hslash \left[ i \omega_n - \frac{\mathcal{E}_{j}}{\hslash} \right]} e^{i \omega_n \delta}.
\end{equation}
From the Matsubara sum $\eqref{eq: primarelazioneesattafrequenzeMatsubara2}$ it follows
\begin{equation}
\dfrac{1}{\beta \hslash} \sum_n \dfrac{e^{i \omega_n \delta}}{\left[ i \omega_n - \frac{\mathcal{E}_{j}}{\hslash} \right]} = - \varepsilon n_{\varepsilon}(\mathcal{E}_j),
\end{equation}
where $n_{\varepsilon}(\mathcal{E}_j)$ is the Fermi-Dirac or Bose-Einstein statistics (see equation $\eqref{eq: distribuzioneBoseFermiunificata}$), then
\begin{align}
\langle \hat{\rho}(\textbf{r}) \rangle_0 &= \left( - \varepsilon \right)^2 \sum_{s=-S}^S \sum_{j} \left| \varphi_{j}(\textbf{r},s) \right|^2 n_{\varepsilon}(\mathcal{E}_j) = \notag \\
&= \sum_{s=-S}^S \sum_{j} \left| \varphi_{j}(\textbf{r},s) \right|^2 n_{\varepsilon}(\mathcal{E}_j),
\end{align}
which shows explicitly that the thermal density is built from the free eigenfunctions and occupation numbers of the non-interacting many-body system. Finally, we can compute the average total number of particles as
\begin{align}
\langle \hat{N} \rangle_0 &= \left\langle \int d\mathbf{r} \, \hat{\rho}(\mathbf{r}) \right\rangle_0 = \notag \\
&= \int d\mathbf{r} \, \langle \hat{\rho}(\mathbf{r}) \rangle_0 = \notag \\
&= \int d\mathbf{r} \sum_{s=-S}^S \sum_j \left| \varphi_j(\mathbf{r}, s) \right|^2 \, n_{\varepsilon}(\mathcal{E}_j) = \notag \\
&= \sum_j n_{\varepsilon}(\mathcal{E}_j).
\end{align}
\newpage
\section{Figures}
\FloatBarrier
\chapter{Finite-temperature Hartree-Fock method for fermionic systems}
In this chapter, we address the finite-temperature Hartree-Fock method applied to fermionic systems, a fundamental approach for describing interacting quantum systems in thermal equilibrium. We will start from the derivation of the Hartree-Fock equations extended to the finite-temperature case, which allow us to account for the statistical effects due to temperature alongside mean-field interactions. To apply this formalism, we will focus on a particularly relevant case: a homogeneous paramagnetic system. For this purpose, we will employ the jellium model, which provides a simple yet realistic framework to analyze the effects of electronic interactions. \newline
Throughout the chapter, we will solve in detail the finite-temperature Hartree-Fock equations applied to the paramagnetic jellium model. These self-consistent equations require the simultaneous determination of electron wave functions and energy eigenvalues, taking into account the average effects of exchange and direct interactions. This calculation will allow us to obtain the energies, that is, the estimate of the system's total energy within the finite-temperature formalism, which reflects not only the statistical distribution of occupied states but also the effects of electronic exchange and direct interactions. \newline
Finally, we will compute the in the jellium model the Hartree-Fock pair correlation function, which describes the conditional probability of finding a pair of electrons at a given relative distance. This function provides a microscopic description of the spatial correlations introduced by the Pauli exclusion principle and the Coulomb interactions as mediated by the Hartree-Fock formalism, highlighting how interactions modify the relative particle distribution compared to an ideal gas of non-interacting fermions.
\section{Hartree-Fock equations at finite temperature}
Here we apply the Hartree-Fock method at the finite temperature by means of Feynman's approach. Let us consider a quantized Hamiltonian of the form $\eqref{eq: HamiltonianascompostainH_0eH_I}$. According to Dyson's expansion, the propagator can be calculated from the eigenenergy $\Sigma^*_{x_1,x_2}(\tau_1,\tau_2)$, which, however, includes a sum of infinite irreducible diagrams. We previously calculated the first-order self-energy analytically, see $\eqref{eq: self-energyprimoordineanalitico}$. The Hartree-Fock method consists of substituting the self-energy of the Dyson equation for its first-order analytically. We will treat the general case, but eventually apply the results to fermions, since bosons at low temperatures undergo the phenomenon of Bose-Einstein condensation. At thermodynamic equilibrium, the expansion $\eqref{eq: self-energyprimoordineanalitico}$ is invariant by time translation, so we can define $\tau = \tau_1 - \tau_2$ and apply a Fourier transform with respect to time as follows
\begin{equation}
\Sigma^{*(1)}_{x_1,x_2}(i \omega_n) = \int_0^{\beta \hslash} d\tau e^{i \omega_n \tau} \Sigma^{*(1)}_{x_1,x_2}(\tau).
\end{equation}
By adapting $\eqref{eq: funzioneGreenMatsubaraserieFourier1}$ to the present case, we obtain
\begin{equation}
G^{(m)(0)}_{x_1,x_2}(-\delta) = \dfrac{1}{\beta \hslash} \sum_{n'} e^{i \omega_{n'} \tau} G^{(m)(0)}_{x_1,x_2}(i \omega_{n'}),
\end{equation}
\begin{equation}
G^{(m)(0)}_{x'_2,x'_2}(-\delta) = \dfrac{1}{\beta \hslash} \sum_{n'} e^{i \omega_{n'} \tau} G^{(m)(0)}_{x'_2,x'_2}(i \omega_{n'}),
\end{equation}
we have
\begin{align}
& \Sigma^{*(1)}_{x_1,x_2}(i \omega_n) = - \hslash U(x_1,x_2) \left( \int_0^{\beta \hslash} d\tau \delta(\tau) \right) \left( \dfrac{1}{\beta \hslash} \sum_{n'} e^{i \omega_{n'} \tau} G^{(m)(0)}_{x_1,x_2}(i \omega_{n'}) \right) \ + \notag \\
&- \varepsilon \hslash \delta(x_1-x_2) \left( \int_0^{\beta \hslash} d\tau \delta(\tau) \right) \left( \int_0^{\beta \hslash} d\tau'_2 \delta(\tau_1-\tau'_2) \right) \int dx'_2 U(x_1,x'_2) \left( \dfrac{1}{\beta \hslash} \sum_{n'} e^{i \omega_{n'} \tau} G^{(m)(0)}_{x'_2,x'_2}(i \omega_{n'}) \right) = \notag \\
&= - \hslash U(x_1,x_2) \left( \dfrac{1}{\beta \hslash} \sum_{n'} e^{i \omega_{n'} \tau} G^{(m)(0)}_{x_1,x_2}(i \omega_{n'}) \right) \ - \varepsilon \hslash \delta(x_1-x_2) \int dx'_2 U(x_1,x'_2) \left( \dfrac{1}{\beta \hslash} \sum_{n'} e^{i \omega_{n'} \tau} G^{(m)(0)}_{x'_2,x'_2}(i \omega_{n'}) \right)  ,
\end{align}
that is, the first order self-energy does not depend on frequency but only on vertices $x_1$ and $x_2$. To include the effects of successive orders of self-energy, in Dyson's expansion we use the first order of self-energy and replace the free propagator with the propagator: we will see that this approach leads to the Hartree-Fock variational method. We apply a Fourier transform to $\eqref{eq: equazioneDysonformaintegrale}$ with respect to time, i.e.,
\begin{equation}
G^{(m)}_{x,x'}(i\omega_n) = G^{(m)(0)}_{x,x'}(i \omega_n) + \int dx_1 \int dx_2 G^{(m)(0)}_{x,x_1}(i \omega_n) \Sigma_{x_1,x_2}^*(i \omega_n)G^{(m)}_{x_2,x'}(i \omega_n).
\label{eq: equazioneDysonformaintegraleinspaziodellefrequenze}
\end{equation}
According to the Hartree-Fock approach, we assume that the propagator has a known shape; however, this Ansatz is not arbitrary. The Hartree-Fock method is a mean-field theory, which replaces the interaction between particles with an effective potential between free particles. In general, different propagators correspond to different effective potentials, allowing us to estimate the form of the eigenfunctions of the Hamiltonian. We assume that the propagator is analogous to the free propagator $\eqref{eq: propagatoreparticelleliberescrittocomeprodottiautostatidivisoenergie}$. The Ansatz for the many-body Green's function is given by
\begin{equation}
G^{(m)}_{x,x'}(i\omega_n) = \sum_{j} \dfrac{\varphi_{j}^{*}(x') \varphi_{j}(x)}{\hslash \left[ i \omega_n - \frac{\mathcal{E}_{j}}{\hslash} \right]},
\label{eq: ansatzperpropagatorevero}
\end{equation}
that is, it has the same structure as the free many-body function, see equation $\eqref{eq: propagatoreparticelleliberescrittocomeprodottiautostatidivisoenergie}$. If we prove that $\eqref{eq: ansatzperpropagatorevero}$ satisfies the $\eqref{eq: equazioneDysonformaintegraleinspaziodellefrequenze}$ where $\left\lbrace \varphi_j \right\rbrace$ and $\left\lbrace \mathcal{E}_j \right\rbrace$ satisfy a self-consistent equation, then we have verified that solving the Dyson equation with first-order self-energy and the propagator in place of the free propagator is equivalent to solve a self-consistent mean-field equation. If the propagator is written as $\eqref{eq: ansatzperpropagatorevero}$, the self-energy at first order changes as follows
\begin{equation}
\Sigma^{*(1)}_{x_1,x_2}(i \omega_n) \rightarrow \tilde{\Sigma}^{*(1)}_{x_1,x_2}(i \omega_n)
\end{equation}
where $\tilde{\Sigma}^{*(1)}_{x_1,x_2}(i \omega_n)$ is obtained by replacing $G^{(m)}$ with $G^{(m)(0)}$ in $\Sigma^{*(1)}_{x_1,x_2}(i \omega_n)$, i.e., 
\begin{align}
\tilde{\Sigma}^{*(1)}_{x_1,x_2}(i \omega_n) &= - \hslash U(x_1,x_2) \left( \dfrac{1}{\beta \hslash} \sum_{n'} e^{i \omega_{n'} \tau} G^{(m)}_{x_1,x_2}(i \omega_{n'}) \right) \ + \notag \\
&- \varepsilon \hslash \delta(x_1-x_2) \int dx'_2 U(x_1,x'_2) \left( \dfrac{1}{\beta \hslash} \sum_{n'} e^{i \omega_{n'} \tau} G^{(m)}_{x'_2,x'_2}(i \omega_{n'}) \right) = \notag \\
&= - \hslash U(x_1,x_2) \left( \dfrac{1}{\beta \hslash} \sum_{n'} e^{i \omega_{n'} \tau} \left( \sum_{j} \dfrac{\varphi_{j}^{*}(x_2) \varphi_{j}(x_1)}{\hslash \left[ i \omega_{n'} - \frac{\mathcal{E}_{j}}{\hslash} \right]} \right) \right) \ + \notag \\
&- \varepsilon \hslash \delta(x_1-x_2) \int dx'_2 U(x_1,x'_2) \left( \dfrac{1}{\beta \hslash} \sum_{n'} e^{i \omega_{n'} \tau} \left( \sum_{j} \dfrac{\left| \varphi_{j}(x'_2) \right|^2}{\hslash \left[ i \omega_{n'} - \frac{\mathcal{E}_{j}}{\hslash} \right]} \right) \right).
\end{align}
Since equation $\eqref{eq: ansatzperpropagatorevero}$ has the same structure as equation $\eqref{eq: propagatoreparticelleliberescrittocomeprodottiautostatidivisoenergie}$, it also satisfies the exact relation $\eqref{eq: primarelazioneesattafrequenzeMatsubara1}$, specifically with Fermi-Dirac statistics, $n_{-1}(\mathcal{E}_j)$. Moreover, consider the case $\tau < 0$, where the exponential factor takes the form $e^{i \omega_n \tau}$ instead of $e^{-i \omega_n \tau}$. Combining these results, we obtain
\begin{align}
\tilde{\Sigma}^{*(1)}_{x_1,x_2}(i \omega_n) &= \varepsilon U(x_1,x_2) \sum_{j} \varphi_{j}^{*}(x_2) \varphi_{j}(x_1) n_{-1}(\mathcal{E}_j) + \left( - \varepsilon \right)^2 \hslash \delta(x_1-x_2) \int dx'_2 U(x_1,x'_2) \sum_{j} \left| \varphi_{j}(x'_2) \right|^2 n_{-1}(\mathcal{E}_j) = \notag \\
&= \varepsilon U(x_1,x_2) \sum_{j} \varphi_{j}^{*}(x_2) \varphi_{j}(x_1) n_{-1}(\mathcal{E}_j) + \hslash \delta(x_1-x_2) \int dx'_2 U(x_1,x'_2) \sum_{j} \left| \varphi_{j}(x'_2) \right|^2 n_{-1}(\mathcal{E}_j).
\end{align}
\begin{theorem}
The Ansatz $\eqref{eq: ansatzperpropagatorevero}$ implies a mean-field theory for the interacting system.
\begin{proof}
We define the operator
\begin{equation}
O_x = i \hslash \omega_n - \hat{\mathcal{H}}_1(x)
\end{equation}
where 
\begin{equation}
\hat{\mathcal{H}}_1(x) = \dfrac{\hat{\textbf{p}}^2}{2m} + V(x) - \mu \hat{N}
\end{equation}
is a grancanonical Hamiltonian, since it includes the chemical potential, i.e., the number of particles is not fixed. Let us consider the non-interacting many-body Green's function $G^{(m)(0)}_{x,x'}(i \omega_n)$, given by equation $\eqref{eq: propagatoreparticelleliberescrittocomeprodottiautostatidivisoenergie}$, and denote by $\mathcal{E}^{(0)}_{\alpha}$ and $\varphi^{(0)}_\alpha(x)$ the eigenvalues and eigenfunctions, respectively, in order to distinguish them from those used in the Ansatz presented here. The eigenfunctions of $\hat{\mathcal{H}}_1$ are $\left\lbrace \varphi^{(0)}_\alpha(x) \right\rbrace$. By definition, $O_x$ is a one-body operator and satisfies the following properties
\begin{equation}
O_x \varphi^{(0)}_\alpha(x) = \left( i \hslash \omega_n - \mathcal{E}^{(0)}_\alpha \right) \varphi^{(0)}_\alpha(x),
\end{equation}
and
\begin{align}
O_x G^{(m)(0)}_{x,x'}(i \omega_n) &= \left( i \hslash \omega_n - \hat{\mathcal{H}}_1(x) \right) \sum_{\alpha} \dfrac{\varphi_{\alpha}^{*(0)}(x') \varphi^{(0)}_{\alpha}(x)}{\hslash \left[ i \omega_n - \frac{\mathcal{E}^{(0)}_{\alpha}}{\hslash} \right]} = \notag \\
&= \left( i \hslash \omega_n - \mathcal{E}^{(0)}_{\alpha} \right) \sum_{\alpha} \dfrac{\varphi_{\alpha}^{*(0)}(x') \varphi^{(0)}_{\alpha}(x)}{\hslash \left[ i \omega_n - \frac{\mathcal{E}^{(0)}_{\alpha}}{\hslash} \right]} = \notag \\
&= \sum_\alpha \varphi_{\alpha}^{*(0)}(x') \varphi^{(0)}_{\alpha}(x) = \notag \\
&= \delta(x-x').
\end{align}
Substituting the self-energy with the first-order energy correction and acting with the operator $O_x$ on both sides of equation $\eqref{eq: equazioneDysonformaintegraleinspaziodellefrequenze}$, we obtain
\begin{equation}
O_x G^{(m)}_{x,x'}(i\omega_n) = O_x G^{(m)(0)}(i \omega_n) + O_x \int dx_1 \int dx_2 G^{(m)(0)}_{x,x_1}(i \omega_n) \tilde{\Sigma}_{x_1,x_2}^*(i \omega_n)G^{(m)}_{x_2,x'}(i \omega_n),
\end{equation}
\begin{equation}
O_x \sum_{j} \dfrac{\varphi_{j}^{*}(x') \varphi_{j}(x)}{\hslash \left[ i \omega_n - \frac{\mathcal{E}_{j}}{\hslash} \right]} = \delta(x-x') + \int dx_1 \int dx_2 O_x G^{(m)(0)}_{x,x_1}(i \omega_n) \tilde{\Sigma}_{x_1,x_2}^*(i \omega_n)\sum_{j} \dfrac{\varphi_{j}^{*}(x') \varphi_{j}(x_2)}{\hslash \left[ i \omega_n - \frac{\mathcal{E}_{j}}{\hslash} \right]},
\end{equation}
\begin{equation}
\sum_{j} \dfrac{\varphi_{j}^{*}(x') O_x \varphi_{j}(x)}{\hslash \left[ i \omega_n - \frac{\mathcal{E}_{j}}{\hslash} \right]} = \delta(x-x') + \int dx_1 \int dx_2 \delta(x-x_1) \tilde{\Sigma}_{x_1,x_2}^*(i \omega_n)\sum_{j} \dfrac{\varphi_{j}^{*}(x') \varphi_{j}(x_2)}{\hslash \left[ i \omega_n - \frac{\mathcal{E}_{j}}{\hslash} \right]},
\end{equation}
\begin{equation}
\sum_{j} \dfrac{\varphi_{j}^{*}(x') O_x \varphi_{j}(x)}{\hslash \left[ i \omega_n - \frac{\mathcal{E}_{j}}{\hslash} \right]} = \delta(x-x') + \int dx_2 \tilde{\Sigma}_{x,x_2}^*(i \omega_n)\sum_{j} \dfrac{\varphi_{j}^{*}(x') \varphi_{j}(x_2)}{\hslash \left[ i \omega_n - \frac{\mathcal{E}_{j}}{\hslash} \right]} .
\end{equation}
We multiply both members of the equation by $\varphi_h(x')$ and integrate with respect to the measure $dx'$, to use the orthonormality of the eigenfunctions. We obtain
\begin{equation}
\dfrac{O_x \varphi_{h}(x)}{\hslash \left[ i \omega_n - \frac{\mathcal{E}_{h}}{\hslash} \right]} = \varphi_{h}(x) + \int dx_2 \tilde{\Sigma}_{x,x_2}^*(i \omega_n) \dfrac{\varphi_{h}(x_2)}{\hslash \left[ i \omega_n - \frac{\mathcal{E}_{h}}{\hslash} \right]} ,
\end{equation}
\begin{equation}
O_x \varphi_{h}(x) = \left[ i \hslash \omega_n - \mathcal{E}_h \right] \varphi_{h}(x) + \int dx_2 \tilde{\Sigma}_{x,x_2}^*(i \omega_n) \varphi_{h}(x_2) ,
\end{equation}
\begin{equation}
\left[ i \hslash \omega_n - \hat{\mathcal{H}}_1(x) \right] \varphi_{h}(x) = \left[ i \hslash \omega_n - \mathcal{E}_h \right] \varphi_{h}(x) + \int dx_2 \tilde{\Sigma}_{x,x_2}^*(i \omega_n) \varphi_{h}(x_2) ,
\end{equation}
\begin{equation}
\mathcal{E}_h \varphi_{h}(x) = \hat{\mathcal{H}}_1(x) \varphi_{h}(x) + \int dx_2 \tilde{\Sigma}_{x,x_2}^*(i \omega_n) \varphi_{h}(x_2).
\label{eq: equazioniautovaloriHFTfinita1}
\end{equation}
That is, for non-interacting particles, $\tilde{\Sigma}_{x,x_2}^*(i \omega_n)=0$ and the previous equation becomes the eigenvalue equation: the interaction modifies the eigenvalue equation via self-energy. We now show that the $\eqref{eq: equazioniautovaloriHFTfinita1}$ corresponds to the set of Hartree-Fock equations. We explicate the self-energy
\begin{equation}
\tilde{\Sigma}_{x,x_2}^*(i \omega_n) = \varepsilon U(x,x_2) \sum_{j} \varphi_{j}^{*}(x_2) \varphi_{j}(x) n_{-1}(\mathcal{E}_j) + \hslash \delta(x-x_2) \int dx'_2 U(x,x'_2) \sum_{j} \left| \varphi_{j}(x'_2) \right|^2 n_{-1}(\mathcal{E}_j),
\end{equation}
and we insert it in $\eqref{eq: equazioniautovaloriHFTfinita1}$, i.e.,
\begin{align}
\mathcal{E}_h \varphi_{h}(x) &= \mathcal{H}_1(x) \varphi_{h}(x) + \varphi_h(x) \left[ \int dx_2 \sum_j \left| \varphi_j(x_2) \right|^2 n_{-1}(\mathcal{E}_j) U(x,x_2) \right] + \notag \\
&+\varepsilon \int dx_2 U(x,x_2) \sum_j \varphi_j^*(x_2) \varphi_j(x) n_{-1}(\mathcal{E}_j) \varphi_h(x_2) ,
\label{eq: equazioniautovaloriHFTfinita2}
\end{align}
which are equal to the Hartree-Fock equations.
\end{proof}
\end{theorem}
We show the meaning of $\eqref{eq: equazioniautovaloriHFTfinita2}$. First of all, the object
\begin{equation}
\varphi_h(x) \left[ \int dx_2 \sum_j \left| \varphi_j(x_2) \right|^2 n_{-1}(\mathcal{E}_j) U(x,x_2) \right]
\end{equation}
is called the direct term. If the equation $\eqref{eq: equazioniautovaloriHFTfinita2}$ contains only the direct term, that is, it takes the form
\begin{equation}
\mathcal{E}_h \varphi_{h}(x) = \hat{\mathcal{H}}_1(x) \varphi_{h}(x) + \varphi_h(x) \left[ \int dx_2 \sum_j \left| \varphi_j(x_2) \right|^2 n_{-1}(\mathcal{E}_j) U(x,x_2) \right] ,
\end{equation}
the equation is solvable, being of the form
\begin{equation}
\mathcal{E}_h \varphi_{h}(x) = \hat{\mathcal{H}}_1(x) \varphi_{h}(x) + \tilde{V}(x) \varphi_h(x) , 
\end{equation}
and the eigenfunctions still satisfy a one-body problem, but the eigenvalues are modified by the mean-field effective potential. The direct term includes the renormalization of energy, which is obtained by replacing each particle with its charge distribution $\left| \varphi_j(x_2) \right|^2 n_{-1}(\mathcal{E}_j)$. It is a term that has a classical equivalent. The object
\begin{equation}
\varepsilon \int dx_2 U(x,x_2) \sum_j \varphi_j^*(x_2) \varphi_j(x) n_{-1}(\mathcal{E}_j) \varphi_h(x_2) 
\end{equation}
is called the exchange term, since it includes a value $\varepsilon$, which depends on the statistical nature of the operators. This term has no classical equivalent; it makes sense only in quantum mechanics.
\section{Solution of the Hartree-Fock equations at finite temperature for translationally invariant systems}
Here, we introduce a set of simplifying assumptions that will reduce the Hartree-Fock equations to a more tractable form and eventually allow for an exact solution. First, assume that the two-body interaction is spin-independent, meaning it depends only on the spatial coordinates, i.e., $U(x_1,x_2) = U(\textbf{r}_1,\textbf{r}_2)$. Consequently, given a solution
\begin{equation}
\varphi_h(x) = \varphi_n(\textbf{r}) \chi_\sigma(s), \ h=(n,\sigma) ,
\end{equation}
the direct and exchange terms become respectively
\begin{align}
& \varphi_h(x) \left[ \int dx_2 \sum_j \left| \varphi_j(x_2) \right|^2 n_{-1}(\mathcal{E}_j) U(x,x_2) \right] = \notag \\
&= \varphi_n(\textbf{r}) \chi_\sigma(s) \left[ \sum_{s_2} \int d^3\textbf{r}_2 \sum_{n'} \sum_{\sigma'} \left| \varphi_{n'}(\textbf{r}_2) \right|^2 \left| \chi_{\sigma'}(s_2) \right|^2 n_{-1}(\mathcal{E}_{n',\sigma'}) U(\textbf{r},\textbf{r}_2) \right] = \notag \\
&= \varphi_n(\textbf{r}) \chi_\sigma(s) \left[ \int d^3\textbf{r}_2 \sum_{n'} \sum_{\sigma'} \left| \varphi_{n'}(\textbf{r}_2) \right|^2 \sum_{s_2} \left| \chi_{\sigma'}(s_2) \right|^2 n_{-1}(\mathcal{E}_{n',\sigma'}) U(\textbf{r},\textbf{r}_2) \right] = \notag \\
&= \varphi_n(\textbf{r}) \chi_\sigma(s) \left[ \int d^3\textbf{r}_2 \sum_{n'} \sum_{\sigma'} \left| \varphi_{n'}(\textbf{r}_2) \right|^2 n_{-1}(\mathcal{E}_{n',\sigma'}) U(\textbf{r},\textbf{r}_2) \right] ,
\end{align}
\begin{align}
& \varepsilon \int dx_2 U(x,x_2) \sum_j \varphi_j^*(x_2) \varphi_j(x) \varphi_h(x_2) n_{-1}(\mathcal{E}_j) = \notag \\
&= \varepsilon \sum_{s_2} \int d^3\textbf{r}_2 U(\textbf{r},\textbf{r}_2) \sum_{n'} \sum_{\sigma'} \varphi_{n'}^*(\textbf{r}_2) \varphi_{n'}(\textbf{r}) \varphi_{n}(\textbf{r}_2) \chi^*_{\sigma'}(s_2) \chi_{\sigma'}(s) \chi_\sigma(s_2) n_{-1}(\mathcal{E}_{n',\sigma'}) = \notag \\
&= \varepsilon \int d^3\textbf{r}_2 U(\textbf{r},\textbf{r}_2) \sum_{n'} \sum_{\sigma'} \varphi_{n'}^*(\textbf{r}_2) \varphi_{n'}(\textbf{r}) \varphi_{n}(\textbf{r}_2) \chi_{\sigma'}(s) n_{-1}(\mathcal{E}_{n',\sigma'}) \left( \sum_{s_2} \chi^*_{\sigma'}(s_2) \chi_\sigma(s_2) \right) = \notag \\
&= \varepsilon \int d^3\textbf{r}_2 U(\textbf{r},\textbf{r}_2) \sum_{n'} \sum_{\sigma'} \varphi_{n'}^*(\textbf{r}_2) \varphi_{n'}(\textbf{r}) \varphi_{n}(\textbf{r}_2) \chi_{\sigma'}(s) n_{-1}(\mathcal{E}_{n',\sigma'}) \delta_{\sigma,\sigma'} = \notag \\
&= \varepsilon \int d^3\textbf{r}_2 U(\textbf{r},\textbf{r}_2) \sum_{n'} \varphi_{n'}^*(\textbf{r}_2) \varphi_{n'}(\textbf{r}) \varphi_{n}(\textbf{r}_2) \chi_{\sigma}(s) n_{-1}(\mathcal{E}_{n',\sigma}),
\end{align}
and Hartree-Fock equations become
\begin{align}
\mathcal{E}_{n,\sigma} \varphi_{n}(\textbf{r}) \chi_\sigma(s) &= \mathcal{H}_1(\textbf{r},s) \varphi_{n}(\textbf{r}) \chi_\sigma(s) + \varphi_n(\textbf{r}) \chi_\sigma(s) \left[ \int d^3\textbf{r}_2 \sum_{n'} \sum_{\sigma'} \left| \varphi_{n'}(\textbf{r}_2) \right|^2 n_{-1}(\mathcal{E}_{n',\sigma'}) U(\textbf{r},\textbf{r}_2) \right] + \notag \\
&+ \varepsilon \int d^3\textbf{r}_2 U(\textbf{r},\textbf{r}_2) \sum_{n'} \varphi_{n'}^*(\textbf{r}_2) \varphi_{n'}(\textbf{r}) \varphi_{n}(\textbf{r}_2) \chi_{\sigma}(s) n_{-1}(\mathcal{E}_{n',\sigma}).
\end{align}
Note that the direct term contains all spins, given the sum with respect to $\sigma'$, whereas the exchange term concerns only the spin of the particle $\varphi_{n}(\textbf{r}) \chi_\sigma(s)$. Furthermore, we assume that the one-body operator also does not depend on the spin variable, i.e., $\hat{\mathcal{H}}_1(\textbf{r},s) = \hat{\mathcal{H}}_1(\textbf{r})$, then the spin eigenfunction $\chi(s)$ can be simplified in both members as follows
\begin{align}
\mathcal{E}_{n,\sigma} \varphi_{n}(\textbf{r}) &= \hat{\mathcal{H}}_1(\textbf{r}) \varphi_{n}(\textbf{r}) + \varphi_n(\textbf{r}) \left[ \int d^3\textbf{r}_2 \sum_{n'} \sum_{\sigma'} \left| \varphi_{n'}(\textbf{r}_2) \right|^2 n_{-1}(\mathcal{E}_{n',\sigma'}) U(\textbf{r},\textbf{r}_2) \right] + \notag \\
&+ \varepsilon \int d^3\textbf{r}_2 U(\textbf{r},\textbf{r}_2) \sum_{n'} \varphi_{n'}^*(\textbf{r}_2) \varphi_{n'}(\textbf{r}) \varphi_{n}(\textbf{r}_2) n_{-1}(\mathcal{E}_{n',\sigma}).
\label{eq: equazioniautovaloriHFTfinita3}
\end{align}
However, the problem remains quite complex, as it still depends on both the two-body interaction \( U \) and the one-body potential \( V \). The Hartree-Fock equation $\eqref{eq: equazioniautovaloriHFTfinita3}$ can be solved exactly only under further simplifying assumptions, namely:
\begin{itemize}
\item the one-body potential is zero, i.e.,
\begin{equation}
V = 0;
\label{eq: condizionecasoparticolareHartreeFock1}
\end{equation} 
\item the two-body interaction is invariant by spatial translation, i.e.,
\begin{equation}
U(\textbf{r}_1,\textbf{r}_2) = U(\textbf{r}_1 - \textbf{r}_2);
\label{eq: condizionecasoparticolareHartreeFock2}
\end{equation}
\item the Fermi-Dirac statistics and the energies do not depend on the spin variable, i.e.,
\begin{equation}
n_{-1}(\mathcal{E}_{n,\sigma}) = n_{-1}(\mathcal{E}_{n}),
\label{eq: condizionecasoparticolareHartreeFock3}
\end{equation}
\begin{equation}
\mathcal{E}_{n,\sigma} = \mathcal{E}_n ,
\label{eq: condizionecasoparticolareHartreeFock4}
\end{equation}
that is, the system is paramagnetic.
\end{itemize}
Consequently, the eigenvalue equation is modified as follows
\begin{align}
\mathcal{E}_{n} \varphi_{n}(\textbf{r}) &= \hat{\mathcal{H}}_1(\textbf{r}) \varphi_{n}(\textbf{r}) + \left( 2S+1 \right) \varphi_n(\textbf{r}) \left[ \int d^3\textbf{r}_2 \sum_{n'} \left| \varphi_{n'}(\textbf{r}_2) \right|^2 n_{-1}\left(\mathcal{E}_{n'}\right) U(\textbf{r}-\textbf{r}_2) \right] + \notag \\
&+ \varepsilon \int d^3\textbf{r}_2 U(\textbf{r}-\textbf{r}_2) \sum_{n'} \varphi_{n'}^*(\textbf{r}_2) \varphi_{n'}(\textbf{r}) \varphi_{n}(\textbf{r}_2) n_{-1}\left(\mathcal{E}_{n'}\right) ,
\label{eq: equazioniautovaloriHFTfinita4}
\end{align}
where in the direct term the sum $\sum_{\sigma'}$ provided the quantity $\left( 2S+1 \right)$. Now, we show that the plane waves $\eqref{eq: ondepiane}$ satisfy $\eqref{eq: equazioniautovaloriHFTfinita4}$. Indeed,
\begin{align}
\mathcal{E}_{\textbf{k}} \dfrac{e^{i \textbf{k} \cdot \textbf{r}}}{\sqrt{V}} &= \hat{\mathcal{H}}_1(\textbf{r}) \dfrac{e^{i \textbf{k} \cdot \textbf{r}}}{\sqrt{V}} + \left( 2S+1 \right) \dfrac{e^{i \textbf{k} \cdot \textbf{r}}}{\sqrt{V}} \left[ \int d^3\textbf{r}_2 \sum_{\textbf{k}'} \dfrac{1}{V} n_{-1}\left(\mathcal{E}_{\textbf{k}'}\right) U(\textbf{r}-\textbf{r}_2) \right] + \notag \\
&+ \varepsilon \int d^3\textbf{r}_2 U(\textbf{r}-\textbf{r}_2) \sum_{\textbf{k}'} \dfrac{1}{V^{\frac{3}{2}}} e^{- i \textbf{k}' \cdot \textbf{r}_2} e^{i \textbf{k}' \cdot \textbf{r}} e^{i \textbf{k} \cdot \textbf{r}_2} n_{-1}\left(\mathcal{E}_{\textbf{k}'}\right) = \notag \\
&= \mathcal{E}^{(0)}_{\textbf{k}} \dfrac{e^{i \textbf{k} \cdot \textbf{r}}}{\sqrt{V}} + \left( 2S+1 \right) \dfrac{e^{i \textbf{k} \cdot \textbf{r}}}{\sqrt{V}} \left[ \int d^3\textbf{r}_2 \sum_{\textbf{k}'} \dfrac{1}{V} n_{-1}\left(\mathcal{E}_{\textbf{k}'}\right) U(\textbf{r}-\textbf{r}_2) \right] + \notag \\
&+ \varepsilon \int d^3\textbf{r}_2 U(\textbf{r}-\textbf{r}_2) \sum_{\textbf{k}'} \dfrac{1}{V^{\frac{3}{2}}} e^{- i \textbf{k}' \cdot \textbf{r}_2} e^{i \textbf{k}' \cdot \textbf{r}} e^{i \textbf{k} \cdot \textbf{r}_2} n_{-1}\left(\mathcal{E}_{\textbf{k}'}\right).
\end{align}
To verify that the plane wave is a solution, we need to modify the exchange term to bring out the object $\frac{e^{i \textbf{k} \cdot \textbf{r}}}{\sqrt{V}}$. Here, the two-body operator satisfies $U(\textbf{r}-\textbf{r}_2) = U(\textbf{r},\textbf{r}_2) = U(\textbf{r}_2,\textbf{r}) = U(\textbf{r}_2-\textbf{r})$, then we set $\textbf{r}_2 - \textbf{r} \rightarrow \textbf{r}_2$, $\textbf{r}_2 \rightarrow \textbf{r}_2 + \textbf{r}$, then the direct term becomes
\begin{align}
\left( 2S+1 \right) \dfrac{e^{i \textbf{k} \cdot \textbf{r}}}{\sqrt{V}} \sum_{\textbf{k}'} \dfrac{1}{V} n_{-1}\left(\mathcal{E}_{\textbf{k}'}\right) \left[ \int d^3\textbf{r}_2  U(\textbf{r}_2) \right] &= \left( 2S+1 \right) \dfrac{e^{i \textbf{k} \cdot \textbf{r}}}{\sqrt{V}} U\left(\textbf{k}=\textbf{0}\right) \sum_{\textbf{k}'} \dfrac{1}{V} n_{-1}\left(\mathcal{E}_{\textbf{k}'}\right) = \notag \\
&= \dfrac{N \left( 2S+1 \right)}{V} \dfrac{e^{i \textbf{k} \cdot \textbf{r}}}{\sqrt{V}} U\left(\textbf{k}=\textbf{0}\right) ,
\end{align}
and the exchange term becomes
\begin{align}
& \varepsilon \int d^3\textbf{r}_2 U(\textbf{r}_2) \sum_{\textbf{k}'} \dfrac{1}{V^{\frac{3}{2}}} e^{- i \textbf{k}' \cdot \left( \textbf{r}_2 + \textbf{r} \right)} e^{i \textbf{k}' \cdot \textbf{r}} e^{i \textbf{k} \cdot \left( \textbf{r}_2 + \textbf{r} \right)} n_{-1}\left(\mathcal{E}_{\textbf{k}'}\right) = \notag \\
&= \varepsilon \int d^3\textbf{r}_2 U(\textbf{r}_2) \sum_{\textbf{k}'} \dfrac{1}{V^{\frac{3}{2}}} e^{- i \textbf{k}' \cdot \textbf{r}_2} e^{- i \textbf{k}' \cdot \textbf{r}} e^{i \textbf{k}' \cdot \textbf{r}} e^{i \textbf{k} \cdot \textbf{r}_2} e^{i \textbf{k} \cdot \textbf{r}} n_{-1}\left(\mathcal{E}_{\textbf{k}'}\right) = \notag \\
&= \varepsilon \int d^3\textbf{r}_2 U(\textbf{r}_2) \sum_{\textbf{k}'} \dfrac{1}{V^{\frac{3}{2}}} e^{- i \textbf{k}' \cdot \textbf{r}_2} e^{i \textbf{k} \cdot \textbf{r}_2} e^{i \textbf{k} \cdot \textbf{r}} n_{-1}\left(\mathcal{E}_{\textbf{k}'}\right) = \notag \\
&= \varepsilon \dfrac{e^{i \textbf{k} \cdot \textbf{r}}}{\sqrt{V}} \sum_{\textbf{k}'} \dfrac{1}{V} n_{-1}\left(\mathcal{E}_{\textbf{k}'}\right) \left[ \int d^3\textbf{r}_2 U(\textbf{r}_2) e^{i \left( \textbf{k} - \textbf{k}' \right) \cdot \textbf{r}_2} \right] = \notag \\
&= \varepsilon \sum_{\textbf{k}'} \dfrac{1}{V} n_{-1}\left(\mathcal{E}_{\textbf{k}'}\right) U\left( \textbf{k} - \textbf{k}' \right) \dfrac{e^{i \textbf{k} \cdot \textbf{r}}}{\sqrt{V}}.
\end{align}
Finally, the plane waves are solutions of the Hartree-Fock equations, and the Hartree-Fock energies are given by
\begin{equation}
\mathcal{E}_{\textbf{k}}^{H.F.} = \mathcal{E}^{(0)}_{\textbf{k}} + \dfrac{N \left(2S+1\right)}{V} U\left(\textbf{k}=\textbf{0}\right) + \varepsilon \sum_{\textbf{k}'} \dfrac{1}{V} n_{-1}\left(\mathcal{E}^{H.F.}_{\textbf{k}'}\right) U\left(\textbf{k}-\textbf{k}'\right) .
\label{eq: equazioneenergieHFtemperaturafinitaT} 
\end{equation}
The direct term contributes a uniform energy shift that is independent of the momentum \( \mathbf{k} \), while the exchange term introduces a nontrivial, momentum-dependent correction to the single-particle energy spectrum.
\section{Hartree-Fock energies in the jellium model}
\begin{figure}[H]
\centering
\includegraphics[scale=0.8]{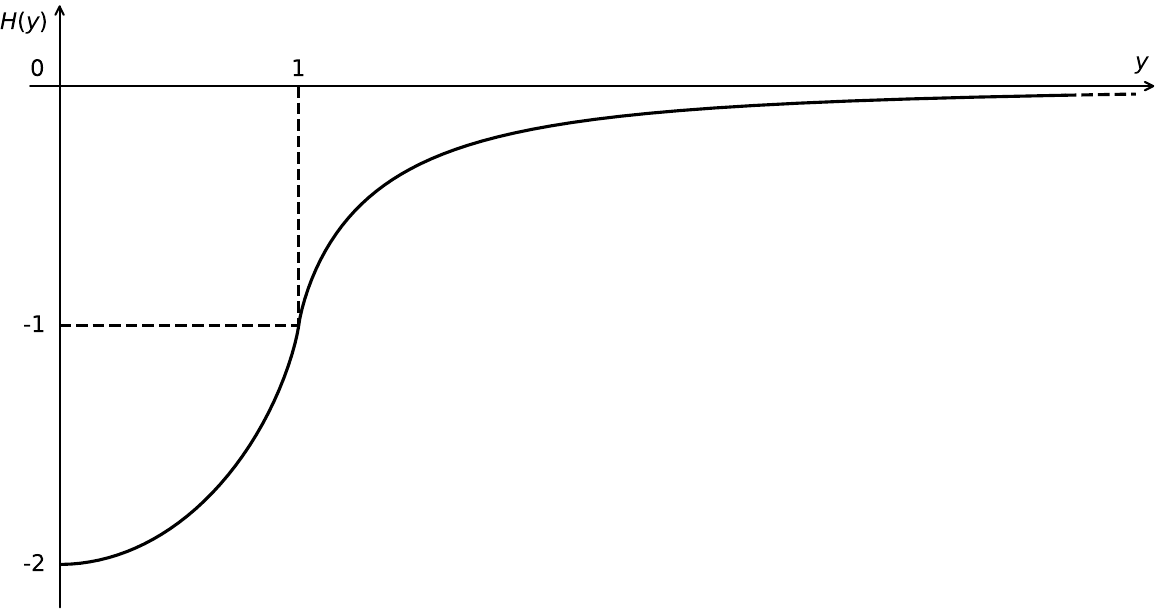}
\caption{Plot of the function $\eqref{eq: H_y_energieHartreeFock}$, defined in terms of the dimensionless variable $y = \frac{k}{k_F}$, which arises in the evaluation of the Hartree-Fock energies $\eqref{eq: energieHartreeFockjellium}$ at zero temperature.}
\label{fig: H_y_energieHartreeFock_figure}
\end{figure}

The equation $\eqref{eq: equazioneenergieHFtemperaturafinitaT}$ can be applied the Hamiltonian of the jellium model in the grand canonical ensemble, i.e., equation $\eqref{eq: hamiltonianamodellojellium3}$. Consequently, $\mathcal{E}^{(0)}_{\textbf{k}} = \frac{\hslash^2 \textbf{k}^2}{2m} - \mu$ and $\varepsilon=-1$. Moreover, since the direct term in the $\eqref{eq: equazioneenergieHFtemperaturafinitaT}$ contains $U\left(\textbf{k}=\textbf{0}\right)$ and the jellium model excludes the zero-momentum component from the sum over momenta, this term vanishes, and we have
\begin{align}
\mathcal{E}_{\textbf{k}}^{H.F.} &= \mathcal{E}^{(0)}_{\textbf{k}} - \sum_{\textbf{k}' \neq \textbf{k}} \dfrac{1}{V} n_{-1}\left(\mathcal{E}^{H.F.}_{\textbf{k}'}\right) U\left(\textbf{k}-\textbf{k}'\right) = \notag \\
&= \mathcal{E}^{(0)}_{\textbf{k}} - \sum_{\textbf{k}' \neq \textbf{k}} \dfrac{1}{V} n_{-1}\left(\mathcal{E}^{H.F.}_{\textbf{k}'}\right) \dfrac{4 \pi q_e^2}{\left|\textbf{k}-\textbf{k}'\right|^2} ,
\end{align}
that, by means of the thermodynamic limit approximation $\eqref{eq: illimitetermodinamico}$ becomes,
\begin{equation}
\mathcal{E}_{\textbf{k}}^{H.F.} = \mathcal{E}^{(0)}_{\textbf{k}} - \pv \int \dfrac{d^3\textbf{k}'}{(2\pi)^3} n_{-1}\left(\mathcal{E}^{H.F.}_{\textbf{k}'}\right) \dfrac{4 \pi q_e^2}{\left|\textbf{k}-\textbf{k}'\right|^2}.
\label{eq: energieHartreeFockjellium}
\end{equation}
Previously, we showed that the Hartree-Fock propagator $\eqref{eq: ansatzperpropagatorevero}$ admits, in a particular case, see equations $\eqref{eq: condizionecasoparticolareHartreeFock1}$, $\eqref{eq: condizionecasoparticolareHartreeFock2}$, $\eqref{eq: condizionecasoparticolareHartreeFock3}$, and $\eqref{eq: condizionecasoparticolareHartreeFock4}$, plane wave solutions. In this case, the Hartree-Fock propagator takes the explicit form
\begin{equation}
G^{(m)}_{x,x'}(i\omega_n) = \sum_{\textbf{k},\sigma} \dfrac{e^{- i \textbf{k} \cdot \textbf{r}}}{\sqrt{V}} \dfrac{e^{i \textbf{k} \cdot \textbf{r}'}}{\sqrt{V}} \dfrac{1}{\hslash \left[ i \omega_n - \frac{\mathcal{E}^{H.F.}_{\textbf{k}}}{\hslash} \right]},
\end{equation}
and summing over spins we have
\begin{equation}
G^{(m)}_{\textbf{r},\textbf{r}'}(i\omega_n) = \sum_{\textbf{k}} \dfrac{e^{i \textbf{k} \cdot (\textbf{r}-\textbf{r}')}}{V} \dfrac{1}{\hslash \left[ i \omega_n - \frac{\mathcal{E}^{H.F.}_{\textbf{k}}}{\hslash} \right]},
\end{equation}
and note that there is no dependence on the spin component $\sigma$, since we are describing a paramagnetic system. We apply a spatial Fourier transform and we get
\begin{equation}
G^{(m)}_{\textbf{k}}(i\omega_n) = \dfrac{1}{\hslash \left[ i \omega_n - \frac{\mathcal{E}^{H.F.}_{\textbf{k}}}{\hslash} \right]}.
\end{equation}
We aim to compute the spectral function associated with the Hartree-Fock propagator: from $\eqref{eq: funzionespettraleparteimmaginariapropagatore}$ we have
\begin{equation}
A^{H.F.}_{\textbf{k}}(\omega) = 2 \pi \delta \left[ \omega - \dfrac{\mathcal{E}^{H.F.}_{\textbf{k}}}{\hslash} \right],
\end{equation}
where we note that the spectral function is identical in form to that of free particles, and in particular the one-to-one correspondence between energies and momenta is preserved, but the energy is not that of a free particle. Since we replace the true propagator with the Hartree-Fock one, we now ask whether the Hartree-Fock spectral function satisfies the same sum rules as the true spectral function $A_{\textbf{k}}(\omega)$, that is, whether it can be substituted for it while providing the same results. It is straightforward to verify that $A^{H.F.}_{\textbf{k}}(\omega)$ satisfies $\eqref{eq: momentonullodifunzionespettrale}$, since the integral of the delta function returns 1, which coincides with the anticommutator between the destruction and creation operators associated with the Hartree-Fock plane wave solution. Then, consider the first moment of the Hartree-Fock spectral function, that is, equation $\eqref{eq: momentoprimodifunzionespettrale}$. Regarding the left-hand side of $\eqref{eq: momentoprimodifunzionespettrale}$, we have
\begin{align}
\int_{-\infty}^{+\infty} \dfrac{d\omega}{2\pi} \omega A^{H.F.}_{\textbf{k}}(\omega) &= \int_{-\infty}^{+\infty} \dfrac{d\omega}{2\pi} \omega 2 \pi \delta \left[ \omega - \dfrac{\mathcal{E}^{H.F.}_{\textbf{k}}}{\hslash} \right] = \notag \\
&= \dfrac{\mathcal{E}^{H.F.}_{\textbf{k}}}{\hslash} = \notag \\
&= \dfrac{\mathcal{E}^{(0)}_{\textbf{k}}}{\hslash} - \dfrac{1}{\hslash} \pv \int \dfrac{d^3\textbf{k}'}{(2\pi)^3} n_{-1}\left(\mathcal{E}^{H.F.}_{\textbf{k}'}\right) \dfrac{4 \pi q_e^2}{\left|\textbf{k}-\textbf{k}'\right|^2},
\label{eq: momentoprimodifunzionespettraleHartreeFocksinistra}
\end{align}
where in the last step we used $\eqref{eq: energieHartreeFockjellium}$. To verify $\eqref{eq: momentoprimodifunzionespettrale}$, we need to compute the right-hand side of $\eqref{eq: momentoprimodifunzionespettrale}$, that is, $\left\langle \left[ \left[ C_{\textbf{k}},\hat{\mathcal{H}} \right],C^\dagger_{\textbf{k}}\right]^{(\varepsilon)} \right\rangle$, where $\varepsilon=-1$, that is, the brackets represent anticommutators, and the thermal average is performed over a Hamiltonian with interaction; here we use the jellium Hamiltonian $\eqref{eq: hamiltonianamodellojellium1}$, adapted to the case with no spin dependence. We have
\begin{align}
\left[ C_{\textbf{k}} , \hat{\mathcal{H}} \right] &= \sum_{\textbf{k}'} \mathcal{E}^{(0)}_{\textbf{k}'} \left[ C_{\textbf{k}}, C^\dag_{\textbf{k}'}C_{\textbf{k}'} \right] + \dfrac{1}{2V} \sum_{\substack{\textbf{k}_1,\textbf{k}_2, \\ \textbf{q} \neq \textbf{0}}} \dfrac{4 \pi q_e^2}{q^2} \left[ C_{\textbf{k}} , C^\dag_{\textbf{k}_1 + \textbf{q}} C^\dag_{\textbf{k}_2 - \textbf{q}} C_{\textbf{k}_2} C_{\textbf{k}_1} \right],
\end{align}
from equation $\eqref{eq: commutatoreoperatorenumeroconoperatoredistruzione}$ it follows
\begin{align}
\sum_{\textbf{k}'} \mathcal{E}^{(0)}_{\textbf{k}'} \left[ C_{\textbf{k}}, C^\dag_{\textbf{k}'}C_{\textbf{k}'} \right] &= \sum_{\textbf{k}'} \mathcal{E}^{(0)}_{\textbf{k}'} \delta_{\textbf{k},\textbf{k}'} C_{\textbf{k}'} = \notag \\
&= \mathcal{E}^{(0)}_{\textbf{k}} C_{\textbf{k}},
\end{align}
Concerning the second commutator, we use the fermionic algebra to move the operator $C_{\textbf{k}}$ into one of the two products. We have
\begin{align}
\left[ C_{\textbf{k}} , C^\dag_{\textbf{k}_1 + \textbf{q}} C^\dag_{\textbf{k}_2 - \textbf{q}} C_{\textbf{k}_2} C_{\textbf{k}_1} \right] &= C_{\textbf{k}} C^\dag_{\textbf{k}_1 + \textbf{q}} C^\dag_{\textbf{k}_2 - \textbf{q}} C_{\textbf{k}_2} C_{\textbf{k}_1} - C^\dag_{\textbf{k}_1 + \textbf{q}} C^\dag_{\textbf{k}_2 - \textbf{q}} C_{\textbf{k}_2} C_{\textbf{k}_1} C_{\textbf{k}} = \notag \\
&= \delta_{\textbf{k},\textbf{k}_1 + \textbf{q}} C^\dag_{\textbf{k}_2 - \textbf{q}} C_{\textbf{k}_2} C_{\textbf{k}_1} - C^\dag_{\textbf{k}_1 + \textbf{q}} C_{\textbf{k}} C^\dag_{\textbf{k}_2 - \textbf{q}} C_{\textbf{k}_2} C_{\textbf{k}_1} + \notag \\
&- C^\dag_{\textbf{k}_1 + \textbf{q}} C^\dag_{\textbf{k}_2 - \textbf{q}} C_{\textbf{k}_2} C_{\textbf{k}_1} C_{\textbf{k}} = \notag \\
&= \delta_{\textbf{k},\textbf{k}_1 + \textbf{q}} C^\dag_{\textbf{k}_2 - \textbf{q}} C_{\textbf{k}_2} C_{\textbf{k}_1} - \delta_{\textbf{k},\textbf{k}_2 - \textbf{q}} C^\dag_{\textbf{k}_1 + \textbf{q}} C_{\textbf{k}_2} C_{\textbf{k}_1} + \notag \\
&+ C^\dag_{\textbf{k}_1 + \textbf{q}} C^\dag_{\textbf{k}_2 - \textbf{q}} C_{\textbf{k}} C_{\textbf{k}_2} C_{\textbf{k}_1} - C^\dag_{\textbf{k}_1 + \textbf{q}} C^\dag_{\textbf{k}_2 - \textbf{q}} C_{\textbf{k}_2} C_{\textbf{k}_1} C_{\textbf{k}} ,
\end{align}
and from $C_{\textbf{k}} C_{\textbf{k}_2} C_{\textbf{k}_1} = - C_{\textbf{k}_2} C_{\textbf{k}} C_{\textbf{k}_1} = C_{\textbf{k}_2} C_{\textbf{k}_1} C_{\textbf{k}}$, we have
\begin{equation}
\left[ C_{\textbf{k}} , C^\dag_{\textbf{k}_1 + \textbf{q}} C^\dag_{\textbf{k}_2 - \textbf{q}} C_{\textbf{k}_2} C_{\textbf{k}_1} \right] = \delta_{\textbf{k},\textbf{k}_1 + \textbf{q}} C^\dag_{\textbf{k}_2 - \textbf{q}} C_{\textbf{k}_2} C_{\textbf{k}_1} - \delta_{\textbf{k},\textbf{k}_2 - \textbf{q}} C^\dag_{\textbf{k}_1 + \textbf{q}} C_{\textbf{k}_2} C_{\textbf{k}_1},
\end{equation}
\begin{align}
& \dfrac{1}{2V} \sum_{\substack{\textbf{k}_1,\textbf{k}_2, \\ \textbf{q} \neq \textbf{0}}} \dfrac{4 \pi q_e^2}{q^2} \left[ C_{\textbf{k}} , C^\dag_{\textbf{k}_1 + \textbf{q}} C^\dag_{\textbf{k}_2 - \textbf{q}} C_{\textbf{k}_2} C_{\textbf{k}_1} \right] = \notag \\
&= \dfrac{1}{2V} \sum_{\substack{\textbf{k}_1,\textbf{k}_2, \\ \textbf{q} \neq \textbf{0}}} \dfrac{4 \pi q_e^2}{q^2} \delta_{\textbf{k},\textbf{k}_1 + \textbf{q}} C^\dag_{\textbf{k}_2 - \textbf{q}} C_{\textbf{k}_2} C_{\textbf{k}_1} - \dfrac{1}{2V} \sum_{\substack{\textbf{k}_1,\textbf{k}_2, \\ \textbf{q} \neq \textbf{0}}} \dfrac{4 \pi q_e^2}{q^2} \delta_{\textbf{k},\textbf{k}_2 - \textbf{q}} C^\dag_{\textbf{k}_1 + \textbf{q}} C_{\textbf{k}_2} C_{\textbf{k}_1} = \notag \\
&= \dfrac{1}{2V} \sum_{\substack{\textbf{k}_2, \\ \textbf{q} \neq \textbf{0}}} \dfrac{4 \pi q_e^2}{q^2} C^\dag_{\textbf{k}_2 - \textbf{q}} C_{\textbf{k}_2} C_{\textbf{k} - \textbf{q}} - \dfrac{1}{2V} \sum_{\substack{\textbf{k}_1, \\ \textbf{q} \neq \textbf{0}}} \dfrac{4 \pi q_e^2}{q^2} C^\dag_{\textbf{k}_1 + \textbf{q}} C_{\textbf{k} + \textbf{q}} C_{\textbf{k}_1} = \notag \\
&= \dfrac{1}{2V} \sum_{\substack{\textbf{k}_2, \\ \textbf{q} \neq \textbf{0}}} \dfrac{4 \pi q_e^2}{q^2} C^\dag_{\textbf{k}_2 - \textbf{q}} C_{\textbf{k}_2} C_{\textbf{k} - \textbf{q}} + \dfrac{1}{2V} \sum_{\substack{\textbf{k}_1, \\ \textbf{q} \neq \textbf{0}}} \dfrac{4 \pi q_e^2}{q^2} C^\dag_{\textbf{k}_1 + \textbf{q}} C_{\textbf{k}_1} C_{\textbf{k} + \textbf{q}},
\end{align}
\begin{align}
\left[ C_{\textbf{k}} , \hat{\mathcal{H}} \right] = \mathcal{E}^{(0)}_{\textbf{k}} C_{\textbf{k}} + \sum_{\substack{\textbf{k}_2, \\ \textbf{q} \neq \textbf{0}}} \dfrac{4 \pi q_e^2}{q^2} C^\dag_{\textbf{k}_2 - \textbf{q}} C_{\textbf{k}_2} C_{\textbf{k} - \textbf{q}} + \sum_{\substack{\textbf{k}_1, \\ \textbf{q} \neq \textbf{0}}} \dfrac{4 \pi q_e^2}{q^2} C^\dag_{\textbf{k}_1 + \textbf{q}} C_{\textbf{k}_1} C_{\textbf{k} + \textbf{q}}. 
\end{align}
We now consider the anticommutator
\begin{equation}
\left\lbrace \left[ C_{\textbf{k}} , \mathcal{H} \right] , C^\dagger_{\textbf{k}} \right\rbrace = \mathcal{E}^{(0)}_{\textbf{k}} \left\lbrace C_{\textbf{k}} , C^\dagger_{\textbf{k}} \right\rbrace + \dfrac{1}{2V} \sum_{\substack{\textbf{k}_2, \\ \textbf{q} \neq \textbf{0}}} \dfrac{4 \pi q_e^2}{q^2} \left\lbrace C^\dag_{\textbf{k}_2 - \textbf{q}} C_{\textbf{k}_2} C_{\textbf{k} - \textbf{q}} , C^\dagger_{\textbf{k}} \right\rbrace + \dfrac{1}{2V} \sum_{\substack{\textbf{k}_1, \\ \textbf{q} \neq \textbf{0}}} \dfrac{4 \pi q_e^2}{q^2} \left\lbrace C^\dag_{\textbf{k}_1 + \textbf{q}} C_{\textbf{k}_1} C_{\textbf{k} + \textbf{q}} , C^\dagger_{\textbf{k}} \right\rbrace,
\end{equation}
where, of course, $\left\lbrace C_{\textbf{k}} , C^\dagger_{\textbf{k}} \right\rbrace = \mathds{1}$, and we need to compute the two anticommutators. Regarding the first one, 
\begin{align}
\left\lbrace C^\dag_{\textbf{k}_2 - \textbf{q}} C_{\textbf{k}_2} C_{\textbf{k} - \textbf{q}} , C^\dagger_{\textbf{k}} \right\rbrace &= C^\dag_{\textbf{k}_2 - \textbf{q}} C_{\textbf{k}_2} C_{\textbf{k} - \textbf{q}} C^\dagger_{\textbf{k}} + C^\dagger_{\textbf{k}} C^\dag_{\textbf{k}_2 - \textbf{q}} C_{\textbf{k}_2} C_{\textbf{k} - \textbf{q}} = \notag \\
&= C^\dag_{\textbf{k}_2 - \textbf{q}} C_{\textbf{k}_2} C_{\textbf{k} - \textbf{q}} C^\dagger_{\textbf{k}} - C^\dag_{\textbf{k}_2 - \textbf{q}} C^\dagger_{\textbf{k}} C_{\textbf{k}_2} C_{\textbf{k} - \textbf{q}} = \notag \\
&= C^\dag_{\textbf{k}_2 - \textbf{q}} C_{\textbf{k}_2} C_{\textbf{k} - \textbf{q}} C^\dagger_{\textbf{k}} - \delta_{\textbf{k},\textbf{k}_2} C^\dag_{\textbf{k}_2 - \textbf{q}}  C_{\textbf{k} - \textbf{q}} + \notag \\
&+ C^\dag_{\textbf{k}_2 - \textbf{q}} C_{\textbf{k}_2} C^\dagger_{\textbf{k}} C_{\textbf{k} - \textbf{q}} = \notag \\
&= C^\dag_{\textbf{k}_2 - \textbf{q}} C_{\textbf{k}_2} C_{\textbf{k} - \textbf{q}} C^\dagger_{\textbf{k}} - \delta_{\textbf{k},\textbf{k}_2} C^\dag_{\textbf{k}_2 - \textbf{q}}  C_{\textbf{k} - \textbf{q}} + \notag \\
&+ \delta_{\textbf{k},\textbf{k} - \textbf{q}} C^\dag_{\textbf{k}_2 - \textbf{q}} C_{\textbf{k}_2} - C^\dag_{\textbf{k}_2 - \textbf{q}} C_{\textbf{k}_2} C_{\textbf{k} - \textbf{q}} C^\dagger_{\textbf{k}} = \notag \\
&= - \delta_{\textbf{k},\textbf{k}_2} C^\dag_{\textbf{k}_2 - \textbf{q}}  C_{\textbf{k} - \textbf{q}} + \delta_{\textbf{k},\textbf{k} - \textbf{q}} C^\dag_{\textbf{k}_2 - \textbf{q}} C_{\textbf{k}_2},
\end{align}
\begin{equation}
\sum_{\substack{\textbf{k}_2, \\ \textbf{q} \neq \textbf{0}}} \dfrac{4 \pi q_e^2}{q^2} \left\lbrace C^\dag_{\textbf{k}_2 - \textbf{q}} C_{\textbf{k}_2} C_{\textbf{k} - \textbf{q}} , C^\dagger_{\textbf{k}} \right\rbrace = - \dfrac{4 \pi q_e^2}{q^2} \sum_{\substack{\textbf{k}_2, \\ \textbf{q} \neq \textbf{0}}}  \delta_{\textbf{k},\textbf{k}_2} C^\dag_{\textbf{k}_2 - \textbf{q}}  C_{\textbf{k} - \textbf{q}} + \sum_{\substack{\textbf{k}_2, \\ \textbf{q} \neq \textbf{0}}} \dfrac{4 \pi q_e^2}{q^2} \delta_{\textbf{k},\textbf{k} - \textbf{q}} C^\dag_{\textbf{k}_2 - \textbf{q}} C_{\textbf{k}_2},
\end{equation}
and in the second term, since $\textbf{q} \neq \textbf{0}$, the delta $\delta_{\textbf{k},\textbf{k} - \textbf{q}}$ is never nonzero, so
\begin{align}
\dfrac{1}{2V} \sum_{\substack{\textbf{k}_2, \\ \textbf{q} \neq \textbf{0}}} \dfrac{4 \pi q_e^2}{q^2} \left\lbrace C^\dag_{\textbf{k}_2 - \textbf{q}} C_{\textbf{k}_2} C_{\textbf{k} - \textbf{q}} , C^\dagger_{\textbf{k}} \right\rbrace &= - \sum_{\substack{\textbf{k}_2, \\ \textbf{q} \neq \textbf{0}}} \dfrac{4 \pi q_e^2}{q^2} \delta_{\textbf{k},\textbf{k}_2} C^\dag_{\textbf{k}_2 - \textbf{q}} C_{\textbf{k} - \textbf{q}} = \notag \\
&= - \dfrac{1}{2V} \sum_{\textbf{q} \neq \textbf{0}} \dfrac{4 \pi q_e^2}{q^2} C^\dag_{\textbf{k} - \textbf{q}} C_{\textbf{k} - \textbf{q}}.
\end{align}
Similarly, for the second anticommutator we have
\begin{align}
\left\lbrace C^\dag_{\textbf{k}_1 + \textbf{q}} C_{\textbf{k}_1} C_{\textbf{k} + \textbf{q}} , C^\dagger_{\textbf{k}} \right\rbrace &= C^\dag_{\textbf{k}_1 + \textbf{q}} C_{\textbf{k}_1} C_{\textbf{k} + \textbf{q}} C^\dagger_{\textbf{k}} + C^\dagger_{\textbf{k}} C^\dag_{\textbf{k}_1 + \textbf{q}} C_{\textbf{k}_1} C_{\textbf{k} + \textbf{q}} = \notag \\
&= C^\dag_{\textbf{k}_1 + \textbf{q}} C_{\textbf{k}_1} C_{\textbf{k} + \textbf{q}} C^\dagger_{\textbf{k}} - C^\dag_{\textbf{k}_1 + \textbf{q}} C^\dagger_{\textbf{k}} C_{\textbf{k}_1} C_{\textbf{k} + \textbf{q}} = \notag \\
&= C^\dag_{\textbf{k}_1 + \textbf{q}} C_{\textbf{k}_1} C_{\textbf{k} + \textbf{q}} C^\dagger_{\textbf{k}} - \delta_{\textbf{k},\textbf{k}_1} C^\dag_{\textbf{k}_1 + \textbf{q}}  C_{\textbf{k} + \textbf{q}} + \notag \\
&+ C^\dag_{\textbf{k}_1 + \textbf{q}} C_{\textbf{k}_1} C^\dagger_{\textbf{k}} C_{\textbf{k} + \textbf{q}} = \notag \\
&= C^\dag_{\textbf{k}_1 + \textbf{q}} C_{\textbf{k}_1} C_{\textbf{k} + \textbf{q}} C^\dagger_{\textbf{k}} - \delta_{\textbf{k},\textbf{k}_1} C^\dag_{\textbf{k}_1 + \textbf{q}}  C_{\textbf{k} + \textbf{q}} + \notag \\
&+ \delta_{\textbf{k},\textbf{k}+\textbf{q}} C^\dag_{\textbf{k}_1 + \textbf{q}} C_{\textbf{k}_1} - C^\dag_{\textbf{k}_1 + \textbf{q}} C_{\textbf{k}_1} C_{\textbf{k} + \textbf{q}}  C^\dagger_{\textbf{k}},
\end{align}
\begin{align}
\dfrac{1}{2V} \sum_{\substack{\textbf{k}_1, \\ \textbf{q} \neq \textbf{0}}} \dfrac{4 \pi q_e^2}{q^2} \left\lbrace C^\dag_{\textbf{k}_1 + \textbf{q}} C_{\textbf{k}_1} C_{\textbf{k} + \textbf{q}} , C^\dagger_{\textbf{k}} \right\rbrace &= - \dfrac{1}{2V} \sum_{\substack{\textbf{k}_1, \\ \textbf{q} \neq \textbf{0}}} \dfrac{4 \pi q_e^2}{q^2} \delta_{\textbf{k},\textbf{k}_1} C^\dag_{\textbf{k}_1 + \textbf{q}} C_{\textbf{k} + \textbf{q}} = \notag \\
&= - \dfrac{1}{2V} \sum_{\textbf{q} \neq \textbf{0}} \dfrac{4 \pi q_e^2}{q^2} C^\dag_{\textbf{k} + \textbf{q}} C_{\textbf{k} + \textbf{q}}.
\end{align}
Combining the results, we have
\begin{equation}
\left\lbrace \left[ C_{\textbf{k}} , \hat{\mathcal{H}} \right] , C^\dagger_{\textbf{k}} \right\rbrace = \mathcal{E}^{(0)}_{\textbf{k}} \mathds{1} - \dfrac{1}{2V} \sum_{\textbf{q} \neq \textbf{0}} \dfrac{4 \pi q_e^2}{q^2} \left( C^\dag_{\textbf{k} - \textbf{q}} C_{\textbf{k} - \textbf{q}} + C^\dag_{\textbf{k} + \textbf{q}} C_{\textbf{k} + \textbf{q}} \right),
\label{eq: momentoprimodifunzionespettraleHartreeFockdestra1}
\end{equation}
which can be further manipulated. Indeed, we set $\textbf{k}' = \textbf{k} - \textbf{q}$ in the first term, from which it follows that $\textbf{q} = \textbf{k} - \textbf{k}'$, and we set $\textbf{k}' = \textbf{k} + \textbf{q}$ in the second term, from which $\textbf{q} = \textbf{k}' - \textbf{k}$, thus
\begin{equation}
- \dfrac{1}{2V} \sum_{\textbf{q} \neq \textbf{0}} \dfrac{4 \pi q_e^2}{q^2} \left( C^\dag_{\textbf{k} - \textbf{q}} C_{\textbf{k} - \textbf{q}} + C^\dag_{\textbf{k} + \textbf{q}} C_{\textbf{k} + \textbf{q}} \right) = - \dfrac{1}{2V} 2 \sum_{\textbf{k}' \neq \textbf{k}} \dfrac{4 \pi q_e^2}{\left| \textbf{k} - \textbf{k}' \right|^2} C^\dag_{\textbf{k}'} C_{\textbf{k}'},
\end{equation}
and $\eqref{eq: momentoprimodifunzionespettraleHartreeFockdestra1}$ can be rewritten as
\begin{equation}
\left\lbrace \left[ C_{\textbf{k}} , \hat{\mathcal{H}} \right] , C^\dagger_{\textbf{k}} \right\rbrace = \mathcal{E}^{(0)}_{\textbf{k}} \mathds{1} - \dfrac{1}{V} \sum_{\textbf{k}' \neq \textbf{k}} \dfrac{4 \pi q_e^2}{\left| \textbf{k} - \textbf{k}' \right|^2} C^\dag_{\textbf{k}'} C_{\textbf{k}'}.
\label{eq: momentoprimodifunzionespettraleHartreeFockdestra2}
\end{equation}
We apply a thermal average to $\eqref{eq: momentoprimodifunzionespettraleHartreeFockdestra2}$ so that
\begin{align}
\left\langle \left\lbrace \left[ C_{\textbf{k}} , \hat{\mathcal{H}} \right] , C^\dagger_{\textbf{k}} \right\rbrace \right\rangle &= \mathcal{E}^{(0)}_{\textbf{k}} \langle \mathds{1} \rangle - \dfrac{1}{V} \sum_{\textbf{k}' \neq \textbf{k}} \dfrac{4 \pi q_e^2}{\left| \textbf{k} - \textbf{k}' \right|^2} \left\langle C^\dag_{\textbf{k}'} C_{\textbf{k}'} \right\rangle = \notag \\
&= \mathcal{E}^{(0)}_{\textbf{k}} - \dfrac{1}{V} \sum_{\textbf{k}' \neq \textbf{k}} n_{-1}\left(\mathcal{E}_{\textbf{k}'}\right) \dfrac{4 \pi q_e^2}{\left| \textbf{k} - \textbf{k}' \right|^2},
\label{eq: momentoprimodifunzionespettraleHartreeFockdestra3}
\end{align}
where we emphasize that the Fermi-Dirac distribution \( n_{-1}(\mathcal{E}_{\textbf{k}'}) \) refers to the eigenvalues \( \mathcal{E}_{\textbf{k}'} \) of the jellium model. Ultimately, by applying the thermodynamic limit approximation $\eqref{eq: illimitetermodinamico}$, we obtain
\begin{equation}
\left\langle \left\lbrace \left[ C_{\textbf{k}} , \hat{\mathcal{H}} \right] , C^\dagger_{\textbf{k}} \right\rbrace \right\rangle = \mathcal{E}^{(0)}_{\textbf{k}} - \pv \int \dfrac{d^3\textbf{k}'}{(2\pi)^3} n_{-1}\left(\mathcal{E}_{\textbf{k}'}\right) \dfrac{4 \pi q_e^2}{\left|\textbf{k}-\textbf{k}'\right|^2}.
\label{eq: momentoprimodifunzionespettraleHartreeFockdestra4}
\end{equation}
We divide the equation $\eqref{eq: momentoprimodifunzionespettraleHartreeFockdestra4}$ by $\hslash$, i.e.,
\begin{equation}
\dfrac{1}{\hslash} \left\langle \left\lbrace \left[ C_{\textbf{k}} , \hat{\mathcal{H}} \right] , C^\dagger_{\textbf{k}} \right\rbrace \right\rangle = \dfrac{\mathcal{E}^{(0)}_{\textbf{k}}}{\hslash} - \dfrac{1}{\hslash} \pv \int \dfrac{d^3\textbf{k}'}{(2\pi)^3} n_{-1}\left(\mathcal{E}_{\textbf{k}'}\right) \dfrac{4 \pi q_e^2}{\left|\textbf{k}-\textbf{k}'\right|^2},
\label{eq: momentoprimodifunzionespettraleHartreeFockdestra5}
\end{equation}
which is not equal to $\eqref{eq: momentoprimodifunzionespettraleHartreeFocksinistra}$. Indeed, $n_{-1}\left( \mathcal{E}_{\mathbf{k}'} \right)$ in $\eqref{eq: momentoprimodifunzionespettraleHartreeFockdestra5}$ is the Fermi-Dirac distribution evaluated at the eigenvalues of the jellium Hamiltonian; on the other hand, equation $\eqref{eq: momentoprimodifunzionespettraleHartreeFocksinistra}$ involves $n_{-1}\left( \mathcal{E}^{H.F.}_{\mathbf{k}'} \right)$, evaluated at the Hartree-Fock eigenvalues. Consequently, the first moment sum rule $\eqref{eq: momentoprimodifunzionespettrale}$ is not satisfied by the Hartree-Fock spectral function $A^{H.F.}_{\textbf{k}}(\omega)$, unlike the zeroth moment sum rule, and this discrepancy worsens as the order of the moments increases. \newline 
Let us then compute the energies $\eqref{eq: energieHartreeFockjellium}$ in a simple case at zero temperature, so that the Fermi-Dirac functions $n_{-1}\left(\mathcal{E}^{H.F.}_{\textbf{k}'}\right)$ are replaced by step functions and the integral can be evaluated in spherical polar coordinates. We have
\begin{align}
\left. \mathcal{E}_{\textbf{k}}^{H.F.} \right|_{T=0} &= \mathcal{E}^{(0)}_{\textbf{k}} - \int_0^{k_F}  \dfrac{2 \pi k'^2}{(2 \pi)^3} dk' \pv \int_{-1}^{+1} d \cos(\theta) \dfrac{4 \pi q_e^2}{k^2 +k'^2 - 2 k k' \cos(\theta)} = \notag \\
&= \mathcal{E}^{(0)}_{\textbf{k}} - \dfrac{q_e^2}{\pi} \int_0^{k_F} dk' \dfrac{k'}{k} \ln \left| \dfrac{k+k'}{k-k'} \right| = \notag \\
&= \mathcal{E}^{(0)}_{\textbf{k}} - \dfrac{q_e^2}{\pi} \left[ 1 + \dfrac{1-y^2}{2y} \ln \left| \dfrac{1+y}{1-y} \right| \right] = \notag \\
&= \mathcal{E}^{(0)}_{\textbf{k}} + \dfrac{q_e^2}{\pi} H(y),
\label{eq: energieHartreeFockjelliumT_0_1}
\end{align}
with
\begin{equation}
y = \dfrac{k}{k_F},
\end{equation}
\begin{equation}
H(y) = - \left[ 1 + \dfrac{1-y^2}{2y} \ln \left| \dfrac{1+y}{1-y} \right| \right],
\label{eq: H_y_energieHartreeFock}
\end{equation}
The asymptotic behaviors of $H(y)$ for the regimes $y \ll 1$, $y = 1$, and $y \gg 1$ correspond, respectively, to points inside the Fermi sphere, on the Fermi surface, and outside the Fermi sphere (see Chapter \ref{The Fermi momentum}). Employing the series expansion
\begin{equation}
\ln \left| \dfrac{1+y}{1-y} \right| = 2 \sum_{n=0}^{+\infty} \dfrac{y^{2n+1}}{2n+1},
\end{equation}
we note that equation $\eqref{eq: H_y_energieHartreeFock}$ satisfies
\begin{equation}
\lim_{y \rightarrow 0^+} H(y) = -2.
\label{eq: H_y_energieHartreeFock_limit_0}
\end{equation}
In addition,
\begin{equation}
\lim_{y \rightarrow +\infty} H(y) = 0,
\label{eq: H_y_energieHartreeFock_limit_infty}
\end{equation}
\begin{equation}
\lim_{y \rightarrow 1} H(y) = -1,
\label{eq: H_y_energieHartreeFock_limit_1}
\end{equation}
as shown in Figure $\eqref{fig: H_y_energieHartreeFock_figure}$. In addition, it can be readily verified that
\begin{equation}
\left. \dfrac{dS(y)}{dy} \right|_{y=1} \rightarrow +\infty
\end{equation}
Specifically, the derivative of the function exhibits a logarithmic divergence at $y = 1$, corresponding to the Fermi surface, indicating a very slow divergence. The Hartree-Fock energies in the jellium model at zero temperature, as given by equation $\eqref{eq: energieHartreeFockjelliumT_0_1}$, can be expressed equivalently as
\begin{equation}
\left. \mathcal{E}_{\textbf{k}}^{H.F.} \right|_{T=0} = \mathcal{E}^{(0)}_{\textbf{k}} + \Sigma(\textbf{k}),
\label{eq: energieHartreeFockjelliumT_0_2}
\end{equation}
where
\begin{equation}
\Sigma(\textbf{k}) = - \pv \int \dfrac{d^3\textbf{k}'}{(2\pi)^3} n_{-1}\left(\mathcal{E}^{H.F.}_{\textbf{k}'}\right) \dfrac{4 \pi q_e^2}{\left|\textbf{k}-\textbf{k}'\right|^2}
\label{eq: energiaselfenergyhartreefockjellium}
\end{equation}
is the self-energy in the Hartree-Fock model. Note that $\eqref{eq: energieHartreeFockjellium}$, based on the assumptions used here, is static energy. To analyze the relative behavior of the spectrum near the Fermi surface, we consider the difference
\begin{equation}
\mathcal{E}_{\textbf{k}}^{H.F.} - \mathcal{E}_{\textbf{k}_F}^{H.F.} = \left[ \mathcal{E}_{\textbf{k}}^{(0)} - \mathcal{E}_{\textbf{k}_F}^{(0)} \right] + \left[ \Sigma(\textbf{k}) - \Sigma(\textbf{k}_F) \right].
\end{equation}
We have
\begin{align}
\mathcal{E}_{\textbf{k}}^{(0)} - \mathcal{E}_{\textbf{k}_F}^{(0)} &= \dfrac{\hslash^2 \textbf{k}^2}{2m} - \dfrac{\hslash^2 \textbf{k}_F^2}{2m} = \notag \\
&= \dfrac{\hslash^2}{2m} (\textbf{k} + \textbf{k}_F) \cdot (\textbf{k} - \textbf{k}_F)
\end{align}
which, in the limit $\textbf{k} \rightarrow \textbf{k}_F$, can be approximated as
\begin{align}
\mathcal{E}_{\textbf{k}}^{(0)} - \mathcal{E}_{\textbf{k}_F}^{(0)} &\sim \dfrac{\hslash^2}{2m} 2 \textbf{k}_F \cdot (\textbf{k} - \textbf{k}_F) = \notag \\
&= \dfrac{\hslash \textbf{k}_F}{m} \cdot \left( \hslash \textbf{k} - \hslash \textbf{k}_F \right) \equiv \notag \\
& \equiv \textbf{v}_F \cdot \left( \hslash \textbf{k} - \hslash \textbf{k}_F \right), 
\end{align}
where we have explicitly defined the Fermi velocity
\begin{equation}
\textbf{v}_F = \dfrac{\hslash \textbf{k}_F}{m}.
\end{equation}
Furthermore, from
\begin{equation}
\Sigma(\textbf{k}) - \Sigma(\textbf{k}_F) \sim \left. \nabla_{\textbf{k}} \Sigma(\textbf{k}) \right|_{\textbf{k}=\textbf{k}_F} \cdot (\textbf{k} - \textbf{k}_F),
\end{equation} 
combining the results, we have
\begin{equation}
\left. \mathcal{E}_{\textbf{k}}^{H.F.} \right|_{T=0} \xrightarrow[\textbf{k} \sim \textbf{k}_F]{} \hslash (\textbf{k} - \textbf{k}_F) \cdot \textbf{v}_F^{H.F.},
\end{equation}
where
\begin{equation}
\textbf{v}_F^{H.F.} = \textbf{v}_F + \left. \nabla_{\textbf{k}} \Sigma(\textbf{k}) \right|_{\textbf{k}=\textbf{k}_F}
\end{equation}
is the Fermi velocity in the Hartree-Fock model. We have shown that near the Fermi surface, a linear dispersion relation holds. From $\eqref{eq: energieHartreeFockjelliumT_0_1}$ it follows
\begin{equation}
\left. \nabla_{\textbf{k}} \Sigma(\textbf{k}) \right|_{\textbf{k}=\textbf{k}_F} = \dfrac{q_e^2}{\pi} \left. \dfrac{dS(y)}{dy} \right|_{y=1},
\end{equation}
and thus 
\begin{equation}
\textbf{v}_F^{H.F.} \xrightarrow[\textbf{k} \to \textbf{k}_F]{} \infty. 
\end{equation}
Since the effective mass of Hartree-Fock quasiparticles can be derived from 
\begin{equation}
\textbf{p}^{H.F.} = m^{H.F.} \textbf{v}^{H.F.}, 
\end{equation}
and the Fermi momentum must remain constant, it follows that 
\begin{equation}
m^{H.F.} \xrightarrow[\textbf{k} \to \textbf{k}_F]{} 0. 
\end{equation}
Consequently, the quasiparticles exhibit a mass tending to zero and a velocity tending to infinity near the Fermi surface. \newline
Note that physical quantities derived within the Hartree-Fock framework, such as the specific heat, do not agree with experimental data. This discrepancy arises because the potential considered here is Coulomb potential, which is long-range, leading to logarithmic divergences not observed experimentally. The reason is that only first-order diagrams have been considered, substituting free propagators with the full Hartree-Fock propagator, while the potential itself remains unrenormalized. To account for the modification of the interaction, it is necessary to consider the medium's response to the interaction, or equivalently, to compute the dielectric function: the effective potential then becomes a short-range potential, as will be shown.
\section{Pair correlation function in the jellium model}
\begin{figure}[H]
\centering
\includegraphics[scale=0.78]{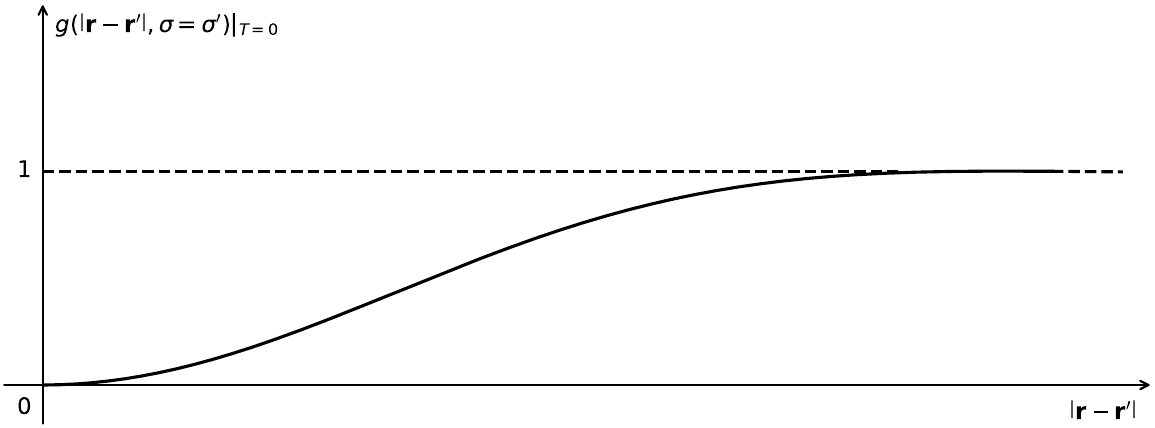}
\caption{Pair correlation function for electrons with parallel spin at zero temperature, as obtained within the Hartree-Fock approximation; see equation $\eqref{eq: funzionecorrelazionecoppiaHartreeFockspinuguali3}$.}
\label{fig: pair_correlation_function_equal_spin}
\end{figure}

We aim to quantify the pair correlation function $\eqref{eq: funzionecorrelazionedicoppia}$ within the Hartree-Fock framework for the jellium model, explicitly expressing the field operators in the plane-wave basis. The expectation value $\left\langle \psi^\dagger(\textbf{r}, s) \psi(\textbf{r}, s) \right\rangle$ represents the particle density per unit volume and per spin. Since we consider the jellium model, the medium is homogeneous, so electrons are evenly distributed between the two spin states, i.e., $\left\langle \psi^\dagger(\textbf{r},s) \psi(\textbf{r},s) \right\rangle = \dfrac{N}{2V}$, and the denominator of $\eqref{eq: funzionecorrelazionedicoppia}$ is
\begin{equation}
\left\langle \psi^\dagger(\textbf{r},s) \psi(\textbf{r},s) \right\rangle \left\langle \psi^\dagger(\textbf{r}',s') \psi(\textbf{r}',s') \right\rangle = \dfrac{N^2}{4 V^2}.
\end{equation}
As for the numerator of $\eqref{eq: funzionecorrelazionedicoppia}$, we have
\begin{align}
\left\langle \psi^\dagger(\textbf{r},s) \psi^\dagger(\textbf{r}',s') \psi(\textbf{r'},s') \psi(\textbf{r},s) \right\rangle &= \sum_{\textbf{k}_1,\sigma_1} \sum_{\textbf{k}_2,\sigma_2} \sum_{\textbf{k}_3,\sigma_3} \sum_{\textbf{k}_4,\sigma_4} \dfrac{e^{- i \textbf{k}_1 \cdot \textbf{r}}}{\sqrt{V}} \dfrac{e^{- i \textbf{k}_2 \cdot \textbf{r}'}}{\sqrt{V}} \dfrac{e^{i \textbf{k}_3 \cdot \textbf{r}'}}{\sqrt{V}} \dfrac{e^{i \textbf{k}_4 \cdot \textbf{r}}}{\sqrt{V}} \chi^*_{\sigma_1}(s) \chi^*_{\sigma_2}(s') \chi_{\sigma_3}(s') \chi_{\sigma_4}(s) \notag \\
& \ \ \ \ \ \left\langle C^\dagger_{\textbf{k}_1,\sigma_1} C^\dagger_{\textbf{k}_2,\sigma_2} C_{\textbf{k}_3,\sigma_3} C_{\textbf{k}_4,\sigma_4} \right\rangle = \notag \\
&= \dfrac{1}{V^2} \sum_{\textbf{k}_1,\textbf{k}_2,\textbf{k}_3,\textbf{k}_4} \sum_{\sigma_1,\sigma_2} e^{- i (\textbf{k}_1 - \textbf{k}_4) \cdot \textbf{r}} e^{- i (\textbf{k}_2 - \textbf{k}_3) \cdot \textbf{r}'} \delta_{\sigma_1,\sigma_4} \delta_{\sigma_2,\sigma_3} \notag \\
& \ \ \ \ \ \left\langle C^\dagger_{\textbf{k}_1,\sigma_1} C^\dagger_{\textbf{k}_2,\sigma_2} C_{\textbf{k}_3,\sigma_3} C_{\textbf{k}_4,\sigma_4} \right\rangle = \notag \\
&= \dfrac{1}{V^2} \sum_{\textbf{k}_1,\textbf{k}_2,\textbf{k}_3,\textbf{k}_4}  e^{- i (\textbf{k}_1 - \textbf{k}_4) \cdot \textbf{r}} e^{- i (\textbf{k}_2 - \textbf{k}_3) \cdot \textbf{r}'} \left\langle C^\dagger_{\textbf{k}_1,\sigma_4} C^\dagger_{\textbf{k}_2,\sigma_3} C_{\textbf{k}_3,\sigma_3} C_{\textbf{k}_4,\sigma_4} \right\rangle,
\end{align}
and by relabeling the dummy spin indices, we obtain
\begin{equation}
\left\langle \psi^\dagger(\textbf{r},s) \psi^\dagger(\textbf{r}',s') \psi(\textbf{r'},s') \psi(\textbf{r},s) \right\rangle = \dfrac{1}{V^2} \sum_{\textbf{k}_1,\textbf{k}_2,\textbf{k}_3,\textbf{k}_4}  e^{- i (\textbf{k}_1 - \textbf{k}_4) \cdot \textbf{r}} e^{- i (\textbf{k}_2 - \textbf{k}_3) \cdot \textbf{r}'} \left\langle C^\dagger_{\textbf{k}_1,\sigma} C^\dagger_{\textbf{k}_2,\sigma'} C_{\textbf{k}_3,\sigma'} C_{\textbf{k}_4,\sigma} \right\rangle.
\end{equation}
Now, since the jellium model describes a system that is invariant under spatial translations, the pair correlation function depends only on the relative coordinate between particles, not on their individual positions, and total momentum is conserved. We therefore consider the term $e^{-i (\mathbf{k}_1 - \mathbf{k}_4) \cdot \mathbf{r}} , e^{-i (\mathbf{k}_2 - \mathbf{k}_3) \cdot \mathbf{r}'}$ and set $\mathbf{r} = \mathbf{r}' + \mathbf{r}''$, yielding
\begin{align}
e^{-i (\mathbf{k}_1 - \mathbf{k}_4) \cdot \mathbf{r}} e^{-i (\mathbf{k}_2 - \mathbf{k}_3) \cdot \mathbf{r}'} &= e^{-i (\mathbf{k}_1 - \mathbf{k}_4) \cdot (\mathbf{r}' + \mathbf{r}'')} \, e^{-i (\mathbf{k}_2 - \mathbf{k}_3) \cdot \mathbf{r}'} = \notag \\
&= e^{-i (\mathbf{k}_1 - \mathbf{k}_4) \cdot \mathbf{r}'} e^{-i (\mathbf{k}_1 - \mathbf{k}_4) \cdot \mathbf{r}''} \, e^{-i (\mathbf{k}_2 - \mathbf{k}_3) \cdot \mathbf{r}'} = \notag \\
&= e^{-i (\mathbf{k}_1 - \mathbf{k}_4) \cdot \mathbf{r}''} e^{-i (\mathbf{k}_1 + \mathbf{k}_2 - \mathbf{k}_3 - \mathbf{k}_4) \cdot \mathbf{r}'}.
\end{align}
Since total momentum must be conserved, we set $\textbf{k}_1 = \textbf{k}_4 + \textbf{q}$ and $\textbf{k}_2 = \textbf{k}_3 - \textbf{q}$, so that $\textbf{k}_3$, $\textbf{k}_4$, and $\textbf{q}$ are independent variables, while $\textbf{k}_1$ and $\textbf{k}_2$ are determined by $\textbf{k}_3$, $\textbf{k}_4$, and $\textbf{q}$, exactly as we did in the jellium model calculation using second quantization. We have
\begin{align}
& \left\langle \psi^\dagger(\textbf{r},s) \psi^\dagger(\textbf{r}',s') \psi(\textbf{r'},s') \psi(\textbf{r},s) \right\rangle = \notag \\
&= \dfrac{1}{V^2} \sum_{\textbf{k}_1,\textbf{k}_2,\textbf{k}_3,\textbf{k}_4}  e^{- i (\textbf{k}_1 - \textbf{k}_4) \cdot \textbf{r}''} e^{-i (\mathbf{k}_1 + \mathbf{k}_2 - \mathbf{k}_3 - \mathbf{k}_4) \cdot \mathbf{r}'} \left\langle C^\dagger_{\textbf{k}_1,\sigma} C^\dagger_{\textbf{k}_2,\sigma'} C_{\textbf{k}_3,\sigma'} C_{\textbf{k}_4,\sigma} \right\rangle = \notag \\
&= \dfrac{1}{V^2} \sum_{\textbf{k}_3,\textbf{k}_4,\textbf{q}} e^{- i \textbf{q} \cdot ( \textbf{r} - \textbf{r}' )} \left\langle C^\dagger_{\textbf{k}_4 + \textbf{q},\sigma} C^\dagger_{\textbf{k}_3 - \textbf{q},\sigma'} C_{\textbf{k}_3,\sigma'} C_{\textbf{k}_4,\sigma} \right\rangle,
\end{align}
and by renaming the dummy momentum indices, we obtain
\begin{equation}
\left\langle \psi^\dagger(\textbf{r},s) \psi^\dagger(\textbf{r}',s') \psi(\textbf{r'},s') \psi(\textbf{r},s) \right\rangle = \dfrac{1}{V^2} \sum_{\textbf{k}_1,\textbf{k}_2,\textbf{q}} e^{- i \textbf{q} \cdot ( \textbf{r} - \textbf{r}' )} \left\langle C^\dagger_{\textbf{k}_1 + \textbf{q},\sigma} C^\dagger_{\textbf{k}_2 - \textbf{q},\sigma'} C_{\textbf{k}_2,\sigma'} C_{\textbf{k}_1,\sigma} \right\rangle.
\end{equation}
Dividing this numerator by the denominator, we obtain
\begin{align}
g(\textbf{r}-\textbf{r}',\sigma,\sigma') &= \dfrac{4V^2}{N^2} \dfrac{1}{V^2} \sum_{\textbf{k}_1,\textbf{k}_2,\textbf{q}} e^{- i \textbf{q} \cdot ( \textbf{r} - \textbf{r}' )} \left\langle C^\dagger_{\textbf{k}_1 + \textbf{q},\sigma} C^\dagger_{\textbf{k}_2 - \textbf{q},\sigma'} C_{\textbf{k}_2,\sigma'} C_{\textbf{k}_1,\sigma} \right\rangle = \notag \\
&= \dfrac{4}{N^2} \sum_{\textbf{k}_1,\textbf{k}_2,\textbf{q}} e^{- i \textbf{q} \cdot ( \textbf{r} - \textbf{r}' )} \left\langle C^\dagger_{\textbf{k}_1 + \textbf{q},\sigma} C^\dagger_{\textbf{k}_2 - \textbf{q},\sigma'} C_{\textbf{k}_2,\sigma'} C_{\textbf{k}_1,\sigma} \right\rangle,
\label{eq:  funzionecorrelazionecoppiaHartreeFock}
\end{align}
where the thermal average is to be taken over the jellium Hamiltonian. We adopt a perturbative approach and compute this average with respect to the free-fermion Hamiltonian, which allows us to apply Wick’s theorem. We make the identifications $1' \equiv C^\dagger_{\textbf{k}_1 + \textbf{q},\sigma}$, $2' \equiv C^\dagger_{\textbf{k}_2 - \textbf{q},\sigma'}$, $3' \equiv C_{\textbf{k}_2,\sigma'}$, $4' \equiv C_{\textbf{k}_1,\sigma}$. We begin by analyzing the term involving different spin components, \( \sigma \neq \sigma' \), corresponding to the direct Hartree-Fock contribution previously computed: operator $1'$ can be contracted only with $4'$, and similarly $2'$ only with $3'$, yielding
\begin{align}
g \left( \textbf{r}-\textbf{r}',\sigma \neq \sigma' \right)^{(0)} &= \dfrac{4}{N^2} \sum_{\textbf{k}_1,\textbf{k}_2,\textbf{q}} e^{- i \textbf{q} \cdot ( \textbf{r} - \textbf{r}' )} \delta_{\textbf{q},\textbf{0}} \left\langle \hat{N}_{\textbf{k}_1,\sigma} \right\rangle \left\langle \hat{N}_{\textbf{k}_2,\sigma'} \right\rangle = \notag \\
&= \dfrac{4}{N^2} \dfrac{N}{2} \dfrac{N}{2} = \notag \\
&= 1.
\label{eq:  funzionecorrelazionecoppiaHartreeFockspindiversi}
\end{align}
Accordingly, to lowest order, two fermions with opposite spin projections are completely uncorrelated, as expected. This result follows directly from the fact that the reference Hamiltonian $\hat{\mathcal{H}}_0$ describes a system of non-interacting particles, where the absence of interactions implies that the connected part of the pair correlation function vanishes. We now evaluate the zeroth-order contribution to the pair correlation function for fermions with identical spin projections. This case corresponds to the exchange term in the Hartree-Fock approximation. For particles with the same spin, the antisymmetry of the fermionic wavefunction leads to nontrivial correlations even in the absence of interactions. In computing the four-point Green's function using Wick's theorem, two distinct contractions contribute to the result. The first contraction, $1'-4'$ and $2'-3'$, reproduces the previously derived direct term. The second contraction, $1'-3'$ and $2'-4'$, yields an additional exchange contribution. This exchange term is accompanied by a minus sign, $\varepsilon = -1$, reflecting the fermionic anticommutation relations. Therefore, the full zeroth-order contribution to the pair correlation function for particles with the same spin includes both the direct and exchange terms. The exchange term reduces the probability of finding two identical fermions at nearby positions, as required by the Pauli exclusion principle and the antisymmetric nature of the many-body wavefunction. We have
\begin{align}
g\left(\textbf{r}-\textbf{r}',\sigma=\sigma'\right)^{(0)} &= 1 - \dfrac{4}{N^2} \sum_{\textbf{k}_1,\textbf{k}_2,\textbf{q}} e^{- i \textbf{q} \cdot ( \textbf{r} - \textbf{r}' )} \delta_{\textbf{k}_1+\textbf{q},\textbf{k}_2} \delta_{\textbf{k}_2 - \textbf{q},\textbf{k}_1} \left\langle C^\dagger_{\textbf{k}_1 + \textbf{q},\sigma}  C_{\textbf{k}_2,\sigma} \right\rangle \left\langle C^\dagger_{\textbf{k}_2 - \textbf{q},\sigma} C_{\textbf{k}_1,\sigma} \right\rangle = \notag \\
&= 1 - \dfrac{4}{N^2} \sum_{\textbf{k}_1,\textbf{k}_2} e^{- i (\textbf{k}_2 - \textbf{k}_1) \cdot ( \textbf{r} - \textbf{r}' )} \left\langle C^\dagger_{\textbf{k}_2,\sigma}  C_{\textbf{k}_2,\sigma} \right\rangle \left\langle C^\dagger_{\textbf{k}_1,\sigma} C_{\textbf{k}_1,\sigma} \right\rangle = \notag \\
&= 1 - \dfrac{4}{N^2} \left( \sum_{\textbf{k}_1} e^{i \textbf{k}_1 \cdot ( \textbf{r} - \textbf{r}' )} \left\langle C^\dagger_{\textbf{k}_1,\sigma} C_{\textbf{k}_1,\sigma} \right\rangle \right) \left( \sum_{\textbf{k}_2} e^{- i \textbf{k}_2 \cdot ( \textbf{r} - \textbf{r}' )} \left\langle C^\dagger_{\textbf{k}_2,\sigma} C_{\textbf{k}_2,\sigma} \right\rangle \right) = \notag \\
&= 1 - \dfrac{4}{N^2} \left( \sum_{\textbf{k}_1} e^{- i \textbf{k}_1 \cdot ( \textbf{r} - \textbf{r}' )} \left\langle C^\dagger_{- \textbf{k}_1,\sigma} C_{-\textbf{k}_1,\sigma} \right\rangle \right) \left( \sum_{\textbf{k}_2} e^{- i \textbf{k}_2 \cdot ( \textbf{r} - \textbf{r}' )} \left\langle C^\dagger_{\textbf{k}_2,\sigma} C_{\textbf{k}_2,\sigma} \right\rangle \right),
\end{align}
where we have rewritten the sum over $-\textbf{k}_1$ instead of $\textbf{k}_1$. Since the jellium model satisfies the relation $C^\dagger_{- \textbf{k}_1,\sigma} C_{-\textbf{k}_1,\sigma} = C^\dagger_{\textbf{k}_1,\sigma} C_{\textbf{k}_1,\sigma}$, and the summation indices are dummy variables, we have
\begin{align}
g \left( \textbf{r}-\textbf{r}',\sigma=\sigma' \right)^{(0)} &= 1 - \dfrac{4}{N^2} \left( \sum_{\textbf{k}} e^{- i \textbf{k} \cdot ( \textbf{r} - \textbf{r}' )} \left\langle \hat{N}_{\textbf{k},\sigma} \right\rangle \right)^2 \notag \\
&= 1 - \dfrac{4}{N^2} \left( \sum_{\textbf{k}} e^{- i \textbf{k} \cdot ( \textbf{r} - \textbf{r}' )} n_{-1}\left(\mathcal{E}_{\textbf{k},\sigma}\right) \right)^2,
\label{eq: funzionecorrelazionecoppiaHartreeFockspinuguali1}
\end{align}
where the Fermi-Dirac statistics \( n_{-1}(\mathcal{E}_{\textbf{k},\sigma}) \) refers to the eigenvalues \( \mathcal{E}_{\textbf{k},\sigma} \) of the jellium model. Now, the pair correlation function is generally less than 1, reflecting the consequences of Fermi-Dirac statistics. We now evaluate $\eqref{eq: funzionecorrelazionecoppiaHartreeFockspinuguali1}$ in the thermodynamic limit approximation $\eqref{eq: illimitetermodinamico}$, in particular at zero temperature, where the Fermi-Dirac functions $n_{-1}\left(\mathcal{E}_{\textbf{k},\sigma}\right)$ reduce to step functions with respect to the Fermi momentum $\textbf{k}_F$. We have
\begin{align}
g\left(\textbf{r}-\textbf{r}',\sigma=\sigma'\right)^{(0)} &= 1 - \dfrac{4}{N^2} \left( \dfrac{V}{(2 \pi)^3} \int d^3 \textbf{k} e^{- i \textbf{k} \cdot ( \textbf{r} - \textbf{r}' )} \Theta(\textbf{k}-\textbf{k}_F) \right)^2 = \notag \\
&= 1 - \dfrac{4}{N^2} \left( \dfrac{V}{(2 \pi)^3} \int_0^{k_F} 2 \pi k^2 dk \int_{-1}^{+1} d\cos(\theta) e^{- i k \left| \textbf{r} - \textbf{r}' \right| \cos(\theta)} \right)^2 = \notag \\
&= 1 - \dfrac{4}{N^2} \left( \dfrac{V}{4 \pi^2} \int_0^{k_F} k^2 dk \dfrac{2 \sin(k \left| \textbf{r} - \textbf{r}' \right|)}{k \left| \textbf{r} - \textbf{r}' \right|} \right)^2 = \notag \\
&= 1 - \dfrac{4}{N^2} \left( \dfrac{V}{2 \pi^2} \dfrac{1}{\left| \textbf{r} - \textbf{r}' \right|} \int_0^{k_F} k \sin \left( k \left| \textbf{r} - \textbf{r}' \right| \right) dk \right)^2,
\label{eq: funzionecorrelazionecoppiaHartreeFockspinuguali2}
\end{align}
then $g\left(\textbf{r}-\textbf{r}',\sigma=\sigma'\right)^{(0)} = g\left(\left| \textbf{r}-\textbf{r}' \right|,\sigma=\sigma' \right)^{(0)}$. From
\begin{equation}
\int_0^{k_F} k \sin \left( k \left| \textbf{r} - \textbf{r}' \right| \right) dk = \dfrac{\sin(k_F \left| \textbf{r} - \textbf{r}' \right|)}{\left| \textbf{r} - \textbf{r}' \right|^2} - \dfrac{k_F \cos(k_F \left| \textbf{r} - \textbf{r}' \right|)}{\left| \textbf{r} - \textbf{r}' \right|},
\end{equation}
it follows
\begin{align}
g(\left| \textbf{r}-\textbf{r}' \right|,\sigma=\sigma')^{(0)} &= 1 - \dfrac{4}{N^2} \left\lbrace \dfrac{V}{2 \pi^2} \dfrac{1}{\left| \textbf{r} - \textbf{r}' \right|^3} \left[ \sin \left( k_F \left| \textbf{r} - \textbf{r}' \right| \right) - k_F \left| \textbf{r} - \textbf{r}' \right| \cos \left( k_F \left| \textbf{r} - \textbf{r}' \right| \right) \right] \right\rbrace^2 = \notag \\
&= 1 - \dfrac{V^2}{N^2 \pi^4} \dfrac{1}{\left| \textbf{r} - \textbf{r}' \right|^6} \left[ \sin \left( k_F \left| \textbf{r} - \textbf{r}' \right| \right) - k_F \left| \textbf{r} - \textbf{r}' \right| \cos \left( k_F \left| \textbf{r} - \textbf{r}' \right| \right) \right]^2 = \notag \\
&= 1 - \dfrac{9}{k_F^6 \left| \textbf{r} - \textbf{r}' \right|^6} \left[ \sin \left( k_F \left| \textbf{r} - \textbf{r}' \right| \right) - k_F \left| \textbf{r} - \textbf{r}' \right| \cos \left( k_F \left| \textbf{r} - \textbf{r}' \right| \right) \right]^2,
\label{eq: funzionecorrelazionecoppiaHartreeFockspinuguali3}
\end{align}
where in the last step we have explicitly expressed the density of non-interacting particles and employed equation $\eqref{eq: relazionedensitafermionimomentoFermid3}$. The function $\eqref{eq: funzionecorrelazionecoppiaHartreeFockspinuguali3}$ is illustrated in Figure $\eqref{fig: pair_correlation_function_equal_spin}$. We note that it satisfies
\begin{align}
\lim_{\left| \textbf{r} - \textbf{r}' \right| \rightarrow +\infty} g \left(\left| \textbf{r}-\textbf{r}' \right|,\sigma=\sigma' \right)^{(0)} &= \lim_{\left| \textbf{r} - \textbf{r}' \right| \rightarrow +\infty} 1 - \dfrac{9}{k_F^6 \left| \textbf{r} - \textbf{r}' \right|^6} \left[ \sin \left( k_F \left| \textbf{r} - \textbf{r}' \right| \right) - k_F \left| \textbf{r} - \textbf{r}' \right| \cos \left( k_F \left| \textbf{r} - \textbf{r}' \right| \right) \right]^2 = \notag \\
&= 1,
\end{align}
as expected, two electrons with the same spin are uncorrelated at large distances. On the other hand,
\begin{equation}
\lim_{\left| \textbf{r} - \textbf{r}' \right| \rightarrow 0^+} g\left( \left| \textbf{r}-\textbf{r}' \right|,\sigma=\sigma' \right)^{(0)} = \lim_{\left| \textbf{r} - \textbf{r}' \right| \rightarrow 0^+} \left\lbrace 1 - \dfrac{9}{k_F^6 \left| \textbf{r} - \textbf{r}' \right|^6} \left[ \sin \left( k_F \left| \textbf{r} - \textbf{r}' \right| \right) - k_F \left| \textbf{r} - \textbf{r}' \right| \cos \left( k_F \left| \textbf{r} - \textbf{r}' \right| \right) \right]^2 \right\rbrace,
\end{equation}
for convenience, we define $x = k_F \left| \textbf{r} - \textbf{r}' \right|$ and we get
\begin{align}
\lim_{x \rightarrow 0^+} \left\lbrace 1 - \dfrac{9}{x^6} \left[ \sin(x) - x \cos(x) \right]^2 \right\rbrace &= \lim_{x \rightarrow 0^+}  \left\lbrace 1 - \dfrac{9}{x^6} \left[ x - \dfrac{x^3}{6} + o(x^3) - x \left( 1 - \dfrac{x^2}{2} + o(x^2) \right) \right]^2 \right\rbrace = \notag \\
&= \lim_{x \rightarrow 0^+} \left\lbrace 1 - \dfrac{9}{x^6} \left[ - \dfrac{x^3}{6} + \dfrac{x^3}{2} + o(x^3) \right]^2 \right\rbrace = \notag \\
&=  \lim_{x \rightarrow 0^+} \left\lbrace 1 - \dfrac{9}{x^6} \left( \dfrac{x^6}{9} + o(x^6) \right) \right\rbrace,
\end{align}
then
\begin{equation}
\lim_{\left| \textbf{r} - \textbf{r}' \right| \rightarrow 0^+} g\left(\left| \textbf{r}-\textbf{r}' \right|,\sigma=\sigma'\right)^{(0)} = 0,
\end{equation}
and this result also follows from the Pauli principle, since it is impossible to find two particles at the same position with the same spin. This consistency was to be expected, given that we considered Slater determinants as eigenstates, and that we performed thermal averages on a Hamiltonian of non-interacting electrons. \newline
From $\eqref{eq:  funzionecorrelazionecoppiaHartreeFock}$ we define
\begin{align}
\tilde{g}(\textbf{r}-\textbf{r}') &= \dfrac{1}{4} \sum_{\sigma,\sigma'} g(\textbf{r}-\textbf{r}',\sigma,\sigma') = \notag \\
&= \dfrac{1}{N^2} \sum_{\sigma,\sigma'} \sum_{\textbf{k}_1,\textbf{k}_2,\textbf{q}} e^{- i \textbf{q} \cdot ( \textbf{r} - \textbf{r}' )} \left\langle C^\dagger_{\textbf{k}_1 + \textbf{q},\sigma} C^\dagger_{\textbf{k}_2 - \textbf{q},\sigma'} C_{\textbf{k}_2,\sigma'} C_{\textbf{k}_1,\sigma} \right\rangle,
\end{align}
which is the probability of finding two electrons at a distance $\textbf{r} - \textbf{r}'$, regardless of their spin. We want to calculate this quantity with respect to the full Hamiltonian, including interactions. We separate the sum over $\textbf{q}$ as follows
\begin{equation}
\tilde{g}(\textbf{r}-\textbf{r}') = \dfrac{1}{N^2} \sum_{\sigma,\sigma'} \sum_{\textbf{k}_1,\textbf{k}_2} \left\langle C^\dagger_{\textbf{k}_1,\sigma} C^\dagger_{\textbf{k}_2,\sigma'} C_{\textbf{k}_2,\sigma'} C_{\textbf{k}_1,\sigma} \right\rangle + \dfrac{1}{N^2} \sum_{\sigma,\sigma'} \sum_{\textbf{k}_1,\textbf{k}_2,\textbf{q} \neq \textbf{0}} e^{- i \textbf{q} \cdot ( \textbf{r} - \textbf{r}' )} \left\langle C^\dagger_{\textbf{k}_1 + \textbf{q},\sigma} C^\dagger_{\textbf{k}_2 - \textbf{q},\sigma'} C_{\textbf{k}_2,\sigma'} C_{\textbf{k}_1,\sigma} \right\rangle,
\end{equation}
and we exploit
\begin{align}
C^\dagger_{\textbf{k}_1,\sigma} C^\dagger_{\textbf{k}_2,\sigma'} C_{\textbf{k}_2,\sigma'} C_{\textbf{k}_1,\sigma} &= - C^\dagger_{\textbf{k}_1,\sigma} C^\dagger_{\textbf{k}_2,\sigma'} C_{\textbf{k}_1,\sigma} C_{\textbf{k}_2,\sigma'} = \notag \\
&= - C^\dagger_{\textbf{k}_1,\sigma} \left( \delta_{\textbf{k}_2,\textbf{k}_1} \delta_{\sigma',\sigma} - C_{\textbf{k}_1,\sigma} C^\dagger_{\textbf{k}_2,\sigma'} \right) C_{\textbf{k}_2,\sigma'} 
\end{align}
in the first term, and we have
\begin{align}
\tilde{g}(\textbf{r}-\textbf{r}') &= - \dfrac{1}{N^2} \sum_{\sigma,\sigma'} \sum_{\textbf{k}_1,\textbf{k}_2} \delta_{\textbf{k}_2,\textbf{k}_1} \delta_{\sigma',\sigma} \left\langle C^\dagger_{\textbf{k}_1,\sigma} C_{\textbf{k}_2,\sigma'} \right\rangle + \dfrac{1}{N^2} \sum_{\sigma,\sigma'} \sum_{\textbf{k}_1,\textbf{k}_2} \left\langle C^\dagger_{\textbf{k}_1,\sigma} C_{\textbf{k}_1,\sigma} C^\dagger_{\textbf{k}_2,\sigma'} C_{\textbf{k}_2,\sigma'}  \right\rangle + \notag \\
&+ \dfrac{1}{N^2} \sum_{\sigma,\sigma'} \sum_{\textbf{k}_1,\textbf{k}_2,\textbf{q} \neq \textbf{0}} e^{- i \textbf{q} \cdot ( \textbf{r} - \textbf{r}' )} \left\langle C^\dagger_{\textbf{k}_1 + \textbf{q},\sigma} C^\dagger_{\textbf{k}_2 - \textbf{q},\sigma'} C_{\textbf{k}_2,\sigma'} C_{\textbf{k}_1,\sigma} \right\rangle = \notag \\
&= - \dfrac{1}{N^2} \sum_{\sigma} \sum_{\textbf{k}_1} \left\langle C^\dagger_{\textbf{k}_1,\sigma} C_{\textbf{k}_1,\sigma} \right\rangle + \dfrac{1}{N^2} \sum_{\sigma,\sigma'} \sum_{\textbf{k}_1,\textbf{k}_2} \left\langle C^\dagger_{\textbf{k}_1,\sigma} C_{\textbf{k}_1,\sigma} C^\dagger_{\textbf{k}_2,\sigma'} C_{\textbf{k}_2,\sigma'}  \right\rangle + \notag \\
&+ \dfrac{1}{N^2} \sum_{\sigma,\sigma'} \sum_{\textbf{k}_1,\textbf{k}_2,\textbf{q} \neq \textbf{0}} e^{- i \textbf{q} \cdot ( \textbf{r} - \textbf{r}' )} \left\langle C^\dagger_{\textbf{k}_1 + \textbf{q},\sigma} C^\dagger_{\textbf{k}_2 - \textbf{q},\sigma'} C_{\textbf{k}_2,\sigma'} C_{\textbf{k}_1,\sigma} \right\rangle = \notag \\
&= - \dfrac{1}{N} + \dfrac{1}{N^2} \sum_{\sigma,\sigma'} \sum_{\textbf{k}_1,\textbf{k}_2} \left\langle C^\dagger_{\textbf{k}_1,\sigma} C_{\textbf{k}_1,\sigma} C^\dagger_{\textbf{k}_2,\sigma'} C_{\textbf{k}_2,\sigma'}  \right\rangle + \notag \\
&+ \dfrac{1}{N^2} \sum_{\sigma,\sigma'} \sum_{\textbf{k}_1,\textbf{k}_2,\textbf{q} \neq \textbf{0}} e^{- i \textbf{q} \cdot ( \textbf{r} - \textbf{r}' )} \left\langle C^\dagger_{\textbf{k}_1 + \textbf{q},\sigma} C^\dagger_{\textbf{k}_2 - \textbf{q},\sigma'} C_{\textbf{k}_2,\sigma'} C_{\textbf{k}_1,\sigma} \right\rangle = \notag \\
&= - \dfrac{1}{N} + \dfrac{1}{N^2} \sum_{\sigma,\sigma'} \sum_{\textbf{k}_1,\textbf{k}_2} \left\langle C^\dagger_{\textbf{k}_1,\sigma} C_{\textbf{k}_1,\sigma} \right\rangle \left\langle C^\dagger_{\textbf{k}_2,\sigma'} C_{\textbf{k}_2,\sigma'}  \right\rangle + \notag \\
&+ \dfrac{1}{N^2} \sum_{\sigma,\sigma'} \sum_{\textbf{k}_1,\textbf{k}_2,\textbf{q} \neq \textbf{0}} e^{- i \textbf{q} \cdot ( \textbf{r} - \textbf{r}' )} \left\langle C^\dagger_{\textbf{k}_1 + \textbf{q},\sigma} C^\dagger_{\textbf{k}_2 - \textbf{q},\sigma'} C_{\textbf{k}_2,\sigma'} C_{\textbf{k}_1,\sigma} \right\rangle,
\end{align}
where in the last step we used that the states are uncorrelated, allowing the factorization of the expectation value as $\left\langle C^\dagger_{\textbf{k}_1 + \textbf{q},\sigma} C^\dagger_{\textbf{k}_2 - \textbf{q},\sigma'} C_{\textbf{k}_2,\sigma'} C_{\textbf{k}_1,\sigma} \right\rangle = \left\langle C^\dagger_{\textbf{k}_1,\sigma} C_{\textbf{k}_1,\sigma} \right\rangle \left\langle C^\dagger_{\textbf{k}_2,\sigma'} C_{\textbf{k}_2,\sigma'}  \right\rangle$. Integrating both sides over position, we obtain
\begin{align}
\int d(\textbf{r}-\textbf{r}') \tilde{g}(\textbf{r}-\textbf{r}') &= - \dfrac{1}{N} \int d(\textbf{r}-\textbf{r}') + \dfrac{1}{N^2} \sum_{\sigma,\sigma'} \sum_{\textbf{k}_1,\textbf{k}_2}  \left\langle C^\dagger_{\textbf{k}_1,\sigma} C_{\textbf{k}_1,\sigma} \right\rangle \left\langle C^\dagger_{\textbf{k}_2,\sigma'} C_{\textbf{k}_2,\sigma'}  \right\rangle \int d(\textbf{r}-\textbf{r}') + \notag \\
&+ \dfrac{1}{N^2} \sum_{\sigma,\sigma'} \sum_{\textbf{k}_1,\textbf{k}_2,\textbf{q} \neq \textbf{0}} \left( \int d(\textbf{r}-\textbf{r}') e^{- i \textbf{q} \cdot ( \textbf{r} - \textbf{r}' )} \right) \left\langle C^\dagger_{\textbf{k}_1 + \textbf{q},\sigma} C^\dagger_{\textbf{k}_2 - \textbf{q},\sigma'} C_{\textbf{k}_2,\sigma'} C_{\textbf{k}_1,\sigma} \right\rangle,
\end{align}
and note that the last term is zero, since it is the integral of an oscillatory term, then
\begin{equation}
\int d(\textbf{r}-\textbf{r}') \tilde{g}(\textbf{r}-\textbf{r}') = - \dfrac{V}{N} + \dfrac{1}{N^2} N^2 V,
\end{equation}
\begin{equation}
\int d(\textbf{r}-\textbf{r}') \tilde{g}(\textbf{r}-\textbf{r}') = - \dfrac{1}{\rho_0} + V,
\end{equation}
and rewriting $V = \int d(\textbf{r} - \textbf{r}') 1$, we finally obtain
\begin{equation}
\int d(\textbf{r}-\textbf{r}') \left[ 1 - \tilde{g}(\textbf{r}-\textbf{r}') \right] = \dfrac{1}{\rho_0},
\end{equation}
which constitutes a sum rule for the pair correlation function.
\newpage
\section{Figures}
\FloatBarrier
\chapter{Phonon propagator}
Here we deal with the description of the Green's function associated with phonons, that is, the quanta of collective lattice vibrations. As with elementary particles, it is possible to define a propagator for phonons as well, which represents the probability amplitude that a phonon created at one space-time point is annihilated at another. In the first part of the chapter, the free phonon propagator will be introduced, corresponding to the harmonic description of the lattice in the absence of interactions with other degrees of freedom (such as electrons, other phonons, or defects). In this approximation, the propagator depends solely on the characteristics of the quantum harmonic system. The lattice propagators will be explicitly derived in both the real-time and imaginary-time formalisms, with particular attention to their analytic properties and frequency dependence. \newline
Subsequently, we will focus on the effects of perturbative interactions that modify phonon propagation. In particular, the concept of phonon self-energy will be introduced, which represents the correction to the free propagator due to scattering processes and coupling with other degrees of freedom. The analysis will be developed within Feynman's diagrammatic formalism, which allows a systematic treatment of interactions through a perturbative expansion of the Hamiltonian. A crucial aspect that will emerge is the connection between the phonon self-energy and the polarization propagator, related to the particle density operator. This link underlies many important phenomena in solid-state physics, such as the renormalization of phonon frequencies, dissipation, and symmetry breaking in the presence of strong interactions.
\section{Free phonon propagator and first-order self-energy for electron-phonon interaction}
\begin{figure}[H]
\centering
\includegraphics[scale=0.64]{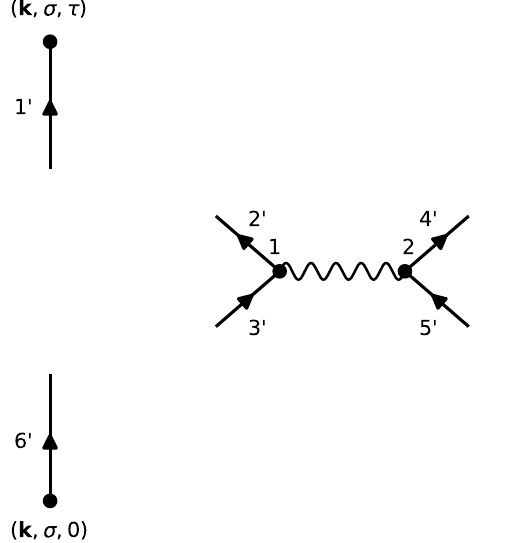}
\caption{Feynman diagram representation of the fermionic operators in the electron-phonon interaction. This diagram arises from the second-order contribution to the fermionic many-body Green's function associated with the electron-phonon interaction; see equation $\eqref{eq: interazioneelettronifononioperatorifermioniciequazione}$.}
\label{fig: interazioneelettronifononioperatorifermionicifigura}
\end{figure}

\begin{figure}[H]
\centering
\includegraphics[scale=0.5]{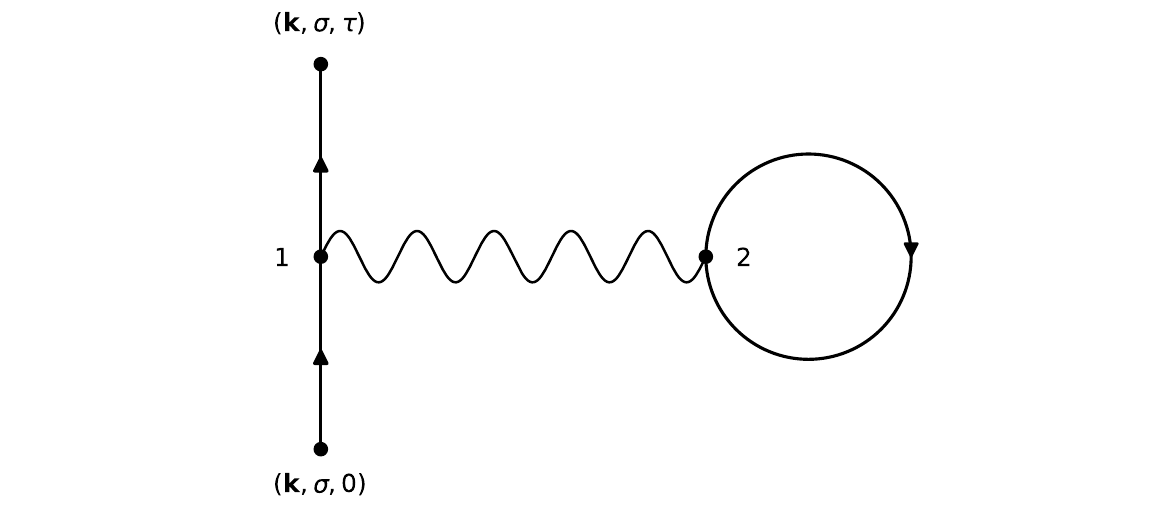}
\caption{Diagrammatic representation of the direct term resulting from a specific contraction in the electron-phonon interaction. It contributes to the second-order correction of the fermionic many-body Green's function associated with the electron-phonon interaction; see equation $\eqref{eq: interazioneelettronifononioperatorifermioniciequazione}$.}
\label{fig: Direct_term_of_phonon_propagator_with_fermionic_operators}
\end{figure}

\begin{figure}[H]
\centering
\includegraphics[scale=0.44]{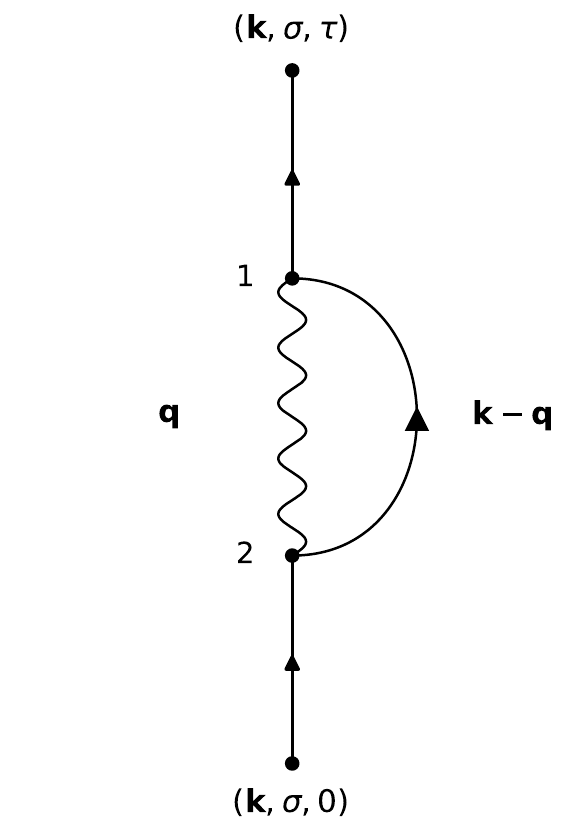}
\caption{Diagrammatic representation of the exchange term resulting from a specific contraction in the electron-phonon interaction. It contributes to the second-order correction of the fermionic many-body Green's function associated with the electron-phonon interaction; see equation $\eqref{eq: interazioneelettronifononioperatorifermioniciequazione}$.}
\label{fig: Exchange_term_of_phonon_propagator_with_fermionic_operators}
\end{figure}

In a previous chapter, we derived the Dyson series for the thermal Green's function of field operators, corresponding to a full Hamiltonian where the free Hamiltonian is given by $\eqref{eq: Hamiltonianasecondaquantizzazioneparticellelibere}$ and the interaction Hamiltonian by $\eqref{eq: operatoreaduecorpisecondaquantizzazione}$, i.e., of the form $\eqref{eq: hamiltonianamateriainterazioneaduecorpi}$. We now consider a full Hamiltonian composed of a free part given by
\begin{equation}
\hat{\mathcal{H}}_0 = \sum_{\textbf{k},\textbf{q}} \mathcal{E}_{\textbf{k}} C^{\dagger}_{\textbf{k},\sigma} C_{\textbf{k},\sigma} + \sum_{\textbf{q}} \hslash \omega_{\textbf{q}} a^{\dagger}_{\textbf{q}} a_{\textbf{q}} ,
\label{eq: hamiltonianaparticellelibereebagnofononico}
\end{equation}
and an interaction part given by $\eqref{eq: interazioneparticellebagnotermico}$, describing a system of fermions or bosons interacting with a bosonic field in a finite volume $V$. In the free Hamiltonian above, the first term accounts for fermions or bosons with dispersion $\mathcal{E}_{\mathbf{k}}$, while the second term describes the free dynamics of bosonic field excitations (e.g., phonons, photons, or a thermal reservoir). A typical physical realization of such a model is that of a system coupled to an external thermal bath. The interaction Hamiltonians in $\eqref{eq: operatoreaduecorpisecondaquantizzazione}$ and $\eqref{eq: interazioneparticellebagnotermico}$ are among the most commonly used in the literature. In what follows, we demonstrate that Feynman's perturbative expansion retains its general structure regardless of the specific interaction Hamiltonian, provided that the interaction potential is suitably adapted.
Previously, note that the Hamiltonian $\eqref{eq: interazioneparticellebagnotermico}$ can be expressed as a function of the (bosonic or fermionic) field operators as follows
\begin{equation}
\hat{\mathcal{H}}_I = \sum_{\textbf{q}} M_{\textbf{q}} \left( a_{\textbf{q}} + a^{\dagger}_{-\textbf{q}} \right) \int dx \hat{\psi}^\dagger(x) e^{i \textbf{q} \cdot \textbf{r}} \hat{\psi}(x).
\label{eq: interazioneparticellebagnotermicoriscritta}
\end{equation}
In addition, since the relation $\hat{\mathcal{H}}_I^\dagger = \hat{\mathcal{H}}_I$ holds, it follows that $M_{\mathbf{q}}^* = M_{-\mathbf{q}}$. \newline
Let us now analyze the meaning of equation $\eqref{eq: interazioneparticellebagnotermico}$. This equation describes a process in which a particle is annihilated at one vertex by the operator \( C_{\mathbf{k},\sigma} \), and another particle is created at another vertex by the operator \( C^{\dagger}_{\mathbf{k}+\mathbf{q},\sigma} \), mediated by bosonic annihilation and creation operators \( a_{\mathbf{q}} \) and \( a^{\dagger}_{\mathbf{q}} \), respectively. We now apply Feynman's perturbation theory to the interaction Hamiltonian given by equation $\eqref{eq: interazioneparticellebagnotermico}$. For \( n = 0 \), we recover the free propagator \( G^{(m)(0)}_{x,x'} \). For \( n = 1 \), the contribution to the interacting Green's function is given by the thermal average
\begin{equation}
\left\langle \hat{T}_\tau\, \hat{\psi}^{(0)}(x,\tau)\, \hat{\mathcal{H}}^{(0)}_I(\tau_1)\, \hat{\psi}^{\dagger (0)}(x',0) \right\rangle_0,
\end{equation}
where $\hat{\mathcal{H}}^{(0)}_I(\tau_1)$ is the interaction Hamiltonian in the interaction picture, and the subscript 0 indicates that the thermal average is taken with respect to the non-interacting Hamiltonian. Such an average factorizes as the product of two thermal averages, one fermionic and one bosonic, since \( \hat{\mathcal{H}}_0 \) governs the dynamics of independent particles, i.e., the fermionic and bosonic sectors are decoupled in the non-interacting Hamiltonian, i.e.,
\begin{align}
& \left\langle \hat{T}_\tau\, \hat{\psi}^{(0)}(x,\tau)\, \hat{\mathcal{H}}^{(0)}_I(\tau_1)\, \hat{\psi}^{\dagger (0)}(x',0) \right\rangle_0 \ = \\
&= \sum_{\textbf{q}_1} M_{\textbf{q}_1} \left\langle \hat{T}_\tau \left[ \hat{\psi}^{(0)}(x,\tau)\, \left( \int dx_1\, \hat{\psi}^{\dagger(0)}(x_1,\tau_1)\, e^{i \mathbf{q}_1 \cdot \mathbf{r}_1}\, \hat{\psi}^{(0)}(x_1,\tau_1) \right)\, \hat{\psi}^{\dagger (0)}(x',0) \right] \right\rangle_{f,0,conn.} \\
&\ \ \ \ \ \left\langle \hat{T}_\tau \left( a_{\mathbf{q}_1}^{(0)}(\tau_1) + a^{\dagger (0)}_{-\mathbf{q}_1}(\tau_1) \right) \right\rangle_{b,0,conn.}.
\end{align}
Consider bosonic thermal average: $a_{\textbf{q}}$ destroys a phonon with momentum $\textbf{q}$ and leads to an $N-1$ particle state, while $a_{-\textbf{q}}^\dagger$ creates a phonon with momentum $- \textbf{q}$ and leads to an $N+1$ particle state, so by orthonormality both of these averages are null, it follows that the first order of the perturbative expansion is null. Then, let us consider the second order of the series for an interaction Hamiltonian given by $\eqref{eq: interazioneparticellebagnotermico}$, i.e.,
\begin{align}
G^{(m)(2)}_{x,x'}(\tau,0) &= G^{(m)(0)}_{x,x'}(\tau,0) - \dfrac{1}{\hslash} \left( - \dfrac{1}{\hslash} \right)^2 \dfrac{1}{2} \int_{0}^{\beta \hslash} d\tau_1 \int_{0}^{\beta \hslash} d\tau_2 \int dx_1 \int dx_2 \sum_{q_1,q_2} M_{\textbf{q}_1} M_{\textbf{q}_2} e^{i \textbf{q}_1 \cdot \textbf{r}_1} e^{i \textbf{q}_2 \cdot \textbf{r}_2} \notag \\
& \left\langle \hat{T}_\tau\,
\hat{\psi}^{(0)}(x,\tau)\,
\hat{\psi}^{\dagger (0)}(x_1,\tau_1)\,
\hat{\psi}^{(0)}(x_1,\tau_1)\,
\hat{\psi}^{\dagger (0)}(x_2,\tau_2)\,
\hat{\psi}^{(0)}(x_2,\tau_2)\,
\hat{\psi}^{\dagger (0)}(x',0)
\right\rangle_{f,0,conn.} \notag \\
&\left\langle \hat{T}_\tau\, 
\left( a^{(0)}_{\mathbf{q}_1}(\tau_1) + a^{\dagger (0)}_{-\mathbf{q}_1}(\tau_1) \right)
\left( a^{(0)}_{\mathbf{q}_2}(\tau_2) + a^{\dagger (0)}_{-\mathbf{q}_2}(\tau_2) \right)
\right\rangle_{b,0,conn.},
\end{align}
where the two spatial integrations come from the operators $\hat{\mathcal{H}}_I^{(0)}(\tau_1)$, $\mathcal{H}_I^{(0)}(\tau_2)$. We compute the bosonic term. Since the system is at thermodynamic equilibrium, it is invariant by time translation, i.e.,
\begin{equation}
\left\langle \hat{T}_{\tau} \left( \hat{a}^{(0)}_{\mathbf{q}_1}(\tau_1 - \tau_2) + \hat{a}^{\dagger(0)}_{-\mathbf{q}_1}(\tau_1 - \tau_2) \right)
\left( \hat{a}_{\mathbf{q}_2} + \hat{a}^{\dagger}_{-\mathbf{q}_2} \right) \right\rangle_{b,0,conn.}.
\end{equation}
Assume $\tau_1-\tau_2>0$, so that the operator $\hat{T}_\tau$ can be removed. The time evolution of the bosonic operators is given by $\eqref{eq: evoluzionetemporaleraprresentazioneHeisenbergtermicaoperatoredistruzione}$, $\eqref{eq: evoluzionetemporaleraprresentazioneHeisenbergtermicaoperatorecreazione}$, that is,
\begin{equation}
a_{\textbf{q}}^{(0)}(\tau) = e^{- \omega_{\textbf{q}} \tau} a_{\textbf{q}},
\end{equation}
\begin{equation}
a_{-\textbf{q}}^{\dagger(0)}(\tau) = e^{\omega_{\textbf{q}} \tau} a^\dagger_{-\textbf{q}},
\end{equation}
where we used the fact that, for a lattice with inversion symmetry, the phonon
frequencies satisfy satisfies \( \omega_{\mathbf{q}} = \omega_{-\mathbf{q}} \). To simplify the notation, we omit the subscripts \( b \) and \( conn. \), and note that the only non-vanishing thermal averages are those of the form \( \langle a_{\mathbf{q}} a^\dagger_{\mathbf{q}'} \rangle_0 \) and \( \langle a^\dagger_{\mathbf{q}} a_{\mathbf{q}'} \rangle_0 \). Therefore, we obtain
\begin{align}
& \left\langle \left( a^{(0)}_{\mathbf{q}_1}(\tau_1 - \tau_2) + a^{\dagger(0)}_{-\mathbf{q}_1}(\tau_1 - \tau_2) \right) \left( a_{\mathbf{q}_2} + a^{\dagger}_{-\mathbf{q}_2} \right) \right\rangle_0 = \notag \\
&= \ \left\langle a^{(0)}_{\mathbf{q}_1}(\tau_1 - \tau_2)\, a^{\dagger}_{-\mathbf{q}_2} \right\rangle_0 + \left\langle a^{\dagger(0)}_{-\mathbf{q}_1}(\tau_1 - \tau_2)\, a_{\mathbf{q}_2} \right\rangle_0 \ = \notag \\
&= \ e^{-\omega_{\mathbf{q}_1}(\tau_1 - \tau_2)}\, \left\langle a_{\mathbf{q}_1}\, a^{\dagger}_{-\mathbf{q}_2} \right\rangle_0 
+ e^{\omega_{\mathbf{q}_1}(\tau_1 - \tau_2)}\, \left\langle a^{\dagger}_{-\mathbf{q}_1}\, a_{\mathbf{q}_2} \right\rangle_0 \ = \notag \\
&= \ e^{-\omega_{\mathbf{q}_1}(\tau_1 - \tau_2)}\, \left\langle \mathds{1} + a^{\dagger}_{-\mathbf{q}_2} a_{\mathbf{q}_1}\,  \right\rangle_0 
+ e^{\omega_{\mathbf{q}_1}(\tau_1 - \tau_2)}\, \left\langle a^{\dagger}_{-\mathbf{q}_1}\, a_{\mathbf{q}_2} \right\rangle_0 \ = \notag \\
&= \ \delta_{\mathbf{q}_1, -\mathbf{q}_2} \left\{ e^{-\omega_{\mathbf{q}_1}(\tau_1 - \tau_2)} \left[ 1 + \dfrac{1}{e^{\beta \hbar \omega_{\mathbf{q}_1}} - 1} \right] + e^{\omega_{\mathbf{q}_1}(\tau_1 - \tau_2)} \dfrac{1}{e^{\beta \hbar \omega_{\mathbf{q}_1}} - 1} \right\},
\end{align}
where we adapted $\eqref{eq: mediatermicaadaggeraconalphadiversi}$ for Bose-Einstein statistics. We define free phonon propagator the object
\begin{align}
D_{\mathbf{q}_1}^{(0)}(\tau_1,\tau_2) &= - \dfrac{1}{\hbar} \delta_{\mathbf{q}_1,-\mathbf{q}_2} \left\langle T_{\tau} \left( a^{(0)}_{\mathbf{q}_1}(\tau_1) + a^{\dagger(0)}_{-\mathbf{q}_1}(\tau_1) \right) 
\left( a^{(0)}_{\mathbf{q}_2}(\tau_2) + a^{\dagger(0)}_{-\mathbf{q}_2}(\tau_2) \right) \right\rangle_0 \ \equiv \notag \\
&\equiv - \dfrac{1}{\hbar} \left\langle T_{\tau} \left( a_{\mathbf{q}_1} + a^{\dagger}_{-\mathbf{q}_1} \right)^{(0)}(\tau_1) 
\left( a_{-\mathbf{q}_1} + a^{\dagger}_{\mathbf{q}_1} \right)^{(0)}(\tau_2) \right\rangle_0.
\label{eq: freephononpropagator}
\end{align}
Why is this object a phononic propagator? The fermionic or bosonic Green's function \( G^{(m)}_{x,x'}(\tau,0) \) creates a fermion or boson at time \( 0 \) and position \( x' \), and annihilates it at position \( x \) and time \( \tau \). Similarly, recalling the structure of equation $\eqref{eq: spostamentoquantizzatofononi}$, the displacement operator \( \mathbf{S}_{\mathbf{n},\mu} \) creates and annihilates bosons within the unit cell via the combination \( \left( a_{\mathbf{q}} + a_{-\mathbf{q}}^\dagger \right) \). The phonon propagator is therefore of the form $\left\langle \hat{T}_\tau\, \mathbf{S}(\tau_1)\, \mathbf{S}^\dagger(\tau_2) \right\rangle$. Up to this point, we have encountered two types of propagators: the fermionic or bosonic propagator \( G^{(m)(0)}_{x,x'}(\tau,0) \) (see equation $\eqref{eq: funzioneGreenMastubaraoperatoricampo}$) in real space, and the free phonon propagator $D_{\mathbf{q}_1}^{(0)}(\tau_1,\tau_2)$ (see equation $\eqref{eq: freephononpropagator}$) in momentum space. Note that \( G \) carries two configuration-space indices, \( x \) and \( x' \), whereas \( D \) depends only on a single momentum \( \mathbf{q} \). This is due to spatial translation symmetry: the propagator depends only on the relative coordinate \( \textbf{r} - \textbf{r}' \), and thus transforms to a single momentum index after Fourier transformation. We now consider the second-order term in the Dyson series for \( G \), and we extract a factor \( \left( -\frac{1}{\hbar} \right) \) in order to reconstruct the phonon propagator, i.e.,
\begin{align}
G^{(m)(2)}_{x,x'}(\tau,0) &= G^{(m)(0)}_{x,x'}(\tau,0) 
+ \left( - \dfrac{1}{\hbar} \right)^2 \dfrac{1}{2} \int_{0}^{\beta \hbar} d\tau_1 \int_{0}^{\beta \hbar} d\tau_2 \int dx_1 \int dx_2 \sum_{\mathbf{q}_1,\mathbf{q}_2} 
M_{\mathbf{q}_1} M_{\mathbf{q}_2} e^{i \mathbf{q}_1 \cdot \mathbf{r}_1} e^{i \mathbf{q}_2 \cdot \mathbf{r}_2} \notag \\
& \left\langle \hat{T}_\tau \hat{\psi}^{(0)}(x,\tau)\hat{\psi}^{\dagger (0)}(x_1,\tau_1)\hat{\psi}^{(0)}(x_1,\tau_1)
\hat{\psi}^{\dagger (0)}(x_2,\tau_2)\hat{\psi}^{(0)}(x_2,\tau_2)\hat{\psi}^{\dagger (0)}(x',0) \right\rangle_{f,0,conn.} \notag \\
&\delta_{\mathbf{q}_1,-\mathbf{q}_2} D_{\mathbf{q}_1}^{(0)}(\tau_1,\tau_2) = \notag \\
&= G^{(m)(0)}_{x,x'}(\tau,0) 
+ \left( - \dfrac{1}{\hbar} \right)^2 \dfrac{1}{2} \int_{0}^{\beta \hbar} d\tau_1 \int_{0}^{\beta \hbar} d\tau_2 \int dx_1 \int dx_2 \sum_{\mathbf{q}} 
|M_{\mathbf{q}}|^2 e^{i \mathbf{q} \cdot (\mathbf{r}_1 - \mathbf{r}_2)} \notag \\
& \left\langle \hat{T}_\tau \hat{\psi}^{(0)}(x,\tau)\hat{\psi}^{\dagger (0)}(x_1,\tau_1)\hat{\psi}^{(0)}(x_1,\tau_1)
\hat{\psi}^{\dagger (0)}(x_2,\tau_2)\hat{\psi}^{(0)}(x_2,\tau_2)\hat{\psi}^{\dagger (0)}(x',0) \right\rangle_{f,0,conn.} 
D_{\mathbf{q}}^{(0)}(\tau_1,\tau_2).
\end{align}
We used the Kronecker delta $\delta_{\mathbf{q}_1, -\mathbf{q}_2}$ to perform the sum over $\mathbf{q}_2$, then renamed the summation index $\mathbf{q}_1 = \mathbf{q}$ and applied the identity $M_{-\mathbf{q}} = M^*_{\mathbf{q}}$. To compare the resulting expression with the first-order Dyson expansion for the field operators, we focus on the structure of the fermionic thermal average. Specifically, we aim to form a "mirrored" operator structure of the type $\hat{\psi}^\dagger(x) \hat{\psi}^\dagger(x') \hat{\psi}(x') \hat{\psi}(x)$. Under the action of the imaginary-time ordering operator $\hat{T}_\tau$, the fermionic field operators can be rearranged using their anticommutation relations. Each exchange of two fermionic operators introduces a factor $\varepsilon = -1$, due to their anticommuting nature. Therefore, we have
\begin{align}
& \hat{T}_\tau \left\lbrace \hat{\psi}^{(0)}(x,\tau)\hat{\psi}^{\dagger (0)}(x_1,\tau_1)\hat{\psi}^{(0)}(x_1,\tau_1)\hat{\psi}^{\dagger (0)}(x_2,\tau_2)\hat{\psi}^{(0)}(x_2,\tau_2)\hat{\psi}^{\dagger (0)}(x',0) \right\rbrace = \notag \\
&= \hat{T}_\tau \left\lbrace \varepsilon \, \hat{\psi}^{(0)}(x,\tau)\hat{\psi}^{\dagger (0)}(x_1,\tau_1)\hat{\psi}^{\dagger (0)}(x_2,\tau_2)\hat{\psi}^{(0)}(x_1,\tau_1)\hat{\psi}^{(0)}(x_2,\tau_2)\hat{\psi}^{\dagger (0)}(x',0) \right\rbrace = \notag \\
&= \hat{T}_\tau \left\lbrace \varepsilon^2 \, \hat{\psi}^{(0)}(x,\tau)\hat{\psi}^{\dagger (0)}(x_1,\tau_1)\hat{\psi}^{\dagger (0)}(x_2,\tau_2)\hat{\psi}^{(0)}(x_2,\tau_2)\hat{\psi}^{(0)}(x_1,\tau_1)\hat{\psi}^{\dagger (0)}(x',0) \right\rbrace,
\end{align}
and we get the mirror operator structure. For $n=1$ we have the first nonzero contribution, indeed the constant
\begin{equation}
\dfrac{1}{2} \left( - \dfrac{1}{\hslash} \right) \left( - \dfrac{1}{\hslash} \right)^n \dfrac{1}{n!} = \dfrac{1}{2} \left( - \dfrac{1}{\hslash} \right)^2 
\end{equation}
belongs to the second order: $\frac{1}{2}$ comes from the order itself, $\frac{1}{n!} = \frac{1}{2}$. From the comparison between the first expansion for $\eqref{eq: operatoreaduecorpisecondaquantizzazione}$ and the second one for $\eqref{eq: interazioneparticellebagnotermico}$, we have
\begin{equation}
U_{eff.}(x_1,x_2,\tau_1,\tau_2) = \sum_{\textbf{q}} |M_{\textbf{q}}|^2 e^{i \textbf{q} \cdot (\textbf{r}_1 - \textbf{r}_2)} D_{\textbf{q}}^{(0)}(\tau_1,\tau_2).
\end{equation}
The interaction between particles is replaced by an effective potential. The effective potential is not instantaneous since there is no $\delta(\tau_1 - \tau_2)$ term. Due to the term $\sum_{\textbf{q}} \hslash \omega_{\textbf{q}} a^\dagger_{\textbf{q}} a_{\textbf{q}}$ in $\hat{\mathcal{H}}_0$, $\omega_{\textbf{q}}$ is a frequency associated with a finite phononic propagation time. Furthermore, the two-body interaction is an electron-electron Coulomb potential, $U_{Coulomb}>0$, while $U_{eff.}<0$, since
\begin{equation}
D_{\textbf{q}}^{(0)}(\tau_1,\tau_2) = - \dfrac{1}{\hslash} \left\langle \hat{T}_{\tau} \left( a_{\textbf{q}} + a_{-\textbf{q}}^{\dagger} \right)^{(0)}(\tau_1) \left( a_{-\textbf{q}} + a_{\textbf{q}}^{\dagger} \right)^{(0)}(\tau_2) \right\rangle_0 < 0.
\end{equation}
The effective potential between electrons is attractive and retarded. Now, since the system is in thermodynamic equilibrium, $D_{\textbf{q}}^{(0)}(\tau_1,\tau_2) = D_{\textbf{q}}^{(0)}(\tau_1-\tau_2)$, and it is invariant under spatial translation, we can expand the field operators in the plane wave basis, thus the second-order spatial Fourier transform of the Green's function is written as
\begin{align}
G^{(m)(2)}_{\textbf{k},\sigma}(\tau) &= G^{(m)(0)}_{\textbf{k},\sigma}(\tau) + \left( - \dfrac{1}{\hslash} \right)^2 \dfrac{1}{2} \int_{0}^{\beta \hslash} d\tau_1 \int_{0}^{\beta \hslash} d\tau_2 \sum_{\textbf{k}_1,\sigma_1} \sum_{\textbf{k}_2,\sigma_2} \sum_{\textbf{q}} |M_{\textbf{q}}|^2 D_{\textbf{q}}^{(0)}(\tau_1-\tau_2) \notag \\
&\left\langle \hat{T}_\tau C^{(0)}_{\textbf{k},\sigma}(\tau) C^{\dagger (0)}_{\textbf{k}_1 + \textbf{q},\sigma_1}(\tau_1) C^{(0)}_{\textbf{k}_1,\sigma_1}(\tau_1) C^{\dagger (0)}_{\textbf{k}_2 - \textbf{q},\sigma_2}(\tau_2) C^{(0)}_{\textbf{k}_2,\sigma_2}(\tau_2) C^{\dagger (0)}_{\textbf{k},\sigma}(0) \right\rangle_{f,0,conn.}, 
\label{eq: interazioneelettronifononioperatorifermioniciequazione}
\end{align}
where the spatial integral of the exponential factor has canceled the volume included in the plane waves. We apply Wick's theorem and compute the contractions, considering only the connected diagrams. We set $1' \equiv C^{(0)}_{\textbf{k},\sigma}(\tau)$, $2' \equiv C^{\dagger (0)}_{\textbf{k}_1 + \textbf{q},\sigma_1}(\tau_1)$, $3' \equiv C^{(0)}_{\textbf{k}_1,\sigma_1}(\tau_1)$, $4' \equiv C^{\dagger (0)}_{\textbf{k}_2 - \textbf{q},\sigma_2}(\tau_2)$, $5' \equiv C^{(0)}_{\textbf{k}_2,\sigma_2}(\tau_2)$, $6' \equiv C^{\dagger (0)}_{\textbf{k},\sigma}(0)$, see Figure $\eqref{fig: interazioneelettronifononioperatorifermionicifigura}$. The contraction $1'-6'$ produces disconnected diagrams, which must be discarded. Furthermore, we observe the presence of two diagrams with the structure "something in–something out–half interaction line", due to the form of the Hamiltonian $\hat{\mathcal{H}}_I$. The contraction of the two half lines at times $\tau_1$, $\tau_2$ gives rise to the phonon propagator; in fact, the two half lines are represented by $\left(a_{\textbf{q}} + a_{-\textbf{q}}^{\dagger}\right)$ and $\left(a_{-\textbf{q}} + a_{\textbf{q}}^{\dagger}\right)$, they compose the phonon propagator and represent the electron-phonon interaction. Ultimately, there are two diagrams, and both appear twice: the direct term diagram (see Figure $\eqref{fig: Direct_term_of_phonon_propagator_with_fermionic_operators}$) from the contractions $1'-2'$, $3'-6'$, $4'-5'$ and $1'-4'$, $2'-3'$, $5'-6'$, and the exchange term diagram (see Figure $\eqref{fig: Exchange_term_of_phonon_propagator_with_fermionic_operators}$) from the contractions $1'-2'$, $3'-4'$, $5'-6'$ and $1'-4'$, $2'-5'$, $3'-6'$. In the direct diagram, due to the loop, an $\varepsilon$ emerges, while in the exchange diagram, one $\varepsilon$ arises from the half loop and one from Wick's theorem, so $\varepsilon^2 = 1$. The analytical expression for the second-order correction to the many-body Green function, $G^{(m)(2)}_{\textbf{k},\sigma}(\tau)$, is given by
\begin{align}
G^{(m)(2)}_{\textbf{k},\sigma}(\tau) &= G^{(m)(0)}_{\textbf{k},\sigma}(\tau) + \left( - \dfrac{1}{\hslash} \right)^2 \int_{0}^{\beta \hslash} d\tau_1 \int_{0}^{\beta \hslash} d\tau_2 \sum_{\textbf{k}_1,\sigma_1} \sum_{\textbf{k}_2,\sigma_2} \sum_{\textbf{q}} |M_{\textbf{q}}|^2 D_{\textbf{q}}^{(0)}(\tau_1-\tau_2) \notag \\
&\bigg[ (- \hslash) G^{(m)(0)}_{\textbf{k},\sigma}(\tau-\tau_1) \delta_{\textbf{k}_1+\textbf{q},\textbf{k}} \delta_{\sigma_1,\sigma} (- \hslash) G^{(m)(0)}_{\textbf{k},\sigma}(\tau_1) \delta_{\textbf{q},\textbf{0}} (- \hslash) (\varepsilon) G^{(m)(0)}_{\textbf{k}_2,\sigma_2}(\tau_2,\tau_2 + \delta) + \notag \\
&+ (- \hslash) G^{(m)(0)}_{\textbf{k},\sigma}(\tau-\tau_1) \delta_{\textbf{k}_1+\textbf{q},\textbf{k}} \delta_{\sigma_1,\sigma} (- \hslash) G^{(m)(0)}_{\textbf{k}_1,\sigma_1}(\tau_1-\tau_2) \delta_{\sigma_1,\sigma_2} \delta_{\textbf{k}_2 - \textbf{q},\textbf{k}_1} (- \hslash) G^{(m)(0)}_{\textbf{k},\sigma}(\tau) \delta_{\textbf{k}_2,\textbf{k}} \delta_{\sigma_2,\sigma} \bigg].
\end{align}
We sum over momenta and spins: the delta functions imply $\textbf{k}_1 = \textbf{k}$, $\sigma_1 = \sigma$ for the direct term and $\textbf{k}_1 = \textbf{k} - \textbf{q}$, $\textbf{k}_2 = \textbf{k}$, $\sigma_1 = \sigma_2 = \sigma$ for the exchange term, and we obtain
\begin{align}
G^{(m)(2)}_{\textbf{k},\sigma}(\tau) &= G^{(m)(0)}_{\textbf{k},\sigma}(\tau) - \hslash \int_{0}^{\beta \hslash} d\tau_1 \int_{0}^{\beta \hslash} d\tau_2 \sum_{\textbf{q}} |M_{\textbf{q}}|^2 D_{\textbf{q}}^{(0)}(\tau_1-\tau_2) \notag \\
&\bigg[ \varepsilon \delta_{\textbf{q},\textbf{0}} \sum_{\textbf{k}_2,\sigma_2} G^{(m)(0)}_{\textbf{k},\sigma}(\tau-\tau_1) G^{(m)(0)}_{\textbf{k}_2,\sigma_2}(\tau_2,\tau_2+\delta) G^{(m)(0)}_{\textbf{k},\sigma}(\tau_1) + \notag \\
&+ G^{(m)(0)}_{\textbf{k},\sigma}(\tau-\tau_1) G^{(m)(0)}_{\textbf{k}-\textbf{q},\sigma}(\tau_1-\tau_2) G^{(m)(0)}_{\textbf{k},\sigma}(\tau_2) \bigg].
\label{eq: funzionediGreensecondoordineconHamiltonianafononica1}
\end{align}
Now, recalling the Dyson algebraic equation $\eqref{eq: equazioneDysonalgebrica}$, we aim to compute the self-energy at the lowest non-trivial order. We observe that, with respect to the interaction Hamiltonian \( \hat{\mathcal{H}}_I \), the contribution we are considering is of second order, but it represents the first non-vanishing term in the perturbative expansion. Moreover, although the Dyson equation involves the improper self-energy \( \Sigma^* \), at this order the only contributing diagrams are irreducible, so we can identify
\begin{equation}
\Sigma^{*(1)} \equiv \Sigma^{(1)}.
\end{equation}
From the comparison between the Dyson equation $\eqref{eq: equazioneDysonalgebrica}$ and the explicit second-order expression for the Green's function $\eqref{eq: funzionediGreensecondoordineconHamiltonianafononica1}$, we obtain
\begin{align}
\Sigma^{(1)}_{\textbf{k},\sigma}(\tau_1 - \tau_2) &= - \varepsilon \hslash \sum_{\textbf{q}} \delta_{\textbf{q},\textbf{0}} |M_{\textbf{q}}|^2 \delta(\tau_1-\tau_2) \int_0^{\beta \hslash} d \tau'_2 D_{\textbf{q}}^{(0)}(\tau_1-\tau'_2) \sum_{\textbf{k}_2,\sigma_2} G^{(m)(0)}_{\textbf{k}_2,\sigma_2}(\tau'_2,\tau'_2+\delta) + \notag \\
&- \hslash \sum_{\textbf{q}} |M_{\textbf{q}}|^2 D_{\textbf{q}}^{(0)}(\tau_1-\tau_2) G^{(m)(0)}_{\textbf{k}-\textbf{q},\sigma}(\tau_1-\tau_2).
\label{eq: selfenergyprimoordineconHamiltonianafononica1}
\end{align}
Note that the contribution to the self-energy originating from the direct term is zero due to the delta $\delta_{\textbf{q},\textbf{0}}$, since $M_{\textbf{q}=\textbf{0}} = 0$. Consequently, the second-order propagator reduces to
\begin{equation}
G^{(m)(2)}_{\textbf{k},\sigma}(\tau) = G^{(m)(0)}_{\textbf{k},\sigma}(\tau) - \hslash \int_{0}^{\beta \hslash} d\tau_1 \int_{0}^{\beta \hslash} d\tau_2 \sum_{\textbf{q}} |M_{\textbf{q}}|^2 D_{\textbf{q}}^{(0)}(\tau_1-\tau_2)  G^{(m)(0)}_{\textbf{k},\sigma}(\tau-\tau_1) G^{(m)(0)}_{\textbf{k}-\textbf{q},\sigma}(\tau_1-\tau_2) G^{(m)(0)}_{\textbf{k},\sigma}(\tau_2),
\label{eq: funzionediGreensecondoordineconHamiltonianafononica2}
\end{equation}
and the first-order self-energy is given by
\begin{equation}
\Sigma^{(1)}_{\textbf{k},\sigma}(\tau_1 - \tau_2) = - \hslash \sum_{\textbf{q}} |M_{\textbf{q}}|^2 D_{\textbf{q}}^{(0)}(\tau_1-\tau_2) G^{(m)(0)}_{\textbf{k}-\textbf{q},\sigma}(\tau_1-\tau_2).
\label{eq: selfenergyprimoordineconHamiltonianafononica2}
\end{equation}
Keeping in mind equation $\eqref{eq: funzioneGreenMatsubaraserieFourier1}$, we apply the temporal Fourier transform and obtain
\begin{align}
\Sigma^{(1)}_{\textbf{k},\sigma}(i \omega_n) &= \int_0^{\beta \hslash} d(\tau_1-\tau_2) e^{i \omega_n (\tau_1 - \tau_2)} \Sigma^{(1)}_{\textbf{k},\sigma}(\tau_1 - \tau_2) = \notag \\
&= - \hslash \sum_{\textbf{q}} |M_{\textbf{q}}|^2 \int_0^{\beta \hslash} d(\tau_1-\tau_2) e^{i \omega_n (\tau_1 - \tau_2)} \dfrac{1}{\beta \hslash} \sum_{n_1} D_{\textbf{q}}^{(0)}(i \omega_{n_1}) e^{- i \omega_{n_1} (\tau_1 - \tau_2)} \notag \\
& \ \ \ \ \ \ \ \ \ \ \dfrac{1}{\beta \hslash} \sum_{n_2} G^{(m)(0)}_{\textbf{k}-\textbf{q},\sigma}(i \omega_{n_2}) e^{- i \omega_{n_2} (\tau_1 - \tau_2)}.
\end{align}
Regarding the time integral, some clarifications are needed: although we are dealing here with electron-phonon interactions, in general $\omega_{n_1}$ is even, as it comes from a bosonic propagator, while $\omega_n$ and $\omega_{n_2}$ are both odd or both even depending on whether the propagators $G$ refer to fermions or bosons, respectively. Consequently, $\omega_n - \omega_{n_2}$ is even regardless of the statistical nature of the particles, and the quantity $\omega_{n_1} - (\omega_n - \omega_{n_2})$ is even. Therefore, the integral is zero unless the overall frequency vanishes, thus producing $\delta_{\omega_{n_2},\omega_n - \omega_{n_1}}$, which implies the presence of a factor $\beta \hslash$, and we have
\begin{equation}
\Sigma^{(1)}_{\textbf{k},\sigma}(i \omega_n) = - \dfrac{1}{\beta} \sum_{\textbf{q}} |M_{\textbf{q}}|^2 \sum_{n_1} D_{\textbf{q}}^{(0)}(i \omega_{n_1}) G^{(m)(0)}_{\textbf{k}-\textbf{q},\sigma}(i (\omega_n - \omega_{n_1})).
\end{equation}
\subsection{Free phonon propagator in momentum and frequency space}
Consider the free phonon propagator $D_{\textbf{q}}^{(0)}(\tau)$ at thermodynamic equilibrium, i.e., equation $\eqref{eq: freephononpropagator}$, which is related to the free Hamiltonian $\eqref{eq: hamiltonianaparticellelibereebagnofononico}$. Assume for simplicity $\tau>0$, then
\begin{align}
D_{\textbf{q}}^{(0)}(\tau) &= - \dfrac{1}{\hslash} \left\langle \left( a_{\textbf{q}} + a_{-\textbf{q}}^\dagger \right)^{(0)}(\tau) \left( a_{-\textbf{q}} + a_{\textbf{q}}^\dagger \right)^{(0)} \right\rangle_0 \ = \notag \\
&= - \dfrac{1}{\hslash} \left\lbrace \left\langle a_{\textbf{q}}^{(0)}(\tau) a_{\textbf{q}}^\dagger \right\rangle_0 + \left\langle a_{-\textbf{q}}^{\dagger (0)}(\tau) a_{-\textbf{q}} \right\rangle_0 \right\rbrace = \notag \\
&= - \dfrac{1}{\hslash} \left\lbrace e^{- \omega_{\textbf{q}} \tau} \left\langle a_{\textbf{q}} a_{\textbf{q}}^\dagger \right\rangle_0 + e^{\omega_{-\textbf{q}} \tau} \left\langle a_{-\textbf{q}}^{\dagger} a_{-\textbf{q}} \right\rangle_0 \right\rbrace = \notag \\
&= - \dfrac{1}{\hslash} \left\lbrace \left( 1 + n_{1}(\mathcal{E}_{\textbf{q}}) \right) e^{- \omega_{\textbf{q}} \tau} + n_{1}(\mathcal{E}_{- \textbf{q}}) e^{\omega_{-\textbf{q}} \tau} \right\rbrace .
\end{align}
Given that $\omega_{\textbf{q}} =\omega_{-\textbf{q}}$, it follows $n_{1}(\mathcal{E}_{\textbf{q}})=n_{1}(\mathcal{E}_{-\textbf{q}})$, i.e.,
\begin{equation}
D_{\textbf{q}}^{(0)}(\tau) = - \dfrac{1}{\hslash} \left\lbrace \left( 1 + n_{1}(\mathcal{E}_{\textbf{q}}) \right) e^{- \omega_{\textbf{q}} \tau} + n_{1}(\mathcal{E}_{\textbf{q}}) e^{\omega_{\textbf{q}} \tau} \right\rbrace.
\label{eq: propagatorefononiconudoall'equilibriotermodinamiconellospaziodiMatsubara}
\end{equation}
We apply a Fourier transform with respect to times, that is
\begin{align}
D_{\textbf{q}}^{(0)}(i \omega_n) &= \int_0^{\beta \hslash} d\tau e^{i \omega_n \tau} D_{\textbf{q}}^{(0)}(\tau) = \notag \\
&= - \dfrac{1}{\hslash} \left\lbrace \left( 1 + n_{1}(\mathcal{E}_{\textbf{q}}) \right) \left( \int_0^{\beta \hslash} d\tau e^{i \omega_n \tau} e^{- \omega_q \tau} \right) + n_{1}(\mathcal{E}_{\textbf{q}}) \left( \int_0^{\beta \hslash} d\tau e^{i \omega_n \tau} e^{\omega_q \tau} \right) \right\rbrace = \notag \\
&= - \dfrac{1}{\hslash} \left\lbrace \dfrac{1 + n_{1}(\mathcal{E}_{\textbf{q}})}{i \omega_n - \omega_{\textbf{q}}} \left( e^{i \omega_n \beta \hslash} e^{- \beta \hslash \omega_{\textbf{q}}} - 1 \right) + \dfrac{n_{1}(\mathcal{E}_{\textbf{q}})}{i \omega_n + \omega_{\textbf{q}}} \left( e^{i \omega_n \beta \hslash} e^{\beta \hslash \omega_q} - 1 \right) \right\rbrace = \notag \\
&= - \dfrac{1}{\hslash} \left\lbrace \dfrac{1 + n_{1}(\mathcal{E}_{\textbf{q}})}{i \omega_n - \omega_{\textbf{q}}} \left( \varepsilon e^{- \beta \hslash \omega_q} - 1 \right) + \dfrac{n_{1}(\mathcal{E}_{\textbf{q}})}{i \omega_n + \omega_{\textbf{q}}} \left( \varepsilon e^{\beta \hslash \omega_{\textbf{q}}} - 1 \right) \right\rbrace = \notag \\
&= - \dfrac{1}{\hslash} \left\lbrace \dfrac{1 + n_{1}(\mathcal{E}_{\textbf{q}})}{i \omega_n - \omega_{\textbf{q}}} \left( e^{- \beta \hslash \omega_{\textbf{q}}} - 1 \right) + \dfrac{n_{1}(\mathcal{E}_{\textbf{q}})}{i \omega_n + \omega_{\textbf{q}}} \left( e^{\beta \hslash \omega_{\textbf{q}}} - 1 \right) \right\rbrace,
\end{align}
and from
\begin{align}
1 + n_{1}(\mathcal{E}_{\textbf{q}}) &= 1 + \dfrac{1}{e^{\beta \hslash \omega_{\textbf{q}}} - 1} = \notag \\
&= \dfrac{e^{\beta \hslash \omega_{\textbf{q}}}}{e^{\beta \hslash \omega_{\textbf{q}}} - 1},
\end{align}
we get
\begin{align}
D_{\textbf{q}}^{(0)}(i \omega_n) &= - \dfrac{1}{\hslash} \left\lbrace \dfrac{1}{i \omega_n - \omega_{\textbf{q}}} \dfrac{e^{\beta \hslash \omega_{\textbf{q}}}}{e^{\beta \hslash \omega_{\textbf{q}}} - 1}\left( e^{- \beta \hslash \omega_{\textbf{q}}} - 1 \right) + \dfrac{1}{i \omega_n + \omega_{\textbf{q}}} \dfrac{1}{e^{\beta \hslash \omega_{\textbf{q}}} - 1} \left( e^{\beta \hslash \omega_{\textbf{q}}} - 1 \right) \right\rbrace = \notag \\
&= - \dfrac{1}{\hslash} \left\lbrace \dfrac{1}{i \omega_n - \omega_{\textbf{q}}} \dfrac{1 - e^{\beta \hslash \omega_q}}{e^{\beta \hslash \omega_{\textbf{q}}} - 1} + \dfrac{1}{i \omega_n + \omega_{\textbf{q}}} \right\rbrace = \notag \\
&= - \dfrac{1}{\hslash} \left\lbrace - \dfrac{1}{i \omega_n - \omega_{\textbf{q}}} + \dfrac{1}{i \omega_n + \omega_{\textbf{q}}} \right\rbrace = \notag \\
&= \dfrac{1}{\hslash} \left\lbrace \dfrac{1}{i \omega_n - \omega_{\textbf{q}}} - \dfrac{1}{i \omega_n + \omega_{\textbf{q}}} \right\rbrace \equiv \notag \\
&\equiv \dfrac{1}{\hslash} \dfrac{2 \omega_{\textbf{q}}}{(i \omega_n)^2-\omega_{\textbf{q}}^2}.
\end{align}
If we set $i \omega_n = \omega + i \delta$ and write
\begin{align}
D_{\textbf{q}}^{(0)}(\omega + i \delta) &= \dfrac{1}{\hslash} \left\lbrace \dfrac{\omega - i \delta - \omega_{\textbf{q}}}{(\omega + i \delta - \omega_{\textbf{q}})(\omega - i \delta - \omega_{\textbf{q}})} - \dfrac{\omega - i \delta + \omega_{\textbf{q}}}{(\omega + i \delta + \omega_{\textbf{q}})(\omega - i \delta + \omega_{\textbf{q}})} \right\rbrace = \notag \\
&= \dfrac{1}{\hslash} \left\lbrace \dfrac{\omega - i \delta - \omega_{\textbf{q}}}{\left( \omega - \omega_{\textbf{q}} \right)^2 + \delta^2} - \dfrac{\omega - i \delta + \omega_{\textbf{q}}}{\left( \omega + \omega_{\textbf{q}} \right)^2 + \delta^2} \right\rbrace ,
\end{align}
then
\begin{equation}
- 2 \hslash \lim_{\delta \rightarrow 0^+} \im \left( D_{\textbf{q}}^{(0)}(\omega + i \delta) \right) = A_{\textbf{q}}(\omega),
\end{equation}
\begin{equation}
A_{\textbf{q}}(\omega) = 2 \pi \left\lbrace \delta(\omega-\omega_{\textbf{q}}) - \delta(\omega+\omega_{\textbf{q}}) \right\rbrace,
\end{equation}
that is, we have derived the spectral function of free phonons. \newline
We now examine the behavior of the free phonon propagator in the high-frequency limit. This regime is relevant in situations where the phonon energy $\hslash \omega_{\textbf{q}}$ becomes much larger than all other energy scales in the problem. In such a limit, the phonon propagator acquires a particularly simple form. Indeed it can be shown that the free phonon propagator $\eqref{eq: propagatorefononiconudoall'equilibriotermodinamiconellospaziodiMatsubara}$ satisfies
\begin{equation}
\lim_{\omega_{\textbf{q}} \rightarrow +\infty} \left( - \hslash \omega_{\textbf{q}} D_{\textbf{q}}^{(0)}(\tau) \right) = 2 \sum_{n \in \mathbb{N}} \delta(\tau - n \beta \hbar).
\label{eq: propagatorefononiconudoall'equilibriotermodinamiconellospaziodiMatsubaralimitealtefrequenze}
\end{equation}
Physically, the high-frequency limit \( \omega_{\textbf{q}} \rightarrow +\infty \) corresponds to the regime in which the phonon dynamics is infinitely faster than the electronic dynamics. In this limit, phonons respond instantaneously to the motion of electrons, and their propagator reduces to a Dirac delta function in imaginary time. This behavior makes the phonon-mediated electron-electron interaction effectively instantaneous (static), analogous to a local interaction. This justifies the use of static approximations in certain effective theories, such as in deriving attractive Hubbard-type models from electron-phonon coupling.
\section{Phonon propagator self-energy}
\begin{figure}[H]
\centering
\includegraphics[scale=0.7]{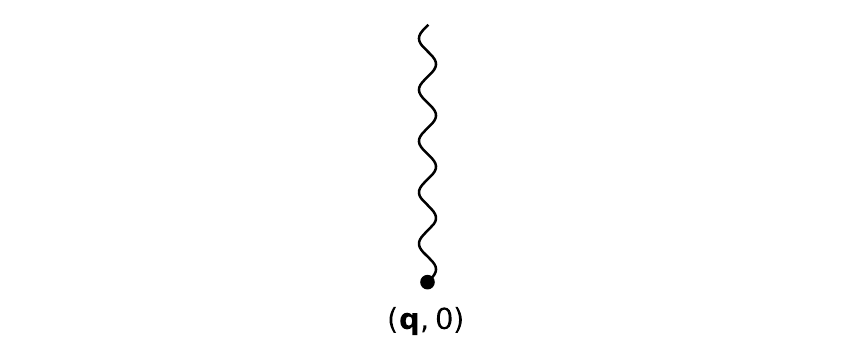}
\caption{Feynman diagram representation of a bosonic annihilation operator, depicted as a vertex corresponding to the absorption of a boson in an interaction process.}
\label{fig: operatorebosonicodistruzionediagrammatica}
\end{figure}

\begin{figure}[H]
\centering
\includegraphics[scale=0.7]{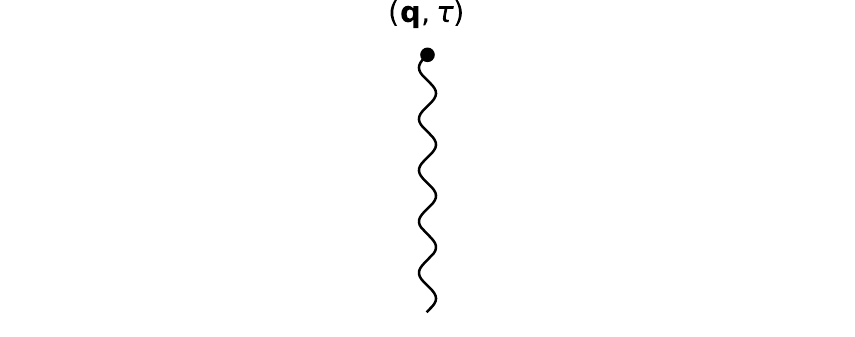}
\caption{Feynman diagram representation of a bosonic creation operator, depicted as a vertex corresponding to the emission of a boson in an interaction process.}
\label{fig: operatorebosonicocreazionediagrammatica}
\end{figure}

\begin{figure}[H]
\centering
\includegraphics[scale=0.7]{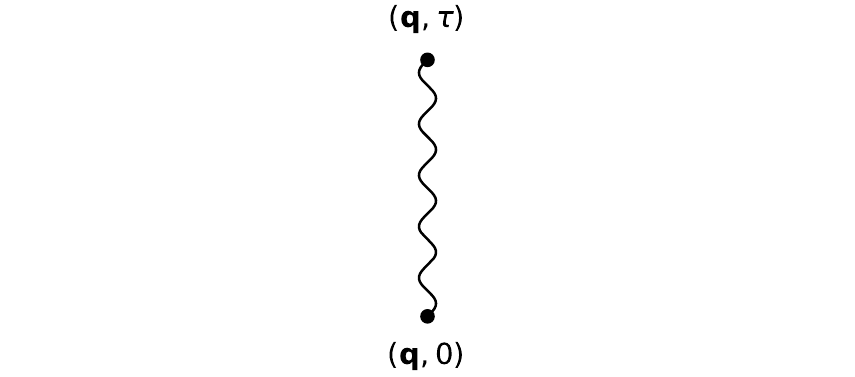}
\caption{Feynman diagram representation of a free phonon propagator, describing the propagation of a phonon between vertices, in the absence of interactions.}
\label{fig: operatorefononicoliberodiagrammatica}
\end{figure}

\begin{figure}[H]
\centering
\includegraphics[scale=0.7]{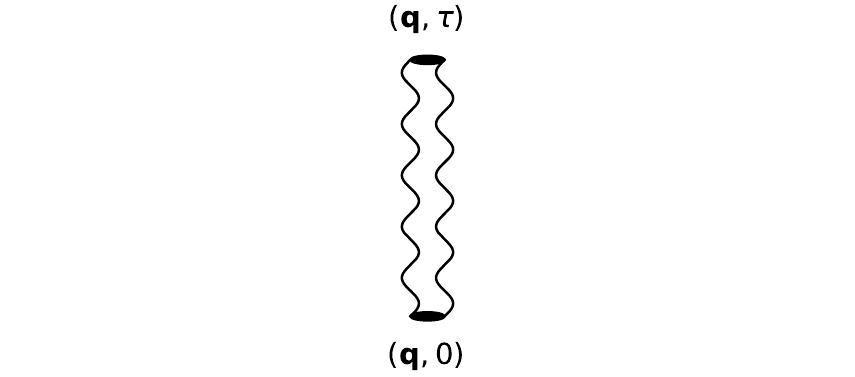}
\caption{Feynman diagram representation of a phonon propagator, describing phonon propagation between interaction vertices.}
\label{fig: operatorefononicodiagrammatica}
\end{figure}

\begin{figure}[H]
\centering
\includegraphics[scale=0.67]{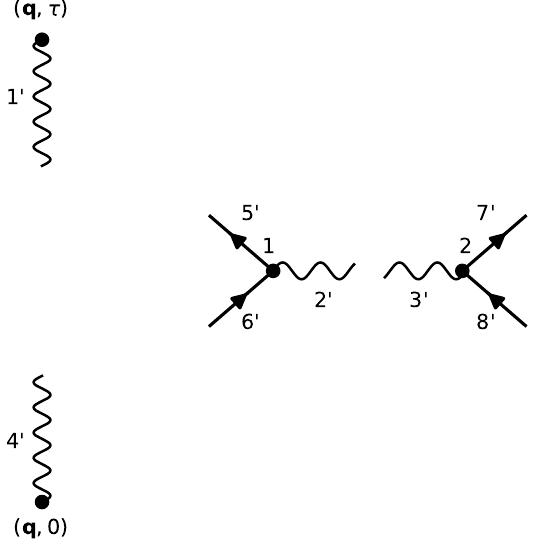}
\caption{Feynman diagram representation of four fermionic and two bosonic operators in the electron-phonon interaction, arising from the second-order expansion of the phonon propagator; see equation $\eqref{eq: svilupposecondoordinepropagatorebosonico2}$.}
\label{fig: interazioneelettronifononioperatorifermioniciebosonicifigura}
\end{figure}

\begin{figure}[H]
\centering
\includegraphics[scale=0.52]{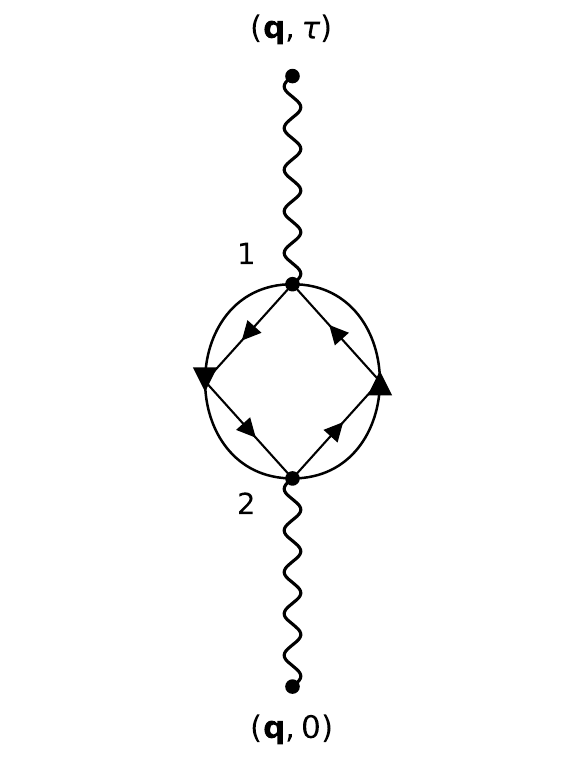}
\caption{Schematic representation of the contractions of four fermionic and two bosonic operators in the electron-phonon interaction, arising from the second-order expansion of the phonon propagator; see equation $\eqref{eq: svilupposecondoordinepropagatorebosonico2}$.}
\label{fig: diagrammacontrazionipropagatorefononicosecondoordine1}
\end{figure}

\begin{figure}[H]
\centering
\includegraphics[scale=0.52]{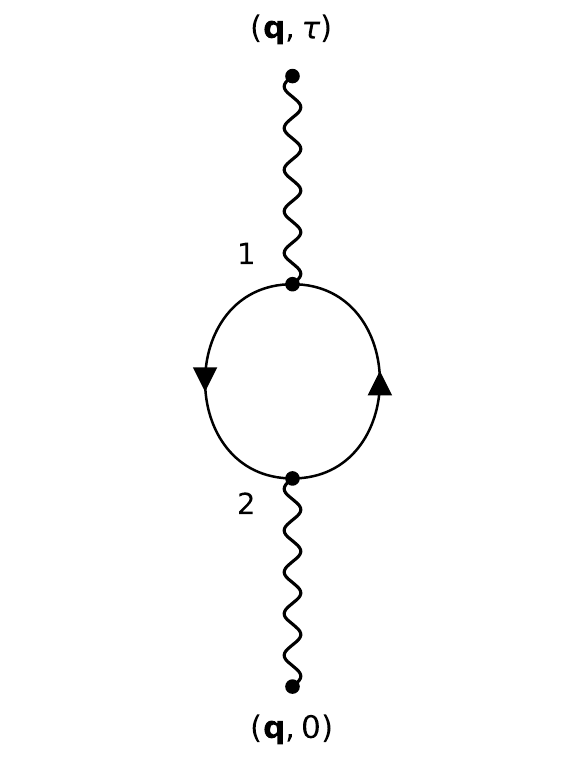}
\caption{Feynman diagram representation of Figure $\eqref{fig: diagrammacontrazionipropagatorefononicosecondoordine1}$.}
\label{fig: diagrammacontrazionipropagatorefononicosecondoordine2}
\end{figure}

\begin{figure}[H]
\centering
\includegraphics[scale=0.56]{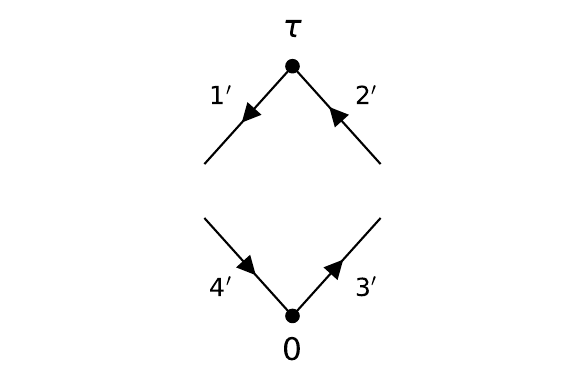}
\caption{Schematic representation of equation $\eqref{eq: propagatoredipolarizzazionespazioimpulsospaziotempiimmaginariprimoordine}$.}
\label{fig: bolladipolarizzazione1}
\end{figure}

\begin{figure}[H]
\centering
\includegraphics[scale=0.56]{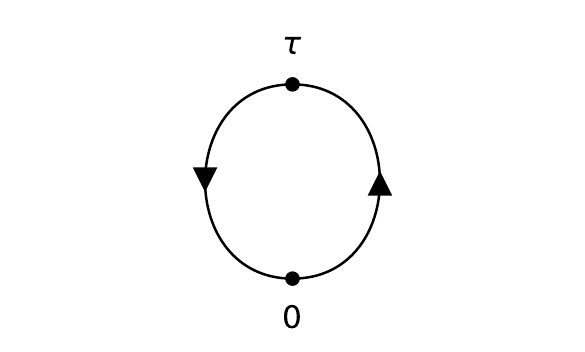}
\caption{Feynman diagram representation of Figure $\eqref{fig: bolladipolarizzazione1}$, i.e., polarization bubble.}
\label{fig: bolladipolarizzazione2}
\end{figure}

\begin{figure}[H]
\centering
\includegraphics[scale=0.7]{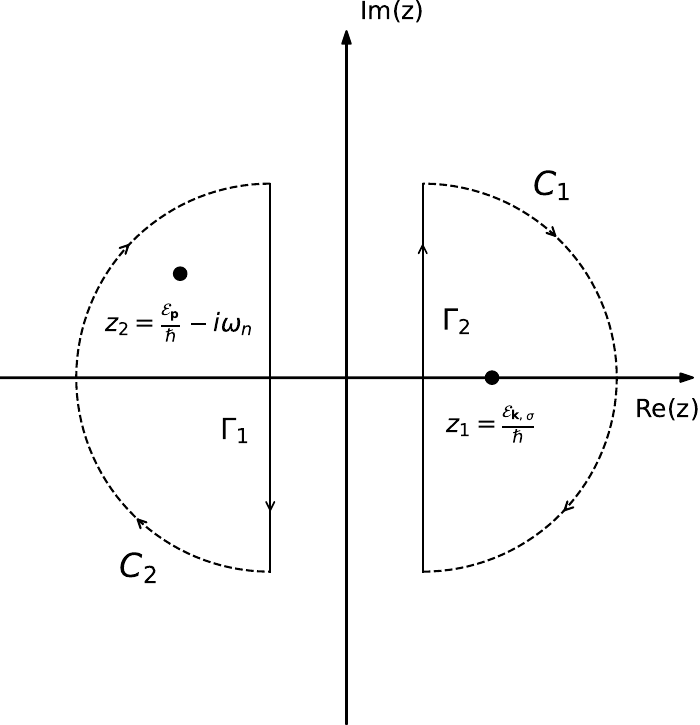}
\caption{Integration contour used for the evaluation of the Matsubara sum in equation $\eqref{eq: secondarelazioneesattafrequenzeMatsubara1}$ via the residue theorem.}
\label{fig: graficointegraleresiduisecondasommamatsubara}
\end{figure}

After defining the free phonon propagator, we consider the phonon propagator, that is
\begin{equation}
D_{\textbf{q}}(\tau) = - \dfrac{1}{\hslash} \left\langle \hat{T}_{\tau} \left( a_{\textbf{q}} + a_{-\textbf{q}}^{\dagger} \right)(\tau) \left( a_{-\textbf{q}} + a_{\textbf{q}}^{\dagger} \right)(0) \right\rangle,
\end{equation}
where the thermal average is taken with respect to the Hamiltonian $\eqref{eq: interazioneparticellebagnotermico}$, and thanks to Feynman's theory we can expand this propagator using the free propagators and the non-interacting Hamiltonian. For convenience, we define
\begin{equation}
A_{\textbf{q}}(\tau) =  \left( a_{\textbf{q}} + a_{-\textbf{q}}^{\dagger} \right)(\tau) ,
\end{equation}
then
\begin{equation}
A^{\dagger}_{\textbf{q}}(\tau) = A_{-\textbf{q}}(\tau),
\end{equation}
\begin{equation}
D_{\textbf{q}}(\tau) = - \dfrac{1}{\hslash} \left\langle \hat{T}_{\tau} A_{\textbf{q}}(\tau) A_{\textbf{q}}^{\dagger}(0) \right\rangle.
\end{equation}
Let us consider the Feynman perturbative expansion. At order zero, the propagator coincides with the free propagator. At order $n=1$, the interaction appears only once, but the term is zero since the interaction Hamiltonian $\hat{\mathcal{H}}_I$ appears only once and the thermal average $\left\langle a_{\textbf{q}} + a_{-\textbf{q}}^{\dagger} \right\rangle_0$ vanishes. The first non-zero order is for $n=2$ and the phonon propagator at second order is written as
\begin{align}
D_{\textbf{q}}^{(2)}(\tau) &= D_{\textbf{q}}^{(0)}(\tau) - \dfrac{1}{\hslash} \left( \dfrac{1}{\hslash} \right)^2 \dfrac{1}{2} \int_0^{\beta \hslash} d(\tau_1) \int_0^{\beta \hslash} d(\tau_2) \sum_{\textbf{q}_1,\textbf{k}_1,\sigma_1} \sum_{\textbf{q}_2,\textbf{k}_2,\sigma_2} M_{\textbf{q}_1} M_{\textbf{q}_2} \notag \\
&\left\langle \hat{T}_\tau A^{(0)}_{\textbf{q}}(\tau) A^{\dagger (0)}_{-\textbf{q}_1}(\tau_1) A^{(0)}_{\textbf{q}_2}(\tau_2) A^{\dagger}_{\textbf{q}}(0)\right\rangle_0 \left\langle \hat{T}_{\tau} C^{\dagger (0)}_{\textbf{k}_1+\textbf{q}_1,\sigma_1}(\tau_1) C^{(0)}_{\textbf{k}_1,\sigma_1}(\tau_1) C^{\dagger (0)}_{\textbf{k}_2+\textbf{q}_2,\sigma_2}(\tau_2) C^{(0)}_{\textbf{k}_2,\sigma_2}(\tau_2) \right\rangle_0 ,
\label{eq: svilupposecondoordinepropagatorebosonico1}
\end{align}
and from $A^{\dagger (0)}_{-\textbf{q}1}(\tau_1) = A^{(0)}_{\textbf{q}_1}(\tau_1)$, we can rewrite
\begin{align}
D_{\textbf{q}}^{(2)}(\tau) &= D_{\textbf{q}}^{(0)}(\tau) - \dfrac{1}{\hslash} \left( \dfrac{1}{\hslash} \right)^2 \dfrac{1}{2} \int_0^{\beta \hslash} d(\tau_1) \int_0^{\beta \hslash} d(\tau_2) \sum_{\textbf{q}_1,\textbf{k}_1,\sigma_1} \sum_{\textbf{q}_2,\textbf{k}_2,\sigma_2} M_{\textbf{q}_1} M_{\textbf{q}_2} \notag \\
&\left\langle \hat{T}_\tau A^{(0)}_{\textbf{q}}(\tau) A^{(0)}_{\textbf{q}_1}(\tau_1) A^{(0)}_{\textbf{q}_2}(\tau_2) A^{\dagger}_{\textbf{q}}(0)\right\rangle_0 \left\langle \hat{T}_{\tau} C^{\dagger (0)}_{\textbf{k}_1+\textbf{q}_1,\sigma_1}(\tau_1) C^{(0)}_{\textbf{k}_1,\sigma_1}(\tau_1) C^{\dagger (0)}_{\textbf{k}_2+\textbf{q}_2,\sigma_2}(\tau_2) C^{(0)}_{\textbf{k}_2,\sigma_2}(\tau_2) \right\rangle_0.
\label{eq: svilupposecondoordinepropagatorebosonico2}
\end{align}
We set $1' \equiv A^{(0)}_{\textbf{q}}(\tau)$, $2' \equiv A^{(0)}_{\textbf{q}_1}(\tau_1)$, $3' \equiv A^{(0)}_{\textbf{q}_2}(\tau_2)$, $4' \equiv A^{\dagger}_{\textbf{q}}(0)$, $5' \equiv C^{\dagger (0)}_{\textbf{k}_1+\textbf{q}_1,\sigma_1}(\tau_1)$, $6' \equiv C^{(0)}_{\textbf{k}_1,\sigma_1}(\tau_1)$, $7' \equiv C^{\dagger (0)}_{\textbf{k}_2+\textbf{q}_2,\sigma_2}(\tau_2)$, $8' \equiv C^{(0)}_{\textbf{k}_2,\sigma_2}(\tau_2)$ and we apply the thermal Wick theorem. The operators $A^{(0)}_{\textbf{q}}(\tau)$ and $A^{\dagger}_{\textbf{q}}(0)$ play the role of destruction and creation field operators, respectively: they are represented by a bosonic line that originates at time $0$, $A^{\dagger}_{\textbf{q}}(0)$, see Figure $\eqref{fig: operatorebosonicodistruzionediagrammatica}$, or terminates at time $\tau$, i.e., $A^{(0)}_{\textbf{q}}(\tau)$, see Figure $\eqref{fig: operatorebosonicocreazionediagrammatica}$. Consequently, similarly to the free and interacting propagators of the field operators, the free phonon propagator is represented as in Figure $\eqref{fig: operatorefononicoliberodiagrammatica}$ and the phonon propagator as in Figure $\eqref{fig: operatorefononicodiagrammatica}$. The interaction vertices are two, $\tau_1$ and $\tau_2$, since we are at second order, and each vertex includes an incoming particle line ($C$) and an outgoing one ($C^{\dagger}$), and a half bosonic line ($2'$ and $3'$), as in Figure $\eqref{fig: interazioneelettronifononioperatorifermioniciebosonicifigura}$. We cannot contract $1'$ with $4'$ since that would yield disconnected diagrams, which are excluded from the Dyson expansion. We must contract $1'$ with one of the two half interaction lines, and both these contractions give the same contribution since both vertices are integrated from $0$ to $\beta \hslash$. We choose to contract $1'-2'$ and $3'-4'$, and this double diagram removes the factor $\frac{1}{2}$ in the second order expansion. The contraction $1'-2'$ produces a free phonon propagator with $\textbf{q}_1 = - \textbf{q}$ up to a factor $- \frac{1}{\hslash}$, and the contraction $3'-4'$ produces a free phonon propagator with $\textbf{q}_2 = \textbf{q}$, up to a factor $- \frac{1}{\hslash}$: the numerical factor $\left(-\frac{1}{\hslash}\right)^2$ in the second order expansion facilitates this algebraic manipulation. Furthermore, the contractions $5'-6'$ and $7'-8'$ are not admissible, as they imply $\textbf{q}_1=0$ and $\textbf{q}_2=0$, respectively, for which the interaction is zero. Therefore, we must contract $5'-8'$ and $6'-7'$. Equation $\eqref{eq: svilupposecondoordinepropagatorebosonico2}$ becomes
\begin{align}
D_{\textbf{q}}^{(2)}(\tau) &= D_{\textbf{q}}^{(0)}(\tau) - \dfrac{1}{\hslash} \int_0^{\beta \hslash} d(\tau_1) \int_0^{\beta \hslash} d(\tau_2) \sum_{\textbf{k}_1,\sigma_1} \sum_{\textbf{k}_2,\sigma_2} \left|M_{\textbf{q}} \right|^2 \notag \\
& \ \ \ \ \ (- \hslash) G^{(0)}_{\textbf{k}_1,\sigma_1}(\tau_1 - \tau_2) \delta_{\textbf{k}_2,\textbf{k}_1 - \textbf{q}} \delta_{\sigma_1,\sigma_2} \varepsilon (- \hslash) G^{(0)}_{\textbf{k}_1 - \textbf{q},\sigma_1}(\tau_2 - \tau_1) D_{\textbf{q}}^{(0)}(\tau - \tau_1) D_{\textbf{q}}^{(0)}(\tau_2),
\label{eq: svilupposecondoordinepropagatorebosonico3}
\end{align}
where $(- \hslash) G^{(0)}_{\textbf{k}_1,\sigma_1}(\tau_1 - \tau_2) \delta_{\textbf{k}_2,\textbf{k}_1 - \textbf{q}} \delta_{\sigma_1,\sigma_2}$ comes from the contraction $5'-8'$ and $\varepsilon (- \hslash) G^{(0)}_{\textbf{k}_1 - \textbf{q},\sigma_1}(\tau_2 - \tau_1)$ comes from the contraction $6'-7'$. The resulting diagram is Figure $\eqref{fig: diagrammacontrazionipropagatorefononicosecondoordine1}$: the contraction $1'-2'$ produces a free phonon propagator with vertices at $\tau$ and $\tau_1$, and in particular at $\tau_1$, there is an incoming and an outgoing line ($C^\dagger$ and $C$); similarly, the contraction $3'-4'$ produces a free phonon propagator with vertices at $\tau_2$ and $0$, and in particular at $\tau_2$ there is an incoming and an outgoing line ($C^\dagger$ and $C$); finally, $G^{(0)}_{\textbf{k}_1-\textbf{q},\sigma_1}(\tau_2 - \tau_1)$ is the propagator of an electron propagating from the internal time $\tau_1$ to the time $\tau_2$, while $G^{(0)}_{\textbf{k}_1,\sigma_1}(\tau_1 - \tau_2)$ is the propagator of an electron propagating from time $\tau_2$ to time $\tau_1$, and together the two propagators form a closed electron loop into which the two pairs of incoming and outgoing lines for each vertex are inserted. In the Feynman diagrammatics, the contractions in Figure $\eqref{fig: diagrammacontrazionipropagatorefononicosecondoordine1}$ are schematized as in Figure $\eqref{fig: diagrammacontrazionipropagatorefononicosecondoordine2}$, where the four fermionic lines are absorbed into the loop. From $\eqref{eq: svilupposecondoordinepropagatorebosonico2}$, summing up we have
\begin{equation}
D_{\textbf{q}}^{(2)}(\tau) = D_{\textbf{q}}^{(0)}(\tau) - \varepsilon \hslash \int_0^{\beta \hslash} d(\tau_1) \int_0^{\beta \hslash} d(\tau_2) \sum_{\textbf{k}_1,\sigma_1} \left|M_{\textbf{q}} \right|^2 G^{(0)}_{\textbf{k}_1-\textbf{q},\sigma_1}(\tau_2 - \tau_1) G^{(0)}_{\textbf{k}_1,\sigma_1}(\tau_1 - \tau_2) D_{\textbf{q}}^{(0)}(\tau - \tau_1) D_{\textbf{q}}^{(0)}(\tau_2).
\label{eq: svilupposecondoordinepropagatorebosonico3}
\end{equation}
To remove the integration with respect to time, we apply a temporal Fourier transform, that is
\begin{align}
D_{\textbf{q}}^{(2)}(i \omega_n) &= \int_0^{\beta \hslash} d\tau e^{i \omega_n \tau} D_{\textbf{q}}^{(2)}(\tau) = \notag \\
&= D_{\textbf{q}}^{(0)}(i \omega_n) - \varepsilon \hslash \int_0^{\beta \hslash} d\tau e^{i \omega_n \tau} \int_0^{\beta \hslash} d(\tau_1) \int_0^{\beta \hslash} d(\tau_2) \sum_{\textbf{k}_1,\sigma_1} \left|M_{\textbf{q}} \right|^2 \dfrac{1}{(\beta \hslash)^4} \notag \\
& \ \ \sum_{n_1,n_4,n_3,n_4} G^{(0)}_{\textbf{k}_1-\textbf{q},\sigma_1}(i \omega_{n_1}) G^{(0)}_{\textbf{k}_1,\sigma_1}(i \omega_{n_2}) D_{\textbf{q}}^{(0)}(i \omega_{n_3}) D_{\textbf{q}}^{(0)}(i \omega_{n_4}) e^{i \omega_{n_1} (\tau_2 - \tau_1)} e^{i \omega_{n_2} (\tau_1 - \tau_2)} e^{i \omega_{n_3} (\tau - \tau_1)} e^{i \omega_{n_4} (\tau_2)}. 
\label{eq: svilupposecondoordinepropagatorebosonico4}
\end{align}
From the integrals with respect to $\tau$, $\tau_1$, $\tau_2$ we respectively have: $n=n_3$, $n_2=n_1+n$, $n_2=n_4 + n_1$ and combining, we get $n_4=n$, furthermore the three integrals provide a numerical factor $\beta \hslash$ each, that is,
\begin{equation}
D_{\textbf{q}}^{(2)}(i \omega_n) = D_{\textbf{q}}^{(0)}(i \omega_n) - \dfrac{\varepsilon}{\beta} \left| M_{\textbf{q}} \right|^2 \sum_{\textbf{k}_1,\sigma_1,n_1} G^{(0)}_{\textbf{k}_1-\textbf{q},\sigma_1}(i \omega_{n_1}) G^{(0)}_{\textbf{k}_1,\sigma_1}(i \omega_{n_1} + i \omega_n) D_{\textbf{q}}^{(0)}(i \omega_{n}) D_{\textbf{q}}^{(0)}(i \omega_{n}) .
\label{eq: svilupposecondoordinepropagatorebosonico5}
\end{equation}
By comparing $\eqref{eq: equazioneDysonalgebrica}$ and $\eqref{eq: svilupposecondoordinepropagatorebosonico5}$, remembering that at the first nonzero order the self-energies $\Sigma$ and $\Sigma^*$ coincide, we have
\begin{equation}
\Sigma^{(1)}_{\textbf{k},\sigma}(i \omega_n) = - \dfrac{\varepsilon}{\beta} \left| M_{\textbf{q}} \right|^2 \sum_{\textbf{k}_1,\sigma_1,n_1} G^{(0)}_{\textbf{k}_1-\textbf{q},\sigma_1}(i \omega_{n_1}) G^{(0)}_{\textbf{k}_1,\sigma_1}(i \omega_{n_1} + i \omega_n).
\label{eq: selfenergyprimoordinepropagatorefononico1}
\end{equation}
Such self-energy is actually the lowest order of another type of propagator, as we will now verify. Consider the particle density operator in second quantization and in momentum space $\eqref{eq: trasformatadiFourieroperatoredensitadifluttuazioneparticelle}$, and define the polarization operator, or polarization propagator, in momentum space and imaginary time as
\begin{equation}
\Pi_{\textbf{q}}(\tau) = - \dfrac{i}{\hslash} \left\langle \left[ \hat{\rho}_{\textbf{q}}(\tau) , \hat{\rho}^{\dagger}_{\textbf{q}}(0) \right] \right\rangle,
\label{eq: propagatoredipolarizzazionespazioimpulsospaziotempiimmaginari}
\end{equation}
which represents a correlation between the particle density at different points and times. Let us consider the free polarization propagator, that is, with respect to the Hamiltonian without interaction, so
\begin{equation}
\Pi^{(0)}_{\textbf{q}}(\tau) = - \dfrac{1}{\hslash} \left\langle \hat{T}_\tau C^{\dagger (0)}_{\textbf{k}_1 + \textbf{q},\sigma_1}(\tau) C^{(0)}_{\textbf{k}_1,\sigma_1}(\tau) C^{\dagger (0)}_{\textbf{k}_2 - \textbf{q},\sigma_2}(0) C^{(0)}_{\textbf{k}_2,\sigma_2}(0) \right\rangle_0,
\label{eq: propagatoredipolarizzazionespazioimpulsospaziotempiimmaginariprimoordine}
\end{equation}
where $\textbf{q}$ must be different from zero, otherwise the interaction is zero, $M_{\textbf{q}=\textbf{0}}=0$. We observe that, in contrast to the free propagator associated with the field operators, this is a higher-order Green's function. This is because the particle density operators involve two field operators each, so the thermal average at first order involves a four-operator expectation value. At each vertex, located at times $0$ and $\tau$, one particle is created and one is annihilated, leading to the contraction of four operators in total. With the identifications $1' \equiv C^{\dagger (0)}_{\textbf{k}_1 + \textbf{q},\sigma_1}(\tau)$, $2' \equiv C^{(0)}_{\textbf{k}_1,\sigma_1}(\tau)$, $3' \equiv C^{\dagger (0)}_{\textbf{k}_2 - \textbf{q},\sigma_2}(0)$, $4' \equiv C^{(0)}_{\textbf{k}_2,\sigma_2}(0)$, the only possible contractions are $1'-4'$, $2'-3'$, see Figure $\eqref{fig: bolladipolarizzazione1}$. Indeed, the contraction $1'-2'$ implies $\textbf{q}=\textbf{0}$, which we have excluded. The only possible diagram then is the one in Figure $\eqref{fig: bolladipolarizzazione2}$, which is called the polarization bubble and analytically the first-order polarization propagator is given by
\begin{equation}
\Pi^{(0)}_{\textbf{q}}(\tau) = - \varepsilon \hslash \sum_{\textbf{k}_1,\sigma_1} G^{(0)}_{\textbf{k}_1 + \textbf{q},\sigma_1}(-\tau) G^{(0)}_{\textbf{k}_1,\sigma_1}(\tau).
\label{eq: propagatoredipolarizzazionespazioimpulsospaziotempiimmaginariprimoordine}
\end{equation}
We apply the temporal Fourier transform and we have
\begin{align}
\Pi^{(0)}_{\textbf{q}}(i \omega_n) &= \int_0^{\beta \hslash} d\tau e^{i \omega_n \tau} \Pi^{(0)}_{\textbf{q}}(\tau) = \notag \\
&= - \varepsilon \hslash \int_0^{\beta \hslash} d\tau e^{i \omega_n \tau} \sum_{\textbf{k}_1,\sigma_1} \dfrac{1}{(\beta \hslash)^2} \sum_{n_1,n_2} G^{(0)}_{\textbf{k}_1 + \textbf{q},\sigma_1}(i \omega_{n_1}) G^{(0)}_{\textbf{k}_1,\sigma_1}(i \omega_{n_2}) e^{- i \omega_{n_1} (- \tau)} e^{- i \omega_{n_2} \tau} = \notag \\
&= - \varepsilon \hslash \int_0^{\beta \hslash} d\tau e^{- i (\omega_n + \omega_{n_1} - \omega_{n_2}) \tau} \sum_{\textbf{k}_1,\sigma_1} \dfrac{1}{(\beta \hslash)^2} \sum_{n_1,n_2} G^{(0)}_{\textbf{k}_1 + \textbf{q},\sigma_1}(i \omega_{n_1}) G^{(0)}_{\textbf{k}_1,\sigma_1}(i \omega_{n_2}) .
\end{align}
The integral with respect to $\tau$ is equal to $\beta \hslash \delta_{n+n_1,n_2}$, so
\begin{equation}
\Pi^{(0)}_{\textbf{q}}(i \omega_n) = - \dfrac{\varepsilon}{\beta} \sum_{\textbf{k}_1,\sigma_1,n_1} G^{(0)}_{\textbf{k}_1+\textbf{q},\sigma_1}(i \omega_{n_1}) G^{(0)}_{\textbf{k}_1,\sigma_1}(i \omega_{n_1} + i \omega_n).
\label{eq: propagatoredipolarizzazionespazioimpulsospaziofrequenzeprimoordine}
\end{equation}
By comparing $\eqref{eq: selfenergyprimoordinepropagatorefononico1}$ and $\eqref{eq: propagatoredipolarizzazionespazioimpulsospaziofrequenzeprimoordine}$, we notice that the two propagators would be proportional, apart from the sign of the momentum $\textbf{q}$; however, the following theorems show that this sign difference is irrelevant for the polarization propagator object. We have
\begin{theorem}\label{thm: secondasommaMatsubara}[Matsubara sum for bosonic and fermionic frequencies] Let $\omega_n$, $\omega_{n_1}$ be Matsubara frequencies, respectively bosonic and fermionic, then they satisfy
\begin{equation}
\dfrac{1}{\beta} \sum_{n_1} \dfrac{1}{\hslash \left[ i \omega_{n_1} - \frac{\mathcal{E}_{\textbf{k}}}{\hslash} \right]} \dfrac{1}{\hslash \left[ i \omega_{n_1} + i \omega_n - \frac{\mathcal{E}_{\textbf{p}}}{\hslash} \right]} = \dfrac{1}{\hslash} \dfrac{n_{-1}(\mathcal{E}_{\textbf{p}}) - n_{-1}(\mathcal{E}_{\textbf{k}})}{i \omega_n + \frac{\mathcal{E}_{\textbf{p}} - \mathcal{E}_{\textbf{k}}}{\hslash}},
\label{eq: secondarelazioneesattafrequenzeMatsubara1}
\end{equation}
where $n_{-1}$ denotes the Fermi-Dirac function.
\begin{proof}
First note that in equation $\eqref{eq: secondarelazioneesattafrequenzeMatsubara1}$ the factor $e^{i \omega_n \delta}$ is absent, unlike the series in $\eqref{eq: seriemodificatapropagatoreMatsubara}$, since this time the general term is proportional to
\begin{equation}
\sum_{n_1} \dfrac{1}{(i \omega_{n_1})^2} \sim \sum_{n_1} \dfrac{1}{n_1^2} ,
\end{equation}
which is a convergent series. We apply the residue method, with calculations very similar to those carried out for the Matsubara sum $\eqref{eq: freephononpropagator}$ (see Figure $\eqref{fig: graficointegraleresiduisecondasommamatsubara}$). The frequencies $\omega_{n_1}$ are fermionic, hence odd, and the frequency $\omega_n$ is bosonic, hence even; consequently, $\omega_{n_1} + \omega_n$ is odd. The Matsubara sum can be computed using the integral
\begin{equation}
I = \int dz \left( - \dfrac{\beta}{2 \pi i} \right) n_{-1}(z) g(z),
\end{equation}
with 
\begin{equation}
g(z) = \dfrac{1}{\hslash \left[ z - \frac{\mathcal{E}_{\textbf{k}}}{\hslash} \right]} \dfrac{1}{\hslash \left[ z + i \omega_n - \frac{\mathcal{E}_{\textbf{p}}}{\hslash} \right] }.
\end{equation}
The poles of $g(z)$ are two: $z_1 = \frac{\mathcal{E}_{\textbf{k}}}{\hslash}$, which we assume to be positive and on the real axis, and $z_2 = \frac{\mathcal{E}_{\textbf{p}}}{\hslash} - i \omega_n$, and both lie outside the contour $\Gamma$. Applying the residue theorem, we get
\begin{equation}
I = \dfrac{1}{\hslash} \left[ \dfrac{n_{-1}(\mathcal{E}_{\textbf{k}})}{\frac{\mathcal{E}_{\textbf{k}}}{\hslash} + i \omega_n - \frac{\mathcal{E}_{\textbf{p}}}{\hslash}} + \frac{n_{-1}(\mathcal{E}_{\textbf{p}} - i \hslash \omega_n)}{\frac{\mathcal{E}_{\textbf{p}}}{\hslash} - i \omega_n - \frac{\mathcal{E}_{\textbf{k}}}{\hslash}} \right] ,
\end{equation}
where the first contribution is clockwise and the second is counterclockwise. Note that the following holds
\begin{align}
n_{-1}(\mathcal{E}_{\textbf{p}} - i \hslash \omega_n) &= \dfrac{1}{e^{\beta (\mathcal{E}_{\textbf{p}} - i \hslash \omega_n)} + 1} = \notag \\
&= \dfrac{1}{e^{\beta \mathcal{E}_{\textbf{p}}} e^{- i \beta \hslash \omega_n} + 1} = \notag \\
&= \dfrac{1}{e^{\beta \mathcal{E}_{\textbf{p}}} + 1} = \notag \\
&= n_{-1}(\mathcal{E}_{\textbf{p}}) ,
\end{align}
where we used that $\omega_n$ is a bosonic frequency, then $e^{- i 2 \pi n}=1$. We get
\begin{equation}
I = \dfrac{1}{\hslash} \left[ \dfrac{n_{-1}(\mathcal{E}_{\textbf{k}})}{\frac{\mathcal{E}_{\textbf{k}}}{\hslash} + i \omega_n - \frac{\mathcal{E}_{\textbf{p}}}{\hslash}} - \frac{n_{-1}(\mathcal{E}_{\textbf{p}}}{\frac{\mathcal{E}_{\textbf{k}}}{\hslash} + i \omega_n - \frac{\mathcal{E}_{\textbf{p}}}{\hslash}} \right] ,
\end{equation}
from which the statement follows.
\end{proof}
\end{theorem}
From $\eqref{eq: secondarelazioneesattafrequenzeMatsubara1}$, we write the propagator $\eqref{eq: propagatoredipolarizzazionespazioimpulsospaziofrequenzeprimoordine}$ as
\begin{equation}
\Pi^{(0)}_{\textbf{q}}(i \omega_n) = \dfrac{1}{\hslash} \sum_{\textbf{k},\sigma} \dfrac{n_{-1}(\mathcal{E}_{\textbf{k}+\textbf{q}}) - n_{-1}(\mathcal{E}_{\textbf{k}})}{i \omega_n + \frac{\mathcal{E}_{\textbf{k}+\textbf{q}} - \mathcal{E}_{\textbf{k}}}{\hslash}}.
\label{eq: propagatorepolarizzazioneprimoordinesommamatsubaraesplicitata}
\end{equation}
We are now able to show the following
\begin{theorem}\label{thm: primoordinepropagatorepolarizzazionespaziomomentispaziofrequenzeproprieta}
The first-order polarization propagator, equation $\eqref{eq: propagatoredipolarizzazionespazioimpulsospaziofrequenzeprimoordine}$, satisfies
\begin{equation}
\Pi^{(0)}_{\textbf{q}}(i \omega_n) = \Pi^{(0)}_{\textbf{q}}(- i \omega_n),
\label{eq: primoordinepropagatorepolarizzazionespaziomomentispaziofrequenzeproprieta1}
\end{equation}
\begin{equation}
\Pi^{(0)}_{\textbf{q}}(i \omega_n) = \Pi^{(0)}_{-\textbf{q}}(i \omega_n).
\label{eq: primoordinepropagatorepolarizzazionespaziomomentispaziofrequenzeproprieta2}
\end{equation}
\begin{proof}
Let us prove $\eqref{eq: primoordinepropagatorepolarizzazionespaziomomentispaziofrequenzeproprieta1}$. We add and subtract $n_{-1}(\mathcal{E}_{\textbf{k}+\textbf{q}})  n_{-1}(\mathcal{E}_{\textbf{k}})$ in the numerator and write
\begin{align}
\Pi^{(0)}_{\textbf{q}}(i \omega_n) &= \dfrac{1}{\hslash} \sum_{\textbf{k},\sigma} \dfrac{n_{-1}(\mathcal{E}_{\textbf{k}+\textbf{q}}) - n_{-1}(\mathcal{E}_{\textbf{k}+\textbf{q}}) + n_{-1}(\mathcal{E}_{\textbf{k}+\textbf{q}}) n_{-1}(\mathcal{E}_{\textbf{k}}) - n_{-1}(\mathcal{E}_{\textbf{k}+\textbf{q}}) n_{-1}(\mathcal{E}_{\textbf{k}})}{i \omega_n + \frac{\mathcal{E}_{\textbf{k}+\textbf{q}} - \mathcal{E}_{\textbf{k}}}{\hslash}} = \notag \\
&= \dfrac{1}{\hslash} \sum_{\textbf{k},\sigma} \frac{n_{-1}(\mathcal{E}_{\textbf{k}+\textbf{q}})  ( 1 - n_{-1}(\mathcal{E}_{\textbf{k}}))}{i \omega_n + \frac{\mathcal{E}_{\textbf{k}+\textbf{q}} - \mathcal{E}_{\textbf{k}}}{\hslash}} - \dfrac{1}{\hslash} \sum_{\textbf{k},\sigma} \frac{n_{-1}(\mathcal{E}_{\textbf{k}}) (1 - n_{-1}(\mathcal{E}_{\textbf{k}+\textbf{q}}))}{i \omega_n + \frac{\mathcal{E}_{\textbf{k}+\textbf{q}} - \mathcal{E}_{\textbf{k}}}{\hslash}},
\end{align}
so we rename the dummy index in the second term as $- \textbf{k}' = \textbf{k} + \textbf{q}$ and we have
\begin{equation}
\Pi^{(0)}_{\textbf{q}}(i \omega_n) = \dfrac{1}{\hslash} \sum_{\textbf{k},\sigma} \dfrac{n_{-1}(\mathcal{E}_{\textbf{k}+\textbf{q}}) ( 1 -n_{-1}(\mathcal{E}_{\textbf{k}}))}{i \omega_n + \frac{\mathcal{E}_{\textbf{k}+\textbf{q}} - \mathcal{E}_{\textbf{k}}}{\hslash}} - \dfrac{1}{\hslash} \sum_{\textbf{k}',\sigma} \frac{n_{-1}(\mathcal{E}_{-\textbf{k}'-\textbf{q}})  (1 - n_{-1}(\mathcal{E}_{-\textbf{k}'})) }{i \omega_n + \frac{\mathcal{E}_{-\textbf{k}'} - \mathcal{E}_{-\textbf{k}'- \textbf{q}}}{\hslash}}. 
\end{equation}
We now recall that we are working within the jellium model in the grand canonical ensemble, that is equation $\eqref{eq: hamiltonianamodellojellium3}$. In this framework, the single-particle energies take the form \( \mathcal{E}_{\textbf{k}} = \frac{\hslash^2 \textbf{k}^2}{2m} - \mu \). It immediately follows that $\mathcal{E}_{-\textbf{k}} = \mathcal{E}_{\textbf{k}}$, $n_{-1}(\mathcal{E}_{-\textbf{k}}) = n_{-1}(\mathcal{E}_{\textbf{k}})$. Using these identities, and by relabeling the dummy summation index \( \textbf{k}' \rightarrow \textbf{k} \) in the second term, we obtain
\begin{align}
\Pi^{(0)}_{\textbf{q}}(i \omega_n) &= \dfrac{1}{\hslash} \sum_{\textbf{k},\sigma} \dfrac{n_{-1}(\mathcal{E}_{\textbf{k}+\textbf{q}}) ( 1 -n_{-1}(\mathcal{E}_{\textbf{k}}))}{i \omega_n + \frac{\mathcal{E}_{\textbf{k}+\textbf{q}} - \mathcal{E}_{\textbf{k}}}{\hslash}} - \dfrac{1}{\hslash} \sum_{\textbf{k},\sigma} \frac{n_{-1}(\mathcal{E}_{\textbf{k}+\textbf{q}}) ( 1 -n_{-1}(\mathcal{E}_{\textbf{k}}))}{i \omega_n + \frac{\mathcal{E}_{\textbf{k}} - \mathcal{E}_{\textbf{k} + \textbf{q}}}{\hslash}} = \notag \\
&= \dfrac{1}{\hslash} \sum_{\textbf{k},\sigma} \frac{n_{-1}(\mathcal{E}_{\textbf{k}+\textbf{q}}) ( 1 - n_{-1}(\mathcal{E}_{\textbf{k}}))}{i \omega_n + \frac{\mathcal{E}_{\textbf{k}+\textbf{q}} - \mathcal{E}_{\textbf{k}}}{\hslash}} - \frac{1}{\hslash} \sum_{\textbf{k},\sigma} \frac{n_{-1}(\mathcal{E}_{\textbf{k}+\textbf{q}}) ( 1 -n_{-1}(\mathcal{E}_{\textbf{k}}))}{i \omega_n - \frac{\mathcal{E}_{\textbf{k}+\textbf{q}} - \mathcal{E}_{\textbf{k}}}{\hslash}} = \notag \\
&= \dfrac{1}{\hslash} \sum_{\textbf{k},\sigma} \dfrac{n_{-1}(\mathcal{E}_{\textbf{k}+\textbf{q}}) ( 1 -n_{-1}(\mathcal{E}_{\textbf{k}}))}{(i \omega_n)^2 + \left( \frac{\mathcal{E}_{\textbf{k}+\textbf{q}} - \mathcal{E}_{\textbf{k}}}{\hslash} \right)^2} 2 \left[ \frac{\mathcal{E}_{\textbf{k}} - \mathcal{E}_{\textbf{k}+\textbf{q}}}{\hslash} \right].
\label{eq: propagatoredipolarizzazionespazioimpulsospaziofrequenzeprimoordinemanipolazione1}
\end{align}
That is, we have shown that $\Pi^{(0)}_{\textbf{q}}(i \omega_n)$ depends on $i \omega_n$ through its square power, so $\eqref{eq: primoordinepropagatorepolarizzazionespaziomomentispaziofrequenzeproprieta1}$ has been proven. \newline
Let us prove $\eqref{eq: primoordinepropagatorepolarizzazionespaziomomentispaziofrequenzeproprieta2}$. Consider the last step of $\eqref{eq: propagatoredipolarizzazionespazioimpulsospaziofrequenzeprimoordinemanipolazione1}$ and set $- \textbf{k}' = \textbf{k} + \textbf{q}$, so that
\begin{equation}
\Pi^{(0)}_{\textbf{q}}(i \omega_n) = \dfrac{1}{\hslash} \sum_{\textbf{k}',\sigma} \dfrac{n_{-1}(\mathcal{E}_{-\textbf{k}'}) ( 1 - n_{-1}(\mathcal{E}_{-\textbf{k}'-\textbf{q}})}{(i \omega_n)^2 + \left( \frac{\mathcal{E}_{-\textbf{k}'} - \mathcal{E}_{-\textbf{k}' - \textbf{q}}}{\hslash} \right)^2} 2 \left[ \frac{\mathcal{E}_{-\textbf{k}' - \textbf{q}} - \mathcal{E}_{-\textbf{k}'}}{\hslash} \right],
\end{equation}
and we rewrite the energies and Fermi-Dirac functions for the opposite momenta, then rename the dummy index $\textbf{k}' \rightarrow \textbf{k}$ as follows
\begin{equation}
\Pi^{(0)}_{\textbf{q}}(i \omega_n) = \dfrac{1}{\hslash} \sum_{\textbf{k},\sigma} \dfrac{n_{-1}(\mathcal{E}_{\textbf{k}}) ( 1 -n_{-1}(\mathcal{E}_{\textbf{k}+\textbf{q}})}{(i \omega_n)^2 + \left( \frac{\mathcal{E}_{\textbf{k}} - \mathcal{E}_{\textbf{k} + \textbf{q}}}{\hslash} \right)^2} 2 \left[ \frac{\mathcal{E}_{\textbf{k} + \textbf{q}} - \mathcal{E}_{\textbf{k}}}{\hslash} \right].
\end{equation}
From the previous equation it follows
\begin{equation}
\Pi^{(0)}_{-\textbf{q}}(i \omega_n) = \dfrac{1}{\hslash} \sum_{\textbf{k},\sigma} \dfrac{n_{-1}(\mathcal{E}_{\textbf{k}}) ( 1 - n_{-1}(\mathcal{E}_{\textbf{k}-\textbf{q}})}{(i \omega_n)^2 + \left( \frac{\mathcal{E}_{\textbf{k}} - \mathcal{E}_{\textbf{k} - \textbf{q}}}{\hslash} \right)^2} 2 \left[ \dfrac{\mathcal{E}_{\textbf{k} - \textbf{q}} - \mathcal{E}_{\textbf{k}}}{\hslash} \right],
\end{equation}
and if we set $\textbf{k}'=\textbf{k}-\textbf{q}$, we have
\begin{equation}
\Pi^{(0)}_{-\textbf{q}}(i \omega_n) = \dfrac{1}{\hslash} \sum_{\textbf{k}',\sigma} \dfrac{n_{-1}(\mathcal{E}_{\textbf{k}'+\textbf{q}}) ( 1 - n_{-1}(\mathcal{E}_{\textbf{k}'}))}{(i \omega_n)^2 + \left( \frac{\mathcal{E}_{\textbf{k}'+\textbf{q}} - \mathcal{E}_{\textbf{k}'}}{\hslash} \right)^2} 2 \left[ \frac{\mathcal{E}_{\textbf{k}'} - \mathcal{E}_{\textbf{k}'+\textbf{q}}}{\hslash} \right],
\end{equation}
and renaming the dummy index $\textbf{k}' \rightarrow \textbf{k}$, we obtain $\eqref{eq: propagatoredipolarizzazionespazioimpulsospaziofrequenzeprimoordinemanipolazione1}$.
\end{proof}
\end{theorem}
Theorem $\ref{thm: primoordinepropagatorepolarizzazionespaziomomentispaziofrequenzeproprieta}$ shows that the first order of the self-energy we calculated above and the first order of the polarization propagator satisfy
\begin{equation}
\Sigma^{(1)}_{\textbf{k},\sigma}(i \omega_n) = \left| M_{\textbf{q}} \right|^2 \Pi^{(0)}_{\textbf{q}}(i \omega_n),
\end{equation}
and this equality can be shown to hold at every order, that is,
\begin{equation}
\Sigma_{\textbf{k},\sigma}(i \omega_n) = \left| M_{\textbf{q}} \right|^2 \Pi_{\textbf{q}}(i \omega_n),
\end{equation}
\begin{equation}
\Sigma^*_{\textbf{k},\sigma}(i \omega_n) = \left| M_{\textbf{q}} \right|^2 \Pi^*_{\textbf{q}}(i \omega_n),
\end{equation}
From the algebraic Dyson equations, adapted for the electron-phonon interaction, namely
\begin{equation}
D_{\textbf{q}}(i \omega_n) = D^{(0)}_{\textbf{q}}(i \omega_n) + D^{(0)}_{\textbf{q}}(i \omega_n) \left| M_{\textbf{q}} \right|^2 \Pi_{\textbf{q}}(i \omega_n) D^{(0)}_{\textbf{q}}(i \omega_n),
\end{equation}
\begin{equation}
D_{\textbf{q}}(i \omega_n) = D^{(0)}_{\textbf{q}}(i \omega_n) + D^{(0)}_{\textbf{q}}(i \omega_n) \left| M_{\textbf{q}} \right|^2 \Pi^*_{\textbf{q}}(i \omega_n) D_{\textbf{q}}(i \omega_n),
\end{equation}
we multiply both sides by $\left| M_{\textbf{q}} \right|^2$
\begin{equation}
\left| M_{\textbf{q}} \right|^2 D_{\textbf{q}}(i \omega_n) = \left| M_{\textbf{q}} \right|^2 D^{(0)}_{\textbf{q}}(i \omega_n) + \left| M_{\textbf{q}} \right|^2 D^{(0)}_{\textbf{q}}(i \omega_n) \Pi_{\textbf{q}}(i \omega_n) \left| M_{\textbf{q}} \right|^2 D^{(0)}_{\textbf{q}}(i \omega_n),
\end{equation}
\begin{equation}
\left| M_{\textbf{q}} \right|^2 D_{\textbf{q}}(i \omega_n) = \left| M_{\textbf{q}} \right|^2 D^{(0)}_{\textbf{q}}(i \omega_n) + \left| M_{\textbf{q}} \right|^2 D^{(0)}_{\textbf{q}}(i \omega_n) \Pi^*_{\textbf{q}}(i \omega_n) \left| M_{\textbf{q}} \right|^2 D_{\textbf{q}}(i \omega_n).
\end{equation}
Note that the quantity
\begin{equation}
U_{eff}(\textbf{q},i \omega_n) = \left| M_{\textbf{q}} \right|^2 D_{\textbf{q}}(i \omega_n)
\end{equation}
it is the effective potential between two particles renormalized by the electron-phonon interaction, and it includes all orders of the perturbative expansion; similarly,
\begin{equation}
U^{(0)}_{eff}(\textbf{q},i \omega_n) = \left| M_{\textbf{q}} \right|^2 D^{(0)}_{\textbf{q}}(i \omega_n)
\end{equation}
is the effective potential without renormalization. The Dyson equations can then be rewritten in a more suggestive form as
\begin{equation}
U_{eff}(\textbf{q},i \omega_n) = U^{(0)}_{eff}(\textbf{q},i \omega_n) + U^{(0)}_{eff}(\textbf{q},i \omega_n) \Pi_{\textbf{q}}(i \omega_n) V^{(0)}_{eff}(\textbf{q},i \omega_n),
\label{eq: equazioneDysonpotenzialeefficacepropagatorepolarizzazione1}
\end{equation}
\begin{equation}
U_{eff}(\textbf{q},i \omega_n) = U^{(0)}_{eff}(\textbf{q},i \omega_n) + U^{(0)}_{eff}(\textbf{q},i \omega_n) \Pi^*_{\textbf{q}}(i \omega_n) V_{eff}(\textbf{q},i \omega_n).
\label{eq: equazioneDysonpotenzialeefficacepropagatorepolarizzazionestar1}
\end{equation}
Equation $\eqref{eq: equazioneDysonpotenzialeefficacepropagatorepolarizzazionestar1}$ can be written as
\begin{equation}
U_{eff}(\textbf{q},i \omega_n) = \dfrac{U^{(0)}_{eff}(\textbf{q},i \omega_n)}{1 - U^{(0)}_{eff}(\textbf{q},i \omega_n) \Pi^*_{\textbf{q}}(i \omega_n)}.
\label{eq: equazioneDysonpotenzialeefficacepropagatorepolarizzazionestar2}
\end{equation}
In particular, the effective potential calculated in the absence of interactions with the material medium, and therefore in vacuum, coincides with the Fourier transform of the Coulomb potential, that is
\begin{equation}
U^{(0)}_{eff}(\textbf{q},i \omega_n) = \dfrac{1}{V} \dfrac{4 \pi q_e^2}{q^2}.
\label{eq: trasformataFourierpotenzialeCoulomb}
\end{equation}
\newpage
\section{Figures}
\FloatBarrier
\chapter{Linear response theory}\label{Linear response theory}
In this chapter, we introduce the theory of linear response, the fundamental theoretical framework used to describe how a quantum system reacts to a weak external perturbation. Linear response constitutes the first useful approximation to connect microscopic properties with experimental observables such as the dielectric function, magnetic susceptibility, and optical conductivity. We will start from the general formulation of linear response within the formalism of Green’s functions and correlation operators, defining the response functions in space-time coordinates and in the frequency domain. The causal principles and analytic properties governing these functions will be analyzed, as well as their close connection to the thermal correlation functions of the system. \newline
Particular attention will be devoted to the dielectric function, which describes the electronic response to an external electric field, the magnetic susceptibility, which characterizes the response to magnetic fields, and the optical conductivity, which quantifies the material’s ability to absorb and transport electromagnetic energy. Special emphasis will be placed on the sum rules of the dielectric function, essential principles that ensure the physical consistency of the response and relate frequency integrals to static observable quantities. The physical significance of the zeros of the dielectric function will also be examined, as they indicate potential instabilities or the presence of collective excitations such as plasmons. \newline
In addition, for the explicit calculation of the dielectric function, three fundamental theoretical approximations will be introduced and discussed. The Hartree-Fock approach explicitly includes electron exchange effects, enhancing the description of correlations at the single-particle level. The Thomas-Fermi approximation provides a semiclassical description of the electronic response at low frequencies and long wavelengths. Finally, the Random Phase Approximation (RPA) captures collective electronic fluctuations and incorporates dynamic screening in a self-consistent manner. \newline
This chapter only aims to provide both the formal tools and practical applications necessary to study and interpret the dynamic and response properties of materials, with particular attention to the fundamental theoretical aspects and widely used computational methodologies in condensed matter physics.
\section{General formalism}
Let $\hat{\mathcal{H}}$ be a time-independent Hamiltonian that describes a system of particles interacting both with each other and with a static external field. The solution to the time-dependent Schrödinger equation is given by $\psi(t) = e^{-i \frac{\hat{\mathcal{H}}}{\hslash} t} \, \tilde{\psi}(0)$. At a certain time $t_0$, a time-dependent external perturbation is switched on, so that the total Hamiltonian becomes
\begin{equation}
\hat{\mathcal{H}}(t) = \hat{\mathcal{H}} + \hat{\mathcal{H}}_{ext}(t),
\end{equation}
with
\begin{equation}
\hat{\mathcal{H}}_{ext}(t) = 0, \ \forall \ t \leq t_0.
\end{equation}
We seek a solution to the Schrödinger equation of the form
\begin{equation}
\psi(t) = e^{- i \frac{\hat{\mathcal{H}}}{\hslash} t} f(t) \tilde{\psi}(0),
\end{equation}
\begin{equation}
f(t) = \mathds{1}, \ \forall \ t \leq t_0,
\end{equation}
where the operator $f(t)$ represents the external perturbation. We substitute $\psi(t)$ into Schrödinger's equation and we get
\begin{equation}
i \hslash \partial_t \left( e^{- i \frac{\hat{\mathcal{H}}}{\hslash} t} f(t) \tilde{\psi}(0) \right) = \left( \hat{\mathcal{H}} + \hat{\mathcal{H}}_{ext}(t) \right) \psi(t),
\end{equation}
\begin{equation}
i \hslash \left( - \dfrac{i}{\hslash} \hat{\mathcal{H}} \, e^{- i \frac{\hat{\mathcal{H}}}{\hslash} t} f(t) \tilde{\psi}(0) + e^{- i \frac{\hat{\mathcal{H}}}{\hslash} t} \dfrac{df(t)}{dt} \tilde{\psi}(0) \right) = \hat{\mathcal{H}} \psi(t) + \hat{\mathcal{H}}_{ext}(t) \psi(t),
\end{equation}
\begin{equation}
\hat{\mathcal{H}} \, e^{- i \frac{\hat{\mathcal{H}}}{\hslash} t} f(t) \tilde{\psi}(0) + i \hslash \, e^{- i \frac{\hat{\mathcal{H}}}{\hslash} t} \dfrac{df(t)}{dt} \tilde{\psi}(0) = \hat{\mathcal{H}} \psi(t) + \hat{\mathcal{H}}_{ext}(t) \psi(t),
\end{equation}
\begin{equation}
\hat{\mathcal{H}} \psi(t) + i \hslash \, e^{- i \frac{\hat{\mathcal{H}}}{\hslash} t} \dfrac{df(t)}{dt} \tilde{\psi}(0) = \hat{\mathcal{H}} \psi(t) + \hat{\mathcal{H}}_{ext}(t) \psi(t),
\end{equation}
\begin{equation}
i \hslash \, e^{- i \frac{\hat{\mathcal{H}}}{\hslash} t} \dfrac{df(t)}{dt} \tilde{\psi}(0) = \hat{\mathcal{H}}_{ext}(t) \psi(t),
\end{equation}
\begin{equation}
i \hslash \, e^{- i \frac{\hat{\mathcal{H}}}{\hslash} t} \dfrac{df(t)}{dt} \tilde{\psi}(0) = \hat{\mathcal{H}}_{\text{ext}}(t) \, e^{- i \frac{\hat{\mathcal{H}}}{\hslash} t} f(t) \tilde{\psi}(0),
\end{equation}
\begin{equation}
i \hslash \, e^{- i \frac{\hat{\mathcal{H}}}{\hslash} t} \dfrac{df(t)}{dt} = \hat{\mathcal{H}}_{ext}(t) \, e^{- i \frac{\hat{\mathcal{H}}}{\hslash} t} f(t),
\end{equation}
and we multiply to the left both members by $e^{i \frac{\mathcal{H}}{\hslash} t}$, then
\begin{equation}
i \hslash \dfrac{df(t)}{dt} = e^{i \frac{\hat{\mathcal{H}}}{\hslash} t} \hat{\mathcal{H}}_{ext}(t) e^{- i \frac{\hat{\mathcal{H}}}{\hslash} t} f(t) ,
\end{equation}
\begin{equation}
i \hslash \dfrac{df(t)}{dt} = \hat{\mathcal{H}}_{ext}^{(0)}(t) f(t).
\end{equation}
We made explicit the Heisenberg representation $\hat{\mathcal{H}}_{ext}^{(0)}(t) = e^{i \frac{\hat{\mathcal{H}}}{\hslash} t} \hat{\mathcal{H}}_{ext}(t) e^{- i \frac{\hat{\mathcal{H}}}{\hslash} t}$, where the subscript $(0)$ indicates that this representation is with respect to the unperturbed Hamiltonian and not the total Hamiltonian. In general, the equation above is an operator equation, then we integrate both members between $t_0$ and $t$ and we get
\begin{equation}
i \hslash \left[ f(t) - f(t_0) \right] = \int_{t_0}^t dt' \hat{\mathcal{H}}_{ext}^{(0)}(t') f(t'),
\end{equation}
\begin{equation}
i \hslash \left[ f(t) - \mathds{1} \right] = \int_{t_0}^t dt' \hat{\mathcal{H}}_{ext}^{(0)}(t') f(t'),
\end{equation}
\begin{equation}
f(t) = \mathds{1} + \frac{1}{i \hslash} \int_{t_0}^t dt' \hat{\mathcal{H}}_{ext}^{(0)}(t') f(t') ,
\end{equation}
which we solve by iterating and we get
\begin{equation}
f(t) = \mathds{1} + \dfrac{1}{i\hslash} \int_{t_0}^t dt' \hat{\mathcal{H}}_{ext}^{(0)}(t') + \dfrac{1}{i\hslash} \int_{t_0}^t dt' \hat{\mathcal{H}}_{ext}^{(0)}(t') \left[ \dfrac{1}{i \hslash} \int_{t_0}^{t'} dt'' \hat{\mathcal{H}}_{ext}^{(0)}(t'') f(t'') \right] .
\end{equation}
In the linear response theory approach, if the modulus of the perturbation is negligible with respect to the initial Hamiltonian, we neglect terms of higher order than prime with respect to $\hat{\mathcal{H}}_{ext}^{(0)}$, i.e.,
\begin{equation}
f(t) \simeq \mathds{1} + \frac{1}{i\hslash} \int_{t_0}^t dt' \hat{\mathcal{H}}_{ext}^{(0)}(t'),
\end{equation}
\begin{equation}
\psi(t) = e^{- i \frac{\hat{\mathcal{H}}}{\hslash} t} \tilde{\psi}(0) + e^{- i \frac{\hat{\mathcal{H}}}{\hslash} t} \dfrac{1}{i \hslash} \int_{t_0}^t dt' \hat{\mathcal{H}}_{ext}^{(0)}(t') \tilde{\psi}(0),
\end{equation}
in other words, within linear response theory, the wave function is expressed as the sum of the unperturbed wave function and a correction induced by the external perturbation. The expectation value of a generic observable can then be evaluated using the perturbed wave function as follows
\begin{align}
\left\langle A(t) \right\rangle &= \left\langle \tilde{\psi}^*(0) A^{(0)}(t) \tilde{\psi}(0)\right\rangle - \dfrac{1}{i \hslash} \left\langle \tilde{\psi}^*(0) \int_{t_0}^t dt' \hat{\mathcal{H}}_{ext}^{(0)}(t') A^{(0)}(t) \tilde{\psi}(0)\right\rangle + \dfrac{1}{i \hslash} \left\langle \tilde{\psi}^*(0) \int_{t_0}^t dt' A^{(0)}(t) \hat{\mathcal{H}}_{ext}^{(0)}(t') \tilde{\psi}(0)\right\rangle,
\end{align}
which is equivalent to
\begin{align}
\langle A \rangle(t) = \langle A \rangle_0(t) - \dfrac{1}{i\hslash} \int_{t_0}^t dt'\, \left\langle \left[ \hat{\mathcal{H}}_{ext}^{(0)}(t'), A^{(0)}(t) \right] \right\rangle_0 ,
\end{align}
where we have neglected the second-order term in $\hat{\mathcal{H}}^{(0)}_{ext}$ and $\langle \ldots \rangle_0$ denotes the expectation value taken with respect to the initial unperturbed state $\tilde{\psi}(t=0)$. We define
\begin{align}
\delta \langle A \rangle(t) \ &= \ \langle A \rangle(t) \ - \langle A \rangle_0(t) = \notag \\
&= - \dfrac{1}{i\hslash} \int_{t_0}^t dt' \left\langle \left[ \hat{\mathcal{H}}_{ext}^{(0)}(t'),A^{(0)}(t)\right]\right\rangle_0 = \notag \\
&= \dfrac{i}{\hslash} \int_{t_0}^t dt' \left\langle \left[\hat{\mathcal{H}}_{ext}^{(0)}(t'),A^{(0)}(t)\right]\right\rangle_0 \ = \notag \\
&= - \dfrac{i}{\hslash} \int_{t_0}^t dt' \left\langle \left[A^{(0)}(t),\hat{\mathcal{H}}_{ext}^{(0)}(t')\right]\right\rangle_0 ,
\end{align}
which quantifies the variation of \(\langle A \rangle\) at time \(t\) due to the external perturbation. It is important to note that, regardless of whether the particles obey Fermi-Dirac or Bose-Einstein statistics, the integrand must involve a commutator. We now rewrite \(\delta \langle A \rangle(t)\) by extending the integration limits to infinity. Since we assume \(\hat{\mathcal{H}}_{ext}(t_0) = 0\), we can safely take \(t_0 \to -\infty\) without affecting the value of the integral. For the upper limit, we introduce a Heaviside function \(\Theta(t - t')\) and take \(t \to \infty\), in accordance with the principle of causality: the system cannot respond to a perturbation before it is applied. We then obtain
\begin{equation}
\delta \langle A \rangle (t) = - \dfrac{i}{\hslash} \int_{-\infty}^{+\infty} dt' \Theta(t-t') \left\langle \left[A^{(0)}(t),\hat{\mathcal{H}}_{ext}^{(0)}(t')\right] \right\rangle_0 .
\end{equation}
Now suppose that the perturbation takes the form $\hat{\mathcal{H}}_{ext}(t) = h(t) B(t)$, where $h(t)$ is an external classical field, that is, a prescribed function of time, which represents the perturbation, and $B(t)$ is the quantum observable to which it is coupled. We get
\begin{align}
\delta \langle A \rangle (t) \ &= - \dfrac{i}{\hslash} \int_{-\infty}^{+\infty} dt' \Theta(t-t') \left\langle \left[A^{(0)}(t),h(t') B^{(0)}(t')\right] \right\rangle = \notag \\
&= - \dfrac{i}{\hslash} \int_{-\infty}^{+\infty} dt' h(t') \ \Theta(t-t') \left\langle \left[A^{(0)}(t),B^{(0)}(t')\right] \right\rangle_0.
\end{align}
Note that $h(t')$ comes out of the commutator, being a function and not an operator. We get the Kubo's formula
\begin{equation}
\delta \langle A \rangle (t) \ = \int_{-\infty}^{+\infty} dt' h(t') G_{AB}^{(r)}(t,t') ,
\label{eq: formuladiKubo}
\end{equation}
which is a manifestation of the fluctuation-dissipation theorem: within the framework of linear response theory, the dissipation of a physical system due to an external perturbation is governed by time-correlated fluctuations of operators. In this context, the many-body Green's function serves as the response function,
as it quantifies how the system reacts to external perturbations.
\begin{remark}[Clarification on the subscript $0$.] It is crucial to emphasize that the retarded Green's function \( G_{AB}^{(r)}(t - t') \) obtained here is computed in the absence of the external perturbation. However, this does not imply a non-interacting system. The thermal average denoted by the subscript $0$, that is, \( \langle \cdot \rangle_0 \), is evaluated with respect to the unperturbed Hamiltonian, which still contains all particle interactions. This distinction is essential to avoid confusion: the "0'' subscript does not mean that interactions are absent, it simply refers to the absence of the external perturbation. Thanks to this, we are justified in introducing the many-body retarded Green's function, which incorporates all internal interactions exactly, while treating the external field as a small perturbation.
\end{remark}
If the external perturbation oscillates with a well-defined frequency \( \omega \), we can express the response in the frequency domain via a Fourier transform
\begin{equation}
\delta \langle A \rangle (\omega) = \int_{-\infty}^{+\infty} dt e^{i \omega t}\int_{-\infty}^{+\infty} dt' h(t') G_{AB}^{(r)}(t,t'),
\end{equation}
which in general is not a convolution, since $G_{AB}^{(r)}(t,t')$ is a function of $t$ and $t'$ individually. Suppose the system is at thermodynamic equilibrium at $T=0$, before the perturbation. Then, $\delta \langle A \rangle (\omega)$ is a convolution, since $G_{AB}^{(r)}(t,t') = G_{AB}^{(r)}(t-t')$; the wave function of the system is the fundamental state $\psi_{fs}$ of $\hat{\mathcal{H}}$, which includes interactions; the retarded Green's function is no longer a thermal average but an expected value of the operator $\left[ A^{(0)}(t),B^{(0)}(t') \right]$ with respect to the state $\psi_{fs}$. Then
\begin{equation}
G_{AB}^{(r)}(t,t') = -\dfrac{i}{\hslash} \, \Theta(t - t') \, \left\langle \psi_{fs} \left| \left[ A^{(0)}(t), B^{(0)}(t') \right] \right| \psi_{fs} \right\rangle_0, \quad T = 0,
\end{equation}
note that at $T=0$ the density matrix $\frac{e^{- \beta \hat{\mathcal{H}}}}{Z}$ is replaced by the identity. Let us consider the first term of the commutator
\begin{equation}
A^{(0)}(t)B^{(0)}(t') = e^{i \frac{\hat{\mathcal{H}}}{\hslash} t} A e^{-i \frac{\hat{\mathcal{H}}}{\hslash} (t-t')} B e^{-i \frac{\hat{\mathcal{H}}}{\hslash} t'}.
\end{equation}
With respect to the fundamental state $\psi_{fs}$ of $\hat{\mathcal{H}}$, the operator $e^{i \frac{\hat{\mathcal{H}}}{\hslash} t}$ acts as $e^{i \frac{\mathcal{E}_{fs}}{\hslash} t}$, i.e.,
\begin{equation}
\left\langle \psi_{fs} \left| A^{(0)}(t) B^{(0)}(t') \right| \psi_{fs} \right\rangle_0 = e^{i \frac{\mathcal{E}_{\text{fs}}}{\hslash} (t - t')} \left\langle \psi_{fs} \left| A \, e^{-i \frac{\hat{\mathcal{H}}}{\hslash} (t - t')} B \right| \psi_{fs} \right\rangle_0.
\end{equation}
Now, the quantity
\begin{equation}
\delta \langle A \rangle (\omega) = \int_{-\infty}^{+\infty} dt e^{i\omega t} \int_{-\infty}^{+\infty} dt' h(t') G_{AB}^{(r)}(t-t')
\end{equation}
is a convolution integral, so
\begin{equation}
\delta \langle A \rangle (\omega) = h(\omega)G_{AB}^{(r)}(\omega).
\end{equation}
However, $T=0$ cannot be reached experimentally. Suppose that before the perturbation the system is in contact with a thermal bath at temperature $T \neq 0$, then the fundamental state $\psi_{fs}$ is replaced by an excited state $\psi_r$. Consequently, the average value of an observable must be computed with respect to the statistical mixture of the states accessible to the system. The coefficients $p_r$, which represent the statistical weights of the states in this mixture, are generally modified by the perturbation, reflecting the change in the system's probability distribution. In particular, from 
\begin{equation}
\delta \langle A \rangle = - \dfrac{i}{\hslash} \int_{t_0}^t dt' \left\langle \psi_r \left| \left[ A^{(0)}(t), \hat{\mathcal{H}}_{ext}^{(0)}(t') \right] \right| \psi_r \right\rangle,
\end{equation}
we observe that if the modified statistical weights were to include the external perturbation $\hat{\mathcal{H}}_{ext}$, higher-order terms in the expansion would appear, which are not allowed within the framework of a linear theory. Therefore, the only admissible choice for the statistical weight is the Maxwell-Boltzmann distribution corresponding to the unperturbed Hamiltonian. Furthermore, since in experiments the temperature is strictly positive ($T > 0$), many-body Green's functions are defined as thermal averages and are thus always computed over excited states, rather than over the ground state. \newline
To recapitulate, the mean value of a physical observable evolves over time to first order as an integral of an average involving the external perturbative Hamiltonian. In the next sections we will see that applying linear response theory to perturbative Hamiltonians containing the operators particle density, spin volume density and current density yields the dielectric function, magnetic susceptibility and optical conductivity, respectively. \newline
For notational convenience, in the remainder of this chapter, thermal averages computed with respect to the unperturbed Hamiltonian, i.e., without the external perturbation but still including all particle interactions, will be denoted simply as \( \langle \ldots \rangle \), omitting the subscript \( 0 \). It is understood that these averages always refer to the interacting system in the absence of the external field.
\section{Dielectric function}\label{Dielectric function}
\subsection{Dielectric function}
Let us consider a system of interacting electrons. The coupling of these electrons, each carrying a charge \( q_e \), to an external, longitudinal electric field is described by the interaction Hamiltonian that involves the particle density operator $\hat{\rho}$ and the scalar potential \( \varphi_{ext}(\mathbf{r},t) \) associated with the external electric field. Since the scalar potential is a classical function of space and time, it commutes with quantum operators and can be factored out of time-ordered or commutator expressions. Accordingly, in the interaction picture, the external perturbation takes the form
\begin{equation}
\hat{\mathcal{H}}_{ext}^{(0)}(t) = \int d^3 \textbf{r} q_e \hat{\rho}^{(0)}(\textbf{r},t) \varphi_{ext}(\textbf{r},t).
\end{equation}
From linear response theory, with $A=\hat{\rho}(\textbf{r})$, we compute the variation in the particle number density as follows
\begin{equation}
\delta \left\langle \hat{\rho}(\mathbf{r}) \right\rangle(t) = \dfrac{i}{\hslash} \int_{-\infty}^{+\infty} dt' \, \Theta(t - t') \int d^3 \mathbf{r}' \, q_e \, \varphi_{ext}(\mathbf{r}', t') \left\langle \left[ \hat{\rho}^{(0)}(\mathbf{r}', t'), \hat{\rho}^{(0)}(\mathbf{r}, t) \right] \right\rangle.
\end{equation}
If we multiply by "-1" and reverse the operators in the commutator, that is,
\begin{equation}
\delta \left\langle \hat{\rho}(\mathbf{r}) \right\rangle(t) = - \dfrac{i}{\hslash} \int_{-\infty}^{+\infty} dt' \Theta(t-t') \int d^3 \textbf{r}' q_e \varphi_{ext}(\textbf{r}',t') \left\langle \left[ \hat{\rho}^{(0)}(\textbf{r},t) , \hat{\rho}^{(0)}(\textbf{r}',t') \right] \right\rangle,
\end{equation}
we recognize the polarization propagator $\eqref{eq: propagatoredipolarizzazionespazioimpulsospaziotempiimmaginari}$ computed in real space and time, namely
\begin{equation}
\Pi(\textbf{r},\textbf{r}',t,t') = - \dfrac{i}{\hslash} \Theta(t-t') \left\langle \left[ \hat{\rho}^{(0)}(\textbf{r},t) , \hat{\rho}^{(0)}(\textbf{r}',t') \right] \right\rangle,
\label{eq: propagatoredipolarizzazionespaziorealetempireali}
\end{equation}
from which we will derive the dielectric function, which can be measured experimentally. We write
\begin{equation}
\delta \left\langle \hat{\rho}(\mathbf{r}) \right\rangle(t) = - \dfrac{i}{\hslash} \int_{-\infty}^{+\infty} dt' \Theta(t-t') \int d^3 \textbf{r}' q_e \varphi_{ext}(\textbf{r}',t') \Pi(\textbf{r},\textbf{r}',t,t'),
\label{eq: formuladiKuboperdensitàdiparticelle1}
\end{equation}
that is, the induced polarization in the medium, meaning the variation of the charged particle density in the medium due to the external perturbation, is related to the polarization Green's function, which does not contain the external electric field, since the densities evolve according to the Hamiltonian $\hat{\mathcal{H}}_0$ that describes only the electron system. Let us assume the system is in thermal equilibrium and invariant under spatial translations, then $\Pi(\textbf{r},\textbf{r}',t,t') = \Pi(\textbf{r}-\textbf{r}',t-t')$. By applying a Fourier transform in space and time, equation $\eqref{eq: formuladiKuboperdensitàdiparticelle1}$ becomes a convolution. In particular, the polarization Green's function is a retarded function, due to the principle of causality. If we substitute the particle density operator in second quantization, that is equation $\eqref{eq: operatorefluttuazionedensitàparticelle2quantizzazione}$, into $\eqref{eq: propagatoredipolarizzazionespaziorealetempireali}$, we obtain
\begin{equation}
\Pi(\textbf{r}-\textbf{r}',t-t') = - \dfrac{i}{\hslash} \Theta(t-t') \sum_{s,s'} \left\langle \left[ \hat{\psi}^\dag(\textbf{r},s,t) \hat{\psi}(\textbf{r},s,t) , \hat{\psi}^\dag(\textbf{r}',s',t') \hat{\psi}(\textbf{r}',s',t') \right] \right\rangle,
\end{equation}
and we expand the field operators in the plane wave basis. The polarization propagator thus contains four sums over momenta vectors $\textbf{k}_1$, $\textbf{k}_2$, $\textbf{k}_3$, $\textbf{k}_4$; we set $\textbf{k}_1 = \textbf{k}_2 + \textbf{q}_1$, $\textbf{k}_3 = \textbf{k}_4 + \textbf{q}_2$, from which it follows that $\textbf{q}_1 = - \textbf{q}_2$, and we obtain
\begin{equation}
\Pi(\textbf{r}-\textbf{r}',t-t') = - \dfrac{i}{\hslash} \, \Theta(t-t') \, \dfrac{1}{V^2} \sum_{\textbf{k}_1,\sigma_1,\textbf{q}_1} \sum_{\textbf{k}_2,\sigma_2} 
\left\langle \left[ \left( C^\dagger_{\textbf{k}_1+\textbf{q}_1,\sigma_1} C_{\textbf{k}_1,\sigma_1} \right)^{(0)}(t), 
\left( C^\dagger_{\textbf{k}_2-\textbf{q}_1,\sigma_2} C_{\textbf{k}_2,\sigma_2} \right)^{(0)}(t') \right] \right\rangle 
e^{- i \textbf{q}_1 \cdot \left( \textbf{r} - \textbf{r}' \right)}.
\end{equation}
We apply a Fourier transform in space and time as follows
\begin{equation}
\Pi_{\textbf{q}}(\omega + i \delta) = \int d^3 \textbf{r} e^{i \textbf{q} \cdot \textbf{r}} \int_{-\infty}^{+\infty} dt e^{i (\omega + i \delta) t} \Pi(\textbf{r},t) ,
\end{equation}
where we have renamed $\textbf{r} - \textbf{r}' \rightarrow \textbf{r}$, $t - t' \rightarrow t$, and the factor $\delta>0$ allows us to integrate in the upper half-plane; thus, the integration is performed first, and the limit $\delta \rightarrow 0^+$ is taken afterwards. The spatial integral yields
\begin{equation}
\int d^3 \textbf{r} e^{i \left( \textbf{q} - \textbf{q}_1 \right) \cdot \textbf{r}} = V \delta_{\textbf{q},\textbf{q}_1}
\end{equation}
and by summing over $\textbf{q}_1$ we have $\textbf{q} = \textbf{q}_1$. In the thermal average, on the other hand, we obtain the spatial Fourier transforms of the particle density, then
\begin{equation}
\Pi_{\textbf{q}}(\omega + i \delta) = - \dfrac{i}{\hslash} \dfrac{1}{V} \int_{-\infty}^{+\infty} dt e^{i (\omega + i \delta) t} \Theta(t) \left\langle \left[ \hat{\rho}^{(0)}_{\textbf{q}}(t) , \hat{\rho}_{\textbf{q}}^{\dagger (0)}(0) \right] \right\rangle, 
\end{equation}
and the variation of the particle density in Fourier space is a convolution, that is
\begin{equation}
\delta \langle \hat{\rho} \rangle (\textbf{q},\omega) = \dfrac{1}{V} q_e \varphi_{ext,\textbf{q}}(\omega) \Pi_{\textbf{q}}(\omega).
\label{eq: variazionedensitàdiparticelleintrasformata}
\end{equation}
Now we can write the dielectric function from the variation in the particle density. The potential $\varphi_{ext}$ is generated by a charge distribution $\rho_{ext}$, which polarizes a system of electrons with charge $q_e$. To describe electric interactions, we make use of the electric field $\textbf{E}$ and the electric displacement field $\textbf{D}$, which satisfy Maxwell's equations. In this regard, we recall the full set of Maxwell's equations in the CGS-Gaussian system, that is
\begin{equation}
\begin{cases}
\nabla \cdot \mathbf{E} = 4 \pi \rho \\[0.8ex]
\nabla \cdot \mathbf{B} = 0 \\[0.8ex]
\nabla \times \mathbf{E} = - \dfrac{1}{c} \dfrac{\partial \mathbf{B}}{\partial t} \\[0.8ex]
\nabla \times \mathbf{B} = \dfrac{1}{c} \dfrac{\partial \mathbf{E}}{\partial t} + \dfrac{4 \pi}{c} \mathbf{J}
\end{cases}
.
\label{eq: equazioniMaxwell}
\end{equation}
Here, $\rho$ and $\textbf{J}$ represent the total charge and current densities. When considering electromagnetic fields in a medium, it is often useful to distinguish between external and induced sources. In this context, the displacement field $\textbf{D}$ satisfies an analogous equation
\begin{equation}
\nabla \cdot \textbf{D} = 4 \pi \rho_{ext},
\end{equation}
where $\rho_{ext}$ is the external charge density. Meanwhile, the electric field $\textbf{E}$ satisfies
\begin{equation}
\nabla \cdot \textbf{E} = 4 \pi (\rho_{ext} + \rho_i),
\end{equation}
where $\rho_i$ is the induced charge density that accounts for the response of the medium. Here, we consider the fields $\textbf{D}$ and $\textbf{E}$ to be slowly varying in time, so that the effects induced by the magnetic field can be neglected. We apply the spatial and temporal Fourier transform and obtain
\begin{equation}
- i \textbf{q} \cdot \textbf{D}_{\textbf{q}}(\omega) = 4 \pi \rho_{ext,\textbf{q}}(\omega), 
\end{equation}
\begin{equation}
- i \textbf{q} \cdot \textbf{E}_{\textbf{q}}(\omega) = 4 \pi \left( \rho_{ext,\textbf{q}}(\omega) + \rho_{i,\textbf{q}}(\omega) \right).
\end{equation}
Given the relations between the fields and the potentials, namely $\textbf{D} = - \nabla \varphi_{ext}$, $\textbf{E} = - \nabla \varphi$, where $\varphi$ is generated by the total charge distribution $\rho_{ext} + \rho_i$, we apply a Fourier transform and obtain
\begin{equation}
\textbf{D}_{\textbf{q}}(\omega) = i \varphi_{ext,\textbf{q}}(\omega) \textbf{q},
\end{equation}
\begin{equation}
\textbf{E}_{\textbf{q}}(\omega) = i \varphi_{\textbf{q}}(\omega) \textbf{q}.
\end{equation}
By comparing the equations for $\textbf{D}_{\textbf{q}}(\omega)$ and $\textbf{E}_{\textbf{q}}(\omega)$, we have
\begin{equation}
\varphi_{ext,\textbf{q}}(\omega) = \dfrac{4 \pi}{q^2} \rho_{ext,\textbf{q}}(\omega),
\end{equation}
\begin{equation}
\varphi_{\textbf{q}}(\omega) = \dfrac{4 \pi}{q^2} \left( \rho_{ext,\textbf{q}}(\omega) + \rho_{i,\textbf{q}}(\omega) \right).
\end{equation}
By definition, the dielectric function $\epsilon$ relates the electric displacement field $\mathbf{D}$ to the electric field $\mathbf{E}$ through the relation 
\begin{equation}
\textbf{E} = \dfrac{\textbf{D}}{\epsilon}, 
\end{equation}
then
\begin{equation}
\varphi_{\textbf{q}}(\omega) = \dfrac{\varphi_{ext,\textbf{q}}(\omega)}{\epsilon_{\textbf{q}}(\omega)},
\label{eq: legametrapotenzialeesternopotenzialetotalefunzionedielettrica}
\end{equation}
\begin{equation}
\dfrac{1}{\epsilon_{\mathbf{q}}(\omega)} = \dfrac{\rho_{\text{ext}, \mathbf{q}}(\omega) + \rho_{i, \mathbf{q}}(\omega)}{\rho_{\text{ext}, \mathbf{q}}(\omega)},
\end{equation}
\begin{equation}
\dfrac{1}{\epsilon_{\textbf{q}}(\omega)} = 1 + \dfrac{\rho_{i,\textbf{q}}(\omega)}{\rho_{ext,\textbf{q}}(\omega)}.
\label{eq: relazionefunzionedielettricapotenziali}
\end{equation}
Ultimately, to calculate the dielectric function, it is necessary to determine the induced charge distribution $\rho_{i,\textbf{q}}(\omega)$, which can be computed using linear response theory. Note that it must be
\begin{equation}
\rho_i = q_e \delta \langle \hat{\rho} \rangle,
\end{equation}
in other words, the induced charge distribution in the material medium is proportional to the electron charge and to the fluctuation in the number of electron, that is why we computed the variation in particle density. We substitute these quantities into equation $\eqref{eq: relazionefunzionedielettricapotenziali}$ and obtain
\begin{equation}
\dfrac{1}{\epsilon_{\textbf{q}}(\omega)} = 1 + \dfrac{1}{V} \dfrac{q_e^2 \varphi_{ext,\textbf{q}}(\omega) \Pi_{\textbf{q}}(\omega)}{\rho_{ext,\textbf{q}}(\omega)}, 
\end{equation}
then
\begin{equation}
\dfrac{1}{\epsilon_{\textbf{q}}(\omega)} = 1 + \dfrac{1}{V} \dfrac{4 \pi q_e^2}{q^2} \Pi_{\textbf{q}}(\omega).
\label{eq: relazionefunzionedielettricapropagatorepolarizzazione}
\end{equation}
Note that if we use the equivalent definition of the dielectric function
\begin{equation}
\dfrac{1}{\epsilon_{\textbf{q}}(\omega)} = \dfrac{U_{eff,\textbf{q}}(\omega)}{U^{(0)}_{eff,\textbf{q}}(\omega)},
\end{equation}
from $\eqref{eq: equazioneDysonpotenzialeefficacepropagatorepolarizzazionestar2}$ we have
\begin{equation}
\dfrac{1}{\epsilon_{\textbf{q}}(\omega)} = \dfrac{U^{(0)}_{eff,\textbf{q}}(\omega)}{U^{(0)}_{eff,\textbf{q}}(\omega) \left( 1 - U^{(0)}_{eff,\textbf{q}}( \omega) \Pi^*_{\textbf{q}}(\omega) \right)},
\end{equation}
\begin{equation}
\epsilon_{\textbf{q}}(\omega) = 1 - U^{(0)}_{eff,\textbf{q}}(\omega) \Pi^*_{\textbf{q}}(\omega),
\end{equation}
and using $\eqref{eq: trasformataFourierpotenzialeCoulomb}$, we get
\begin{equation}
\epsilon_{\textbf{q}}(\omega) = 1 - \dfrac{1}{V} \dfrac{4 \pi q_e^2}{q^2} \Pi^*_{\textbf{q}}(\omega).
\label{eq: relazionefunzionedielettricapropagatorepolarizzazionestar}
\end{equation}
In the following sections, we will investigate both the general properties of the dielectric function \( \epsilon_{\mathbf{q}}(\omega) \) and various theoretical models used to approximate it in many-body systems. Understanding the behavior of \( \epsilon_{\mathbf{q}}(\omega) \) is essential, as it governs how electric fields are screened in a material and determines the effective interactions between charged particles. Moreover, the study of the dielectric function is not limited to theoretical interest: it is also directly accessible through experimental techniques. One of the most prominent examples is Electron Energy Loss Spectroscopy (EELS), which measures the energy lost by fast electrons as they pass through a material. This energy loss is associated with the excitation of collective modes, such as plasmons (see below), and is directly related to the imaginary part of the inverse dielectric function,
\[
\text{EELS} \propto \operatorname{Im}\left[ -\dfrac{1}{\epsilon_{\mathbf{q}}( \omega)} \right].
\]
As such, EELS provides valuable momentum- and frequency-resolved information about the screening properties and charge dynamics of quantum systems. In a typical EELS experiment, a beam of high-energy electrons (in the range of 10--300~keV) is directed through a thin sample. As the electrons interact with the material, they may lose energy by exciting internal degrees of freedom such as plasmons or interband transitions. By analyzing the energy distribution of the transmitted electrons with high resolution, one obtains the energy loss spectrum. Moreover, by varying the scattering angle, it is possible to determine the momentum transfer \( \mathbf{q} \), thereby mapping out the full dispersion \( \omega(\mathbf{q}) \) of collective excitations. This makes EELS a powerful tool to probe the dielectric response over a broad range of energies and wavevectors.
\subsection{Zeros of the dielectric function}
\begin{figure}[H]
\centering
\includegraphics[scale=0.75]{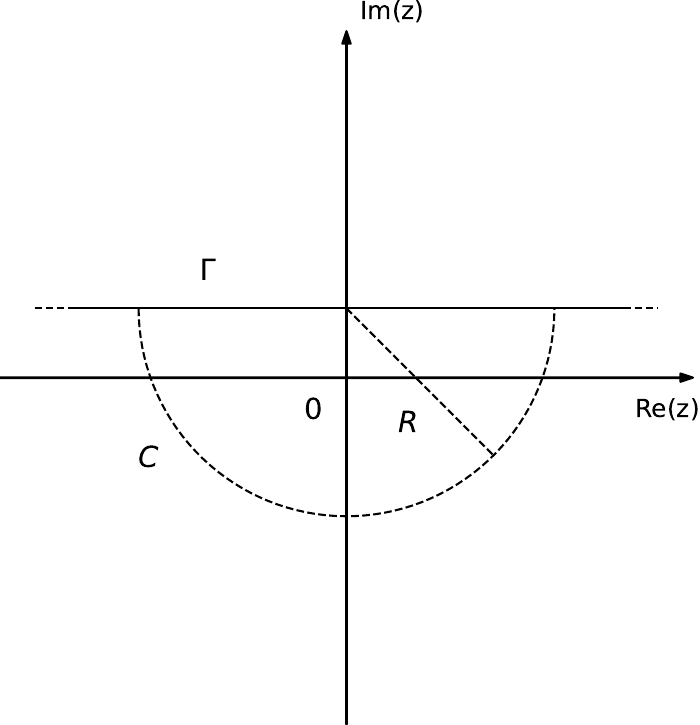}
\caption{Integration contour used for the evaluation of the variation in particle density in equation $\eqref{eq: caricapervariazionedensitacaricaspazioposizionietempi}$ via the residue theorem.}
\label{fig: integrale_residui_carica_per_variazione_densita_particelle}
\end{figure}

From $\eqref{eq: variazionedensitàdiparticelleintrasformata}$ it follows
\begin{equation}
q_e \delta \langle \hat{\rho} \rangle(\textbf{q},\omega) = \dfrac{1}{V} q^2_e \varphi_{ext,\textbf{q}}(\omega) \Pi_{\textbf{q}}(\omega),
\label{eq: caricapervariazionedensitadiparticelleintrasformata1}
\end{equation}
furthermore, since we can invert equations $\eqref{eq: relazionefunzionedielettricapropagatorepolarizzazione}$ and $\eqref{eq: relazionefunzionedielettricapropagatorepolarizzazionestar}$, we can write the polarization propagator as follows
\begin{equation}
\Pi_{\textbf{q}}(\omega) = \dfrac{q^2 V}{4 \pi q_e^2} \left( \dfrac{1}{\epsilon_{\textbf{q}}(\omega)} - 1 \right),
\label{eq: propagatorepolarizzazioneinfunzionedifunzionedielettrica}
\end{equation}
\begin{equation}
\Pi^*_{\textbf{q}}(\omega) = \dfrac{q^2 V}{4 \pi q_e^2} \left( \epsilon_{\textbf{q}}(\omega) - 1 \right),
\label{eq: propagatorepolarizzazionestarinfunzionefunzionedielettrica}
\end{equation}
and equation $\eqref{eq: caricapervariazionedensitadiparticelleintrasformata1}$ can be written from equation $\eqref{eq: propagatorepolarizzazioneinfunzionedifunzionedielettrica}$ as
\begin{align}
q_e \delta \langle \hat{\rho} \rangle (\textbf{q},\omega) &= \dfrac{1}{V} q^2_e \varphi_{ext,\textbf{q}}(\omega) \Pi_{\textbf{q}}(\omega) = \notag \\
&= \dfrac{1}{V} q^2_e \varphi_{ext,\textbf{q}}(\omega) \dfrac{q^2 V}{4 \pi q_e^2} \left( \dfrac{1}{\epsilon_{\textbf{q}}(\omega)} - 1 \right) = \notag \\
&= \dfrac{q^2}{4 \pi} \varphi_{ext,\textbf{q}}(\omega) \left( \dfrac{1}{\epsilon_{\textbf{q}}(\omega)} - 1 \right) ,
\label{eq: caricapervariazionedensitadiparticelleintrasformatainfunzionefunzionedielettrica1}
\end{align}
or equivalently, if we use equation $\eqref{eq: propagatorepolarizzazionestarinfunzionefunzionedielettrica}$ in equation $\eqref{eq: caricapervariazionedensitadiparticelleintrasformatainfunzionefunzionedielettrica1}$, then equation $\eqref{eq: caricapervariazionedensitadiparticelleintrasformata1}$ can be written as
\begin{align}
q_e \delta \langle \hat{\rho} \rangle (\textbf{q},\omega) &= \dfrac{q^2}{4 \pi} \varphi_{ext,\textbf{q}}(\omega) \left( \dfrac{1}{\epsilon_{\textbf{q}}(\omega)} - 1 \right) = \notag \\
&= \dfrac{q^2}{4 \pi} \varphi_{ext,\textbf{q}}(\omega) \left( \dfrac{1 - \epsilon_{\textbf{q}}(\omega)}{\epsilon_{\textbf{q}}(\omega)} \right) = \notag \\
&= \dfrac{q^2}{4 \pi} \varphi_{ext,\textbf{q}}(\omega) \left( - \dfrac{\epsilon_{\textbf{q}}(\omega) - 1}{\epsilon_{\textbf{q}}(\omega)} \right) = \notag \\
&= \dfrac{q^2}{4 \pi} \varphi_{ext,\textbf{q}}(\omega) \dfrac{\dfrac{1}{V} \dfrac{4 \pi q_e^2}{q^2} \Pi^*_{\textbf{q}}(\omega)}{\epsilon_{\textbf{q}}(\omega)} = \notag \\
&= \dfrac{q_e^2}{V} \varphi_{ext,\textbf{q}}(\omega)  \dfrac{\Pi^*_{\textbf{q}}(\omega)}{\epsilon_{\textbf{q}}\omega)}.
\label{eq: caricapervariazionedensitadiparticelleintrasformatainfunzionefunzionedielettrica2}
\end{align}
Here, we aim to study the meaning of the zeros of the dielectric function. Let us consider an impulsive external perturbation, that is of the form
\begin{equation}
\varphi_{ext}(\textbf{r},t) = \varphi_0 e^{- i \textbf{q} \cdot \textbf{r}} \delta(t) ,
\end{equation}
and we compute the spatial and temporal Fourier transform, that is
\begin{align}
\varphi_{ext,\textbf{k}}(\omega) &= \int d^3 \textbf{r} e^{i \textbf{k} \cdot \textbf{r}} \int dt e^{i \omega t} \varphi_0 e^{- i \textbf{q} \cdot \textbf{r}} = \notag \\
&= \delta(\textbf{k}-\textbf{q}) \varphi_0.
\end{align}
Then, we consider the spatial and temporal Fourier transform of $\eqref{eq: caricapervariazionedensitadiparticelleintrasformatainfunzionefunzionedielettrica2}$, namely
\begin{equation}
q_e \delta \langle \hat{\rho} \rangle (\textbf{r},t) = \int \dfrac{d^3 \textbf{k}}{(2 \pi)^3} e^{- i \textbf{k} \cdot \textbf{r}} \int \dfrac{d \omega}{2 \pi} e^{- i \omega t} \dfrac{q_e^2}{V} \varphi_{ext,\textbf{k}}(\omega)  \dfrac{\Pi^*_{\textbf{k}}(\omega)}{\epsilon_{\textbf{k}}(\omega)},
\end{equation} 
and we compute it in the case of the impulsive field described above, i.e.,
\begin{align}
q_e \delta \langle \hat{\rho} \rangle (\textbf{r},t) &= \int \dfrac{d^3 \textbf{k}}{(2 \pi)^3} e^{- i \textbf{k} \cdot \textbf{r}} \int \dfrac{d \omega}{2 \pi} e^{- i \omega t} \dfrac{q_e^2}{V} \delta(\textbf{k}-\textbf{q}) \varphi_0 \dfrac{\Pi^*_{\textbf{k}}(\omega)}{\epsilon_{\textbf{k}}(\omega)} = \notag \\
&= \dfrac{e^{- i \textbf{q} \cdot \textbf{r}}}{(2 \pi)^3} \varphi_0 \int_{\Gamma} dz e^{- i z t} \dfrac{\Pi^*_{\textbf{q}}(z)}{\epsilon_{\textbf{q}}(z)} ,
\label{eq: caricapervariazionedensitacaricaspazioposizionietempi}
\end{align}
where the integral is over a line $\Gamma$ parallel to the real axis contained in the upper half-plane, and we integrate using the residue theorem. Consider the contour $\Gamma + C$, see Figure $\eqref{fig: integrale_residui_carica_per_variazione_densita_particelle}$. The factor $e^{- i z t}$ ensures convergence along the semicircle. Applying Jordan's lemma, in the limit $R \rightarrow \infty$, the contribution is given by the poles of the integrand function in the lower half-plane, namely by the zeros of the function $\frac{1}{\epsilon_{\textbf{q}}(z)}$. Suppose that a zero is of the form $z_{\textbf{q}} = \Omega_{\textbf{q}} - i \gamma_{\textbf{q}}$, with $\gamma_{\textbf{q}}>0$. From the residue theorem, we obtain, up to constants, that the variation of particle density, following an impulsive perturbation, evolves as
\begin{equation}
q_e \delta \langle \hat{\rho} \rangle (\textbf{r},t) \propto e^{- i \textbf{q} \cdot \textbf{r}} e^{- i \Omega_{\textbf{q}} t} e^{- i \omega_{\textbf{q}} t}.
\end{equation}
We have shown that the Fourier transform of an impulsive perturbation includes all the oscillation frequencies of the material medium, with a decay time that directly depends on the zeros of the function $\frac{1}{\epsilon_{\textbf{q}}(z)}$, and we will now calculate these frequencies. \newline
We compute these frequencies semiclassically. After the impulsive perturbation, the particle density becomes
\begin{equation}
\rho(\textbf{r},t) = \rho_0 + \delta \rho(\textbf{r},t) ,
\end{equation}
where $\rho_0$ is the particle density at equilibrium and the quantity $\delta \rho(\textbf{r},t)$ measures the induced variation. At thermodynamic equilibrium, the density $\rho_0$ is balanced by that of the positive ions. The external perturbation implies electron motion, and thus charge imbalances, which in turn create an electric field that satisfies
\begin{equation}
\dive \textbf{E} = 4 \pi q_e \delta \rho.
\label{eq: primaMaxwellelementovolumevariazionedensitacarica}
\end{equation}
We will apply Newton’s law $\textbf{F}=m\textbf{a}$, following an Eulerian approach to motion. We are interested in studying the quantities related to a fixed infinitesimal volume, such as the acting force and velocity, as time passes. Since we work within a continuous model for the material medium, the mass contained in an infinitesimal macroscopical volume is given by
\begin{align}
dm &= \rho_m dV = \notag \\
&= m_e \rho dV,
\end{align}
where $\rho_m$ is the electronic mass density, which is obviously related to the electronic particle density $\rho$. With respect to the infinitesimal volume $dV$, Newton’s equation is written as
\begin{equation}
q_e \textbf{E} \rho(\textbf{r},t) dV = m_e dV \dfrac{d \textbf{v}}{dt} \rho(\textbf{r},t),
\end{equation}
\begin{equation}
q_e \textbf{E} \rho(\textbf{r},t) = m_e \dfrac{d \textbf{v}}{dt} \rho(\textbf{r},t),
\end{equation}
where the Eulerian velocity $\textbf{v}(x,y,z,t)$ indicates the velocity of the infinitesimal volume. At time $t+dt$ the volume has moved, and the velocity must be calculated at the new point, that is, it is a velocity $\textbf{v}(x+v_x dt,y+v_y dt, z + v_z dt, t+ dt)$. From the Taylor expansion we have
\begin{equation}
\textbf{v}(x+v_x dt,y+v_y dt, z + v_z dt, t+ dt) = \textbf{v}(x,y,z,t) + \dfrac{\partial \textbf{v}}{\partial t} dt + \dfrac{\partial \textbf{v}_x}{\partial x} v_x dt + \dfrac{\partial \textbf{v}_y}{\partial y} v_y dt + \dfrac{\partial \textbf{v}_z}{\partial z} v_z dt,
\end{equation}
thus, the acceleration is given by
\begin{align}
\textbf{a} &= \dfrac{\textbf{v}(x+v_x dt,y+v_y dt, z + v_z dt, t+ dt) - \textbf{v}(x,y,z,t)}{dt} = \notag \\
&= \dfrac{\partial \textbf{v}}{\partial t} + \textbf{v} \cdot \grad \textbf{v}.
\end{align}
Since we work within the framework of linear response theory, we consider first order with respect to velocity, so the total time derivative of velocity can be replaced by a partial time derivative. Under the assumption of linear theory, moreover, $\rho_e \sim \rho_0$, then
\begin{equation}
q_e \textbf{E} \rho_0 = m_e \dfrac{\partial \textbf{v}}{\partial t} \rho_0,
\end{equation}
\begin{equation}
q_e \textbf{E} = m_e \dfrac{\partial \textbf{v}}{\partial t}.
\label{eq: forzaNewtonsuelementovolumevariazionedensitacarica}
\end{equation}
Now, from the continuity equation 
\begin{equation}
\dfrac{d \rho}{dt} + \rho \dive \textbf{v} = 0, 
\end{equation}
and considering a first-order approximation, i.e., linearizing the equation around a uniform equilibrium state, we obtain
\begin{align}
\dfrac{d \rho}{dt} &\sim \dfrac{\partial \rho}{\partial t} = \notag \\
&= \dfrac{\partial \rho_0}{\partial t} + \dfrac{\partial \delta \rho}{\partial t} = \notag \\
&= \dfrac{\partial \delta \rho}{\partial t} ,
\end{align}
and $\rho \sim \rho_0$, and continuity equation becomes
\begin{equation}
\dfrac{\partial \delta \rho}{\partial t} + \rho_0 \dive \textbf{v} = 0.
\label{eq: equazionecontinuitavariazionedensitacarica}
\end{equation}
We differentiate $\eqref{eq: equazionecontinuitavariazionedensitacarica}$ with respect to time, i.e.,
\begin{equation}
\dfrac{\partial^2 \delta \rho}{\partial t^2} + \rho_0 \dive \dfrac{\partial \textbf{v}}{\partial t} = 0,
\end{equation}
from $\eqref{eq: forzaNewtonsuelementovolumevariazionedensitacarica}$ we have
\begin{equation}
\dfrac{\partial^2 \delta \rho}{\partial t^2} + \dfrac{q_e \rho_0}{m_e} \dive \textbf{E} = 0,
\end{equation}
and from $\eqref{eq: primaMaxwellelementovolumevariazionedensitacarica}$ we have
\begin{equation}
\dfrac{\partial^2 \delta \rho}{\partial t^2} + \dfrac{4 \pi q_e^2 \rho_0}{m_e} \delta \rho = 0,
\end{equation}
which is a harmonic motion equation for $\delta \rho$. We have derived the frequency at which the induced charge density oscillates, namely the plasma frequency
\begin{equation}
\omega_p^2 = \dfrac{4 \pi q_e^2 \rho_0}{m_e},
\label{eq: frequenzadiplasma}
\end{equation}
which is associated with quasiparticles called plasmons. In particular, such quasiparticles satisfy bosonic statistics, despite originating from a system of fermions. Indeed, the dielectric function is related to the polarization propagator, which includes a commutator of particle density operators, which are bosonic operators.
\subsection{Sum rules for the dielectric function}
From spectral function theory, we know that the polarization propagator must satisfy
\begin{equation}
\lim_{\delta \rightarrow 0^+} - 2 \hslash \im \Pi_{\textbf{q}}(\omega + i \delta) = A_{\textbf{q}}(\omega),
\end{equation}
where $A_{\textbf{q}}(\omega)$ is the spectral function of the polarization propagator, which from $\eqref{eq: funzionespettrale}$ can be written in terms of the eigenstates of the complete Hamiltonian as
\begin{equation}
A_{\textbf{q}}(\omega) = \dfrac{2 \pi}{Z} \sum_{n,m} \left( e^{- \beta \mathcal{E}_n} - e^{- \beta \mathcal{E}_m} \right) \delta \left[ \omega - \frac{\mathcal{E}_m - \mathcal{E}_n}{\hslash} \right] \left| \langle n|\hat{\rho}_{\textbf{q}}|m \rangle \right|^2 .
\label{eq: funzionespettralepropagatorepolarizzazione}
\end{equation}
\begin{theorem}
The spectral function of the polarization propagator, that is, equation $\eqref{eq: funzionespettralepropagatorepolarizzazione}$, satisfies
\begin{itemize}
\item[1)]
\begin{equation}
A_{\textbf{q}}(-\omega) = - A_{\textbf{q}}(\omega);
\label{eq: funzionespettralepropagatorepolarizzazionedisparirispettoafrequenze}
\end{equation}
\item[2)]
\begin{equation}
\int_{-\infty}^{+\infty} \dfrac{d\omega}{2\pi} A_{\textbf{q}}(\omega) = 0;
\label{eq: funzionespettralepropagatorepolarizzazionemomentozero}
\end{equation}
\item[3)]
\begin{equation}
\int_{-\infty}^{+\infty} \dfrac{d\omega}{2\pi} \omega A_{\textbf{q}}(\omega) = \dfrac{\hslash q^2 N}{m_e}.
\label{eq: funzionespettralepropagatorepolarizzazionemomentoprimo}
\end{equation}
\end{itemize}
\begin{proof}
\item[1)] From
\begin{equation}
A_{\textbf{q}}(-\omega) = \dfrac{2 \pi}{Z} \sum_{n,m} \left( e^{- \beta \mathcal{E}_n} - e^{- \beta \mathcal{E}_m} \right) \delta \left[ - \omega - \frac{\mathcal{E}_m - \mathcal{E}_n}{\hslash} \right] \left| \langle n|\hat{\rho}_{\textbf{q}}|m \rangle \right|^2,
\end{equation}
we use the property $\delta(x) = \delta(-x)$, and we write
\begin{equation}
A_{\textbf{q}}(-\omega) = \dfrac{2 \pi}{Z} \sum_{n,m} \left( e^{- \beta \mathcal{E}_n} - e^{- \beta \mathcal{E}_m} \right) \delta \left[ \omega + \frac{\mathcal{E}_m - \mathcal{E}_n}{\hslash} \right] \left| \langle n|\hat{\rho}_{\textbf{q}}|m \rangle \right|^2,
\end{equation}
and we invert the order of the indices $n$, $m$ as follows
\begin{equation}
A_{\textbf{q}}(-\omega) = \dfrac{2 \pi}{Z} \sum_{n,m} \left( e^{- \beta \mathcal{E}_m} - e^{- \beta \mathcal{E}_n} \right) \delta \left[ \omega + \frac{\mathcal{E}_n - \mathcal{E}_m}{\hslash} \right] \left| \langle m|\hat{\rho}_{\textbf{q}}|n \rangle \right|^2.
\end{equation}
Note that the delta can be rewritten as
\begin{equation}
\delta \left[ \omega + \frac{\mathcal{E}_n - \mathcal{E}_m}{\hslash} \right] = \delta \left[ \omega - \frac{\mathcal{E}_m - \mathcal{E}_n}{\hslash} \right],
\end{equation}
and given that $\hat{\rho}_{\textbf{q}}$ is Hermitian, it satisfies
\begin{equation}
\left| \langle m|\hat{\rho}_{\textbf{q}}|n \rangle \right|^2 = \left| \langle n|\hat{\rho}_{\textbf{q}}|m \rangle \right|^2,
\end{equation}
from which it follows
\begin{equation}
A_{\textbf{q}}(-\omega) = \dfrac{2\pi}{Z} \sum_{n,m} \left( e^{- \beta \mathcal{E}_m} - e^{- \beta \mathcal{E}_n} \right) \delta \left[ \omega - \frac{\mathcal{E}_m - \mathcal{E}_n}{\hslash} \right] \left| \langle n|\hat{\rho}_{\textbf{q}}|m \rangle \right|^2.
\end{equation}
which is the thesis.
\item[2)] From $\eqref{eq: momentonullodifunzionespettrale}$ with $A=\hat{\rho}_{\textbf{q}}$, $B=\hat{\rho}^{\dagger}_{\textbf{q}}$, we have
\begin{equation}
\int_{-\infty}^{+\infty}\frac{d\omega}{2\pi} A_{\textbf{q}}(\omega) = \left\langle \left[\hat{\rho}_{\textbf{q}},\hat{\rho}^{\dagger}_{\textbf{q}}\right] \right\rangle,
\end{equation}
and from $\eqref{eq: trasformatadiFourieroperatoredensitadifluttuazioneparticellecommutatore1}$ the thesis follows.
\item[3)] From $\eqref{eq: momentoprimodifunzionespettrale}$ with $A=\hat{\rho}_{\textbf{q}}$, $B=\hat{\rho}^{\dagger}_{\textbf{q}}$, we have
\begin{equation}
\int_{-\infty}^{+\infty} \dfrac{d\omega}{2\pi} \omega A_{\textbf{q}}(\omega) = \dfrac{1}{\hslash} \left\langle \left[ \left[ \hat{\rho}_{\textbf{q}},\hat{\mathcal{H}} \right],\hat{\rho}^{\dagger}_{\textbf{q}} \right] \right\rangle,
\end{equation}
where $\hat{\mathcal{H}}$ is the jellium Hamiltonian; in particular, we employ the form given in equation $\eqref{eq: hamiltonianamodellojellium2}$ of the jellium model, with the obvious identifications
\begin{equation}
\hat{\mathcal{H}}_0 = \sum_{\textbf{k},\sigma} \mathcal{E}^{(0)}_{\textbf{k}} C^\dag_{\textbf{k},\sigma}C_{\textbf{k},\sigma},
\end{equation}
\begin{equation}
\hat{\mathcal{H}}_I = \dfrac{1}{2V} \sum_{\textbf{q} \neq \textbf{0}} \dfrac{4 \pi q_e^2}{q^2} \left( \hat{N} - \hat{\rho}_{\textbf{q}} \hat{\rho}^{\dagger}_{\textbf{q}} \right),
\end{equation}
where $\hat{N}$ is the total number operator and $\mathcal{E}^{(0)}_{\textbf{k}}= \frac{\hslash^2 \textbf{k}^2}{2m_e} - \mu$, and we compute $\left[ \hat{\rho}_{\textbf{q}},\hat{\mathcal{H}}\right] = \left[ \hat{\rho}_{\textbf{q}},\hat{\mathcal{H}}_0 \right] + \left[ \hat{\rho}_{\textbf{q}},\hat{\mathcal{H}}_I \right]$. The object $\left[ \hat{\rho}_{\textbf{q}},\hat{\mathcal{H}}_0\right]$ is obtained from $\eqref{eq: trasformatadiFourieroperatoredensitadifluttuazioneparticellecommutatore2}$, with $\mathcal{E}_{\textbf{k}}=\mathcal{E}^{(0)}_{\textbf{k}}$, while the second object decomposes as
\begin{align}
\left[ \hat{\rho}_{\textbf{q}}, \hat{\mathcal{H}}_I \right] &= \dfrac{1}{2V} \sum_{\textbf{q}' \neq \textbf{0}} \dfrac{4 \pi q_e^2}{q'^2} \left[ \hat{\rho}_{\textbf{q}} , \left( \hat{N} - \hat{\rho}_{\textbf{q}'} \hat{\rho}^{\dagger}_{\textbf{q}'} \right) \right] = \notag \\
&= \dfrac{1}{2V} \sum_{\textbf{q}' \neq \textbf{0}} \dfrac{4 \pi q_e^2}{q'^2} \left( \left[ \hat{\rho}_{\textbf{q}} , \hat{N} \right] - \left[ \hat{\rho}_{\textbf{q}} , \hat{\rho}_{\textbf{q}'} \hat{\rho}^{\dagger}_{\textbf{q}'} \right] \right).
\end{align}
and from $\eqref{eq: trasformatadiFourieroperatoredensitadifluttuazioneparticellecommutatore4}$, $\eqref{eq: trasformatadiFourieroperatoredensitadifluttuazioneparticellecommutatore5}$ it is zero, and the first moment can be rewritten as
\begin{equation}
\int_{-\infty}^{+\infty} \dfrac{d\omega}{2\pi} \omega A_{\textbf{q}}(\omega) = \dfrac{1}{\hslash} \left\langle \left[ \left[ \hat{\rho}_{\textbf{q}}, \hat{\mathcal{H}}_0 \right], \hat{\rho}^{\dagger}_{\textbf{q}} \right] \right\rangle.
\end{equation}
From $\eqref{eq: trasformatadiFourieroperatoredensitadifluttuazioneparticellecommutatore3}$, with $\mathcal{E}_{\textbf{k}}=\mathcal{E}^{(0)}_{\textbf{k}}$, we have
\begin{equation}
\left\langle \left[ \left[ \hat{\rho}_{\textbf{q}}, \hat{\mathcal{H}}_0 \right], \hat{\rho}^{\dagger}_{\textbf{q}} \right] \right\rangle = \sum_{\textbf{k},\sigma} \left\langle \left( \mathcal{E}^{(0)}_{\textbf{k}-\textbf{q}} - 2 \mathcal{E}^{(0)}_{\textbf{k}} + \mathcal{E}^{(0)}_{\textbf{k}+\textbf{q}} \right) C^\dag_{\textbf{k},\sigma} C_{\textbf{k},\sigma} \right\rangle,
\end{equation}
and from
\begin{align}
\mathcal{E}^{(0)}_{\textbf{k}-\textbf{q}} - 2 \mathcal{E}^{(0)}_{\textbf{k}} + \mathcal{E}^{(0)}_{\textbf{k}+\textbf{q}} &= \dfrac{\hslash^2 (\textbf{k}-\textbf{q})^2}{2m_e} - 2 \dfrac{\hslash^2 \textbf{k}^2}{2m_e} + \dfrac{\hslash^2 (\textbf{k}+\textbf{q})^2}{2m_e} = \notag \\
&= \hslash^2 \dfrac{\textbf{k}^2 - 2 \textbf{k} \cdot \textbf{q} + \textbf{q}^2 - 2 \textbf{k}^2 + \textbf{k}^2 + 2 \textbf{k} \cdot \textbf{q} + \textbf{q}^2}{2m_e} = \notag \\
&= \dfrac{\hslash^2 q^2}{2m_e} ,
\end{align}
it follows
\begin{align}
\left\langle \left[ \left[ \hat{\rho}_{\textbf{q}}, \hat{\mathcal{H}}_0 \right], \hat{\rho}^{\dagger}_{\textbf{q}} \right] \right\rangle &= \dfrac{\hslash^2 q^2}{m_e} \sum_{\textbf{k}, \sigma} \left\langle C^{\dagger}_{\textbf{k}, \sigma} C_{\textbf{k}, \sigma} \right\rangle = \notag \\
&= \dfrac{\hslash^2 q^2 N}{m_e} ,
\label{eq: mediacommutatorecommutatorerhoqH0conrhodaggerq}
\end{align}
from which the thesis follows.
\end{proof}
\end{theorem}
Thanks to equation $\eqref{eq: funzionespettralepropagatorepolarizzazionemomentoprimo}$, we are now able to prove a sum rule for the dielectric function.
\begin{theorem}[The f-sum rule for the dielectric function]
The dielectric function satisfies the f-sum rule
\begin{equation}
\int_{-\infty}^{+\infty} \dfrac{d \omega}{2 \pi} \omega \im \left[ \dfrac{1}{\varepsilon_{\textbf{q}}(\omega)} \right] = - \dfrac{\omega_p^2}{2},
\end{equation}
where $\omega_p$ is the plasma frequency $\eqref{eq: frequenzadiplasma}$.
\begin{proof}
We multiply the imaginary part of equation $\eqref{eq: relazionefunzionedielettricapropagatorepolarizzazione}$ by $(-2 \hslash)$, that is
\begin{equation}
- 2 \hslash \im \left[ \dfrac{1}{\epsilon_{\textbf{q}}(\omega)} \right] = \dfrac{1}{V} \dfrac{4 \pi q_e^2}{q^2} \im \left[ - 2 \hslash \Pi_{\textbf{q}}(\omega) \right],
\end{equation}
and given that the imaginary part of the polarization propagator is related to the spectral function, we have
\begin{equation}
- 2 \hslash \im \left[ \dfrac{1}{\epsilon_{\textbf{q}}(\omega)} \right] = \dfrac{1}{V} \dfrac{4 \pi q_e^2}{q^2} A_{\textbf{q}}(\omega),
\end{equation}
\begin{equation}
A_{\textbf{q}}(\omega) = - 2 \hslash \dfrac{V q^2}{4 \pi q_e^2} \im \left[ \dfrac{1}{\epsilon_{\textbf{q}}(\omega)} \right].
\end{equation}
We substitute such equation into $\eqref{eq: funzionespettralepropagatorepolarizzazionemomentoprimo}$ and we have
\begin{equation}
\int_{-\infty}^{+\infty} \dfrac{d\omega}{2\pi} \omega \left( - 2 \hslash \dfrac{V q^2}{4 \pi q_e^2} \im \left[ \dfrac{1}{\epsilon_{\textbf{q}}(\omega)} \right] \right) = \dfrac{\hslash q^2 N}{m_e},
\end{equation}
\begin{equation}
\int_{-\infty}^{+\infty} \dfrac{d\omega}{2\pi} \omega \im \left[ \dfrac{1}{\epsilon_{\textbf{q}}(\omega)} \right] = - \dfrac{4 \pi q_e^2 N}{2m_e V},
\end{equation}
which is the thesis.
\end{proof}
\end{theorem}
We will now derive another sum rule for the dielectric function. Let us consider the polarization propagator in momentum and real-time space, that is
\begin{equation}
\Pi_{\textbf{q}}(t) = - \dfrac{i}{\hslash} \Theta(t) \left\langle \left[ \hat{\rho}^{(0)}_{\textbf{q}}(t) , \hat{\rho}^{\dagger}_{\textbf{q}}(0) \right] \right\rangle,
\label{eq: propagatoredipolarizzazionespazioimpulsospaziotempireali}
\end{equation}
and suppose we know the eigenstates of the interacting Hamiltonian, that is, we can write
\begin{equation}
\Pi_{\textbf{q}}(t) = \dfrac{1}{\hslash} \sum_{n,m} \dfrac{e^{- \beta \mathcal{E}_n} - e^{- \beta \mathcal{E}_m}}{Z} \left| \langle n|\hat{\rho}_{\textbf{q}}|m \rangle \right|^2.
\end{equation}
We apply a Fourier transform in frequency space, inserting as usual an arbitrarily small factor $\delta$ to ensure convergence, and we have
\begin{align}
\Pi_{\textbf{q}}(\omega + i \delta) &= \int_{-\infty}^{+\infty} dz e^{i (\omega + i \delta) t} \Pi_{\textbf{q}}(t) = \notag \\
&= \dfrac{1}{\hslash} \sum_{n,m} \dfrac{e^{- \beta \mathcal{E}_n} - e^{- \beta \mathcal{E}_m}}{Z} \left| \langle n|\hat{\rho}_{\textbf{q}}|m \rangle \right|^2 \dfrac{1}{\omega + i \delta + \frac{\mathcal{E}_n - \mathcal{E}_m}{\hslash}},
\end{align}
consequently, from $\eqref{eq: relazionefunzionedielettricapropagatorepolarizzazione}$ the dielectric function can be written as
\begin{equation}
\dfrac{1}{\epsilon_{\textbf{q}}(\omega + i \delta)} = 1 + \dfrac{1}{V} \dfrac{4 \pi q_e^2}{q^2} \dfrac{1}{\hslash} \sum_{n,m} \dfrac{e^{- \beta \mathcal{E}_n} - e^{- \beta \mathcal{E}_m}}{Z} \left| \langle n|\hat{\rho}_{\textbf{q}}|m \rangle \right|^2 \dfrac{1}{\omega + i \delta + \frac{\mathcal{E}_n - \mathcal{E}_m}{\hslash}}.
\label{eq: reciprocofunzionedielettricainfunzionediautostatidiH}
\end{equation}
If we rewrite the previous equation as
\begin{equation}
\dfrac{1}{\epsilon_{\textbf{q}}(\omega + i \delta)} = 1 + \dfrac{1}{V} \dfrac{4 \pi q_e^2}{q^2} \dfrac{1}{\hslash \omega} \sum_{n,m} \dfrac{e^{- \beta \mathcal{E}_n} - e^{- \beta \mathcal{E}_m}}{Z} \left| \langle n|\hat{\rho}_{\textbf{q}}|m \rangle \right|^2 \dfrac{1}{1 + \frac{i \delta}{\omega} + \frac{\mathcal{E}_n - \mathcal{E}_m}{\hslash \omega}},
\end{equation}
and if we write
\begin{equation}
\dfrac{1}{1 + \frac{i \delta}{\omega} + \frac{\mathcal{E}_n - \mathcal{E}_m}{\hslash \omega}} = \dfrac{1}{1 + \frac{i \delta + \frac{\mathcal{E}_n - \mathcal{E}_m}{\hslash}}{\omega}} ,
\end{equation}
we observe that in the high-frequency limit $\omega \rightarrow +\infty$, we can use the following approximation
\begin{equation}
\dfrac{1}{1 + \frac{i \delta + \frac{\mathcal{E}_n - \mathcal{E}_m}{\hslash}}{\omega}} = 1 - \dfrac{i \delta + \frac{\mathcal{E}_n - \mathcal{E}_m}{\hslash}}{\omega},
\end{equation}
then
\begin{equation}
\left. \dfrac{1}{\epsilon_{\textbf{q}}(\omega + i \delta)} \right|_{\omega \rightarrow \infty} = 1 + \dfrac{1}{V} \dfrac{4 \pi q_e^2}{q^2} \dfrac{1}{\hslash \omega} \sum_{n,m} \dfrac{e^{- \beta \mathcal{E}_n} - e^{- \beta \mathcal{E}_m}}{Z} \left| \langle n|\hat{\rho}_{\textbf{q}}|m \rangle \right|^2 \left( 1 - \dfrac{i \delta}{\omega} - \dfrac{\mathcal{E}_n - \mathcal{E}_m}{\hslash \omega} \right).
\end{equation}
As for the sum
\begin{equation}
\left( 1 - \dfrac{i \delta}{\omega} \right) \sum_{n,m} \dfrac{e^{- \beta \mathcal{E}_n} - e^{- \beta \mathcal{E}_m}}{Z} \left| \langle n|\hat{\rho}_{\textbf{q}}|m \rangle \right|^2 - \sum_{n,m} \dfrac{e^{- \beta \mathcal{E}_n} - e^{- \beta \mathcal{E}_m}}{Z} \left| \langle n|\hat{\rho}_{\textbf{q}}|m \rangle \right|^2 \dfrac{\mathcal{E}_n - \mathcal{E}_m}{\hslash \omega},
\end{equation}
we show that the first term is zero: indeed, the general term of the series is antisymmetric, as can be seen by swapping the indices $n$ and $m$,
\begin{equation}
A_{mn} = \dfrac{e^{- \beta \mathcal{E}_m} - e^{- \beta \mathcal{E}_n}}{Z} \left| \langle m \left| \hat{\rho}_{\mathbf{q}} \right| n \rangle \right|^2 = - \dfrac{e^{- \beta \mathcal{E}_n} - e^{- \beta \mathcal{E}_m}}{Z} \left| \langle n \left| \hat{\rho}_{\mathbf{q}} \right| m \rangle \right|^2 = - A_{nm},
\end{equation}
then it is a sum of antisymmetric terms over a pair of symmetric indices, so it is zero, and the dielectric function in the limit $\omega \rightarrow +\infty$ can be rewritten as
\begin{equation}
\left. \dfrac{1}{\epsilon_{\textbf{q}}(\omega)} \right|_{\omega \rightarrow \infty} = 1 - \dfrac{1}{V} \dfrac{4 \pi q_e^2}{q^2} \dfrac{1}{(\hslash \omega)^2} \sum_{n,m} \dfrac{e^{- \beta \mathcal{E}_n} - e^{- \beta \mathcal{E}_m}}{Z} \left| \langle n|\hat{\rho}_{\textbf{q}}|m \rangle \right|^2 \left( \mathcal{E}_n - \mathcal{E}_m \right).
\end{equation}
From
\begin{align}
\left| \left\langle n \left| \hat{\rho}_{\textbf{q}} \right| m \right\rangle \right|^2 \left( \mathcal{E}_n - \mathcal{E}_m \right) &= \left\langle n \left| \hat{\rho}_{\textbf{q}} \right| m \right\rangle \left\langle m \left| \hat{\rho}^\dagger_{\textbf{q}} \right| n \right\rangle \left( \mathcal{E}_n - \mathcal{E}_m \right) = \notag \\
&= \left\langle n \left| \left( \mathcal{E}_n - \mathcal{E}_m \right) \hat{\rho}_{\textbf{q}} \right| m \right\rangle \left\langle m \left| \hat{\rho}^\dagger_{\textbf{q}} \right| n \right\rangle = \notag \\
&= \left\langle n \left| \hat{\mathcal{H}}_0 \hat{\rho}_{\textbf{q}} - \hat{\rho}_{\textbf{q}} \hat{\mathcal{H}}_0 \right| m \right\rangle \left\langle m \left| \hat{\rho}^\dagger_{\textbf{q}} \right| n \right\rangle = \notag \\
&= \left\langle n \left| \left[ \hat{\mathcal{H}}_0 , \hat{\rho}_{\textbf{q}} \right] \right| m \right\rangle \left\langle m \left| \hat{\rho}^\dagger_{\textbf{q}} \right| n \right\rangle = \notag \\
&= - \left\langle n \left| \left[ \hat{\rho}_{\textbf{q}} , \hat{\mathcal{H}}_0 \right] \right| m \right\rangle \left\langle m \left| \hat{\rho}^\dagger_{\textbf{q}} \right| n \right\rangle ,
\end{align}
we get
\begin{equation}
\left. \dfrac{1}{\epsilon_{\textbf{q}}(\omega)} \right|_{\omega \rightarrow \infty} = 1 + \dfrac{1}{V} \dfrac{4 \pi q_e^2}{q^2} \dfrac{1}{(\hslash \omega)^2} \sum_{n,m} \dfrac{e^{- \beta \mathcal{E}_n} - e^{- \beta \mathcal{E}_m}}{Z} \left\langle n|\left[ \rho_{\textbf{q}} , \hat{\mathcal{H}}_0 \right]|m \right\rangle \left\langle m|\hat{\rho}^\dagger_{\textbf{q}}|n \right\rangle,
\end{equation}
then, we manipulate the sum as follows
\begin{align}
& \sum_{n,m} \dfrac{e^{- \beta \mathcal{E}_n} - e^{- \beta \mathcal{E}_m}}{Z} \left\langle n \left| \left[ \hat{\rho}_{\textbf{q}} , \hat{\mathcal{H}}_0 \right] \right| m \right\rangle \left\langle m \left| \hat{\rho}^\dagger_{\textbf{q}} \right| n \right\rangle \ = \notag \\
&= \sum_{n,m} \dfrac{e^{- \beta \mathcal{E}_n}}{Z} \left\langle n \left| \left[ \hat{\rho}_{\textbf{q}} , \hat{\mathcal{H}}_0 \right] \right| m \right\rangle \left\langle m \left| \hat{\rho}^\dagger_{\textbf{q}} \right| n \right\rangle - \sum_{n,m} \dfrac{e^{- \beta \mathcal{E}_m}}{Z} \left\langle n \left| \left[ \hat{\rho}_{\textbf{q}} , \hat{\mathcal{H}}_0 \right] \right| m \right\rangle \left\langle m \left| \hat{\rho}^\dagger_{\textbf{q}} \right| n \right\rangle = \notag \\
&= \sum_{n,m} \dfrac{e^{- \beta \mathcal{E}_n}}{Z} \left\langle m \left| \hat{\rho}^\dagger_{\textbf{q}} \right| n \right\rangle \left\langle n \left| \left[ \hat{\rho}_{\textbf{q}} , \hat{\mathcal{H}}_0 \right] \right| m \right\rangle - \sum_{n,m} \dfrac{e^{- \beta \mathcal{E}_m}}{Z} \left\langle n \left| \left[ \hat{\rho}_{\textbf{q}} , \hat{\mathcal{H}}_0 \right] \right| m \right\rangle \left\langle m \left| \hat{\rho}^\dagger_{\textbf{q}} \right| n \right\rangle = \notag \\
&= \sum_m \left\langle m \left| \hat{\rho}^\dagger_{\textbf{q}} \left( \sum_n \dfrac{e^{- \beta \mathcal{E}_n}}{Z} \left| n \right\rangle \left\langle n \right| \right) \left[ \hat{\rho}_{\textbf{q}} , \hat{\mathcal{H}}_0 \right] \right| m \right\rangle - \sum_n \left\langle n \left| \left[ \hat{\rho}_{\textbf{q}} , \hat{\mathcal{H}}_0 \right] \left( \sum_m \dfrac{e^{- \beta \mathcal{E}_m}}{Z} \left| m \right\rangle \left\langle m \right| \right) \hat{\rho}^\dagger_{\textbf{q}} \right| n \right\rangle = \notag \\
&= \sum_m \left\langle m \left| \hat{\rho}^\dagger_{\textbf{q}} \left[ \hat{\rho}_{\textbf{q}} , \hat{\mathcal{H}}_0 \right] \right| m \right\rangle - \sum_n \left\langle n \left| \left[ \hat{\rho}_{\textbf{q}} , \hat{\mathcal{H}}_0 \right] \hat{\rho}^\dagger_{\textbf{q}} \right| n \right\rangle,
\end{align}
and since the summation indices are dummy indices, we have constructed a thermal average of $\langle[[\hat{\rho}_{\textbf{q}},\mathcal{H}_0],\hat{\rho}^{\dagger}_{\textbf{q}}]\rangle$, from which it follows
\begin{equation}
\left. \dfrac{1}{\epsilon_{\textbf{q}}(\omega)} \right|_{\omega \rightarrow \infty} = 1 + \dfrac{1}{V} \dfrac{4 \pi q_e^2}{q^2} \dfrac{1}{(\hslash \omega)^2} \left\langle\left[\left[ \hat{\rho}_{\textbf{q}},\hat{\mathcal{H}}_0\right], \hat{\rho}^{\dagger}_{\textbf{q}}\right] \right\rangle.
\end{equation}
From $\eqref{eq: mediacommutatorecommutatorerhoqH0conrhodaggerq}$ we have
\begin{align}
\left. \dfrac{1}{\epsilon_{\textbf{q}}(\omega)} \right|_{\omega \rightarrow \infty} &= 1 + \dfrac{1}{V} 4 \pi q_e^2 \dfrac{1}{\omega^2} \dfrac{N}{m_e} = \notag \\
&= 1 + \dfrac{\omega_p^2}{\omega^2}.
\label{eq: reciprocofunzionedielettricalimitegrandiomega}
\end{align}
If we write
\begin{equation}
\left. \dfrac{1}{\epsilon_{\textbf{q}}(\omega)} \right|_{\omega \rightarrow \infty} = \dfrac{\omega^2 + \omega_p^2}{\omega^2},
\end{equation}
we equivalently have
\begin{align}
\left. \epsilon_{\textbf{q}}(\omega) \right|_{\omega \rightarrow \infty} &= \dfrac{\omega^2}{\omega^2 + \omega_p^2} = \notag \\
&= \dfrac{1}{1 + \frac{\omega_p^2}{\omega^2}} ,
\end{align}
and in the limit $\omega \rightarrow \infty$ that we are considering here, we can use the following approximation
\begin{equation}
\left. \epsilon_{\textbf{q}}(\omega) \right|_{\omega \rightarrow \infty} = 1 - \dfrac{\omega_p^2}{\omega^2}.
\end{equation}
In other words, the dielectric function has a zero at the plasma frequency, meaning a resonance phenomenon occurs at the plasma frequency.
\begin{remark}[Landau's interpretation of the zeros of the dielectric function]
Starting from expression $\eqref{eq: relazionefunzionedielettricapropagatorepolarizzazionestar}$, we consider a frequency of the form $\omega = \Omega_{\textbf{q}} - i \gamma_{\textbf{q}}$, with $\gamma_{\textbf{q}} > 0$, that is, we consider the lower complex half-plane, where we expect the dielectric function to have zeros, as they correspond to poles of the polarization propagator. We set $\epsilon_{\textbf{q}}(\Omega_{\textbf{q}} - i \gamma_{\textbf{q}}) = 0$, from $\eqref{eq: relazionefunzionedielettricapropagatorepolarizzazionestar}$ it follows
\begin{equation}
1 - \dfrac{1}{V} \dfrac{4 \pi q_e^2}{q^2} \left[ \re \Pi^*_{\textbf{q}}\left( \Omega_{\textbf{q}} - i \gamma_{\textbf{q}} \right) + i \im \Pi^*_{\textbf{q}}\left( \Omega_{\textbf{q}} - i \gamma_{\textbf{q}} \right) \right] = 0 .
\end{equation}
Landau considers a Taylor expansion of the polarization propagator for $\gamma_{\textbf{q}} \rightarrow 0$. In particular, since the propagator is analytic at points that are zeros of the dielectric function, the complex derivative can be taken along any direction; thus, we differentiate the propagator with respect to the real axis and write
\begin{equation}
1 - \dfrac{1}{V} \dfrac{4 \pi q_e^2}{q^2} \left[ \Pi^*_{\textbf{q}}\left( \Omega_{\textbf{q}} \right) - i \gamma_{\textbf{q}} \left. \dfrac{\partial \Pi^*_{\textbf{q}} \left( \omega \right)}{\partial \omega} \right|_{\omega=\Omega_{\textbf{q}}} \right] = 0,
\label{eq: sviluppoTaylorfunzionedielettrica}
\end{equation}
which implies two equations for the real and imaginary parts. Regarding the real part of $\eqref{eq: sviluppoTaylorfunzionedielettrica}$, neglecting terms of order higher than first in $\gamma_{\textbf{q}}$, we have
\begin{equation}
1 - \dfrac{1}{V} \dfrac{4 \pi q_e^2}{q^2} \re \Pi^*_{\textbf{q}}\left( \Omega_{\textbf{q}} \right) = 0.
\end{equation}
From the imaginary part equation, we have
\begin{equation}
\gamma_{\textbf{q}} = \dfrac{\im \Pi^*_{\textbf{q}}\left( \Omega_{\textbf{q}} \right)}{\left. \dfrac{\partial \re \Pi^*_{\textbf{q}} \left( \omega \right)}{\partial \omega} \right|_{\omega=\Omega_{\textbf{q}}}},
\end{equation}
That is, the imaginary part of the polarization propagator is responsible for the damping, similarly to what we showed for the many-body propagator. If the imaginary part is zero, the damping is zero and the associated particles have an infinite lifetime.
\end{remark}
\subsection{Hartree-Fock approximation}
\begin{figure}[H]
\centering
\includegraphics[scale=0.8]{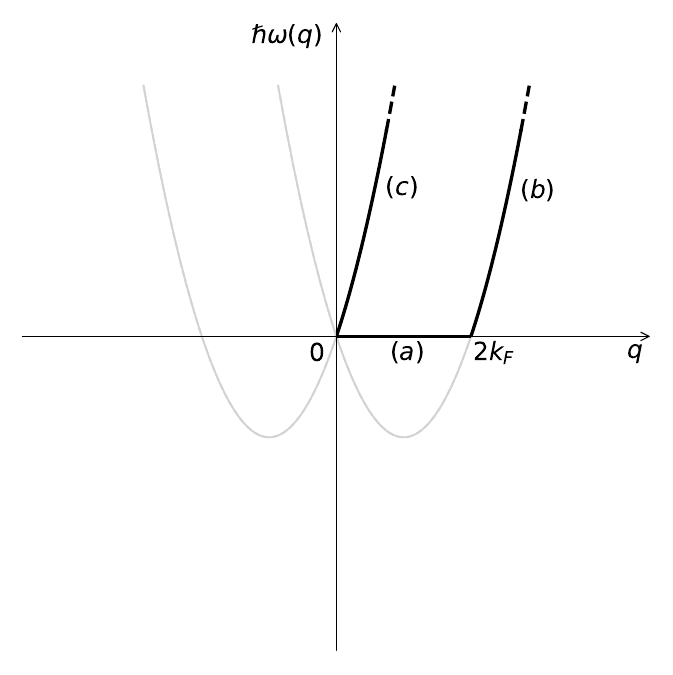}
\caption{The hole-particle processes in the Hartree-Fock approximation for the dielectric function occur for momentum values between curves (a), (b), and (c).}
\label{fig: grafico_processi_buca_particella_Hartree_Fock_1}
\end{figure}

A first approximation for calculating the dielectric function is the Hartree-Fock approximation, which consists of replacing the polarization propagator with its first-order term, that is
\begin{equation}
\left. \left[ q_e \delta \langle \rho \rangle (\textbf{q},\omega) \right] \right|_{(H.F.)} = \dfrac{1}{V} q^2_e \varphi_{ext,\textbf{q}}(\omega) \Pi^{(0)}_{\textbf{q}}(\omega),
\end{equation}
That is, from $\eqref{eq: relazionefunzionedielettricapropagatorepolarizzazione}$, this means performing a first order expansion in terms of $\frac{1}{\epsilon_{\textbf{q}}(\omega)}$. Since the dielectric function is written as in $\eqref{eq: reciprocofunzionedielettricainfunzionediautostatidiH}$, the Hartree-Fock approach consists of replacing the eigenfunctions of the complete Hamiltonian with those corresponding to the non-interacting Hamiltonian. Now, suppose $T=0$, consequently
\begin{equation}
\dfrac{e^{- \beta \mathcal{E}_n}}{Z} =
\begin{cases}
1, \ \text{ground state} \\
0, \ \text{otherwise}
\end{cases},
\end{equation}
and we need to compute quantities of the form
\begin{equation}
\left| \left\langle n \left| \hat{\rho}_{\textbf{q}} \right|0 \right\rangle \right|^2.
\end{equation}
Note that the ground state is a Slater determinant of plane waves with maximum momentum equal to the Fermi momentum $\textbf{k}_F$ (see Chapter \ref{The Fermi momentum}), and the particle density operator destroys the plane wave of momentum $\textbf{k}$ and creates one of momentum $\textbf{k} + \textbf{q}$. The operator $C_{\textbf{k},\sigma}$ contained in $\hat{\rho}_{\textbf{q}}$ can act on the ground state $|0\rangle$ only if the new momentum $\textbf{k}$ has magnitude less than $\textbf{k}_F$, while the operator $C^\dagger_{\textbf{k}+\textbf{q},\sigma}$ can act on the state $|n \rangle$ only if the new momentum is $\textbf{k} + \textbf{q}$. In other words, in the Hartree-Fock approximation, hole-particle processes occur, which are fixed-particle-number excitations. Let $\mathcal{E}_m$ be the energy of the initial state, which in our case is the ground state, so $\mathcal{E}_m = 0$, and $\mathcal{E}_n$ the energy of the final state, with the energy difference given by
\begin{align}
\Delta \mathcal{E}(\textbf{q}) &= \mathcal{E}_{\textbf{k}+\textbf{q}} - \mathcal{E}_{\textbf{k}} = \notag \\
&= \dfrac{\hslash^2 (\textbf{k}+\textbf{q})^2}{2m_e} - \dfrac{\hslash^2 \textbf{k}^2}{2m_e} = \notag \\
&= \dfrac{\hslash^2}{2m_e} \left[ \textbf{q}^2 + 2 \textbf{k} \cdot \textbf{q} \right],
\end{align}
Such energy is the energy required to create a hole-particle pair and occurs at a fixed transferred momentum $\textbf{q}$, and for the constraints discussed previously, it must hold that
\begin{equation}
\max \left\lbrace 0, \dfrac{\hslash^2}{2m_e} \left( \textbf{q}^2 - 2 \textbf{k}_F \cdot \textbf{q} \right) \right\rbrace \leq \Delta \mathcal{E}(\left| \textbf{q} \right|) \leq \dfrac{\hslash^2}{2m_e} \left( \textbf{q}^2 + 2 \textbf{k}_F \cdot \textbf{q} \right).
\end{equation}
The graph of the energy $\hslash \omega$ as a function of the modulus of $\textbf{q}$ is shown in Figure $\eqref{fig: grafico_processi_buca_particella_Hartree_Fock_1}$; in particular, it is the planar region enclosed between arcs $(b)$ and $(c)$, including segment $(a)$. Arc $(c)$ corresponds to the parabola $\hslash \omega(q) = \frac{\hslash^2}{2m_e} \left( \textbf{q}^2 + 2 \textbf{k}_F \cdot \textbf{q} \right)$, which has zeros at $\textbf{q}=0$ and $\textbf{q}=-2 \textbf{k}_F$; arc $(b)$ corresponds to the parabola $\hslash \omega(q) = \frac{\hslash^2}{2m} \left( \textbf{q}^2 - 2 \textbf{k}_F \cdot \textbf{q} \right)$, which has zeros at $\textbf{q}=0$ and $\textbf{q}=2 \textbf{k}_F$; segment $(a)$ corresponds to processes in which the energy variation is zero because the particle is annihilated on the Fermi surface and recreated with the same momentum, also on the Fermi surface. Note that the excitations produced in the Hartree-Fock approximation are single-particle excitations and not collective. Moreover, the Hartree-Fock approximation contains no singularities, whereas we have seen previously that the zeros of the dielectric function correspond to resonance frequencies, the plasma frequencies, which are frequencies of collective excitations.
\subsection{Thomas-Fermi approximation}
Here we study the Thomas-Fermi approximation for the dielectric function, which consists of considering an external perturbation constant in time, that is, of the form
\begin{equation}
\varphi_{ext,\textbf{q}}(\omega) = 2 \pi \delta(\omega) \varphi_{ext,\textbf{q}},
\end{equation}
and we aim to study the variation of particle density, that is, the polarization propagator. If we divide $\eqref{eq: caricapervariazionedensitadiparticelleintrasformatainfunzionefunzionedielettrica1}$ by the charge $q_e$, we obtain
\begin{equation}
\delta \langle \hat{\rho} \rangle (\textbf{q},\omega) = \dfrac{q^2}{4 \pi q_e} \varphi_{ext,\textbf{q}}(\omega) \left( \dfrac{1}{\epsilon_{\textbf{q}}(\omega)} - 1 \right),
\end{equation}
and by substituting the Thomas-Fermi perturbation, we have
\begin{equation}
\delta \langle \hat{\rho} \rangle (\textbf{q},\omega) = \dfrac{q^2}{2 q_e} \delta(\omega) \varphi_{ext}(\textbf{q}) \left( \dfrac{1}{\epsilon_{\textbf{q}}(\omega)} - 1 \right).
\end{equation}
We aim to compute this density variation in the time domain, so we apply a temporal Fourier transform as follows
\begin{align}
\delta \langle \hat{\rho} \rangle (\textbf{q},t) &= \int_{-\infty}^{+\infty} \dfrac{d \omega}{2 \pi} e^{- i \omega t} \delta \langle \hat{\rho} \rangle (\textbf{q},\omega) = \notag \\
&= \int_{-\infty}^{+\infty} \dfrac{d \omega}{2 \pi} e^{- i \omega t} \dfrac{q^2}{2 q_e} \delta(\omega) \varphi_{ext,\textbf{q}} \left( \dfrac{1}{\epsilon_{\textbf{q}}(\omega)} - 1 \right),
\end{align}
then
\begin{equation}
\delta \langle \hat{\rho} \rangle (\textbf{q}) = \dfrac{q^2}{4 \pi q_e} \varphi_{ext,\textbf{q}} \left( \dfrac{1}{\epsilon_{\textbf{q}}(0)} - 1 \right) .
\label{eq: variazionedensitaparticelleThomasFermi1}
\end{equation}
As expected, this variation is time-independent, reflecting the fact that it results from a static (i.e., zero-frequency) perturbation in the frequency domain. The response of the medium is therefore stationary. Since the static dielectric function $\epsilon_{\textbf{q}}(0)$ is not known explicitly, Thomas and Fermi propose an Ansatz for the particle density variation $\delta \langle \hat{\rho} \rangle (\textbf{q})$. In the absence of perturbation, the particle density is calculated as
\begin{align}
\rho_0 &= \dfrac{1}{V} \sum_{\textbf{k},\sigma} \langle C^{\dagger}_{\textbf{k},\sigma} C_{\textbf{k},\sigma}\rangle_0 = \notag \\
&= \dfrac{1}{V} \sum_{\textbf{k},\sigma} n_{-1} \left(\mathcal{E}^{(0)}_{\textbf{k}} \right),
\end{align}
where $n_{-1}\left( \mathcal{E}^{(0)}_{\textbf{k}} \right)$ corresponds to Fermi-Dirac statistics, with energies $\mathcal{E}^{(0)}_{\textbf{k}} = \frac{\hslash^2 \textbf{k}^2}{2m_e} - \mu$, and represents a zeroth-order approximation for the particle density, as expected. As before, we work within the framework of the jellium model, see equation $\eqref{eq: hamiltonianamodellojellium2}$. When an external perturbation \( \varphi_{\text{ext}} \) is applied, it induces a charge density within the material. As a result, the system becomes polarized and produces its own electric field, which contributes to the total electrostatic potential \( \varphi \). Thomas and Fermi proposed an Ansatz for this perturbation in real space, focusing on a charged particle Hamiltonian of the form
\begin{equation}
\hat{\mathcal{H}} = \dfrac{\hat{\textbf{p}}^2}{2m_e} + q_e \varphi(\textbf{r}),
\end{equation}
and it is assumed that the new particle density becomes
\begin{align}
\rho(\textbf{r}) &= \dfrac{1}{V} \sum_{\textbf{k},\sigma} n_{-1}\left( \mathcal{E}_{\textbf{k}}^{(0)} + q_e \varphi(\textbf{r}) \right) = \notag \\
&= \dfrac{1}{V} \sum_{\textbf{k},\sigma} \dfrac{1}{e^{\beta \left( \mathcal{E}^{(0)}_{\textbf{k}} + q_e \varphi(\textbf{r}) \right)} + 1} .
\label{eq: densitaparticelleThomasFermi}
\end{align}
Note that this approximation assumes knowledge of the particle density at each position and for each conjugate momentum, which indeed violates quantum mechanics: the Thomas-Fermi approach is semi-classical. Furthermore, since the system is at thermodynamic equilibrium, the chemical potential must be constant within each macroscopic volume element; otherwise, there would be charge flows from one point to another. Since we are interested in the variation of particle density, we expand $\eqref{eq: densitaparticelleThomasFermi}$ around $\varphi=0$ using a first-order Taylor polynomial, that is
\begin{equation}
\rho(\textbf{r}) = \rho_0 + \dfrac{1}{V} \sum_{\textbf{k},\sigma} \dfrac{\partial n_{-1}\left( \mathcal{E}^{(0)}_{\textbf{k}},\varphi(\textbf{r}) \right)}{\partial \mathcal{E}^{(0)}_{\textbf{k}}} q_e \varphi(\textbf{r}),
\end{equation}
then
\begin{equation}
\delta \langle \rho \rangle (\textbf{r)} = \dfrac{1}{V} \sum_{\textbf{k},\sigma} \dfrac{n_{-1}\left( \mathcal{E}^{(0)}_{\textbf{k}},\varphi(\textbf{r}) \right)}{\partial \mathcal{E}^{(0)}_{\textbf{k}}} q_e \varphi(\textbf{r}).
\end{equation}
We apply a Fourier transform in momentum space and obtain
\begin{align}
\delta \langle \rho \rangle (\textbf{q)} &= \dfrac{1}{V} \sum_{\textbf{k},\sigma} \dfrac{\partial n_{-1}\left( \mathcal{E}^{(0)}_{\textbf{k}},\varphi(\textbf{r}) \right)}{\partial \mathcal{E}^{(0)}_{\textbf{k}}} q_e \varphi_{\textbf{q}} = \notag \\
&= \dfrac{1}{V} \sum_{\textbf{k},\sigma} \dfrac{\partial n_{-1}\left( \mathcal{E}^{(0)}_{\textbf{k}},\varphi(\textbf{r}) \right)}{\partial \mathcal{E}^{(0)}_{\textbf{k}}} q_e \dfrac{\varphi_{ext,\textbf{q}}}{\epsilon_{\textbf{q}}(0)},
\label{eq: variazionedensitaparticelleThomasFermi2}
\end{align}
where we have substituted $\eqref{eq: legametrapotenzialeesternopotenzialetotalefunzionedielettrica}$ in the case where the external potential does not depend on frequency and the dielectric function is considered only for $\omega=0$, that is, in the static case we are addressing. Equating $\eqref{eq: variazionedensitaparticelleThomasFermi1}$ and $\eqref{eq: variazionedensitaparticelleThomasFermi2}$, we have
\begin{equation}
\dfrac{q^2}{4 \pi q_e} \varphi_{ext,\textbf{q}} \left( \dfrac{1}{\epsilon_{\textbf{q}}(0)} - 1 \right) = \dfrac{1}{V} \sum_{\textbf{k},\sigma} \dfrac{\partial n_{-1}\left( \mathcal{E}^{(0)}_{\textbf{k}},\varphi(\textbf{r}) \right)}{\partial \mathcal{E}^{(0)}_{\textbf{k}}} q_e \dfrac{\varphi_{ext,\textbf{q}}}{\epsilon_{\textbf{q}}(0)},
\end{equation}
\begin{equation}
\dfrac{q^2}{4 \pi q_e} \left( \dfrac{1-\epsilon_{\textbf{q}}(0)}{\epsilon_{\textbf{q}}(0)} \right) = \dfrac{1}{V} \sum_{\textbf{k},\sigma} \dfrac{\partial n_{-1}\left( \mathcal{E}^{(0)}_{\textbf{k}},\varphi(\textbf{r}) \right)}{\partial \mathcal{E}^{(0)}_{\textbf{k}}} q_e \dfrac{1}{\epsilon_{\textbf{q}}(0)},
\end{equation}
\begin{equation}
\dfrac{q^2}{4 \pi q_e^2} \left( 1-\epsilon_{\textbf{q}}(0) \right) = \dfrac{1}{V} \sum_{\textbf{k},\sigma} \dfrac{\partial n_{-1}\left( \mathcal{E}^{(0)}_{\textbf{k}},\varphi(\textbf{r}) \right)}{\partial \mathcal{E}^{(0)}_{\textbf{k}}},
\end{equation}
\begin{equation}
1-\epsilon_{\textbf{q}}(0) = \dfrac{4 \pi q_e^2}{q^2} \dfrac{1}{V} \sum_{\textbf{k},\sigma} \dfrac{\partial n_{-1}\left( \mathcal{E}^{(0)}_{\textbf{k}},\varphi(\textbf{r}) \right)}{\partial \mathcal{E}^{(0)}_{\textbf{k}}},
\end{equation}
and finally we get
\begin{equation}
\epsilon_{\textbf{q}}(0) = 1 - \dfrac{4 \pi q_e^2}{q^2} \dfrac{1}{V} \sum_{\textbf{k},\sigma} \dfrac{\partial n_{-1}\left( \mathcal{E}^{(0)}_{\textbf{k}},\varphi(\textbf{r}) \right)}{\partial \mathcal{E}^{(0)}_{\textbf{k}}}.
\label{eq: funzionedielettricaThomasFermi1}
\end{equation}
Ultimately, to compute $\eqref{eq: funzionedielettricaThomasFermi1}$, we need to estimate the quantity
\begin{equation}
\dfrac{1}{V} \sum_{\textbf{k},\sigma} \dfrac{\partial n_{-1}\left( \mathcal{E}^{(0)}_{\textbf{k}},\varphi(\textbf{r}) \right)}{\partial \mathcal{E}^{(0)}_{\textbf{k}}}.
\end{equation}
First of all, note that the energies do not depend on spin, so the sum over $\sigma$ yields the spin degeneracy $g=2$,
\begin{equation}
\dfrac{2}{V} \sum_{\textbf{k}} \dfrac{\partial n_{-1}(\textbf{k})}{\partial \mathcal{E}^{(0)}_{\textbf{k}}}.
\end{equation}
We will compute this quantity in the limits of low and high temperatures relative to the Fermi temperature, i.e., $T \ll T_F$ and $T \gg T_F$. At low temperatures, the Fermi-Dirac function can be approximated by a Heaviside theta function $\Theta(\textbf{k}_F-\textbf{k})=\Theta(\mathcal{E}_{\textbf{k}_F} - \mathcal{E}^{(0)}_{\textbf{k}})$, which equals $1$ below the Fermi level and $0$ otherwise, and its derivative with respect to energy is $- \delta \left( \mathcal{E}^{(0)}_{\textbf{k}} \right)$. In the high-temperature limit, the Fermi function is replaced by the Maxwell-Boltzmann distribution $e^{- \beta \mathcal{E}^{(0)}_{\textbf{k}}}$, whose derivative with respect to energy is $- \beta e^{- \beta \mathcal{E}^{(0)}_{\textbf{k}}}$. We have
\begin{equation}
\dfrac{2}{V} \sum_{\textbf{k}} \dfrac{\partial n_{-1}\left( \mathcal{E}^{(0)}_{\textbf{k}},\varphi(\textbf{r}) \right)}{\partial \mathcal{E}^{(0)}_{\textbf{k}}} 
=
- \dfrac{2}{V} \sum_{\textbf{k}}
\begin{cases}
\delta \left( \mathcal{E}^{(0)}_{\textbf{k}} \right), \ T \ll T_F \\
\beta e^{- \beta \mathcal{E}^{(0)}_{\textbf{k}}}, \ T \gg T_F
\end{cases}
.
\end{equation}
For $T \ll T_F$, we have
\begin{align}
- \dfrac{2}{V} \sum_{\textbf{k}} \delta \left( \mathcal{E}^{(0)}_{\textbf{k}} \right) &\simeq 2 \int \dfrac{d^3 \textbf{k}}{(2 \pi)^3} \delta \left( \mathcal{E}^{(0)}_{\textbf{k}} \right) = \notag \\
&= 2 \int_{-\infty}^{+\infty} \dfrac{4 \pi k^2 dk}{(2 \pi)^3} \delta \left[ \dfrac{\hslash^2 k^2}{2m_e} - \mathcal{E}_F \right] ,
\end{align}
where we have used the thermodynamic limit approximation $\eqref{eq: illimitetermodinamico}$ and spherical polar coordinates. We observe that within the integral appears a composite function of the momentum, the function
\begin{equation}
f(k) = \dfrac{\hslash^2 k^2}{2m_e} - \mathcal{E}_F,
\end{equation}
which has zeros at
\begin{equation}
k_F = \pm \sqrt{\dfrac{2 m_e \mathcal{E}_F}{\hslash^2}},
\end{equation}
and whose first derivative is
\begin{equation}
f'(k) = \dfrac{\hslash^2 k}{m_e}.
\end{equation}
By the property of the delta function of a composite argument, that is,
\begin{align}
\delta(k) &= \frac{1}{|f'(k_F)|} \delta(k - k_F) + \frac{1}{|f'(-k_F)|} \delta(k + k_F) = \notag \\
&= \frac{m_e}{\hbar^2 k_F} \left( \delta(k - k_F) + \delta(k + k_F) \right),
\end{align}
we have
\begin{align}
- \frac{2}{V} \sum_{\mathbf{k}} \delta \left( \mathcal{E}^{(0)}_{\mathbf{k}} \right) &= 2 \int_{-\infty}^{+\infty} \frac{4 \pi k^2 \, dk}{(2 \pi)^3} \frac{m_e}{\hbar^2 k_F} \left( \delta(k - k_F) + \delta(k + k_F) \right) = \notag \\
&= \frac{m_e}{\pi^2 \hbar^2 k_F} \left( \int k^2 \delta(k - k_F) \, dk + \int k^2 \delta(k + k_F) \, dk \right) = \notag \\
&= \frac{2 m_e k_F^2}{\pi^2 \hbar^2 k_F} = \notag \\
&= \frac{2 m_e k_F}{\pi^2 \hbar^2}.
\end{align}
Now, since we are considering low temperatures, a first-order approximation for the particle density variation, and working in models where electronic interactions are neglected, such as the jellium model, we can use $\eqref{eq: relazionedensitafermionimomentoFermid3}$ and proceed as follows
\begin{align}
- \dfrac{2}{V} \sum_{\textbf{k}} \delta \left( \mathcal{E}^{(0)}_{\textbf{k}} \right) &= \frac{2 m_e k_F}{\pi^2 \hbar^2} \frac{3 k_F^2}{3 k_F^2} = \notag \\
&= 3 \frac{2 m_e}{\hbar^2 k_F^2} \frac{k_F^2}{3 \pi^2} = \notag \\
&= \frac{3 \rho_0}{2 \mathcal{E}_F}.
\end{align}
For $T \gg T_F$, the Maxwell-Boltzmann distribution, including the spin degeneracy $g=2$, yields the number of particles $N$, thus
\begin{equation}
- \dfrac{2}{V} \sum_{\textbf{k}} \beta e^{- \beta \mathcal{E}^{(0)}_{\textbf{k}}} = - \beta \rho_0.
\end{equation} 
The dielectric function in the Thomas-Fermi model satisfies
\begin{equation}
\epsilon_{\textbf{q}}(0) = 1 + \dfrac{6 \pi q_e^2}{q^2} \dfrac{\rho_0}{\mathcal{E}_F}, \ T \ll T_F,
\label{eq: funzionedielettricaThomasFermi2bassetemperature}
\end{equation}
\begin{equation}
\epsilon_{\textbf{q}}(0) = 1 + \dfrac{4 \pi q_e^2}{q^2} \beta \rho_0, \ T \gg T_F,
\label{eq: funzionedielettricaThomasFermi2altetemperature}
\end{equation}
or equivalently we can write in compact form as follows
\begin{equation}
\left. \epsilon_{\textbf{q}}(0) \right|_{T.F.} = 1 + \dfrac{q^2_{T.F.}}{q^2},
\label{eq: funzionedielettricaThomasFermi2compatta}
\end{equation}
with
\begin{equation}
q^2_{T.F.}
=
\begin{cases}
\dfrac{6 \pi q_e^2 \rho_0}{\mathcal{E}_F}, \ \ T \ll T_F \\
4 \pi q_e^2 \beta \rho_0, \ T \gg T_F
\end{cases}
.
\label{eq: q2_TF}
\end{equation}
Let us consider as the external perturbation the potential produced by a charged particle $Q$ placed at a distance from the material medium, that is,
\begin{equation}
\varphi_{ext}(\textbf{r}) = \dfrac{Q}{\textbf{r}},
\end{equation}
\begin{equation}
\varphi_{ext,\textbf{q}} = \dfrac{4 \pi Q}{q^2},
\end{equation}
which is a stationary perturbation that induces a redistribution of charges and consequently it implies an induced total potential $\varphi_{\textbf{q}}$ that we want to determine. From $\eqref{eq: legametrapotenzialeesternopotenzialetotalefunzionedielettrica}$ and $\eqref{eq: funzionedielettricaThomasFermi2compatta}$, we have
\begin{align}
\varphi_{\textbf{q}}(\omega) &= \dfrac{\varphi_{ext,\textbf{q}}(\omega)}{\epsilon_{\textbf{q}}(\omega)} = \notag \\
&= \dfrac{4 \pi Q}{q^2} \dfrac{1}{1 + \frac{q^2_{T.F.}}{q^2}} = \notag \\
&= \dfrac{4 \pi Q}{q^2 + q^2_{T.F.}}.
\end{align}
We note that from $\eqref{eq: trasformataFourierpotenzialeYukawa}$ the total potential is a Yukawa potential,
\begin{equation}
\varphi_{ext}(r) = \frac{Q}{r} e^{-q_{T.F.} r},
\end{equation}
that is, the Thomas-Fermi approximation transforms a long-range external perturbation (Coulomb potential) into a short-range interaction (Yukawa potential).
\subsection{Random Phase Approximation}
\begin{figure}[H]
\centering
\includegraphics[scale=0.8]{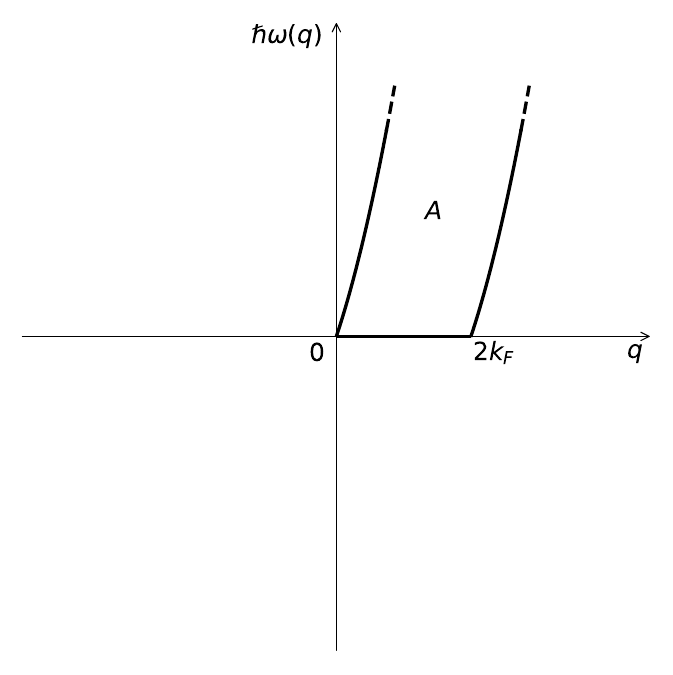}
\caption{The hole-particle processes in the Hartree-Fock approximation for the dielectric function occur for momentum values belonging to the $A$ region, that is, in the region of the plane enclosed by the segment and the two arcs of the parabola (bold curve).}
\label{fig: grafico_processi_buca_particella_Hartree_Fock_2}
\end{figure}

\begin{figure}[H]
\centering
\includegraphics[scale=0.8]{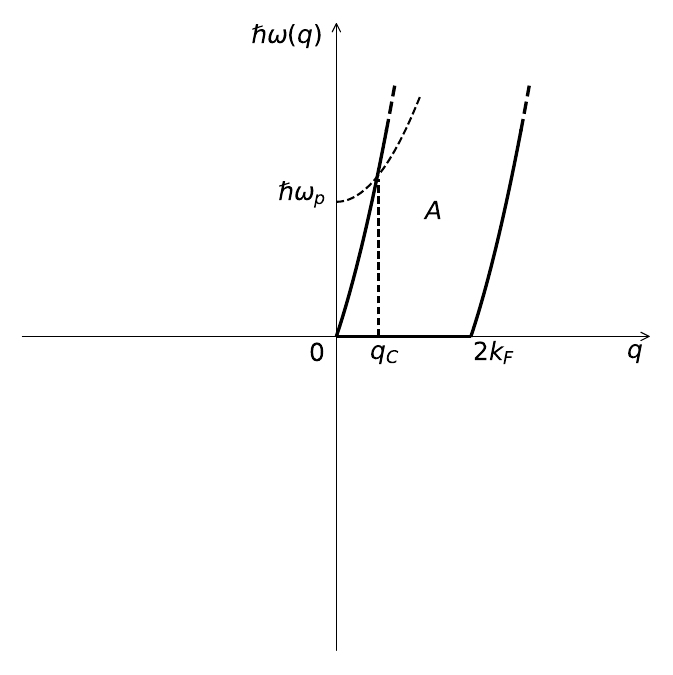}
\caption{Region A is the energy region where particle-hole processes are allowed, while the dashed curve corresponds to equation $\eqref{eq: relazionedispersioneplasmoni4}$, i.e., the plasmon dispersion relation.}
\label{fig: grafico_processi_buca_particella_Hartree_Fock_3}
\end{figure}

\begin{figure}[H]
\centering
\includegraphics[scale=0.6]{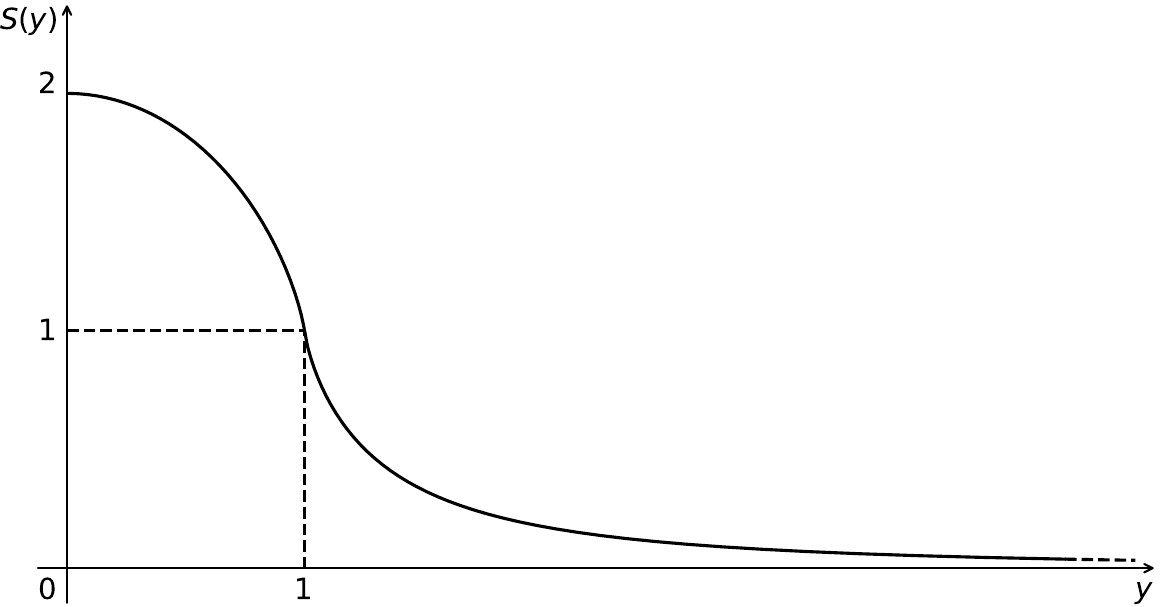}
\caption{The function $\eqref{eq: S_y_RPA_approximation_equation}$ appears in the RPA approximation equation of the dielectric function for $\omega=0$, $T=0$, and corresponds to the opposite of the function defined in equation $\eqref{eq: H_y_energieHartreeFock}$, i.e., Figure $\eqref{fig: H_y_energieHartreeFock_figure}$.}
\label{fig: S_y_RPA_approximation_figure}
\end{figure}

Unlike the Hartree-Fock approximation, which performs an expansion in terms of $\frac{1}{\epsilon_{\textbf{q}}(\omega)}$, the Random Phase Approximation (R.P.A.) method performs an expansion in terms of $\epsilon_{\textbf{q}}(\omega)$. In particular, given $\eqref{eq: relazionefunzionedielettricapropagatorepolarizzazionestar}$, the random phase approximation consists in replacing the polarization propagator with the lowest order, that is
\begin{equation}
\left. \epsilon_{\textbf{q}}(\omega) \right|_{R.P.A.} = 1 - \dfrac{1}{V} \dfrac{4 \pi q_e^2}{q^2} \Pi^{(0)}_{\textbf{q}}(\omega),
\label{eq: funzionedielettricaRPA1}
\end{equation}
where $\Pi^{(0)}_{\textbf{q}}(\omega)$ is the polarization bubble and is given by $\eqref{eq: propagatorepolarizzazioneprimoordinesommamatsubaraesplicitata}$, which we write as
\begin{equation}
\Pi^{(0)}_{\textbf{q}}(\omega + i \delta) = \dfrac{1}{\hslash} \sum_{\textbf{k},\sigma} \dfrac{n_{-1}(\mathcal{E}_{\textbf{k}+\textbf{q}}) - n_{-1}(\mathcal{E}_{\textbf{k}})}{\omega + i \delta + \frac{\mathcal{E}_{\textbf{k}+\textbf{q}} - \mathcal{E}_{\textbf{k}}}{\hslash}},
\end{equation}
that is, we have introduced a convergence factor $\delta>0$. Since we are dealing with electrons and the energies are of the form $\mathcal{E}_{\textbf{k}} = \frac{\hslash^2 \textbf{k}^2}{2m} - \mu$, we can eliminate the sum over $\sigma$ and introduce a spin degeneracy $g=2$ as follows
\begin{equation}
\Pi^{(0)}_{\textbf{q}}(\omega + i \delta) = \dfrac{2}{\hslash} \sum_{\textbf{k}} \dfrac{n_{-1}(\mathcal{E}_{\textbf{k}+\textbf{q}}) - n_{-1}(\mathcal{E}_{\textbf{k}})}{\omega + i \delta + \frac{\mathcal{E}_{\textbf{k}+\textbf{q}} - \mathcal{E}_{\textbf{k}}}{\hslash}}.
\label{eq: propagatorepolarizzazioneprimoordinesommamatsubaraesplicitatasenzasommaspin}
\end{equation}
Here we are interested in the static case. Since the spectral function is antisymmetric with respect to $\omega$, see $\eqref{eq: funzionespettralepropagatorepolarizzazionedisparirispettoafrequenze}$, the imaginary part of the polarization propagator is also antisymmetric, and therefore it vanishes at $\omega=0$, that is
\begin{equation}
\im \Pi^{(0)}_{\textbf{q}}(\omega=0) = 0.
\end{equation}
The polarization propagator for $\omega=0$ has only a real part, and in the thermodynamic limit the sum is replaced by a principal value integral as follows
\begin{equation}
\re \Pi^{(0)}_{\textbf{q}}(\omega=0) = 2 V \pv \int \dfrac{d^3 \textbf{k}}{(2 \pi)^3} \dfrac{n_{-1}(\mathcal{E}_{\textbf{k}+\textbf{q}}) - n_{-1}(\mathcal{E}_{\textbf{k}})}{\mathcal{E}_{\textbf{k}+\textbf{q}} - \mathcal{E}_{\textbf{k}}},
\end{equation}
from which the R.P.A. approximation of dielectric function is
\begin{equation}
\left. \epsilon_{\textbf{q}}(0) \right|_{R.P.A.} = 1 - \dfrac{4 \pi q_e^2}{q^2} 2 \pv \int \dfrac{d^3 \textbf{k}}{(2 \pi)^3} \dfrac{n_{-1}(\mathcal{E}_{\textbf{k}+\textbf{q}}) - n_{-1}(\mathcal{E}_{\textbf{k}})}{\mathcal{E}_{\textbf{k}+\textbf{q}} - \mathcal{E}_{\textbf{k}}}.
\label{eq: funzionedielettricaRPA2}
\end{equation}
Note that in the limit $\textbf{q} \rightarrow \textbf{0}$, the integrand function becomes $\frac{\partial n_{-1}(\mathcal{E}_{\textbf{k})}}{\partial \mathcal{E}_{\textbf{k}}}$, and thus the dielectric function at infinitesimal momenta is that calculated by Thomas-Fermi. We then need to quantify this function for finite $\textbf{q}$. Now, assuming $T=0$, the Fermi-Dirac functions become step functions. A famous approximation in the scientific literature at zero temperature is the following
\begin{equation}
\left[ n_{-1}(\mathcal{E}_{\textbf{k}+\textbf{q}}) - n_{-1}(\mathcal{E}_{\textbf{k}}) \right]_{T=0} \approx - 2 \left[ n_{-1}(\mathcal{E}_{\textbf{k}}) \right]_{T=0},
\end{equation}
then
\begin{equation}
\left. \epsilon_{\textbf{q}}(\omega) \right|_{R.P.A., T=0} = 1 + \dfrac{4 \pi q_e^2}{q^2} 2 \pv \int \dfrac{d^3 \textbf{k}}{(2 \pi)^3} \dfrac{2 n_{-1}(\mathcal{E}_{\textbf{k}})}{\mathcal{E}_{\textbf{k}+\textbf{q}} - \mathcal{E}_{\textbf{k}}},
\label{eq: funzionedielettricaRPA3}
\end{equation}
where the Fermi-Dirac function is replaced by a Heaviside function $\Theta(\textbf{k}_F - \textbf{k})$. We have
\begin{align}
\left. \epsilon_{\textbf{q}}(\omega) \right|_{R.P.A., \ T=0} &= 1 - \dfrac{4 \pi q_e^2}{q^2} 4 \int_0^{k_F} \dfrac{2 \pi k^2 dk}{(2 \pi)^3} \pv \int_{-1}^{+1} d \cos(\theta) \dfrac{1}{\dfrac{\hslash^2 q^2}{2m_e} + \dfrac{2 \hslash^2 k q \cos(\theta)}{2m_e}} = \notag \\
&= 1 + \dfrac{4 \pi q_e^2}{q^2} \dfrac{4}{(2 \pi)^2} \dfrac{2m_e}{\hslash^2} \int_0^{k_F} \dfrac{2 \pi k^2 dk}{(2 \pi)^3} \pv \int_{-1}^{+1} d \cos(\theta) \dfrac{1}{q^2 + 2 k q \cos(\theta)}.
\end{align}
By evaluating the well-known integral, we find
\begin{equation}
\left. \epsilon_{\textbf{q}}(\omega) \right|_{R.P.A., \ T=0} = 1 + \dfrac{q_{T.F.}^2(T=0)}{2 q^2} S(y),
\label{eq: funzionedielettricaRPA4}
\end{equation}
with
\begin{equation}
y = \dfrac{q}{2 k_F},
\end{equation}
\begin{equation}
S(y) = 1 + \dfrac{1-y^2}{2y} \ln \left| \dfrac{1+y}{1-y} \right|,
\label{eq: S_y_RPA_approximation_equation}
\end{equation}
which is the opposite of function $\eqref{eq: H_y_energieHartreeFock}$. Consequently, from $\eqref{eq: H_y_energieHartreeFock_limit_0}$, $\eqref{eq: H_y_energieHartreeFock_limit_infty}$, $\eqref{eq: H_y_energieHartreeFock_limit_1}$, equation $\eqref{eq: S_y_RPA_approximation_equation}$ satisfies
\begin{equation}
\lim_{y \rightarrow 0^+} S(y) = 2,
\label{eq: S_y_RPA_approximatrion_limit_0}
\end{equation}
\begin{equation}
\lim_{y \rightarrow +\infty} S(y) = 0,
\label{eq: S_y_RPA_approximatrion_limit_infty}
\end{equation}
\begin{equation}
\lim_{y \rightarrow 1} S(y) = 1,
\label{eq: S_y_RPA_approximatrion_limit_1}
\end{equation}
as shown in Figure $\eqref{fig: S_y_RPA_approximation_figure}$. In particular, for $q \rightarrow 0^+$, i.e., $y \rightarrow 0^+$, from $\eqref{eq: S_y_RPA_approximatrion_limit_0}$ and $\eqref{eq: funzionedielettricaRPA4}$ it follows that
\begin{align}
\left. \epsilon_{\textbf{q}}(0) \right|_{R.P.A., \ T=0, \ q \rightarrow 0^+} &= 1 + \dfrac{q_{T.F.}^2(T=0)}{q^2} = \notag \\
&= \left. \epsilon_{\textbf{q}}(0) \right|_{T.F.}.
\end{align}
Now, if we consider an external perturbation given by a charged particle at a distance, we ask what the induced total potential is in the R.P.A. approximation. Using the residue theorem, it is possible to show that this potential in real space has the form
\begin{equation}
\varphi(r) = \dfrac{\cos(2 k_F r)}{r^3},
\end{equation}
which is the inverse Fourier transform of
\begin{equation}
\varphi(q) = \dfrac{4 \pi Q}{q^2} \left( 1 + \dfrac{q_{T.F.}^2(T=0)}{2 q^2} S(y) \right)^{-1}.
\end{equation}
Note that, unlike the Thomas-Fermi approximation where the potential decays exponentially, here the induced potential decays following a power law and includes oscillations, known as Friedel oscillations. \newline
Let us now consider arbitrary frequencies, i.e., finite frequencies. The imaginary part of $\eqref{eq: funzionedielettricaRPA2}$ is equal to the imaginary part of the integral, which can be easily shown to be antisymmetric with respect to the frequency $\omega$. Furthermore, this imaginary part is responsible for the hole-particle processes, which are allowed for energies within the region $A$ of Figure $\eqref{fig: grafico_processi_buca_particella_Hartree_Fock_2}$, as seen in the Hartree-Fock approximation. It follows that the imaginary part of the dielectric function in the R.P.A. model must be zero for energies within the region $\mathbb{R}^2 \setminus A$ of Figure $\eqref{fig: grafico_processi_buca_particella_Hartree_Fock_2}$. Thus, we focus on energies belonging to this region, where the dielectric function has only a real part. Since hole-particle excitations are not allowed, we are investigating whether the R.P.A. approximation generates new phenomena. We write $\eqref{eq: funzionedielettricaRPA2}$ for arbitrary $\omega$, that is, from $\eqref{eq: propagatorepolarizzazioneprimoordinesommamatsubaraesplicitatasenzasommaspin}$, $\eqref{eq: funzionedielettricaRPA1}$ we have
\begin{equation}
\left. \epsilon_{\textbf{q}}(\omega + i \delta) \right|_{R.P.A.} = 1 - \dfrac{4 \pi q_e^2}{q^2} \dfrac{2}{\hslash} \pv \int \dfrac{d^3 \textbf{k}}{(2 \pi)^3} \dfrac{n_{-1}(\mathcal{E}_{\textbf{k}+\textbf{q}}) - n_{-1}(\mathcal{E}_{\textbf{k}})}{\omega + i \delta + \frac{\mathcal{E}_{\textbf{k}+\textbf{q}} - \mathcal{E}_{\textbf{k}}}{\hslash}}.
\label{eq: funzionedielettricaRPA5}
\end{equation}
and we manipulate as follows
\begin{align}
\left. \epsilon_{\textbf{q}}(\omega + i \delta) \right|_{R.P.A.} &= 1 - \dfrac{4 \pi q_e^2}{q^2} \dfrac{2}{\hslash} \left[ \pv \int \dfrac{d^3 \textbf{k}}{(2 \pi)^3} \dfrac{n_{-1}(\mathcal{E}_{\textbf{k}+\textbf{q}})}{\omega + i \delta + \frac{\mathcal{E}_{\textbf{k}+\textbf{q}} - \mathcal{E}_{\textbf{k}}}{\hslash}} - \pv \int \dfrac{d^3 \textbf{k}}{(2 \pi)^3} \dfrac{n_{-1}(\mathcal{E}_{\textbf{k}})}{\omega + i \delta + \frac{\mathcal{E}_{\textbf{k}+\textbf{q}} - \mathcal{E}_{\textbf{k}}}{\hslash}} \right] = \notag \\
&= 1 - \dfrac{4 \pi q_e^2}{q^2} \dfrac{2}{\hslash} \left[ \pv \int \dfrac{d^3 \textbf{k}'}{(2 \pi)^3} \dfrac{n_{-1}(\mathcal{E}_{\textbf{k}'})}{\omega + i \delta + \frac{\mathcal{E}_{\textbf{k}'} - \mathcal{E}_{\textbf{k}'-\textbf{q}}}{\hslash}} - \pv \int \dfrac{d^3 \textbf{k}}{(2 \pi)^3} \dfrac{n_{-1}(\mathcal{E}_{\textbf{k}})}{\omega + i \delta + \frac{\mathcal{E}_{\textbf{k}+\textbf{q}} - \mathcal{E}_{\textbf{k}}}{\hslash}} \right],
\end{align}
from which we rename the dummy variable and factorize as follows
\begin{align}
\left. \epsilon_{\textbf{q}}(\omega + i \delta) \right|_{R.P.A.} &= 1 - \dfrac{4 \pi q_e^2}{q^2} \dfrac{2}{\hslash} \pv \int \dfrac{d^3 \textbf{k}}{(2 \pi)^3} n_{-1}(\mathcal{E}_{\textbf{k}}) \left[ \dfrac{1}{\omega + i \delta + \frac{\mathcal{E}_{\textbf{k}} - \mathcal{E}_{\textbf{k}-\textbf{q}}}{\hslash}} - \dfrac{1}{\omega + i \delta + \frac{\mathcal{E}_{\textbf{k}+\textbf{q}} - \mathcal{E}_{\textbf{k}}}{\hslash}} \right] = \notag \\
&= 1 - \dfrac{4 \pi q_e^2}{q^2} \dfrac{2}{\hslash \omega} \pv \int \dfrac{d^3 \textbf{k}}{(2 \pi)^3} n_{-1}(\mathcal{E}_{\textbf{k}}) \left[ \dfrac{1}{1 + \frac{i \delta}{\omega} + \frac{\mathcal{E}_{\textbf{k}} - \mathcal{E}_{\textbf{k}-\textbf{q}}}{\hslash \omega}} - \dfrac{1}{1 + \frac{i \delta}{\omega} + \frac{\mathcal{E}_{\textbf{k}+\textbf{q}} - \mathcal{E}_{\textbf{k}}}{\hslash \omega}} \right].
\end{align}
Thereafter, assume $T=0$, so that the Fermi-Dirac functions can be replaced by step functions, and then we consider the limit for small $\textbf{q}$. From the calculations, we obtain
\begin{equation}
\left. \epsilon_{\textbf{q}}(\omega) \right|_{R.P.A., \ T=0, \ \omega \neq 0} = 1 - \dfrac{4 \pi q_e^2 \rho_0}{m_e \omega^2} \left( 1 + \dfrac{9}{5} \dfrac{\hslash^2}{m_e^2} \dfrac{q^2 k_F^2}{\omega^2} \right) .
\end{equation}
If we set this dielectric function equal to zero, we obtain
\begin{equation}
\dfrac{4 \pi q_e^2 \rho_0}{m \omega^2} \left( 1 + \dfrac{9}{5} \dfrac{\hslash^2}{m_e^2} \dfrac{q^2 k_F^2}{\omega^2} \right) = 1.
\end{equation}
We set
\begin{equation}
A = \frac{4 \pi q_e^2 \rho_0}{m_e}, 
\end{equation}
\begin{equation}
B = \frac{9}{5} \frac{\hbar^2}{m_e^2} q^2 k_F^2,
\end{equation}
then
\begin{equation}
\frac{A}{\omega^2} \left(1 + \frac{B}{\omega^2}\right) = 1,
\end{equation}
\begin{equation}
\frac{A}{\omega^2} + \frac{AB}{\omega^4} = 1 ,
\end{equation}
\begin{equation}
A \omega^2 + AB = \omega^4,
\end{equation}
\begin{equation}
\omega^4 - A \omega^2 - AB = 0.
\end{equation}
Whose solution is given by
\begin{equation}
\omega^2 = \frac{A + \sqrt{A^2 + 4AB}}{2},
\end{equation}
then
\begin{align}
\omega &= \sqrt{ \dfrac{A + \sqrt{A^2 + 4AB} }{2} } = \notag \\
&= \sqrt{\frac{\frac{4\pi q_e^2 \rho_0}{m_e} + \sqrt{ \left( \frac{4\pi q_e^2 \rho_0}{m_e} \right)^2 + 4 \frac{4\pi q_e^2 \rho_0}{m_e} \frac{9}{5} \frac{\hbar^2}{m^2_e} q^2 k_F^2}}{2}} = \notag \\
&= \sqrt{ \frac{\omega_p^2 + \sqrt{\omega_p^4 + \frac{36}{5} \omega_p^2 \frac{\hbar^2}{m^2_e} q^2 k_F^2}}{2}} ,
\label{eq: relazionedispersioneplasmoni1}
\end{align}
where we have made explicit the plasma frequency $\eqref{eq: frequenzadiplasma}$. We now compute a first-order expansion of the frequency $\eqref{eq: relazionedispersioneplasmoni1}$ in $q^2$. We have
\begin{equation}
\delta = \dfrac{36}{5} \omega_p^2 \dfrac{\hbar^2}{m^2_e} q^2 k_F^2 \ll \omega_p^2 ,
\end{equation}
\begin{align}
\sqrt{\omega_p^4 + \omega_p^2 \delta} &= \omega_p^2 \sqrt{ 1 + \dfrac{\delta}{\omega_p^2} } = \notag \\
&\approx \omega_p^2 \left( 1 + \dfrac{1}{2} \dfrac{\delta}{\omega_p^2} \right) = \notag \\
&= \omega_p^2 + \dfrac{1}{2} \delta,
\end{align}
\begin{align}
\omega &\approx \sqrt{ \frac{ \omega_p^2 + \left( \omega_p^2 + \frac{1}{2} \delta \right) }{2} } = \notag \\
&= \sqrt{ \dfrac{2 \omega_p^2 + \frac{1}{2} \delta}{2} } = \notag \\
&= \sqrt{ \omega_p^2 + \dfrac{1}{4} \delta } = \notag \\
&\approx \omega_p \left( 1 + \dfrac{1}{2} \dfrac{1}{4} \dfrac{\delta}{\omega_p^2} \right) = \notag \\
&= \omega_p \left( 1 + \dfrac{1}{8} \dfrac{\delta}{\omega_p^2} \right) = \notag \\
&= \omega_p \left( 1 + \dfrac{1}{8} \dfrac{36}{5} \dfrac{\hbar^2}{m^2_e} \dfrac{q^2 k_F^2}{\omega_p^2} \right) = \notag \\
&= \omega_p \left( 1 + \dfrac{9}{10} \dfrac{\hbar^2}{m^2_e \omega_p^2} q^2 k_F^2 \right) .
\label{eq: relazionedispersioneplasmoni2}
\end{align}
If we invert the first of $\eqref{eq: q2_TF}$ as follows
\begin{equation}
\rho_0 = \dfrac{\mathcal{E}_F q^2_{TF}(T=0)}{6 \pi q_e^2},
\end{equation}
and we manipulate as follows
\begin{align}
\omega &= \omega_p \left( 1 + \dfrac{9}{10} \dfrac{\hbar^2 k_F^2}{2 m_e} \dfrac{1}{m_e \omega_p^2} q^2 \right) = \notag \\
&= \omega_p \left( 1 + \dfrac{9}{10} \mathcal{E}_F \dfrac{1}{m_e} \dfrac{1}{\omega_p^2} q^2 \right) = \notag \\
&= \omega_p \left( 1 + \dfrac{9}{10} \mathcal{E}_F \dfrac{1}{m_e} \dfrac{m_e}{4 \pi q_e^2 \rho_0} q^2 \right) = \notag \\
&= \omega_p \left( 1 + \dfrac{9}{10} \mathcal{E}_F \dfrac{1}{4 \pi q_e^2 \rho_0} q^2 \right) = \notag \\
&= \omega_p \left( 1 + \dfrac{9}{10} \mathcal{E}_F \dfrac{1}{4 \pi q_e^2 \frac{\mathcal{E}_F q^2_{\text{TF}}(T=0)}{6 \pi q_e^2}} q^2 \right) ,
\label{eq: relazionedispersioneplasmoni3}
\end{align}
we finally get
\begin{equation}
\omega = \omega_p \left( 1 + \dfrac{27}{20} \dfrac{q^2}{q^2_{TF}} \right).
\label{eq: relazionedispersioneplasmoni4}
\end{equation}
As we anticipated, the semiclassical and quantum calculations of the plasma frequency coincide, but the quantum treatment has additionally provided the equation $\eqref{eq: relazionedispersioneplasmoni4}$, which is a dispersion relation at small momenta $q$ for quasiparticles, the plasmons, which are collective excitations of an electron gas. The equation $\eqref{eq: relazionedispersioneplasmoni4}$ is shown in Figure $\eqref{fig: grafico_processi_buca_particella_Hartree_Fock_3}$, together with the allowed region for hole-particle processes. In the region where the imaginary part of the dielectric function is zero, hole-particle processes do not occur, and the plasmons correspond to a dielectric function that has only a nonzero real part: from the discussion on Landau damping, we deduce that in this region the plasmons have infinite lifetime. Note that the plasmon dispersion relation intersects with the energy region of hole-particle processes; consequently, for sufficiently large values of $q$, starting from a critical momentum $q_C$, plasmons are no longer stable because they can decay into hole-particle processes.
\section{Magnetic susceptibility}
Here we aim to study the response of a material medium to a weak magnetic field. The following Hamiltonian is written in the framework of first quantization, and describes the coupling between a single spin and a magnetic field directed along the $z$-axis, that is,
\begin{equation}
\mathcal{H}_I = - \dfrac{g \mu_B}{\hslash} \hat{S}_z H_z ,
\end{equation}
where $\mu_B = \frac{\hslash |q_e|}{2 m_e c}$ is the Bohr magneton, and $g$ is a constant known as the gyromagnetic factor. For a system of $N$ non-interacting spins, we generalize as follows
\begin{equation}
\mathcal{H}_I = - \dfrac{g \mu_B}{\hslash} \sum_{i=1}^N \hat{S}_{i,z} H_z. 
\end{equation}
In the thermodynamic limit the interaction hamiltonian in first quantization takes the form
\begin{equation}
\mathcal{H}_I = - \dfrac{g \mu_B}{\hslash} \int d^3 \textbf{r} s_z(\textbf{r}) H_z(\textbf{r}),
\end{equation}
where $s_z(\textbf{r})$ is a spin density, that is it has dimensions of spin per unit volume, and the discrete and continuous expressions are equivalent only if
\begin{equation}
s_z(\textbf{r}) = \sum_{i=1}^N \delta(\textbf{r}-\textbf{r}_i) \hat{S}_z(\textbf{r}) ,
\end{equation}
and we express such a density in second quantization. Since it is a one-body operator, we have
\begin{align}
\hat{s}_z(\textbf{r}) &= \int dx' \hat{\psi}^\dag(x') \delta(\textbf{r} - \textbf{r}') \hat{S}_z \hat{\psi}(x') = \notag \\
&= \sum_{s} \hat{\psi}^\dag(\textbf{r},s) \hat{S}_z \hat{\psi}(\textbf{r},s) .
\end{align}
Let us assume that the material medium is spatially translationally invariant. In this case, we can expand by means of $\eqref{eq: operatorecampodistruzionebaseondepianeperspin}$ $\eqref{eq: operatorecampocreazionebaseondepianeperspin}$, and from $\eqref{eq: equazioneautovaloriSzconbasechi}$, we have
\begin{align}
\hat{s}_z(\textbf{r}) &= \sum_{s} \hat{\psi}^\dag(\textbf{r},s) \delta(\textbf{r} - \textbf{r}') \hat{S}_z \hat{\psi}(\textbf{r},s) = \notag \\
&= \dfrac{1}{V} \sum_{s} \sum_{\textbf{k}_1,\textbf{k}_2,\sigma_1,\sigma_2} e^{i (\textbf{k}_2 - \textbf{k}_1) \cdot \textbf{r}} C^{\dagger}_{\textbf{k}_1,\sigma_1} \chi_{\sigma_1}(s) \hat{S}_z \chi_{\sigma_2}(s) C_{\textbf{k}_2,\sigma_2} = \notag \\
&= \dfrac{1}{V} \sum_{s} \sum_{\textbf{k}_1,\textbf{k}_2,\sigma_1,\sigma_2} e^{i (\textbf{k}_2 - \textbf{k}_1) \cdot \textbf{r}} C^{\dagger}_{\textbf{k}_1,\sigma_1} \chi_{\sigma_1}(s) \hslash \sigma_2 \chi_{\sigma_2}(s) C_{\textbf{k}_2,\sigma_2} = \notag \\
&= \dfrac{1}{V} \sum_{\textbf{k}_1,\textbf{k}_2,\sigma_1,\sigma_2} e^{i (\textbf{k}_2 - \textbf{k}_1) \cdot \textbf{r}} \hslash \sigma_2 C^{\dagger}_{\textbf{k}_1,\sigma_1} \delta_{\sigma_1,\sigma_2} C_{\textbf{k}_2,\sigma_2} = \notag \\
&= \dfrac{1}{V} \sum_{\textbf{k}_1,\textbf{k}_2,\sigma} e^{i (\textbf{k}_2 - \textbf{k}_1) \cdot \textbf{r}} \hslash \sigma C^{\dagger}_{\textbf{k}_1,\sigma} C_{\textbf{k}_2,\sigma} ,
\end{align}
where in particular we used the completeness relation $\sum_s \chi_{\sigma_1}(s) \chi_{\sigma_2}(s) = \delta_{\sigma_1,\sigma_2}$, and then performed a sum over $\sigma_2$ renaming the dummy index $\sigma_1 \rightarrow \sigma$. Now that the spin density is expressed in second quantization, we compute its spatial Fourier transform, that is
\begin{align}
\hat{s}_z(\textbf{q}) &= \int d^3 \textbf{r} e^{i \textbf{q} \cdot \textbf{r}} \hat{s}_z(\textbf{r}) = \notag \\
&= \dfrac{1}{V} \sum_{\textbf{k}_1,\textbf{k}_2,\sigma} \int d^3 \textbf{r} e^{i (\textbf{q} + \textbf{k}_2 - \textbf{k}_1) \cdot \textbf{r}} \hslash \sigma C^{\dagger}_{\textbf{k}_1,\sigma} C_{\textbf{k}_2,\sigma} = \notag \\
&= \dfrac{1}{V} \hslash \sum_{\textbf{k}_1,\textbf{k}_2,\sigma} V \delta_{\textbf{q} + \textbf{k}_2 - \textbf{k}_1} \sigma C^{\dagger}_{\textbf{k}_1,\sigma} C_{\textbf{k}_2,\sigma} = \notag \\
&= \hslash \sum_{\textbf{k},\sigma} \sigma C^{\dagger}_{\textbf{k} + \textbf{q},\sigma} C_{\textbf{k},\sigma} ,
\end{align}
where we used
\begin{equation}
\int d^3 \textbf{r} e^{i (\textbf{q} + \textbf{k}_2 - \textbf{k}_1)} = V \delta_{\textbf{q} + \textbf{k}_2 - \textbf{k}_1},
\end{equation}
\begin{equation}
\textbf{k}_1 = \textbf{k}_2 + \textbf{q},
\end{equation}
and finally, we renamed the dummy index $\textbf{k}_2 \rightarrow \textbf{k}$. We observe that $\hat{s}_z(\textbf{q})$ has a structure very similar to the Fourier transform of the particle density. We sum over $\sigma = \mp \frac{1}{2}$, and we obtain
\begin{equation}
\hat{s}_z(\textbf{q}) = \dfrac{\hslash}{2} \sum_{\textbf{k}} \left[ C^{\dagger}_{\textbf{k} + \textbf{q},\frac{1}{2}} C_{\textbf{k},\frac{1}{2}} - C^{\dagger}_{\textbf{k} + \textbf{q},-\frac{1}{2}} C_{\textbf{k},-\frac{1}{2}} \right] ,
\end{equation}
which has an immediate interpretation: an electron with momentum $\textbf{k}$ and spin up is annihilated and created with unchanged spin and momentum $\textbf{k} + \textbf{q}$, and the term is preceded by a positive sign; alternatively, an electron with momentum $\textbf{k}$ and spin down is annihilated and created with unchanged spin and momentum $\textbf{k} + \textbf{q}$, and the term is preceded by a negative sign.
\begin{remark}[Spin volume densities along the $x$ and $y$ axes]
Let us write the spatial spin densities corresponding to the $x$ and $y$ components. The difference compared to the $z$ component case treated above lies in the action of the operators $\hat{S}_x$ and $\hat{S}_y$ on the eigenfunction $\chi_{\sigma}(s)$, see $\eqref{eq: equazioneautovaloriSxconbasechi}$, $\eqref{eq: equazioneautovaloriSyconbasechi}$. The spin volume density along $x$ in second quantization is then given by
\begin{align}
\hat{s}_x(\textbf{r}) &= \sum_{s} \hat{\psi}^\dag(\textbf{r},s) \delta(\textbf{r} - \textbf{r}') \hat{S}_x \hat{\psi}(\textbf{r},s) = \notag \\
&= \dfrac{1}{V} \sum_{s} \sum_{\textbf{k}_1,\textbf{k}_2,\sigma_1,\sigma_2} e^{i (\textbf{k}_2 - \textbf{k}_1) \cdot \textbf{r}} C^{\dagger}_{\textbf{k}_1,\sigma_1} \chi_{\sigma_1}(s) \hat{S}_x \chi_{\sigma_2}(s) C_{\textbf{k}_2,\sigma_2} = \notag \\
&= \dfrac{1}{V} \sum_{s} \sum_{\textbf{k}_1,\textbf{k}_2,\sigma_1,\sigma_2} e^{i (\textbf{k}_2 - \textbf{k}_1) \cdot \textbf{r}} C^{\dagger}_{\textbf{k}_1,\sigma_1} \chi_{\sigma_1}(s) \hslash \left| \sigma_2 \right| \chi_{-\sigma_2}(s) C_{\textbf{k}_2,\sigma_2} = \notag \\
&= \dfrac{1}{V} \sum_{\textbf{k}_1,\textbf{k}_2,\sigma_1,\sigma_2} e^{i (\textbf{k}_2 - \textbf{k}_1) \cdot \textbf{r}} \hslash \left| \sigma_2 \right| C^{\dagger}_{\textbf{k}_1,\sigma_1} \delta_{\sigma_1,-\sigma_2} C_{\textbf{k}_2,\sigma_2} = \notag \\
&= \dfrac{1}{V} \sum_{\textbf{k}_1,\textbf{k}_2,\sigma} e^{i (\textbf{k}_2 - \textbf{k}_1) \cdot \textbf{r}} \hslash \left| \sigma \right| C^{\dagger}_{\textbf{k}_1,\sigma} C_{\textbf{k}_2,-\sigma},
\end{align}
and its Fourier transform is
\begin{align}
\hat{s}_x(\textbf{q}) &= \int d^3 \textbf{r} e^{i \textbf{q} \cdot \textbf{r}} \hat{s}_x(\textbf{r}) = \notag \\
&= \dfrac{1}{V} \sum_{\textbf{k}_1,\textbf{k}_2,\sigma} \int d^3 \textbf{r} e^{i (\textbf{q} + \textbf{k}_2 - \textbf{k}_1) \cdot \textbf{r}} \hslash \left| \sigma \right| C^{\dagger}_{\textbf{k}_1,\sigma} C_{\textbf{k}_2,-\sigma} = \notag \\
&= \dfrac{1}{V} \hslash \sum_{\textbf{k}_1,\textbf{k}_2,\sigma} V \delta_{\textbf{q} + \textbf{k}_2 - \textbf{k}_1} \left| \sigma \right|  C^{\dagger}_{\textbf{k}_1,\sigma} C_{\textbf{k}_2,\sigma} = \notag \\
&= \hslash \sum_{\textbf{k},\sigma} \left| \sigma \right| C^{\dagger}_{\textbf{k} + \textbf{q},\sigma} C_{\textbf{k},-\sigma},
\end{align}
that is
\begin{equation}
\hat{s}_x(\textbf{q}) = \dfrac{\hslash}{2} \sum_{\textbf{k}} \left[ C^{\dagger}_{\textbf{k} + \textbf{q},-\frac{1}{2}} C_{\textbf{k},\frac{1}{2}} + C^{\dagger}_{\textbf{k} + \textbf{q},\frac{1}{2}} C_{\textbf{k},-\frac{1}{2}} \right].
\end{equation}
The spin volume density along $y$ in second quantization is then given by
\begin{align}
\hat{s}_y(\textbf{r}) &= \sum_{s} \hat{\psi}^\dag(\textbf{r},s) \delta(\textbf{r} - \textbf{r}') \hat{S}_y \hat{\psi}(\textbf{r},s) = \notag \\
&= \dfrac{1}{V} \sum_{s} \sum_{\textbf{k}_1,\textbf{k}_2,\sigma_1,\sigma_2} e^{i (\textbf{k}_2 - \textbf{k}_1) \cdot \textbf{r}} C^{\dagger}_{\textbf{k}_1,\sigma_1} \chi_{\sigma_1}(s) \hat{S}_y \chi_{\sigma_2}(s) C_{\textbf{k}_2,\sigma_2} = \notag \\
&= \dfrac{1}{V} \sum_{s} \sum_{\textbf{k}_1,\textbf{k}_2,\sigma_1,\sigma_2} e^{i (\textbf{k}_2 - \textbf{k}_1) \cdot \textbf{r}} C^{\dagger}_{\textbf{k}_1,\sigma_1} \chi_{\sigma_1}(s) i \hslash \sigma_2 \chi_{-\sigma_2}(s) C_{\textbf{k}_2,\sigma_2} = \notag \\
&= \dfrac{1}{V} i \hslash \sum_{\textbf{k}_1,\textbf{k}_2,\sigma_1,\sigma_2} e^{i (\textbf{k}_2 - \textbf{k}_1) \cdot \textbf{r}} \sigma_2 C^{\dagger}_{\textbf{k}_1,\sigma_1} \delta_{\sigma_1,-\sigma_2} C_{\textbf{k}_2,\sigma_2} = \notag \\
&= - \dfrac{1}{V} i \hslash \sum_{\textbf{k}_1,\textbf{k}_2,\sigma} e^{i (\textbf{k}_2 - \textbf{k}_1) \cdot \textbf{r}} \sigma C^{\dagger}_{\textbf{k}_1,\sigma} C_{\textbf{k}_2,-\sigma},
\end{align}
and its Fourier transform is
\begin{align}
\hat{s}_y(\textbf{q}) &= \int d^3 \textbf{r} e^{i \textbf{q} \cdot \textbf{r}} \hat{s}_y(\textbf{r}) = \notag \\
&= - \dfrac{1}{V} i \hslash \sum_{\textbf{k}_1,\textbf{k}_2,\sigma} \int d^3 \textbf{r} e^{i (\textbf{q} + \textbf{k}_2 - \textbf{k}_1) \cdot \textbf{r}} \sigma C^{\dagger}_{\textbf{k}_1,\sigma} C_{\textbf{k}_2,-\sigma} = \notag \\
&= - \dfrac{1}{V} i \hslash \sum_{\textbf{k}_1,\textbf{k}_2,\sigma} V \delta_{\textbf{q} + \textbf{k}_2 - \textbf{k}_1} \sigma C^{\dagger}_{\textbf{k}_1,\sigma} C_{\textbf{k}_2,\sigma} = \notag \\
&= - i \hslash \sum_{\textbf{k},\sigma} \sigma C^{\dagger}_{\textbf{k} + \textbf{q},\sigma} C_{\textbf{k},-\sigma} ,
\end{align}
that is
\begin{equation}
\hat{s}_y(\textbf{q}) = i \dfrac{\hslash}{2} \sum_{\textbf{k}} \left[ C^{\dagger}_{\textbf{k} + \textbf{q},-\frac{1}{2}} C_{\textbf{k},\frac{1}{2}} - C^{\dagger}_{\textbf{k} + \textbf{q},\frac{1}{2}} C_{\textbf{k},-\frac{1}{2}} \right].
\end{equation}
Note that, unlike the spin volume density along $x$, in second quantization the spin volume densities along $y$ and $z$ annihilate an electron with momentum $\textbf{k}$ and spin down (up) and create one with momentum $\textbf{k} + \textbf{q}$ and spin up (down), and vice versa: these densities flip the spin, as in first quantization.
\end{remark}
Starting from $\hat{s}_z(\textbf{r})$, we define the magnetization density as
\begin{equation}
\hat{M}_z(\textbf{r}) = \dfrac{g \mu_B}{\hslash} \hat{s}_z(\textbf{r}) ,
\end{equation}
and its Fourier transform, obviously proportional to $\hat{s}_z(\textbf{q})$, is
\begin{equation}
\hat{M}_z(\textbf{q}) = \dfrac{g \mu_B}{2} \sum_{\textbf{k}} \left[ C^{\dagger}_{\textbf{k} + \textbf{q},\frac{1}{2}} C_{\textbf{k},\frac{1}{2}} - C^{\dagger}_{\textbf{k} + \textbf{q},-\frac{1}{2}} C_{\textbf{k},-\frac{1}{2}} \right].
\end{equation}
Here we aim to apply linear response theory. As a first step, we recall that in this framework the relevant Hamiltonian is the one describing the unperturbed system. In our case, this Hamiltonian is expressed in the formalism of second quantization and, using the definition of the magnetization volume density, it takes the form
\begin{equation}
\hat{\mathcal{H}}_I  = - \int d^3 \textbf{r} \hat{M}_z(\textbf{r}) H_z(\textbf{r}).
\end{equation}
Now we aim to quantify the change in magnetization density, that is $\delta \langle \hat{M}_z \rangle(\mathbf{q}, \omega)$, as a function of the external magnetic field. Through calculations analogous to those performed to calculate $\delta \langle \rho \rangle (\textbf{q},\omega)$ in the section $\eqref{Dielectric function}$, we obtain
\begin{equation}
\delta \langle \hat{M}_z \rangle(\mathbf{q}, \omega) = - \dfrac{1}{V} H_z(\textbf{q},\omega) G_M(\textbf{q},\omega),
\end{equation}
where
\begin{equation}
G_M(\textbf{q},\omega + i \delta) = - \dfrac{i}{\hslash} \dfrac{1}{V} \int_{-\infty}^{+\infty} dt e^{i (\omega + i \delta) t} \Theta(t) \left\langle \left[ \hat{M}^{(0)}_z(\textbf{q},t) , \hat{M}_z^{\dagger (0)}(\textbf{q},0) \right] \right\rangle.
\end{equation}
It is an equivalent of the polarization propagator, and it is called the magnetization propagator. Note that the variable $\delta$ is introduced to compute the integral in the upper complex half-plane and then take the limit $\delta \rightarrow 0^+$. We define the magnetic susceptibility as
\begin{equation}
\chi(\textbf{q},\sigma) = - G_M(\textbf{q},\omega) ,
\end{equation}
which includes a thermal average over the Hamiltonian without interaction with the field, but in which the particles interact among themselves. Recall that at the lowest order of the density-density Green's function expansion one obtains the polarization bubble, similarly, the lowest order of the spin-spin Green's function produces the so-called magnetization bubble. The thermal average contained in the magnetic susceptibility is of the form
\begin{equation}
\left\langle C^{\dagger}_{\textbf{k} + \textbf{q},\sigma} C_{\textbf{k},\sigma} C^{\dagger}_{\textbf{k}' - \textbf{q},\sigma'} C_{\textbf{k}',\sigma'} \right\rangle_0,
\end{equation}
note that from Wick's theorem, the contraction $\left\langle C^{\dagger}_{\textbf{k} + \textbf{q},\sigma} C_{\textbf{k},\sigma} \right\rangle$ implies $\textbf{q}=\textbf{0}$, which is not admissible, so the unique contraction is $\left\langle C^{\dagger}_{\textbf{k} + \textbf{q},\sigma} C_{\textbf{k}',\sigma'} \right\rangle$, which implies $\textbf{k}' = \textbf{k}+\textbf{q}$, $\sigma = \sigma'$, and therefore the lowest order of the spin-spin Green's function is equal to the polarization bubble. In other words, the polarization bubble at lowest order describes both the dielectric response of a system to a weak longitudinal electric field and the magnetic response to a weak external magnetic field.
\section{Optical conductivity}
Here we aim to derive Ohm's law at the microscopic level
\begin{equation}
\textbf{J} = \sigma \textbf{E},
\label{eq: leggediOhm}
\end{equation}
or equivalently, to derive the optical conductivity $\sigma$. For now, we begin with a classical treatment of the problem. This will allow us to develop the physical intuition and identify the key quantities involved. We will then proceed to reformulate the discussion within the frameworks of first and second quantization. At this stage, we adopt a classical approach: let $q_e$ be an electron in an electromagnetic field; its Hamiltonian is of the form
\begin{equation}
\mathcal{H} = \dfrac{1}{2m_e} \left( \textbf{p} - \dfrac{q_e}{c} \textbf{A} \right)^2 + q_e \varphi + V ,
\end{equation}
where $\varphi$ and $\textbf{A}$ are respectively the scalar and vector potentials. The fields are related to the potentials by
\begin{equation}
\textbf{E} = - \dfrac{1}{c} \dfrac{\partial \textbf{A}}{\partial t} - \nabla \varphi,
\end{equation}
\begin{equation}
\textbf{B} = \nabla \times \textbf{A},
\end{equation}
and we also know that any other scalar and vector potentials must satisfy
\begin{equation}
\textbf{A}' = \textbf{A} + \nabla f,
\end{equation}
\begin{equation}
\varphi' = \varphi - \dfrac{1}{c} \dfrac{\partial f}{\partial t},
\end{equation}
where $f$ is the gauge function. Among the gauge functions, we choose the one for which $\dive \textbf{A} = 0$, which implies $\dive \textbf{A}' = 0$; thus, the vector potential we use is purely transverse, i.e., $\textbf{A} = \textbf{A}_T$. In the Coulomb gauge, the equations to calculate $\varphi$ are decoupled, and one solution is the Coulomb potential
\begin{equation}
\varphi(\textbf{r},t) = \int d^3 \textbf{r} \dfrac{\rho(\textbf{r},t)}{\left| \textbf{r} - \textbf{r}' \right|} ,
\end{equation}
with $\rho(\textbf{r},t)$ the charge density. Here, we aim to study the response of a material medium to a weak transverse electromagnetic field. Under the assumption that the medium is far from the sources of the field, the following approximations hold
\begin{itemize}
\item the electromagnetic wave can be approximated as a plane wave;
\item the Coulomb scalar potential is negligible, since $\left| \textbf{r} - \textbf{r}' \right| \rightarrow \infty$.
\end{itemize}
Under the above assumptions we have
\begin{equation}
\textbf{E} = - \dfrac{1}{c} \dfrac{\partial \textbf{A}}{\partial t},
\end{equation}
\begin{equation}
\mathcal{H} = \dfrac{1}{2 m_e} \left( \textbf{p} - \dfrac{q_e}{c} \textbf{A} \right)^2 + V.
\end{equation}
We also recall that the velocity of the particle, as a consequence of the electromagnetic field, becomes
\begin{equation}
\textbf{v} = \dfrac{1}{m_e} \left( \textbf{p} - \dfrac{q_e}{c} \textbf{A} \right),
\end{equation}
then the current density vector can be written as
\begin{equation}
\textbf{J} = q_e \rho \textbf{v},
\end{equation}
with $\rho$ the particle density. In quantum mechanics, for an $N$-particle Hilbert space, the current density vector is written as
\begin{equation}
\textbf{J}_{\textbf{A}} = \dfrac{q_e}{m_e} \sum_{i=1}^N \left( \textbf{p}_i - \dfrac{q_e}{c} \textbf{A}(\textbf{r}_i) \right) \delta(\textbf{r} - \textbf{r}_i) ,
\end{equation}
where we have made explicit the dependence of the current density on the vector potential. We write the current density as the sum of the contribution with zero vector potential and the contribution with nonzero vector potential, that is,
\begin{equation}
\textbf{J}_{\textbf{A}} = \textbf{J}_{\textbf{A}=\textbf{0}} - \dfrac{q_e^2}{m_e c} \textbf{A}(\textbf{r}) \rho(\textbf{r}).
\label{eq: densitadicorrenteinfunzionedelpotenzialevettore}
\end{equation}
We now apply linear response theory to the current density. To this end, we explicitly switch to the formalism of first quantization, rewriting the Hamiltonian as in equation $\eqref{eq: HamiltonianascompostainH_0eH_I}$, where
\begin{equation}
\mathcal{H}_0 = \dfrac{\hat{\textbf{p}}^2}{2 m_e} + V,
\end{equation}
\begin{equation}
\mathcal{H}_I = - \dfrac{q_e^2}{2m_e c^2} \textbf{A} ^2 - \dfrac{q_e}{2m_e c} \left( \hat{\textbf{p}} \cdot \textbf{A} + \textbf{A} \cdot \hat{\textbf{p}} \right).
\end{equation}
Under the assumption of a weak field, we can neglect the term $\textbf{A}^2$; moreover, in general the operators $\textbf{p}$ and $\textbf{A}$ do not commute, but in the Coulomb gauge they do commute, so
\begin{align}
\mathcal{H}_I &= - \dfrac{q_e}{mc} \hat{\textbf{p}} \cdot \textbf{A} = \notag \\
&= - \dfrac{q_e}{c} \textbf{v}_{\textbf{A}=\textbf{0}} \cdot \textbf{A}.
\end{align}
We generalize this interaction term between the particles and the external field to the case of $N$ particles, as follows
\begin{equation}
\mathcal{H}_I^{(0)}(t) = - \dfrac{1}{c} \int d^3 \textbf{r} \textbf{J}_{\textbf{A}=\textbf{0}}(\textbf{r},t) \cdot \textbf{A}(\textbf{r},t) ,
\end{equation}
where we have replaced the current density evaluated in the absence of external perturbation. We note that in linear response theory we must use the Heisenberg representation of the Hamiltonian operator, that is,
\begin{equation}
\mathcal{H}_I^{(0)}(t) = - \dfrac{1}{c} \int d^3 \textbf{r} \textbf{J}^{(0)}_{\textbf{A}=\textbf{0}}(\textbf{r},t) \cdot \textbf{A}(\textbf{r},t) ,
\end{equation}
where we note that since the vector potential is a semiclassical field, the Heisenberg representation does not apply to $\textbf{A}$. Note that the interaction Hamiltonian contains the perturbation, the potential $\textbf{A}$, at the lowest order, consistent with linear response theory. From linear response theory, we have
\begin{equation}
\langle \textbf{J}_{\textbf{A}}(\textbf{r}) \rangle (t) = \langle \textbf{J}_{\textbf{A}}(\textbf{r}) \rangle(0) + \dfrac{i}{\hslash} \int_{-\infty}^{+\infty} dt' \Theta(t-t') \left\langle \left[ - \dfrac{1}{c} \int d^3 \textbf{r}' \textbf{J}^{(0)}_{\textbf{A}=\textbf{0}}(\textbf{r}',t') \cdot \textbf{A}(\textbf{r}',t') , \textbf{J}^{(0)}_{\textbf{A}=\textbf{0}}(\textbf{r},t) \right] \right\rangle.
\label{eq: formulaKubodensitadicorrente1}
\end{equation}
Note that the time evolution in the Heisenberg picture of the current density, that is, the object
\begin{equation}
\textbf{J}^{(0)}_{\textbf{A}}(\textbf{r},t) = \textbf{J}^{(0)}_{\textbf{A}=\textbf{0}}(\textbf{r},t) - \dfrac{q_e^2}{m_e c} \textbf{A}(\textbf{r}) \rho(\textbf{r}) 
\end{equation}
has been replaced in the commutator by $\textbf{J}^{(0)}_{\textbf{A}=\textbf{0}}(\textbf{r},t)$, since the commutator $\left[ \textbf{J}^{(0)}_{\textbf{A}=\textbf{0}}(\textbf{r}',t') \cdot \textbf{A}(\textbf{r}',t') , \textbf{A}(\textbf{r}) \rho(\textbf{r}) \right]$, up to constants, produces operators of second order in $\textbf{A}$, which are neglected here. The thermal average of the current density $\eqref{eq: densitadicorrenteinfunzionedelpotenzialevettore}$ is given by
\begin{align}
\langle \textbf{J}_{\textbf{A}}\rangle (0) &= \langle \textbf{J}_{\textbf{A}=\textbf{0}} \rangle - \left\langle \dfrac{q_e^2}{m_e c} \textbf{A}(\textbf{r},t=0) \rho(\textbf{r}) \right\rangle = \notag \\
&= - \dfrac{q_e^2}{m_e c} \langle \textbf{A}(\textbf{r},t=0) \rho(\textbf{r}) \rangle = \notag \\
&= - \dfrac{q_e^2}{m_e c} \textbf{A}(\textbf{r},t=0) \rho(\textbf{r}) .
\label{eq: mediatermicadensitacorrente}
\end{align}
In the absence of any external perturbation, the current density within the material medium vanishes, and therefore its expectation value is also zero. Let us consider a field of the form
\begin{equation}
\textbf{E}(t) = E_0 e^{- i \omega t} \hat{\textbf{i}},
\end{equation}
that is, a plane wave traveling along the $\hat{\textbf{i}}$ axis. Note that the dependence of $\textbf{E}$ on $\textbf{r}$ has been omitted; we will explain why. We have implicitly assumed $\textbf{q} = \textbf{0}$, then from $q = \frac{2 \pi}{\lambda}$, we are assuming that the wavelength of the incident wave is very large: in solid-state physics, the system excitations are relevant for energies associated with wavelengths on the order of $10^3 \ \angstrom$, whereas the lattice constant in the material medium is on the order of $1 \angstrom$. In other words, the photon momentum plays no role. The vector potential associated with the electric field is given by
\begin{equation}
\textbf{A}(t) = \dfrac{c}{i \omega} \textbf{E}(t) ,
\end{equation}
and since the vector potential also does not depend on position, it is irrotational, and thus $\textbf{B} = \textbf{0}$. In other words, we are studying the case of a transverse electric field perturbing a material medium. We integrate with respect to position and divide both sides of equation $\eqref{eq: formulaKubodensitadicorrente1}$, adapted for a transverse field, by the volume, that is,
\begin{align}
\dfrac{1}{V} \int d^3 \textbf{r} \langle \textbf{J}_{\textbf{A}}(\textbf{r})\rangle (t) &= \dfrac{1}{V} \int d^3 \textbf{r} \langle \textbf{J}_{\textbf{A}}(\textbf{r}) \rangle (0) + \notag \\
&+ \dfrac{i}{\hslash} \dfrac{1}{V} \int d^3 \textbf{r} \int_{-\infty}^{+\infty} dt' \Theta(t-t') \left\langle \left[ - \dfrac{1}{c} \int d^3 \textbf{r}' \textbf{J}^{(0)}_{\textbf{A}=\textbf{0}}(\textbf{r}',t') \cdot \textbf{A}(t') , \textbf{J}^{(0)}_{\textbf{A}=\textbf{0}}(\textbf{r},t) \right] \right\rangle.
\label{eq: formulaKubodensitadicorrente2}
\end{align}
We have
\begin{equation}
\dfrac{1}{V} \int d^3 \textbf{r} \langle \textbf{J}_{\textbf{A}}(\textbf{r}) \rangle (t) = \dfrac{1}{V} \langle \textbf{J}_{\textbf{q}=\textbf{0}} \rangle(t),
\end{equation}
and from $\eqref{eq: mediatermicadensitacorrente}$ it follows
\begin{align}
\dfrac{1}{V} \int d^3 \textbf{r} \langle \textbf{J}_{\textbf{A}}(\textbf{r}) \rangle(0) &= - \dfrac{1}{V} \dfrac{q_e^2}{m c} \int d^3 \textbf{r} \textbf{A}(\textbf{r},t=0) \rho(\textbf{r}) = \notag \\
&= - \dfrac{q_e^2}{i m \omega} \textbf{E}(t) \int d^3 \textbf{r} \dfrac{\rho(\textbf{r})}{V} = \notag \\
&= - \dfrac{q_e^2}{i m \omega} \dfrac{N}{V} \textbf{E}(t) = \notag \\
&= - \dfrac{q_e^2 \rho}{i m \omega} \textbf{E}(t) ,
\end{align}
where we have made explicit that $N = \int d^3 \textbf{r} \rho(\textbf{r})$. Moreover, in the last term, only $\textbf{J}^{(0)}_{\textbf{A}=\textbf{0}}(\textbf{r},t)$ depends on $\textbf{r}$, and only $\textbf{J}^{(0)}_{\textbf{A}=\textbf{0}}(\textbf{r}',t')$ depends on $\textbf{r}'$. Therefore, by performing these integrals, we obtain their respective Fourier transforms in momentum space for $\textbf{q} = \textbf{0}$. Ultimately, equation $\eqref{eq: formulaKubodensitadicorrente2}$ becomes
\begin{align}
\dfrac{1}{V} \langle \textbf{J}_{\textbf{q}=\textbf{0}} \rangle (t) &= - \dfrac{q_e^2 \rho}{i m \omega} \textbf{E}(t) + \dfrac{i}{\hslash} \dfrac{1}{V} \int_{-\infty}^{+\infty} dt' \Theta(t-t') \left( - \dfrac{1}{c} \right) \left\langle \left[ \textbf{J}^{(0)}_{\textbf{q}=\textbf{0},\textbf{A}=\textbf{0}}(t') \cdot \dfrac{c}{i \omega} \textbf{E}(t') , \textbf{J}^{(0)}_{\textbf{q}=\textbf{0},\textbf{A}=\textbf{0}}(t) \right] \right\rangle = \notag \\
&= - \dfrac{q_e^2 \rho}{i m \omega} \textbf{E}(t) - \dfrac{1}{\hslash \omega} \dfrac{1}{V} \int_{-\infty}^{+\infty} dt' \Theta(t-t') \left\langle \left[ \textbf{J}^{(0)}_{\textbf{q}=\textbf{0},\textbf{A}=\textbf{0}}(t') \cdot E_0 e^{- i \omega t'} \hat{\textbf{i}} , \textbf{J}^{(0)}_{\textbf{q}=\textbf{0},\textbf{A}=\textbf{0}}(t) \right] \right\rangle = \notag \\
&= - \dfrac{q_e^2 \rho}{i m \omega} \textbf{E}(t) - \dfrac{1}{\hslash \omega} \dfrac{1}{V} \int_{-\infty}^{+\infty} dt' \Theta(t-t') e^{- i \omega t'} \left\langle \left[ \textbf{J}^{(0)}_{\textbf{q}=\textbf{0},\textbf{A}=\textbf{0}}(t') , \textbf{J}^{(0)}_{\textbf{q}=\textbf{0},\textbf{A}=\textbf{0}}(t) \right] \right\rangle \cdot E_0 \hat{\textbf{i}}.
\label{eq: formulaKubodensitadicorrente3}
\end{align}
We observe that $e^{- i \omega t'} = e^{i \omega (t-t')} e^{- i \omega t}$, which allows us to factor out the electric field and rewrite the expression as
\begin{equation}
\dfrac{1}{V} \langle \textbf{J}_{\textbf{q}=\textbf{0}} \rangle (t) = - \dfrac{q_e^2 \rho}{i m \omega} \textbf{E}(t) - \dfrac{1}{\hslash \omega} \dfrac{1}{V} \int_{-\infty}^{+\infty} dt' \Theta(t-t') e^{i \omega (t-t')} \left\langle \left[ \textbf{J}^{(0)}_{\textbf{q}=\textbf{0},\textbf{A}=\textbf{0}}(t') , \textbf{J}^{(0)}_{\textbf{q}=\textbf{0},\textbf{A}=\textbf{0}}(t) \right] \right\rangle \cdot \textbf{E}(t),
\label{eq: formulaKubodensitadicorrente4}
\end{equation}
after which we invert the commutator,
\begin{equation}
\dfrac{1}{V} \langle \textbf{J}_{\textbf{q}=\textbf{0}} \rangle (t) = - \dfrac{q_e^2 \rho}{i m \omega} \textbf{E}(t) + \dfrac{1}{\hslash \omega} \dfrac{1}{V} \int_{-\infty}^{+\infty} dt' \Theta(t-t') e^{i \omega (t-t')} \left\langle \left[ \textbf{J}^{(0)}_{\textbf{q}=\textbf{0},\textbf{A}=\textbf{0}}(t) , \textbf{J}^{(0)}_{\textbf{q}=\textbf{0},\textbf{A}=\textbf{0}}(t') \right] \right\rangle \cdot \textbf{E}(t),
\label{eq: formulaKubodensitadicorrente4}
\end{equation}
we exploit the condition of thermodynamic equilibrium,
\begin{equation}
\dfrac{1}{V} \langle \textbf{J}_{\textbf{q}=\textbf{0}}\rangle (t) = - \dfrac{q_e^2 \rho}{i m \omega} \textbf{E}(t) + \dfrac{1}{\hslash \omega} \dfrac{1}{V} \int_{-\infty}^{+\infty} dt' \Theta(t-t') e^{i \omega (t-t')} \left\langle \left[ \textbf{J}^{(0)}_{\textbf{q}=\textbf{0},\textbf{A}=\textbf{0}}(t-t') , \textbf{J}^{(0)}_{\textbf{q}=\textbf{0},\textbf{A}=\textbf{0}}(0) \right] \right\rangle \cdot \textbf{E}(t),
\label{eq: formulaKubodensitadicorrente5}
\end{equation}
we rename the dummy variable in the integral as $t - t' \rightarrow t''$,
\begin{equation}
\dfrac{1}{V} \langle \textbf{J}_{\textbf{q}=\textbf{0}} \rangle (t) = - \dfrac{q_e^2 \rho}{i m \omega} \textbf{E}(t) - \dfrac{1}{\hslash \omega} \dfrac{1}{V} \int_{-\infty}^{+\infty} (-dt'') \Theta(t'') e^{i \omega t''} \left\langle \left[ \textbf{J}^{(0)}_{\textbf{q}=\textbf{0},\textbf{A}=\textbf{0}}(t'') , \textbf{J}^{(0)}_{\textbf{q}=\textbf{0},\textbf{A}=\textbf{0}}(0) \right] \right\rangle \cdot \textbf{E}(t),
\label{eq: formulaKubodensitadicorrente6}
\end{equation}
and finally we set $t''=t$, then we obtain
\begin{equation}
\dfrac{1}{V} \langle \textbf{J}_{\textbf{q}=\textbf{0}} \rangle (t) = - \dfrac{q_e^2 \rho}{i m \omega} \textbf{E}(t) + \dfrac{1}{\hslash \omega} \dfrac{1}{V} \int_{-\infty}^{+\infty} dt \Theta(t) e^{i \omega t} \left\langle \left[ \textbf{J}^{(0)}_{\textbf{q}=\textbf{0},\textbf{A}=\textbf{0}}(t) , \textbf{J}^{(0)}_{\textbf{q}=\textbf{0},\textbf{A}=\textbf{0}}(0) \right] \right\rangle \cdot \textbf{E}(t).
\label{eq: formulaKubodensitadicorrente7}
\end{equation}
Ohm's law $\eqref{eq: leggediOhm}$ has been derived in frequency space and for $\textbf{q} = \textbf{0}$, that is,
\begin{equation}
\dfrac{1}{V} \textbf{J}_{\textbf{q}=\textbf{0}}(\omega) = \sigma(\omega) \textbf{E}(\omega),
\end{equation}
where we set
\begin{equation}
\sigma(\omega) = \dfrac{q_e^2 \rho}{i m \omega} + \dfrac{i}{\omega} G_{\textbf{J}}(\textbf{q}=\textbf{0},\omega + i \delta),
\end{equation}
\begin{equation}
G_{\textbf{J}}(\textbf{q}=\textbf{0},\omega + i \delta) = - \dfrac{i}{\hslash} \dfrac{1}{V} \int_{-\infty}^{+\infty} dt \Theta(t) e^{- i (\omega + i \delta) t} \left\langle \left[ \textbf{J}^{(0)}_{\textbf{q}=\textbf{0},\textbf{A}=\textbf{0}}(t) , \textbf{J}^{(0)}_{\textbf{q}=\textbf{0},\textbf{A}=\textbf{0}}(0) \right] \right\rangle.
\end{equation}
Again, the integral is evaluated in the upper complex half-plane, followed by taking the limit $\delta \rightarrow 0$. If the electric field is weak, the optical conductivity is an intrinsic property of a material medium and is calculated as the current density autocorrelation in the absence of external perturbation. Since calculating the optical conductivity is very complex, dedicated models have been developed for insulating and conducting materials, using semiclassical approaches. In particular, in the following, we present two simple models of linear optical response: the Lorentz and Drude models. These provide intuitive insight into the behavior of the conductivity function. A fully microscopic calculation of the optical conductivity is considerably more complex and requires the Mori-Zwanzig formalism, which lies beyond the scope of this text.
\subsection{Lorentz and Drude models of optical conductivity}
The simplest model of an insulator is the Lorentz model: atoms oscillate around their equilibrium positions, and electrons are bound to the atomic nuclei by elastic forces with constant $k_{el}$. Moreover, the electrons move in a medium where there is viscous friction, which according to Stokes' law is directly proportional to the electron’s velocity and the viscous friction coefficient $\gamma$. We also assume the presence of an external electric field of the type introduced earlier. A single electron then satisfies the differential equation
\begin{equation}
m_e \ddot{\textbf{r}} = - k_{el} \textbf{r} - m_e \gamma \dot{\textbf{r}} + q_e E_0 e^{- i \omega t} \hat{\textbf{i}},
\label{eq: equazionedifferenzialeLorentz}
\end{equation}
whose solution is given by the sum of the general integral of the associated homogeneous equation and a particular solution of the full equation. The solution of the associated homogeneous equation is transient, meaning it tends to zero at large times, and since we are interested in stationary solutions, we directly consider a particular solution. Let us assume that this particular solution is of the form
\begin{equation}
\textbf{r}(t) = \textbf{r}(0) e^{- i \omega t},
\end{equation}
that is, it has the same oscillation frequency as the external perturbation, and we substitute this solution into the differential equation. We obtain
\begin{equation}
- m_e \omega^2 \textbf{r}(0) e^{- i \omega t} = - k_{el} \textbf{r}(0) e^{- i \omega t} + i m_e \omega \gamma \textbf{r}(0) e^{- i \omega t} + q_e E_0 e^{- i \omega t} \hat{\textbf{i}} ,
\end{equation}
which implies
\begin{equation}
\textbf{r}(0) = \dfrac{q_e E_0 \hat{\textbf{i}}}{k_{el} - i m_e \omega \gamma - m_e \omega^2}.
\end{equation}
The current density in the Lorentz model is given by
\begin{equation}
\textbf{J} = n q_e (- i \omega) \textbf{r}(0) e^{-i \omega t}
\end{equation}
which, from Ohm's law $\eqref{eq: leggediOhm}$, implies
\begin{equation}
n q_e (- i \omega) \textbf{r}(0) e^{-i \omega t} = \sigma E_0 e^{- i \omega t} \hat{\textbf{i}},
\end{equation}
and by substituting $\textbf{r}(0)$, we have
\begin{equation}
n q_e (- i \omega) \dfrac{q_e E_0 \hat{\textbf{i}}}{k_{el} - i m_e \omega \gamma - m_e \omega^2} = \sigma E_0 \hat{\textbf{i}},
\end{equation}
and
\begin{align}
\sigma(\omega) &= - i n q_e^2 \dfrac{\omega}{k_{el} - i m_e \omega \gamma - m_e \omega^2} = \notag \\
&= - i \dfrac{n q_e^2}{m_e} \dfrac{\omega}{\omega_0^2 - i \omega \gamma - \omega^2} = \notag \\
&= i \dfrac{n q_e^2}{m_e} \dfrac{\omega}{\omega^2 - \omega_0^2 + i \omega \gamma} = \notag \\
&=  i \frac{n q_e^2}{m_e \gamma} \frac{\gamma \frac{\omega}{\omega_0^2}}{\frac{\omega^2}{\omega_0^2} - 1 + i \frac{\omega \gamma}{\omega_0^2}} = \notag \\
&= i \frac{n q_e^2}{m_e \gamma} \frac{\gamma \frac{\omega}{\omega_0^2}}{\left( \frac{\omega^2}{\omega_0^2} - 1 \right)^2 + \frac{\omega^2 \gamma^2}{\omega_0^4}} \left( \frac{\omega^2}{\omega_0^2} - 1 - i \dfrac{\omega \gamma}{\omega_0^2} \right),
\label{eq: conducibilitaotticaLorentz}
\end{align}
where we have defined the electron's natural frequency as
\begin{equation}
\omega_0^2 = \dfrac{k_{el}}{m_e}.
\end{equation}
Let us consider the real part
\begin{equation}
\re \sigma(\omega) = \dfrac{n q_e^2}{m_e \gamma} \frac{\gamma^2 \frac{\omega^2}{\omega_0^4}}{\left[ \frac{\omega^2}{\omega_0^2} - 1 \right]^2 + \frac{\omega^2 \gamma^2}{\omega_0^4}}
\label{eq: parterealeconducibilitaotticaLorentz}
\end{equation}
of the Lorentz optical conductivity. It is zero at $\omega=0$ and also vanishes in the infinite frequency limit. Moreover, it has a maximum at $\omega = \omega_0$, which is given by
\begin{equation}
\re \sigma(\omega_0) = \dfrac{n q_e^2}{m_e \gamma}.
\end{equation}
Let us consider the imaginary part
\begin{equation}
\im \sigma(\omega) = \dfrac{n q_e^2}{m_e \gamma} \frac{\gamma \frac{\omega}{\omega_0^2}}{\left[ \frac{\omega^2}{\omega_0^2} - 1 \right]^2 + \frac{\omega^2 \gamma^2}{\omega_0^4}} \left( \frac{\omega^2}{\omega_0^2} - 1 \right)
\label{eq: parteimmaginariaconducibilitaotticaLorentz}
\end{equation}
of the Lorentz optical conductivity. The imaginary part is also zero at $\omega = 0$. This is a distinctive feature of an insulator: indeed, an electric field with zero angular frequency is a static field and obviously does not produce any current density. Furthermore, the imaginary part is negative for $\omega < \omega_0$, positive for $\omega > \omega_0$, zero at $\omega = \omega_0$, and tends to zero at very high frequencies. \newline
Now, let us consider the case where the material medium is a conductor. The simplest model is the Drude model, which is very similar to the Lorentz model, except that the electrons are free to move. Consequently, the differential equation to study is similar to $\eqref{eq: equazionedifferenzialeLorentz}$, but with $k_{el}=0$. By performing calculations analogous to those in the Lorentz model, we obtain the expression for the optical conductivity in the Drude model
\begin{equation}
\sigma(\omega) = i \dfrac{n q_e^2}{m_e \gamma} \dfrac{\gamma}{\omega^2 + \gamma^2} \left( \omega - i \gamma \right).
\end{equation}
Let us consider the real part
\begin{equation}
\re \sigma(\omega) = \dfrac{n q_e^2}{m_e \gamma} \dfrac{\gamma^2}{\omega^2 + \gamma^2}
\label{eq: parterealeconducibilitaotticaDrude}
\end{equation}
of the Drude optical conductivity. This function is a Lorentzian with parameter $\gamma$; in particular, it is nonzero at $\omega = 0$, which is due to the resistance that the material medium opposes to the moving electrons. This resistance arises from electron-phonon interactions, meaning that the Hamiltonian on which we want to calculate the optical conductivity in a conducting material must include the electron-phonon interaction. A maximum at zero frequency is a distinctive feature of a conducting material. Let us consider the imaginary part
\begin{equation}
\im \sigma(\omega) = \dfrac{n q_e^2}{m_e \gamma} \dfrac{\gamma \omega}{\omega^2 + \gamma^2}
\label{eq: parteimmaginariaconducibilitaotticaDrude}
\end{equation}
of the Drude optical conductivity. It is zero at $\omega = 0$, positive, attains a maximum at $\omega = \gamma$, and then decreases as the frequency increases. \newline
Importantly, in formal terms, the Lorentz model for bound electrons in an insulator is mathematically equivalent to a driven series RLC electrical circuit. In this analogy, the electron mass \( m_e \) plays the role of the inductance \( L \), accounting for the inertial response; the elastic restoring force with constant \( k_{el} \) corresponds to the reciprocal of the capacitance \( C^{-1} \); and the damping due to viscous friction, characterized by the coefficient \( \gamma \), is identified with the resistance \( R \). The external oscillating electric field acts like a time-dependent voltage source applied to the circuit. As a result, the frequency-dependent optical conductivity \( \sigma(\omega) \) mirrors the complex admittance of the circuit. Importantly, just as in an RLC circuit, only the real part of the optical conductivity corresponds to energy dissipation, it reflects the irreversible conversion of electromagnetic energy into heat due to friction. The imaginary part, by contrast, is associated with the reactive (non-dissipative) response of the system, related to energy temporarily stored in the bound motion of the electrons or in the electric and magnetic fields of the circuit. This formal correspondence provides a powerful physical intuition for interpreting optical responses in solids in terms of familiar circuit elements. \newline
On the other hand, the Drude model is formally equivalent to that of a series RL circuit driven by an alternating voltage \(V(t)\). In this analogy, the electron mass \(m_e\) plays the role of the inductance \(L\), representing the electron's inertia which resists changes in current (electron velocity). The damping coefficient \(\gamma\), which quantifies the friction due to scattering, corresponds directly to the resistance \(R\), responsible for dissipating electrical energy into heat. The external electric field \(\textbf{E}(t)\), which drives the electrons, is analogous to the voltage source \(V(t)\) in the circuit, supplying energy to maintain oscillations. The frequency-dependent optical conductivity \(\sigma(\omega)\) of the Drude model corresponds to the complex admittance of the RL circuit, where the real part \(\mathrm{Re}[\sigma(\omega)]\) represents the dissipative conductance that converts electromagnetic energy irreversibly into heat, and the imaginary part \(\mathrm{Im}[\sigma(\omega)]\) represents the reactive susceptance, associated with energy temporarily stored in the electrons' kinetic motion and released back to the field.  This analogy not only provides a clear physical picture of how electrons respond to time-varying electric fields but also offers a useful framework to understand how resistive losses and inertial effects shape the frequency-dependent response of metals, just as resistance and inductance govern the behavior of an RL circuit. \newline
Finally, to better understand the physical significance of the imaginary parts of the Lorentz and Drude optical conductivities, let us consider their phase relationship with the driving electric field. Ohm's law can be rewritten as
\begin{align}
\textbf{J} &= \re \left[ \sigma \right] \textbf{E} + i \im \left[ \sigma \right] \textbf{E} = \notag \\
&= \re \left[ \sigma \right] \textbf{E} + e^{i \frac{\pi}{2} \sgn(\im \sigma)} \left| \im \left[ \sigma \right] \right| \textbf{E} ,
\end{align}
showing that the current density consists of two components: an in-phase part proportional to the real part of the conductivity, and an out-of-phase part proportional to the imaginary part, which leads or lags the electric field by \( \pm \frac{\pi}{2} \). This phase difference is characteristic of reactive behavior, corresponding to energy temporarily stored and released rather than dissipated. The sign of \(\mathrm{Im}[\sigma]\) thus indicates whether the system behaves more inductively or capacitively, which directly connects back to the effective circuit analogies of the Lorentz (RLC) and Drude (RL) models.
\newpage
\section{Figures}
\FloatBarrier
\begin{figure}[H]
\centering
\includegraphics[scale=0.6]{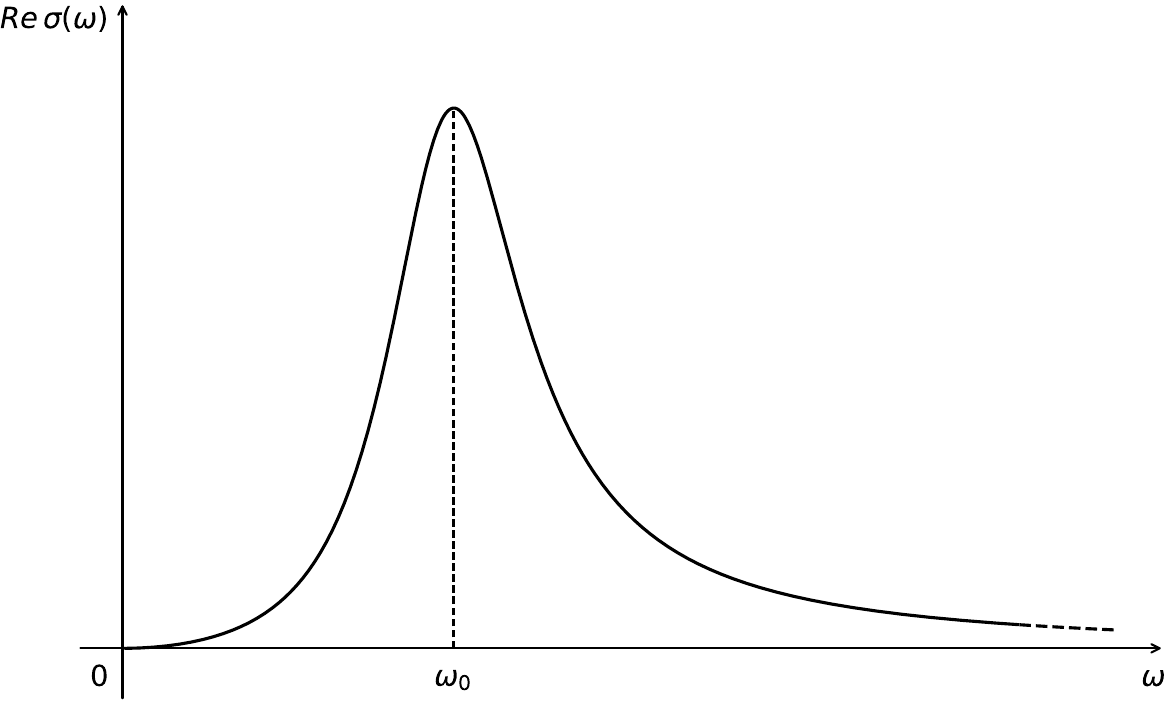}
\caption{Real part of the optical conductivity in the Lorentz model, see equation $\eqref{eq: parterealeconducibilitaotticaLorentz}$.}
\label{fig: sigmareLorentz}
\end{figure}
\begin{figure}[H]
\centering
\includegraphics[scale=0.6]{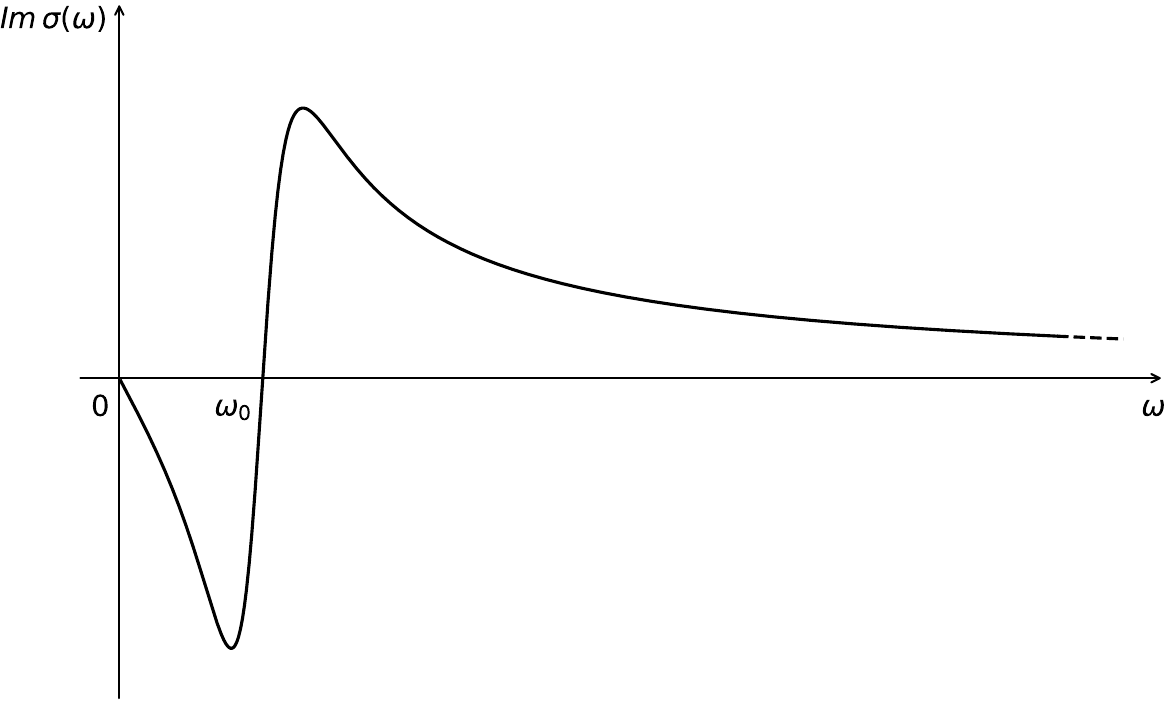}
\caption{Imaginary part of the optical conductivity in the Lorentz model, see equation $\eqref{eq: parteimmaginariaconducibilitaotticaLorentz}$.}
\label{fig: sigmaimLorentz}
\end{figure}
\begin{figure}[H]
\centering
\includegraphics[scale=0.6]{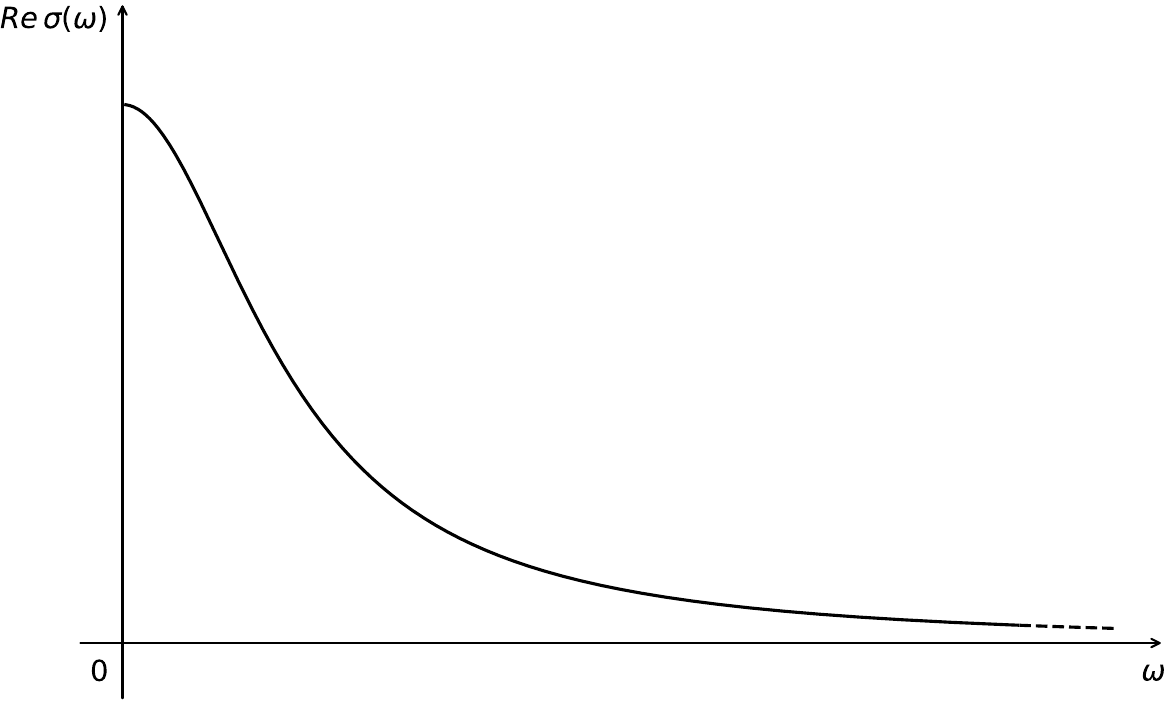}
\caption{Real part of the optical conductivity in the Drude model, see equation $\eqref{eq: parterealeconducibilitaotticaDrude}$.}
\label{fig: sigmareDrude}
\end{figure}
\begin{figure}[H]
\centering
\includegraphics[scale=0.6]{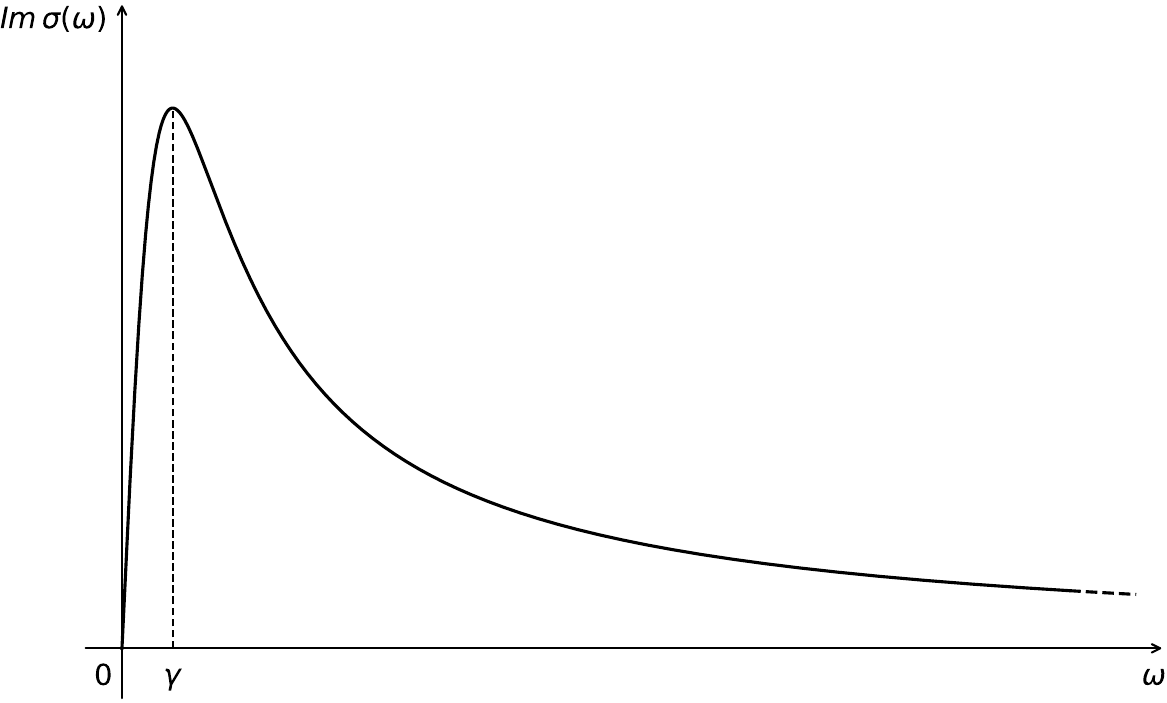}
\caption{Imaginary part of the optical conductivity in the Drude model, see equation $\eqref{eq: parteimmaginariaconducibilitaotticaDrude}$.}
\label{fig: sigmaimDrude}
\end{figure}

\part{Further Bosonic Quantum Fields}

\chapter{Bose-Einstein condensation of non-interacting bosons}
Bose-Einstein condensation (BEC) is a macroscopic quantum phase transition that occurs when a dilute gas of indistinguishable bosons is cooled to temperatures near absolute zero, leading to the macroscopic occupation of the ground state and the emergence of a coherent quantum phase. Originally predicted by S.N. Bose and A. Einstein in the 1920s and experimentally realized in 1995 in ultracold atomic gases, BEC represents a paradigmatic example of quantum statistics manifesting at the macroscopic scale. \newline
From a classical viewpoint, the behavior of dilute gases is described by Maxwell-Boltzmann statistics, in which particles are treated as distinguishable and can independently populate available energy levels without constraints. This framework, however, fails to capture the collective behavior of quantum particles at low temperatures, particularly the indistinguishability and wavefunction symmetry inherent to bosons. \newline
Quantum mechanically, bosons follow Bose-Einstein statistics, which allow multiple particles to occupy the same quantum state. Below a critical temperature, a non-negligible fraction of the particles condenses into the lowest energy state, forming a Bose-Einstein condensate. This phase is characterized by long-range coherence and quantum degeneracy, with no classical analog. \newline
The onset of condensation can be quantitatively understood through the occupation number distribution: above the critical temperature, particles populate a wide range of energy states, while below it, the ground state occupation diverges with system size. Theoretical treatments involve both thermodynamic approaches, employing the grand canonical ensemble, chemical potential, and density of states, and quantum field-theoretic formulations, often in the framework of second quantization, where the bosonic field operators exhibit macroscopic eigenvalues. \newline
BECs play a central role in various domains of modern physics, including quantum gases, superfluidity, quantum simulation, and precision metrology. Their high degree of tunability and coherence make them ideal platforms for probing fundamental aspects of quantum many-body physics, symmetry breaking, and low-temperature phase transitions, as well as for engineering quantum technologies. \newline
Throughout this chapter, we consider a system of non-interacting bosons confined in a cubic box with periodic boundary conditions. The system is consistently described by the Hamiltonian given in equation $\eqref{eq: Hamiltonianasecondaquantizzazioneparticellelibere}$. In this framework, the single-particle energies are shifted by the chemical potential \( \mu \), and we define \( \mathcal{E}'_i = \mathcal{E}_i - \mu \), where \( \mathcal{E}_i = \frac{\hslash^2 \mathbf{k}_i^2}{2m} \). 
\section{Qualitative study at the low and high temperature limit of non-interacting bosons}
In this section, we analyze the behavior of a gas of non-interacting bosons in the limiting regimes of low and high temperatures. Given a system of non-interacting bosons, let us consider the average number \( \langle \hat{N}_i \rangle \) of bosons in a given state \( i \), i.e.,
\begin{align}
\langle \hat{N}_i \rangle &= n_{1}\left( \mathcal{E}_i - \mu \right) = \notag \\
&= \dfrac{1}{e^{\beta \mathcal{E}_i} e^{- \beta \mu} - 1} = \notag \\
&= \dfrac{e^{\beta \mu} e^{- \beta \mathcal{E}_i}}{1 - e^{\beta \mu} e^{- \beta \mathcal{E}_i}},
\end{align}
where $n_{1}\left( \mathcal{E}_i - \mu \right)$ is the Bose-Einstein statistics and we used $\eqref{eq: mediatermicanumerodioccupazioneedistribuzioneFermiBose}$. In addition, $\langle \hat{N}_i \rangle$ is nonnegative, then it follows
\begin{equation}
e^{\beta \mathcal{E}_i} e^{- \beta \mu} - 1 \geq 0,
\end{equation}
\begin{equation}
\mu \leq \mathcal{E}_i.
\end{equation}
Without loss of generality, we set the ground-state energy to zero, \( \mathcal{E}_0 = 0 \), so that \( \mathcal{E}_i \geq 0 \) for all \( i \). \newline
As \( T \to 0^+ \), consider any excited state \( i \) with \( \mathcal{E}_i > 0 \). Then,
\begin{equation}
\left. \langle \hat{N}_i \rangle \right|_{T=0} = 0, \ \forall \ i \neq 0.
\end{equation}
The total average number of particles $\langle \hat{N} \rangle$ is conserved, then it is equal to the average number of bosons in the fundamental state, i.e., $\langle \hat{N}_0 \rangle$, and we have
\begin{align}
\langle \hat{N} \rangle &= \langle \hat{N}_0 \rangle = \notag \\
&= \dfrac{e^{\beta \mu} e^{- \beta \mathcal{E}_0}}{1 - e^{\beta \mu} e^{- \beta \mathcal{E}_0}} = \notag \\
&= \dfrac{e^{\beta \mu}}{1 - e^{\beta \mu}} = \notag \\
&= \dfrac{\alpha}{1 - \alpha} ,
\end{align}
where we defined the fugacity as
\begin{equation}
\alpha = e^{\beta \mu}.
\end{equation}
From the fugacity we get $\mu$ as follows
\begin{equation}
\langle \hat{N} \rangle - \alpha \langle \hat{N} \rangle = \alpha,
\end{equation}
\begin{equation}
\alpha \left( \langle \hat{N} \rangle + 1 \right) = \langle \hat{N} \rangle,
\end{equation}
\begin{equation}
\alpha = \dfrac{\langle \hat{N} \rangle}{\langle \hat{N} \rangle+1} ,
\end{equation}
\begin{equation}
\beta \mu = \ln \dfrac{\langle \hat{N} \rangle}{\langle \hat{N} \rangle+1},
\end{equation}
\begin{equation}
\mu = k_B T \ln \dfrac{\langle \hat{N} \rangle}{\langle \hat{N} \rangle+1}.
\end{equation}
At the thermodynamic limit, the chemical potential tends to $0$, but since it is defined as non-positive, at most null, we deduce that
\begin{equation}
\lim_{T \rightarrow 0^+} \lim_{\langle \hat{N} \rangle \rightarrow \infty} \mu = 0.
\end{equation}
This reflects the macroscopic occupation of the ground state at low temperature, i.e., the onset of BEC. \newline
In the opposite regime \( T \to \infty \), the Bose-Einstein distribution approaches the Maxwell-Boltzmann distribution. Given that \( \beta \to 0^+ \), we require
\begin{equation}
e^{\beta (\mathcal{E}_i - \mu)} \gg 1,
\end{equation}
which implies
\begin{equation}
\beta \mu \to -\infty.
\end{equation}
In this limit, the average occupation number becomes
\begin{align}
\left. \langle \hat{N}_i \rangle \right|_{T \to +\infty} &\sim e^{\beta \mu} e^{- \beta \mathcal{E}_i} \sim \notag \\
&\sim e^{\beta \mu},
\end{align}
since
\begin{equation}
\lim_{\beta \to 0^+} e^{- \beta \mathcal{E}_i} = 1.
\end{equation}
This shows that at infinite temperature, the particles are evenly distributed over the available excited states, and quantum statistical effects are suppressed.
\section{Quantitative study at the low and high temperature limit of non-interacting bosons}
\subsection{Average number of bosons in a box}
\begin{figure}[H]
\centering
\includegraphics[scale=1]{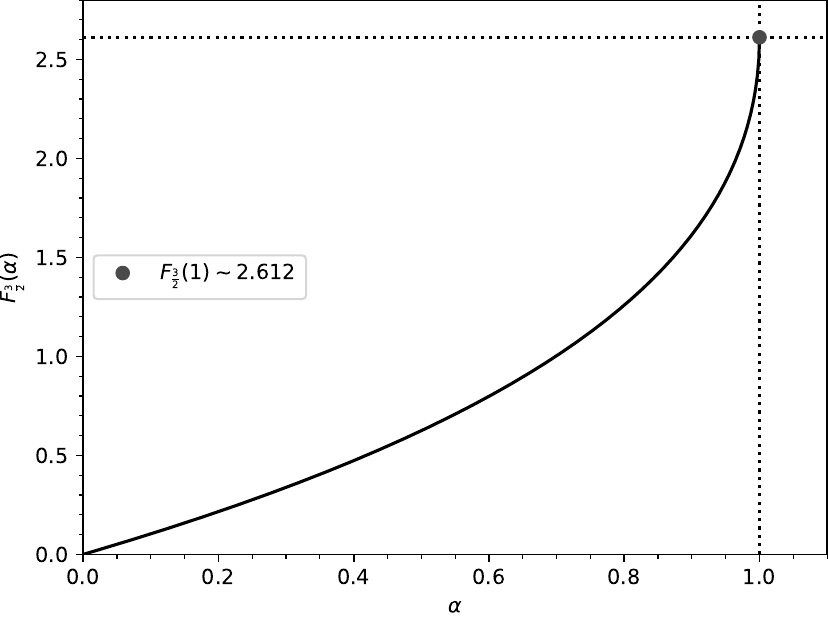}
\caption{The polylogarithmic function $F_{\frac{3}{2}}$ as a function of the fugacity $\alpha = e^{\beta\mu}$ in the range $[0,1]$, including its maximum. This function arises in the expression for the average number of non-interacting bosons, written in terms of the thermal de Broglie wavelength, i.e., equation $\eqref{eq: numeromediobosoninoninteragentisbagliatoinfunzionedilunghezzadondadebroglietermica}$.}
\label{fig: polylogaritmicfunction}
\end{figure}

Previously, we qualitatively discussed the behavior of a system of non-interacting bosons in the low-temperature (quantum) and high-temperature (classical) regimes. We now aim to quantitatively analyze these results. To begin, consider the total average number of non-interacting bosons
\begin{align}
\langle \hat{N} \rangle &= \sum_i \dfrac{e^{\beta \mu} e^{- \beta \mathcal{E}_i}}{1-e^{\beta \mu} e^{- \beta \mathcal{E}_i}} = \notag \\
&= \sum_i \dfrac{\alpha e^{- \beta \mathcal{E}_i}}{1 - \alpha e^{- \beta \mathcal{E}_i}}.
\end{align}
From $\mu \leq 0$, we have $\alpha \leq 1$. Moreover, if $\mathcal{E}_i > 0$, then $e^{- \beta \mathcal{E}_i} < 1$, so in particular $\alpha e^{- \beta \mathcal{E}_i} < 1$. The function
\begin{equation}
\dfrac{1}{1 - \alpha e^{- \beta \mathcal{E}_i}}
\end{equation}
can be seen as the sum of the geometric series
\begin{equation}
\sum_{j=0}^N \left( \alpha e^{- \beta \mathcal{E}_i} \right)^j ,
\end{equation}
and the average number of particles is written as
\begin{align}
\langle \hat{N} \rangle &= \sum_i \sum_{j=0}^{\infty} \alpha e^{- \beta \mathcal{E}_i} \left( \alpha e^{- \beta \mathcal{E}_i} \right)^j = \notag \\
&= \sum_i \sum_{j=0}^{\infty} \alpha^{j+1} e^{- (j+1) \beta \mathcal{E}_i} = \notag \\
&= \sum_i \sum_{j=1}^{\infty} \alpha^j e^{- j \beta \mathcal{E}_i}.
\end{align}
Since the eigenvalues of the bosonic Hamiltonian in a box are independent of the spin degree of freedom, a spin degeneracy factor $g$ arises, and we get
\begin{equation}
\langle N \rangle = g \sum_{\textbf{k}} \sum_{j=1}^{\infty} \alpha^j e^{- j \beta \mathcal{E}_{\textbf{k}}}.
\end{equation}
Recall that the chemical potential is contained in the definition of the fugacity \( \alpha \). Now, we aim to rewrite the sum over momentum states as an integral over energy, introducing the density of states \( \rho(\mathcal{E}) \). Accordingly, the sum over \( \textbf{k} \) is replaced by an integral with respect to the measure \( d\mathcal{E} \), namely
\begin{equation}
\langle \hat{N} \rangle = g \sum_{j=1}^{\infty} \int d\mathcal{E} \rho(\mathcal{E}) \alpha^j e^{- j \beta \mathcal{E}} ,
\end{equation} 
where the density of states must be defined as follows
\begin{equation}
\rho(\mathcal{E}) = \sum_{\textbf{k}} \delta(\mathcal{E}-\mathcal{E}_{\textbf{k}}).
\end{equation}
In the thermodynamic limit approximation $\eqref{eq: illimitetermodinamico}$, the density of states $\rho(\mathcal{E}) = \sum_{\textbf{k}} \delta(\mathcal{E}-\mathcal{E}_{\textbf{k}})$ can be rewritten in integral form as
\begin{align}
\rho(\mathcal{E}) &= \dfrac{V}{(2 \pi)^3} \int d^3\textbf{k} \delta(\mathcal{E}-\mathcal{E}_{\textbf{k}}) = \notag \\
&= \dfrac{V}{(2 \pi)^3} \int d^3\textbf{k} \delta\left(\mathcal{E}-\dfrac{\hslash^2 \textbf{k}^2}{2m}\right) ,
\end{align}
which in spherical polar coordinates becomes
\begin{align}
\rho(\mathcal{E}) &= \dfrac{V}{(2 \pi)^3} 4 \pi \int_0^\infty dk k^2 \delta\left(\mathcal{E}-\dfrac{\hslash^2 k^2}{2m}\right) = \notag \\
&= \dfrac{V}{2 \pi^2} \int_0^\infty dk k^2 \delta\left(\mathcal{E}-\dfrac{\hslash^2 k^2}{2m}\right).
\end{align}
Now, given a decomposition
\begin{equation}
\delta(f(x)) = \sum_i \dfrac{\delta(x-x_i)}{\left| f'(x_i) \right|},
\end{equation}
where $x_i$ is a zero of $f$ but not of $f'$, we recognize the identifications
\begin{align}
f(x_i) & \longrightarrow \mathcal{E}(k_i), \\
x_i & \longrightarrow k_{\pm} = \pm \sqrt{\dfrac{2 m \mathcal{E}}{\hslash^2}} , \\
\left| f'(x) \right| & \longrightarrow \left| \dfrac{d \mathcal{E}}{dk} \right| = \dfrac{\hslash^2 k}{m} , \\
\left| f'(x_i) \right| & \longrightarrow \left| \dfrac{d \mathcal{E}}{dk}(k_i) \right| = \dfrac{\hslash^2}{m} \sqrt{\dfrac{2 m \mathcal{E}}{\hslash^2}} .
\end{align}
Since the integration is with respect to the positive semi-axis, we are only interested in $k_+$ and the density of states transforms as follows
\begin{align}
\rho(\mathcal{E}) &= \dfrac{V}{2 \pi^2} \int_0^\infty dk k^2 \frac{\delta \left( k-\sqrt{\frac{2 m \mathcal{E}}{\hslash^2}} \right)}{\frac{\hslash^2}{m} \sqrt{\frac{2 m \mathcal{E}}{\hslash^2}}} = \notag \\
&= \dfrac{V}{2 \pi^2} \dfrac{2 m \mathcal{E}}{\hslash^2} \dfrac{1}{\frac{\hslash^2}{m} \sqrt{\frac{2 m \mathcal{E}}{\hslash^2}}} = \notag \\
&= \dfrac{V}{2 \pi^2} \sqrt{\dfrac{2 m \mathcal{E}}{\hslash^2}} \dfrac{m}{\hslash^2}.
\label{eq: densitàstatibosoninoninteragenti}
\end{align} 
The average number of bosons becomes
\begin{equation}
\langle \hat{N} \rangle = g \sum_{j=1}^\infty \alpha^j \int_0^\infty d\mathcal{E} \dfrac{V}{2 \pi^2} \sqrt{\dfrac{2 m \mathcal{E}}{\hslash^2}} \dfrac{m}{\hslash^2} e^{- j \beta \mathcal{E}}.
\end{equation}
We set $x = j \beta \mathcal{E}$, then
\begin{equation}
\langle \hat{N} \rangle = g \sum_{j=1}^\infty \alpha^j \int_0^\infty \dfrac{dx}{j \beta} \dfrac{V}{2 \pi^2} \sqrt{\dfrac{2 m \mathcal{E}}{\hslash^2}} \dfrac{m}{\hslash^2} e^{-x} ,
\end{equation}
and multiply and divide by $j \beta$ as follows
\begin{align}
\langle \hat{N} \rangle &= g \sum_{j=1}^\infty \alpha^j \int_0^\infty \dfrac{dx}{(j \beta)^{\frac{3}{2}}} \dfrac{V}{2 \pi^2} \sqrt{\dfrac{2 m j \beta \mathcal{E}}{\hslash^2}} \dfrac{m}{\hslash^2} e^{-x} = \notag \\
&= g \sum_{j=1}^\infty \dfrac{\alpha^j}{j^{\frac{3}{2}}} \int_0^\infty dx (k_B T)^{\frac{3}{2}} \dfrac{V}{2 \pi^2} \sqrt{\dfrac{2 m}{\hslash^2}} \dfrac{m}{\hslash^2} \sqrt{x} e^{-x} = \notag \\
&= g \sum_{j=1}^\infty \dfrac{\alpha^j}{j^{\frac{3}{2}}} \left( \dfrac{V}{\sqrt{2}} \dfrac{1}{\pi^2} \left( \dfrac{m k_B T}{\hslash^2} \right)^{\frac{3}{2}} \right) \left[ \int_0^\infty dx \sqrt{x} e^{-x} \right].
\end{align}
Clearly, $\langle \hat{N} \rangle$ is a dimensionless quantity, all physical dimensions are instead contained in $V \left( \frac{m k_B T}{\hslash^2} \right)^{\frac{3}{2}}$, then
\begin{equation}
\left[ \dfrac{\hslash^2}{m k_B T} \right] = [\text{length}]^2 ,
\end{equation}
and we recall de Broglie thermal wavelength
\begin{align}
\lambda_T &\equiv \sqrt{\dfrac{2 \pi \hslash^2}{m k_B T}} \equiv \notag \\
&\equiv \sqrt{\dfrac{h^2}{2 \pi m k_B T}} ,
\end{align}
finally we multiply and divide $\langle \hat{N} \rangle$ by $(2 \pi)^{\frac{3}{2}}$ and obtain
\begin{align}
\langle \hat{N} \rangle &= g \sum_{j=1}^\infty \dfrac{\alpha^j}{j^{\frac{3}{2}}} \left( \dfrac{V}{\sqrt{2}} \dfrac{1}{\pi^2} (2 \pi)^{\frac{3}{2}} \left( \dfrac{m k_B T}{2 \pi \hslash^2} \right)^{\frac{3}{2}} \right) \left[ \int_0^\infty dx \sqrt{x} e^{-x} \right] = \notag \\
&= g \sum_{j=1}^\infty \dfrac{\alpha^j}{j^{\frac{3}{2}}} \dfrac{V}{\lambda_T^3} \dfrac{2}{\sqrt{\pi}} \left[ \int_0^\infty dx \sqrt{x} e^{-x} \right] = \notag \\
&= g  \sum_{j=1}^\infty \dfrac{\alpha^j}{j^{\frac{3}{2}}} \dfrac{V}{\lambda_T^3} \dfrac{2}{\sqrt{\pi}} \Gamma\left(\dfrac{3}{2}\right),
\end{align}
where we recognize the Gamma function (see Chapter $\ref{Euler gamma function}$) evalueted at $\frac{3}{2}$. In particular, from
\begin{align}
\Gamma\left(\dfrac{3}{2}\right) &= \int_0^\infty dx \sqrt{x} e^{-x} = \notag \\
&= \dfrac{\sqrt{\pi}}{2} ,
\end{align}
we get
\begin{equation}
\langle \hat{N} \rangle = g \sum_{j=1}^\infty \dfrac{\alpha^j}{j^{\frac{3}{2}}} \dfrac{V}{\lambda_T^3}.
\label{eq: numeromediobosoninoninteragentisbagliatoinfunzionedilunghezzadondadebroglietermica}
\end{equation}
The function
\begin{equation}
F_{\frac{3}{2}}(\alpha) = \sum_{j=1}^\infty \dfrac{\alpha^j}{j^{\frac{3}{2}}}, \ \alpha \in [0,1]
\end{equation}
is known in the literature as polylogarithmic function, see Figure $\eqref{fig: polylogaritmicfunction}$, and it is monotonically increasing with respect to fugacity, reaching its maximum ($2.6$ approximately) at $\alpha=1$. Here, we are interested in emphasizing that for any value of the chemical potential $\mu$ the polylogarithmic function is a finite quantity, so the average number of bosons is a power law of the form
\begin{equation}
\langle \hat{N} \rangle \propto \dfrac{V}{\lambda_T^3}.
\end{equation}
Indeed, at the low temperature limit, the thermal de Broglie wavelength becomes large, i.e., 
\begin{equation}
\dfrac{1}{\lambda_T^3} \sim T^{\frac{3}{2}} .
\end{equation}
That is, below a certain temperature there is no longer any value of \(\alpha\) for which equation $\eqref{eq: numeromediobosoninoninteragentisbagliatoinfunzionedilunghezzadondadebroglietermica}$ can be satisfied. This would appear to imply a violation of particle-number conservation. The issue lies in having replaced the sum over \(\textbf{k}\) with an integral. This leads to
\begin{equation}
\rho(\mathcal{E}) \propto \sqrt{\mathcal{E}} ,
\end{equation}
where the weight associated with \(\mathcal{E}_0\) is seen to vanish; however, we know that the single-particle ground state is the only one that is macroscopically occupied at \(T = 0\). Equation $\eqref{eq: numeromediobosoninoninteragentisbagliatoinfunzionedilunghezzadondadebroglietermica}$ describes the population of excited states, so the replacement of the series with the integral in the calculation of $\rho(\mathcal{E})$ is allowed for $\mathcal{E}>0$. The $\eqref{eq: numeromediobosoninoninteragentisbagliatoinfunzionedilunghezzadondadebroglietermica}$ must be corrected as
\begin{align}
\langle \hat{N}_e \rangle &= \sum_{i \neq 0} \langle \hat{N}_i \rangle = \notag \\
&= g F_{\frac{3}{2}}(\alpha) \dfrac{V}{\lambda_T^3}, \ T \geq T_C.
\label{eq: numeromediobosoninoninteragentistatieccitatiinfunzionedilunghezzadondadebroglietermica}
\end{align}
with
\begin{equation}
\langle \hat{N} \rangle = \langle \hat{N}_0 \rangle + \langle \hat{N}_e \rangle.
\label{eq: NtotaleugualeN0Neccitati}
\end{equation}
The temperature $T_C$ is the lowest temperature at which equation $\eqref{eq: numeromediobosoninoninteragentisbagliatoinfunzionedilunghezzadondadebroglietermica}$ can be satisfied without introducing the correction factor $\left\langle \hat{N}_0 \right\rangle$. At this critical temperature, $\alpha$ attains its maximum value; that is, $\mu = 0$. Assuming that the chemical potential \( \mu \) remains constant and strictly negative, let us compare two temperatures \( T \) and \( T' \), with \( T' < T \) (hence \( \beta' > \beta \)), we have $\beta' \mu < \beta \mu$, then
\begin{equation}
\alpha' = e^{\beta' \mu} < e^{\beta \mu} = \alpha,
\end{equation}
\begin{equation}
F_{\frac{3}{2}}(\alpha') < F_{\frac{3}{2}}(\alpha) .
\end{equation}
Together with $\lambda_{T'} > \lambda_{T}$, the equation $\eqref{eq: numeromediobosoninoninteragentisbagliatoinfunzionedilunghezzadondadebroglietermica}$ provides
\begin{equation}
\langle \hat{N}' \rangle = g F_{\frac{3}{2}}(\alpha') \dfrac{V}{\lambda_{T'}^3} \ < \ g F_{\frac{3}{2}}(\alpha) \dfrac{V}{\lambda_T^3} = \langle \hat{N} \rangle,
\end{equation}
which implies that the occupation numbers of excited states increase as the temperature decreases. We manipulate the expression of the average number of bosons to bring out the dependence on the critical temperature. For $T=T_C$, the $\eqref{eq: numeromediobosoninoninteragentistatieccitatiinfunzionedilunghezzadondadebroglietermica}$ can be written as
\begin{equation}
\dfrac{\langle \hat{N} \rangle \lambda_{T_C}^3}{g V} = F_{\frac{3}{2}}(1) .
\label{eq: lunghezzadondadideBroglietermicaallatemperaturacritica}
\end{equation} 
A comparison of equations $\eqref{eq: NtotaleugualeN0Neccitati}$ and $\eqref{eq: lunghezzadondadideBroglietermicaallatemperaturacritica}$ yields
\begin{equation}
\langle \hat{N} \rangle - \langle \hat{N}_0 \rangle = \left( \dfrac{T}{T_C} \right)^{\frac{3}{2}} \langle \hat{N} \rangle ,
\end{equation}
\begin{equation}
\langle \hat{N}_0 \rangle = \left[ 1 - \left( \dfrac{T}{T_C} \right)^{\frac{3}{2}} \right] \langle \hat{N} \rangle, 
\end{equation}
which is the average number of bosons in the fundamental state as a function of temperature. Obviously, the number of bosons in the fundamental state increases as the temperature decreases, and BEC is complete when $T=0$. Finally, if we write $\eqref{eq: lunghezzadondadideBroglietermicaallatemperaturacritica}$ as
\begin{equation}
\dfrac{\langle \hat{N} \rangle \lambda_{T_C}^3}{V} = g F_{\frac{3}{2}}(1) ,
\end{equation}
that is, we note that BEC becomes observable when the De Broglie thermal wavelength is comparable to the average spatial distance between bosons.
\subsection{Average energy}
For convenience, and only within a part of this section, we simplify the notation by writing \( \mathcal{E}'_i = \mathcal{E}_i - \mu \). In the grancanonical ensemble formalism, the related gran partition function $Z$ must be computed as a trace of states of Fock space as follows
\begin{align}
Z &= \sum_{N=0}^\infty \sum_{\text{permanents}} \left\langle N_0,\ldots,N_\infty \left| e^{- \beta \hat{\mathcal{H}}} \right| N_0,\ldots,N_\infty \right\rangle,
\end{align}
i.e., it is a sum over all subspaces with fixed number $N$ of particles, and in each of these spaces one must sum over all possible states, which we write in the occupation-number basis $\left| N_0,\ldots,N_\infty \right\rangle$. Since the action of the Hamiltonian on that state returns the eigenvalues $\sum_{\textbf{k},\sigma} \mathcal{E}'_{\textbf{k}} N_{\textbf{k},\sigma}$, and the sum over permanents, for a fixed number $N$ of particles are equal to the sums over the occupation numbers, we get
\begin{align}
Z &= \sum_{N=0}^\infty \sum_{\text{permanents}} e^{- \beta \sum_{\textbf{k},\sigma} \mathcal{E}'_{\textbf{k}} N_{\textbf{k},\sigma}} = \notag \\
&= \sum_{N_1=0}^\infty \sum_{N_2=0}^\infty \ldots \sum_{N_\infty = 0}^\infty e^{- \beta \sum_{\textbf{k},\sigma} \mathcal{E}'_{\textbf{k}} N_{\textbf{k},\sigma}} = \notag \\
&= \prod_{\textbf{k},\sigma} \sum_{N=0}^\infty e^{- N \beta \mathcal{E}'_{\textbf{k}}}.
\end{align}
Since $\mathcal{E}'_{\textbf{k}} > 0, \ \forall \ \textbf{k}$, we have $e^{- N \beta \mathcal{E}'_{\textbf{k}}} < 1$, then
\begin{equation}
\sum_{N=0}^\infty e^{- N \beta \mathcal{E}'_{\textbf{k}}} = \dfrac{1}{1 - e^{- \beta \mathcal{E}'_{\textbf{k}}}} ,
\end{equation}
\begin{equation}
Z = \prod_{\textbf{k},\sigma} \dfrac{1}{1 - e^{- \beta \mathcal{E}'_{\textbf{k}}}}. 
\end{equation}
From the statistical mechanics, the grand potential $\Omega(V,T,\mu)$ is connected to the grand partition function by means of $\Omega = - k_B T \ln Z$, then the grand potential of a set of non-interacting bosons is given by
\begin{align}
\Omega &= - k_B T \ln \prod_{\textbf{k},\sigma} \dfrac{1}{1 - e^{- \beta \mathcal{E}'_{\textbf{k}}}} = \notag \\
&= - k_B T \sum_{\textbf{k},\sigma} \ln \dfrac{1}{1 - e^{- \beta \mathcal{E}'_{\textbf{k}}}} = \notag \\
&= - g k_B T \sum_{\textbf{k}} \ln \dfrac{1}{1 - e^{- \beta \mathcal{E}'_{\textbf{k}}}} = \notag \\
&= g k_B T \sum_{\textbf{k}} \ln \left( 1 - e^{- \beta \mathcal{E}'_{\textbf{k}}} \right) = \notag \\
&= g k_B T \ln \left( 1 - e^{\beta \mu} \right) + g k_B T \sum_{\textbf{k} \neq \textbf{0}} \ln \left( 1 - e^{- \beta \mathcal{E}'_{\textbf{k}}} \right) ,
\end{align}
where in the last step we separated the ground state from the excited states. In the thermodynamic limit, the terms of excited states can be written in a integral form by means of the density of states $\eqref{eq: densitàstatibosoninoninteragenti}$ as follows
\begin{align}
g k_B T \sum_{\textbf{k} \neq \textbf{0}} \ln \left( 1 - e^{- \beta \mathcal{E}'_{\textbf{k}}} \right) & \simeq g k_B T \int_0^\infty d\mathcal{E} \rho(\mathcal{E}) \ln \left( 1 - e^{- \beta (\mathcal{E} - \mu)} \right) = \notag \\
&= g k_B T \dfrac{V}{2 \pi^2} \sqrt{\dfrac{2m}{\hslash^2}} \dfrac{m}{\hslash^2} \int_0^\infty d\mathcal{E} \sqrt{\mathcal{E}} \ln \left( 1 - e^{- \beta (\mathcal{E} - \mu)} \right),
\end{align}
where we made explicit $\mathcal{E}'_{\textbf{k}} = \mathcal{E}_{\textbf{k}} - \mu$, and the discrete variable \(\mathcal{E}_{\textbf{k}}\) is replaced by the continuous measure \(d\mathcal{E}\). For simplicity, we omit the numerical constants and perform integration by parts, i.e.,
\begin{align}
\dfrac{2}{3} \int_0^\infty d\mathcal{E} \dfrac{3}{2} \sqrt{\mathcal{E}} \ln \left( 1 - e^{- \beta (\mathcal{E} - \mu)} \right) &= \dfrac{2}{3} \left[ \mathcal{E}^{\frac{3}{2}} \ln \left( 1 - e^{- \beta (\mathcal{E} - \mu)} \right)  \right]_0^\infty - \dfrac{2}{3} \int_0^\infty d\mathcal{E} \dfrac{\mathcal{E}^{\frac{3}{2}} (- e^{-\beta \mathcal{E}} e^{\beta \mu}) (- \beta)}{1 - e^{- \beta \mathcal{E}} e^{\beta \mu}} = \notag \\
&= - \dfrac{2}{3} \beta \int_0^\infty d\mathcal{E} \dfrac{\mathcal{E}^{\frac{3}{2}} e^{-\beta \mathcal{E}} e^{\beta \mu}}{1 - e^{- \beta \mathcal{E}} e^{\beta \mu}} = \notag \\
&= - \dfrac{2}{3} \beta \int_0^\infty d\mathcal{E} \dfrac{\mathcal{E}^{\frac{3}{2}}}{e^{\beta (\mathcal{E} - \mu)} - 1}.
\end{align}
Now we insert the numerical constants and multiply and divide by $2$
\begin{equation}
- g k_B T \dfrac{V}{2 \pi^2} \sqrt{\dfrac{2m}{\hslash^2}} \dfrac{m}{\hslash^2} \dfrac{2}{3} \beta \int_0^\infty d\mathcal{E} \dfrac{\mathcal{E}^{\frac{3}{2}}}{e^{\beta (\mathcal{E} - \mu)} -1} = - \dfrac{2}{3} g \dfrac{V}{(2 \pi)^2} \left( \dfrac{2m}{\hslash^2} \right)^{\frac{3}{2}} \int_0^\infty d\mathcal{E} \dfrac{\mathcal{E}^{\frac{3}{2}}}{e^{\beta (\mathcal{E} - \mu)} - 1},
\end{equation}
then the grand potential of a system of non-interacting bosons is given by
\begin{equation}
\Omega = g k_B T \ln \left( 1 - e^{\beta \mu} \right) - \dfrac{2}{3} g \dfrac{V}{(2 \pi)^2} \left( \dfrac{2m}{\hslash^2} \right)^{\frac{3}{2}} \int_0^\infty d\mathcal{E} \dfrac{\mathcal{E}^{\frac{3}{2}}}{e^{\beta (\mathcal{E} - \mu)} - 1}.
\end{equation}
The pressure of the system is defined as
\begin{align}
p &= - \left. \dfrac{\partial \Omega}{\partial V} \right|_{T,\mu} \ = \notag \\
&= \dfrac{2}{3} g \dfrac{1}{(2 \pi)^2} \left( \dfrac{2m}{\hslash^2} \right)^{\frac{3}{2}} \int_0^\infty d\mathcal{E} \dfrac{\mathcal{E}^{\frac{3}{2}}}{e^{\beta (\mathcal{E} - \mu)} - 1},
\end{align}
which does not depend on the condensate term. Now we compute the average value of the internal energy $U$, which is related to the grand partition function as
\begin{equation}
U = - \dfrac{\partial}{\partial \beta} \ln Z \mid_{\beta \mu},
\end{equation}
with 
\begin{equation}
\ln Z = - g \ln \left( 1 - e^{\beta \mu} \right) + g \sum_{\textbf{k} \neq \textbf{0}} \ln \left( 1 - e^{- \beta \mathcal{E}'_{\textbf{k}}} \right).
\end{equation}
It follows
\begin{align}
U &= - g \sum_{\textbf{k} \neq \textbf{0}} \dfrac{(- \mathcal{E}_{\textbf{k}}) e^{- \beta \mathcal{E}_{\textbf{k}}} e^{\beta \mu}}{1 - e^{- \beta \mathcal{E}_{\textbf{k}}} e^{\beta \mu}} = \notag \\
&= g \sum_{\textbf{k} \neq \textbf{0}} \dfrac{\mathcal{E}_{\textbf{k}} e^{- \beta \mathcal{E}_{\textbf{k}}} e^{\beta \mu}}{1 - e^{- \beta \mathcal{E}_{\textbf{k}}} e^{\beta \mu}} = \notag \\
&= g \sum_{\textbf{k} \neq \textbf{0}} \dfrac{\mathcal{E}_{\textbf{k}}}{e^{\beta (\mathcal{E}_{\textbf{k}} - \mu)} - 1},
\end{align}
and by means of the density of states $\eqref{eq: densitàstatibosoninoninteragenti}$, we have
\begin{equation}
U = g \dfrac{V}{(2 \pi)^2} \left( \dfrac{2m}{\hslash^2} \right)^{\frac{3}{2}} \int_0^\infty d\mathcal{E} \dfrac{\mathcal{E}^{\frac{3}{2}}}{e^{\beta (\mathcal{E} - \mu)} - 1}.
\label{eq: energiaquantisticainsiemedibosoninoninteragenti}
\end{equation}
Comparing the expressions of $U$ and $p$, the gas of non-interacting bosons satisfies
\begin{equation}
pV = \dfrac{2}{3} U.
\label{eq: pV=3/2Ugasdibosoni}
\end{equation}
\section{A comparison with classical physics scenario}
\subsection{Average number of bosons at high temperatures}
At high temperatures, Bose-Einstein statistics can be replaced by Maxwell-Boltzmann statistics, i.e.,
\begin{align}
\langle \hat{N}_{\textbf{k}} \rangle &= \dfrac{1}{e^{\beta \mathcal{E}_{\textbf{k}}} e^{- \beta \mu} - 1} \approx \notag \\
&\approx e^{- \beta \mathcal{E}_{\textbf{k}}} e^{\beta \mu},
\end{align}
then
\begin{align}
\langle \hat{N} \rangle &= \sum_{\textbf{k},\sigma} \langle \hat{N}_{\textbf{k}} \rangle \approx \notag \\
&\approx g e^{\beta \mu} \sum_{\textbf{k}} e^{- \beta \mathcal{E}_{\textbf{k}}} ,
\end{align}
and in the thermodynamic limit approximation $\eqref{eq: illimitetermodinamico}$, we get
\begin{align}
\langle \hat{N} \rangle &\approx g \dfrac{V}{(2 \pi)^3} e^{\beta \mu} \int d^3\textbf{k} e^{- \beta \mathcal{E}_{\textbf{k}}} = \notag \\
&= g \dfrac{V}{(2 \pi)^3} 4 \pi e^{\beta \mu} \int_0^\infty dk k^2 e^{- \beta \mathcal{E}_{\textbf{k}}}.
\end{align}
We define the dimensionless variable
\begin{equation}
x = \hslash k \sqrt{\dfrac{\beta}{2m}},
\label{eq: variabileadimensionalexcondensazioneBoseEinstein1}
\end{equation}
then
\begin{equation}
x^3 = \hslash^3 k^3 \left( \dfrac{\beta}{2m} \right)^{\frac{3}{2}},
\end{equation}
\begin{equation}
3 x^2 dx = 3 \hslash^3 k^2 \left( \dfrac{\beta}{2m} \right)^{\frac{3}{2}} dk,
\end{equation}
\begin{equation}
x^2 dx = \hslash^3 k^2 \left( \dfrac{\beta}{2m} \right)^{\frac{3}{2}} dk.
\end{equation}
We multiply and divide the integral by 
\begin{equation}
\dfrac{(2 m k_B T)^{\frac{3}{2}}}{\hslash^3} \equiv \left( \dfrac{2 m k_B T}{\hslash^2} \right)^{\frac{3}{2}},
\end{equation}
and we get
\begin{align}
\langle \hat{N} \rangle &= g \dfrac{V}{(2 \pi)^3} 4 \pi e^{\beta \mu} \left( \dfrac{2 m k_B T}{\hslash^2} \right)^{\frac{3}{2}} \int_0^\infty dk e^{- \beta \frac{\hslash^2 k^2}{2m}} \hslash^3 k^2 \left( \dfrac{1}{2 m k_B T} \right)^{\frac{3}{2}} \equiv \notag \\
&\equiv g \dfrac{V}{(2 \pi)^3} 4 \pi e^{\beta \mu} \left( \dfrac{2 m k_B T}{\hslash^2} \right)^{\frac{3}{2}} \int_0^\infty dx e^{-x^2} x^2 = \notag \\
&= g \dfrac{V}{(2 \pi)^3} 4 \pi \dfrac{\sqrt{\pi}}{4} e^{\beta \mu} \left( \dfrac{2 m k_B T}{\hslash^2} \right)^{\frac{3}{2}} = \notag \\
&= g \dfrac{V}{2^3 \pi^{\frac{3}{2}}} e^{\beta \mu} \left( \dfrac{2 m k_B T}{\hslash^2} \right)^{\frac{3}{2}} = \notag \\
&= g V e^{\beta \mu} \left( \dfrac{m k_B T}{2 \pi \hslash^2} \right)^{\frac{3}{2}}. 
\label{eq: numeromediobosonisenzainterazionealtetemperature} 
\end{align}
It follows
\begin{equation}
e^{\beta \mu} = \dfrac{\langle \hat{N} \rangle}{V} \dfrac{1}{g} \left( \dfrac{2 \pi \hslash^2}{m k_B T} \right)^{\frac{3}{2}},
\end{equation}
\begin{equation}
e^{\beta \mu} = \dfrac{\langle \hat{N} \rangle}{V} \dfrac{\lambda_T^3}{g},
\end{equation}
\begin{equation}
\beta \mu = \ln \left[ \dfrac{\langle \hat{N} \rangle}{V} \dfrac{\lambda_T^3}{g} \right].
\label{eq: betamucasoclassico}
\end{equation}
By recalling the definition of the thermal de Broglie wavelength, we arrive at the following limiting expressions
\begin{equation}
\lim_{T \rightarrow +\infty} \beta \mu = - \infty,
\end{equation} 
\begin{equation}
\lim_{T \rightarrow 0^+} \beta \mu = + \infty.
\end{equation}
From equation $\eqref{eq: betamucasoclassico}$, we infer the existence of a critical temperature $T_C$ at which the chemical potential vanishes, $\mu(T_C) = 0$, mirroring the behavior observed in the quantum statistical description. This temperature marks the onset of BEC in the quantum case. It is important to note, however, that the classical and quantum descriptions of a non-interacting Bose gas coincide only in the high-temperature limit. At lower temperatures, quantum effects become significant, and the classical approximation fails to capture the emergence of collective phenomena such as BEC.
\subsection{Average energy in the high-temperature regime}
In the high-temperature regime and in the thermodynamic limit approximation $\eqref{eq: illimitetermodinamico}$, we obtain
\begin{align}
U &= \sum_{\textbf{k},\sigma} \mathcal{E}_{\textbf{k}} \langle \hat{N}_{\textbf{k}} \rangle = \notag \\
&= g \sum_{\textbf{k}} \mathcal{E}_{\textbf{k}} \dfrac{1}{e^{\beta \mathcal{E}_{\textbf{k}}} e^{- \beta \mu} - 1} \approx \notag \\
& \approx g \sum_{\textbf{k}} \mathcal{E}_{\textbf{k}} e^{- \beta \mathcal{E}_{\textbf{k}}} e^{\beta \mu} \approx \notag \\
&\approx g \dfrac{V}{(2 \pi)^3} e^{\beta \mu} \int d^3\textbf{k} \dfrac{\hslash^2 k^2}{2m} e^{- \beta \frac{\hslash^2 k^2}{2m}} = \notag \\
&= g \dfrac{V}{(2 \pi)^3} 4 \pi e^{\beta \mu} \int_0^\infty dk \dfrac{\hslash^2 k^4}{2m} e^{- \beta \frac{\hslash^2 k^2}{2m}}.
\end{align}
Again, from $\eqref{eq: variabileadimensionalexcondensazioneBoseEinstein1}$, we get
\begin{equation}
x^5 = \hslash^5 k^5 \left( \dfrac{\beta}{2m} \right)^{\frac{5}{2}},
\end{equation}
\begin{equation}
5 x^4 dx = 5 \hslash^5 k^4 \left( \dfrac{\beta}{2m} \right)^{\frac{5}{2}} dk,
\end{equation}
\begin{equation}
x^4 dx = \hslash^5 k^4 \left( \dfrac{\beta}{2m} \right)^{\frac{5}{2}} dk.
\end{equation}
Multiplying and dividing the integral by $\frac{\hslash^3 \beta^{\frac{5}{2}}}{(2m)^{\frac{3}{2}}}$, we get
\begin{align}
U &= g \dfrac{V}{(2 \pi)^3} 4 \pi e^{\beta \mu} \dfrac{(2m)^{\frac{3}{2}}}{\hslash^3 \beta^{\frac{5}{2}}} \int_0^\infty dk \hslash^5 k^4 \left( \dfrac{\beta}{2m} \right)^{\frac{5}{2}} e^{- \beta \frac{\hslash^2 k^2}{2m}} = \notag \\
&= g \dfrac{V}{(2 \pi)^3} 4 \pi e^{\beta \mu} \dfrac{(2m)^{\frac{3}{2}}}{\hslash^3 \beta^{\frac{5}{2}}} \int_0^\infty dx x^4 e^{-x^2} \equiv \notag \\
&\equiv g \dfrac{V}{(2 \pi)^3} 4 \pi e^{\beta \mu} \dfrac{(2m)^{\frac{3}{2}} (k_B T)^{\frac{5}{2}}}{\hslash^3} \int_0^\infty dx x^4 e^{-x^2} = \notag \\
&= g \dfrac{V}{(2 \pi)^3} 4 \pi \dfrac{3 \sqrt{\pi}}{8} e^{\beta \mu} \dfrac{(2m)^{\frac{3}{2}} (k_B T)^{\frac{5}{2}}}{\hslash^3}.
\end{align}
Now, we manipulate physical and numerical quantities, excluding $V$ and $g$, as follows
\begin{align}
\dfrac{m^{\frac{3}{2}} (k_B T)^{\frac{5}{2}}}{\hslash^3} &= k_B T \dfrac{m^{\frac{3}{2}} (k_B T)^{\frac{3}{2}}}{\hslash^3} = \notag \\
&= k_B T \left( \dfrac{m k_B T}{\hslash^2} \right)^{\frac{3}{2}} ,
\end{align}
\begin{align}
\dfrac{(2^{\frac{3}{2}}) (4) (3)}{(8) (8)} \dfrac{\pi^{\frac{1}{2}} \pi}{\pi^2} &= 3 \dfrac{2^{\frac{3}{2}}}{2^4} \dfrac{1}{\pi^{\frac{3}{2}}} = \notag \\
&= 3 \dfrac{2^3}{(2^4)(2^{\frac{3}{2}})} \dfrac{1}{\pi^{\frac{3}{2}}} = \notag \\
&= \dfrac{3}{2} \dfrac{1}{(2\pi)^{\frac{3}{2}}} ,
\end{align}
and we write
\begin{align}
U &= \dfrac{3}{2} g k_B T V e^{\beta \mu} \left( \dfrac{m k_B T}{2 \pi \hslash^2} \right)^{\frac{3}{2}} = \notag \\
&= \dfrac{3}{2} g k_B T V e^{\beta \mu} \dfrac{1}{\lambda_T^3} = \notag \\
&= \dfrac{3}{2} \langle \hat{N} \rangle k_B T ,
\label{eq: principioequipartizioneenergia}
\end{align}
where we included the average number $\eqref{eq: numeromediobosonisenzainterazionealtetemperature}$ of bosons without interaction at high temperatures, i.e., we obtain the energy equipartition principle.
\section{The specific heat of a non interacting boson system}
Given the energy expression $\eqref{eq: energiaquantisticainsiemedibosoninoninteragenti}$, derived within a quantum framework and valid at all temperatures, we now focus on its behavior in the low-temperature regime. As previously established, BEC occurs below a critical temperature $T_C$, at which point the bosonic gas undergoes a phase transition and a macroscopic number of particles occupy the ground state. We have
\begin{align}
U &= g \dfrac{V}{(2 \pi)^2} \left( \dfrac{2m}{\hslash^2} \right)^{\frac{3}{2}} \int_0^\infty d\mathcal{E} \dfrac{\mathcal{E}^{\frac{3}{2}}}{e^{\beta (\mathcal{E} - \mu)} - 1} = \notag \\
&= g \dfrac{V}{(2 \pi)^2} \left( \dfrac{2m}{\hslash^2} \right)^{\frac{3}{2}} \int_0^\infty d\mathcal{E} \dfrac{\mathcal{E}^{\frac{3}{2}}}{e^{\beta \mathcal{E}} - 1} ,
\end{align}
given that for temperatures much lower than the critical temperature, $T < T_C$, it follows that $e^{- \beta \mu} \rightarrow 1$. If we rewrite $\eqref{eq: variabileadimensionalexcondensazioneBoseEinstein1}$ as
\begin{equation}
x= \sqrt{\beta \mathcal{E}} ,
\label{eq: variabileadimensionalexcondensazioneBoseEinstein2}
\end{equation}
from
\begin{equation}
x^5 = \beta^{\frac{5}{2}} \mathcal{E}^{\frac{5}{2}},
\end{equation}
\begin{equation}
5 x^4 dx = \dfrac{5}{2} \beta^{\frac{5}{2}} \mathcal{E}^{\frac{3}{2}} d \mathcal{E},
\end{equation}
\begin{equation}
x^4 dx = \dfrac{1}{2} \beta^{\frac{5}{2}} \mathcal{E}^{\frac{3}{2}} d \mathcal{E},
\end{equation}
we multiply and divide $U$ by $\frac{\beta^{\frac{5}{2}}}{2}$ and we get
\begin{align}
U &= \dfrac{2}{\beta^{\frac{5}{2}}} g \dfrac{V}{(2 \pi)^2} \left( \dfrac{2m}{\hslash^2} \right)^{\frac{3}{2}} \int_0^\infty dx \dfrac{x^4}{e^{\sqrt{x}} - 1} \equiv \notag \\
&\equiv 2 g \dfrac{V}{(2 \pi)^2} \left( \dfrac{2m}{\hslash^2} \right)^{\frac{3}{2}} (k_B T)^{\frac{5}{2}} \int_0^\infty dx \dfrac{x^4}{e^{\sqrt{x}} - 1}.
\end{align}
It is now evident that the internal energy \( U \) of a system of non-interacting bosons exhibits a power-law dependence on temperature in the low-temperature regime:
\begin{equation}
U \propto T^{\frac{5}{2}}, \quad T \ll T_C.
\end{equation}
In the high-temperature limit, the Bose-Einstein distribution approaches the Maxwell-Boltzmann statistics. According to the equipartition theorem, the internal energy then scales linearly with temperature
\begin{equation}
U \propto T, \quad T \gg T_C.
\end{equation}
As a result, the specific heat at constant volume, \( C_V = \frac{\partial U}{\partial T} \), behaves as follows
\begin{equation}
C_V \propto
\begin{cases}
T^{\frac{3}{2}}, & T \ll T_C, \\
\frac{3}{2} R, & T \gg T_C,
\end{cases}
\end{equation}
where \( R \) is the ideal gas constant. According to this simplified model, \( C_V(T) \) features a cusp at the critical temperature \( T = T_C \).
\newpage
\section{Figures}
\FloatBarrier
\chapter{Bogoljubov's theory for interacting bosonic systems}
In this chapter, we develop Bogoljubov's theory for interacting bosonic systems in a self-contained and rigorous manner, starting from the physical assumptions and proceeding all the way to observable consequences. \newline
The central idea of Bogoljubov's approach is to leverage the macroscopic occupation of the ground state by condensed bosons to systematically linearize the interaction terms. This leads to an effective approximation of the Hamiltonian, which takes a diagonalizable form. This, in turn, allows for the explicit determination of the spectrum of elementary excitations and of the ground state of the system. The method is valid in the regime of finite density and weak interactions, typically at zero temperature. We begin by introducing the second-quantized Hamiltonian for bosons with two-body interactions, and we apply Bogoljubov's key assumption that the zero-momentum component of the bosonic field can be treated as a classical quantity. We then analyze the conditions under which this approximation is valid, and show how it leads to a quadratic approximation of the Hamiltonian, expressible in terms of bosonic creation and annihilation operators. \newline
At this stage, we introduce the Bogoljubov canonical transformations specific to bosonic systems, which serve as a powerful tool for approximately diagonalizing the Hamiltonian in the presence of weak interactions. These transformations are constructed so as to preserve the fundamental bosonic commutation relations, and they allow us to express the Hamiltonian in terms of new creation and annihilation operators corresponding to non-interacting bosonic quasiparticles. We derive these transformations explicitly, highlighting both their mathematical structure and physical significance, and we use them to obtain the energy spectrum of the quasiparticles. \newline
Once the diagonal form of the Hamiltonian has been obtained, we proceed to analyze the ground state of the system in this new representation. In particular, we compute the correction to the condensate density resulting from interactions, which lead to a depletion of the condensate even at zero temperature. We also determine the modified vacuum energy, which incorporates the contribution from quantum fluctuations around the Bose-Einstein condensed state. These results offer a more accurate and realistic description of the interacting bosonic system compared to the ideal non-interacting BEC, and they establish a direct connection between the microscopic theory and observable macroscopic phenomena.
\section{Bogoljubov assumption}
In the previous chapter, we showed that a description of the specific heat of a bosonic system requires the inclusion of inter-particle interactions, especially near the condensation transition. In this chapter, we begin our study of the interacting Bose gas starting from the simplest regime, i.e., the zero-temperature limit. At \( T = 0 \), one expects that a large macroscopic part of bosons within a Bose-Einstein condensate occupy a one-particle quantum state. Let us denote by \( \varphi_0 = |0\rangle \) the ground-state eigenfunction of the one-particle Hamiltonian, which defines the condensate mode. The many-body wavefunction \( |\phi_N\rangle \) of the condensate composed of \( N \) bosons can then be represented as a fully symmetric product state, given by
\begin{equation}
|\phi_{N}\rangle = \dfrac{a^{\dagger \ N}_0}{\sqrt{N!}} |0\rangle,
\end{equation}
where $a^\dagger_0$ creates a boson in the state $\varphi_0$. The operator number satisfies
\begin{equation}
a_0^\dagger a_0 |\phi_{N}\rangle = N |\phi_{N}\rangle.
\end{equation}
Instead, in the thermodynamic limit the operator $a_0 a_0^\dagger$ satisfies
\begin{align}
a_0 a_0^\dagger |\phi_{N} \rangle &= (N+1) |\phi_{N} \rangle \sim \notag \\
&\sim N |\phi_{N} \rangle, 
\end{align}
and similarly for the annihilation operator
\begin{align}
a_0 |\phi_{N} \rangle &= \sqrt{N} |\phi_{N-1}\rangle \sim \notag \\
&\sim \sqrt{N} |\phi_{N}\rangle.
\end{align}
In other words, we are assuming that in the condensate the bosonic operators commute. Considering the operators as classical functions, we can replace them by their eigenvalues, that is, by $\sqrt{N}$. This is the basic assumption of Bogoljubov's theory. In addition, given an Hamiltonian of interacting boson particles of the form $\eqref{eq: hamiltonianamateriainterazioneaduecorpi}$, with 
\begin{equation}
\mathcal{H}_0 = \dfrac{\hat{\textbf{p}}^2}{2m} + V(x) - \mu,
\end{equation}
Bogoljubov considered a system composed of ultracold atoms that is strongly dilute. The ultracold condition is essential for the realization and study of BEC, while the diluteness implies that the characteristic range of the interatomic interactions is much smaller than the average interparticle spacing. Consequently, the pairwise interaction potential can be effectively approximated by a contact potential, i.e.,
\begin{equation}
U(x,x') = U \delta(\left| x-x' \right|),
\end{equation}
and our starting point is the second-quantized Hamiltonian
\begin{equation}
\hat{\mathcal{H}} = \int dx \hat{\psi}^\dagger(x) \left[ \dfrac{\hat{\textbf{p}}^2}{2m} + V(x) - \mu \right] \hat{\psi}(x) + \dfrac{U}{2} \int dx \hat{\psi}^\dagger(x) \hat{\psi}^\dagger(x) \hat{\psi}(x) \hat{\psi}(x).
\end{equation}
Bogoljubov assumed
\begin{equation}
\hat{\psi}(x) = \varphi_0(x) a_0 + \hat{\phi}(x),
\end{equation}
where $\hat{\phi}(x)$ denotes the excited states and the BEC state must satisfy
\begin{equation}
\int dx \varphi_0^*(x) \hat{\phi}(x) = 0,
\end{equation}
\begin{equation}
\int dx \left| \varphi_0(x) \right|^2 = 1.
\end{equation}
Our goal is to determine the ground state $\varphi_0(x)$. Since the ground state corresponds to a stable equilibrium configuration, we expand the excited state up to second order around the ground state. At zero temperature, if the bosons were non-interacting, then the number of particles in the condensate would satisfy $N_0 = N$. However, because the bosons interact, we expect $N_0 \neq N$, although both remain of the same order of magnitude. Accordingly, we replace the operator $a_0$ by the variable $\sqrt{N_0}$, i.e.,
\begin{equation}
\hat{\psi}(x) = \sqrt{N_0} \varphi_0(x) + \hat{\phi}(x).
\end{equation}
where $\sqrt{N_0} \varphi_0(x)$ is a c-number, i.e., it is not an operator. In other words, the macroscopic part of the field is treated as a number, and only the fluctuations are quantized.
\section{Bogoljubov's Hamiltonian}
Our objective is to express the Hamiltonian as a second-order expansion in the field operators around the ground state. For this purpose, we introduce the Bogoljubov field operator and retain only those terms that are of order two or lower in $\hat{\phi}$, $\hat{\phi}^\dagger$, i.e.,
\begin{align}
\hat{\mathcal{H}} &= \int dx \left( \sqrt{N_0} \varphi_0^*(x) + \hat{\phi}^\dagger(x) \right) \left[ \dfrac{\hat{\textbf{p}}^2}{2m} + V(x) - \mu \right] \left( \sqrt{N_0} \varphi_0(x) + \hat{\phi}(x) \right) + \notag \\
&+ \dfrac{U}{2} \int dx \left( \sqrt{N_0} \varphi_0^*(x) + \hat{\phi}^\dagger(x) \right) \left( \sqrt{N_0} \varphi_0^*(x) + \hat{\phi}^\dagger(x) \right) \left( \sqrt{N_0} \varphi_0(x) + \hat{\phi}(x) \right) \left( \sqrt{N_0} \varphi_0(x) + \hat{\phi}(x) \right).
\end{align}
The contributions in the first integral are
\begin{equation}
N_0 \int dx \varphi_0^*(x) \left[ \dfrac{\hat{\textbf{p}}^2}{2m} + V(x) - \mu \right] \varphi_0(x),
\end{equation}
\begin{equation}
\int dx \hat{\phi}^\dagger(x) \left[ \dfrac{\hat{\textbf{p}}^2}{2m} + V(x) - \mu \right] \hat{\phi}(x),
\end{equation}
\begin{equation}
\sqrt{N_0} \int dx \varphi_0^*(x) \left[ \dfrac{\hat{\textbf{p}}^2}{2m} + V(x) - \mu \right] \hat{\phi}(x),
\end{equation}
\begin{equation}
\sqrt{N_0} \int dx \hat{\phi}^\dagger(x) \left[ \dfrac{\hat{\textbf{p}}^2}{2m} + V(x) - \mu \right] \varphi_0(x).
\end{equation}
On the other hand, the contributions in the second integral can be simplified by recognizing the following identification
\begin{equation}
\int dx (a+b)(c+d)(e+f)(g+h),
\end{equation}
where
\begin{equation}
aceg \equiv \dfrac{U N_0^2}{2} \int dx \varphi_0^*(x) \varphi_0^*(x) \varphi_0(x) \varphi_0(x),
\end{equation}
\begin{equation}
aceh \equiv \dfrac{U N_0^\frac{3}{2}}{2} \int dx \varphi_0^*(x) \varphi_0^*(x) \varphi_0(x) \hat{\phi}(x),
\end{equation}
\begin{equation}
acfg \equiv \dfrac{U N_0^\frac{3}{2}}{2} \int dx \varphi_0^*(x) \varphi_0^*(x) \hat{\phi}(x) \varphi_0(x),
\end{equation}
\begin{equation}
acfh \equiv \dfrac{U N_0}{2} \int dx \varphi_0^*(x) \varphi_0^*(x) \hat{\phi}(x) \hat{\phi}(x),
\end{equation}
\begin{equation}
adeg \equiv \dfrac{U N_0^\frac{3}{2}}{2} \int dx \varphi_0^*(x) \hat{\phi}^\dagger(x) \varphi_0(x) \varphi_0(x),
\end{equation}
\begin{equation}
adeh \equiv \dfrac{U N_0}{2} \int dx \varphi_0^*(x) \hat{\phi}^\dagger(x) \varphi_0(x) \hat{\phi}(x),
\end{equation}
\begin{equation}
adfg \equiv \dfrac{U \sqrt{N_0}}{2} \int dx \varphi_0^*(x) \hat{\phi}^\dagger(x) \hat{\phi}(x) \varphi_0(x),
\end{equation}
\begin{equation}
adfh \equiv \dfrac{U \sqrt{N_0}}{2} \int dx \varphi_0^*(x) \hat{\phi}^\dagger(x) \hat{\phi}(x) \hat{\phi}(x),
\end{equation}
\begin{equation}
bdfh \equiv \dfrac{U}{2} \int dx \hat{\phi}^\dagger(x) \hat{\phi}^\dagger(x) \hat{\phi}(x) \hat{\phi}(x),
\end{equation}
\begin{equation}
bdfg \equiv \dfrac{U \sqrt{N_0}}{2} \int dx \hat{\phi}^\dagger(x) \hat{\phi}^\dagger(x) \hat{\phi}(x) \varphi_0(x),
\end{equation}
\begin{equation}
bdeg \equiv \dfrac{U N_0}{2} \int dx \hat{\phi}^\dagger(x) \hat{\phi}^\dagger(x) \varphi_0(x) \varphi_0(x),
\end{equation}
\begin{equation}
bdeh \equiv \dfrac{U \sqrt{N_0}}{2} \int dx \hat{\phi}^\dagger(x) \hat{\phi}^\dagger(x) \varphi_0(x) \hat{\phi}(x),
\end{equation}
\begin{equation}
bceg \equiv \dfrac{U N_0^\frac{3}{2}}{2} \int dx \hat{\phi}^\dagger(x) \varphi^*_0(x) \varphi_0(x) \varphi_0(x),
\end{equation}
\begin{equation}
bceh \equiv \dfrac{U N_0}{2} \int dx \hat{\phi}^\dagger(x) \varphi^*_0(x) \varphi_0(x) \hat{\phi}(x),
\end{equation}
\begin{equation}
bcfg \equiv \dfrac{U N_0}{2} \int dx \hat{\phi}^\dagger(x) \varphi^*_0(x) \hat{\phi}(x) \varphi_0(x),
\end{equation}
\begin{equation}
bcfh \equiv \dfrac{U \sqrt{N_0}}{2} \int dx \hat{\phi}^\dagger(x) \varphi_0^*(x) \hat{\phi}(x) \hat{\phi}(x).
\end{equation}
In order to express the Hamiltonian as a quadratic expansion in the fluctuation field $\hat{\phi}$ around the condensate wavefunction $\varphi_0$, all terms of order higher than second must be neglected, specifically, the contributions corresponding to $adfh$, $bdfh$, $bdfg$, $bdeh$ and $bcfh$. Consequently, the contributions to the second integral are given by
\begin{equation}
aceg \equiv \dfrac{U N_0^2}{2} \int dx \left| \varphi_0(x) \right|^2 \left| \varphi_0(x) \right|^2,
\end{equation}
\begin{equation}
aceh \equiv \dfrac{U N_0^\frac{3}{2}}{2} \int dx \varphi_0^*(x) \left| \varphi_0(x) \right|^2 \hat{\phi}(x),
\end{equation}
\begin{equation}
acfg \equiv \dfrac{U N_0^\frac{3}{2}}{2} \int dx \varphi_0^*(x) \left| \varphi_0(x) \right|^2 \hat{\phi}(x),   
\end{equation}
\begin{equation}
acfh \equiv \dfrac{U N_0}{2} \int dx \left[ \varphi_0^*(x) \right]^2 \hat{\phi}(x) \hat{\phi}(x),
\end{equation}
\begin{equation}
adeg \equiv \dfrac{U N_0^\frac{3}{2}}{2} \int dx \varphi_0(x) \left| \varphi_0(x) \right|^2 \hat{\phi}^\dagger(x),
\end{equation}
\begin{equation}
adeh \equiv \dfrac{U N_0}{2} \int dx \left| \varphi_0(x) \right|^2 \hat{\phi}^\dagger(x) \hat{\phi}(x),
\end{equation}
\begin{equation}
adfg \equiv \dfrac{U \sqrt{N_0}}{2} \int dx \left| \varphi_0(x) \right|^2 \hat{\phi}^\dagger(x) \hat{\phi}(x),
\end{equation}
\begin{equation}
bdeg \equiv \dfrac{U N_0}{2} \int dx \left[ \varphi_0(x)\right]^2 \hat{\phi}^\dagger(x) \hat{\phi}^\dagger(x),
\end{equation}
\begin{equation}
bceg \equiv \dfrac{U N_0^\frac{3}{2}}{2} \int dx \varphi_0(x) \left| \varphi_0(x) \right|^2 \hat{\phi}^\dagger(x),
\end{equation}
\begin{equation}
bceh \equiv \dfrac{U N_0}{2} \int dx \left| \varphi_0(x) \right|^2 \hat{\phi}^\dagger(x) \hat{\phi}(x),
\end{equation}
\begin{equation}
bcfg \equiv \dfrac{U N_0}{2} \int dx \left| \varphi_0(x) \right|^2 \hat{\phi}^\dagger(x) \hat{\phi}(x),
\end{equation}
and from $aceh \equiv acfg$, $adeg \equiv bceg$, $adeh \equiv adfg \equiv bceh \equiv bcfg$, we have
\begin{align}
\int dx (a+b)(c+d)(e+f)(g+h) &\simeq \int dx \ aceg + 2 \int dx \ aceh + \int dx \ acfh + 2 \int dx \ adeg + \notag \\
&+ 4 \int dx \ adeh + \int dx \ bdeg. 
\end{align}
By combining the results obtained from the integrals contained in the Hamiltonian, we have
\begin{align}
\hat{\mathcal{H}} &= N_0 \int dx \varphi_0^*(x) \left[ \dfrac{\hat{\textbf{p}}^2}{2m} + V(x) - \mu \right] \varphi_0(x) + \int dx \hat{\phi}^\dagger(x) \left[ \dfrac{\hat{\textbf{p}}^2}{2m} + V(x) - \mu \right] \hat{\phi}(x) + \notag \\
&+ \sqrt{N_0} \int dx \varphi_0^*(x) \left[ \dfrac{\hat{\textbf{p}}^2}{2m} + V(x) - \mu \right] \hat{\phi}(x) + \sqrt{N_0} \int dx \hat{\phi}^\dagger(x) \left[ \dfrac{\hat{\textbf{p}}^2}{2m} + V(x) - \mu \right] \varphi_0(x) + \notag \\
&+ \dfrac{U N_0^2}{2} \int dx \left| \varphi_0(x) \right|^2 \left| \varphi_0(x) \right|^2 + U N_0^\frac{3}{2} \int dx \varphi_0^*(x) \left| \varphi_0(x) \right|^2 \hat{\phi}(x) + \dfrac{U N_0}{2} \int dx \left[ \varphi_0^*(x) \right]^2 \hat{\phi}(x) \hat{\phi}(x) + \notag \\
&+ U N_0^\frac{3}{2} \int dx \varphi_0(x) \left| \varphi_0(x) \right|^2 \hat{\phi}^\dagger(x) + 2 U N_0 \int dx \left| \varphi_0(x) \right|^2 \hat{\phi}^\dagger(x) \phi(x) + \dfrac{U N_0}{2} \int dx \left[ \varphi_0(x)\right]^2 \hat{\phi}^\dagger(x) \hat{\phi}^\dagger(x).
\end{align}
The state $\varphi_0$ represents the ground state and thus corresponds to a minimum of the system's energy. Therefore, the sum of all first-order terms in $\hat{\phi}$ and $\hat{\phi}^\dagger$ must vanish, allowing the Hamiltonian to be expanded up to second order as follows
\begin{align}
\hat{\mathcal{H}} &= N_0 \int dx \varphi_0^*(x) \left[ \dfrac{\hat{\textbf{p}}^2}{2m} + V(x) - \mu \right] \varphi_0(x) + \int dx \hat{\phi}^\dagger(x) \left[ \dfrac{\hat{\textbf{p}}^2}{2m} + V(x) - \mu \right] \hat{\phi}(x) + \notag \\
&+ \dfrac{U N_0^2}{2} \int dx \left| \varphi_0(x) \right|^2 \left| \varphi_0(x) \right|^2 + \dfrac{U N_0}{2} \int dx \left[ \varphi_0^*(x) \right]^2 \hat{\phi}(x) \hat{\phi}(x) + \notag \\
&+ 2 U N_0 \int dx \left| \varphi_0(x) \right|^2 \hat{\phi}^\dagger(x) \hat{\phi}(x) + \dfrac{U N_0}{2} \int dx \left[ \varphi_0(x)\right]^2 \hat{\phi}^\dagger(x) \hat{\phi}^\dagger(x),
\label{eq: sviluppoquadraticoHamiltonianabosoniconinterazioni}
\end{align}
which holds if and only if
\begin{align}
& \sqrt{N_0} \int dx \varphi_0^*(x) \left[ \dfrac{\hat{\textbf{p}}^2}{2m} + V(x) - \mu \right] \hat{\phi}(x) + \sqrt{N_0} \int dx \hat{\phi}^\dagger(x) \left[ \dfrac{\hat{\textbf{p}}^2}{2m} + V(x) - \mu \right] \varphi_0(x) + \notag \\
& \ \ \ + U N_0^\frac{3}{2} \int dx \varphi_0^*(x) \left| \varphi_0(x) \right|^2 \hat{\phi}(x) + U N_0^\frac{3}{2} \int dx \varphi_0(x) \left| \varphi_0(x) \right|^2 \hat{\phi}^\dagger(x) = 0.
\end{align} 
Note that the equation above is of the form $z + z^* = 0$, with
\begin{align}
z &\equiv \sqrt{N_0} \int dx \hat{\phi}^\dagger(x) \left[ \dfrac{\hat{\textbf{p}}^2}{2m} + V(x) - \mu \right] \varphi_0(x) + U N_0^\frac{3}{2} \int dx \varphi_0(x) \left| \varphi_0(x) \right|^2 \hat{\phi}^\dagger(x) = \notag \\
&= \sqrt{N_0} \int dx \hat{\phi}^\dagger(x) \left[ \dfrac{\hat{\textbf{p}}^2}{2m} + V(x) - \mu + U N_0 \left| \varphi_0(x) \right|^2 \right] \varphi_0(x).
\end{align}
Since $0 = z + z^* = 2 \re z$, the real part of $z$ must vanish. In our case, the Hamiltonian is generally complex, so an alternative particular solution is given by
\begin{equation}
\int dx \hat{\phi}^\dagger(x) \left[ \dfrac{\hat{\textbf{p}}^2}{2m} + V(x) - \mu + U N_0 \left| \varphi_0(x) \right|^2 \right] \varphi_0(x) = 0, \ \forall \ \hat{\phi}^\dagger(x),
\end{equation}
then
\begin{equation}
\left[ \dfrac{\hat{\textbf{p}}^2}{2m} + V(x) - \mu + U N_0 \left| \varphi_0(x) \right|^2 \right] \varphi_0(x) = 0,
\label{eq: Gross-Pitaevskij}
\end{equation}
which is known as Gross-Pitaevsky equation and must be satisfied. Bogoljubov demonstrated that a Hamiltonian of the form $\eqref{eq: sviluppoquadraticoHamiltonianabosoniconinterazioni}$ can be exactly diagonalized in a particular case. Under the assumptions of ultracold atoms interacting via a two-body contact potential, and further assuming that $V(x)=0$, the condensate state corresponds to a plane wave with zero momentum, that is
\begin{equation}
\varphi_0 = \dfrac{1}{\sqrt{V}} ,
\label{eq: ondapianamomentonulloBogoljubov}
\end{equation}
which satisfies the $\eqref{eq: Gross-Pitaevskij}$. Indeed,
\begin{equation}
\left[ - \mu + U N_0 \left| \dfrac{1}{\sqrt{V}} \right|^2 \right] \dfrac{1}{\sqrt{V}} = 0,
\end{equation}
\begin{equation}
- \mu + \dfrac{U N_0}{V} = 0,
\end{equation}
\begin{equation}
\mu = \dfrac{U N_0}{V}.
\end{equation}
Note that we have omitted the momentum operator, since when acting on $\varphi_0$ it yields the zero eigenvalue. Given the form of the solution in equation $\eqref{eq: ondapianamomentonulloBogoljubov}$, we expand the excitation field operator in terms of plane waves $\eqref{eq: ondepiane}$ as follows
\begin{equation}
\hat{\phi}(\textbf{r}) = \sum_{\textbf{k} \neq \textbf{0}} \dfrac{e^{i \textbf{k} \cdot \textbf{r}}}{\sqrt{V}} a_{\textbf{k}}.
\end{equation}
Note we have neglected the spin degree of freedom, which contributes only a numerical degeneracy factor to each term in the equations. We then introduce the Bogoljubov field operator
\begin{align}
\hat{\psi}(\textbf{r}) &= \sqrt{N_0} \varphi_0 + \hat{\phi}(\textbf{r}) = \notag \\
&= \sqrt{\dfrac{N_0}{V}} + \sum_{\textbf{k} \neq \textbf{0}} \dfrac{e^{i \textbf{k} \cdot \textbf{r}}}{\sqrt{V}} a_{\textbf{k}}
\end{align}
in equation $\eqref{eq: sviluppoquadraticoHamiltonianabosoniconinterazioni}$, replacing $x$ by $\textbf{r}$, and we write
\begin{align}
\hat{\mathcal{H}} &= - N_0 \mu \int d\textbf{r} \dfrac{1}{V} + \int d\textbf{r} \left( \sum_{\textbf{k} \neq \textbf{0}} \dfrac{e^{- i \textbf{k} \cdot \textbf{r}}}{\sqrt{V}} a^\dagger_{\textbf{k}} \right) \left[ \dfrac{\hat{\textbf{p}}^2}{2m} - \mu \right] \left( \sum_{\textbf{k}' \neq \textbf{0}} \dfrac{e^{i \textbf{k}' \cdot \textbf{r}}}{\sqrt{V}} a_{\textbf{k}'} \right) + \notag \\
&+ \dfrac{U N_0^2}{2} \int d\textbf{r} \dfrac{1}{V^2} + \dfrac{U N_0}{2} \int d\textbf{r} \dfrac{1}{V} \left[ \sum_{\textbf{k} \neq \textbf{0}} \dfrac{e^{i \textbf{k} \cdot \textbf{r}}}{\sqrt{V}} a_{\textbf{k}} \right]^2 + \notag \\
&+ 2 U N_0 \int d\textbf{r} \dfrac{1}{V} \left| \sum_{\textbf{k} \neq \textbf{0}} \dfrac{e^{i \textbf{k} \cdot \textbf{r}}}{\sqrt{V}} a_{\textbf{k}} \right|^2 + \dfrac{U N_0}{2} \int d\textbf{r} \dfrac{1}{V} \left[ \sum_{\textbf{k} \neq \textbf{0}} \dfrac{e^{- i \textbf{k} \cdot \textbf{r}}}{\sqrt{V}} a^\dagger_{\textbf{k}} \right]^2 = \notag \\
&= - \dfrac{U N_0^2}{V} + \sum_{\textbf{k} \neq \textbf{0}} \sum_{\textbf{k}' \neq \textbf{0}} \int d\textbf{r} \dfrac{e^{- i \textbf{k} \cdot \textbf{r}}}{\sqrt{V}} \dfrac{\hslash^2 \textbf{k}'^2}{2m} \dfrac{e^{i \textbf{k}' \cdot \textbf{r}}}{\sqrt{V}} a^\dagger_{\textbf{k}} a_{\textbf{k}'} - \dfrac{U N_0}{V} \int d\textbf{r} \left( \sum_{\textbf{k} \neq \textbf{0}} \dfrac{e^{- i \textbf{k} \cdot \textbf{r}}}{\sqrt{V}} a^\dagger_{\textbf{k}} \right) \left( \sum_{\textbf{k}' \neq \textbf{0}} \dfrac{e^{i \textbf{k}' \cdot \textbf{r}}}{\sqrt{V}} a_{\textbf{k}'} \right) + \dfrac{U N_0^2}{2V} + \notag \\
&+ \dfrac{U N_0}{2V} \int d\textbf{r} \left( \sum_{\textbf{k} \neq \textbf{0}} \dfrac{e^{i \textbf{k} \cdot \textbf{r}}}{\sqrt{V}} a_{\textbf{k}} \right) \left( \sum_{\textbf{k}' \neq \textbf{0}} \dfrac{e^{i \textbf{k}' \cdot \textbf{r}}}{\sqrt{V}} a_{\textbf{k}'} \right) + \dfrac{2 U N_0}{V} \int d\textbf{r} \left( \sum_{\textbf{k} \neq \textbf{0}} \dfrac{e^{- i \textbf{k} \cdot \textbf{r}}}{\sqrt{V}} a^\dagger_{\textbf{k}} \right) \left( \sum_{\textbf{k}' \neq \textbf{0}} \dfrac{e^{i \textbf{k}' \cdot \textbf{r}}}{\sqrt{V}} a_{\textbf{k}'} \right) + \notag \\
&+ \dfrac{U N_0}{2V} \int d\textbf{r} \left( \sum_{\textbf{k} \neq \textbf{0}} \dfrac{e^{- i \textbf{k} \cdot \textbf{r}}}{\sqrt{V}} a^\dagger_{\textbf{k}} \right) \left( \sum_{\textbf{k}' \neq \textbf{0}} \dfrac{e^{- i \textbf{k}' \cdot \textbf{r}}}{\sqrt{V}} a^\dagger_{\textbf{k}'} \right) =
\end{align}
\begin{align}
&= - \dfrac{U N_0^2}{2V} + \sum_{\textbf{k} \neq \textbf{0}} \sum_{\textbf{k}' \neq \textbf{0}} \int d\textbf{r} \dfrac{e^{- i (\textbf{k}-\textbf{k}') \cdot \textbf{r}}}{V} \dfrac{\hslash^2 \textbf{k}'^2}{2m} a^\dagger_{\textbf{k}} a_{\textbf{k}'} - \dfrac{U N_0}{V} \sum_{\textbf{k} \neq \textbf{0}} \sum_{\textbf{k}' \neq \textbf{0}} \int d\textbf{r} \dfrac{e^{- i (\textbf{k}-\textbf{k}') \cdot \textbf{r}}}{V} a^\dagger_{\textbf{k}} a_{\textbf{k}'} + \notag \\
&+ \dfrac{U N_0}{2V} \sum_{\textbf{k} \neq \textbf{0}} \sum_{\textbf{k}' \neq \textbf{0}} \int d\textbf{r} \dfrac{e^{i (\textbf{k}+\textbf{k}') \cdot \textbf{r}}}{V} a_{\textbf{k}} a_{\textbf{k}'} + \dfrac{2 U N_0}{V} \sum_{\textbf{k} \neq \textbf{0}} \sum_{\textbf{k}' \neq \textbf{0}} \int d\textbf{r} \dfrac{e^{- i (\textbf{k}-\textbf{k}') \cdot \textbf{r}}}{V} a^\dagger_{\textbf{k}} a_{\textbf{k}'} + \notag \\
&+ \dfrac{U N_0}{2V} \sum_{\textbf{k} \neq \textbf{0}} \sum_{\textbf{k}' \neq \textbf{0}} \int d\textbf{r} \dfrac{e^{- i (\textbf{k}+\textbf{k}') \cdot \textbf{r}}}{V} a^\dagger_{\textbf{k}} a^\dagger_{\textbf{k}'} = \notag \\
&= - \dfrac{U N_0^2}{2V} + \sum_{\textbf{k} \neq \textbf{0}} \sum_{\textbf{k}' \neq \textbf{0}} \delta(\textbf{k}-\textbf{k}') \dfrac{\hslash^2 \textbf{k}'^2}{2m} a^\dagger_{\textbf{k}} a_{\textbf{k}'} - \dfrac{U N_0}{V} \sum_{\textbf{k} \neq \textbf{0}} \sum_{\textbf{k}' \neq \textbf{0}} \delta(\textbf{k}-\textbf{k}') a^\dagger_{\textbf{k}} a_{\textbf{k}'} + \dfrac{U N_0}{2V} \sum_{\textbf{k} \neq \textbf{0}} \sum_{\textbf{k}' \neq \textbf{0}} \delta(\textbf{k}+\textbf{k}') a_{\textbf{k}} a_{\textbf{k}'} + \notag \\
&+ \dfrac{2 U N_0}{V} \sum_{\textbf{k} \neq \textbf{0}} \sum_{\textbf{k}' \neq \textbf{0}} \delta(\textbf{k}-\textbf{k}') a^\dagger_{\textbf{k}} a_{\textbf{k}'} + \dfrac{U N_0}{2V} \sum_{\textbf{k} \neq \textbf{0}} \sum_{\textbf{k}' \neq \textbf{0}} \delta(\textbf{k}+\textbf{k}') a^\dagger_{\textbf{k}} a^\dagger_{\textbf{k}'} = \notag \\
&= - \dfrac{U N_0^2}{2V} + \sum_{\textbf{k} \neq \textbf{0}} \dfrac{\hslash^2 \textbf{k}^2}{2m} a^\dagger_{\textbf{k}} a_{\textbf{k}} - \dfrac{U N_0}{V} \sum_{\textbf{k} \neq \textbf{0}} a^\dagger_{\textbf{k}} a_{\textbf{k}} + \dfrac{U N_0}{2V} \sum_{\textbf{k}' \neq \textbf{0}} a_{-\textbf{k}'} a_{\textbf{k}'} + \dfrac{2 U N_0}{V} \sum_{\textbf{k} \neq \textbf{0}} a^\dagger_{\textbf{k}} a_{\textbf{k}} + \dfrac{U N_0}{2V} \sum_{\textbf{k} \neq \textbf{0}} a^\dagger_{\textbf{k}} a^\dagger_{-\textbf{k}}.
\end{align}
After appropriately renaming the dummy indices, we finally obtain
\begin{equation}
\hat{\mathcal{H}} = - \dfrac{U N_0^2}{2V} + \sum_{\textbf{k} \neq \textbf{0}} \left( \dfrac{\hslash^2 \textbf{k}^2}{2m} + \dfrac{U N_0}{V} \right) a^\dagger_{\textbf{k}} a_{\textbf{k}} + \dfrac{U N_0}{2V} \sum_{\textbf{k} \neq \textbf{0}} \left( a^\dagger_{\textbf{k}} a^\dagger_{-\textbf{k}} + a_{-\textbf{k}} a_{\textbf{k}} \right) ,
\label{eq: hamiltonianadiBogoljubov}
\end{equation}
which is known as the Bogoljubov's Hamiltonian. The quadratic terms of the form $a^\dagger_{\textbf{k}'} a^\dagger_{-\textbf{k}'}$ and $a_{-\textbf{k}'} a_{\textbf{k}'}$ will be analyzed in detail in the following section.
\section{Bogoljubov's bosonic transformations}\label{Bogoljubov's bosonic transformations}
Here, we derive the transformations necessary to diagonalize the bosonic Hamiltonian $\eqref{eq: hamiltonianadiBogoljubov}$, commonly known as the Bogoljubov bosonic transformations. We can make use of the bosonic algebra and exploit the Jacobi identity $\eqref{eq: Jacobiidentity}$, and we have
\begin{align}
\left[ a_{\textbf{k}} , a^\dagger_{\textbf{k}'} a^\dagger_{-\textbf{k}'} \right] &= \left[ a_{\textbf{k}} , a^\dagger_{\textbf{k}'} \right] a^\dagger_{-\textbf{k}'} + a^\dagger_{\textbf{k}'} \left[ a_{\textbf{k}} , a^\dagger_{-\textbf{k}'} \right] = \notag \\
&= \delta_{\textbf{k},\textbf{k}'} a^\dagger_{-\textbf{k}'} + \delta_{\textbf{k},-\textbf{k}'} a^\dagger_{\textbf{k}'},
\end{align}
\begin{align}
\left[ a_{\textbf{k}} , \sum_{\textbf{k}' \neq \textbf{0}} a^\dagger_{\textbf{k}'} a^\dagger_{-\textbf{k}'} \right] &= \sum_{\textbf{k}' \neq \textbf{0}} \left[ a_{\textbf{k}} , a^\dagger_{\textbf{k}'} a^\dagger_{-\textbf{k}'} \right] = \notag \\
&= \sum_{\textbf{k}' \neq \textbf{0}} \left( \delta_{\textbf{k},\textbf{k}'} a^\dagger_{-\textbf{k}'} + \delta_{\textbf{k},-\textbf{k}'} a^\dagger_{\textbf{k}'} \right) = \notag \\
&= 2 a^\dagger_{-\textbf{k}},
\end{align}
and
\begin{align}
\left[ a^\dagger_{\textbf{k}} , a_{-\textbf{k}'} a_{\textbf{k}'} \right] &= \left[ a^\dagger_{\textbf{k}} , a_{-\textbf{k}'} \right] a_{\textbf{k}'} + a_{-\textbf{k}'} \left[ a^\dagger_{\textbf{k}} , a_{\textbf{k}'} \right] = \notag \\
&= - \delta_{\textbf{k},-\textbf{k}'} a_{\textbf{k}'} - \delta_{\textbf{k},\textbf{k}'} a_{-\textbf{k}'},
\end{align}
\begin{align}
\left[ a^\dagger_{\textbf{k}} , \sum_{\textbf{k}' \neq \textbf{0}} a_{-\textbf{k}'} a_{\textbf{k}'} \right] &= \sum_{\textbf{k}' \neq \textbf{0}} \left[ a^\dagger_{\textbf{k}} , a_{-\textbf{k}'} a_{\textbf{k}'} \right] = \notag \\
&= \sum_{\textbf{k}' \neq \textbf{0}} \left( - \delta_{\textbf{k},-\textbf{k}'} a_{\textbf{k}'} - \delta_{\textbf{k},\textbf{k}'} a_{-\textbf{k}'} \right) = \notag \\
&= - 2 a_{-\textbf{k}}.
\end{align}
Bogoljubov approach involves considering transformations that map $a_{\textbf{k}}$ and $a^\dagger_{-\textbf{k}}$ into a linear combination of new creation and annihilation operators related to a momentum and its opposite. We define
\begin{equation}
b_{\textbf{k}} = u_{\textbf{k}} a_{\textbf{k}} + v_{\textbf{k}} a^\dagger_{-\textbf{k}}, \ \textbf{k} \neq \textbf{0} ,
\label{eq: operatoredistruzioneBogoljubov}
\end{equation}
\begin{equation}
b^\dagger_{-\textbf{k}} = v^*_{-\textbf{k}} a_{\textbf{k}} + u^*_{-\textbf{k}} a^\dagger_{-\textbf{k}}, \ \textbf{k} \neq \textbf{0} ,
\label{eq: operatorecreazioneBogoljubov}
\end{equation}
or equivalently in matrix form
\begin{equation}
\begin{pmatrix}
b_{\textbf{k}} \\
b^\dagger_{-\textbf{k}}
\end{pmatrix}
=
\begin{pmatrix}
u_{\textbf{k}} & v_{\textbf{k}} \\
v^*_{-\textbf{k}} & u^*_{-\textbf{k}}
\end{pmatrix}
\begin{pmatrix}
a_{\textbf{k}} \\
a^\dagger_{-\textbf{k}}
\end{pmatrix} 
, \ \textbf{k} \neq \textbf{0} ,
\label{eq: trasformazioneBogoljubovbosoni}
\end{equation}
where
\begin{equation}
\begin{pmatrix}
u_{\textbf{k}} & v_{\textbf{k}} \\
v^*_{-\textbf{k}} & u^*_{-\textbf{k}}
\end{pmatrix}
, \ \textbf{k} \neq \textbf{0}
\label{eq: matriceinizialetrasformazionebosonicaBogoljubov}
\end{equation}
is the Bogoljubov's bosonic matrix. Then, the goal is to compute the complex coefficients $u_{\textbf{k}}$, $u^*_{-\textbf{k}}$, $v_{\textbf{k}}$, $v^*_{-\textbf{k}}$. As a first assumption, Bogoljubov requires that the transformations $\eqref{eq: operatoredistruzioneBogoljubov}$, $\eqref{eq: operatorecreazioneBogoljubov}$ are canonical, i.e., they satisfy a bosonic algebra
\begin{equation}
\left[ b_{\textbf{k}'} , b^\dagger_{\textbf{k}} \right] = \delta_{\textbf{k}',\textbf{k}} ,
\label{eq: primovincolooperatoriBogoljubov}
\end{equation}
which implies
\begin{equation}
\left[ u_{\textbf{k}'} a_{\textbf{k}'} + v_{\textbf{k}'} a^\dagger_{-\textbf{k}'} , v^*_{\textbf{k}} a_{-\textbf{k}} + u^*_{\textbf{k}} a^\dagger_{\textbf{k}} \right] = \delta_{\textbf{k}',\textbf{k}},
\end{equation}
\begin{equation}
\left[ u_{\textbf{k}'} a_{\textbf{k}'} , v^*_{\textbf{k}} a_{-\textbf{k}} \right] + \left[ u_{\textbf{k}'} a_{\textbf{k}'} , u^*_{\textbf{k}} a^\dagger_{\textbf{k}} \right] + \left[ v_{\textbf{k}'} a^\dagger_{-\textbf{k}'} , v^*_{\textbf{k}} a_{-\textbf{k}} \right] + \left[ v_{\textbf{k}'} a^\dagger_{-\textbf{k}'} , u^*_{\textbf{k}} a^\dagger_{\textbf{k}} \right] = \delta_{\textbf{k}',\textbf{k}} ,
\end{equation}
\begin{equation}
u_{\textbf{k}'} u^*_{\textbf{k}} \delta_{\textbf{k}',\textbf{k}} - v_{\textbf{k}'} v^*_{\textbf{k}} \delta_{-\textbf{k}',-\textbf{k}} = \delta_{\textbf{k}',\textbf{k}},
\end{equation}
\begin{equation}
u_{\textbf{k}'} u^*_{\textbf{k}} \delta_{\textbf{k}',\textbf{k}} - v_{\textbf{k}'} v^*_{\textbf{k}} \delta_{\textbf{k}',\textbf{k}} = \delta_{\textbf{k}',\textbf{k}},
\end{equation}
and for $\textbf{k} = \textbf{k}'$, we have
\begin{equation}
| u_{\textbf{k}} |^2 - | v_{\textbf{k}} |^2 = 1.
\label{eq: primacondizionecoefficientitrasformazioneBogoljubov}
\end{equation}
As a second assumption, Bogoljubov requires that the Hamiltonian $\eqref{eq: hamiltonianadiBogoljubov}$ becomes diagonal with respect the new operators, i.e.,
\begin{equation}
\hat{\mathcal{H}} = \tilde{\mathcal{E}} + \sum_{\textbf{k} \neq \textbf{0}} \mathcal{E}_{\textbf{k}} b^\dagger_{\textbf{k}} b_{\textbf{k}} ,
\end{equation}
and consequently, $\eqref{eq: hamiltonianadiBogoljubov}$ must satisfy
\begin{equation}
\left[ \hat{\mathcal{H}} , b_{\textbf{k}} \right] = - \mathcal{E}_{\textbf{k}} b_{\textbf{k}}.
\label{eq: secondovincolooperatoriBogoljubov}
\end{equation}
Obviously, involving the commutator of the Hamiltonian with the creation operator $b^\dagger_{-\textbf{k}}$ is equally correct. From the definition of $b_{\textbf{k}}$, $\eqref{eq: secondovincolooperatoriBogoljubov}$ implies the study of the commutators between $\eqref{eq: hamiltonianadiBogoljubov}$ and $a_{\textbf{k}}$, $a^\dagger_{-\textbf{k}}$. From $\eqref{eq: commutatoreoperatorenumeroconoperatoredistruzione}$ and $\eqref{eq: commutatoreoperatorenumeroconoperatorecreazione}$ we have
\begin{equation}
\left[ a^\dagger_{\textbf{k}'} a_{\textbf{k}'} , a_{\textbf{k}} \right] = - \delta_{\textbf{k}',\textbf{k}} a_{\textbf{k}},
\label{eq: commutatoreoperatorenumeromomentokconoperatoredistruzionemomentokprimo}
\end{equation}
\begin{equation}
\left[ a^\dagger_{\textbf{k}'} a_{\textbf{k}'} , a^\dagger_{\textbf{k}} \right] = \delta_{\textbf{k}',\textbf{k}} a^\dagger_{\textbf{k}},
\label{eq: commutatoreoperatorenumeromomentokconoperatorecreazionemomentokprimo}
\end{equation}
respectively, and from Jacobi identity $\eqref{eq: Jacobiidentity}$ we have
\begin{align}
\left[ a^\dagger_{-\textbf{k}} , a_{-\textbf{k}'} a_{\textbf{k}'} \right] &= \left[ a^\dagger_{-\textbf{k}} , a_{-\textbf{k}'} \right] a_{\textbf{k}'} + a_{-\textbf{k}'} \left[ a^\dagger_{-\textbf{k}} , a_{\textbf{k}'} \right] = \notag \\
&= - \delta_{-\textbf{k},-\textbf{k}'} a_{\textbf{k}'} - \delta_{-\textbf{k},\textbf{k}'} a_{-\textbf{k}'}.
\label{eq: commutazioneoperatoriquadraticidistruzioneconcreazione2}
\end{align}
Now let us compute $\eqref{eq: secondovincolooperatoriBogoljubov}$, that is
\begin{equation}
\left[ \sum_{\textbf{k}' \neq \textbf{0}} \left( \dfrac{\hslash^2 \textbf{k}'^2}{2m} + \dfrac{U N_0}{V} \right) a^\dagger_{\textbf{k}'} a_{\textbf{k}'} + \dfrac{U N_0}{2V} \sum_{\textbf{k}' \neq \textbf{0}} \left( a^\dagger_{\textbf{k}'} a^\dagger_{-\textbf{k}'} + a_{-\textbf{k}'} a_{\textbf{k}'} \right) , b_{\textbf{k}} \right] = - \mathcal{E}_{\textbf{k}} b_{\textbf{k}},
\end{equation}
\begin{equation}
\left[ \sum_{\textbf{k}' \neq \textbf{0}} \left( \dfrac{\hslash^2 \textbf{k}'^2}{2m} + \dfrac{U N_0}{V} \right) a^\dagger_{\textbf{k}'} a_{\textbf{k}'} + \dfrac{U N_0}{2V} \sum_{\textbf{k}' \neq \textbf{0}} \left( a^\dagger_{\textbf{k}'} a^\dagger_{-\textbf{k}'} + a_{-\textbf{k}'} a_{\textbf{k}'} \right) , u_{\textbf{k}} a_{\textbf{k}} + v_{\textbf{k}} a^\dagger_{-\textbf{k}} \right] = - \mathcal{E}_{\textbf{k}} \left( u_{\textbf{k}} a_{\textbf{k}} + v_{\textbf{k}} a^\dagger_{-\textbf{k}} \right),
\end{equation}
\begin{align}
& \sum_{\textbf{k}' \neq \textbf{0}} \left( \dfrac{\hslash^2 \textbf{k}'^2}{2m} + \dfrac{U N_0}{V} \right) \left[ a^\dagger_{\textbf{k}'} a_{\textbf{k}'} , u_{\textbf{k}} a_{\textbf{k}} \right] + \sum_{\textbf{k}' \neq \textbf{0}} \left( \dfrac{\hslash^2 \textbf{k}'^2}{2m} + \dfrac{U N_0}{V} \right) \left[ a^\dagger_{\textbf{k}'} a_{\textbf{k}'} , v_{\textbf{k}} a^\dagger_{-\textbf{k}} \right] + \notag \\
& \ \ \ + \dfrac{U N_0}{2V} \sum_{\textbf{k}' \neq \textbf{0}} \left[ a^\dagger_{\textbf{k}'} a^\dagger_{-\textbf{k}'} , u_{\textbf{k}} a_{\textbf{k}} \right] + \dfrac{U N_0}{2V} \sum_{\textbf{k}' \neq \textbf{0}} \left[ a_{-\textbf{k}'} a_{\textbf{k}'} , v_{\textbf{k}} a^\dagger_{-\textbf{k}} \right] = - \mathcal{E}_{\textbf{k}} \left( u_{\textbf{k}} a_{\textbf{k}} + v_{\textbf{k}} a^\dagger_{-\textbf{k}} \right),
\end{align}
\begin{align}
& - \sum_{\textbf{k}' \neq \textbf{0}} \left( \dfrac{\hslash^2 \textbf{k}'^2}{2m} + \dfrac{U N_0}{V} \right) u_{\textbf{k}} \delta_{\textbf{k}',\textbf{k}} a_{\textbf{k}} + \sum_{\textbf{k}' \neq \textbf{0}} \left( \dfrac{\hslash^2 \textbf{k}'^2}{2m} + \dfrac{U N_0}{V} \right) v_{\textbf{k}} \delta_{\textbf{k}',\textbf{k}} a^\dagger_{-\textbf{k}} + \dfrac{U N_0}{2V} \sum_{\textbf{k}' \neq \textbf{0}} u_{\textbf{k}} \left( - \delta_{\textbf{k},\textbf{k}'} a^\dagger_{-\textbf{k}'} - \delta_{\textbf{k},-\textbf{k}'} a^\dagger_{\textbf{k}'} \right) + \notag \\
&+ \dfrac{U N_0}{2V} \sum_{\textbf{k}' \neq \textbf{0}} v_{\textbf{k}} \left( \delta_{-\textbf{k},-\textbf{k}'} a_{\textbf{k}'} + \delta_{-\textbf{k},\textbf{k}'} a_{-\textbf{k}'} \right) = - \mathcal{E}_{\textbf{k}} \left( u_{\textbf{k}} a_{\textbf{k}} + v_{\textbf{k}} a^\dagger_{-\textbf{k}} \right),
\end{align}
\begin{align}
& - \left( \dfrac{\hslash^2 \textbf{k}^2}{2m} + \dfrac{U N_0}{V} \right) u_{\textbf{k}} a_{\textbf{k}} + \left( \dfrac{\hslash^2 \textbf{k}^2}{2m} + \dfrac{U N_0}{V} \right) v_{\textbf{k}} a^\dagger_{-\textbf{k}} - \dfrac{U N_0}{2V} u_{\textbf{k}} \left( a^\dagger_{-\textbf{k}} + a^\dagger_{-\textbf{k}} \right) + \dfrac{U N_0}{2V} v_{\textbf{k}} \left( a_{\textbf{k}} + a_{\textbf{k}} \right) = \notag \\
&= - \mathcal{E}_{\textbf{k}} \left( u_{\textbf{k}} a_{\textbf{k}} + v_{\textbf{k}} a^\dagger_{-\textbf{k}} \right),
\end{align}
\begin{equation}
- \left( \dfrac{\hslash^2 \textbf{k}^2}{2m} + \dfrac{U N_0}{V} \right) u_{\textbf{k}} a_{\textbf{k}} + \left( \dfrac{\hslash^2 \textbf{k}^2}{2m} + \dfrac{U N_0}{V} \right) v_{\textbf{k}} a^\dagger_{-\textbf{k}} - \dfrac{U N_0}{V} u_{\textbf{k}} a^\dagger_{-\textbf{k}} + \dfrac{U N_0}{V} v_{\textbf{k}} a_{\textbf{k}} = - \mathcal{E}_{\textbf{k}} \left( u_{\textbf{k}} a_{\textbf{k}} + v_{\textbf{k}} a^\dagger_{-\textbf{k}} \right).
\end{equation}
By comparing the coefficients of $a_{\textbf{k}}$ and $a^\dagger_{-\textbf{k}}$, we obtain
\begin{equation}
\begin{cases}
- \left( \dfrac{\hslash^2 \textbf{k}^2}{2m} + \dfrac{U N_0}{V} \right) u_{\textbf{k}} + \dfrac{U N_0}{V} v_{\textbf{k}} = - \mathcal{E}_{\textbf{k}} u_{\textbf{k}} \\ 
\left( \dfrac{\hslash^2 \textbf{k}^2}{2m} + \dfrac{U N_0}{V} \right) v_{\textbf{k}} - \dfrac{U N_0}{V} u_{\textbf{k}} = - \mathcal{E}_{\textbf{k}} v_{\textbf{k}}
\end{cases} ,
\end{equation}
\begin{equation}
\begin{cases}
\left( \dfrac{\hslash^2 \textbf{k}^2}{2m} + \dfrac{U N_0}{V} \right) u_{\textbf{k}} - \dfrac{U N_0}{V} v_{\textbf{k}} = \mathcal{E}_{\textbf{k}} u_{\textbf{k}} \\ 
- \dfrac{U N_0}{V} u_{\textbf{k}} + \left( \dfrac{\hslash^2 \textbf{k}^2}{2m} + \dfrac{U N_0}{V} \right) v_{\textbf{k}} = - \mathcal{E}_{\textbf{k}} v_{\textbf{k}}
\end{cases} ,
\label{eq: sistemaequazioniproblemaautovalorigeneralizzatoBogoljubov}
\end{equation}
or in matrix form
\begin{equation}
\begin{pmatrix}
\dfrac{\hslash^2 \textbf{k}^2}{2m} + \dfrac{U N_0}{V} & - \dfrac{U N_0}{V} \\
- \dfrac{U N_0}{V} & \dfrac{\hslash^2 \textbf{k}^2}{2m} + \dfrac{U N_0}{V}  
\end{pmatrix}
\begin{pmatrix}
u_{\textbf{k}} \\
v_{\textbf{k}}
\end{pmatrix}
=
\begin{pmatrix}
\mathcal{E}_{\textbf{k}} & 0 \\
0 & - \mathcal{E}_{\textbf{k}}
\end{pmatrix}
\begin{pmatrix}
u_{\textbf{k}} \\
v_{\textbf{k}}
\end{pmatrix} ,
\end{equation}
which is a generalized eigenvalue equation. We set
\begin{equation}
\begin{cases}
u_{\textbf{k}} + v_{\textbf{k}} = s_{\textbf{k}} \\
u_{\textbf{k}} - v_{\textbf{k}} = d_{\textbf{k}} \\
\end{cases} ,
\end{equation}
then we add member by member the equations of $\eqref{eq: sistemaequazioniproblemaautovalorigeneralizzatoBogoljubov}$, i.e.,
\begin{equation}
\left( \dfrac{\hslash^2 \textbf{k}^2}{2m} + \dfrac{U N_0}{V} \right) (u_{\textbf{k}}+v_{\textbf{k}}) - \dfrac{U N_0}{V} (u_{\textbf{k}}+v_{\textbf{k}}) = \mathcal{E}_{\textbf{k}} (u_{\textbf{k}}-v_{\textbf{k}}),
\end{equation}
\begin{equation}
\dfrac{\hslash^2 \textbf{k}^2}{2m} (u_{\textbf{k}}+v_{\textbf{k}}) = \mathcal{E}_{\textbf{k}} (u_{\textbf{k}}-v_{\textbf{k}}),
\end{equation}
and we subtract member by member the equations of $\eqref{eq: sistemaequazioniproblemaautovalorigeneralizzatoBogoljubov}$ as follows
\begin{equation}
\left( \dfrac{\hslash^2 \textbf{k}^2}{2m} + \dfrac{U N_0}{V} \right) (u_{\textbf{k}}-v_{\textbf{k}}) + \dfrac{U N_0}{V} (u_{\textbf{k}}-v_{\textbf{k}}) = \mathcal{E}_{\textbf{k}} (u_{\textbf{k}}+v_{\textbf{k}}),
\end{equation}
\begin{equation}
\left( \dfrac{\hslash^2 \textbf{k}^2}{2m} + 2 \dfrac{U N_0}{V} \right) (u_{\textbf{k}}-v_{\textbf{k}}) = \mathcal{E}_{\textbf{k}} (u_{\textbf{k}}+v_{\textbf{k}}).
\end{equation}
Let us compute the solutions of the equations
\begin{equation}
\dfrac{\hslash^2 \textbf{k}^2}{2m} s_{\textbf{k}} = \mathcal{E}_{\textbf{k}} d_{\textbf{k}},
\label{eq: Bogoljubovs_ked_k1}
\end{equation}
\begin{equation}
\left( \dfrac{\hslash^2 \textbf{k}^2}{2m} + 2 \dfrac{U N_0}{V} \right) d_{\textbf{k}} = \mathcal{E}_{\textbf{k}} s_{\textbf{k}}.
\label{eq: Bogoljubovs_ked_k2}
\end{equation}
We multiply both members of $\eqref{eq: Bogoljubovs_ked_k1}$ by 
\begin{equation}
\dfrac{\hslash^2 \textbf{k}^2}{2m} + 2 \dfrac{U N_0}{V},
\end{equation}
and we have
\begin{equation}
\left( \dfrac{\hslash^2 \textbf{k}^2}{2m} + 2 \dfrac{U N_0}{V} \right) \dfrac{\hslash^2 \textbf{k}^2}{2m} s_{\textbf{k}} = \left( \dfrac{\hslash^2 \textbf{k}^2}{2m} + 2 \dfrac{U N_0}{V} \right) \mathcal{E}_{\textbf{k}} d_{\textbf{k}},
\end{equation}
then we use $\eqref{eq: Bogoljubovs_ked_k2}$ and we get
\begin{equation}
\left( \dfrac{\hslash^2 \textbf{k}^2}{2m} + 2 \dfrac{U N_0}{V} \right) \dfrac{\hslash^2 \textbf{k}^2}{2m} s_{\textbf{k}} = \mathcal{E}^2_{\textbf{k}} s_{\textbf{k}}.
\end{equation}
A possible solution of the above equation is $s_{\textbf{k}} = 0$, but this must be discarded, as it would imply 
\begin{equation}
u_{\textbf{k}} = - v_{\textbf{k}} ,
\end{equation}
\begin{equation}
u^2_{\textbf{k}} = v^2_{\textbf{k}} ,
\end{equation}
\begin{equation}
u^2_{\textbf{k}} - v^2_{\textbf{k}} = 0,
\end{equation}
which contradicts the assumption $\eqref{eq: primacondizionecoefficientitrasformazioneBogoljubov}$. Since $s_{\textbf{k}} \neq 0$, it must be
\begin{equation}
\mathcal{E}^2_{\textbf{k}} = \left( \dfrac{\hslash^2 \textbf{k}^2}{2m} + 2 \dfrac{U N_0}{V} \right) \dfrac{\hslash^2 \textbf{k}^2}{2m},
\end{equation}
\begin{equation}
\mathcal{E}_{\textbf{k}} = \sqrt{\mathcal{E}_{\textbf{k}}^0 \left( \mathcal{E}_{\textbf{k}}^0 + 2 \dfrac{U N_0}{V} \right)}.
\label{eq: energiespettroBogoljubov}
\end{equation}
Moreover, given
\begin{align}
1 &= u^2_{\textbf{k}} - v^2_{\textbf{k}} = \notag \\
&= \left( u_{\textbf{k}}+v_{\textbf{k}} \right) \left( u_{\textbf{k}}-v_{\textbf{k}} \right) = \notag \\
&= s_{\textbf{k}} d_{\textbf{k}},
\end{align}
we insert $d_{\textbf{k}} = s^{-1}_{\textbf{k}}$ into the equation $\eqref{eq: Bogoljubovs_ked_k1}$ and we get
\begin{equation}
\dfrac{\hslash^2 \textbf{k}^2}{2m} s^2_{\textbf{k}} = \mathcal{E}_{\textbf{k}} ,
\end{equation}
\begin{equation}
s_{\textbf{k}} = \sqrt{\dfrac{\mathcal{E}_{\textbf{k}}}{\mathcal{E}^0_{\textbf{k}}}}, 
\end{equation}
\begin{equation}
d_{\textbf{k}} = \sqrt{\dfrac{\mathcal{E}^0_{\textbf{k}}}{\mathcal{E}_{\textbf{k}}}}, 
\end{equation}
\begin{equation}
2 u_{\textbf{k}} = s_{\textbf{k}} + d_{\textbf{k}},
\end{equation}
\begin{equation}
2 v_{\textbf{k}} = s_{\textbf{k}} - d_{\textbf{k}},
\end{equation}
which imply
\begin{equation}
u_{\textbf{k}} = \dfrac{1}{2} \left( \sqrt{\dfrac{\mathcal{E}_{\textbf{k}}}{\mathcal{E}^0_{\textbf{k}}}} + \sqrt{\dfrac{\mathcal{E}^0_{\textbf{k}}}{\mathcal{E}_{\textbf{k}}}} \right) ,
\end{equation}
\begin{equation}
v_{\textbf{k}} = \dfrac{1}{2} \left( \sqrt{\dfrac{\mathcal{E}_{\textbf{k}}}{\mathcal{E}^0_{\textbf{k}}}} - \sqrt{\dfrac{\mathcal{E}^0_{\textbf{k}}}{\mathcal{E}_{\textbf{k}}}} \right),
\end{equation}
that is, we have a solution for the coefficients $u_{\textbf{k}}$, $v_{\textbf{k}}$. Note that such solutions satisfy
\begin{equation}
u_{-\textbf{k}} = u_{\textbf{k}},
\end{equation}
\begin{equation}
v_{-\textbf{k}} = v_{\textbf{k}}.
\end{equation}
Now we want to write the old operators as a function of Bogoljubov bosonic operators. For this purpose, it is necessary to compute the inverse Bogoljubov's bosonic transformation, that is, the inverse of the matrix $\eqref{eq: matriceinizialetrasformazionebosonicaBogoljubov}$. We have
\begin{theorem}
If the coefficients of the Bogoljubov's bosonic matrix $\eqref{eq: matriceinizialetrasformazionebosonicaBogoljubov}$ satisfy
\begin{equation}
| u_{\textbf{k}} |^2 - | v_{-\textbf{k}} |^2 = 1,
\label{eq: condizionecoefficientibosoniciBogoljubovperaveretrasformazioneinversa}
\end{equation}
then such a matrix is invertible and its inverse is given by the matrix
\begin{equation}
\begin{pmatrix}
u^*_{\textbf{k}} & -v_{-\textbf{k}} \\
-v^*_{\textbf{k}} & u_{-\textbf{k}}
\end{pmatrix}
, \ \textbf{k} \neq \textbf{0}.
\label{eq: matriceinizialetrasformazionebosonicainversaBogoljubov}
\end{equation}
\begin{proof}
From
\begin{equation}
\left[ b_{\textbf{k}} , b_{-\textbf{k}} \right] = 0 ,
\end{equation}
\begin{equation}
\left[ u_{\textbf{k}} a_{\textbf{k}} + v_{\textbf{k}} a^\dagger_{-\textbf{k}} , u_{-\textbf{k}} a_{-\textbf{k}} + v_{-\textbf{k}} a^\dagger_{\textbf{k}} \right] = 0,
\end{equation}
\begin{equation}
u_{\textbf{k}} v_{-\textbf{k}} \left[ a_{\textbf{k}} , a^\dag_{\textbf{k}} \right] + v_{\textbf{k}} u_{-\textbf{k}} \left[ a^\dag_{-\textbf{k}} , a_{-\textbf{k}} \right] = 0,
\end{equation}
\begin{equation}
u_{\textbf{k}} v_{-\textbf{k}} - v_{\textbf{k}} u_{-\textbf{k}} = 0,
\end{equation}
we define
\begin{equation}
r_{\textbf{k}} = \dfrac{v_{\textbf{k}}}{u_{\textbf{k}}},
\end{equation}
which satisfies
\begin{equation}
r_{\textbf{k}} = r_{-\textbf{k}},
\end{equation}
thanks to
\begin{equation}
\dfrac{v_{\textbf{k}}}{u_{\textbf{k}}} = \dfrac{v_{-\textbf{k}}}{u_{-\textbf{k}}}.
\end{equation}
We divide the equation $\eqref{eq: primacondizionecoefficientitrasformazioneBogoljubov}$ by $\left| u_{\textbf{k}} \right|^2$ as follows
\begin{equation}
\dfrac{\left| u_{\textbf{k}} \right|^2}{\left| u_{\textbf{k}} \right|^2} - \dfrac{\left| v_{\textbf{k}} \right|^2}{\left| u_{\textbf{k}} \right|^2} = \dfrac{1}{\left| u_{\textbf{k}} \right|^2},
\end{equation}
then
\begin{equation}
1 - \left| r_{\textbf{k}} \right|^2 = \dfrac{1}{\left| u_{\textbf{k}} \right|^2},
\end{equation}
\begin{equation}
\left| u_{\textbf{k}} \right|^2 \left( 1 - \left| r_{\textbf{k}} \right|^2 \right) = 1,
\end{equation}
\begin{equation}
\left| u_{\textbf{k}} \right|^2 = \dfrac{1}{1 - \left| r_{\textbf{k}} \right|^2}.
\end{equation}
Since $u_{\textbf{k}} \in \mathbb{C}$, its solutions are given by
\begin{equation}
u_{\textbf{k}} = \dfrac{e^{i \gamma_{\textbf{k}}}}{\sqrt{1-\left| r_{\textbf{k}} \right|^2}} ,
\end{equation}
\begin{equation}
u_{-\textbf{k}} = \dfrac{e^{i \gamma_{-\textbf{k}}}}{\sqrt{1-\left| r_{\textbf{k}} \right|^2}} ,
\end{equation}
with $\gamma_{\textbf{k}}, \ \gamma_{-\textbf{k}} \in \mathbb{R}$. We insert such solutions of $\left| u_{\textbf{k}} \right|^2$ in $\eqref{eq: primacondizionecoefficientitrasformazioneBogoljubov}$, i.e.,
\begin{equation}
\dfrac{1}{1-\left| r_{\textbf{k}} \right|^2} - \left| v_{\textbf{k}} \right|^2 = 1,
\end{equation}
\begin{equation}
\dfrac{1 - \left| v_{\textbf{k}} \right|^2 + \left| v_{\textbf{k}} \right|^2 \left| r_{\textbf{k}} \right|^2}{1-\left| r_{\textbf{k}} \right|^2} = 1,
\end{equation}
\begin{equation}
1 - \left| v_{\textbf{k}} \right|^2 + \left| v_{\textbf{k}} \right|^2 \left| r_{\textbf{k}} \right|^2 = 1-\left| r_{\textbf{k}} \right|^2,
\end{equation}
\begin{equation}
- \left| v_{\textbf{k}} \right|^2 + \left| v_{\textbf{k}} \right|^2 \left| r_{\textbf{k}} \right|^2 = -\left| r_{\textbf{k}} \right|^2 ,
\end{equation}
\begin{equation}
\left| v_{\textbf{k}} \right|^2 \left( 1 - \left| r_{\textbf{k}} \right|^2 \right) = \left| r_{\textbf{k}} \right|^2.
\end{equation}
\begin{equation}
\left| v_{\textbf{k}} \right|^2 = \dfrac{\left| r_{\textbf{k}} \right|^2}{ \left( 1 - \left| r_{\textbf{k}} \right|^2 \right)}.
\end{equation}
We choose for the solution of $v_{\textbf{k}}$ the one having the same phase as $u_{\textbf{k}}$, i.e.,
\begin{equation}
v_{\textbf{k}} = \dfrac{e^{i \gamma_{\textbf{k}}} r_{\textbf{k}}}{\sqrt{1-\left| r_{\textbf{k}} \right|^2}},
\end{equation}
\begin{equation}
v_{-\textbf{k}} = \dfrac{e^{i \gamma_{-\textbf{k}}} r_{\textbf{k}}}{\sqrt{1-\left| r_{\textbf{k}} \right|^2}}.
\end{equation}
Now, we multiply $\eqref{eq: operatoredistruzioneBogoljubov}$ by $u^*_{-\textbf{k}}$ and $\eqref{eq: operatorecreazioneBogoljubov}$ by $v_{\textbf{k}}$, i.e.,
\begin{equation}
u^*_{-\textbf{k}} b_{\textbf{k}} = u^*_{-\textbf{k}} u_{\textbf{k}} a_{\textbf{k}} + u^*_{-\textbf{k}} v_{\textbf{k}} a^\dagger_{-\textbf{k}} ,
\end{equation}
\begin{equation}
v_{\textbf{k}} b^\dagger_{-\textbf{k}} = v_{\textbf{k}} v^*_{-\textbf{k}} a_{\textbf{k}} + v_{\textbf{k}} u^*_{-\textbf{k}} a^\dagger_{-\textbf{k}},
\end{equation}
and we subtract the second equation from the first equation as follows
\begin{equation}
u^*_{-\textbf{k}} b_{\textbf{k}} - v_{\textbf{k}} b^\dagger_{-\textbf{k}} = \left( u^*_{-\textbf{k}} u_{\textbf{k}} - v_{\textbf{k}} v^*_{-\textbf{k}} \right) a_{\textbf{k}},
\end{equation}
\begin{equation}
a_{\textbf{k}} = \dfrac{u^*_{-\textbf{k}} b_{\textbf{k}} - v_{\textbf{k}} b^\dagger_{-\textbf{k}}}{u^*_{-\textbf{k}} u_{\textbf{k}} - v_{\textbf{k}} v^*_{-\bar{k}}}.
\label{eq: operatoredistruzioneinvertitoinfunzionedioperatoriBogoljuboviniziale}
\end{equation}
Now, the denominator of the above equation can be manipulated as
\begin{align}
u^*_{-\textbf{k}} u_{\textbf{k}} - v^*_{-\textbf{k}} v_{\textbf{k}} &= \dfrac{e^{- i \gamma_{-\textbf{k}}}}{\sqrt{1-\left| r_{\textbf{k}} \right|^2}} \dfrac{e^{i \gamma_{\textbf{k}}}}{\sqrt{1-\left| r_{\textbf{k}} \right|^2}} - \dfrac{e^{- i \gamma_{-\textbf{k}}} r^*_{\textbf{k}}}{\sqrt{1-\left| r_{\textbf{k}} \right|^2}} \dfrac{e^{i \gamma_{\textbf{k}}} r_{\textbf{k}}}{\sqrt{1-\left| r_{\textbf{k}} \right|^2}} = \notag \\
&= \dfrac{e^{i (\gamma_{\textbf{k}} - \gamma_{-\textbf{k}})}}{1-\left| r_{\textbf{k}} \right|^2} - \dfrac{e^{i (\gamma_{\textbf{k}} - \gamma_{-\textbf{k}})} \left| r_{\textbf{k}} \right|^2}{1-\left| r_{\textbf{k}} \right|^2} = \notag \\
&= \dfrac{e^{i (\gamma_{\textbf{k}} - \gamma_{-\textbf{k}})} \left( 1 - \left| r_{\textbf{k}} \right|^2 \right)}{1 - \left| r_{\textbf{k}} \right|^2} = \notag \\
&= e^{i (\gamma_{\textbf{k}} - \gamma_{-\textbf{k}})},
\end{align}
then
\begin{align}
\dfrac{u^*_{-\textbf{k}}}{u^*_{-\textbf{k}} u_{\textbf{k}} - v^*_{-\textbf{k}} v_{\textbf{k}}} &= \dfrac{e^{-i \gamma_{-\bar{k}}}}{\sqrt{1 - \left| r_{\textbf{k}} \right|^2} e^{i (\gamma_{\textbf{k}} - \gamma_{-\textbf{k}})}} = \notag \\
&= \dfrac{e^{-i \gamma_{\textbf{k}}}}{\sqrt{1 - \left| r_{\textbf{k}} \right|^2}} = \notag \\
&= u^*_{\textbf{k}},
\end{align}
\begin{align}
\dfrac{v_{\textbf{k}}}{u^*_{-\textbf{k}} u_{\textbf{k}} - v^*_{-\textbf{k}} v_{\textbf{k}}} &= \dfrac{e^{i \gamma_{\textbf{k}}} r_{\bar{k}}}{\sqrt{1 - \left| r_{\textbf{k}} \right|^2} e^{i (\gamma_{\textbf{k}} - \gamma_{-\textbf{k}})}} = \notag \\
&= \dfrac{e^{i \gamma_{-\textbf{k}}} r_{\textbf{k}}}{\sqrt{1 - \left| r_{\textbf{k}} \right|^2}} = \notag \\
&= v_{-\textbf{k}}.
\end{align}
We insert the two results above in $\eqref{eq: operatoredistruzioneinvertitoinfunzionedioperatoriBogoljuboviniziale}$, i.e.,
\begin{align}
a_{\textbf{k}} &= \dfrac{u^*_{-\textbf{k}}}{u^*_{-\textbf{k}} u_{\textbf{k}} - v_{\textbf{k}} v^*_{-\textbf{k}}} b_{\textbf{k}} - \dfrac{v_{\textbf{k}}}{u^*_{-\textbf{k}} u_{\textbf{k}} - v_{\textbf{k}} v^*_{-\textbf{k}}} b^\dagger_{-\textbf{k}} = \notag \\
&= u^*_{\textbf{k}} b_{\textbf{k}} - v_{-\textbf{k}} b^\dagger_{-\textbf{k}},
\end{align}
which together with its complex conjugate provide
\begin{equation}
\begin{cases}
a_{\textbf{k}} = u^*_{\textbf{k}} b_{\textbf{k}} - v_{-\textbf{k}} b^\dagger_{-\textbf{k}} \\
a^\dagger_{\textbf{k}} = u_{\textbf{k}} b^\dagger_{\textbf{k}} - v^*_{-\textbf{k}} b_{-\textbf{k}} 
\end{cases},
\label{eq: operatoriinfunzionidioperatoridiBogoljubov}
\end{equation}
or equivalently
\begin{equation}
\begin{cases}
a_{\textbf{k}} = u^*_{\textbf{k}} b_{\textbf{k}} - v_{-\textbf{k}} b^\dagger_{-\textbf{k}} \\
a^\dagger_{-\textbf{k}} = u_{-\textbf{k}} b^\dagger_{-\textbf{k}} - v^*_{\textbf{k}} b_{\textbf{k}} 
\end{cases},
\end{equation}
where the momentum in the creation operator $a^\dagger_{\textbf{k}}$ has been inverted. Such transformations are given in matrix form as
\begin{equation}
\begin{pmatrix}
a_{\textbf{k}} \\
a^\dagger_{-\textbf{k}}
\end{pmatrix}
=
\begin{pmatrix}
u^*_{\textbf{k}} & -v_{-\textbf{k}} \\
-v^*_{\textbf{k}} & u_{-\textbf{k}}
\end{pmatrix}
\begin{pmatrix}
b_{\textbf{k}} \\
b^\dagger_{-\textbf{k}}
\end{pmatrix} 
, \ \textbf{k} \neq \textbf{0}.
\label{eq: trasformazioneinversaBogoljubovbosoni}
\end{equation}
Let us verify that the operators $\eqref{eq: operatoriinfunzionidioperatoridiBogoljubov}$ satisfy a commutative algebra, indeed
\begin{align}
\left[ a_{\textbf{k}'} , a^\dagger_{\textbf{k}} \right] &= \left[ u^*_{\textbf{k}'} b_{\textbf{k}'} - v_{-\textbf{k}'} b^\dagger_{-\textbf{k}'} , u_{\textbf{k}} b^\dagger_{\textbf{k}} - v^*_{-\textbf{k}} b_{-\textbf{k}} \right] = \notag \\
&= \left[ u^*_{\textbf{k}'} b_{\textbf{k}'} , u_{\textbf{k}} b^\dagger_{\textbf{k}} \right] + \left[ v_{-\textbf{k}'} b^\dagger_{-\textbf{k}'} , v^*_{-\textbf{k}} b_{-\textbf{k}} \right] = \notag \\
&= u^*_{\textbf{k}'} u_{\textbf{k}} \delta_{\textbf{k}',\textbf{k}} - v_{-\textbf{k}'} v^*_{-\textbf{k}} \delta_{-\textbf{k}',-\textbf{k}},
\end{align}
that for $\textbf{k} \neq \textbf{k}'$ is null, while for $\textbf{k}=\textbf{k}'$ given that $\delta_{-\textbf{k}',-\textbf{k}} = \delta_{\textbf{k}',\textbf{k}}$ becomes
\begin{align}
\left[ a_{\textbf{k}} , a^\dagger_{\textbf{k}} \right] &= u^*_{\textbf{k}} u_{\textbf{k}} - v_{-\textbf{k}} v^*_{-\textbf{k}} = \notag \\
&= \left| u_{\textbf{k}} \right|^2 - | v_{-\textbf{k}} |^2,
\end{align}
which is equal to the identity if the $\eqref{eq: condizionecoefficientibosoniciBogoljubovperaveretrasformazioneinversa}$ holds. Note that $\eqref{eq: condizionecoefficientibosoniciBogoljubovperaveretrasformazioneinversa}$ is similar to $\eqref{eq: primacondizionecoefficientitrasformazioneBogoljubov}$ except for a sign of the momentum of $v$. However, an Ansatz for coefficients $u_{\textbf{k}}$ and $v_{\textbf{k}}$ (equations $\eqref{eq: condizionecoefficientitrasformazioneBogoljubovbosoni1}$, $\eqref{eq: condizionecoefficientitrasformazioneBogoljubovbosoni2}$) will be set in this section, according to which $\eqref{eq: primacondizionecoefficientitrasformazioneBogoljubov}$ and $\eqref{eq: condizionecoefficientibosoniciBogoljubovperaveretrasformazioneinversa}$ become equal.
\end{proof}
\end{theorem}
Now, let us insert the above solutions into $\eqref{eq: hamiltonianadiBogoljubov}$.
\begin{align}
\hat{\mathcal{H}} &= - \dfrac{U N_0^2}{2V} + \sum_{\textbf{k} \neq \textbf{0}} \left( \dfrac{\hslash^2 \textbf{k}^2}{2m} + \dfrac{U N_0}{V} \right) \left[ \left( u_{\textbf{k}} b^\dagger_{\textbf{k}} - v^*_{-\textbf{k}} b_{-\textbf{k}} \right) \left( u^*_{\textbf{k}} b_{\textbf{k}} - v_{-\textbf{k}} b^\dagger_{-\textbf{k}} \right) \right] + \notag \\
&+ \dfrac{U N_0}{2V} \sum_{\textbf{k} \neq \textbf{0}} \left[ \left( u_{\textbf{k}} b^\dagger_{\textbf{k}} - v^*_{-\textbf{k}} b_{-\textbf{k}} \right) \left( u_{-\textbf{k}} b^\dagger_{-\textbf{k}} - v^*_{\textbf{k}} b_{\textbf{k}} \right) + \left( u^*_{-\textbf{k}} b_{-\textbf{k}} - v_{\textbf{k}} b^\dagger_{\textbf{k}} \right) \left( u^*_{\textbf{k}} b_{\textbf{k}} - v_{-\textbf{k}} b^\dagger_{-\textbf{k}} \right) \right] = \notag \\
&= - \dfrac{U N_0^2}{2V} + \sum_{\textbf{k} \neq \textbf{0}} \left( \dfrac{\hslash^2 \textbf{k}^2}{2m} + \dfrac{U N_0}{V} \right) \left[ \left| u_{\textbf{k}} \right|^2 b^\dagger_{\textbf{k}} b_{\textbf{k}}  + \left| v_{-\textbf{k}} \right|^2 b_{-\textbf{k}} b^\dagger_{-\textbf{k}} - u_{\textbf{k}} v_{-\textbf{k}} b^\dagger_{\textbf{k}} b^\dagger_{-\textbf{k}} - u^*_{\textbf{k}} v^*_{-\textbf{k}} b_{-\textbf{k}} b_{\textbf{k}} \right] + \notag \\
&+ \dfrac{U N_0}{2V} \sum_{\textbf{k} \neq \textbf{0}} \bigg[ \left( u_{\textbf{k}} u_{-\textbf{k}} b^\dagger_{\textbf{k}} b^\dagger_{-\textbf{k}} - u_{\textbf{k}} v^*_{\textbf{k}} b^\dagger_{\textbf{k}} b_{\textbf{k}} - u_{-\textbf{k}} v^*_{-\textbf{k}} b_{-\textbf{k}} b^\dagger_{-\textbf{k}} + v^*_{-\textbf{k}} v^*_{\textbf{k}} b_{-\textbf{k}} b_{\textbf{k}} \right) + \notag \\
&+ \left( u^*_{-\textbf{k}} u^*_{\textbf{k}} b_{-\textbf{k}} b_{\textbf{k}} - u^*_{-\textbf{k}} v_{-\textbf{k}} b_{-\textbf{k}} b^\dagger_{-\textbf{k}} - u^*_{\textbf{k}} v_{\textbf{k}} b^\dagger_{\textbf{k}} b_{\textbf{k}} + v_{\textbf{k}} v_{-\textbf{k}} b^\dagger_{\textbf{k}} b^\dagger_{-\textbf{k}} \right) \bigg].
\end{align}
We would like to simplify the form of the above Hamiltonian. Let us assume
\begin{equation}
u_{\textbf{k}} , \ v_{\textbf{k}} \in \mathbb{R},
\end{equation}
\begin{equation}
u_{-\textbf{k}} = u_{\textbf{k}},
\label{eq: condizionecoefficientitrasformazioneBogoljubovbosoni1}
\end{equation}
\begin{equation}
v_{-\textbf{k}} = v_{\textbf{k}},
\label{eq: condizionecoefficientitrasformazioneBogoljubovbosoni2}
\end{equation}
and write
\begin{align}
\hat{\mathcal{H}} &= - \dfrac{U N_0^2}{2V} + \sum_{\textbf{k} \neq \textbf{0}} \left( \dfrac{\hslash^2 \textbf{k}^2}{2m} + \dfrac{U N_0}{V} \right) \left[ u_{\textbf{k}}^2 b^\dagger_{\textbf{k}} b_{\textbf{k}} +  v_{\textbf{k}}^2 b_{-\textbf{k}} b^\dagger_{-\textbf{k}} - u_{\textbf{k}} v_{\textbf{k}} \left( b^\dagger_{\textbf{k}} b^\dagger_{-\textbf{k}} + b_{-\textbf{k}} b_{\textbf{k}} \right) \right] + \notag \\
&+ \dfrac{U N_0}{2V} \sum_{\textbf{k} \neq \textbf{0}} \bigg[ \left( u^2_{\textbf{k}} b^\dagger_{\textbf{k}} b^\dagger_{-\bar{k}} - u_{\textbf{k}} v_{\textbf{k}} b^\dagger_{\textbf{k}} b_{\textbf{k}} - u_{\textbf{k}} v_{\textbf{k}} b_{-\textbf{k}} b^\dagger_{-\textbf{k}} + v^2_{\textbf{k}} b_{-\textbf{k}} b_{\textbf{k}} \right) + \notag \\
&+ \left( u^2_{\textbf{k}} b_{-\textbf{k}} b_{\textbf{k}} - u_{\textbf{k}} v_{\textbf{k}} b_{-\textbf{k}} b^\dagger_{-\textbf{k}} - u_{\textbf{k}} v_{\textbf{k}} b^\dagger_{\bar{k}} b_{\textbf{k}} + v^2_{\textbf{k}} b^\dagger_{\textbf{k}} b^\dagger_{-\textbf{k}} \right) \bigg].
\end{align}
We use the bosonic algebra $b_{-\textbf{k}} b^\dagger_{-\textbf{k}} = \mathds{1} + b^\dagger_{-\textbf{k}} b_{-\textbf{k}}$ and set $- \textbf{k} \rightarrow \textbf{k}$ in the number operators, as follows
\begin{align}
\hat{\mathcal{H}} &= - \dfrac{U N_0^2}{2V} + \sum_{\textbf{k} \neq \textbf{0}} \left( \dfrac{\hslash^2 \textbf{k}^2}{2m} + \dfrac{U N_0}{V} \right) \left[ \left( u_{\textbf{k}}^2 + v^2_{\textbf{k}} \right) b^\dagger_{\textbf{k}} b_{\textbf{k}} + v^2_{\textbf{k}} - u_{\textbf{k}} v_{\textbf{k}} \left( b^\dagger_{\textbf{k}} b^\dagger_{-\textbf{k}} + b_{-\textbf{k}} b_{\textbf{k}} \right) \right] + \notag \\
&+ \dfrac{U N_0}{2V} \sum_{\textbf{k} \neq \textbf{0}} \bigg[ \left( u^2_{\textbf{k}} b^\dagger_{\textbf{k}} b^\dagger_{-\textbf{k}} - u_{\textbf{k}} v_{\textbf{k}} b^\dagger_{\textbf{k}} b_{\textbf{k}} - u_{\textbf{k}} v_{\textbf{k}} b^\dagger_{\textbf{k}} b_{\textbf{k}} - u_{\textbf{k}} v_{\textbf{k}} + v^2_{\textbf{k}} b_{-\textbf{k}} b_{\textbf{k}} \right) + \notag \\
&+ \left( u^2_{\textbf{k}} b_{-\textbf{k}} b_{\textbf{k}} - u_{\textbf{k}} v_{\textbf{k}} b^\dagger_{\textbf{k}} b_{\textbf{k}} - u_{\textbf{k}} v_{\textbf{k}} - u_{\textbf{k}} v_{\textbf{k}} b^\dagger_{\textbf{k}} b_{\textbf{k}} + v^2_{\textbf{k}} b^\dagger_{\textbf{k}} b^\dagger_{-\textbf{k}} \right) \bigg] = \notag \\
&= - \dfrac{U N_0^2}{2V} + \sum_{\textbf{k} \neq \textbf{0}} \left( \dfrac{\hslash^2 \textbf{k}^2}{2m} + \dfrac{U N_0}{V} \right) \left[ \left( u_{\textbf{k}}^2 + v^2_{\textbf{k}} \right) b^\dagger_{\textbf{k}} b_{\textbf{k}} - u_{\textbf{k}} v_{\textbf{k}} \left( b^\dagger_{\textbf{k}} b^\dagger_{-\textbf{k}} + b_{-\textbf{k}} b_{\textbf{k}} \right) + v^2_{\textbf{k}} \right] + \notag \\
&+ \dfrac{U N_0}{2V} \sum_{\textbf{k} \neq \textbf{0}} \left[ \left( u^2_{\textbf{k}} + v^2_{\textbf{k}} \right) \left( b^\dagger_{\textbf{k}} b^\dagger_{-\textbf{k}} + b_{-\textbf{k}} b_{\textbf{k}} \right) - 4 u_{\textbf{k}} v_{\textbf{k}} b^\dagger_{\textbf{k}} b_{\textbf{k}} - 2 u_{\textbf{k}} v_{\textbf{k}} \right].
\end{align}
We define the constant
\begin{equation}
\tilde{\mathcal{E}} = - \dfrac{U N_0^2}{2V} + \sum_{\textbf{k} \neq \textbf{0}} \left[ \left( \dfrac{\hslash^2 \textbf{k}^2}{2m} + \dfrac{U N_0}{V} \right) v^2_{\textbf{k}} - \dfrac{U N_0}{V} u_{\textbf{k}} v_{\textbf{k}} \right],
\end{equation}
and the Hamiltonian becomes
\begin{align}
\hat{\mathcal{H}} &= \tilde{\mathcal{E}} + \sum_{\textbf{k} \neq \textbf{0}} \left[ \left( \dfrac{\hslash^2 \textbf{k}^2}{2m} + \dfrac{U N_0}{V} \right) \left( u^2_{\textbf{k}} + v^2_{\textbf{k}} \right) - 2 \dfrac{U N_0}{V} u_{\textbf{k}} v_{\textbf{k}} \right] b^\dagger_{\textbf{k}} b_{\textbf{k}} + \notag \\
&- \sum_{\textbf{k} \neq \textbf{0}} \left[ \left( \dfrac{\hslash^2 \textbf{k}^2}{2m} + \dfrac{U N_0}{V} \right) u_{\textbf{k}} v_{\textbf{k}} - \dfrac{U N_0}{2 V} \left( u^2_{\textbf{k}} + v^2_{\textbf{k}} \right) \right] \left( b^\dagger_{\textbf{k}} b^\dagger_{-\textbf{k}} + b_{-\textbf{k}} b_{\textbf{k}} \right).
\end{align}
If we set
\begin{equation}
F \equiv \dfrac{\hslash^2 \textbf{k}^2}{2m} + \dfrac{U N_0}{V},
\end{equation}
\begin{equation}
T \equiv \dfrac{U N_0}{V},
\end{equation}
then the Hamiltonian
\begin{equation}
\hat{\mathcal{H}} = \tilde{\mathcal{E}} + \sum_{\textbf{k} \neq \textbf{0}} \left[ F \left( u^2_{\textbf{k}} + v^2_{\textbf{k}} \right) - T \left( 2 u_{\textbf{k}} v_{\textbf{k}} \right) \right] b^\dagger_{\textbf{k}} b_{\textbf{k}} - \sum_{\textbf{k} \neq \textbf{0}} \left[ F u_{\textbf{k}} v_{\textbf{k}} - \dfrac{T}{2} \left( u^2_{\textbf{k}} + v^2_{\textbf{k}} \right) \right] \left( b^\dagger_{\textbf{k}} b^\dagger_{-\textbf{k}} + b_{-\textbf{k}} b_{\textbf{k}} \right)
\end{equation}
shows a similarity between the coefficient of $b^\dagger_{\textbf{k}} b_{\textbf{k}}$ and that of $\left( b^\dagger_{\textbf{k}} b^\dagger_{-\textbf{k}} + b_{-\textbf{k}} b_{\textbf{k}} \right)$. Indeed, first we multiply and divide by $2$ the number operator and manipulate as follows
\begin{equation}
\dfrac{1}{2} \sum_{\textbf{k} \neq \textbf{0}} 2 b^\dagger_{\textbf{k}} b_{\textbf{k}} = \dfrac{1}{2} \sum_{\textbf{k} \neq \textbf{0}} \left( b^\dagger_{\textbf{k}} b_{\textbf{k}} + b^\dagger_{-\textbf{k}} b_{-\textbf{k}} \right) ,
\end{equation}
and we write
\begin{equation}
\hat{\mathcal{H}} = \tilde{\mathcal{E}} + \dfrac{1}{2} \sum_{\textbf{k} \neq \textbf{0}} \left[ F \left( u^2_{\textbf{k}} + v^2_{\textbf{k}} \right) - T \left( 2 u_{\textbf{k}} v_{\textbf{k}} \right) \right] \left( b^\dagger_{\textbf{k}} b_{\textbf{k}} + b^\dagger_{-\textbf{k}} b_{-\textbf{k}} \right) - \dfrac{1}{2} \sum_{\textbf{k} \neq \textbf{0}} \left[ F \left( 2 u_{\textbf{k}} v_{\textbf{k}} \right) - T \left( u^2_{\textbf{k}} + v^2_{\textbf{k}} \right) \right] \left( b^\dagger_{\textbf{k}} b^\dagger_{-\textbf{k}} + b_{-\textbf{k}} b_{\textbf{k}} \right).
\end{equation}
Consequently, the hamiltonian is diagonal if
\begin{align}
F \left( 2 u_{\textbf{k}} v_{\textbf{k}} \right) - T \left( u^2_{\textbf{k}} + v^2_{\textbf{k}} \right) &\equiv \left( \dfrac{\hslash^2 \textbf{k}^2}{2m} + \dfrac{U N_0}{V} \right) 2 u_{\textbf{k}} v_{\textbf{k}} - \dfrac{U N_0}{V} \left( u^2_{\textbf{k}} + v^2_{\textbf{k}} \right) = \notag \\
&= 0, \ \forall \ \textbf{k} \neq \textbf{0}.
\label{eq: condizionehamiltonianaBogoljubovdiagonale}
\end{align}
Now, given $\eqref{eq: primacondizionecoefficientitrasformazioneBogoljubov}$, we set the Ansatz 
\begin{equation}
u_{\textbf{k}} = \cosh(\theta_{\textbf{k}}),
\end{equation}
\begin{equation}
v_{\textbf{k}} = \sinh(\theta_{\textbf{k}}),
\end{equation}
and we insert
\begin{equation}
2 \cosh(\theta_{\textbf{k}}) \sinh(\theta_{\textbf{k}}) = \sinh(2 \theta_{\textbf{k}}),
\end{equation}
\begin{equation}
\cosh^2(\theta_{\textbf{k}}) + \sinh^2(\theta_{\textbf{k}}) = \cosh(2 \theta_{\textbf{k}}),
\end{equation}
into the equation $\eqref{eq: condizionehamiltonianaBogoljubovdiagonale}$, then
\begin{equation}
F \sinh(2 \theta_{\textbf{k}}) = T \cosh(2 \theta_{\textbf{k}}),
\end{equation}
\begin{equation}
F \sinh(2 \theta_{\textbf{k}}) = T \sqrt{1 + \sinh^2(2 \theta_{\textbf{k}})},
\end{equation}
\begin{equation}
F^2 \sinh^2(2 \theta_{\textbf{k}}) = T^2 \left( 1 + \sinh^2(2 \theta_{\textbf{k}}) \right),
\end{equation}
\begin{equation}
\sinh^2(2 \theta_{\textbf{k}}) = \dfrac{T^2}{F^2-T^2},
\end{equation}
\begin{align}
\cosh^2(2 \theta_{\textbf{k}}) &= 1 + \sinh^2(2 \theta_{\textbf{k}}) = \notag \\
&= \dfrac{F^2 - T^2 + T^2}{F^2-T^2} = \notag \\
&= \dfrac{F^2}{F^2 - T^2},
\end{align}
\begin{equation}
\begin{cases}
\sinh(2 \theta_{\textbf{k}}) = \dfrac{T}{\sqrt{F^2-T^2}} \\ 
\cosh(2 \theta_{\textbf{k}}) = \dfrac{F}{\sqrt{F^2-T^2}}
\end{cases}.
\label{eq: soluzionicondizionehamiltonianaBogoljubovdiagonale}
\end{equation}
In conclusion, 
\begin{align}
\hat{\mathcal{H}} &= \tilde{\mathcal{E}} + \dfrac{1}{2} \sum_{\textbf{k} \neq \textbf{0}} \left[ F \left( u^2_{\textbf{k}} + v^2_{\textbf{k}} \right) - T \left( 2 u_{\textbf{k}} v_{\textbf{k}} \right) \right] \left( b^\dagger_{\textbf{k}} b_{\textbf{k}} + b^\dagger_{-\textbf{k}} b_{-\textbf{k}} \right) = \notag \\
&= \tilde{\mathcal{E}} + \dfrac{1}{2} \sum_{\textbf{k} \neq \textbf{0}} \left[ F \cosh(2 \theta_{\textbf{k}}) - T \sinh(2 \theta_{\textbf{k}}) \right] \left( b^\dagger_{\textbf{k}} b_{\textbf{k}} + b^\dagger_{-\textbf{k}} b_{-\textbf{k}} \right) = \notag \\
&= \tilde{\mathcal{E}} + \dfrac{1}{2} \sum_{\textbf{k} \neq \textbf{0}} \left[ F \dfrac{F}{\sqrt{F^2-T^2}} - T \dfrac{T}{\sqrt{F^2-T^2}} \right] \left( b^\dagger_{\textbf{k}} b_{\bar{k}} + b^\dagger_{-\textbf{k}} b_{-\textbf{k}} \right) = \notag \\
&= \tilde{\mathcal{E}} + \dfrac{1}{2} \sum_{\textbf{k} \neq \textbf{0}} \left[ \dfrac{F^2}{\sqrt{F^2-T^2}} - \dfrac{T^2}{\sqrt{F^2-T^2}} \right] \left( b^\dagger_{\textbf{k}} b_{\textbf{k}} + b^\dagger_{-\textbf{k}} b_{-\textbf{k}} \right) = \notag \\
&= \tilde{\mathcal{E}} + \dfrac{1}{2} \sum_{\textbf{k} \neq \textbf{0}} \left[ \sqrt{F^2-T^2} \right] \left( b^\dagger_{\bar{k}} b_{\textbf{k}} + b^\dagger_{-\textbf{k}} b_{-\textbf{k}} \right) = \notag \\
&= \tilde{\mathcal{E}} + \dfrac{1}{2} \sum_{\textbf{k} \neq \textbf{0}} \left[ \sqrt{\left( \dfrac{\hslash^2 \textbf{k}^2}{2m} + \dfrac{U N_0}{V} \right)^2 - \left( \dfrac{U N_0}{V} \right)^2} \right] \left( b^\dagger_{\textbf{k}} b_{\textbf{k}} + b^\dagger_{-\textbf{k}} b_{-\textbf{k}} \right) = \notag \\
&= \tilde{\mathcal{E}} + \dfrac{1}{2} \sum_{\textbf{k} \neq \textbf{0}} \left[ \sqrt{\left( \dfrac{\hslash^2 \textbf{k}^2}{2m} \right)^2 + \dfrac{U^2 N^2_0}{V^2} + 2 \dfrac{\hslash^2 \textbf{k}^2}{2m} \dfrac{U N_0}{V} - \dfrac{U^2 N_0^2}{V^2}} \right] \left( b^\dagger_{\textbf{k}} b_{\textbf{k}} + b^\dagger_{-\textbf{k}} b_{-\textbf{k}} \right) = \notag \\
&= \tilde{\mathcal{E}} + \dfrac{1}{2} \sum_{\textbf{k} \neq \textbf{0}} \left[ \sqrt{\dfrac{\hslash^2 \textbf{k}^2}{2m} \left( \dfrac{\hslash^2 \textbf{k}^2}{2m} + 2 \dfrac{U N_0}{V} \right)} \right] \left( b^\dagger_{\textbf{k}} b_{\textbf{k}} + b^\dagger_{-\textbf{k}} b_{-\textbf{k}} \right) = \notag \\
&= \tilde{\mathcal{E}} + \dfrac{1}{2} \sum_{\textbf{k} \neq \textbf{0}} \mathcal{E}_{\textbf{k}} \left( b^\dagger_{\textbf{k}} b_{\textbf{k}} + b^\dagger_{-\textbf{k}} b_{-\textbf{k}} \right) = \notag \\
&= \tilde{\mathcal{E}} + \dfrac{1}{2} \sum_{\textbf{k} \neq \textbf{0}} \mathcal{E}_{\textbf{k}} \left( 2 b^\dagger_{\textbf{k}} b_{\textbf{k}} \right) ,
\end{align}
and we finally obtain the diagonalized Bogoljubov Hamiltonian
\begin{equation}
\hat{\mathcal{H}} = \tilde{\mathcal{E}} + \sum_{\textbf{k} \neq \textbf{0}} \mathcal{E}_{\textbf{k}} b^\dagger_{\textbf{k}} b_{\textbf{k}},
\label{eq: hamiltonianaBogoljubovdiagonalizzata}
\end{equation}
with
\begin{equation}
\mathcal{E}_{\textbf{k}} = \sqrt{\dfrac{\hslash^2 \textbf{k}^2}{2m} \left( \dfrac{\hslash^2 \textbf{k}^2}{2m} + 2 \dfrac{U N_0}{V} \right)}.
\end{equation}
Bogoljubov's bosonic transformations are given by
\begin{align}
\begin{pmatrix}
b_{\textbf{k}} \\
b^\dagger_{-\textbf{k}}
\end{pmatrix}
&=
\begin{pmatrix}
u_{\textbf{k}} & v_{\textbf{k}} \\
v_{\textbf{k}} & u_{\textbf{k}}
\end{pmatrix}
\begin{pmatrix}
a_{\textbf{k}} \\
a^\dagger_{-\textbf{k}}
\end{pmatrix} 
\equiv \notag \\
&\equiv
\begin{pmatrix}
\cosh(\theta_{\textbf{k}}) & \sinh(\theta_{\textbf{k}}) \\
\sinh(\theta_{\textbf{k}}) & \cosh(\theta_{\textbf{k}})
\end{pmatrix}
\begin{pmatrix}
a_{\textbf{k}} \\
a^\dagger_{-\textbf{k}}
\end{pmatrix}
, \ \textbf{k} \neq \textbf{0} ,
\end{align}
\begin{align}
\begin{pmatrix}
a_{\textbf{k}} \\
a^\dagger_{-\textbf{k}}
\end{pmatrix}
&=
\begin{pmatrix}
u_{\textbf{k}} & -v_{\textbf{k}} \\
-v_{\textbf{k}} & u_{\textbf{k}}
\end{pmatrix}
\begin{pmatrix}
b_{\textbf{k}} \\
b^\dagger_{-\textbf{k}}
\end{pmatrix}
\equiv \notag \\
&\equiv
\begin{pmatrix}
\cosh(\theta_{\textbf{k}}) & - \sinh(\theta_{\textbf{k}}) \\
- \sinh(\theta_{\textbf{k}}) & \cosh(\theta_{\textbf{k}})
\end{pmatrix}
\begin{pmatrix}
b_{\textbf{k}} \\
b^\dagger_{-\textbf{k}}
\end{pmatrix}
, \ \textbf{k} \neq \textbf{0}.
\end{align}
Note that
\begin{align}
\begin{pmatrix}
u_{\textbf{k}} & v_{\textbf{k}} \\
v_{\textbf{k}} & u_{\textbf{k}}
\end{pmatrix}
\begin{pmatrix}
u_{\textbf{k}} & -v_{\textbf{k}} \\
-v_{\textbf{k}} & u_{\textbf{k}}
\end{pmatrix}
&= 
\begin{pmatrix}
u^2_{\textbf{k}}-v^2_{\textbf{k}} & - u_{\textbf{k}} v_{\textbf{k}} \\
u_{\textbf{k}} v_{\textbf{k}} - u_{\textbf{k}} v_{\textbf{k}} & - v^2_{\textbf{k}} + u^2_{\textbf{k}}
\end{pmatrix}
= \notag \\
&=
\begin{pmatrix}
1 & 0 \\
0 & 1
\end{pmatrix},
\end{align}
\begin{align}
&
\begin{pmatrix}
\cosh(\theta_{\textbf{k}}) & \sinh(\theta_{\textbf{k}}) \\
\sinh(\theta_{\textbf{k}}) & \cosh(\theta_{\textbf{k}})
\end{pmatrix}
\begin{pmatrix}
\cosh(\theta_{\textbf{k}}) & - \sinh(\theta_{\textbf{k}}) \\
- \sinh(\theta_{\textbf{k}}) & \cosh(\theta_{\textbf{k}})
\end{pmatrix}
= \notag \\
&=
\begin{pmatrix}
\cosh^2(\theta_{\textbf{k}}) - \sinh^2(\theta_{\textbf{k}}) & - \cosh(\theta_{\textbf{k}}) \sinh(\theta_{\textbf{k}}) + \sinh(\theta_{\textbf{k}}) \cosh(\theta_{\textbf{k}}) \\
\sinh(\theta_{\textbf{k}}) \cosh(\theta_{\textbf{k}}) - \cosh(\theta_{\textbf{k}}) \sinh(\theta_{\textbf{k}}) & - \sinh^2(\theta_{\textbf{k}}) + \cosh^2(\theta_{\textbf{k}})
\end{pmatrix}
= \notag \\
&=
\begin{pmatrix}
1 & 0 \\
0 & 1
\end{pmatrix},
\end{align}
as it must be.
\begin{remark}[Low and high momentum limits of the Bogoljubov bosonic spectrum] Let us examine the behavior of the eigenvalues $\eqref{eq: energiespettroBogoljubov}$ in the limits of high and low momenta. Note that for low momenta we have
\begin{align}
\mathcal{E}_{\textbf{k}} & \underset{k \to 0}{\sim} 2 \dfrac{U N_0}{V} \mathcal{E}^0_{\textbf{k}} = \notag \\
& \ \ = \sqrt{\dfrac{U N_0 \hslash^2 k^2}{mV}} = \notag \\
& \ \ = \hslash k \sqrt{\dfrac{U N_0}{mV}},
\end{align}
i.e., in Bogoljubov's theory, for small momenta, the interactions between bosons imply a linear dispersion relation. From
\begin{equation}
\omega = k c,
\end{equation}
\begin{equation}
\mathcal{E} = \hslash k c,
\end{equation}
we note an analogy with the dispersion relation for small momenta, where the velocity of the bosons is given by
\begin{equation}
c = \sqrt{\dfrac{U N_0}{mV}},
\label{eq: velocitàspettroBogoljubovpiccolimomenti}
\end{equation}
\begin{align}
\left[ \dfrac{U N_0}{m V} \right] &= \dfrac{\left[ E \right]}{\left[ m \right] \left[ L \right]^3} = \notag \\
&= \dfrac{\left[ m \right] \left[ L \right]^2}{\left[ m \right] \left[ T \right]^2} = \notag \\
&= \dfrac{\left[ L \right]^2}{\left[ T \right]^2}.
\end{align}
On the other hand, for high moments, bosons can be regarded as free particles.
\end{remark}
\section{Fundamental state of the Bogoljubov's bosonic Hamiltonian}
Now, since the Hamiltonian is diagonal, we can compute its fundamental state $|\Psi \rangle$, which satisfies
\begin{equation}
b_{\textbf{k}} |\Psi\rangle = 0, \ \forall \ \textbf{k}.
\end{equation}
To do this, consider a single harmonic oscillator Hamiltonian, i.e., $\mathcal{H}_1 = \hslash \omega a^\dagger a$, and we define the operator
\begin{equation}
S = z \left( a a - a^\dagger a^\dagger \right), \ z \in \mathbb{R}.
\end{equation}
The parameter \( z \) is chosen such that the transformation \( e^{S} \) is unitary, which requires \( S^\dagger = -S \). In order to analyze the transformed operator \( e^S \mathcal{H}_1 e^{-S} \), we define
\begin{equation}
h(z) = e^{S(z)} a e^{-S(z)},
\end{equation}
and we compute the first derivative of $h(z)$ with respect to the $z$, i.e.,
\begin{align}
\dfrac{dh(z)}{dz} &= \dfrac{dS(z)}{dz} e^{S(z)} a e^{-S(z)} - e^{S(z)} a e^{-S(z)} \dfrac{dS(z)}{dz} = \notag \\ 
&= e^{S(z)} \dfrac{dS(z)}{dz} a e^{-S(z)} - e^{S(z)} a \dfrac{dS(z)}{dz} e^{-S(z)} = \notag \\
&= e^{S(z)} \left( \dfrac{dS(z)}{dz} a - a \dfrac{dS(z)}{dz} \right) e^{-S(z)} = \notag \\
&= e^{S(z)} \left\lbrace \left( a a - a^\dagger a^\dagger \right) a - a \left( a a - a^\dagger a^\dagger \right) \right\rbrace e^{-S(z)} = \notag \\
&= e^{S(z)} \left( a a a - a^\dagger a^\dagger a - a a a + a a^\dagger a^\dagger \right) e^{-S(z)} = \notag \\
&= e^{S(z)} \left( - a^\dagger a^\dagger a + a a^\dagger a^\dagger \right) e^{-S(z)} ,
\end{align}
and we use the bosonic algebra as follows
\begin{align}
\dfrac{dh(z)}{dz} &= e^{S(z)} \left( - a^\dagger \left( - \mathds{1} + a a^\dagger \right) + \left( a^\dagger a + \mathds{1} \right) a^\dagger \right) e^{-S(z)} = \notag \\
&= e^{S(z)} \left( a^\dagger - a^\dagger a a^\dagger + a^\dagger a a^\dagger + a^\dagger \right) e^{-S(z)} ,
\end{align}
then
\begin{equation}
\dfrac{dh(z)}{dz} = 2 e^{S(z)} a^\dagger e^{-S(z)}. 
\end{equation}
Now we compute the second derivative of $h(z)$ with respect to $z$, i.e.,
\begin{align}
\dfrac{d^2h(z)}{dz^2} &= 2 e^{S(z)} \left( \dfrac{dS(z)}{dz} a^\dagger - a^\dagger \dfrac{dS(z)}{dz} \right) e^{-S(z)} = \notag \\
&= 2 e^{S(z)} \left\lbrace \left( a a - a^\dagger a^\dagger \right) a^\dagger - a^\dagger \left( a a - a^\dagger a^\dagger \right) \right\rbrace e^{-S(z)} = \notag \\
&= 2 e^{S(z)} \left( a a a^\dagger - a^\dagger a^\dagger a^\dagger - a^\dagger a a + a^\dagger a^\dagger a^\dagger \right) e^{-S(z)} = \notag \\
&= 2 e^{S(z)} \left( a a a^\dagger - a^\dagger a a \right) e^{-S(z)} = \notag \\
&= 2 e^{S(z)} \left( a \left( \mathds{1} + a^\dagger a \right) - \left( - \mathds{1} + a a^\dagger \right) a \right) e^{-S(z)} = \notag \\
&= 2 e^{S(z)} \left( a + a a^\dagger a + a - a a^\dagger a \right) e^{-S(z)},
\end{align}
then
\begin{align}
\dfrac{d^2h(z)}{dz^2} &= 4 e^{S(z)} a e^{-S(z)} \equiv \notag \\
&\equiv 4 h(z) ,
\end{align}
which is a second-order differential equation and has solution
\begin{equation}
h(z) = c_1 e^{2z} + c_2 e^{-2z}.
\end{equation}
The system of initial conditions
\begin{equation}
\begin{cases}
h(0) = a \\
h'(0) = 2 a^\dagger
\end{cases},
\end{equation}
implies
\begin{equation}
\begin{cases}
c_1 + c_2 = a \\
2 c_1 - 2 c_2 = 2 a^\dagger
\end{cases},
\end{equation}
\begin{equation}
\begin{cases}
c_1 + c_2 = a \\
c_1 - c_2 = a^\dagger
\end{cases}.
\end{equation}
We add the equations and have
\begin{equation}
c_1 = \dfrac{a+a^\dagger}{2} ,
\end{equation}
then we subtract the equations and have
\begin{equation}
c_2 = \dfrac{a-a^\dagger}{2}.
\end{equation}
We insert $c_1$ and $c_2$ in $h(z)$ as follows
\begin{align}
h(z) &= \dfrac{a+a^\dagger}{2} e^{2z} + \dfrac{a-a^\dagger}{2} e^{-2z} = \notag \\
&= \dfrac{e^{2z} a}{2} + \dfrac{e^{2z} a^\dagger}{2} + \dfrac{e^{-2z} a}{2} - \dfrac{e^{-2z} a^\dagger}{2} = \notag \\
&= \dfrac{e^{2z}+e^{-2z}}{2} a + \dfrac{e^{2z}-e^{-2z}}{2} a^\dagger = \notag \\
&= \cosh(2z) a + \sinh(2z) a^\dagger \equiv \notag \\
&\equiv b.
\end{align}
Similarly, we define the operator
\begin{equation}
l(z) = e^{S(z)} a^\dagger e^{-S(z)},
\end{equation}
and we get 
\begin{align}
\dfrac{dl(z)}{dz} &= \dfrac{dS(z)}{dz} e^{S(z)} a^\dagger e^{-S(z)} - e^{S(z)} a^\dagger e^{-S(z)} \dfrac{dS(z)}{dz} = \notag \\ 
&= e^{S(z)} \dfrac{dS(z)}{dz} a^\dagger e^{-S(z)} - e^{S(z)} a^\dagger \dfrac{dS(z)}{dz} e^{-S(z)} = \notag \\
&= e^{S(z)} \left( \dfrac{dS(z)}{dz} a^\dagger - a^\dagger \dfrac{dS(z)}{dz} \right) e^{-S(z)} = \notag \\
&= e^{S(z)} \left\lbrace \left( a a - a^\dagger a^\dagger \right) a^\dagger - a^\dagger \left( a a - a^\dagger a^\dagger \right) \right\rbrace e^{-S(z)} = \notag \\
&= e^{S(z)} \left( a a a^\dagger - a^\dagger a^\dagger a^\dagger - a^\dagger a a + a^\dagger a^\dagger a^\dagger \right) e^{-S(z)} = \notag \\
&= e^{S(z)} \left( a a a^\dagger - a^\dagger a a \right) e^{-S(z)} = \notag \\
&= e^{S(z)} \left( a \left( \mathds{1} + a^\dagger a \right) - \left( - \mathds{1} + a a^\dagger \right) a \right) e^{-S(z)} = \notag \\
&= e^{S(z)} \left( a + a a^\dagger a + a - a a^\dagger a \right) e^{-S(z)} = \notag \\
&= 2 e^{S(z)} a e^{-S(z)},
\end{align}
\begin{equation}
\dfrac{dl(z)}{dz} = 2 e^{S(z)} a e^{-S(z)}. 
\end{equation}
\begin{align}
\dfrac{d^2l(z)}{dz^2} &= 2 e^{S(z)} \left( \dfrac{dS(z)}{dz} a - a \dfrac{dS(z)}{dz} \right) e^{-S(z)} = \notag \\
&= 2 e^{S(z)} \left\lbrace \left( a a - a^\dagger a^\dagger \right) a - a \left( a a - a^\dagger a^\dagger \right) \right\rbrace e^{-S(z)} = \notag \\
&= 2 e^{S(z)} \left( a a a - a^\dagger a^\dagger a - a a a + a a^\dagger a^\dagger \right) e^{-S(z)} = \notag \\
&= 2 e^{S(z)} \left( - a^\dagger a^\dagger a + a a^\dagger a^\dagger \right) e^{-S(z)} = \notag \\
&= 2 e^{S(z)} \left( - a^\dagger \left( - \mathds{1} + a a^\dagger \right) + \left( \mathds{1} + a^\dagger a \right) a^\dagger \right) e^{-S(z)} = \notag \\
&= 2 e^{S(z)} \left( a^\dagger - a^\dagger a a^\dagger + a^\dagger + a^\dagger a a^\dagger \right) e^{-S(z)},
\end{align}
\begin{align}
\dfrac{d^2l(z)}{dz^2} &= 4 e^{S(z)} a^\dagger e^{-S(z)} \equiv \notag \\
&\equiv 4 l(z),
\end{align}
\begin{equation}
l(z) = c_1 e^{2z} + c_2 e^{-2z},
\end{equation}
\begin{equation}
\begin{cases}
l(0) = a^\dagger \\
l'(0) = 2 a
\end{cases},
\end{equation}
\begin{equation}
\begin{cases}
c_1 + c_2 = a^\dagger \\
2 c_1 - 2 c_2 = 2 a
\end{cases},
\end{equation}
\begin{equation}
\begin{cases}
c_1 + c_2 = a^\dagger \\
c_1 - c_2 = a
\end{cases},
\end{equation}
\begin{equation}
c_1 = \dfrac{a+a^\dagger}{2},
\end{equation}
\begin{equation}
c_2 = \dfrac{a^\dagger-a}{2}.
\end{equation}
We insert $c_1$ and $c_2$ in $l(z)$ as follows 
\begin{align}
l(z) &= \dfrac{a+a^\dagger}{2} e^{2z} + \dfrac{a^\dagger-a}{2} e^{-2z} = \notag \\
&= \dfrac{e^{2z} a}{2} + \dfrac{e^{2z} a^\dagger}{2} + \dfrac{e^{-2z} a^\dagger}{2} - \dfrac{e^{-2z} a}{2} = \notag \\
&= \dfrac{e^{2z}-e^{-2z}}{2} a + \dfrac{e^{2z}+e^{-2z}}{2} a^\dagger = \notag \\
&= \sinh(2z) a + \cosh(2z) a^\dagger \equiv \notag \\
&\equiv b^\dagger,
\end{align}
given that $l(z)$ is the adjoint of $h(z)$. Note that if we set $u=\cosh$ and $v=\sinh$, as we assumed for the Bogoljubov's bosonic transformations, then $b$ and $b^\dagger$ become $\eqref{eq: operatoredistruzioneBogoljubov}$ and $\eqref{eq: operatorecreazioneBogoljubov}$, respectively. Let us verify that such a transformation is canonical, i.e.,
\begin{align}
\left[ b,b^\dagger \right] &= \left[ \cosh(2z) a + \sinh(2z) a^\dagger , \sinh(2z) a + \cosh(2z) a^\dagger \right] = \notag \\
&= \cosh(2z)^2 \left[ a , a^\dagger \right] + \sinh^2(2z) \left[ a^\dagger , a \right] = \notag \\
&= \cosh^2(2z) - \sinh^2(2z) = \notag \\
&= 1.
\end{align}
We now compute how \( e^{-S} \) acts on the position and momentum operators in one spatial dimension, that is, on \( \hat{x} \) and \( \hat{p} \). We have
\begin{align}
\hat{x}' &= e^{S(z)} \hat{x} e^{-S(z)} = \notag \\
&= \sqrt{\dfrac{\hslash}{2 m \omega}} e^{S(z)} \left( a + a^\dagger \right) e^{-S(z)} = \notag \\
&= \sqrt{\dfrac{\hslash}{2 m \omega}} \left( \cosh(2z) a + \sinh(2z) a^\dagger + \cosh(2z) a^\dagger + \sinh(2z) a \right) = \notag \\
&= \sqrt{\dfrac{\hslash}{2 m \omega}} \left( \cosh(2z) + \sinh(2z) \right) \left( a + a^\dagger \right) = \notag \\
&= \sqrt{\dfrac{\hslash}{2 m \omega}} \left( \dfrac{e^{2z}+e^{-2z}}{2} + \dfrac{e^{2z}-e^{-2z}}{2} \right) \left( a + a^\dagger \right) = \notag \\
&= \sqrt{\dfrac{\hslash}{2 m \omega}} e^{2z} \left( a + a^\dagger \right) = \notag \\
&= e^{2z} \hat{x},
\end{align}
\begin{align}
\hat{p}' &= e^{S(z)} \hat{p} e^{-S(z)} = \notag \\
&= i \sqrt{\dfrac{m \hslash \omega}{2}} e^{S(z)} \left( a^\dagger - a \right) e^{-S(z)} = \notag \\
&= i \sqrt{\dfrac{m \hslash \omega}{2}} \left( \cosh(2z) a^\dagger + \sinh(2z) a - \cosh(2z) a - \sinh(2z) a^\dagger \right) = \notag \\
&= i \sqrt{\dfrac{m \hslash \omega}{2}} \left( \cosh(2z) - \sinh(2z) \right) \left( a^\dagger - a \right) = \notag \\
&= i \sqrt{\dfrac{m \hslash \omega}{2}} \left( \dfrac{e^{2z}+e^{-2z}}{2} - \dfrac{e^{2z}-e^{-2z}}{2} \right) \left( a^\dagger - a \right) = \notag \\
&= i \sqrt{\dfrac{m \hslash \omega}{2}} e^{-2z} \left( a^\dagger - a \right) = \notag \\
&= e^{-2z} \hat{p}.
\end{align}
Recall that the fundamental state $|0\rangle$ of $\mathcal{H}_1$ is an eigenstate of the annihilation operator $a$. In particular, in the spaces of positions and momenta, the fundamental state is a Gaussian function. The fundamental state associated to new operators is still a Gaussian function: it has a larger width in the space of positions and a smaller width in the space of momenta. Note that $\hat{x}'$ and $\hat{p}'$ satisfy the uncertainty principle. Let us study the transformed vacuum state
\begin{align}
|\tilde{0} \rangle &\equiv e^{-S(z)} |0 \rangle = \notag \\
&= e^{- z \left( a a - a^\dagger a^\dagger \right)} |0\rangle.
\end{align}
From Jacobi identity $\eqref{eq: Jacobiidentity}$, we get
\begin{align}
\left[ a a , a^\dagger a^\dagger \right] &= a \left[ a , a^\dagger a^\dagger \right] + \left[ a, a^\dagger a^\dagger \right] a = \notag \\
&= a a^\dagger \left[ a , a^\dagger \right] + a \left[ a , a^\dagger \right] a^\dagger + a^\dagger \left[ a , a^\dagger \right] a + \left[ a , a^\dagger \right] a^\dagger a = \notag \\
&= 2 a a^\dagger + 2 a^\dagger a = \notag \\
&= 2 + 2 a^\dagger a + 2 a^\dagger a = \notag \\
&= 2 + 4 a^\dagger a.
\end{align}
Accordingly, we recall that by definition the fundamental state $|\tilde{0} \rangle$ of the transformed Hamiltonia must satisfy
\begin{equation}
b |\tilde{0} \rangle = 0,
\end{equation}
which implies
\begin{equation}
\left( \cosh(2z) a + \sinh(2z) a^\dagger \right) |\tilde{0} \rangle = 0.
\label{eq: operatoretrasformatodistruzioneBogoljubovsustatofondamentale}
\end{equation}
Now, we want to write the eigenstate of the fundamental state of the transformed Hamiltonian as a function of the eigenfunctions of the original Hamiltonian by means of a series expansion. Since the new eigenstates are even functions in the space of positions (they are Gaussians), only even powers of the creation operator $a^\dagger$ can be considered in the expansion of $|\tilde{0} \rangle$. We therefore expect a solution of the form
\begin{equation}
|\tilde{0} \rangle = c_0 |0 \rangle + c_2 a^\dagger a^\dagger |0 \rangle + c_4 a^\dagger a^\dagger a^\dagger a^\dagger |0 \rangle + \ldots \ .
\end{equation}
In place of the series we consider only a quadratic term $a^\dagger a^\dagger$ and impose for the transformed fundamental state the following Ansatz
\begin{equation}
|\tilde{0} \rangle = \mathcal{N}(z) \exp \left\lbrace - \dfrac{1}{2} f(z) a^\dagger a^\dagger \right\rbrace |0 \rangle,
\end{equation}
then we must compute $\mathcal{N}(z)$ and $f(z)$. The equation $\eqref{eq: operatoretrasformatodistruzioneBogoljubovsustatofondamentale}$ gives
\begin{equation}
\left( \cosh(2z) a + \sinh(2z) a^\dagger \right) \exp \left\lbrace - \dfrac{1}{2} f(z) a^\dagger a^\dagger \right\rbrace |0 \rangle = 0,
\end{equation}
\begin{equation}
\cosh(2z) a \exp \left\lbrace - \dfrac{1}{2} f(z) a^\dagger a^\dagger \right\rbrace |0 \rangle + \sinh(2z) a^\dagger \exp \left\lbrace - \dfrac{1}{2} f(z) a^\dagger a^\dagger \right\rbrace |0\rangle = 0.
\end{equation}
Since the action of the annihilation operator $a$ on the vacuum state is the null state, the argument of the first exponential can be replaced with a commutator as follows
\begin{equation}
\cosh(2z) \left[ a , \exp \left\lbrace - \dfrac{1}{2} f(z) a^\dagger a^\dagger \right\rbrace \right] |0 \rangle + \sinh(2z) a^\dagger \exp \left\lbrace - \dfrac{1}{2} f(z) a^\dagger a^\dagger \right\rbrace |0 \rangle = 0. 
\end{equation}
If $\left[ \left[ A,B \right] , B \right] = 0$, then $\left[ A,e^B \right] = \left[ A , B \right] e^B$, which is valid in our case, with $A=a$ and $B=- \frac{1}{2} f(z) a^\dagger a^\dagger$, so the above equation becomes
\begin{equation}
\cosh(2z) \left[ a , - \dfrac{1}{2} f(z) a^\dagger a^\dagger \right] \exp \left\lbrace - \dfrac{1}{2} f(z) a^\dagger a^\dagger \right\rbrace |0 \rangle + \sinh(2z) a^\dagger \exp \left\lbrace - \dfrac{1}{2} f(z) a^\dagger a^\dagger \right\rbrace |0 \rangle = 0,
\end{equation}
\begin{equation}
\left( \cosh(2z) \left[ a , - \dfrac{1}{2} f(z) a^\dagger a^\dagger \right] + \sinh(2z) a^\dagger \right) \exp \left\lbrace - \dfrac{1}{2} f(z) a^\dagger a^\dagger \right\rbrace |0 \rangle = 0.
\end{equation}
From
\begin{align}
\left[ a , - \dfrac{1}{2} f(z) a^\dagger a^\dagger \right] &= - \dfrac{1}{2} f(z) \left[ a , a^\dagger a^\dagger \right] = \notag \\
&= - \dfrac{1}{2} f(z) \left\lbrace a^\dagger \left[ a , a^\dagger \right] + \left[ a , a^\dagger \right] a^\dagger \right\rbrace = \notag \\
&= - f(z) a^\dagger,
\end{align}
we have
\begin{equation}
\left( - \cosh(2z) f(z) + \sinh(2z) \right) a^\dagger \exp \left\lbrace - \dfrac{1}{2} f(z) a^\dagger a^\dagger \right\rbrace |0\rangle = 0,
\end{equation}
\begin{equation}
- \cosh(2z) f(z) + \sinh(2z) = 0,
\end{equation}
\begin{equation}
f(z) = \tanh(2z),
\end{equation}
then
\begin{equation}
|\tilde{0} \rangle = \mathcal{N}(z) \exp \left\lbrace - \dfrac{1}{2} \tanh(2z) a^\dagger a^\dagger \right\rbrace |0 \rangle.
\end{equation}
First we multiply both members of the above equation by $\langle 0|$ and we then expand the exponential operator explicitly as a power series as follows
\begin{align}
\left\langle 0 \middle| \tilde{0} \right\rangle &= \mathcal{N}(z) \left\langle 0 \left| \exp \left\lbrace - \dfrac{1}{2} \tanh(2z)\, a^\dagger a^\dagger \right\rbrace \right| 0 \right\rangle = \notag \\
&= \mathcal{N}(z) \lim_{N \rightarrow \infty} \left\langle 0 \left| \left( 1 - \dfrac{1}{2} \tanh(2z) \dfrac{a^\dagger a^\dagger}{N} \right)^N \right| 0 \right\rangle = \notag \\
&= \mathcal{N}(z) \lim_{N \rightarrow \infty} \left\langle 0 \left| \left( 1 - \dfrac{1}{2} \tanh(2z) \dfrac{a a}{N} \right)^N \right| 0 \right\rangle = \notag \\
&= \mathcal{N}(z),
\end{align}
then, by integrating over the one-dimensional spatial coordinate \( x \), we obtain
\begin{align}
\mathcal{N}(z) &= \langle 0|\tilde{0} \rangle = \notag \\
&= \int_{-\infty}^{+\infty} dx \langle 0|x \rangle \langle x|\tilde{0} \rangle = \notag \\
&= \int_{-\infty}^{+\infty} dx \left( \psi_0(x) \right)^* \tilde{\psi}_0(x),
\end{align}
where the fundamental state of the harmonic oscillator is given by
\begin{equation}
\psi_0(x) = N_0 \exp \left( - \dfrac{m \omega}{2 \hslash} x^2 \right),
\end{equation}
with
\begin{equation}
N_0 = \left( \dfrac{m \omega}{\pi \hslash} \right)^{\frac{1}{4}}.
\end{equation}
We have
\begin{align}
\mathcal{N}(z) &= \left( \dfrac{m \omega}{\pi \hslash} \right)^{\frac{1}{4}} \left( \dfrac{\tilde{m} \tilde{\omega}}{\pi \hslash} \right)^{\frac{1}{4}} \int_{-\infty}^{+\infty} dx \exp \left( - \dfrac{m \omega + \tilde{m} \tilde{\omega}}{2 \hslash} x^2 \right) = \notag \\
&= \left( m \omega \tilde{m} \tilde{\omega} \right)^{\frac{1}{4}} \dfrac{1}{\sqrt{\pi \hslash}} \dfrac{\sqrt{2 \pi \hslash}}{\sqrt{m \omega + \tilde{m} \tilde{\omega}}} = \notag \\
&= \left( \dfrac{1}{2} \dfrac{m \omega + \tilde{m} \tilde{\omega}}{\sqrt{m \omega \tilde{m} \tilde{\omega}}} \right)^{-\frac{1}{2}} = \notag \\
&= \left( \dfrac{1}{2} \left[ \sqrt{\dfrac{m \omega}{\tilde{m} \tilde{\omega}}} + \sqrt{\dfrac{\tilde{m} \tilde{\omega}}{m \omega}} \right] \right)^{-\frac{1}{2}} ,
\end{align}
where $\tilde{m}$ and $\tilde{\omega}$ are the mass and frequency of the quasiparticle of the transformed Hamiltonian, respectively. We define
\begin{equation}
e^{2z} = \sqrt{\dfrac{m \omega}{\tilde{m} \tilde{\omega}}},
\end{equation}
then
\begin{align}
\mathcal{N}(z) &= \left( \dfrac{1}{2} \left[ e^{2z} + e^{-2z} \right] \right)^{-\frac{1}{2}} = \notag \\
&= \dfrac{1}{\sqrt{\cosh(2z)}} ,
\end{align}
\begin{equation}
|\tilde{0} \rangle = \dfrac{1}{\sqrt{\cosh(2z)}} \exp \left\lbrace - \dfrac{1}{2} \tanh(2z) a^\dagger a^\dagger \right\rbrace |0 \rangle.
\end{equation}
We have obtained the operatorial identity
\begin{equation}
e^{- z \left( a a - a^\dagger a^\dagger \right)} |0 \rangle = \dfrac{1}{\sqrt{\cosh(2z)}} \exp \left\lbrace - \dfrac{1}{2} \tanh(2z) a^\dagger a^\dagger \right\rbrace |0\rangle.
\end{equation}
Regarding the Bogoljubov transformed bosonic Hamiltonian, the property
\begin{equation}
b_{\textbf{k}} |\psi \rangle = 0, \ \forall \ \textbf{k} \neq \textbf{0}
\label{eq: condizionestatofondamentaleBogoljubov}
\end{equation}
is verified if the fundamental state has the structure
\begin{equation}
|\psi \rangle = \dfrac{\left(a_0^\dagger\right)^{N_0}}{\sqrt{N_0 !}} |0\rangle \exp \left\lbrace - \sum_{\substack{\textbf{q}_1 \neq \textbf{0}, \\ \textbf{q}_2 \neq \textbf{0}}} \dfrac{M_{\textbf{q}_1,\textbf{q}_2}}{2} a^\dagger_{\textbf{q}_1} a^\dagger_{\textbf{q}_2} \right\rbrace |0 \rangle,
\label{eq: statofondamentaleBogoljubov}
\end{equation}
where
\begin{equation}
\dfrac{\left(a_0^\dagger\right)^{N_0}}{\sqrt{N_0 !}} |0 \rangle
\end{equation}
is the eigenfunction related to the condensate. Now we want to prove that $\eqref{eq: statofondamentaleBogoljubov}$ satisfies the $\eqref{eq: condizionestatofondamentaleBogoljubov}$. Given
\begin{equation}
\left( u_{\textbf{k}} a_{\textbf{k}} + v_{\textbf{k}} a^\dagger_{-\textbf{k}} \right) \dfrac{\left(a_0^\dagger\right)^{N_0}}{\sqrt{N_0 !}} |0 \rangle \exp \left\lbrace - \sum_{\substack{\textbf{q}_1 \neq \textbf{0}, \\ \textbf{q}_2 \neq \textbf{0}}} \dfrac{M_{\textbf{q}_1,\textbf{q}_2}}{2} a^\dagger_{\textbf{q}_1} a^\dagger_{\textbf{q}_2} \right\rbrace |0 \rangle,
\label{eq: azionedell'operatoredidistruzionediBogoljubovsullostatofondamentale}
\end{equation}
note that the operator $a^\dagger_{-\textbf{k}}$ commutes both with $\left(a_0^\dagger\right)^{N_0}$ and with $a^\dagger_{\textbf{q}} a^\dagger_{\textbf{q}'}$. On the other hand, $a_{\textbf{k}}$ commutes with $\left(a_0^\dagger\right)^{N_0}$ but not with the exponential, because of $a^\dagger_{\textbf{k}} a^\dagger_{\textbf{q}'}$ and $a^\dagger_{\textbf{q}} a^\dagger_{\textbf{k}}$ in the series. Then, given
\begin{equation}
a_{\textbf{k}} \exp \left\lbrace - \sum_{\substack{\textbf{q}_1 \neq \textbf{0}, \\ \textbf{q}_2 \neq \textbf{0}}} \dfrac{M_{\textbf{q}_1,\textbf{q}_2}}{2} a^\dagger_{\textbf{q}_1} a^\dagger_{\textbf{q}_2} \right\rbrace |0 \rangle,
\end{equation}
we insert the identity
\begin{equation}
\mathds{1} = \exp \left\lbrace - \sum_{\substack{\textbf{q}_1 \neq \textbf{0}, \\ \textbf{q}_2 \neq \textbf{0}}} \dfrac{M_{\textbf{q}_1,\textbf{q}_2}}{2} a^\dagger_{\textbf{q}_1} a^\dagger_{\textbf{q}_2} \right\rbrace \exp \left\lbrace \sum_{\substack{\textbf{q}_1 \neq \textbf{0}, \\ \textbf{q}_2 \neq \textbf{0}}} \dfrac{M_{\textbf{q}_1,\textbf{q}_2}}{2} a^\dagger_{\textbf{q}_1} a^\dagger_{\textbf{q}_2} \right\rbrace
\end{equation}
to the left of $a_{\textbf{k}}$ as follows
\begin{equation}
\exp \left\lbrace - \sum_{\substack{\textbf{q}_1 \neq \textbf{0}, \\ \textbf{q}_2 \neq \textbf{0}}} \dfrac{M_{\textbf{q}_1,\textbf{q}_2}}{2} a^\dagger_{\textbf{q}_1} a^\dagger_{\textbf{q}_2} \right\rbrace \exp \left\lbrace \sum_{\substack{\textbf{q}_1 \neq \textbf{0}, \\ \textbf{q}_2 \neq \textbf{0}}} \dfrac{M_{\textbf{q}_1,\textbf{q}_2}}{2} a^\dagger_{\textbf{q}_1} a^\dagger_{\textbf{q}_2} \right\rbrace a_{\textbf{k}} \exp \left\lbrace - \sum_{\substack{\textbf{q}_1 \neq \textbf{0}, \\ \textbf{q}_2 \neq \textbf{0}}} \dfrac{M_{\textbf{q}_1,\textbf{q}_2}}{2} a^\dagger_{\textbf{q}_1} a^\dagger_{\textbf{q}_2} \right\rbrace,
\end{equation}
and we study the transformed of $a_{\textbf{k}}$, i.e.,
\begin{align}
\tilde{a}_{\textbf{k}} &\equiv \exp \left\lbrace \sum_{\substack{\textbf{q}_1 \neq \textbf{0}, \\ \textbf{q}_2 \neq \textbf{0}}} \dfrac{M_{\textbf{q}_1,\textbf{q}_2}}{2} a^\dagger_{\textbf{q}_1} a^\dagger_{\textbf{q}_2} \right\rbrace a_{\textbf{k}} \exp \left\lbrace - \sum_{\substack{\textbf{q}_1 \neq \textbf{0}, \\ \textbf{q}_2 \neq \textbf{0}}} \dfrac{M_{\textbf{q}_1,\textbf{q}_2}}{2} a^\dagger_{\textbf{q}_1} a^\dagger_{\textbf{q}_2} \right\rbrace \equiv \notag \\
&\equiv e^{T} a_{\textbf{k}} e^{-T},
\end{align}
with
\begin{equation}
T = \sum_{\substack{\textbf{q}_1 \neq \textbf{0}, \\ \textbf{q}_2 \neq \textbf{0}}} \dfrac{M_{\textbf{q}_1,\textbf{q}_2}}{2} a^\dagger_{\textbf{q}_1} a^\dagger_{\textbf{q}_2}.
\end{equation}
From the conjugation formula
\begin{equation}
e^{T} a_{\textbf{k}} e^{-T} = a_{\textbf{k}} + \left[ T , a_{\textbf{k}} \right] + \dfrac{1}{2!} \left[ T , \left[ T , a_{\textbf{k}} \right] \right] + \ldots
\end{equation}
we compute
\begin{align}
\left[ T , a_{\textbf{k}} \right] &= \dfrac{1}{2} \left( \sum_{\textbf{q}_1 \neq \textbf{0}} M_{\textbf{q}_1,\textbf{k}} \left[ a^\dagger_{\textbf{q}_1} a^\dagger_{\textbf{k}} , a_{\textbf{k}} \right] + \sum_{\textbf{q}_2 \neq \textbf{0}} M_{\textbf{k},\textbf{q}_2} \left[ a^\dagger_{\textbf{k}} a^\dagger_{\textbf{q}_2} , a_{\textbf{k}} \right] \right) = \notag \\
&= \dfrac{1}{2} \bigg( \sum_{\textbf{q}_1 \neq \textbf{0}} M_{\textbf{q}_1,\textbf{k}} a^\dagger_{\textbf{q}_1} \left[ a^\dagger_{\textbf{k}} , a_{\textbf{k}} \right] + \sum_{\textbf{q}_1 \neq \textbf{0}} M_{\textbf{q}_1,\textbf{k}} \left[ a^\dagger_{\textbf{q}_1} , a_{\textbf{k}} \right] a^\dagger_{\textbf{k}} + \notag \\
&+ \sum_{\textbf{q}_2 \neq \textbf{0}} M_{\textbf{k},\textbf{q}_2} a^\dagger_{\textbf{k}} \left[ a^\dagger_{\textbf{q}_2} , a_{\textbf{k}} \right] + \sum_{\textbf{q}_2 \neq \textbf{0}} M_{\textbf{k},\textbf{q}_2} \left[ a^\dagger_{\textbf{k}} , a_{\textbf{k}} \right] a^\dagger_{\textbf{q}_2} \bigg) = \notag \\
&= \dfrac{1}{2} \left( \sum_{\textbf{q}_1 \neq \textbf{0}} M_{\textbf{q}_1,\textbf{k}} \left( - a^\dagger_{\textbf{q}_1} - \delta_{\textbf{q}_1,\textbf{k}} a^\dagger_{\textbf{k}} \right) + \sum_{\textbf{q}_2 \neq \textbf{0}} M_{\textbf{k},\textbf{q}_2} \left( - \delta_{\textbf{q}_2,\textbf{k}} a^\dagger_{\textbf{k}} - a^\dagger_{\textbf{q}_2} \right) \right) = \notag \\
&= \dfrac{1}{2} \left( - \sum_{\textbf{q}_1 \neq \textbf{0}} M_{\textbf{q}_1,\textbf{k}} a^\dagger_{\textbf{k}} - M_{\textbf{k},\textbf{k}} a^\dagger_{\textbf{k}} - \sum_{\textbf{q}_2 \neq \textbf{0}} M_{\textbf{k},\textbf{q}_2} a^\dagger_{\textbf{q}_2} - M_{\textbf{k},\textbf{k}} a^\dagger_{\textbf{k}} \right).
\end{align}
Since $M_{\textbf{q}_1,\textbf{q}_2}$ quantifies the interaction between two particles with momenta $\textbf{q}_1$ and $\textbf{q}_2$, assume that $M$ is a symmetric matrix with the diagonal elements equal to zero, i.e.,
\begin{equation}
M_{\textbf{k},\textbf{k}} = 0, \ \forall \ \textbf{k} \neq \textbf{0},
\end{equation}
\begin{equation}
M_{\textbf{k},\textbf{q}_2} \equiv M_{\textbf{q}_2,\textbf{k}},
\end{equation}
then we rename the dummy indices $\textbf{q}_1 \equiv \textbf{q}_2 \equiv \textbf{q}$ and write
\begin{equation}
\left[ T , a_{\textbf{k}} \right] = - \sum_{\textbf{q} \neq \textbf{0}} M_{\textbf{q},\textbf{k}} a^\dagger_{\textbf{q}}.
\end{equation}
Since the other terms in $e^{T} a_{\textbf{k}} e^{-T}$ involve commutators of $[T, a_{\textbf{k}}]$ with $T$ itself, the only non-zero term is the one we calculated, i.e.,
\begin{align}
\tilde{a}_{\textbf{k}} &= e^T a_{\textbf{k}} e^{-T} = \notag \\
&= a_{\textbf{k}} + \left[ T , a_{\textbf{k}} \right] = \notag \\
&= a_{\textbf{k}} - \sum_{\textbf{q} \neq \textbf{0}} M_{\textbf{q},\textbf{k}} a^\dagger_{\textbf{q}}.
\end{align}
So, from $\eqref{eq: azionedell'operatoredidistruzionediBogoljubovsullostatofondamentale}$, $a^\dagger_{-\textbf{k}}$ it follows
\begin{equation}
\dfrac{\left(a_0^\dagger\right)^{N_0}}{\sqrt{N_0 !}} |0 \rangle \exp \left\lbrace - \sum_{\substack{\textbf{q}_1 \neq \textbf{0}, \\ \textbf{q}_2 \neq \textbf{0}}} \dfrac{M_{\textbf{q}_1,\textbf{q}_2}}{2} a^\dagger_{\textbf{q}_1} a^\dagger_{\textbf{q}_2} \right\rbrace \left( u_{\textbf{k}} \tilde{a}_{\textbf{k}} + v_{\textbf{k}} a^\dagger_{-\textbf{k}} \right)|0 \rangle = 0,
\end{equation}
\begin{equation}
\dfrac{\left(a_0^\dagger\right)^{N_0}}{\sqrt{N_0 !}} |0\rangle \exp \left\lbrace - \sum_{\substack{\textbf{q}_1 \neq \textbf{0}, \\ \textbf{q}_2 \neq \textbf{0}}} \dfrac{M_{\textbf{q}_1,\textbf{q}_2}}{2} a^\dagger_{\textbf{q}_1} a^\dagger_{\textbf{q}_2} \right\rbrace \left( u_{\textbf{k}} a_{\textbf{k}} - u_{\textbf{k}} \sum_{\textbf{q} \neq \textbf{0}} M_{\textbf{q},\textbf{k}} a^\dagger_{\textbf{q}} + v_{\textbf{k}} a^\dagger_{-\textbf{k}} \right)|0 \rangle = 0,
\end{equation}
\begin{equation}
\dfrac{\left(a_0^\dagger\right)^{N_0}}{\sqrt{N_0 !}} |0 \rangle \exp \left\lbrace - \sum_{\substack{\textbf{q}_1 \neq \textbf{0}, \\ \textbf{q}_2 \neq \textbf{0}}} \dfrac{M_{\textbf{q}_1,\textbf{q}_2}}{2} a^\dagger_{\textbf{q}_1} a^\dagger_{\textbf{q}_2} \right\rbrace \left( - u_{\textbf{k}} \sum_{\textbf{q} \neq \textbf{0}} M_{\textbf{q},\textbf{k}} a^\dagger_{\textbf{q}} + v_{\textbf{k}} a^\dagger_{-\textbf{k}} \right)|0 \rangle = 0,
\end{equation}
where in the last step the annihilation operator has been removed, since it acts on the vacuum state. A possible solution for the above equation is given by
\begin{equation}
- u_{\textbf{k}} \sum_{\textbf{q} \neq \textbf{0}} M_{\textbf{q},\textbf{k}} a^\dagger_{\textbf{q}} + v_{\textbf{k}} a^\dagger_{-\textbf{k}} = 0,
\end{equation}
which is verified if and only if only
\begin{equation}
M_{\textbf{q},\textbf{k}} = \dfrac{v_{\textbf{k}}}{u_{\textbf{k}}} \delta_{\textbf{q},-\textbf{k}}.
\end{equation}
We have shown that $\eqref{eq: statofondamentaleBogoljubov}$ satisfies $\eqref{eq: condizionestatofondamentaleBogoljubov}$ with $M_{\textbf{q},\textbf{q}} = 0$, $M_{\textbf{q}_1,\textbf{q}_2} = M_{\textbf{q},-\textbf{q}}$, then the fundamental state of a Bogoljubov's bosonic Hamiltonian is
\begin{equation}
|\psi \rangle = \dfrac{\left(a_0^\dagger\right)^{N_0}}{\sqrt{N_0 !}} |0 \rangle \exp \left\lbrace - \sum_{\textbf{q} \neq \textbf{0}} \dfrac{M_{\textbf{q},-\textbf{q}}}{2} a^\dagger_{\textbf{q}} a^\dagger_{-\textbf{q}} \right\rbrace |0 \rangle.
\end{equation}
Note that $M_{\textbf{q},-\textbf{q}} = \frac{v_{\textbf{q}}}{u_{\textbf{q}}}$ explicitly incorporates the effects of interactions, as expected.
\section{Number density of bosons}
Given the annihilation field operator of the Bogoljubov's theory, i.e.,
\begin{equation}
\hat{\psi}(\textbf{r}) = \sqrt{\dfrac{N_0}{V}} + \sum_{\textbf{k} \neq \textbf{0}} \dfrac{e^{i \textbf{k} \cdot \textbf{r}}}{\sqrt{V}} \left( u_{\textbf{k}} b_{\textbf{k}} - v_{\textbf{k}} b^\dagger_{-\textbf{k}} \right),
\end{equation}
let us compute its thermal average with respect to the diagonalized Hamiltonian $\eqref{eq: hamiltonianaBogoljubovdiagonalizzata}$. Note that the interaction term does not contribute to the thermal average, as it does not conserve particle number. In particular,
\begin{equation}
\left\langle u_{\textbf{k}} b_{\textbf{k}} - v_{\textbf{k}} b^\dagger_{-\textbf{k}} \right\rangle = 0,
\end{equation}
since the operator \( u_{\textbf{k}} b_{\textbf{k}} - v_{\textbf{k}} b^\dagger_{-\textbf{k}} \) does not commute with the total number operator and therefore has vanishing expectation value. Then
\begin{equation}
\langle \hat{\psi}(\textbf{r}) \rangle = \sqrt{\dfrac{N_0}{V}},
\end{equation}
that is, the Bogoljubov field operator is the sum of its mean and a fluctuation term contained in iterations. The number density of bosons within an infinitesimal volume centered at $\textbf{r}$ is
\begin{align}
\hat{\rho}(\textbf{r}) &= \langle \hat{\psi}^\dagger(\textbf{r}) \hat{\psi}(\textbf{r}) \rangle = \notag \\
&= \left\langle \left( \sqrt{\dfrac{N_0}{V}} + \sum_{\textbf{k} \neq \textbf{0}} \dfrac{e^{-i \textbf{k} \cdot \textbf{r}}}{\sqrt{V}} \left( u_{\textbf{k}} b^\dagger_{\textbf{k}} - v_{\textbf{k}} b_{-\textbf{k}} \right) \right) \left( \sqrt{\dfrac{N_0}{V}} + \sum_{\textbf{k}' \neq \textbf{0}} \dfrac{e^{i \textbf{k}' \cdot \textbf{r}}}{\sqrt{V}} \left( u_{\textbf{k}'} b_{\textbf{k}'} - v_{\textbf{k}'} b^\dagger_{-\textbf{k}'} \right) \right) \right\rangle ,
\end{align}
and from $\langle u_{\textbf{k}} b_{\textbf{k}} - v_{\textbf{k}} b^\dagger_{-\textbf{k}} \rangle = 0$ we get
\begin{equation}
\hat{\rho}(\textbf{r}) = \dfrac{N_0}{V} + \sum_{\textbf{k} \neq \textbf{0}} \sum_{\textbf{k}' \neq \textbf{0}} \left\langle \dfrac{e^{-i \textbf{k} \cdot \textbf{r}}}{\sqrt{V}} \dfrac{e^{i \textbf{k}' \cdot \textbf{r}}}{\sqrt{V}} \left( u_{\textbf{k}} b^\dagger_{\textbf{k}} - v_{\textbf{k}} b_{-\textbf{k}} \right) \left( u_{\textbf{k}'} b_{\textbf{k}'} - v_{\textbf{k}'} b^\dagger_{-\textbf{k}'} \right) \right\rangle.
\end{equation}
Moreover, the thermal average of the product of two general wave functions related to two different momenta is null, because of the integral of a product including an oscillating term, then
\begin{equation}
\hat{\rho}(\textbf{r}) = \dfrac{N_0}{V} + \dfrac{1}{V} \sum_{\textbf{k} \neq \textbf{0}} \left\langle \left( u_{\textbf{k}} b^\dagger_{\textbf{k}} - v_{\textbf{k}} b_{-\textbf{k}} \right) \left( u_{\textbf{k}} b_{\textbf{k}} - v_{\textbf{k}} b^\dagger_{-\textbf{k}} \right) \right\rangle.
\end{equation}
The thermal averages of the products of two creation operators and two annihilation operators are zero, then the number density operator becomes
\begin{align}
\hat{\rho}(\textbf{r}) &= \dfrac{N_0}{V} + \dfrac{1}{V} \sum_{\textbf{k} \neq \textbf{0}} \left\langle u^2_{\textbf{k}} b^\dagger_{\textbf{k}} b_{\textbf{k}} + v^2_{\textbf{k}} b_{-\textbf{k}} b^\dagger_{-\textbf{k}} \right\rangle = \notag \\
&= \dfrac{N_0}{V} + \dfrac{1}{V} \sum_{\textbf{k} \neq \textbf{0}} u^2_{\textbf{k}} \left\langle b^\dagger_{\textbf{k}} b_{\textbf{k}} \right\rangle + \dfrac{1}{V} \sum_{\textbf{k} \neq \textbf{0}} v^2_{\textbf{k}} \left\langle b_{\textbf{k}} b^\dagger_{\textbf{k}} \right\rangle = \notag \\
&= \dfrac{N_0}{V} + \dfrac{1}{V} \sum_{\textbf{k} \neq \textbf{0}} u^2_{\textbf{k}} \left\langle b^\dagger_{\textbf{k}} b_{\textbf{k}} \right\rangle + \dfrac{1}{V} \sum_{\textbf{k} \neq \textbf{0}} v^2_{\textbf{k}} \left\langle \mathds{1} + b^\dagger_{\textbf{k}} b_{\textbf{k}} \right\rangle = \notag \\
&= \dfrac{N_0}{V} + \dfrac{1}{V} \sum_{\textbf{k} \neq \textbf{0}} u^2_{\textbf{k}} \langle \hat{N}_{\textbf{k}} \rangle + \dfrac{1}{V} \sum_{\textbf{k} \neq \textbf{0}} v^2_{\textbf{k}} \left( 1 + \langle \hat{N}_{\textbf{k}} \rangle \right) \equiv \notag \\
&\equiv \dfrac{N_0}{V} + \dfrac{1}{V} \sum_{\textbf{k} \neq \textbf{0}} \left( u^2_{\textbf{k}} + v^2_{\textbf{k}} \right) \langle \hat{N}_{\textbf{k}} \rangle + \dfrac{1}{V} \sum_{\textbf{k} \neq \textbf{0}} v^2_{\textbf{k}} ,
\end{align}
where we renamed the dummy index $- \textbf{k} \rightarrow \textbf{k}$ and we used the bosonic algebra. At finite temperature, the number density of bosons has two additional terms to that of the condensate. What about low temperatures? We have
\begin{align}
\lim_{T \rightarrow 0^+} \langle \hat{N}_{\textbf{k}} \rangle &= \lim_{T \rightarrow 0^+} \dfrac{1}{e^{\beta \mathcal{E}_{\textbf{k}}} - 1} = \notag \\
&= 0,
\end{align}
\begin{equation}
\lim_{T \rightarrow 0^+} \dfrac{1}{V} \sum_{\textbf{k} \neq \bar{0}} \left( u^2_{\textbf{k}} + v^2_{\textbf{k}} \right) \langle \hat{N}_{\textbf{k}} \rangle = 0,
\end{equation}
then
\begin{equation}
\lim_{T \rightarrow 0^+} \hat{\rho}(\textbf{r}) = \dfrac{N_0}{V} + \dfrac{1}{V} \sum_{\textbf{k} \neq \textbf{0}} v^2_{\textbf{k}},
\end{equation}
that is, at low temperatures the number density of bosons includes also an additional term. Given
\begin{equation}
\begin{cases}
u^2_{\textbf{k}} - v^2_{\textbf{k}} = 1 \\
u^2_{\textbf{k}} + v^2_{\textbf{k}} = \left( u_{\textbf{k}} + v_{\textbf{k}} \right)^2 - 2 u_{\textbf{k}} v_{\textbf{k}}
\end{cases},
\end{equation}
from
\begin{align}
\left( u_{\textbf{k}} + v_{\textbf{k}} \right)^2 &= \left[ \dfrac{1}{2} \left( \sqrt{\dfrac{\mathcal{E}_{\textbf{k}}}{\mathcal{E}^0_{\textbf{k}}}} + \sqrt{\dfrac{\mathcal{E}^0_{\textbf{k}}}{\mathcal{E}_{\textbf{k}}}} \right) + \dfrac{1}{2} \left( \sqrt{\dfrac{\mathcal{E}_{\textbf{k}}}{\mathcal{E}^0_{\textbf{k}}}} - \sqrt{\dfrac{\mathcal{E}^0_{\textbf{k}}}{\mathcal{E}_{\textbf{k}}}} \right) \right]^2 = \notag \\
&= \left[ \dfrac{1}{2} 2 \sqrt{\dfrac{\mathcal{E}_{\textbf{k}}}{\mathcal{E}^0_{\textbf{k}}}} \right]^2 = \notag \\
&= \dfrac{\mathcal{E}_{\textbf{k}}}{\mathcal{E}^0_{\textbf{k}}},
\end{align}
\begin{align}
2 u_{\textbf{k}} v_{\textbf{k}} &= 2 \dfrac{1}{4} \left( \sqrt{\dfrac{\mathcal{E}_{\textbf{k}}}{\mathcal{E}^0_{\textbf{k}}}} + \sqrt{\dfrac{\mathcal{E}^0_{\textbf{k}}}{\mathcal{E}_{\textbf{k}}}} \right) \left( \sqrt{\dfrac{\mathcal{E}_{\textbf{k}}}{\mathcal{E}^0_{\textbf{k}}}} - \sqrt{\dfrac{\mathcal{E}^0_{\textbf{k}}}{\mathcal{E}_{\textbf{k}}}} \right) = \notag \\
&= \dfrac{1}{2} \left( \dfrac{\mathcal{E}_{\textbf{k}}}{\mathcal{E}^0_{\textbf{k}}} - \dfrac{\mathcal{E}^0_{\textbf{k}}}{\mathcal{E}_{\bar{k}}} \right) = \notag \\
&= \dfrac{1}{2} \dfrac{\mathcal{E}^2_{\textbf{k}} - \mathcal{E}^0_{\textbf{k}}}{\mathcal{E}^0_{\textbf{k}} \mathcal{E}_{\textbf{k}}} = \notag \\
&= \dfrac{1}{2} \dfrac{\mathcal{E}^0_{\textbf{k}} \left( \mathcal{E}^0_{\textbf{k}} + 2 \frac{U N_0}{V} \right) - \left( \mathcal{E}^0_{\textbf{k}} \right)^2}{\mathcal{E}^0_{\textbf{k}} \mathcal{E}_{\textbf{k}}} = \notag \\
&= \dfrac{1}{2} \dfrac{\mathcal{E}^0_{\textbf{k}} + 2 \frac{U N_0}{V} - \mathcal{E}^0_{\textbf{k}}}{\mathcal{E}_{\textbf{k}}} = \notag \\
&= \dfrac{U N_0}{\mathcal{E}_{\textbf{k}} V},
\end{align}
we have
\begin{align}
\left( u_{\textbf{k}} + v_{\textbf{k}} \right)^2 - 2 u_{\textbf{k}} v_{\textbf{k}} &= \dfrac{\mathcal{E}_{\textbf{k}}}{\mathcal{E}^0_{\textbf{k}}} - \dfrac{U N_0}{\mathcal{E}_{\textbf{k}} V} = \notag \\
&= \dfrac{\mathcal{E}^2_{\textbf{k}} - \mathcal{E}^0_{\textbf{k}} \frac{U N_0}{V}}{\mathcal{E}^0_{\textbf{k}} \mathcal{E}_{\textbf{k}}} = \notag \\
&= \dfrac{\mathcal{E}^0_{\textbf{k}} \left( \mathcal{E}^0_{\textbf{k}} + 2 \frac{U N_0}{V} \right) - \mathcal{E}^0_{\textbf{k}} \frac{U N_0}{V}}{\mathcal{E}^0_{\textbf{k}} \mathcal{E}_{\textbf{k}}} = \notag \\
&= \dfrac{\mathcal{E}^0_{\textbf{k}} + 2 \frac{U N_0}{V} - \frac{U N_0}{V}}{\mathcal{E}_{\textbf{k}}} = \notag \\
&= \dfrac{\mathcal{E}^0_{\textbf{k}} + \frac{U N_0}{V}}{\mathcal{E}_{\textbf{k}}} ,
\end{align}
and the system becomes
\begin{equation}
\begin{cases}
u^2_{\textbf{k}} - v^2_{\textbf{k}} = 1 \\
u^2_{\textbf{k}} + v^2_{\textbf{k}} = \dfrac{\mathcal{E}^0_{\textbf{k}} + \frac{U N_0}{V}}{\mathcal{E}_{\textbf{k}}}
\end{cases}.
\end{equation}
By subtracting the second equation from the first equation, we get
\begin{equation}
2 v^2_{\textbf{k}} = \dfrac{\mathcal{E}^0_{\textbf{k}} + \frac{U N_0}{V}}{\mathcal{E}_{\textbf{k}}} - 1,
\end{equation}
\begin{equation}
v^2_{\textbf{k}} = \dfrac{1}{2} \left( \dfrac{\mathcal{E}^0_{\textbf{k}} + \frac{U N_0}{V}}{\mathcal{E}_{\textbf{k}}} - 1 \right),
\end{equation}
which we insert in the number density of bosons at low temperatures, i.e.,
\begin{equation}
\lim_{T \rightarrow 0^+} \hat{\rho}(\textbf{r}) = \dfrac{N_0}{V} + \dfrac{1}{2V} \sum_{\textbf{k} \neq \textbf{0}} \left( \dfrac{\mathcal{E}^0_{\textbf{k}} + \frac{U N_0}{V}}{\mathcal{E}_{\textbf{k}}} - 1 \right).
\end{equation}
We modify the dispersion relation as follows
\begin{align}
\mathcal{E}_{\textbf{k}} &= \sqrt{\left( \dfrac{\hslash^2 \textbf{k}^2}{2m} \right)^2 + \dfrac{\hslash^2 \textbf{k}^2}{2m} 2 \dfrac{U N_0}{V}} \equiv \notag \\
&\equiv \sqrt{\dfrac{\hslash^2 \textbf{k}^2 U N_0}{mV}} \sqrt{\dfrac{\hslash^2 \textbf{k}^2 V}{4 m U N_0} + 1} ,
\end{align}
and we note that
\begin{align}
\left[ \dfrac{\hslash^2 V}{4 m U N_0} \right] &= \dfrac{\left[ E \right]^2 \left[ T \right]^2 \left[ L \right]^3}{\left[ m \right] \left[ E \right] \left[ L \right]^3} = \notag \\
&= \dfrac{\left[ E \right] \left[ T \right]^2}{\left[ m \right]} = \notag \\
&= \dfrac{\left[ m \right] \left[ L \right]^2 \left[ T \right]^2}{\left[ m \right] \left[ T \right]^2} = \notag \\
&= \left[ L \right]^2 ,
\end{align}
consequently, the object
\begin{align}
l &= \dfrac{\hslash}{m} \sqrt{\dfrac{mV}{2 U N_0}} \equiv \notag \\
&\equiv \dfrac{\hslash}{\sqrt{2} mc}
\end{align}
has the dimensions of a length. The dispersion relation becomes
\begin{equation}
\mathcal{E}_{\textbf{k}} = \hslash k c \sqrt{\dfrac{l^2 k^2}{2} + 1} ,
\end{equation}
where we used the velocity $\eqref{eq: velocitàspettroBogoljubovpiccolimomenti}$. From the identity
\begin{align}
l^2 k^2 &= k^2 \dfrac{\hslash^2 V}{2 m U N_0} = \notag \\
&= \dfrac{\mathcal{E}^0_{\textbf{k}} V}{U N_0},
\end{align}
we write the non-condensate term as
\begin{align}
\dfrac{1}{2V} \sum_{\textbf{k} \neq \textbf{0}} \left( \dfrac{\mathcal{E}^0_{\textbf{k}} + \frac{U N_0}{V}}{\mathcal{E}_{\textbf{k}}} - 1 \right) 
&= \dfrac{1}{2V} \sum_{\textbf{k} \neq \textbf{0}} \left( \frac{\frac{U N_0}{V} \left( \frac{\mathcal{E}^0_{\textbf{k}} V}{U N_0} + 1 \right)}{\hslash k c \sqrt{\frac{l^2 k^2}{2} + 1}} - 1 \right) = \notag \\
&= \dfrac{1}{2V} \sum_{\textbf{k} \neq \textbf{0}} \left( \frac{\frac{U N_0}{V} \left( l^2 k^2 + 1 \right)}{\hslash k \sqrt{\frac{U N_0}{m V}} \sqrt{\frac{l^2 k^2 + 2}{2}}} - 1 \right) = \notag \\
&= \dfrac{1}{2V} \sum_{\textbf{k} \neq \textbf{0}} \left( \frac{\sqrt{\frac{2 m U N_0}{\hslash^2 k^2 V}} \left( l^2 k^2 + 1 \right)}{\sqrt{l^2 k^2 + 2}} - 1 \right) = \notag \\
&= \dfrac{1}{2V} \sum_{\textbf{k} \neq \textbf{0}} \left( \frac{l^2 k^2 + 1}{l k \sqrt{l^2 k^2 + 2}} - 1 \right).
\end{align}
In the thermodynamic limit approximation $\eqref{eq: illimitetermodinamico}$, given that the series excludes the zero-momentum term, the integral must be interpreted as a Cauchy principal value, i.e.,
\begin{align}
\dfrac{1}{2V} \sum_{\textbf{k} \neq \textbf{0}} \left( \dfrac{\mathcal{E}^0_{\textbf{k}} + \frac{U N_0}{V}}{\mathcal{E}_{\textbf{k}}} - 1 \right) &\simeq \dfrac{1}{2V} \dfrac{V}{\left( 2 \pi \right)^3} \pv \int d^3\textbf{k} \left( \frac{l^2 k^2 + 1}{l k \sqrt{l^2 k^2 + 2}} - 1 \right) = \notag \\
&= \dfrac{1}{2 \left( 2 \pi \right)^3} \pv \int d^3\textbf{k} \left( \frac{l^2 k^2 + 1}{l k \sqrt{l^2 k^2 + 2}} - 1 \right).
\end{align}
For convenience, we multiply and divide by \( l^3 \), and rewrite as
\begin{equation}
\dfrac{1}{l^3} \dfrac{1}{2 \left( 2 \pi \right)^3} \pv \int d^3\left(l \textbf{k}\right) \left( \frac{l^2 k^2 + 1}{l k \sqrt{l^2 k^2 + 2}} - 1 \right).
\end{equation}
Since the integrand depends only on the modulus of \( \mathbf{k} \), we switch to spherical coordinates, and we set $y = l k$, then
\begin{align}
\dfrac{1}{l^3} \dfrac{1}{2 \left( 2 \pi \right)^3} \pv \int d^3y \left( \frac{y^2 + 1}{y \sqrt{y^2 + 2}} - 1 \right) &= \dfrac{2^{\frac{3}{2}} m^3 c^3}{\hslash^3} \dfrac{1}{2^4 \pi^3} 4 \pi \pv \int_0^{+\infty} dy y^2 \left( \frac{y^2 + 1}{y \sqrt{y^2 + 2}} - 1 \right) = \notag \\
&= \dfrac{m^3 c^3}{\sqrt{2} \hslash^3 \pi^2} \pv \int_0^{+\infty} dy y^2 \left( \frac{y^2 + 1}{y \sqrt{y^2 + 2}} - 1 \right).
\end{align}
The number density of bosons at low temperatures does not depend on the position and it is given by
\begin{equation}
\langle \hat{\rho} \rangle = \dfrac{N_0}{V} + \dfrac{m^3 c^3}{\sqrt{2} \hslash^3 \pi^2} \pv \int_0^{+\infty} dy \ y^2 \left( \frac{y^2 + 1}{y \sqrt{y^2 + 2}} - 1 \right).
\end{equation}
The total number density at low temperatures consists of two contributions: the condensate density \( \frac{N_0}{V} \) and the non-condensate density, described by the second term. Importantly, this latter term depends on the interaction strength through the sound velocity \( c \), with a characteristic power-law behavior
\begin{equation}
\langle \hat{\rho}_{\text{non-condensate}} \rangle \propto c^3 \propto U^{\frac{3}{2}}.
\end{equation}
This non-integer dependence on the interaction parameter \( U \) reveals a key insight: Bogoljubov's prediction goes beyond standard perturbation theory, which would only yield integer powers of \( U \). In other words, the result embodies the effect of an infinite resummation of Feynman diagrams, accounting for collective many-body phenomena that cannot be captured by expanding to finite order in \( U \). This highlights the non-perturbative nature of the Bogoljubov approximation and its ability to capture long-wavelength quantum correlations in an interacting Bose gas.
\chapter{Quantization of the electromagnetic field: photons}\label{Quantization of the electromagnetic field: photons}
This final chapter presents a semirelativistic framework for the quantization of the electromagnetic field, in which the field is treated quantum mechanically while the charged particles are described within a nonrelativistic approximation. This approach provides a consistent theoretical setting for the study of systems where the dynamics of photons is essential, yet relativistic effects on matter can be safely neglected. The resulting formulation is therefore well suited to a wide range of physical situations of interest in atomic, molecular, and condensed-matter physics, where light–matter interactions play a fundamental role. \newline
We begin by revisiting the classical description of the electromagnetic field in terms of scalar and textbftor potentials, a formulation that naturally incorporates gauge invariance. A central step in the construction is the decomposition of these potentials into longitudinal and transverse components, which allows one to clearly distinguish between unphysical gauge-dependent degrees of freedom and the physically relevant ones. In particular, the longitudinal component of the textbftor potential does not correspond to independent dynamical degrees of freedom and can be eliminated by an appropriate choice of gauge, most commonly the Coulomb gauge. The transverse component, on the other hand, describes propagating modes of the electromagnetic field and is directly associated with physical radiation. \newline
On this basis, we analyze the Lagrangian describing a system of charged particles interacting with the electromagnetic field and derive the corresponding Hamiltonian formulation. The resulting Hamiltonian contains the kinetic energy of the particles, the energy stored in the electromagnetic field, and the interaction terms that couple the particle currents to the field. The derivation requires a careful treatment of the constraints arising from gauge invariance, including both primary and secondary constraints, which naturally guide the selection of a consistent gauge and ensure a well-defined Hamiltonian dynamics. \newline
The final part of the chapter is devoted to the quantization of the electromagnetic field. Since only the transverse component of the textbftor potential represents true physical degrees of freedom, it is promoted to a quantum operator and expanded in terms of its normal modes. Each transverse mode is associated with bosonic creation and annihilation operators, corresponding to photons with well-defined momentum and polarization. This leads to a photon-based description of the electromagnetic field, which provides the foundation for the quantum theory of light–matter interactions. Within this formalism, one can describe fundamental processes such as absorption and emission of radiation, as well as more complex collective phenomena in interacting many-body quantum systems.
\section{Electromagnetic field and potentials}
We begin by reviewing some general properties of the electromagnetic (e.m.) field in preparation for its quantization. To this end, it is useful to recall the notions of longitudinal and transverse components of a textbftor field. Given a textbftor field $\textbf{V}(\textbf{r})$, we define the longitudinal and transverse components of the field, $\textbf{V}^{(L)}(\textbf{r})$ and $\textbf{V}^{(T)}(\textbf{r})$ respectively, such that
\begin{equation}
\textbf{V}(\textbf{r}) = \textbf{V}^{(L)}(\textbf{r}) + \textbf{V}^{(T)}(\textbf{r}) ,
\end{equation}
\begin{equation}
\nabla \cdot \textbf{V}^{(T)}(\textbf{r}) = 0,
\label{eq: divergenzacampotrasversale}
\end{equation}
\begin{equation}
\nabla \times \textbf{V}^{(L)}(\textbf{r}) = \textbf{0}.
\label{eq: rotorecampolongitudinale}
\end{equation}
That is, the longitudinal component is irrotational (curl-free), while the transverse component is divergence-free. This decomposition is unique under suitable boundary conditions and plays a central role in the analysis of electromagnetic fields. Now, we show how to derive the components $\textbf{V}^{(L)}(\textbf{r})$, $\textbf{V}^{(T)}(\textbf{r})$. Let $f(\textbf{r})$ and $\textbf{u}(\textbf{r})$ be the divergence and the rotor of the field $\textbf{V}(\textbf{r})$, respectively; from the definitions of the longitudinal and transverse components of the field it follows
\begin{align}
\nabla \cdot \textbf{V}(\textbf{r}) &= \nabla \cdot \textbf{V}^{(L)}(\textbf{r}) = \notag \\
&= f(\textbf{r}),
\end{align}
\begin{align}
\nabla \times \textbf{V}(\textbf{r}) &= \nabla \times \textbf{V}^{(T)}(\textbf{r}) = \notag \\
&= \textbf{u}(\textbf{r}),
\end{align}
from which, applying the Fourier transform, we have
\begin{equation}
- i \textbf{q} \cdot \textbf{V}^{(L)}(\textbf{q}) = f(\textbf{q}),
\label{eq: menoiqVLqugualefq}
\end{equation}
\begin{equation}
- i \textbf{q} \times \textbf{V}^{(T)}(\textbf{q}) = \textbf{u}(\textbf{q}).
\label{eq: qvettorcomponentetrasversalecampo}
\end{equation}
Furthermore, from $\eqref{eq: divergenzacampotrasversale}$, $\eqref{eq: rotorecampolongitudinale}$, the Fourier transforms of the components $\textbf{V}^{(T)}(\textbf{r})$ and $\textbf{V}^{(L)}(\textbf{r})$ are given by
\begin{equation}
- i \textbf{q} \cdot \textbf{V}^{(T)}(\textbf{q}) = 0,
\end{equation}
\begin{equation}
- i \textbf{q} \times \textbf{V}^{(L)}(\textbf{q}) = \textbf{0},
\end{equation}
then 
\begin{equation}
\textbf{V}^{(L)}(\textbf{q}) \parallel \textbf{q}, 
\end{equation}
\begin{equation}
\textbf{V}^{(T)}(\textbf{q}) \perp \textbf{q}.
\end{equation}
Consequently, we can set $\textbf{V}^{(L)}(\textbf{q}) = a \textbf{q}$. From the equation $\eqref{eq: menoiqVLqugualefq}$ we derive 
\begin{equation}
f(\textbf{q}) = - i a q^2,
\end{equation}
\begin{equation}
a = i \dfrac{f(\textbf{q})}{q^2},
\end{equation}
then
\begin{equation}
\textbf{V}^{(L)}(\textbf{q}) = i \dfrac{f(\textbf{q})}{q^2} \textbf{q}.
\end{equation}
From $\eqref{eq: menoiqVLqugualefq}$, it follows
\begin{align}
\textbf{V}^{(L)}(\textbf{q}) &= i \dfrac{f(\textbf{q})}{q^2} \textbf{q} = \notag \\
&= i \dfrac{\left(- i \textbf{q} \cdot \textbf{V}(\textbf{q}) \right)}{q^2} \textbf{q} = \notag \\
&= \dfrac{\textbf{q} \cdot \textbf{V}(\textbf{q})}{q^2} \textbf{q}.
\label{eq: trasformatacomponentelongitudinalecampoinfunzioneditrasformatacampo}
\end{align}
To determine the trasversal component, we multiply both members of $\eqref{eq: qvettorcomponentetrasversalecampo}$ textbftorially by $\textbf{q}$ and we employ the identity $\mathbf{A} \times (\mathbf{B} \times \mathbf{C}) = (\mathbf{A} \cdot \mathbf{C}) \mathbf{B} - (\mathbf{A} \cdot \mathbf{B}) \mathbf{C}$ as follows
\begin{align}
- i \textbf{q} \times \left( \textbf{q} \times \textbf{V}^{(T)}(\textbf{q}) \right) &= - i \left[ \left( \mathbf{q} \cdot \mathbf{V}^{(T)} \right) \mathbf{q} - q^2 \mathbf{V}^{(T)} \right] = \notag \\
&= i q^2 \textbf{V}^{(T)}(\textbf{q}) = \notag \\
&= \textbf{q} \times \textbf{u}(\textbf{q}) ,
\end{align}
then
\begin{equation}
\textbf{V}^{(T)}(\textbf{q}) = - i \dfrac{\textbf{q} \times \textbf{u}(\textbf{q})}{q^2}.
\end{equation}
From $\eqref{eq: rotorecampolongitudinale}$ we have
\begin{align}
\textbf{u}(\textbf{q}) &= - i \textbf{q} \times \textbf{V}^{(T)}(\textbf{q}) = \notag \\
&= - i \textbf{q} \times \textbf{V}(\textbf{q}),
\end{align}
then
\begin{align}
\textbf{V}^{(T)}(\textbf{q}) &= - \dfrac{i}{q^2} \textbf{q} \times \left( - i \textbf{q} \times \textbf{V}(\textbf{q}) \right) = \notag \\
&= - \dfrac{\textbf{q} \times \left( \textbf{q} \times \textbf{V}(\textbf{q}) \right)}{q^2}.
\label{eq: componentetrasversalevettoreinfunzionedelvettore}
\end{align}
Ultimately, we wrote the Fourier transforms of the longitudinal and transverse components of a textbftor field as a function of the Fourier transform of the field itself. \newline
We now apply the decomposition of textbftor fields into longitudinal and transverse components to Maxwell's equations, expressed in the CGS-Gaussian unit system; see the system of equations in $\eqref{eq: equazioniMaxwell}$. In particular, from $\nabla \cdot \textbf{E}^{(T)}=0$, $\nabla \times \textbf{E}^{(L)}=\textbf{0}$, $\nabla \cdot \textbf{B}^{(T)}=0$, $\nabla \times \textbf{B}^{(L)}=\textbf{0}$, it follows
\begin{equation}
\begin{cases}
\nabla \cdot \mathbf{E}^{(L)} = 4 \pi \rho \\[0.8ex]
\nabla \cdot \mathbf{B}^{(L)} = 0 \\[0.8ex]
\nabla \times \mathbf{E}^{(T)} = - \dfrac{1}{c} \dfrac{\partial \mathbf{B}}{\partial t} \\[0.8ex]
\nabla \times \mathbf{B}^{(T)} = \dfrac{1}{c} \dfrac{\partial \mathbf{E}}{\partial t} + \dfrac{4 \pi}{c} \mathbf{J}
\end{cases}
.
\label{eq: equazioniMaxwell2}
\end{equation}
The divergence and the rotor of the longitudinal component of the magnetic field are zero, consequently this component is zero, consequently the magnetic field is a transverse field, i.e.,
\begin{equation}
\textbf{B} = \textbf{B}^{(T)}.
\end{equation}
In contrast, the electric field has both longitudinal and transverse components. Recalling that the divergence of a rotor is zero, we apply the divergence to Maxwell's fourth equation, obtaining
\begin{align}
0 &= \nabla \cdot (\nabla \times \textbf{B}) = \notag \\
&= \dfrac{1}{c} \dfrac{\partial \nabla \cdot \textbf{E}}{\partial t} + \dfrac{4 \pi}{c} \nabla \cdot \textbf{J} = \notag \\
&= \dfrac{1}{c} \dfrac{\partial\rho}{\partial t} + \dfrac{4 \pi}{c} \nabla \cdot \textbf{J},
\end{align}
then
\begin{equation}
\dfrac{\partial\rho}{\partial t} + \nabla \cdot \textbf{J} = 0.
\end{equation}
In this way the continuity law is derived. We aim to show that the longitudinal component of 
\begin{equation}
\dfrac{1}{c} \dfrac{\partial \textbf{E}}{\partial t} + \dfrac{4 \pi}{c} \textbf{J}
\end{equation}
vanishes. Indeed, by definition, the rotor of the longitudinal component is zero; moreover, the divergence of the longitudinal component is
\begin{equation}
\nabla \cdot \left( \dfrac{1}{c} \dfrac{\partial \textbf{E}^{(L)}}{\partial t} + \dfrac{4 \pi}{c} \textbf{J}^{(L)} \right) = \dfrac{4 \pi}{c} \left( \dfrac{\partial\rho}{\partial t} + \nabla \cdot \textbf{J} \right) = 0,
\end{equation}
where we used $\nabla \cdot \textbf{J}^{(L)} = \nabla \cdot \textbf{J}$, since $\nabla \cdot \textbf{J}^{(T)}=0$. Consequently, both the rotor and the divergence of the longitudinal component of the considered field are zero, so the considered field is transverse, and we can rewrite Maxwell's equations $\eqref{eq: equazioniMaxwell2}$ as
\begin{equation}
\begin{cases}
\nabla \cdot \mathbf{E}^{(L)} = 4 \pi \rho \\[0.8ex]
\nabla \cdot \mathbf{B}^{(L)} = 0 \\[0.8ex]
\nabla \times \mathbf{E}^{(T)} = - \dfrac{1}{c} \dfrac{\partial \mathbf{B}^{(T)}}{\partial t} \\[0.8ex]
\nabla \times \mathbf{B}^{(T)} = \dfrac{1}{c} \dfrac{\partial \mathbf{E}^{(T)}}{\partial t} + \dfrac{4 \pi}{c} \mathbf{J}^{(T)}
\end{cases}
.
\label{eq: equazioniMaxwell3}
\end{equation}
Thus, the first and second Maxwell equations refer to the longitudinal components of the fields, while the third and fourth Maxwell equations refer to the transverse components of the fields. \newline
Let us now introduce the textbftor potential and the scalar potential. Since the divergence of the magnetic field is zero and since the divergence of a rotor is zero, the magnetic field can be written as the rotor of a textbftor field $\textbf{A}$, called the textbftor potential. We have
\begin{align}
\textbf{B} &= \nabla \times \textbf{A} = \notag \\
&= \nabla \times \left( \textbf{A}^{(L)} + \textbf{A}^{(T)} \right) = \notag \\
&= \nabla \times \textbf{A}^{(T)}.
\end{align}
From Maxwell's third equation we derive
\begin{align}
\nabla \times \textbf{E} &= - \dfrac{1}{c} \dfrac{\partial \textbf{B}}{\partial t} = \notag \\
&= - \dfrac{1}{c} \dfrac{\partial \left( \nabla \times \textbf{A} \right)}{\partial t} ,
\end{align}
then
\begin{equation}
\nabla \times \left( \textbf{E} + \dfrac{1}{c} \dfrac{\partial \textbf{A}}{\partial t} \right) = \textbf{0}.
\end{equation}
Since the rotor of a gradient is zero, we can write the quantity 
\begin{equation}
\textbf{E} + \dfrac{1}{c} \dfrac{\partial \textbf{A}}{\partial t}
\end{equation}
as the gradient of a scalar function $\varphi$, which we call a scalar potential, and we get
\begin{equation}
\textbf{E} = - \dfrac{1}{c} \dfrac{\partial \textbf{A}}{\partial t} - \nabla \varphi.
\label{eq: campoEinfunzionedipotenzialevettoreepotenzialescalare}
\end{equation}
In particular, since the gradient is a longitudinal textbftor, because the rotor of a gradient is zero, we get
\begin{equation}
\textbf{E}^{(L)} = - \dfrac{1}{c} \dfrac{\partial \textbf{A}^{(L)}}{\partial t} - \nabla \varphi,
\end{equation}
\begin{equation}
\textbf{E}^{(T)} = - \dfrac{1}{c} \dfrac{\partial \textbf{A}^{(T)}}{\partial t}.
\label{eq: ETinfunzionediderivatatemporaleAT}
\end{equation}
By introducing the potentials $\textbf{A}$, $\varphi$, Maxwell's second and third equations are automatically satisfied, so it is necessary to determine the potentials from the first and fourth equations. Substituting the fields $\textbf{E}$, $\textbf{B}$ as a function of the potentials $\textbf{A}$, $\varphi$ into Maxwell's first and fourth equations, we find
\begin{equation}
\nabla^2 \varphi = - \dfrac{1}{c} \dfrac{\partial \nabla \cdot \textbf{A}}{\partial t} - 4 \pi \rho,
\end{equation}
\begin{equation}
\left( \dfrac{1}{c^2} \dfrac{\partial^2}{\partial t^2} - \nabla^2 \right) \textbf{A} = \dfrac{4\pi}{c} \textbf{J} - \nabla \left( \nabla \cdot \textbf{A} \right) - \dfrac{1}{c} \nabla \left( \dfrac{\partial \varphi}{\partial t} \right).
\end{equation}
Since the rotor of a gradient is zero, we can transform the textbftor potential $\textbf{A}$ by the addition of the gradient of a scalar function while leaving the magnetic field unchanged. Under the transformationl $\textbf{A} \rightarrow \textbf{A}' = \textbf{A} + \nabla f$, the magnetic field is unchanged. Imposing that the electric field is unchanged under the transformation $(\textbf{A},\varphi) \rightarrow (\textbf{A}',\varphi')$, we get
\begin{equation}
\textbf{E}' = - \dfrac{1}{c} \dfrac{\partial \textbf{A}'}{\partial t} - \nabla \varphi' = - \dfrac{1}{c} \dfrac{\partial \textbf{A}}{\partial t} - \dfrac{1}{c} \dfrac{\partial \nabla f}{\partial t} - \nabla \varphi' ,
\end{equation}
and from $\eqref{eq: campoEinfunzionedipotenzialevettoreepotenzialescalare}$ we have
\begin{equation}
\varphi' = \varphi - \dfrac{1}{c} \dfrac{\partial f}{\partial t}.
\end{equation}
Accordingly, the gauge transformation that leaves the electromagnetic field unchanged between two reference systems is
\begin{equation}
\begin{cases}
\textbf{A}' = \textbf{A} + \nabla f \\
\varphi' = \varphi - \dfrac{1}{c} \dfrac{\partial f}{\partial t}
\end{cases}
.
\end{equation}
Since the rotor of the textbftor potential is fixed by the equation $\nabla \times \textbf{A} = \textbf{B}$, to uniquely fix the gauge, i.e., equivalently to define the textbftor potential, one must assign $\nabla \cdot \textbf{A}$. We observe in particular that the gauge transformation for the textbftor potential consists of the addition of the gradient of the gauge function, which is a longitudinal field, consequently the transverse component of the textbftor potential is unaffected by the gauge transformation. \newline
If we set $\nabla \cdot \textbf{A}=0$, we get Coulomb gauge. In this case, we have $\nabla \cdot \textbf{A}=\nabla \cdot \textbf{A}^{(L)}=0$, $\nabla \times \textbf{A}^{(L)}=\textbf{0}$, so the textbftor potential is a transverse field. In Coulomb gauge, Maxwell's first equation becomes
\begin{align}
\nabla \cdot \textbf{E}^{(L)} &= \nabla \cdot (- \nabla \varphi) = \notag \\
&= 4 \pi \rho ,
\end{align}
then
\begin{equation}
\nabla^2 \varphi = - 4 \pi \rho,
\end{equation}
which is the Poisson equation for the scalar potential, which has as its solution an instantaneous scalar potential, that is, $\varphi(t)$ depends on $\rho(t)$. We observe that the quantity that has physical meaning is the electromagnetic field, so there is no contradiction in having a scalar potential that propagates instantaneously: e.m. field must be retarded. In the Coulomb gauge, from $\textbf{E}^{(L)} = - \nabla \varphi$ and equation $\eqref{eq: ETinfunzionediderivatatemporaleAT}$, the longitudinal component of the electric field depends only on the scalar field and the transverse component of the electric field depends only on the textbftor potential. \newline
If we set 
\begin{equation}
\nabla \cdot \textbf{A} + \dfrac{1}{c} \dfrac{\partial \varphi}{\partial t} = 0, 
\end{equation}
we get Lorenz gauge. In this case the equations of $\textbf{A}$, $\varphi$ are decoupled, and in particular we have
\begin{equation}
\begin{cases}
\left( \dfrac{1}{c^2} \dfrac{\partial^2}{\partial t^2} - \nabla^2 \right) \varphi = 4 \pi \rho \\
 \left( \dfrac{1}{c^2} \dfrac{\partial^2}{\partial t^2} - \nabla^2 \right) \textbf{A} = \dfrac{4 \pi}{c} \textbf{J}
\end{cases}
.
\end{equation}
Now, let us compute the Fourier transforms of electric field and magnetic field. The Fourier transforms of Maxwell's equations $\eqref{eq: equazioniMaxwell}$ are given by
\begin{equation}
\begin{cases}
- i \mathbf{q} \cdot \mathbf{E}_{\mathbf{q}} = 4 \pi \rho_{\mathbf{q}} \\[0.8ex]
- i \mathbf{q} \cdot \mathbf{B}_{\mathbf{q}} = 0 \\[0.8ex]
- i \mathbf{q} \times \mathbf{E}_{\mathbf{q}} = - \dfrac{1}{c} \dfrac{\partial \mathbf{B}_{\mathbf{q}}}{\partial t} \\[0.8ex]
- i \mathbf{q} \times \mathbf{B}_{\mathbf{q}} = \dfrac{1}{c} \dfrac{\partial \mathbf{E}_{\mathbf{q}}}{\partial t} + \dfrac{4 \pi}{c} \mathbf{J}_{\mathbf{q}}
\end{cases}
,
\label{eq: equazioniMaxwell4}
\end{equation}
and the Fourier transform of the continuity equation is given by
\begin{equation}
- i \textbf{q} \cdot \textbf{J}_{\textbf{q}} + \dfrac{\partial \rho_{\textbf{q}}}{\partial t}=0.
\label{eq: equazionecontinuitaspazioq}
\end{equation}
Since $\textbf{E}(\textbf{r})$, $\textbf{B}(\textbf{r})$ are real, their Fourier transforms are complex fields. In particular,
\begin{equation}
\textbf{E}(\textbf{r}) = \dfrac{1}{(2 \pi)^3} \int d\textbf{q} \textbf{E}_{\textbf{q}} e^{- i \textbf{q} \cdot \textbf{r}},
\end{equation}
\begin{equation}
\textbf{E}^*(\textbf{r}) = \dfrac{1}{(2 \pi)^3} \int d\textbf{q} \textbf{E}^*_{\textbf{q}} e^{i \textbf{q} \cdot \textbf{r}}, 
\end{equation}
then $\textbf{E}(\textbf{r})=\textbf{E}^*(\textbf{r})$ implies
\begin{equation}
\textbf{E}^*_{\textbf{q}} = \textbf{E}_{- \textbf{q}}.
\label{eq: vincolosuEqstar}
\end{equation}
Similarly, we get
\begin{equation}
\textbf{B}^*_{\textbf{q}} = \textbf{B}_{- \textbf{q}}.
\label{eq: vincolosuBqstar}
\end{equation}
From equation $\eqref{eq: vincolosuEqstar}$, $\eqref{eq: vincolosuBqstar}$ we get
\begin{equation}
\textbf{A}^*_{\textbf{q}} = \textbf{A}_{- \textbf{q}},
\label{eq: vincolosuAqstar}
\end{equation}
\begin{equation}
\varphi^*_{\textbf{q}} = \varphi_{- \textbf{q}}.
\label{eq: vincolosuvarphiqstar}
\end{equation}
In the space of wave textbftors, fields are related to potentials by
\begin{equation}
\textbf{E}_{\textbf{q}} = - \dfrac{1}{c} \dfrac{\partial \textbf{A}_{\textbf{q}}}{\partial t} + i \varphi_{\textbf{q}} \textbf{q},
\label{eq: Einfunzionedipotenzialispazioq}
\end{equation}
\begin{equation}
\textbf{B}_{\textbf{q}} = - i \textbf{q} \times \textbf{A}_{\textbf{q}}.
\label{eq: Binfunzionedipotenzialespazioq}
\end{equation}
From equation $\eqref{eq: trasformatacomponentelongitudinalecampoinfunzioneditrasformatacampo}$, we get 
\begin{equation}
\textbf{E}^{(L)}_{\textbf{q}}=\dfrac{\textbf{E} \cdot \textbf{q}}{q^2} \textbf{q}. 
\end{equation}
Combining this expression with the Fourier-transformed version of the first of Maxwell's equations, we obtain
\begin{equation}
\textbf{E}^{(L)}_{\textbf{q}} = \dfrac{i 4 \pi \rho_{\textbf{q}}}{q^2} \textbf{q}.
\label{eq: trasformataFouriercomponentelongitudinaleE}
\end{equation}
Now, by the convolution theorem, the Fourier transform of $\int d\textbf{r} f(\textbf{r}) g(\textbf{r}-\textbf{r}')$ is given by $f(\textbf{q}) g(\textbf{q})$ while Parseval equality implies
\begin{align}
\int d\textbf{r} f^*(\textbf{r}) g(\textbf{r}) &= \int d\textbf{r} \int \dfrac{d\textbf{q}}{(2 \pi)^3} f^*_{\textbf{q}} e^{i \textbf{q} \cdot \textbf{r}} \int \dfrac{d \textbf{q}'}{(2 \pi)^3} g_{\textbf{q}'} e^{- i \textbf{q}' \cdot \textbf{r}} = \notag \\
&= \int \dfrac{d\textbf{q}}{(2 \pi)^3} \int d\textbf{q}' \delta(\textbf{q}-\textbf{q}') f^*_{\textbf{q}} g_{\textbf{q}'} = \notag \\
&= \int \dfrac{d \textbf{q}}{(2 \pi)^3} f^*_{\textbf{q}} g_{\textbf{q}}.
\end{align}
Thus, let us consider the Fourier transform of the longitudinal electric field. Recall that the Fourier transform of $\frac{1}{r}$ is given by
\begin{equation}
\dfrac{4 \pi}{q^2} = \int d\textbf{r} \dfrac{e^{i \textbf{q} \cdot \textbf{r}}}{r}.
\end{equation}
In addition, recall that given a field $\textbf{V}(\textbf{r})=\nabla h(\textbf{r})$, then $\textbf{V}_{\textbf{q}}=- i h_{\textbf{q}} \textbf{q}$, consequently the quantity 
\begin{equation}
i \dfrac{4 \pi}{q^2} \textbf{q} 
\end{equation}
is the Fourier transform of 
\begin{equation}
- \nabla \left( \dfrac{1}{r} \right) = \dfrac{\textbf{r}}{r^3}. 
\end{equation}
From the convolution theorem and from $\eqref{eq: trasformataFouriercomponentelongitudinaleE}$ we derive
\begin{equation}
\textbf{E}^{(L)}(\textbf{r},t) = \int d \textbf{r}' \rho(\textbf{r}',t) \dfrac{\textbf{r}-\textbf{r}'}{\left| \textbf{r} - \textbf{r}' \right|^3}.
\end{equation}
Therefore, we solved Maxwell's first equation independently of potentials, that is, independently of gauge: the solution of the equation implies the longitudinal component of the electric field is instantaneous, representing the electrostatic field generated by the charge distribution $\rho(\textbf{r},t)$. Again, there is no contradiction in having found an instantaneous field component, because the magnitude of physical interest is the electric field, which, on the other hand, is retarded overall, i.e., the transverse component of the electric field balances the retarded longitudinal component. \newline
Let us now analyze the energy of the electromagnetic field, which is given by the spatial integral of the energy density, i.e.,
\begin{equation}
U = \dfrac{1}{8 \pi} \int d\textbf{r} \left[ \textbf{E}(\textbf{r}) \cdot \textbf{E}(\textbf{r}) + \textbf{B}(\textbf{r}) \cdot \textbf{B}(\textbf{r}) \right].
\end{equation}
The fields $\textbf{E}(\textbf{r})$, $\textbf{B}(\textbf{r})$ are real, then we can write
\begin{equation}
U = \dfrac{1}{8 \pi} \int d\textbf{r} \left[ \textbf{E}^*(\textbf{r}) \cdot \textbf{E}(\textbf{r}) + \textbf{B}^*(\textbf{r}) \cdot \textbf{B}(\textbf{r}) \right].
\end{equation}
Now we can use the Parseval equation derived earlier and write the energy of the electromagnetic field as a function of the Fourier transforms of the electric and magnetic fields as
\begin{equation}
U = \dfrac{1}{8 \pi} \int \dfrac{d\textbf{q}}{(2 \pi)^3} \left[ \textbf{E}^*_{\textbf{q}} \cdot \textbf{E}_{\textbf{q}} + \textbf{B}^*_{\textbf{q}} \cdot \textbf{B}_{\textbf{q}} \right].
\end{equation}
Recalling that the magnetic field is transverse while the electric field has both longitudinal and transverse components, we get
\begin{equation}
U = \dfrac{1}{8 \pi} \int \dfrac{d\textbf{q}}{(2 \pi)^3} \left[ \left| \textbf{E}^{(L)}_{\textbf{q}} \right|^2 + \left| \textbf{E}^{(T)}_{\textbf{q}} \right|^2 + \left| \textbf{B}^{(T)}_{\textbf{q}} \right|^2 \right].
\end{equation}
Specifically, regarding the contribution to the energy of the longitudinal component of the electric field, we have
\begin{align}
U^{(L)} &= \dfrac{1}{8 \pi} \int \dfrac{d\textbf{q}}{(2 \pi)^3} \textbf{E}^{(L) *}_{\textbf{q}} \cdot \textbf{E}^{(L)}_{\textbf{q}} = \notag \\
&= \dfrac{1}{8 \pi} \int \dfrac{d\textbf{q}}{(2 \pi)^3} (4 \pi)^2 \dfrac{\rho^*_{\textbf{q}} \rho_{\textbf{q}}}{q^2}.
\end{align}
From
\begin{equation}
\dfrac{4\pi}{q^2} \rho(\textbf{q}) = \int d\textbf{r}' \dfrac{\rho(\textbf{r}')}{\left| \textbf{r} - \textbf{r}' \right|},
\end{equation}
and from Parseval identity, we get
\begin{equation}
U^{(L)} = \dfrac{1}{2} \int d\textbf{r} \rho(\textbf{r}) \int d\textbf{r}' \dfrac{\rho(\textbf{r}')}{\left| \textbf{r} - \textbf{r}' \right|}.
\end{equation}
Consequently, regardless of the gauge, the contribution to the energy of $\textbf{E}^{(L)}$ is equal to the electrostatic energy of the charge distribution.
\section{Lagrangian of a system of charged particles in an electromagnetic field}
In order to proceed to the quantization of the electromagnetic field, we must first study the Lagrangian of a system of charges in an electromagnetic field, and then move on to the related Hamiltonian and finally to its quantization. \newline
The Lagrangian of a particle with charge $Q$ in an electromagnetic field is
\begin{equation}
\mathcal{L} = \dfrac{1}{2} m \textbf{v}^2 + Q \left( \dfrac{\textbf{A} \cdot \textbf{v}}{c} - \varphi \right).
\end{equation}
Given a system consisting of charges and the electromagnetic field, we assume that the relative Lagrangian is given by
\begin{equation}
\mathcal{L} = \dfrac{1}{2} \sum_{\alpha} m_{\alpha} \textbf{v}^2_{\alpha} + \dfrac{1}{8\pi} \int d\textbf{r} \left( \textbf{E}^* \cdot \textbf{E} - \textbf{B}^* \cdot \textbf{B} \right) + \sum_{\alpha} Q_{\alpha} \left( \dfrac{\textbf{A}(\textbf{r}_\alpha,t) \cdot \textbf{v}}{c} - \varphi(\textbf{r}_{\alpha},t) \right).
\end{equation}
Since the charge density and current density are respectively written as
\begin{equation}
\rho(\textbf{r}) = \sum_{\alpha} Q_{\alpha} \delta(\textbf{r} - \textbf{r}_{\alpha}),
\end{equation}
\begin{equation}
\textbf{J}(\textbf{r}) = \sum_{\alpha} Q_{\alpha} \textbf{v}_{\alpha} \delta(\textbf{r} - \textbf{r}_{\alpha}),
\end{equation}
we can write the Lagrangian as
\begin{equation}
\mathcal{L} = \dfrac{1}{2} \sum_{\alpha} m_{\alpha} \textbf{v}^2_{\alpha} + \dfrac{1}{8\pi} \int d\textbf{r} \left( \textbf{E}^* \cdot \textbf{E} - \textbf{B}^* \cdot \textbf{B} \right) + \dfrac{1}{c} \int d\textbf{r} \left( \textbf{J} \cdot \textbf{A} - c \rho \varphi \right).
\label{eq: Lagrangianaparticellecaricheincampoelettromagneticoforma1}
\end{equation}
To prove that $\eqref{eq: Lagrangianaparticellecaricheincampoelettromagneticoforma1}$ is the right form of the Lagrangian, we must show that the associated Lagrange equations are the Lorentz equation for charges and the Maxwell equations for fields. We focus in particular on the fields. We write the Lagrangian $\eqref{eq: Lagrangianaparticellecaricheincampoelettromagneticoforma1}$ by applying Fourier transforms as follows
\begin{equation}
\mathcal{L} = \dfrac{1}{2} \sum_{\alpha} m_{\alpha} \textbf{v}^2_{\alpha} + \dfrac{1}{8\pi} \int \dfrac{d\textbf{q}}{(2\pi)^3} \left( \textbf{E}^*_{\textbf{q}} \cdot \textbf{E}_{\textbf{q}} - \textbf{B}^*_{\textbf{q}} \cdot \textbf{B}_{\textbf{q}} \right) + \dfrac{1}{c} \int \dfrac{d\textbf{q}}{(2 \pi)^3} \left( \textbf{J}^*_{\textbf{q}} \cdot \textbf{A}_{\textbf{q}} - c \rho^*_{\textbf{q}} \varphi_{\textbf{q}} \right) ,
\label{eq: Lagrangianaparticellecaricheincampoelettromagneticoforma2}
\end{equation}
where, since charge density and current density are real quantities, we substituted $\rho(\textbf{r}) \rightarrow \rho^*(\textbf{r})$, $\textbf{J}(\textbf{r}) \rightarrow \textbf{J}^*(\textbf{r})$ and then we used the Parseval identity. Recall that by writing the fields as a function of potentials, Maxwell's second and third equations are automatically satisfied, so we must verify that Lagrange's equations lead to the first and fourth equations. 
Note that the Lagrangian is written as a function of the variables $\textbf{A}_{\textbf{q}}(t)$, $\varphi_{\textbf{q}}(t)$: let us consider as Lagrangian variables $\textbf{A}^*_{\textbf{q}}(t)$, $\varphi^*_{\textbf{q}}(t)$. If we derive the Lagrangian with respect to $\varphi^*_{\textbf{q}}$, since $\mathcal{L}$ does not depend on $\dot{\varphi}^*_{\textbf{q}}$, we obtain 
\begin{equation}
\left( - \dfrac{1}{c} \dot{\textbf{A}}_{\textbf{q}} + i \varphi_{\textbf{q}} \textbf{q} \right) \dfrac{- i \textbf{q}}{4 \pi} - \rho_{\textbf{q}} = 0 ,
\end{equation}
which is the Maxwell's first equation, see $\eqref{eq: equazioniMaxwell4}$. Note that in calculating the derivative one must remember that $\int d \textbf{q}$ is invariant under the change of variable $\textbf{q} \rightarrow - \textbf{q}$. To obtain Maxwell's fourth equation, we need to consider the field $\textbf{A}^*_{\textbf{q}}$ as the Lagrangian variable. In this case the Lagrangian depends on $\dot{\textbf{A}}^*_{\textbf{q}}$, because the electric field depends on the derivative of the textbftor potential, and in particular the canonical momentum conjugate to $\textbf{A}^*_{\textbf{q}}$ is given by
\begin{equation}
\dfrac{\partial \mathcal{L}}{\partial \dot{\textbf{A}}^*_{\textbf{q}}} = - \dfrac{1}{4 \pi c} \dot{\textbf{E}}_{\textbf{q}}.
\end{equation}
This equation will be fundamental to the quantization of the electromagnetic field. The term of the magnetic field in the Lagrangian is
\begin{align}
\textbf{B}^*_{\textbf{q}} \cdot \textbf{B}_{\textbf{q}} &= \left( i \textbf{q} \times \textbf{A}^*_{\textbf{q}} \right) \cdot (- i \textbf{q} \times \textbf{A}_{\textbf{q}}) = \notag \\
&= \textbf{A}^*_{\textbf{q}} \cdot \left[ \left( \textbf{q} \times \textbf{A}_{\textbf{q}} \right) \times \textbf{q} \right] = \notag \\
&= - \textbf{A}^*_{\textbf{q}} \cdot \left[ \left( \textbf{q} \times \textbf{q} \right) \times \textbf{A}_{\textbf{q}} \right],
\end{align}
then
\begin{equation}
\dfrac{\partial \mathcal{L}}{\partial \textbf{A}^*_{\textbf{q}}} = - \dfrac{1}{4 \pi} \left[ \left( \textbf{q} \times \textbf{A}_{\textbf{q}} \right) \times \textbf{q} \right] + \dfrac{1}{c} \textbf{J}_{\textbf{q}}.
\end{equation}
Consequently, Lagrange's equation takes the form
\begin{equation}
\dfrac{1}{4 \pi c} \dot{\textbf{E}}_{\textbf{q}} - \dfrac{1}{4 \pi} \left[ \left( \textbf{q} \times \textbf{A}_{\textbf{q}} \right) \times \textbf{q} \right] + \dfrac{1}{c} \textbf{J}_{\textbf{q}} = 0,
\end{equation}
\begin{equation}
\dfrac{1}{4 \pi c} \dot{\textbf{E}}_{\textbf{q}} + \dfrac{1}{c} \textbf{J}_{\textbf{q}} = \dfrac{1}{4 \pi} \dfrac{1}{(-i)} \textbf{B}_{\textbf{q}} \times \textbf{q},
\end{equation}
\begin{equation}
\dfrac{1}{4 \pi c} \dot{\textbf{E}}_{\textbf{q}} + \dfrac{1}{c} \textbf{J}_{\textbf{q}} = - i \dfrac{1}{4 \pi} \textbf{q} \times \textbf{B}_{\textbf{q}},
\end{equation}
and multiplying by $4\pi$ we obtain Maxwell's fourth equation. Having derived Maxwell's first and fourth equations, the Lagrangian $\eqref{eq: Lagrangianaparticellecaricheincampoelettromagneticoforma1}$ is related to a system of charged particles in an electromagnetic field. \newline
Let us now show that the Lagrangian $\eqref{eq: Lagrangianaparticellecaricheincampoelettromagneticoforma2}$ can be made independent of $\textbf{A}^{(L)}_{\textbf{q}}$ and $\varphi_{\textbf{q}}$. From Maxwell's first equation, see $\eqref{eq: equazioniMaxwell4}$, writing $\textbf{E}_{\textbf{q}}$ as a function of potentials, we get
\begin{equation}
i \textbf{q} \cdot \left( - \dfrac{1}{c} \dot{\textbf{A}}_{\textbf{q}} + i \varphi_{\textbf{q}} \textbf{q} \right) = - 4 \pi \rho_{\textbf{q}},
\end{equation}
\begin{equation}
q^2 \varphi_{\textbf{q}} = 4 \pi \rho_{\textbf{q}} - i \dfrac{\textbf{q} \cdot \dot{\textbf{A}}_{\textbf{q}}}{c},
\end{equation}
that is, the scalar potential is then written as a function of the textbftor potential as
\begin{equation}
\varphi_{\textbf{q}} = \dfrac{4 \pi \rho_{\textbf{q}}}{q^2} - i \dot{A}^{(L)}_{\textbf{q}} \dfrac{1}{cq}.
\end{equation}
Substituting in the electric field, we find
\begin{align}
\textbf{E}_{\textbf{q}} &= - \dfrac{1}{c} \left( \dot{\textbf{A}}^{(L)}_{\textbf{q}} +\dot{\textbf{A}}^{(T)}_{\textbf{q}} \right) + i \left( \dfrac{4 \pi \rho_{\textbf{q}}}{q^2} - i \dot{A}^{(L)}_{\textbf{q}} \dfrac{1}{cq} \right) \textbf{q} = \notag \\
&= - \dfrac{1}{c} \dot{\textbf{A}}^{(L)}_{\textbf{q}} - \dfrac{1}{c} \dot{\textbf{A}}^{(T)}_{\textbf{q}} + i \dfrac{4 \pi \rho_{\textbf{q}}}{q^2} \textbf{q} + \dfrac{1}{cq} \dot{A}^{(L)}_{\textbf{q}} \textbf{q} = \notag \\
&= - \dfrac{1}{c} \dot{\textbf{A}}^{(T)}_{\textbf{q}} + i \dfrac{4 \pi \rho_{\textbf{q}}}{q^2} \textbf{q} .
\end{align}
In particular, the term that appears in the Lagrangian is
\begin{equation}
\textbf{E}^*_{\textbf{q}} \cdot \textbf{E}_{\textbf{q}} = \dfrac{1}{c^2} \dot{\textbf{A}}^{(T) *}_{\textbf{q}} \dot{\textbf{A}}^{(T)}_{\textbf{q}} + \dfrac{(4 \pi)^2 \rho^*_{\textbf{q}} \rho_{\textbf{q}}}{q^2}.
\end{equation}
Let us now analyze the magnetic field-dependent Lagrangian term. Recall that in general the transverse component of a textbftor field $\textbf{V}$ is given by $\eqref{eq: componentetrasversalevettoreinfunzionedelvettore}$, then we can write
\begin{align}
\textbf{B}^*_{\textbf{q}} \cdot \textbf{B}_{\textbf{q}} &= q^2 \textbf{A}^*_{\textbf{q}} \cdot \textbf{A}^{(T)}_{\textbf{q}} = \notag \\
&= q^2 \textbf{A}^{(T)*}_{\textbf{q}} \cdot \textbf{A}^{(T)}_{\textbf{q}}.
\end{align}
Substituting in the Lagrangian, we find
\begin{align}
\mathcal{L} &= \dfrac{1}{2} \sum_{\alpha} m_{\alpha} \textbf{v}^2_{\alpha} + \dfrac{1}{8\pi} \int \dfrac{d\textbf{q}}{(2 \pi)^3} \left( \dfrac{1}{c^2} \dot{\textbf{A}}^{(T)*}_{\textbf{q}} \cdot \dot{\textbf{A}}^{(T)}_{\textbf{q}} - q^2 \textbf{A}^{(T)*}_{\textbf{q}} \cdot \textbf{A}^{(T)}_{\textbf{q}} \right) + \dfrac{1}{2} \int \dfrac{d\textbf{q}}{(2 \pi)^3} \dfrac{4 \pi \rho^*_{\textbf{q}} \rho_{\textbf{q}}}{q^2} + \notag \\
&+ \dfrac{1}{c} \int \dfrac{d \textbf{q}}{(2 \pi)^3} \left( \textbf{J}^*_{\textbf{q}} \cdot \textbf{A}_{\textbf{q}} - c \dfrac{4 \pi \rho^*_{\textbf{q}} \rho_{\textbf{q}}}{q^2} + \dfrac{i}{q} \dot{A}^{(L)}_{\textbf{q}} \rho_{\textbf{q}} \right) = \notag \\
&= \dfrac{1}{2} \sum_{\alpha} m_{\alpha} \textbf{v}^2_{\alpha} + \dfrac{1}{8\pi} \int \dfrac{d\textbf{q}}{(2 \pi)^3} \left( \dfrac{1}{c^2} \dot{\textbf{A}}^{(T)*}_{\textbf{q}} \cdot \dot{\textbf{A}}^{(T)}_{\textbf{q}} - q^2 \textbf{A}^{(T)*}_{\textbf{q}} \cdot \textbf{A}^{(T)}_{\textbf{q}} \right) - \dfrac{1}{2} \int \dfrac{d\textbf{q}}{(2 \pi)^3} \dfrac{4 \pi \rho^*_{\textbf{q}} \rho_{\textbf{q}}}{q^2} + \notag \\
&+ \dfrac{1}{c} \int \dfrac{d \textbf{q}}{(2 \pi)^3} \left( \textbf{J}^*_{\textbf{q}} \cdot \textbf{A}_{\textbf{q}} + \dfrac{i}{q} \dot{A}^{(L)}_{\textbf{q}} \rho_{\textbf{q}} \right).
\end{align}
Now we analyze the term $\textbf{J}^*_{\textbf{q}} \cdot \textbf{A}_{\textbf{q}} = J^{(L) *}_{\textbf{q}} A^{(L)}_{\textbf{q}} + J^{(T) *}_{\textbf{q}} A^{(T)}_{\textbf{q}}$. From equation $\eqref{eq: equazionecontinuitaspazioq}$ we get
\begin{equation}
J^{(L)}_{\textbf{q}} = - i \dfrac{\dot{\rho}_{\textbf{q}}}{q} ,
\end{equation}
\begin{equation}
J^{(L) *}_{\textbf{q}} = i \dfrac{\dot{\rho}^*_{\textbf{q}}}{q},
\end{equation}
\begin{equation}
\textbf{J}^*_{\textbf{q}} \cdot \textbf{A}_{\textbf{q}} = i \dfrac{\dot{\rho}^*_{\textbf{q}}}{q} A^{(L)}_{\textbf{q}} + J^{(T) *}_{\textbf{q}} A^{(T)}_{\textbf{q}}.
\end{equation}
in the last term of the Lagrangian then appears the quantity
\begin{equation}
i \dfrac{\dot{\rho}^*_{\textbf{q}}}{q} A^{(L)}_{\textbf{q}} + J^{(T) *}_{\textbf{q}} A^{(T)}_{\textbf{q}} + \dfrac{i}{q} \dot{A}^{(L)}_{\textbf{q}} \rho_{\textbf{q}} = J^{(T) *}_{\textbf{q}} A^{(T)}_{\textbf{q}} + \dfrac{d}{dt} \left( i \dfrac{\rho_{\textbf{q}}}{q} A^{(L)}_{\textbf{q}} \right).
\end{equation}
The lagrangian depends on $A^{(L)}$ only through a total derivative with respect to time: two lagrangians that differ by a total derivative with respect to time lead to the same equations of motion, so we can ignore the term dependent on the longitudinal component of the textbftor potential. Note this result is very important, because the longitudinal component of the textbftor potential changes with respect to the gauge, while the transverse component remains unchanged, so it represents a Lagrangian coordinate without physical meaning. We finally write the lagrangian as
\begin{equation}
\mathcal{L} = \dfrac{1}{2} \sum_{\alpha} m_{\alpha} \textbf{v}^2_{\alpha} + \dfrac{1}{8\pi} \int \dfrac{d\textbf{q}}{(2 \pi)^3} \left( \dfrac{1}{c^2} \dot{\textbf{A}}^{(T)*}_{\textbf{q}} \cdot \dot{\textbf{A}}^{(T)}_{\textbf{q}} - q^2 \textbf{A}^{(T)*}_{\textbf{q}} \cdot \textbf{A}^{(T)}_{\textbf{q}} \right) - \dfrac{1}{2} \int \dfrac{d\textbf{q}}{(2 \pi)^3} \dfrac{4 \pi \rho^*_{\textbf{q}} \rho_{\textbf{q}}}{q^2} + \dfrac{1}{c} \int \dfrac{d \textbf{q}}{(2 \pi)^3} \textbf{J}^{(T) *}_{\textbf{q}} \cdot \textbf{A}^{(T)}_{\textbf{q}}.
\label{eq: Lagrangianaparticellecaricheincampoelettromagneticoforma3}
\end{equation}
\section{Hamiltonian of a system of charged particles in an electromagnetic field}
Having determined the Lagrangian for a system of charged particles interacting with an electromagnetic field, we now proceed to derive the corresponding Hamiltonian. Since the Lagrangian $\mathcal{L}$ depends only on the transverse component of the textbftor potential, we omit the superscript ${}^{(T)}$ in what follows for notational simplicity, then
\begin{equation}
\mathcal{L} = \dfrac{1}{2} \sum_{\alpha} m_{\alpha} \textbf{v}^2_{\alpha} + \dfrac{1}{8\pi} \int \dfrac{d\textbf{q}}{(2 \pi)^3} \left( \dfrac{1}{c^2} \dot{\textbf{A}}^*_{\textbf{q}} \cdot \dot{\textbf{A}}_{\textbf{q}} - q^2 \textbf{A}^*_{\textbf{q}} \cdot \textbf{A}_{\textbf{q}} \right) - \dfrac{1}{2} \int \dfrac{d\textbf{q}}{(2 \pi)^3} \dfrac{4 \pi \rho^*_{\textbf{q}} \rho_{\textbf{q}}}{q^2} + \dfrac{1}{c} \int \dfrac{d \textbf{q}}{(2 \pi)^3} \textbf{J}^*_{\textbf{q}} \cdot \textbf{A}_{\textbf{q}}.
\label{eq: Lagrangianaparticellecaricheincampoelettromagneticoforma4}
\end{equation}
Let us define two orthogonal versors $\textbf{e}_{{\textbf{q}},1}$, $\textbf{e}_{{\textbf{q}},2}$ such that
\begin{equation}
\begin{cases}
\textbf{e}_{{\textbf{q}},s} \cdot \textbf{q} = 0 \\
\textbf{e}_{-{\textbf{q}},s} = \textbf{e}_{{\textbf{q}},s}
\end{cases}
, \ s \in \left\lbrace 1,2 \right\rbrace, \ \forall \ \textbf{q} ,
\end{equation}
where $s$ is the polarization index. We then write the transverse field component of the textbftor potential as
\begin{equation}
\textbf{A}^{(T)}_{\textbf{q}} = A_{{\textbf{q}},1} \textbf{e}_{{\textbf{q}},1} + A_{{\textbf{q}},2} \textbf{e}_{{\textbf{q}},2}, 
\end{equation}
with $A^*_{{\textbf{q}},s}=A_{-{\textbf{q}},s}$. With this form of $\textbf{A}^{(T)}_{\textbf{q}}$, the second term of the Lagrangian becomes
\begin{equation}
\dfrac{1}{c^2} \left( \dot{\textbf{A}}^*_{\textbf{q}} \cdot \dot{\textbf{A}}_{\textbf{q}} - c^2 q^2 \textbf{A}^*_{\textbf{q}} \cdot \textbf{A}_{\textbf{q}} \right) = \dfrac{1}{c^2} \left( \sum_{s=1}^2 \dot{A}^*_{{\textbf{q}},s} \dot{A}_{{\textbf{q}},s} - c^2 q^2 A^*_{{\textbf{q}},s} A_{{\textbf{q}},s} \right).
\end{equation}
The canonical momentum conjugate to the Lagrangian variable $A^*_{{\textbf{q}},s}$ is given by
\begin{align}
\Pi_{{\textbf{q}},s} &= \dfrac{\partial \mathcal{L}}{\partial \dot{A}^*_{{\textbf{q}},s}} = \notag \\
&= \dfrac{1}{4 \pi c^2} \dot{A}_{{\textbf{q}},s}.
\end{align}
Since the transverse component of the electric field is $\textbf{E}_{\textbf{q}} = - \dfrac{1}{c} \dot{\textbf{A}}_{\textbf{q}} = - 4 \pi c \bm{\Pi}_{\textbf{q}}$, we have 
\begin{align}
\bm{\Pi}_{\textbf{q}} &= - \dfrac{\textbf{E}_{\textbf{q}}}{4 \pi c} = \notag \\
&= \dfrac{1}{4 \pi c^2} \dot{\textbf{A}}_{\textbf{q}}.
\end{align}
If we focus on the second term of the Lagrangian, which describes the free e.m. field, we get
\begin{equation}
\dfrac{\partial \mathcal{L}}{\partial A^*_{{\textbf{q}},s}} = - \dfrac{q^2}{4 \pi} A_{{\textbf{q}},s},
\end{equation}
\begin{equation}
\dfrac{d \Pi_{{\textbf{q}},s}}{dt} = \dfrac{1}{4 \pi c^2} \ddot{A}_{{\textbf{q}},s},
\end{equation}
then
\begin{equation}
\ddot{A}_{{\textbf{q}},s} = - \omega^2_{\textbf{q}} A_{{\textbf{q}},s},
\end{equation}
with $\omega_{\textbf{q}} = c^2 q^2$. In other words, the Lagrangian of free fields is associated with infinite independent harmonic oscillators with frequencies. \newline
Let us now compute the Hamiltonian of the system. In general, given a Lagrangian $\mathcal{L}(q_h,\dot{q}_h)$, the associated Hamiltonian is $\mathcal{H}= \sum_h p_h \dot{q}_h(q_h,p_h) - \mathcal{L}(q_h,\dot{q}_h(q_h,p_h))$. In our case, the Lagrangian variables are $\textbf{r}_{\alpha}$ and $\textbf{A}_{\textbf{q}}$, so we have to write the variables  $\dot{\textbf{r}}_{\alpha}$ and $\dot{\textbf{A}}^*_{\textbf{q}}$ as a function of the Hamiltonian variables. Recall that the last term of the Lagrangian depends on $\dot{\textbf{r}}_{\alpha}$, indeed 
\begin{align}
\dfrac{1}{c} \int \dfrac{d \textbf{q}}{(2 \pi)^3} \textbf{J}^*_{\textbf{q}} \cdot \textbf{A}_{\textbf{q}} &= \dfrac{1}{c} \int d\textbf{r} \textbf{J}(\textbf{r}) \cdot \textbf{A}(\textbf{r}) = \notag \\
&= \dfrac{1}{c} \sum_{\alpha} q_{\alpha} \dot{\textbf{r}}_{\alpha} \cdot \textbf{A}(\textbf{r}_{\alpha}),
\end{align}
then
\begin{align}
p_{\alpha,i} &= \frac{\partial \mathcal{L}}{\partial \dot{r}_{\alpha,i}} = \notag \\
&= m_{\alpha} \dot{r}_{\alpha,i} + \frac{Q_{\alpha}}{c} A_i(\textbf{r}_{\alpha}),
\end{align}
\begin{equation}
\dot{\textbf{r}}_{\alpha} = \dfrac{\textbf{p}_{\alpha}}{m_\alpha} - \dfrac{Q_{\alpha}}{m_\alpha c} \textbf{A}(\textbf{r}_{\alpha}).
\end{equation}
Using the relation between the Lagrangian and the Hamiltonian, we find
\begin{equation}
\mathcal{H} = \sum_{\alpha} \textbf{p}_{\alpha} \cdot \dot{\textbf{r}}_{\alpha} + \int \dfrac{d \textbf{q}}{(2 \pi)^3} \bm{\Pi}_{\textbf{q}} \cdot \textbf{A}^*_{\textbf{q}} - \mathcal{L}.
\end{equation}
The term describing particles in the electromagnetic field is obtained by adding $\sum_{\alpha} \textbf{p}_{\alpha} \cdot \dot{\textbf{r}}_{\alpha}$ with the first and last terms of the Lagrangian, i.e., the one that depends on the current density, and we obtain
\begin{align}
\sum_{\alpha} \textbf{p}_{\alpha} \cdot \dot{\textbf{r}}_{\alpha} - \dfrac{1}{2} \sum_{\alpha} m_{\alpha} \dot{\textbf{r}}^2_{\alpha} - \dfrac{1}{c} \sum_{\alpha} Q_{\alpha} \dot{\textbf{r}}_{\alpha} \cdot \textbf{A}(\textbf{r}_{\alpha}) &= \sum_{\alpha} \left( \textbf{p}_{\alpha} - \dfrac{Q_{\alpha}}{c} \textbf{A}(\textbf{r}_{\alpha}) \right) \cdot \dot{\textbf{r}}_{\alpha} - \dfrac{1}{2} \sum_{\alpha} m_{\alpha} \dot{\textbf{r}}^2_{\alpha} = \notag \\
&= \sum_{\alpha} \left( \textbf{p}_{\alpha} - \dfrac{Q_{\alpha}}{c} \textbf{A}(\textbf{r}_{\alpha}) \right) \cdot \left( \dfrac{\textbf{p}_{\alpha}}{m_{\alpha}} - \dfrac{Q_{\alpha}}{m_{\alpha} c} \textbf{A}(\textbf{r}_{\alpha}) \right) + \notag \\
&- \dfrac{1}{2} \sum_{\alpha} m_{\alpha} \left( \dfrac{\textbf{p}_{\alpha}}{m_{\alpha}} - \dfrac{Q_{\alpha}}{m_{\alpha} c} \textbf{A}(\textbf{r}_{\alpha}) \right)^2 = \notag \\
&= \sum_{\alpha} \dfrac{1}{2 m_{\alpha}} \left( \textbf{p}_{\alpha} - \dfrac{Q_{\alpha}}{c} \textbf{A}(\textbf{r}_{\alpha}) \right)^2.
\end{align}
Substituting $\dot{\textbf{A}}_{\textbf{q}}$ as a function of $\bm{\Pi}_{\textbf{q}}$, we finally get
\begin{equation}
\mathcal{H} = \sum_{\alpha} \dfrac{1}{2 m_{\alpha}} \left( \textbf{p}_{\alpha} - \dfrac{Q_{\alpha}}{c} \textbf{A}_{\alpha}(\textbf{r}_{\alpha})  \right)^2 + \dfrac{1}{2} \int \dfrac{d \textbf{q}}{(2 \pi)^3} \dfrac{4 \pi}{q^2} \rho^*_{\textbf{q}} \rho_{\textbf{q}} + \int \dfrac{d \textbf{q}}{(2 \pi)^3} \dfrac{1}{2} \left[ 4 \pi c^2 \bm{\Pi}^*_{\textbf{q}} \cdot \bm{\Pi}_{\textbf{q}} + \dfrac{q^2}{4 \pi} \textbf{A}^*_{\textbf{q}} \cdot \textbf{A}_{\textbf{q}} \right].
\label{eq: Hamiltonianaparticellecaricheincampoelettromagneticoforma1}
\end{equation}
\section{Quantization of the Hamiltonian of a system of charged particles in an electromagnetic field}
After deriving the Hamiltonian for a system of charges in an electromagnetic field, we can proceed to its quantization. In the quantized theory, the classical field amplitudes \( A_{\textbf{q},s} \) and their conjugate momenta \( \Pi_{\textbf{q},s} \) are promoted to operators. Now, we must specify the commutation relations they satisfy. As is standard in canonical quantization, we impose that each pair of conjugate variables satisfies a commutation relation of the form
\begin{equation}
\left[ \hat{A}^\dagger_{{\textbf{q}},s} , \hat{\Pi}_{{\textbf{q}}',s'} \right] = i \hslash \delta_{ss'} \delta({\textbf{q}}-{\textbf{q}}') ,
\end{equation}
with $\hat{A}^\dagger_{{\textbf{q}},s} = \hat{A}_{{-\textbf{q}},s}$. Since the free e.m. field is described by a Hamiltonian of infinite independent harmonic oscillators, we define the annihilation and creation operators as follows
\begin{equation}
a_{{\textbf{q}},s} = \dfrac{1}{2} \left( \sqrt{\dfrac{2 \tilde{m} \omega_{\textbf{q}}}{\hbar}} \hat{A}_{{\textbf{q}},s} + i \sqrt{\dfrac{2}{\hbar \tilde{m} \omega_{\textbf{q}}}} \hat{\Pi}_{{-\textbf{q}},s} \right),
\end{equation}
\begin{equation}
a^\dagger_{{\textbf{q}},s} = \dfrac{1}{2} \left( \sqrt{\dfrac{2 \tilde{m} \omega_{\textbf{q}}}{\hbar}} \hat{A}_{{-\textbf{q}},s} - i \sqrt{\dfrac{2}{\hbar \tilde{m} \omega_{\textbf{q}}}} \hat{\Pi}_{{\textbf{q}},s} \right),
\end{equation}
where $q \in \left\lbrace 1,2 \right\rbrace$ and $\tilde{m} = \frac{1}{4 \pi c^2}$ plays the role of an effective mass in the fields Hamiltonian. It is verified that the  operators $a_{{\textbf{q}},s}$, $a^\dagger_{{\textbf{q}},s}$ are bosonic operators, since they satisfy the algebra $\left[ a_{{\textbf{q}},s} , a^\dagger_{{\textbf{q}}',s'} \right] = \delta({\textbf{q}}-{\textbf{q}}') \delta_{ss'}$. We then write the Hamiltonian as a function of the creation and annihilation operators. Reversing the relations above, we have
\begin{equation}
\hat{A}_{\textbf{q},s} = \sqrt{\dfrac{2 \pi \hbar c^2}{\omega_{\textbf{q}}}} \left( a_{\textbf{q},s} + a_{-\textbf{q},s}^{\dagger} \right),
\end{equation}
\begin{equation}
\hat{\Pi}_{\textbf{q},s} = i \sqrt{\dfrac{\hbar \omega_{\textbf{q}}}{8 \pi c^2}} \left( a_{\textbf{q},s}^{\dagger} - a_{-\textbf{q},s} \right),
\end{equation}
\begin{equation}
\hat{\mathcal{H}} = \sum_{\alpha} \dfrac{1}{2 m_{\alpha}} \left( \hat{\textbf{p}}_{\alpha} - \dfrac{Q_{\alpha}}{c} \hat{\textbf{A}}_{\alpha}(\hat{\textbf{r}}_{\alpha})  \right)^2 + \dfrac{1}{2} \int \dfrac{d \textbf{q}}{(2 \pi)^3} \dfrac{4 \pi}{q^2} \rho^*_{\textbf{q}} \rho_{\textbf{q}} + \int \dfrac{d \textbf{q}}{(2 \pi)^3} \sum_{s=1}^2 \hslash \omega_{\textbf{q}} \left( a^\dagger_{{\textbf{q}},s} a_{{\textbf{q}},s} + \dfrac{1}{2} \right),
\label{eq: quantizzazionecampoem}
\end{equation}
which is the Hamiltonian describing a system of non-relativistic charged particles interacting with the quantized electromagnetic field. The first term accounts for the minimal coupling between each charged particle and the quantized transverse textbftor potential \( \hat{\textbf{A}} \), and governs the dynamics of charges in the presence of radiation. The second term describes the instantaneous Coulomb interaction between charges, expressed in Fourier space via the charge density \( \rho_{\textbf{q}} \). Finally, the third term represents the Hamiltonian of the free electromagnetic field in the Coulomb gauge, quantized as a collection of independent harmonic oscillators for each wavetextbftor \( \textbf{q} \) and polarization state \( s \). The bosons associated with the transverse modes of the electromagnetic field are called photons. \newline
Now, we write the quantized textbftor potential \( \hat{\textbf{A}}(\textbf{r}, t) \) and the fields \( \hat{\textbf{E}}(\textbf{r}, t) \), \( \hat{\textbf{B}}(\mathbf{r}, t) \) as functions of the creation and annihilation operators. From
\begin{equation}
\hat{\textbf{A}}_{\textbf{q}} = \sqrt{\dfrac{2 \pi \hbar c^2}{\omega_{\textbf{q}}}} \sum_{s=1}^2 \left( a_{{\textbf{q}},s} + a^\dagger_{{-\textbf{q}},s} \right) \textbf{e}_{{\textbf{q}},s}, 
\end{equation}
we obtain in the Heisenberg picture (see Eq. \ref{eq: Heisenbergpicture})
\begin{align}
\hat{\textbf{A}}(\textbf{r},t) &= \int \dfrac{d \textbf{q}}{(2 \pi)^3} e^{i \textbf{q} \cdot \textbf{r}} \hat{\textbf{A}}_{\textbf{q}}(t) = \notag \\
&= \int \dfrac{d \textbf{q}}{(2 \pi)^3} e^{i \textbf{q} \cdot \textbf{r}} \sqrt{\dfrac{2 \pi \hbar c^2}{\omega_{\textbf{q}}}} \sum_{s=1}^2 \textbf{e}_{{\textbf{q}},s} \left( a_{{\textbf{q}},s}(t) + a^\dagger_{-{\textbf{q}},s}(t) \right).
\end{align}
The textbftor potential is parallel to $\textbf{e}_{{\textbf{q}},s}$, then it is transverse. The time evolution factor is that which characterizes an operator evolving according to a Hamiltonian of a harmonic oscillator. We also observe that the textbftor potential is written as a superposition of progressive and regressive plane waves whose amplitude is quantized, since it is given by an operator. As for the e.m. field, from $\eqref{eq: ETinfunzionediderivatatemporaleAT}$ we have
\begin{equation}
\hat{\textbf{E}}^{(T)}(\textbf{r},t) = \int \dfrac{d \textbf{q}}{(2 \pi)^3} e^{i \textbf{q} \cdot \textbf{r}} i \sqrt{2 \pi \hbar \omega_{\textbf{q}}} \sum_{s=1}^2 \textbf{e}_{{\textbf{q}},s} \left( a_{{\textbf{q}},s}(t) + a^\dagger_{-{\textbf{q}},s}(t) \right),
\end{equation}
and from $\eqref{eq: Binfunzionedipotenzialespazioq}$, we have
\begin{equation}
\hat{\textbf{B}}(\textbf{r},t) = \int \dfrac{d \textbf{q}}{(2 \pi)^3} e^{i \textbf{q} \cdot \textbf{r}} \sqrt{\dfrac{2 \pi \hbar c^2}{\omega_{\textbf{q}}}} \sum_{s=1}^2 \left( i \textbf{q} \times \textbf{e}_{{\textbf{q}},s} \right) \left( a_{{\textbf{q}},s}(t) + a^\dagger_{-{\textbf{q}},s}(t) \right).
\end{equation}
Recall that in classical mechanics the momentum of the electromagnetic field is defined by $\textbf{p} = \int \dfrac{d \textbf{r}}{4 \pi c} \textbf{E} \times \textbf{B}$. Substituting the fields, we find
\begin{align}
\hat{\textbf{p}}^{(T)} &= \left( 2 \pi \hbar c i^2 \right) \int \dfrac{d \textbf{r}}{4 \pi c} \int \dfrac{d \textbf{q}}{(2 \pi)^3} \int \dfrac{d \textbf{q}'}{(2 \pi)^3} \sum_{s=1}^2 \sum_{s'=1}^2 \notag \\
&e^{i (\textbf{q} + \textbf{q}') \cdot \textbf{r}} \left( \textbf{e}_{{\textbf{q}},s} \times \left( \textbf{q}' \times \textbf{e}_{\textbf{q}',s'} \right) \right) \left( a_{{\textbf{q}},s}(t) + a^\dagger_{-{\textbf{q}},s}(t) \right) \left( a_{{\textbf{q}}',s'}(t) + a^\dagger_{-{\textbf{q}}',s'}(t) \right).
\end{align}
The integral with respect to spatial coordinates gives $\delta({\textbf{q}}+{\textbf{q}}')$, then
\begin{align}
\hat{\textbf{p}}^{(T)} &= \dfrac{1}{2} \int \dfrac{d \textbf{q}}{(2 \pi)^3} \sum_{s=1}^2 \sum_{s'=1}^2 \left( \textbf{e}_{{\textbf{q}},s} \times \left( \textbf{q} \times \textbf{e}_{-{\textbf{q}},s'} \right) \right) \hbar \left( a_{{\textbf{q}},s}(t) + a^\dagger_{-{\textbf{q}},s}(t) \right) \left( a_{-{\textbf{q}},s'}(t) + a^\dagger_{{\textbf{q}},s'}(t) \right) = \notag \\
&= \dfrac{1}{2} \int \dfrac{d \textbf{q}}{(2 \pi)^3} \sum_{s=1}^2 \sum_{s'=1}^2 \hbar \delta_{s,s'} \textbf{q} \left( a_{{\textbf{q}},s}(t) + a^\dagger_{-{\textbf{q}},s}(t) \right) \left( a_{-{\textbf{q}},s'}(t) + a^\dagger_{{\textbf{q}},s'}(t) \right) = \notag \\
&= \int \dfrac{d \textbf{q}}{(2 \pi)^3} \sum_{s=1}^2 \hbar \textbf{q} a^\dagger_{{\textbf{q}},s} a_{{\textbf{q}},s} .
\end{align}
Consequently the momentum $\hat{\textbf{p}}^{(T)}$ is the sum of $\hslash \textbf{q}$ as many times as the number of photons with momentum $\textbf{q}$ and for each of the two polarizations $s$. \newline
To summarize, we have found that the e.m. field can be described as a field of bosonic particles, photons, each carrying energy \( \hslash \omega_{\textbf{q}} \) and momentum \( \hslash \textbf{q} \). The quantization of the number of photons follows from the fact that the number operator has a discrete spectrum. Moreover, recalling the relativistic expression for the energy of a particle, i.e., $U= \sqrt{p^2 c^2 + m^2 c^4}$, and noting that each photon has energy \( \hslash \omega_{\textbf{q}} = \hslash q c \), we conclude that the rest mass of the photon is zero. It is important to stress that photons correspond only to the transverse component of the textbftor potential. To see this explicitly, consider the linear momentum associated with the longitudinal component of the electric field. From the first of Maxwell’s equations (see equation $\eqref{eq: equazioniMaxwell4}$), we get
\begin{align}
\hat{\textbf{p}}^{(L)} &= \int \dfrac{d \textbf{r}}{4 \pi c} \int \dfrac{d \textbf{q}}{(2 \pi)^3} \int \dfrac{d \textbf{q'}}{(2 \pi)^3} e^{- i \textbf{q} \cdot \textbf{r}} e^{i \textbf{q}' \cdot \textbf{r}} \rho_{\textbf{q}} \dfrac{4 \pi i^2}{q^2} \textbf{q} \times \left( \textbf{q}' \times \hat{\textbf{A}}_{-{\textbf{q}}'} \right) = \notag \\
&= \dfrac{1}{c} \int \dfrac{d \textbf{q}}{(2 \pi)^3} \dfrac{\rho_{\textbf{q}}}{q^2} \textbf{q} \times \left( - \textbf{q} \times \hat{\textbf{A}}_{-{\textbf{q}}} \right) = \notag \\
&= \dfrac{1}{c} \int \dfrac{d \textbf{q}}{(2 \pi)^3} \rho_{\textbf{q}} \hat{\textbf{A}}^*_{\textbf{q}} = \notag \\
&= \dfrac{1}{c} \int d\textbf{r} \rho(\textbf{r}) \hat{\textbf{A}}(\textbf{r},t) = \notag \\
&= \dfrac{1}{c} \sum_{\alpha} Q_{\alpha} \hat{\textbf{A}}(\textbf{r}_{\alpha},t),
\end{align}
which coincides exactly with the second term in the expression for the total momentum of particles in the electromagnetic field (see $\eqref{eq: quantizzazionecampoem}$). This shows that the longitudinal field is entirely determined by the presence of sources and does not contribute independent quantum excitations. In contrast, the transverse components of the field describe free oscillations and are responsible for the quantum excitations of the electromagnetic field, that is, the photons. \newline
Now, let us consider the transverse electric field and that the temperature is zero, in other words, let us consider the fundamental state of photons, then the mean value of the electric field is zero since it depends on the mean value of $a^\dagger$ and $a$. This implies that in the fundamental state of the Hamiltonian of the free electromagnetic field there are no oscillations. This result is not inconsistent with classical physics; in fact, to compare a system described by a quantum harmonic oscillator with a classical one, we must consider the system in a coherent state, which is the state most similar to classical states. Recall that given a coherent state $|\psi_{c,s}(\alpha) \rangle$, we have that $\langle \hat{N} \rangle= \langle \psi_{c,s}(\alpha)|a^\dagger a|\psi_{c,s}(\alpha) \rangle = \left| \alpha \right|^2$, where $\langle \hat{N} \rangle$ denotes the average number of quanta in the state $\alpha$. We place $\alpha$ as $\alpha = \sqrt{\langle \hat{N} \rangle} e^{i \varphi}$ and we consider the Hamiltonian of free fields for a fixed $\textbf{q}$ and a fixed polarization $s$, and evaluate the mean value of the electric field over a coherent state. Recalling that $|\psi_{c,s}(\alpha) \rangle=e^{-S(\alpha)}|0 \rangle$ and that $e^{-S} a e^S = a + a^\dagger$, $e^{-S} a e^S = a^\dagger - a$, we obtain
\begin{equation}
\left\langle \hat{\textbf{E}} \right\rangle_{{\textbf{q}},s} \propto \textbf{e}_{{\textbf{q}},s} e^{i \textbf{q} \cdot \textbf{r}} \sqrt{\langle \hat{N} \rangle} e^{i \varphi} e^{- i \omega_{\textbf{q}} t} + \text{c.c.}
\end{equation}
The electric field is described by a plane wave whose intensity is proportional to \( \langle \hat{N} \rangle \), the average number of photons. Varying the intensity of the electromagnetic wave corresponds to changing its average photon number.
\begin{remark}[Blackbody radiation]
The quantization of the electromagnetic field leads naturally to the derivation of the blackbody radiation spectrum. Consider a free electromagnetic field confined within a cavity. The quantized field can be described by the Hamiltonian 
\begin{equation}
\hat{\mathcal{H}} = \sum_{\textbf{q},s} \mathcal{E}_{\textbf{q},s} \, a^\dagger_{\textbf{q},s} a_{\textbf{q},s}, 
\end{equation}
where \( a^\dagger_{\textbf{q},s} \) and \( a_{\textbf{q},s} \) denote the creation and annihilation operators for photons with wavetextbftor \( \textbf{q} \) and polarization state \( s \), and \( \mathcal{E}_{\textbf{q},s} = \hslash \omega_{\textbf{q}} \) is the energy of each mode. At thermal equilibrium at temperature \( T \), the average number of photons in each mode is given by the Bose-Einstein distribution. Therefore, the average value of the Hamiltonian becomes
\begin{align}
\left\langle \hat{\mathcal{H}} \right\rangle &= \sum_{\textbf{q},s} \hslash \omega_{\textbf{q}} \, \langle \hat{N}_{\textbf{q},s} \rangle = \notag \\
&= \sum_{\textbf{q},s} \dfrac{\hslash \omega_{\textbf{q}}}{e^{\beta \hslash \omega_{\textbf{q}}} - 1}.
\end{align}
By transforming the discrete sum into an integral over the density of states in momentum space, and considering the two possible transverse polarizations of each photon, one obtains the energy density per unit frequency
\begin{equation}
u(\omega) = \dfrac{\hslash \omega^3}{\pi^2 c^3} \dfrac{1}{e^{\beta \hslash \omega} - 1},
\end{equation}
which is Planck's law of blackbody radiation. This result not only explains the observed spectrum of blackbody radiation but also resolves the ultraviolet catastrophe predicted by classical physics.
\end{remark}
\subsection{Absorption, stimulated emission and spontaneous emission}
We see how the quantization of the e.m. field implies the stimulated emission and absorption processes, and in particular justifies the phenomenon of spontaneous emission, that is, the instability of excited states. \newline
Consider a hydrogenoid atom, described by the Hamiltonian
\begin{equation}
\hat{\mathcal{H}} = \dfrac{\hat{\textbf{p}}^2}{2m_e} - \dfrac{q_e^2}{r} + \sum_{\textbf{q},s} \hslash \omega_\textbf{q} a^\dagger_{\textbf{q},s} a_{\textbf{q},s} + \dfrac{q_e}{m_e c} \hat{\textbf{p}} \cdot \hat{\textbf{A}}
\end{equation}
where we neglected the term proportional to $\hat{\textbf{A}}^2$. In other words, we consider the Hamiltonian as the sum of the Hamiltonian of the unperturbed system, which describes the system of the electron of the hydrogenoid atom and the e.m. field, i.e., free photons, and the perturbation Hamiltonian 
\begin{equation}
\hat{\mathcal{H}}_I = \dfrac{q_e}{m_e c} \hat{\textbf{p}} \cdot \hat{\textbf{A}}. 
\end{equation}
In the occupation number representation, the eigenstates of the free field are written as
\begin{equation}
|N_{\textbf{q}_1,s_1},N_{\textbf{q}_2,s_2},\ldots \rangle = 
\dfrac{\left( a^\dagger_{\textbf{q}_1,s_1} \right)^{N_{\textbf{q}_1,s_1}}}{\sqrt{N_{\textbf{q}_1,s_1}!}} 
\dfrac{\left( a^\dagger_{\textbf{q}_2,s_2} \right)^{N_{\textbf{q}_2,s_2}}}{\sqrt{N_{\textbf{q}_2,s_2}!}} 
\cdots |0 \rangle.
\end{equation}
Using Fermi's golden rule, the transition probability is
\begin{equation}
W = \dfrac{2 \pi}{\hslash} \left| \langle i|\hat{\mathcal{H}}_I|f \rangle \right|^2 \delta(U_f - U_i),
\end{equation}
and consider an initial and a final state characterized respectively by the energies
\begin{equation}
U_i = U_{N_1} + \left( \hslash \omega_{\textbf{q}_1} N_{\textbf{q}_1,s_1} + \ldots \right),
\end{equation}
\begin{equation}
U_f = U_{N_2} + \left( \hslash \omega_{\textbf{q}_1} N_{\textbf{q}_1,s_1} + \ldots \right).
\end{equation}
The interaction term depends on the textbftor potential $\hat{\textbf{A}}$, which in turn depends on $(a^\dagger + a)$, consequently the only permissible transitions are those such that the final and initial states differ by the energy of a single photon, i.e.,
\begin{equation}
U_{n_2} = U_{n_1} \pm \hslash \omega_q.
\end{equation}
Consequently, considering only the one-photon interaction term we find that radiation allows transitions between electronic states that differ by the energy equal to the energy of a photon. This gives rise to the processes of stimulated absorption and stimulated emission. \newline
Let us now focus on the phenomenon of spontaneous emission. The textbftor potential contains a term involving the operator \( a^\dagger \), so even if we start from an initial state with zero photons, the system can evolve into a state with one photon. Consequently, the interaction between the electron and the electromagnetic field, described by the term \( \hat{\textbf{p}} \cdot \hat{\textbf{A}} \), which depends on \( a^\dagger \), allows the creation of a photon and enables a transition from a higher to a lower electronic energy level. For this reason, excited electronic states are unstable, as they can decay spontaneously, i.e., without the presence of initial photons, via spontaneous emission processes.

\part{Toolkit of Group Theory and Symmetries in Physics}

\chapter{Foundations of group theory}\label{Foundations of group theory}
Group theory is a fundamental mathematical tool in theoretical physics, particularly in quantum mechanics, where it provides a systematic framework for analyzing many structural properties of physical systems. By studying the structure and representations of groups, one can classify quantum states, determine selection rules, predict spectral degeneracies, and construct operators compatible with the fundamental characteristics of the system. Beyond its role in describing symmetries, group theory thus emerges as an effective language and a powerful technique for addressing concrete problems across a wide range of quantum contexts. \newline
In this chapter, we introduce the abstract concept of a group, starting from the algebraic definition. We then move on to continuous groups, or Lie groups, which are equipped with a differentiable structure defined using tools from mathematical analysis. This approach not only highlights the key properties of Lie groups relevant for our purposes, but also makes the subject accessible to students without a specialist background in advanced geometry or topology. Lie groups describe the most important continuous symmetries in physics, such as rotations, translations, and gauge transformations. We will further explore not only the structure of Lie groups, but also that of the corresponding Lie algebras, which represent their infinitesimal approximation and constitute a powerful tool for computation and classification. Lie algebras, defined as vector spaces endowed with a commutator operation satisfying antisymmetry and the Jacobi identity, allow one to study continuous symmetries locally by analyzing the infinitesimal generators of the transformations. \newline
Particular attention will be given to the groups \(SO(3)\) and \(SU(2)\), which play a fundamental role in quantum physics. The group \(SO(3)\) describes rotations in three-dimensional space and is intimately connected to the conservation of angular momentum, a key principle in both classical and quantum mechanics. Its elements correspond to physical rotations that can be visualized as movements of rigid bodies or coordinate transformations preserving distances and orientations. On the other hand, \(SU(2)\) is the double cover of \(SO(3)\), meaning that it "wraps around" the rotation group in a two-to-one manner. Although \(SU(2)\) and \(SO(3)\) share the same Lie algebra, reflecting their identical local or infinitesimal structure, they differ in their global topological properties. This subtle distinction is essential in quantum theory because \(SU(2)\) naturally accommodates spin-$\frac{1}{2}$ particles and other quantum spin states that do not have classical analogs. We will explore how the representations of \(SU(2)\) provide the mathematical framework to describe spin states and how these representations underpin the quantum mechanical treatment of angular momentum. Understanding this connection sheds light on the profound ways in which symmetry principles govern the behavior of quantum systems and highlight the interplay between abstract mathematical structures and physical phenomena.
\section{Lie group}
\begin{definition}[Group]
Let $G$ be a set and let $\cdot : G \times G \rightarrow G$ be a binary operation, i.e., given two elements $g$ and $g'$ of G, the product $gg' :=g \cdot g'$ is equal to another element, say $g''$, in the set. The operation is also called multiplication or composition. The couple ($G$,\textquotedblleft $\cdot$\textquotedblright) is said to be a group if it satisfies the following requirements:
\begin{itemize}
\item {Associativity}. The composition is associative, i.e.,
\begin{equation}
\left( g g' \right) g'' = g \left( g' g'' \right), \ \forall \ g, g', g'' \in G. 
\end{equation}
\item {Identity}. There exists and is unique an element e in G such that 
\begin{equation}
e g = g e = e, \ \forall \ g \in G.
\end{equation}
\item {Inverse}. For each g in G there exists and is unique an element, which we denote by $g^{-1}$, such that
\begin{equation}
g g^{-1} = g^{-1} g = e.
\end{equation}
\end{itemize} 
\end{definition}
If there is no danger of confusion, the composition dot symbol is usually omitted.
\begin{definition}[Lie group]
Lie groups are groups whose elements are analytic functions dependent on $n$ real variables. Given a Lie group $G$, the vector $\bar{a} = (a_{1} , \ldots , a_{n})$, with $a_i \in \mathbb{R}$, identifies the element $g(\bar{a}) \in G$.
\end{definition}
If $\dim G = n$, then each group element can be parametrized by $n$ real coordinates. Without loss of generality, let us assume that the identity of a Lie group is in biunivocal correspondence with the $n$-dimensional null vector. Since the elements of the Lie group are analytic functions, the inverse of one element and the product of two elements can be computed by a Taylor series expansion of the elements of the group.
\begin{definition}[\(GL(V)\)]
For a finite-dimensional vector space \(V\), the general linear group \(GL(V)\) is the group of all invertible linear operators \(V \to V\), equipped with the composition of maps.
\end{definition}

\begin{definition}[Homomorphism]
Given two groups $G$, $G'$, a map
\begin{equation}
f: G \rightarrow G'
\end{equation}
is a homomorphism if it preserves the multiplication, i.e.,
\begin{equation}
f(g_{1}g_{2}) = f(g_{1}) f(g_{2}).
\end{equation}
\end{definition}
Two groups are said to be homomorphic if they are connected by a homomorphism. If the homomorphism is bijective, the homomorphism is said to be isomorphic, and two groups connected by an isomorphism are said to be isomorphic. By definition, we have the following results.
\begin{theorem}
Let $G$ and $G'$ be isomorphic groups and let $e$ and $e'$ be the identities in $G$ and in $G'$ respectively, then
\begin{equation}
f(e) = e' ,
\end{equation}
that is, the identity $e \in G$ is mapped into the identity $e' \in G'$ by the isomorphism.
\begin{proof}
From
\begin{align}
g' &= f(g) = \notag \\
&= f(e g) = \notag \\
&= f(e) f(g) = \notag \\
&= f(e) g',
\end{align}
and
\begin{align}
g' &= f(g) = \notag \\
&= f(g e) = \notag \\
&= f(g) f(e) = \notag \\
&= g' f(e),
\end{align}
the assertion follows.
\end{proof}
\end{theorem}
\begin{theorem} 
Let $f : G \to G'$ be a group homomorphism and let $g \in G$. Then
\begin{equation}
f(g^{-1}) = f(g)^{-1}.
\end{equation}
\begin{proof}
Since $f$ is a homomorphism, one has
\begin{align}
f(g^{-1}) f(g) &= f(g^{-1} g) = \notag \\
&= f(e) = \notag \\
&= e',
\end{align}
where $e'$ denotes the identity element of $G'$. Similarly,
\begin{align}
f(g) f(g^{-1}) &= f(g g^{-1}) = \notag \\
&= f(e) = \notag \\
&= e'.
\end{align}
Therefore, $f(g^{-1})$ is the inverse of $f(g)$, and the result follows.
\end{proof}
\end{theorem}

\begin{definition}[Representation of a group]
Given a group \(G\), let \(GL(V)\) denote the group of invertible linear operators on \(V\). If the map
\begin{equation}
T : G \rightarrow GL(V)
\end{equation}
is a homomorphism, i.e.,
\begin{equation}
T(g_{1}g_{2}) = T(g_{1}) T(g_{2}), \ \forall \ g_{1}, g_{2} \in G
\end{equation}
and the operator $T$ is called a representation of the group $G$.
\end{definition}
\begin{definition}[Projective representation]
Let \(G\) be a Lie group and let \(V\) be a vector space. A projective representation of \(G\) on \(V\) is a map
\begin{equation}
S : G \longrightarrow \mathrm{GL}(V)
\end{equation}
such that for all \(g_1, g_2 \in G\) one has
\begin{equation}
S(g_1)\, S(g_2) = \omega(g_1,g_2)\, S(g_1 g_2),
\end{equation}
where \(\omega(g_1,g_2)\) is a nonzero complex scalar, called a multiplier or factor system. If \(\omega(g_1,g_2) = 1\) for all \(g_1, g_2 \in G\), the representation is an ordinary representation.
\end{definition}
\begin{definition}[Faithful representation]
Let \(G\) be a Lie group and let \(V\) be a vector space. A representation
\begin{equation}
S : G \longrightarrow \mathrm{GL}(V)
\end{equation}
is called faithful if it is injective, that is, if
\begin{equation}
S(g) = \mathrm{Id}_V \;\Rightarrow\; g = e,
\end{equation}
where \(e\) denotes the identity element of \(G\).
\end{definition}
\begin{definition}[Fully reducible representation]
Let \(G\) be a Lie group and let \(V\) be a vector space. A representation
\begin{equation}
S : G \longrightarrow \mathrm{GL}(V)
\end{equation}
is called fully reducible if \(V\) can be written as a direct sum of invariant subspaces,
\begin{equation}
V = \bigoplus_{\alpha} V_{\alpha},
\end{equation}
where each \(V_{\alpha}\) is invariant under the action of \(G\), and the restriction of \(S\) to each \(V_{\alpha}\) is irreducible.
\end{definition}
\subsection{SO(3) and SU(2) Lie groups}
\begin{itemize}
\item {Spatial translation group $T(3)$.}\label{Spatial translation group T(3)} \newline
Let $\textbf{r}$ and $\textbf{r}'$ be two points in 3-dimensional space $\mathbb{R}^3$, which are connected by a spatial translation of a vector $\textbf{a}$, i.e.,
\begin{equation}
\textbf{r}' = \textbf{r} + \textbf{a}.
\end{equation}
Given three points $\textbf{r}_1$, $\textbf{r}_2$ and $\textbf{r}_3$ connected as follows
\begin{equation}
\textbf{r}_1 = \textbf{r}_2 + \textbf{a}_{12},
\end{equation}
\begin{equation}
\textbf{r}_2 = \textbf{r}_3 + \textbf{a}_{23},
\end{equation}
we have
\begin{align}
\textbf{r}_1 &= \textbf{r}_2 + \textbf{a}_{12} = \notag \\
&= \textbf{r}_3 + \textbf{a}_{23} + \textbf{a}_{12} = \notag \\
&= \textbf{r}_3 + \textbf{a}_{13},
\end{align}
with
\begin{equation}
\textbf{a}_{13} = \textbf{a}_{12} + \textbf{a}_{23}.
\end{equation}
Consequently, the couple $(\mathbb{R}^3,+)$ forms a group, denoted $T(3)$ and known as the spatial translations group. Given $g_1, g_2 \in T(3)$, with $g_1 =\textbf{a}_1$, $g_2 = \textbf{a}_2$, we have
\begin{align}
g_1 g_2 \textbf{r} &= g_1 ( g_2 \textbf{r} ) = \notag \\
&= g_1 ( \textbf{r} + \textbf{a}_1 ) = \notag \\
&= \textbf{r} + \textbf{a}_1 + \textbf{a}_2 = \notag \\
&= g \textbf{r} ,
\end{align}
then the composition $\cdot$ of two spatial translations is defined as
\begin{equation}
g_1 \cdot g_2 = \textbf{a}_1 + \textbf{a}_2.
\end{equation}
\item {The rotation group $SO(3)$.}
Let $\textbf{r}$ and $\textbf{r}'$ be two points in 3-dimensional space connected by a rotational matrix $R$, i.e.,
\begin{equation}
\textbf{r}' = R \textbf{r} ,
\end{equation}
with $R^{T}=R^{-1}$ and $\det R = 1$. Given three points $\textbf{r}_1$, $\textbf{r}_2$ and $\textbf{r}_3$ connected as follows
\begin{equation}
\textbf{r}_1 = R_{12} \textbf{r}_2,
\end{equation}
\begin{equation}
\textbf{r}_2 = R_{23} \textbf{r}_3,
\end{equation}
we get
\begin{align}
\textbf{r}_1 &= R_{12} \textbf{r}_2 = \notag \\
&= R_{12} \left( R_{23} \textbf{r}_3 \right) = \notag \\
&= R_{12} R_{23} \textbf{r}_3 = \notag \\
&= R_{13} \textbf{r}_3 ,
\end{align}
with
\begin{equation}
R_{13} = R_{12} R_{23}.
\end{equation}
Note that
\begin{align}
R_{13}^T &= \left( R_{12} R_{23} \right)^T = \notag \\
&= R_{23}^T R_{12}^T = \notag \\
&= R_{23}^{-1} R_{12}^{-1} = \notag \\
&= R_{13}^{-1},
\end{align}
and
\begin{align}
\det R_{13} &= \det \left( R_{12} R_{23} \right) = \notag \\
&= \det R_{12} \det R_{23} = \notag \\
&= 1,
\end{align}
i.e., $R_{13}$ is a rotational matrix. Each rotational matrix can be parameterized by means of Euler angles, which are three real numbers, then the set of the rotational matrices together with the composition of the multiplication of matrices form a Lie group, denoted $SO(3)$ and known as the rotation group:
\begin{equation}
SO(3)=\lbrace RR^{T}=R^{T}R= \mathds{1}, \ \det R = 1 \rbrace. 
\end{equation}
Note that the group $SO(3)$ is also known as the special orthogonal group, since the orthogonal group, denoted $O(3)$, is the group of the $3 \times 3$ real matrices $A$ which satisfy $A^T=A^{-1}$, i.e.,
\begin{equation}
O(3)=\lbrace AA^{T}=A^{T}A=\mathds{1}, \ \det A = \pm 1 \rbrace.
\end{equation} 
\item {The Euclidean group $E(3)$.} \newline
Let $\textbf{r}$ and $\textbf{r}'$ be two points in 3-dimensional space connected by a rototranslation, i.e.,
\begin{equation}
\textbf{r}' = R \textbf{r} + \textbf{a} ,
\end{equation}
where $\textbf{a}$ is the spatial translation vector and $R$ is a rotation matrix. It is immediate to verify that any rototranslation can be written as
\begin{equation}
v'_{i} = \sum_{j=1}^4 A_{ij} v_{j} ,
\end{equation}
with
\begin{equation}
A = 
\begin{pmatrix}
& & & a_{1} \\
& R & & a_{2} \\
& & & a_{3} \\
0 & 0 & 0 & 1
\end{pmatrix}.
\label{eq: matricerototraslazione}
\end{equation}
Indeed, given two rototranslation matrices $A^{(1)}$, $A^{(2)}$, the product
\begin{align}
A^{(2)} A^{(1)} &= 
\begin{pmatrix}
R_{11}^{(2)} & R_{12}^{(2)} & R_{13}^{(2)} & a_1^{(2)} \\
R_{21}^{(2)} & R_{22}^{(2)} & R_{23}^{(2)} & a_2^{(2)} \\
R_{31}^{(2)} & R_{32}^{(2)} & R_{33}^{(2)} & a_3^{(2)} \\
0 & 0 & 0 & 1
\end{pmatrix}
\begin{pmatrix}
R_{11}^{(1)} & R_{12}^{(1)} & R_{13}^{(1)} & a_1^{(1)} \\
R_{21}^{(1)} & R_{22}^{(1)} & R_{23}^{(1)} & a_2^{(1)} \\
R_{31}^{(1)} & R_{32}^{(1)} & R_{33}^{(1)} & a_3^{(1)} \\
0 & 0 & 0 & 1
\end{pmatrix}
= \notag \\
&=
\begin{pmatrix}
\sum_{m=1}^3 R_{1m}^{(2)} R_{m1}^{(1)} & \sum_{m=1}^3 R_{1m}^{(2)} R_{m2}^{(1)} & \sum_{m=1}^3 R_{1m}^{(2)} R_{m3}^{(1)} & \sum_{m=1}^3 R_{1m} a^{(1)}_m + a^{(2)}_1 \\ \\
\sum_{m=1}^3 R_{2m}^{(2)} R_{m1}^{(1)} & \sum_{m=1}^3 R_{2m}^{(2)} R_{m2}^{(1)} & \sum_{m=1}^3 R_{2m}^{(2)} R_{m3}^{(1)} & \sum_{m=1}^3 R_{2m} a^{(1)}_m + a^{(2)}_2 \\ \\
\sum_{m=1}^3 R_{3m}^{(2)} R_{m1}^{(1)} & \sum_{m=1}^3 R_{3m}^{(2)} R_{m2}^{(1)} & \sum_{m=1}^3 R_{3m}^{(2)} R_{m3}^{(1)} & \sum_{m=1}^3 R_{3m} a^{(1)}_m + a^{(2)}_3 \\ \\
0 & 0 & 0 & 1
\end{pmatrix}
\end{align}
is still a rototranslation matrix, where the rotation matrix element is given by
\begin{equation}
R_{ij} = \sum_{m=1}^3 R_{im}^{(2)} R_{mj}^{(1)},
\end{equation}
and the component of the translation vector is given by
\begin{equation}
a_i = \sum_{m=1}^3 R_{im}^{(2)} a_m^{(1)} + a_i^{(2)}.
\end{equation}
Consequently, rototranslations together with the matrix multiplication form a group, which is denoted $E(3)$ and known as the Euclidean group. A generical element of the Euclidean group is a couple of the form $g=(R,\textbf{a}) \in E(3)$, where $R$ is a rotation matrix parameterized by $3$ variables and $\textbf{a}$ is a translation vector, that is, the Euclidean group has dimension $6$. Now we ask what is the form of multiplication in $E(3)$, in other words, what is the composition of two rototranslations, say $g_1=(R_1,\textbf{a}_1)$ and $g_2=(R_2,\textbf{a}_2)$. Let us apply the element group $g_2 g_1 \in E(3)$ to a point $\textbf{r}$ as follows
\begin{align}
g_2 g_1 \textbf{r} &= g_2 (g_1 \textbf{r}) = \notag \\
&= g_2 (R_1 \textbf{r} + \textbf{a}_1) = \notag \\
&= R_2 (R_1 \textbf{r} + \textbf{a}_1) + \textbf{a}_2 = \notag \\
&= R_2 R_1 \textbf{r} + R_2 \textbf{a}_1 + \textbf{a}_2 = \notag \\
&= g \textbf{r} ,
\end{align}
then the composition $\cdot$ of two rototranslations is defined as
\begin{equation}
g_2 g_1 = (R_2R_1,R_2\textbf{a}_1+\textbf{a}_2).
\end{equation}
Obviously, spatial translations and rotations with respect to an axis are subgroups of the Euclidean group. Indeed, the group of spatial translations can be seen as a rototranslation of the vector $\textbf{a}$ and a null rotation $\mathds{1}$, that is, it is characterized by the multiplication
\begin{equation}
\left( \mathds{1}, \textbf{a})(\mathds{1},\textbf{a}' \right) = \left( \mathds{1},\textbf{a}+\textbf{a}' \right).
\end{equation}
On the other hand, the group of rotations with respect to an axis can be seen as a rototranslation with translation vector $\textbf{a}=\textbf{0}$ and a rotation $R$, that is, it is characterized by the multiplication
\begin{equation}
\left( R,\textbf{0})(R',\textbf{0} \right) = \left( RR',\textbf{0} \right).
\end{equation}
\end{itemize}
The special unitary group of degree $n$, denoted $SU(n)$, is the Lie group of $n \times n$ unitary matrices with unitary determinant. Here we study the case $n=2$, i.e.,
\begin{equation}
SU(2)=\lbrace UU^{\dagger}=U^{\dagger}U=\mathds{1}, \ \det U = 1 \rbrace. 
\end{equation}
In particular, we show a useful characterization of the $SU(2)$ group, i.e.,
\begin{theorem}
Any element $M \in SU(2)$ can be written as
\begin{equation}
M = a \mathds{1} + i b \sigma_x + i c \sigma_y + i d \sigma_z
\label{eq: primaformaelementiSU(2)}
\end{equation}
where $\sigma_x$, $\sigma_y$ and $\sigma_z$ are Pauli matrices and the real coefficients $a$, $b$, $c$ and $d$ satisfy
\begin{equation}
a^2 + b^2 + c^2 + d^2 = 1.
\label{eq: vincolocoefficientielementiSU(2)}
\end{equation}
\begin{proof}
Given a $2 \times 2$ matrix $M$, that is of the form $\eqref{eq: genericamatriceMcomplessa2x2}$, its inverse matrix is
\begin{align}
M^{-1} 
&= \dfrac{1}{\det M}
\begin{pmatrix}
M_{22} & -M_{12} \\
-M_{21} & M_{11}
\end{pmatrix} 
= \notag \\
&=
\begin{pmatrix}
M_{22} & -M_{12} \\
-M_{21} & M_{11}
\end{pmatrix} ,
\end{align}
and from $M^{-1}=M^\dagger$, we get
\begin{equation}
M_{22} = M_{11}^* ,
\end{equation}
\begin{equation}
M_{21} = - M_{12}^*.
\end{equation}
Consequently, if $M \in SU(2)$, then
\begin{equation}
M =
\begin{pmatrix}
M_{11} & M_{12} \\
-M_{12}^* & M_{11}^*
\end{pmatrix}.
\label{eq: proprietàmatriciSU(2)}
\end{equation}
Since the set of the identity matrix and the Pauli matrices is a basis for the $2 \times 2$ complex matrix space, see theorem $\ref{A basis of M2x2C}$, then $M$ satisfies the $\eqref{eq: primaformaelementiSU(2)}$, i.e.,
\begin{equation}
\begin{pmatrix}
a & 0 \\
0 & a 
\end{pmatrix}
+
\begin{pmatrix}
0 & ib \\
ib & 0
\end{pmatrix}
+ 
\begin{pmatrix}
0 & c \\
-c & 0
\end{pmatrix}
+ 
\begin{pmatrix}
id & 0 \\
0 & id
\end{pmatrix}
=
\begin{pmatrix}
a+id & c+ib \\
-c+ib & a-id
\end{pmatrix}
\end{equation}
if and only if the $\eqref{eq: vincolocoefficientielementiSU(2)}$ holds.
\end{proof}
\end{theorem}
\subsection{Homomorphism 2:1 from $SU(2)$ on $SO(3)$}
A useful application in physics is given by
\begin{theorem}
There exists a $2:1$ homomorphism $R$ from $SU(2)$ on $SO(3)$, i.e.,
\begin{equation}
R : U,-U \in SU(2) \rightarrow R(U)=R(-U) \in SO(3).
\end{equation}
\begin{proof}
The proof is divided into the following steps:
\begin{itemize}
\item[1.] there exists $R$ that maps $U \in SU(2)$ to a real $3 \times 3$ matrix $R(U)$ such that
\begin{equation}
R(U_{1})R(U_{2}) = R(U_{1} U_{2});
\end{equation}
\item[2.] the map $R$ is in $SO(3)$, that is, $R^{T}(U)R(U) = \mathds{1}$, $\det \left( R(U) \right) = +1$;
\item[3.] the map $R$ is on $SO(3)$, i.e., it is a homomorphism;
\item[4.] the homomorphism $R$ is $2:1$, that is, if $R(U_{1}) = R(U_{2})$, then $U_{1} = \pm U_{2}$.
\end{itemize}
\begin{itemize}
\item[1.] Let $\sigma_{1}$, $\sigma_{2}$, $\sigma_{3}$ be the Pauli matrices, i.e., $\sigma^{\dagger}_{i}=\sigma_{i}$, $\Tr \sigma_{i} = 0$. The set of three such matrices together with the identical matrix forms a basis for the space of $2 \times 2$ complex matrices $M$ (see Chapter \ref{Complex 2 times 2 matrices}). In particular, if $M$ is a hermitian null-trace matrix, we have $\alpha_{0} = 0$, $\alpha_{i} = \alpha^{\dagger}_{i}$ for $i=1,2,3$, i.e., any $2 \times 2$ hermitian null-trace matrix is a linear combination of real coefficients of the matrices $\sigma_{i}$. Now, given $\sigma'_{i} = U \sigma_{i} U^{\dagger}, \ U \in SU(2)$, since $\Tr \sigma'_{i} = 0$, $\sigma'^{\ \dagger}_{i} = \sigma'_{i}$, the matrix $\sigma'_{i}$ can be written as a linear combination of the Pauli matrices, i.e.,
\begin{equation}
U \sigma_{i} U^{\dagger} = \sum_{j=1}^3 R_{ji}(U) \sigma_{j}.
\end{equation}
The $9$ real coefficients $R_{ji}(U)$ are uniquely defined by $U$ due to the linear independence of the set $\left\lbrace \sigma_{i} \right\rbrace$. Consequently,
\begin{equation}
R: U \rightarrow R(U)
\end{equation}
is a map that for each $U$ associates a real $3 \times 3$ matrix $R(U)$. Given $V \in SU(2)$, we have
\begin{equation}
VU \sigma_{i} U^{\dagger}V^{\dagger} = (VU)\sigma_{i}(VU)^{\dagger},
\end{equation}
\begin{equation}
V \left( \sum_{k=1}^3 R_{ki}(U) \sigma_{k} \right) V^{\dagger} = \sum_{j=1}^3 R_{ji}(VU) \sigma_{j},
\end{equation}
\begin{equation}
\sum_{j=1}^3 \left( \sum_{k=1}^3 R_{ki}(U) R_{jk}(V) \right) \sigma_{j} = \sum_{j=1}^3 R_{ji}(VU) \sigma_{j} ,
\end{equation}
\begin{equation}
\sum_{k=1}^3 R_{jk}(V) R_{ki}(U) = R_{ji}(VU),
\end{equation}
\begin{equation}
R(V)R(U) = R(VU).
\end{equation}
\item[2.]
From the identity $\sigma_{i} \sigma_{j} = \delta_{ij} \mathds{1} + i \sum_{k=1}^3 \epsilon_{ijk} \sigma_{k}$, we have
\begin{align}
U \sigma_{i} \sigma_{j} U^{\dagger} &= U \sigma_{i} U^{\dagger} U \sigma_{j} U^{\dagger} = \notag \\
&= \sum_{l=1}^3 R_{li}(U) \sigma_{l} \sum_{m=1}^3 R_{mj}(U) \sigma_{m} = \notag \\
&= \sum_{l=1}^3 \sum_{m=1}^3 R_{li}(U) R_{mj}(U) \sigma_{l} \sigma_{m} = \notag \\
&= \sum_{l=1}^3 \sum_{m=1}^3 R_{li}(U) R_{mj}(U) \left( \delta_{lm} \mathds{1} + i \sum_{n=1}^3 \epsilon_{lmn} \sigma_{n} \right) =  \notag \\
&= \sum_{l=1}^3 \sum_{m=1}^3 \delta_{lm} R_{li}(U) R_{mj}(U) \mathds{1} + i \sum_{l=1}^3 \sum_{m=1}^3 \sum_{n=1}^3 R_{li}(U) R_{mj}(U) \epsilon_{lmn} \sigma_{n} = \notag \\
&= \sum_{m=1}^3 R_{mi}(U) R_{mj}(U) \mathds{1} +  i \sum_{l=1}^3 \sum_{m=1}^3 \sum_{n=1}^3 R_{li}(U) R_{mj}(U) \epsilon_{lmn} \sigma_{n} = \notag \\
&= \sum_{m=1}^3 R_{im}^{T}(U) R_{mj}(U) \mathds{1} + i \sum_{l=1}^3 \sum_{m=1}^3 \sum_{n=1}^3 R_{li}(U) R_{mj}(U) \epsilon_{lmn} \sigma_{n} = \notag \\
&= \left( R^{T} R \right)_{ij} \mathds{1} + i \sum_{l=1}^3 \sum_{m=1}^3 \sum_{n=1}^3 R_{li}(U) R_{mj}(U) \epsilon_{lmn} \sigma_{n}.
\end{align}
On the other hand, we have
\begin{equation}
U \sigma_{i} \sigma_{j} U^{\dagger} = \delta_{ij} \mathds{1} + i \sum_{k=1}^3 \epsilon_{ijk} R_{nk} \sigma_{k} ,
\end{equation}
which, by comparison, implies $R^{T} R = \mathds{1}$, that is $R(U) \in O(3)$, and
\begin{equation}
\sum_{l=1}^3 \sum_{m=1}^3 R_{li}(U) R_{mj}(U) \epsilon_{lmn} = \sum_{k=1}^3 \epsilon_{ijk} R_{nk}(U).
\end{equation}
Multiplying both members of the above equation by $R_{ns}(U)$ and using $R^{T} R = \mathds{1}$, we have
\begin{equation}
\sum_{l=1}^3 \sum_{m=1}^3 \sum_{n=1}^3 R_{li}(U) R_{mj}(U) R_{ns}(U) \epsilon_{lmn} = \epsilon_{ijs},
\end{equation} 
\begin{equation}
\det \left( R(U) \right) = 1 ,
\end{equation}
that implies $R(U) \in SO(3)$.
\item[3.] Given $S \in SO(3)$, which we parameterize by means of Euler angles as $S = S_{3}(\gamma) S_{2}(\beta) S_{3}(\alpha)$, we show that there exists $U_{3}(\alpha)$ such that $R\left(U_{3}(\alpha)\right)=S_{3}(\alpha)$, similarly for $U_{2}(\beta)$ and $U_{3}(\gamma)$. With the composition of the row-by-column product of the matrices, it follows that $R\left( U_{3}(\gamma) U_{2}(\beta) U_{3}(\alpha) \right) = S$, i.e., $R$ is a homomorphism on $SO(3)$. Recall
\begin{equation}
S_{3}(\alpha) =
\begin{pmatrix}
\cos{(\alpha)} & \sin{(\alpha)} & 0 \\
-\sin{(\alpha)} & \cos{(\alpha)} & 0 \\
0 & 0 & 1
\end{pmatrix}.
\end{equation}
Given $U_{3}(\alpha) \sigma_{i} U_{3}^{\dagger}(\alpha) = \sum_{j=1}^3 S_{3}(\alpha)_{ji} \sigma_{j}$, which implies
\begin{equation}
U_{3}(\alpha) \sigma_{1} U_{3}^{\dagger}(\alpha) = \cos{(\alpha)} \sigma_{1} - \sin{(\alpha)} \sigma_{2},
\end{equation}
\begin{equation}
U_{3}(\alpha) \sigma_{2} U_{3}^{\dagger}(\alpha) = \sin{(\alpha)} \sigma_{1} + \cos{(\alpha)} \sigma_{2},
\end{equation}
\begin{equation}
U_{3}(\alpha) \sigma_{3} U_{3}^{\dagger}(\alpha) = \sigma_{3},
\end{equation}
it is easily verified that a solution is given by
\begin{align}
U_{3}(\alpha) &= e^{i \frac{\alpha}{2} \sigma_3} = \notag \\
&= \cos{\left(\dfrac{\alpha}{2}\right)} + i \sigma_{3} \sin{\left(\dfrac{\alpha}{2}\right)} = \notag \\
&= 
\begin{pmatrix}
e^{i \frac{\alpha}{2}} & 0 \\
0 & e^{- i \frac{\alpha}{2}}
\end{pmatrix} \in SU(2),
\end{align}
then
\begin{align}
U_{3}(\alpha) \sigma_{1} U_{3}^{\dagger}(\alpha) &= \left[ \cos{\left(\dfrac{\alpha}{2}\right)} + i \sigma_{3} \sin{\left(\dfrac{\alpha}{2}\right)} \right] \sigma_{1} \left[ \cos{\left(\dfrac{\alpha}{2}\right)} - i \sigma_{3} \sin{\left(\dfrac{\alpha}{2}\right)} \right] = \notag \\
&= \cos^{2}{\left(\dfrac{\alpha}{2}\right)} \sigma_{1} - i \sigma_{1} \sigma_{3} \sin{\left(\dfrac{\alpha}{2}\right)} \cos{\left(\dfrac{\alpha}{2}\right)} + i \sigma_{3} \sigma_{1} \sin{\left(\dfrac{\alpha}{2}\right)} \cos{\left(\dfrac{\alpha}{2}\right)} + \sigma_{3} \sigma_{1} \sigma_{3} \sin^{2}{\left(\dfrac{\alpha}{2}\right)} = \notag \\
&= \cos^{2}{\left(\dfrac{\alpha}{2}\right)} \sigma_{1} - i \sigma_{1} \sigma_{3} \sin{\left(\dfrac{\alpha}{2}\right)} \cos{\left(\dfrac{\alpha}{2}\right)} - i \sigma_{1} \sigma_{3} \sin{\left(\dfrac{\alpha}{2}\right)} \cos{\left(\dfrac{\alpha}{2}\right)} + \sigma_{3} \sigma_{1} \sigma_{3} \sin^{2}{\left(\dfrac{\alpha}{2}\right)} = \notag \\
&= \cos^{2}{\left(\dfrac{\alpha}{2}\right)} \sigma_{1} - i \sigma_{1} \sigma_{3} 2 \sin{\left(\dfrac{\alpha}{2}\right)} \cos{\left(\dfrac{\alpha}{2}\right)} + \sigma_{3} \sigma_{1} \sigma_{3} \sin^{2}{\left(\dfrac{\alpha}{2}\right)} = \notag \\
&= \cos^{2}{\left(\dfrac{\alpha}{2}\right)} \sigma_{1} + \sigma_{3} \sigma_{1} \sigma_{3} \sin^{2}{\left(\dfrac{\alpha}{2}\right)} - i \sigma_{1} \sigma_{3} 2 \sin{\left(\dfrac{\alpha}{2}\right)} \cos{\left(\dfrac{\alpha}{2}\right)} = \notag \\
&= \cos^{2}{\left(\dfrac{\alpha}{2}\right)} \sigma_{1} + \sigma_{3} \sigma_{1} \sigma_{3} \sin^{2}{\left(\dfrac{\alpha}{2}\right)} - i \sigma_{1} \sigma_{3} \sin{\left(\alpha\right)} = \notag \\
&= \cos^{2}{\left(\dfrac{\alpha}{2}\right)} \sigma_{1} + \sigma_{3} \sigma_{1} \sigma_{3} \sin^{2}{\left(\dfrac{\alpha}{2}\right)} - i \left(i \epsilon_{132} \sigma_{2} \right) \sin{\left(\alpha\right)} = \notag \\
&= \cos^{2}{\left(\dfrac{\alpha}{2}\right)} \sigma_{1} + \sigma_{3} \sigma_{1} \sigma_{3} \sin^{2}{\left(\dfrac{\alpha}{2}\right)} + ( - \sigma_{2}) \sin{\left(\alpha\right)} = \notag \\
&= \cos^{2}{\left(\dfrac{\alpha}{2}\right)} \sigma_{1} - \sin^{2}{\left(\dfrac{\alpha}{2}\right)} \sigma_{1} - \sin{\left(\alpha\right)} \sigma_{2} = \notag \\
&= \cos{\left(\alpha\right)} \sigma_{1} - \sin{\left(\alpha\right)} \sigma_{2} ,
\end{align}
and similarly for $U_{2}(\beta)$ and $U_{3}(\gamma)$.
\item[4.] Let $U$, $V \in SU(2)$ be such that
\begin{equation}
U \sigma_{i} U^{\dagger} = V \sigma_{i} V^{\dagger} ,
\end{equation} 
then $U = \pm V$ and
\begin{equation}
V^{\dagger} U \sigma_{i} U^{\dagger} U = V^{\dagger} V \sigma_{i} V^{\dagger} U,
\end{equation}
\begin{equation}
V^{\dagger} U \sigma_{i} = \sigma_{i} V^{\dagger} U,
\end{equation}
\begin{equation}
\left[ V^{\dagger}U , \sigma_{i} \right] = 0,
\end{equation}
so, for every hermitian null-trace matrix $M$, we have
\begin{equation}
\left[ V^{\dagger}U , M \right] = 0,
\end{equation}
consequently,
\begin{equation}
V^{\dagger} U = \alpha \mathds{1}.
\end{equation}
If we compute the determinants of both members of the above equation, we have $\alpha^{2} = 1$, then $\alpha = \pm 1$ and $U$ and $-U$ map to the matrix $R(U)$.
\end{itemize}
\end{proof} 
\end{theorem}
\subsection{Germ of a Lie group}
Since the elements of a Lie group are analytic functions, given a Lie group $G$, the germ of the group is defined as the set
\begin{equation}
U_{\epsilon} = \lbrace g(\bar{a}) \in G : |\bar{a}|<\epsilon, \ \forall \ \epsilon>0 \rbrace,
\end{equation}
where $|\bar{a}|$ denotes the Euclidean norm. If $g \in U_{\epsilon}$, then $g^{-1} \in U_{\epsilon}$. We have
\begin{theorem}
The group elements close to the identity are constructed by means of the germ of the group. 
\begin{proof}
Consider the sets of the form
\begin{equation}
U^{n}_{\epsilon} = \left\{ g \in G \ :\ g = g_1 g_2 \cdots g_n,\ \text{with } g_i \in U_\epsilon \text{ for all } i = 1,\dots,n \right\},
\end{equation}
and define
\begin{equation}
G_0 = \bigcup_{n \in \mathbb{N}} U^n_\epsilon.
\end{equation}
We now show that $G_0$ is a subgroup of $G$.
\begin{itemize}
\item{Closure under multiplication.} Let $g, g' \in G_0$. Then there exist $m, m' \in \mathbb{N}$ such that
\[
g = g_1 g_2 \cdots g_m,\quad g_i \in U_\epsilon,
\qquad
g' = g'_1 g'_2 \cdots g'_{m'},\quad g'_j \in U_\epsilon.
\]
It follows that
\[
gg' = g_1 g_2 \cdots g_m g'_1 g'_2 \cdots g'_{m'} \in U^{m + m'}_\epsilon \subset G_0,
\]
so $G_0$ is closed under multiplication.
\item{Closure under inversion.} Let $g \in G_0$. Then there exists $m \in \mathbb{N}$ such that
\[
g = g_1 g_2 \cdots g_m,\quad g_i \in U_\epsilon.
\]
Since $U_\epsilon$ is symmetric (i.e., closed under inversion), we have $g_i^{-1} \in U_\epsilon$ for each $i$. Therefore,
\[
g^{-1} = g_m^{-1} \cdots g_2^{-1} g_1^{-1} \in U^m_\epsilon \subset G_0.
\]
\item{Identity element.} The identity $e$ belongs to $U^1_\epsilon \subset G_0$.
\end{itemize}
\end{proof}
\end{theorem}
Given a group $G$ and an element $g$ in a neighborhood of identity, i.e.,
\begin{equation}
g(\epsilon \bar{v}) = \mathds{1} + \epsilon \sum_{\alpha}^n v_{\alpha} T_{\alpha},
\end{equation}
any element of the connected component of the identity of the Lie group can be obtained as a product of such infinitesimal elements. Indeed, given
\begin{equation}
\epsilon = \dfrac{1}{N} > 0 ,
\end{equation}
we have
\begin{align}
g(\epsilon \bar{v}) \ldots g(\epsilon \bar{v}) &= g^{N}(\epsilon \bar{v}) = \notag \\
&= \left( \mathds{1} + \dfrac{1}{N} \sum_{\alpha=1}^n v_{\alpha} T_{\alpha} \right)^{N},
\end{align}
and we finally get
\begin{equation}
\lim_{N \rightarrow \infty} \left( \mathds{1} + \dfrac{1}{N} \sum_{\alpha=1}^n v_{\alpha} T_{\alpha} \right)^{N} = e^{\sum_{\alpha=1}^n v_{\alpha} T_{\alpha}}.
\end{equation}
The group $G_{0}$ that is composed by union of the powers of the germs is the whole group $G$ if $G$ is connected, otherwise it constitutes only its connected component. For example in $O(3)$ the germ reconstructs $SO(3)$ but not $O(3) \setminus SO(3)=PSO(3)$ since the elements of $PSO(3)$ are not connected with continuity to the identity.
\section{Lie algebra}\label{Lie algebra}
A mathematical structure that encapsulates the local properties of a Lie group is its Lie algebra. By definition, a Lie algebra is a real vector space $L$ of dimension $n$, equipped with a binary operation called the Lie bracket, denoted by $[\cdot,\cdot]$ which satisfies the following properties:
\begin{itemize}
\item bilinearity: the bracket is linear in both arguments;
\item antisymmetry: for all \( x, y \in L \), we have \( [x, y] = -[y, x] \);
\item Jacobi identity: see equations $\eqref{eq: Jacobiidentity}$, $\eqref{eq: Jacobiidentity1}$. 
\end{itemize}
An example of Lie brackets is provided by the commutator.
\begin{definition}[Representation of a Lie algebra]
Given a finite vector space $W$, $\dim(W)=k$, an application $T$ that maps an element of the algebra $L$ with an invertible linear operator on $W$, i.e.,
\begin{equation}
T : L \rightarrow GL(W),
\end{equation}
which is a homomorphism with respect to composition of the matrix commutator in $GL(W)$ is said a representation of the Lie algebra $L$. 
\end{definition}
The invertible linear operators acting on $W$ that are images of the Lie algebra $L$ through the homomorphism $T$ form a closed vector space with respect to the Lie bracket.
\begin{definition}[Homomorphism between Lie algebras]
The Lie algebras $L$ and $L'$ are homomorphic if there exists a linear map 
\begin{equation}
T: L \rightarrow L',
\end{equation}
which preserves Lie brackets, i.e.,
\begin{equation}
T: [x,y] \rightarrow [T(x),T(y)]=T\left( [x,y] \right).
\end{equation}
\end{definition}
\subsection{Structure constants}
Let $L$ be a Lie algebra with basis $\lbrace e_{\alpha} \rbrace$, since $[e_{\alpha},e_{\beta}] \in L$, we can write 
\begin{equation}
[e_{\alpha},e_{\beta}] = \sum_\gamma C_{\alpha \beta \gamma} e_{\gamma},
\end{equation}
where $C_{\alpha \beta \gamma} \in \mathbb{R}$ are known as the structure constants of a Lie algebra. First note that because of the antisymmetry of the Lie brackets, they satisfy
\begin{equation}
C_{\alpha \beta \gamma} = - C_{\beta \alpha \gamma}.
\end{equation}
In addition, we have
\begin{theorem}
The structure constants satisfy
\begin{equation}
\sum_\delta \sum_\epsilon \left( C_{\alpha \beta \delta} C_{\delta \gamma \epsilon} + C_{\gamma \alpha \delta} C_{\delta \beta \epsilon} + C_{\beta \gamma \delta} C_{\delta \alpha \epsilon} \right) e_{\epsilon} = 0.
\label{eq: vincolocostantidistruttura}
\end{equation}
\begin{proof}
From Jacobi identity $\eqref{eq: Jacobiidentity1}$, we get
\begin{align}
0 &= \left[ [e_{\alpha} , e_{\beta}] , e_{\gamma} \right] + \left[ [e_{\gamma} , e_{\alpha}] , e_{\beta} \right] + \left[ [e_{\beta} , e_{\gamma}] , e_{\alpha} \right] = \notag \\
&= \left[ \sum_\delta C_{\alpha \beta \delta} e_{\delta} , e_{\gamma} \right] + \left[ \sum_\delta C_{\gamma \alpha \delta} e_{\delta} , e_{\beta} \right] + \left[ \sum_\delta C_{\beta \gamma \delta} e_{\delta} , e_{\alpha} \right],
\end{align}
then
\begin{align}
\left[ \sum_\delta C_{\alpha \beta \delta} e_{\delta} , e_{\gamma} \right] &= \sum_\delta C_{\alpha \beta \delta} \left[ e_{\delta} , e_{\gamma} \right] = \notag \\
&= \sum_\delta \sum_\epsilon C_{\alpha \beta \delta} C_{\delta \gamma \epsilon} e_{\epsilon},
\end{align}
\begin{align}
\left[ \sum_\delta C_{\gamma \alpha \delta} e_{\delta} , e_{\beta} \right] &= \sum_\delta C_{\gamma \alpha \delta} \left[ e_{\delta} , e_{\beta} \right] = \notag \\
&= \sum_\delta \sum_\epsilon C_{\gamma \alpha \delta} C_{\delta \beta \epsilon} e_{\epsilon},
\end{align}
\begin{align}
\left[ \sum_\delta C_{\beta \gamma \delta} e_{\delta} , e_{\alpha} \right] &= \sum_\delta C_{\beta \gamma \delta} \left[ e_{\delta} , e_{\alpha} \right] = \notag \\
&= \sum_\delta \sum_\epsilon C_{\beta \gamma \delta} C_{\delta \alpha \epsilon} e_{\epsilon},
\end{align}
from which the equation $\eqref{eq: vincolocostantidistruttura}$ follows.
\end{proof}
\end{theorem}
Given any two elements of a Lie algebra, its structure constants allow us to calculate the action of Lie brackets as a combination of the basis of the Lie algebra. 
\begin{theorem}
Given a Lie algebra $L$ and two bases $\lbrace e_i \rbrace$, $\lbrace e'_j \rbrace$, the structure constants in such a bases are connected by
\begin{equation}
C'_{\alpha \beta \tau} = \sum_\gamma \sum_\delta \sum_\epsilon A_{\alpha \gamma} A_{\beta \delta} \left( A \right)^{-1}_{\epsilon \tau} C_{\gamma \delta \epsilon}.
\label{eq: leggitrasformazionecostantidistruttura}
\end{equation}
\begin{proof}
Since the bases $\lbrace e_{\beta} \rbrace$ and $\lbrace e'_{\alpha} \rbrace$ are connected by a linear transformation $A$ as follows
\begin{equation}
e'_{\alpha} = \sum_\beta A_{\alpha \beta} e_{\beta},
\end{equation}
\begin{equation}
e_{\beta} = \sum_\alpha \left( A \right)^{-1}_{\beta \alpha} e'_{\alpha},
\end{equation}
and the Lie bracket provides
\begin{equation}
\left[ e'_{\alpha} , e'_{\beta} \right] = \sum_\gamma C'_{\alpha \beta \tau} e'_{\tau},
\end{equation}
we have
\begin{align}
\left[ e'_{\alpha} , e'_{\beta} \right] &= \left[ \sum_\gamma A_{\alpha \gamma} e_{\gamma} , \sum_\delta A_{\beta \delta} e_{\delta} \right] = \notag \\
&= \sum_\gamma \sum_\delta A_{\alpha \gamma} A_{\beta \delta} \left[ e_{\gamma} , e_{\delta} \right] = \notag \\
&= \sum_\gamma \sum_\delta \sum_\epsilon A_{\alpha \gamma} A_{\beta \delta} C_{\gamma \delta \epsilon} e_{\epsilon} = \notag \\
&= \sum_\gamma \sum_\delta \sum_\epsilon \sum_\tau A_{\alpha \gamma} A_{\beta \delta} \left( A \right)^{-1}_{\epsilon \tau} C_{\gamma \delta \epsilon} \ e'_{\tau},
\end{align}
from which the equation $\eqref{eq: leggitrasformazionecostantidistruttura}$ follows.
\end{proof}
\end{theorem}
\subsection{The Lie algebra of the Lie group of dimension $n$ of invertible matrices of dimension k}
\begin{theorem}
Let $G$ be the Lie group of dimension $n$ of invertible matrices of dimension $k$, then the span of the infinitesimal generators of $G$ together with the Lie brackets of the commutator between matrices is a Lie algebra.
\begin{proof}
By definition, the elements $g \equiv A \in G$ are invertible matrices of dimension $k$ that depend on $n$ real parameters with respect to which they are analytic functions, so a generic matrix element is written as $A_{ij}(\bar{a})$, $i, j \in (1,\ldots,k)$, $\bar{a}=(a_1,\ldots,a_n)$, $a_i \in \mathbb{R}$. Consider the a neighborhood of the identity matrix. Since the identity is mapped by the null vector $\bar{0}$, i.e., $A_{ij}(\bar{0}) = \delta_{ij}$, the elements belonging to the neighborhood of the identity are such that the parameters $a_\alpha$ are infinitesimal. Since the elements $g \in G$ are analytic functions with respect to the parameterization vector, an element can be approximated by a first-order Taylor polynomial. Given an infinitesimal parameterization of the form $\bar{a}_\epsilon = \epsilon \bar{v}$, $a_{\epsilon,\alpha} = \epsilon v_{\alpha}$, $0<\epsilon<1$, $\alpha \in \lbrace 1,\ldots,n \rbrace$, we have
\begin{align}
A_{ij}(\epsilon \bar{v}) &\simeq A_{ij}(\bar{0}) + \epsilon  \sum_{\alpha=1}^n v_{\alpha} \left. \dfrac{\partial A_{ij}}{\partial a_{\alpha}} \right|_{\bar{a}=\bar{0}} \ = \notag \\
&= \delta_{ij} + \epsilon \sum_{\alpha=1}^n v_{\alpha} \left. \dfrac{\partial A_{ij}}{\partial a_{\alpha}} \right|_{\bar{a}=\bar{0}} \ = \notag \\
&= \delta_{ij} + \epsilon \sum_{\alpha=1}^n v_{\alpha} (T_{\alpha})_{ij},
\end{align}
where we defined the matrices
\begin{equation}
(T_{\alpha})_{ij} = \left. \dfrac{\partial A_{ij}}{\partial a_{\alpha}} \right|_{\bar{a}=\bar{0}}.
\end{equation}
We set
\begin{align}
T_{ij} &= \sum_{\alpha=1}^n v_{\alpha} (T_{\alpha})_{ij} = \notag \\
&= \sum_{\alpha=1}^n v_{\alpha} \left. \dfrac{\partial A_{ij}}{\partial a_{\alpha}} \right|_{\bar{a}=\bar{0}},
\end{align}
then
\begin{equation}
A_{ij}(\epsilon \bar{v}) = \delta_{ij} + \epsilon T_{ij}.
\end{equation}
By definition, the matrices $\lbrace T_{\alpha} \rbrace$ are $n$ matrices $k \times k$ and form a linearly independent system. Otherwise, there would exist at least one nonzero $\lambda_{\beta}$ for which $A_{ij} = \sum_{\alpha=1}^n \lambda_{\alpha} (T_{\alpha})_{ij} = \delta_{ij}$, which contradicts the hypothesis of biunivocal correspondence between $A_{ij}$ matrices and the parameters $(a_{1},\ldots,a_{n})$. Then the system of matrices $T_{\alpha}$ is a system of generators. Let $L$ be the vector space generated by these matrices, that is, $L=\Span \lbrace T_{\alpha} \rbrace$. We show that $L$ together with the Lie bracket of the matrix commutator is a Lie algebra. The matrices $\lbrace T_{\alpha} \rbrace$ of the group $G$ are called infinitesimal generators of the Lie algebra $L$. Previously, we show that two different systems of generators span the same space $L$. Let $\bar{b}=(b_{1},\ldots,b_{n})$ be a new parameterization of the group $G$, and let it be writable as a function of the old parameterization $\bar{a}$, i.e., $\bar{b}=\bar{b}(\bar{a})$, $\bar{b}(\bar{0})=\bar{0}$. Let $\bar{b}_\epsilon = \bar{w}$, $b_{\epsilon,\alpha} = \epsilon w_{\alpha}$, $0<\epsilon<1$, $\alpha \in \lbrace 1,\ldots,n \rbrace$ a new infinitesimal parameterization, then
\begin{align}
A_{ij}(\epsilon \bar{w}) & \simeq A_{ij}(\bar{0}) + \epsilon \sum_{\alpha=1}^n v_{\alpha} w_{\alpha} \left. \dfrac{\partial A_{ij}}{\partial b_{\alpha}} \right|_{\bar{b}=\bar{0}} = \notag \\
&= \delta_{ij} + \epsilon \sum_{\alpha=1}^n w_{\alpha} \left. \dfrac{\partial A_{ij}}{\partial b_{\alpha}} \right|_{\bar{b}=\bar{0}} = \notag \\
&= \delta_{ij} + \epsilon \sum_{\alpha=1}^n v_{\alpha} w_{\alpha} \left( \left. \dfrac{\partial a_{\gamma}}{\partial b_{\alpha}} \right|_{\bar{b}=\bar{0}} \right) \left( \left. \dfrac{\partial A_{ij}}{\partial a_{\gamma}} \right|_{\bar{a}=\bar{0}} \right).
\end{align}
On the other hand, in the new parameterization must be
\begin{equation}
A_{ij}(\epsilon \bar{w}) = \delta_{ij} + \epsilon \sum_{\alpha=1}^n v_{\alpha} w_{\alpha} (T'_{\alpha})_{ij},
\end{equation}
then
\begin{equation}
(T'_{\alpha})_{ij} = \left( \left. \dfrac{\partial a_{\gamma}}{\partial b_{\alpha}} \right|_{\bar{b}=\bar{0}} \right) (T_{\gamma})_{ij} ,
\end{equation}
that is, the infinitesimal generators with respect to the parameterization $\bar{b}$ are linear combination of the generators of the parameterization $\bar{a}$, i.e., they belong to the same space $L$. Now, we show that $L$ together with the matrix commutator is a Lie algebra. Consider two distinct elements in a neighborhood of the identity, say $g_{1} = A_{ij}(\epsilon \bar{v})$, $g_{2} = A_{ij}(\epsilon' \bar{w})$. We set
\begin{equation}
B(\epsilon,\epsilon') = g^{-1}_{2}(\epsilon') \ g^{-1}_{1}(\epsilon) \ g_{2}(\epsilon') \ g_{1}(\epsilon)
\end{equation}
and we perform a Taylor expansion of \( B_{ij} \) in \( \epsilon \) and \( \epsilon' \). The leading-order term is given by
\begin{equation}
B_{ij}(\epsilon,\epsilon') \simeq \delta_{ij} + o(\epsilon \epsilon').
\end{equation}
To understand why the first nonzero term is a function of the product $\epsilon \epsilon'$, note that
\begin{equation}
B(\epsilon,0)=B(0,\epsilon')= \mathds{1}, \ \forall \ \epsilon, \epsilon' ,
\end{equation}
then, if we set one between $\epsilon$ and $\epsilon'$ equal to zero, the expansion of $B$ with respect to the other parameter must return zero terms. Now, since
\begin{equation}
g_{1} = \mathds{1} + \epsilon \sum_{\delta=1}^n v_{\delta} T_{\delta} + o(\epsilon^{2}),
\end{equation}
\begin{equation}
g_{2} = \mathds{1} + \epsilon' \sum_{\gamma=1}^n w_{\gamma} T_{\gamma} + o(\epsilon'^{2}),
\end{equation}
\begin{equation}
g^{-1}_{1} = \mathds{1} - \epsilon \sum_{\beta=1}^n v_{\beta} T_{\beta} + o(\epsilon^{2}),
\end{equation}
\begin{equation}
g^{-1}_{2} = \mathds{1} - \epsilon' \sum_{\alpha=1}^n w_{\alpha} T_{\alpha} + o(\epsilon'^{2}),
\end{equation}
we have
\begin{align}
B(\epsilon,\epsilon') &= \left(\mathds{1} - \epsilon' \sum_{\alpha=1}^n w_{\alpha} T_{\alpha}\right) \left(\mathds{1} - \epsilon \sum_{\beta=1}^n v_{\beta} T_{\beta}\right) \left(\mathds{1} + \epsilon' \sum_{\gamma=1}^n w_{\gamma} T_{\gamma}\right) \left(\mathds{1} + \epsilon \sum_{\delta=1}^n v_{\delta} T_{\delta}\right).
\end{align}
We compute the products in $B(\epsilon,\epsilon')$ and ignore terms of higher order than the product $\epsilon \epsilon'$, i.e.,
\begin{equation}
B(\epsilon,\epsilon') = \mathds{1} + \epsilon \epsilon' \sum_{\alpha=1}^n w_{\alpha} T_{\alpha} \sum_{\beta=1}^n v_{\beta} T_{\beta} - \epsilon \epsilon' \sum_{\alpha=1}^n w_{\alpha} T_{\alpha} \sum_{\delta=1}^n v_{\delta} T_{\delta} - \epsilon \epsilon' \sum_{\beta=1}^n v_{\beta} T_{\beta} \sum_{\gamma=1}^n w_{\gamma} T_{\gamma} + \epsilon \epsilon' \sum_{\gamma=1}^n w_{\gamma} T_{\gamma} \sum_{\delta=1}^n v_{\delta} T_{\delta}.
\end{equation}
The second and third terms cancel each other, hence
\begin{align}
B(\epsilon,\epsilon') &= \mathds{1} - \epsilon \epsilon' \sum_{\beta=1}^n v_{\beta} T_{\beta} \sum_{\gamma=1}^n w_{\gamma} T_{\gamma} + \epsilon \epsilon' \sum_{\gamma=1}^n w_{\gamma} T_{\gamma} \sum_{\delta=1}^n v_{\delta} T_{\delta} = \notag \\
&= \mathds{1} + \epsilon \epsilon' \sum_{\gamma=1}^n \sum_{\beta=1}^n w_\gamma v_\beta \left[ T_\gamma , T_\beta \right] ,
\end{align}
where we renamed the dummy index $\delta \rightarrow \beta$. If $\epsilon$ and $\epsilon'$ are infinitesimal, the quantity $\epsilon \epsilon' \sum_{\gamma=1}^n \sum_{\beta=1}^n w_\gamma v_\beta \left[ T_\gamma , T_\beta \right]$ belongs to a neighborhood of the identity, so it belongs to the group, consequently, this object can be written as a linear combination of the matrices $T_{\tau}$, so there must exist coefficients $C_{\tau}(\bar{v},\bar{w})$ such that
\begin{equation}
\sum_{\gamma=1}^n \sum_{\beta=1}^n w_\gamma v_\beta \left[ T_\gamma , T_\beta \right] = \sum_{\tau=1}^n C_{\tau}(\bar{v},\bar{w}) T_{\tau}, \ \forall \ \bar{v}, \bar{w}.
\end{equation}
Since the vectors $\bar{v}$ and $\bar{w}$ are arbitrary, we set their components equal to Kronecker deltas, i.e., $w_{\gamma,\alpha} = \delta_{\gamma \alpha}$, $v_{\beta,\delta} = \delta_{\beta \delta}$, then
\begin{align}
\sum_{\tau=1}^n C_{\alpha \beta \tau} T_{\tau} &= \sum_{\gamma=1}^n \sum_{\beta=1}^n \delta_{\gamma \alpha} \delta_{\beta \delta} \left[T_{\gamma},T_{\delta}\right] = \notag \\
&= \left[T_{\alpha},T_{\beta}\right].
\end{align}
We have shown that the commutator of the generators $\lbrace T_{\alpha} \rbrace$ of $L$ is itself a generator, then the vector space $L$ together with the application of the commutator of matrices is a Lie algebra, in particular, it is the Lie algebra of the Lie group of dimension $n$ of invertible matrices of dimension $k$.
\end{proof}
\end{theorem}
\subsection{Lie algebra of $SO(3)$}
Let $\delta \omega$ be an infinitesimal angle, then an infinitesimal rotation between two bases $\left\lbrace \textbf{e}_i \right\rbrace$ e $\left\lbrace \textbf{e}'_i \right\rbrace$ can be written as
\begin{align}
\textbf{e}'_{i} &= \textbf{e}_{i} + \delta \bm{\omega} \times \textbf{e}_{i} = \notag \\
&= \textbf{e}_{i} + \sum_{k=1}^3 \delta \omega_{k} \textbf{e}_{k} \times \textbf{e}_{i} = \notag \\
&= \textbf{e}_{i} + \sum_{k=1}^3 \sum_{l=1}^3 \delta \omega_{k} \epsilon_{kil} \textbf{e}_{l} = \notag \\
&= \sum_{l=1}^3 \delta_{il} \textbf{e}_{l} + \sum_{k=1}^3 \sum_{l=1}^3 \delta \omega_{k} \epsilon_{kil} \textbf{e}_{l} = \notag \\
&= \sum_{l=1}^3 \left( \delta_{il} + \sum_{k=1}^3 \epsilon_{kil} \delta \omega_{k} \right) \textbf{e}_{l} = \notag \\
&= \sum_{l=1}^3 R_{il} \textbf{e}_{l} ,
\end{align}
where
\begin{equation}
R_{il} = \delta_{il} + \sum_{k=1}^3 \epsilon_{kil} \delta \omega_{k}
\label{eq: rotazioneinfinitesima}
\end{equation}
is the matrix element of an infinitesimal rotation and $\epsilon_{kil}$ is the Levi-Civita symbol. Note that from $\eqref{eq: rotazioneinfinitesima}$ it follows that the matrix element of the inverse of an infinitesimal rotational matrix is given by
\begin{equation}
(R)^{-1}_{li} = \delta_{li} - \sum_{k=1}^3 \epsilon_{kli} \delta \omega_{k} .
\label{eq: rotazioneinfinitesimainversa}
\end{equation}
By definition, the infinitesimal generators of a rotation are
\begin{align}
(T_{k})_{li} &= \left. \dfrac{\partial (R)^{-1}_{li}}{\partial w_{k}} \right|_{\textbf{w}=\textbf{0}} \ = \notag \\
&= - \epsilon_{kli},
\end{align}
and from $k,l,i \in (1,2,3)$, the infinitesimal generators of the group of rotations are $3$ ($n=3$) square matrices of dimension $3$ ($k=3$). The Lie algebra of $SO(3)$ is called $L_{SO(3)}$ or $so(3)$ and satisfies
\begin{equation}
[T_{k},T_{l}] = \sum_{m=1}^3 \epsilon_{klm} T_{m}.
\end{equation}
\subsection{Lie algebra of $E(3)$}
Consider the elements $A \in E(3)$ infinitely close to the identity, i.e.,
\begin{equation}
A = \mathds{1} + \sum_{\alpha=1}^3 \delta w_{\alpha} \widetilde{I}_{\alpha} + \sum_{\alpha=1}^3 \delta a_{\alpha} \widetilde{\pi}_{\alpha},
\label{eq: rototraslazioneinfinitesima}
\end{equation}
where $\delta w_{\alpha}$ and $\delta a_{\alpha}$ are the components of infinitesimal three-dimensional vectors of rotations and translations, respectively. The objects
\begin{equation}
\widetilde{I}_{\alpha} =
\begin{pmatrix}
I_{\alpha} & 0 \\[1mm]
0 & 0
\end{pmatrix}
\end{equation}
denote the \( 4 \times 4 \) matrices that include the infinitesimal generators \( I_{\alpha} \) of three-dimensional rotations, as well as the objects
\begin{equation}
(\widetilde{\pi}_{\alpha})_{ij} = \delta_{i\alpha} \, \delta_{j4}, \quad i,j = 1,\dots,4
\end{equation}
denote the $4 \times 4$ matrices representing the infinitesimal generators of translations, where only the element in the \( \alpha \)-th row and 4-th column is nonzero. \newline
It is easily verified that such a generators satisfy
\begin{equation}
\begin{cases}
\left[ \widetilde{I}_{\alpha} , \widetilde{I}_{\beta} \right] = \epsilon_{\alpha \beta \gamma} \widetilde{I}_{\gamma} \\
\left [\widetilde{\pi}_{\alpha},\widetilde{\pi}_{\beta} \right]=0 \\
\left[ \widetilde{I}_{\alpha} , \widetilde{\pi}_{\beta} \right] = \epsilon_{\alpha \beta \gamma} \widetilde{\pi}_{\gamma}
\end{cases}.
\end{equation}
\subsection{Lie algebra of $SU(2)$}
Given a neighborhood of the identity of $U \in SU(2)$, i.e., $U = \mathds{1} + \epsilon T$, $U^{\dagger} = \mathds{1} + \epsilon T^{\dagger}$, what conditions must the matrix $T$ satisfy in order for $U \in SU(2)$? From
\begin{align}
\mathds{1} &= U U^{\dagger} = \notag \\
&= (\mathds{1} + \epsilon T)(\mathds{1} + \epsilon T^{\dagger}) \simeq \notag \\
&\simeq \mathds{1} + \epsilon (T+T^{\dagger}) ,
\end{align}
it follows $\epsilon (T+T^{\dagger}) = 0$, then $T = - T^{\dagger}$, that is, the unitary condition for the $U$ matrix implies that the $T$ matrix must be antihermitian, i.e.,
\begin{equation}
U =
\begin{pmatrix}
1 + \epsilon T_{11} & \epsilon T_{12} \\
\epsilon T_{21} & 1 + \epsilon T_{22}
\end{pmatrix}.
\end{equation}
Let us compute the determinant of $U$ and stop at the first order with respect to $\epsilon$, i.e.,
\begin{align}
\det U &= 1 + \epsilon T_{11} + \epsilon T_{22} = \notag \\
&= 1 + \epsilon \Tr T,
\end{align}
then $\Tr T = 0$. The object $T$ is antihermitian null-trace $2 \times 2$ matrix, so it can be written as a linear combination of the Pauli matrices with pure imaginary numbers, i.e.,
\begin{equation}
T = \dfrac{i}{2} \sum_{\alpha=1}^3 v_{\alpha} \sigma_{\alpha}, \ v_{\alpha} \in \mathbb{R},
\end{equation}
where the factor $\frac{1}{2}$ was chosen by convention in the literature. We have
\begin{align}
U &= \mathds{1} + \epsilon \sum_\alpha^3 v_{\alpha} \left( \dfrac{i}{2} \sigma_{\alpha} \right) = \notag \\
&= \mathds{1} + \epsilon \sum_{\alpha=1}^3 v_{\alpha} T_{\alpha},
\end{align}
with
\begin{equation}
T_{\alpha} = \dfrac{i}{2} \sigma_{\alpha},
\end{equation}
tthat is, the infinitesimal generators are proportional to the Pauli matrices, then they satisfy
\begin{equation}
[T_{\alpha},T_{\beta}] = - \sum_{\gamma=1}^3 \epsilon_{\alpha \beta \gamma} T_{\gamma} ,
\end{equation}
which is the Lie algebra of the group $SU(2)$, and is called $L_{SU(2)}$ or $su(2)$.
\subsection{The algebras of $SO(3)$ and $SU(2)$ are isomorphic}
It has been shown that there is a $2:1$ homomorphism between $SU(2)$ and $SO(3)$. Locally there is a $1:1$ correspondence between the elements of $SU(2)$ and $SO(3)$ ones. Since Lie algebras are derived from the germ of the identity and the germs of $SU(2)$ and $SO(3)$ are equivalent, then the two Lie algebras are equal. Locally the two groups have the same properties, globally, their topology is different. 
\section{From a Lie algebra to a Lie group}
Given a Lie algebra $L$ on a real vector space $V$ of dimension $n$.  A natural question arises: does there exist a Lie group $G$ whose Lie algebra is precisely $L$, i.e., such that the elements of $L$ correspond to the infinitesimal generators of $G$? By definition, the object $T \in V$ can be written as a linear combination of the basis of the infinitesimal generators, i.e., $T = \sum_\alpha^n \lambda_{\alpha} T_{\alpha}$. From the matrix exponential $\ref{Matrix exponential}$, we define an exponential map as follows
\begin{equation}
\exp : \ \bar{\lambda} = (\lambda_1, \ldots, \lambda_n) \in \mathbb{R}^n \rightarrow \exp\left( \sum_{\alpha}^n \lambda_\alpha T_\alpha \right) .
\label{eq: mappaesponenziale}
\end{equation}
We can now show
\begin{theorem}[Local Lie group of a Lie algebra]
If $|\bar{\lambda}|<\epsilon$, the exponential map $\eqref{eq: mappaesponenziale}$ maps the elements of $L$ to a Lie group and it is a $1:1$ application.
\begin{proof}
Consider the set
\begin{equation}
U_{\epsilon} = \lbrace e^{\sum_\alpha^n \lambda_{\alpha} T_{\alpha}} : |\overline{\lambda}| < \epsilon \rbrace,
\end{equation}
where \( |\bar{\lambda}| \) denotes the Euclidean norm. First note that the identity $e$ is in $U_{\epsilon}$; second, if $g \in U_{\epsilon}$, then from $g^{-1} = e^{- \sum_\alpha^n \lambda_{\alpha} T_{\alpha}}$ and $|-\bar{\lambda}|=|\bar{\lambda}|<\epsilon$ it follows \( g^{-1} \in U_{\epsilon} \), i.e., \( U_\epsilon \) is a neighborhood of the identity in the local Lie group generated by \( L \), i.e., it is a germ. We call $G$ the Lie group of such a germ. Consequently, from any set of matrices with well-defined structure constants, by means of the exponential map, we can construct a group that has $T_{\alpha}$ as generators. Given $g_{1},g_{2} \in U_{\epsilon}$, with $g_{1}=g(\bar{\lambda_{1}})$, $g_{2}=g(\bar{\lambda_{2}})$, the product $g=g(\bar{\lambda})=g_{1}g_{2}$ may not be in $U_{\epsilon}$, however, for continuity, it must be
\begin{equation}
\exists \ \delta < \epsilon , \ U_{\delta} \subset U_{\epsilon} : g_{1}, g_{2} \in U_{\delta} \Rightarrow g_{2} g_{1} \in U_{\epsilon}.
\end{equation}
Since $g_{2} g_{1} \in U_{\epsilon}$, its canonical coordinates are unique, so there exists $\phi_{\alpha}=\phi_{\alpha}(\bar{\lambda}_{1},\bar{\lambda}_{2})$ such that $e^{\sum_\alpha^n \phi_{\alpha} T_{\alpha}} = g_{2} g_{1}$. Since $G$ is a Lie group, the functions $\phi^{\alpha}(\overline{\lambda}_{1},\overline{\lambda}_{2})$ must be analytic. Now, we show that the Taylor's expansion of such functions is uniquely determined by the structure constants $C_{\alpha \beta \gamma}$ of the Lie algebra $L$. Let $T_{1}$ and $T_{2}$ be the matrices related to $g_{1}$ and $g_{2}$, respectively, then let us compute the element $T \in L$ such that
\begin{align}
g_{2} g_{1} &= e^{T_{2}} e^{T_{1}} = \notag \\
&= e^{T}.
\end{align}
We multiply both members to the right by $e^{-T}$ as follows
\begin{equation}
e^{T_{2}} e^{T_{1}} e^{- T} = \mathds{1}.
\end{equation}
We parameterize $g_{1}$ and $g_{2}$ as follows
\begin{equation}
g_{1}(s) = e^{s T_{1}(s)},
\end{equation}
\begin{equation}
g_{2}(s) = e^{s T_{2}(s)},
\end{equation}
with $s \in [0,1]$. We set an Ansatz
\begin{equation}
\tilde{T}(s) = s (T_{2} + T_{1}) + T(s) ,
\end{equation}
then
\begin{equation}
e^{s T_{2}} e^{s T_{1}} e^{- s (T_{2} + T_{1}) - T(s)} = \mathds{1}, \ \forall s \in [0,1].
\end{equation}
A differential equation for $T(s)$ can be derived, i.e.,
\begin{align}
0 &= e^{sT_{2}} T_{2} e^{sT_{1}} e^{-s(T_{1} + T_{2}) - T(s)} + e^{sT_{2}} e^{sT_{1}} T_{1} e^{-s(T_{1} + T_{2}) - T(s)} + e^{sT_{2}} e^{sT_{1}} \left( -T_{2} -T_{1} - \dfrac{dT(s)}{ds} \right) e^{-s(T_{2} + T_{1}) - T(s)} = \notag \\
&= e^{sT_{2}} T_{2} e^{sT_{1}} e^{-s(T_{1} + T_{2}) - T(s)} - e^{sT_{2}} e^{sT_{1}} \left( T_{2} + \dfrac{dT(s)}{ds} \right) e^{-s(T_{2} + T_{1}) - T(s)} = \notag \\
&= e^{sT_{2}} e^{sT_{1}} \left( e^{-sT_{1}} T_{2} e^{sT_{1}} - T_{2} - \dfrac{dT(s)}{ds} \right) e^{-s(T_{2} + T_{1}) -T(s)} ,
\end{align}
where in the last step we manipulated the term $T_{2}$ by means of the identity $e^{-sT_{1}} e^{sT_{1}}$. It must be
\begin{align}
\dfrac{dT(s)}{ds} &= e^{-sT_{1}} T_{2} e^{sT_{1}} - T_{2} = \notag \\
&= U(s) - T_{2}.
\end{align}
Let us calculate $U(s)$, we differentiate $U(s)$ with respect to $s$ as follows
\begin{equation}
\dfrac{dU(s)}{ds} = \left[ U(s) , T_{1} \right],
\end{equation}
then, note that $U(0)=T_{2}$, which is the initial condition to the Cauchy problem
\begin{equation}
\begin{cases}
U(s) = T_{2} + \int_{0}^{s} \left[ U(\tau) , T_{1} \right] d\tau \\
U(0) = T_{2} 
\end{cases} .
\end{equation}
Consider the function $U(s)$ as a series expansion with respect to $s$ and proceed iteratively, i.e.,
\begin{equation}
U(0) = T_{2},
\end{equation}
\begin{equation}
U(1) = T_{2} + s [T_{2},T_{1}],
\end{equation}
and we insert $U(1)$ into $U(s)$, then integrate. By performing an iteration, we get
\begin{equation}
U(s) = \sum_{n=1}^{\infty} \dfrac{s^{n}}{n!} \ T_{1}^{n} \lbrace T_{2} \rbrace,
\end{equation}
with
\begin{equation}
T_{1}^{n} \lbrace T_{2} \rbrace = \underbrace{[[[ \ldots}_{n \ times} T_{2} , T_{1}],T_{1}], \ldots ,T_{1}].
\end{equation}
Since $U(s)$ is an iterated commutator, it belongs to the Lie algebra, that is, $U(s) \in L$. Now, we can solve the differential equation as follows
\begin{align}
\dfrac{dT(s)}{ds} &= U(s) - T_{2} = \notag \\
&= T_{2} + \sum_{n=1}^{\infty} \dfrac{s^{n}}{n!} \ T_{1}^{n} \lbrace T_{2} \rbrace - T_{2} = \notag \\
&= \sum_{n=1}^{\infty} \dfrac{s^{n}}{n!} \ T_{1}^{n} \lbrace T_{2} \rbrace, 
\end{align}
which implies
\begin{equation}
T(s) = \sum_{n=1}^{\infty} \dfrac{s^{n+1}}{(n+1)!} \ T_{1}^{n} \lbrace T_{2} \rbrace ,
\end{equation}
with
\begin{equation}
T(0) = 0,
\end{equation}
then the object $\tilde{T}$ can be written as
\begin{align}
\tilde{T} &= \hat{T}(s=1) = \notag \\
&= T_{1} + T_{2} + \sum_{n=1}^{\infty} \dfrac{1}{(n+1)!} \ T_{1}^{n} \lbrace T_{2} \rbrace .
\end{align}
Considering only the first term of the series, we have
\begin{align}
\tilde{T} &= \sum_\alpha \left( \lambda_{1,\alpha} T_{\alpha} + \lambda_{2,\alpha} T_{\alpha} \right) + \dfrac{1}{2} [T_{2},T_{1}] +o(|\overline{\lambda}|^{3}) = \notag \\
&= \sum_\alpha \left( \lambda_{1,\alpha} T_{\alpha} + \lambda_{2,\alpha} T_{\alpha} \right) + \dfrac{1}{2} \left[ \sum_\beta \lambda_{2,\beta} T_{\beta},\sum_\gamma \lambda_{1,\gamma} T_{\gamma} \right] + o(|\overline{\lambda}|^{3}) = \notag \\
&= \sum_\alpha \left( \lambda_{1,\alpha} T_{\alpha} + \lambda_{2,\alpha} T_{\alpha} \right) + \dfrac{1}{2} \sum_\beta \sum_\gamma \sum_\alpha \lambda_{2,\beta} \lambda_{1,\gamma} C_{\beta \gamma \alpha} T_{\alpha} + o(|\overline{\lambda}|^{3}).
\end{align}
Regarding the second order term, we have
\begin{align}
\dfrac{1}{6} \left [ \left[ T_{2},T_{1} \right],T_{1} \right] &= \dfrac{1}{6} \left[ \left[ \sum_\epsilon \lambda_{2,\epsilon} T_{\epsilon} , \sum_\tau \lambda_{1,\tau} T_{\tau} \right] , \sum_\gamma \lambda_{1,\gamma} T_{\gamma} \right] = \notag \\
&= \dfrac{1}{6} \sum_\epsilon \sum_\tau \sum_\gamma \lambda_{2,\epsilon} \lambda_{1,\tau} \lambda_{1,\gamma} \left[ \left[ T_{\epsilon} , T_{\tau} \right] , T_{\gamma} \right] = \notag \\
&= \dfrac{1}{6} \sum_\epsilon \sum_\tau \sum_\gamma \lambda_{2,\epsilon} \lambda_{1,\tau} \lambda_{1,\gamma} \left[ \sum_\xi C_{\epsilon \tau \xi} T_{\xi} , T_{\gamma} \right] = \notag \\
&= \dfrac{1}{6} \sum_\epsilon \sum_\tau \sum_\gamma \sum_\xi  \lambda_{2,\epsilon} \lambda_{1,\tau} \lambda_{1,\gamma} C_{\epsilon \tau \xi} \left[ T_{\xi} , T_{\gamma} \right] = \notag \\
&= \dfrac{1}{6} \sum_\epsilon \sum_\tau \sum_\gamma \sum_\xi \sum_\alpha \lambda_{2,\epsilon} \lambda_{1,\tau} \lambda_{1,\gamma} C_{\epsilon \tau \xi} C_{\xi \gamma \alpha} T_\alpha ,
\end{align}
which turns out to be uniquely determined by the structure constants of the Lie algebra. The same occurs for every order of the series, then the local $1:1$ correspondence between Lie algebra $L$ and Lie group $G$ is proved.
\end{proof}
\end{theorem}
We have thus shown that, in a neighborhood of the identity, the Lie algebra $L$ uniquely determines the local structure of a Lie group $G$, with a one-to-one correspondence between elements of the algebra and elements of the group via the exponential map. In particular, the group multiplication law near the identity is fully characterized by the structure constants of $L$, as demonstrated through the expansion involving iterated commutators. This proves the existence of a local Lie group associated with any given Lie algebra, whose group law is analytic and consistent with the algebraic structure of $L$.
\subsection{Connectivity of a Lie group}
Here we discuss the global properties of the exponential map associated with a basis of a Lie algebra. It is well known that the Lie groups $SU(2)$ and $SO(3)$ are locally isomorphic; that is, they share the same Lie algebra and hence identical structure constants. However, they differ at the global level, as their group manifolds and matrix representations are not isomorphic. The fundamental distinction between these groups lies in their topological connectivity properties. 
Let $G$ be a Lie group, which can be viewed as a smooth geometric object with a group structure. Consider a fixed point in $G$, typically the identity element, and closed curves (loops) that start and end at this point. Two such curves are said to be equivalent if one can be continuously deformed into the other without breaking or detaching from the base point; otherwise, they are not equivalent. This notion partitions all curves into equivalence classes. We define the connectivity of the Lie group as the number of equivalence classes of closed curves based at the identity. A common geometric visualization of \(SO(3)\) is provided by the axis--angle representation, in which rotations are parametrized by vectors whose direction determines the rotation axis and whose magnitude corresponds to the rotation angle. In this representation, rotations by an angle \(\pi\) lie on a sphere, and antipodal points on this sphere represent the same rotation. This description is purely illustrative and does not endow \(SO(3)\) with a manifold structure. We define a continuous deformation as a family of curves parameterized continuously with respect to \(t\), where \(t=0\) corresponds to the identity element and \(t=1\) to the curve itself. To make the classification of loops more intuitive, we introduce the notion of "jump" curves as a visual tool to distinguish between topological classes of loops in \(SO(3)\). This terminology is informal, but it captures the essential idea of curves crossing the boundary where antipodal points are identified. We distinguish between 0-jump curves (\(C_0\)) and one-jump curves (\(C_1\)), where a "jump" corresponds to a crossing of the spherical surface associated with the identification of antipodal points.
More precisely, a curve is said to have a "jump" each time it passes through the spherical surface, which identifies two antipodal points as the same rotation. An example of a \(C_0\) curve is one that starts from the identity and returns to it without crossing the spherical surface. In contrast, a \(C_1\) curve starts at the identity, reaches the spherical surface (crossing it once), and then continues from the identified antipodal point to return to the origin. It can be shown that curves with an even number of jumps belong to the equivalence class \(C_0\), while curves with an odd number of jumps belong to \(C_1\). Therefore, there are exactly two equivalence classes of closed curves in \(SO(3)\), implying that
\begin{equation}
\mathrm{conn}[SO(3)] = 2,
\end{equation}
on the other hand, it can be shown that
\begin{equation}
\mathrm{conn}[SU(2)] = 1.
\end{equation}
\subsection{Universal covering group}
Before introducing the notion of a universal covering group, we recall two basic definitions.
\begin{definition}[Kernel of a group homomorphism]
If
\begin{equation}
T : G \longrightarrow H
\end{equation}
is a group homomorphism, then the kernel of \(T\) is given by
\begin{equation}
\ker(T) = \{\, g \in G \mid T(g) = e_H \,\},
\end{equation}
where \(e_H\) denotes the identity element of \(H\).
\end{definition}
\begin{definition}[Left coset]
Let \(G\) be a group and let \(H \subset G\) be a subgroup. For any element \(g \in G\), the left coset of \(H\) with respect to \(g\) is defined as
\begin{equation}
gH = \left\lbrace gh \mid h \in H \right\rbrace.
\end{equation}
\end{definition}
Given two groups $\overline{G}$ and $G$, we say that $\overline{G}$ is the \emph{universal covering group} of $G$ if:
\begin{itemize}
\item $\overline{G}$ is simply connected, i.e., $\mathrm{conn}[\overline{G}] = 1$;
\item there exists a group homomorphism
\begin{equation}
T : \overline{G} \rightarrow G
\end{equation}
such that $\ker(T)$ is a discrete subgroup with $h_1, \ldots, h_k$ distinct elements, and
\begin{equation}
k = \mathrm{conn}[G].
\end{equation}
\end{itemize}
Any Lie group that is not simply connected admits a universal covering group. The universal covering group of a Lie group is itself a Lie group and is unique up to isomorphism. Given a Lie algebra, there exists a unique simply connected Lie group whose Lie algebra is generated by it. To summarize, the process of constructing Lie groups from a given Lie algebra proceeds as follows:
\begin{itemize}
\item [a.] Apply the exponential map to the generators of the Lie algebra $L$, yielding a simply connected Lie group $\overline{G}$:
\begin{equation}
\exp : L \rightarrow \overline{G};
\end{equation}
\item [b.] Construct group homomorphisms $T_i : \overline{G} \rightarrow G_i$ such that $\ker(T_i)$ is a discrete normal subgroup of $\overline{G}$;
\item [c.] Define the quotient group
\begin{equation}
G_i = \overline{G} / \ker(T_i),
\end{equation}
which is a Lie group with the same Lie algebra as $\overline{G}$.
\end{itemize}
In this way, we identify all possible Lie groups sharing a given Lie algebra. Among these, the simply connected one is unique (up to isomorphism). All such groups share the same universal covering group. \newline
We now prove point (c).
\begin{itemize}
\item {Normal subgroup.} Let $f : \overline{G} \rightarrow G$ be a group homomorphism. The kernel of $f$, denoted $\ker(f)$, is the set of elements in $\overline{G}$ that are mapped to the identity element $e \in G$. That is,
\begin{equation}
\ker(f) = \lbrace \overline{g} \in \overline{G} \mid f(\overline{g}) = e \rbrace.
\end{equation}
We now show that:
\begin{itemize}
\item[i)] $\ker(f)$ is a subgroup of $\overline{G}$, that is,
\begin{equation}
\overline{e} \in \ker(f), \quad \overline{h}_1, \overline{h}_2 \in \ker(f) \Rightarrow \overline{h}_1 \overline{h}_2 \in \ker(f);
\end{equation}
\item[ii)] $\ker(f)$ is a normal (invariant) subgroup of $\overline{G}$, i.e., for all $\overline{g} \in \overline{G}$ and $\overline{h} \in \ker(f)$,
\begin{equation}
\overline{g} \, \overline{h} \, \overline{g}^{-1} \in \ker(f).
\end{equation}
Indeed, since $f$ is a homomorphism, we get
\begin{align}
f(\overline{g} \, \overline{h} \, \overline{g}^{-1}) &= f(\overline{g}) f(\overline{h}) f(\overline{g}^{-1}) = \notag \\
&= f(\overline{g}) e f(\overline{g})^{-1} = \notag \\
&= f(\overline{g}) f(\overline{g})^{-1} = \notag \\
&= f(\overline{e}) = \notag \\
&= e,
\end{align}
then the conjugate element lies in the kernel.
\end{itemize}
\item {Quotient group.} The kernel $\ker(f)$ induces an equivalence relation on $\overline{G}$: two elements $\overline{g}_1, \overline{g}_2 \in \overline{G}$ are equivalent if and only if
\begin{equation}
\overline{g}_1 \overline{g}_2^{-1} \in \ker(f),
\end{equation}
i.e.,
\begin{equation}
\overline{g}_1 \sim \overline{g}_2 \quad \Leftrightarrow \quad \exists \, \overline{h} \in \ker(f) \text{ such that } \overline{g}_1 = \overline{h} \overline{g}_2.
\end{equation}
This partitions $\overline{G}$ into left cosets of $\ker(f)$. The set of all such cosets forms the quotient group
\begin{equation}
Q = \overline{G} / \ker(f),
\end{equation}
whose elements are the equivalence classes in $\overline{G}$ under this relation.
To verify that $Q$ is a group, consider two cosets $g_1 = \overline{h}_1 \overline{g}_1$, $g_2 = \overline{h}_2 \overline{g}_2$, with $\overline{h}_1, \overline{h}_2 \in \ker(f)$. Then, the product of the two cosets is
\begin{align}
g_1 g_2 &= (\overline{h}_1 \overline{g}_1)(\overline{h}_2 \overline{g}_2) = \notag \\
&= \overline{h}_1 (\overline{g}_1 \overline{h}_2 \overline{g}_1^{-1}) (\overline{g}_1 \overline{g}_2),
\end{align}
where we used the associativity of the group product. Since $\ker(f)$ is normal, $\overline{g}_1 \overline{h}_2 \overline{g}_1^{-1} \in \ker(f)$, and the product lies in the same coset class:
\begin{equation}
g_1 g_2 \in Q.
\end{equation}
If $\ker(f)$ is discrete and $\overline{G}$ is simply connected, then the quotient group $G = \overline{G}/\ker(f)$ is a Lie group with the same Lie algebra as $\overline{G}$.
\end{itemize}
The group $SU(2)$ is simply connected and is the universal covering group of $SO(3)$. Its only nontrivial discrete normal subgroup is
\begin{equation}
\overline{H} = \lbrace \mathds{1}, -\mathds{1} \rbrace,
\end{equation}
and the corresponding quotient is
\begin{equation}
SU(2) / \overline{H} = SO(3).
\end{equation}
There are no other discrete normal subgroups of $SU(2)$, so the only Lie groups with the Lie algebra of $SU(2)$ are $SU(2)$ and $SO(3)$.
\section{From Lie group representations to Lie algebra representations}
Let $G'$ be the representation of the Lie group $G$ by the homomorphism $S$, and in particular $G'$ is a group of linear operators on a vector space $V$. If $V$ has finite dimension, we can take $G'$ as a group of invertible matrices; if $V$ has infinite dimension, then $G'$ is a group of invertible linear operators on $V$. We have
\begin{theorem} 
Given a Lie group and its representation, $G$ and $G'$ respectively, let $L$ and $L'$ be the respective Lie algebras, then a representation of $G$ induces a representation of $L$. 
\begin{proof}
Let $G$ be a group of matrices and $g \in G$ an element close to the identity, i.e.,
\begin{equation}
g = \mathds{1} + \epsilon T, \ T \in L.
\end{equation}
Now, the representation is a homomorphism, the identity of $G$ is mapped to the identity of $G'$, and the homomorphism is continuous, so elements of $G$ near the identity $e$ go into elements of $G'$ near the identity $e'=S(e)$
\begin{equation}
S(g) = \mathds{1}' + \epsilon T', \ T' \in L'.
\end{equation}
Consequently 
\begin{equation}
L \ni T \rightarrow T' \in L'
\end{equation}
is a map. We show that such a map:
\begin{itemize}
\item [i)] is linear;
\item [ii)] is a homomorphism between Lie algebras.
\end{itemize}
\begin{itemize}
\item [i)]
Given
\begin{equation}
g_{1} = \mathds{1} + \epsilon_{1} T_{1},
\end{equation}
\begin{equation}
g_{2} = \mathds{1} + \epsilon_{2} T_{2} ,
\end{equation}
\begin{equation}
g'_{1} = \mathds{1}' + \epsilon_{1} T'_{1},
\end{equation}
\begin{equation}
g'_{2} = \mathds{1}' + \epsilon_{2} T'_{2} ,
\end{equation}
$S$ is a homomorphism, then 
\begin{equation}
S(g_{1} g_{2}) = S(g_{1}) S(g_{2}),
\end{equation}
\begin{equation}
g_{1} g_{2} \simeq \mathds{1} + \epsilon_{1} T_{1} + \epsilon_{2} T_{2} ,
\end{equation}
which is mapped through $S$ in 
\begin{equation}
g'_{1} g'_{2} \simeq \mathds{1}' + \epsilon_{1} T'_{1} + \epsilon_{2} T'_{2} ,
\end{equation}
then the map
\begin{equation}
s: \epsilon_{1} T_{1} + \epsilon_{2} T_{2} \rightarrow \epsilon_{1} T'_{1} + \epsilon_{2} T'_{2} 
\end{equation}
is a linear map.
\item [ii)] We show that $s$ is a homomorphism and in particular that it preserves commutators. We introduce the quantities 
\begin{equation}
B(\epsilon_{1},\epsilon_{2}) = g^{-1}_{1} g^{-1}_{2} g_{1} g_{2},
\end{equation}
\begin{equation}
B'(\epsilon_{1},\epsilon_{2}) = g'^{-1}_{1} g'^{-1}_{2} g'_{1} g'_{2}.
\end{equation}
At the first order respect to $\epsilon_{i}$ it is worth
\begin{equation}
B(\epsilon_{1},\epsilon_{2}) = \mathds{1} + \epsilon_{1} \epsilon_{2} \left( T_{1} T_{2} - T_{2} T_{1} \right),
\end{equation}
\begin{equation}
B'(\epsilon_{1},\epsilon_{2}) = \mathds{1}' + \epsilon_{1} \epsilon_{2} \left( T'_{1} T'_{2} - T'_{2} T'_{1} \right),
\end{equation}
then 
\begin{equation}
S: B \rightarrow B'
\end{equation}
implies that
\begin{equation}
s: [T_{1},T_{2}] \rightarrow [T'_{2},T'_{1}]
\end{equation}
is a homomorphism of Lie algebras.
\end{itemize}
\end{proof}
\end{theorem}
\chapter{Symmetries}
In this chapter, we explore the central role of symmetries in quantum mechanics, beginning with a rigorous proof of Wigner's theorem. This fundamental result establishes that every symmetry transformation, which preserves the structure of quantum probabilities, must be represented in the state space by either a unitary or antiunitary operator. Wigner's theorem thus provides a crucial link between physical symmetries and their mathematical representations, laying the groundwork for a systematic study of invariant properties in quantum systems. \newline
Next, we introduce the operator forms associated with the primary spatial symmetries: translations and rotations. The spatial translation operator describes uniform shifts of a system through space, while the rotation operator encodes changes in spatial orientation in three dimensions. These operators act not only on quantum states but also on physical observables, such as position, linear momentum, and angular momentum, transforming them according to well-defined rules. Thanks to the group theory framework developed in the previous chapter, these properties and their operator structures can be demonstrated quickly and elegantly. In particular, the connection to the infinitesimal generators of these transformations, namely the linear momentum and angular momentum operators, becomes transparent, highlighting their role as observable quantities conserved due to the underlying symmetries. \newline
Finally, we present concrete applications, including the conservation of linear momentum derived from invariance under spatial translations, and the invariance of certain Hamiltonians under time-reversal symmetry. The latter has profound implications for the dynamics and spectral properties of quantum systems. Together, these theoretical foundations provide essential insight into how symmetries govern and constrain the physical behavior of quantum systems.
\section{Symmetry transformation and Wigner's theorem}
Given two observers \( O \) and \( O' \), their measuring instruments are said to be \emph{identical} if they yield the same measurement outcomes, and their physical descriptions are considered equivalent. Mathematically, this corresponds to the existence of a symmetry transformation relating the two measurement processes. Such a transformation leaves invariant a property of the physical system, regardless of the observer. Let \( A, B \) be observables acting on the Hilbert space \( H \) associated with observer \( O \), and \( A', B' \) be observables acting on the Hilbert space \( H' \) associated with observer \( O' \). Assume these observables have nondegenerate spectra, then
\begin{equation}
A \, |\varphi_{r}\rangle = \lambda_{r} \, |\varphi_{r}\rangle,
\end{equation}
\begin{equation}
B \, |\psi_{s}\rangle = \mu_{s} \, |\psi_{s}\rangle,
\end{equation}
\begin{equation}
A' \, |\varphi'_{r}\rangle = \lambda_{r} \, |\varphi'_{r}\rangle,
\end{equation}
\begin{equation}
B' \, |\psi'_{s}\rangle = \mu_{s} \, |\psi'_{s}\rangle.
\end{equation}
In quantum mechanics, the act of measurement is related not only to the numerical values of the spectrum but also to the measurement probabilities associated with the eigenstates. Specifically, if observer \( O \) first measures the observable \( A \) and immediately afterward measures the observable \( B \), the probability of obtaining the eigenvalue \(\mu_s\) following the eigenvalue \(\lambda_r\) is given by the transition (or superposition) probability $|\langle \psi_s | \varphi_r \rangle|^2$. If the two physical descriptions are equivalent, this probability must be preserved under the transformation, so that
\begin{equation}
|\langle \psi_s | \varphi_r \rangle|^2 = |\langle \psi'_s | \varphi'_r \rangle|^2.
\label{eq: ipotesitrasformazionesimmetria}
\end{equation}
Since state vectors are defined only up to an overall phase factor, it is natural to consider equivalence classes of vectors of fixed norm \( \|\psi\|^2 = \langle \psi | \psi \rangle \) rather than the vectors themselves. We denote by \( r = \|\psi\|^2 \) the norm squared (or "radius") of the equivalence class of vectors of the form \( e^{i \alpha} |\psi \rangle \), with \( \alpha \in \mathbb{R} \), and call any vector \( |\psi \rangle \) a representative of its equivalence class of radius \( r \).
Consequently, a symmetry transformation establishes a one-to-one correspondence between the equivalence classes (radii) of $H$ and those of $H'$ such that equation $\eqref{eq: ipotesitrasformazionesimmetria}$ holds. We now proceed to characterize such symmetry transformations.
\begin{theorem}[Wigner's theorem]
Given a symmetry transformation, that is, a map between rays of Hilbert spaces \(H\) and \(H'\) of dimension \(n\) that preserves the overlap function
\begin{equation}
\left| \langle \psi | \varphi \rangle \right|^{2} \ = \ \left| \langle \psi' | \varphi' \rangle \right|^{2},
\end{equation}
such a map can be represented by an operator
\begin{equation}
\Omega : H \rightarrow H',
\end{equation}
either unitary and linear, i.e.,
\begin{equation}
\langle \Omega \psi | \Omega \varphi \rangle = \langle \psi | \varphi \rangle,
\end{equation}
\begin{equation}
\Omega \left( c_{1} |\psi_{1}\rangle + c_{2} |\psi_{2}\rangle \right) = c_{1} \Omega |\psi_{1}\rangle + c_{2} \Omega |\psi_{2}\rangle,
\end{equation}
or antiunitary and antilinear, i.e.,
\begin{equation}
\langle \Omega \psi | \Omega\varphi \rangle = \langle \psi | \varphi \rangle^{*},
\end{equation}
\begin{equation}
\Omega \left(c_{1} |\psi_{1}\rangle + c_{2} |\psi_{2}\rangle \right) = c^{*}_{1} \Omega |\psi_{1}\rangle + c^{*}_{2} \Omega |\psi_{2}\rangle.
\end{equation}
\begin{proof}
We divide the proof into several steps.
\begin{itemize}
\item {Step 1}. Let $\lbrace |i \rangle \rbrace$ be a basis in $H$ and let $\lbrace |i' \rangle \rbrace$ be the vectors transformed through a symmetry transformation. We show that the symmetry transforms complete orthonormal systems into complete orthonormal systems, i.e., the superscripted system is a basis in $H'$. From the conservation of the overlap function it follows that
\begin{align}
|\langle i' | j' \rangle|^{2} &= |\langle i | j \rangle|^{2} = \notag \\
&= \delta_{ij},
\end{align}
then the orthonormality is preserved. Now, suppose that $\lbrace |i' \rangle \rbrace$ is not complete, then there exists a vector $|a' \rangle$ orthogonal to all superscripted versors 
\begin{equation}
0 = |\langle a'|i' \rangle|^{2},
\end{equation}
and the transformed of $|a' \rangle$, i.e., $|a \rangle$, should satisfy the equality
\begin{equation}
0 = |\langle a|i \rangle|^{2},
\end{equation}
which is an absurdity because by hypothesis $\lbrace |i \rangle \rbrace$ is complete. So, the system $\lbrace |i' \rangle \rbrace$ is complete and orthonormal, i.e., it is a basis.
\item {Step 2}. We need to establish a convention for fixing the phase of vectors in \( H' \). Specifically, we choose, for each basis vector in \( H' \), a representative that satisfies
\begin{equation}
|1 \rangle + |n \rangle \longrightarrow |1' \rangle + |n' \rangle , \ \forall \ n.
\end{equation} 
Consider
\begin{equation}
|\psi_{n} \rangle = \dfrac{1}{\sqrt{2}} \ \left[|1 \rangle +|n \rangle \right] , \ n \neq 1,
\end{equation}
then we expand the transformed $|\psi'_{n}\rangle$ respect to the basis $\lbrace |i' \rangle \rbrace$, i.e.,
\begin{equation}
|\psi'_{n} \rangle = \sum_{l} \ c_{nl} |l' \rangle.
\end{equation}
From the conservation of the overlap function, it follows 
\begin{align}
|c_{nn}|^{2} &= |c_{n1}|^{2} = \notag \\
&= \dfrac{1}{\sqrt{2}} ,
\end{align}
\begin{equation}
|c_{nl}|^{2} = 0, \ n \neq l, \ n \neq 1,
\end{equation}
we arbitrarily set 
\begin{equation}
c_{nn} = c_{nl} = 1,
\end{equation}
then
\begin{equation}
|1\rangle + |n \rangle \longrightarrow |1' \rangle + |n' \rangle, \ \forall \ n.
\end{equation} 
\item {Step 3}. Let us see how any state with $c_{n}$ real components is transformed under the symmetry transformation. Given
\begin{equation}
|u \rangle = \sum_{n} \ c_{n} |n \rangle,
\end{equation}
and its transformed
\begin{equation}
|u' \rangle = \sum_{n'} \ c'_{n} |n' \rangle,
\end{equation}
we get
\begin{equation}
|\langle \psi_{n}|u \rangle |^{2} = |c_{1}+c_{n}|^{2}, 
\end{equation}
\begin{equation}
|\langle \psi'_{n}|u' \rangle|^{2} = |c'_{1}+c'_{n}|^{2} ,
\end{equation}
which together with 
\begin{align}
|\langle n|u \rangle|^{2} &= |c_{n}|^{2} = \notag \\
&= |c'_{n'}|^{2} = \notag \\
&= |\langle n'|u' \rangle|^{2}, \ \forall \ n
\end{align}
implies
\begin{equation}
\begin{cases}
|c_{n}|^{2} = |c'_{n'}|^{2}, \ \forall \ n \\
|c_{1}|^{2} = |c'_{1'}|^{2}, \\
|c_{1}+c_{n}|^{2} = |c'_{1'}+c'_{n'}|^{2}, \ \forall \ n \neq 1
\end{cases}.
\end{equation}
If we chose the representative of $|u' \rangle$ such that
\begin{equation}
c'_{1'} = c_{1},  
\end{equation}
\begin{equation}
c'_{n'} = c_{n},
\end{equation}
the last equality of the system is valid if and only if all the coefficients are real. Note that in the case of real components the operator is exclusively unitary.
\item {Step 4}. We generalize step $3$ to a symmetry transformation of states with any complex components. Given $j$, $k$, consider the states
\begin{equation}
|j\rangle + |j+1\rangle + \ldots + |j+k\rangle \longrightarrow |j'\rangle + |j'+1\rangle + \ldots + |j'+k'\rangle,
\end{equation}
where the images are written by choosing each representative with real components. Consider the scalar product of $|j \rangle + |j+1 \rangle + \ldots + |j+k \rangle$ by 
\begin{equation}
|u\rangle = \sum_{n} c_{n} |n\rangle,
\end{equation}
then
\begin{equation}
\langle u| \left( |j\rangle + |j+1\rangle + \ldots + |j+k\rangle \right) = \sum_{s=j}^{j+n} c_{s},
\end{equation}
and similarly
\begin{equation}
\langle u'| \left( |j'\rangle + |j'+1\rangle + \ldots + |j'+k'\rangle \right) = \sum_{s'=j'}^{j'+n'} c_{s'} .
\end{equation}
We represent the two summations as broken lines $\Gamma$ and $\Gamma'$ in the complex plane. Each segment with a given index in $\Gamma$ corresponds to a segment in $\Gamma'$ with the same modulus, due to the correspondence between the radii. By conservation of probability, the total lengths of $\Gamma$ and $\Gamma'$ must be equal. This condition can be satisfied in only two possible ways:
\begin{itemize}
\item The coefficients \( c_i \) and \( c'_{i'} \) differ by a global phase rotation:
\begin{equation}
c'_{i'} = e^{i\alpha} c_i,
\end{equation}
which, of course, preserves the squared moduli and hence the length of the broken line.
\item The broken line \( \Gamma' \) is obtained from \( \Gamma \) by a combination of rotation and reflection in the complex plane. As a result, each coefficient transforms as
\begin{equation}
c'_{i'} = e^{i\beta} c_i^*,
\end{equation}
which also preserves the moduli and the total length.
\end{itemize}
Since the coefficients are defined up to a phase, we can choose the representatives such that $\alpha=\beta=0$ so there are only two cases, i.e.,
\begin{equation}
c'_{i'} = c_{i} ,
\end{equation}
and
\begin{equation}
c'_{i'} = c^{*}_{i}.
\end{equation}
It remains to be shown that the two possibilities are mutually exclusive, that is, the choice of a transformation for the coefficients of a state involves all vectors of the $H$ space.
\item {Step 5}. Given
\begin{equation}
\lvert j \rangle + e^{i \alpha} \lvert k \rangle \quad , \quad \alpha \neq n \pi
\end{equation}
Let us assume that this vector is transformed through a linear transformation, i.e.,
\begin{equation}
c'_n = c_n.
\end{equation}
According to step 2, then
\begin{equation}
\lvert j \rangle + e^{i \alpha} \lvert k \rangle \longrightarrow \lvert j' \rangle + e^{i \alpha} \lvert k' \rangle.
\end{equation}
Given a vector \( \lvert u \rangle \), assume that it transforms through an antilinear transformation, and consider the scalar product
\begin{equation}
\left( \langle j \rvert + e^{-i \alpha} \langle k \rvert \right) \lvert u \rangle = c_j + e^{-i \alpha} c_k,
\end{equation}
\begin{equation}
\left( \langle j' \rvert + e^{-i \alpha} \langle k' \rvert \right) \lvert u' \rangle = c'_{j'} + e^{-i \alpha} c'_{k'} = c_j^* + e^{-i \alpha} c_k^*.
\end{equation}
where in the last step we used the antilinearity condition. From the conservation of the transition probability we have
\begin{equation}
|c_{j} + e^{- i \alpha} c_{k}|^{2} = |c^{*}_{j} + e^{- i \alpha} c^{*}_{k}|^{2},
\end{equation}
\begin{equation}
c_{j} c^{*}_{k} e^{i \alpha} + c^{*}_{j} c_{k} e^{- i \alpha} = c^{*}_{j} c_{k} e^{i \alpha} + c_{j} c^{*}_{k} e^{- i \alpha},
\end{equation}
and let $\gamma$ be the phase between $c_{j}$ and $c_{k}$, meaning that
\begin{equation}
\cos{(\gamma+\alpha)} = \cos{(\gamma-\alpha)}.
\end{equation}
We note that it must be $\gamma \neq n \pi$, since otherwise it would be $c_{i}=c^{*}_{i}$ and we are assuming instead that the vectors have components with non-zero imaginary part. In addition, the condition
\begin{equation}
\cos{(\gamma+\alpha)} = \cos{(\gamma-\alpha)} 
\end{equation}
implies
\begin{equation}
\begin{cases}
\gamma + \alpha = \alpha - \gamma + 2 n \pi \\
\gamma + \alpha = \gamma - \alpha + 2 n \pi 
\end{cases},
\end{equation}
\begin{equation}
\begin{cases}
\gamma = n \pi \\ 
\alpha = n \pi
\end{cases},
\end{equation}
which contradicts the assumptions about angles, so $|u \rangle$ also transforms linearly. Similarly, it is proved in the complementary case, and the theorem is thus proved.
\end{itemize}
\end{proof}
\end{theorem}
\section{Wigner's operator for rotations}
Given two reference systems at rest, rotated with respect to each other by any angle, and given the measurement equivalence principle for the rotations of the references, Wigner's theorem ensures the existence of an operator $\hat{\Omega}$ between the states of the measurements. Consider three observers at rest, say $O$, $O'$ and $O''$, rotated with respect to each other by arbitrary angles around a common direction. The matrix $R_{1}$ maps $O$ to $O'$ and the matrix $R_{2}$ maps $O'$ to $O''$, and if we apply Wigner's theorem to each rotation, then there exist $\hat{\Omega}_{1}$ and $\hat{\Omega}_{2}$ unitary or antiunitary operators. Since $O$ and $O''$ are also connected by rotation and the rotations form a group, $O''$ is connected to $O$ by the composition $R_{2} R_{1}$. Therefore, there must exist a map $\hat{\Omega}^{*}$ between the Hilbert spaces of $O$ and $O''$. Given the state $|\psi \rangle$ of the observer $O$, by the action of the operator $\hat{\Omega}_{1}$ on $|\psi \rangle$ one obtains the state vector $|\psi' \rangle$ of the observer $O'$, then, by the action of the operator $\hat{\Omega}_{2}$ on the state $|\psi' \rangle$ one obtains the state vector $|\psi' \rangle$ of the observer $O''$. Consequently, we have
\begin{align}
|\psi'' \rangle &= \hat{\Omega}_{2} |\psi' \rangle = \notag \\
&= \hat{\Omega}_{2} \hat{\Omega}_{1} |\psi \rangle,
\end{align}
\begin{equation}
\hat{\Omega}_{2} \hat{\Omega}_{1} = e^{i \gamma} \hat{\Omega}^{*}.
\end{equation}
On the other hand,
\begin{equation}
\hat{\Omega}(R_{2}) \hat{\Omega}(R_{1}) = e^{i \gamma} \hat{\Omega}(R_{2} R_{1}) ,
\end{equation}
that is, the operators $\hat{\Omega}$ provide a projective representation of the group of rotations $SO(3)$ through the homomorphism
\begin{equation}
R \rightarrow \hat{\Omega}(R).
\end{equation}
\begin{theorem}
The $\hat{\Omega}(R)$ representation of $SO(3)$ is unitary.
\begin{proof}
An antiunitary matrix cannot become the identity matrix with continuity. The group of rotations is a connected group: any rotation can gradually recur to the identity with continuity, and since the trivial rotation, i.e., the identity, must be matched by $\hat{\Omega}(0) = e^{i \gamma} \mathds{1}$, $\hat{\Omega}(R)$ must be a unitary operator.
\end{proof}
\end{theorem}
\section{Mathematical derivation of the angular momentum operator}
Here, the angular momentum operator will be deduced from the general theory of rotation symmetry. Consider a Hilbert space $H$, on which a local representation of the rotations is provided by a unitary operator $\hat{R}$, with $\hat{R}=\hat{\Omega}(R)$. Given an infinitesimal rotation $\eqref{eq: rotazioneinfinitesima}$, the infinitesimal rotation operator $\hat{R}$ has the form
\begin{equation}
\hat{R} = \mathds{1} + \sum_{\alpha=1}^3 \delta w_{\alpha} \hat{I}_{\alpha},
\label{eq: operatorerotazioneinfinitesima1}
\end{equation}
where $\hat{I}_{\alpha}$ are the infinitesimal generators of the representation of the rotations. Given a representation of the group $SO(3)$, the associated infinitesimal generators define a representation of its Lie algebra $L_{SO(3)}$. As a consequence, they satisfy the corresponding commutation relations
\begin{equation}
\left[ \hat{I}_{\alpha},\hat{I}_{\beta} \right] = \sum_{\gamma=1}^3 \epsilon_{\alpha \beta \gamma} \hat{I}_{\gamma}.
\end{equation}
Since $\hat{R}$ is unitary, that is, $\hat{R} \hat{R}^{\dagger} = \mathds{1}$, it follows
\begin{align}
\mathds{1} &= \left( \mathds{1} + \sum_{\alpha=1}^3 \delta w_{\alpha} \hat{I}_{\alpha} \right) \left( \mathds{1} + \sum_{\alpha=1}^3 \delta w_{\alpha} \hat{I}^{\dagger}_{\alpha} \right) = \notag \\
&= \mathds{1} + \sum_{\alpha=1}^3 \delta w_{\alpha} \hat{I}^{\dagger}_{\alpha} + \sum_{\alpha=1}^3 \delta w_{\alpha} \hat{I}_{\alpha} + \sum_{\alpha=1}^3 \sum_{\beta=1}^3 \delta w_{\alpha} \delta w_{\beta} \hat{I}_{\alpha} \hat{I}^{\dagger}_{\beta} \simeq \notag \\
& \simeq \mathds{1} + \sum_{\alpha=1}^3 \delta w_{\alpha} \left( \hat{I}_{\alpha} + \hat{I}^{\dagger}_{\alpha} \right) ,
\end{align}
i.e., if and only if
\begin{equation}
\hat{I}_{\alpha} = - \hat{I}^{\dagger}_{\alpha},
\end{equation}
that is, if and only if the operators $\hat{I}_{\alpha}$ are anti-Hermitian. Note that from $\eqref{eq: operatorerotazioneinfinitesima1}$ it follows
\begin{align}
\hat{R}^\dagger &= \mathds{1} + \sum_{\alpha=1}^3 \delta w_{\alpha} \hat{I}^\dagger_{\alpha} = \notag \\
&= \mathds{1} - \sum_{\alpha=1}^3 \delta w_{\alpha} \hat{I}_{\alpha} ,
\label{eq: operatorerotazioneaggiuntainfinitesima1}
\end{align}
where we used $\hat{I}^\dagger_{\alpha} = - \hat{I}_{\alpha}$. Now, we define the hermitian operators $\hat{J}_{\alpha}$ as
\begin{equation}
\hat{I}_{\alpha} = - \dfrac{i}{\hslash} \hat{J}_{\alpha}.
\end{equation}
They satisfy
\begin{equation}
- \dfrac{1}{\hbar^{2}} \left[\hat{J}_{\alpha},\hat{J}_{\beta}\right] = - \dfrac{i}{\hbar} \sum_{\gamma=1}^3 \epsilon_{\alpha \beta \gamma} \hat{J}_{\gamma} ,
\end{equation}
\begin{equation}
\dfrac{1}{\hbar^{2}} \left[\hat{J}_{\alpha},\hat{J}_{\beta}\right] = \dfrac{i}{\hbar} \sum_{\gamma=1}^3 \epsilon_{\alpha \beta \gamma} \hat{J}_{\gamma} ,
\end{equation}
\begin{equation}
\left[\hat{J}_{\alpha},\hat{J}_{\beta}\right] = i \hbar \sum_{\gamma=1}^3 \epsilon_{\alpha \beta \gamma} \hat{J}_{\gamma}.
\label{eq: algebramomentoangolare}
\end{equation}
From the general theory of symmetry, with the only requirement that the operator $\hat{J}_{\alpha}$ be Hermitian and satisfy the algebra of rotation group, we derived the general definition of angular momentum, i.e., the $\eqref{eq: algebramomentoangolare}$. Note that equations $\eqref{eq: operatorerotazioneinfinitesima1}$ and $\eqref{eq: operatorerotazioneaggiuntainfinitesima1}$ become
\begin{equation}
\hat{R} = \mathds{1} - \dfrac{i}{\hslash} \sum_{\alpha=1}^3 \delta w_{\alpha} \hat{J}_{\alpha},
\label{eq: operatorerotazioneinfinitesima2}
\end{equation}
\begin{align}
\hat{R}^\dagger &= \mathds{1} + \dfrac{i}{\hslash} \sum_{\alpha=1}^3 \delta w_{\alpha} \hat{J}_{\alpha},
\label{eq: operatorerotazioneaggiuntainfinitesima2}
\end{align}
respectively. We know that, locally, any rotation can be expressed as the exponential of a linear combination of the generators of the Lie algebra $L_{SO(3)}$ associated with the $SO(3)$ group, i.e.,
\begin{equation}
R_{\textbf{u}}(\delta \varphi) = e^{\delta \varphi \textbf{u} \cdot \textbf{I}},
\end{equation}
then, such a rotation is represented by the unitary operator
\begin{equation}
\hat{R}_{\textbf{u}}(\delta \varphi) = e^{- \frac{i}{\hslash} \delta \varphi \textbf{u} \cdot \hat{\textbf{J}}}.
\end{equation}
For finite rotations, we have
\begin{equation}
R_{\textbf{u}}(\varphi) = \lim_{N \rightarrow \infty} \left[ R_{\textbf{u}}\left(\dfrac{\varphi}{N}\right) \right]^{N}.
\label{eq: rotazionefinita}
\end{equation}
Recall, however, that exponentiating the generators of a Lie algebra (in this case $L_{SO(3)}$) does not give a representation of the group ($SO(3)$) but that of its universal covering ($SU(2)$).
\section{Law of transformations of observables under finite and infinitesimal rotations}
Let $A$ and $A'$ be a generic observable measured by $O$ and $O'$ respectively, the principle of equivalence of measures implies
\begin{align}
|\langle \psi | A | \psi \rangle|^{2} &= |\langle \psi' | A' | \psi' \rangle|^{2} = \notag \\
&= |\langle \psi | \hat{R}^{\dagger} A' \hat{R} | \psi \rangle|^{2}, \quad \forall \, |\psi\rangle,
\end{align}
then
\begin{equation}
A = \hat{R}^{\dagger} A' \hat{R},
\label{eq: leggetrasformazioneoperatoresottorotazione1}
\end{equation}
\begin{equation}
A' = \hat{R} A \hat{R}^{\dagger}.
\label{eq: leggetrasformazioneoperatoresottorotazione2}
\end{equation}
\subsection{Scalar operators}
Let $S$ be the operator associated with an observable that is a scalar quantity, that is, $S=S'$. Then
\begin{equation}
S = \hat{R} S \hat{R}^{\dagger} ,
\end{equation}
from which multiplying to the right by $\hat{R}$, we have for scalar observables the following characterization
\begin{equation}
S \hat{R} = \hat{R} S,
\end{equation}
then
\begin{equation}
\left[ \hat{R} , S \right] = 0.
\end{equation}
\subsection{Vector operators}
Here, we emphasize that there are two ways to characterize the transformation law of a vector under a rotation. An observer studies two equivalent rotated physical systems or a physical system is studied by means of two rotated spatial reference systems.
\begin{itemize}
\item{\textit{Active rotations}.}
Given a vector $\textbf{V}$, the rotated vector is the vector $\textbf{V}'$ whose components are obtained by applying the rotation matrix $R$ as follows
\begin{equation}
V'_{i} = \sum_{j=1}^3 R_{ij} V_{j},
\label{eq: rotazioneantiorariaattiva}
\end{equation}
or equivalently
\begin{equation}
\textbf{V}' = R \textbf{V},
\label{eq: vettorecolonnarotazioneattiva}
\end{equation}
where $\textbf{V}'$ and $\textbf{V}$ are column vectors.
\item{\textit{Passive rotations}.}
Given the components of a vector $\textbf{V}$ with respect to two rotated spatial reference systems, i.e.,
\begin{equation}
\textbf{e}'_{i} = \sum_{j=1}^3 R_{ij} \textbf{e}_{j},
\end{equation}
\begin{equation}
\textbf{e}_{j} = \sum_{j=1}^3 R^{-1}_{ji} \textbf{e}'_{j},
\end{equation}
\begin{equation}
\textbf{V} = \sum_{j=1}^3 V_{j} \textbf{e}_{j},
\end{equation}
\begin{equation}
\textbf{V} = \sum_{i=1}^3 V'_{i} \textbf{e}'_{i}.
\end{equation}
The components $V_{j}$ and $V'_{i}$ must satisfy
\begin{equation}
\sum_{i=1}^3 V'_{i} \textbf{e}'_{i} = \sum_{j=1}^3 V_{j} \textbf{e}_{j},
\end{equation}
then
\begin{equation}
\sum_{i=1}^3 V'_{i} \textbf{e}'_{i} = \sum_{j=1}^3 V_{j} \sum_{i=1}^3  R^{-1}_{ji} \textbf{e}'_{i},
\end{equation}
\begin{equation}
V'_{i} = \sum_{j=1}^3 V_{j} R_{ji}^{-1} ,
\end{equation}
or equivalently
\begin{equation}
\textbf{V}'^T = \textbf{V}^T R^{-1} .
\label{eq: vettorerigarotazionepassiva}
\end{equation}
where $\textbf{V}'^T$ and $\textbf{V}^T$ are row vectors.
\end{itemize}
Given the position vector operator $\hat{\textbf{r}}$, we postulate that its coordinates transform under rotation analogously to the associated classical dynamical quantities, according to the interpretation of passive rotations, as both observers measure the same state in different references, rotated with respect to each other. If an observer $O$ measures the state
\begin{equation}
\hat{r}_{i} |\textbf{r} \rangle = r_{i} |\textbf{r}\rangle, \ i \in \lbrace 1,2,3 \rbrace ,
\end{equation}
then the observer $O'$ measures
\begin{align}
\hat{r}'_{i} |\mathbf{r}'\rangle \ &= r'_{i} |\mathbf{r}'\rangle \ = \notag \\
&= \sum_{j=1}^3 r_j R^{-1}_{ji} |\mathbf{r}'\rangle \ = \notag \\
&= \sum_{j=1}^3 R^{-1}_{ji} r_j |\mathbf{r}'\rangle \ = \notag \\
&= \sum_{j=1}^3 R^{-1}_{ji} \hat{r}_{j} |\mathbf{r}'\rangle.
\end{align}
Since the previous relation must hold for every state \( |\mathbf{r}'\rangle \), it follows that it is an operator identity, i.e.,
\begin{equation}
\hat{r}'_{i} = \sum_{j=1}^3 R^{-1}_{ji} \hat{r}_{j},
\end{equation}
\begin{equation}
\hat{R} \hat{r}_{i} \hat{R}^{\dagger} = \sum_{j=1}^3 R^{-1}_{ji} \hat{r}_{j}.
\end{equation}
We use the above equation as the definition of the transformation of a vector operator $\lbrace K^{i} \rbrace$ under finite rotations, i.e.,
\begin{equation}
\hat{R} K_{i} \hat{R}^{\dagger} = \sum_{j=1}^3 R^{-1}_{ji} K_{j}.
\label{eq: leggetrasformazioneoperatorevettoriale}
\end{equation}
Let us now derive the transformation rule for a vector operator under an infinitesimal rotation. According to equation $\eqref{eq: leggetrasformazioneoperatorevettoriale}$, from equations $\eqref{eq: rotazioneinfinitesimainversa}$, $\eqref{eq: operatorerotazioneinfinitesima2}$ and $\eqref{eq: operatorerotazioneaggiuntainfinitesima2}$, we get
\begin{equation}
\sum_{j=1}^3 \left( \delta_{ji} - \sum_{\alpha=1}^3 \delta w_{\alpha} \epsilon_{\alpha j i} \right) K_{j} = \left( \mathds{1} - \dfrac{i}{\hslash} \sum_{\alpha=1}^3 \delta w_{\alpha} \hat{J}_{\alpha} \right) K_{i} \left( \mathds{1} + \dfrac{i}{\hslash} \sum_{\alpha=1}^3 \delta w_{\alpha} \hat{J}_{\alpha} \right),
\end{equation}
\begin{equation}
\sum_{j=1}^3 \delta_{ji} K_{j} - \sum_{j=1}^3 \sum_{\alpha=1}^3 \delta w_{\alpha} \epsilon_{\alpha j i} K_{j} = \left( K_{i} - \dfrac{i}{\hslash} \sum_{\alpha=1}^3 \delta w_{\alpha} \hat{J}_{\alpha} K_{i} \right) \left( \mathds{1} + \dfrac{i}{\hslash} \sum_{\alpha=1}^3 \delta w_{\alpha} \hat{J}_{\alpha} \right),
\end{equation}
\begin{equation}
K_{i} - \sum_{j=1}^3 \sum_{\alpha=1}^3 \delta w_{\alpha} \epsilon_{\alpha j i} K_{j} = K_{i} - \dfrac{i}{\hslash} \sum_{\alpha=1}^3 \delta w_{\alpha} \hat{J}_{\alpha} K_{i} + \dfrac{i}{\hslash} \sum_{\alpha=1}^3 \delta w_{\alpha} K_{i} \hat{J}_{\alpha} + \dfrac{1}{\hslash^{2}} \sum_{\alpha=1}^3 \sum_{\beta=1}^3 \delta w_{\alpha} \delta w_{\beta} \hat{J}_{\alpha} K_{i} \hat{J}_{\beta},
\end{equation}
\begin{equation}
K_{i} - \sum_{j=1}^3 \sum_{\alpha=1}^3 \delta w_{\alpha} \epsilon_{\alpha j i} K_{j} \simeq K_{i} - \dfrac{i}{\hslash} \sum_{\alpha=1}^3 \delta w_{\alpha} \hat{J}_{\alpha} K_{i} + \dfrac{i}{\hslash} \sum_{\alpha=1}^3 \delta w_{\alpha} K_{i} \hat{J}_{\alpha},
\end{equation}
\begin{equation}
- \sum_{j=1}^3 \sum_{\alpha=1}^3 \delta w_{\alpha} \epsilon_{\alpha j i} K_{j} = - \dfrac{i}{\hslash} \sum_{\alpha=1}^3 \delta w_{\alpha} \hat{J}_{\alpha} K_{i} + \dfrac{i}{\hslash} \sum_{\alpha=1}^3 \delta w_{\alpha} K_{i} \hat{J}_{\alpha},
\end{equation}
\begin{equation}
- \sum_{j=1}^3 \sum_{\alpha=1}^3 \delta w_{\alpha} \epsilon_{\alpha j i} K_{j} = - \dfrac{i}{\hslash} \sum_{\alpha=1}^3 \delta w_{\alpha} \hat{J}_{\alpha} K_{i} + \dfrac{i}{\hslash} \sum_{\alpha=1}^3 \delta w_{\alpha} K_{i} \hat{J}_{\alpha}.
\end{equation}
Ultimately, since infinitesimal increments are arbitrary, it must be
\begin{equation}
- \sum_{j=1}^3 \epsilon_{\alpha j i} K_{j} = \dfrac{i}{\hslash} \left( K_{i} \hat{J}_{\alpha} - \hat{J}_{\alpha} K_{i} \right),
\end{equation}
\begin{equation}
- \sum_{j=1}^3 \epsilon_{\alpha j i} K_{j} = \dfrac{i}{\hslash} \left[ K_{i} , \hat{J}_{\alpha} \right],
\end{equation}
\begin{equation}
i \hslash \sum_{j=1}^3 \epsilon_{\alpha j i} K_{j} = \left[ K_{i} , \hat{J}_{\alpha} \right],
\end{equation}
\begin{equation}
- i \hslash \sum_{j=1}^3 \epsilon_{\alpha i j} K_{j} = \left[ K_{i} , \hat{J}_{\alpha} \right],
\end{equation}
\begin{equation}
i \hslash \sum_{j=1}^3 \epsilon_{\alpha i j} K_{j} = \left[ \hat{J}_{\alpha} , K_{i} \right].
\end{equation}
Since the quantity \(\sum_{j=1}^3 \epsilon_{\alpha i j} K_{j}\) is nonzero only when the indices \(\alpha\), \(i\), and \(j\) are all distinct, it follows that
\begin{equation}
\left[ \hat{J}_{\alpha} , K_{i} \right] = i \hslash \sum_{\gamma=1}^3 \epsilon_{\alpha i j} K_{j},
\end{equation}
which is the transformation of a vector operator under an infinitesimal rotation.
\section{Angular momentum operator}
Here we will work with irreducible sets of observables.
\subsection{Spinless particle}
For a spinless particle we refer to the irreducible system of observables $(\hat{r_{\alpha}},\hat{p}_{\alpha})$, $\alpha = 1,2,3$ on the space $\mathcal{L}^{2}(\mathbb{R}^{3})$. Since the momentum is a vector operator, in the case of finite or infinitesimal rotation it must transform as the position operator does. Due to irreducibility of the system $(\hat{r_{\alpha}},\hat{p}_{\alpha})$, from the commutators $\left[ \hat{J}_{\alpha} , \hat{x}_{\beta} \right]$, $\left[ \hat{J}_{\alpha} , \hat{p}_{\beta} \right]$, the difference of any two $\hat{\textbf{J}}$, $\hat{\textbf{J}}'$ can be calculated. Indeed, suppose that $\hat{\textbf{J}}$ and $\hat{\textbf{J}}'$ satisfy the same commutator, then it must be
\begin{equation}
\left[ \hat{J}_{\alpha} - \hat{J}'_{\alpha} , \hat{r}_{\beta} \right] = 0,
\end{equation}
\begin{equation}
\left[ \hat{J}_{\alpha} - \hat{J}'_{\alpha} , \hat{p}_{\beta} \right] = 0,
\end{equation}
and Schur's lemma implies
\begin{equation}
\hat{J}_{\alpha} - \hat{J}'_{\alpha} = c_{\alpha} \mathds{1},
\end{equation}
\begin{equation}
\hat{J}'_{\alpha} = \hat{J}_{\alpha} - c_{\alpha} \mathds{1}.
\end{equation}
Consequently, the finite rotation operator $\hat{R}$ satisfies
\begin{align}
\hat{R}_{\hat{\textbf{u}}}(\varphi) &= e^{- \frac{i}{\hslash} \varphi \hat{\textbf{J}}' \cdot \textbf{u}} = \notag \\
&= e^{- \frac{i}{\hslash} \varphi \left(\hat{\textbf{J}} - \textbf{c} \mathds{1} \right) \cdot \textbf{u}} = \notag \\
&= e^{\frac{i}{\hslash} \varphi \textbf{c} \cdot \textbf{u}} e^{- \frac{i}{\hslash} \varphi \hat{\textbf{J}} \cdot \textbf{u}},
\end{align}
i.e., the operators of the rotations obtained by exponentiating the generators $\hat{\textbf{J}}'$ differ from the operators associated with the generators $\hat{\textbf{J}}$ by a phase factor. In general, the unitary operators representing the action of a rotation provide projective representations of the group of rotations, that is, they are defined up to a phase factor, so $\hat{\textbf{J}}$ and $\hat{\textbf{J}}'$ are physically equivalent. It is easy to verify that a solution $\hat{\textbf{J}}$ is provided by the definition of classical orbital angular momentum
\begin{equation}
\hat{\textbf{J}} = \hat{\textbf{L}} = \hat{\textbf{r}} \times \hat{\textbf{p}}.
\end{equation}
For a spinless particle, the generators of the infinitesimal rotations are the components of the particle's orbital angular momentum.
\subsection{Particle with spin}
Given a particle with a fixed spin, its irreducible set is given by $(\hat{r_{\alpha}},\hat{p}_{\alpha}, \hat{S}_{\alpha})$, $\alpha = 1,2,3$. How do spin operators transform under rotations? By analogy with other vector observables, assume
\begin{equation}
\left[ \hat{J}_{\alpha} , \hat{r}_{\beta} \right] = i \hslash \sum_{\gamma=1}^3 \epsilon_{\alpha \beta \gamma} \hat{r}_{\gamma},
\end{equation}
\begin{equation}
\left[ \hat{J}_{\alpha} , \hat{p}_{\beta} \right] = i \hslash \sum_{\gamma=1}^3 \epsilon_{\alpha \beta \gamma} \hat{p}_{\gamma},
\end{equation}
\begin{equation}
\left[ \hat{J}_{\alpha} , \hat{S}_{\beta} \right] = i \hslash \sum_{\gamma=1}^3 \epsilon_{\alpha \beta \gamma} \hat{S}_{\gamma}.
\end{equation}
For particles with spin, the total angular momentum operator is given by
\begin{equation}
\hat{J}_{\alpha} = \hat{L}_{\alpha} + \hat{S}_{\alpha},
\end{equation}
or equivalently
\begin{equation}
\hat{\textbf{J}} = \hat{\textbf{L}} + \hat{\textbf{S}}.
\end{equation}
\section{Action of rotations on the state vector}
Here we study the action of the rotation operator on a generic state vector. In general, the exponential map of the infinitesimal generators of the Lie algebra $so(3)$ provides elements of the universal covering group of $SO(3)$, i.e., $SU(2)$. We want to understand whether and under what conditions the exponential map provides a faithful representation of the $SO(3)$ group.
\subsection{Spinless particle}
Under an infinitesimal rotation $R$, according to the viewpoint of passive rotations, a vector $V$ transforms as
\begin{align}
\textbf{V}' &= R^{-1} \textbf{V} = \notag \\
&= \textbf{V} - \delta \bm{\omega} \times \textbf{V} = \notag \\
&= \textbf{V} - \delta \varphi \ \textbf{u} \times \textbf{V} ,
\end{align}
then let us compute the action of the operator of an infinitesimal rotation
\begin{align}
\hat{R}_{\textbf{u}}(\delta \varphi) &= e^{- \frac{i}{\hslash} \delta \varphi \textbf{u} \cdot \hat{\textbf{J}}} \simeq \notag \\
& \simeq \hat{\mathds{1}} - \dfrac{i}{\hslash} \delta \varphi \textbf{u} \cdot \hat{\textbf{J}} \equiv \notag \\
& \equiv \hat{\mathds{1}} - \dfrac{i}{\hslash} \delta \varphi \textbf{u} \cdot \hat{\textbf{L}}
\end{align}
on the state vector $\psi(\textbf{r})$. The object 
\begin{align}
\left( \hat{R}_{\textbf{u}}(\delta \varphi) \psi \right) (\textbf{r}) &= \left( \left( \mathds{1} - \dfrac{i}{\hslash} \delta \varphi \ \textbf{u} \cdot \hat{\textbf{L}} \right) \psi \right) (\textbf{r}) = \notag \\
&= \left( \left( \mathds{1} - \dfrac{i}{\hslash} \delta \varphi \ \textbf{u} \cdot \left( \textbf{r} \times \dfrac{\hslash}{i} \nabla \right) \right) \psi \right) (\textbf{r}) = \notag \\
&= \left( \left( \mathds{1} - \delta \varphi \ \textbf{u} \cdot \left( \textbf{r} \times \nabla \right) \right) \psi \right) (\textbf{r}) 
\end{align}
is the first-order Taylor polynomial of the function $\psi (\textbf{r} - \delta \varphi \ \textbf{u} \times \textbf{r})$, indeed
\begin{align}
\psi (\textbf{r} - \delta \varphi \ \textbf{u} \times \textbf{r}) &= \psi(\textbf{r}) - \delta \varphi \left( \textbf{u} \times \textbf{r} \right) \cdot \nabla \psi(\textbf{r}) + o(\delta \varphi^{2}) = \notag \\
&= \psi(\textbf{r}) - \delta \varphi \left( \textbf{u} \cdot \left( \textbf{r} \times \nabla \psi(\textbf{r}) \right) \right) + o(\delta \varphi^{2}),
\end{align}
which is the wave function evaluated on the position vector transformed according to the inverse of the rotation matrix $R$, i.e.,
\begin{align}
\left( \hat{R}_{\textbf{u}}(\delta \varphi) \psi \right) (\textbf{r}) &= \psi (\textbf{r} - \delta \varphi \ \textbf{u} \times \textbf{r}) = \notag \\
&= \psi \left( R^{-1}_{\textbf{u}}(\delta \varphi) \textbf{r} \right).
\end{align}
From $\eqref{eq: rotazionefinita}$, the operator of a finite rotation is
\begin{align}
\hat{R}_{\textbf{u}}(\varphi) &= \lim_{N \rightarrow \infty} \left[ \mathds{1} - \dfrac{i}{\hslash} \dfrac{\varphi}{N} \textbf{u} \cdot \hat{\textbf{J}} \right]^{N} = \notag \\
&= e^{- \frac{i}{\hslash} \varphi \textbf{u} \cdot \hat{\textbf{J}}}.
\end{align}
Finally, the action of a finite rotation on the state vector is given by
\begin{align}
\left( \hat{R}_{\textbf{u}}(\varphi) \psi \right) (\textbf{r}) &= e^{- \frac{i}{\hslash} \varphi \textbf{u} \cdot \hat{\textbf{J}}} \psi(\textbf{r}) = \notag \\
&= e^{- \frac{i}{\hslash} \varphi \textbf{u} \cdot \hat{\textbf{L}}} \psi(\textbf{r}) = \notag \\
&= \left( \lim_{N \rightarrow + \infty} \left( \mathds{1} - \dfrac{i}{\hslash} \dfrac{\varphi}{N} \textbf{u} \cdot \hat{\textbf{L}} \right)^{N} \psi \right) (\textbf{r}) = \notag \\
&= \lim_{N \rightarrow + \infty} \psi \left( R^{-1}_{\textbf{u}} \left( \left( \dfrac{\varphi}{N} \right)^{N} \right) \textbf{r} \right) = \notag \\
&= \psi \left( R^{-1}_{\textbf{u}} \left( \varphi \right) \textbf{r} \right).
\label{eq: rotazionevettoredistato}
\end{align}
Equation $\eqref{eq: rotazionevettoredistato}$ shows that, under a finite rotation, the wave function of a spinless particle transforms like a scalar field.
\subsubsection{Fully reducible faithful unitary representation}
We study whether the representation of rotations for a spinless particle is projective or faithful. For this purpose, we compute the equation $\eqref{eq: rotazionevettoredistato}$ for $\varphi = 2 \pi$, i.e.,
\begin{align}
\left( \hat{R}_{\textbf{u}}(2 \pi) \psi \right) (\textbf{r}) &= \psi \left( R^{-1}_{\textbf{u}} \left( 2 \pi \right) \textbf{r} \right) = \notag \\
&= \psi (\textbf{r}),
\end{align}
that is, for a spinless particle, the operator $\hat{R}$ provides a faithful unitary representation of the group $SO(3)$. Now, we ask whether this representation is completely reducible or irreducible. In other words, can the matrices representing rotations be brought, in a suitable basis, into block diagonal form? The answer is affirmative: the representation is completely reducible, consisting of infinitely many blocks labeled by the orbital angular momentum quantum number. Let us see why. Any wave function can be expressed as
\begin{align}
\psi(\textbf{r}) &= \sum_{nlm} c_{nlm} v_{nl}(r) Y_{lm}(\theta,\phi) = \notag \\
&= \sum_{nlm} c_{nlm} \psi_{nlm} (\textbf{x}),
\end{align}
with
\begin{equation}
\psi_{nlm}(\textbf{x}) = v_{nl}(r) Y_{lm}(\theta,\phi).
\end{equation}
Fixed an arbitrary orbital quantum number $l$, if the radial components satisfy
\begin{equation}
\int v_{nl}(r) v_{n'l}(r) r^{2} dr = \delta_{n,n'},
\end{equation}
then, since the spherical harmonics functions are a complete orthonormal system, i.e.,
\begin{equation}
\int Y^{*}_{lm}(r) Y_{l'm'}(r) d\Omega = \delta_{l,l'} \delta_{m,m'},
\end{equation}
the set $\lbrace \psi_{nlm} \rbrace$ is an orthonormal complete system. In addition, note that 
\begin{equation}
\left[\hat{\textbf{L}}^{2},\hat{L}_{i}\right]=0
\end{equation}
implies
\begin{equation}
\left[\hat{\textbf{L}}^{2},\hat{R}_{\textbf{u}}(\varphi)\right]=0,
\end{equation}
where $\hat{R}_{\textbf{u}}(\varphi)$ is the exponential function of the operator $\hat{L}_{i}$. The operator $\hat{\textbf{L}}^{2}$ acts on the state vector as
\begin{equation}
\hat{\textbf{L}}^{2} |\psi_{nlm} \rangle = \hslash^{2} l(l+1) |\psi_{nlm} \rangle,
\end{equation}
and since $\hat{\mathbf{L}}^2$ and the rotation operator $\hat{R}_{\mathbf{u}}(\varphi)$ commute, the action of a rotation on a state with orbital angular momentum $l$ yields another state with the same $l$. As a result, the Hilbert space of wave functions can be decomposed into a direct sum of subspaces with fixed $l$: these subspaces are invariant under the action of rotation operators. Therefore, the representation provided by $\hat{R}_{\mathbf{u}}(\varphi)$ is completely reducible. The number of such invariant blocks is infinite, due to the infinite number of allowed values of the principal quantum number $n$, with each $l$ being constrained by $n$. For fixed $n$ and $l$, the Hilbert space $H$ can be written as a direct sum of $(2l + 1)$-dimensional subspaces $H_{nl}$, that is,
\begin{equation}
H = \oplus \ H_{nl},
\end{equation}
\begin{equation}
H_{nl} = |\psi \rangle,
\end{equation}
\begin{equation}
|\psi \rangle = \sum_{m=-l}^{l} c_{m} |\psi_{nlm} \rangle , 
\end{equation}
where each $H_{nl}$ is invariant under the rotation associated with the orbital angular momentum $l$.  
\subsection{Particle with spin}
In Schrödinger's representation, the dynamical evolution of a particle with spin is described by $\psi = \psi_{o} \otimes \psi_{s}$, and its total angular momentum is given by
\begin{equation}
\hat{\textbf{J}} = \hat{\textbf{L}} + \hat{\textbf{S}},
\end{equation}
with $\hat{\textbf{L}}$ and $\hat{\textbf{S}}$ acting on the orbital and spin spaces of the wave function, respectively. Since $[\hat{\textbf{L}},\hat{\textbf{S}}]=0$, $\hat{R}_{\textbf{u}}(\varphi)$ can be manipulated as follows
\begin{align}
\hat{R}_{\textbf{u}}(\varphi) &= e^{- \frac{i}{\hslash} \varphi \textbf{u} \cdot \left( \hat{\textbf{L}}+\hat{\textbf{S}} \right) } = \notag \\
&= e^{- \frac{i}{\hslash} \varphi \textbf{u} \cdot \hat{\textbf{L}}} e^{- \frac{i}{\hslash} \varphi \textbf{u} \cdot \hat{\textbf{S}}},
\end{align}
then
\begin{align}
\hat{R}_{\textbf{u}}(\varphi) \psi &= \hat{R}_{\textbf{u}}(\varphi) ( \psi_{o} \otimes \psi_{s} ) = \notag \\
&= \left( e^{- \frac{i}{\hslash} \varphi \textbf{u} \cdot \hat{\textbf{L}}} \ \psi_{o} \right) \otimes \left( e^{- \frac{i}{\hslash} \varphi \textbf{u} \cdot \hat{\textbf{S}}} \ \psi_{s} \right).
\end{align}
Let us compute $e^{- \frac{i}{\hslash} \varphi \textbf{u} \cdot \hat{\textbf{S}}} \ \psi_{s}$. We will consider the case $s=\frac{1}{2}$.  Let us compute the matrix elements of the exponentials of the spin operators, i.e.,
\begin{align}
\left\langle i \middle| e^{- \frac{i}{\hslash} \varphi \mathbf{u} \cdot \hat{\mathbf{S}}} \middle| j \right\rangle &= \left\langle i \middle| e^{- \frac{i}{\hslash} \varphi \mathbf{u} \cdot \hat{\mathbf{S}}} \middle| j \right\rangle = \notag \\
&= \left( e^{- \frac{i}{2} \varphi \mathbf{u} \cdot \bm{\sigma}} \right)_{ij},
\end{align}
and we expand the exponential in a series of powers as follows
\begin{equation}
e^{- \frac{i}{2} \varphi \textbf{u} \cdot \bm{\sigma}} = \sum_{n=0}^{+ \infty} \dfrac{1}{n!} \left( - \dfrac{i \varphi}{2} \right)^{n} (\textbf{u} \cdot \bm{\sigma})^{n}.
\end{equation}
From the $\eqref{eq: Pauliidentity}$, with $\textbf{a}=\textbf{b}=\textbf{u}$, we have
\begin{equation}
\left( \textbf{u} \cdot \bm{\sigma} \right)^{2} = \mathds{1},
\end{equation}
the series can be decomposed into the sum of even and odd terms as follows
\begin{align}
e^{- \frac{i}{2} \varphi \textbf{u} \cdot \bm{\sigma}} &= \sum_{n=0}^{+ \infty} \dfrac{1}{n!} \left( -\dfrac{i \varphi}{2} \right)^{n} (\textbf{u} \cdot \bm{\sigma})^{n} = \notag \\
&= \sum_{n=0}^{+ \infty} \left[ \dfrac{1}{(2n+1)!} \left( - \dfrac{i \varphi}{2} \right)^{2n+1} (\textbf{u} \cdot \bm{\sigma})^{2n+1} + \dfrac{1}{(2n)!} \left( - \dfrac{i \varphi}{2} \right)^{2n} (\textbf{u} \cdot \bm{\sigma})^{2n} \right] = \notag \\
&= \sum_{n=0}^{+ \infty} \left[ \dfrac{1}{(2n+1)!} \left( - \dfrac{i \varphi}{2} \right)^{2n+1} (\textbf{u} \cdot \bm{\sigma})^{2n} (\textbf{u} \cdot \bm{\sigma}) + \dfrac{1}{(2n)!} \left( - \dfrac{i \varphi}{2} \right)^{2n} \mathds{1} \right] = \notag \\
&= \sum_{n=0}^{+ \infty} \left[ \dfrac{1}{(2n+1)!} \left( - \dfrac{i \varphi}{2} \right)^{2n+1}  (\textbf{u} \cdot \bm{\sigma}) + \dfrac{1}{(2n)!} \left( - \dfrac{i \varphi}{2} \right)^{2n} \mathds{1} \right] = \notag \\
& = \cos{\left( \dfrac{\varphi}{2} \right)} \mathds{1} - i \sin{\left( \dfrac{\varphi}{2} \right)} (\textbf{u} \cdot \bm{\sigma}).
\label{eq: rotazionespin1/2}
\end{align}
The $\eqref{eq: rotazionespin1/2}$ is the rotation of the spin $s=\frac{1}{2}$ by a finite angle $\varphi$ around the direction $\textbf{u}$.
\subsubsection{Projective unitary representation}
Here we study whether the representation of the rotations of a particle with spin is projective or faithful. Let us compute the $\eqref{eq: rotazionevettoredistato}$ for $\varphi = 2 \pi$, i.e.,
\begin{equation}
\hat{R}_{\textbf{u}}(2 \pi) ( \psi_{o} \otimes \psi_{s} ) = \psi_{o} \otimes \left( e^{- \frac{i}{\hslash} 2 \pi \textbf{u} \cdot \hat{\textbf{S}}} \ \psi_{s} \right) ,
\end{equation}
where $\textbf{u} \cdot \hat{\textbf{S}}$ is the projection of the spin $\hat{\textbf{S}}$ along the direction $\textbf{u}$. Since the spin component in the irreducible set $(\hat{r_{\alpha}},\hat{p}_{\alpha}, \hat{S}_{\alpha})$ is arbitrary, we can arrange the axes of the spatial reference system such that 
\begin{equation}
\hat{S}_{z} = \textbf{u} \cdot \hat{\textbf{S}} ,
\end{equation}
and we use the eigenfunctions of $\hat{S}_z$, denoted by $\chi_\sigma(s)$, and the corresponding eigenvalue equation, as follows
\begin{align}
e^{- \frac{i}{\hbar} 2 \pi \hat{S}_{z}} |\psi_{s} \rangle &= \sum_{\sigma=-S}^S \ e^{- \frac{i}{\hbar} 2 \pi \hat{S}_{z}} c_{\sigma} \chi_\sigma(s) = \notag \\
&= \sum_{\sigma=-S}^S \ e^{- \frac{i}{\hbar} 2 \pi \hbar \sigma} c_{\sigma} \chi_\sigma(s) = \notag \\
&= \sum_{\sigma=-S}^S \ e^{- i 2 \pi \sigma} c_{\sigma} \chi_\sigma(s).
\end{align}
If spin $S$ is integer, $\sigma$ is integer; if $S$ is semi-integer, $\sigma$ is semi-integer, so
\begin{equation}
e^{- \frac{i}{\hbar} 2 \pi \hat{S}_{z}} \left| \psi_{s} \right\rangle \ = 
\begin{cases}
+\left| \psi_{s} \right\rangle , \ \text{if } S \text{ is integer} \\
- \left| \psi_{s} \right\rangle , \ \text{if } S \text{ is half-integer} \\
\end{cases},
\end{equation}
\begin{equation}
\begin{cases}
S \text{ integer} \ \Rightarrow \ \hat{R}_{\vec{u}}(2\pi) = \mathds{1} \otimes \mathds{1} \ \Rightarrow \text{faithful representation} \\
S \text{ half-integer} \ \Rightarrow \ \hat{R}_{\vec{u}}(2\pi) = -\mathds{1} \otimes \mathds{1} \ \Rightarrow \text{double-valued representation}
\end{cases}.
\end{equation}
\begin{remark}
Since the unitary operators are defined up to a phase factor, the operators $\mathds{1}$ and $- \mathds{1}$ are physically equivalent, the measurements being provided by the square moduli. If, on the other hand, wave functions were classical fields, only faithful representations of spin would have physical meaning, i.e., a correspondence of state vectors in classical physics would be possible only for $j$ integers.
\end{remark}
\section{The angular momentum and momentum operators as infinitesimal generators of $E(3)$}
Let \( O \) and \( O' \) be two spatial reference frames connected by a roto-translation. Then, to each element \( g = (R, \mathbf{a}) \in E(3) \) we associate an operator \( \hat{\Omega}\left(R_{\mathbf{u}}(\varphi), \mathbf{a} \right) \) such that, if \( |\psi\rangle \) is the state described in frame \( O \), the transformed state \( \hat{\Omega} |\psi\rangle \) corresponds to its description in frame \( O' \). Just as for rotations, the operator \( \hat{\Omega} \) must be unitary also for translations. Given an infinitesimal roto-translation as in equation $\eqref{eq:  rototraslazioneinfinitesima}$, the corresponding infinitesimal transformation operator \( \hat{\Omega} \) takes the form
\begin{equation}
\hat{\Omega} = \mathds{1} + \sum_{\alpha=1}^{3} \delta w_{\alpha} \hat{\widetilde{I}}_{\alpha} + \sum_{\alpha=1}^{3} \delta a_{\alpha} \hat{\widetilde{\pi}}_{\alpha},
\label{eq: operatorerototraslazioneinfinitesima1}
\end{equation}
the constraint $\hat{\Omega} \hat{\Omega}^{\dagger} = \hat{\Omega}^{\dagger} \hat{\Omega} = \mathds{1}$ implies
\begin{equation}
\hat{\widetilde{I}}_{\alpha} = - \hat{\widetilde{I}}_{\alpha}^{\dagger},
\end{equation}
\begin{equation}
\hat{\pi}_{\alpha} = - \hat{\widetilde{\pi}}_{\alpha}^{\dagger},
\end{equation}
then
\begin{equation}
\hat{\Omega}^\dagger = \mathds{1} - \sum_{\alpha=1}^{3} \delta w_{\alpha} \hat{\widetilde{I}}_{\alpha} - \sum_{\alpha=1}^{3} \delta a_{\alpha} \hat{\widetilde{\pi}}_{\alpha}.
\label{eq: operatorerototraslazioneinfinitesimaaggiunto1}
\end{equation}
We define the two hermitian operators by means of
\begin{equation}
\hat{\widetilde{I}}_{\alpha} = - \dfrac{i}{\hslash} \hat{J}_{\alpha},
\end{equation}
\begin{equation}
\hat{\widetilde{\pi}}_{\alpha} = - \dfrac{i}{\hslash} \hat{P}_{\alpha},
\end{equation}
which satisfy the algebras
\begin{equation}
\begin{cases}
\left[ \hat{J}_{\alpha},\hat{J}_{\beta} \right] = i \hslash \sum_{\gamma=1}^3 \epsilon_{\alpha \beta \gamma} \hat{J}_{\gamma} \\
\left[ \hat{P}_{\alpha},\hat{P}_{\beta} \right] = 0 \\
\left[ \hat{J}_{\alpha},\hat{P}_{\beta} \right] = i \hslash \sum_{\gamma=1}^3 \epsilon_{\alpha \beta \gamma} \hat{P}_{\gamma}
\end{cases}.
\end{equation}
Independently of a physical system, we define the momentum $\hat{\textbf{P}}$ and the angular momentum $\hat{\textbf{J}}$ operators as the infinitesimal generators of translations and rotations, respectively. Note that the equations $\eqref{eq: operatorerototraslazioneinfinitesima1}$, $\eqref{eq: operatorerototraslazioneinfinitesimaaggiunto1}$ become
\begin{equation}
\hat{\Omega} = \mathds{1} - \dfrac{i}{\hslash} \sum_{\alpha=1}^{3} \delta w_{\alpha} \hat{J}_{\alpha} - \dfrac{i}{\hslash} \sum_{\alpha=1}^{3} \delta a_{\alpha} \hat{P}_{\alpha},
\label{eq: operatorerototraslazioneinfinitesima2}
\end{equation}
\begin{equation}
\hat{\Omega}^{\dagger} = \mathds{1} + \dfrac{i}{\hslash} \sum_{\alpha=1}^{3} \delta w_{\alpha} \hat{J}_{\alpha} + \dfrac{i}{\hslash} \sum_{\alpha=1}^{3} \delta a_{\alpha} \hat{P}_{\alpha},
\label{eq: operatorerototraslazioneinfinitesimaaggiunto2}
\end{equation}
respectively. The action of infinite infinitesimal rotations and infinite infinitesimal translations lead to the operator $\hat{\Omega}\left(R_{\hat{\textbf{u}}}(\varphi),\textbf{a}\right)$ of a finite rototranslation, i.e.,
\begin{equation}
\hat{\Omega}(R_{\textbf{u}}(\varphi),\textbf{a}) = e^{- \frac{i}{\hslash} \varphi \textbf{u} \cdot \hat{\textbf{J}} - \frac{i}{\hslash} \textbf{a} \cdot \hat{\textbf{P}}}.
\label{eq: operatorerototraslazione}
\end{equation}
\section{Angular momentum and momentum of a spinless particle}
Let $(\hat{r}_i, \hat{p}_j)$ be an irreducible set of operators on the Hilbert space $\mathcal{H}$ of a quantum system. To determine how the total angular momentum $\hat{\mathbf{J}}$ and the momentum $\hat{\mathbf{P}}$ depend on the operators $(\hat{r}_i, \hat{p}_j)$, we must analyze how these operators transform under rototranslations. It is important to adopt the passive interpretation of rotations. Let $O$ and $O'$ be two spatial reference frames related by a rototranslation. Their respective orthonormal bases are connected through a rotation matrix $R$, such that $e'_{i} = \sum_{j=1}^3 R_{ij} e_{j}$, $e_{j} = \sum_{i=1}^3 R^{-1}_{ji} e'_{i}$. Accordingly, for a generic point $P$, its coordinates in the two reference frames are related as follows
\begin{align}
\sum_{i=1}^3 r'_{i} e'_{i} &= O'P = \notag \\
&= OP - OO' = \notag \\
&= \sum_{j=1}^3 r_{j} e_{j} - \sum_{j=1}^3 a_{j} e_{j} = \notag \\
&= \sum_{j=1}^3 (r_{j} - a_{j}) e_{j} = \notag \\
&= \sum_{j=1}^3 \sum_{i=1}^3 (r_{j} - a_{j}) R^{-1}_{ji} e'_{i} ,
\end{align}
and, by comparison, we have
\begin{equation}
r'_{i} = \sum_{j=1}^3 (r_{j} - a_{j}) R^{-1}_{ji}.
\end{equation}
Let us postulate that the position operator transforms under rototranslation as the classical function does, i.e.,
\begin{align}
\hat{r}'_{i} |\textbf{r}'\rangle \ &= r'_i |\textbf{r}'\rangle = \notag \\
&= \sum_{j=1}^3 (r_{j} - a_{j}) R^{-1}_{ji} |\textbf{r}'\rangle = \notag \\
&= \sum_{j=1}^3 R^{-1}_{ji} (r_{j} - a_{j}) |\textbf{r}'\rangle = \notag \\
&= \sum_{j=1}^3 R^{-1}_{ji} \left( \hat{r}_{j} - a_{j} \mathds{1} \right) |\textbf{r}'\rangle,
\end{align}
then
\begin{equation}
\hat{r}'_{i} = \sum_{j=1}^3 R^{-1}_{ji} \left( \hat{r}_{j} - a_{j} \mathds{1} \right) ,
\end{equation}
\begin{equation}
\hat{\Omega} \hat{r}_{i} \hat{\Omega}^{\dagger} = \sum_{j=1}^3 R^{-1}_{ji} \left( \hat{r}_{j} - a_{j} \mathds{1} \right).
\label{eq: rotazioneoperatoreposizionespinless}
\end{equation}
In classical physics, the momentum of a particle does not vary under a translation, so we postulate that the action of a rototranslation on $\hat{\textbf{p}}$ reduces to a rotation, i.e.,
\begin{equation}
\hat{p}'_{i} = \sum_{j=1}^3 R^{-1}_{ji} \hat{p}_{j},
\end{equation}
\begin{equation}
\hat{\Omega} \hat{p}_{i} \hat{\Omega}^{\dagger} = \sum_{j=1}^3 R^{-1}_{ji} \hat{p}_{j}.
\label{eq: rotazioneoperatoremomentospinless}
\end{equation}
The infinitesimal generators $\hat{\textbf{J}}$ and $\hat{\textbf{P}}$ can now be calculated from the system given by $\eqref{eq: rotazioneoperatoreposizionespinless}$, $\eqref{eq: rotazioneoperatoremomentospinless}$, where $R^{-1}_{ji}$, $\hat{\Omega}$ and $\hat{\Omega}^\dagger$ are given by $\eqref{eq: rotazioneinfinitesimainversa}$, $\eqref{eq: operatorerototraslazioneinfinitesima2}$ and $\eqref{eq: operatorerototraslazioneinfinitesimaaggiunto2}$, respectively, and the translation vector $\textbf{a}$ is replaced by the infinitesimal vector $\delta \textbf{a}$. From $\eqref{eq: rotazioneoperatoremomentospinless}$ we have
\begin{equation}
\left[ \hat{J}_{\alpha} , \hat{p}_{i} \right] = i \hslash \sum_{j=1}^3 \epsilon_{\alpha i j} \hat{p}_{j}.
\end{equation}
In addition, since the translations leave the momentum unchanged, we have
\begin{equation}
\left[ \hat{p}_{\alpha} , \hat{P}_{\beta} \right] = 0, \ \forall \ \alpha, \beta.
\end{equation}
From $\hat{\Omega} \hat{r}_{i} \hat{\Omega}^{\dagger} = \sum_{j=1}^3 R^{-1}_{ji} \left( \hat{r}_{j} - \delta a_{j} \mathds{1} \right)$, we have
\begin{align}
& \left( \mathds{1} - \dfrac{i}{\hslash} \sum_{\alpha=1}^{3} \delta w_{\alpha} \hat{J}_{\alpha} - \dfrac{i}{\hslash} \sum_{\alpha=1}^{3} \delta a_{\alpha} \hat{P}_{\alpha} \right) \hat{r}_{i} \left( \mathds{1} + \dfrac{i}{\hslash} \sum_{\alpha=1}^{3} \delta w_{\alpha} \hat{J}_{\alpha} + \dfrac{i}{\hslash} \sum_{\alpha=1}^{3} \delta a_{\alpha} \hat{P}_{\alpha} \right) = \notag \\
& \ \ \ \ \ \ \ \ \ \ \ \ = \sum_{j=1}^3 \left( \delta_{ji} - \sum_{\alpha=1}^3 \delta w_{\alpha} \epsilon_{\alpha j i} \right) \left( \hat{r}_{j} - \delta a_{j} \mathds{1} \right),
\end{align}
\begin{align}
& \left( \hat{r}_{i} - \dfrac{i}{\hslash} \sum_{\alpha=1}^{3} \delta w_{\alpha} \hat{J}_{\alpha} \hat{r}_{i} - \dfrac{i}{\hslash} \sum_{\alpha=1}^{3} \delta a_{\alpha} \hat{P}_{\alpha} \hat{r}_{i} \right) \left( \mathds{1} + \dfrac{i}{\hslash} \sum_{\alpha=1}^{3} \delta w_{\alpha} \hat{J}_{\alpha} + \dfrac{i}{\hslash} \sum_{\alpha=1}^{3} \delta a_{\alpha} \hat{P}_{\alpha} \right) = \notag \\
& \ \ \ \ \ \ \ \ \ \ \ \ = \sum_{j=1}^3 \delta_{ji} \hat{r}_{j} - \sum_{j=1}^3 \delta_{ji} \delta a_{j} \mathds{1} - \sum_{j=1}^3 \sum_{\alpha=1}^3 \delta w_{\alpha} \epsilon_{\alpha j i} \hat{r}_{j} + \sum_{j=1}^3 \sum_{\alpha=1}^3 \delta w_{\alpha} \delta a_{j} \epsilon_{\alpha j i},
\end{align}
\begin{align}
& \hat{r}_{i} + \dfrac{i}{\hslash} \sum_{\alpha=1}^{3} \delta w_{\alpha} \hat{r}_{i} \hat{J}_{\alpha} + \dfrac{i}{\hslash} \sum_{\alpha=1}^{3} \delta a_{\alpha} \hat{r}_{i} \hat{P}_{\alpha} - \dfrac{i}{\hslash} \sum_{\alpha=1}^{3} \delta w_{\alpha} \hat{J}_{\alpha} \hat{r}_{i} + \dfrac{1}{\hslash^2} \sum_{\alpha,\beta=1}^3 \delta w_{\alpha} \delta w_{\beta} \hat{J}_{\alpha} \hat{r}_{i} \hat{J}_{\beta} + \notag \\
& \ \ \ + \dfrac{1}{\hslash^2} \sum_{\alpha,\beta=1}^3 \delta w_{\alpha} \delta a_{\beta} \hat{J}_{\alpha} \hat{r}_{i} \hat{P}_{\beta} - \dfrac{i}{\hslash} \sum_{\alpha=1}^{3} \delta a_{\alpha} \hat{P}_{\alpha} \hat{r}_{i} + \dfrac{1}{\hslash^2} \sum_{\alpha,\beta=1}^3 \delta a_{\alpha} \delta w_{\beta} \hat{P}_{\alpha} \hat{r}_{i} \hat{J}_{\beta} + \notag \\
& \ \ \ \ \ \ + \dfrac{1}{\hslash^2} \sum_{\alpha,\beta=1}^3 \delta a_{\alpha} \delta a_{\beta} \hat{P}_{\alpha} \hat{r}_{i} \hat{P}_{\beta} = \hat{r}_{i} - \delta a_{i} \mathds{1} - \sum_{j=1}^3 \sum_{\alpha=1}^3 \delta w_{\alpha} \epsilon_{\alpha j i} \hat{r}_{j} + \sum_{j=1}^3 \sum_{\alpha=1}^3 \delta w_{\alpha} \delta a_{j} \epsilon_{\alpha j i},
\end{align}
\begin{equation}
\hat{r}_{i} + \dfrac{i}{\hslash} \sum_{\alpha=1}^{3} \delta w_{\alpha} \hat{r}_{i} \hat{J}_{\alpha} + \dfrac{i}{\hslash} \sum_{\alpha=1}^{3} \delta a_{\alpha} \hat{r}_{i} \hat{P}_{\alpha} - \dfrac{i}{\hslash} \sum_{\alpha=1}^{3} \delta w_{\alpha} \hat{J}_{\alpha} \hat{r}_{i} - \dfrac{i}{\hslash} \sum_{\alpha=1}^{3} \delta a_{\alpha} \hat{P}_{\alpha} \hat{r}_{i} = \hat{r}_{i} - \delta a_{i} \mathds{1} - \sum_{j=1}^3 \sum_{\alpha=1}^3 \delta w_{\alpha} \epsilon_{\alpha j i} \hat{r}_{j},
\end{equation}
where in the last step we neglected second-order terms with respect to infinitesimal increments. Then
\begin{equation}
\dfrac{i}{\hslash} \sum_{\alpha=1}^{3} \delta w_{\alpha} \hat{r}_{i} \hat{J}_{\alpha} + \dfrac{i}{\hslash} \sum_{\alpha=1}^{3} \delta a_{\alpha} \hat{r}_{i} \hat{P}_{\alpha} - \dfrac{i}{\hslash} \sum_{\alpha=1}^{3} \delta w_{\alpha} \hat{J}_{\alpha} \hat{r}_{i} - \dfrac{i}{\hslash} \sum_{\alpha=1}^{3} \delta a_{\alpha} \hat{P}_{\alpha} \hat{r}_{i} = - \delta a_{i} \mathds{1} - \sum_{j=1}^3 \sum_{\alpha=1}^3 \delta w_{\alpha} \epsilon_{\alpha j i} \hat{r}_{j},
\end{equation}
\begin{equation}
\dfrac{i}{\hslash} \sum_{\alpha=1}^{3} \delta w_{\alpha} \left[ \hat{r}_{i} , \hat{J}_{\alpha} \right] + \dfrac{i}{\hslash} \sum_{\alpha=1}^{3} \delta a_{\alpha} \left[ \hat{r}_{i} , \hat{P}_{\alpha} \right] = - \sum_{j=1}^3 \sum_{\alpha=1}^3 \delta w_{\alpha} \epsilon_{\alpha j i} \hat{r}_{j} - \delta a_{i} \mathds{1}.
\end{equation}
By comparison we get
\begin{equation}
\dfrac{i}{\hslash} \left[ \hat{r}_{i} , \hat{J}_{\alpha} \right] = - \sum_{j=1}^3 \epsilon_{\alpha j i} \hat{r}_{j},
\end{equation}
\begin{equation}
\dfrac{i}{\hslash} \sum_{\alpha=1}^{3} \delta a_{\alpha} \left[ \hat{r}_{i} , \hat{P}_{\alpha} \right] = - \delta a_{i} \mathds{1}.
\end{equation}
Regarding the first equation, in analogy with the treatment of the angular momentum, we obtain
\begin{equation}
\left[ \hat{r}_{i} , \hat{J}_{\alpha} \right] = i \hslash \sum_{j=1}^3 \epsilon_{\alpha j i} \hat{r}_{j},
\end{equation}
\begin{equation}
\left[ \hat{r}_{i} , \hat{J}_{\alpha} \right] = - i \hslash \sum_{j=1}^3 \epsilon_{\alpha i j} \hat{r}_{j},
\end{equation}
\begin{equation}
- \left[ \hat{J}_{\alpha} , \hat{r}_{i} \right] = - i \hslash \sum_{j=1}^3 \epsilon_{\alpha i j} \hat{r}_{j},
\end{equation}
\begin{equation}
\left[ \hat{J}_{\alpha} , \hat{r}_{i} \right] = i \hslash \sum_{j=1}^3 \epsilon_{\alpha i j} \hat{r}_{j},
\end{equation}
where the sum over $j$ has been eliminated because the Levi-Civita symbol is nonzero only for $\alpha \neq i \neq j$. Concerning the second equation of the system, note that for it to be satisfied for any choice of increments $\delta a_\alpha$, the commutator $\left[ \hat{r}_{i} , \hat{P}_{\alpha} \right]$ must vanish for $i \neq \alpha$, then
\begin{equation}
\left[ \hat{r}_{i} , \hat{P}_{\alpha} \right] = i \hslash \delta_{i,\alpha}.
\end{equation}
Ultimately, the solutions satisfy
\begin{equation}
\begin{cases}
\left[ \hat{J}_{\alpha} , \hat{r}_{i} \right] = i \hslash \sum_{j=1}^3 \epsilon_{\alpha i j} \hat{r}_{j} \\
\left[ \hat{r}_{i} , \hat{P}_{\alpha} \right] = i \hslash \delta_{i,\alpha} \\
\left[ \hat{p}_{i} , \hat{J}_{\alpha} \right] = i \hslash \sum_{j=1}^3 \epsilon_{i \alpha j} \hat{p}_{j} \\
\left[ \hat{p}_{\alpha} , \hat{P}_{\beta} \right] = 0
\end{cases} .
\end{equation}
i.e., for a particle with no spin and subject to infinitesimal rototranslation, we derive in a general way the identifications
\begin{align}
\hat{J}_{\alpha} &= \hat{L}_{\alpha} = \notag \\
&= \sum_{\gamma=1}^3 \epsilon_{\alpha \beta \gamma} \hat{r}_{\beta} \hat{p}_{\gamma},
\end{align}
and
\begin{equation}
\hat{P}_{\alpha} = \hat{p}_{\alpha}.
\label{eq: momentogeneratoreoperatoretraslazione}
\end{equation}
\section{Translation operator}
Starting from equation $\eqref{eq: operatorerototraslazione}$, and using $\eqref{eq: momentogeneratoreoperatoretraslazione}$, we can define the spatial translation operator $\hat{T}(\textbf{a})$ as the operator corresponding to a roto-translation with zero rotation (i.e., with the identity as the rotation matrix) and translation vector $\textbf{a}$, in other words
\begin{align}
\hat{T}(\textbf{a}) &\equiv \hat{\Omega}(\mathds{1},\textbf{a}) = \notag \\
&= e^{- \frac{i}{\hslash} \textbf{a} \cdot \hat{\textbf{P}}} = \notag \\
&= e^{- \frac{i}{\hslash} \textbf{a} \cdot \hat{\textbf{p}}}.
\label{eq: operatoretraslazionespaziale}
\end{align}
The definition of the translation operator leads to a fundamental result of quantum mechanics: the theorem of invariance under spatial translations. Although rooted in the formal structure of quantum theory, this theorem plays a central role in many-body physics, where translational symmetry is extensively exploited to simplify calculations. We have
\begin{theorem}[Invariance under spatial translations] Spatial translation invariance of the Hamiltonian implies the conservation of the momentum operator.
\begin{proof}
Let $\mathcal{H}$ be the Hamiltonian of the system, then the symmetry under spatial translations implies
\begin{equation}
\left[ \hat{T}(\textbf{a}), \mathcal{H} \right] = 0, \ \forall \ \textbf{a},
\end{equation}
\begin{equation}
\hat{T}^\dagger(\textbf{a}) \mathcal{H} \hat{T}(\textbf{a}) = \mathcal{H}.
\end{equation}
We differentiate both sides with respect to $\textbf{a}$, and evaluate at $\textbf{a}=\textbf{0}$, that is
\begin{align}
\left. \nabla_{\textbf{a}} \left( \hat{T}^\dagger(\textbf{a}) \mathcal{H} \hat{T}(\textbf{a}) \right) \right|_{\textbf{a}=\textbf{0}}
&= \left. \left( \nabla_{\textbf{a}} \hat{T}(\textbf{a}) \mathcal{H} + \mathcal{H} \nabla_{\textbf{a}} \hat{T}(\textbf{a}) \right) \right|_{\textbf{a}=\textbf{0}} = \notag \\
&= \frac{i}{\hslash} \hat{\textbf{p}} \mathcal{H} - \frac{i}{\hslash} \mathcal{H} \hat{\textbf{p}} = \notag \\
&= \frac{i}{\hslash} [\hat{\textbf{p}}, \mathcal{H}].
\end{align}
Since the derivative of the left-hand side vanishes, because it equals the derivative of $\mathcal{H}$, a constant operator, we get $\left[ \hat{\textbf{p}}, \mathcal{H} \right] = 0$, then, by the Heisenberg equation of motion
\begin{align}
\frac{d\hat{\textbf{p}}}{dt} &= \dfrac{i}{\hslash} \left[ \mathcal{H}, \hat{\textbf{p}} \right] = \notag \\
&= 0,
\end{align}
the statement follows.
\end{proof}
\end{theorem}
Ultimately, we show that a state vector transforms under a spatial translation as does a scalar field, indeed
\begin{align}
\left( \hat{T}(\textbf{a}) \psi \right) (\textbf{r}) &= e^{- \frac{i}{\hslash} \textbf{a} \cdot \hat{\textbf{p}}} \psi(\textbf{r}) = \notag \\
&= \left( \lim_{N \rightarrow + \infty} \left( \mathds{1} - \dfrac{i}{\hslash} \dfrac{\textbf{a}}{N} \cdot \hat{\textbf{p}} \right)^{N} \psi \right) (\textbf{r}) = \notag \\
&= \left( \lim_{N \rightarrow + \infty} \left( \mathds{1} - \dfrac{\textbf{a}}{N} \cdot \nabla \right)^{N} \psi \right) (\textbf{r}) = \notag \\
&= \lim_{N \rightarrow + \infty} \psi \left( \left( \mathds{1} - \dfrac{\textbf{a}}{N} \cdot \nabla \right)^{N} \textbf{r} \right) = \notag \\
&= \psi(\textbf{r}-\textbf{a}).
\label{eq: azioneoperatoretraslazionespazialesufunzionedonda}
\end{align}
\section{Ordinary $SO(3)$ matrix representations of operators}
Let $\lbrace A_{1} , \ldots , A_{n} \rbrace$ be a set of linearly independent operators that transform, under rotations, into linear combinations of the set itself by means of $n \times n$ matrices $M(R)$, i.e.,
\begin{equation}
\hat{\Omega}(R) A_{i} \hat{\Omega}^{\dagger}(R) = \sum_{j=1}^n A_{j} M_{ji}(R),
\end{equation}
that is, any linear combination with complex coefficients of the $\lbrace A_{i} \rbrace$ will transform under rotation into an element of the n-dimensional space $\Span \lbrace A_{1} , \ldots , A_{n} \rbrace$. We have
\begin{theorem}
If $\hat{\Omega}(R)$ is a projective representation of the rotation group, then the matrices $M(R)$ define an ordinary representation of $SO(3)$ acting on the operators, i.e.,
\begin{equation}
M(R_{2}) M(R_{1}) = M(R_{2} R_{1}).
\end{equation}
\begin{proof}
The rotation $R_{2} R_{1}$ satisfies
\begin{equation}
\hat{\Omega}(R_{2} R_{1}) A_{i} \hat{\Omega}^{\dagger}(R_{2} R_{1}) = \sum_{k=1}^n A_{k} M_{ki} (R_{2} R_{1}),
\end{equation}
and since $\hat{\Omega}(R)$ is a projective representation, we have
\begin{align}
e^{i \varphi(R_{2},R_{1})} \hat{\Omega}(R_{2}) \hat{\Omega}(R_{1}) A_{i} e^{- i \varphi(R_{2},R_{1})} \hat{\Omega}^{\dagger}(R_{1}) \hat{\Omega}^{\dagger}(R_{2}) &= \hat{\Omega}(R_{2}) \left( \hat{\Omega}(R_{1}) A_{i} \hat{\Omega}^{\dagger}(R_{1}) \right) \hat{\Omega}^{\dagger}(R_{2}) = \notag \\
&= \hat{\Omega}(R_{2}) \left( \sum_{j=1}^n A_{j} M_{ji}(R_{1}) \right) \hat{\Omega}^{\dagger}(R_{2}) = \notag \\
&= \sum_{j=1}^n \sum_{k=1}^n A_{k} M_{kj}(R_{2}) M_{ji}(R_{1}),
\end{align}
then
\begin{equation}
\sum_{j=1}^n \sum_{k=1}^n A_{k} M_{kj}(R_{2}) M_{ji}(R_{1}) = \sum_{k=1}^n A_{k} M_{ki} (R_{2} R_{1}),
\end{equation}
and since $\lbrace A_{i} \rbrace$ is a linearly independent set, the assertion follows.
\end{proof}
\end{theorem}
Given a vector operator \( \mathbf{K} \), it transforms under rotations as
\begin{align}
\hat{\Omega}(R) K_{i} \hat{\Omega}^{\dagger}(R) &= \sum_{j=1}^3 R^{-1}_{ji} K_{j} = \notag \\
&= \sum_{j=1}^3 R^{T}_{ji} K_{j} = \notag \\
&= \sum_{j=1}^3 R_{ij} K_{j} .
\end{align}
The matrices \( M(R) \) form a faithful representation of the three-dimensional rotation matrices \( R \in SO(3) \), and in this case they coincide with them, i.e.,
\begin{equation}
M_{ij}(R) \equiv R_{ij}, \ i, j = 1,2,3.
\end{equation}
\section{Time reversal symmetry}
\subsection{Time reversal in classical mechanics}
Given the Hamiltonian of a single particle in a conservative potential, i.e., $\mathcal{H} = \frac{\textbf{p}^2}{2m} + V(\textbf{r})$, where \( V(\textbf{r}) \) is the potential associated with a conservative force field, the Hamiltonian is an even function of the conjugate momentum, that is, $\mathcal{H}(\textbf{r}, \textbf{p}) = \mathcal{H}(\textbf{r}, -\textbf{p})$. Indeed, under the substitution \( \textbf{p} \rightarrow -\textbf{p} \), both the potential term \( V(\textbf{r}) \) and the kinetic energy remain unchanged. Consequently, if \( \textbf{r}(t) \), \( \textbf{p}(t) \) are solutions of the Hamilton equations of motion,
\begin{equation}
\dfrac{d{\textbf{r}}(t)}{dt} = \dfrac{\partial \mathcal{H}}{\partial \textbf{p}},
\end{equation}
\begin{equation}
\dfrac{d\textbf{p}(t)}{dt} = - \dfrac{\partial \mathcal{H}}{\partial \textbf{r}},
\end{equation}
then also the functions
\begin{equation}
\textbf{r}'(t) = \textbf{r}(-t),
\label{eq: trasformazioneinversionetemporaler}
\end{equation}
\begin{equation}
\textbf{p}'(t) = - \textbf{p}(-t)
\label{eq: trasformazioneinversionetemporalep}
\end{equation}
satisfy the same equations of motion. In other words, \( \textbf{r}'(t) \) and \( \textbf{p}'(t) \) describe a valid physical trajectory obtained by time reversal: the particle retraces its path in reverse, with the momentum inverted at each point. Indeed,
\begin{align}
\dfrac{d\textbf{p}'(t)}{dt} &= - \dfrac{d\textbf{p}(-t)}{dt} = \notag \\
&= \dfrac{d\textbf{p}(t)}{dt} = \notag \\
&= - \dfrac{\partial \mathcal{H}}{\partial \textbf{r}} = \notag \\
&= - \dfrac{\partial \mathcal{H}}{\partial \textbf{r}'} \dfrac{\partial \textbf{r}'_i}{\partial \textbf{r}} = \notag \\
&= - \dfrac{\partial \mathcal{H}}{\partial \textbf{r}'},
\end{align}
and
\begin{align}
\dfrac{d\textbf{r}'(t)}{dt} &= \dfrac{d\textbf{r}(-t)}{dt} = \notag \\
&= - \dfrac{d\textbf{r}(t)}{dt} = \notag \\
&= - \dfrac{\partial \mathcal{H}}{\partial \textbf{p}} = \notag \\
&= \dfrac{\partial \mathcal{H}}{\partial (-\textbf{p})},
\end{align}
where the last step is due to the fact that $\mathcal{H}$ is even with respect to $\textbf{p}$, then
\begin{equation}
\dfrac{d\textbf{r}'(t)}{dt} = \dfrac{\partial \mathcal{H}}{\partial \textbf{p}'}.
\end{equation}
The solutions $\textbf{r}'(t)$, $\textbf{p}'(t)$ have considerable physical significance. If a particle moves from $A$ at time $t-t_0$ to $B$ at time $t_0$, its dynamics is described by the solutions $\textbf{r}(t)$, $\textbf{p}(t)$, that is,
\begin{align}
A \ \ \ & \xrightarrow[\text{+t}]{} \ \ \ B \notag \\
\left( \textbf{r}(t_0-t),\textbf{p}(t_0-t) \right) \ \ \ & \xrightarrow[\text{+t}]{} \ \ \ \left( \textbf{r}(t_0),\textbf{p}(t_0) \right).
\end{align}
The functions $\textbf{r}'(t)$, $\textbf{p}'(t)$ are solutions of the particle dynamics. In particular, given $\eqref{eq: trasformazioneinversionetemporaler}$, $\eqref{eq: trasformazioneinversionetemporalep}$, if the motion starts at the time $-t_0$ (note that we reverse the sign of time) and we set 
\begin{equation}
B \equiv \left(\textbf{r}(t_0),\textbf{p}(t_0)\right) \xrightarrow  B \equiv \left(\textbf{r}(-t_0),-\textbf{p}(t_0)\right)
\end{equation}
since the functions are equal at each instant, the evolution described by $\textbf{r}'(t)$, $\textbf{p}'(t)$ predicts that the particle moves from $B$ to $A$, in the time interval $t$, with reversed momentum, i.e.,
\begin{align}
B \ \ \ & \xrightarrow[\text{+t}]{} \ \ \ A \notag \\
\left( \textbf{r}(-t_0),\textbf{p}(-t_0) \right) \ \ \ & \xrightarrow[\text{+t}]{} \ \ \ \left( \textbf{r}(t-t_0),\textbf{p}(t-t_0) \right).
\end{align}
Therefore, 
\begin{equation}
A \equiv \left( \textbf{r}(t_0-t),\textbf{p}(t_0-t) \right) \equiv \left( \textbf{r}(t-t_0),-\textbf{p}(t-t_0) \right),
\end{equation}
\begin{equation}
B \equiv \left( \textbf{r}(t_0),\textbf{p}(t_0) \right) \equiv \left( \textbf{x}(-t_0),-\textbf{p}(-t_0) \right).
\end{equation}
Due to the inversion of the sign of time, the functions \( \textbf{r}'(t) \) and \( \textbf{p}'(t) \) are referred to as the time-reversed (or temporal inverses) of \( \textbf{r}(t) \) and \( \textbf{p}(t) \), respectively. In fact, the initial point, the final point, and the direction along the trajectory are all reversed under this transformation.
\subsection{Time reversal in quantum mechanics}
Here, we will study a formal theory of symmetry by time reversal in quantum mechanics from which we will also derive the properties of the time reversal operator. For this purpose, it is sufficient to restrict the discussion to the Hilbert space of a spinless particle. Let $\hat{\Theta}$ be the time reversal operator
\begin{equation}
\hat{\Theta}: \ |\alpha\rangle \longrightarrow \hat{\Theta} |\alpha\rangle.
\end{equation}
Now, given a state $|\alpha \rangle$ at time $t=0$, its time evolution in an infinitesimal time interval $\delta t$ is given by
\begin{equation}
|\alpha,\delta t \rangle= \left( \mathds{1} - i \dfrac{\mathcal{H}}{\hslash} \delta t \right) |\alpha \rangle.
\end{equation}
On the other hand, if $\hat{\Theta}$ acts on the state $|\alpha \rangle$ at $t=0$, after a time $\delta t$, that state becomes
\begin{equation}
\hat{\Theta} |\alpha \rangle \longrightarrow \left( \mathds{1} - i \dfrac{\mathcal{H}}{\hslash} \delta t \right) \hat{\Theta} |\alpha \rangle .
\end{equation}
If the system is invariant under time reversal, then the previous state is equal to the action of the time reversal operator $\hat{\Theta}$ on the state evolved from time \(0\) to time \(-\delta t\), that is, \(\hat{\Theta} |\alpha, -\delta t \rangle\). Therefore,
\begin{equation}
\left( \mathds{1} - i \dfrac{\mathcal{H}}{\hslash} \delta t \right) \hat{\Theta} |\alpha \rangle = \hat{\Theta} \left( \mathds{1} - i \dfrac{\mathcal{H}}{\hslash} (- \delta t) \right) |\alpha \rangle,
\end{equation}
\begin{equation}
- i \mathcal{H} \hat{\Theta} |\alpha \rangle = \hat{\Theta} i \mathcal{H} |\alpha \rangle .
\end{equation}
According to Wigner's theorem on symmetries, the time reversal operator is either unitary or antiunitary. If we assume that it is unitary, and thus linear, then the presence of the imaginary unit \( i \) in the previous equation becomes irrelevant, and we obtain
\begin{equation}
- \mathcal{H} \hat{\Theta} |\alpha \rangle = \hat{\Theta} \mathcal{H} |\alpha \rangle, 
\end{equation}
which, however, leads to physically incorrect conclusions. Indeed, if \(\mathcal{E}_\alpha\) is the eigenvalue associated with the state \(|\alpha\rangle\), the time-reversed state satisfies
\begin{align}
\mathcal{H} \hat{\Theta} |\alpha \rangle &= - \hat{\Theta} \mathcal{H} |\alpha \rangle = \notag \\
&= - \mathcal{E}_\alpha \hat{\Theta} |\alpha \rangle,
\end{align}
that is, the opposite of a state is associated with the opposite of energy, which makes no sense. Thus, the time reversal operator is antiunitary, in particular is antilinear
\begin{equation}
- i \mathcal{H} \hat{\Theta} |\alpha \rangle = \hat{\Theta} i \mathcal{H} |\alpha \rangle, 
\end{equation}
\begin{equation}
- i \mathcal{H} \hat{\Theta} |\alpha \rangle = -i \hat{\Theta} \mathcal{H} |\alpha \rangle,
\end{equation}
\begin{equation}
\left[ \mathcal{H} , \hat{\Theta} \right] = 0.
\end{equation}
\subsection{Transformation of operators under time reversal}
In this section, we study how physical observables transform under the action of the time reversal operator. First, let us specialize equation $\eqref{eq: unacaratterizzazionediosservabilisottoazioneoperatoriantiunitari}$ to our case as follows 
\begin{equation}
\langle \beta | A | \alpha \rangle \ = \ \langle \tilde{\alpha} | \hat{\Theta} A \hat{\Theta}^{-1} | \tilde{\beta} \rangle
\label{eq: osservabilepariodisparisottoinversionetemporale1}
\end{equation}
for a physical observable $A$ and an antiunitary operator $\hat{\Theta}$, then $A$ is even or odd depending on whether under time reversal, it goes into itself or its opposite, i.e.,
\begin{equation}
\hat{\Theta} A \hat{\Theta}^{-1} = \pm A.
\label{eq: osservabilepariodisparisottoinversionetemporale2}
\end{equation}
The equations $\eqref{eq: osservabilepariodisparisottoinversionetemporale1}$, $\eqref{eq: osservabilepariodisparisottoinversionetemporale2}$ set a constraint on the phase factors of the elements of the $A$ operator taken with respect to the temporally inverted state vectors, indeed
\begin{align}
\langle \beta | A | \alpha \rangle \ &= \langle \tilde{\alpha} | \hat{\Theta} A \hat{\Theta}^{-1} | \tilde{\beta} \rangle = \notag \\
&= \pm \langle \tilde{\alpha} | A | \tilde{\beta} \rangle = \notag \\
&= \pm \langle \tilde{\beta} | A | \tilde{\alpha} \rangle^*,
\end{align}
and in particular, the expectation values of the observable satisfy
\begin{equation}
\langle \alpha | A | \alpha \rangle = \pm \langle \tilde{\alpha} | A | \tilde{\alpha} \rangle^{*} .
\label{eq: criterioditrasformazioneosservabilesottoinversionetemporale}
\end{equation}
Equation $\eqref{eq: criterioditrasformazioneosservabilesottoinversionetemporale}$ expresses the condition that an observable operator must satisfy under time reversal.  
For the momentum operator $\hat{\textbf{p}}$, it follows from the classical analysis that its expectation value in the time-reversed state is the negative of that in the original state, i.e., it changes sign,
\begin{equation}
\langle \alpha | \hat{\textbf{p}} | \alpha \rangle \ = \ - \langle \tilde{\alpha} | \hat{\textbf{p}} | \tilde{\alpha} \rangle^{*} ,
\end{equation}
that is, the momentum operator is an odd operator, i.e.,
\begin{equation}
\hat{\Theta} \hat{\textbf{p}} \hat{\Theta}^{-1} = - \hat{\textbf{p}}, 
\end{equation}
then
\begin{align}
\hat{\mathbf{p}} \, \hat{\Theta} |\mathbf{p}'\rangle &= \left(- \hat{\Theta} \hat{\mathbf{p}} \hat{\Theta}^{-1}\right) \hat{\Theta} |\mathbf{p}'\rangle = \notag \\
&= - \hat{\Theta} \hat{\mathbf{p}} |\mathbf{p}'\rangle = \notag \\
&= - \hat{\Theta} \mathbf{p}' |\mathbf{p}'\rangle = \notag \\
&= - \mathbf{p}' \, \hat{\Theta} |\mathbf{p}' \rangle ,
\end{align}
This confirms that \(\hat{\Theta} |\mathbf{p}'\rangle\) is an eigenstate of the transformed momentum operator with eigenvalue \(-\mathbf{p}'\). For the position operator, instead, we have
\begin{equation}
\langle \alpha | \hat{\mathbf{r}} | \alpha \rangle = \langle \tilde{\alpha} | \hat{\mathbf{r}} | \tilde{\alpha} \rangle^*,
\end{equation}
\begin{equation}
\hat{\Theta} \hat{\textbf{r}} \hat{\Theta}^{-1} = \hat{\textbf{r}},
\end{equation}
and
\begin{align}
\hat{\textbf{r}} \, \hat{\Theta} |\textbf{r}'\rangle &= \left( \hat{\Theta} \, \hat{\textbf{r}} \, \hat{\Theta}^{-1} \right) \hat{\Theta} |\textbf{r}'\rangle = \notag \\
&= \hat{\Theta} \, \hat{\textbf{r}} |\textbf{r}'\rangle = \notag \\
&= \hat{\Theta} \, \textbf{r}' |\textbf{r}'\rangle = \notag \\
&= \textbf{r}' \, \hat{\Theta} |\textbf{r}'\rangle.
\end{align}
\subsection{An application of the time reversal operator to quantum many-body theory}
Given a Hamiltonian quadratic in both the position and momentum operators of many independent particles, such a Hamiltonian is invariant under time-reversal symmetry. Under the action of the time-reversal operator, the momentum operator transforms as follows
\begin{equation}
\hat{\Theta} \hat{\textbf{k}} \hat{\Theta}^{-1} = - \hat{\textbf{k}}.
\end{equation}
Then, the relation
\begin{equation}
\hat{\Theta} \mathcal{H}(\mathbf{k}) \hat{\Theta}^{-1} = \mathcal{H}(-\mathbf{k})
\end{equation}
implies that the eigenvalues of the Hamiltonian must satisfy the symmetry
\begin{equation}
\mathcal{E}_{\mathbf{k}} = \mathcal{E}_{-\mathbf{k}}.
\end{equation}
Indeed, if $|\psi_{\mathbf{k}}\rangle$ is an eigenstate of $\mathcal{H}(\mathbf{k})$ with eigenvalue $\mathcal{E}_{\mathbf{k}}$, then
\begin{equation}
\mathcal{H}(\mathbf{k}) |\psi_{\mathbf{k}}\rangle = \mathcal{E}_{\mathbf{k}} |\psi_{\mathbf{k}}\rangle.
\end{equation}
Applying the time-reversal operator $\hat{\Theta}$ and using the transformation property of the Hamiltonian, we obtain
\begin{align}
\mathcal{H}(-\mathbf{k}) \left( \hat{\Theta} |\psi_{\mathbf{k}}\rangle \right)
&= \hat{\Theta} \mathcal{H}(\mathbf{k}) |\psi_{\mathbf{k}}\rangle
= \notag \\
&= \mathcal{E}_{\mathbf{k}} \left( \hat{\Theta} |\psi_{\mathbf{k}}\rangle \right).
\end{align}
Therefore, $\hat{\Theta} |\psi_{\mathbf{k}}\rangle$ is an eigenstate of $\mathcal{H}(-\mathbf{k})$ with the same eigenvalue $\mathcal{E}_{\mathbf{k}}$. From this it follows that the eigenvalues of the Hamiltonian are symmetric with respect to momentum inversion. From a physical point of view, the time-reversal operator reverses the direction of momentum but does not change the energy, which is a scalar quantity. Consequently, the energy associated with a state of momentum $\mathbf{k}$ must be equal to that of the state with inverted momentum $-\mathbf{k}$.

\part{Appendices}

\appendix
\chapter{Complex $2 \times 2$ matrices}\label{Complex 2 times 2 matrices}
Let $M_2(\mathbb{C})$ denote the vector space of complex $2 \times 2$ matrices, then
\begin{theorem}[A basis of $M_2(\mathbb{C})$]\label{A basis of M2x2C}
Given the vector space $M_2(\mathbb{C})$ with elements
\begin{align}
M &= 
\begin{pmatrix}
M_{11} & M_{12} \\
M_{21} & M_{22}
\end{pmatrix} 
= \notag \\
&= M_{11}
\begin{pmatrix}
1 & 0 \\
0 & 0
\end{pmatrix}
+ M_{12}
\begin{pmatrix}
0 & 1 \\
0 & 0
\end{pmatrix}
+ M_{21}
\begin{pmatrix}
0 & 0 \\
1 & 0
\end{pmatrix}
+ M_{22}
\begin{pmatrix}
0 & 0 \\
0 & 1
\end{pmatrix} ,
\label{eq: genericamatriceMcomplessa2x2}
\end{align}
a basis is given by the set of the identity with the Pauli matrices.
\begin{proof}
The set $\left\lbrace \mathds{1},\sigma_1,\sigma_2,\sigma_3 \right\rbrace$ is a basis of $M_2(\mathbb{C})$ if and only if 
\begin{itemize}
\item [1.] the set $\left\lbrace \mathds{1},\sigma_1,\sigma_2,\sigma_3 \right\rbrace$ is linearly independent;
\item [2.] each element $M \in M_2(\mathbb{C})$ can be written as
\begin{equation}
M = \alpha_0 \mathds{1} + \alpha_1 \sigma_1 + \alpha_2 \sigma_2 + \alpha_3 \sigma_3, \ \{\alpha_j\}_{j=0}^{3} \in \mathbb{C}.
\end{equation}
\end{itemize}
\begin{itemize}
\item [1.] Let $\{\alpha_j\}_{j=0}^{3} \in \mathbb{C}$ such that
\begin{equation}
\alpha_0 \mathds{1} + \alpha_1 \sigma_1 + \alpha_2 \sigma_2 + \alpha_3 \sigma_3 = 0,
\end{equation}
then
\begin{equation}
\begin{pmatrix}
\alpha_0 & 0 \\
0 & \alpha_0
\end{pmatrix}
+
\begin{pmatrix}
0 & \alpha_1 \\
\alpha_1 & 0 
\end{pmatrix}
+
\begin{pmatrix}
0 & - i \alpha_2 \\
i \alpha_2 & 0 
\end{pmatrix}
+
\begin{pmatrix}
\alpha_3 & 0 \\
0 & - \alpha_3
\end{pmatrix}
= 
\begin{pmatrix}
0 & 0 \\
0 & 0
\end{pmatrix},
\end{equation}
\begin{equation}
\begin{pmatrix}
\alpha_0 + \alpha_3 & \alpha_1 - i \alpha_2 \\
\alpha_1 + i \alpha_2 & \alpha_0 - \alpha_3
\end{pmatrix}
=
\begin{pmatrix}
0 & 0 \\
0 & 0
\end{pmatrix},
\end{equation}
\begin{equation}
\alpha_0 = \alpha_1 = \alpha_2 = \alpha_3 = 0.
\end{equation}
\item [2.] From 
\begin{equation}
\begin{pmatrix}
\alpha_0 + \alpha_3 & \alpha_1 - i \alpha_2 \\
\alpha_1 + i \alpha_2 & \alpha_0 - \alpha_3
\end{pmatrix}
=
\begin{pmatrix}
M_{11} & M_{12} \\
M_{21} & M_{22}
\end{pmatrix},
\end{equation}
we get
\begin{equation}
\begin{cases}
\alpha_0 + \alpha_3 = M_{11} \\
\alpha_1 - i \alpha_2 = M_{12} \\
\alpha_1 + i \alpha_2 = M_{21} \\
\alpha_0 - \alpha_3 = M_{22}
\end{cases},
\end{equation}
which has solutions given by
\begin{equation}
\alpha_0 = \dfrac{1}{2} \left( M_{11}+M_{22} \right),
\label{eq: coefficientisviluppomatrice2x2_1} 
\end{equation}
\begin{equation}
\alpha_1 = \dfrac{1}{2} \left( M_{12} + M_{21} \right),
\label{eq: coefficientisviluppomatrice2x2_2}
\end{equation}
\begin{equation} 
\alpha_2 = \dfrac{i}{2} \left( M_{12}-M_{21} \right),
\label{eq: coefficientisviluppomatrice2x2_3} 
\end{equation}
\begin{equation}
\alpha_3 = \dfrac{1}{2} \left( M_{11}-M_{22} \right).
\label{eq: coefficientisviluppomatrice2x2_4}
\end{equation}
\end{itemize}
\end{proof}
\end{theorem}
Now we are able to prove
\begin{theorem}[Characterization of traceless Hermitian $2 \times 2$ matrices] Let $M \in M_2(\mathbb{C})$ be a traceless and hermitian matrix, then
\begin{equation}
M = \sum_{i=1}^3 \alpha_i \sigma_i , \ \alpha_i \in \mathbb{R}.
\end{equation}
\begin{proof}
Since the matrix is hermitian, i.e., $M = M^\dagger$, it satisfies
\begin{equation}
\begin{pmatrix}
M_{11} & M_{12} \\
M_{21} & M_{22}
\end{pmatrix}
=
\begin{pmatrix}
M_{11}^* & M_{21}^* \\
M_{12}^* & M_{22}^*
\end{pmatrix},
\end{equation}
then
\begin{equation}
M =
\begin{pmatrix}
M_{11} & M_{12} \\
M_{12}^* & M_{22}
\end{pmatrix}
, \ M_{11}, M_{22} \in \mathbb{R}, \ M_{12} \in \mathbb{C}.
\label{eq: matricecomplessa2x2hermitiana}
\end{equation}
On the other hand, the null trace condition implies
\begin{equation}
M_{11} + M_{22} = 0.
\label{eq: traccianullamatrice2x2} 
\end{equation}
From $\eqref{eq: coefficientisviluppomatrice2x2_1}$, $\eqref{eq: coefficientisviluppomatrice2x2_2}$, $\eqref{eq: coefficientisviluppomatrice2x2_3}$, $\eqref{eq: coefficientisviluppomatrice2x2_4}$, $\eqref{eq: matricecomplessa2x2hermitiana}$, $\eqref{eq: traccianullamatrice2x2}$, we have
\begin{equation}
\alpha_0 = 0,
\end{equation}
\begin{align}
\alpha_1 &= \dfrac{1}{2} \left( M_{12} + M_{21} \right) = \notag \\
&= \dfrac{1}{2} \left( M_{12} + M_{12}^* \right) = \notag \\
&= \dfrac{2 \re(M_{12})}{2} = \notag \\
&= \re(M_{12}) \in \mathbb{R},
\end{align}
\begin{align}
\alpha_2 &= \dfrac{i}{2} \left( M_{12}-M_{21} \right) = \notag \\
&= \dfrac{i}{2} \left( M_{12}-M_{12}^* \right) = \notag \\
&= \dfrac{i}{2} \left( - 2 i \im(M_{12}) \right) = \notag \\
&= \im(M_{12}) \in \mathbb{R},
\end{align}
\begin{equation}
\alpha_3 = \dfrac{1}{2} \left( M_{11}-M_{22} \right) \in \mathbb{R},
\end{equation}
from which the statement follows.
\end{proof}
\end{theorem}
\chapter{Matrix exponential}\label{Matrix exponential}
\begin{definition}
Let $A \in \mathbb{C}^{n \times n}$ be a square matrix. The matrix exponential $e^A$ is defined via the power series, i.e.,
\begin{align}
e^A &= \sum_{k=0}^{\infty} \frac{1}{k!} A^k = \notag \\
&= \mathds{1} + A + \frac{1}{2!}A^2 + \frac{1}{3!}A^3 + \cdots
\end{align}
\end{definition}
Note that this series converges absolutely for any matrix $A \in \mathbb{C}^{n \times n}$, since the operator norm of each term is bounded by the corresponding term in the scalar exponential series:
\begin{equation}
\left\| \frac{1}{k!} A^k \right\| \leq \frac{1}{k!} \| A \|^k.
\end{equation}
Therefore, by comparison with the convergent series $\sum_{k=0}^\infty \frac{1}{k!} \|A\|^k = e^{\|A\|}$, the matrix exponential is well defined for all square matrices. The matrix exponential plays a fundamental role in many areas of mathematics and physics, particularly in the study of continuous symmetry transformations, solutions of linear differential equations, and the representation theory of Lie groups. It satisfies several important properties, of which we highlight two below.
\begin{itemize}
\item Baker–Campbell–Hausdorff (BCH) formula. For any two square matrices (or operators) \(A\) and \(B\), the following relation holds
\begin{equation}
e^A e^B = \exp\left( A + B + \frac{1}{2}[A,B] + \frac{1}{12}[A,[A,B]] - \frac{1}{12}[B,[A,B]] + \cdots \right),
\label{eq: BCHformula}
\end{equation}
where \([A,B] = AB - BA\) denotes the commutator. The series continues with higher-order nested commutators and converges under appropriate conditions. This formula allows expressing the product of exponentials as the exponential of a single matrix, capturing the non-commutativity of \(A\) and \(B\).
\item Glauber formula (special case of BCH). For two operators \(A\) and \(B\) whose commutator \([A,B]\) is a complex scalar (i.e., a c-number) or proportional to the identity operator, and such that all higher nested commutators vanish, the exponential of their sum can be factorized as
\begin{equation}
e^{A+B} = e^{-\frac{1}{2}[A,B]} e^{A} e^{B}.
\label{eq: Glauberformula}
\end{equation}
This formula is a special case of the BCH formula and applies, for example, to bosonic creation and annihilation operators \(a^\dagger\) and \(a\), where $\left[ a, a^\dagger \right] = \mathds{1}$.
\item Additivity for commuting matrices. If \( A \) and \( B \) commute, i.e., \( [A, B] = AB - BA = 0 \), then the exponential of their sum factorizes:
\begin{equation}
e^{A+B} = e^A e^B.
\end{equation}
This property follows directly from the Glauber formula, and is crucial in physics when combining generators of commuting symmetries (e.g., momentum components in translationally invariant systems).
\item Differentiation rule. Let \( A \) be a constant matrix and consider the one-parameter family of exponentials \( e^{tA} \), where \( t \in \mathbb{R} \). Then the derivative with respect to \( t \) is
\begin{align}
\dfrac{d}{dt} e^{tA} &= A e^{tA} = \notag \\
&= e^{tA} A.
\end{align}
\end{itemize}
\chapter{Unitary and antiunitary operators}
\begin{definition}
An operator $U$ is said to be linear if 
\begin{equation}
U \left( c_1 |\alpha \rangle + c_2 |\beta \rangle \right) = c_1 U |\alpha \rangle + c_2 U |\beta \rangle, \ \forall |\alpha \rangle, \ |\beta \rangle.
\end{equation}
$U$ is unitary if it satisfies $U U^\dagger = U^\dagger U = \mathds{1}$ and if it preserves the scalar product, i.e.,
\begin{equation}
\left\langle \alpha \middle| U^\dagger U \middle| \beta \right\rangle = \left\langle \alpha \mid \beta \right\rangle.
\end{equation}
An operator $\Theta$ is called antilinear if 
\begin{equation}
\Theta \left( c_1 \left| \alpha \right\rangle + c_2 \left| \beta \right\rangle \right)
= c_1^* \Theta \left| \alpha \right\rangle + c_2^* \Theta \left| \beta \right\rangle, \quad \forall \left| \alpha \right\rangle, \left| \beta \right\rangle,
\end{equation}
i.e., constants and antilinear operators do not commute. 
\end{definition}
To exploit the properties of antilinearity, recall that every bra can be seen as a linear application between the space of kets and the complex field, so the bra $\langle \alpha| \Theta$ will act as an antilinear application on a generic ket $|\beta \rangle$. In the scalar product we consider the action of $\Theta$ on the ket and apply a complex conjugate as follows
\begin{equation}
\left( \left\langle \alpha \right| \Theta \right) \left| \beta \right\rangle = \left[ \left\langle \alpha \middle| \left( \Theta \left| \beta \right\rangle \right) \right\rangle \right]^*.
\end{equation}
An antilinear operator $\Theta$ is said to be antiunitary if it satisfies $\Theta \Theta^\dagger = \Theta^\dagger \Theta = \mathds{1}$ which coincides with the first property of unitary operators and depends on the fact that the product of two antiunitary operators is a unitary operator. If an antiunitary operator acts on a bra and its ket, the scalar product becomes
\begin{align}
\left( \left\langle \alpha \right| \Theta^\dagger \right) \left( \Theta \left| \beta \right\rangle \right) &= \left[ \left\langle \alpha \middle| \Theta^\dagger \Theta \middle| \beta \right\rangle \right]^* = \notag \\
&= \left\langle \beta \middle| \alpha \right\rangle.
\label{eq: conservazionemodulooperatoreantiunitario}
\end{align}
that is, only the square modulus of the scalar product is preserved. \newline
Now, we want to show that an antiunitary operator $\Theta$ is equivalent to the nonunique product of a unitary operator $U$ and an operator $K$ of complex conjugation, defined as 
\begin{equation}
K c |\alpha \rangle = c^* K |\alpha \rangle.
\end{equation} 
Let us now explicate the action of $K$ on the state vectors. Given the basis $|a \rangle$ and the expansion
\begin{align}
\left| \alpha \right\rangle &= \sum_a \left| a \right\rangle \left\langle a \middle| \alpha \right\rangle = \notag \\
&= \sum_a \left\langle a \middle| \alpha \right\rangle \left| a \right\rangle,
\end{align}
the transformed vector is
\begin{align}
K \left| \alpha \right\rangle &= K \sum_a \left\langle a \middle| \alpha \right\rangle \left| a \right\rangle = \notag \\
&= \sum_a K \left\langle a \middle| \alpha \right\rangle \left| a \right\rangle = \notag \\
&= \sum_a \left\langle a \middle| \alpha \right\rangle^* K \left| a \right\rangle = \notag \\
&= \sum_a \left\langle a \middle| \alpha \right\rangle^* \left| a \right\rangle,
\end{align}
where we chose the standard basis $|a \rangle$ on which of course $K$ acts as the identity operator. If the basis is changed, the action of $K$ on a state vector changes. Consequently, the unitary operator $U$ in the product $\Theta = UK$ is not unique but depends on the basis. Now, we verify that the factorization $\Theta = UK$ is an antiunitary operator. First,
\begin{align}
\Theta \left( c_1 \left| \alpha \right\rangle + c_2 \left| \beta \right\rangle \right) &= U K \left( c_1 \left| \alpha \right\rangle + c_2 \left| \beta \right\rangle \right) = \notag \\
&= c_1^* U K \left| \alpha \right\rangle + c_2^* U K \left| \beta \right\rangle = \notag \\
&= c_1^* \Theta \left| \alpha \right\rangle + c_2^* \Theta \left| \beta \right\rangle.
\end{align}
To prove the $\eqref{eq: conservazionemodulooperatoreantiunitario}$, we follow an approach frequent in the scientific literature, i.e., we consider the action of $\Theta$ on the kets and calculate the changes of the bras by complex conjugation of the kets as follows
\begin{align}
\left| \tilde{\alpha} \right\rangle &= \Theta \left| \alpha \right\rangle = \notag \\
&= \sum_{a_1} \left\langle a_1 | \alpha \right\rangle^* \, U \left| a_1 \right\rangle = \notag \\
&= \sum_{a_1} \left\langle \alpha | a_1 \right\rangle \, U \left| a_1 \right\rangle,
\end{align}
\begin{align}
\left| \tilde{\beta} \right\rangle &= \Theta \left| \beta \right\rangle = \notag \\
&= \sum_{a_2} \left\langle \beta | a_2 \right\rangle \, U \left| a_2 \right\rangle,
\end{align}
\begin{equation}
\left\langle \tilde{\beta} \right| = \sum_{a_2} \left\langle a_2 \middle| \beta \right\rangle \left\langle a_2 \right| U^\dagger,
\end{equation}
\begin{align}
\left\langle \tilde{\beta} \middle| \tilde{\alpha} \right\rangle &= \sum_{a_1, a_2} \left\langle a_2 \middle| \beta \right\rangle \left\langle a_2 \middle| U^\dagger U \middle| a_1 \right\rangle \left\langle \alpha \middle| a_1 \right\rangle = \notag \\
&= \sum_{a_1, a_2} \left\langle a_2 \middle| \beta \right\rangle \left\langle a_2 \middle| a_1 \right\rangle \left\langle \alpha \middle| a_1 \right\rangle = \notag \\
&= \sum_{a_1} \left\langle a_1 \middle| \beta \right\rangle \left\langle \alpha \middle| a_1 \right\rangle = \notag \\
&= \sum_{a_1} \left\langle \alpha \middle| a_1 \right\rangle \left\langle a_1 \middle| \beta \right\rangle = \notag \\
&= \left\langle \alpha \middle| \beta \right\rangle = \notag \\
&= \left\langle \beta \middle| \alpha \right\rangle^*.
\end{align}
\begin{theorem}[A useful characterization of anti-unitary operators] Let $O$ be a linear operator, then
\begin{equation}
\left\langle \beta \middle| O \middle| \alpha \right\rangle
= \left\langle \tilde{\alpha} \middle| \Theta O^\dagger \Theta^{-1} \middle| \tilde{\beta} \right\rangle,
\end{equation}
where $\Theta$ is an antiunitary operator.
\begin{proof}
We define
\begin{equation}
\left| \tilde{\alpha} \right\rangle = \Theta \left| \alpha \right\rangle,
\end{equation}
\begin{equation}
\left| \tilde{\beta} \right\rangle = \Theta \left| \beta \right\rangle,
\end{equation}
\begin{equation}
\left| \tilde{\gamma} \right\rangle = O^\dagger \left| \beta \right\rangle,
\end{equation}
\begin{equation}
\left| \tilde{\gamma} \right\rangle = \Theta O^\dagger \left| \beta \right\rangle,
\end{equation}
then
\begin{align}
\left\langle \beta \right| O \left| \alpha \right\rangle &= \left\langle \gamma \right| \alpha \rangle = \notag \\
&= \left\langle \tilde{\alpha} \right| \tilde{\gamma} \rangle = \notag \\
&= \left\langle \tilde{\alpha} \right| \Theta O^\dagger \left| \beta \right\rangle = \notag \\
&= \left\langle \tilde{\alpha} \right| \Theta O^\dagger \Theta^{-1} \Theta \left| \beta \right\rangle = \notag \\
&= \left\langle \tilde{\alpha} \right| \Theta O^\dagger \left| \beta \right\rangle = \notag \\
&= \langle \tilde{\alpha} | \Theta O^\dagger \Theta^{-1} | \tilde{\beta} \rangle.
\end{align}
\end{proof}
\end{theorem}
In the physically relevant case of observables, where \( A = A^\dagger \), the above equation takes the following form
\begin{equation}
\left\langle \beta \right| A \left| \alpha \right\rangle = \langle \tilde{\alpha} | \Theta A \Theta^{-1} | \tilde{\beta} \rangle.
\label{eq: unacaratterizzazionediosservabilisottoazioneoperatoriantiunitari}
\end{equation}
\chapter{Trace of an operator}
\begin{definition}[Trace of a linear operator on a finite-dimensional Hilbert space]
Let $A$ be a linear operator acting on the finite-dimensional complex Hilbert space $H$, $\dim(H)=r$, and $\lbrace \varphi_{i} \rbrace$ a complete orthonormal system, the trace of an operator $A$ is defined as 
\begin{equation}
\Tr(A) = \sum_{i=1}^{r} \langle \varphi_{i}|A|\varphi_{i} \rangle .
\end{equation}
\end{definition}
It follows from the definition that the trace is a linear operator, being a finite sum of diagonal elements with respect to the basis $\varphi_{i}$. For a Hermitian operator, the related trace is a real number.
\begin{theorem}
The trace of an operator is independent of the choice of complete orthonormal basis.
\begin{proof}
Given a basis $\lbrace \varphi'_{i} \rbrace$, we have
\begin{align}
\Tr(A) &= \sum_{i=1}^{r} \langle \varphi'_{i}|A|\varphi'_{i} \rangle = \notag \\
&= \sum_{i} \sum_{j} \sum_{k} \langle \varphi'_{i}|\varphi_{j} \rangle  \langle \varphi_{j}|A|\varphi_{k} \rangle \langle \varphi_{k}|\varphi'_{i} \rangle = \notag \\
&= \sum_{i} \sum_{j} \sum_{k} \ \delta_{ij} \delta_{ki} \langle \varphi_{j}|A|\varphi_{k} \rangle = \notag \\
&= \sum_{j} \langle \varphi_{j}|A|\varphi_{j} \rangle .
\end{align}
\end{proof}
\end{theorem}
\begin{theorem}[Cyclic property of the trace] Given a set of any $n$ operators $\left\lbrace A_1,\ldots,A_n\right\rbrace$, the trace satisfies
\begin{equation}
\Tr \left(A_1 A_2 \ldots A_n\right) = \Tr \left( A_n A_1 \ldots A_{n-1} \right) = \ldots = \Tr \left( A_2 \ldots A_n A_1 \right).
\label{eq: proprietàciclicatraccia}
\end{equation}
which is known as the cyclic property of the trace operator.
\begin{proof}
Since the proof does not depend on the number of the operators, for sake of simplicity consider the case $n=3$. By definition, the trace of the product $A_1A_2A_3$ is given in the bracket notation by
\begin{equation}
\Tr \left(ABC\right) = \sum_n \langle n|ABC|n \rangle ,
\end{equation}
then we insert the identities $\sum_m |m \rangle \langle m|$, $\sum_p |p \rangle  \langle p|$ as follows
\begin{align}
\Tr \left(ABC\right) &= \sum_n \sum_m \sum_p \langle n|A|m \rangle \langle m|B|p \rangle \langle p|C|n \rangle \equiv \notag \\
&\equiv \sum_n \sum_m \sum_p A_{nm} B_{mp} C_{pn}.
\end{align}
Note that row-by-column products between operators can be arranged in several equivalent ways, i.e.,
\begin{align}
\sum_n \sum_m \sum_p A_{nm} B_{mp} C_{pn} &= \sum_p \sum_m \sum_n C_{pn} A_{nm} B_{mp} = \notag \\
&= \sum_m \sum_n \sum_p B_{mp} C_{pn} A_{nm} ,
\end{align}
which is the thesis.
\end{proof}
\end{theorem}
\begin{theorem}
Given a projection operator on the space $H' \subset H$, $\left(\dim(H’)=g<r \right)$, i.e.,
\begin{equation}
P^{g} = \sum_{j=1}^{g} |\varphi_{j} \rangle \langle \varphi_{j}|
\end{equation}
we have
\begin{equation}
\Tr\left(P^{g}\right) = g.
\end{equation}
\begin{proof}
From definition it follows
\begin{align}
\Tr(P^{g}) &= \sum_{i=1}^{g} \langle \varphi_{i}|P^{g}|\varphi_{i} \rangle = \notag \\
&= \sum_{i} \sum_{j} \langle \varphi_{i}|\varphi_{j}\rangle \langle \varphi_{j}|\varphi_{i} \rangle = \notag \\
&= \sum_{i} \sum_{j} \langle \varphi_{j}|\varphi_{i} \rangle^{*}  \langle \varphi_{j}|\varphi_{i} \rangle = \notag \\
&= \sum_{j=1}^{g} |\langle \varphi_{j}|\varphi_{j} \rangle |^{2} = \notag \\
&= g.
\end{align}
\end{proof}
\end{theorem}
Finally, for infinitely dimensional Hilbert spaces, the sum contained in the trace becomes a series, so the definition is adapted as follows.
\begin{definition}[A trace class operator]
The trace class operators are the linear operators defined on infinite-dimensional Hilbert spaces for which the series defined by the trace is absolutely convergent.
\end{definition}
\chapter{The thermodynamic limit}\label{The thermodynamic limit}
The formalism of the continuous Fourier transform allows for a smooth transition from systems defined in finite, periodic volumes to the description of extended systems, enabling the replacement of discrete mode sums with continuous integrals over momentum or frequency space. We now present a heuristic derivation of this formulation.
\begin{theorem}[Fourier series expansion]\label{Fourier series expansion}
Let \( f(x) \) be a square summable function defined on the interval \(\left[-\frac{L}{2}, \frac{L}{2}\right]\) and periodic of period \(L\). Such a function can be computed at any point on the real axis by means of its periodic extension, which is given by its Fourier series, i.e., \( f(x) \) can be written as follows
\begin{equation}
f(x) = \sum_{n=-\infty}^{\infty} c_n \frac{e^{-i \frac{2 \pi n}{L} x}}{\sqrt{L}},
\end{equation}
with
\begin{equation}
c_n = \int_{-\frac{L}{2}}^{+\frac{L}{2}} f(x) \frac{e^{i \frac{2 \pi n}{L} x}}{\sqrt{L}} \, dx, \ \forall n \in \mathbb{Z}.
\label{eq: svilupposerieFourier}
\end{equation}
\end{theorem}
From a physical point of view, Fourier series is the expansion of a periodic function in the spatial plane wave basis 
\begin{equation}
\left\lbrace \dfrac{e^{i k_n x}}{\sqrt{L}} \right\rbrace,
\end{equation} 
which satisfy
\begin{equation}
e^{i k_n x} = e^{i k_n (x+L)}.
\end{equation} 
Of course, the same interpretation applies to periodic functions with temporal period $T$. In this case, the theorem $\ref{Fourier series expansion}$ is adapted by substituting $k = \frac{2 \pi}{L}$ and $k_n = \frac{2 \pi n}{L}$ with
\begin{equation}
\omega = \dfrac{2 \pi}{T},
\end{equation}
\begin{equation}
\omega_n = \dfrac{2 \pi n}{T},
\end{equation}
respectively, so that a function periodic in time is expanded in the basis of temporal plane waves 
\begin{equation}
\left\lbrace \dfrac{e^{i \omega_n t}}{\sqrt{T}} \right\rbrace,
\end{equation} 
which satisfy
\begin{equation}
e^{i \omega_n t} = e^{i \omega_n (t+T)}.
\end{equation} 
Now we are able to introduce a widely used substitution in statistical physics, known as the thermodynamic limit approximation,that is
\begin{equation}
\sum_{\textbf{k}} \ldots \rightarrow \dfrac{V}{(2 \pi)^3} \int d^3 \textbf{k} \ldots
\label{eq: illimitetermodinamico}
\end{equation}
which is valid at the limits $N \rightarrow \infty$, $V \rightarrow \infty$. Indeed, consider a sum with respect to the wave vector $\textbf{k}$ and recall that 
\begin{equation}
k_i = \dfrac{n_i}{2 \pi}, \ n_i \in \mathbb{Z}, \ i=x,y,z. 
\end{equation}
Now, if the sum can be decomposed into three sums with respect to the coordinate axes, i.e.,
\begin{equation}
\sum_{\textbf{k}} \ldots = \sum_{k_x} \sum_{k_y} \sum_{k_z} \ldots
\end{equation}
and the limits $N \rightarrow \infty$, $V \rightarrow \infty$ are valid, each of these sums can be approximated by an integral. Then, we multiply and divide by the volume $V=L_xL_yL_z$ of the physical system as follows
\begin{equation}
V \left( \dfrac{1}{L_x} \dfrac{1}{L_y} \dfrac{1}{L_z} \sum_{k_x} \sum_{k_y} \sum_{k_z} \ldots \right).
\end{equation}
Consequently, the discrete quantity $\textbf{k}$ is replaced by the measure $d^3 \textbf{k}$, the sum gives way to an integral extended over $\mathbb{R}^3$, and a factor $(2 \pi)^{-1}$ arises for each coordinate axes and the $\eqref{eq: illimitetermodinamico}$ has been proved.
\chapter{Euler gamma function}\label{Euler gamma function}
The Euler gamma function is defined as
\begin{equation}
\Gamma(z) = \int_0^\infty t^{z-1} e^{-t} dt, \ z \in \mathbb{C}, \ \re(z) > 0.
\label{eq: funzionegammaEulero}
\end{equation}
\begin{theorem}
The Euler gamma function satisfies the following recurrence relation
\begin{equation}
z \Gamma(z) = \Gamma(z+1).
\label{eq: equazionericorrenzaGammaEulero}
\end{equation}
\begin{proof}
By definition, we have
\begin{align}
z \Gamma(z) &= z \int_{0}^{\infty} t^{z-1} \ e^{-t} \ dt = \notag \\
&= \int_{0}^{\infty} z t^{z-1} \ e^{-t} \ dt = \notag \\
&= \int_{0}^{\infty} \left( \dfrac{d}{dt} t^{z} \right) \ e^{-t} \ dt = \notag \\
&= \left[ t^{2} e^{-t} \right]_{0}^{\infty} - \int_{0}^{\infty} t^{z} \dfrac{d}{dt} e^{-t} dt = \notag \\
&= \int_{0}^{\infty} t^{z+1} \ e^{-t} \ dt ,
\end{align}
which is equal to the Euler gamma function computer for $z+1$.
\end{proof}
\end{theorem}
In the particular case $z=n \in \mathbb{N}$, the $\eqref{eq: equazionericorrenzaGammaEulero}$ gives
\begin{align}
\Gamma(n+1) &= n \Gamma(n) = \notag \\
&= n (n-1) \Gamma(n-1) = \notag \\
&\ \vdots \notag \\
&= n! \Gamma(1) ,
\end{align}
with
\begin{align}
\Gamma(1) &= \int_{0}^{\infty} t^{0} \ e^{-t} \ dt = \notag \\
&= \int_{0}^{\infty} e^{-t} \ dt = \notag \\
&= \left[ - e^{-t} \right]_{0}^{\infty} = \notag \\
&= 1.
\end{align}
In other words, the Euler gamma function is a generalization of the factorial function, that is
\begin{equation}
\Gamma(n+1) = n!, \ \forall \ n \in \mathbb{N}.
\end{equation}
Consequently, for large values of $z \in \mathbb{N}$ we can use the Stirling approximation, i.e.,
\begin{align}
\Gamma(z+1) \ &= \ z! \ \simeq \notag \\
&\simeq \ z^{z} \ e^{-z} \ \sqrt{2 \pi z} \ = \notag \\
&= \ z^{z+\frac{1}{2}} \ e^{-z} \ \sqrt{2 \pi} ,
\end{align}
with
\begin{equation}
z \in \mathbb{N} , \ z \rightarrow \infty.
\end{equation}
It can be shown that this approximation is valid even for complex numbers, i.e.,
\begin{equation}
\Gamma(z+1) \ \simeq \ z^{z-\frac{1}{2}} \ e^{-z} \ \sqrt{2 \pi}
\end{equation}
with
\begin{equation}
z \in \mathbb{C}, \ \re(z) > 0, \ \re(z) \rightarrow \infty.
\end{equation}
In statistical physics, a widely used equation derived from the Stirling approximation is given by
\begin{align}
\ln{\Gamma(z+1)} & \simeq \ln{\left( z^{z+\frac{1}{2}} \ e^{-z} \ \sqrt{2 \pi} \right)} = \notag \\
&= \left( z + \dfrac{1}{2} \right) \ln{z} - z + \ln{\sqrt{2 \pi}} \ = \notag \\
&= z \ln{z} + \dfrac{1}{2} \ln{z} - z + \ln{\sqrt{2 \pi}} \simeq \notag \\
&\simeq z \ln{z} - z ,
\end{align}
with
\begin{equation}
z \in \mathbb{R}^{+}, \ z \rightarrow \infty.
\end{equation}
\chapter{The volume of a hypersphere}
We aim to compute the volume of a hypersphere in arbitrary dimension. Since in many problems of statistical physics (both classical and quantum) the dimension is related to the number of degrees of freedom of the system, typically represented by the spatial degrees $d = 3N$, where $N$ is the number of particles, we will in particular consider a hypersphere in the momentum subspace of phase space, where each of the $3N$ degrees of freedom corresponds to a cartesian momentum component of the $N$ particles. We have
\begin{theorem}
Let $B^{3N}(R) = \left\lbrace x \in \mathbb{R}^{3N} : \|x\| \le R \right\rbrace$ be the hypersphere of radius $R = \sqrt{2m \mathcal{E}}$ in the momentum phase space of $N$ particles in $3$ dimensions. Here, the radius $R$ is set by the maximal total kinetic energy $\mathcal{E}$ of the system, assuming each particle has mass $m$. Then, the volume of this hypersphere is
\begin{equation}
V_{3N} = \dfrac{\pi^{\frac{3N}{2}}}{\Gamma\left(\frac{3N}{2} + 1\right)} R^{3N}.
\label{eq: volumeipersfera}
\end{equation}
\begin{proof}
By dimensional analysis and rotational invariance, the volume must scale as a constant $B_{3N}$, depending on the space dimension, multiplied by a power of the radius $R$ dictated by the dimensionality of the space, namely,
\begin{equation}
V_{3N} = B_{3N} R^{3N}.
\end{equation}
Since the hypersphere is defined in momentum space, we have
\begin{equation}
0 \leq \sum_{i=1}^{3N} \dfrac{p^{2}_{i}}{2m} \leq \mathcal{E},
\end{equation}
\begin{equation}
0 \leq \sum_{i=1}^{3N} p^{2}_{i} \leq 2 m \mathcal{E},
\end{equation}
and since the phase space is constructed from orthogonal coordinates, this condition defines a ball in $\mathbb{R}^{3N}$ endowed with the Euclidean norm. Accordingly, consider a Gaussian integral in cartesian coordinates
\begin{align}
I &= \int_{0 \leq \sum_{i=1}^{3N} p^{2}_{i} \leq 2m \mathcal{E}} e^{- \sum_{i=1}^{3N} p_{i}^{2}} d^{3N} p = \notag \\
&= \int_{- \sqrt{2m \mathcal{E}}}^{+\sqrt{2m \mathcal{E}}} e^{- p_{1}^{2}} dp_{1} \int_{- \sqrt{2m \mathcal{E}}}^{+\sqrt{2m \mathcal{E}}} e^{- p_{2}^{2}} dp_{2} \ldots \int_{- \sqrt{2m \mathcal{E}}}^{+\sqrt{2m \mathcal{E}}} e^{- p_{3N}^{2}} dp_{3N}.
\end{align}
In the thermodynamic limit, the contribution from the boundary of the integration domain is negligible, allowing the extension of the integration limits to infinity, i.e.,
\begin{align}
I & \simeq \int_{- \infty}^{+ \infty} e^{- p_{1}^{2}} dp_{1} \int_{- \infty}^{+ \infty} e^{- p_{2}^{2}} dp_{2} \ldots \int_{- \infty}^{+ \infty} e^{- p_{3N}^{2}} dp_{3N} = \notag \\
&= \left[ \sqrt{\pi} \right]^{3N} = \notag \\
&= \pi^{\frac{3N}{2}}.
\end{align}
Let us express this Gaussian integral in a hyperspherical coordinate system, with one radial dimension $p$ and $3N - 1$ angular dimensions. Note that in $\mathbb{R}^{3}$ the surface ($V_2)$ and volume ($V_3$) elements satisfy
\begin{equation}
dV_{2} = 2 \pi R dR,
\end{equation}
\begin{equation}
dV_{3} = 4 \pi R^{2} dR ,
\end{equation}
and since the $(3N-1)$-dimensional hypersurface is obtained by differentiating the $3N$-dimensional volume with respect to the radius,
\begin{equation}
V_{d-1} = \frac{d}{dR} V_d,
\end{equation}
we have in particular
\begin{equation}
V_{3N-1} = 3N B_{3N} R^{3N-1}.
\end{equation}
It follows that the hypervolume element in hyperspherical coordinates is
\begin{equation}
dV_{3N} = 3N B_{3N} R^{3N-1} dR,
\end{equation}
after integration over $3N-1$ angular variables. Consequently, from Cartesian to hyperspherical coordinates, the hypervolume element is transformed as
\begin{equation}
dV_{3N} = \prod_{i=1}^{3N} dp_{i} ,
\end{equation}
\begin{equation}
dV_{3N} = 3N B_{3N} p^{3N-1} dp,
\end{equation}
and the Gaussian integral is written as
\begin{align}
I \ &= \ \pi^{\frac{3N}{2}} \ = \notag \\
&= \ 3N B_{3N} \int_{0}^{\sqrt{2m \mathcal{E}}} e^{-p^{2}} p^{3N-1} dp \ \simeq \notag \\
&\simeq \ 3N B_{3N} \int_{0}^{\infty} e^{-p^{2}} p^{3N-1} dp.
\end{align}
Let us compute $B_{3N}$. We have
\begin{align}
\pi^{\frac{3N}{2}} \ &= \ 3N B_{3N} \int_{0}^{\infty} e^{-p^{2}} p^{3N-1} dp \ = \notag \\
&= \ 3N B_{3N} \int_{0}^{\infty} \dfrac{2p}{2p} e^{-p^{2}} p^{3N-1} dp ,
\end{align}
and from
\begin{equation}
t=p^{2},
\end{equation}
\begin{equation}
dt = 2 p dp,
\end{equation}
it follows
\begin{align}
\pi^{\frac{3N}{2}} \ &= \ 3N B_{3N} \int_{0}^{\infty} \dfrac{2p}{2p} e^{-p^{2}} p^{3N-1} dp \ = \notag \\
&= \ \dfrac{3N}{2} B_{3N} \int_{0}^{\infty} e^{-p^{2}} \dfrac{p^{3N-1}}{p} 2pdp \ = \notag \\
&= \ \dfrac{3N}{2} B_{3N} \int_{0}^{\infty} e^{-p^{2}} p^{3N-2} 2pdp \ = \notag \\
&= \dfrac{3N}{2} B_{3N} \int_{0}^{\infty} e^{-p^{2}} p^{2\left( \frac{3N}{2}-1\right)} 2pdp \ = \notag \\
&= \ \dfrac{3N}{2} B_{3N} \int_{0}^{\infty} e^{-t} \ t^{\left(\frac{3N}{2}-1\right)} \ dt,
\end{align}
and recalling the definition of the Euler Gamma function $\Gamma(z)$ from $\eqref{eq: funzionegammaEulero}$, we find
\begin{equation}
\pi^{\frac{3N}{2}} \ = \ \dfrac{3N}{2} B_{3N} \ \Gamma\left( \dfrac{3N}{2} \right),
\end{equation}
\begin{equation}
B_{3N} = \dfrac{\pi^{\frac{3N}{2}}}{\frac{3N}{2} \ \Gamma\left( \frac{3N}{2} \right)}.
\end{equation}
From the property $\eqref{eq: equazionericorrenzaGammaEulero}$ of Gamma function we have
\begin{equation}
B_{3N} = \dfrac{\pi^{\frac{3N}{2}}}{\Gamma\left( \frac{3N}{2} + 1 \right)},
\end{equation}
from which the thesis follows.
\end{proof}
\end{theorem}
\chapter{The Fermi momentum}\label{The Fermi momentum}
The Fermi momentum $k_F$ is a fundamental quantity in the description of electrons in a solid, especially at zero temperature. It represents the maximum magnitude of the wavevector of electrons occupied in their ground state, corresponding to the Fermi energy $\mathcal{E}_F$: all states with momentum less than or equal to $k_F$ are occupied, while those with higher momentum are empty. The following theorem provides a general relation, valid in an arbitrary $d$-dimensional space, between the particle number density $\rho_0$ of fermionic particles and the Fermi momentum $k_F$, where the subscript $0$ emphasizes that we are considering non-interacting fermions. 
\begin{theorem}[Particle density of non-interacting particles as a function of Fermi momentum in $d$-dimensions]
Consider a system of non-interacting fermions, with spin degeneracy $g$, confined in a volume $V \subset \mathbb{R}^d$, at zero temperature. The number density $\rho_0 = \frac{N}{V}$ is related to the magnitude of the Fermi momentum $\mathbf{k}_F$ by the relation
\begin{equation}
\rho_0 = g \dfrac{1}{(2\pi)^d} \dfrac{\pi^{\frac{d}{2}}}{\Gamma\left(\frac{d}{2} + 1 \right)} k_F^d.
\end{equation}
\begin{proof}
At zero temperature, all states with energy less than or equal to the Fermi energy are occupied. Since the energy depends only on the magnitude of the wavevector $\textbf{k}$, the occupied states correspond to all vectors $\textbf{k} \in \mathbb{R}^d$ such that $k \leq k_F$. In the thermodynamic limit, momentum space becomes continuous, and the number of states in a volume element $d^d k$ in momentum space is
\begin{equation}
dN = g \frac{V}{(2\pi)^d} d^d k,
\end{equation}
where the factor $\frac{V}{(2\pi)^d}$ comes from the periodic quantization in a volume $V$, and $g$ is the number of spin states accessible for each $\textbf{k}$. The total number of particles is obtained by integrating over all states inside the Fermi sphere, that is
\begin{equation}
N = g \frac{V}{(2\pi)^d} \int_{k \leq k_F} d^d k.
\end{equation}
The volume of the hypersphere in $\mathbb{R}^d$ is given by equation $\eqref{eq: volumeipersfera}$ with $d=3N$ and radius $R=k_F$, i.e.,
\begin{equation}
\int_{k \leq k_F} d^d k = \dfrac{\pi^{\frac{d}{2}}}{\Gamma\left(\frac{d}{2} + 1 \right)} k_F^d,
\end{equation}
then we obtain
\begin{equation}
N = g \dfrac{V}{(2\pi)^d} \dfrac{\pi^{\frac{d}{2}}}{\Gamma\left(\frac{d}{2} + 1 \right)} k_F^d,
\end{equation}
and dividing both sides by the volume $V$, the thesis follows.
\end{proof}
\end{theorem}
Such relation is exact for free electron models, such as the ideal Fermi gas (an ideal gas that obeys Fermi statistics) and the jellium model, where the complexity of electron interactions and the crystal's periodic potentials are neglected, allowing electrons to be treated as free particles confined within a finite volume. In particular, from the previous theorem, for the physically relevant cases $d = 1, 2, 3$, we have the following
\begin{theorem}[Particle density of non-interacting particles as a function of Fermi momentum in 1D, 2D, and 3D dimensions]
The relation between the particle density $\rho_0$ and the magnitude of the Fermi momentum $k_F$ for non-interacting fermions with spin degeneracy $g$ in spatial dimensions $d = 1, 2, 3$ is
\begin{equation}
\rho_0 = \frac{g}{2\pi} k_F,
\label{eq: relazionedensitafermionimomentoFermid1}
\end{equation}
\begin{equation}
\rho_0 = \frac{g}{4\pi} k_F^2,
\label{eq: relazionedensitafermionimomentoFermid2}
\end{equation}
\begin{equation}
\rho_0 = \frac{g}{6\pi^2} k_F^3.
\label{eq: relazionedensitafermionimomentoFermid3}
\end{equation}
\end{theorem}
\chapter{Operational aspects of tensor products}
\begin{definition}[Tensor product between vectors] Let $V$ and $W$ be finite-dimensional vector spaces over a field, with dimensions $n$ and $m$, respectively. The tensor product $V \otimes W$ of two vector spaces $V$ and $W$ (over the same field) is a vector space to which is associated a bilinear map $V \times W \rightarrow V \otimes W$ that maps a pair $(v,w)$, $v \in V$, $w \in W$ to an element of $V \otimes W$ denoted $v \otimes w$. Let $e_1, \ldots, e_n$, $f_1, \ldots , f_m$, be bases for $V$ and $W$ respectively; then the tensor product basis $\left\lbrace e_i \otimes f_j \right\rbrace_{i,j}^{n,m}$ of $V \otimes W$ can be constructed as follows
\begin{align}
\left\lbrace e_i \otimes f_j \right\rbrace_{i,j=1}^{n,m} &=
\begin{pmatrix}
e_1 \\
\vdots \\
e_n
\end{pmatrix}
\begin{pmatrix}
f_1 & \ldots & f_m
\end{pmatrix} 
= \notag \\ 
&=
\begin{pmatrix}
e_1 f_1 & e_1 f_2 & \ldots & e_1 f_m \\
e_2 f_1 & e_2 f_2 & \ldots & e_2 f_m \\
\vdots & \vdots & \ddots & \vdots \\
e_n f_1 & e_n f_2 & \ldots & e_n f_m
\end{pmatrix}
,
\end{align}
with this definition we have that $\dim(V \otimes W) = nm$.
\end{definition}
\begin{definition}[Tensor product between matrices] Let $A$ and $B$ be finite-dimensional matrices over a field, with dimensions $n \times m$ and $p \times q$, respectively. Given
\begin{equation}
A =
\begin{pmatrix}
a_{11} & \ldots & a_{1m} \\
\vdots & \ddots & \vdots \\
a_{n1} & \ldots & a_{nm}
\end{pmatrix} ,
\end{equation}
\begin{equation}
B =
\begin{pmatrix}
b_{11} & \ldots & b_{1q} \\
\vdots & \ddots & \vdots \\
b_{p1} & \ldots & b_{pq}
\end{pmatrix} ,
\end{equation}
the tensor product can be constructed as follows
\begin{align}
A \otimes B &= 
\begin{pmatrix}
a_{11} & \ldots & a_{1m} \\
\vdots & \ddots & \vdots \\
a_{n1} & \ldots & a_{nm}
\end{pmatrix}
\otimes
\begin{pmatrix}
b_{11} & \ldots & b_{1q} \\
\vdots & \ddots & \vdots \\
b_{p1} & \ldots & b_{pq}
\end{pmatrix}
= \notag \\
&=
\begin{pmatrix}
a_{11} \begin{pmatrix}
b_{11} & \ldots & b_{1q} \\
\vdots & \ddots & \vdots \\
b_{p1} & \ldots & b_{pq}
\end{pmatrix} 
& \ldots &
a_{1m} \begin{pmatrix}
b_{11} & \ldots & b_{1q} \\
\vdots & \ddots & \vdots \\
b_{p1} & \ldots & b_{pq}
\end{pmatrix} \\
\vdots & \ddots & \vdots \\
a_{n1} \begin{pmatrix}
b_{11} & \ldots & b_{1q} \\
\vdots & \ddots & \vdots \\
b_{p1} & \ldots & b_{pq}
\end{pmatrix}
& \ldots &
a_{nm} \begin{pmatrix}
b_{11} & \ldots & b_{1q} \\
\vdots & \ddots & \vdots \\
b_{p1} & \ldots & b_{pq}
\end{pmatrix} 
\end{pmatrix} = \notag \\
&=
\begin{pmatrix}
a_{11} b_{11} & a_{11} b_{12} & \ldots & a_{11} b_{1q} & \ldots & \ldots & a_{1m} b_{11} & a_{1m} b_{12} & \ldots & a_{1m} b_{1q} \\
a_{11} b_{21} & a_{11} b_{22} & \ldots & a_{11} b_{2q} & \ldots & \ldots & a_{1m} b_{21} & a_{1m} b_{22} & \ldots & a_{1m} b_{2q} \\
\vdots & \vdots & \ddots & \vdots & & & \vdots & \vdots & \ddots & \vdots \\
a_{11} b_{p1} & a_{11} b_{p2} & \ldots & a_{11} b_{pq} & \ldots & \ldots & a_{1m} b_{p1} & a_{1m} b_{p2} & \ldots & a_{1m} b_{pq} \\
\vdots & \vdots & \ddots & \vdots & \ddots & & \vdots & \vdots & \ddots & \vdots \\
\vdots & \vdots & \ddots & \vdots & & \ddots & \vdots & \vdots & \ddots & \vdots \\
a_{n1} b_{11} & a_{n1} b_{12} & \ldots & a_{n1} b_{1q} & \ldots & \ldots & a_{nm} b_{11} & a_{nm} b_{12} & \ldots & a_{nm} b_{1q} \\
a_{n1} b_{21} & a_{n1} b_{22} & \ldots & a_{n1} b_{2q} & \ldots & \ldots & a_{nm} b_{21} & a_{nm} b_{22} & \ldots & a_{nm} b_{2q} \\
\vdots & \vdots & \ddots & \vdots & & & \vdots & \vdots & \ddots & \vdots \\
a_{n1} b_{p1} & a_{n1} b_{p2} & \ldots & a_{n1} b_{pq} & \ldots & \ldots & a_{nm} b_{p1} & a_{nm} b_{p2} & \ldots & a_{nm} b_{pq}
\end{pmatrix},
\label{eq: prodottotensorialematrici}
\end{align}
which is a matrix of dimension $np \times mq$. Explicitly, each block \( a_{ij} b \) is a scalar multiple of the entire matrix \( b \). Thus, the resulting matrix has size \( np \times mq \), where each scalar \( a_{ij} \) multiplies every entry of \( b \) in its corresponding block. Accordingly, if \( T = A \otimes B \), the generic entry of \( T \) can be labeled as
\begin{equation}
\left\lbrace t_{ij,lk} \right\rbrace, \ 1 \leq i \leq n, \ 1 \leq j \leq m, \ 1 \leq l \leq p, \ 1 \leq k \leq q,
\label{eq: genericoelementomatriceprodottotensorialematrici}
\end{equation}
i.e., the row and column indices are ordered lexicographically according to the structure of the tensor product.
\end{definition}
\chapter{Jacobi identity}
The Jacobi identity is a property of a binary operation that describes how the order of evaluation, the placement of parentheses in a multiple product, affects the result of the operation.
\begin{theorem}[Jacobi identity]
Given the operators $A$, $B$ and $C$, let $\left[ \cdot , \cdot \right]$ the commutator parentheses, then
\begin{equation}
\left[ A, BC \right] = \left[ A, B \right] C + B \left[ A, C \right].
\label{eq: Jacobiidentity}
\end{equation}
\begin{proof}
We exploit the commutator as follows
\begin{equation}
\left[ A, BC \right] = ABC - BCA,
\end{equation}
and we add and substract the operator $BAC$ and obtain
\begin{align}
\left[ A, BC \right] &= ABC - BAC + BAC - BCA = \notag \\
&= \left( ABC - BAC \right) + \left( BAC - BCA \right) = \notag \\
&= \left( AB - BA \right) C + B \left( AC - CA \right) ,
\end{align}
from which the thesis follows.
\end{proof}
\end{theorem}
Note that equation $\eqref{eq: Jacobiidentity}$ can be equivalently rewritten as
\begin{equation}
\left[ A, \left[ B, C \right] \right] + \left[ B, \left[ C, A \right] \right] + \left[ C, \left[ A,B \right] \right] = 0.
\label{eq: Jacobiidentity1}
\end{equation}
Moreover, because the identity is formulated in terms of commutators, it usually plays a central role in many-body theory applications where bosonic operators, obeying canonical commutation relations, are involved.
\chapter{Breit-Wigner formula}
\begin{theorem}
The following result holds
\begin{equation}
\lim_{y \to 0^+} \frac{1}{\pi} \frac{y}{x^2 + y^2} = \delta(x)
\label{eq: BreitWignerformula}
\end{equation}
in the sense of distributions, for every continuous and bounded test function $\varphi(x)$ we have
\begin{equation}
\lim_{y \to 0^+} \int_{-\infty}^{+\infty} \frac{1}{\pi} \frac{y}{x^2 + y^2} \varphi(x) \, dx = \varphi(0).
\end{equation}
\begin{proof}
We will proceed step by step.
\begin{itemize}
\item{Normalization.} We verify that the function
\begin{equation}
f_y(x) = \frac{1}{\pi} \frac{y}{x^2 + y^2}
\end{equation}
is normalized, then
\begin{equation}
\int_{-\infty}^{+\infty} f_y(x) \, dx = \int_{-\infty}^{+\infty} \frac{1}{\pi} \frac{y}{x^2 + y^2} \, dx.
\end{equation}
We make the change of variable $t = \frac{x}{y}$, from which it follows that $dx = y\, dt$, then
\begin{align}
\int_{-\infty}^{+\infty} f_y(x) \, dx &= \int_{-\infty}^{+\infty} \frac{1}{\pi} \frac{y}{y^2 t^2 + y^2} y \, dt = \notag \\
&= \int_{-\infty}^{+\infty} \frac{1}{\pi} \frac{1}{t^2 + 1} \, dt = \notag \\
&= 1.
\end{align}
\item{Distributional limit}. Let us consider the integral
\begin{align}
I_y &= \int_{-\infty}^{+\infty} f_y(x) \varphi(x) \, dx = \notag \\
&= \int_{-\infty}^{+\infty} \frac{1}{\pi} \frac{y}{x^2 + y^2} \varphi(x) \, dx.
\end{align}
We make again the change of variable $t = \frac{x}{y}$ and obtain
\begin{align}
I_y &= \int_{-\infty}^{+\infty} \frac{1}{\pi} \frac{y}{y^2 t^2 + y^2} \varphi(y t) y \, dt = \notag \\
&= \int_{-\infty}^{+\infty} \frac{1}{\pi} \frac{1}{t^2 + 1} \varphi(y t) \, dt.
\end{align}
\item{The limit}. Since $\varphi$ is a continuous function, if $y \to 0^+$, we have
\begin{equation}
\varphi(y t) \to \varphi(0),
\end{equation}
and by applying the dominated convergence theorem, we obtain
\begin{align}
\lim_{y \to 0^+} I_y &= \varphi(0) \int_{-\infty}^{+\infty} \frac{1}{\pi} \frac{1}{t^2 + 1} \, dt = \notag \\
&= \varphi(0),
\end{align}
which is the thesis.
\end{itemize}
\end{proof}
\end{theorem}
\chapter{Kramers-Krönig relations}\label{Kramers-Krönig relations}
\FloatBarrier
\begin{figure}[H]
\centering
\includegraphics[scale=0.6]{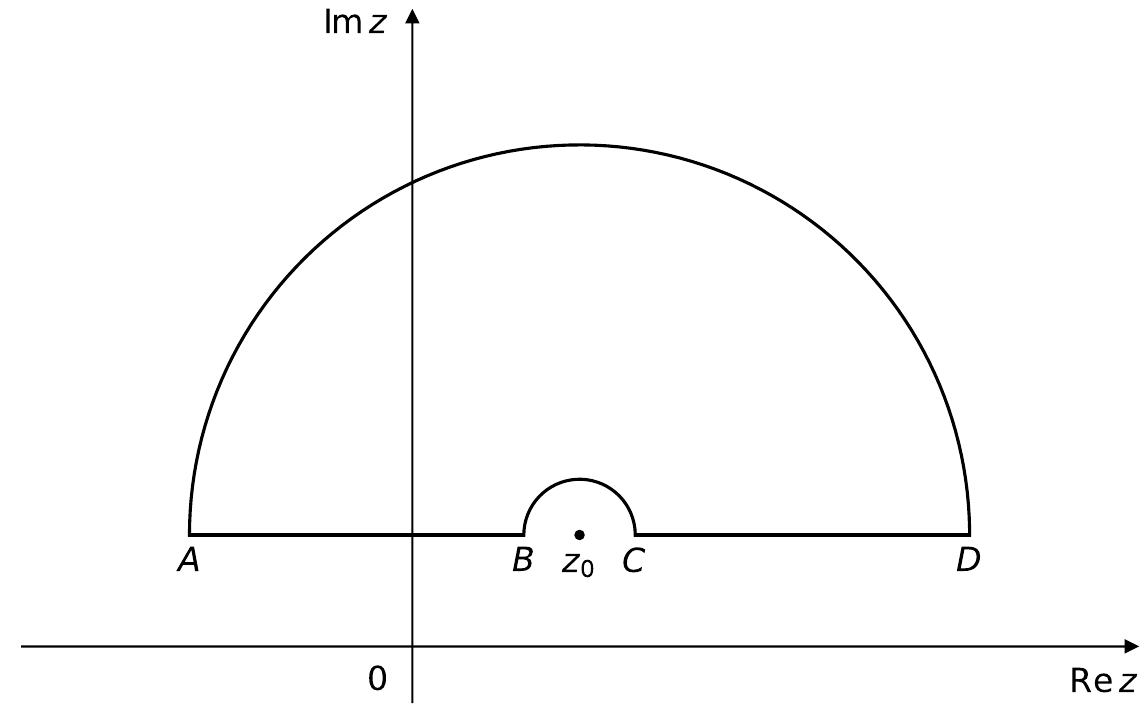}
\caption{The integration contour $\Gamma$ appearing in equation $\eqref{eq: curvagammarelazioniKramersKronig}$, used in the derivation of the Kramers-Krönig relations.}
\label{fig: curvagammarelazioniKramersKronig}
\end{figure}
The definition of an integral in the sense of the Cauchy principal value ($\pv$), or simply the Cauchy principal value, is used in cases where an improper integral has a non-integrable singularity within the domain of integration. Formally, it is defined as follows
\begin{definition}[Cauchy principal value]\label{def: valoreprincipaleCauchy}
Let $f(x)$ be a function with a singularity at a point $c \in (a, b)$. The Cauchy principal value of the integral
\begin{equation}
\int_a^b f(x)\,dx
\end{equation}
is defined as
\begin{equation}
\mathrm{P.V.} \int_a^b f(x)\,dx = \lim_{\epsilon \to 0^+} \left( \int_a^{c - \epsilon} f(x)\,dx + \int_{c + \epsilon}^b f(x)\,dx \right),
\end{equation}
and if this limit exists, the integral is said to admit a principal value.
\end{definition}
Thanks to Definition $\ref{def: valoreprincipaleCauchy}$, we can now prove a useful relation in theoretical physics. We have
\begin{theorem}[Kramers-Krönig relations]
Let $k(z)$ be an analytic function in the upper half-plane and such that
\begin{equation}
\lim_{|z| \to \infty} k(z) = 0,
\end{equation}
then, for every $x_0 \in \mathbb{R}$, the function $k(z)$ satisfies the Kramers-Krönig relations
\begin{equation}
\re k(x_0) = \frac{1}{\pi} \, \pv \int_{-\infty}^{+\infty} \frac{\im k(x)}{x - x_0} \, dx,
\label{eq: KK_real}
\end{equation}
\begin{equation}
\im k(x_0) = - \frac{1}{\pi} \, \pv \int_{-\infty}^{+\infty} \frac{\re k(x)}{x - x_0} \, dx.
\label{eq: KK_imag}
\end{equation}
\end{theorem}
\begin{proof}
Let $z_0 = x_0 + i y_0$ with $y_0 > 0$. Let us consider the function
\begin{equation}
F(z) = \frac{k(z)}{z - z_0}.
\label{eq: curvagammarelazioniKramersKronig}
\end{equation}
Since $k(z)$ is analytic for $\im z > 0$ and tends to zero as $|z| \to \infty$, the function $F(z)$ is analytic in the upper half-plane except for a simple pole at $z = z_0$. We define a closed contour $\Gamma$ given by the union of
\begin{itemize}
\item two horizontal segments parallel to the real axis at height $y_0$, i.e.,
\begin{equation}
AB: x \in [x_0 - R, x_0 - \varepsilon],
\end{equation}
\begin{equation}
CD: x \in [x_0 + \varepsilon, x_0 + R],
\end{equation}
with $R \to +\infty$ and $\varepsilon \to 0^+$,
\item two semicircles in the upper half-plane, i.e.,
\begin{equation}
\text{small semicircle of radius } \varepsilon \text{ around } z_0,
\end{equation}
\begin{equation}
\text{big semicircle of radius } R \text{ closed in the upper half-plane.}
\end{equation}
\end{itemize}
Since $F$ is analytic inside $\Gamma$ except for the pole at $z_0$, by Cauchy's theorem we have
\begin{equation}
\oint_{\Gamma} F(z) \, dz = 0.
\end{equation}
Let us compute the individual contributions. 
\begin{itemize}
\item{Integrals on the segments}  
\begin{equation}
\int_{x_0 - R}^{x_0 - \varepsilon} \frac{k(x + i y_0)}{x - x_0} \, dx + \int_{x_0 + \varepsilon}^{x_0 + R} \frac{k(x + i y_0)}{x - x_0} \, dx = \pv \int_{-\infty}^{+\infty} \frac{k(x + i y_0)}{x - x_0} \, dx.
\end{equation}
\item{Integral on the semicircle of radius $\varepsilon$}  
Using the parametrization $z = z_0 + \varepsilon e^{i\theta}$, $\theta \in [\pi,0]$,
\begin{equation}
\int_{\pi}^0 i k(z_0 + \varepsilon e^{i\theta}) \, d\theta \xrightarrow[\varepsilon \to 0^+]{} -i \pi k(z_0).
\end{equation}
\item{Integral on the semicircle of radius $R$}  
Using the parametrization $z = z_0 + R e^{i\theta}\), \(\theta \in [0, \pi]$,
\begin{equation}
\int_0^\pi i k(z_0 + R e^{i\theta}) \, d\theta
\xrightarrow[R \to \infty]{} 0,
\end{equation}
by the hypothesis $\lim_{|z| \to \infty} k(z) = 0$. \newline
\end{itemize}
Putting together the contributions and taking the limits \(R \to \infty\), \(\varepsilon \to 0^+\), we get
\begin{equation}
\pv \int_{-\infty}^{+\infty} \frac{k(x + i y_0)}{x - x_0} \, dx = i \pi k(x_0 + i y_0),
\end{equation}
from which, separating the real and imaginary parts of $k$, we obtain the Kramers-Krönig relations.
\end{proof}
\chapter{Residue theorem}
The residue theorem, supported by tools such as Jordan's lemma, provides an elegant and powerful method for evaluating complex and real integrals, especially those involving functions with isolated singularities. Its power lies in the fact that, instead of integrating directly along the contour, it suffices to know the residues of the function at the interior singular points.
\begin{definition}[Simple pole] An isolated singularity $ z_0 $ of a complex function $ f(z) $ is called a simple pole if, in a neighborhood of $ z_0 $, the function can be written in the form
\begin{equation}
f(z) = \frac{g(z)}{z - z_0},
\end{equation}
where $ g(z) $ is complex differentiable in a neighborhood of $ z_0 $ and $ g(z_0) \neq 0 $.
\end{definition}
\begin{definition}[Residue at a simple pole] Let $ f(z) $ be a complex function with an isolated singularity at $ z_0 $. If $ z_0 $ is a simple pole of $ f $, then in a neighborhood of $ z_0 $ the function can be expressed by the Laurent series expansion
\begin{equation}
f(z) = \frac{a_{-1}}{z - z_0} + a_0 + a_1 (z - z_0) + \cdots,
\end{equation}
where the coefficient $ a_{-1} $ of the $(z - z_0)^{-1}$ term is nonzero. The residue of $ f $ at $ z_0 $ is precisely this coefficient $ a_{-1} $, which can be computed conveniently by the formula
\begin{equation}
\operatorname{Res}_{z_0}(f) = \lim_{z \to z_0} (z - z_0) f(z).
\end{equation}
\end{definition}
In practice, multiplying $ f(z) $ by $ (z - z_0) $ and taking the limit as $ z \to z_0 $ yields a finite number that captures the most significant part of the singularity of $ f $ at $ z_0 $. This value is fundamental for computing complex integrals via the residue theorem, as we show.
\begin{theorem}\label{thm: teoremaresidui}
Let $ \Omega \subset \mathbb{C} $ be an open set, and let $ z_1, z_2, \ldots, z_n $ be isolated singularities of a complex-valued function $ f $. Let $ \gamma $ be a positively oriented, simple closed curve lying in $ \Omega \setminus \{z_1, \ldots, z_n\} $, and suppose that each $ z_k $ lies in the interior of the region enclosed by $ \Gamma $. Assume that the function $ f : \Omega \setminus \{z_1, \ldots, z_n\} \to \mathbb{C} $ is complex-differentiable at every point of $ \Omega \setminus \{z_1, \ldots, z_n\} $. Then the integral of $ f $ around $ \Gamma $ is given by
\begin{equation}
\oint_{\Gamma} f(z) \, dz = 2\pi i \sum_{k=1}^{n} I_{z_k}(\Gamma) \, \operatorname{Res}_{z_k}(f),
\label{eq: formulateoremaresidui}
\end{equation}
where $ \operatorname{Res}_{z_k}(f) $ denotes the residue of $ f $ at the singularity $ z_k $, and $ I_{z_k}(\Gamma) $ is the winding number of the curve $ \gamma $ around $ z_k $. \newline
In the special case where $ \gamma $ is positively oriented, that is, traversed in the counterclockwise direction so that the interior region lies on the left while moving along the curve, and encloses each $ z_k $ exactly once, we have $ I_{z_k}(\Gamma) = 1 $, and the formula simplifies to
\begin{equation}
\oint_{\Gamma} f(z) \, dz = 2\pi i \sum_{k=1}^{n} \operatorname{Res}_{z_k}(f) .
\label{eq: formulateoremaresiduisemplificata}
\end{equation}
\end{theorem}
\begin{lemma}\label{lmm: lemmaJordan}
Let $f(z)$ be continuous on the upper semicircular arc $C_R^+$ of radius $R$ centered at the origin, and let $a > 0$. Assume there exists a constant $M$ such that
\begin{equation}
|f(z)| \leq \frac{M}{|z|} \quad \text{for } |z| = R.
\end{equation}
Then, for the arc $C_R^+$ traversed counterclockwise, we get
\begin{equation}
\lim_{R \to \infty} \int_{C_R^+} f(z) e^{i a z} \, dz = 0.
\end{equation}
\end{lemma}
\end{document}